\documentclass[a4paper,12pt]{book} 
\usepackage{geometry}
\usepackage[usenames,dvipsnames,svgnames,table]{xcolor}
\usepackage{amsmath}
\usepackage{gensymb}
\usepackage{amssymb}
\usepackage{graphics}
\usepackage{graphicx}
\usepackage{epsfig}
\usepackage{fancyhdr}
\usepackage{wasysym}
\usepackage{latexsym}
\usepackage{amsfonts}
\usepackage{subfigure}
\usepackage{float}
\usepackage{color,colortbl}
\usepackage{cite}
\usepackage{notoccite} 
\usepackage{soul}
\usepackage{multirow}
\usepackage{etoolbox}
\usepackage{authblk}
\usepackage{blindtext}
\usepackage{dcolumn}
\usepackage{bm}
\usepackage{qcircuit}
\usepackage{braket}
\usepackage{physics}
\usepackage{braket}
\usepackage{varwidth}
\usepackage{empheq}
\usepackage[]{algorithm2e}
\usepackage{algorithmic}
\usepackage{framed}
\usepackage{enumitem}
\usepackage[utf8]{inputenc}
\usepackage[english]{babel}
\usepackage{url}
\usepackage{hyperref}
\usepackage[font={footnotesize}]{caption}
\usepackage{enumitem}
\usepackage{multirow}
\usepackage{textcomp}
\usepackage[T1]{fontenc}
\usepackage{listings}
\usepackage{blindtext}
\usepackage{subfiles}
\usepackage{fullpage}
\usepackage{makeidx}
\usepackage{fancyhdr}
\usepackage[calcwidth]{titlesec}
\usepackage[font={footnotesize},labelfont={bf}]{caption}
\usepackage{setspace}%
\providecommand{\U}[1]{\protect\rule{.1in}{.1in}}
\usepackage[symbol]{footmisc}
\graphicspath{{images/}{../images/fig1}}
\graphicspath{{images/}{../images/fig2}}
\graphicspath{{images/}{../images/fig3}}
\graphicspath{{images/}{../images/fig4}}
\definecolor{anti-flashwhite}{rgb}{0.95, 0.95, 0.96}
\hypersetup{
	colorlinks=true,
	linktocpage=false,
	hyperindex=true,
	linkcolor=blue,
	filecolor=cyan,
	citecolor=green,      
	urlcolor=magenta,
	pdftitle={Quantum-Computation},
	pdfauthor={Bhupesh~Bishnoi},
	bookmarks=true,
	bookmarksopen=true,
	pdfpagemode=UseOutlines,
	pdfstartview={FitH},
}
\titleformat{\section}[hang]{\bfseries}
{\Large\thesection}{12pt}{\Large}[{\titlerule[0.5pt]}]
\titleformat{\chapter}[display]
{\normalfont\Large\filcenter}
{\titlerule[1pt]
	\vspace{1pt}
	\titlerule
	\vspace{1pc}
	\bfseries\LARGE\MakeUppercase{\chaptertitlename} \thechapter}
{1pc}
{\titlerule
	\vspace{1pc}
	\Huge\bfseries}
\makeindex

\begin{document}

\title{\textbf{{\Huge Quantum-Computation}}}
\author{\bf{Bhupesh~Bishnoi\thanks{\texttt{\bf{bishnoi[At]ieee[Dot]org}}}}}
\date{}
\makeatletter
\def\@fnsymbol#1{\ensuremath{\ifcase#1\or \dagger \else\@ctrerr\fi}}
\makeatother

\maketitle

\noindent \textbf{Copyright Notice:}

\bigskip

\noindent The Work, \textbf{``Quantum-Computation''} is to be published by relevant publication press. The Work is on four part:
\begin{enumerate}
\item \textbf{Quantum-Computation and Applications}
\item \textbf{Quantum-Computation Algorithms}
\item \textbf{Quantum-Computation and Quantum-Communication}
\item \textbf{Universal Quantum-Computation}	
\end{enumerate}

This book could be useful for self-learning or as a research reference book. This book is available under a Creative Commons Attribution-NonCommercial-NoDerivatives:CC BY-NC-ND
license. This means that you can copy and redistribute the material in any medium or format of this draft, pre-publication copy as you wish,  
as long as you attribute the author, you do not use it for
commercial purposes, and you must give appropriate credit, provide a link to the license, and indicate if changes were made. You may do so in any reasonable manner, but not in any way that suggests the licensor endorses you or your use. You may not use the material for commercial purposes. If you remix, transform, or build upon the material, you may not distribute the modified material.(see
\begin{quotation}
	\texttt{https://creativecommons.org/licenses/by-nc-nd/4.0/}
\end{quotation}
for a readable summary of the terms of the license). These requirements can be
waived if you obtain permission from the present author. By releasing the draft, pre-publication copy of the book
under this license, I expect and encourage new developments in the theory, and the latest open problems. It might also be a helpful starting point for a book on a related topic.  

\bigskip

\noindent \copyright\ in the Work, Bhupesh~Bishnoi, 2020

\bigskip

\noindent NB: The copy of the work, as displayed on this web site, is a draft, pre-publication copy only. The final, published version of the work can be purchased through relevant publication press and other standard distribution channels. This draft copy is made available for personal use only and must not be sold or re-distributed.

\bigskip\bigskip\noindent Bhupesh~Bishnoi\newline Tsukuba, Ibaraki,
Japan\newline June 2020
\newline \texttt{{bishnoi[At]ieee[Dot]org}}
\newline \texttt{{bishnoi[At]computer[Dot]org}}
	
\tableofcontents

\part{Quantum-Computation and Applications}
\chapter{Quantum-Computation and Applications}

\maketitle

In this research notebook on quantum computation and applications for quantum engineers, researchers, and scientists, we will discuss and summarized the core principles and practical application areas of quantum computation. We first discuss the historical prospect from which quantum computing emerged from the early days of computing before the dominance of modern microprocessors. And the re-emergence of that quest with the sunset of Moore's law in the current decade. The mapping of computation onto the behavior of physical systems is a historical challenge vividly illustrate by considering how quantum bits may be realized with a wide variety of physical systems, spanning from atoms to photons, using semiconductors and superconductors. The computing algorithms also change with the underline variety of physical systems and the possibility of encoding the information in the quantum systems compared to the ordinary classical computers because of these new abilities afforded by quantum systems. We will also consider the emerging engineering, science, technology, business, and social implications of these advancements. We will describe a substantial difference between quantum and classical computation paradigm. After we will discuss and understand engineering challenges currently faced by developers of the real quantum computation system. We will evaluate the essential technology required for quantum computers to be able to function correctly. Later on, discuss the potential business application, which can be touch by these new computation capabilities. We utilize the IBM Quantum Experience to run the real-world problem, although on a small scale.

\section{The era to Quantum Computation and Information}
First and foremost, quantum computation and information have been a scientific curiosity implemented in the academic laboratory and a few governments and industrial laboratories around the world \cite{nielsen_quantum_2011,chuang_quantum_2014,preskill_quantum_2019,vazirani_quantum_1997, watrous_theory_2011,vazirani_quantum_2007,aaronson_quantum_2006,shor_quantum_2003,chuang_quantum_2006,aaronson_quantum_2010,harrow_quantum_2018,childs_quantum_2008,cleve_introduction_2007}. It started as a discipline in physics. But over the last five years to ten years it is transitioning from scientific curiosity to technical reality \cite{russo_coming_2018,noauthor_next_nodate,noauthor_here_nodate}. In that transition, many problems need to be addressed. Many of them are falling on the engineering side. So, going forward, people who come from traditional engineering disciplines will have significant roles to play. There is a new term that is being coined called ``quantum engineering.'' Moreover, it is bridging quantum science and the traditional engineering disciplines. Both will have to pivot to meet the needs of building quantum testbeds, which will lead to future quantum systems. In terms of background, many people today, whether trained as quantum physicists or trained as quantum engineers, are discussing a lot about the other discipline \cite{harrow_why_2012}. The hope is that as we move forward that we begin to define the engineering aspects of quantum information, that we begin to abstract it in the way that engineers abstract concepts from physics to make it an engineering and systems problem, and as this happens in the current decade, people will be able to contribute more without knowing the underlying, in-depth quantum physics. So, the motivation is to begin bridge quantum science and quantum engineering. In this sense, quantum computers are becoming available commercially for larger people outside of the academic laboratory. Now, it becomes more the domain of computer sciences rather than physicists and mathematicians. Traditionally physicists and mathematicians have been at the forefront of this field and pushing it. Now the goal is to bring computer scientists and electrical engineers up to the same level as we go forwards to tackle systems engineering challenges. With the advent of access to these computers will certainly bring in more computer scientists, software programmers, and electrical engineers in the domain. However, it does not mean that physicists and mathematicians stop doing work in quantum science and theory. so, the physicists and mathematicians will continue to be actively involved in this new computational development. Also, related to cost-effective ways to build a quantum computing experience, in-house familiarity, and expertise, there are examples of online quantum computers available to open source to use and operate. These include the IBM Quantum Experience\cite{sisodia_circuit_2018}. Rigetti has an online quantum computer\cite{cervera-lierta_exact_2018}. D-Wave has an online quantum computer. Google is also preparing an online quantum computer\cite{mcclean_openfermion_2019}. The impact and advantage of these open-source availability are many discussions started about the algorithms that might run on these existing systems and have an impact on the society with the current state of the art quantum hardware capability. We also get to interface with faculty and graduate students at the university and institutions. Access to talent is a huge part of the development. if we have some good ideas to explore and would be the solution, then we can engage directly with places like IBM, Rigetti, D-Wave, and Google to implement on their systems. So, there is a natural way via academia to ramp into the quantum information field. It starts with education, and we hope those are some useful ideas and a beautiful way to go ahead. With the current state of art quantum computation technological capability, it looks like there is no big relation between we can calculate today and the future quantum computational potential. So, should we need to study and worried about quantum computing today? There is some area which will affect today's world, one is information security, and it is important to everyone. we want to keep information secure not just today but for the next few decades. So, we write down and share the information, e.g., business strategy, trade-secret, government communique, personal information with our partner, but do not want to reveal everyone else and get hacked. So, we encrypt the information and share on electronic channel worldwide. currently, state of the art encryption scheme is strong, and we do not have a quantum computational capability yet, so decryption of information is not possible for others, this is true, but in future, we will have the quantum computational capability and with that, today shared and communicated information can be decrypted in the future. So, how long do we want today's information to be secure in the future. We need to understand the implications of quantum computing and now realize our information encryption vulnerability. So, that is one important aspect. Another important aspect is that most economic problems can be boiled down to optimization problems. We need to optimize that if it is doing routing and layout of an electrical circuit in a complicated chip. It could be as simple as home deliveries. we want to pick the shortest path possible. There are many examples of optimization problems. we will be able to identify certain problems within our organization, which might benefit from quantum computing, which is manifestly not related to factoring but more closely related to optimization \cite{farhi_quantum_2014} or quantum simulation or quantum emulation. These are problems that we want to think about and explore the solution because, in the future, with quantum computers, it will have an impact on our life.

Quantum computing is not going to replace classical computing. We will always need a classical computer to run beside it and control it. At least as we understand it today, quantum computers solve specific problems that we know of today much more efficiently than classical computers. However, those problems are not necessarily the types that we would want to run on our classical computer. It may be better said that a quantum computer can run any algorithm that we can run on our classical computer. However, it may not do so, any better than we can currently do with our classical computer. So, for the foreseeable future, we believe that quantum computers will, for example, remain in the cloud, or we could think of them as these large mainframe-type computers. we use it for problems, like quantum simulation. in the nearer term, particularly before we can make fully error-corrected, fully error-resistant quantum computers, we will be using qubits that are faulty. We, therefore, cannot, by themselves, do a very, very long computation. so, it is likely that we will first see quantum computers that are viewed as accelerators for a core processor with a classical computer \cite{smart_experimental_2019,hegade_experimental_2019}. The classical computer is running and orchestrating an overall algorithm. Nevertheless, it pings or pulls the quantum node periodically to get an answer to a sub-problem of a small part of the problem, takes that answer back, and then incorporates it into a larger algorithm that is being run. So, indeed, It is true that quantum computers, very likely, will be working alongside classical computers, at least for the foreseeable future. On classical digital computers, digital words and gates are applied several billion times a second. what is the analogous fundamental operation on a quantum computer? One starts with qubits. Then how are operations commanded? In classical computers, arbitrary Boolean logic can be performed with a universal set of one-bit and two-bit gates. There is not a unique arrangement of single and two-bit gates that will give us a universality. However, with just a handful of gates or even one gate, one can perform universal Boolean logic. then in computers, there is a clock that runs at gigahertz rates and just clocks the application of these gates throughout some algorithm or computation. As in a quantum computer, it is a bit different, and there are also some similarities. let us start with the similarities. First, quantum computers also have the concept of universality and the fact that there is a combination of single-qubit and two-qubit gates, which can reach, as we say, any point in the Hilbert space. We can take any quantum state and turn it into any other quantum state spanned by the qubits that we have. so, that is the concept of universality. Again, it is a handful of gates, as we have seen, not unique. However, with one and two-qubit gates, we can perform arbitrary quantum logic. Now, how it works, is that we will take a massive quantum superposition state of $ N $ qubits. That gets fed into a computer, or that is the starting point of the computer. Then through single-qubit and two-qubit gates, according to a prescription that the algorithm designer decides we implement and set up quantum interference to occur in such a way that by the end of the computation, ideally, all of the probability amplitude resides in one of these states that is, in fact, the answer to our problem. It encodes the answer to the problem so that when we measure with a very high probability or even unit probability, the system collapses onto that single state. The probability that we measure a given state goes as the magnitude squared of the coefficient in front. That is why it must approach unit value. When we make that measurement with high probability, we will get the right answer. so, that is how these two computers work. Now, in a classical computer, running faster and faster is always desirable. It is also true that we do want to have a fast gate speed in a quantum computer. For example, if one type of quantum computer can run at 100 megahertz and another type of quantum computer can do the same calculation in the same way. However, it only runs at 100 kilohertz, then, the one that is running at 100 megahertz will run 1,000 times faster. we can celebrate that a quantum computer can run or can solve problems exponentially faster. so, something that might take the age of the universe now it only takes a day on the 100-megahertz quantum computer. Nevertheless, on the 100-kilohertz quantum computer, it would take 1,000 days. In comparison to it found the age of the universe, it is a fantastic achievement. However, 1,000 days that is around three years. so, from a human time frame timescale, that is not as useful as something that runs in a day. So, there are two aspects. There is the exponential improvement or the strong polynomial improvement, the quantum enhancement, that happens \cite{bremner_classical_2011}. However, on top of that, it is still important that we do not forget the prefactor, that we run quickly, and we have a faster clock and gating. The next challenge is addressing large-scale quantum computing from sub-modules. so, would it be possible to use quantum networking or other methods\cite{shang_continuous-variable_2019,liao_satellite-relayed_2018}? Like a cluster computer does, to simulate a system of 1,000 qubits, but by only using 20 such sub-units with every 50 qubits, and this is the concept of modularity. It is an active area of research. It is expected that we will add something that looks like modularity in any large qubits system that we build. In classical computer systems, we always see some degree of modularity in building those systems because they are complicated. As in the early stages of quantum computing, we are working with qubits that are individually faulty. There is a whole research line on trying to address that through fault tolerance and error correction \cite{edmunds_measuring_2017}. However, that is going to take some time. In the meantime, we are looking at these noisy systems, and this is the concept of noisy, intermediate-scale quantum computing, which is quantum computing using the qubits we have today or in the foreseeable future, which are quite good, but not good enough to run a computation from start to finish in its entirety. so, these NISQ \cite{preskill_quantum_2018}, quantum computers very likely will be clusters of, 50 to 100-qubits and very likely will be a co-processor that runs alongside a classical computer where the classical computer is going to pull these quantum subsystems or ask a question to this sub-module many times and get answers back and aggregate these answers in order to address and solve an overall problem. That is certainly one way we can imagine modularity being incorporated into a quantum computer. There are other examples of this in trapped ions, superconducting qubits, for example, to take the trapped ion example, a single ion trap can hold on the order of 100 qubits\cite{delehaye_single-ion_2018,huntemann_single-ion_2016}. However, then to go to 1,000 qubits, that is beyond what any single trap would be able to accommodate and hold. So, the idea would be that we build 100-qubit ion traps and then communicate between them using lasers and entanglement to run a larger scale computer built out of these smaller modules. A similar concept is being developed at QCI for superconducting qubits, where we have a microwave cavity. Inside this cavity, we can encode multiple qubits worth of information, whether that is several modes within that cavity, or the information is being stored in complicated \texttt{Ket} states. However, then these resonators would then be coherently coupled to one another to make a larger scale quantum computer. So, modularity will be a part of future quantum computing as we build the larger systems.

Now, other than quantum system hardware, there is also active development in the quantum programming languages. There is much activity going on into the development of the software stack. Each of the major industrial players who are looking at quantum computing is developing it, IBM has their Qiskit \cite{larose_overview_2019}, Google or Rigetti \cite{mcclean_openfermion_2019}, and Microsoft, all are developing software platforms and addressing various levels in the software stack, whether that is software which immediately controls the physical qubit layer, the hardware layer, or whether It is software that is interpreting a program that is written by a user who is trying to implement a quantum algorithm and translating that program and compiling that into the types of instructions that a quantum computer needs in order to operate. So, there is a lot of research activity in the programming domain, not just in the industry, but also in the university. There are academic programs that are focused on these very tasks. We expect that these research activities will continue both because there is much work to do, but also because quantum computers themselves are going to mature with every new generation.
Moreover, as they mature, the quantum programming languages will also mature \cite{killoran_strawberry_2019}. There is another aspect of software development, which is not directly related to implementing an algorithm but is equally important and related to the development of very basic hardware iteration. The electronic design automation (EDA) type design software that is needed to design quantum computers. There are existing simulators to simulate the electromagnetic of the superconducting chip or semiconducting chip or whether It is layout optimizer or whether it is taking a GDS file design of the quantum circuit and then simulating that quantum circuit to understand if it is at least electromagnetically behaving the way that It is supposed to. All of these are essential aspects of designing a real system. So, that is another area where software development is being done. It would be fair to say that with the launch of the IBM Quantum System One, the first commercially available quantum computer, that quantum computing has emerged from the laboratory and will more become the domain of computer scientists, software programmers, electrical engineer off course with physicists and mathematicians. The IBM quantum system is the first, commercially available universal gate model type quantum computer. The D-Wave system has also been around for a few years. That is also commercially available to perform quantum annealing.

We will start with an introduction to the types of quantum computing devices that exist. We will also look at the history of classical electronic computing. Then we compared that to where quantum computing is today. We look at quantum gates, single-qubit gates, two-qubit gates, and how they are used in universal quantum algorithms. We discuss quantum interference and quantum parallelism and how that underlies the power of a quantum computer. We look at examples of quantum simulation \cite{king_observation_2018} or emulation, quantum annealing devices\cite{gabor_assessing_2019}, and the universal gate model quantum computer. We will also discuss qubit modality and their performance. Thus, we start with the DiVincenzo criteria for quantum computers. We will discuss qubit robustness and the coherence time. We will also discuss how the gate time is critical and introduced the metric for qubits called the gate fidelity. Then, we compare different modalities against one another. We will also investigate several of them, including defect centers, ion traps, superconducting qubits, semiconducting qubits. We will focus on trapped ions and the superconducting qubits. We are looking at the promises of quantum computing, and the promise of quantum communication, and looking at quantum advantage and algorithms. We look at circuit models and look at the Deutsch-Jozsa quantum algorithm. At the end, we will discuss about various industry player in the quantum computation domain to discuss about their research and perspectives on quantum computing, we discuss IBM \cite{abenojar_ibm_nodate,noauthor_ibm_nodate,noauthor_qiskit_nodate,noauthor_qiskitopenqasm_2020},Google\cite{mcclean_openfermion_2019,noauthor_quantumlib_nodate,noauthor_quantumlibopenfermion_2020},Microsoft\cite{svore_q_2018,soeken_programming_2018,noauthor_quantum_nodate},IonQ\cite{noauthor_ionq_nodate},Rigetti\cite{noauthor_home_nodate,noauthor_rigetti_nodate},QCI\cite{circuits_quantum_nodate},and D-WaveSystems\cite{noauthor_d-wave_nodate}. In the next, we are going to look at circuit models and discuss the Deutsch-Jozsa quantum algorithm. We will then apply the Deutsch-Jozsa algorithm and run it on the IBM Quantum Experience.

\section{Quantum Computing}
We discuss quantum computers every day in the news and the popular press. There is excitement about quantum computing. It is said that quantum computers will solve certain types of problems. Ones of tremendous importance to humankind are problems that today are prohibitive or even impossible to solve with current computers. We will discuss pharmaceuticals and drug discovery. We are gaining a better understanding of new materials like high-temperature superconductors, New methods for machine learning \cite{shiba_convolution_2019,schuld_machine_2019, bassman_towards_2020,havlicek_supervised_2019}, artificial intelligence, optimization problems, and financial services in technology. Quantum computers will even challenge and change the way we securely communicate information. It certainly sounds like a fantastic and exciting future, which leads us to a few fundamental questions. What exactly is a quantum computer, and what is its suitable application? More importantly, when will we have one? Quantum computers are not just smaller, faster versions of classical computers. They are fundamentally different. Whereas in the digital computer world, a bit, which is one fundamental element of information, is a zero or a one. In a quantum computer, we can have a quantum bit, or qubit, that is in a superposition of zero and one. We can design and control them. We are engineering and manipulating quantum mechanics. So, when We have a quantum computer is a fascinating and nuanced question.
Moreover, the answer will, therefore, be finicky. We have been saying that quantum computers are ten years away. We have been saying that for decades. Depending on the definition, we already have quantum computers, but they are just small. It is not decades away or a century away from that quantum age has now arrived. Quantum computers are not merely faster, smaller versions of the conventional computers we have today. Nor are they another incremental step in the evolution of Moore's Law. Instead, quantum computers stand for a new, fundamentally different type of computing paradigm, one that carries tremendous advantage for certain types of problems of importance. Quantum computing could transform industries where there are significant optimization problems. We have a lot of discrete or binary decisions to make to figure out, do we do this first or that first. Another way to understand the difference between classical and quantum computers is to look at quantum systems of quantum simulation. A quantum processor is a suitable tool for modelling other quantum systems. Biomolecule systems and tons of other systems that we use, material systems fundamentally work on based on those quantum mechanical properties. We need a quantum machine to simulate quantum effects. When we can manipulate individual molecules and understand what is going on in those molecules, how they bond, then We will be able to have an excellent handle on generating new things and novel materials that might be very useful. Still, we are just at the very beginning of quantum computing development. Assembling and testing the prototype processors. It is a bit like being in the 1950s at the dawn of transistor-based computing.
Furthermore, just as integrated circuits led to an information processing revolution last century, driving economic growth and productivity, many people today believe that quantum computing will have a similar impact on this century. Quantum computing and quantum algorithms present fundamentally new programming and algorithm design paradigms. How do we fundamentally unlock new ideas in computing? We are still discussing a lot about how to improve the individual components and connect them. We are looking to enable the increased complexity and functionality of these qubits. We are here at the very beginning of the new revolution. We find that it is tremendously exciting. The goal here is to separate the promise from the hype and to technologically understand the basics of the quantum computation working principle and its applications. We will begin by focusing on those basics.

Quantum computers are not merely smaller, faster versions of today's computers. Instead, they represent a fundamentally new paradigm for processing information. They can exceed the performance of conventional computers for problems of importance to humankind and businesses alike, in areas such as
\begin{itemize}
\item Cybersecurity,
\item Materials science,
\item Chemistry,
\item Pharmaceuticals,
\item Machine learning,
\item Optimization, and more.
\end{itemize}

\section{An overview of Quantum Computer}
\begin{figure}[H] \centering{\includegraphics[scale=.3]{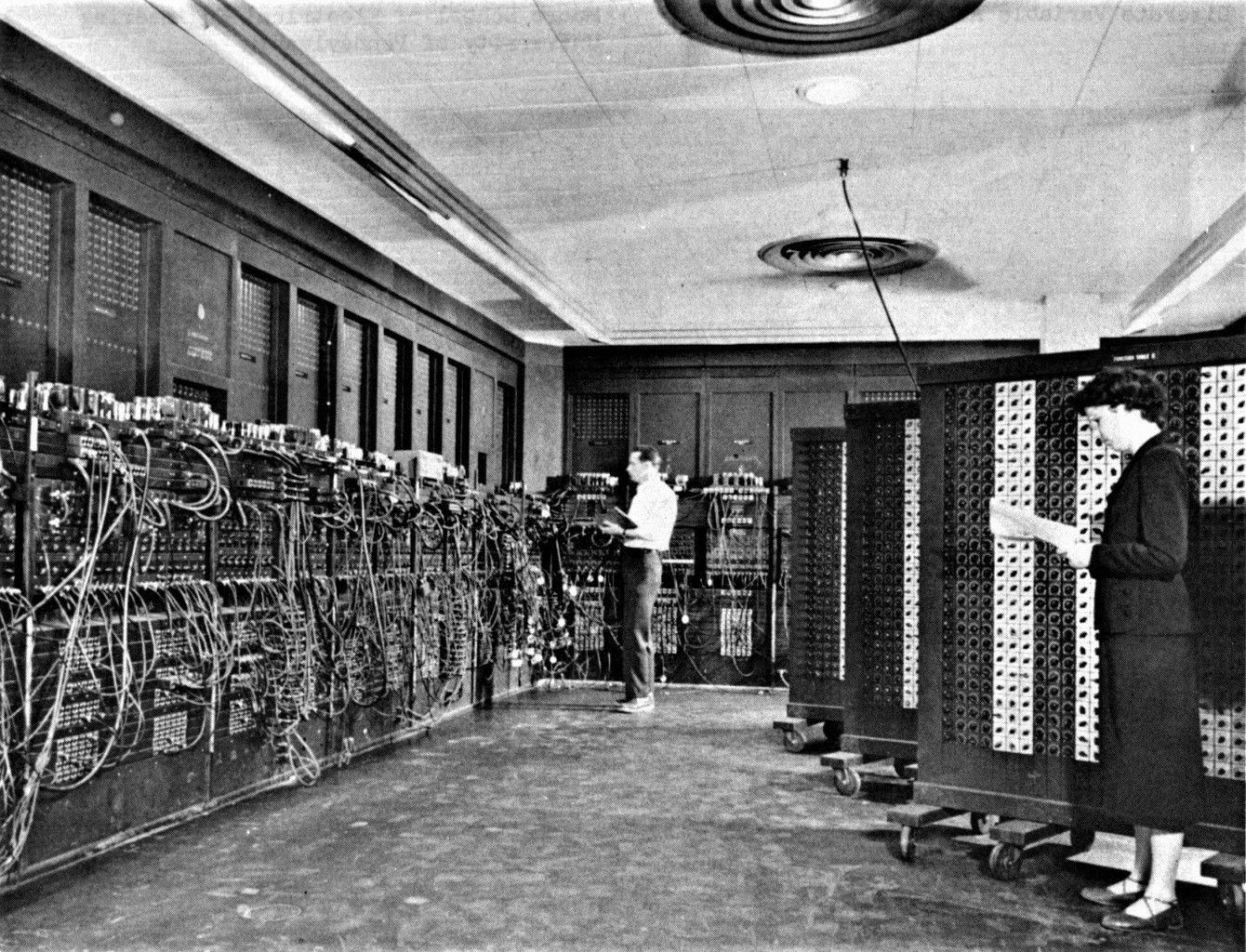}}
\caption{ENIAC; Source:U.S. Army}
\label{fig1_1}
\end{figure}

Classical computers have changed dramatically over the past 80 years, from the room-filling vacuum-tube-based computers like ENIAC (Figure \ref{fig1_1})

Currently, quantum computers are at the research and prototyping stage, looking more like an ENIAC than a laptop or tablet. They often occupy an entire laboratory space with a variety of machines and tools to house and operate the core of the quantum computer. A portion of this infrastructure surrounding the quantum computing ``core'' is necessary to shield the quantum computer from sources of electromagnetic noise, mechanical vibration, heat, and other noise sources, which tend to degrade performance. Another portion, comprising conventional ``classical'' computers, electronics, and optical systems, is used to control the quantum computer, implement an algorithm, and read out the result.

\begin{figure}[H] \centering{\includegraphics[scale=.7]{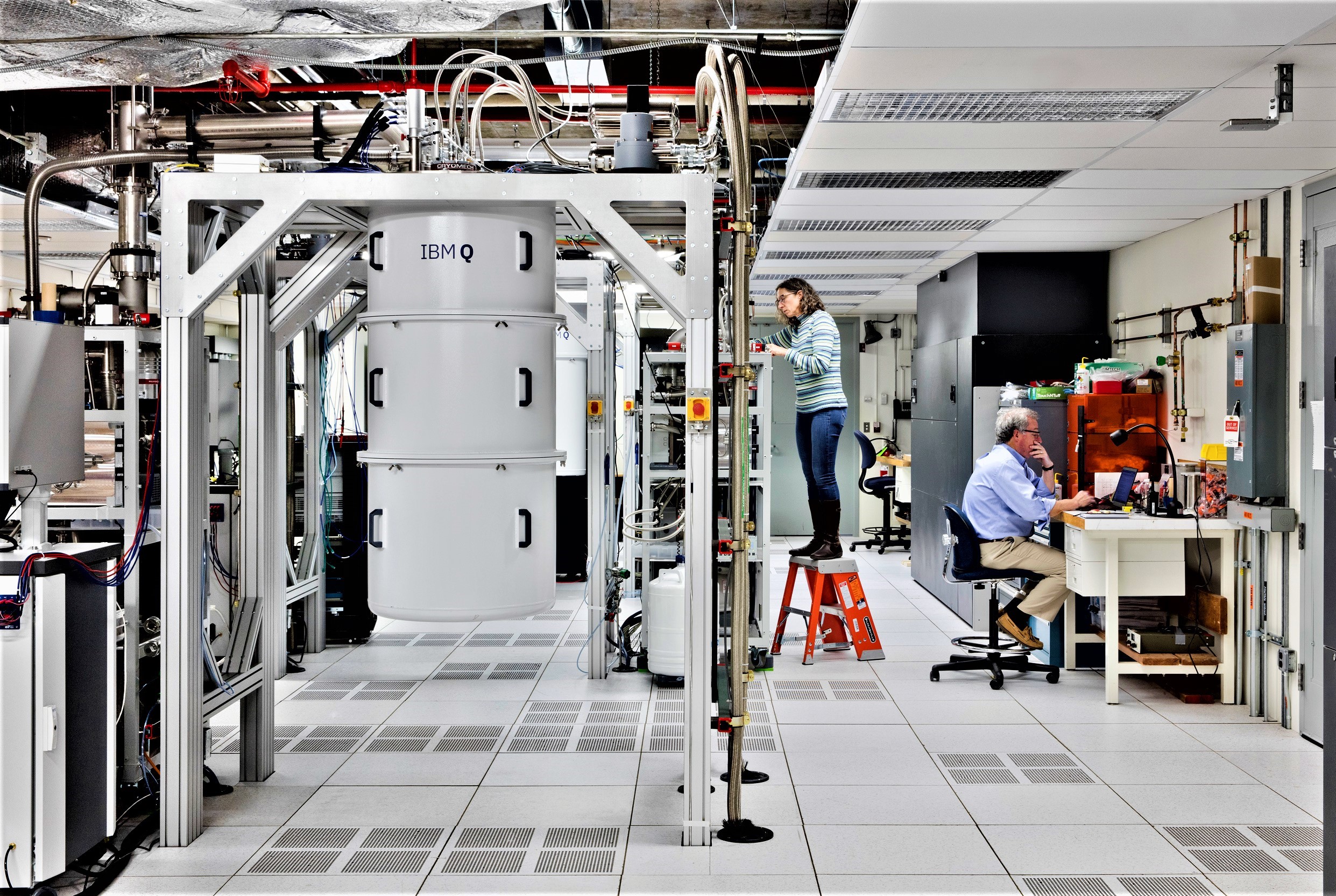}}
    \caption{A research-grade superconducting quantum processor. The processor is located inside the white dilution refrigerator hanging from the support structure. The refrigerator cools the processor to the milliKelvin temperatures required to operate it;Source:IBM}
    \label{fig1_2}
\end{figure}

\begin{figure}[H] \centering{\includegraphics[scale=.4]{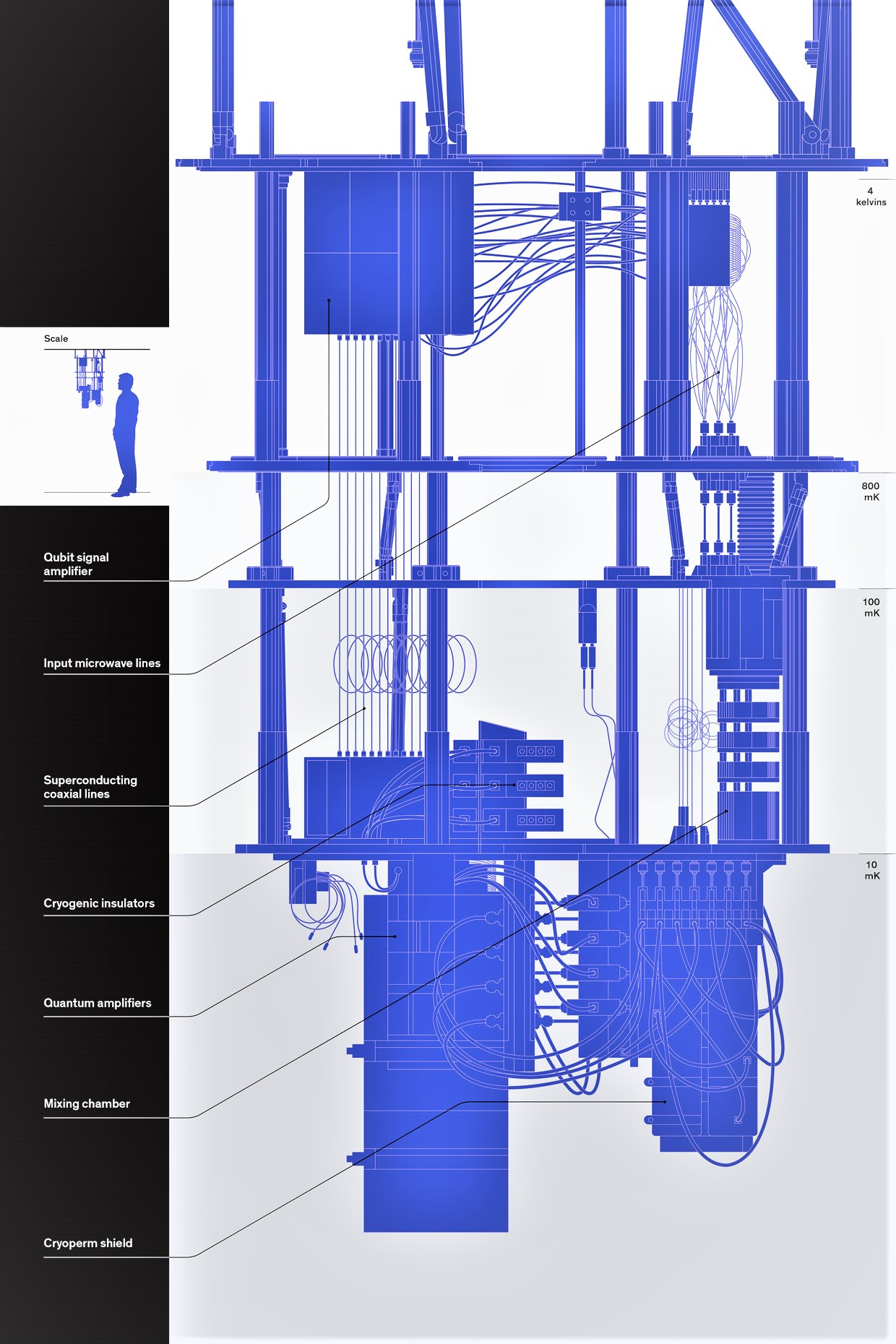}}
	\caption{Dilution refrigerator;Source:IBM}
	\label{fig1_2_1}
\end{figure}

\begin{figure}[H] \centering{\includegraphics[scale=.7]{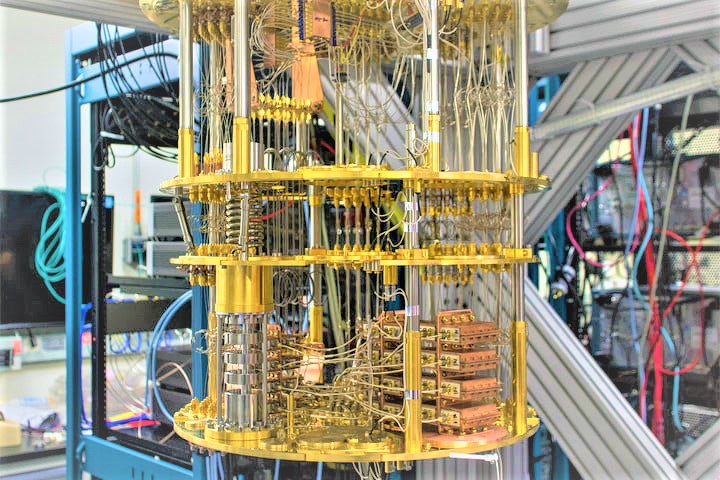}}
	\caption{superconducting quantum processor;Source:IBM}
	\label{fig1_2_2}
\end{figure}

\begin{figure}[H] \centering{\includegraphics[scale=.8]{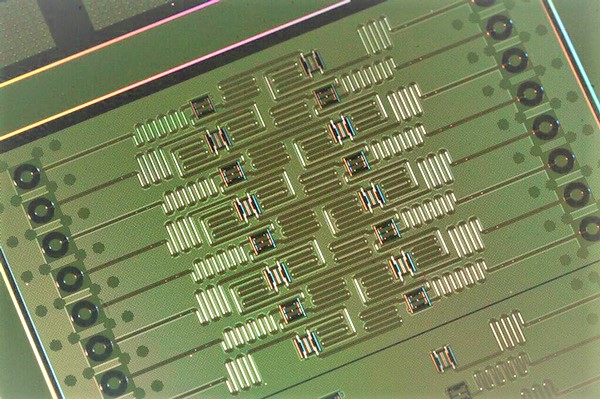}}
	\caption{superconducting qubit citcuit;Source:IBM}
	\label{fig1_2_3}
\end{figure}

\begin{figure}[H] \centering{\includegraphics[scale=1.1]{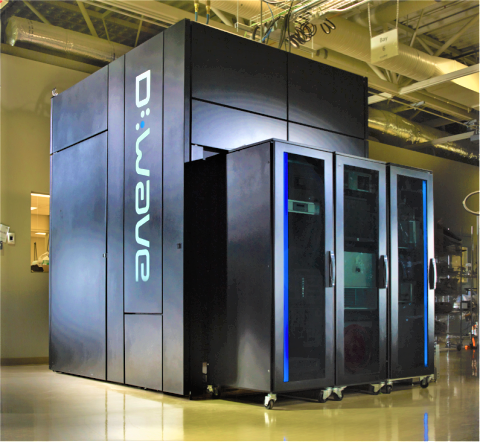}}
	\caption{superconducting quantum annealer;Source:D-Wave}
	\label{fig1_2_4}
\end{figure}

In the picture above, we see a large research-grade ``dilution refrigerator'' used to house and cool a prototype superconducting quantum processor. Refrigeration is required to cool the quantum chip to its operating temperature of less than 20 milliKelvin, a temperature more than 100 times colder than outer space. The refrigerator also serves to reduce the thermal load and noise that would otherwise degrade performance, arising in large part from the room-temperature electronics connected to the chip through various types of electrical cabling. To the left of the refrigerator are racks of such electronics, including arbitrary waveform generators, microwave signal generators, and current sources used to control the processor.

\begin{figure}[H] \centering{\includegraphics[scale=.4]{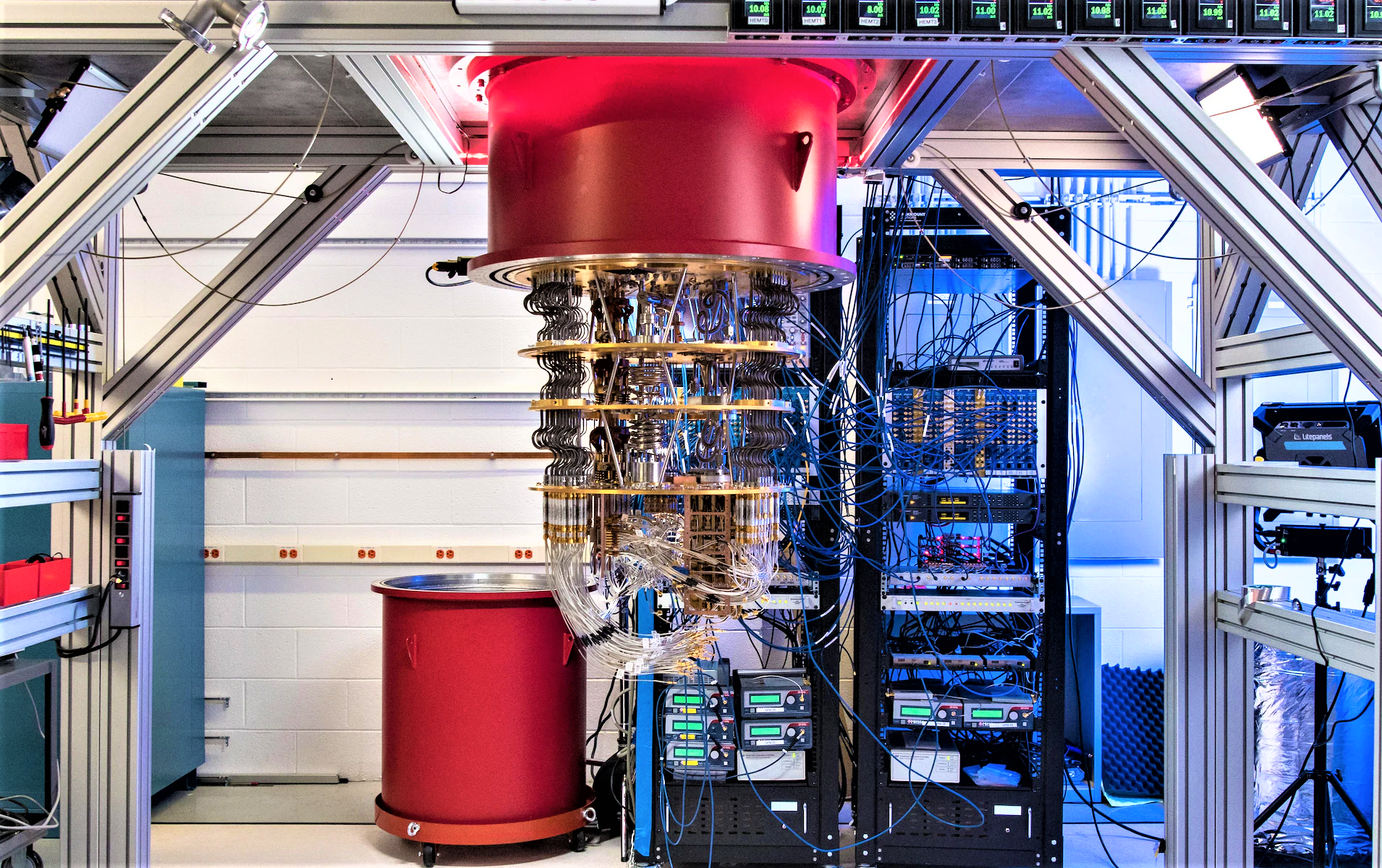}}
	\caption{Google System;Source:Google}
	\label{fig1_2_5}
\end{figure}

\begin{figure}[H] \centering{\includegraphics[scale=.7]{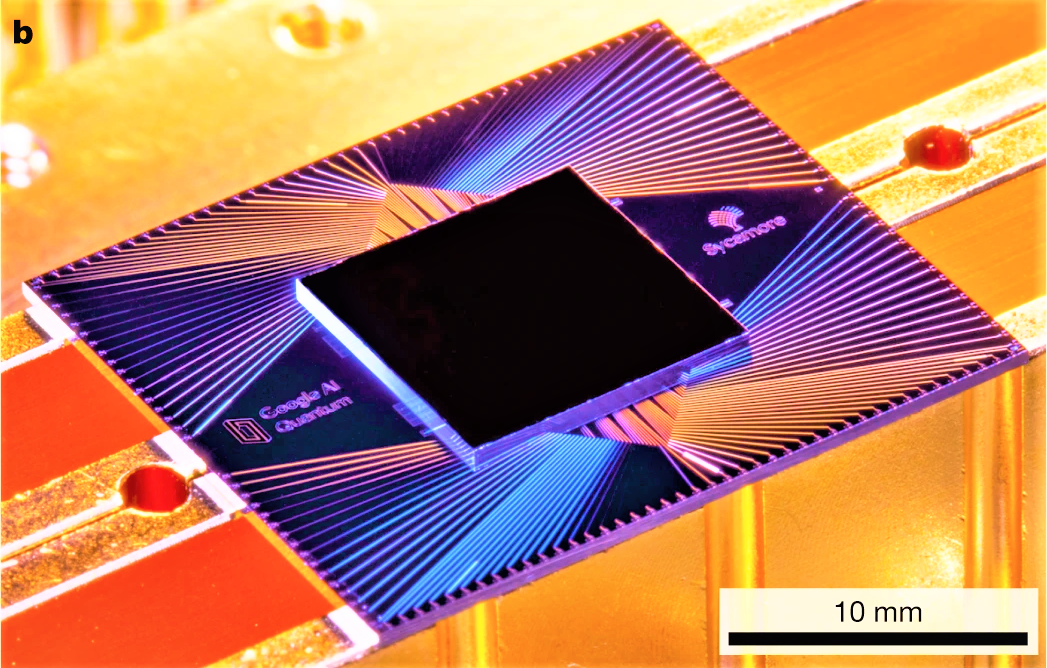}}
	\caption{Google superconducting quantum processor;Source:Google}
	\label{fig1_2_6}
\end{figure}

In the figure below, we see an optical table, on which stands a large black box housing the optical system used to control and measure a trapped-ion quantum computer\cite{stuart_chip-integrated_2019}. The trapped ion computer ``core'' itself may reside in a cryogenic chamber at a temperature around 3-4 Kelvin, a temperature similar to outer space to obtain and maintain an ultra-high vacuum (although this is not required). High-stability lasers send light through a variety of mirrors, beamsplitters, optical modulators, and the like to address and read the individual ions that comprise the quantum computer.

\begin{figure}[H] \centering{\includegraphics[scale=.2]{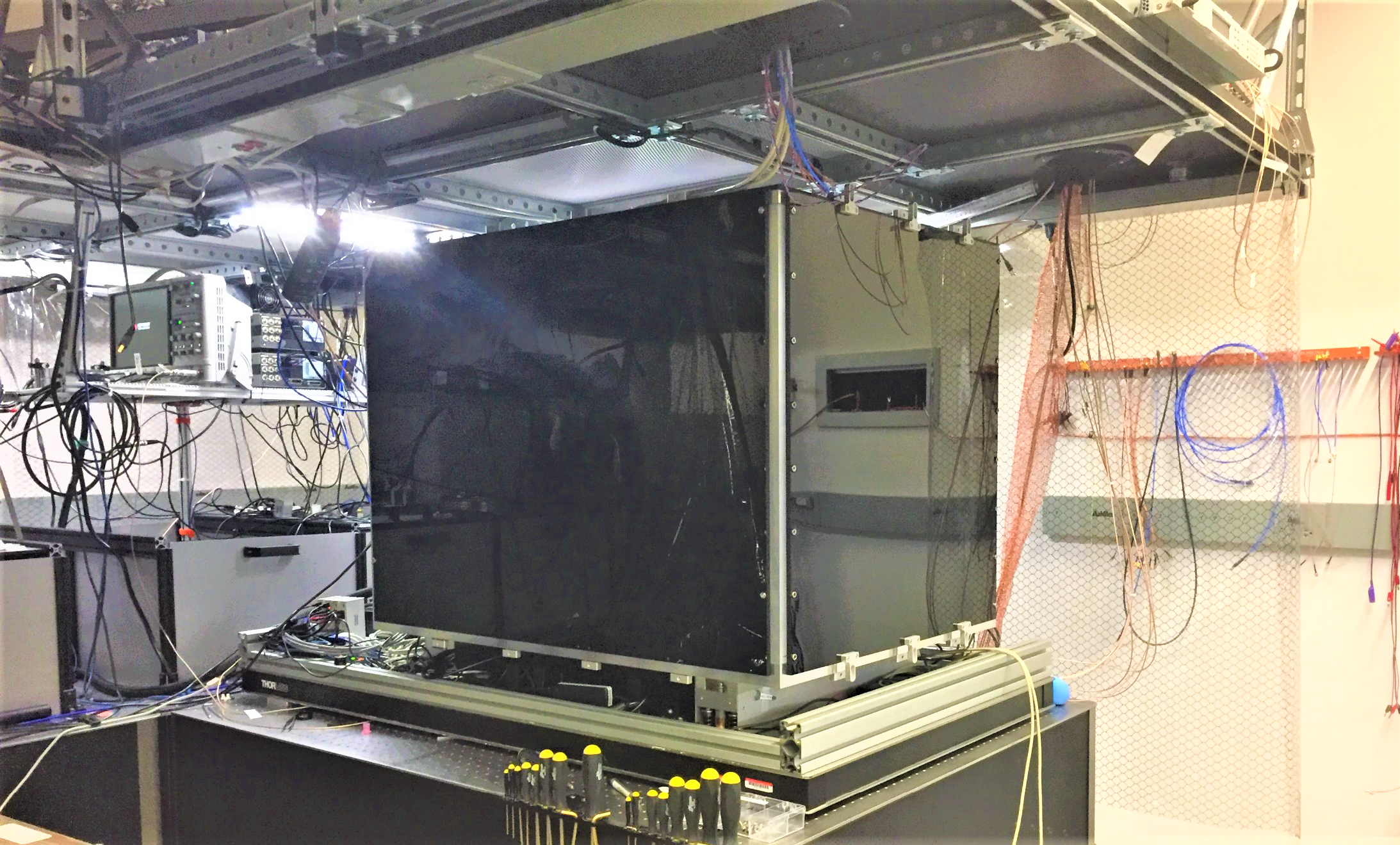}}
    \caption{Trapped-ion quantum computer; Source: IonQ} 
    \label{fig1_3}
\end{figure}

\begin{figure}[H] \centering{\includegraphics[scale=.6]{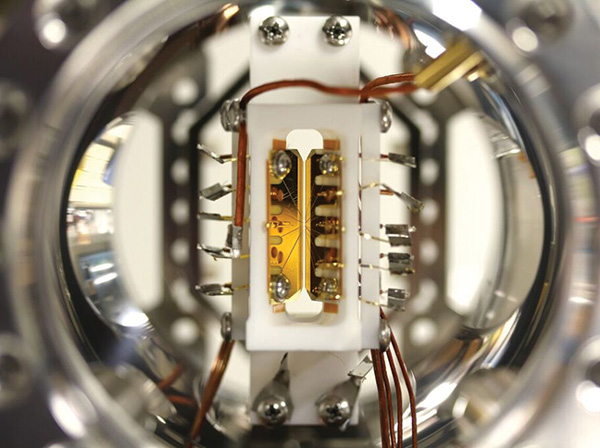}}
	\caption{Trapped-ion quantum computer; Source: Joint Quantum Institute }
	\label{fig1_3_1}
\end{figure}

\begin{figure}[H] \centering{\includegraphics[scale=.4]{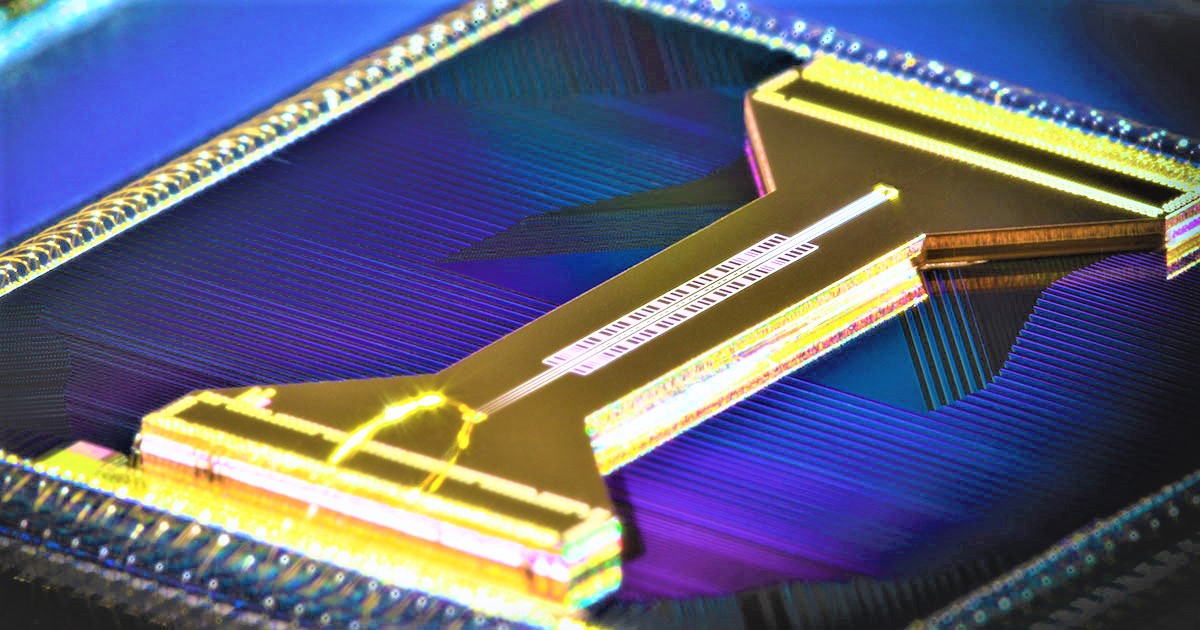}}
	\caption{Honeywell's Ion Trap Quantum; Source: Honeywell }
	\label{fig1_3_2}
\end{figure}

\begin{figure}[H] \centering{\includegraphics[scale=.4]{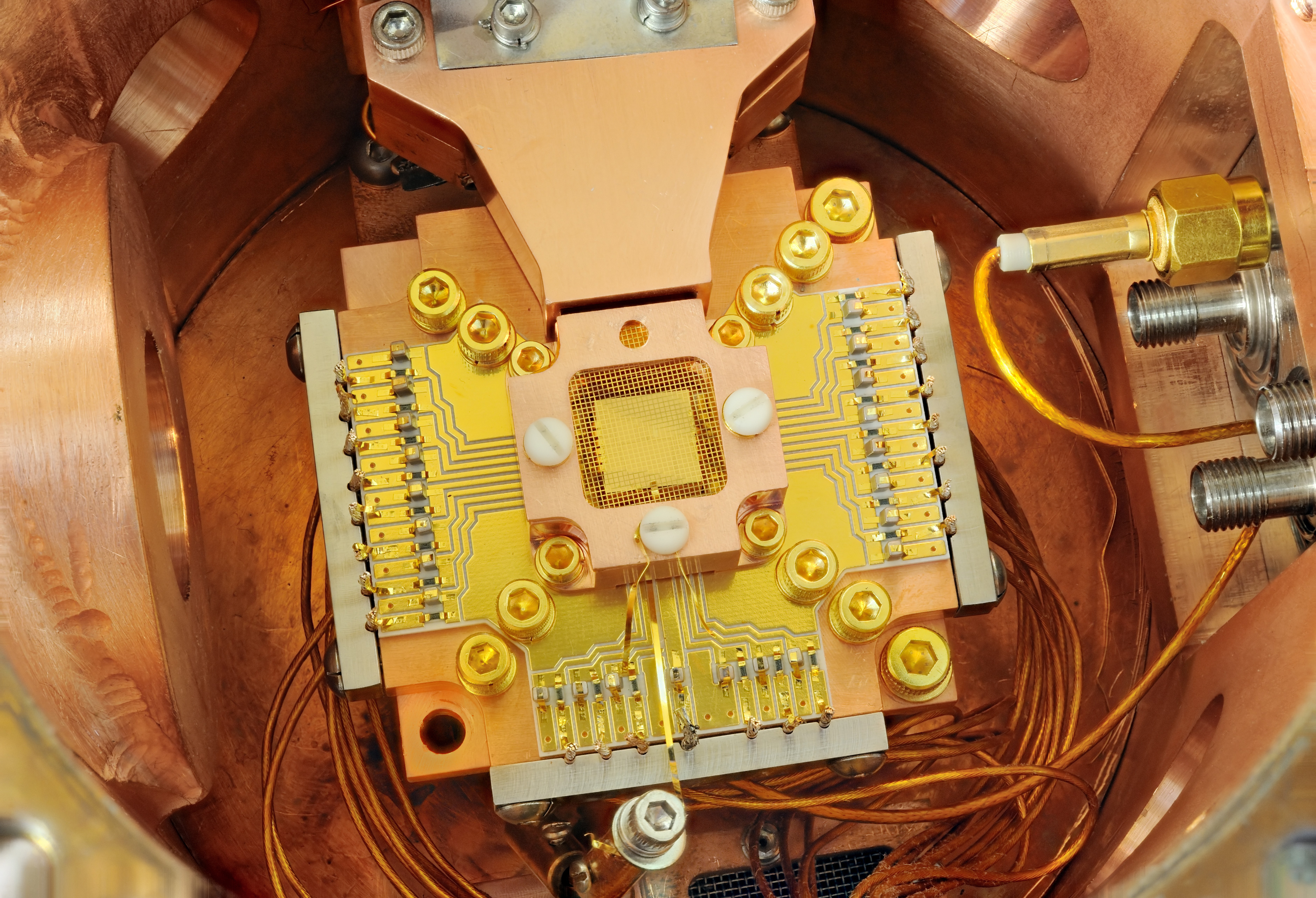}}
	\caption{Trapped ion quantum computer; Source: NIST }
	\label{fig1_3_3}
\end{figure}

\begin{figure}[H] \centering{\includegraphics[scale=.3]{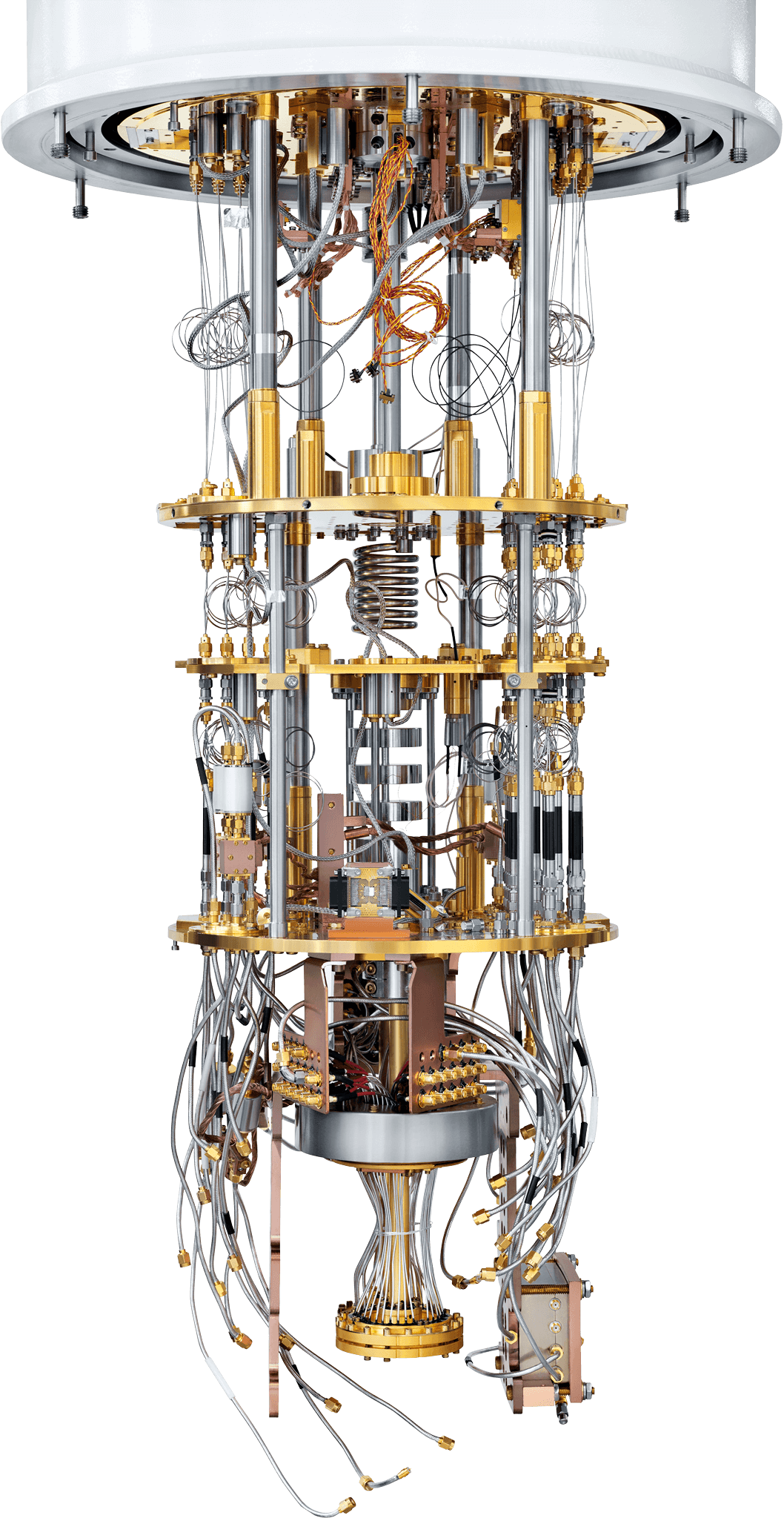}}
	\caption{QPU (quantum processing unit) ; Source: Rigetti }
	\label{fig1_3_4}
\end{figure}

\begin{figure}[H] \centering{\includegraphics[scale=.6]{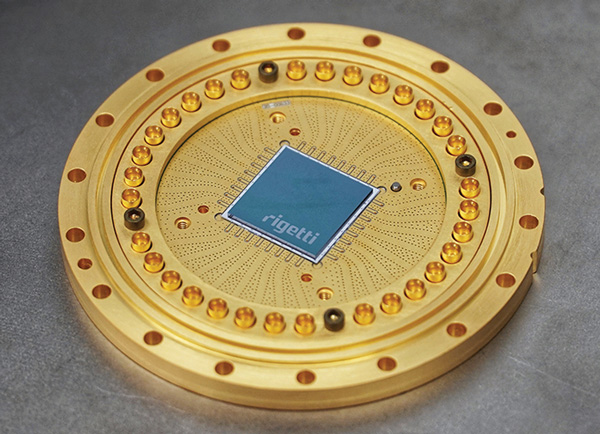}}
	\caption{gold-plated copper disk with a silicon chip; Source: Rigetti }
	\label{fig1_3_5}
\end{figure}

\begin{figure}[H] \centering{\includegraphics[scale=.6]{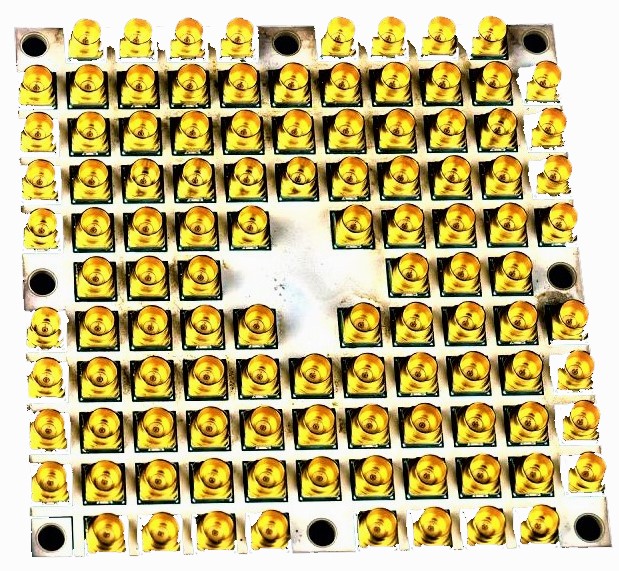}}
	\caption{49 qubits superconducting quantum processor Tangle Lake; Source: Intel}
	\label{fig1_3_6}
\end{figure}

\section{Physical and Conceptual Models of Classical Computation}
To understand quantum computing, we first need to understand how information is processed today and, more generally, what constitutes ``a computer'' and ``computation.''\\
The most common approach to classical information processing today uses a conventional electronic computer comprising a memory and a transistor-based computational processing unit. However, this is not the only physical manifestation of a classical computer. For example, human beings and our brains process information with different physical methods and architectures \cite{linke_experimental_2017}. As these two examples suggest, there are many physical models of classical computation.\\ 
\textbf{Physical Models of Classical Computation:}\\
\textbf{Mechanical:} A computational system built with primarily mechanical components is a mechanical computer. Adding machines used for bookkeeping during the first part of the last century is an example of such computers. The input system is composed of numbered key buttons. After entering the numbers, a user pulls the crank, gear-wheels start turning, and the sum is mechanically computed and displayed. As in the case of an adding machine, mechanical computers are generally designed to implement application-specific tasks.\\
\textbf{Electrical:} Electrical computers use electrical elements that switch on/off electrical currents or voltages. Today's personal computers are in this class, and they use transistors as the fundamental switching elements. Transistors enable the construction of a universal classical computer, one that can, in principle, tackle any computable problem. However, it may not be able to do so efficiently (in a reasonable amount of time or using a reasonable amount of physical hardware).\\
\textbf{Optical:} Systems that use photons the fundamental particles of light to perform computation are optical computers. The gates used to perform logic with photons can be engineered using nonlinear optical materials. As of today, existing optical computers tend to be application-specific.\\
\textbf{Biological:} Biological molecules, for example, proteins or DNA, can be used to process information. The individual, necessary elements for a fully operational biological computer, such as biological transistors, have been demonstrated. However, present biological, computational systems are hybrids that require the addition of electrical or mechanical components to operate.\\
There are also several conceptual models and architectures of classical computation, which in principle, any of the above physical systems can be used to process information.\\ 
\textbf{Conceptual Models and Architectures of Classical Computation:}\\
\textbf{Turing Machine:} A Turing machine comprises a memory tape and reads/write head. The memory tape is divided into discrete cells that store data. The head successively manipulates the cells. According to a set of rules, cell data may be altered depending on the cell's prior information, which is also accessible by the head. This scheme provides an architecture to construct universal computational systems.\\
\textbf{Cellular Automaton:} A cellular automaton comprises an array of cells, each connected to several of its neighboring cells. After the cells are set to initial values, the cellular automaton evolves according to a set of rules that governs how the state of each cell changes in response to the states of its neighboring cells. Depending on the rules and connectivity of the cells, the result may be a uniform, oscillating, chaotic, or another intricate pattern. This computational concept is used to simulate or mimic the behavior of biological or chemical systems. For example, a global function emerges from a large number of seeming independent agents that interact with one another in a specified way.\\
\textbf{Von Neumann Architecture:} Architectures comprising a central processing unit and a memory unit are called von Neumann architectures. The central processing unit contains a controller, an arithmetic logic unit, and registers. The controller manages the computational processes: it requests data from memory, stores it in a register, directs the arithmetic logic unit to process the data in the register, and then sends the result back to the memory. This model is used for most present-day computational devices.

We are all familiar with laptop computers, desktop computers, even servers in the cloud that we interact with daily. In this section, we will refer to these as classical computers to contrast them with quantum computers. We will be discussing in detail later in the section. Classical computers use transistor-based integrated circuits-computer chips to process information and solve problems, whether for implementing a financial transaction, simulating a weather pattern, developing a CAD design, or even just drafting an email to a colleague. However, it is important to remember that there are many alternative ways in which one can process information. So, before we get started, let us discuss about what constitutes a computer and computation. There are many models of classical computation, and we would like to illustrate a wide variety of models, some of which we might find unexpected. One model is mechanical. So, for example, the thermostats on the walls, are driven by pneumatic pressure, and they actuate a little switch. So, they are a little kind of computer that controls temperature, but it is an analog computer. So mechanical computers do not need to be digital. They can come in all kinds of shapes and forms. the one that is the most famous in history is Babbage's difference engine. We can have electrical circuits, which are ways of building a universal classical computer. We can also have optical computers that are made from information carriers that are photons and not electrons. Biological. In many ways, we and We are walking, discussing computers. this idea of biology and biological systems as computers is currently going through a renaissance because of the notions of neural networks \cite{verdon_quantum_2019,schuld_circuit-centric_2020} and deep learning and these kinds of networks of neurons that act as computation. We also want to make a distinction between these models and some conceptual models. These are models by which we might realize computation, and these are the conceptual models that we might want to realize, one of them, the Turing machine. It is a machine that has a head and tape, and the tape has slots on it, which may have ones and zeros. it is something which has an extent to left and right, which goes off, in principle, to infinity. the tape is a kind of memory. the thing which is ostensibly doing the computation is a head that can read and write to this tape, but the only thing inside this head is a finite state machine. So, there are different states, and there are transitions between the states which happen to depend on what is read at what time and the earlier state that it used to exist. Turing machines like this come in many different flavors. There are probabilistic Turing machines, and there are universal Turing machines. So, given a certain kind of structure of a finite state machine here, we find that this Turing machine, then, can simulate any other Turing machine. Here are another model cellular automata. here, the model of computation is a world which is a grid in two dimensions in n dimensions where the point is that we have some kind of state in a local cell of this grid, and it undergoes transitions based on the state of its neighbors. we may have a local Cartesian neighborhood. we may have super Cartesian, but depending on what we are surrounded by, we change our state. we change ourselves to become empty or filled or different colors and so forth. these rules of patterns and pattern changes can give rise to computation. There is Von Neumann architecture. It is, again, a conceptual model of computation. here, the idea is to split memory from something which does arithmetic. So, we have an arithmetic logic unit, for example, some registers. Memory reads out data and feeds it into this ALU. Then the ALU feeds data back into the memory, and this is the model that is used by all processors today, including the ones on all our phones. There is DNA-based computation. This is the idea that we have strands of bases in AGCT, and then, A and T associate each other, which is called ligation. then G and C also ligate together. when we have two different strands of DNA, they will pattern match other strands in the right locations to produce base-pair ligations. this has been shown to allow a kind of computation using polymerase chain reaction tools. If we have a beaker, for example, with just one DNA strand, with PCR, we can amplify the number of DNA strands there so we can detect a certain DNA sequence. Thus, it is elegant that we can think of using such tools to do computation. we want us to be open to such models of computation, especially today, because we are starting to reconsider what it means to do computation. In many ways, we are at the end of the road of silicon today. We cannot rely on Moore's Law much longer to provide increasing scaling. That is exponential of capability and size, power, and weight to build the computers. We need to look at different physical mechanisms to build computation. that is why all these different approaches, where we represent information in different ways, is so appealing to think about because the next thing beyond silicon, could be something different very different that utilize these ideas. we might discover that it is already happening all around us, or within us, if we only know how to think about it in the right way. So, although this section is about quantum computation, ostensibly, what We want to think about is how we are also thinking about a broader question of what is the physics of computation? How do we think of physical mechanisms as doing computation? Moreover, how do we think we might be able to exploit physical mechanisms and biological mechanisms that exist to realize the computations that we want to achieve?

\section{Origins of Quantum Computing}

How long has the idea of quantum computing been around? When did it start, and what have been the key milestones in its development? Before answering these questions for quantum computing, it is worth looking back at the history of classical electronic computing and how those technologies developed over the past 100 years.

The development of classical computers did not jump directly from the vacuum tube to laptops and smartphones. Commercial demand for intermediate products throughout the 1900s incentivized companies to develop and advance the technologies that, over time, led to the ubiquitous classical computers we have today. Early examples of such ``off-ramp'' products include:

\textbf{Radar:} Before being replaced by transistors, vacuum tubes were used to modulate radar signals.

\textbf{Frequency mixers:} Some of the first research on transistors developed out of an attempt to build frequency mixers for radio receivers during World War II. It was the starting point for Bell Lab's work on transistors.

\textbf{Transistor radios:} The development of the bipolar junction transistor lead to the creation of transistor radios sold by companies like Texas Instruments, IDEA, and Sony. Unlike vacuum-tube radios, which could not output sound while the tubes were warming up, transistor radios could turn on and output sound immediately.

\textbf{Amplifiers:} Transistors were used (and are still used today) in all manner of products requiring electrical amplification, including sound speakers, hearing aids, radios, and telephones.

Along these lines, it will be challenging to sustain intense commercial interest and funding for quantum computing technology development if the first useful quantum computer is a 1,000,000-qubit fault-tolerant machine that is still 20-30 years in the future. Nearer-term commercial applications of quantum information \cite{nielsen_quantum_2011} technologies will be needed to seed and maintain the virtuous cycle of technology development needed to realize large-scale quantum computers. Some of these quantum information technologies ``off-ramp'' applications could be, for example,

\textbf{Noisy, intermediate-scale quantum simulation:} Small, error-prone quantum computers may find use in simulating small-scale quantum systems perhaps as a co-processor to a classical computer. 

\textbf{Noisy, intermediate-scale Optimization:} Noisy, error-prone quantum computers may also have applications to optimization or classification problems. For more examples of noisy, intermediate-scale quantum (NISQ) applications of quantum computing \cite{preskill_quantum_2018}.

Besides, the various components needed to build these quantum systems will likely generate new business opportunities and expand existing ones. For example, the optics, electronics, refrigeration, software, and services are likely `` dual-use'' beyond solely quantum computing. These products will benefit from the enhancements required for quantum computing and, in this enhanced state, support customers with applications beyond solely quantum computing.

Before diving into quantum computing, revisiting the history of classical electronic computing in the last century is worthwhile. We will see a few takeaways from this history that are relevant to the current and future development of quantum computing. Lee De Forest invented the first three-terminal triode vacuum tube in 1906. Vacuum tubes, much like the transistors that would later follow, are essentially faucets for electricity. The application of a small voltage on one terminal effectively opens a valve, which allows current to flow between the other two terminals. As such, vacuum tubes were used primarily as amplifiers for radio receivers, but they could also be used as on/off switches to implement logic gates. Thus, it was about 40 years after that first invention. We had the first large scale computers based on vacuum tubes, such as the electronic numerical integrator and the computer, or ENIAC, developed at the University of Pennsylvania in the 1940s. Also, around that time in 1947, the first transistor was invented at Bell Laboratories, and the first fully transistor-based computer soon followed. That computer, the transistor experimental computer number 0, or TX0, was built at MIT and Lincoln Laboratory in the mid-1950s, and it featured discrete transistors and magnetic core memory. Quite different from the computers we know and use today. Shortly after that, in 1959, the first integrated circuits using silicon were demonstrated. However, still, it was a good 20 or 30 years before we had the types of integrated circuit chips and memory chips that we now use in the computers daily. The first commercially available monolithic processor, the Intel 4004, appeared in 1971. It was a 4-bit processor, featured 2100 transistors, and clocked in at around 740 kilohertz. Within a year or two, however, Intel came out with another processor, the 8008. An 8-bit processor with nearly double the number of transistors. This doubling of the number of transistors approximately every two years was exemplary of what became known as Moore's Law. by the 1990s, following Moore's Law, the number of transistors had increased into the millions. today, we have computer chips with five billion transistors or more with multicore processors and graphical processing units, GPUs with close to 20 billion transistors. Although performance increases had previously simply followed from this Moore's law type scaling, these straightforward improvements have significantly waned over the past decade. Nonetheless, with the development of high k dielectrics, low resistance interconnects, multicore processors, 3D integration, and the like, we can expect continued improvements in the performance of classical processors for years to come. In contrast to these 100 plus years of classical computing development, quantum computing is much more recent. In the early 1980s, Richard Feynman suggested that if we want to simulate a quantum system, a task that is very hard for a classical computer, we should use a quantum system to perform that simulation \cite{feynman_simulating_1982,hey_feynman_2018}. He was suggesting we should build a quantum computer. He also noted that it is a fascinating problem, because it is not so easy, and he was right. Over the next decade or so, researchers thought about what kinds of algorithms could potentially give a quantum advantage for real-world problems of significance, and how fragile quantum states could ever be used to implement such an algorithm. The answers started in the mid-1990s, including two significant milestones in the history of quantum computing. The first was the discovery of Shor's algorithm, developed by Peter Shor \cite{shor_algorithms_1994}. Shor's algorithm was not the first quantum algorithm to show the quantum advantage. However, it was the first that also addressed an important practical problem, namely the factorization of large numbers. Now that is an important problem because the difficulty of factorization is a pillar for the present-day encryption schemes that protect the information. Essentially factoring large numbers is a very challenging problem on a classical computer, which is why it is used for public-key encryption. Peter Shor showed that factorization could be done efficiently on a quantum computer. Also, around that time, Peter Shor and his colleagues Robert Calder bank and Andrew Steane developed the first quantum error-correcting codes \cite{steane_error_1996}, which, once fully implemented, will enable quantum computers to continue to operate robustly in the presence of errors. Since then, researchers have focused on both the underlying physics, as well as the hardware that we can use to build quantum computers. Starting at the single-qubit level, researchers have explored numerous qubit modalities, from superconductors to trapped ions, semiconductors, and more. Today we have processors operating with 10 to 20 qubits and reports of 50 qubits being available soon. There is also a marked transition from prototype demonstrations in the early 2000s to where we are today, which is engineering larger and larger quantum systems. We are even now seeing examples of cloud quantum computers \cite{dumitrescu_cloud_2018} on the web that can be used by people worldwide to try out algorithms, and We will use one in this section. So, what does this all mean? Well, we think there are a couple of takeaways from this brief historical discussion. The first is that technology development takes time. It took over a century to get from the first triode vacuum tubes to the computers we have today. That development is not over. It continues today. there are many changes along the way. The right approach to building a classical computer changed many times over the years. There was no single right answer. The right technology in the 1940s was different from the 1980s, and that was also different from today. Nonetheless, in hindsight, all these steps were crucial to the overall development. Similarly, we can expect that going forward, and quantum computing will likely go through technology evolution. The best qubit modalities today are not necessarily the ones that will excel in the future. However, the observation is that in the absence of effort, we should not expect the right technology to appear if we wait long enough. Technology is developed; it is not bestowed. The road to future quantum computers, whatever they may end up looking like, is paved with the technologies we develop today. Finally, we should not underestimate the crucial role that commercialization played in the development of classical computing technologies. From the very beginning, transistors had commercial applications that generated revenue long before computers were available, including radio amplifiers, hearing aids. Although governments played a key role in seeding the development of transistor-based computers and are playing an equally crucial role in the development of quantum computers today, it was commercial development of transistors and computer chips that ultimately enabled the virtuous cycle of development that made possible Moore's law like scaling that led to the computers we use today. a major challenge for quantum computing is to identify these kinds of commercially useful applications. For qubits and small-scale quantum processors that can kickstart a similar virtuous development cycle. One that will be needed if we are to realize large scale quantum computers.

\section{How is a Quantum Computer Different}
How is a quantum computer different than a classical computer? In this section, we will compare and contrast classical and quantum computers to gain a better understanding of the unique ways in which a quantum computer represents information.

So how is a quantum computer different? We can begin to answer that question by comparing it with a classical computer. Classical computers are the computing devices that we use every day at work and home, and they process information using transistors, each of which can store one bit of information. We will call this a classical bit, which is binary, and it can take on one of two states. It can either be in state 0, let us say the absence of a voltage on the transistor gate, or it can be in state 1, the presence of a voltage on that gate. These are discrete, robust states, and when we measure the state of that transistor, we will see either a 0 or a 1, depending on where that bit was set. We can contrast that with a quantum computer, which is built from logical elements called qubits, which is short for quantum bits. A qubit is binary in the sense that it is realized using a quantum coherent two-state system, and so it can be set in state 0 or state 1, but because it is quantum mechanical, it can do much more. A qubit can also be at a quantum superposition state. It is a single state, but it carries aspects of both state 0 and state 1 simultaneously, and this is a manifestly quantum mechanical effect. We can represent a qubit state on what called a Bloch Sphere, which for this discussion, we can think of as the planet Earth, where state 0 is at the North Pole, and state 1 is at the South Pole. In this representation, a classical bit can be either at the North Pole or the South Pole, but that is it.

\begin{figure}[H] \centering{\includegraphics[scale=1.1]{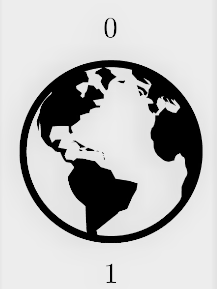}}
	\caption{The two logical states of a classical system correspond to north and south on a globe}
	\label{fig1_5}
\end{figure}

In contrast, a qubit can exist anywhere on its surface. Now, a qubit can also be at the North Pole or the South Pole. That is fine, but when it is anywhere else, the qubit is in a superposition state, again, a single state that takes on aspects of state 0 and state 1 simultaneously, as shown in the notation here. Superposition states result in probabilistic measurements, meaning that if we had said an equal superposition of 0 and 1, and we measure the qubit, we have a 50/50 chance of measuring in state 0 and a 50/50 chance of getting state 1. If we identically prepare that same state and measure it, and do that repeatedly, Half of the time, we will get a 0, and Half of the time, and We will get a 1. As a result, quantum computers rely on encoding information in fundamentally different ways than classical computers. So, on a classical computer N, classical bits represent a single N-bit state. For example, if N equals 3, we have 3 classical bits, and they can represent the state 000 or 001, all the way up to 111. There are eight different combinations, but the three classical bits can represent only one of them at a time. Thus, when we want to process the information on a classical computer, we pick one of those states as the input, we process the information, and we get a result as an output. However, if we also want to process information using a different input state, we have two choices. We can either process in parallel by using additional copies of the hardware, or with added time, we can process sequentially on the same piece of hardware. It is classical parallelism, and in both cases, we needed more resources, either more hardware or more time. The qubits in a quantum computer, on the other hand, can be set into a single superposition state that simultaneously carries aspects of all these 2 to the N components. So, for example, with three qubits, a quantum computer can represent aspects of all eight different components in a single quantum superposition state. Consequently, we have a quantum version of parallelism, and importantly, we also have quantum interference between those constituent components. Quantum parallelism and quantum interference make a quantum computer different, and in the next section, we will give examples of how they work in a quantum processor.

In transistor-based classical computers, a transistor represents a classical binary bit that can store one bit of information. In one of two distinct logical states, classical bits are found in logic state 0 or logic state 1. ``State 0'' corresponds to the transistor switch being ``off'' (e.g., no voltage is applied to the transistor gate, and so no current flows in the transistor channel), and ``state 1'' corresponds to the transistor switch being ``on'' (e.g., a voltage is applied to the gate, and so a current flows through the transistor channel). These discrete states are robust and can be measured with near certainty.

The fundamental elements of quantum computers are ``quantum bits'', typically referred to as ``qubits.'' Qubits are quantum-mechanical two-level systems. They are binary in the sense that they can be initialized in classical states 0 or 1. However, as quantum mechanical objects, qubits can also be prepared in a quantum superposition state: a single quantum state that embodies aspects of both state 0 and state 1.

Quantum superposition states are succinctly represented using Dirac notation \cite{dirac_fundamental_1925, dirac_quantum_1926,dirac_mathematical_1978,dirac_basis_1929,dirac_quantum_1929,dirac_theory_1926}. In this notation, quantum states are expressed as ``kets,'' where $\vert 0\rangle$ and $\vert 1\rangle$ represent the quantum states 0 and 1 respectively. A qubit that is in a superposition of these two states is then written as $\vert \psi \rangle =a\vert 0 \rangle +b\vert 1\rangle$. The coefficients a and b are called ``probability amplitudes,'' and they are related to the relative ``weighting'' of the two states in the superposition. An obvious special case occurs when either is zero, in which case the state$ \vert \psi \rangle$ is no longer in a superposition. For example, if $a=0$, then $\vert \psi \rangle =\vert 1\rangle$. More generally, both a and b can be complex numbers and therefore must satisfy $\vert a\vert ^{2}+\vert b\vert ^{2}=1$, a normalization to unity that ensures the ``weights'' being compared are of a standard, consistent size. This is analogous to the convention that probabilities are set to sum to 1.

Both classical and quantum bits can be visualized on a ``Bloch sphere,'' a tool which can be thought of as the planet Earth, as pictured in Fig. 5. By convention, the ``north pole'' of the sphere represents state 0, and the ``south pole'' represents state 1. A classical bit is either at the north pole or the south pole, but nowhere else. In contrast, a qubit may exist anywhere on the surface of the sphere. When the qubit state is anywhere except for the north and south poles, it is in a superposition state.

\begin{figure}[H] \centering{\includegraphics[scale=0.5]{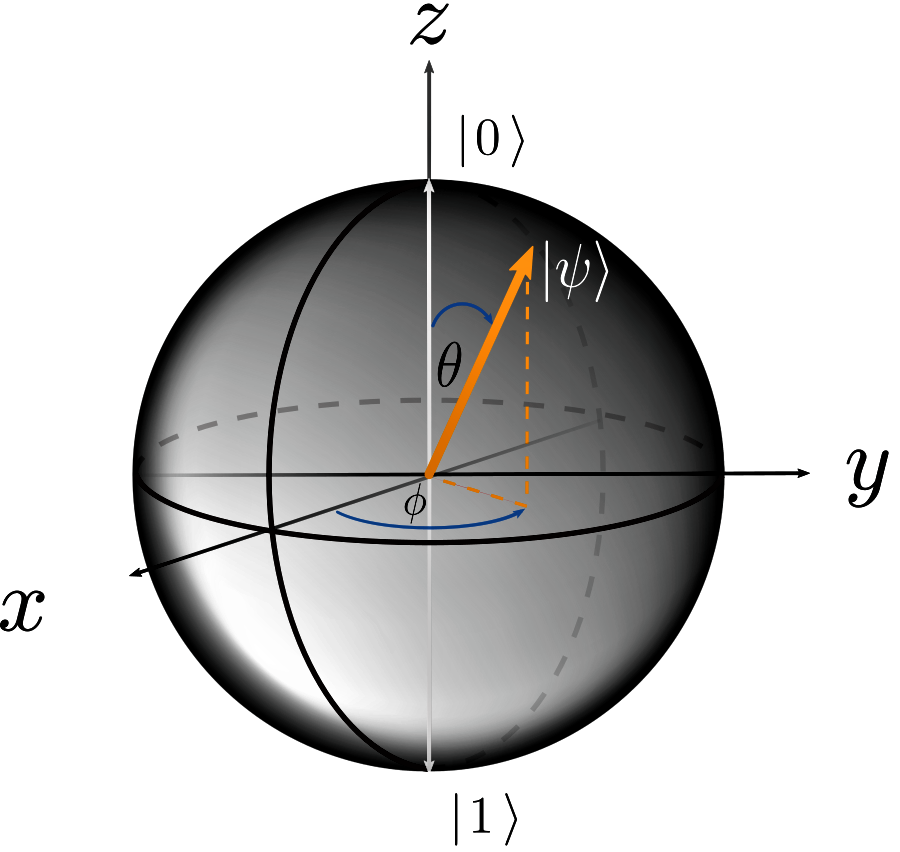}}
    \caption{The Bloch Sphere}
    \label{fig1_6}
\end{figure}

To better illuminate the connection between the state of a qubit and the Bloch sphere, the coefficients of the state $\vert \psi \rangle$ can be expressed as $\vert \psi \rangle = \cos \left( \theta/2 \right) \vert 0 \rangle + e^{i \phi } \sin \left( \theta/2 \right) \vert 1 \rangle$. It is a straightforward exercise to confirm that these coefficients satisfy the normalization condition mentioned above. The two angles $\theta$ and $ \phi $, determine the point on the surface of the Bloch sphere corresponding to the state $\vert \psi \rangle$, as pictured in Fig. 6. Intuitively, these angles relate to the globe in Fig. 5, because $\theta$ moves the state $\vert \psi \rangle$ in the north-south direction and corresponds to the qubit ``latitude,'' while $\phi$ moves the state along the east-west direction and corresponds to its ``longitude.''

Ideal projective measurement of a qubit occurs along a single axis of the Bloch sphere, for example, the z-axis (which on the globe would be the line connecting the north and south poles). It is called the measurement basis, and measurement will yield a classical result either ``state 0'' or ``state 1'' along this axis. The measurement process itself is probabilistic, and the probability of obtaining either $\vert 0\rangle$ or $\vert 1\rangle$ is related to the qubit's projection onto the measurement basis. As an example, consider when the qubit is an equal superposition of states $\vert 0\rangle$ and$ \vert 1\rangle$. It occurs whenever $\theta =\pi /2$ and corresponds to the states along the equator of the globe. In these cases, the state, when measured along the z-axis, is equally likely to result in the outcome $\vert 0\rangle$ or $\vert 1\rangle$, because their probability amplitudes are the same. Intuitively, any point on the equator when projected onto the z-axis is at the center of the Earth, equally ``far'' from the north and south pole.

As a result, quantum computers rely on encoding information in fundamentally different ways than classical computers\cite{setia_superfast_2019}. For N bits, there are $2^ N$ possible classical states. However, a classical computer can represent only one of these N-bit states at a time. Processing multiple N-bit states can either be performed sequentially in time or parallel using additional copies of the hardware. It is classical parallelism. In contrast, the qubits in a quantum computer can be set into a single superposition state that may simultaneously carry aspects of all $ 2^ N $ components. As we will see shortly, this enables two uniquely quantum mechanical effects: quantum parallelism and quantum interference.

In the context of the Bloch sphere, we have discussed about making measurements along what we call the qubits quantization axis, or the z-axis. the question was, basically, are there ways in which we can make measurements along different axes of the Bloch sphere? And the answer is yes. it depends a bit on the modality that we are discussing about. But generally speaking, we could always keep our measurement apparatus measuring along the z-axis. right before we make that measurement for example, if we wanted to measure the x-axis what we could do is a rotation that basically brings the x-axis to the z-axis and then make our measurement. so, we are still measuring along the z-axis. But by doing this rotation right before the measurement, we are effectively rotating the qubit states along x up to the z-axis and then making a measurement. so, this is a common way to do state tomography \cite{thew_qudit_2002} and process tomography measurements when It is more convenient to just have our measurement basis be just along the x-axis, just along the z-axis is to do a rotation right before measurement to measure effectively in these different bases.

Bloch Sphere, The same way that a person's position can be defined by a point on the Earth, a quantum state can be defined by a point on the Bloch sphere. While a point on the globe always refers to position, a point on the Bloch sphere refers to a qubit's state. For example, a qubit state can be spin-up (north pole), spin-down (south pole), or in a superposition of spin-up and spin-down (anywhere else). Identify the qubit states on the Bloch sphere.

\begin{figure}[H] \centering{\includegraphics[scale=.5]{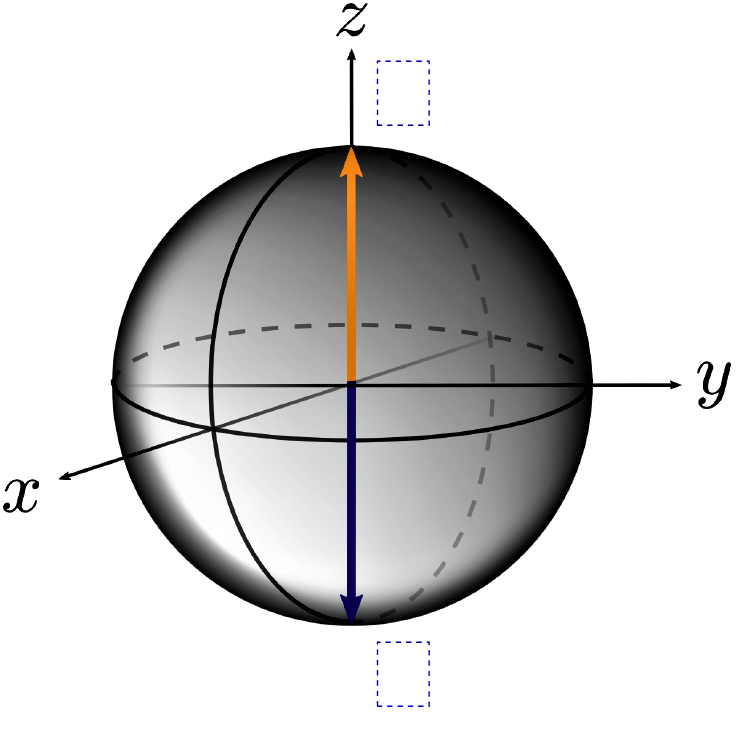}}\caption{Bloch Sphere}\label{fig1_7}
\end{figure}

On the Bloch Sphere, the z-axis runs from the south to the north pole of the sphere. Axes x-axis and y-axis are perpendicular to one another in the plane of the equator of the sphere. The north pole corresponds to state 0 (orange vector), and the south pole corresponds to state 1 (blue vector). The equator corresponds to equal superposition states. $ (|0> + |1>)/\sqrt(2) $ is at the surface of the sphere in the +x direction. $ (|0> - |1>)/\sqrt(2) $ is at the surface of the sphere in the -x-direction.

\section{Dirac Notation}

We discussed that a quantum bit qubit for short is the name given to a quantum-mechanical two-level system. The particular physical system we viewed was the spin of an electron in an atom, where the two states were spin-up and spin-down. Although qubits are realized using physical systems, it is advantageous to think of them as mathematical objects \cite{rue_mathematical_nodate}, because it will be easier to work with them using mathematics \cite{laforest_mathematics_nodate}. The approach is technology agnostic, independent of a particular physical system. In this unit, we will discuss basic concepts from linear algebra, necessary for understanding how quantum states and gates operate using Dirac notation\cite{dirac_fundamental_1925, dirac_quantum_1926,dirac_mathematical_1978,dirac_basis_1929,dirac_quantum_1929,dirac_theory_1926}. At the end of this unit, we will see the concept of measurement at a very high level; more details about it will be given in the following section.

As an introductory example, consider the state space representation of four light bulbs. In this classical system, each light bulb is a classical two-state system and can be either: OFF $\to$ state 0, or ON $\to$ state 1. It means that our classical system can be in $2^4=16$ possible configurations. Suppose that for some reason, we decide to communicate our ATM PIN (which is 1248) to our neighbor in an extremely insecure manner using light bulbs. To do this, we would first write each digit of the ATM number using binary representation, and then turn ON/OFF the lights according to the predefined state-space definition:

1=0001 $\to$ OFF OFF OFF ON,

2=0010 $\to$ OFF OFF ON OFF,

4=0100 $\to$ OFF ON OFF OFF,

8=1000 $\to$ ON OFF OFF OFF.

To send the decimal number 1, we will keep the first three bulbs OFF and the last one ON. To send the number 2, we will keep OFF the first 2 bulbs, ON the third bulb, and OFF the fourth bulb.

Quantum bits and classical bits both represent two-state systems, as in the section, qubits have unique quantum-mechanical properties. Thus, to represent the state of a qubit, people use a standard notation called Dirac notation, or ``bra''-``ket'' (read: bracket) notation. The representation uses vectors, which can then be manipulated using linear algebra concepts, such as matrix multiplication. If it has been a while, the following text and links will serve as a refresher.

1. States 0 and 1 are represented as kets $\left\vert 0\right\rangle$ and $\left\vert 1\right\rangle$ (the ket in bra-ket), and correspond to column vectors. In particular, ket $\left\vert 0\right\rangle$ and ket$ \left\vert 1\right\rangle$ are usually written as:

\begin{equation}\label{eq1_01}
\left\vert 0\right\rangle = \left(\begin{array}{c} 1\\ 0 \end{array}\right)
\end{equation}
\begin{equation}\label{eq1_02}
\left\vert 1\right\rangle = \left(\begin{array}{c} 0\\ 1 \end{array}\right)
\end{equation}

2. Bras (the bra in bra-ket) are the Hermitian conjugate of kets. Operationally, a Hermitian conjugate is found by transposing a vector (or matrix) and taking the complex conjugate of each element. Since the states $\vert 0\rangle$ and $\vert 1 \rangle$, as written above, contain only real numbers, the Hermitian conjugate is equivalent to the transpose and results in the following row vectors $\displaystyle \left\langle 0\right\vert \displaystyle = \displaystyle \left(\begin{array}{c c} 1 & 0 \end{array}\right), \displaystyle \left\langle 1\right\vert \displaystyle = \displaystyle \left(\begin{array}{c c} 0 & 1 \end{array}\right).$

The use of the Hermitian conjugate may be more evident after the next point.

3. The inner product between two states, say $\left\vert \phi \right\rangle$ and $\left\vert \psi \right\rangle$, is written as the bracket (as in, bra-ket) $\left\langle \phi \right\vert \left. \psi \right\rangle$, and in general results in a complex number. This is evident through an example. Consider the quantum state
\begin{equation}\label{eq1_03}
\begin{split}
\left\vert \psi\right\rangle & = \alpha\vert 0\rangle + \beta \vert 1\rangle\\ & =\alpha\left(\begin{array}{c} 1 \\ 0 \end{array}\right) + \beta\left(\begin{array}{c} 0 \\ 1 \end{array}\right) \\ & =\left(\begin{array}{c} \alpha\\ \beta \end{array}\right),
\end{split}
\end{equation}

3. Taking the inner product of two vectors shows that $\langle 0 \vert \psi \rangle =\alpha$ and$ \langle 1 \vert \psi \rangle =\beta$. In this example, the inner product of the general state$ \vert \psi \rangle$ with each of the basis states$ \vert 0\rangle$ and $\vert 1\rangle$ returns a number which corresponds to the ``probability amplitude'' of $ \vert \psi \rangle$ in each of those states. Hence, the Hermitian conjugate is a mathematical tool used to calculate the projection of one state onto another. Finally, it should be noted that since the inner product is a complex number in general, it can be decomposed into a product of its modulus (it is magnitude) represented as $\left\vert \left\langle \psi \right\vert \left. \phi \right\rangle \right\vert $and a phase factor, $e^{i\theta }$, where $ \theta$ is the angle between the vectors representing the states $\left\vert \psi \right\rangle$ and $\left\vert \phi \right\rangle$.

4. The norm or ``length'' of the vector representing a state $\left\vert \psi \right\rangle$ is given by the square root of the inner product:\\ $\left\vert \left\langle \psi \right\vert \left. \psi \right\rangle \right\vert =\sqrt {\left\langle \psi \right\vert \left. \psi \right\rangle }. \sqrt {\left\langle \psi \right\vert \left. \psi \right\rangle }$

5. Physical states represented in ket notation have a norm equal to one, that is $\left\langle \psi \right\vert \left. \psi \right\rangle =1$. Checking and ensuring that the norm of a state has unit value is procedure called ``normalization''. Since $\left\vert 0\right\rangle$ and $\left\vert 1\right\rangle$ are physical states with unit norm, they must also satisfy the following condition (since $ 1^2 = 1)$:
\begin{equation}\label{eq1_04}
\left\vert \left\langle 0\left\vert \right. 0 \right\rangle \right\vert =\sqrt{\left\langle 0\left\vert \right. 0 \right\rangle} , \; \;  \left\vert \left\langle 1\left\vert \right. 1 \right\rangle \right\vert =\sqrt{\left\langle 1\left\vert \right. 1 \right\rangle}.
\end{equation}

States $\left\vert 0\right\rangle$ and $\left\vert 1\right\rangle$ are orthogonal, i.e $\left\langle 0\left\vert \right.1 \right\rangle =\left\langle 1\left\vert \right. 0 \right\rangle =0.$ This means there is no projection of state $\left\vert 0\right\rangle$ on to state $\left\vert 1\right\rangle$ and visa versa. They are independent vectors, and so there is no way to write $\left\vert 0\right\rangle$ in terms of $\left\vert 1\right\rangle$ or vice versa; this is called linear independence.

When a quantum state is the sum of linearly independent states, such as $ \left\vert 0\right\rangle $ and $ \left\vert 1\right\rangle $, it is said to be in a superposition state. This is the case for the state $ \vert \psi \rangle =\alpha \vert 0\rangle + \beta \vert 1\rangle $ defined above. The coefficients $ \alpha $ and $ \beta $ are referred to as probability amplitudes and, as we have discussed, are in general complex numbers. The hermitian conjugate of $ \left\vert \psi \right\rangle $ is $ \displaystyle \left\langle \psi \right\vert \displaystyle = \displaystyle \alpha ^*\left\langle 0\right\vert +\beta ^*\left\langle 1\right\vert , \displaystyle = \displaystyle \left(\begin{array}{c c} \alpha ^* & \beta ^* \end{array}\right) $, where $  \alpha ^* $ and $ \beta ^* $ are the complex conjugates of $  \alpha  $ and $ \beta $ respectively.

To better understand the probability amplitudes $ \alpha $ and $ \beta $ of $ \vert \psi \rangle $ represent, Let us think more about the superposition concept. A light bulb is either ON or OFF, and that is it. When we look at it, or ``measure'' it, we know precisely which state it had been in and continues to be in. On the other hand, while a quantum system can certainly be in the classical states $ \left\vert 0\right\rangle$ or $ \left\vert 1\right\rangle $, it can also be in a superposition state: a single state that carries aspects of both $ \left\vert 0\right\rangle $ and $ \left\vert 1\right\rangle $. What does this mean?

Let us take a qubit prepared in the superposition state $ \vert \psi \rangle =\alpha \left\vert 0\right\rangle + \beta \left\vert 1\right\rangle $. When this qubit is measured, quantum mechanics tells us that the qubit state will be projected onto our measurement basis. In the examples in the section, we are measuring along the z-axis, that is, the axis which represents states  $ \left\vert 0\right\rangle  $ and $\left\vert 1\right\rangle $. Measurements must give us a classical result, and so any given measurement will result in one of the classical states: either state $ \left\vert 0\right\rangle $ or state $ \left\vert 1\right\rangle $. We never measure a superposition state directly. However, if we identically prepare and measure the state $ \left\vert \psi \right\rangle $ many times, we will find that we will obtain state $ \left\vert 0\right\rangle $ with probability $ \left\vert \alpha \right\vert ^2 $ and state $ \left\vert 1\right\rangle $ with probability $ \left\vert \beta \right\vert ^2 $. We call the coefficients $ \alpha $ and $ \beta $ probability amplitudes, since their magnitude squared will yield the probability that we measure their respective states. As shown in the example in (3) above, these probability amplitudes can be found by projecting the vector representing state $ \left\vert \psi \right\rangle $ onto the vectors representing the states $ \left\vert 0\right\rangle $ and $ \left\vert 1\right\rangle $.

\begin{figure}[H] \centering{\includegraphics[scale=1]{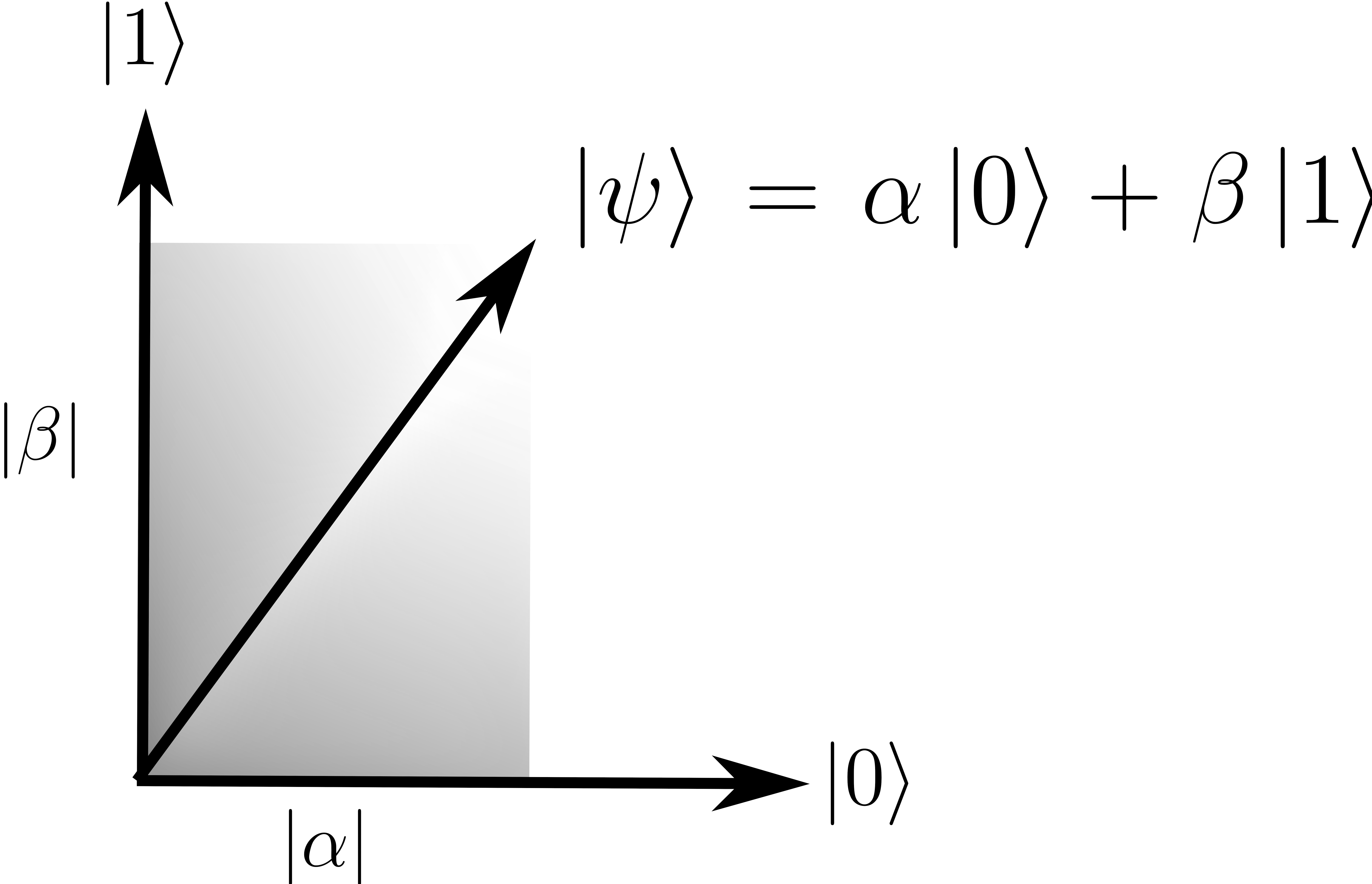}}
\caption{0 and 1 states}
\label{fig1_8}
\end{figure}

This is represented in the last figure, which shows the projection of the superposition state $ \left\vert \psi \right\rangle =\alpha \left\vert 0\right\rangle + \beta \left\vert 1\right\rangle $ on to the measurement axis corresponding to states $ \left\vert 0\right\rangle$ and $\left\vert 1\right\rangle $. Notice that the closer the state $ \left\vert \psi \right\rangle $ is to $ \left\vert 0\right\rangle $, the larger the projection $ \left\vert \alpha \right\vert $ and thus the probability $ \left\vert \alpha \right\vert ^2  $of measuring state$  \left\vert 0\right\rangle $. In fact, when $ \left\vert \psi \right\rangle $ coincides with $ \left\vert 0\right\rangle $ the value of $ \left\vert \alpha \right\vert $ becomes equal to 1, and we will measure state $ \left\vert 0\right\rangle $ with certainty (probability $ \left\vert \alpha \right\vert ^2 $ equal to 1).

To summarize, a superposition state $ \left\vert \psi \right\rangle = \alpha \left\vert 0\right\rangle + \beta \left\vert 1\right\rangle $ satisfies the normalization condition $ \left\vert \alpha \right\vert ^2+\left\vert \beta \right\vert ^2=1 $ (this ensures that the probabilities of measuring all states add to unity), and the probability of measuring the states $ \left\vert 0\right\rangle$ and $\left\vert 1\right\rangle $ are $ p(0)=\left\vert \left\langle 0 \right\vert \left.\psi \right\rangle \right\vert ^2=|\alpha |^2 $and$ p(1)=\left\vert \left\langle 1 \right\vert \left.\psi \right\rangle \right\vert ^2=|\beta |^2 $ respectively.

\section{Bloch Sphere}

The Bloch sphere is a useful tool for visualizing single-qubit states. Using Dirac notation, as we have discussed, one can write an arbitrary single-qubit state $ \vert \psi \rangle $ as
\begin{equation}\label{eq1_05}
\lvert \psi \rangle =\alpha \lvert 0 \rangle +\beta \lvert 1 \rangle ,
\end{equation}

where $ \alpha $ and $ \beta $ are the probability amplitudes and $ \vert \alpha \vert ^{2}+\vert \beta \vert ^{2}=1 $. In general, probability amplitudes are complex numbers, and can therefore always be written as the product of a real number and a complex exponential phase factor. For example, the probability amplitudes $ \alpha $ and $ \beta $ can be expressed as
\begin{equation}\label{eq1_06}
\alpha = |\alpha | (\cos \phi _{\alpha } + i\sin \phi _{\alpha }) = |\alpha | e^{i\phi _{\alpha }} \rightarrow ae^{i\phi _ a},
\end{equation}

\begin{equation}\label{eq1_07}
\beta = |\beta | (\cos \phi _{\beta } + i\sin \phi _{\beta }) = |\beta | e^{i\phi _{\beta }} \rightarrow be^{i\phi _ b},
\end{equation}

where $ a=|\alpha |$ and $ b=|\beta | $ are the magnitudes of $ \alpha $ and $ \beta, $ and $ \phi _{a}=\phi _{\alpha }=\arg (\alpha ) $ and $  \phi _{b}=\phi _{\beta }=\arg (\beta ) $ are the arguments $ \alpha $ and $ \beta $ referred to as ``phases''. Using this convention, the single-qubit state $ \vert \psi \rangle $ becomes

\begin{equation}\label{eq1_08}
\begin{split}
\lvert \psi \rangle & = ae^{i\phi _ a} \vert 0 \rangle +be^{i\phi _ b} \vert 1 \rangle \\ &
= e^{i\phi_a} \left( a\vert 0 \rangle +be^{i(\phi_b - \phi_a) }\vert 1 \rangle  \right) \\&
\equiv e^{i\phi_a} \left( a\vert 0 \rangle +be^{i\phi }\vert 1 \rangle  \right)
\end{split}
\end{equation}

where we have factored out the phase $ e^{i\phi _ a} $, referred to as the global phase, and defined a relative phase $ \phi =\phi _ b-\phi _ a $ with $ \phi $ defined from 0 to $ 2\pi $.

We do this because it is only relative phases that play a role in quantum interference or the values of physical observables based on measurements. Any phases that sit out front may be omitted without harm.

To see this explicitly, remember that the probability of measuring the state $ \vert \psi \rangle $ in another state $ \vert \mu \rangle $ is given by $ p(\mu )=\vert \langle \mu \vert \psi \rangle \vert ^{2} $. Defining $ \vert \mu \rangle =e^{i\phi _ g} \left( c\vert 0 \rangle +de^{i\phi _ r }\vert 1 \rangle \right) $, a straight forward calculation yields:

\begin{equation}\label{eq1_09}
\begin{split}
p(\mu) & =\vert \langle \mu\vert\psi\rangle\vert^{2} \\
& =\left| \left( e^{-i\phi_g} \left( c\langle 0 \vert +de^{-i\phi_r }\langle 1 \vert  \right) \right) \left(e^{i\phi_a} \left( a\lvert 0 \rangle +be^{i\phi }\lvert 1 \rangle  \right)\right)\right|^{2}\\
& =\left| e^{i(\phi_{a}-\phi_{g})}\left( c\langle 0 \vert +de^{-i\phi_r }\langle 1 \vert \right)\left( a\lvert 0 \rangle +be^{i\phi }\lvert 1 \rangle  \right)\right|^{2}\\
& =\left| e^{i(\phi_{a}-\phi_{g})}\left( ac+bd e^{i(\phi-\phi_r)} \right)\right|^{2}\\
& =\left| e^{i(\phi_{a}-\phi_{g})}\right|^{2} \left|\left( ac+bd e^{i(\phi-\phi_r)} \right)\right|^{2}\\
& = \left|\left( ac+bd e^{i(\phi-\phi_r)} \right)\right|^{2}.\\
\end{split}
\end{equation}

Two important properties of complex numbers are used in this calculation. First, for any two complex numbers w and z, it is always true that $ (wz)^{*}=w^{*}z^{*} $. And, second, $ \vert z \vert ^{2}=z^{*} z $ where $ * $ denotes the complex conjugate. The latter property is useful because it means that $ \vert e^{ix}\vert ^{2}=e^{-ix} e^{ix} =1  $ for any real x, and this leads to the removal of the global phases when calculating the measurement probability. This absence of both $ \phi _{a} $ and $  \phi _{g} $ from the final result, regardless of their value, indicates that the global phase has no physical relevance. Therefore, it is conventional to omit these phase factors from calculations.

Removing the global phase from $ \vert \psi \rangle  $reduces the number of variables needed to specify a state from four $ (a,b,\phi _{a},\phi _{b})  $to three $ (a,b,\phi) $. One further degree of freedom can be removed by directly incorporating the normalization condition $ \lvert \alpha \rvert ^2+\lvert \beta \rvert ^2=1  $into the coefficients. Following convention, this is performed by parameterizing a and b using the trigonometric functions
\begin{equation}\label{eq1_10}
\begin{split}
a & =\cos \left(\theta /2\right),\\
b & =\sin \left(\theta /2\right),
\end{split}
\end{equation}

where $ \theta $ goes from $ 0 $ to $ \pi $. The reason for selecting trigonometric functions is due to the natural geometric interpretation of the angle $ \theta $, as will be discussed in more detail below. Therefore, the state of a single qubit can be represented in complete generality by
\begin{equation}\label{eq1_11}
\lvert \psi \rangle = \cos \left(\theta /2\right)\vert 0 \rangle +\sin \left(\theta /2\right)e^{i\phi } \vert 1 \rangle
\end{equation}
which has only two free variables.

Geometrically, $ \theta $ and $ \phi $ can be mapped to a point on a sphere, referred to as the ``Bloch sphere'', using a spherical coordinate system. The angle $ \theta $ is called the ``polar angle'', and it is measured from the positive z-axis to the Bloch vector representing the state $ \lvert \psi \rangle $. The angle $ \phi $ is called the ``azimuthal angle,'' It is measured from the positive x-axis to the projection of state $ \lvert \psi \rangle $ onto the x-y plane (see the figure for the correct orientation).\\
Let us consider the polar and azimuthal angles for a few standard quantum states.\\
First, Let us consider the ``poles'' where the z axis meets the surface of the sphere, corresponding to $ \theta =0 $ and $ \theta =\pi $ and representing the states $ \lvert \psi \rangle =\vert 0\rangle $ and $ \lvert \psi \rangle = \vert 1\rangle $ respectively. Note that for $ \theta =\pi, $ corresponding to $ \lvert \psi \rangle = e^{i \phi } \lvert 1 \rangle \rightarrow \lvert 1 \rangle $, the angle $ \phi $ becomes a global phase factor and is therefore not needed.\\
$ - (\theta =0,\phi ) \rightarrow \vert 0\rangle $: this is the point where the z axis meets the north pole in the positive-z direction $ (z=+1) $. \\
$ - (\theta =\pi ,\phi ) \rightarrow \vert 1\rangle: $ this is the point where the z axis meets the south pole in the negative-z direction $ (z=-1) $.\\
Next, Let us consider the equal superposition states $ \lvert \psi \rangle = \vert 0 \rangle + e^{i\phi } \vert 1 \rangle $ on the ``equator'' in the x-y plane. These states all share $ \theta =\pi /2 $, and are uniquely identified on the equator by the angle $ \phi $. Let us further look at four specific examples as we work our way around the equator:\\
$ - (\theta =\pi /2,\phi =0) \rightarrow \frac{1}{\sqrt {2}}\left(\vert 0\rangle + \vert 1\rangle \right): $ this is the point where the x axis meets the equator in the positive-x direction $ (x=+1) $.\\
$ - (\theta =\pi /2,\phi =\pi /2) \rightarrow \frac{1}{\sqrt {2}}\left(\vert 0\rangle + i\vert 1\rangle \right): $ this is the point where the y axis meets the equator in the positive-y direction $ (y=+1) $.\\
$ - (\theta =\pi /2,\phi =\pi ) \rightarrow \frac{1}{\sqrt {2}}\left(\vert 0\rangle - \vert 1\rangle \right): $ this is the point where the x axis meets the equator in the negative-x direction $ (x=-1) $.\\
$ - (\theta =\pi /2,\phi =3\pi /2) \rightarrow \frac{1}{\sqrt {2}}\left(\vert 0\rangle - i\vert 1\rangle \right): $ this is the point where the y axis meets the equator in the negative-y direction $ (y=-1) $.

\begin{figure}[H] \centering{\includegraphics[scale=0.15]{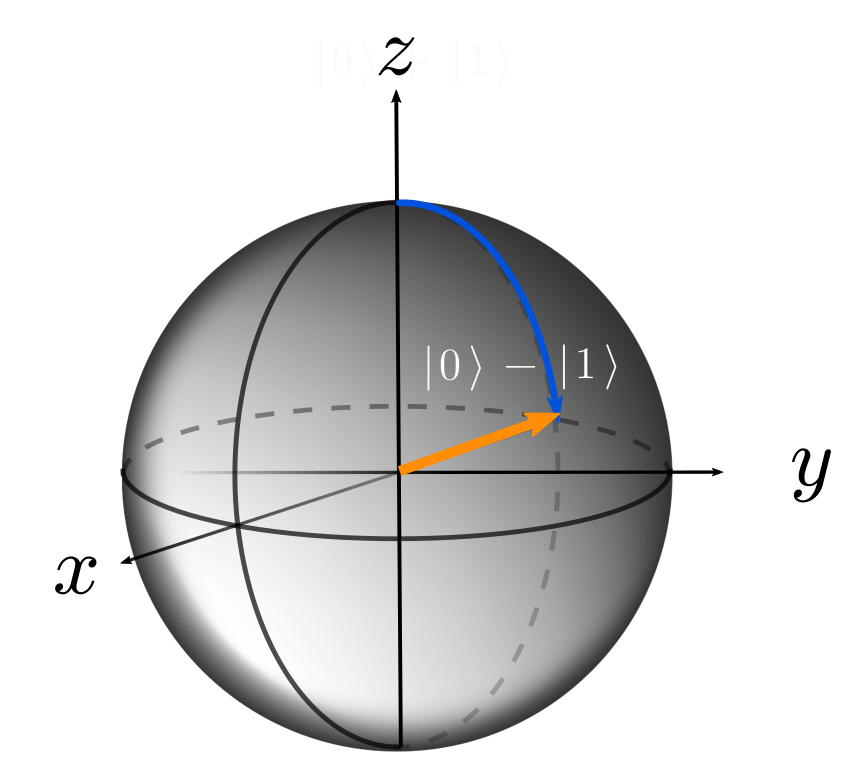}}\caption{Bloch Sphere}\label{fig1_10}
\end{figure}

\section{Quantum Parallel and Interference}

How does a quantum computer process information? How can quantum logic operations lead to quantum advantage? In this section, we will be introduced to two quantum-mechanical phenomena quantum parallelism and quantum interference that are fundamental to quantum information processing \cite{nielsen_quantum_2011,mermin_quantum_2007,national_academies_of_sciences_quantum_2018}. We will be given visceral, intuitive examples that allow us to ``see'' directly how the quantum versions of parallelism and interference efficiently manipulate the weighting coefficients probability amplitudes within a quantum superposition state, process quantum information.

Quantum parallelism and quantum interference are what make a quantum computer different than a classical computer. However, what is these quantum effects? Furthermore, how do they work in a quantum computer? To gain some insight, we will consider an example of a small quantum computer with three qubits. Here we have three atoms. Each of which has an electron with a spin.
Moreover, each of these spins can either be pointed up, which We will call spin-up, or pointed down, which We will call spin-down. We will use these three electron spins as the qubits. There are eight different spin combinations that we can have, from all three pointed up, to all three pointed down. We place these qubits in a single quantum superposition state, comprising all eight of these spin configurations. It takes eight complex numbers, C1 through C8, to specify the weighting of each of these components. The superposition state can then be represented as a state register with all eight spin configurations and their respective coefficients. Let us now imagine that we want to perform an operation that flips the spin of Atom 1. We can do this by applying an electromagnetic pulse with the right strength and the right duration, such that it rotates the spin of Atom 1 by 180 degrees. It is called a $  \pi $ pulse.
Furthermore, it acts to flip the spin. Spin-up rotates to spin-down, and spin-down rotates to spin-up. So, when we flip the spin in Atom 1, it acts to flip the spin on each of the spins in the configurations that make up the superposition state. For example, the coefficients C1 through C4, associated initially with spin-up in Atom 1, are now associated with spin-down. Similarly, the coefficients C5 through C8, associated initially with spin-down, are now associated with spin-up. It happens simultaneously across all the spin configurations that make up the quantum superposition state, even though we are performing only a single operation on a single qubit, and this is an example of quantum parallelism. Let us now look at quantum interference between these states. In this case, we will address Atom 3. We will consider a type of pulse called a $ \pi/2 $ pulse. A $ \pi/2 $ pulse does takes a spin-up and rotates it to a superposition state of up plus down. If we have taken quantum mechanics before, we remember that there is a normalization factor 1 over square root of 2 sitting in front. It maintains the length of the vector on the Bloch sphere. However, here we will omit those factors as they are not crucial for this discussion. So, a $ \pi/2 $ pulse rotates a spin-up to a superposition of up plus down. We can visualize that on the Bloch sphere. The spin-up is pointed at the North Pole, and we rotate it to the equator by rotating it $ \pi/2 $, or 90 degrees.
We will associate the direction of the vector that it is now pointing with a plus sign. Thus, the superposition state is up plus down. In the state space, for the moment, let us just look at coefficient C5. C5 is associated initially with a spin-up on Atom 3. After the $ \pi/2 $ rotation, it is now associated with both a spin-up and a spin-down. Next, let us look at what happens to spin-down. A $ \pi/2 $ pulse will rotate a spin-down pointed at the South Pole up to the equator, but now in the opposite direction. We will associate this new direction with a minus sign.
Thus, the resulting superposition state is up to minus down. In the state space, the coefficient C6, associated initially with a spin-down in Atom 3, is now associated with both up and down, but with a minus sign for the spin-down. So, we find plus 6 for spin-up and minus 6 for spin-down. So, what does it all mean? Well, if C5 equals C6, for example, then C5 minus C6 is zero.
Moreover, there is no longer any weighting to the up-up-down state. It is an example of destructive quantum interference. At the same time, there is constructive quantum interference that increases the weighting of the state with C5 plus C6.
Furthermore, this is also an example of quantum parallelism because this quantum interference process also happens simultaneously to all the other states in the register. So, quantum parallelism and quantum interference form the foundation for how a quantum computer processes information. As, with even a single gate operation, quantum parallelism and quantum interference allow us to simultaneously manipulate and change the values of the many weighting coefficients that comprise a superposition state. At a fundamental level, we can efficiently implement quantum algorithms on a quantum computer\cite{giri_review_2017}. 

Quantum parallelism and quantum interference are two quantum mechanical concepts that distinguish a quantum computer from a classical computer.\\
\textbf{Quantum Parallelism:}

Let us revisit the concept of quantum parallelism introduced in the section. We looked at three qubits; here, Let us look at two qubits.

Suppose we have two qubits, realized by two separate electrons and their associated spins. Each electron spin can either be pointed up the ``spin-up'' state $ \lvert \uparrow \rangle $ or it can be pointed down, the ``spin-down state$  \lvert \downarrow \rangle $.'' As qubits, they can also be in superpositions states of $ \lvert \uparrow \rangle $ and $ \lvert \downarrow \rangle $.

A system of $  N=2 $ spins can be found in $  2^ N=4 $ possible spin configurations. An equal superposition of these configurations results in four complex probability amplitudes (weighting factors) $ c_ i $:
\begin{equation}\label{eq1_12}
\lvert \Psi \rangle =c_1 \lvert \downarrow \downarrow \rangle +c_2 \lvert \downarrow \uparrow \rangle +c_3 \lvert \uparrow \downarrow \rangle +c_4 \lvert \uparrow \uparrow \rangle
\end{equation}

A $ \pi $-pulse applied to the first qubit (left-most spin in the bra-ket) will flip its spin. This rotation is implemented using an electromagnetic pulse with a precise amplitude and duration such that it rotates the spin by 180 degrees.
\begin{equation}\label{eq1_13}
\lvert \Psi \rangle \xrightarrow {\pi \text {-pulse on left spin}}\lvert \Psi '\rangle = c_3 \lvert \downarrow \downarrow \rangle +c_4 \lvert \downarrow \uparrow \rangle +c_1 \lvert \uparrow \downarrow \rangle +c_2 \lvert \uparrow \uparrow \rangle
\end{equation}

As we can see, a single $ \pi $-pulse on a single qubit effectively shuffles the individual probability amplitudes amongst all of the $ 2^ N=4 $ spin configurations making up a quantum superposition state. It is an example of quantum parallelism.\\

\textbf{Quantum Interference:}

Let us now explore what happens when a $ \pi/2 $-pulse applied to the second qubit. There are two cases to consider:\\

If the second qubit is in the spin-up state, a $ \pi/2 $ pulse applied along the y-axis will rotate the spin from the north pole down to the equator. This aligns the spin with the +x direction, creating the equal superposition state $ (\lvert \uparrow \rangle + \lvert \downarrow \rangle )/\sqrt {2} $ with a $ ``+" $ sign.\\

If the second qubit is instead in the spin-down state, a $ \pi/2 $ pulse applied along the y-axis will rotate the spin in the same counter-clockwise direction, bringing it from the south pole up to the equator. This aligns the spin in the x-direction, creating the equal superposition state $ (\lvert \uparrow \rangle - \lvert \downarrow \rangle )/\sqrt {2}, $ this time with a corresponding $ ``-" $ sign.

The resulting state is:
\begin{equation}\label{eq1_14}
\begin{split}
\lvert \Psi \rangle \xrightarrow {\pi \text {/2-pulse}}\lvert \Psi ''\rangle =  \\ & \frac{1}{\sqrt {2}}(c_2-c_1) \lvert \downarrow \downarrow \rangle +\frac{1}{\sqrt {2}}(c_2+c_1) \lvert \downarrow \uparrow \rangle \\& +\frac{1}{\sqrt {2}}(c_4-c_3) \lvert \uparrow \downarrow \rangle +\frac{1}{\sqrt {2}}(c_4+c_3) \lvert \uparrow \uparrow \rangle
\end{split}
\end{equation}

The probability amplitudes now add and subtract one another. Suppose there are two coefficients with equal values, for example, $ c_3=c_4 $. In this case, there is a complete cancellation of the probability amplitude for $ \lvert \uparrow \downarrow \rangle.$ Such a reduction of the probability amplitude is called ``destructive quantum interference.'' On the other hand, there is a doubling of the probability amplitude in front of $  \lvert \uparrow \uparrow \rangle. $ Such an enhancement of the probability amplitude is called ``constructive quantum interference.'' Furthermore, since the constructive and destructive quantum interference happens across the entire state space, this is also an example of quantum parallelism.

\section{Quantum Gates} 

What are quantum logic gates? How are they visualized? In this section, we will discuss the single-qubit and two-qubit gates. We will see an example of each X gate and the CNOT gate and contrast them with their classical analogs. A small set of such single-qubit and two-qubit gates forms a universal gate set that can be used to implement any algorithm on a circuit-model quantum computer.

Classical computers can perform arbitrary Boolean logic with a small set of single-bit and two-bit gates. For example, the NOT gate, combined with the AND gate, is considered universal in that it can, in principle, implement any classical algorithm that uses binary logic. Similarly, quantum algorithms can be run on quantum computers using a small, universal set of single and two-qubit gates. Let us begin with an example of a single-qubit gate called the X-gate and its classical analog, the NOT gate. A NOT gate takes one bit as its input, and it inverts it. For example, a 0 at the input is inverted to state 1, and a 1 at the input is inverted to state 0. The quantum analog of this gate is called the X-gate, which takes a quantum state as its input, in this case, state 0, and rotates it to state 1 or it takes state 1 and rotates it to state 0. We can represent this operation on the Bloch Sphere with state 0 at the North Pole and state 1 at the South Pole. We see an envelope of the pulse that we are using to drive the X-gate. We call this a $ \pi $ pulse because it will rotate the Bloch vector representing the qubit's state from the North Pole to the South Pole, a rotation of 180 degrees. The red arrow that comes in and out of the Bloch Sphere screen represents the envelope of the pulse that we are using to drive this operation.
Moreover, visually, much like the spokes of an umbrella will rotate around the umbrella's central axis when we twist its handle. The Bloch vector rotates around the axis to which we are applying the pulse. In this case, since this pulse is applied along the x-axis of the Bloch Sphere, therefore we call the operation an X-gate. Now, as we show it here, we are rotating from the North Pole to the South Pole from state 0 to state 1. In this configuration, this is simply a classical operation, but the X-gate can do much more. The X-gate can take as its input any superposition state, that is any starting point on the Bloch Sphere and rotate it around the x-axis by 180 degrees. It is a quantum mechanical operation.
The qubit starts in a superposition state, and it ends in a superposition state. What the X-gate essentially does is take the input quantum state, $ \alpha $ 0 plus $ \beta $ 1, and swaps the coefficients to generate an output state, $ \beta $ 0 plus $ \alpha $ 1. The X-gate is one of a handful of standard single-qubit gates that rotate the qubit state around a few different axes on the Bloch Sphere. Let us now consider an example of a two-qubit gate, and the controlled-NOT gate, or CNOT-gate, and its classical analog, the exclusive OR, or the XOR-gate. XOR takes two bits as inputs, bit x, and y. We will call bit x the control bit and bit y the target bit. The truth table for XOR shows the output states of all four possible input state combinations. For example, when the control bit x is in its 0 states, the target bit y, whether a 0 or a 1, remains unchanged, and its value is just passed to the output. However, when control bit x is set to state 1, the target bit y is inverted. 0 becomes 1, and 1 becomes 0. The quantum analog of this gate is the CNOT-gate, and it takes as inputs qubit x and qubit y. Again, we will call x the control bit and y the target bit. When qubit x is in state 0, qubit y remains unchanged. When qubit x is in state 1, qubit y undergoes a $ \pi $ rotation, a rotation of 180 degrees. It means that the rotation of qubit y depends on the state of qubit x. We can consider an interesting example where x, the control qubit, is in an equal superposition of 0 and 1, and y, the target qubit, is in state 0. To determine the output, let us take it one piece at a time. When x is in state 0, y remains unchanged. When x is in state 1, the qubit y undergoes a rotation of 180 degrees that flips state 0 to state 1. The resulting output state, 0, 0, plus 1, 1, is a fascinating state because it cannot be factorized into an x component cross a y component. This type of state is entangled, and an entangled state is a manifestly quantum mechanical state. Universal quantum computation can be built from a small subset of these types of single and two-qubit gates \cite{kitaev_classical_2002}. A universal gate set allows us to perform any type of quantum algorithm on a gate model quantum computer\cite{de_ridder_quantum_2019}.

Classical computers perform arbitrary Boolean logic with a small set of single-bit and two-bit logic gates. The NOT gate (a single-bit gate) in conjunction with the AND gate (a two-bit gate) are together one example of a universal gate set. Such a universal gate set can, in principle, implement any arbitrary classical algorithm based on boolean logic.

The quantum analogue of a NOT gate is the X-gate. A NOT gate inverts its input  (NOT 0 $\rightarrow$ 1, and NOT 1 $\rightarrow$  0) . Similarly, the X-gate would swap states $ \lvert 0\rangle$ and $ \lvert 1\rangle $. The swap can be visualized on the Bloch sphere as a 180-degree rotation around the x-axis (this is why it is called an  ``X-gate'' ). Because the rotation is 180-degrees, the signal we send to the qubit in order to perform an X-gate is referred to as a $ \pi $-pulse. In general, the X-gate can be applied to any arbitrary quantum superposition state, and it acts to swap the probability amplitudes on the states $ \lvert 0\rangle $ and $ \lvert 1\rangle $:
\begin{equation}\label{eq1_15}
\frac{1}{\sqrt {2}}(\alpha \lvert 0\rangle + \beta \lvert 1\rangle ) \xrightarrow {\text {X gate}} \frac{1}{\sqrt {2}}(\beta \lvert 0\rangle + \alpha \lvert 1\rangle )
\end{equation}

In addition to the X-gate, one may rotate the qubit state around the y-axis or the z-axis. A $ \pi $-rotation around the y-axis is called a Y-gate, and a $ \pi $-rotation around the z-axis is called a Z-gate.

Not all classical gates have a direct quantum analog. It is because quantum circuits must be reversible \cite{pednault_breaking_2018}. In a reversible circuit, one can precisely reconstruct the input state(s) given the output state(s). For example, a NOT gate is reversible, because, given the output 0/1, we know the input was 1/0. However, the AND gate is not reversible, because, for example, given the output 0, we cannot tell if the input had been 00, 01, or 10. The reason quantum circuits have to be reversible has to do with the fact that coherent quantum states ideally always undergo unitary evolution. So a quantum gate can be ``undone'' by applying the inverse of that unitary evolution.

There is a quantum analog to the classical exclusive-OR gate (abbreviated XOR), called the controlled-NOT gate (abbreviated CNOT). The CNOT gate is a ``conditional gate'' comprising two qubits: a ``control qubit'' and a ``target qubit''. When the control qubit is in state $ \lvert 0\rangle $, the target qubit remains unchanged. However, when the control qubit is in state $ \lvert 1\rangle $, an X-gate is applied to the target qubit: it undergoes a $ \pi $-rotation around the x-axis of the Bloch sphere.

Consider an interesting example where the control qubit is in an equal superposition of $ \lvert 0\rangle $ and$  \lvert 1\rangle $, and the target qubit is in state$  \lvert 0\rangle $:
\begin{equation}\label{eq1_16}
\frac{1}{\sqrt {2}}(\lvert 0\rangle +\lvert 1\rangle ) \otimes \lvert 0\rangle \xrightarrow {\text {CNOT}} \frac{1}{\sqrt {2}}(\lvert 0\rangle \otimes \lvert 0\rangle +\lvert 1\rangle \otimes \lvert 1\rangle ).
\end{equation}

(Note: We are here introducing ``tensor notation,'' \cite{orus_practical_2014} where $ \lvert x\rangle \otimes \lvert y \rangle $ indicates a two-qubit state with the first qubit in-state $ \lvert x \rangle, $ and the second qubit in-state $ \lvert y \rangle $. Occasionally, we may drop the $ \otimes $ and write a two-quit state as simply$  \lvert x\rangle \lvert y \rangle. $) The resulting output state is remarkable because it can be no longer separable into two single-qubit components such as $ (\ldots )_ x \otimes (\ldots )_ y $. It is known as an entangled state, and it is a manifestly quantum mechanical state.

Universal quantum computation can be built from a small subset of these types of single and two-qubit gates. A universal gate set enables us to perform any type of algorithm quantum or classical on a gate model quantum computer. However, universality does not imply quantum advantage. Many algorithms can be implemented on a quantum computer in principle but feature no quantum advantage.

\section{List of Classical Gates}
Classical computing is performed using a universal set of Boolean logic gates. The table below introduces several examples of single-bit and two-bit gates. Subsets of these gates form a universal gate set. For example, the NOT gate and the AND gate together can implement any Boolean logic function.
\begin{figure}[H] \centering{\includegraphics[scale=0.25]{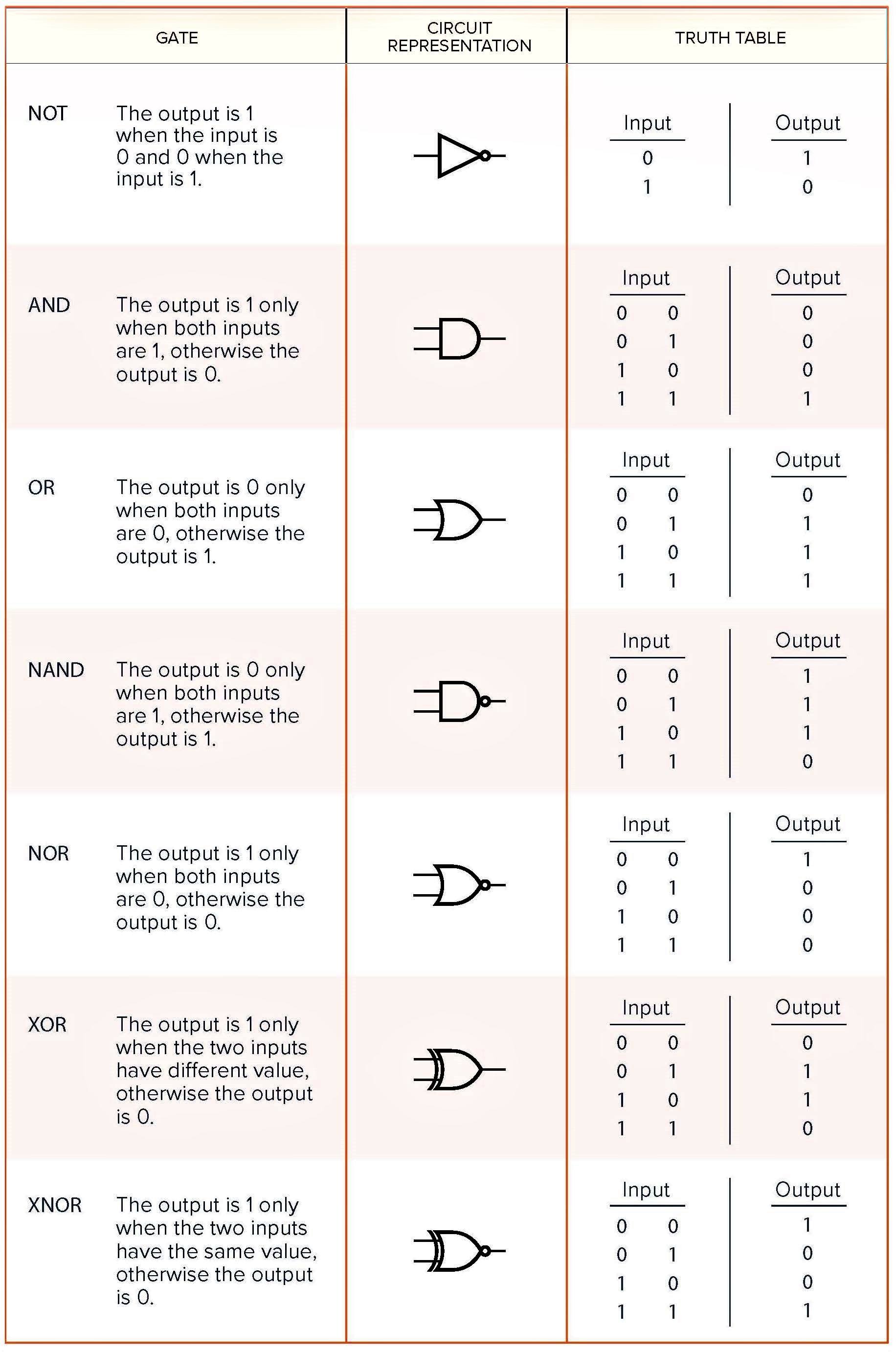}}
    \caption{Classical Gates}
    \label{fig1_4}
\end{figure}

The table above comprises a set of single-bit and two-bit Boolean logic gates. Logic gates perform Boolean functions on the inputs to yield output. The second column contains the graphical symbols used to represent the gates introduced in column one, and the third column contains the truth tables for these gates. Truth tables comprise all possible combinations of input states, $ 2^1 $ for single-bit gates and $ 2^2 = 4 $ for two-bit gates, and the corresponding output state.

The most basic Boolean logic gate is the NOT gate, a single-bit gate, which inverts the input bit, logic $ \rightarrow $ logic 1, and vice versa. Single-bit gates alone are insufficient to form a universal gate set; universal computation requires some form of logical operation on multiple bits, such as a two-bit gate.

The table presents six two-bit Boolean logic gates. The AND gate outputs a logic 1 only if the two inputs are both in the logic 1 state. A NAND gate is an inverted AND gate, that is, an AND gate followed by a NOT gate. The OR gate outputs logic state 1 if at least one input bit is 1. The NOR gate is an OR followed by a NOT gate. The exclusive OR, the XOR gate, outputs a logic 1 only if the two input bits differ. An inverted XOR gate, an XOR gate followed by a NOT gate, is called an XNOR gate and outputs a logic 1 if the two input bits are the same.

As these six two-bit gates foretell, gates are not unique and can be generated using combinations of other single-bit and two-bit gates.

\section{Single-Qubit Gates}

Universal quantum computing is performed using a small set of single-qubit and two-qubit gates. The table below introduces several single-qubit gates used in the circuit model of quantum computation.
\begin{figure}[H] \centering{\includegraphics[scale=0.48]{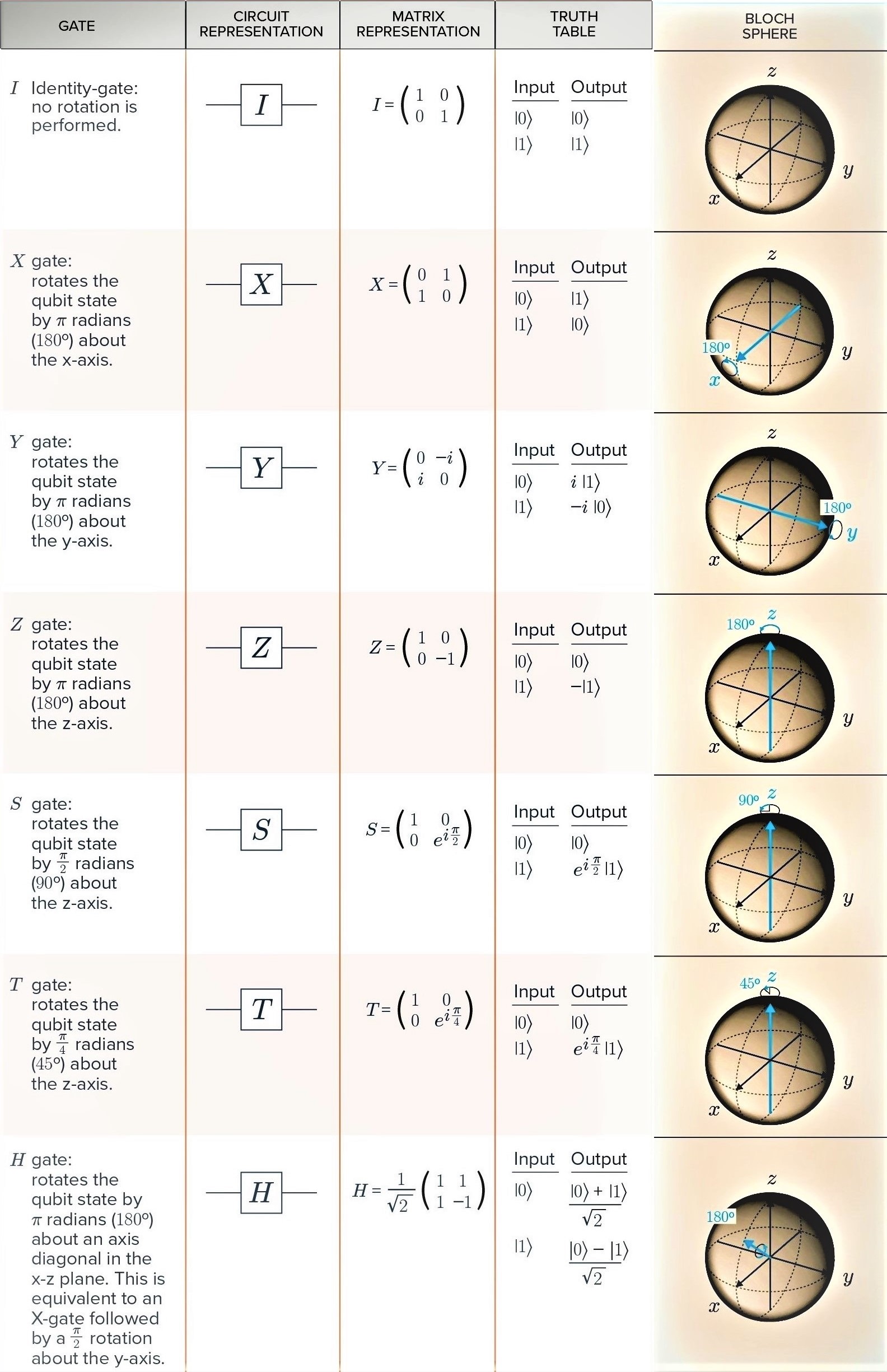}}
    \caption{Single Qubit}
    \label{fig1_13}
\end{figure}

The first column of the table introduces the gate and a short definition of its action. The second column shows the graphical representation of each gate, and the third column shows the corresponding matrix representation. The fourth column provides the truth table for the input states $ \lvert 0\rangle $ and $ \lvert 1\rangle $. The last column indicates the rotation on the Bloch sphere that the gate performs. We will walk through the I, X, Z, and Y gates to provide a sense of how single-qubit gates work, in general.

The first gate is ``identity'', an operation that does not alter the input state and is used to represent a lossless quantum channel\cite{shor_capacities_2003}. It is represented by the identity matrix.
\begin{equation}\label{eq1_17}
I\lvert 0 \rangle = \left(\begin{array}{c c} 1 & 0 \\ 0 & 1 \end{array}\right) \left(\begin{array}{c}1\\ 0\end{array}\right)= \left(\begin{array}{c}1\\ 0\end{array}\right) = \lvert 0 \rangle
\end{equation}

\begin{equation}\label{eq1_18}
I\lvert 1 \rangle = \left(\begin{array}{c c} 1 & 0 \\ 0 & 1 \end{array}\right) \left(\begin{array}{c}0\\ 1\end{array}\right)= \left(\begin{array}{c}0\\ 1\end{array}\right) = \lvert 1 \rangle
\end{equation}

There are two items to note here. First, we use ``operator notation'' to mathematically represent the application of a gate operation on a qubit state. For example, the identity operation applied to qubit state $ \lvert 0 \rangle $ is written: $ I\lvert 0 \rangle $. we may sometimes see a ``hat'' added to an operator, e.g., $ \hat{I} $, to make its role clear. The corresponding matrix and vector can then replace the operator and state, respectively, to calculate the gate's action. Second, we note that the identity operator leaves the state of the qubit unchanged.

The ``X-gate'' can be visualized as performing a rotation around the x-axis on the Bloch sphere. As discussed previously, the X-gate is the quantum analog of the classical NOT gate. See below how the X-gate acts on states $ \lvert 0 \rangle $ and $ \lvert 1 \rangle: $
\begin{equation}\label{eq1_19}
X\lvert 0 \rangle = \left(\begin{array}{c c} 0 & 1 \\ 1 & 0 \end{array}\right) \left(\begin{array}{c}1\\ 0\end{array}\right)= \left(\begin{array}{c}0\\ 1\end{array}\right) = \lvert 1 \rangle
\end{equation}

\begin{equation}\label{eq1_20}
X\lvert 1 \rangle = \left(\begin{array}{c c} 0 & 1 \\ 1 & 0 \end{array}\right) \left(\begin{array}{c}0\\ 1\end{array}\right)= \left(\begin{array}{c}1\\ 0\end{array}\right) = \lvert 0 \rangle .
\end{equation}

The X-gate is said to perform a ``bit flip'', because the qubit states $ \lvert 0 \rangle $ and $ \lvert 1 \rangle $ are ``flipped'' to$  \lvert 1 \rangle  $and $ \lvert 0 \rangle $ respectively. As in the section, more generally, the X-gate swaps the probability amplitudes in a quantum state: X $ (\alpha \lvert 0 \rangle + \beta \lvert 1 \rangle ) \rightarrow \beta \lvert 0 \rangle + \alpha \lvert 1 \rangle $.

The Z-gate ``adds a phase'' of -1 to the $ \lvert 1 \rangle $ state and leaves $ \lvert 0 \rangle $ unchanged.
\begin{equation}\label{eq1_21}
Z\lvert 0 \rangle = \left(\begin{array}{c c} 1 & 0 \\ 0 & -1 \end{array}\right) \left(\begin{array}{c}1\\ 0\end{array}\right)= \left(\begin{array}{c}1\\ 0\end{array}\right) = \lvert 0 \rangle
\end{equation}

\begin{equation}\label{eq1_22}
Z\lvert 1 \rangle = \left(\begin{array}{c c} 1 & 0 \\ 0 & -1 \end{array}\right) \left(\begin{array}{c}0\\ 1\end{array}\right)= -\left(\begin{array}{c}0\\ 1\end{array}\right) = -\lvert 1 \rangle
\end{equation}

Although the net result is to multiple that state $ \lvert 1 \rangle $ by -1, the language ``adds a phase'' is often used. The terminology arises from the fact that the Z-gate will rotate a state about the Z-axis by $ \pi $ radians (180 degrees). Phases are additive, and so applying a $ \pi $ phase shift to a starting phase $ \phi _0 $ results in $ e^{i (\phi _0 + \pi )} = e^{i \phi}  e^{i \pi } = -e^{i \phi } $. Thus, ``adding a phase'' in the exponent leads to a factor $ e^{i\pi} = -1. $

The Y-gate is a rotation around the y axis by $  \pi $ radians (180 degrees). This operation may also be written in terms of the X-gate, the Z-gate, and a global phase. That is, the Y-gate may be viewed as performing both a bit flip and a phase flip, as well as introducing a global phase factor of i:
\begin{equation}\label{eq1_23}
Y = iXZ = i\left(\begin{array}{c c} 0 & 1 \\ 1 & 0 \end{array}\right) \left(\begin{array}{c c} 1 & 0 \\ 0 & -1 \end{array}\right) = \left(\begin{array}{c c} 0 & -i \\ i & 0 \end{array}\right)
\end{equation}

Applying the Y-gate to qubit states $ \lvert 0 \rangle $ and $ \lvert 1 \rangle $ yields:
\begin{equation}\label{eq1_24}
Y\lvert 0 \rangle = \left(\begin{array}{c c} 0 & -i \\ i & 0 \end{array}\right) \left(\begin{array}{c}1\\ 0\end{array}\right)= \left(\begin{array}{c}0\\ i\end{array}\right) = i\lvert 1 \rangle
\end{equation}

\begin{equation}\label{eq1_25}
Y\lvert 1 \rangle = \left(\begin{array}{c c} 0 & -i \\ i & 0 \end{array}\right) \left(\begin{array}{c}0\\ 1\end{array}\right)= -\left(\begin{array}{c}i\\ 0\end{array}\right) = -i\lvert 0 \rangle
\end{equation}

Rotations around the z-axis by an arbitrary angle $  \phi  $ causes state $ \lvert 1 \rangle $ to acquire a phase $ e^{i\phi }  $ and leaves state $ \lvert 0 \rangle $ unaffected. Rotations around the z-axis by the specific angles $ \pi/2 $ and $ \pi/4 $ are referred to as the S-gate and T-gate, respectively.

Finally, the Hadamard gate, or H-gate for short, induces a $  \pi $ rotation around an axis exactly in between the x-axis and z-axis of the Bloch sphere. For example, it takes states located on the z-axis and rotates them on to the x-axis, creating equal superposition states: $ H\lvert 0 \rangle = \frac{1}{\sqrt {2}}(\lvert 0 \rangle + \lvert 1 \rangle$ and $H \lvert 1 \rangle = \frac{1}{\sqrt {2}}(\lvert 0\rangle - \lvert 1 \rangle. $

\section{Two-Qubit Gates}

Two-qubit gates are unitary quantum operations on two qubits. When these gates cannot be written as the product of two single-qubit gates, they are called ``entangling gates''. They may also be ``controlled'' or ``conditional'' gates. The word controlled arises from the conditional implementation of these gates, in which a unitary operation $ U $ is conditionally applied to a target qubit depending on the state of the control qubit. A two-qubit controlled gate, denoted here as $ U_{A, B}^ c,  $with the first qubit as control (A), and the second qubit as target (B) can be written in Dirac notation as
\begin{equation}\label{eq1_26}
U_{A,B}^ c = \lvert 0\rangle \langle 0\rvert_{A} \otimes I_ B+\lvert 1\rangle \langle 1\rvert _{A}\otimes U_{B},
\end{equation}

Where $ I_ B $ and $ U_ B $ are single-qubit operations applied to qubit B, depending on the state of qubit A. If qubit A is in state $ \lvert 0 \rangle $, then the identity operation is applied to qubit B, and its state remains unchanged. However, when qubit A is in state $ \lvert 1 \rangle $, the unitary single-qubit operation $ U_ B $ is applied to qubit B:

\begin{equation}\label{eq1_27}
\lvert 0\rangle _ A \otimes \lvert 0\rangle _ B  \to  \lvert 0\rangle _ A \otimes I_ B\lvert 0\rangle _ B,
\end{equation}

\begin{equation}\label{eq1_28}
\lvert 0\rangle _ A \otimes \lvert 1\rangle _ B  \to  \lvert 0\rangle _ A \otimes I_ B\lvert 1\rangle _ B,
\end{equation}

\begin{equation}\label{eq1_29}
\lvert 1\rangle _ A \otimes \lvert 0\rangle _ B  \to  \lvert 1\rangle _ A \otimes U_ B\lvert 0\rangle _ B,
\end{equation}

\begin{equation}\label{eq1_30}
\lvert 1\rangle _ A \otimes \lvert 1\rangle _ B  \to  \lvert 1\rangle _ A \otimes U_ B\lvert 1\rangle _ B.
\end{equation}
The specific operation $ U_ B $ depends on the type of two-qubit gate.

In quantum circuits, a controlled gate is indicated by a vertical line that connects two qubits: a solid black circle indicates the control qubit, and the target qubit is indicated by a symbol representing a single-qubit unitary operation U that is conditionally applied, as shown in the circuit below.

\begin{figure}[H] \centering{\includegraphics[scale=3]{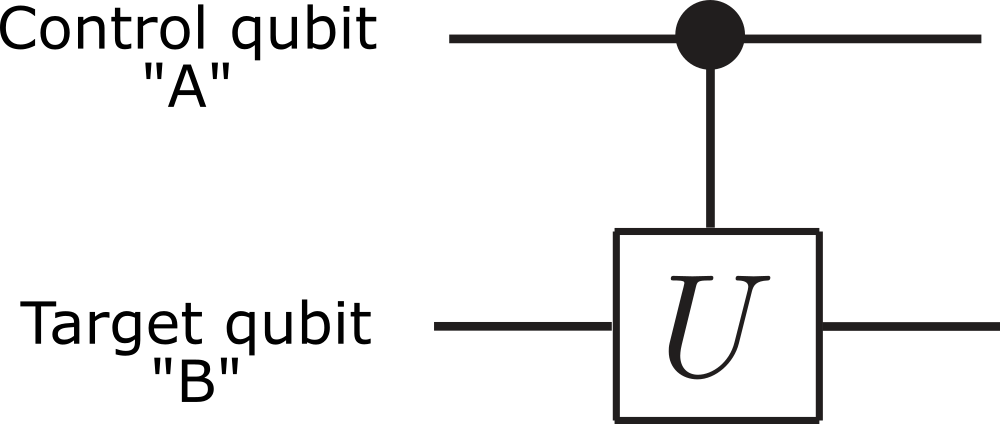}}
\caption{control unitary AB}
\label{fig1_16}
\end{figure}

Note that the roles of target and control qubit may be swapped \cite{paler_influence_2019}. A controlled unitary, $ U_{B,A}^ c $, with the B as the control qubit and A as the target qubit is denoted:
\begin{equation}\label{eq1_31}
U_{B,A}^ c = I_ A\otimes \lvert 0\rangle \langle 0\rvert _{B} +U_ A\otimes \lvert 1\rangle \langle 1\rvert _{B},
\end{equation}

Where $ I_ A $ represents the identity gate operating on qubit A, and $ U_ A  $ is the single-qubit gate that operates conditionally on qubit A. The quantum circuit of this controlled gate is shown in the following figure.

\begin{figure}[H] \centering{\includegraphics[scale=4]{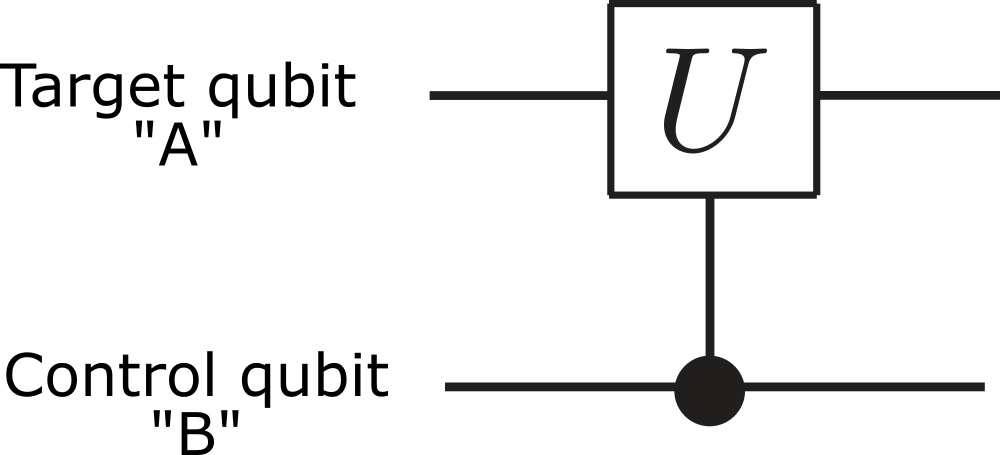}}
    \caption{control unitary BA}
    \label{fig1_17}
\end{figure}

Two common examples of conditional gates are the controlled-phase gate referred to as the CZ-gate and the controlled-NOT gate most commonly referred to as the CNOT-gate (it could also be called a CX-gate). Both gates take their name from the single-qubit operation U that is conditionally applied to the target qubit.

In the case of the controlled-phase gate, the conditional single-qubit operation is the Z-gate. The word ``phase'' refers to the phase factor $ (\exp {i \pi }=-1) $ that results from an application of the Z-gate on state $ \lvert 1 \rangle $,
\begin{equation}\label{eq1_32}
\begin{split}
Z\lvert 0\rangle & = \lvert 0\rangle, \\ 
Z\lvert 1\rangle & = -\lvert 1\rangle.
\end{split}
\end{equation}

If qubit A is the control and B is the target, the controlled-phase gate can be written in Dirac notation as
\begin{equation}\label{eq1_33}
Z_{A,B}^ c = \lvert 0\rangle \langle 0\rvert _{A} \otimes I_ B+\lvert 1\rangle \langle 1\rvert _{A}\otimes Z_{B},
\end{equation}

and its quantum circuit is given by

\begin{figure}[H] \centering{\includegraphics[scale=1]{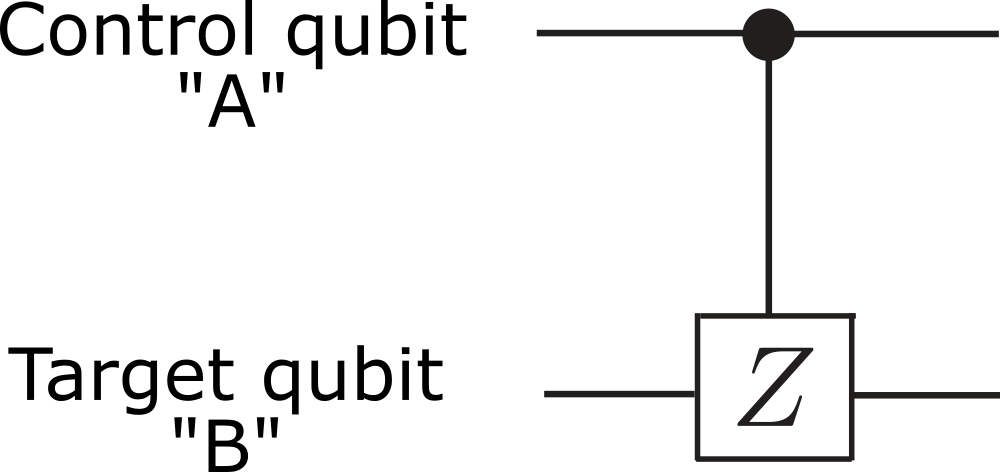}}\caption{CZ gate}\label{fig1_18}
\end{figure}

In the case of the CNOT gate, the single-qubit operation is the X-gate. The word ``NOT'' comes from the effect of the X-gate on the state $ \lvert 0\rangle $ and $ \lvert 1\rangle $,

\begin{equation}\label{eq1_33_1}
\begin{split}
X\lvert 0\rangle & = \lvert 1\rangle, \\ 
X\lvert 1\rangle & = \lvert 0\rangle.
\end{split}
\end{equation}

If qubit A is the control and B is the target, CNOT operator and its corresponding quantum circuit are:
\begin{equation}\label{eq1_34}
CNOT_{A,B} = \lvert 0\rangle \langle 0\rvert _{A} \otimes I_ B+\lvert 1\rangle \langle 1\rvert _{A}\otimes X_{B}.
\end{equation}

\begin{figure}[H] \centering{\includegraphics[scale=4]{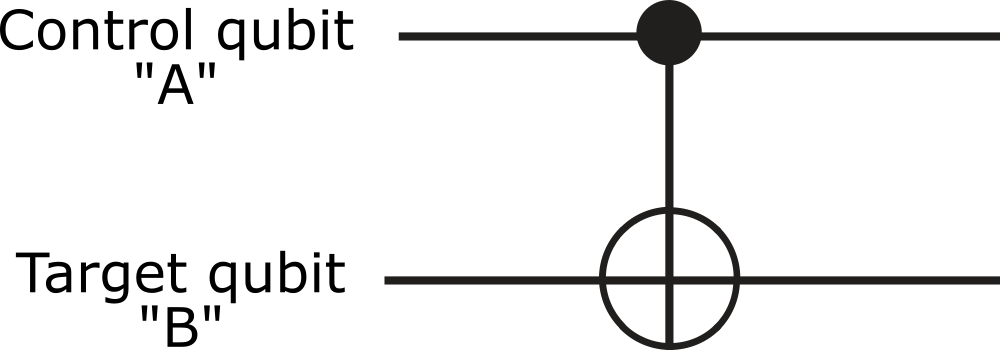}}\caption{CNOT AB}\label{fig1_19}
\end{figure}

Similarly, if the role of the control and target qubits are interchanged, the CNOT operator and its quantum circuit are:
\begin{equation}\label{eq1_35}
CNOT_{B,A} = I_ A \otimes \lvert 0\rangle \langle 0\rvert _{B} +X_{A}\otimes \lvert 1\rangle \langle 1\rvert _{B}.
\end{equation}

\begin{figure}[H] \centering{\includegraphics[scale=4]{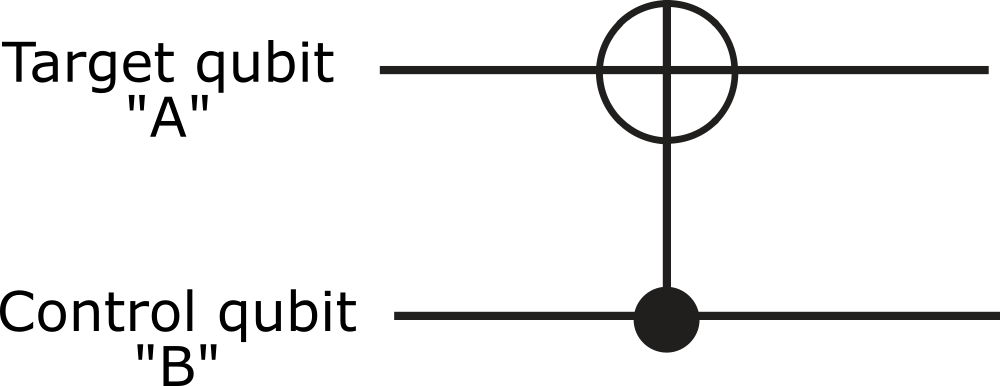}}\caption{CNOT BA}\label{fig1_20}
\end{figure}

A CNOT gate can be implemented using a Z-gate by applying a Hadamard gate, H, before and after a Z-gate, since X=HZH. The figures below represent this identity.

\begin{figure}[H] \centering{\includegraphics[scale=.6]{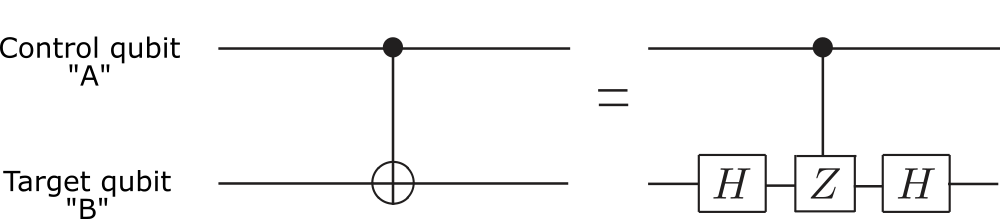}}\caption{CNOT CZ Identity}\label{fig1_21}
\end{figure}

The table below shows a summary of the two controlled-gates CZ and CNOT. Note that the sub-indexes A and B in $ cZ_{A, B} $ are usually omitted. In their absence, it is generally assumed that the first qubit is the ``control qubit,'' and the second qubit is the ``target qubit.'' It is not only valid for the controlled-phase gate (CZ) but also the controlled-unitary gate (cU) and the controlled-NOT gate (CNOT).

\begin{figure}[H] \centering{\includegraphics[scale=.5]{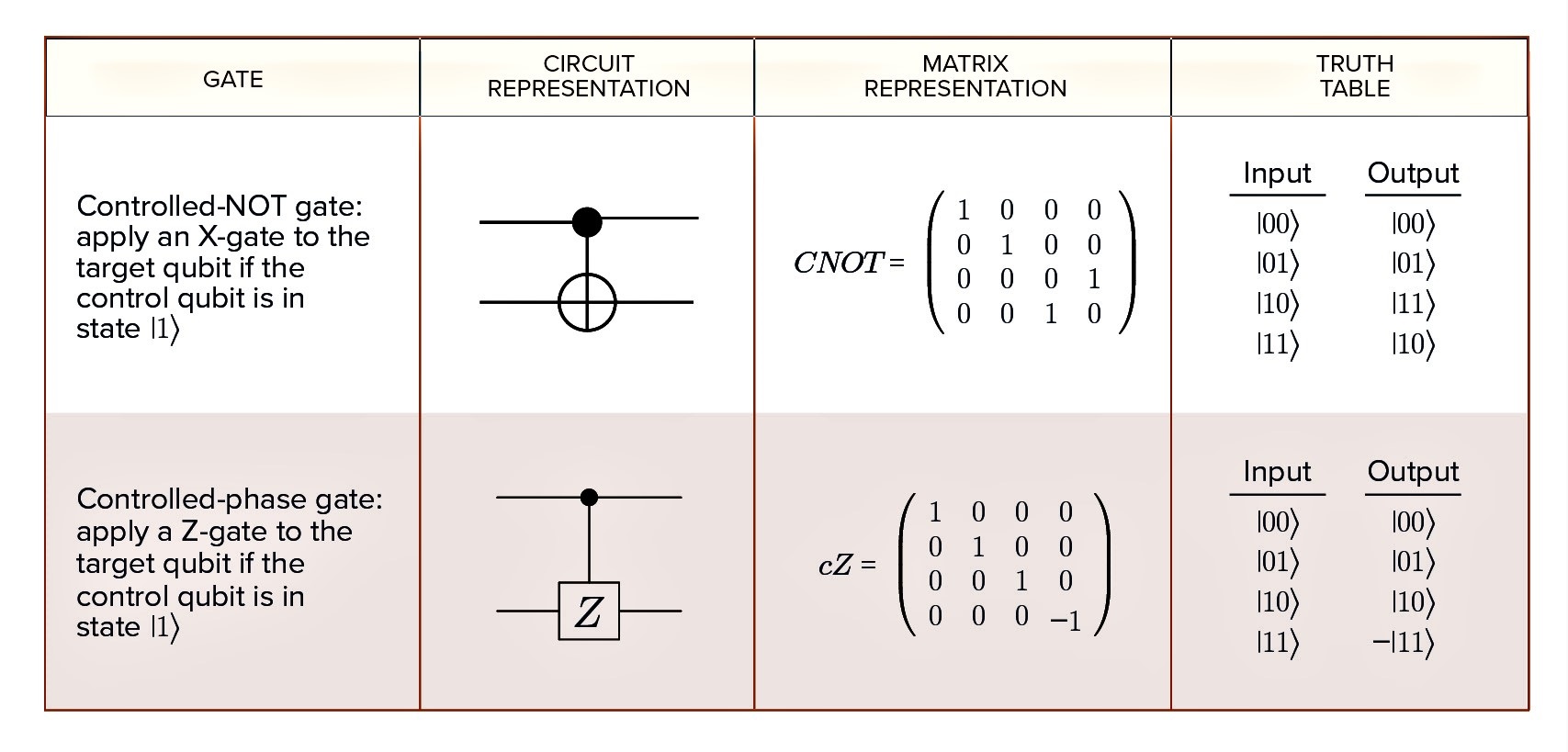}}\caption{Two Qubit}\label{fig1_22}
\end{figure}

\section{How Universal Algorithm Works}
How is a universal quantum algorithm implemented on a quantum computer? In this section, we will have an intuitive introduction to how a universal quantum computer uses single-qubit and two-qubit gates to implement an algorithm.

Let us get an intuitive picture of how a quantum algorithm works. A universal quantum algorithm is built from a small set of single and coupled qubit gates. The input to a quantum computer is a massive superposition state, in general. Then, we apply the types of single-qubit gates that we have just discussed. The single-qubit gate operates on all the states simultaneously through quantum parallelism. It is followed by quantum interference, which modifies the coefficients in front of those states. We will also apply the types of coupled qubit gates that we have discussed, for example, the rotation of qubit y-depends on the state of qubit-x. Again, through quantum parallelism, these operations apply to the entire state space, followed by quantum interference, which will again modify the coefficients.
Moreover, the goal of an algorithm designer is to ensure that, by the end of the algorithm, one of these coefficients has a value that is unity or very close to unity. It corresponds to the state that gives us the answer to the problem. It is a crucial point because we need to measure the output from the quantum computer to find the answer to the problem. The measurement process itself, in a quantum system, is probabilistic and leads to a classical result. So, although the output of the quantum computer is a massive superposition state, in general, the measurement process will project those qubits on to only one of the classical states that makes up the superposition. The probability that we get any given state corresponds to the magnitude squared of that coefficient. So, having a coefficient that is close to one or unity will give us a very high probability, the correct answer.

A quantum algorithm comprises a sequence of single-qubit and two-qubit gates. Quantum parallelism and quantum interference are used by the algorithm designer to take an input state typically a massive superposition state and, in a step-by-step manner, modify the weighting coefficients (probability amplitudes) until the quantum mechanical state evolves into an output state that encodes the answer to the problem. Since a projective quantum measurement will yield a single, classical state, it is imperative that the algorithm results in a final state with a probability amplitude near unity such that near-unity probability, the measurement will lead to the correct answer.

\section{Universal Quantum Algorithm}

What types of algorithms can be run on a universal quantum computer? What are the advantages, requirements, and challenges associated with universal quantum computation?  In this section, we present a high-level introduction to universal quantum algorithms and quantum computation.

There are a couple of different kinds of quantum computers. The most general is a Universal Fault-Tolerant Quantum Computer. The gate model algorithm that we discussed earlier runs on this type of computer. Building a Universal, Fault-Tolerant Quantum Computer\cite{kivlichan_improved_2019}, we think, is one of the most significant scientific and technological endeavors today. Such a machine, when we build it, will have all the power that quantum has to offer. A Universal Fault-Tolerant Quantum Computer is a holy grail, in some sense, as it means we will be able to program, implement, and reliably run complex, large-scale quantum algorithms. We use a small set of single and two-qubit gates to implement universal quantum computation. In principle, this computer can run any type of algorithm-quantum or classical. However, there are only specific algorithms that are known to exhibit a quantum advantage. Something that We will refer to as quantum speed up. One example is Shor's Factoring Algorithm. Shor's algorithm is used to break public-key cryptography. For example, RSA encryption a scheme that is used for secure communications\cite{jonsson_pkcs_nodate}. The security of RSA encryption is based on the premise that factoring a large integer number into two smaller prime numbers is a challenging computation for a classical computer to perform. The computational complexity scales poorly with the length of the encryption key. So, for example, if we are dissatisfied with our current security level, just double the length of our public key, and it will become exponentially harder for a classical computer to break that encryption scheme.
In contrast, Shor's algorithm can perform the same tasks, with only a marginal increase in difficulty, as the key's length has increased. Other examples include Grover's algorithm for searching an unsorted database \cite{grover_fast_1996}, or sampling solutions to linear equations. For these types of algorithms, quantum speedup of some degree exists over known classical algorithms when run on a universal quantum computer\cite{pednault_breaking_2018}. The actual amount of speed up for these algorithms we will discuss later in the section. The advantage of a Gate Model Quantum computer is that it is universal. In principle, it can run any algorithm. Now, in practice, not all algorithms will exhibit a quantum speedup, but many will present. The challenge lies in making a computer that is both large enough and operates for long enough to complete an algorithm and obtain the answer. Even though the fundamental building blocks, the qubits themselves, are prone to errors. In a quantum computer, each operation has a probability of being noisy, That is of adding an error to the computation. We need a way to combat the build-up of all these errors so that the output we get out of the algorithm, or the computation is reliable. We will see later in the section; such errors can be overcome by something called Quantum Error Correction. However, it will require additional resources to implement it.

The power of a universal computer is that it can implement any algorithm that can be expressed in terms of a circuit comprised of logical gates. A universal, fault-tolerant quantum computer will enable people to program, implement, and reliably run arbitrarily complex quantum algorithms. In fact, a universal quantum computer can run any type of algorithm quantum or classical. Determining when it is advantageous to implement an algorithm on a quantum computer, as opposed to a classical computer, is one of the principal aims of quantum information science. The study of how efficiently a computer, whether a classical or a quantum computer, can solve a given problem results in the formal segregation of algorithms into what are called ``computational complexity classes.'' This efficiency is determined based on how the computational resources needed to solve a problem, grow, or ``scale'' with the size of the problem instance. For example, how the size of the memory or the number of computational steps scales with the problem size.

Two important computational complexity classes are P and NP\cite{aaronson_computational_2010,orus_tensor_2019}. An example of a problem in P is the multiplication of two numbers of length n, which requires $ n^2 $ time steps to complete. For example, this means multiplying the binary numbers 01 and 10 requires 4-time steps since each number has n=2 digits, but multiplying 100 and 110 requires 9-time steps, since n=3 for these numbers. This time scaling, and any time scaling of $ a \times n^ b $ where both a and b are constants, is referred to as ``polynomial scaling'' and is considered efficient.

These problems should be compared with those in the complexity class NP, where the time required to solve a problem is exponential in the number of problem inputs and is therefore not considered to be efficiently solvable. An example of such scaling would be $ 2^ n $. One common point of confusion is that time steps required to solve both P and NP problems can be comparable for small n. For example, when n=1,2,3, the given polynomial scaling requires time steps of $ n^2=1,4,9, $ as compared to the exponential scaling, which requires $ 2^ n=2,4,8 $ time steps. However, as n increases, these two scalings rapidly diverge, and this is why they are categorized into different complexity classes. Considering n=10, these two scalings are already an order of magnitude apart, with $ 10^2=100  $and $ 2^{10}=1024 $. Go to n=100, and $ 100^2=10,000 $, whereas $ 2^{100}>1\times 10^{30}  $that is, 10 followed by 30 zeros! That is the power of exponential scaling. Finally, it is important to note that problems are classified into these categories based on the best-known algorithm for solving them and are therefore subject to change as discoveries are made.

One of the main reasons for the current interest in quantum computers is that specific problems can be solved in polynomial time on a quantum computer even though the best known classical algorithm for the same problem requires exponential time. A prominent example of this is Shor's algorithm. Shor's algorithm effectively factors large integer numbers into two constituent prime numbers, a computationally hard task for classical computers. The difficulty of Factorization is that it is used as the foundation for public-key cryptography\cite{barbeau_secure_2019}, for example, RSA encryption. It is even using the best-known classical algorithm, the resources required to break RSA encryption using a classical computer increase exponentially with the length of the public key. Hence, the problem is considered to be in NP (although it has not yet been formally proven to be NP). Practically, this means that doubling the number of bits in an RSA encryption key makes it exponentially more difficult for a classical computer to break the scheme. In contrast, Shor's algorithm can perform the same task efficiently on a quantum computer, with only a marginal increase in difficulty as the length of the key increases.

\section{Quantum Simulation Emulation}
Quantum computers can simulate a quantum system using both digital gates and analog evolution. They can also emulate a quantum system by tailoring the qubits and their connectivity. In this section, we will take a look at quantum simulations \cite{bloch_quantum_2012} and a specific example, nitrogen fixation. Another type of quantum computer is a quantum simulator \cite{zhang_observation_2017}, a processor that simulates a physical system's behavior, a chemical reaction, or a biological process. Simulations can be implemented using single and two-qubit gates as on a universal quantum computer, and we will call these digital simulations. Alternatively, the qubits themselves, the way that they connect, and the strengths of those connections can be designed to emulate a system's behavior. We will call this an analog simulator. There are also examples of hybrid simulators \cite{chen_hybrid_2019}that use aspects of both digital and analog approaches \cite{bauman_downfolding_2019}. Nature is quantum mechanical. So, with quantum computers, we get this ability to take a step closer to being able to computationally model aspects of physical systems. So, we think this is a pretty fantastic potential first application of quantum computers. As the first business application, with the knowledge of today, we would say the simulation of molecules in materials \cite{bassman_towards_2020} is the most likely, which, in turn, can have a broad impact on science, technology, and society. One example is the simulation of chemical reaction mechanisms. It is estimated that somewhere between 1\% and 2\% of worldwide energy production is used to produce ammonia for agricultural fertilizer. The critical step is a process called nitrogen fixation. In industry, nitrogen fixation is performed using the Haber Bosch process, which requires both high pressure and high temperature.
In contrast, there exist bacteria that use an enzyme called molybdenum nitrogenase, which, even at room temperature, can catalyze atmospheric nitrogen into ammonia. How do the bacteria do it? we do not know precisely. We do know the critical component is an iron-molybdenum cofactor, a chemical compound that acts as a helper molecule for the enzyme. However, it is not known how the reaction works in detail. Simulations with classical computers, exist but only provide approximate answers. If we could study that process and understand it, we could potentially engineer a catalyst that makes this reaction efficient. Now to study this with classical computers, this is the lifetime of the universe time scale to get a solution.
On the other hand, quantum simulation of the chemical reaction steps would provide the reaction energies involved in the nitrogen fixation. This type of quantum chemistry simulation has many applications\cite{hempel_quantum_2018,mccaskey_quantum_2019,lanyon_towards_2010}. For example, developing new pharmaceutical drugs or tailoring a material to have unique properties. For these types of simulations, it has been shown that a quantum speedup can exist over known classical algorithms.

Quantum computers can be applied to many types of simulation problems. The challenge is that many of these problems require large numbers of well-behaved qubits, and such large-scale quantum computers are still likely a decade or more away. Before we get to the point where we can simulate systems entirely on quantum computers, there may be room for hybrid classical-quantum systems to hold a quantum advantage over purely classical algorithms\cite{pednault_breaking_2018}. One example is called a variational quantum eigensolver, another hybrid quantum-classical algorithm is Variational Quantum Fidelity Estimation\cite{cerezo_variational_2020},variational quantum factoring (VQF) algorithm\cite{anschuetz_variational_2019}.

A variational quantum eigensolver \cite{kandala_hardware-efficient_2017,thornton_quantum_2019,endo_variational_2019,mcardle_variational_2019,yuan_theory_2019,mcclean_theory_2016,peruzzo_variational_2014,li_efficient_2017} determines the lowest-energy state of a quantum system, a difficult task of importance to quantum chemistry. The protocol goes as follows:

1. The quantum processor is set so that it simulates the dynamics of the quantum system of interest

2. The classical computer proposes a trial ground state, which is then loaded into the quantum processor

3. The quantum processor calculates the energy of that state and passes it back to the classical computer

4. The classical computer takes that value for the energy and uses it to generate a new ground trial state, one which should give a lower energy

5. This process repeats until the VQE cannot find a new trial state that gives yet lower energy. This state is then likely the lowest-energy state of the system.

A vital aspect of this process is that the classical computer is not merely being used to initialize a quantum computation; the classical and quantum computers are passing information back and forth throughout the simulation. Therefore, the quantum computer is acting as a co-processor. It need not be coherent over the entire duration of the simulation, but only over the period of time, it takes to simulate the energy of a single trial state. This approach may allow useful computations to be done in the nearer term using less reliable qubits.

The VQE method has been shown to successfully determine the molecular spectra \cite{omalley_scalable_2016}of a hydrogen molecule $ H_2$. To simulate the hydrogen molecule comprising two electrons, a quantum processor with 2 superconducting qubits is sufficient \cite{gambetta_building_2017}. It has also been used to simulate other small molecules, such as $ BeH_2 $. All of these simulations to date are prototype problems that demonstrate the mechanics of the quantum simulation, but the problems themselves are easily solved on a classical computer. As the number of qubits increases, VQE can be applied to larger, more challenging simulations.

\section{Quantum Annealing}

A quantum annealer is a different type of computer that addresses classical optimization problems by mapping them onto a set of interconnected qubits and then searching for a solution (or solutions) that minimizes the total energy. In this section, we will get a high-level introduction to quantum annealing computers and how they work.

The third type of quantum computer is called a quantum annealer, and it is used solely to solve classical optimization problems. Optimization problems are those that we face every day. What is the best way for us to be able to drive home when there is traffic? What is the best way to route aircraft around an airport? All of those are optimization problems. If we are planning a very complicated mission, space mission with many moving parts that involve materiel, people, gasoline, if we can optimize and improve the efficiency of something like that, quantum computing could have a considerable impact. Quantum annealers do not use digital gates. Instead, an optimization problem is encoded directly into the qubits and their connectivity to one another, and by the strength of those connections. Finding the qubit states that then minimize their total energy is equivalent to optimizing and finding an encoding problem. However, how do we anneal these qubits into states that minimize their energy? So, the name quantum annealing is related to making a sword, where we take a piece of metal, and then we heat it, and in the old days, a swordsmith would beat the sword into some shape, and they would cool it, quench it in water heat it up again, and then beat it into shape and do that multiple times.
Furthermore, that was called an annealing process. So, heat it, cool it down, heat it, cool it down. Quantum computing uses a similar technique. To perform annealing, we start by setting the qubit states and their couplings to one another in a configuration where we know the ground state. Let us call this the starting configuration or its starting Hamiltonian. We then slowly change the qubits and their couplings from those in the starting Hamiltonian to those in the encoded problem Hamiltonian \cite{bravyi_tapering_2017}. Now, if we make these changes slowly enough, it is likely that We will remain in or very near the ground state of the system so that by the end of this evolution, the qubits are in their ground state, and that represents the solution to the optimization problem. If we have a problem that we can craft as a landscape like the Alps, for example, what the computer does is, without adding or subtracting, it would find the lowest valley or valleys in that landscape. It is probabilistic, and again, does not add or subtract, but rather, it is not a three-dimensional version of the Alps. However, it is an n-dimensional energy landscape, and it is finding a low energy solution to that problem. The challenge is that for problems of unusual size and complexity, it is almost certain that no matter how slowly we change these parameters, the annealer will, at some point, leave its ground state. It is due to the more significant number of energy states and their proximity to one another and as a result, quantum annealers must find additional mechanisms to return the computer to the ground state, for example, through quantum tunneling or by introducing excess relaxation, a loss of energy that returns the computer to its ground state. If we hit a barrier like a mountain range, as we described earlier, what the machine does is rather than we have to put more energy to climb the mountainside and go down the other side, we use the quantum mechanical property of tunneling. We tunnel through the barrier to get to what should be a lower energy solution in the next valley if we will.
On the other hand, add too much relaxation and the computer may not even be quantum mechanical anymore. As a result, it is currently unknown if a quantum annealer can exhibit quantum enhancement for a general class of optimization problems. It may just represent another type of classical computer. Although, as a classical computer, it would not scale well with the size of the problem. However, it may still be interesting if it were substantially faster than any of today's transistor-based classical computers. Volkswagen, in March of 2017, announced the optimization of about 500 taxis going from downtown Beijing to the airport, always congestion used the machine to see if we could come up with optimized routes for each of those taxis to get everyone moving more quickly. Worked, but it is a prototype application. Could not yet handle all the vehicles in Beijing, for example. There are technical challenges facing quantum computing. For the quantum annealing computers, how do we scale them up to be able to start attacking more real-world problems instead of subsets of real-world problems? There has a keen interest in quantum annealers because classical optimization problems are everywhere, from supply transport optimization to sensor and satellite tasking, pattern recognition\cite{leymann_towards_2019}, needle in a haystack problem. Many problems can be reduced to optimization, and so there is an extensive application pull to understand quantum annealers better.

The third type of quantum computer is a quantum annealer, an ``application-specific'' computer that is used solely to solve classical optimization problems. Quantum annealers do not use digital gates. Instead, an optimization problem is encoded directly into the qubits and their interactions (qubit interaction is referred to as coupling). Minimizing the total energy of the system is equivalent to optimizing and finding an encoding problem.

To quickly explain some physics jargon: Every quantum system has some amount of energy that is determined by the state of each object (e.g., each qubit) in the system. The state of each individual object, in turn, determines the global state of the entire system, and each global state corresponds to the system having particular total energy. The Hamiltonian is a function that matches each state to an energy value. So, for our purposes, the Hamiltonian is a function that tells us ``if our qubits are in the state $ \vert \psi \rangle $, our system has energy $ E_\psi $. The state that gives the lowest energy for a system is called the ground state.

The idea behind quantum annealers is that if we have a quantum system governed by one Hamiltonian, we can transfer it to another Hamiltonian by changing the system's parameters.

Quantum annealing starts with a qubit state configuration for which the ground state is well known. It is the starting configuration or starting Hamiltonian. By gradually turning off the starting Hamiltonian, while turning on the problem Hamiltonian, one can transition to the configuration that encodes the problem. If this transition is performed slowly enough and the system has no noise, then the computer will remain; it is ground state throughout the evolution. Measuring the state of the computer in the final configuration yields the answer to the problem. As described here, this is an ``adiabatic quantum computer.''\cite{farhi_quantum_2000, aharonov_adiabatic_2005}

The challenge is that for problems of unusual size and complexity, it is almost certain that no matter how slowly the coupling parameters are changed, the annealer will leave its ground state. There are two primary reasons for this: first, the ``minimum gap'' between the ground state energy and nearby excited-state energy levels gets very small as the size of the problem increases; and second, there is always noise in the system due to non-zero temperature. Although operation in a cryogenic environment reduces noise, it does not eliminate it \cite{bardin_28nm_2019}. as the minimum gap gets smaller eventually, even cryogenic temperatures begin to look relatively ``warm.''

Consequently, quantum annealers must find additional mechanisms such as quantum tunneling or excess relaxation to keep the computer in the ground state. On the other hand, too much relaxation will make the computer manifestly classical. As a result, it is presently unknown if a quantum annealer can exhibit quantum enhancement for a general class of optimization problems or whether it is merely another type of classical computer. Nevertheless, there is intense research into quantum annealers today, because the application pull is strong. Many real-world can be cast as optimization problems (e.g., financial portfolio optimization \cite{rebentrost_quantum_2018}, product distribution, routing of autonomous vehicles\cite{irie_quantum_2019}), and so there is a strong motivation to understand quantum annealers better. Quantum annealing is a method for finding solutions to classical optimization problems. The quantum annealer encodes the problem on to a set of qubits and biases those qubits into a starting state with a known ground state (that is not the solution to the problem). The system is then slowly evolved towards a set of bias points that do represent the problem being solved, and the ground state of the system is the answer to the problem. However, a quantum annealer is likely to leave its ground state during this evolution. The concept is that, though a combination of quantum tunneling and relaxation processes, the system should find its way back to the ground state (or a lower-energy state) to find the encoded (approximate) solution to the optimization problem \cite{paini_approximate_2019}.

\section{The DiVincenzo Criteria for Quantum Computers}
There are a variety of physical systems that have been proposed as the qubits for quantum computers. What properties must a given technology possess in order to build a quantum computer? In this section, we will discuss the DiVincenzo Criteria\cite{divincenzo_two-bit_1995,divincenzo_physical_2000}, the minimum set of requirements a qubit technology must have to be considered a viable candidate for quantum computing, and for communicating quantum information.  

There are numerous examples of quantum mechanical two-level systems in nature that could potentially serve as qubits. For example, the electronic states of an ion, or the electron spin of a phosphorus atom implanted in silicon \cite{yoneda_quantum_2020} or a nuclear spin of a defect in diamond. We can even manufacture electrical circuits that behave as quantum mechanical two-level systems or artificial atoms. However, what makes a useful qubit? Furthermore, how do we assess these qubit modalities and compare them to one another?
Furthermore, how can we determine if a modality is well-suited for quantum computing? We can begin to answer these questions by first drawing on the intuition from classical computing. What makes an excellent classical logic element? Furthermore, why did we end up with transistors? If we want to build a computer at a scale large enough to perform new problems, we should start with a technology that scales well, one where we know how to define and characterize the logic elements and where we can manufacture them in large numbers. These devices must also accurately represent and process classical information. We need to set their states to provide input to the computer, and we need to be able to measure the result to get an answer. Finally, these devices must be robust against failure to complete the computation and reliably obtain the answer. Transistors satisfy all these requirements very well, so it is no wonder that today's computers are built from transistors and not, vacuum tubes, or mechanical switches. So, what makes a useful qubit? Around the year 2000, David DiVincenzo, then a researcher at IBM, articulated five necessary conditions that any qubit technology must at least possess for a suitable physical implementation for large scale quantum computation \cite{divincenzo_physical_2000}. First, it should be a scalable physical system with well-defined and characterized qubits. Second, we must be able to set the input state and, third, measure the resulting output state. Fourth, we must be able to perform a universal set of gate operations.
For example, the single and two-qubit gates that we discussed previously that are needed to run an algorithm. Fifth, the qubits must robustly represent quantum information. In many ways, these requirements are like those for classical computers. It is only in how the requirements are met that we can identify differences—for example, performing a universal set of one-qubit and two-qubit gates rather than universal Boolean logic. Alternatively, to robustly represent quantum information, qubits must have long coherence times\cite{herbschleb_ultra-long_2019}, a concept that loosely translates to the meantime to failure for a transistor. In addition to these five requirements for the qubit technology, David DiVincenzo added two conditions related to the communication of quantum information between qubits. So, continuing from number five, number six is that the technology must support the interconversion of quantum information between a stationary qubit and a flying qubit.
Moreover, seven, there must be a way to transfer flying qubits faithfully between two locations. These two requirements describe a quantum version of an interconnect that is, the means to take the quantum information encoded in one qubit, convert it to an object that can move, like a photon, provide the means to guide a photon without loss to another qubit at a distant location, and then hand back that quantum information. Again, analogous to requirements for routing signals within a classical computer but following the rules of quantum mechanics. These criteria, today referred to as the DiVincenzo Criteria, articulate the basic requirements that any qubit technology must possess if it is a viable physical implementation for quantum computation.

Several quantum mechanical two-level systems potentially could serve as qubits. For example, the electronic states of an ion, the electron spin of a phosphorus atom implanted in silicon, a nuclear spin of a crystal defect in diamond, and even manufactured ``artificial atoms,'' electrical circuits can be described as quantum mechanical two-level systems. What are the basic requirements any one of these modalities must at least possess to be a candidate for a quantum computing technology?

In 2000, David DiVincenzo, then a researcher at IBM, articulated five fundamental requirements for any qubit technology to be a suitable physical implementation for large scale quantum computation (see the paper here \cite{divincenzo_physical_2000}).

In addition to these five criteria for qubit technology, David DiVincenzo added two conditions related to the communication of quantum information between qubits.

\begin{figure}[H] \centering{\includegraphics[scale=.5]{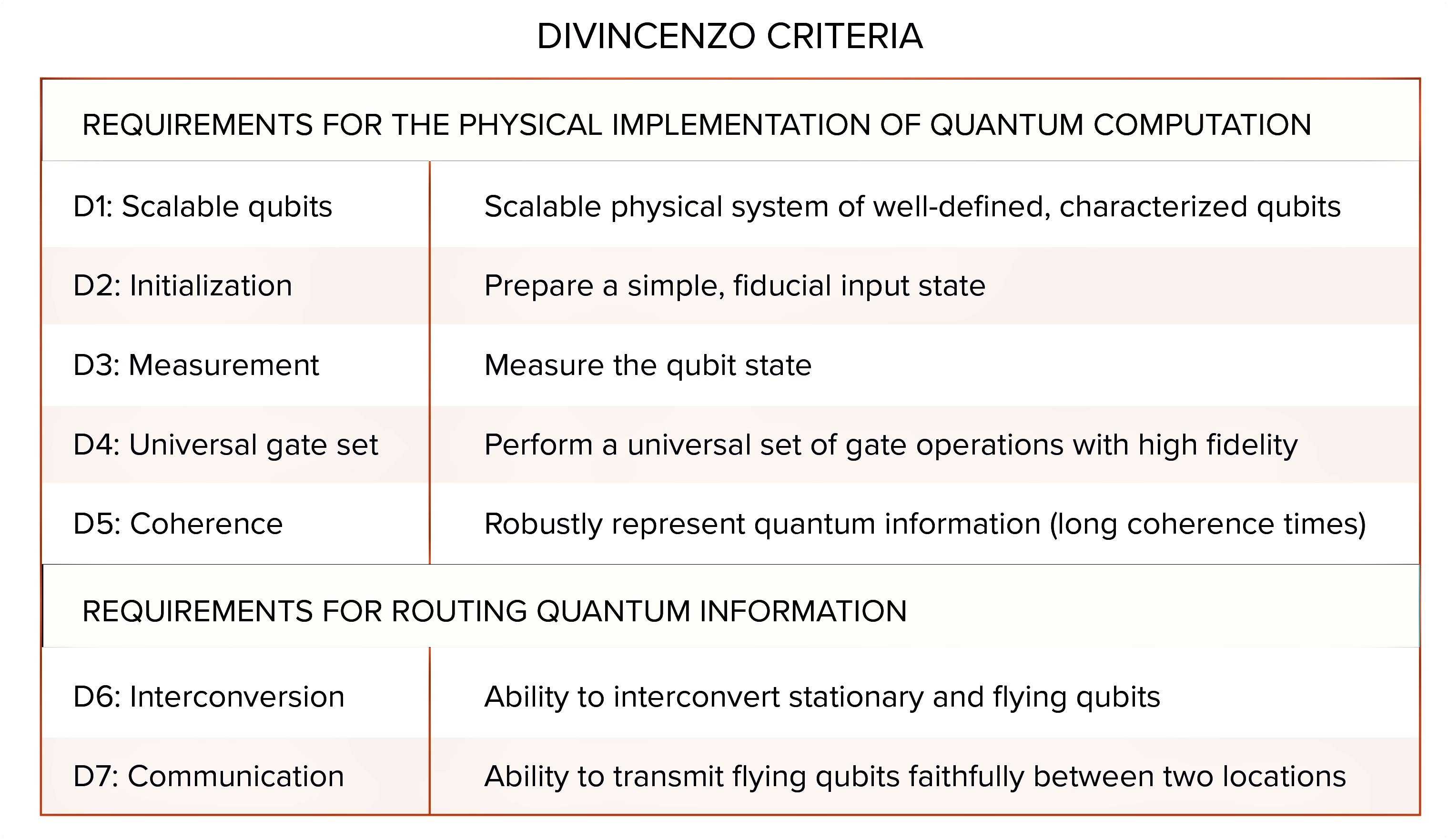}}
    \caption{DiVincenzo Criteria}\label{fig1_26}
\end{figure}

These criteria, known as the DiVincenzo Criteria, are, in many ways, adapted from the conditions for classical operational computers and summarize the fundamental requirements qubit technologies need to fulfill at a minimum to be a candidate quantum computing technology.

\section{Qubit Coherence and Gate Time}

In this section, we will discuss two types of errors that can occur in qubits energy relaxation and decoherence and their corresponding characteristic lifetimes $ T_1 $ and $ T_2 $. We will also consider the clock speed at which qubit operations can be performed\cite{klimov_fluctuations_2018}.

The average number of operations that can be performed within a qubit lifetime is a proxy for a more rigorous metric called gate fidelity. Implementing more operations before an error occurs is good. As we will see in this section, longer coherence times do not necessarily translate to more operations per gate time, as the gate operations themselves are generally slower in long-lived qubit modalities. However, just as with classical computers, there is a distinct advantage to faster clock speed, as that means we obtain results faster.  

The DiVincenzo criteria articulated the requirements that a qubit technology must have to be a viable candidate for the physical implementation of a quantum computer. In this section, we will build on two of those criteria, related to qubit robustness and quantum gates, to define metrics that will allow us to compare qubit modalities with one another. To do this, let us first look, in more detail, at the qubit coherence time, the analog of the meantime to failure for a transistor. Quantum computers, like classical computers, must be built from robust elements. The coherence time is one metric that quantifies the robustness of a quantum bit. Essentially, it is the amount of time, on average, that a qubit state is maintained before the quantum state is lost. As an illustration, let us consider a qubit that we set into a quantum state psi, and consider what happens to that qubit over time. At first, the state is well-defined, we just put it in that state, we are confident that we did a good job, and so we know what the state is. Over time, however, the qubit begins to interact with its environment. When it does so, the qubit experiences noise that alters the qubit state in ways that we did not anticipate. Intuitively, we can imagine that the state begins to blur. as time goes on, and the qubit is subject to more of this environmental noise. Eventually, we can no longer recognize the state, and the quantum information is fully lost. The qubit is still there, but it is no longer in the state that we would have expected it to be in after this amount of time. The noise has altered the behavior of the qubit, and it is now in some other state. Let us now look at how we quantify qubit lifetimes, by looking at errors on the Bloch Sphere. There are two fundamental ways in which a qubit loses quantum information. The first is energy relaxation. Imagine that we put the qubit in its excited state, state 1. Unfortunately, it probably will not stay there, due to noise at the qubit frequency, the qubit will, at some point, lose its energy to the environment and return to the ground state. This loss of energy is called energy relaxation, and on average, it occurs after a time called $ T_1 $. The second way a qubit can lose quantum information is through a loss of phase coherence. Imagine that we intentionally set the qubit at one point on the equator of the Bloch Sphere. It likely will not stay in that direction forever, due to interactions with the environment. there are two ways that phase coherence can be lost. First, the qubit state may move around on the equator, due to the environmental noise. If we repeat the experiment often, the noise will sometimes drive the Bloch vector east along the equator, sometimes to the west, and over time, the Bloch vector fans out more and more. It does this until, eventually, we can no longer tell which direction, or equivalently, which phase, the Bloch vector has. the average time it takes for this to happen is called the pure dephasing time, $ T_{\phi } $. Now there is another way things can go wrong on the equator. Remember that, on the equator, the qubit is in a superposition state of 0 and 1, with 1 being the excited state. If that component of the superposition state loses its energy to the environment, then the 1 state flips to 0, and the superposition state is lost. Essentially, the qubit is relaxed from the equator to the north pole. Energy relaxation is also a phase breaking process since once the Bloch vector points to the north pole, we can no longer tell which way it had been pointing on the equator, that phase information is lost to the environment. Thus, the average amount of time that a qubit remains coherent is related to both the dephasing time, $ T_{\phi } $, and the energy relaxation time, $ T_1 $, which together give a time $ T_2 $, over which phase coherence is lost. Thus, a qubit loses its quantum information by two mechanisms, energy relaxation and loss of phase coherence, characterized by the times $ T_1 $ and $ T_2 $. Now another important metric for quantum computers, just as with classical computers, is the clock speed, the time required to perform a quantum operation. It is called the gate time, and although it generally differs for single and two-qubit gates, we can use a typical time, or conservatively, use the slowest gate time, to define the clock speed with which we can operate the quantum computer. as with all computers, faster is better. Even if we have an exponential speedup from a quantum algorithm, it still takes some time to perform that algorithm. All else being equal, a faster clock speed will translate to obtaining the answer more quickly. A key figure of merit, then, is the number of gates one can perform within the qubit lifetime. The more gates one can implement before an error occurs, the larger an algorithm one can run. This metric also illustrates an interesting trade-off with qubits. In general, qubits with long lifetimes have less interaction with their environment. That is good because they have less sensitivity to noise. However, the trade-off is that they respond more slowly, even when we intentionally try to control them. It is because They are not just weakly interacting with their environment. They are also weakly interacting with the control fields. Similarly, qubits that more strongly interact with their environment may have shorter coherence times, but they generally respond faster. The number of gates one can perform, on average, before an error occurs, may not differ much between these two cases. However, one of these qubit modalities may have a much faster clock speed than the other one, and that is a good thing. This figure of merit, the average number of gates one can perform before an error occurs, is a proxy for a more general and rigorous concept, called gate fidelity, which We will discuss next.we would like to understand if a quantum processor will have the same limitation of classical processors regarding the gigahertz and clock speed hardware. It is possible to have a or is it possible to have a quantum CPU with much more than with many more gigahertz than is typical nowadays at 3 gigahertz? that the clock speed matters whether we are discussing about classical or quantum computers from the point of view of it will take a certain amount of time to solve a problem. if we want to solve that problem faster, then we need to run the computer faster. But we think making a comparison between classical clock speeds and quantum clock speeds is not really the right comparison because, as we know, a classical computer can really get bogged down on some of these hard problems that a quantum computer can solve very efficiently. so, a problem like factorization, which might take the age of the universe on a classical computer so, just completely impractical on 100-megahertz quantum computer might be able to do this factorization could be done in about an hour or a couple of hours. that is for, a 2,000 or 4,000-bit number. So, that has nothing to do with clock speed. That has to do with quantum computing being a fundamentally different computing paradigm than classical computing. But then what we discuss and we think this is an important point, is that if we have a quantum computer and it can do that calculation in a day at, 100 megahertz, fantastic. It did not take the age of the universe. It took a day much more efficient, exponentially more efficient. But if we run on another quantum computer that is 1,000 times slower, then that will take 1,000 days. so, there, the speed actually matters. So, within quantum computing paradigm \cite{bo_ewald_introduction_2019,maslov_outlook_2019}, comparing different computers, it makes sense to discuss about gate speed. But we also want to emphasize that It is not just gate speed. So, it is gate speed, assuming that all else is equal, meaning that we have two computers and they have the same connectivity, the same overhead for error correction. If they are identical in and along all of those axes, then clearly, running one 1,000 times faster is going to give we an answer at 1,000 times sooner. But gate speed is not the only thing that determines how fast we achieve a solution. the degree to which qubits can be connected to one another for example, do we only have in array of qubits, we only able to connect to a nearest neighbour or do we have an all-to-all connectivity? Can every qubit discuss through a two-qubit gate directly to any other qubit in the system? That is a huge difference. That makes a tremendous difference in the execution speed. So, it is not just clock speed that matters. But clock speed does matter, as explained. 

There are two processes by which a qubit loses quantum information. The first is called energy relaxation, and it is characterized by time $ T_1 $. It is characteristic time it takes for a qubit in its excited state to relax back to its ground state by emitting energy to the environment. It can be visualized on the Bloch sphere as a qubit prepared in the excited state (south pole) switching to the ground state (north pole) through energy loss. $ T_1 $ also characterizes the time over which a qubit will absorb energy from the environment. However, most leading qubit technologies today operate in a regime where such energy absorption processes are negligible compared with energy decay.

The second way a qubit loses information is through decoherence. There are two ways for a qubit to lose coherence. The first is that due to environmental noise, a Bloch vector on the equator might move along the equator away from it is the original position in a manner we cannot predict, and over time, we lose track of which way the vector is pointing. It is called dephasing, and it occurs over a characteristic time $ T_{\phi } $. The second is that the excited-state portion of the qubit superposition state could lose its energy to the environment, and the qubit ``falls off'' the equator and back to the ground state (north pole). It is also a phase-breaking process since once the vector is at the north pole, we can no longer tell which way it had been pointing on the equator, and it occurs over the same characteristic relaxation time, $ T_1 $. Together, these two dephasing mechanisms lead to the decoherence time $ T_2 $, which is a function of the dephasing time and the energy relaxation time: $ 1/T_2 = 1/T_{\phi } + 1/2T_{1} $.

Mitigating the resulting errors can be quite challenging on quantum computers, much more so than correcting classical errors on conventional computers. For example, one can periodically error-correct a bit by simply measuring and resetting its state on a classical computer. For example, a classical bit might be stored as a voltage, where a voltage of 5V denotes a 1, and 0V denotes a 0. Even though we cannot assume that a bit set to 1 will stay at 5V forever, we can periodically measure it is voltage and reset it to 5V if it starts to decrease.

In contrast, because of the way measurements affect quantum systems, this option is not directly available to us in quantum computers. It is not possible to measure a state and then reset it precisely, because making that measurement disrupts the quantum state of the system in a manner that is not entirely reversible. It is why we work very hard to shield qubits from environmental noise in the first place. Still, there are quantum versions of error mitigation and error correction that can fix residual errors, provided the underlying qubits are ``good enough''.

\section{Gate Fidelity}

How does one confirm that a gate operation is working as intended? In this section, we will discuss the gate fidelity of a quantum operation. Gate fidelity quantifies the quality of gate operation, and it is used to compare qubit modalities of varying types. 

In the last section, we focused on two of the DiVincenzo criteria, qubit coherence, and quantum gates. We developed an intuitive figure of merit, the average number of gate operations that can be performed before an error occurs. Such metrics are important because they allow us to compare different qubit modalities, even when those modalities have remarkably different properties. However, the definition we introduced only accounted for errors due to qubit decoherence. That is fine for qubits dominated by decoherence, but other sources of errors limit some qubit modalities. For example, control errors, imperfections in the pulses that are used to drive a gate operation. For these qubits, even if their coherence times were practically infinite, their gates would still be subject to control errors. So, we need a more rigorous way to characterize the robustness of gate operation, one that is sensitive to a broad set of error sources. that leads us to this section's subject, a more general concept called gate fidelity. Gate fidelity is a rigorous means to define how well a gate operation works. Essentially, it is a measure of how closely the actual gate operation matches, on average, a theoretically ideal version of that operation. For example, if we apply an X-gate to a qubit prepared in state 0, how close do we get to state 1? Now intuitively, errors could occur along any direction of the Bloch Sphere, and so we need to check for errors along with all directions after the operation is complete. Until this point in the section, we have only measured the qubit along the Z-direction. In principle, we can measure the qubit along any axis that passes through the center of the Bloch Sphere, including the x-axis or the y-axis. Besides, gate errors may be manifest differently, depending on the starting point of the qubit, so we need to check how the gate performs against different initial states, and not just the north pole, for example. Now at this point, we may be asking ourselves, how is this even feasible? The qubit state can be anywhere on the surface of the Bloch Sphere, representing an infinite number of possible initial and final states. How can we possibly check them all? Well, we do not have to. That is because we can define a basis set that spans the entire qubit state space. To draw an analogy, think of global positioning. the position anywhere on the globe can be represented by projecting it onto a convenient basis, in this case, a coordinate system that spans the globe. We often use latitude and longitude, but we could equally use Cartesian coordinates, x-, y-, and z. Similarly, we can define a basis that spans the entire qubit state space, and then use it to specify any qubit state at the input. We can also use it to characterize any state at the output, by projecting back onto this basis. It works because quantum mechanics is described by linear algebra. It is enough to understand projections to and from the basis set to know what will happen globally. We will see in the section later, and such projective measurement plays a crucial role in the digitization of errors and the ability to correct them. Although the coefficients of a superposition state will dictate the probability of obtaining a 0 or 1 when the measurement is done, for any given measurement, we will get a 0 or a 1. In this way, quantum states and their errors, which appear to be analog, can be digitized through projective measurement. Now we discuss previously that n-qubits could represent $ 2^{N} $ states. Here, the basis can be represented in a matrix form, and the number of matrix elements that span the state space is the square of the number of states. So, $ 2^{N} $, quantity squared. For a single qubit, n equals 1, and we need four elements. For two qubits n equals 2, and the number quickly increases to 16. So, determining the gate fidelity is essentially a standard black box type problem. We have a gate operation, and theoretically, we know how it should perform. However, due to errors, its actual operation is not precisely known. That is the black box. Thus, we need to characterize, on average, how well the gate operation performs. To do this, we probe the gate operation using the input basis states, perform the actual gate operation, and then, for each input, project the resulting output state onto the entire basis set. We then compare the results to what we would expect from a theoretically ideal operation to determine the gate fidelity. The described approach is called process tomography \cite{mohseni_quantum-process_2008}, and it represents a complete description of the errors during a gate operation. It also requires many steps to implement, as the product of input and output elements is $ 2^{4N} $. Now, although there are $ 2^{2N}$-th constraints, due to the properties of these matrices, this only reduces the total number of measurements by a small amount. Thus, implementing process tomography scales very poorly with the number of qubits, and becomes impractical as n gets large. This is an issue, because although the gates only act on one or two qubits, the errors may leak to nearby qubits, qubit 3, qubit 4, or qubit 5. Thus, to account for this, the basis we choose must encompass all n qubits. Besides, process tomography is sensitive to all errors, including initialization errors, when preparing the input state, and measurement errors at the output, which, although certainly real errors, are unrelated to the quality of the gate operation itself. As a result, an alternative approach, called randomized benchmarking \cite{villalonga_flexible_2019,corcoles_process_2013,sheldon_characterizing_2016,magesan_characterizing_2012}, has also been developed. Randomized benchmarking essentially interleaves the gate operation being characterized by a random assortment of other gate operations\cite{mavadia_experimental_2018}. Although half of the gates are chosen at random, we know what they are, and so we can predict the expected output state, assuming all the gates were ideal. It is then compared to the same experiment, but with only the random assortment of gates, to see how much the error rate has changed when adding the interleaved gate. This change in error rate is then attributed to the interleave gate itself. This approach is repeated for increasing numbers of pulses to obtain a refined estimate for the average error per gate and, thereby, the gate fidelity. Randomized benchmarking is much more efficient than process tomography, and it is also insensitive to initialize and measurement errors. However, it only provides a net error rate without revealing specific error channels. However, it is measured, a gate fidelity of 100\% means that the actual operation perfectly matches the ideal operation. For example, no matter where the qubit starts on the Bloch Sphere, the actual x-gate would perfectly rotate the input state to the correct output state. Now, as we might expect, it is generally not possible to achieve a perfect gate fidelity. There will always be some level of error, whether due to qubit decoherence during the operation, imperfections in the control pulse itself. The goal is to see how many nines, two nines, 99\%, three nines, 99.9\%, that one can achieve. The higher the fidelity, the closer it comes to an ideal gate. We will see later in the section, achieving high fidelity is critical, because it translates directly to two important aspects of quantum error correction. First, the fidelity must at least reach a minimum value, called the threshold, for the error correction to give us a net improvement in the error rate. second, once above this threshold, the higher the fidelity, the less the resource overhead required to implement the error correction.

Gate fidelity is a metric that characterizes how well a gate operation works. Intuitively, we are asking the question: how close does the actual implemented gate operation come to an ideal, theoretically calculated one?

To answer this question, we need to consider all potential errors that may impact the gate operation and the resulting qubit evolution. These errors can, in principle, arise from any direction on the Bloch sphere, so we need to check the fidelity for different starting states of the qubit. Similarly, we need to assess the qubit's final state in all directions along which errors may occur.

To do this, we define a basis set for the qubit that spans all of the possible qubit states and qubit operations. Global positioning is an analogous concept, where we can identify our position on the planet earth using a basis of longitude and latitude, or, equivalently, we could use the Cartesian coordinates x-, y-, and z. Similarly, since quantum mechanics is governed by states and operators that follow linear algebra rules, we can define a basis for the qubit states and operations.

Qubit state tomography is the projection of a qubit state onto this basis\cite{thew_qudit_2002}, and it fully characterizes a qubit state (its position on the Bloch sphere, for example). Similarly, process tomography fully characterizes a gate operation. Since a gate may be applied to any qubit input state, we must first prepare the qubit in each of the basis states that span the possible input state-space. Then, for each of these inputs, the gate operation is applied, and the resulting output state must then be projected and measured against the entire set of basis states at the output. This process is then repeated for each input state. Ultimately, the net results are then compared to a theoretical gate operation, and the ``distance'' between the two results quantifies the gate errors.

The number of elements in the input matrix that represents the basis states for N qubits is $ 2^{2N} $, and there is the same number of elements in the output matrix. Since the gate operation must be applied to each input state, and the result in turn projected on to each of the output states, the total number of input-output combinations is $ 2^{2N}\times 2^{2N}=2^{4N} $. For a single qubit, this is 16; for two qubits, the number quickly increases to 256. while there are a few constraints that will reduce these totals by $ 2^{2N} $, the net number of combinations that must be measured is only reduced to 12 and 240, respectively. It is another example of the power of exponential growth, but, in this case, the rapid expansion in the number of states works against us. We must make all of these measurements to characterize the gate entirely. Besides, process tomography is sensitive to state preparation and measurement errors commonly referred to as SPAM errors. While SPAM errors are real errors, they are independent of the gate operation and should not be included in the gate fidelity.

An alternative approach is called randomized benchmarking. Randomized benchmarking characterizes a gate operation's aggregate performance by interleaving it with a random (but known) assortment of other gates\cite{mcgeoch_principles_2019,mcgeoch_practical_2019}. The random gates are first characterized by themselves to assess a baseline level of error, including SPAM errors. Then, the same measurement is performed with the desired gate operation interleaved into the sequence. The outputs are then compared, and the net additional error is attributed to the addition of the desired gate operations. This process is repeated while increasing the number of gates to refine an estimate for the net error per gate. In this way, randomized benchmarking provides an aggregate fidelity that is somewhat more efficient to implement and is ideally independent of SPAM errors. The trade-off means that it averages all error channels without quantifying each individually (as is possible with process tomography).

Gate fidelity is a crucial metric that allows us to compare remarkably different qubit technologies on an equal footing. Gate fidelity is a benchmark that determines the feasibility, efficacy, and resource overhead required for implementing quantum error correction.

\section{Qubit Modalities: Electron and Nuclear Spins}

There are several physical manifestations of qubits. In the next three sections, we will discuss several qubit modalities. In this first section, we will be introduced to physical qubit modalities based on electron and nuclear spins. In the second section, neutral atoms and trapped ions. In the third section, superconducting qubits and other modalities.

Until this point in the section, we have generally discussed qubits as quantum mechanical two-level systems. However, how do we physically realize a qubit in practice? Moreover, what are the leading physical realizations today? In this section, we will have a general introduction to several candidate qubit technologies. Later, we will present an in-depth look at two of the leading qubit modalities. Let us start with electron spin. As in the introduction to quantum parallelism and quantum interference, the electron spin has two states, spin up and spin down. The question is how to isolate a single electron. In this first example, the electron spin is trapped in a quantum dot, a small region of semiconductor material where a single electron can be trapped. We start with a two-dimensional sheet of electrons called a two-dimensional electron gas, which can be realized at the interface of a slab of silicon and a slab of silicon germanium. Metallic gates are then defined on the surface of the device to define the quantum dot region electrostatically, and by applying the appropriate gate voltages, a single electron spin can be trapped there. Combinations of microwaves and baseband pulses are then used to implement quantum gates. Now we will find several key properties of quantum dot qubits shown here. We will not read through them all, as we can find them in the table associated with this section. In the next section, we will make comparisons between qubit modalities based on these numbers. Now, in addition to the single dot shown here, there are also more complex designs, including double dots, triple dots, and even dots based on CMOS devices. Each of which adds complexity to gain certain advantages. In general, the main attraction of quantum dots is that they leverage advanced silicon fabrication technology\cite{hardy_single_2019}, two, they are relatively small in area and so in principle can be integrated into large numbers. Three, much like CMOS, they are controlled using gate voltages. The main challenge is that multiple gates are required to define a single qubit. Thus, 3D integration techniques will certainly be required to realize large scale circuits. Another example of an electron spin qubit is in phosphorus-doped silicon \cite{vandersypen_quantum_2019}. Here phosphorus atoms are implanted in the silicon substrate, and the spin-off of its outermost electron serves as the qubit. As with quantum dots, electrostatic gates control the qubits using a combination of the baseband, RF, and microwave pulses. The main advantages of phosphorus-doped silicon are that it also leverages silicon fabrication technologies. The electron spin qubit has very long coherence times\cite{vandersypen_quantum_2019}. The primary challenge is that the dopants must be close to one another, around 10 nanometers, to have a sufficiently large two-qubit coupling to implement a two-qubit gate. It makes it extremely challenging to place the gates and route the wires required to control the qubits. It also means that crosstalk between control lines and qubits will be significant. Again, 3D integration will be required here. A second challenge is that the phosphorus dopants must be implanted in the silicon with very high precision. This remains an outstanding challenge. Phosphorus dopants also have a nuclear spin that can be used as a qubit controlled using radiofrequency or RF pulses. The main advantage of nuclear spin qubits is their extremely long coherence times, which arise because the spins are largely decoupled from their environment. The trade, though, is that the gate times are rather slow. A second system that affords both an electron spin and a nuclear spin is the nitrogen-vacancy center in diamond. Diamond is formed from a tetrahedral lattice of carbon atoms. Occasionally, however, a defect interrupts this lattice, and one example is called a nitrogen-vacancy \cite{childress_diamond_2013}, in which a nitrogen atom is injected into the lattice and causes the carbon vacancy to form. The result is an extra pair of electrons that form a spin 1 system, which is the lowest 2 spin levels that can be used as the qubit. Alternatively, the nuclear spin of the nitrogen atom, or the surrounding carbon atoms, may also be used as the qubit. The electron spin is addressed by a combination of microwave pulses to implement the quantum control and lasers to initialize and measure the qubit. The nuclear spin has a magnetic field dependent splitting that is controlled at radio frequencies. The advantage of NV centers is that they are rather well-suited to the interconversion and communication of quantum information. They also have reasonably long coherence times. They can even be operated at room temperature, albeit with some lower coherence. The primary challenge for NV centers is their scalability. While few qubits spin clusters local to a single NV center have been demonstrated, currently, it is not possible to place high coherence nitrogen vacancies in precise locations to create large qubit arrays.

\section{Qubit Modalities: Atomic States}
Another qubit technology is based on the internal states of atoms, or neutral atoms, as they are called, to contrast them with the ions that we will discuss next. Neutral atoms can be trapped by cross propagating optical beams, which combine to form an egg carton like potential. The qubit states are hyperfine states, resulting from an interaction between the electron spin and the nuclear spin. Such hyperfine transitions are driven at very well-defined microwave frequencies, commonly used for atomic clocks\cite{delehaye_single-ion_2018,huntemann_single-ion_2016}. These are highly stable qubits, and as such, their coherence times are very long. Thus, the gate fidelity in neutral atoms is generally limited by control errors. The main advantage of neutral atoms is their long coherence times, and the Ability to trap these neutral atoms in two dimensional, and even three dimensional, arrays. Arrays with up to 49 qubits have been demonstrated, although not yet adequately controlled. There are a few challenges, however. One is the high laser power required to trap and control these neutral atoms. Another is that loading the trap is a stochastic process. Atom re-arrangement can later be made to fill in the gaps, but it requires extra steps.
Furthermore, third, neutral atoms will require integrated optics to ultimately be scalable, something that is not yet been implemented, although concepts exist for its implementation. The next example, trapped ions, is a leading qubit modality today. Trapped ions have been used as atomic clocks for decades. These systems are stable and very well-characterized. Ion qubits generally start as atoms with two electrons in their outermost shell. Then, one of those electrons is removed through ionization. The qubit is realized as either an optical transition between orbital states of this outermost electron or a microwave transition between hyperfine states. A primary advantage is that many of the DiVincenzo criteria are satisfied for trapped-ion qubits \cite{bruzewicz_trapped-ion_2019}. To date, arrays of 10 to 20 trapped-ion qubits have been demonstrated, and surface traps in silicon are now being developed and used to both capture and control these ions in a scalable manner\cite{schafer_fast_2018}. The primary challenge is the 3D integration of optical and electrical technologies into the surface traps to make them scalable. We will discuss more trapped ions and their relative advantages and challenges at the end of this section.

A trapped-ion is a charged particle that is confined in all three spatial dimensions by electromagnetic fields. The initial goal of trapping ions stemmed primarily from the desire to perform high-precision spectroscopy and mass spectrometry on the ions. The hope was that by suspending them free in space, the ions could be interrogated for long periods of time (which improves precision), while at the same time being as weakly perturbed by their confinement as possible (which improves accuracy). However, it turns out that it is not trivial to construct such a trap due to a well-known result from electromagnetism, known as Earnshaw's theorem, which states that three-dimensional trapping cannot be achieved simply by electrostatic fields alone.

Despite this complication, two ingenious methods were developed in the late 1950s to trap ions: the Penning trap and the Paul trap, invented by Hans Dehmelt and Wolfgang Paul, respectively. The Penning trap gets around Earnshaw's theorem by utilizing a combination of static electric and magnetic fields, and the Paul trap does so by utilizing radiofrequency electric fields. For this work, Dehmelt and Paul shared the 1989 Nobel Prize in physics.

In 1975, David Wineland, who studied electrons confined in Penning traps with Dehmelt and who would receive a Nobel Prize of his own in 2012, first proposed with Dehmelt a method to cool the motion of an atomic ion in a trap to very low temperature with the use of lasers. The ultimate motivation of this laser cooling was to realize even higher precision atomic spectroscopy due to the reduction of spectroscopic shifts that result from the motion. Wineland subsequently moved on from his postdoc to the National Institute for Standards and Technology (NIST), and in 1978, he along with his colleagues there, implemented laser cooling of a cloud of trapped magnesium ions. At nearly the same time, Dehmelt and colleagues at the University of Heidelberg in Germany demonstrated the cooling of barium ions using the same method.

The demonstration of laser cooling of atomic ions, and the high-accuracy and precision spectroscopy it enabled, led to another idea: using laser-cooled trapped ions as a frequency standard, or atomic clock. In 1984, Wineland and his colleagues at NIST demonstrated this concept using beryllium ions in a Penning trap. In order to realize this high-performance clock, ion trapping and cooling of the motion were not the only essential things. Instead, having two internal (electronic) quantum states of the ion that had long coherence times was essential. This is because the certainty in the splitting between these two levels sets the accuracy and precision of the clock, and this certainty is degraded by decoherence. In addition to long coherence times, it was necessary to be able to prepare accurately, or initialize the internal quantum state of the ion before interrogation by the radiation field used as the clock's master oscillator, as well as to read out the final state of the ion after this interrogation had occurred.

In 1994, Peter Shor developed an algorithm that could be run on a quantum computer and perform the cryptanalytical task of factorization of numbers into their constituent primes exponentially faster than classical ones \cite{shor_algorithms_1994}. Later that year, motivated to some degree by the importance of factoring algorithms to the security of many modern cryptosystems, Ignacio Cirac and Peter Zoller at the University of Innsbruck laid out a framework for physically implementing a quantum computer using cold ions confined in a Paul trap\cite{roos_viewpoint_2012}. They recognized that most of the critical ingredients for quantum computing with ions were already in place due to work on atomic clocks, namely ion-motion cooling, high coherence, high-fidelity internal ion state preparation and readout, and the ability to put ions in superpositions of two internal states accurately. This last technique had been demonstrated as part of the method of clock interrogation, and it was evident that it represented the presence and control of a qubit in the trapped-ion system \cite{pino_demonstration_2020}. The chief innovation of Cirac and Zoller was their development of a method to use the motion of the ions in the trap, coupled to one another via their Coulomb repulsion, as a way to implement a so-called controlled-NOT (CNOT) gate \cite{chen_demonstration_2008}. The CNOT gate creates quantum entanglement between the ions \cite{monz_14-qubit_2011}, which is essential for the implementation of Shor's algorithm, by universal quantum computing.

Over the next few years, the CNOT gate and ion-ion entanglement were demonstrated experimentally by Wineland's team at NIST and by a group at the University of Innsbruck led by Rainer Blatt. Since then, motivated by the success of these early demonstrations and by the general interest in quantum computing, there has been an explosion of groups around the world developing and implementing techniques and technology for more sophisticated control of trapped-ion qubits. This includes a proof-of-principle demonstration of Shor's algorithm by Blatt's group using five trapped ions to factor the number fifteen. While this work has mostly been carried out by academic research groups, it has recently begun to be explored in the private sector as well, with a startup company called IonQ focused on developing a commercially-available trapped-ion quantum computer. One of IonQ's founders, Chris Monroe, worked on Wineland's team to implement the first trapped-ion CNOT gate.The Honeywell has also recently reported an effort to develop quantum computers with trapped ions. The field of trapped-ion quantum computing also relies heavily on other specialty industries, such as high precision optics and laser companies, since laser light is one of the primary means of trapped-ion quantum control, and it must be highly stable in both its absolute frequency as well as its direction of propagation.

The future appears bright for trapped ions since they offer a platform for high-coherence qubits and high-precision, universal quantum control. The main challenges going forward will be scaling systems of trapped ions to a large enough that they can move beyond the science experiments and proof-of-principle demonstrations of today to quantum processors that can outperform classical computers for practically useful tasks. In order to achieve this, there will be a need for the further development of ion control technology and techniques that do not suffer performance degradation as the system size grows. Besides, the development of quantum algorithms that make efficient use of trapped-ion qubits will be required. New academic research groups and industrial efforts will likely spring up to meet these challenges.

\section{Qubit Modalities: Superconducting Qubits}
The next example, superconducting qubits, are manufactured artificial atoms. Unlike the previous examples based on spins and naturally occurring atoms, superconducting qubits are electrical circuits that behave like atoms. Necessarily superconducting qubits are nonlinear oscillators built from inductors and capacitors \cite{goto_bifurcation-based_2016,goto_quantum_2019}. The inductor is realized by a Josephson Junction \cite{abdo_active_2019,bergeal_analog_2010,abdo_josephson_2011,}, a nonlinear inductor that makes the resonator anharmonic. We will discuss this in the following section. Anharmonic oscillators feature in the addressable two-level system that We will call the qubit. The qubit states can either be states of phase across the Josephson Junction, the flux in a superconducting loop, or even charge on the Junction island. The main advantages of superconducting qubits are that the gates are fast compared with the other qubits, and They are manufactured on silicon wafers using materials and tools common to CMOS foundries.
To date, arrays of 10 to 20 qubits have been demonstrated, including the cloud-based quantum processors that We will use in this section. The main challenge is the integration of control and readout technologies that maintain qubit coherence, even at millikelvin temperatures. We will discuss the challenges, advantages, and the current state of the art in the following section. Finally, several other qubit modalities are being pursued that are at various stages of development and maturity. The most developed and mature include linear optics quantum computing, where the presence or absence of a photon constitutes the qubit. Quantum information is processed using linear optical components like beam splitters\cite{bouland_generation_2014}, phase shifters, mirrors, and interferometers. Effective nonlinear interactions are achieved using single-photon sources, photodetectors, and the like. The main challenge with linear optical quantum computing is that it is tough to make photons interact with one another. Besides, high fidelity memory is a big challenge, and many schemes rely on probabilistic sources and gates to process information. Another type of quantum computer based on neutral atoms is a quantum emulator. A quantum emulator uses Rydberg atoms \cite{greenland_coherent_2010} or Bose-Einstein condensates to emulate a condensed matter or atomic system using tailored atomic energies and coupling terms\cite{kshetrimayum_simple_2017}. A more exotic qubit modality that is generating significant interest today is based on something called a Majorana fermion \cite{he_chiral_2017}, a fermion that is its own anti-particle. Several efforts worldwide are trying to realize a Majorana qubit\cite{zhang_quantized_2018}, using a combination of superconducting and semiconductor materials\cite{xu_coherent_2016,antipov_effects_2018}, which feature a strong spin-orbit interaction. If successfully realized, Majorana qubits have been theoretically shown to exhibit topological protection, resilience to noise \cite{wallman_noise_2016}. That is not unlike the resilience afforded by quantum error correction. The challenge is that They are challenging to realize. To date, there has been no definitive demonstration of a Majorana qubit featuring this topological protection. Other qubit candidates include molecular ions, which are trapped and controlled in a manner like atomic ions, and electrons on liquid helium, which are free electrons that form an electron lattice on the surface of liquid helium and can be controlled using electron spin resonance techniques.
In superconducting qubits, it appears that there really is only a move up and down in an energy level. how is this useful for quantum computation? From the Bloch sphere, it appears that spin and superposition play an important role in storing information that would be used in the final answer after the gate operations. How is this working? Whether it is the superconducting qubit or trapped ion qubit, whether it is a physical electron spin in a semiconductor qubit or what we call a pseudospin, a system that behaves like a spin, in all of these cases, a qubit has two energy levels that we care about, state 0 and state 1. But because it is a spin-1/2 system or a pseudospin-1/2 system, a system that behaves like a spin-1/2 system, quantum mechanically, the states of that system are described within this Bloch sphere picture that we have discussed about. so, classically, states 0 and 1 would be the North Pole and the South Pole. the excited state would have an energy that is generally higher by some value than the ground state energy is state 0. But the point is that we can put the qubit into superpositions of these states, 0 and 1. when they are in a superposition of 0 and 1, the state or the Bloch vector is pointing to any position on the earth, any position except the North and South Pole. Now, it is important to remember that It is a bit hard to describe these states in the vernacular because the words we use describe the world around us, which is classical. so, we often say that the qubit is in state 0 and state 1 simultaneously. There is some truth to that. But the thing to keep in mind, and this is what is hard to describe in words, is that It is one state. The qubit is in a state. that state could be state 0, that state could be state 1 those are classical states or it could be in a superposition state. It is a single state. But it carries aspects of both 0 and 1. so, with some weighting coefficients that we call probability amplitudes could be $ \alpha 0 + \beta 1 $  and so, it is a single state that carries aspects of both 0 and 1. So, even though we say, look, It is in a superposition of 0 and 1, that does not imply that It is in both states and that there are somehow two states. It is a single state. But it carries aspects of both 0 and 1. that is true whether It is a superconducting qubit, which, due to its design, is a pseudospin-1/2 system, or whether It is, an electron spin, literally an electron spin, that is being electrostatically trapped in a semiconducting material.

\section{Comparing Qubit Modalities}
In this section, we will compare the different qubit modalities that we just discussed. We will first do this in the context of the DiVincenzo criteria, and then compare the gate fidelity and the clock rate for a single qubit and two-qubit gates. Let us begin with a stoplight chart and the DiVincenzo criteria. Now, this is undoubtedly going to be a subjective assessment. With all the ongoing rapid technology development, the actual coloring will undoubtedly change over time. Still, it is useful to take a snapshot of the current state of the art to gain insight into the relative strengths and challenges that different modalities face, and that is the purpose here. The stoplight chart assigns colors to indicate progress. Green will indicate that a modality has made the requisite demonstrations and has sufficiently matured to the point that we can envision it proceeding to 100 plus qubits. Yellow indicates that concepts or first demonstrations may exist, but that the technology is not ready to scale to the 100-qubit level. Red indicates that no realistic concepts are currently known that would enable reaching 100 qubits. However, we are not going to consider any of these qubit modalities. They are still too nascent for the purposes here, so we will not see any red. Starting with the first DiVincenzo criteria, D1, scalable systems, and most mature are neutral atoms, trapped ions, and superconducting qubits. All have demonstrated systems of 10 to tens of qubits, with 50 qubit prototypes coming to a line soon. Silicon germanium and doped silicon qubits are yellow because they will require high wire counts to address their qubits.
Moreover, those wires need to be routed with high density, due to the relatively small qubit geometries. All these issues will need to be addressed with 3D integration, which is only just beginning to be conceived for these modalities. Besides, both doped silicon and NV centers are currently limited by the inability to implant high coherence dopants or defects with the precision required to reach the 100-qubit level. Next, for initialization D2, most technologies are doing well. The one exception is NV centers, where the strong laser used to initialize the qubit will occasionally remove an electron. This makes the defect charge-neutral, effectively destroying the qubit until it is reset. For measurement D3, the measurement of neutral atoms is extremely slow and would certainly limit the ultimate clock speed of an error-corrected system. That is why this one is colored yellow.
In terms of D4, universal gates, the doped silicon community has not yet reported a 2-qubit gate fidelity. For neutral atoms and silicon-germanium qubits, although their two-qubit gates have been demonstrated, they still have a relatively low fidelity, around 80\%, which is why these are colored yellow. In terms of coherence, D5, all these technologies, grade well. For D6 and D7, the interconversion and communication of quantum information, there has been no published work yet for silicon germanium or doped silicon approaches. However, it is anticipated that conversion to microwave photons should follow from the same approaches used for superconducting qubits. Now, as we can see, both trapped ions and superconducting qubits have made significant progress toward meeting the DiVincenzo criteria. It is one reason they are viewed today as leading candidates. Let us turn now to the gate fidelity and gate speed of these modalities. The number of operations before an error occurs directly related to the fidelity, and they are plotted here against the gate speed for single and two-qubit gates. The upper right corner represents the best performance.
Additionally, it is desirable to be above the dashed red line, which is representative of the most lenient threshold for quantum error correction. What we see is that trapped ions have the highest single qubit fidelity. On the other hand, they are about 500 times slower than a superconducting qubit gate. For two-qubit gates, superconducting qubits and trapped ions have similar levels of gate fidelity. However, again, the superconducting qubit gate is about 1,000 times faster than the ion traps. Some technologies, like doped silicon, have measured and reported single-qubit gate fidelities, but They are still working on their two-qubit gate fidelities. Again, we can see that both trapped ions and superconducting qubits are leading modalities today. In the case studies, we will take a detailed look at each of these technologies.

The following table provides a snapshot of the current state of technology for several leading qubit modalities as of 2020. Entries labeled ``TBD'' have not been reported for a given qubit modality.

\begin{figure}[H] \centering{\includegraphics[scale=.5]{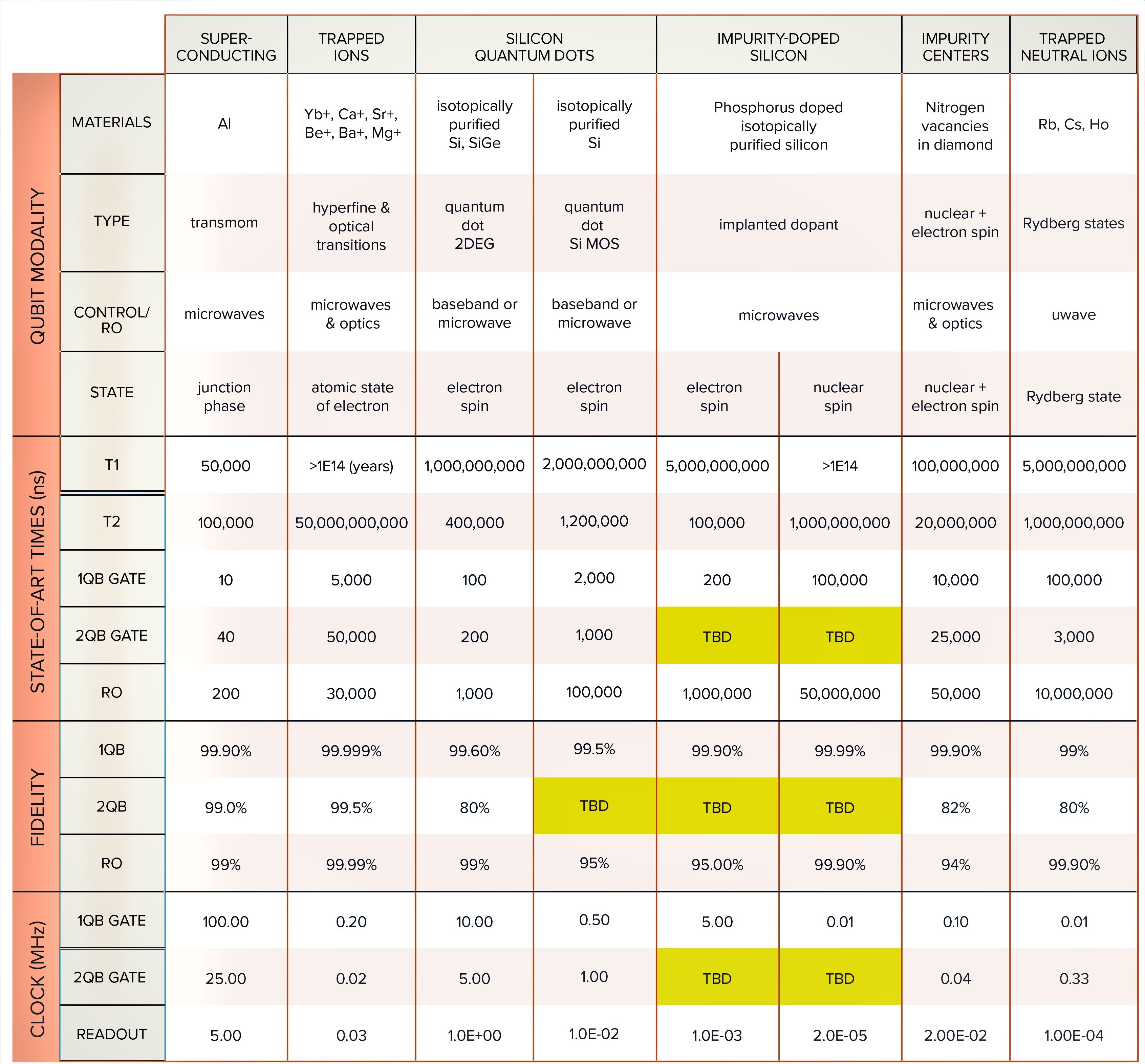}}\caption{Quibit Modality}\label{fig1_27}
\end{figure}

The table below is the reproduction of the ``stoplight chart'' from the section. It provides us with an overview of several leading qubit technologies and their performance concerning the seven DiVincenzo Criteria.

\begin{figure}[H] \centering{\includegraphics[scale=.5]{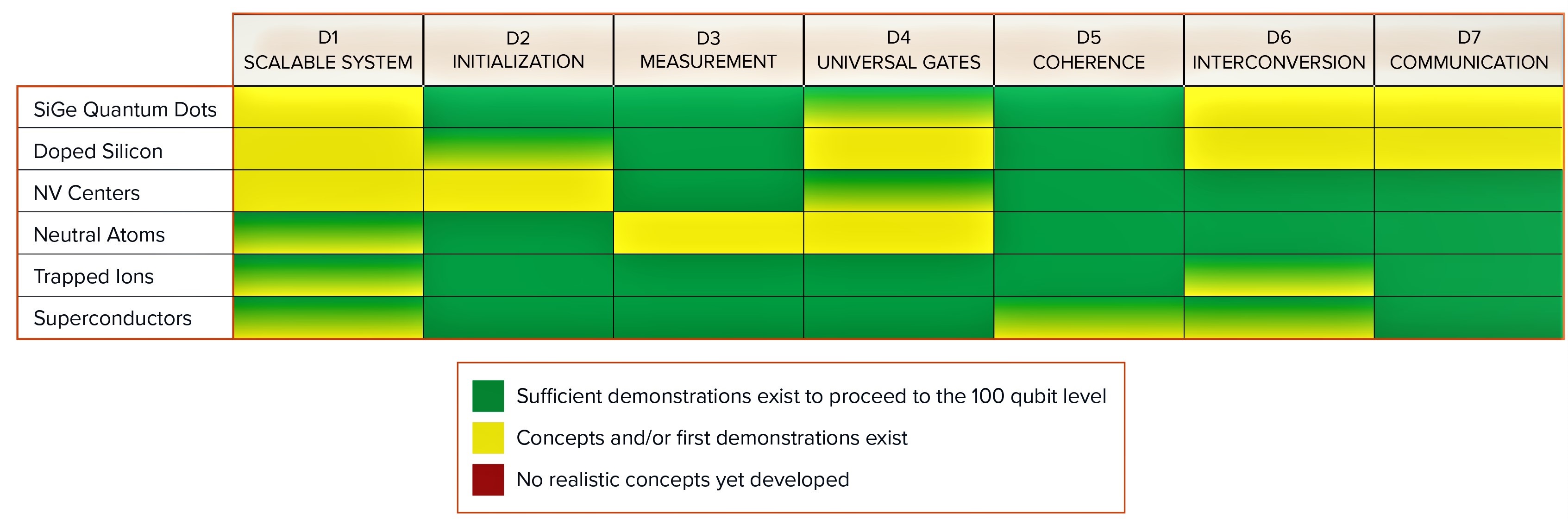}}\caption{DiVincenzo Criteria}\label{fig1_28}
\end{figure}

This evaluation is somewhat subjective, but it serves to place these technologies on a common footing to compare progress concerning one another. It indicates where each technology excels and where improvements are needed.

The DiVincenzo criteria, which is, what is the minimum requirement for a particle or a system to be used as a qubit? And among all of the elementary particles in the standard model, which ones can or cannot be used and why? So, we believe, the DiVincenzo criteria specify what is the minimum again, minimum requirement for a technology or modality to be considered for quantum computing in a scalable sense. so there are many engineering problems beyond that to actually truly scale to a large size. But the DiVincenzo criteria look, whatever technology we have, it has to at least meet these requirements. so, for example, it has to be a uniquely identifiable and addressable two-level system. we need to have a universal set of gates typically, a handful of single and two-qubit gates. so, these were the DiVincenzo criteria. It has to have a long coherence time, for example. so, any technology that we might imagine, we wonder if this would be a good candidate to be used in quantum computing, would start with those five criteria. Now, if we look at the elementary particles in the standard model. we think that we can take out some examples of ones that would be good examples and one that would be a bad example. So, first, a good example is the one that is currently being used, which is the electron and its spin. So, the electron is a spin-1/2. whether this is an electron that arises from doping a semiconductor, with phosphorus the phosphorus dopant has one extra electron and that electron is then used as the quantum bit of information or whether that electron is, say starts out in a two-dimensional electron gas and, through electrostatic gating, we either deplete most of the electrons away and leave just one that is a depletion mode semiconductor quantum dot, or There is another version, which is an accumulation mode, and we start with no electrons and then we apply a positive voltage to a gate so, that we pull up one and only one electron however we gather that electron or isolate that electron, it is a spin-1/2. It is been shown that these spin-1/2 electrons are controllable. We can address it with either electric or magnetic fields. So, the spin is magnetic. But through spin-orbit coupling, for example, one can use an electric field to drive a rotation of the spin. There are single-qubit gates. contemporary research now demonstrating two-qubit gates with, doped silicon\cite{broome_two-electron_2018}. so, we think that the electron is a good example of an elementary particle that would make a good qubit. Some other examples electron and helium surface. People have done experiments where they try to isolate single electrons on the surface with the helium. It is less mature. But that would be another example. So, what is an example of a particle which may not be such a great qubit? And we can think of one off the top of my head, which is a neutrino. So, a neutrino we might think, They do not interact with anything. so, their coherence time is presumably infinite. true enough. But it is not enough to have a long coherence time. we also have to have a reasonable gate time. It is actually the number of gates that we can apply within that coherence time that matters, how many gates per coherence time. How many gates can we apply before decoherence sets in? That is the metric. It is related to the fidelity of the qubit. so, a neutrino, although it lasts forever as we know, it is very hard to detect these neutrinos because It is very hard to couple to them and to control them. So, even if we could grab onto a neutrino, which we think is very difficult, and hold it in isolation, where we could operate on it, It is very challenging to actually perform the analog of quantum operations on such a neutrino because it interacts so, weakly with its environment. so, this is an example of a particle which would not make a good qubit simply because the gate times would be far too long so, perfectly isolated, but useless for quantum computing. NV centers and doped silicon modalities are still working to meet criterion D1, as they are currently limited by the inability to implant high-coherence dopants or defects with high precision reliably.
Besides, NV centers are still addressing criterion D2, as the strong laser used to initialize the qubits will occasionally remove an electron and make the defect charge neutral. The doped silicon modality qubits are working towards criteria D4, D6, and D7. Two-qubit gates are just now being demonstrated, and definitive fidelity measurements are being performed. The interconversion and communication of quantum information have yet to be demonstrated.

\section{Trapped Ions: Introduction}

We will explore todays leading trapped ions qubit modality engineering, science, and technological aspects. Trapped ions were the first qubit technology, primarily due to their historical use in atomic clocks and precision measurements. Ions start as neutral atoms, the chemical elements in the periodic table. For example, we will discuss strontium in this section. A strontium ion is formed when its outermost electron is removed, and a process called ionization. With one less electron, the atom now has a net positive charge, so it is no longer charged neutral, and We will call this an ion. Because the ion is charged, it can be trapped or held in place using oscillatory electromagnetic fields. Although these fields used to be applied using large electrodes arranged in a three-dimensional configuration, today, they are implemented using surface traps electrodes manufactured on silicon wafers that hold the ions just above the wafer surface. Trapped ion qubits are one of the leading modalities being developed to realize a quantum computer, and it has several advantages. From a business perspective, trapped ions are attractive because they leverage a substantial existing technology base, one that goes back three decades to the development of atomic clocks and mass spectrometers. Besides, trapped ions today are fundamentally silicon-based technology. The surface traps themselves are fabricated on silicon wafers, and they use standard semiconductor fabrication tools and techniques. Also, all the critical control and readout circuitry can be integrated with existing CMOS electronics. For example, the electrode voltage is used to hold and move the ions, or the integrated photonic waveguides and gradings used to control the ions optically.
Furthermore, even the photodetectors used to read out the ions are all compatible with existing CMOS foundries and materials. In this sense, they are lithographically scalable to large numbers of qubits. Today, there are a growing number of commercial efforts pursuing or supporting the development of trapped ion qubits, including a startup company called IonQ. From an engineering perspective, the surface traps used to capture and hold ions are designed using standard CAD software and layout tools. The electrode control electronics are CMOS circuits, and They are designed in the same way as conventional CMOS electronics.
Similarly, the photonic waveguides and couplers are designed and integrated directly into the silicon. Scientifically, state of the art trapped ion quantum processors has demonstrated several prototype quantum algorithms, such as Shor's algorithm at the 10-qubit level, and larger-scale circuits at the several tens of qubit level are expected soon. Although trapped ions operate more slowly than superconducting qubits, they have demonstrated a more significant degree of connectivity between the ions. It may help offset the relatively slow speed by making problem embedding more efficient. Finally, from a technology standpoint, trapped ion qubits leverage both microwave and data communications technology. The lasers used in trapped ion experiments have an application to atomic clocks, precision navigation, even biology, and optogenetics. Developing compact instrumentation, for example, size, weight, power, even cost, will benefit an array of technologies even beyond trapped ions.

\section{Trapped Ions: How They Work}

In the following section, we will introduce how trapped ions work, from the ionization of neutral atoms to how they are captured using surface traps. MIT Lincoln Laboratory, is developing technologies to enable practical quantum information processing, including quantum computation, simulation, and sensing, \cite{stuart_chip-integrated_2019,bruzewicz_trapped-ion_2019,sedlacek_evidence_2018,niffenegger_integrated_2020}. Starting with neutral atoms with two electrons in the outermost level, we can remove one of these electrons, producing an ion with a positive charge that we can hold onto very tightly using electromagnetic fields. In machines like these, we can trap individual atoms, the smallest amount of an element in the case, strontium, or things like calcium. That is one atom. That is something like smaller than a tenth of a nanometer. we manipulate them with lasers and electromagnetic fields such that we can hold onto them for many hours and manipulate their quantum states to do things like quantum information processing. So, in an individual trapped ion, the amount of time that the quantumness survives, or the coherence time, can be made to be very long. In systems like this, where people regularly manipulate trapped ions, this coherence time can be seconds, even minutes. this allows us to do lots of coherent quantum operations between larger numbers of qubits in the amount of time available to do anything with those quantum systems. Other advantages are that the error rates in the quantum operations that we perform single qubit, double qubit operations that are keys to quantum information processing\cite{behnia_tachyon_2018}, and the preparation and readout of the quantum states can be done with very low error rate in trapped ion systems. Because these atomic systems are very clean, single quantum systems that we can control very well, we can theoretically predict exactly how They are going to behave. We know where they are, and we can manipulate their states very, very cleanly. We have defined the qubit state within a single ion, but how do we hold on to one atom? Utilizing the positive charge of the ion, we can trap it in a combination of static and dynamic electric fields. By applying a radio frequency electric field to two diametrically opposed rods out of a four-rod square configuration, we can produce a quadripolar field in two dimensions. At any one moment in time, this field will tend to push an ion toward the center of the trap in one dimension and push it away in the other. However, since this field oscillates in time, for the right combination of frequency and amplitude, the ion can effectively be pushed toward the center in all directions as if it were in a two-dimensional bowl. In the third dimension, we can apply a combination of static voltages to trap the ion. This is the same technology used for mass spectrometry, and we see potential offshoots of ion quantum information processing in improvement to small sample sensing and identification. A promising approach to large-scale ion control and manipulation is based on chip-based traps, as shown here. By patterning metal using standard microfabrication techniques onto an insulating substrate, we can produce a flat version of the four-rod trap structure We just described. We will describe in a bit, and ions are trapped in space above the center of the trap where the quadripolar trapping field is produced. To hold onto an individual ion for a long time, a very low-pressure environment is required, as molecules in the air have enough energy to knock the ions out of the trap. Ultra-high vacuum conditions with background pressures more than 12 orders of magnitude less than standard atmospheric pressure can be maintained in specialized chambers. These extremely low pressures are produced in the laboratory. It employs a cryopump, in which various surfaces in the chamber are made very cold, within a few degrees of absolute zero, utilizing refrigeration. Much as water condenses from the air onto a cold beverage container on a humid day, cryo-pumping results in condensation of almost all the air's constituents quickly and dramatically lowering the pressure in the closed vacuum chamber. Ions can be maintained for hours to days in UHV systems like this. It is important to point out that the ions themselves are not cooled in this manner. They are cooled too much lower temperatures, eventually to only a few tens of millionths of a degree above absolute zero, using lasers. Perhaps counterintuitively, it is possible to remove motional energy from atoms by allowing them to absorb light at a very particular frequency. They re-emit light at a slightly higher frequency, leading to a reduction in motional energy. This laser cooling process developed 30 years ago, is a key enabling technology for quantum computing with atomic systems. This animation shows the process by which we get individual ions into the trap. Inside the vacuum system, we heat a piece of metal to a few hundred degrees celsius, producing hot atomic vapor. Using a combination of laser beams and a magnetic field gradient, we cool approximately a million neutral atoms into what is known as a magneto-optical trap. These cold atoms can then be accelerated toward the ion trap chip using another laser beam. With the radio frequency trapping potential applied to the trap electrodes, neutral and ionized atoms, using other lasers within the trapping volume, will feel the trapping fields and be localized within the trap, where they are cooled with yet another set of lasers, these resonant with transitions in the ion. The result is a single atom held in space approximately 50 micrometers from the surface of the chip. We can image the ions using the light they scatter during cooling, producing images like that shown here.

\section{Trapped Ions: Qubit Operations}

In the following section, we introduces how to control and measure the states of trapped ions to implement universal quantum computation. It is done using pulses of coherent laser light, of carefully controlled amplitude and phase, applied to the trapped ion. Each ion encodes roughly one qubit, and multiple ions can be trapped next to each other, to allow multi-qubit systems to be controlled and measured.

Once we have the ions, how do we manipulate the qubits? Single-qubit gates are brought about by applying laser pulses resonant with the qubit transition for a certain amount of time. If the qubit starts in the ground state as a function of the laser pulse duration, the qubit state coherently oscillates between the ground state and the excited state, performing what is known as Rabi oscillations. By choosing the appropriate time to stop the Rabi oscillations, we could perform a flip of the quantum state from 0 to 1, producing a quantum version of the classical NOT gate, or the inverter. In this way, we perform a $ \pi $ pulse.
Interestingly, if we use a laser pulse half as long as a $ \pi $ pulse to perform a $ \pi/2 $ pulses, that is a gate with no classical analog, we produce in the ion qubit a superposition state, 0 and 1 at the same time, where the electron is effectively in two states at once. These manipulations form the basis of single-qubit operations on ion qubits. Along with single-qubit gates, 2-qubit gates are required to do arbitrary quantum computations. We bring about these operations utilizing the strong Coulomb interaction between two positively charged ions. Two ions in the same trap share vibrational modes due to their interaction, much like two balls with a spring connecting them. These modes can be excited, depending on their internal qubit states, using laser beams. Thus, the internal qubit states can be entangled through the quantum vibrational mode channel used as a quantum bus.
The basic 2-qubit gate is a controlled-NOT gate, somewhat analogous to a classical exclusive OR gate. It can be produced using this bus interaction in combination with a few single-qubit gates, we just described. If one ion starts in an equal superposition state, the state after application of a controlled-NOT gate will be a maximally entangled state, capable of effects like Einstein called ``spooky action at a distance.'' Not only is this spooky action the way entangled systems work, as verified many times in laboratories around the world, it is also a fundamental component that large-scale quantum computers will rely on for their operation. Finally, after a series of single and 2-qubit operations, as we would perform to carry out an algorithm, the quantum state of the ion qubit must be measured. Ions allow a particularly useful mechanism for high-fidelity state readout when compared with other systems\cite{harty_high-fidelity_2014,ballance_high-fidelity_2016,wang_high-fidelity_2020,gaebler_high-fidelity_2016}. By illuminating the ion with light resonant with an auxiliary transition, we cause the ion to absorb and re-emit light on this transition if it is measured in zero states.
In contrast, the light will be off-resonant if it is measured to be in the 1 state, and the ion will be dark. The ion state may have been in a superposition before the measurement, but during illumination, the state will be projected to 0 or 1 and remain there. It forms a quantum non-demolition measurement, allowing us to scatter many photons and get useful statistics on the ion state. By setting a threshold on the number of photons we detect from the ion during measurement using a single photon-sensitive detector, we can measure the ion's state with very high fidelity. Therefore, we have established techniques to perform all the operations required for quantum computing using trapped ions, and the properties of the ions themselves allow for very low error in these operations. Researchers in the field of trapped ion quantum computing have demonstrated basic quantum computing primitives in a few ion experiments that approach or surpass the fidelity levels we think we need for useful large-scale systems. Coherence times for a single ion are about 10 minutes long for quantum properties to persist, even becoming comparable to human time scales. Single-qubit gates, with errors at the one-in-a-million-one level, have been achieved using microwave fields.
Furthermore, 2-qubit gate fidelities are at the 99.9 percent level in experiments performed by a few different groups. Besides, a few tens of ions have been trapped and individually manipulated as well, though not simultaneously with the highest fidelity gates. The remaining challenge is to maintain the exquisite level of quantum control that is possible while scaling systems to many ions. It must include providing control and readout capability for arrays of ions without simply multiplying up the number and size of the bulk optics and external electronics setups used today equipment that can take up a small room. It is an exciting area of current research.

\section{Trapped Ions: Chip-Scale Integration Technology}

In the following section, we introduces photonic integration technologies that are being developed to engineer larger-scale surface traps for multi-qubit trapped-ion processors of the future\cite{bruzewicz_trapped-ion_2019,bruzewicz_dual-species_2019,loh_brillouin_2020}. One of the most significant technical challenges for trapped ion quantum computing systems is providing all of the laser controls needed. It is precisely focused on the many ions held just about a hundred micrometers above the surface of a microfabricated chip. we explains how this can be accomplished by integrating optical waveguides into the chip, and lithographically patterning optical gratings onto the chip's surface, such that light traveling within the chip can exit, then come to a focus, on single atomic ions above the chip.  This laser light needs to come in many different colors, ranging from blue wavelengths to red and far-infrared.  Surprisingly, the technology exists to guide all these colors within the integrated optics. As we discussed previously, all the required operations for performing quantum computing with trapped ions have been demonstrated in research groups around the world. However, this has been achieved in only a few-qubit system, that is, with around 1 to 10 ions. Perhaps the most significant obstacle to realizing a practically useful quantum computer is the current lack of ability to control and measure huge numbers of ions. we think We will need thousands to millions, with the same exquisite precision demonstrated in the few-ion systems. It is the direction, the direction of so-called scalability, in which the field is looking. There are four key things we need to do with the trapped ion qubits. We would like to go through them and discuss the technology being developed to address how we might do each one on a large scale. That is, on many ion qubits. First, ions need to be trapped or held in fixed positions. we already discussed that this is done with a combination of both static and oscillating voltages applied to metal electrodes placed around the positions. We would like the ions to be trapped in. These electrodes are typically centimeter to millimeter scale and are made in standard machine shops. It works well for few-ion systems, but to trap huge arrays of ions finely detailed and complex electrode structures are required. this is something challenging to achieve with machining techniques. We would like to utilize micro or nanofabrication techniques, like those employed for making classical computer chips, where metal layers are deposited on wafer substrates, like silicon or sapphire, and are subsequently deposited on wafer substrates finely patterned using what is called optical lithography. Fortunately, we have figured out a way to use these techniques to make ion traps. It turns out it works to unfold the electrodes normally placed around the ion onto a plane that lies below it. By applying a similar set of voltages to these planar electrodes, the ion can be trapped above the surface of the plane, with the surface to ion distance scale being set by the size and spacing of the electrodes. Since these planar electrodes can be fabricated using microchip techniques, they can be made small, and in an arbitrary pattern. It allows us to produce large numbers of zones arrayed in two dimensions for large numbers of ions to be trapped in, as well as to reduce the size of the electrodes to the micrometer scale. In these types of traps, which we call surface electrode traps, ions are typically trapped a few tens to 100 microns above the chip surface. An important additional benefit of these microchip surface electrode traps is that they provide a platform for integration of a host of other key ion control technologies. We have the potential to put anything we can currently put in classical computer chips and more into these ion traps. The second key thing we need to do with the ions is to control their internal quantum states. That is, perform the actual quantum gates or quantum operations. Here, we are including the initialization steps of cooling the ions' motions and setting the internal electronic states in which the ions begin a computation. As we already discussed, this is done primarily with lasers. For the types of ions, we plan to use, we require about a dozen different laser wavelengths, ranging from the near-ultraviolet to the near-infrared, pretty much over the whole visible spectrum. These layers are directed and tightly focused on the ions housed inside a vacuum chamber, through the chamber windows using large numbers of bulk optics, like mirrors and lenses.  It is a pretty effective way to control a few ions, but it is nearly impossible to imagine how to use bulk optics to address a large ion array. We need to find a way to focus a dozen different colored laser beams on each ion in a highly controlled manner. For example, we will want the ability to hit one ion that resides in the middle of a large two-dimensional ion array with one laser beam without hitting any of the other ions, which would lead to operation error. A dream would be to plug a fiber for each color of light we need into the chip and have that light routed around the chip, much like a metal trace or wire routes electrical signals. This light will be split up into many paths of the chip, and then directed and focused out of the chip plane to each ion location. Perhaps surprisingly, this technology exists. It is called integrated photonics, and we are beginning to incorporate it into ion traps. we can think of integrated photonics as basically fiber optics on a chip. These tiny fibers called waveguides are made by depositing the right materials, which must be transparent over the visible spectrum, onto the trap chips. These materials are then patterned using the same techniques used to pattern the ion trap electrodes. These waveguide patterns define the paths that the laser light travels along, and we can design them to split the light from one path into many branches. To get the light out of the chip and onto the ions, we can pattern periodic gaps in the waveguide material that act like a diffractive grating. This grating will bounce and focus the light in the vertical direction. This out-coupling process works in reverse. it turns out to be a very effective way of getting the light into the chip from a fiber optic cable coupled to the laser source. Lincoln Laboratory have demonstrated quantum control of trapped ions with light integrated into a surface electrode ion-trap chip in collaboration with MIT. In this case, the waveguides run below the trap's metal electrodes, and we pattern holes in the electrodes so that the light can get through to reach the ions.

\section{Trapped Ions: Leveraging CMOS for Integration}

In the following section, We introduces CMOS integration technologies that are being developed to control the ion trap electrodes needed to hold and shuttle ions, and integrated photodetectors for readout in larger-scale surface traps for multi-qubit trapped-ion processors of the future. The third thing we need to do with the ions is to measure or read out their quantum states. As explained earlier, this is done by counting photons that are emitted from the ions when a readout laser illuminates them. This readout needs to be done fast, and the speed is determined by the number of photons we can collect from the ion and how quickly and efficiently the photon detector can give us a click when the photon hits it. Ions emit light isotopically, or in all directions, so it is not very easy to collect it all. For few-ion systems, the light collection is typically done with a huge lens, which, due to its size, is located outside the vacuum chamber. It can collect a few percents of the total emitted light and send it to a detector, such as a photomultiplier tube that converts individual arriving photons into short electrical pulses that we can count on electronics. It works well because we only must collect from a tiny region in space where the few ions reside. For large arrays of ions, this technique is highly inefficient. Instead, we are now working to eliminate the large collection lens and integrate the photon detectors into the ion trap chips with a detector right below each ion\cite{mueller_simulating_2011}. Since the detector is located only a few tens of micrometers away from the ion, they can be made very small and still collect the same amount of light as a big lens placed far away. At the same time, these detectors can be arrayed around the ion trap chip in huge numbers to match the size and pitch of the ion qubit array. It can, in principle, be a very efficient and scalable way to collect and detect light in a trapped-ion quantum processor. The detectors that we are using for this purpose are known as Avalanche Photodiodes or APDs. They are routinely fabricated by academic research groups and by industry using the same facilities and techniques for microchip technology. there is, therefore, a clear path to incorporating them in ion traps. We could detect photons emitted from trapped ions using these integrated APDs. The fourth thing we need to do with the ions is moving them around. We call this shuttling, and it provides a capability that is unique to the ion qubit modality. To create the large entangled states required for practical quantum algorithms, we must find some way to move and distribute quantum information around our quantum computer. Ions can be shuttled by changing the voltages applied to the trap electrodes. The voltages are typically generated with electronics located outside the vacuum system and brought to the electronics via long wires that are connected to the chip around the edges. It is done routinely and is straightforward for few-ion systems because there are few electrodes and corresponding voltage sources required for such a small scale. However, for large arrays of ions, we will require lots of electrodes, and therefore lots of voltages and wires. Think about 10 per ion. Once again, we lean on integration to solve this scalability problem. Lincoln Laboratory have begun to develop tunable voltage sources integrated into ion traps, with an individual voltage source below each electrode and connected through on-chip wiring. Commercial CMOS foundry facilities can be used to fabricate these integrated electronic devices because they are built from the same transistor technology used in classical computer chips. These individual voltage sources called Digital to Analog Converters, or DACs, can all be programmed and varied at high speed with digital signals that come down just a few wires connected to the trap chip, like the USB communication protocol. We would also like to point out that, as we go forward, there are other electronic devices that we can think about integrating, things like circuits that detect the electrical pulses from the photon counters, or even classical computing processors can store and manipulate classical information. As we might expect, we ultimately need to integrate these different control and measurement technologies together into one chip. It means that we need to use a fabrication process that is compatible with it all. It is not a trivial concern since the performance of the required devices is very sensitive to the fabrication techniques and the materials used. We keep this important constraint in mind as we develop the scalable technology, and make sure we are keeping everything compatible. All the technologies We have discussed are completely compatible with advanced CMOS fabrication techniques and materials. It means that we are truly poised to take advantage of all the incredible scalable technology that has been developed for classical computers, propelled by billions of dollars of investment over decades for the quantum computer. It is the vision of what a scalable trapped ion quantum computer will look like. Ions are trapped above the plane of a surface electrode ion-trap chip. They are individually addressed and controlled with laser light delivered by integrated photonics, and light emitted by the ions is detected with integrated single-photon APDs below the trap electrodes. The ions are shuttled around by varying the voltages on these trap electrodes using integrated electronic circuits. The movie shows a quantum computing algorithm, and we may think to ourselves, this does not look like a computer. We would say that we are trying to build something that computes in a fundamentally different and transformative way. we would argue that we should not expect it to look like anything We have ever seen.\\
\textbf{Measuring Ions:}

The state of an ion is measured by illuminating it with light that resonates with an auxiliary transition, depending on the qubit's state. It means the transition will ``light up'' or ``stay dark.'' In the following figure, the arrow pointing up indicates that the electron transitions to a higher-energy auxiliary state. The arrow pointing down indicates that the electron goes back to the initial state. If the electron goes from the auxiliary state back to the ground state, it will emit light (light up); if not, it will stay dark.

\section{Superconducting Qubits: Introduction}

In this section, we will explore the leading qubit modality superconducting qubits. We will take a closer look at superconducting qubits. Superconducting qubits are electrical circuits built from inductors, capacitors, and Josephson Tunnel Junctions that exhibit quantum mechanical behavior when cooled to millikelvin temperatures. It is often said that superconducting qubits are ``artificial atoms.'' Furthermore, what meant by that is that these circuits can be designed to have properties like atoms. For example, the energy level separation between state 0 and state 1, or their sensitivity to environmental noise, can all be determined by design.
Furthermore, this is quite different from qubits based on elements that we find in the periodic table, where we just go with what nature provides. Today, superconducting qubits are one of the leading modalities being developed to realize a quantum computer, and there are several advantages. From a business perspective, superconducting qubits are attractive because they leverage a substantial existing technology base. For example, superconducting qubits are fundamentally silicon technology. They are fabricated on silicon wafers. They use standard semiconductor fabrication tools and techniques.
Furthermore, materials like aluminum or titanium nitride are fully compatible with CMOS foundries. In this sense, they are graphically scalable to large numbers of qubits. Many major corporations are pursuing superconducting qubits, including Google, IBM, and Intel, as well as startup companies like D-Wave \cite{mcgeoch_practical_2019}and Rigetti. From an engineering perspective, superconducting qubits are designed in much the same way as classical transistor-based circuits. They use the same kinds of CAD software for layout and simulation and have many of the same considerations as larger-scale computer chips, for example, when routing wires to and from the devices. Scientifically, state-of-the-art superconducting quantum processors are pushing 50 to 100 qubits.
Furthermore, this is at a level where even the most massive foreseeable classical computers, let alone the ones we have today, simply would not be able to simulate the quantum computer's behavior or its output. When this happens, we will have reached a point that is often referred to as quantum supremacy\cite{arute_quantum_2019,markov_quantum_2018}, a point where a quantum computer has performed a task that could not be calculated precisely on a classical computer. Finally, from a technology standpoint, superconducting qubits leverage and rely on existing technologies, such as microwave control electronics and microwave packaging \cite{randall_efficient_2015,paraoanu_microwave-induced_2006}, often using the same frequency bands used in cell phone electronics.

\section{Superconducting Qubits: How They Work}

We will introduce how quantum mechanical artificial atoms can be built from superconducting electrical circuits, including inductors, capacitors, and Josephson junctions. We will also introduce dilution refrigerators, which cool these artificial atoms to near absolute zero degrees Kelvin, and why they are needed for superconducting quantum computing \cite{rosenberg_3d_2017,rosenberg_3d_2019}. Superconducting circuits are a quantum computing modality where quantum information is stored in superpositions of charge and current \cite{song_10-qubit_2017}. In a superconducting metal, as we cool it down below some transition temperature, the electrons pair up into what known as Cooper pairs after one of the inventors of the most successful theory of superconductivity, the BCS theory. The B is for Bardeen, and the S is for Schrieffer. Superconductors have many applications. there is a large market for them far beyond for quantum computing. Superconductors have zero resistance at DC frequencies. So, they are useful for applications where we have large currents, for example, to create large magnetic fields. So, if we get an MRI, the large magnetic fields will probably be generated by currents going around in superconductors. If we tried to make those fields out of regular wire, we would have so much heating that it would not work. Superconductors are also sometimes used for maglev trains and low power classical digital circuits. One of the simplest types of the superconducting circuit is just a linear harmonic oscillator made up of some inductor, L, and a capacitor, C. In an LC oscillator, if we put in energy at the right frequency, it will cycle between charging up the capacitor and putting current through the inductor. How fast this happens depends on the values of the capacitor and the inductor. However, for the circuits we are discussing, it happens a few billion times a second or at gigahertz frequencies. It is also the same frequency that many consumer electronics work at, like our Wi-Fi network at home or our cell phone, which means that we get to use much equipment that is being developed over the years to meet growing consumer and commercial needs. A linear harmonic oscillator has equally spaced energy levels. the energy spectra are quantized, meaning we can only have a discrete number of excitations. we can observe this quantization provided our circuit is cold enough that the energy associated with the temperature is much less than the energy level spacing. Our materials are good enough that we do not lose much energy every oscillation period. It is one of the reasons we need to use superconductors. If the metal were normal, we would lose too much energy per oscillation period. However, there is a problem with using a simple linear oscillator as a qubit. If we have a qubit, we want to be able to move it around the Bloch Sphere into superpositions of 0 and 1. However, in a linear oscillator, all the energy levels are equally spaced. It means that if we are in the 1 state and try to put in the energy to make the qubit go to the 0 states, we can end up in the 2 states or some higher state instead. Thus, we cannot think of the system as a qubit anymore, because a qubit should only have two energy levels. The way to make a linear oscillator into something that can be used as a qubit is to introduce a nonlinear element. That will make it, so the energy splitting between the 0 and the 1 state is different from the energy levels between the other states. The most common way of doing this is to use a Josephson Junction or JJ. Brian Josephson predicted JJ's in 1962. A JJ is just a very thin insulating barrier between two superconductors. If the barrier is thin enough, the superconducting electrons can tunnel through the barrier without losing any energy. The most important thing about a Josephson Junction for the purposes is that the inductance depends on the current going through the junction. It is different from a loop of wire where the inductance is just a fixed value. the nonlinear dependence of the inductance changes the energy spectra, so the levels are no longer equally spaced. It means we can uniquely address one energy level that we are going to make into the qubit. One of the DiVincenzo Criteria is that we must be able to initialize the system. If we want to be able to put the system in the ground state and have it stay there, thermal excitations must not excite the qubit out of the ground state. That means we need to be cold. Exactly how cold this is depending on what the energy levels are. A typical superconducting qubit frequency of 5 gigahertz corresponds to about 250 millikelvins. So, we need to be much colder than that, around 10 millikelvin or so. we can get to these temperatures using dilution fridges. It is a three-case stage, 3 Kelvin, so 3 degrees above absolute zero. We get to that temperature by using a pulse tube cooler, which we might be able to discuss in the background. then to get colder, we use a dilution refrigerator, which uses a mixture of helium-3 and helium-4. Helium-3 is just a lighter isotope of helium-4. the basic concept behind a dilution refrigerator is that it is like the way we cool a coffee cup. we blow across the top of it. what we are doing is removing the vapor. what must happen is more coffee has to come out of the liquid and into the vapor, and that takes energy, and that is how our coffee cools. So, we can use the same trick with helium-3 and helium-4. Dilution fridges used to be very specialized, hard to use, and required liquid helium. Now, in part driven by the interest in quantum computing, companies are making automated systems that are much easier to operate and to maintain. They also do not require a continuous source of liquid helium to run, only electricity.

\section{Superconducting Qubits: ``Artificial Atoms''}

In the this section, we introduces artificial atoms as they are used for superconducting quantum processing, including their coupling to resonators for control and readout, coherence times, and single-qubit and two-qubit gates. One of the neat things about superconducting qubits is that they are macroscopic things, but they act a lot like atoms. They have discrete energy levels, and we can couple them to cavities just like we can with atoms. With atoms, we have cavity quantum electrodynamics, or cavity QED, that describes how light in a cavity interacts with atoms. With superconducting qubits, we have circuit QED, which describes how light in a cavity interacts with superconducting circuits\cite{devoret_superconducting_2013}. The math is the same. However, unlike with atoms where we are limited to whatever is on the periodic table, with superconducting qubits, we can engineer the energy levels to be whatever we want by changing the values of the capacitors and the sizes and numbers of the Josephson Junctions. Therefore, superconducting circuits are sometimes called artificial atoms. Because They are like atoms, but we can engineer their energy levels. Here is a picture of a fabricated superconducting qubit. The two squares of metal make up the capacitor. In between, they are a loop of superconducting material that has JJs in it. In this qubit, it is easiest to think of the encoded information in currents moving clockwise and counterclockwise. To control the qubit, we can send in a pulse of microwave energy at the qubit is frequency \cite{randall_efficient_2015}. The phase and length of the control pulse will determine how much the qubit rotates around the x and y-axes of the Bloch sphere. So, this instrument is called an arbitrary waveform generator.
Moreover, we can think of as mostly the central command for the experiments. So, these are what we are going to load the qubit pulses, the entangling pulses, and the readout pulses. So, it is going to send, for example, our INQ signals to our qubit. It could also send an INQ signal to our readouts. So, this would be enough to control, at a minimum, a single qubit. By multiplexing, that is, by sending signals at multiple frequencies down the same lines, we can use a unit like this to control multiple qubits simultaneously. Typical single-qubit control pulses are tens of nanoseconds, which is very short compared to the best coherence times of 100 microseconds, and fidelities greater than 99.9\% have been achieved.
Now, we also need to be able to do two-qubit gates between two superconducting qubits. There are lots of ways to a couple of superconducting qubits to each other, including coupling them directly to each other through a capacitive, or inductive interaction, or mediating the coupling with a resonator or another qubit. In many coupling schemes, it is necessary to change the qubits' energy levels, which we can do by changing the magnetic flux through the loop. Two qubit gate times range from tens to hundreds of nanoseconds. Gate fidelities greater than 99\% have been demonstrated. Now, at some point in the quantum algorithm, we also need to be able to get information out of the qubit. One common way of doing this is to couple it to a linear resonator, like the harmonic oscillator we discussed before. In this picture, we see a qubit coupled to a readout resonator. Previously, we discussed using an inductor and a capacitor to make a resonator. However, in this case, we are just taking a microwave transmission line with a distributed inductance and capacitance and is also a resonator. The qubit interacts with a resonator just like an atom interacts with a cavity.
Moreover, we end up shifting the resonator frequency by a different amount if the qubit is in the zero states or the one state. So, by sending a pulse of microwave energy to interrogate the resonator, we can discuss the qubit's state. Readout times vary but can be as short as hundreds of nanoseconds.

\section{Superconducting Qubits: Manufacturing Artificial Atoms}

In this section, we will introduces how superconducting qubits can be manufactured with high coherence using modern CMOS-compatible toolsets.\cite{wang_coherent_2019,sung_non-gaussian_2019,yan_flux_2016}. We often say that superconducting qubits are artificial atoms. What we mean by that is that we can build electrical circuits that behave much like the natural atoms on the periodic table. For these artificial atoms, we can design all their properties, their energy level spacing, their sensitivity to noise. It is quite different from qubits based on natural atoms, where we are limited to what we are given by nature. Thus, a key advantage of superconducting qubits is that we design these superconducting quantum circuits and their properties. A second advantage is that superconducting qubits are silicon technology. We manufacture superconducting qubits on silicon wafers using the same tools photolithography, metal deposition, and metal etch. That is used by industry to make Complementary Metal-Oxide-Semiconductor, or CMOS, transistors. The active elements we use, Josephson Junctions, have thin oxide barriers, just like transistors' thin gates. Even the metals we use aluminum, titanium nitride, niobium are all compatible with CMOS fabrication. Thus, in an authentic sense, superconducting qubits are silicon technology.
Moreover, this affords us lithographic scalability. It is a straightforward path to increasingly complex circuitry with many interconnected qubits. There are a few differences in the specific process temperatures and other processing parameters used for CMOS and superconducting qubits. This is related primarily to the need for pristine materials and fabrication in order to maintain high qubit coherence. Superconducting qubits have a variety of ways they can lose quantum information, or what we call decoherence. Several examples of decoherence channels are illustrated here, including charges fluctuating on the device surface \cite{motzoi_simple_2009,muller_interacting_2015}, trapped magnetic vortices, and stray magnetic or electrical fields. Many of these channels can be enhanced and suppressed by the materials and fabrication choices we make in manufacturing superconducting qubits, as well as by their design. Research within the superconducting qubit community for the last 20 years has focused on identifying and mitigating decoherence sources to improve qubit performance with tremendous success. We have seen more than five orders of magnitude improvement in qubit coherence within these 20 years. In 2018, multiple major modalities of superconducting qubits with unique parameters tuned for different applications had achieved coherence times that exceed the most lenient thresholds for quantum error correction. This is one of the reasons superconducting qubits are at the forefront of demonstrations today.

\section{Superconducting Qubits: Fabrication}

In the next section, we describes the tool chain used for fabricating superconducting qubits, and the process flow, in this section.  The three main steps include the first deposition of superconducting material, a second large-scale patterning of the pristine material, then a third fine-scale patterning of the actual qubit.

Patterning superconducting qubits requires state of the art fabrication technologies. For example, MIT Lincoln Laboratory fabricate superconducting qubits in the 70,000 square foot micro-electronics laboratory. Within this clean room have a full suite of 90 nanometers CMOS tools, and also have dedicated superconducting deposition and etch tools. These dedicated superconducting tools are essential for limiting magnetic contamination sources, which is one of the decoherence within qubits. When we are fabricating superconducting qubits, we first deposit pristine materials. We then work to keep them as pristine as possible by minimizing, processing, and choosing the parameters carefully. Many processing steps have the potential to introduce sources of fabrication induced loss, and We have systematically studied ways to mitigate this potential. There are three main steps in the baseline superconducting qubit fabrication process. First, we prepare substrates and deposit a pristine layer of the superconductor, often aluminum or titanium nitride. Second, we pattern this pristine material into essentially everything for the superconducting qubit circuit, other than the qubit loop. This can include readout and control circuitry, such as resonators that interact with the qubit, the device ground plane, and shunt capacitors, which can store energy from the qubits. Third, we add the qubit loops, which contain Josephson Junctions. Josephson Junctions are thin oxide barriers sandwiched between two layers of superconductor. We often use aluminum as this qubit loop superconducting material. After we fabricate the qubit wafers, we conduct extensive testing of the devices. We then wire-bonded package some chips from the wafers to cool down and measure in the dilution fridges. We fabricate the devices on either 200-millimeter manufacturing style wafers or smaller 50-millimeter prototyping wafers. For the 50-millimeter prototyping wafers, we focus on quickly turning around new designs for rapid testing. For the 200-millimeter manufacturing scale wafers, we focus on the high yield of defect-free complex designs. We now will walk through these main process steps in more detail and highlight at each stage some of the key considerations, starting with the preparation of the silicon substrate and deposition of the high-quality base metal. Silicon substrates have a native surface oxide that is about 1 and 1/2 nanometers thick. The silicon oxide can contain dangling bonds and is a source of loss for superconducting qubits. In order to remove this oxide, we first do a wet chemical etch, and then we load the wafers immediately into the molecular beam epitaxy or MBE system. Using the MBE, we further prepare the silicon surface by annealing at high temperature and reconstructing the top monolayers of silicon. This is a molecular beam epitaxy deposition system, and this is critical in making superconducting qubits because it forms an essential part of the qubit, the very base metal. the metal we typically use is aluminum. This forms the capacitors, and the wiring, and the connections to the Josephson Junctions, which form the qubit itself. The aluminum comes from these effusion cells, which is a fancy tool, but all it means is that we have a little cone-shaped crucible where we put our aluminum in, and it is melted through these wires around the circuitry. the specific amount of power that goes into it controls the temperature of the aluminum down to under a tenth of a degree. That means we have precise control over the vapor pressure of the aluminum, which means we control how fast the aluminum film is deposited into the tool. we can monitor, in situ, the growth of these aluminum films using those high energy RHEED gun, which stands for reflective high energy electron diffraction What this gives us is a two-dimensional diffraction view of the surface of the wafer. So, we can see the silicon wafer, two, the desorption of the hydrogen on the silicon wafer, and three, the deposition in real-time of the aluminum on the wafer. Next, we pattern the high-quality metal into the shunt capacitor's control and readout circuitry and the ground plane. We pattern a layer of optical resist onto the superconducting base metal to define the features of interest. The pattern is transferred into the underlying superconducting material using either a wet chemical etches or a plasma etch process. Afterward, we strip the resist mask using a chemical stripper. Sources of loss within superconducting qubits can be attributed either to the wafers' materials or to the fabrication process. As a proxy for assessing the loss within superconducting qubits, we can use coplanar waveguide resonators. Lincoln Lab  is useing quarter-wave resonators that are capacitively coupled to a center feed line to deposit the same metal as the qubit base metal, and pattern, and etch using identical processes. The standard chip layout has five resonators that are each spaced 200 megahertz apart in frequency by varying the length of the resonators. When cooled to milli-Kelvin temperatures, which have passed the superconducting transition point, we can look at a loss within these resonators as a function of photon number. We assess performance at the single-photon limit, where on average, a single photon is in the resonator cavity. As of 2018, we typically see single-photon quality factors of 500,000 to one million for aluminum and one to 2.3 million for titanium nitride.

\section{Superconducting Qubits: High Coherence Qubit Loops and Josephson Junctions}

We will describes how high-coherent superconducting qubits are lithographically patterned and fabricated using modern semiconductor fabrication methods.  The section includes introduction of cleanroom, where the fabrication process and the actual equipment employed in the process. Next, We will look at the fabrication of high coherence qubit loops and Josephson Junctions. Starting from the patterned base metal, the next step is to lithographically define the region where we will deposit the qubit loops and the embedded Josephson Junctions. We start by depositing a stack of three layers of material. First, we spin on a layer of methyl methacrylate resist that we use as a spacer layer. Second is a layer of germanium, which we use as a hard mask. For the prototyping process on top, we coat a layer of ZEP electron beam or e-beam resist. We pattern the ZEP using an electron beam lithography system to define the qubit loops. Lincoln Lab use a Raith e-beam system. The 100 kilovolts system has a 50-megahertz clock speed, which enables reasonable write times of the nanometer-scale patterns. We say reasonable because we raster an electron beams back and forth across the wafer's surface rather than exposing full dye at a time as we would do in photolithography. This is a slower process than photolithography, but we gain patterning flexibility for rapid prototyping. Using the e-beam system patterning lines down to less than 10 nanometers on the system are demonstrated. For Josephson Junctions, we routinely pattern sub-150 nanometer features. Alternately, we also can pattern the qubit loops using stepper photolithography. For manufacturing scale wafers, Lincoln Lab use a 193-nanometer wavelength ASML scanner, which enables to pattern features smaller than 100 nanometers. Optical lithography enables orders of magnitude speedup in write time compared to e-beam lithography, since now we are exposing much larger write areas at a time, rather than rastering a nanometer-scale beam across the wafer. After the resist is exposed, by either e-beam lithography or photolithography, we transfer the pattern into the germanium hard mask using a plasma etch. We use plasma etching to pattern features on a smaller scale than what we can do with wet chemistry. we use gas, an ionized gas, and electric field to etch small metal features, silicon features, or oxide features to make the layers in our integrated circuit. After etching the germanium, we etch the spacer methyl methacrylate layer using an oxygen plasma etch. This exposes the silicon substrate and metal contact regions. Besides, this oxygen plasma simultaneously removes the ZEP top layer of resist. In addition to defining the open area for the qubit loops, we also pattern small germanium bridges, where We will pattern the Josephson Junctions. To release the bridges, we over-etch and undercut the methyl methacrylate. Zooming in on a cross-sectional view of the freestanding germanium bridge, we can schematically show the shadow evaporation process used to define the Josephson Junctions. All the shadow evaporation steps happen in situ within different chambers of the same vacuum system. Since we will be making superconducting contact between the qubit loop and the underlying base metal capacitive shunts, we first must prepare that base layer by removing the metal's native oxide. To do this, we sputter away the oxide using argon ions. We then transfer the wafer into the deposition module and put down the first layer of aluminum. Afterward, we move the wafer into the oxidation chamber, flow in oxygen, and allow it to oxidize the aluminum for a specific amount of time to reach the target oxide thickness. The target oxide thickness depends on the desired qubit parameter for the Josephson Junction critical current. Next, we move the wafer back into the deposition chamber, tilt the wafer to the opposite angle, and deposit a second layer of aluminum. We are now located at the PLASSYS shadow evaporation system, a tool that we use to fabricate one of the critical components of the superconducting qubits. This is where we deposit the Josephson Junctions and the associated qubit loops that surround them. So, what we do in here is four steps. First, we load a wafer into the load-lock. Then we load it into the etch chamber once it has pumped down to vacuum. In this etch chamber, we first ion mills the surface as a method to prepare it before we put down the Josephson Junction layers. Now in this etch chamber, we are preparing that top surface by removing the top layer, about a nanometer of aluminum oxide before we come in and deposit these junctions and qubit loops. So, then we move from the etch module over to the deposition module. In this module, we first put down the bottom layer of what will define the Josephson Junctions and qubit loops. This connects directly to that MBE material. From there, we move into the oxidation module. This module's job in life is to be able to put down very pristine, very uniform aluminum oxide layers. It flows in oxygen into the chamber when we have this exposed aluminum film, and it oxidizes to a precise thickness that We have tuned. From there, we move back into the deposition module. by tilting the stage differently, we can use the e-beam defined shadow mask to create a small layer of overlap between the bottom layer of aluminum. that is now being oxidized and a top layer of aluminum. That layer of overlap is specifically the Josephson Junction for the superconducting qubit loop. From there, the process is complete, and we move back out to the load -lock and take the wafers out. After shadow evaporation, we remove the wafer from the deposition system and lift off the resist stack in a chemical solvent. Once the resist is removed, wafer fabrication is complete. The completed wafer contains the superconducting qubits with integrated Josephson Junctions, as well as base metal pattern and capacitive shunts, the ground plane, and readout and control circuitry.

\section{Superconducting Qubits: Testing}

In this section, we introduces the use of data-driven process monitoring for assessing fabrication yield and device parameter spreads. The next stage is room temperature testing. we have to do extensive automated testing on every wafer that we fabricate. We conduct data-driven process monitoring to assess device performance and drive the process development. We test each component of the superconducting qubit system. Two examples include checking the critical current density of the Josephson Junctions and measuring the contact resistance between the base metal and the qubit loop shadow evaporated metal. Each wafer is tested using an automated wafer probing station. After we load the wafer, we do thousands of four-point probe measurements using a 26-pin probe card combined with a switch matrix. At each probing location, we apply current to a four-wire test structure and measure the voltage drop. The switch matrix re-assigns the current end voltage locations for all test structures accessible to the 26-pin probe card. Then, the probe card is lifted and transferred to a new position where the measurements continue. After testing is complete, the results are automatically databased. Besides, some devices can be selected for further cryogenic testing, either at liquid helium temperature of 4.2 Kelvin or at millikelvin temperatures in one of the dilution fridges. As one example of room temperature testing, we measure the resistance of Josephson Junctions with varying junction lengths ranging from 100 nanometers to three microns. We plot the inverse of the resistance, which is the conductance, as a function of the junction length. We use the slope of that plot to extract out the low temperature critical current density, JC, of the wafer. On the prototyping 50-millimeter wafers, we measure the critical current density at six identically patterned process monitor chip locations spread across the wafer. In this example, we targeted a critical current density of three micrograms per micron squared. We met that target average critical current density and had less than 2\% cross-wafer variation. In a second example of the test structures, we measure contact resistance between the metal layers. We show a false-color scanning electron microscope image of contact between the base MBE aluminum metal in blue and the top shadow evaporated Josephson Junction metal in orange.
Additionally, we also check for continuity of millimeter long snaking lines and isolation between interdigitated combs. We use these measurements to check for any particle defects and consistency of lithographic patterning. After testing, the last step is to select wafers for dicing and packaging. Wafers are diced using an automated water-cooled dicing saw. The qubit chips are then packaged, and wire bonded to make connections from the outside cabling to the chip circuitry. From there, the chips are loaded into one of the dilution fridges for measurement.

\section{Superconducting Qubits: Why 3D Integration?}

We will describe in this section why three-dimensional Integration is needed for superconducting qubit chips and illustrate conceptually how this can be achieved using through-silicon vias and stacked wafer technology\cite{rosenberg_3d_2017,rosenberg_3d_2019}. So far, people have been working with small arrays of qubits, up to around a few tens of qubits. At some point, however, we run into a fundamental problem of where to put everything. We want many qubits that are coupled together. However, if we look back at the picture, we can see that the actual qubit is only a small portion of what needs to go on the chips. We have bias lines, control lines, readout resonators, for example. those can all be bigger than the qubit. Now, we can imagine making some of these smaller. However, eventually, if we have a large array of qubits, it will be hard even to get the wiring to the qubits in the center if we are working in two dimensions. most of the demonstrations to date have been limited to geometries where we can laterally access the qubits on the surface of a chip. This problem is not unique to superconducting qubits. Many applications have this same issue. Consider, for example, a large-scale imager with lots of pixels. we need to be able to get our signals out of the pixels. However, we also want to be able to put many pixels on a chip to fill a 2D array. The one-way industry has solved this issue by moving to 3D Integration, where signals are brought vertically instead of laterally. We think it is necessary to move to 3D Integration to make large scale superconducting qubit circuits. However, we need to be careful when doing so because the qubits are affected by things that are not an issue with classical electronics, which is what 3D integration was developed. For example, one way to efficiently route wires in 3D is to build up a multi-layer stack of metal, with a dielectric in between so wiring on different layers can cross each other. However, we know that having lossy dielectric materials near a qubit can cause it to lose its quantum state. So, we have developed a new idea for how to get the benefits of 3D Integration, efficient wiring, and the ability to bring in signals vertically without affecting the qubit. Here is the proposed scheme, which has three separate silicon chips that are held together with indium bump bonds. Within the three stacks, each of the layers is fabricated separately. We have seen that there are process incompatibilities between the processes in separate layers of these three stacks. For example, we must stay at a relatively low temperature once We have deposited shadow evaporated Josephson Junctions on the qubit layer, the interposer layer. However, we need to have processed at a higher temperature for the readouts and interconnect wafer. By separately fabricating each layer, and combining them at the end of the process, we can combine the best processes and optimize the capabilities of the full stack. The top layer of the three stacks contains the qubits. Here we show two examples, capacitively shunted flux qubits\cite{yan_flux_2016}, that we use for quantum annealing applications \cite{yarkoni_boosting_2019,das_colloquium_2008,susa_quantum_2018,kadowaki_quantum_1998,pino_quantum_2018}, and transmons, another type of qubit that we use for gate-based quantum computing. As of 2020, state of the art single qubit coherence times for each of these styles are on the order of 50 to 100 microseconds. Simultaneously work is going on to increase qubit coherence further and retain that coherence as we move into the third dimension with increasing interconnect complexity. The second tier in the three stacks is the interposer wafer, where Through Silicon Vias, or TSVs, that are lined with superconducting metal, route signals between the two sides of the wafer. The interposer provides three key benefits. First, it provides an isolated mode volume for each of the high coherence qubits to retain coherence times comparable to those in a planar device geometry. Second, the interposer provides a nearby surface for inductive or capacitive coupling across the vacuum gap between the qubit plane and the interposer plane. This interposer surface can be used for control and readout circuitry, or for coupling qubits that bridge two devices on the qubit plane. The third benefit of the interposer is that it connects the qubits with the readouts and interconnect multi-layer wafer on the bottom layer of the three stacks, while still having the qubits keep a healthy distance from the lossy dielectrics on that multi-layer interconnect module. Moving down to the multi-layer readout and interconnect layer on niobium devices. We previously fabricated tri-layer Josephson Junctions for superconducting qubits, which turned out to have lower coherence than the aluminum shadow evaporated Josephson Junctions that we use today. We also currently fabricate fully planarized multi-layer niobium devices with integrated tri-layer Josephson Junctions for both Single Flux Quantum, or SFQ circuits, as well as near quantum, limited traveling wave parametric amplifiers, or TWPAs. For the three-stack configuration, we envision embedding Josephson Junctions within the multi-layer niobium wiring that could be used for active circuit components, such as on-chip TWPAs. Last, after the fabrication of each of these three separate wafers, we need to assemble the complete circuit. To do this, we are developing a double bump bonding approach with indium thermal compression bonding. Indium, which is superconducting at the millikelvin operational temperatures, can be used both for the mechanical stability of the wafer stack and for electrical connectivity. We align the chips, using a precision bump bonder system. We use the feature in feature alignment marks, where part of the alignment structure is on each wafer, both during alignment, and as a check afterward of how well we align the two surfaces. We have seen that we routinely achieve better than one-micron alignment between the chips. In addition to lateral alignment, tilt control is also critical for consistent inductive or capacitive coupling across the vacuum gap. Using multiple techniques, we have demonstrated tilt control between wafers of less than 250 micro radians.

\section{Superconducting Qubits: How 3D Integration Works}

In this section, we will focus on the fabrication procedure used for superconducting qubit systems. How through-silicon vias allow a three-dimensional circuit structure to be realized\cite{rosenberg_3d_2017,rosenberg_3d_2019}. Unlike most traditional silicon fabrication processes used for conventional CPUs, multiple wafers are employed for this process.  This enables chips made with superconducting qubits to be stacked together with other chips, e.g., for readout, and potentially, classical control circuitry.  Such high levels of Integration are likely needed for future quantum computing systems.

Let us now take a deeper dive into the fabrication of the superconducting TSV interposer wafer. First, we fabricate the TSVs, fill them with superconducting material, and pattern the metal on the wafer surface. Next, we mount the interposer wafer to a temporary carrier wafer, flip over the TSV wafer, and remove the excess wafer thickness down to the TSVs. Afterward, we add metal to this revealed surface. Fabricate control circuitry or qubits onto that surface. This revealed side is the surface that ultimately will be near the qubit layer qubits. After all the processing is complete, we dice the wafer and release the individual chips from the temporary carrier wafer. In the end, the chips are available for bump bonding into the three stacks. After the wafers are fabricated, but before we dice and release them, we do extensive room temperature and cold temperature testing to confirm that the TSVs are superconducting. Here, we see a 200-millimeter TSV wafer, which has 52 identical dies. On each die, we have several TSV test structures in addition to the active circuits that we are using for the qubit stacks. We use the process control monitor chips to assess the individual metal properties, as well as four-point probe structures, to look at both single TSVs and chains of TSVs. As one example, we have links of 400 TSVs in a series that we can probe to assess the resistance.
When we look at the room temperature resistance of these 400 TSV chains, we see that, on average, we have 37 ohms of resistance per link. Of even more interest to us at room temperature is that we see the standard deviation is only two ohms, which shows a high degree of uniformity across the wafer. Afterward, we took a subset of these devices and cooled them down in the dilution fridge to assess the superconducting transition temperature. The devices go superconducting around 1.6 Kelvin, and that the midpoint is 3.1 Kelvin. It is well above the millikelvin operational temperature. In advance of having the full three stacks available for testing, in 2016 and 2017, we also conducted several experiments looking at components of the eventual three stack. Since qubits are so sensitive to materials and processes, the first experiment we did was to take a regular single-chip qubit and design a flip-chip version of it where all the inductors and capacitors had the same values. However, the qubit chip was flipped and bonded to another chip.
We wanted to make sure that the extra processing and the presence of another chip bonded to the qubit chip did not affect the coherence time. Here is what the circuit looked like for the single-chip qubit. The chip has six superconducting flux qubits, and each coupled to a bias line and a readout resonator. For the flip-chip version, we took all the control and readout circuitry and moved it to another chip. The only things left on the qubit chip were the qubits themselves. Then we bonded the two chips together. The figure on the right shows an infrared image looking through both chips and showing that structures on one chip are well aligned to those on the other chip.
We found that the relaxation and coherence times of the single-chip and flipped chip qubits were virtually identical. This is interesting for a couple of reasons. First, it shows that 3D Integration does not significantly degrade qubit performance, which is essential. Second, we have demonstrated that we can move all the control and readout circuitry to another chip, which we are planning to do for the full 3D integration scheme. In this experiment, we stuck with the same general design because we wanted to isolate the effect of 3D Integration. In future work, we are planning to shrink the resonator and other components so they can fit under or between qubits in a 2D array. We are excited to be continuing to develop further and further demonstrations using this three-stack scalable architecture. Within both digital quantum computers and simulators and analog quantum systems, there is a strong need for high connectivity between qubits and significant control and readout complexity. Although there are differences between digital and analog quantum algorithms, both systems have similar 3D scalability needs that are requiring significant engineering developments. We are planning to use this three-stack hardware to scale to testbeds with 100 qubits or more. We will then use what we discuss from these systems as a stepping-stone to future large-scale demonstrations of quantum computers.

\begin{figure}[H] \centering{\includegraphics[scale=0.6]{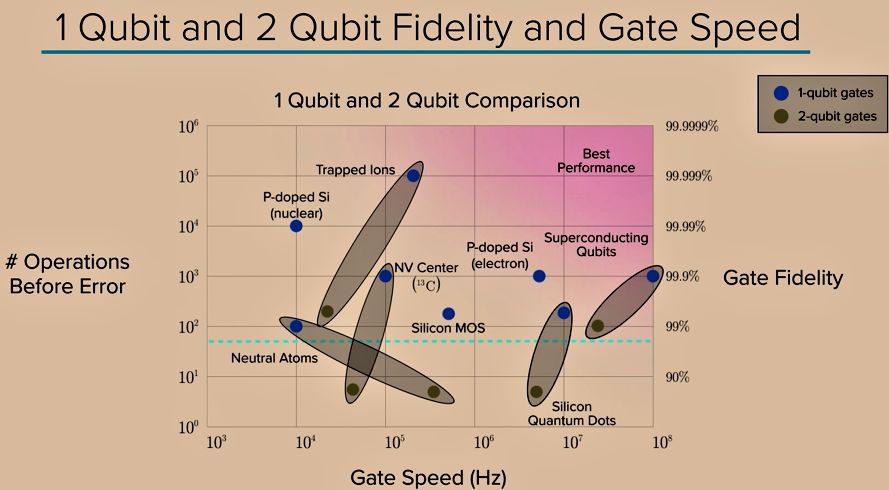}}\caption{1-Qubit and 2-Qubit fidelity and gate speed}\label{fig1_31}
\end{figure}

\begin{figure}[H] \centering{\includegraphics[scale=0.5]{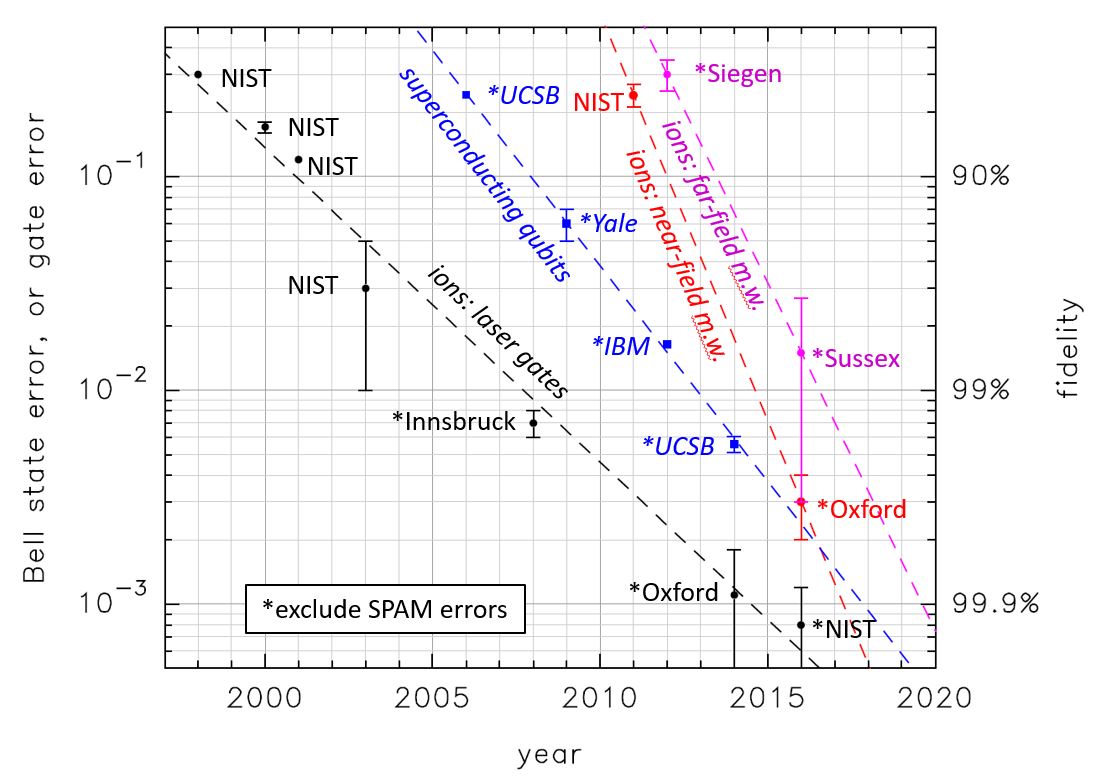}}\caption{1-Qubit and 2-Qubit fidelity and gate error}\label{fig1_31_1}
\end{figure}

\section{The Promise of Quantum Computers}
we have discussed how a quantum computer differs from a classical computer, and we have seen examples of the tremendous potential of quantum computing. In this section, we will turn to a more practical question: How can one realize the promise of quantum computing and quantum communication? Let us begin with the promise of quantum computation. 

In this section, we will look at quantum algorithms and quantum communication. We will start by considering the promise of these technologies, what is needed to realize that promise, how to get there, and where we are today. Let us begin with quantum algorithms and the promise of quantum computers. As we discussed earlier, quantum computers are not smaller, faster versions of classical computers, nor are they an incremental step in the evolution of Moore's law. Instead, quantum computing represents a fundamentally new approach to processing information.
Furthermore, it is the only computing model we know of today. That is qualitatively different from existing computers. This difference and the potential promise of a quantum computer is reflected in the fact that simulating a quantum computer using a classical computer requires exponential overhead. This means that, as we add one qubit to the quantum computer, the size of the classical computer required to simulate it will grow exponentially. Now we do not often experience exponential growth in the everyday lives, and so, to gain an intuition for how powerful it can be, we may remember the old riddle where our employer asks us how we would like to be paid this month, and we have two choices. Either We will start us with one penny, and then each day for one month, We will double the number of pennies and keep the pennies from the last day. Alternatively, we can have \$1 million. Now, it is entirely reasonable to think through the first few days of pennies. For example, two pennies on day one, four pennies on day two, double that to eight pennies on day three.
Moreover, we begin to think that \$1 million is starting to look pretty good. However, in fact, if we take that \$1 million, we have lost a lot of money because the number of pennies is growing exponentially as 2 to the n-th power, where n is the number of days. So, on the 31st day, the last day of the month, we have 2 to the 31 pennies. That is equivalent to about \$21 million. That is the power of exponential growth. Returning to computing, if we consider the amount of memory, we would need to store all the available states in the state space of a quantum computer We will similarly see the immensity of exponential growth. For example, representing all the quantum states of just 30 qubits would require at least a very powerful laptop.
Increase that to just 40 qubits, and now we need a supercomputer. Double that to just 80 qubits, and we would need all the classical computers we have on Earth. Then double that again to 160 qubits, and now we need all the silicon atoms on Earth. Even small increases in qubits translate to exponential growth in the amount of memory required to represent all 2 to the n numbers that n qubits can store. So, there is a big difference between classical and quantum computers.
Furthermore, this suggests the potential for substantial quantum advantage. However, what is needed to realize this in practice? To realize a quantum advantage, we need to identify the intersection of three essential requirements. First, we would like to identify a useful problem, one that has practical importance. Second, this problem should not have a known fast classical algorithm to solve it. Otherwise, we might as well just use a classical computer. Then third, the problem must have a known fast quantum algorithm, one that offers a substantial speedup. Now, this sounds great, but in fact, this promise has been mostly empty, except for a small area of overlap that we know about today.
Nonetheless, there are useful problems with a quantum advantage, and this improvement can be quantified using a mathematical expression. One place is in the prefactor of that expression \cite{smolin_pretending_2013}, and one place is in the exponential. The prefactor A can be a large fixed number, or it can scale polynomials with the number of qubits n. The exponential scaling of a problem of size n is in the exponential factor. Either of these two types of improvement can indicate quantum advantage. In the following sections, we will compare several algorithms and their respective speedup.

In the section above, an example of the power of exponential growth. In the salary example, if we earn two pennies on the first day, and double that amount each day for 31 days, then we will receive $ 2^{31} $ pennies on the last day of the month, equivalent to over \$ 21M dollars. This growth rate is termed exponential because the output $ y=2^{x} $ depends exponentially on the number of inputs x (in this case, the number of days). In computer science, algorithms can be classified according to the scaling of the resources required to run an algorithm as the number of inputs the problem size grows. An algorithm is said to scale exponentially if the resources (time, space) needed to run the algorithm grow following an expression such as $ y=c^{x} $, where c is a constant greater than 1 and x is the problem size. Two other examples of growth functions are linear and polynomial. An algorithm is said to be linear if it can be solved by using$  y=ax  $resources, where a is a constant, and polynomial if it uses $ y=ax^c $ resources where c is a constant greater than or equal to 1. The linear case is itself an example of polynomial scaling with $ y=c=1 $.

\section{Understanding Quantum Algorithms}
Realizing the promise of quantum computing relies on developing new quantum algorithms that offer a quantum advantage. In this section, we will review several algorithms that are known today and present several different approaches to the development of new quantum algorithms.  

In the last section, the promise of quantum computing and its potential for quantum advantage. However, although a quantum computer can do anything a classical computer can do, it often does not improve. we only have a small set of useful algorithms at the disposal today. So, what can we do to realize the promise of quantum computing? Quantum advantage starts with having useful quantum algorithms. Developing new algorithms is very challenging, in large part because quantum computers are based on quantum mechanics. Thus, developing an algorithm for a quantum computer is quite different from doing so on a classical computer. To understand how quantum algorithms work and what they can do. There are currently a couple of different approaches. One is to develop algorithms based on mathematical theorems and their proofs. Shor's algorithm and Grover's algorithm are both excellent examples. Based on either intuition or insight into the problem, a quantum algorithm is first proposed. then, it is theoretically determined if the algorithm is efficient or not. For example, proving that an N-bit number can be factored with high probability, in a time that scales polynomials with the size of that and that N-bit number, rather than exponentially. Again, this type of development generally requires insight into a structure of the problem, that lends itself to enhancement on a quantum computer. A second approach is to use classical computers to simulate quantum computers' behavior, to glean some insight into algorithmic primitives that can then be used to realize quantum enhancement in larger systems. Now, as we might imagine, simulating a quantum computer with a classical computer is not very efficient. After all, if we could easily simulate a quantum computer, we would not need to build one. Currently, classical supercomputers, with significant effort, can simulate about 50 qubits. Various probabilistic and Monte-Carlo techniques have been developed to further increase the qubit number, in exchange for accuracy or completeness in the solution. Again, these kinds of classical simulations can provide insights into how a quantum algorithm works at small scales\cite{chen_classical_2018}. So, it can be leveraged in quantum algorithms run on larger scale quantum computers. Beyond small scale simulations, a third approach is just to throw caution to the wind, build a quantum system, and see what happens. The D-Wave system is a great example of this approach. D-Wave has built a quantum annealer with more than 2000 qubits \cite{stollenwerk_flight_2019}, a fantastic engineering achievement\cite{mcgeoch_practical_2019}. Although there is, yet, no theoretical or experimental evidence pointing to quantum enhancement in quantum annealers, at least for machines and problems studied so far, there remains a lot that we do not know. as we discussed previously, because optimization problems are so important, there is a tremendous application pull to develop quantum-enhanced optimization tools\cite{mehta_quantum_2019}. Thus, getting a real machine into the hands of engineers, computer scientists, and algorithm designers is a great way to start discussing what is and is not possible with the various quantum computing approaches. So, developing quantum algorithms is very challenging, and tremendous efforts have been applied to their development. it is because of this that we have useful, or potentially useful, algorithms today. Examples in the user now category are Shor's factoring algorithm \cite{shor_polynomial-time_1997}and Grover's search algorithm. Shor's algorithm can be used to break public-key cryptosystems and afford exponential speedup over known classical algorithms. Doing so at a meaningful scale requires a medium-sized quantum computer, with thousands of logical qubits \cite{kapit_very_2016}. Although we do not have such quantum computers yet, they are foreseeable. research today is oriented towards developing new, quantum-resistant public-key cryptography standards, as well as a new generation of quantum algorithms to break them. Grover's algorithm is related to a general class of search, collision, and optimization algorithms that afford a quantum advantage over known classical algorithms \cite{kirke_application_2019}. However, currently, more research is needed to help make these algorithms practical. There are a couple of reasons for this. One is the data loading problem. For example, we can efficiently load large amounts of data into a quantum computer so it can be efficiently searched. Also, there is currently a rather limited range of problem sizes that admit benefit. Namely, problems that are large enough to make use of the quantum advantage, and yet not so large that it becomes practically prohibitive to operate on a human time scale. For example, a quantum enhancement that reduces the runtime from a few thousand years to a few decades is obviously a fantastic improvement. However, very few people would find that useful. Although Shor's and Grover's algorithms are perhaps the most well-known examples of a quantum algorithm, it is widely anticipated that quantum simulation will have the greatest economic impact in the future. The simulation would apply broadly to a range of problems, from solid-state and nuclear physics to material science and chemistry. these are problems of importance to both science and industry. Classical simulations of quantum systems are generally limited to very small problem sizes, or an array of simplifying assumptions, such as no entanglement, or semi-classical dynamics\cite{chen_classical_2018}. Exact solutions are generally intractable because the number of variables needed to describe a fully quantum system grows exponentially with the problem size. It has been shown that quantum simulations can efficiently model quantum systems. the hope is that some of these problems may see the quantum advantage, on a smaller scale, or even error-prone quantum computers. Whether or not this is achievable remains an open question. Other quantum algorithms in the ``may be useful" category include sampling solutions to linear equations \cite{chen_hybrid_2019}, adiabatic optimization problems, and machine learning\cite{riste_demonstration_2017}. In these cases, the ultimate degree of quantum advantage will depend on several factors, including input-output complexity, such as the loading problem. We mentioned earlier and making a meaningful connection to applications. Besides, to offer a useful advantage, these algorithms must also outperform existing heuristics. Classical computing methods that, while not guaranteed to give an exact or optimal answer, do give very high-quality solutions. That is often enough for many applications. In summary, developing quantum algorithms that exhibit quantum advantage is at the heart of realizing the promise of a quantum computer. in the next section, we will look at the degree and type of quantum enhancement these algorithms provide.
In the context of D-Wave quantum annealing versus gate model quantum computing\cite{streif_solving_2019,pelofske_solving_2019}. so, what is  the potential of D-Wave's quantum annealing machine versus, IBM or others' quantum gate machines? And more specifically, which model will be a better choice for businesses in the immediate future? The universal gate model machines we know if we can build and engineer that system,  there are problems that will have a quantum advantage when run on a quantum computer. But to get to that point requires fault tolerance and error correction. Now, in contrast to that, a quantum annealer addresses specifically classical optimization problems. it is not theoretically known whether a quantum annealer will afford quantum advantage or not over classical computers. There is no theory that says it will, nor is there a theory that says that it cannot. so, that is an open question. Now, it is true that there are many, many problems that can be broken down into optimization problems. Optimization problems are ubiquitous \cite{inagaki_coherent_2016}. so, there is a huge application pull to develop computers, whatever they may be, that can solve optimization problems better than we can currently do. that is really the motivation for quantum annealing\cite{vyskocil_embedding_2019,susa_exponential_2018}. we simply do not know if There is going to be quantum advantage there. Now, it could be that a quantum annealer is just a much better classical computer. if that is the case, it still has value. we think we still hope that There is a region of the parameter space where quantum annealers do show quantum enhancement. What we can say is that the quantum annealing in particular, the system that the D-Wave company has built, is really a fantastic engineering achievement. They very quickly, over a decade, went from just a few qubits up to systems with more than 2,000 qubits. They co-locate cryogenic electronics to control that machine. That is a fantastic achievement. But as a system engineer will know that to do that, they had to restrict the flexibility of that machine to a degree that it can scale quickly to those large sizes. so, the reason we bring that up is that there is some hope that, even though quantum enhancement has not been observed yet for a general class of problems the phase space that has been tested so, far is rather narrow. For example, only zz-type coupling interactions have been demonstrated. The degree of connectivity is limited to one specific type. the coherence times of the qubits that are used in the D-Wave machine are rather low when compared to the coherence times that we have on gate model machines. so, by moving to different types of coupling for example, xx-or yy-, sometimes called non-stochastic type couplings or to moving towards much higher coherence qubits, there is a hope that this may actually be a region of parameter space where quantum enhancement can be found.

\section{Quantum Advantage}

In this section, we will compare the resource scaling for several algorithms when implemented on a classical computer and a quantum computer. An improvement (reduction) in the resource-scaling is a quantifiable metric for the degree of quantum advantage.

In the last section, we introduced several algorithms, including those like Shor's and Grover's algorithm, that are useful now if we can build a quantum computer that is large enough and robust enough to run the algorithm. We also looked at quantum simulation, a very promising set of algorithms which, when realized, has the potential for substantial economic impact, and may even find utility on today's smaller scale, non-error-corrected machines.  we looked at a few potentially useful algorithms currently under development. For each of these examples, let us now look at the resource requirements for implementing an algorithm on both a classical and a quantum computer. It will then show the degree of quantum advantage one can obtain by using the quantum computer. We will also identify the leading limitations as They are currently known for realizing this advantage. Let us take them in order of their applicability as we see it today. At the top of the list is a quantum simulation\cite{bauman_quantum_2019} with application to quantum chemistry and material science\cite{motta_low_2018,babbush_low-depth_2018}. The classical resources scale as $ 2^ N $-th power for simulation of N atoms, whereas the quantum resources scale as N to a constant power C, where C generally ranges from 2 to 6. Now, by resources, we mean both the time required to reach a solution, referred to as a temporal resource, and the number of logic or memory elements needed to implement the problem, often referred to as a spatial resource. To quantify the resource requirements, we quote a mathematical expression for the scaling law. We are looking at how the resource requirements grow as the problem gets larger, as parameterized by the size N, for example, the number of atoms being simulated. Presumably, as N gets larger, the problem gets harder. However, the question is, how does it get harder? Does it require exponentially more resources, such as $ 2^ N $-th power, or a polynomial scaling, like N to an integer power? For quantum simulation, we see that the classical resources scale exponentially with N, whereas the quantum resources only scale polynomials. It means that there is an exponential advantage to using a quantum computer. To simulate the system dynamics for a time $ t $, both a classical and a quantum computer would require a similar number of time steps. For example, if we want to simulate the dynamics of a reaction for one second, we would divide it into a similar number of time slices. However, the quantum advantage is that there is an exponential reduction in the amount of memory needed to perform the simulation on a quantum computer. Thus, the quantum advantage is exponential. The main limitation is in determining the mapping of a physical problem onto the qubits, their couplings, and the gate operations needed to implement a quantum simulation. Next, factoring and related number-theoretic algorithms also have an exponential scaling in the number of classical resources, going as 2 to the n-th power and the number of digits being processed. In contrast, the quantum resource requirements scale polynomial, as N to the third power. Thus, the quantum advantage is again exponential. The main limitation, or perhaps uncertainty, is that the best-known classical algorithm has not been proven to be optimal. So, there may still be a more efficient classical algorithm yet to be discovered\cite{tang_quantum-inspired_2019}. if so, then the degree of quantum advantage might also change. Sampling solutions to linear systems of equations \cite{harrow_quantum_2009} is the next algorithm. In this case, we solve linear algebraic problems of the type $ ax = b $, where $ a $ is matrix, $ b $ is a known vector, and $ x $ is an unknown vector. The classical resources scale exponentially as $ 2^ N $-th power, whereas the quantum resources scale only approximately as $ N $. so, this algorithm also exhibits exponential advantage. The main limitation is related to a variety of restrictions on operating conditions. For example, a requirement for a sparse matrix A. The classical resources required for optimization problems also scale exponentially, as $ 2^ N $-th power, where $ N $ is again related to the size of the problem. However, in this case, the corresponding quantum resources, and the quantum advantage, are not well-defined. It is in part because one generally cannot determine if the resulting answer is indeed optimal. We can only tell if it is better than a solution We have had previously. Thus, we can only derive empirical evidence that a quantum optimization algorithm provides a quantum advantage. this is currently the main limitation of this algorithm. Finally, there is Grover's search algorithm for unsorted or unstructured data. Here, the scale of the classical resources with the number of data elements $ N $, whereas the quantum resources go as the $ \sqrt{N} $. Thus, the quantum advantage is the $ \sqrt{N} $, a polynomial enhancement. The main limitation of this type of search algorithm is the data loading problem. Namely, how can we efficiently load in a large amount of unsorted or unstructured data that needs to be searched? In general, we can see that, for many, if not all these algorithms, there can be a substantial quantum advantage when the problem size $ N  $ becomes large. for those with an exponential advantage, classical computers, and Moore's law-like scaling will never be able to catch a quantum computer's performance \cite{soloviev_beyond_2017}.

Developing applications that demonstrate a clear quantum advantage is challenging. However, there are several areas where the quantum advantage is known to exist (if we can build a large enough quantum computer to run the problem at scale).\\
\textbf{Simulation of quantum systems: } Quantum simulations of quantum systems afford an exponential advantage in the amount of space (memory) over classical simulations. Essentially, quantum systems generally comprise a very large number of variables to monitor during a simulation, and these can be handled more efficiently on a quantum computer.\\
\textbf{Shor's factoring algorithm: }  has an exponential temporal advantage over the fastest known classical algorithm for factoring. While it has not been formally proven that an efficient classical algorithm does not exist, it is believed that it does not.\\
\textbf{ Linear systems of equations:}  An exponential quantum advantage has been proven for linear systems that satisfy certain properties. From a matrix A and a vector b, the quantum algorithm can find a vector x such that $Ax = b$  as long as A is sparse and has a low condition number. \\
\textbf{Grover's search algorithm:} has a proven polynomial speedup over the best possible classical search algorithms. This advantage is still subject to the data loading problem: the data to be searched must also be loaded into the quantum computer efficiently. Overcoming this loading problem is an active area of research.

It is challenging to prove that a quantum computer offers a quantum speedup for optimization problems. There are two main reasons for this. First, although one can generally identify a candidate solution and confirm that it is better than a previous solution, knowing (or proving) that the solution is optimal is generally not possible. Second, near-optimal solutions are often sufficient in practice for most problems. For example, in finding the shortest route home over a distance measured in kilometers, solutions that differ by a few meters are practically interchangeable with the true optimal solution. Proving speedup is thus confounded with the notion of ``good enough,'' making such proofs challenging. There is currently intense research to find quantum algorithms that provide a quantum advantage for finding higher-quality solutions in less time than is possible on a classical computer.

The following table compares the resource scaling for several algorithms that feature a quantum advantage. The resource scaling with system size N is shown for both classical and quantum versions of the algorithm when performed on a classical and quantum computing. The resulting degree of quantum advantage is shown along with current known limitations to the quantum algorithm implementation. 

\begin{figure}[H] \centering{\includegraphics[scale=.5]{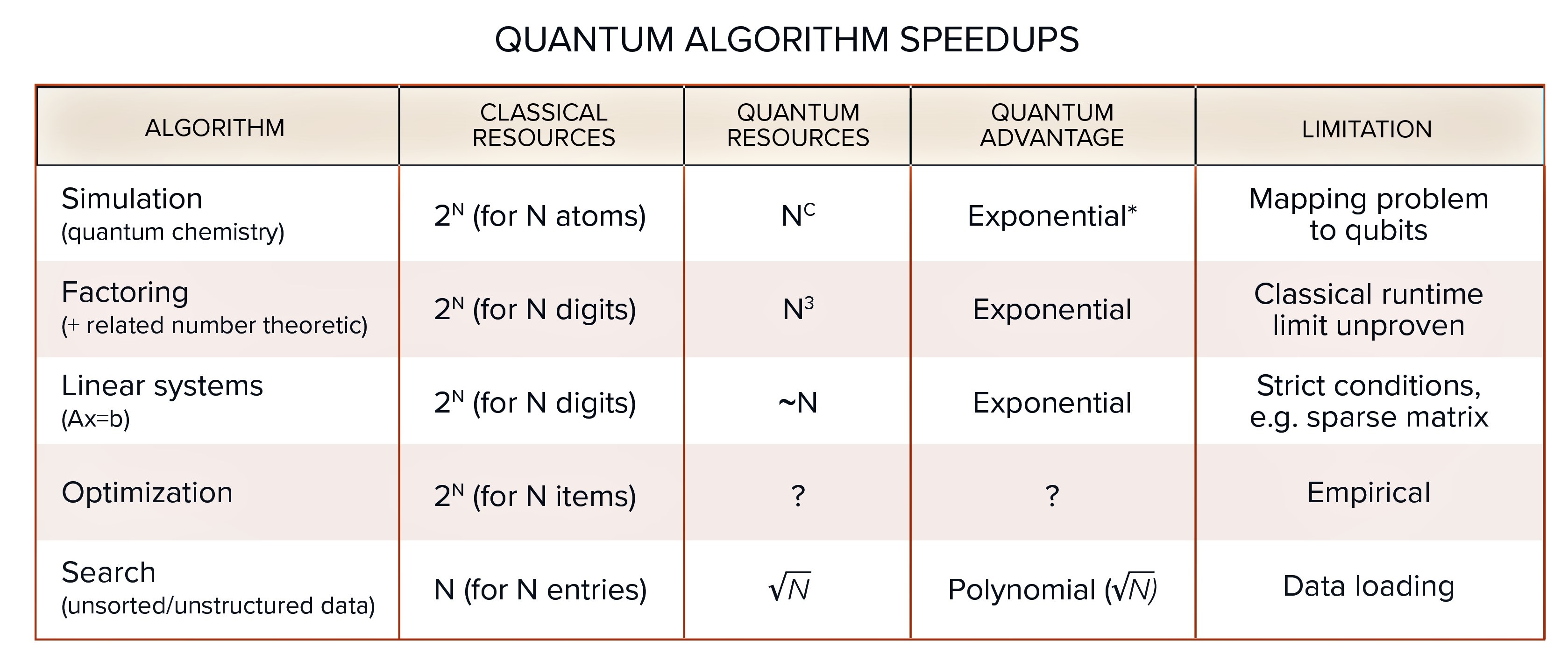}}\caption{Quantum Speedups}\label{fig1_33}
\end{figure}

\section{The Promise of Quantum Communication}

Quantum communication is qualitatively different from classical communication. In this section, we will discuss the promise of quantum communication and how to realize this potential by looking at what criteria must be met to realize the quantum advantage.

Just as quantum mechanics can dramatically enhance the way we process information, it also has the potential to enhance the way we communicate information. So, in this section, let us look at the promise of quantum communication. Communication is broadly defined as the conveyance of information, and it encompasses sending that information as well as receiving it. It certainly refers to verbal communication between people as well as nonverbal forms such as written text and signatures. It often comes in an encoded form. For example, the digital zeros and ones that are created by sampling the voices on a cell phone or representing the words of an email that are then routed from one computer to another. Even information on a chip must be shuttled around, communicated between a processor and its memory.
Moreover, this concept can be extended to include several spatially distributed computing nodes, all of which are available to attack a problem but must be coordinated and require communication between these different nodes. All of these are examples of classical communication. However, quantum communication is different, and it represents a fundamentally different means to encode, convey, and authenticate information. These differences and the potential promise of quantum communication are reflected in three basic facts or theorems. The first is that classical information encoded in qubits can be transmitted two times faster than by classical means. This is achieved using a concept known as superdense coding, and it is a provable quantum enhancement. Two bits of classical information can be transmitted using a single qubit when that qubit is part of an entangled pair of qubits called a Bell state\cite{noauthor_big_nodate,noauthor_big_nodate-1}. Now, two qubits, but We only needed to transmit one of them to send two classical bits. That is superdense coding. The second is that quantum bits cannot be copied or cloned.
Moreover, it is, again, a provable consequence of quantum mechanics. Now, if We prepared a qubit in each state, we know that state and so We can prepare a second qubit in the same state. However, if We have a qubit in an unknown state, we cannot make an exact copy of it, and that is called the no-cloning theorem. The third fact is that attempts to intercept or measure quantum bits can be detected. The no-cloning theorem cannot duplicate those bits. Thus, an eavesdropper must measure in order to glean any information.
Moreover, it is the measurement process that always leaves a signature that is detectable in some way. What is needed to leverage these three facts and realize the promise of quantum communication? To realize the quantum advantage, we need to identify problems at the intersection of three key applications. The first is to find a problem that can leverage the enhanced capacity to encode information that quantum mechanics enabled. Second, there are problems related to authentication—the ability to confirm an individual's identity or the truth of an attribute.
Furthermore, third are problems associated with collaboration, the collaboration between people or agents such as secret sharing or game theory\cite{khan_nash_2019}. For example, enhance means to auction or to vote without attribution. Alternatively, using quantum mechanics to verifiably share public goods in a way that avoids certain rational yet self-destructive tendencies such as the tragedy of the commons. Realizing the promise of quantum communication will require finding problems that exhibit quantum advantage. In the next section, we will look at several such problems.

Communication describes the conveyance of information between a sender and a receiver. Some familiar examples of classical communication include verbal or written language, encoded information transmitted from one location to another via smartphones or the internet, and fetch calls made by a classical processor to its memory unit. Classical communication beyond speaking or writing is generally mediated by classical states of light or electricity, incapable of assuming quantum superposition.

Alternatively, the field of quantum communication considers the advantages that can potentially be found by communicating via the transmission of quantum states. Such quantum communication is fundamentally different from classical communication with respect to encoding, conveying, and authenticating information. These differences, and the potential promise of quantum communication, are reflected in the following three basic concepts.

The first concept is called superdense coding, also referred to as quantum dense coding (QDC). QDC is a fundamental quantum communications protocol that conveys two bits of classical information through the transmission of a single qubit. This protocol makes use of quantum entanglement as a resource and requires that the two users\cite{wong-campos_demonstration_2017}, traditionally referred to as Alice and Bob, have access to pre-shared entangled qubits.

As an overview, the protocol presumes that Alice and Bob have already pre-shared entangled qubits, such that each has one qubit from an entangled pair. Alice wants to send information to Bob, and so she proceeds to encode her information by performing one of four operations to the qubit in her possession. She then sends this qubit to Bob, so that Bob now has both entangled qubits. Bob then performs a joint measurement on each of the qubits now in his possession. By doing this, he is able to determine which of the four operations Alice performed during the encoding phase, and hence reveal two bits of classical information (since $ \log _{2}(4)=2) $. Thus, Alice manipulates and sends one qubit to Bob, and in doing so, effectively transmits two classical bits of information. 

The important feature of quantum mechanics, which enables this protocol to work, is the presence of entanglement, which provides access to a larger space of possible states. More specifically, if Alice and Bob did not share entanglement between their qubits, then the most information Alice could encode on the qubit she transmits to Bob is $ \log _{2}=1 $ bit, since the state space is spanned by two orthogonal states (usually written as $ \vert 0\rangle$ and $\vert 1\rangle) $  it would just be a single, classical bit. However, due to the nonlocal nature of entanglement, Alice can hold just one of the two entangled qubits and still have access to the larger state space of the entangled qubits one that is spanned by the four orthogonal Bell states. Whichever Bell state Alice and Bob initially shared, Alice can change it to any of the other four through the application of one of the four following unitary operations to her local qubit: Identity, X-gate, Y-gate, or Z-gate.

To see how this works, Let us look at the four distinct Bell states for qubits A and B and label each with a unique binary number (00, 01, 10, 11):
\begin{equation}\label{eq1_36}
\begin{split}
\lvert \Phi ^+ \rangle & = \frac{1}{\sqrt {2}}\left(\lvert 0 \rangle _ A \lvert 0 \rangle _ B + \lvert 1 \rangle _ A \lvert 1 \rangle _ B \right) ~ ~ ~ \rightarrow ~ ~ ~ 00 \\
\lvert \Phi ^- \rangle & = \frac{1}{\sqrt {2}}\left(\lvert 0 \rangle _ A \lvert 0 \rangle _ B - \lvert 1 \rangle _ A \lvert 1 \rangle _ B \right) ~ ~ ~ \rightarrow ~ ~ ~ 01 \\
\lvert \Psi ^+ \rangle & = \frac{1}{\sqrt {2}}\left(\lvert 0 \rangle _ A \lvert 1 \rangle _ B + \lvert 1 \rangle _ A \lvert 0 \rangle _ B \right) ~ ~ ~ \rightarrow ~ ~ ~ 10 \\
\lvert \Psi ^- \rangle & = \frac{1}{\sqrt {2}}\left(\lvert 0 \rangle _ A \lvert 1 \rangle _ B - \lvert 1 \rangle _ A \lvert 0 \rangle _ B \right) ~ ~ ~ \rightarrow ~ ~ ~ 11
\end{split}
\end{equation}

As an example, we assume that Alice and Bob initially share a $ \vert \Phi ^{+}\rangle $ state. If Alice wishes to send 00, then she need only send her qubit to Bob (the Identity operation). However, if Alice wishes to send the bits 10, then she must first perform an X-gate on her qubit, indicated with the subscript; \\ $ A: (X\otimes I) \lvert \Phi ^+ \rangle = \frac{1}{\sqrt {2}}\left(\lvert 0 \rangle _ A \lvert 1 \rangle _ B + \lvert 1 \rangle _ A \lvert 0 \rangle _ B \right)=\vert \Psi ^{+}\rangle $, with similar relations for the $ Y- $ and $ Z- $ gates. Once the encoding is complete, Alice transmits her qubit to Bob. With possession of both qubits, Bob is able to perform a measurement known as a Bell-state measurement. This measurement allows him to determine which of the four Bell states the two qubits are in, so he can decode the message.

While QDC offers a simple, clear indication of how classical and quantum communications differ, there are several important challenges to its real-world implementation. The first is that distributing a Bell state between distant users, a requirement to begin QDC is not trivial in practice. One of the most promising methods currently for this is to use photons entangled in such degrees of freedom as polarization, frequency, or time of arrival. Unfortunately, a single photon is extremely fragile. Therefore, it is likely to be corrupted or lost during transmission through free-space or optical fiber over distances of tens or hundreds of kilometers. A second challenge is how to store those pre-shared entangled photons before they are needed for communication. A third challenge is that the most straightforward ways to perform Bell-state measurements as currently do not distinguish between all four Bell states. For example, a setup might not be able to distinguish between the $ \vert \Psi ^{\pm }\rangle $ states, and hence only $ \log _{2}$ (3) bits are conveyed with a single qubit. For these reasons, the rate at which information can be conveyed using QDC is at present significantly less than the rate at which classical communication can be performed in practice, making QDC an important protocol to study from a theoretical perspective, but unlikely to be of practical use any time soon.

The second concept is the no-cloning theorem of quantum mechanics, which states that an arbitrary unknown quantum bit cannot be copied. This is a fundamental consequence of the linearity of quantum mechanics. It is important to stress that this theorem only applies to unknown quantum states, for if the state is known, it can be identically prepared over and over again. However, given a qubit in an unknown state, there is no means to copy it and end up with two qubits in the same unknown state. This concept has far-reaching consequences, one of the most direct of which is that quantum states used for quantum communication cannot be amplified, since an amplifier should essentially make many copies of a state and sends them along the transmission path to amplify the signal. Efforts to overcome this limitation have resulted in significant research into what are called quantum repeaters. These devices potentially enable longer distance transmission of quantum states, but at a heavy technological price.

The third concept is that attempts to intercept or measure a quantum state are detectable, a pillar of secure quantum communication. The no-cloning theorem rules out duplication of an unknown qubit state. Therefore, a potential eavesdropper has no choice but to measure some aspect of a qubit to glean its information, and this measurement process always leaves a detectable signature. A sender and receiver, therefore, know when an eavesdropper is present.

Quantum communication holds the potential for quantum advantage. This promise is found at the intersection of three broad applications: source authentication, secure communication, and collaboration, each of which will be described in the next section.

\section{Understanding Quantum Communication}
The quantum advantage potential for quantum communication can be understood in the context of the number of participating parties, from two-user communication to multi-user distributed applications. In this section, we will discuss several such examples that illustrate the promise of quantum communication.  

In the last section, we discussed the promise of quantum communication and its potential for quantum advantage. That advantage was based on three tenets of quantum mechanics, namely quantum-enhanced channel capacity, a concept known as superdense coding, the no-cloning theorem, which forbids copying an unknown quantum state, and a form of communication non-disturbance, by which any attempt to intercept or measure quantum bits can be detected. We can begin to understand how these tenets enable the promise of quantum communication by considering the number of participants as simultaneously participating in that quantum communication. The first example is the point to point secure communication using quantum key distribution. Secure communication relies on the use of a private key, essentially a string of bits used by a cryptographic algorithm. Quantum key distribution provides a means to transmit and share such a secret key securely. The secret key can then be used to encrypt and decrypt information. Any attempt to intercept that key can be detected, and when this happens, the compromised bits are simply discarded and replaced with additional secure bits. In this way, quantum mechanics enables two parties to communicate securely. Besides, We will discuss later, and quantum mechanics can also be used to generate true randomness, a resource for many cryptographic applications, and a step above existing classical pseudo-random number generators.
Second are applications involving a few participants. One important example is quantum secret sharing amongst two or more people. Let us say Alice wishes to send a secret to both Bob and Charlie. She would like them to learn that information simultaneously so that neither Bob nor Charlie can have an advantage over one another by learning the secret first. Using quantum communication to distribute an entangled state shared between Bob and Charlie, Alice has created a situation where Bob and Charlie must coordinate to uncover what that secret is. They can coordinate on an open classical channel, but quantum mechanics will ensure that they receive the information simultaneously. Finally, the third set of applications involves multiple distributed participants or multiple distributed computing nodes that, in concert, perform a quantum algorithm. One example is quantum scheduling, which uses Grover's Algorithm to search for a time when n distributed people are available.
Another example is related to distributed computing systems, such as the leader election problem. In this application, a unique leader or a master node must be chosen from amongst all the distributed computing nodes. In both cases, it is known that quantum communication and quantum computing together provide an advantage over classical approaches. Despite the promise and numerous examples of problems where quantum communication gives an advantage, there are only a few practical applications to date. One application that is available now, even commercialized\cite{mohseni_commercialize_2017}, is quantum key distribution or QKD. In QKD schemes, Alice and Bob want to share a private key, and any attempts by Eve to intercept it can be detected. There are demonstrations of QKD using photons over 100 kilometers of optical fiber and even a demonstration of transmitting signals between two points on earth using satellite quantum communication \cite{liao_satellite-relayed_2018}. QKD even serves as a foil to quantum codebreaking by Shor's Algorithm. We have already discussed, Shor's Algorithm can efficiently break RSA public key encryption. It is also efficient against Diffie-Hellman key exchange and elliptic curve cryptographic schemes. However, messages encrypted using a one-time pad are, in principle, secure. One-time pad schemes use each bit in a key only once and then discard it. They, therefore, need to be continually renewed, and this can be done securely with QKD. The main issue with QKD security is that, as with most cryptographic schemes, there are many channels for the attack. For example, one must generally control access to the hardware used to implement a QKD scheme\cite{zhuang_floodlight_2016}. Another limitation is signal attenuation. Classical communication links use repeaters to regenerate and amplify a signal. However, due to the no-cloning theorem, this is not straightforward with quantum communication. Instead, We will discuss later in the section, a quantum version of repeaters has been proposed that can effectively extend communication lengths using quantum teleportation and distributed entanglement\cite{podoshvedov_efficient_2019}. In conjunction with quantum memory, distributed entanglement would enable a quantum communication link to be established\cite{friis_observation_2018}. Other applications that may be useful include quantum secret sharing, auctioning or voting without attribution, verifiable quantum digital signatures, and message authentication. For all these examples, developing robust protocols and hardware systems that are immune to compromise will be necessary to realize the promise of quantum communication schemes fully.

\section{Fundamentals of Quantum Communication}

\subsection{Non-Cloning Theorem}

The no-cloning theorem states that it is impossible to replicate with certainty an arbitrary quantum bit. We can illustrate this statement by trying to come up with an operator $ \hat{U}_{clone} $ that can clone a qubit. Let us assume we have found a unitary operator $ \hat{U}_{clone} $ that can clone single-qubit states $ \lvert 0 \rangle $ and $ \lvert 1 \rangle. $
\begin{equation}\label{eq1_37}
\hat{U}_{clone}\lvert 0 \rangle \rightarrow \lvert 0 \rangle \lvert 0 \rangle ~ ~ ~ ~ ~ \& ~ ~ ~ ~ ~ \hat{U}_{clone}\lvert 1 \rangle \rightarrow \lvert 1 \rangle \lvert 1 \rangle
\end{equation}

However, if we apply this to an equal superposition state, following the rules of linear algebra, we find:
\begin{equation}\label{eq1_38}
\begin{split}
\hat{U}_{clone} \left(\frac{1}{\sqrt {2}} (\lvert 0 \rangle + \lvert 1 \rangle ) \right) \rightarrow \frac{1}{\sqrt {2}}(\lvert 0 \rangle \lvert 0 \rangle +\lvert 1 \rangle \lvert 1 \rangle ) \\
~ ~ ~ ~ ~ ~ ~ ~ ~ ~ ~ ~ ~ ~ ~ ~ ~ ~ ~ ~ ~ ~ ~ ~ ~ ~ ~ ~ \neq \left( \frac{1}{\sqrt {2}} (\lvert 0 \rangle + \lvert 1 \rangle ) \right) \left(\frac{1}{\sqrt {2}} (\lvert 0 \rangle + \lvert 1 \rangle ) \right)
\end{split}
\end{equation}

As this example suggests, it is impossible to define a $ \hat{U}_{clone} $ that can clone any arbitrary quantum state. This is a consequence of linear algebra and the nature of quantum mechanics, and it is called the no-cloning theorem.

\subsection{Non-Disturbance}

Non-disturbance describes the subsequent alteration of a quantum state due to measurement. The measurement process projects the quantum state on a particular predefined basis and causes an identifiable change. This process can be illustrated using the Bloch sphere. Suppose our predefined measurement basis is the z-axis. A measurement of a qubit in a superposition state will project it onto the z-axis. The resulting quantum state after projection (aligned with the z-axis) differs from the original superposition state before the measurement (it was not aligned with the z-axis). To conclude, attempted measurements of an arbitrary quantum state alter the quantum state and are, therefore, detectable.

\subsection{Quantum Key Distribution}

Quantum Key Distribution (QKD) is a communication method that uses unique features of quantum mechanics to exchange a secret key to encrypt and decrypt messages securely. In the underlying case, the keys are generated via a quantum channel between two parties, Alice and Bob. The principles of non-disturbance enable the detection of potential eavesdroppers on their quantum channels\cite{gyongyosi_survey_2018}. If the measured level of anomalies in their quantum measurement is below a certain threshold, the key is considered secure. Once a secure key is generated, it can, in turn, be used to encode and securely transmit information. The method of generating a secure key using quantum mechanics is known as quantum key distribution.

\section{Industry Perspective: Introduction}

In this section, we will discuss several leading industrial and start-up efforts targeting quantum computing. To get the discussion started, we asked several baseline questions. First, what technological approach are we employing to realize quantum computers, and why? Second, what is the business application we have in mind for the quantum computing systems we are developing? Third, what are the major technical hurdles we are facing? Fourth, how will our approach contribute to the advancement of scientific knowledge?
Furthermore, fifth, what other industries are either vertically or horizontally integrated with respect to our efforts? We also add in other perspectives and thoughts not captured in these questions. The results are very interesting and informative. Thus, with that brief introduction, let us discuss the major commercial players in quantum computing today.

\section{Industry Perspective: IBM}

So, if we want to make a qubit, first we must decide what kind of qubit we want to make at IBM, we like to use superconducting circuit qubits. So, as the name already suggests, it is a circuit approach to qubits, and we use many superconductors. the favorite superconductors here are niobium and aluminum. we also make them on the silicon wafers. So, there is much expertise that we have locally and how to process and make them on the silicon platform. one ingredient that all the different superconducting circuits share is a Josephson junction, which is a little bit of an unusual circuit element. However, the way we make it is, we have two pieces of superconductors that are very weakly coupled. We just separate two superconducting electrodes by a very thin tunnel barrier. That is just an insulating layer. it is like 1/1000 of a hair thick. So very thin. This Josephson junction behaves like a non-linear inductor, but then to make a qubit, we need a second element, a capacitor. we make like of regular linear capacitor out of superconducting material. Then, tying these two together gives us an LC resonator that can behave as a qubit if we tune the parameters correctly. So now we have a qubit. In order to make a useful qubit processor, then we need to tie these different qubits together. We use microwave buses to couple them. then lastly, we provide input and output through a readout resonator to each of the qubits that then couples to the outside. Right over here is one of the dilution refrigeration systems, which cool down the superconducting qubit devices. There are several different plates inside here, which all sit at different temperatures. That gets down to 15 millikelvins. That is colder than outer space itself. The sound we hear is a pulse 2 compressor, which essentially is pumping on a closed cycle of helium 4, which helps us get the system cold. We have much other equipment that we use in order to run the processors. So, there is a lot of microwave hardware, different passive components including filters and attenuators, and co-ax cables, which allow us to send the signals down to address the qubits and to read them out to us the controllability. This is a four-qubit package. We have a five-qubit device that is right now inside of the fridge, but the general idea for how we package up these devices and cool them down is the same. this is a printed circuit board to which we mount the qubit chips. we wire bond to them to connect to essentially these co-axial pins. These co-axial pins connect to cables inside the fridge, allowing us to send the signals down and take signals back and read them out. We are not going to have quantum computers in our pocket. We will have quantum systems in the cloud that we will be able to access. for many people, day-to-day basis, they may not even know that their information is coming from a quantum computer in the future. However, they would benefit from the value that quantum computers have created. By the time we have 50 qubits or so, that system no conventional classical computer, that We have ever built or could ever build, could emulate what that 50-qubit quantum system will have. we are going to see it in the years to come. A central question in general that one must answer here is considering the amount of effort and the challenge that it is to build a quantum computer, why do we want to do this? Moreover, the answer is that there are applications, or there are algorithms for which we know these algorithms vastly outperform their classical counterparts. So, we may be familiar with Shor's factoring algorithm or Grover's search and so forth, just to name a few. What is important to keep in mind here is that these algorithms require a universal fault-tolerant quantum computer\cite{chow_implementing_2014,corcoles_demonstration_2015,kandala_error_2019,colless_computation_2018}. At the current stage, we are still quite off from that. Several challenges must be overcome. The hardware must improve. We must develop better error-correcting codes so that the demands of the hardware become less. Nevertheless, at the same time, there has been steady and considerable progress in the experiments, with increased coherence rates and increased gate fidelities. We now enter a regime where it becomes challenging for classical computers to simulate these devices. Now, once we build increasingly large devices, we can ask ourselves, well, if this device is hard to simulate, is there already something meaningful that We can do with this? Furthermore, this is the question of trying to find near-term applications for quantum computers. Considering that we do not have a fault-tolerant computer, we must start making some concessions. Number one is we are going to have to look at algorithms or heuristics that do not come with a clean analysis, as they do for the cases as mentioned above. The other thing is that we must develop algorithms that are closer to the hardware that we are working on. lastly, we must think about errors constantly while developing those algorithms. We are currently focusing on three areas of applications: quantum simulation, quantum heuristics for optimization, and machine learning with quantum computers. One must keep this dichotomy to near-term and long-term in mind because many of the current algorithms that have been developed rely on having a fault-tolerant quantum computer. here, it is the task or goal basically to develop applications that are more closely tailored to the actual hardware that we are dealing with. Work is being done to develop error-correcting codes for use with quantum computers. Does the IBM Cube use an error-correcting code, or does it perform computations on a best-effort basis? If it does, what are the error-correcting codes it uses? And how does it implement that? So, many groups, including IBM, but groups worldwide, are focused on developing and demonstrating error-detection and error-correcting codes. today, there have been, we would say, very small-scale demonstrations of error detection by IBM and Google delved. there has been some demonstrations of error detection and correction, for example, at Yale, again, on smaller scale systems. But if we are asking about the online quantum computer that IBM's using today, the answer is that it does not use error correction, active error correction, in this sense. As we will discuss there is a hierarchy to an architecture for building a quantum computer. at the bottom of that architecture, are the physical qubits. those qubits are they are good. But they are faulty. They have error rates of $ 10^{-3} $, to $ 10^{-4} $. to really run a computation at scale for a period of a day 24 hours, let us say we really need error rates at $ 10^{-15} $, to $ 10^{-20} $. So, to achieve that, we need to do active error correction. But in between the hardware and the active error correction, there is another layer that is been developed. this is used by groups worldwide. this is called passive error suppression or error mitigation \cite{kandala_error_2019}. what this is, essentially, dynamical decoupling. by applying the same types of gates, single and two-qubit gates, mostly single-qubit gates, that we use to control and run an algorithm, control the computer and run an algorithm, those same gates can be used to mitigate certain types of errors basically, coherent errors, errors where we do not lose information to the environment, but the qubit is dephasing in some way. those types of errors can be mitigated using dynamical decoupling. so, by applying these pulses, mitigating these coherent errors\cite{endo_mitigating_2019}, we can make the fundamental physical qubits last even longer. They have even better error rates. that is important because error correction then, active error correction, is very expensive. the better the qubits are to begin with, the less overhead one needs to implement full-op error correction. so, what we would say is that we are not doing active error correction today, generally. But what we are doing, we would say, quite broadly is doing this passive error suppression. we want to call it error correction. So, we will call it passive error suppression or error mitigation \cite{endo_practical_2018}. Another part is, what is the status of error-correcting code development as it pertains to trapped ion qubits and semiconducting qubits? just to reiterate, we think in both cases, these technologies or the researchers who are pursuing these technologies are first working on error detection. So, to perform error correction, we first have to identify if and when an error occurred and on which qubits. the magic of quantum error correction, is that through specially designed syndrome measurements, we can make measurements which do not project the quantum information that we are trying to preserve. But they project onto a space where that information we gather is was there an error or not. So, we never discuss, for example, whether the qubit was in state 0 or state 1. We discussed that there was a bit flip error and that a 0 became a 1 or 1 became a 0. we discussed which qubit it happened on. so, with that information, we can go back and flip the qubit back to its correct value. But we never discussed what the correct value was. We never discuss whether it was supposed to be a 0 or supposed to be a 1. We only discussed that a bit flip error occurred, and we should flip it back. So, that intuitively the type of or the way that error correction works and to do that, we first have to detect if there was an error. We have to perform a syndrome measurement. So, many groups are working on doing those syndrome measurements for different types of error-correcting codes. Some of them are relatively simple codes, such as the bit flip code or the phase flip code. Some of them are more mature codes, like the surface code \cite{litinski_game_2019}. So, the surface code\cite{jones_layered_2012}, for example, requires what is called a wait for parity check \cite{farhi_limit_1998}. researchers, are trying to demonstrate that detection with high efficiency. then once their detection is mastered, then we can run error correction. We can go back in and correct for those errors once we discuss where they are.

\section{Industry Perspective: Google}

As with any information technology, quantum computing requires an entire stack. So, we go from application software to programming environments, all the way down to the hardware that realizes the actual quantum operations. for the hardware, we are currently banking on superconducting qubits since it is a technology suite that is complete. Thus, we know how to initialize a qubit. We know how to manipulate a qubit, or several qubits, by gate operations. we know how to measure those qubits. we know to do all those operations with rather high fidelity. However, we also keep a keen eye on competing technologies, such as silicon qubits or topologically protected qubits. However, at this point, we feel good that with superconducting technology, We will be able to build a small error-corrected quantum computer. On the software side, the most natural way to offer users access to a quantum processor is through a cloud interface. For example, a superconducting qubit the processors, they sit in a refrigerator that keeps them at about 10 millikelvins. it is too bulky to put this under our desk. However, it is no problem to have those in a data center. we abstract this complexity away from the user by offering a cloud API through which we access these processors. the roadmap for developing applications looks as follows. We will first start by passing a waypoint, which is known as quantum supremacy. Quantum supremacy roughly means the point at which a quantum computer, or some quantum device, surpasses the equivalent available technology on classical computers or classical devices. We are particularly interested in computational quantum supremacy, which means that a quantum computer will surpass for a well-defined computational task what we can do with a state-of-the-art supercomputer and state of the art classical algorithms. We call it the way post because the quantum supremacy benchmark task is not necessarily tied to something useful. So, it is important to realize that once we have achieved quantum supremacy, we are not yet in the phase of error-corrected quantum computing. To do error correction, like in classical error correction, we need redundancy. Unfortunately, the ratio of physical qubits to logical qubits to have one well-protected logical qubit is very high. It is about 1,000 to one. So, in order to have 1,000 logical qubits, we would need about a million physical qubits. we are still several years away from that.
So, about quantum supremacy. it says, it looks like quantum supremacy has not been proved yet. Why do we need to invest our resources in quantum technology-related projects now before quantum supremacy has been proven? we would say that the aspect of quantum computing that there is a quantum speed-up that would, if we could build that system, demonstrate quantum supremacy that has been proven mathematically, that there are problems for which a quantum computer, if and when we build it, will provide quantum speed-up. Now, this second part of that question related to quantum supremacy has not been demonstrated yet. we think that we are close. we will demonstrate on probably a computer on the order of 100 qubits a calculation of some type that cannot be efficiently simulated on a classical computer. we think that that will happen. that will be a concrete demonstration of a problem for which there is quantum supremacy \cite{harrow_quantum_2017, boixo_characterizing_2018}. we think that will happen within a year or two. Many companies are focused on doing exactly that, which is making a demonstration of quantum supremacy. Now, will that be for a problem that is useful that we will then use in our company to solve some problem that we have? That remains to be seen. We do not know yet. It could be that quantum supremacy is first demonstrated on a problem which may be of importance to, fundamental physics. It may be a problem that demonstrates some aspect of physics, but still needs to be translated into a practical problem for businesses. But we think that the concept of quantum supremacy is much more important than that. The concept is that we have demonstrated, in a real physical system, this concept that there are tasks that a quantum computer can perform and come up with an answer, and we can check that It is the right answer, but that a classical computer just cannot do efficiently. it will be a concrete demonstration. Even though we know mathematically, we have proven that this should be the case and is the case, we think having that physical hardware and demonstrating it directly is important. Nevertheless, we are optimistic that there are several areas in which we can develop algorithms that can already do something useful before error correction. There are probably three applications that We would encourage investigation. One is the quantum simulation. So, this is simulating quantum systems that cannot be simulated by even the world's biggest supercomputers. The second area would be on discrete optimization. So, this is optimizing very complicated optimization problems. the third is machine learning. In terms of verticals, the verticals that We have seen much interest in would be automotive, aerospace, chemistry, pharmaceuticals, and, more broadly, just defense. the reason why We would say that these are interesting, especially automotive and aerospace, is that both these kinds of industries have both complicated optimization problems, but they also have chemistry-related problems\cite{reiher_elucidating_2017}. For example, accelerating the development of batteries for electric cars. Right now, there are many issues with the battery technology we use. For example, it uses a rare material, such as cobalt. So, the number of batteries with today's technologies that we can build. It is estimated to be 80 to 90 million batteries only, and then all cobalt reserves are exhausted. So, we need to find alternatives. there are just a lot of areas within the electrochemistry of a battery that, with quantum simulation, we possibly can better understand and devise better materials. Quantum processors are finicky beasts. They are cooled to below outer space temperatures. they need to be robust against noise and any interference that happens. so, one of the biggest challenges is just isolating these systems and keeping them in their quantum state. There are two main sources of noise for the systems. there is coherent noise and incoherent noise. we work to reduce both through excellent fabrication, also through the control electronics and the readout as well. Also, scalability is a big challenge for these systems. We are operating in a 20 to the 50-qubit range today. We are looking to scale up to 1,000, eventually a million, qubits. we must invent new technologies as we hit each of those certain milestones. What gotten we to one or two or 10 qubits is not going to get us to 1,000. so that scalability with the packaging, the electronics, the control is key for us to build a system that can scale up to a large enough size that we can run error correction on these chips, which we need. We all know how important computers are in our lives. We think about quantum computing. We have made a change into a realm where the basic rules are different. So, we think understanding the power of quantum computing is important for science, in general. In recent years, everyone has been very hyped, for a good reason, about machine learning and artificial intelligence. people have also been very excited about the possibility of using quantum computers to speed up computation. So, there is a natural question, can we put these things together? Furthermore, to that end, we have personally been thinking about specific setups for quantum computers to run machine learning tasks. We hope that we will discover some areas in classical machine learning where the quantum computers' additional power accelerates or speeds up the machine learning tasks or allows for more reliable predictions, or perhaps learns to learn with fewer data coming in. Another area, which We think is sort of an obvious approach to use quantum computers, is if our data itself is not classical. So, we all know we can store classical data, like images, on classical computers. However, if We gave us a quantum system and We just told we it was a quantum state, that cannot be represented by a long bit string. The quantum state is too complicated. But nonetheless, quantum states exist in nature. So, we could ask, if We presented a quantum state to a machine a quantum computer could it learn to distinguish some attribute of that state which a conventional computer could not do? Moreover, the conventional computer might not be able to do it because it cannot even accept the input. Alternatively, even if somehow or another, we can encode the input,  the quantum computer is just more naturally inclined to be able to answer questions about quantum states. So, we think that is a whole other area for quantum machine learning\cite{kerenidis_quantum_2017}, which needs to be explored. For more information about Google's Quantum AI research,\cite{foxen_demonstrating_2020,arute_hartree-fock_2020,mcclean_decoding_2020,takeshita_increasing_2020}.

\section{Industry Perspective: Microsoft}

Microsoft, are building a full quantum stack, and the approach we are taking harnessing topological qubits that perform longer or more complex computations at a lower cost and higher accuracy. What it allows us to do is store and process quantum information in a more protected way. It is stored in a topological state of matter, these Majorana quasiparticles. What remarkable is that it has built-in fault tolerance and error correction. The actual value of the qubit is stored in a joint parity state, say over a nanowire, where that state is stored across two wires to form a qubit.
Moreover, the computation then occurs by moving these quasiparticles around one another. It becomes what we call a braid in space-time, and this braid in space-time is the actual computation. Now it turns out these braids can be implemented through measurement only. So, we do not ever move the quasiparticles, but we can do joint measurements of these quasiparticles, and through sequences of joint measurements, we can enact and implement quantum computation. So, while other quantum approaches have more qubits right now, by number, in the short term, right, we believe the topological approach will deliver orders of magnitude, better accuracy, and scalability in the long term. When We think about the first business application, we think about how We will bring quantum to Microsoft's cloud, Azure, and allow it to be integrated directly with people and organizations, apps, data, and infrastructure that is already existing.
Moreover, we see users in Azure using quantum in concert with advanced classical conventional processing and storage. A quantum system, it will be unlike any other system built to date, and there are a handful of challenges that we must address as we seek to scale this system up. One challenge is the development of a robust, scalable qubit. We have been working on that advancement, with the topological qubit. Another challenge as we work up the stack is the development of the control and readout platform. We have been developing a state-of-the-art cryogenic compute and control system that cools matter to the coldest temperatures inside this dilution refrigerator, while also using an extremely small amount of energy. Throughout this full-stack, then, as we work up even further, we also need software to enable programming the quantum system and to enable running an actual application on the quantum system. We are building out not only a quantum development kit that includes a quantum focus programming language called $ Q\# $, but we are also building the full runtime environment for running and controlling the quantum system. Within this runtime, there are several challenges to address, everything from optimizing compilation, layout, and scheduling of the algorithm on the actual device. We must be able to tune and calibrate and verify that the qubits and the operations and the compute. Then, at the highest level, we also face the challenge of developing quantum applications. Quantum computing and quantum algorithms present fundamentally new programming and algorithm design paradigms.
How do we get speedups over classical approaches with these new paradigms? How do we fundamentally unlock new ideas in computing? We think we are only scratching the surface in terms of the potential of quantum computers. We already believe that what We have done in quantum computing has contributed to scientific knowledge. In 2012, the team member, Leo Kouwenhoven, and his team in Delft, they demonstrated evidence of the Majorana quasiparticle that underlies the topological qubit. This is not only fundamentally new physics, that really pushes the frontier of the scientific knowledge, but it is also a giant step forward toward a qubit that enables scaling more efficiently. In addition, we think other areas have pushed the scientific knowledge forward. It is in the area of studying quantum methods, quantum algorithms, and quantum programs. So, we have been able to bring new ideas from quantum computing over to classical methods and algorithms and programs. In turn, this has resulted in not only improvements in quantum algorithms, but We have seen improvements in classical algorithms from learning about quantum algorithms. So, for example, we have discovered remarkable connections between synthesizing gate sequences for a quantum computer, like single-qubit rotation operations, decomposing those into a sequence of universal discrete single-qubit operations. We have discovered that there is a connection there with algebraic number theory, which enables proofs of previously opened conjectures. It is truly amazing that quantum computing has the potential to really revolutionize the world to solve problems that we really cannot imagine solving with conventional classical computing. This is a fundamental, game-changing work. It is so exciting to be a part of that.

\section{Industry Perspective: IonQ}

We use individual atoms as the quantum bits. These are atomic ions, they are charged. We hold them in a vacuum chamber, and They are away from any surface. They are not part of the solid. So, they are nearly perfectly isolated from the outside environment, and this makes them perfect quantum testbeds to do quantum operations, quantum gates and to build up a large-scale quantum computer. There have been many results over the last few decades on manipulating these individual atoms and making small scale entangled states, small algorithms. They have largely been demonstration experiments, and they exploit the beautiful coherence properties of individual atoms and that each qubit is the same as its neighbor. So, when these qubits are idle, their quantum memories are as perfect as we can get because They are individual atomic clocks. There are challenges in scaling the system up to go beyond 50 or so qubits would require some sophisticated control engineering to not only trap these ions in a single device but also to think about moving entanglement not only between that device but to another device that has as many qubits. The challenge now in the implementation of quantum computing is to invent and perfect the controller system around these ions. This means that we must develop laser systems and optical control systems to move information around. While it is a challenge and this is unconventional computer hardware, the big benefit of having optical controllers from the outside is that the system is entirely reconfigurable. That means that the atoms themselves are not wired together explicitly; we wire them from the outside. So, we can change the pattern of wiring between qubits, and this leads to great opportunities in the programmability of the quantum computer. One key challenge facing the entire quantum computing industry is finding near-term commercially viable applications. There are very well-known applications for large scale quantum computers. However, today we are focused on developing flexible and fully reprogrammable quantum systems to help drive this technology development. While we expect early applications to be those that map naturally to quantum processes, like quantum chemistry or material simulation, even if niche, these applications will help drive the technology development and will allow us to continue to attract top talent. One of the biggest hurdles for scaling trapped ion quantum computation to larger scales is the sophistication of the laser control that is needed to manipulate these qubits. Currently, we are using multiple channels of laser beams that are controlled by these acousto-optic modulators. Traditionally, these types of optics were constructed on a big optical table using individual optical components, as we would see in an atomic physics lab. As the system gets more sophisticated, the complexity of that optical system gets out of control and, therefore, leading to instabilities, meaning things drift over time, very difficult to service if something goes wrong or some parts of it get misaligned. Because it is so spread out over such a large area, it becomes very sensitive to environmental conditions, such as temperature, the humidity of the room. It turns out that there are significant advances in optical integration technologies that were mostly invented in other nearby industries, such as optical lithography, telecommunications where optics are used to send data signals very far away, and also imaging systems where people are building very large scale high-resolution images. So, we are starting to adopt the technological progress of these areas in these other fields to try to design the control systems for the qubits that are much more scalable, has a lot of multiplexity, and also stable with the precision that we need to control the qubits. So, the ultimate solution, we believe, is to modularize the system where we take a large-scale quantum computer and break them up into a large number of smaller modules, each containing a manageable number,  on the order of 100 qubits. We then try to make sure that those modules can be manufactured in vast numbers with a more reliable and compact geometry. Then assemble many, many of those modules and connect them together using a quantum interface in the form of optical communication or photonic interfaces, and that is how we are going to build a large-scale system. We find most exciting is not just the development and continued optimization of well-known algorithms today and applications today, but especially the discovery of new solutions, new capabilities. While we are in the very early days of quantum computing, if the development of classical computing is any indication, it is going to be very exciting next 5, 10, even 50 years.

\section{Industry Perspective: Rigetti}

Rigetti Computing is a full-stack quantum computing company building hybrid quantum-classical computing solutions. So, let unpack both of those. On the full stack side, this means we do everything from making the own design software to making the own chips in the case, superconducting qubit chips. However, we are packaging them in fridges, building control electronics, building software on top, and building that all into a cloud-deployed quantum programming environment, which, for us, is called Forest. So that is the first part. The reason we make this integrated approach is that a lot of the stuff at each of the layers has already been developed in academia, and now the big challenge is to integrate them and engineer them at scale and for a lower cost. The second bit is that we build hybrid quantum-classical computers, and this is to recognize that we are not going to have big, perfect quantum computers in the next few years. We are going to have small, noisy ones. So that means we should not treat quantum computers as standalone devices, standalone compute. They are co-processors that work in concert with classical computing resources. So, a classical computer can do things like tune around the noise in the quantum programs, or to assemble an ensemble of short quantum programs into the answer to a larger algorithm. So, from the ground-up, we have built the control racks, the instruction set, the programming environments, all to work in this hybrid model\cite{shaydulin_hybrid_2019}. So, we are building full-stack hybrid superconducting qubit systems. There are three sets of areas that we are most interested in beginning with. Those are quantum simulation, optimization, and machine learning. So fundamentally, in quantum simulation, it is because we have these quantum systems that we can apply all the techniques We have discussed in classical approaches to solving computational quantum chemistry and map them into how to do quantum programming for quantum simulation. Then in optimization and ML, it is a little more nascent, but there we are using the fact that we have, basically, a large linear algebra resource and trying to figure out how to sample and exploit from it. So, we are excited about both those areas of development. We think the most important business challenge is not something that affects just Rigetti Computing. It affects the whole industry, and it is education. If we must have a Ph.D. in quantum computing to build the technology or use the technology, this is never going to change the world. So, we need to figure out a way to make a new kind of community. It is a quantum computer engineering community, and there needs to be a shared corpus. There needs to be shared the basic content. The basic textbooks are still to be written for this quantum computer engineering community. We need to both do that on a technical level, and then translate that into something on the business side that people come to understand. One way to think about quantum computers is as a way of exploring this new reality that is out there. We have known about quantum physics for 110 years, and it is played its way subtly into how the technology works in a lot of ways, but we never really grasped it. A quantum computer is like a vessel that we can use to go into this new area of exploration to see what we can do with large, well-controlled quantum systems. That is a very, very fundamental question.
On the one hand, if it turns out that we cannot build quantum computers, this would break the models of physics, which is almost as interesting as the stuff we can do by applying them in a powerful way. There is so much foundational stuff to do. There are so many foundational concepts to discover. There is just a lot of ways on how to create an impact on how the field will evolve and grow. There is just not a lot of historical opportunities to be at the forefront of something like that. It is like we are going back to the 50s or 60s and saying, this computing thing that might happen. How could we be involved in that? Furthermore, quantum computers can happen faster because we have all regular computing to help us.

\section{Industry Perspective: QCI}

We have been pioneering a couple of new approaches for how to make quantum computations more robust and more easily scalable. This is what we call hardware efficient error correction, which means we need fewer parts to build into our quantum computer. Something called the modular architecture, which is a way of assembling smaller pieces into a more powerful whole, entangles them. The machines we see in the lab, are used to cool the special superconducting quantum circuits to temperatures a few thousandths of a degree above absolute zero, where we can prolong their quantum behavior, and where all the elements inside become superconducting so that there is no electrical loss in the computer at all. There are several other players in this field, including Google and IBM.
Moreover, they also use superconducting qubits and a circuit QED approach. They have decided to make their quantum processors out of a relatable technology, in the form of big 2D grids of these devices. QCI takes a different approach. We believe that just like supercomputers should be, and are, built out of networks of simple processors, quantum computers should not just be large grids of qubits. Thus, we are making the architecture modular, in the sense that the goal is to build a quantum computer out of reliable, easy to understand, and easily characterize units that are connected in a reconfigurable way. Thus, we feel we have a real head start, and a much easier path to rapidly scaling to large scale and useful quantum computers. What is special about the quantum system is that they offer possibilities that no classical system will ever offer. They operate on using, not the conventional logic we are used to, which is the Greek logic, but the logic of nature. There are aspects of it which we cannot predict as of now. We mean, there is an event and breakthrough that will make this new science available in fields that are not even thinkable as of now. So just a bit like asking the makers of the first classical vacuum tube computer to predict the internet.

\section{Industry Perspective: D-Wave Systems}

D-Wave uses superconductor integrated circuits to build annealing quantum computers. In terms of annealing quantum computing, this is kind of the most interesting part of what D-Wave has chosen to do because it is very different from what most of the quantum computing community has chosen to focus on, which is called gate model quantum computing. We are trying to harness quantum mechanical tunneling. So, the qubits have the possibility of tunneling back and forth between the zero and the one state. We can control how easy it is for them to tunnel back and forth between these two states. So, we start the algorithm, the quantum annealing algorithm, where all the qubits in the system are tuned so that it is easy to tunnel back and forth between zero and one.
Furthermore, the idea with quantum annealing is to start in that configuration and progressively make it more difficult to tunnel back and forth between the zero and the one state. So, at the beginning of the algorithm, the system, each one of the qubits is in a superposition of zero. One and the system is, in a sense, in a superposition of all possible answers that the final processor could encode. Some of those answers in that superposition are going to be the better or best answers for our set of constraints. We just do not know which ones yet. The algorithm works because we start with our dimmer switch turned down in this way, and we gradually turn up the strength of the constraints that we wanted to program on the processor. We are stating the problem more and more strongly as we go.
Moreover, as we are doing this, we are making it more and more difficult for the qubits to tunnel back and forth between the zero and the one-state—each of them. The idea with this, and again, just at sort of a hand-wavy level, is that we progressively select out, preferentially select out, those combinations of zeroes and ones, which are going to end up being the better answers. The lower energy cost-function answers out of this large massive superposition. Annealing quantum computing is relevant in areas of optimization, sampling, and machine learning. It cannot run Grover's algorithm, which is a prescription for a gate model algorithm, but it naturally solves an optimization problem at its core. Everything in a superconducting integrated circuit is made with Josephson's junctions. The current generations of processors have about 135,000 of those junctions on them. So, scaling the technology to be able to yield at that level has been challenging. However, also, the actual architecture of the processor to function, to be programmed, measured, and annealing properly at that kind of scale has been a big challenge. D-Wave has had to tackle the problem of bringing classical control circuitry directly into the quantum processors' environment to control and manipulate them and read them out properly. However, we are not there yet. So, there is still is a little way to go to mature technology. There is a whole world in front of us. What we find are the major challenges still facing us are continuing to scale the processors and crossing into a territory where we can run real commercial applications. So, D-Wave was the first company to produce a commercial quantum computer. We are the first to be able to go through these rapid development cycles, get customer feedback. Then the physicists, engineers, and software developers can improve the architectures as sort of, as pointed by customers working on actual applications and making them better. People discuss, in the digital computing world, that Moore's Law is coming to an end. In the part of the quantum computing market, we have gone from 120 qubits to 500 to 1000 to 2000, and then machines with four of five thousand qubits are on the horizon in the next few years as well.

\section{Limits of Classical Computing: A Brief Introduction to Computational Complexity}

We have discussed the promise of quantum computers and their potential for quantum advantage. Some problems are very hard to compute on a classical computer, and for some of these problems, a quantum computer is much more efficient. However, what makes a problem hard in the first place? we have probably discussed problems being categorized as $ P $ or $ NP $, $ NP- $complete\cite{lucas_ising_2014}. Nevertheless, what do these categories mean, and how do computer scientists classify a problem as an easy problem or a hard one? In this deep dive, We will discuss about the computational complexity of classical computing\cite{aaronson_computational_2010}, how it is defined, and how problems are classified, including a discussion of several standard problems that are classified as $ P $, and $ NP $, and $ NP-$complete. Let us describe a little bit about universality in the language of the circuit model of computation. That is what we naturally know best and is what we will build quantum computation upon as a language. So, in the circuit model, for example, we may have AND gates and NOT gates. we want to claim that all Boolean circuits, circuits that compute Boolean functions, can be composed of AND and NOT. this is easy to see, but first, let us look at the family of Boolean functions. They will be functions that take, if there are $ n $ bits of input, say $ x_0 $ to $ x_{n-1} $, and the result of this is going to be either 0 or 1, so for example, in the notation, We will say that we may have $ f $ of $ x_{1}x_{2} $ be $ x_{1}x_{2} $. this product represents an AND gate. Alternatively, for example, $ f $ of $ x $ is equal to $ \bar x $ . This is how we will represent a NOT gate? Now, the question of universality dovetails with another question. we have not finished the universality description yet. However, we want us to keep this in mind of complexity because it is not useful if We say something about universality without saying how much cost it takes to accomplish something. So, the key idea which we will not have time to go into much depth on, but We hope we already know something about, which is that some math problems are harder than others. we do not mean it is hard to multiply two million-digit numbers, and that is harder than multiplying two 10-digit numbers. No, that is not the point. The point is They are scaling with respect to the size of the problem. this leads to a statement which we hope that all of we already know of in some form, which is called the Strong Church-Turing Thesis. we will write this up for us because we think the words of this are instrumental in understanding the perspective. So, we say, any model of computation can be simulated on a specific kind of machine. the machine and model We would choose will be this one here, the Turing machine. However, we are going to choose a specific variant of it, the probabilistic Turing machine. we need to say something about the cost of this kind of simulation. the essence of that thesis is that this simulation cost comes with, at most, a polynomial increase in the number of elementary operations required. It is an excellent thesis. It defines equivalence between models, whether they be electrical and optical or electrical and DNA or quantum and DNA or quantum and classical. for even that to be possible, conceptually is remarkable. we have not defined a lot of the technical terms in this, like simulation and the overhead costs, but We hope we will appreciate that as we go along. So now, in order to highlight this statement of equivalence, let us make sure we are aware of one of the greatest motivating factors for quantum computing, which is the difference between two of the most important classes of mathematical problems. we will use this fact that many problems can be expressed as decision problems. So, for example, is the number $ M $ prime? Furthermore, the answer to this is yes or no. Is this a hard problem or an easy problem? This was not known for many years. It was then realized that we could answer this question with some randomness. we would not know it for certain. This is Rabin's primality testing algorithm. then some 15 years ago, somebody in India proved that we could do it deterministically. So, it moved from this probabilistic model, which people had to use previously for this question, to something which did not need probability. So today, there is a deterministic primality testing algorithm. This problem is called primality. That is one example. Here is another one. It is called factoring. Given a composite integer m, but not just the number that we are going to try to factor, also and a number l, an integer l, which is less than $ m $, does $ m $ have a factor that is non-trivial which is less than l? So, we need to bound the sides of the range of numbers we are going to consider as being answers to the problem. again, this is a yes or no question. we have this distinction. If the time taken to answer this question, needed to answer this question, it is polynomial in the size of the question. Here, for example, for factoring, this is the number of bits of $ m $, the number of digits in the number, not just the number itself, then we say that the problem is polynomial in complexity. We say it is in this class that We will call $ P $. Now, let us break down the class on whether this is answered by a yes or by a no. If the yes instances of the problem are easily checked, and We will use the word verified as a technical term with the aid of a witness, which is a short description piece of information that enables somebody else who is not skilled at the arts but can be very reliable, to verify our claim. Sometimes, we discuss Merlin and Arthur \cite{terhal_adaptive_2004}. We say that the Merlin is very clever and can come up with proofs, but we need the Arthur, who is not clever but can be very, very reliable to take that proof and verify the claim. So, there are two parties to this, the verifier and the prover. the verifier takes this witness. If this is true, then the problem, even if it is challenging, is the fact that we can verify that the proof of it means we will say that the problem is in this class called $ NP $. in some senses, we want to say this is non-polynomial, but the main point is the distinction between $ P $ and $ NP $. for the sake of completeness, we have another parallel to this: the mirror image. If no instances, so this is the answer is yes, and the answer is no with the same language, then we say that the problem is in a different class called $ co-NP $. we do this not because We want to say anything more about P versus $ co-NP $ or $ NP $ versus $ co-NP $, but just to share with we a trivial fact in a moment. So, the reason We show this is because so much of the motivation for computation today, and much of quantum computation comes from this question of $ NP $ versus $ P $. We think that the $ P $ problems are easy. The $ NP $ problems are the hard ones and meaningful ones to do. There is this plot that we can make, which shows that, if this is the space of all mathematical problems, then $ P $ is a subset of something we might say is $ NP $, but there is also this extra area over here, which is the hardest of the $ NP $ problems. we call these $ NP- $complete problems. they are defined as such because, if we can solve any of the $ NP- $complete problems, then it gives us a polynomial-time algorithm to solve any other problems in the $ NP $ regime. So, where is quantum computation in all this map? Well, we are not going to answer that for we today, but We hope that, through this, we will start to appreciate where quantum computation is relative to this landscape of the field of all problems and their complexities. Quantum computation sits a kind of difference in this landscape. It has a complexity class, typically of something we call $ BQP $ \cite{aaronson_bqp_2009,aaronson_forrelation_2014,raz_oracle_2019}. part of the reason is that the model is slightly different and does not fit directly into either of these classes, because sometimes the output is quantum mechanical. There are errors involved, and some things We will come to later. Good. We hope that many of us already knew about most of this, but We also hope that it starts to connect some things for us. An example of an $ NP- $complete problem and now We going to go back to the concept of universality is, so an example is a problem called $ 3-SAT $. this is about the satisfiability of Boolean functions, which looks something like this. So, suppose we have, again, a function of bits $ x_{0} $ through $ x_{n-1} $. The formulas that are involved in $ 3-SAT $ look something like $ x_{1} + x_{3} + x_{9}$, OR'd together, AND'd with there is a multiplication here another term, like $ x_{4} + \bar{x_{7}} + x_{11}$, and so forth and so on, where each one of these terms just has three bits. our goal is to say, does there exist an assignment of zeros and ones to the $ x $ inputs such that the output is equal to 1? It seems very simple. It is straightforward to write this problem on the board. However, if we can solve this problem fast in time. That is polynomial with $ n $ here, and then it turns out we can quickly solve all the rest of these problems. if we can solve this problem fast, we can solve problems like the optimum way to pack boxes into a FedEx truck, or the optimum way to route a packet in for information from San Francisco to Boston. we know, it is remarkable how powerful such a simple problem can be. However, we can show that, in practice, we cannot solve most of the instances of this problem as well as we would like to. Are there simple problems that are neither $ P $ nor $ NP $? Are there simple problems that are neither P nor NP? Another interesting class that sits outside of this class is counting the number of solutions to a problem. That is the class called sharp$ -P $. so forth, because we might count the number of things to count the number. However, we know this is how we get a career as a theoretical computer scientist. It is very effective.

\section{Computational Complexity: Topics on Complexity Theory}

So, what is complexity theory? Theoretical computer scientists love to come up with complexity classes\cite{bernstein_quantum_1993}. So, $ P $ is polynomial-time computation. $ NP $, things we can prove in polynomial time. So, who here does not know about $ NP $? So let us explain $ NP $ a little bit. The most famous $ NP- $complete problem may be the traveling salesman problem. Here we have a graph, distances between, let us call them, cities, which are the nodes of the graph. The question, is there a path visiting all cities less than length l? So, this might be 1, 2, 3, 5, or no 2, 3, 1, 1, 2, 3. So this would be a path of length 12.
Moreover, there are distances We mean there are roads between the cities. We have not put on here at distances along those roads. If these were the only roads, it would be an easy problem. If there are only six vertices, it is an easy problem. So, it is easy to prove. So, if We want to prove that there is a path, we just give us the path, and we add up the distances and check it.
Moreover, we can prove it. However, it is hard, or it may be hard to prove no. So how can We prove that there is no path shorter than distance 12 on that route? We could run through all paths and check all their distances, but that takes a very long time. If We know, and if We were trying to convince us that we have done this, would there be a better way to convince us other than just say, well, we ran through all the paths and checked them?
Furthermore, there are better ways for the traveling salesman problem, but they all seem to be an exponential time in the worst case. Or the exponential time in the number of cities. Now, some problems are $ NP- $Complete, which is as hard as any $ NP $ problem. In other words, if We could solve an $ NP- $Complete the problem in polynomial time in the length of its input, we could solve any problem in $ NP $ in polynomial time the length of its input. We do that by reducing one problem to another.
Furthermore, at some point in the 1990s, we think computer scientists came up with interactive proofs. There are three classes of interactive proofs. $ IP $, no bound on the number of rounds, $ AM $, and here $ ``A" $ stands for Arthur, and $ ``M" $ stands for Merlin. Merlin has infinite computing power, and Arthur only has a polynomial amount of computing power. So, $ A $ sends $ M $ message, $ M $ returns the message. So that is two rounds. Then there is also an $ MA $. $ M $ gives proof. $ A $ verifies it with coin flips. So, $ NP $, Merlin gives Arthur a proof, and Arthur verifies it deterministically. $ MA $, Merlin gives Arthur a proof and Arthur verifies it, but probabilistically.
Moreover, $ AM $, Arthur sends Merlin a message, and Merlin returns a message. The classic example of $ AM $ is graph non-isomorphism. Are these two graphs isomorphic and $ NP $. We are given two graphs, and We guess these have 5, 6, 1, 2, 3, 4, 5 6. So, the question is, is there a way to relabel the vertices so that they are the same?
Moreover, the guessed answer is an $ NP $, and an isomorphism is in $ NP $. Merlin just gives Arthur a relabeling. So, Merlin says 1 in this graph corresponds to 5 in this graph. 3 in this graph corresponds to 4 in this graph. Non-isomorphism is an $ AM $. Arthur takes one of the graphs. He re-numbers the vertices at random, and he sends it to Merlin.
Furthermore, Merlin is supposed to Merlin says which graph it is. Now, if the two graphs are isomorphic, we know, Merlin has the scrambled graph, but a random scrambling of this graph looks exactly like a random scrambling of this graph. So, he cannot tell them apart. If Merlin has an infinitely powerful computer, he can try all possible permutations and tell them apart if the two graphs are not isomorphic. So that says that non-isomorphism is an Arthur-Merlin.

\section{Classical Circuit Model}

This section we will program and execute the Deutsch-Jozsa algorithm on a cloud-based quantum computer using the IBM Quantum Experience\cite{lesovik_arrow_2019}. The Deutsch-Jozsa algorithm is programmed and implemented using a quantum circuit model of quantum computation. To get started, let us first review an example of a circuit model for classical computation. 

In this section, we will run an actual quantum algorithm, the Deutsch Jozsa algorithm, on a real quantum computer. Of the section, we will first discuss in detail how that works. To do this, we will need to introduce the algorithm using a circuit model of quantum computation \cite{nielsen_quantum_2011}. However, before we do that, let us first examine a circuit model in the context of classical computation. The circuit model for classical computation begins with input data prepared at the initialization stage. We can think of the computational bits in the input register being reset to all zeros and then set to the initial values that will be input to the computer. Now, this reset step may not be necessary for a classical computer. However, we will include it here for comparison with the quantum circuit model that We will discuss later. The initialized bits are then inputted to a computational stage. Here, a series of logical operations are implemented using classical Boolean logic gates to compute a series of functions.
Furthermore, once this is complete, the output for this stage encodes the result of the computation. For example, let us consider a full adder circuit comprising a series of Boolean logic gates. The full adder takes as inputs the bits B1 through B3. Where B1 and B2 are the binary numbers we want to add, and B3 holds any carry in that we may have from a previous calculation. Here we do not have a carry forward, so its value is zero. An output of this circuit is the sum, B1 plus B2 modulo 2, where the modulo 2 arises because this is binary arithmetic. The second output is the arithmetic carry out that may result from the addition. These two values are then assigned to bits B4 and B5 and stored in the output register. In this case, we add 1 plus 1, which is 2 with decimal numbers.
Furthermore, for the binary addition performed by this circuit, the addition is modulo 2, 1 plus 1 equal 0, and has a carry forward of 1. Finally, the results from this addition problem are obtained in a measurement stage, where the values of bits B4 and B5 are the binary representation of the answer, which may then be converted back to a decimal result. This type of initialized, compute, and measure process is the foundation for a universal classical machine. To access this machine, a user interface provides the input data and the program instruction set to calculate the desired function. This instruction set is then applied to the physical hardware using a controller layer. This layer takes the input data, sets the input bits, implements the physical logic gates according to the instruction set, measures the output, and then sends the resulting data back to the user interface. This computational model illustrates the basic building blocks that we need to implement an algorithm using a quantum circuit model. The main distinction will be the role of quantum mechanics and its impact on the initialization, compute, and measurement stages.

The blueprint for implementing an algorithm on a classical computer is the classical circuit mode. It describes how a set of input states is manipulated by a processing core and subsequently stored in an output register.

For example, a 2-bit$\times$2-bit binary multiplier uses a series of Boolean logic gates to multiply two input numbers, each taking an integer value between 0 and 3. The computation starts by first resetting the input register and then initializing it to the values of the two inputs represented as binary numbers. After that, the input states are then processed by a gate sequence that implements multiplication. Finally, the resulting output bit sequence, the computed result is placed in the output register where it can be readout.

All possible 2-bit inputs (a and b) and resulting outputs (c) are summarized in the following table:
\begin{table}[H]
\centering
\caption{2-bit inputs (a and b) and resulting outputs (c)}
\label{tab:1_1:Table 1}
\resizebox{\textwidth}{!}{
\begin{tabular}{|c|c|c|c|c|}\hline
decimal & binary & decimal output & binary composition & binary output \\\hline
        $ a \cdot b $     &   $ [a_1a_0]\cdot [b_1b_0]  $     &   $ c $      &      $ c_32^3+c_22^2+c_12^1+c_02^0] $                &       $ [c_3c_2c_1c_0] $           \\ \hline
        $ 0 \cdot 0 $    &        $(00) \cdot (00)  $    &       $ 0 $             &        $ 0\cdot 2^3+0\cdot 2^2 + 0\cdot 2^1 +0\cdot 2^0 $                &        $ 0000  $           \\ \hline
        $ 0 \cdot 1 $    &        $(00) \cdot (01)  $    &       $ 0 $             &        $ 0\cdot 2^3+0\cdot 2^2 + 0\cdot 2^1 +0\cdot 2^0 $                &        $ 0000 $           \\\hline
        $ 0 \cdot 2 $    &        $(00) \cdot (10)  $    &       $ 0 $             &        $ 0\cdot 2^3+0\cdot 2^2 + 0\cdot 2^1 +0\cdot 2^0 $                &        $ 0000 $           \\\hline
        $ 0 \cdot 3 $    &        $(00) \cdot (11)  $    &       $ 0 $             &        $ 0\cdot 2^3+0\cdot 2^2 + 0\cdot 2^1 +0\cdot 2^0 $                &        $ 0000 $           \\\hline
        $ 1 \cdot 0 $    &        $(01) \cdot (00)  $    &       $ 0 $             &        $ 0\cdot 2^3+0\cdot 2^2 + 0\cdot 2^1 +0\cdot 2^0 $                &        $ 0000 $           \\\hline
        $ 2 \cdot 0 $    &        $(10) \cdot (00)  $    &       $ 0 $             &        $ 0\cdot 2^3+0\cdot 2^2 + 0\cdot 2^1 +0\cdot 2^0 $                &        $ 0000 $           \\\hline
        $ 3 \cdot 0 $    &        $(11) \cdot (00)  $    &       $ 0 $             &        $ 0\cdot 2^3+0\cdot 2^2 + 0\cdot 2^1 +0\cdot 2^0 $                &        $ 0000 $           \\\hline
        $ 1 \cdot 1 $    &        $(01) \cdot (01)  $    &       $ 1 $             &        $ 0\cdot 2^3+0\cdot 2^2 + 0\cdot 2^1 +1\cdot 2^0 $                &        $ 0001 $           \\\hline
        $ 1 \cdot 2 $    &        $(01) \cdot (10)  $    &       $ 2 $             &        $ 0\cdot 2^3+0\cdot 2^2 + 1\cdot 2^1 +0\cdot 2^0 $                &        $ 0010 $           \\\hline
        $ 2 \cdot 1 $    &        $(10) \cdot (01)  $    &       $ 2 $             &        $ 0\cdot 2^3+0\cdot 2^2 + 1\cdot 2^1 +0\cdot 2^0 $                &        $ 0010 $           \\\hline
        $ 1 \cdot 3 $    &        $(01) \cdot (11)  $    &       $ 3 $             &        $ 0\cdot 2^3+0\cdot 2^2 + 1\cdot 2^1 +1\cdot 2^0 $                &        $ 0011 $           \\\hline
        $ 3 \cdot 1 $    &        $(11) \cdot (01)  $    &       $ 3 $             &        $ 0\cdot 2^3+0\cdot 2^2 + 1\cdot 2^1 +1\cdot 2^0 $                &        $ 0011 $           \\\hline
        $ 2 \cdot 2 $    &        $(10) \cdot (10)  $    &       $ 4 $             &        $ 0\cdot 2^3+1\cdot 2^2 + 0\cdot 2^1 +0\cdot 2^0 $                &        $ 0100 $           \\\hline
        $ 2 \cdot 3 $    &        $(10) \cdot (11)  $    &       $ 6 $             &        $ 0\cdot 2^3+1\cdot 2^2 + 1\cdot 2^1 +0\cdot 2^0 $                &        $ 0110 $           \\\hline
        $ 3 \cdot 2 $    &        $(11) \cdot (10)  $    &       $ 6 $             &        $ 0\cdot 2^3+1\cdot 2^2 + 1\cdot 2^1 +0\cdot 2^0 $                &        $ 0110 $           \\\hline
        $ 3 \cdot 3 $    &        $(11) \cdot (11)  $    &       $ 9 $             &        $ 1\cdot 2^3+0\cdot 2^2 + 0\cdot 2^1 +1\cdot 2^0 $                &        $ 1001 $           \\\hline
\end{tabular}}
\end{table}

A 2-bit$\times$2-bit multiplier can be constructed using Boolean logic gates of the following type (note: this is not a unique construction):

AND gate: outputs logic-state 1 solely if both input bits are in logic state-1, and outputs 0 otherwise.

XOR gate: outputs logic-state 1 if the input bits differ, and outputs 0 otherwise.

Analyzing the output bits $ c_0 $ through $ c_3 $ yields the following conclusions:
\begin{itemize}
\item $ c_0 $ equals 1 only if both a and b are non-zero and odd. Therefore, an AND gate applied to the bits $ a_0 $ and $ b_0 $ is sufficient to compute $ c_0 $:\\
$ c_0 = a_0~ AND~ b_0 $ 
\item $ c_1 $ will equal 1 if not all input bits and either $  a_0 $ and $  b_1 $ or $ b_0 $ and $ a_1 $ are in logic-state 1:\\
$ c_1 = (a_0~ AND~ b_1)~ XOR~ (a_1~ AND~ b_0) $
\item $ c_2 $ will equal 1 only if both primary bits $ a_1 $ and $ b_1 $  but not all involved bits  are in logic-state 1:\\
$ c_2 = ((a_0~ AND~ b_0)~ AND~ (a_1~ AND~ b_1))~ XOR~ (a_1~ AND~ b_1) $
\item $ c_3 $ equals 1 only if all bits composing a and b are initialized in logic state 1:\\
$ c_3 = (a_0~ AND~ b_0)~ AND~ (a_1~ AND~ b_1) $
\end{itemize}

The result stored in the four-bit output register is obtained via a measurement of the register and, for the users convenience, can be converted back to a decimal result. The sequence of initialization, computation, and measurement summarizes the working principle of universal classical machines. Although universal quantum machines \cite{goto_universal_2016} follow the principles of quantum mechanics and therefore require a different approach for implementing individual computational steps, the concepts derived from the classical circuit models serve as the basis for designing quantum circuit models.

\section{Quantum Circuit Model}

In this section, we will discuss the quantum circuit model and its similarities and differences with the classical circuit model. The quantum circuit model applies to gate-model algorithms, such as the Deutsch-Jozsa algorithm that we will implement in the lab practicum.

In the last section, we introduced a circuit model for classical computation. This model included an initialization stage to set the input bits, and a compute stage that implemented classical logic gates to compute a function and a measurement stage of extracting the output. In this section, we will introduce an analogous circuit model for quantum computation. The basic structure is the same. Initialize, compute, and measure. We will start with an initialization stage where the qubits are first to reset to a state $ \lvert 0 \rangle  $ and then initialized to their input values. We will assume here that the initial state is also $ \lvert 0 \rangle  $, $ \lvert 0 \rangle  $, $ \lvert 0 \rangle  $. The initialized qubits are then inputted to the computation stage, where a series of quantum logic operations will be performed. In general, one of the first steps in this block is to create an equal superposition state. This is done by applying single qubit Hadamard gates to the input register. To see how this works, let us first consider just a single qubit in state $ \lvert 0 \rangle  $. The Hadamard gate applied to this qubit takes state $ \lvert 0 \rangle  $ and rotates it to an equal superposition state,  $ \lvert 0+1 \rangle  $, and the normalization constant is $ \frac{1}{\sqrt{2}}$. Now, what happens when we have two qubits each initialized in state zero? Applying a Hadamard gate to each one independently rotates each into an equal superposition state $ \lvert 0+1 \rangle  $.
Moreover, when we take the tensor product effectively multiplying out the terms, we find an equal superposition state of 2 qubits, $ \lvert 00 \rangle + \lvert 01 \rangle + \lvert 10 \rangle + \lvert 11 \rangle$. now, the normalization is the square of $ \frac{1}{\sqrt{2}}$, which is $ \frac{1}{2}$. Similarly, applying a Hadamard gate to three qubits, each initialized in state zero, results in an equal superposition state of three qubits. We have seen before, this state comprises eight terms, from state $ \lvert 000 \rangle$ to state $ \lvert 111 \rangle$. the normalization is now $ \frac{1}{2\sqrt{2}}$ . Each coefficient now has this value. Creating a large, equal superposition state sets the stage for quantum parallelism and quantum interference to occur during quantum logic operations.
The logic operations themselves are the single-qubit and two-qubit gates we discussed earlier. For example, a single qubit $ X- $ gate or another Hadamard gate or a two-qubit CNOT gate, chosen to implement a function or algorithm. As discussed previously, the quantum parallelism and quantum interference that occurred during these operations changed the coefficients' values in the superposition state. Finally, the qubits are then measured to determine the answer. As we also discuss earlier in the section, the measurement process projects the qubits onto the measurement basis. This effectively collapses the massive superposition state onto a single classical state with a probability that is the magnitude squared of its coefficient. In doing so, the measurement leads to a single classical binary result, either a $ 0 $ or a $ 1 $ for each qubit. Now, if we do not perform any logic operations and simply measure the equal superposition state from the input, each coefficient has the same value, and we have an equal probability, one eighth, of measuring any one of these states. Thus, as we discussed earlier in the section, to be successful, a quantum algorithm is designed such that after applying a designated sequence of quantum logic gates, ideally, all the probability amplitude resides in one of the coefficients.
Moreover, this coefficient sits in front of the state that encodes the answer to the problem. Thus, when we make a projective measurement, the probability is a unity that we obtain this result. Using this basic quantum circuit model, we initialize, compute, and measure, and we can implement a universal quantum algorithm. In the next section, we will apply this model to a specific example, the Deutsch-Jozsa algorithm.

The structure of the classical circuit model initialization, computation, and measurement are directly applicable to quantum circuit models. The initialization stage starts with the qubits being set to their starting states, often the ground state  $ \lvert 0 \rangle  $ for each qubit. The initialized qubits are subsequently processed by a sequence of quantum logic operations in the computation stage.

In general, the first step of the computational stage is to create a massive superposition state to set the stage for quantum parallelism and quantum interference during the algorithm. For example, to create an equal superposition state from three qubits initialized in $ \lvert 000 \rangle $, we may apply Hadamard gates to each qubit individually, as shown below.
\begin{equation}\label{eq1_39}
\begin{split}
\displaystyle H\otimes H\otimes H \lvert 000 \rangle    \displaystyle & =    \displaystyle \left( \frac{\lvert 0 \rangle +\lvert 1 \rangle }{\sqrt {2}}\right)\left( \frac{\lvert 0 \rangle +\lvert 1 \rangle }{\sqrt {2}}\right)\left( \frac{\lvert 0 \rangle +\lvert 1 \rangle }{\sqrt {2}}\right),         (1) \\
\displaystyle & =    \displaystyle \frac{1}{\sqrt {2}^3}\left(\lvert 000 \rangle +\lvert 001\rangle +\lvert 010 \rangle +\lvert 011 \rangle \right.         (2) \\ &
\displaystyle +\left.\lvert 100 \rangle +\lvert 101 \rangle +\lvert 110 \rangle + \lvert 111 \rangle \right)         (3)
\end{split}
\end{equation}

After creating this equal superposition state, a sequence of single and two-qubit gates is applied to implement a particular function or algorithm. Quantum algorithms use quantum parallelism and quantum interference to ideally coalesce all of the probability amplitude into the coefficient of the state that encodes the answer; by bringing this coefficient's value to unity, the probability of measuring this state which goes as the magnitude squared of the coefficient is also unity. The values of the other coefficients and thus the probability of measuring states that do not contain the solution ideally decrease to zero. Sometimes, as in the Deutsch-Jozsa algorithm, the algorithm reaches a final answer in one step. In other algorithms, such as Shor's factoring algorithm or Grover's search algorithm, the algorithm repeats a cycle of operations that gradually modify the coefficients. The system converges to a solution after some number of iterations. In either case, at the end of the computation, the quantum system is in a state that comprises the solution state with coefficient near $ 1 $, and all other coefficients near $ 0 $. 

Upon completion of the computation stage, the qubits at the output are measured to determine the answer. Since the solution state has a coefficient at or near 1, we are very likely to measure that state when we perform our measurement. Because of the probabilistic nature of this measurement, and the potential for imperfect coalescence to the solution state, the measurement may need to be performed several times to ensure that we did not project the wrong answer by chance. In practice, the entire process may need to be run multiple times even for algorithms that, in theory, have the correct coefficient converging to 1 because there is always noise in the system that can lead a small probability that the algorithm gives an incorrect result.

\section{The Deutsch-Jozsa Algorithm}

In this section, we will discuss how the Deutsch-Jozsa algorithm works in detail. The section will begin with a general discussion of the N-qubit algorithm. It then presents the key mathematical steps required to implement the algorithm for a single data qubit. 

In this section, we will introduce the Deutsch-Jozsa Algorithm. Deutsch-Jozsa was one of the first quantum algorithms that exhibited a provable, exponential speedup over a classical algorithm. In this section, we will discuss how it works and walks through its key steps. by the end of this section, we will implement it on a real cloud-based quantum computer. To get started, imagine that we have an unknown function $ f $. It takes $ N $ Bits as input, and it outputs a single result, either a $ 0 $ or a $ 1 $. Now, all we know about this function is that it has a unique property. It is either a constant function, or it is a balanced function. What do we mean by that? A constant function takes any input, and no matter what that input may be, the function always produces the same result. For example, independent of the input, the function always outputs a zero, or no matter the input, it always outputs a one. A balanced function, on the other hand, will take those inputs, and for half of them, we do not specify which half, but for half of them, it will output a zero, and for the other half, it will output a one. In this sense, the output is balanced. The Deutsch-Jozsa problem is the following. Determine whether a function $ f $ is constant or balanced based on queries that we make of the function $ f $. we can send any input state to the function, and we get a result back, and we can query the function as many times as we like. Now for $ N $ Bits, there are two to the $ n- $th power different input states from which we can choose. we will need to run the function for at least half of them to determine with certainty if the function is constant or balanced. we will have to do it for half plus one. That is because even if we get all zeros for the first half of the states, we will need to try one more to see if the second half remains zeros, a constant function, or all one's balanced function. Thus, a deterministic classical algorithm will take $ 2^{{\frac{N}{2}}+1} $  step. it always works. Now, if we have $ N $ Qubits instead, we can create an equal superposition state of all $ N $ Qubits, and we will determine the answer in just one step. it always works. that is an exponential speed up. To see how this works in detail, let us simplify the problem to just one bit or one qubit. It is the same approach as for $ N $ Bits or $ N $ Qubits, but it will be much easier to follow if we just take $ n $ equals to one. In this case, for a constant function, we have $ f $ of $ 0 $ equals $ f $ of $ 1 $, and the output is either a $ 0 $ for both, or it is a $ 1 $ for both. for a balanced function, one of the outputs is $ 0 $, and the other is $ 1 $. In the truth table, we can make the following observation, that if we take the exclusive or $ f $ of $ 0 $ and $ f $ of $ 1 $, the result is $ 0 $ for a constant function, and $ 1 $ for a balanced function. In this case, classically, it takes two steps to implement the algorithm for $ n $ equal one. again, quantum-mechanically, it only takes one step. Now, that may not seem like a big speedup. It is only twice as fast, but, for $ N $ equal one, it is an exponential speed up. this generalizes to any number $ N $ that we may choose. Again, we are going with $ N $ equal one because it will be easier to see how the algorithm works. To implement the Deutsch-Jozsa Algorithm for $ N $ equal $ 1 $, we need two qubits. One is the data qubit, that is, the $ N $ equal one qubit, and the other is a helper qubit or an ancilla qubit. we will use the quantum circuit model to implement this algorithm. First, we initialize the qubits into their starting states. The data qubit is prepared in state zero and the helper qubit in state one. As we proceed through the algorithm, we will indicate the position at each stage of the algorithm and the state of these two qubits. For example, after initialization, we see that the data qubit is in state zero, and the helper qubit is in state one. To keep it all straight, we will use yellow to highlight the data qubit state and green to highlight the helper qubit state. We first create an equal superposition state by applying a Hadamard gate to both qubits. This results in a superposition state for each qubit. There is a plus sign for the data qubit since it started in state zero and a minus sign for the helper qubit since it started in state one. Next, we will just expand out the terms in the data qubit. The zero-state tensor product, the helper qubit superposition state plus one state tensor product, the helper qubit state. This state then inputs into the quantum circuit block. We will just call it $ U_ f $. The data qubit is $ x $, and the helper qubit is $ y $.  $ U_ f $ implements the set of logical operations. it does two things. First, it implements the unknown function $ f $, and second, it outputs the exclusive or of bits $ y $ and $ f $ of $ x $. We will show later how this can be implemented with single-qubit and two-qubit operations. For now, though, let us just figure out what happens. At the output of $ U_ f $, we have the helper qubit. it takes on the state $ y $, $ x $, or $ f $ of $ x $. So, where the data qubit is zero, the helper qubit is $ x  $ ord with $ f $ of $ 0 $. where the data qubit is one, the helper qubit state is $ x $ ord with $ f $ of $ 1 $. Now we make an observation. we should try this on scratch paper to convince ourselves that the expression can be written in this way with a minus $ 1 $ to the $ f $ of $ 0 $ power and a minus $ 1 $ to the $ f $ of $ 1 $ power. To check, there are four cases to consider. For example, if $ f $ of $ x $ equals $ 0 $ for any $ x $, the $ x $ or expressions on the left maintain the superposition $ 0-1 $. However, if $ f $ of $ x $ equals $ 1 $ for any $ x $, then the $ x $ or expressions on the left result is $ 1-0 $. to return this to the superposition state $ 0-1 $, we need to multiply by a $ -1 $. It is achieved by taking  $ -1 $ to the $ fx $ power. when $ fx $ equals $ 1 $, this is the minus one that we are looking for. The expression also works when $ f $ of $ 0 $ and $ f $ of $ 1 $ are not equal, but take on the values $ 0 $ and $ 1 $, the third case, or $ 1 $ and $ 0 $, the fourth case. The fact that this works for all four cases simultaneously is an example of quantum parallelism. In the next step, we factor out the helper qubit term $ 0-1 $ divided by root 2. we move the $ -1 $ to the $ f $ of $ 0 $ power and the $ -1 $ to the $ f $ of $ 1 $ power over to the data qubit. Next, we apply Hadamard gates to the data and helper qubits. On the data qubit, state zero rotates to $ 0+1 $, and state $ 1 $ rotates to  $0- 1 $. For the helper qubit,  $ 0-1 $ rotates to state $ 1 $. Next, we just note that  $ -1 $ to the $ f $ of $ x $ power is the same as $ e $ to the power $- \pi $ $ f $ of $ x $. then, we just rearrange terms to collect the coefficients of state zero and the coefficients of state one. this is an example of where quantum interference occurs, changing the coefficients values depending on whether the function is constant or balanced. For example, if $ f $ of 0 equals $ f $ of $ 1 $, a constant function, then $ b $ equals 0, and $ a $ is either plus or minus $ 1 $. Thus, a measurement will yield state 0 with unity probability. we observe that state zero is equivalent to state $ f  $of 0, $ x $ ord with $ f $ of $ 1 $. In contrast, if $ f $ of 0 does not equal $ f $ of 1, a balanced function, then an equal 0 and $ b $ is plus or minus $ 1 $. In this case, a measurement will yield state one with unity probability. we observe that state one can also be replaced with the state f of 0, $ x $, or $ f $ of $ 1 $. Thus, if we measure a 0 on the qubit, the function is constant, but if we measure a 1 on the qubit, the function is balanced, and we will always measure a state one on the helper qubit. Thus, we were able to determine whether the function was constant or balanced with one evaluation of the quantum algorithm. the same is true if we had instead used $ N $ data qubits. One evaluation always works. In contrast, the classical case must evaluate half of the 2 to the $ N $ states plus one to get a deterministic answer. Thus, the quantum speedup is one step versus 2 to the $ N $ minus 1 plus 1 step. Lastly, we can look at ways to implement the logical operation $ U _ f $. We will not discuss this through all of them here. we can find them in the text unit following this section. However, four cases are corresponding to the two constant functions 0, 0, and 1, 1, and the two balanced functions 0, 1, and 1, 0. We encourage us to work through each of these cases. Now that we understand how the Deutsch-Jozsa Algorithm works within the quantum circuit model, we are ready to write a quantum computer program that will implement it on a quantum computer. We will see how that generally is done in the next section, and then more specifically, in the lab practicum.

To ensure that we understand each step in the computation stage of the Deutsch-Jozsa Algorithm, Let us a review of the section for the 1-qubit case.\\
\textbf{Step 0:} To start with, we have a data qubit initialized in the state $ \lvert 0 \rangle $, and a ``helper'' (or ``ancilla'') qubit initialized in $ \lvert 1 \rangle. $
\begin{equation}\label{40}
\lvert \Psi _0 \rangle = \lvert 0 \rangle \lvert 1_ a \rangle
\end{equation}

We will use the subscript ``a'' to designate the ancilla qubit. Also note that If we wanted to run the algorithm on an N-bit number, we would need N data qubits, but still only 1 ancilla qubit.\\
\textbf{Step 1:} The first step of the computation stage is to put both the data and ancilla qubits in a superposition state by applying a Hadamard gate to each.
\begin{equation}\label{eq1_41}
\lvert \Psi _1 \rangle = H \otimes H\lvert \Psi _0\rangle
\end{equation}
\begin{equation}\label{eq1_42}
\lvert \Psi _1 \rangle = H\rangle \lvert 0 \rangle \otimes H \lvert 1_ a \rangle
\end{equation}
\begin{equation}\label{eq1_43}
\lvert \Psi _1 \rangle = \Big(\frac{\lvert 0 \rangle + \lvert 1 \rangle }{\sqrt {2}}\Big) \otimes \Big(\frac{\lvert 0_ a \rangle -\lvert 1_ a \rangle }{\sqrt {2}}\Big)
\end{equation}
\begin{equation}\label{eq1_44}
\lvert \Psi _1 \rangle = \frac{1}{\sqrt {2}} \Big[ \lvert 0 \rangle \Big(\frac{\lvert 0_ a \rangle -\lvert 1_ a \rangle }{\sqrt {2}}\Big) + \lvert 1 \rangle \Big(\frac{\lvert 0_ a \rangle -\lvert 1_ a \rangle }{\sqrt {2}}\Big)\Big]
\end{equation}
\textbf{Step 2:} Next, we apply the unitary operation $ U_ f $, which is implemented by a sequence of quantum logic gates (we will not worry about exactly which logic gates here). $ U_ f $ has the property that if the data qubits are in-state $ \lvert x \rangle $, then it puts the ancilla qubit into the state $ 1 \oplus f(x) $, where f is the function that is either balanced or constant, and $ \oplus $ denotes addition modulo 2 (note that this is equivalent to XOR for a single bit).
\begin{equation}\label{eq1_45}
\lvert \Psi _2 \rangle = U_ f\lvert \Psi _1\rangle
\end{equation}

\begin{equation}\label{eq1_46}
\lvert \Psi _2 \rangle = \frac{1}{\sqrt {2}} \Big[ \lvert 0 \rangle \Big(\frac{\lvert 0_ a \oplus f(0) \rangle -\lvert 1_ a \oplus f(0) \rangle }{\sqrt {2}}\Big) + \lvert 1 \rangle \Big(\frac{\lvert 0_ a \oplus f(1) \rangle -\lvert 1_ a \oplus f(1) \rangle }{\sqrt {2}}\Big)\Big]
\end{equation}

\begin{equation}\label{eq1_47}
\lvert \Psi _2 \rangle = \frac{1}{2} \Big[\lvert 0 \rangle \lvert 0_ a \oplus f(0) \rangle -\lvert 0 \rangle \lvert 1_ a \oplus f(0) \rangle + \lvert 1 \rangle \lvert 0_ a \oplus f(1) \rangle - \lvert 1 \rangle \lvert 1_ a \oplus f(1) \rangle \Big]
\end{equation}

We can rewrite this as:
\begin{equation}\label{eq1_48}
\lvert \Psi _2 \rangle = \frac{1}{\sqrt {2}} \Big[ \lvert 0 \rangle (-1)^{f(0)} \Big(\frac{\lvert 0_ a \rangle -\lvert 1_ a \rangle }{\sqrt {2}}\Big) + \lvert 1 \rangle (-1)^{f(1)} \Big(\frac{\lvert 0_ a \rangle -\lvert 1_ a \rangle }{\sqrt {2}}\Big)\Big]
\end{equation}
So, we can see that the effect of applying $ U_ f $ to our quantum state was the insertion of these $ (-1)^{f(x)} $ factors. These factors will lead to constructive and destructive interference enhancing and suppressing terms in the output state.\\ 
\textbf{Step 3:} Once again, apply Hadamard gates to both qubits.
\begin{equation}\label{eq1_49}
\lvert \Psi _3 \rangle = H \otimes H\lvert \Psi _2\rangle
\end{equation}
\begin{equation}\label{eq1_50}
\lvert \Psi _3 \rangle = \frac{1}{\sqrt {2}} \Big[(-1)^{f(0)} \Big(\frac{\lvert 0 \rangle +\lvert 1 \rangle }{\sqrt {2}}\Big) \lvert 1_ a \rangle + (-1)^{f(1)} \Big(\frac{\lvert 0 \rangle -\lvert 1 \rangle }{\sqrt {2}}\Big) \lvert 1_ a \rangle \Big]
\end{equation}
If $ f(0) = f(1) $, the $ \lvert 1 \rangle $ terms on the data qubit will cancel out, leaving only$  \lvert 0 \rangle $ terms. But if $ f(0) \neq f(1) $, then the $ \lvert 0 \rangle $ terms will cancel out (because one will get a plus sign and the other will get a negative sign) and we will be left with only $ \lvert 1 \rangle $ terms in the data qubit. So, we can now measure the data qubit, knowing that if we measure $ \lvert 0 \rangle $ then f(x) is constant, while if we measure $ \lvert 1 \rangle $, then f(x) is balanced.

If f(x) only operates on a 1-bit number (as in this example), then it is extremely easy to just measure f(x) on all the possible inputs classically (because there are only two possible inputs). However, imagine f(x) operates on a 50-bit number. We would then have to apply the function to $ 2^{50}/2+1 \approx 10^{14} $ inputs to know with 100\% probability that f(x) is constant. In contrast, the quantum algorithm allows us to determine this is only one application of the function.

Even so, the problem is a bit contrived. Although not a deterministic answer, one can establish classically whether the function is balanced or constant with high likelihood after several trials. For example, if one repeatedly outputs a 0, then one gradually builds confidence that the function is likely constant. There are not any real-world problems in which we have a function that we know is either constant or balanced. To obtain the answer with certainty for any a priori unknown function, a classical computer will need to make $N/2+1$ queries, whereas the Deutsch-Jozsa algorithm only needs to make one. Thus, although the Deutsch-Jozsa algorithm does not address a practical problem, it is an easy-to-follow demonstration of how quantum computers can give exponential speedup over the best possible classical algorithms.

\section{Introduction to Quantum Software Program General overview}
Up to this point, we have focused primarily on quantum computing algorithms and hardware. However, how do we efficiently design a quantum processor? How do we validate that it is working properly? Moreover, how do we program it? As with classical computers, there is an array of software that is used for these purposes. In this section, we will discuss the similarities and differences between the classical and quantum computing software stacks.  

Like classical computers, quantum computers have a software stack used in the development of programs, analysis of designs, and testing of constructive programs and circuits. The function and examples of these quantum software tools are a topic of this section\cite{larose_overview_2019}. Let us start by contrasting quantum software tools to classical ones to perform similar types of functions. To create a classical circuit design, one may use a schematic capture CAD tool. A quantum analog to this is the circuit composer from IBM's quantum experience. For program-based designs, for example, HDL circuit designs are written in VHDL or Verilog, or high-level computer programs are written in a language like c++, one uses a compiler to translate the program to a lower level format. Numerous quantum compilers exist that perform the same function. Some examples include Quipper developed by Dalhousie University and Q sharp from Microsoft. These quantum compilers typically produce an intermediate format known as QASM \cite{cross_open_2017} or quantum assembly language. Finally, to translate a program or circuit to a format that a machine can execute, one uses an assembler. A similar function is required for quantum circuits. Each hardware platform will have specific gates that it can execute, and the connectivity that it provides is specific to the technology's topology constraints. Mapping from QASM programs to hardware is something that IBM's QISKit software performs \cite{cross_open_2017}, in this case, for superconductor-based technology. For classical circuit designs, a circuit simulation tool like Spice is commonly used to determine circuit properties like power consumption, timing and to verify the correctness of the circuit. One of the main concerns of quantum circuits is how noise or errors impact the operation of the circuit. Quantum simulation tools exist at multiple levels of abstraction that model error in the operation of quantum circuits. Quantum assimilation tools are also used to calculate the fidelity of gates operating under the control waveforms and to verify circuits' correctness. The last category of tools as those used for testing of fabricated chips, PCBs. For classical chips, one may use hardware and software tools, such as JTAG and boundary-scan, to verify the operation of a chip. For quantum circuits, quantum characterization, validation, and verification, or QCVV tools provide a similar function. These tools evaluate results obtained from experimental runs to calculate the Fidelity devices and to verify their correctness. Let us now discuss the main steps one would use to program a quantum computer. Program generation, hardware-specific circuit mapping, and hardware control and execution. The program generation phase is hardware agnostic, whereas the other two phases deal with aspects specific to the architecture of the hardware platform \cite{smith_practical_2017}. Another difference between quantum computing systems and classical ones is that the control and processing systems are typically separate. There is a classical system for control and a quantum device for processing. In many cases, these two systems are physically separated. For example, a superconducting quantum computer may use room temperature instruments for control, whereas the quantum circuits require millikelvin temperatures, and must be located within a dilution refrigerator. Let us now look at the three main methods for program generation, the first being schematic capture. One of the front ends of IBM's quantum experience, the composer tool, is an example of a schematic capture tool, where one uses a GUI to construct a circuit consisting of one and two-qubit gates and measurements. The tool knows of underlying hardware constraints and prevents the user from violating these. In many cases, it is convenient to specify the quantum circuit as a program, which allows for the specification of larger-scale circuits. This leads to the motivation behind the second method for program generation, high-level quantum languages. Examples of this are tools like Quipper, Q Sharp, and Scaffold. High-level programs written in these languages are compiled into QASM circuits or can be displayed as circuit diagrams. The last method for program generation that we will discuss is problem specific generation. Google's Open Fermion software tool is an example of this method. This package is designed specifically for formulating simulations of quantum chemistry. An advantage of this approach is that it incorporates the steps and the known methods used for the class of problems, without requiring the user to know detailed domain knowledge or the methods used to solve these problems. A user specifies a problem at a high level, and the package generates a quantum circuit that can be mapped to a hardware-specific platform or simulation systems. Thus, we have seen the type of software tools commonly used to program a quantum computer and how these tools are like tools used for classical computers. In the next section, we will look at other types of software used in the control, design, and testing of quantum computers.

\section{Introduction to Quantum Software Analysis Testing}

In this second section on software, we discuss of instrument control then transitions to software used for circuit validation, verification, and benchmarking, and software tools used to model qubit performance in the presence of noise at various levels of complexity. 

As discussed previously, a quantum computer requires a classical control system. For small to medium-sized quantum computers, the control can be realized using commercially available instruments, such as arbitrary waveform generators, microwave sources, and laser systems. Software is required to program and synchronize the various instruments required. Labor is one example of this type of software. This software also provides a result visualization and logging functions. Quantum computer control systems that are custom designs require firmware and software control software. IBM's QISKit for superconducting circuits and ARTIQ for ion traps are examples of these types of software control systems.
Additionally, and especially for small scale experiments, many experimental groups use homegrown software control systems. We now discuss hardware testing and validation of quantum systems, which is typically performed with the aid of analysis techniques and software known as QCVV, Quantum Characterization, Verification, and Validation. These techniques take the results of a sequence of experiments and produce descriptive metrics, like gate fidelity or operator descriptions of the underlying gates. Two of the most popular techniques in use today are randomized benchmarking and gate set tomography. In randomized benchmarking, an experiment consists of a long sequence of a repeated target gate interspersed with random other gates. The random gates essentially randomize the error seen in the target gate and allow us to calculate the average fidelity of this target gate. This procedure also mitigates the impact that imperfect state preparation and measurement have on the fidelity. One of the disadvantages of randomized benchmarking is that it only provides a single metric for the gate, namely, the fidelity, which may not be enough to help diagnose the cause of error in the system.
Gate set tomography goes beyond randomized benchmarking by providing process map descriptions of the quantum gates. These process maps are operator descriptions of both the gate and the error seen in the experiment. Both randomized benchmarking and gate set tomography require a large amount of data and are, therefore, limited to the analysis of small quantum systems. Developing scalable QCVV techniques is an ongoing area of research, which will be increasingly important as larger quantum systems emerge. As mentioned earlier, noise and error are a major concern in today's quantum devices and computers. One of the main uses of quantum simulation software is to understand the impact of this noise. One can apply modeling and simulation at many different levels of abstraction, ranging from finite element models of materials to devices up to quantum circuit models. QuTiP is a Python-based package that provides libraries useful for modeling and simulating open quantum systems, and systems interacting with unwanted degrees of freedom in the environment. Static modeling can also be applied to devices and circuits to obtain important metrics affecting their performance as qubits. These metrics include the coherence time of the qubits and the sensitivity of the qubit to specific types of noise. Simulating the performance of quantum error correction circuits is another important use of modeling and simulation. Here, one uses simulation to determine the logical error rate of the circuit and determine the scaling of this logical error rate as a function of the individual physical devices. One final example of simulation does not involve error at all. Simulation is also used to understand the computational advantage that quantum computers have over classical ones. Quantum supremacy circuits have been proposed as circuits that are difficult to simulate classically but are easy for a quantum computer to execute. Several groups have developed parallel simulators that run on high-performance computing systems and use these simulators to determine how large a circuit is required to demonstrate a quantum advantage \cite{jmiszczak_list_2015}. Thus, in summary, quantum computers have a software stack that serves the same purpose as the software stack used for classical computers\cite{calude_-quantizing_2007}. These quantum software tools are important for programming, design, and testing of today's quantum computers. It will be even more important to the goal of realizing large scale quantum computers.

There exists an array of software tools used in the design, programming, and validation of classical and quantum computers. On the one hand, circuits are designed using CAD software, including circuit layout tools and conceptual circuit schematic-capture. On the other hand, programs are written and compiled through several layers of abstraction, from the high-level program codes we use to ``write programs'' to the compilation of instruction sets and low-level hardware-specific implementations\cite{booth_comparing_2018}. Besides, the software is used to make predictions of performance for differing hardware architectures. Once assembled, hardware needs to be benchmarked with the assistance of software. All together, these various software tools comprise the software stack. There are similarities and differences between the classical and quantum versions of a software stack.

The path from the underlying quantum-physical operations (unitary evolutions) introduced in the previous section to actual physical quantum systems is highly dependent on the hardware\cite{franson_limitations_2018}. Aside from high-level descriptions of quantum gate sequences, there are currently no set standards and, just as many qubit modalities continue to compete, the development of the quantum computing software stack today is a very diverse undertaking. It is linked to the efforts of several academic groups, larger corporations, and small start-up companies, many of which have only formed in the last 2 years. This page of a volunteer-run wiki provides a comprehensive list of a large number of available software platforms, ordered by type, and the (classical) programming language by which they are realized.

Programming a quantum computer comprises three primary types of software:

1. Program generation software

2. Circuit mapping software

3. Control and execution software

Program generation is generally technology agnostic, whereas circuit mapping, control, and execution all require knowledge of the hardware architecture. Program generation can be further divided into three categories:\\
1. schematic capture (generally related to the quantum circuit model)\\
2. high-level programming languages (abstracting away subroutines such as phase estimation or period-finding via the Quantum Fourier Transform) and lower-level assembly languages (abstracting away from specific hardware controls and subroutines)\\
3. problem-specific platforms (e.g., to generate auxiliary inputs to quantum chemistry calculations)

In general, computer code may be abstracted multiple times before the high-level language used by human beings to ``code'' a quantum computer is compiled and assembled into the instructions that are ultimately used at the hardware level. One example of this reduction is the quantum assembly language (QASM) that we will use to program the Deutsch-Jozsa algorithm directly. QASM is very generic and close to the quantum circuit model.

As this reduction occurs, the hardware-agnostic codes must eventually be implemented on physical circuits using hardware-specific control and execution. Whereas the control and execution systems are typically co-located (and, in many cases, the same technology base) for classical computers, quantum processors are generally controlled by classical hardware. These are different technologies and often reside in different locations. For example, a superconducting quantum computer is controlled with classical electronic instruments at room temperature, while the quantum information is processed inside the dilution refrigerator at cryogenic temperatures\cite{nielsen_quantum_2011}.

\section{IBMQ Experience and QASM Programming}
In this section, We introduces the IBM Quantum Experience quantum computing system, a graphical user interface called the composer, and a programming language called QASM. 

We think one of the most exciting developments in quantum computing in recent years has been the ubiquitous access to real quantum devices. Just a couple of years ago, students would discuss concepts and quantum information and quantum computation without any real means for hands-on experimentation. However, in 2016 with the launch of the IBM Q Experience, the first generation of quantum computers came online. They have generated much excitement, with nearly 80,000 unique users running more than three million experiments. this section will introduce us to some of the resources that will help us get started with quantum programming. Let us start by looking at the composer, which is a simple graphical user interface for building and executing quantum programs. This space has two parts. In the top part, we see information about the devices that are available for experiments. On the left, we see a schematic of the chip where each white square is one of the qubits. right next to that is a diagram of how the qubits are connected to each other. This connectivity diagram influences the operations we can do on this device, as We will see in a moment. One important thing that we must get used to when working with quantum computers is that They are always noisy and imperfect \cite{bremner_achieving_2017}. So, on the right, we see a lot of information about the noise level on each of these qubits and each of the gates. We do not have to believe these numbers. we can design and run certain experiments that will let us measure the exact errors. There is a lot of more interesting data about each of these chips. For example, we see that this one is sitting in a dilution refrigerator with a temperature of around 21 millikelvin, which is colder than the surface of Pluto. At the bottom of the page is the composer itself. we can see there are five wires, each representing one of the qubits in the quantum computer. on the panel on the right, we have the gates available for building the quantum circuit. As we know, the circuit model is a simple yet powerful model for quantum computation. A quantum circuit is a recipe for how to transform the state of several qubits by applying various gates. It can be shown that any quantum computation can be done by just operating on one or two qubits at a time. the very small set of gates is enough for universal quantum computation. So as simple as this interface seems, we can do quite complex computations with it. So, let us do a simple entanglement experiment. All the qubits initially start as being in the zero states. To create an entangled pair of qubits, we first put one of the qubits in a superposition state by applying a Hadamard gate to it. Next, we toggle the second qubit conditioned on the first qubit by applying a CNOT gate. This leaves the qubits to go in a state of 0, 0 plus 1, 1, a state that cannot be described in terms of the state of each qubit individually. It is an entangled state. However, how can we know what state in the qubits are? The qubit is state-space inaccessible to us. the only way we can get any information is to measure the qubits. So, let us add two final measurement operations to this circuit. However, each time we measure the qubits, we read a normal classical bitstream, such as 0, 0, or 1, 1. So, to infer the qubits' state right before the measurement, we have to repeat this experiment many times and approximate the probability distribution that existed with the entangled states. So now that we are done building the circuit let us execute it. we see that we have the option of either sending our circuit to a simulator or real hardware \cite{villalonga_flexible_2019}. First, let us do a simulation to make sure the circuit is working as expected. here is what we get, a histogram of all the measurement outcomes. By default, in the IBM Q Experience, each circuit will be executed about 1,000 times. As we expected, roughly $50\%$ of the time, we measured both qubits in state 0. the other $50\%$ of the time, we measured both qubits in state 1. This is an indication of the state correlations that arise as a result of quantum entanglement. Now let us submit the circuit to a real quantum computer. By clicking run, our circuit will be sent to the IBM Research Labs in New York, where they will be translated to the language of qubit manipulations, namely, control pulses. After we receive the measurement results, we see that, indeed, the same outcome is achieved. However, there are some imperfections here in the result. this is exactly due to the noisy qubits and gates that We discussed earlier. A composer is a great tool for just playing around and visualizing our circuits. However, we may prefer to build or alter our circuits more quickly. we can do this by switching to the QASM Editor, which gives us a textual representation of the circuit. we can also import QASM from the file. QASM, or quantum assembly, is a circuit description language. The graphical interface we were using previously was being translated to QASM code under the hood. QASM is a hardware-agnostic language, meaning it can be translated to any physical chip or even a different quantum computing technology altogether. It expresses data dependencies without explicit timings for the instructions, which will be decided at a later stage. Here we see the same circuit we just built written in QASM. The first line imports a standard library of gates for us to use. This is exactly akin to the menu of gates we had previously. The next two lines define the qubits as a quantum register of size 5 and the classical bits, which will be used to hold the final measurement results. Finally, we have the Hadamard, CNOT, and measurement operations. Let us suppose that we want to repeat the entangling experiment, but this time on two different qubits, let us qubit 2 and 3. This seems trivial. We just change the indices on Q and C. However. We get an error message saying that a CNOT is not allowed from Q2 to Q3. That is right. This is due to the connectivity graph of the qubits in this chip. We see that in order to do a CNOT operation between these two qubits, we need to designate Q3 as control and Q2 as a target. Luckily, there is an easy transformation that can be used to flip a CNOT. This circuit identity is achieved by sandwiching the CNOT that we have between two layers of Hadamard's, giving us the flipped CNOT. Now we repeat the experiment on the real chip again. We see the same result as we expected. If we look closely, however, we see that the accuracy of the results is slightly degraded compared to the previous experiment. This is because the new circuit uses more gates, leading to a higher chance of errors accumulating in the circuit. To conclude, it is easy to use the IBM Q Experience to program a quantum computer. All the many layers of technology that goes into building a working quantum computing stack are conveniently abstracted. Quantum information science is no longer only done in the lab\cite{preskill_quantum_2012}. Now we can control a real quantum computer remotely. While these devices are still small and noisy, in a way, they force us to be cleverer in using them. For example, we must consider the connectivity of the qubits and try to use fewer gates in order to preserve information fidelity. We will see many more interesting examples of quantum information science if we visit the IBM Q Experience user guide, and each one accompanied it with its composer circuit and QASM code. we also encourage us to get involved in the Community Forum, which is a great place to discuss anything related to quantum computing and discussing it from each other.

In the following sections, we will discuss Open QASM \cite{cross_open_2017} and how this programming language can be used to demonstrate several simple quantum circuits and, ultimately, the Deutsch-Jozsa quantum algorithm on the IBM Quantum Experience quantum computer.

The IBM Q Experience platform allows anyone with access to the internet to write and run quantum algorithms on a real quantum computer. In this section, we will receive the necessary resources to implement algorithms on the IBM platform. However, we encourage to register on the IBM page to access the composer's graphical interface (alternatively, if we are comfortable writing text-based programs, the interface here may be just fine).

The IBM Q composer allows one to express quantum circuits and quantum algorithms in a simple graphical way, and on the IBM QE site, we will be able to save the results of our programs and access them later. To introduce us to the use of the platform, IBM has a user beginner's guide. If we want to discuss how to implement more complicated algorithms, we recommend we visit their full user guide. For researchers, it is now possible to request exclusive access.

Once in the composer, we can create a new experiment and choose if we want to run our program on a real quantum computer or to simulate it using a classical computer. When we decide to run our code on the quantum computer, we will have to choose the number of times that we want to run the code and which ``backend'' we will use. There are four possible backends on which we can run our codes. We will only use either the classical numerical simulator or one of the five-qubit backends, ibmqx2 \cite{shukla_complete_2018} (may be unavailable due to maintenance) and ibmqx4 (when available)\cite{shukla_complete_2018}. We will discuss more backend options in the following sections.

\begin{figure}[H] \centering{\includegraphics[scale=.4]{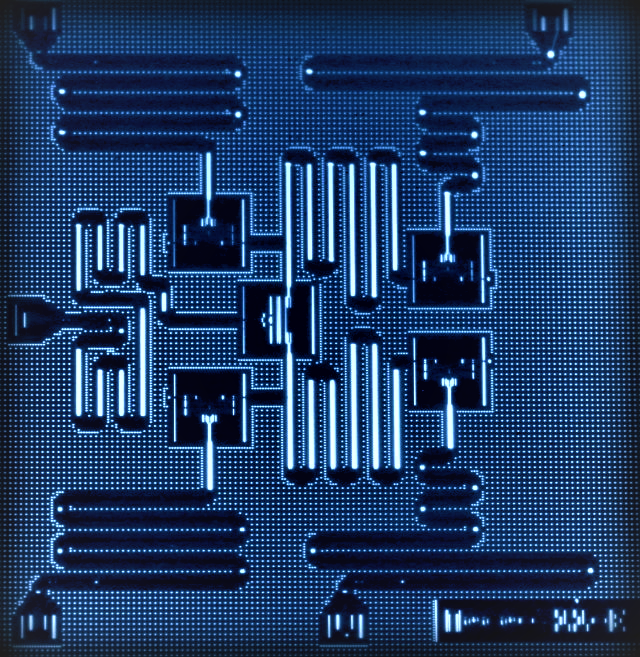}}\caption{IBM QEX2}\label{fig1_34}
\end{figure}

The IBM Q Experience platform not only allows us to use the composer, but it also allows us to write our code using Open QASM (Quantum Assembly Language). This is a simple text language that has elements of C and assembly languages. By using QASM, we can describe any quantum circuit, in principle, since it has built-in a universal set of quantum gates. In this section, we will discuss how to express quantum circuits and write quantum algorithms using QASM. If we want to discuss the theory behind QASM, we recommend that we read the following paper written by IBM researchers: Open Quantum Assembly Language \cite{cross_open_2017}.

Here, we will receive detailed instructions on how to use QASM, we will guide us step by step on how to create our first programs, and we will teach us how to read our results. In this first set of exercises, we will discuss syntax and common mistakes, measurements, single-qubit gates, two-qubit gates, and backends. We will have the opportunity to practice our knowledge and to prove our skills in the end-of-section test. Besides the results that the IBM platform gives us, we will also make a theoretical analysis of the quantum state evolution, so we will easily see the correspondence between the analytical and experimental results.

The numerical simulation employed here is based on the IBM QISKit engine, an open-source package that allows the inclusion of gate and measurement noise, for added realism.  This noise model is based on recent calibration data from the 5-qubit IBM QE machines. The output histograms are statistically similar to those that we would obtain from real quantum computers.

By default, each exercise begins with numerical simulations to give us the most rapid feedback for our check-our-understanding answer submission.  With select exercises, once our submission is deemed correct, our QASM program will automatically be submitted for execution on a real quantum computer.   Once it is run, we should see a plot comparing the simulated and real quantum computer's results, side-by-side, like this:

\begin{figure}[H] \centering{\includegraphics[scale=.5]{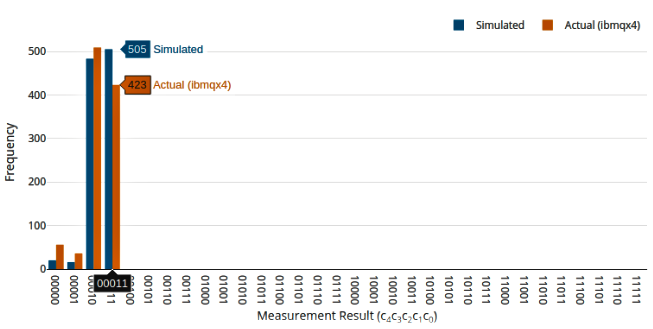}}\caption{Real QC Results Histogram}\label{fig1_35}
\end{figure}

We will need to be patient, however, because these run requests must be queued for execution since hardware availability is limited. Each program must be run multiple times to generate the output statistics needed for a pedagogical explanation.

\section{Syntax of QASM}

In the figure below, each horizontal line represents the evolution of a qubit with time proceeding from left to right. The five qubits are labeled in order as: q[0], q[1], q[2], q[3] and q[4]. The qubits are always initialized in the quantum state:

\begin{equation}\label{51}
\lvert 0\rangle =\left(\begin{array}{c}1 \\ 0 \end{array} \right)
(1)
\end{equation}

\begin{figure}[H] \centering{\includegraphics[scale=1]{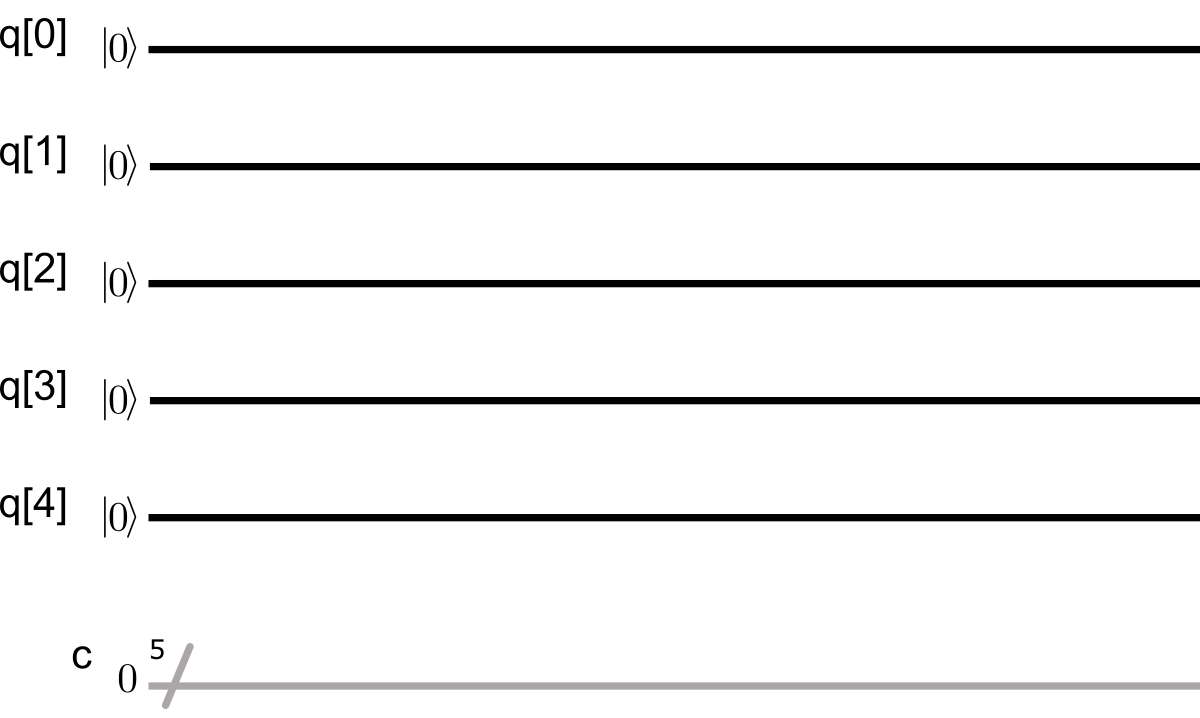}}\caption{Circuit}\label{fig1_36}
\end{figure}

Classical bits used to store measurement results are indicated by the letter ``c'' and the number ``5'' indicates that there are five classical bits in the register being represented with the grey line.

The figure below shows a graphical representation of adding a measurement block to the first qubit. The vertical grey line indicates that the measurement result will be stored in the classical bit register. The number below the grey arrowhead indicates into which classical bit the result will be stored. In this case, the result is stored in bit c[0].

\begin{figure}[H] \centering{\includegraphics[scale=1]{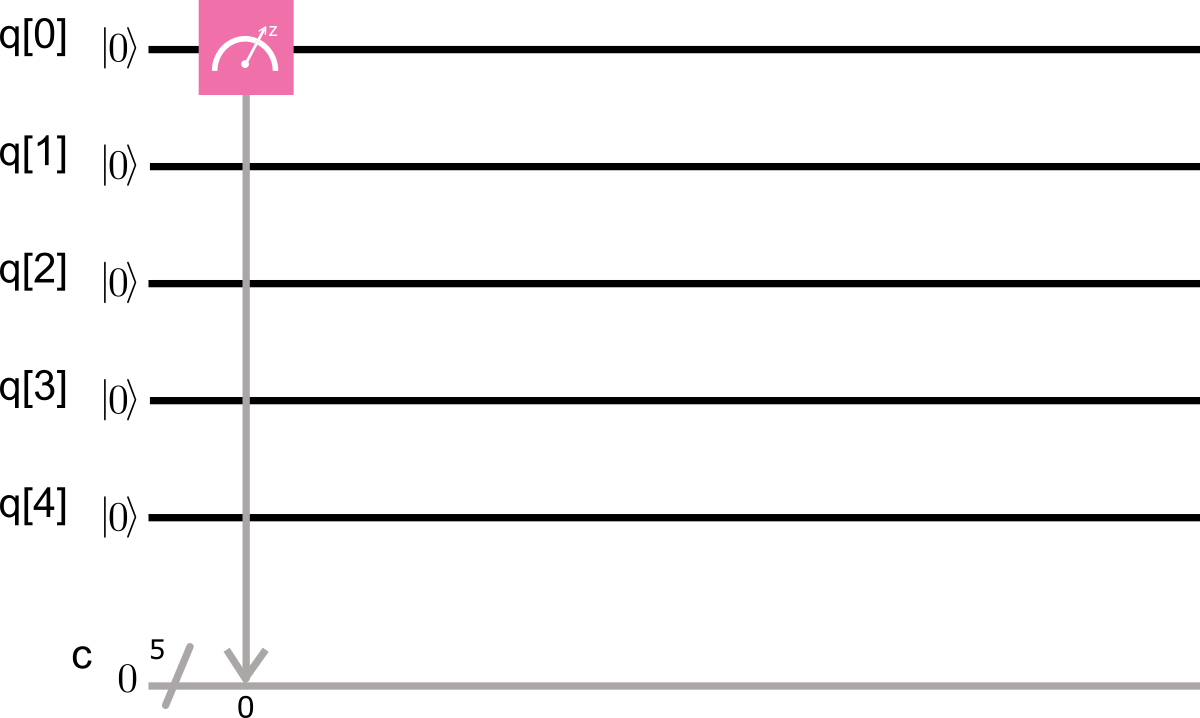}}\caption{Circuit}\label{fig1_37}
\end{figure}

\begin{lstlisting}
The following QASM code implements the quantum circuit above:
include ``qelib1.inc'';
qreg q[5];
creg c[5];

// This is a comment
measure q[0] -> c[0];
\end{lstlisting}

Let us take a look at what each line of the previous code does.

Line 1: Includes all gates in the quantum gate library. By adding this, we can then use the X-gate, Y-gate, Z-gate, CNOT gate. Notice that each line has to end in a semicolon ``;'' (without quotes).

Line 2: Defines a quantum register of five qubits. In this example, we are using five qubits, but we can elect to use any number of qubits from one to five. Note that even if we define a two-qubit register, the quantum circuits depicted here and at the IBM quantum experience site will always graphically illustrate all five qubits by default. This is because the physical quantum computer being used has 5 qubits. Nonetheless, we can choose to work with only a subset of those five qubits.  

Line 3: Defines a classical register of five bits. This register is used to store the quantum measurement results. As with the quantum register, even if we define a two-bit classical register, the quantum circuits will indicate all five bits by the grayline. Nonetheless, we can choose to work with only a subset of those five bits.

Line 4: we can insert blank lines to help us organize our sections.

Line 5: we can also insert comments using the ``//'' (two back-slashes, without quotes). This helps other readers understand our intention in each line. Text that appears after the ``//'' (without quotes) is not evaluated.

Line 6: Measures the first qubit q[0] and stores the measurement result in bit c[0]. We could alternatively store the information on c[1], but, for organizational purposes, we recommend a direct numerical mapping q[0] to c[0].

\begin{enumerate}[wide, labelwidth=!, labelindent=0pt]   
\item  Syntax Error I, Identify on which line there is an error:\\
\begin{lstlisting}
include``qelib1.inc'';
qreg q[5];
creg c[5];
measure q[0] -> c[0];
\end{lstlisting}
Solution:\\
A space is needed between include and ``qelib1.inc''.

\item  Syntax Error II, Identify on which line there is an error:\\
\begin{lstlisting}
include ``qelib1.inc'';
qreg q[5];
creg c[5]
measure q[0] -> c[0];
\end{lstlisting}
Solution:\\
There is a missing ; at the end of the line.

\item  Syntax Error III, Identify on which line there is an error:\\
\begin{lstlisting}
include ``qelib1.inc'';
qreg q{5};
creg c[5];
measure q[0] -> c[0];
\end{lstlisting}
Solution:\\
Parentesis should be square [] not {}.

\item  Syntax Error IV, Identify on which line there is an error:\\
\begin{lstlisting}
include ``qelib1.inc'';
qreg q[5];\\ this is a comment
creg c[5];
measure q[0] -> c[0];
\end{lstlisting}
Solution:\\
Comments sould be added after // not $\\$.
\item  Syntax Error V, Identify in which line there is an error:\\

\begin{lstlisting}
include ``qelib1.inc'';
qreg q[2];
creg c[2];
measure q[0] -> c[2];
\end{lstlisting}

Solution:\\
The classical register index should be less than 2. In line 3, creg $ c[2] $ defined a classical register with two indexes $ c[0] $ and $ c[1] $.

\section{Measurements}

As we discussed in the previous section, the quantum measurement is probabilistic. When we perform a measurement, we project the state of the qubit onto either $ \lvert 0\rangle $ or $ \lvert 1\rangle  $ with a probability that is the magnitude squared of their respective coefficients (their probability amplitudes). To understand how QASM deals with measurements, Let us go back to the first example,\\

\begin{lstlisting}
include ``qelib1.inc'';
qreg q[5];
creg c[5];

// This is a comment
measure q[0] -> c[0];
\end{lstlisting}

In this example, the first qubit q[0] is measured, and the result of the measurement is stored in classical bit c[0]. The analytical probabilities of projecting q[0] onto states $ \lvert 0\rangle $ and $ \lvert 1\rangle $ are given by $ p(q[0],\lvert 0\rangle )=\lvert \langle 0\lvert 0\rangle \rvert ^{2}=1 $ and $ p(q[0],\lvert 1\rangle )=\lvert \langle 0\lvert 1\rangle \rvert ^{2}=0 $.

The IBM QE platform either simulates QASM code or runs it on a real quantum computer, producing numerical results that can be presented as a table. The following figure shows the tabular result of simulating the previous code over 10 ``shots''  identically prepared and executed experiments. In this example, the qubit was projected 10 times onto state $ \lvert 0\rangle $. Note that because the qubit is only projected on to one state in this example, the only label is 0.

\begin{figure}[H] \centering{\includegraphics[scale=.17]{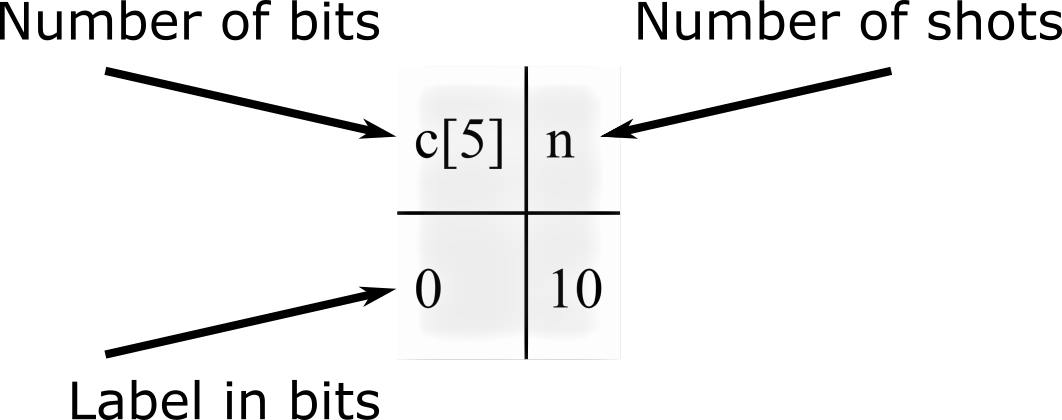}}\caption{Distribution}\label{fig1_38}
\end{figure}

The results are different when the code is run on a real (imperfect) quantum computer. The figure below shows the result of executing the code 8192 times. In this case, there are two different labels because qubit q[0] was projected to state $ \lvert 0\rangle $ but was sometimes projected to state $ \lvert 1\rangle $. Specifically, it was projected on to $ \lvert 0\rangle $ 8182 times and onto $ \lvert 1\rangle $ 10 times.

\begin{figure}[H] \centering{\includegraphics[scale=.5]{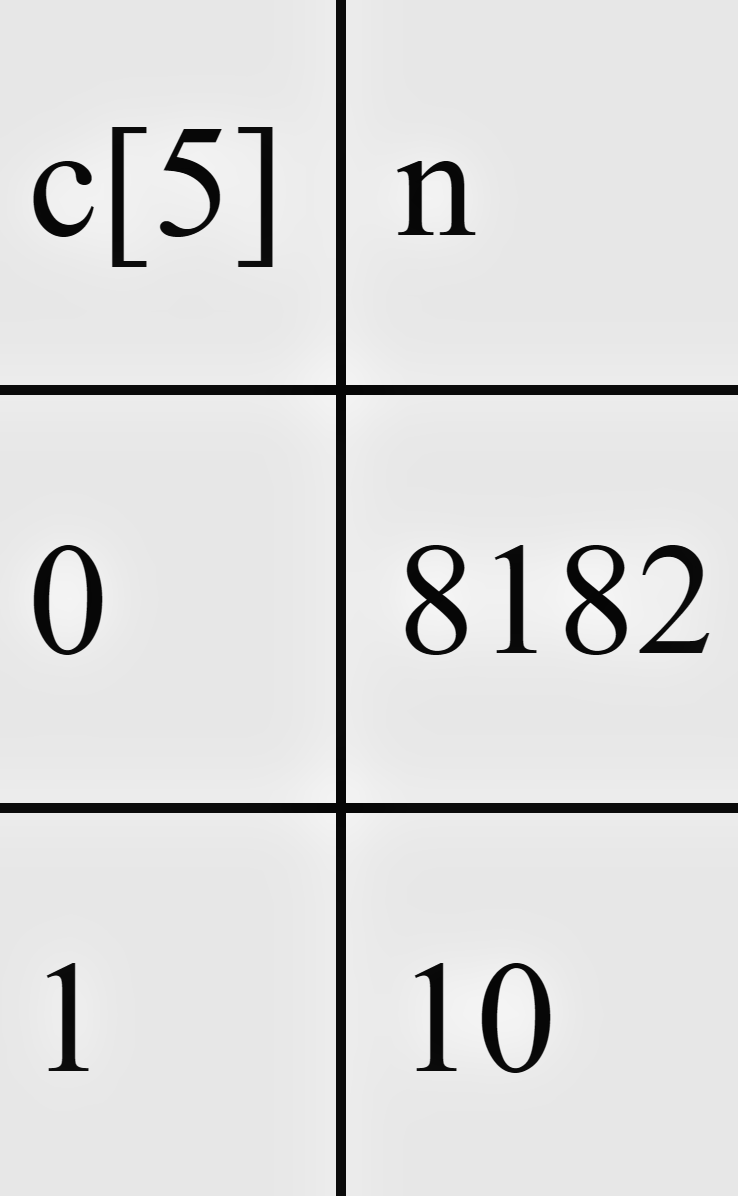}}\caption{Distribution}\label{fig1_39}
\end{figure}

\item  Measuring I, Which option creates a classical register with 3 indexes?\\
creg c[0];\\
creg c[1]\\
creg c[2]\\
creg c[3]\\
Solution:\\
For the classical register to have 3 bits, it must be defined as ``creg $ c[3] $'', this gives we indices c[0], c[1] and c[2]. The first qubit is labeled q[0], the second is labeled q[1], and so on. To store the second qubit in c[1] we must assign ``$ q[1] -> c[1] $'' 

\item  Measuring II, Which quantum circuit corresponds to the graphical representation of the following code\\

\begin{lstlisting}
include ``qelib1.inc'';
qreg q[5];
creg c[5];
measure q[1] -> c[3];
measure q[3] -> c[1];
\end{lstlisting}

\begin{figure}[H] \centering{\includegraphics[scale=1]{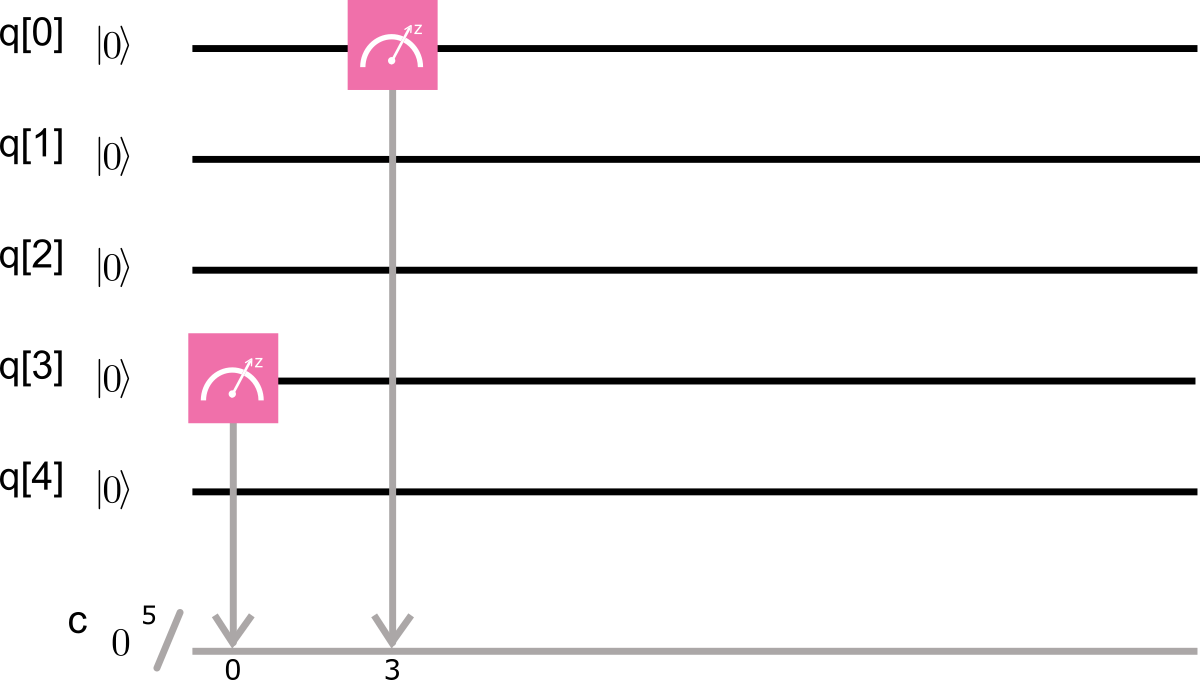}}\caption{MC 1c}\label{fig1_40}
\end{figure}

\begin{figure}[H] \centering{\includegraphics[scale=1]{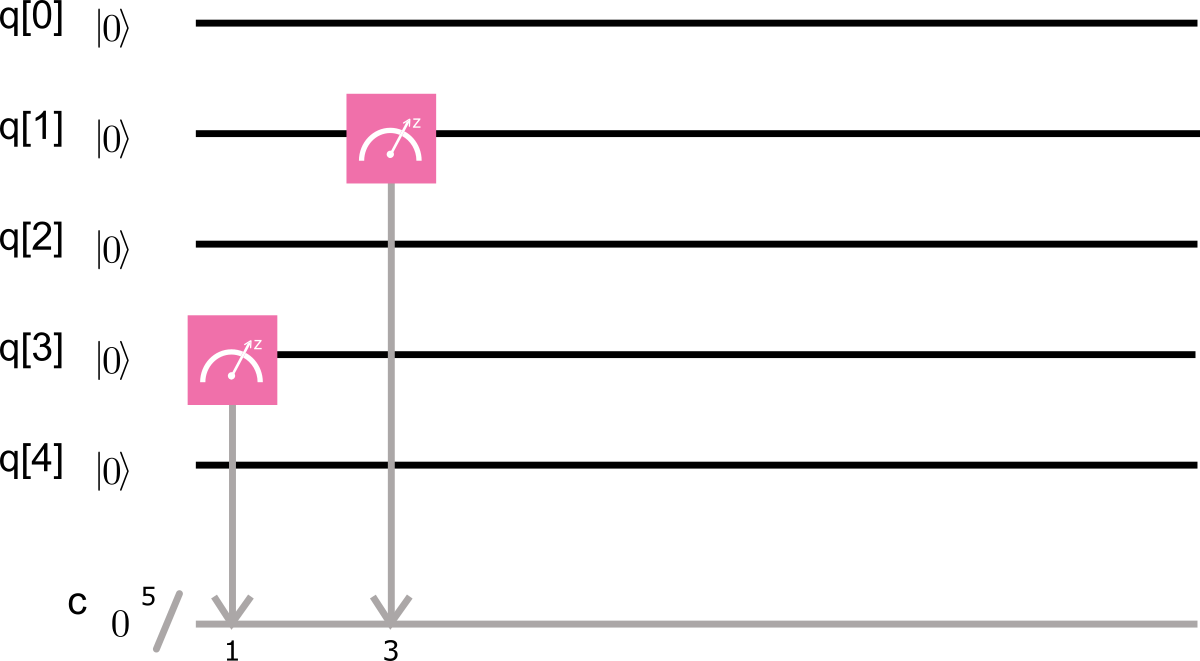}}\caption{MC 1b}\label{fig1_41}
\end{figure}

\begin{figure}[H] \centering{\includegraphics[scale=1]{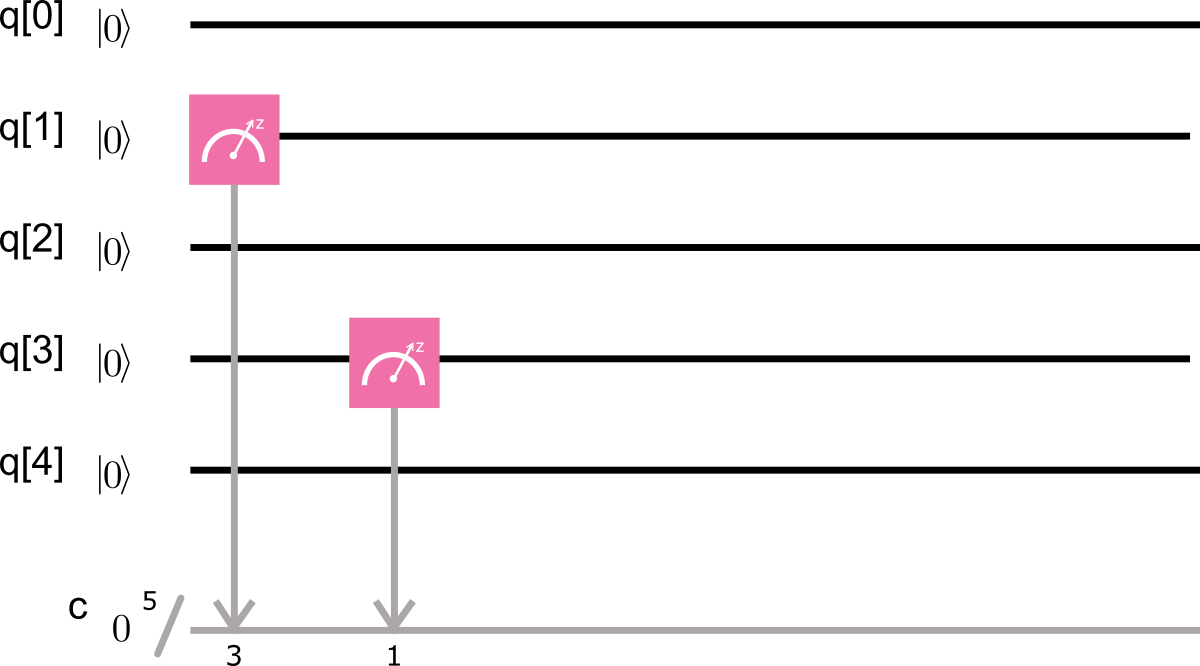}}\caption{MC 1a}\label{fig1_42}
\end{figure}

\begin{figure}[H] \centering{\includegraphics[scale=1]{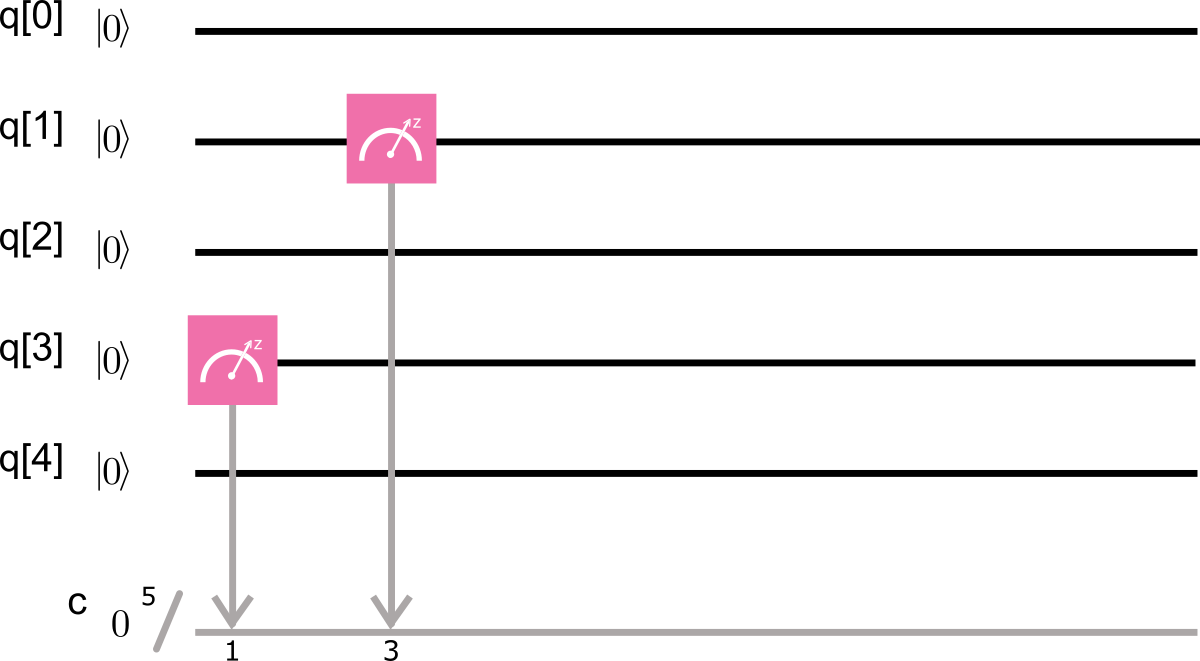}}\caption{MC 1d}\label{fig1_43}
\end{figure}

To better understand how the measurement results are stored in the classical register, Let us consider the quantum circuit below.

\begin{figure}[H] \centering{\includegraphics[scale=1]{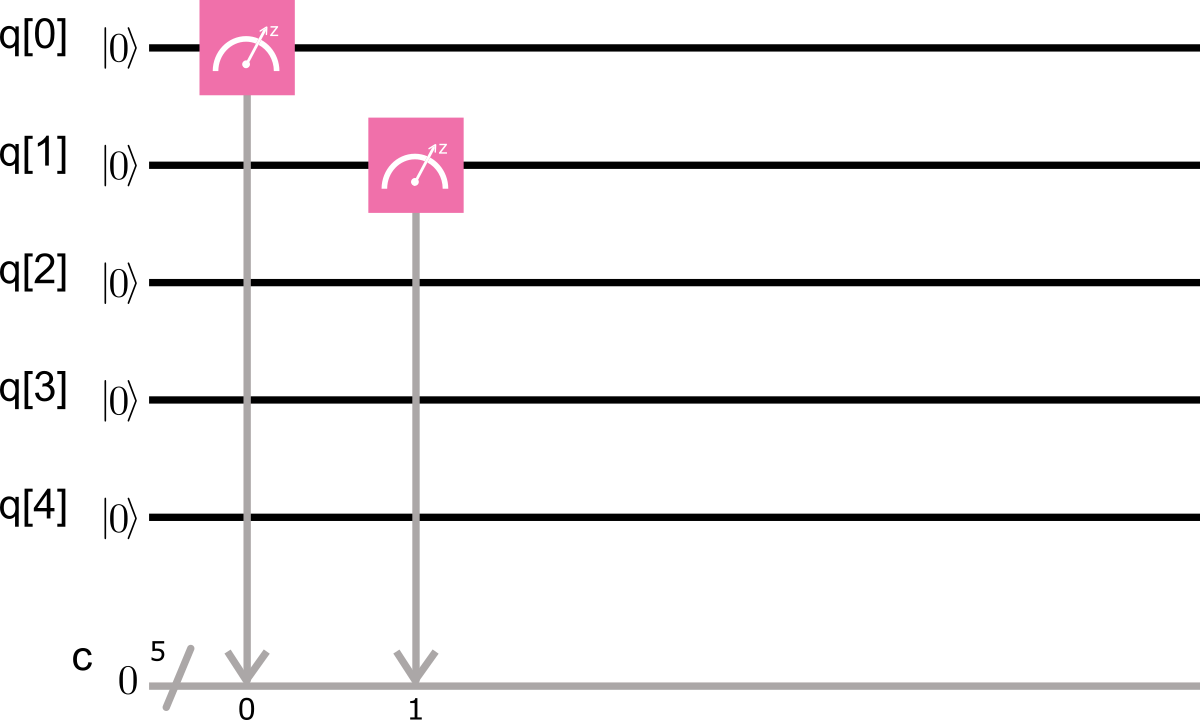}}\caption{circuit}\label{fig1_44}
\end{figure}

The QASM code that generates this circuit is\\

\begin{lstlisting}
include ``qelib1.inc''; 
qreg q[5];
creg c[5];
measure q[0] -> c[0];
measure q[1] -> c[1];
\end{lstlisting}

Similar to the previous example, the analytical probabilities of projecting qubits q[0] and q[1] onto states $ \lvert 0\rangle $ and $ \lvert 1\rangle $ are respectively given by $ p(q[0],\lvert 0\rangle )=\lvert \langle 0\lvert 0\rangle \rvert ^{2}=1, p(q[0],\lvert 1\rangle )=\lvert \langle 0\lvert 1\rangle \rvert ^{2}=0, and p(q[1],\lvert 0\rangle )=\lvert \langle 0\lvert 0\rangle \rvert ^{2}=1, p(q[1],\lvert 1\rangle )=\lvert \langle 0\lvert 1\rangle \rvert ^{2}=0. $

The following figure shows the result of simulating the previous code with 10 shots. Qubits q[0] and q[1] were projected 10 times into state $ \lvert 0\rangle  \lvert 0\rangle $. Note that the label in the left column is ``00'' because two qubits are being measured.

\begin{figure}[H] \centering{\includegraphics[scale=.5]{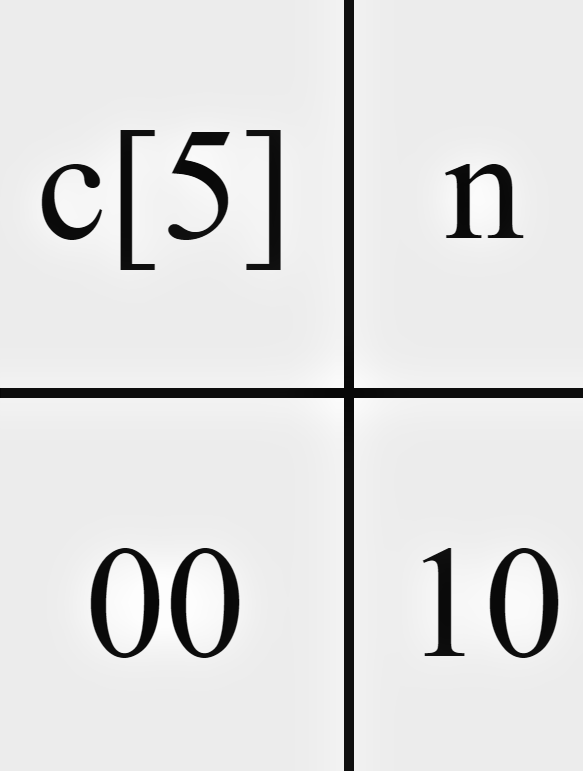}}\caption{Distribution Updated}\label{fig1_45}
\end{figure}

The figure below shows the result of running the previous code in a real quantum computer 1024 times. Since the qubit q[0]measurement was stored in c[0], and q[1] in c[1], the labels are given in the order c[1]c[0]. The number n=1014 at the right of label ``00'' indicates that qubits q[0] and q[1] were projected onto state $ \lvert 00\rangle $ 1014 times. The number n=1 at the right of label ``01'' indicates that qubits q[0] and q[1] were projected onto state $ \lvert 01\rangle $ 1 time. The number n=9 at the right of label ``10'' indicates that qubits q[0] and q[1] were projected onto state $ \lvert 10\rangle $ 9 times. Note that there is no label ``11''; this is because the qubits were never projected on to state $ \lvert 11\rangle $.

\begin{figure}[H] \centering{\includegraphics[scale=.5]{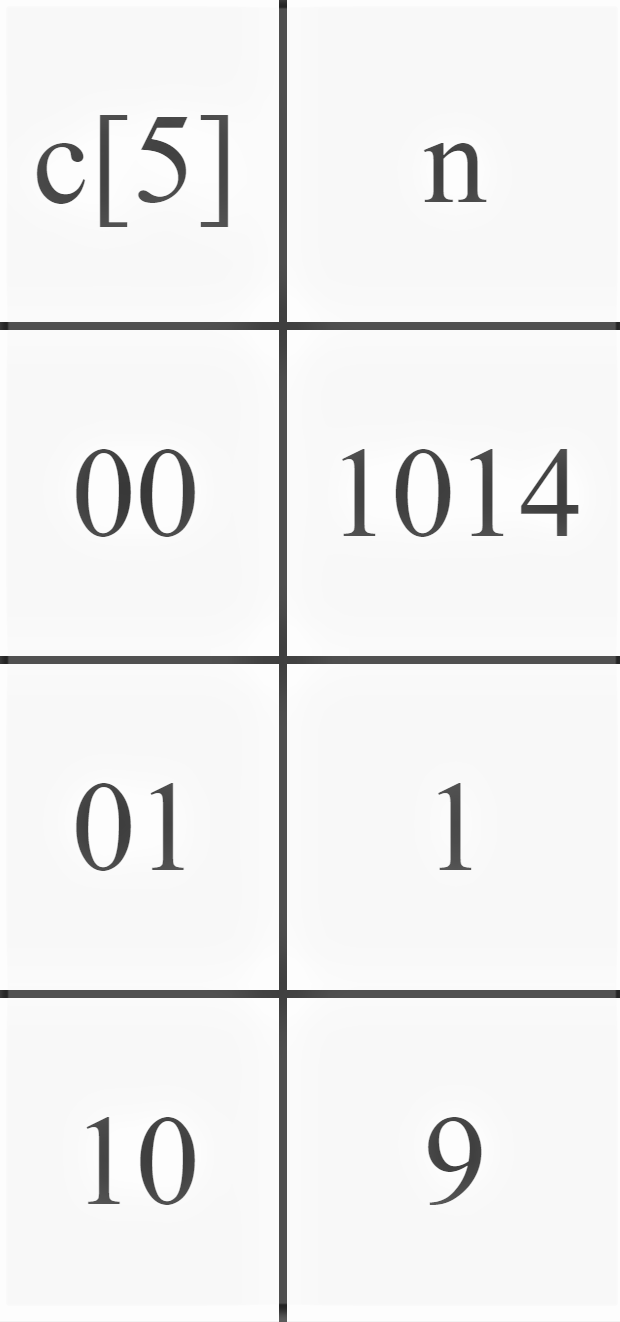}}\caption{Distribution Updated}\label{fig1_46}
\end{figure}

The following table shows a summary of the previous results. We can see that the measurement results are in general but not perfect agreement with the analytical probabilities.

\begin{table}[H]
\centering
\caption{Results}
\label{tab:1_1:Table 2}
\resizebox{\textwidth}{!}{
\begin{tabular}{|c|c|c|c|}\hline
Analytical Probabilities & Quantum state $ \lvert q[0]q[1] \rangle $ & Result Label c[1]c[0] & Projection Frequency n \\\hline
1    &  $ \lvert00\rangle $             &   00           &  1014                    \\\hline
0    &  $ \lvert01\rangle $             &   10           &  9                    \\\hline
0    &  $ \lvert10\rangle $             &   01           &  1                    \\\hline
0    &  $ \lvert11\rangle $             &   11           &  0                  \\ \hline 
\end{tabular}}
\end{table}

To understand how the labeling of the results works can be complicated, so let us analyze the following example.

\begin{figure}[H] \centering{\includegraphics[scale=1]{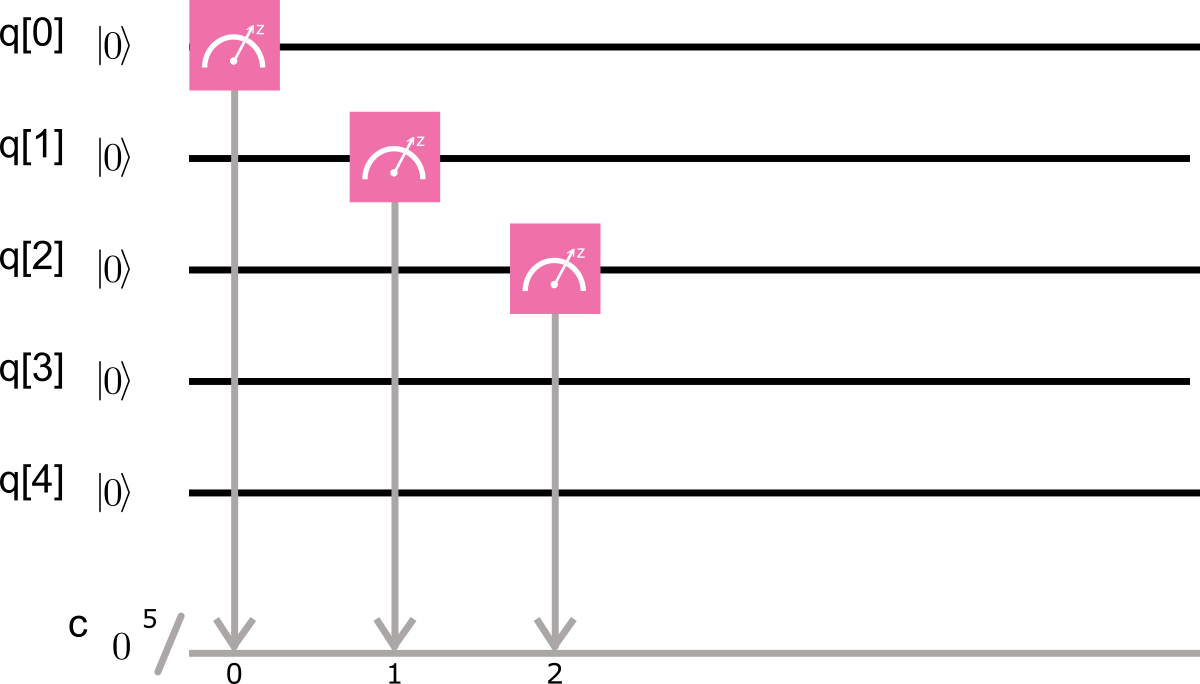}}\caption{Circuit}\label{fig1_47}
\end{figure}

The code that generates this quantum circuit is\\
\begin{lstlisting}
include ``qelib1.inc''; 
qreg q[5];
creg c[5];
measure q[0] -> c[0];
measure q[1] -> c[1];
measure q[2] -> c[2];
\end{lstlisting}
Since the q[0], q[1] and q[2] measurements are stored in c[0], c[1] and c[2] respectively, the projection results will be stored as in Table 2. If instead the measurements are stored in the following way,\\
\begin{lstlisting}
include ``qelib1.inc''; 
qreg q[5];
creg c[5];
measure q[0] -> c[2];
measure q[1] -> c[1];
measure q[2] -> c[0];
\end{lstlisting}
then the projection results will be stored as shown in Table 3

\begin{table}[H]
    \centering
    \caption{q[0] $->$ c[0]; q[1] $->$ c[1]; q[2] $->$ c[2]}
    \label{tab:1_1:Table 3}
    \resizebox{\textwidth}{!}{
        \begin{tabular}{|c|c|}\hline
            Label c[2]c[1]c[0] & Quantum state $ \lvert q[0]q[1]q[2]\rangle $ \\\hline
            000             &  $ \lvert000\rangle $    \\\hline
            001             &   $ \lvert100\rangle $       \\\hline
            010             &  $ \lvert010\rangle $      \\\hline
            011             &   $ \lvert110\rangle $        \\ \hline 
            100             &  $ \lvert001\rangle $ \\\hline
            101             &  $ \lvert101\rangle $ \\\hline
            110             &  $ \lvert011\rangle $ \\\hline
            111             &  $ \lvert111\rangle $ \\\hline
\end{tabular}}
\end{table}

\begin{table}[H]
    \centering
    \caption{q[0] $->$ c[2]; q[1] $->$ c[1]; q[2] $->$ c[0]}
    \label{tab:1_1:Table 4}
    \resizebox{\textwidth}{!}{
        \begin{tabular}{|c|c|}\hline
            Label c[2]c[1]c[0] & Quantum state $ \lvert q[0]q[1]q[2]\rangle $ \\\hline
            000             &  $ \lvert000\rangle $    \\\hline
            100             &   $ \lvert100\rangle $       \\\hline
            010             &  $ \lvert010\rangle $      \\\hline
            110             &   $ \lvert110\rangle $        \\ \hline 
            001             &  $ \lvert001\rangle $ \\\hline
            101             &  $ \lvert101\rangle $ \\\hline
            011             &  $ \lvert011\rangle $ \\\hline
            111             &  $ \lvert111\rangle $ \\\hline
    \end{tabular}}
\end{table}

Consider the quantum circuit and measurement results below and respond to the following questions.

\begin{table}[H]
    \centering
    \caption{q[0] $->$ c[2]; q[1] $->$ c[1]; q[2] $->$ c[0]}
    \label{tab:1_1:Table 5}
    \resizebox{\textwidth}{!}{
        \begin{tabular}{|c|c|}\hline
            c[5] & Quantum state  n \\\hline
            000             &  7939    \\\hline
            001             &   113       \\\hline
            010             &  135      \\\hline
            011             &   3        \\ \hline 
            100             &  11 \\ \hline
        \end{tabular}}
\end{table}

\begin{figure}[H] \centering{\includegraphics[scale=1]{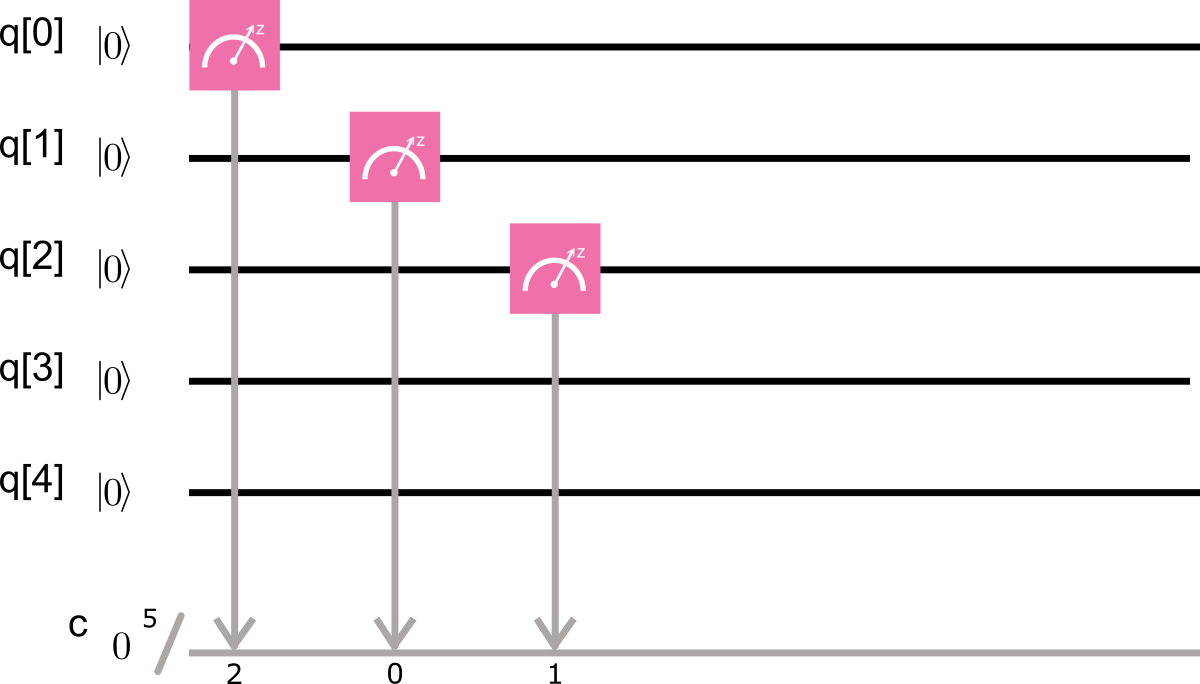}}\caption{Assessment}\label{fig1_48}
\end{figure}

\item  Measuring III, Which QASM code generates this quantum circuit?\\

\begin{lstlisting}
include ``qelib1.inc''; 
qreg q[5];
creg c[5];
measure q[0] -> c[0];
measure q[1] -> c[1];
measure q[2] -> c[2];
\end{lstlisting}

\begin{lstlisting}
include ``qelib1.inc''; 
qreg q[5];
creg c[5];
measure q[0] -> c[2];
measure q[1] -> c[0];
measure q[2] -> c[1];
\end{lstlisting}

\begin{lstlisting}
include ``qelib1.inc''; 
qreg q[5];
creg c[5];
measure q[0] -> c[2];
measure q[1] -> c[1];
measure q[2] -> c[0];
\end{lstlisting}

\item  Measuring IV: How many times was the code run?
Solution:\\
The sum of all the measurements gives the number of times the experiment is run; 7939+113+135+3+11=8201

\item  Measuring V, How many times were the qubits projected to state $ \lvert 001\rangle $ 
Solution:\\
The corresponding label to state $ \lvert 001\rangle $ is 010. This label shows n=135.

\item  Measuring VI, How many times were the qubits projected to state $ \lvert 010\rangle $?
Solution:
The corresponding label to state $ \lvert 001\rangle $ is 001. This label shows n=113.

\item  Measuring VII, How many times were the qubits projected to state $ \lvert 011\rangle $?
Solution:\\
The corresponding label to state $ \lvert 011\rangle $ is 011. This label shows n=3.

\item  Measuring VIII, How many times were the qubits projected to state $ \lvert 100\rangle $?
Solution:\\
The corresponding label to state $ \lvert 100\rangle $ is 100. This label shows n=11.

\section{Single-Qubit Gates}

Up until this point, we have discussed how to measure qubits using QASM and the IBM quantum computer. The IBM platform also enables us to perform single-qubit and two-qubit operations. Table 1 shows the predefined single-qubit gates and the QASM line to apply them on qubit q[0]. We can extrapolate this concept to other qubits.

\begin{table}[H]
\centering
\caption{single-qubit gates}
\label{tab:1_1:Table 6}
    \begin{tabular}{|c|c|}  \hline
        Gate & QASM line     \\ \hline
         $ X=\left( \begin{array}{cc} 0 & 1 \\ 1 & 0 \end{array} \right) $    &  x q[0];             \\\hline
         $ Y=\left( \begin{array}{cc} 0 & -i \\ i & 0 \end{array} \right) $    &  y q[0];              \\\hline
         $Z=\left( \begin{array}{cc} 1 & 0 \\ 0 & -1 \end{array} \right)  $    &  z q[0];             \\\hline
         $H=\frac{1}{\sqrt{2}}\left( \begin{array}{cc} 1 & 1 \\ 1 & -1 \end{array} \right)      $    &  h q[0];              \\\hline
        $S=\left( \begin{array}{cc} 1 & 0 \\ 0 & e^{i\frac{\pi}{2}} \end{array} \right)      $     &  s q[0];              \\\hline
        $S^{\dagger}=\left( \begin{array}{cc} 1 & 0 \\ 0 & -e^{i\frac{\pi}{2}} \end{array} \right)  $     &  sdg q[0];            \\\hline
        $T=\left( \begin{array}{cc} 1 & 0 \\ 0 & e^{i\frac{\pi}{4}} \end{array} \right)      $     &  t q[0];              \\
        $ T^{\dagger}=\left( \begin{array}{cc} 1 & 0 \\ 0 & -e^{i\frac{\pi}{4}} \end{array} \right)     $     &  tdg q[0];          \\\hline
    \end{tabular}
\end{table}

In the figure below, an X-gate is applied to the first qubit q[0], and then q[0] and q[1] are measured. The final state of the two-qubit system is given by $ \lvert 10\rangle $.

\begin{figure}[H] \centering{\includegraphics[scale=1]{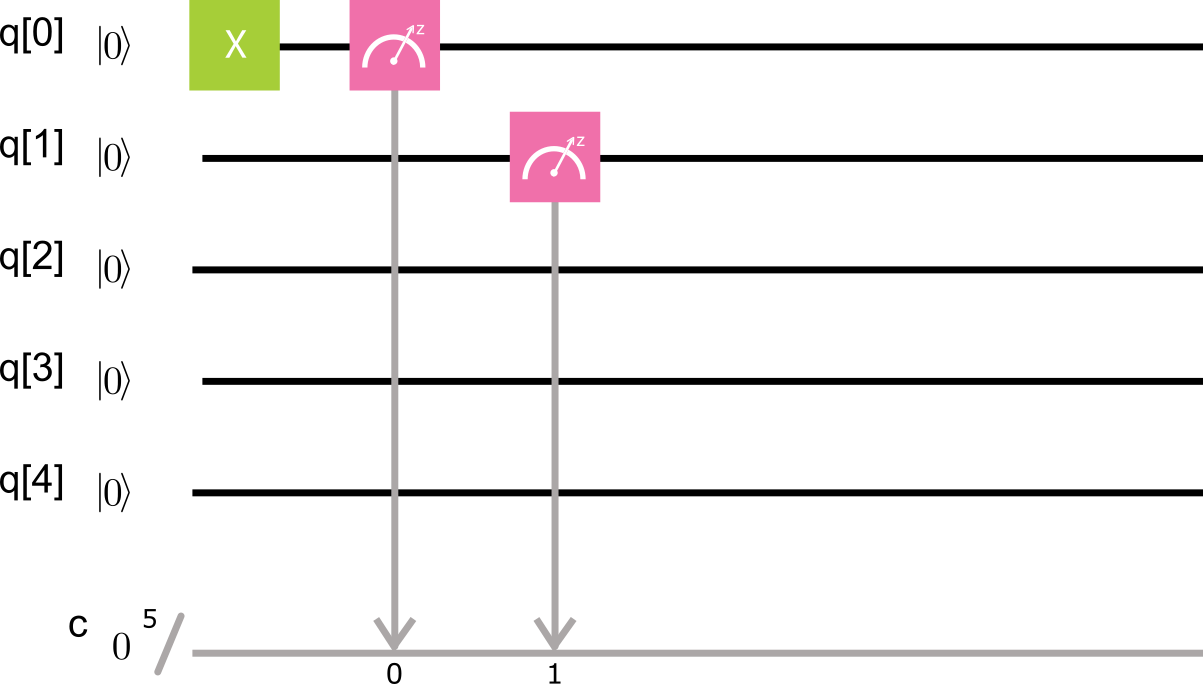}}\caption{Circuit}\label{fig1_49}
\end{figure}

The code below generates the above quantum circuit. Note that there are two blank lines, 4 and 6, this is just to make the code look cleaner. In this way, we will have a first section in which we define our registers, a second section in which we apply the quantum gates, and a third section in which we measure.\\

\begin{lstlisting}
include ``qelib1.inc'';
qreg q[5];
creg c[5];

x q[0];

measure q[0] -> c[0];
measure q[1] -> c[1];
\end{lstlisting}

Table 2 shows the results of running the code 1024 times on the perfect quantum simulator. Notice that the two qubits were projected onto state $ \lvert 10\rangle $ every time.

\begin{table}[H]
\centering
\caption{Results of running the code 1024 times on the perfect quantum simulator}
\label{tab:1_1:Table 7}
\begin{tabular}{|c|c|}  \hline
        c[5] & n     \\ \hline
        01 & 1024             \\\hline
\end{tabular}
\end{table}

Table 3 shows the results of running the code 1024 times on a real IBM quantum computer. In this case, notice that the two qubits were projected onto state $ \lvert 00\rangle $ 70 times, onto state $ \lvert 01\rangle 2  $ times, onto state $ \lvert 10\rangle $ 941 times, and onto state, $ \lvert 11\rangle $ 11 times. The difference between the results in tables 2 and 3 are not only given by the errors in the measurements, but also by the error in the X-gate.

\begin{table}[H]
    \centering
    \caption{Results of running the code 1024 times on a real IBM quantum computer}
    \label{tab:1_1:Table 8}
    \begin{tabular}{|c|c|}  \hline
        c[5] & n     \\ \hline
        00 & 70             \\\hline
        01 & 941             \\\hline
        10 & 2             \\\hline
        11 & 11             \\\hline
    \end{tabular}
\end{table}

\item  Single-qubit Gates I
)
Which code corresponds to the following quantum circuit

\begin{figure}[H] \centering{\includegraphics[scale=1]{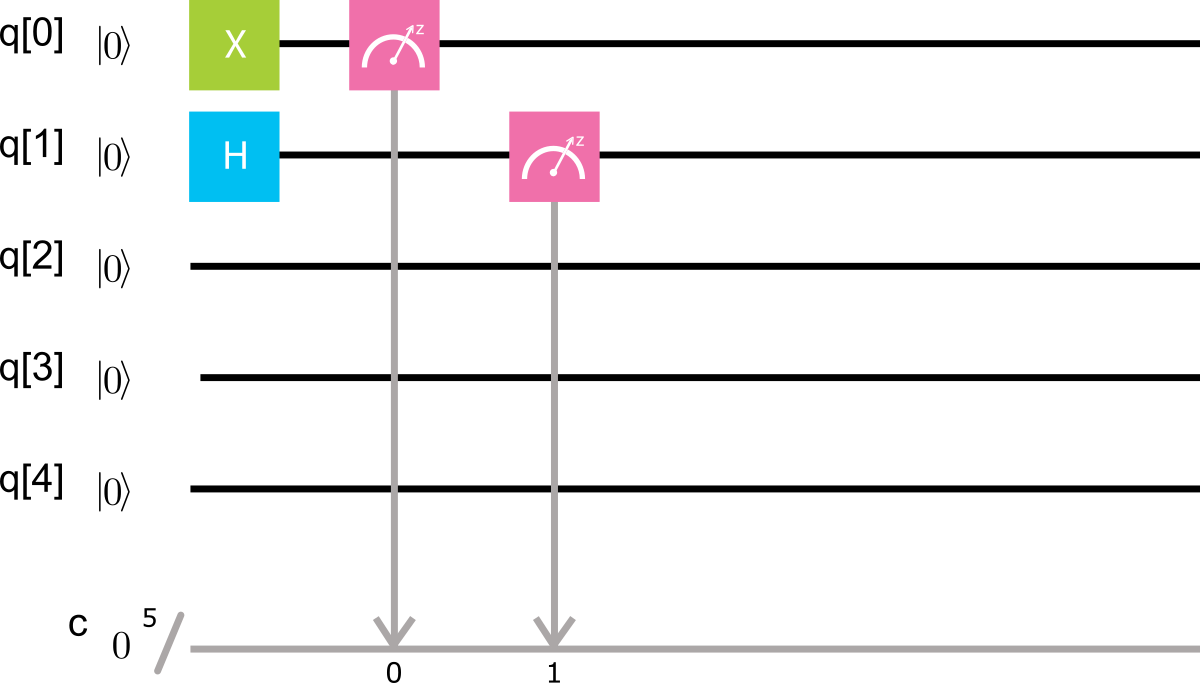}}\caption{Circuit}\label{fig1_50}
\end{figure}

\begin{lstlisting}
include ``qelib1.inc'';

qreg q[5];
creg c[5];

x q[0];
h q[1];
measure q[0] -> c[0];
measure q[1] -> c[1];
\end{lstlisting}

\begin{lstlisting}
include ``qelib1.inc'';

qreg q[5];
creg c[5];

x q[1];
h q[2];
measure q[1] -> c[1];
measure q[2] -> c[2];
\end{lstlisting}

\begin{lstlisting}
include ``qelib1.inc'';

qreg q[5];
creg c[5];

x q[0];
h q[1];
measure q[0] -> c[1];
measure q[1] -> c[0];
\end{lstlisting}

\begin{lstlisting}
include ``qelib1.inc'';

qreg q[5];
creg c[5];
x q[1];
h q[2];
measure q[1] -> c[2];
measure q[2] -> c[1];
\end{lstlisting}
Solution:\\
Since an X-gate is applied on qubit q[0], line 6 should be ``x q[0]''. Similarly, line 7 should be ``h q[1]'', as this is applying a Hadamard gate on qubit q[1].

\item  Single-qubit Gates II

Which the following code generates a quantum circuit

\begin{lstlisting}
include ``qelib1.inc'';

qreg q[5];
creg c[5];

x q[0];
x q[2];
measure q[1] -> c[1];
measure q[3] -> c[3];
\end{lstlisting}

\begin{figure}[H] \centering{\includegraphics[scale=1]{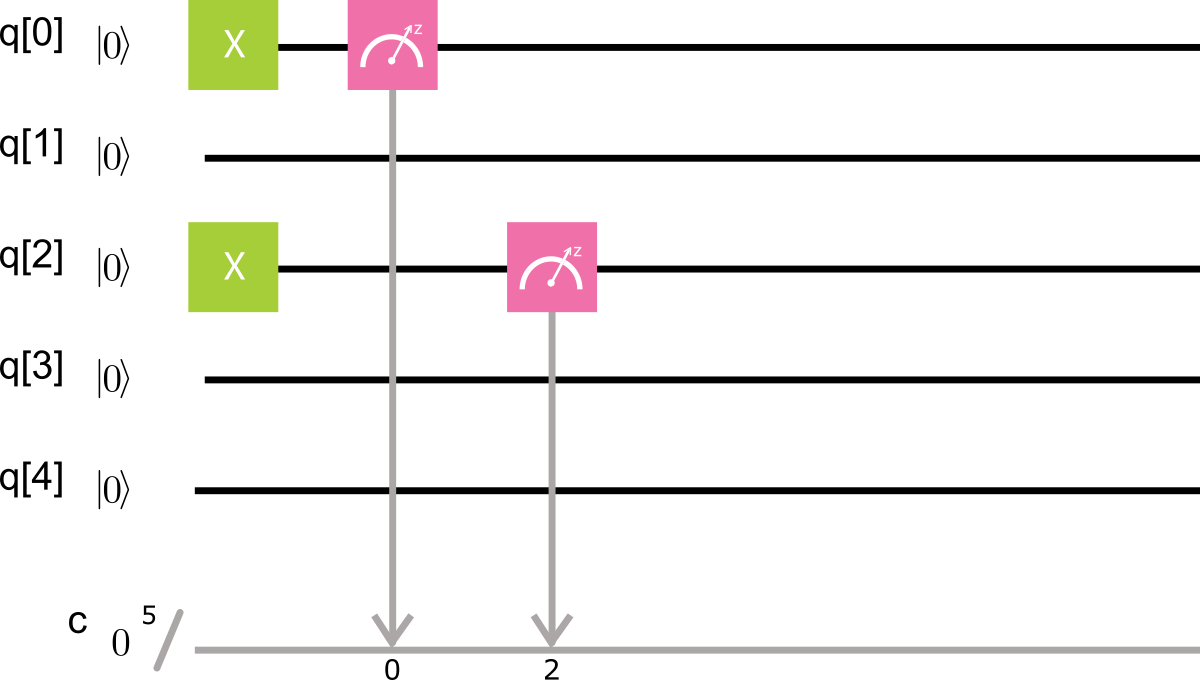}}\caption{IBM C15a}\label{fig1_51}
\end{figure}

\begin{figure}[H] \centering{\includegraphics[scale=1]{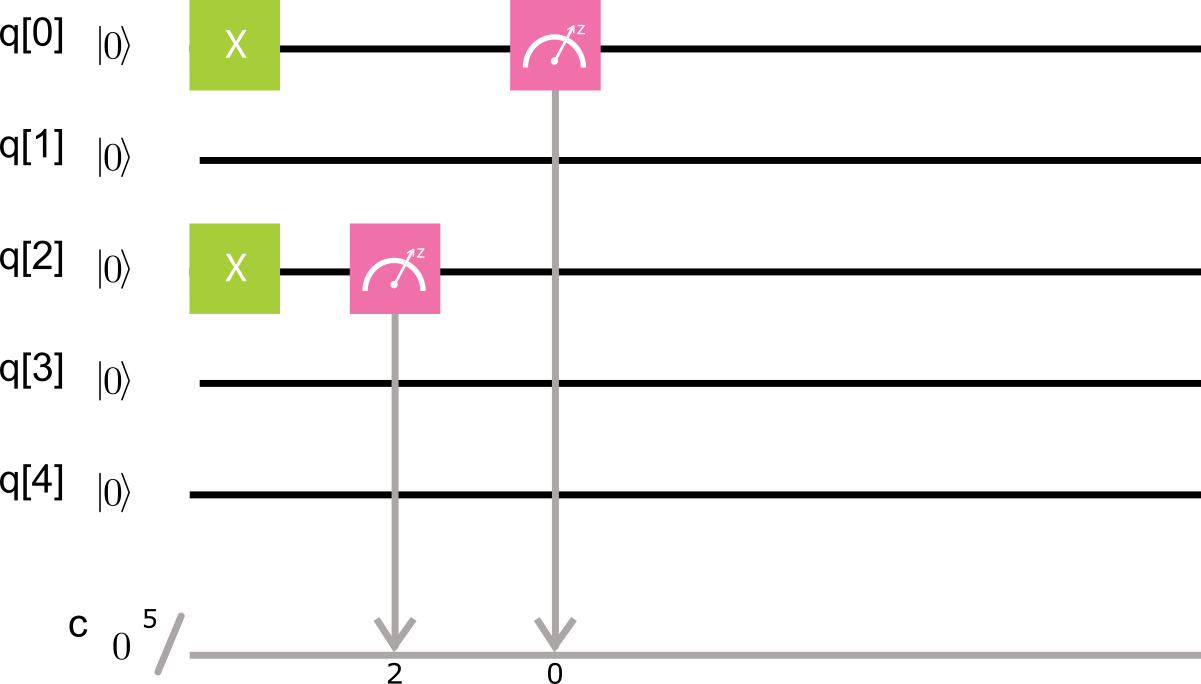}}\caption{IBM C15b}\label{fig1_52}
\end{figure}

\begin{figure}[H] \centering{\includegraphics[scale=1]{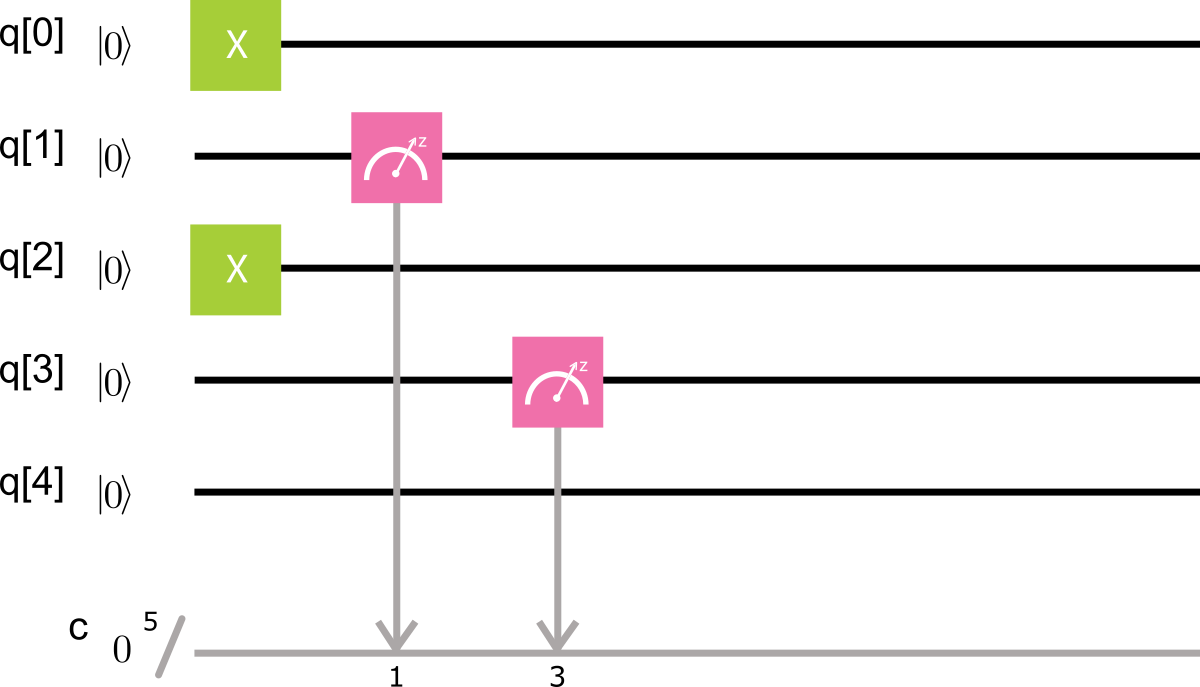}}\caption{IBM C15c}\label{fig1_53}
\end{figure}

\begin{figure}[H] \centering{\includegraphics[scale=1]{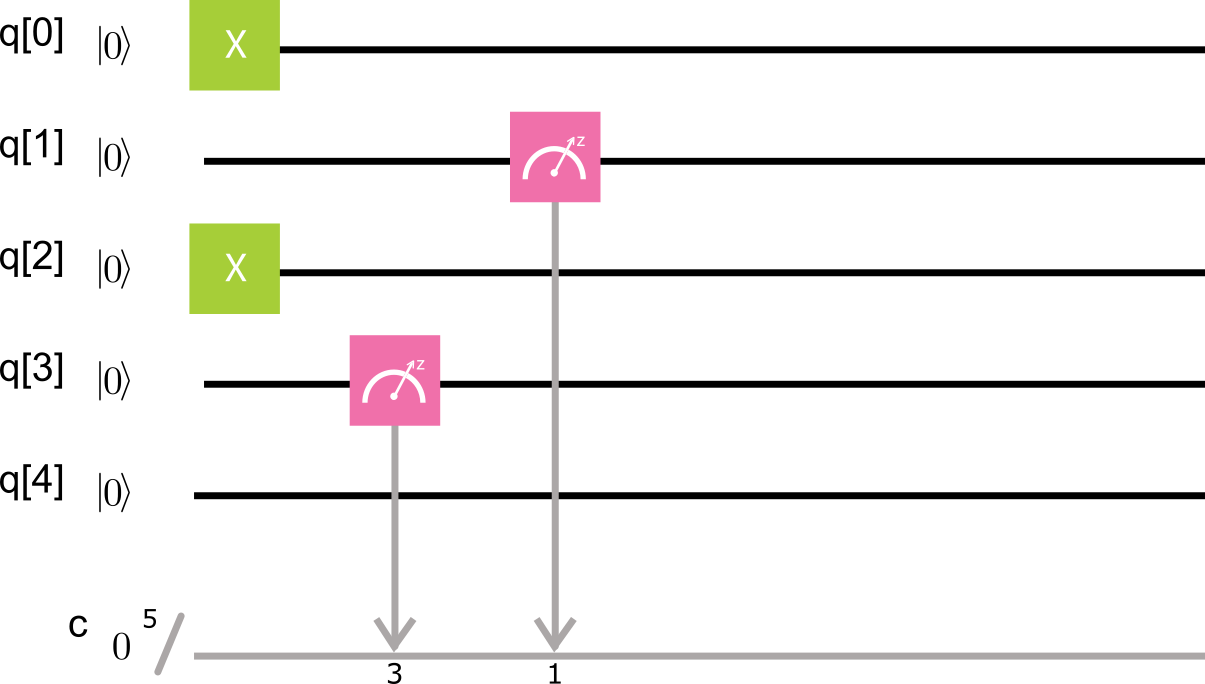}}\caption{IBM C15d}\label{fig1_54}
\end{figure}

\item First QASM Code, In the following activity we will run QASM code. Notice that the measurement results will not be given in a tabular format presented as previously, but instead we will see a histogram. Write in the console the following QASM lines:\\
\begin{lstlisting}
include ``qelib1.inc'';
qreg q[5];
creg c[5];
h q[1];
h q[2];
measure q[1] -> c[1];
measure q[2] -> c[2];
\end{lstlisting}

\begin{figure}[H] \centering{\includegraphics[scale=.5]{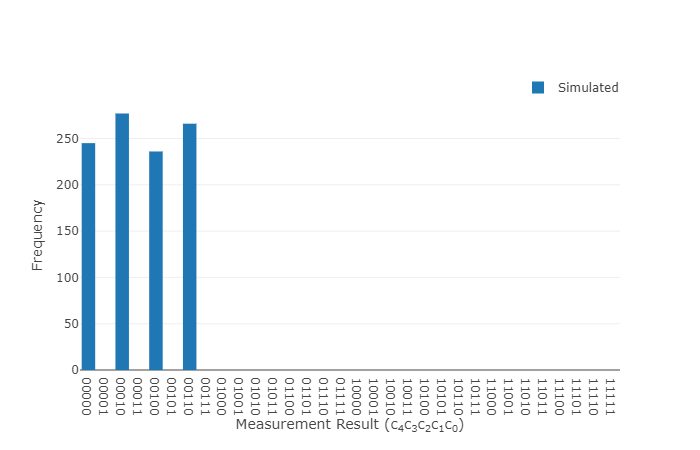}}\caption{Plot}\label{fig1_55}
\end{figure}

\item  Two Single-Qubit Gates, Write the code that generates the following quantum circuit

\begin{figure}[H] \centering{\includegraphics[scale=1]{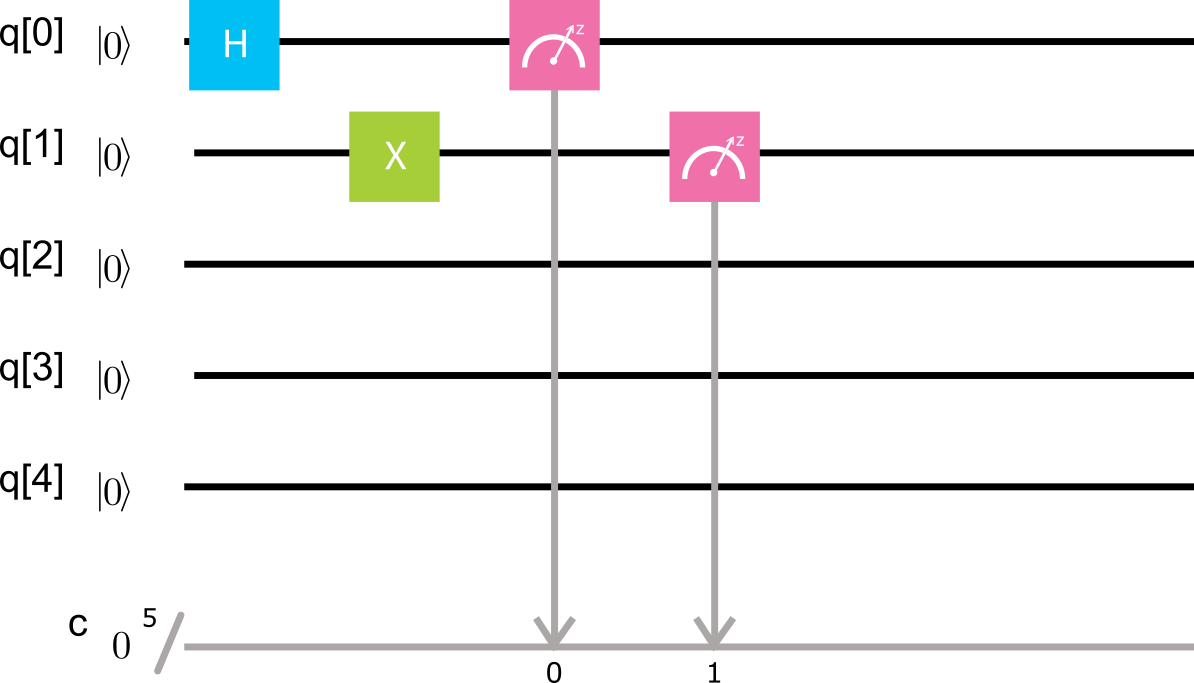}}\caption{IBM QE2}\label{fig1_56}
\end{figure}

\begin{lstlisting}
include ``qelib1.inc'';
qreg q[5];
creg c[5];
h q[0];
x q[1];
measure q[0] -> c[0];
measure q[1] -> c[1];
\end{lstlisting}

Solution:\\
Initially, qubits q[0] and q[1] are in the ground state $ \lvert 00\rangle $. A Hadamard gate is applied on q[0], and an X-gate on q[1]. This leaves both qubits in the final state $ \frac{\lvert 0\rangle +\lvert 1\rangle }{\sqrt{2}}\lvert 0\rangle $. In a perfect simulation, the two-qubit system would project on states $ \lvert 01\rangle $ and $ \lvert 11\rangle $ with probability 0.5 each.

\begin{figure}[H] \centering{\includegraphics[scale=.5]{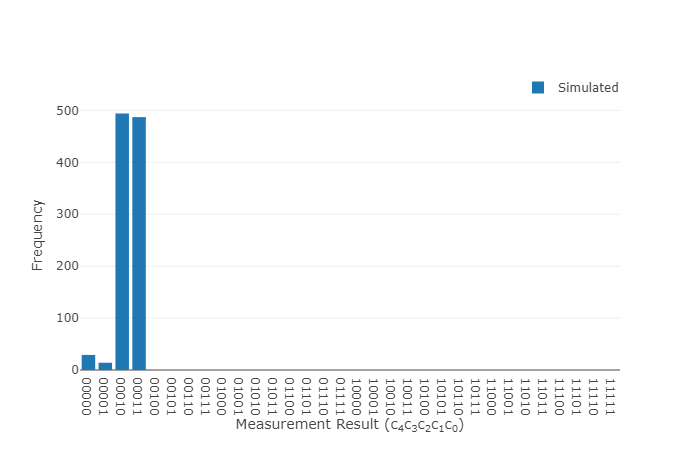}}\caption{Plot}\label{fig1_57}
\end{figure}

\section{Two-Qubit Gates and Backends}

Controlled NOT Gate

As we discussed, a universal gate set includes two-qubit gates. A commonly used two-qubit gate is the controlled-NOT operation or CNOT gate. Independent of which qubit is used as the control qubit and which as the target qubit, the CNOT gate will always apply an X-gate onto the target qubit if the control qubit is in the $ \lvert 1\rangle  $state. When the first qubit is defined as control and the second as a target, the matrix form of the CNOT is given by
\begin{equation}\label{eq1_52}
CNOT=\left( \begin{array}{cccc} 1 & 0 & 0 & 0\\ 0 & 1 & 0 & 0 \\ 0 & 0 & 0 & 1\\ 0 & 0 & 1 & 0 \end{array} \right)
\end{equation}
Notice that not all combinations of control and target qubits are allowed in the IBM QE. The combinations that we can use are determined by the topology of the IBM Q experience backend.

\subsection{ibmqx2 Backend}

The first backend that we will show is called ibmqx2. The figure below shows its topology.

\begin{figure}[H] \centering{\includegraphics[scale=.2]{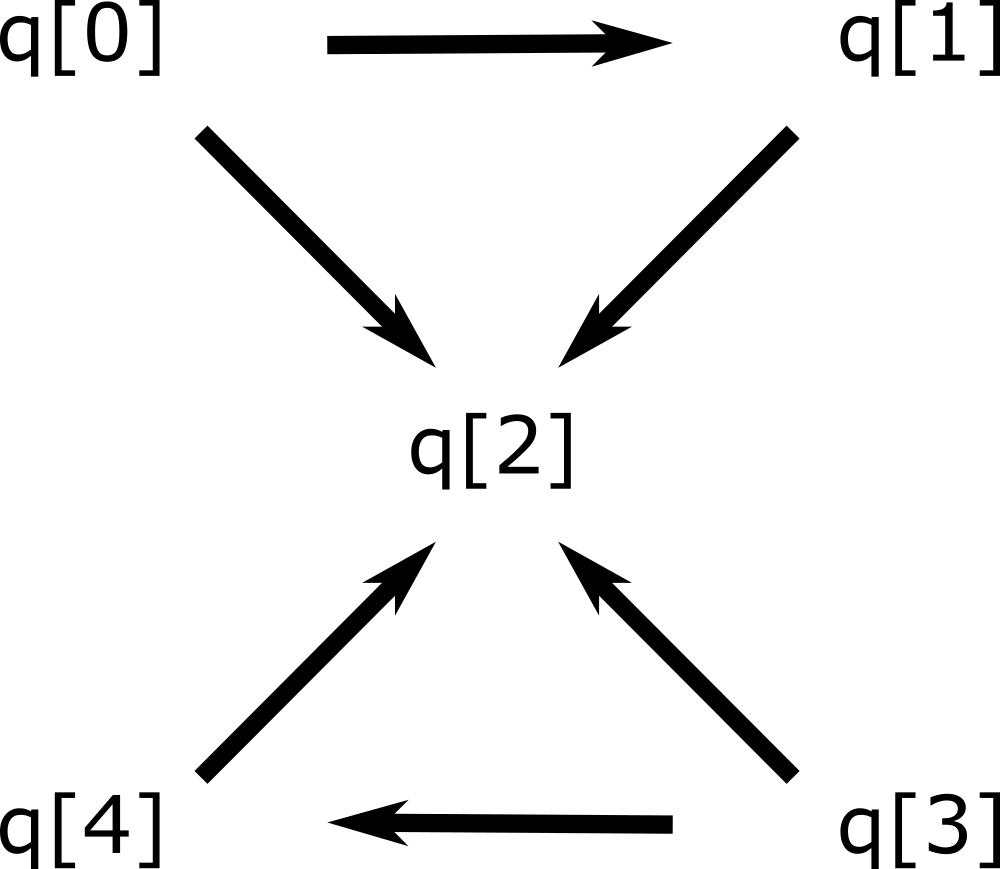}}\caption{Topology}\label{fig1_58}
\end{figure}

The following table shows all the allowed combinations for applying a CNOT-gate in QASM using the ibmqx2 backend.

\begin{table}[H]
    \centering
    \caption{CNOT-gate in QASM}
    \label{tab:1_1:Table 9}
    \begin{tabular}{|c|c|c|}  \hline
        Control qubit & Target qubit   & 
        QASM line  \\ \hline
    q[0] & q[1] & cx q[0],q[1] \\ \hline
    
    q[0] & q[2] & cx q[0],q[2] \\ \hline
    
    q[1] & q[2] & cx q[1],q[2] \\ \hline
    
    q[3] & q[2] & cx q[3],q[2] \\ \hline
    
    q[3] & q[4] & cx q[3],q[4] \\ \hline
    
    q[4] & q[2] & cx q[4],q[2] \\ \hline
    \end{tabular}
\end{table}

\item  CNOT Gate. The following code generates the quantum circuit below.\\
\begin{lstlisting}
include ``qelib1.inc'';
qreg q[5];
creg c[5];
x q[0];
cx q[0],q[1];
measure q[0] -> c[0];
measure q[1] -> c[1];
\end{lstlisting}

\begin{figure}[H] \centering{\includegraphics[scale=1]{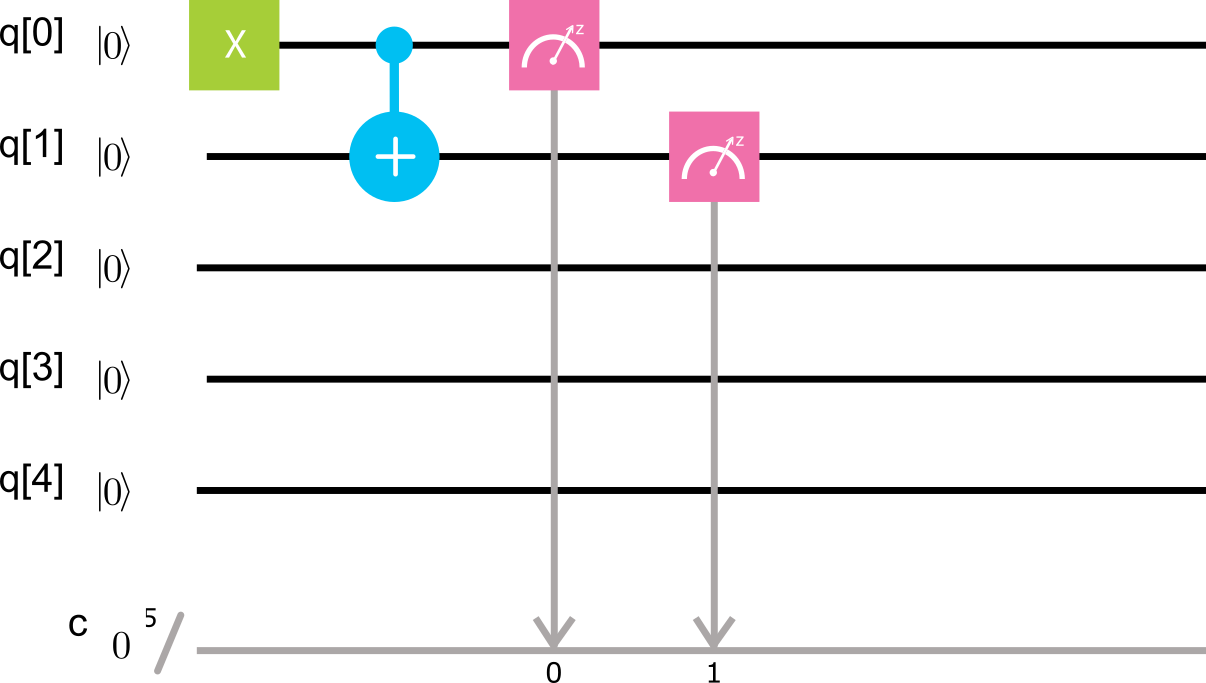}}\caption{CNIT2}\label{fig1_59}
\end{figure}

In the console, write the code that generates the following circuit.

\begin{figure}[H] \centering{\includegraphics[scale=1]{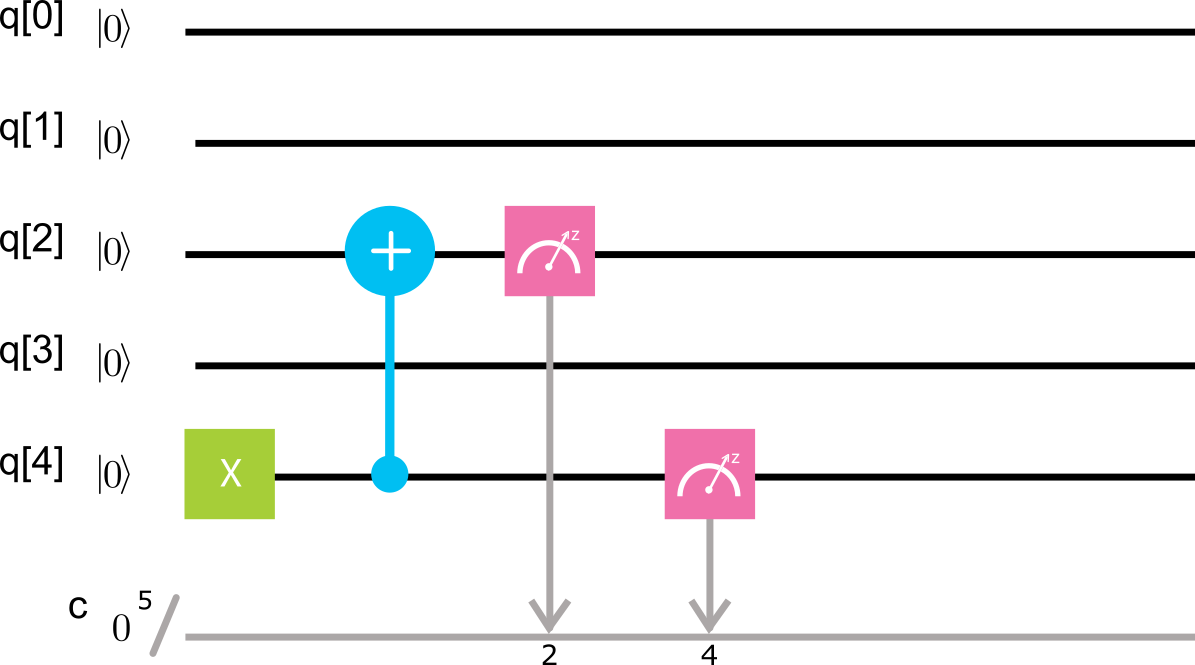}}\caption{CNOT3}\label{fig1_60}
\end{figure}
\begin{lstlisting}
include ``qelib1.inc'';
qreg q[5];
creg c[5];
x q[4];
cx q[4],q[2];
measure q[2] -> c[2];
measure q[4] -> c[4];
\end{lstlisting}
Solution:\\
An X-gate is applied on qubit q[4], followed by a CNOT gate with q[4] as control qubit and q[2] as target. The final state of the five qubit system is given by $ \lvert 00101\rangle $. The only non-zero analytical probability is $ p(q[2]q[4],\lvert 01\rangle )=1 $. This is, if the code is run in an perfect quantum computer, the system should always project onto $ \lvert 00101\rangle $

\begin{figure}[H] \centering{\includegraphics[scale=.6]{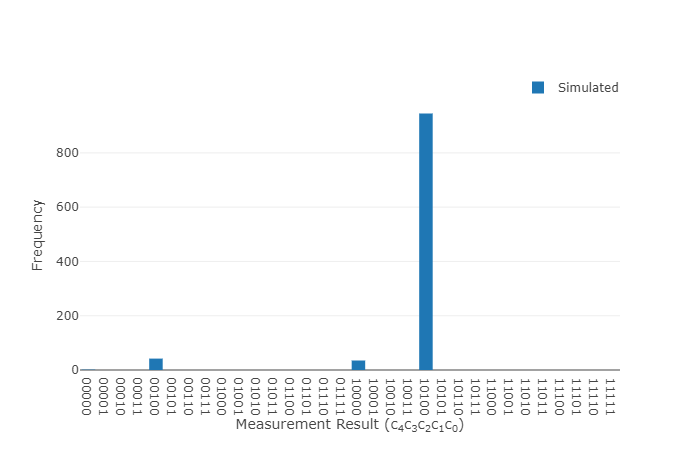}}\caption{Plot}\label{fig1_61}
\end{figure}

\item  Bell State Creation, In the console, write the code that generates the following circuit

\begin{figure}[H] \centering{\includegraphics[scale=1]{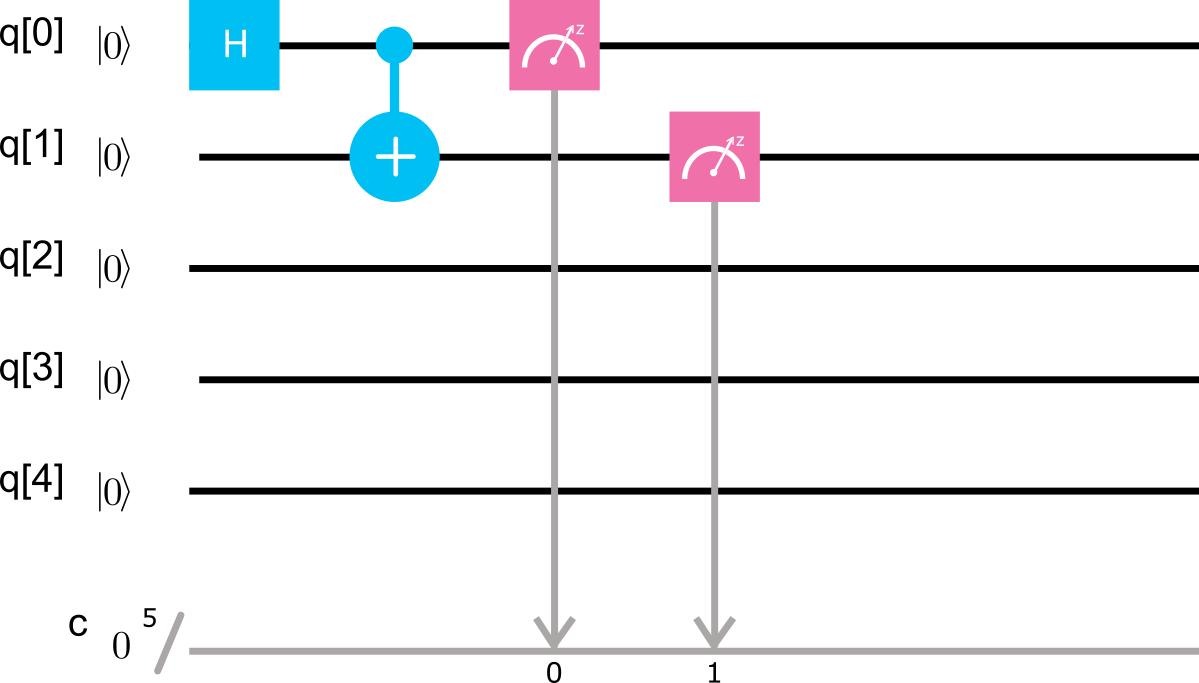}}\caption{IBM QE4}\label{fig1_62}
\end{figure}
\begin{lstlisting}
include ``qelib1.inc'';
qreg q[5];
creg c[5];
h q[0];
cx q[0],q[1];
measure q[0] -> c[0];
measure q[1] -> c[1];
\end{lstlisting}

Solution:\\
A Hadamard gate is applied on qubit q[0], followed by a CNOT gate with q[0] as control qubit and q[1] as target. The final state of the five qubit system is given by $ \frac{\lvert 00\rangle+ \lvert 11\rangle}{\sqrt{2}} $. The non-zero analytical probability are $ p(q[0]q[1],\lvert 00\rangle )=0.5 $, and $ p(q[0]q[1],\lvert 11\rangle )=0.5 $.

\begin{figure}[H] \centering{\includegraphics[scale=.6]{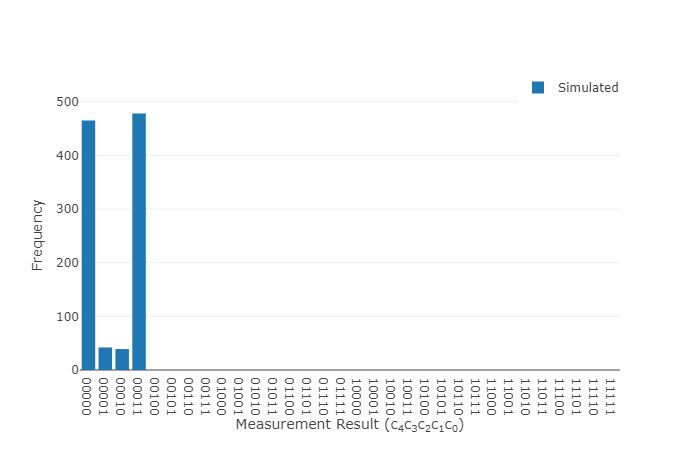}}\caption{Plot}\label{fig1_63}
\end{figure}

\subsection{ibmqx4 Backend}

The second backend that we will show is called ibmqx4. The figure below shows its topology.

\begin{figure}[H] \centering{\includegraphics[scale=.18]{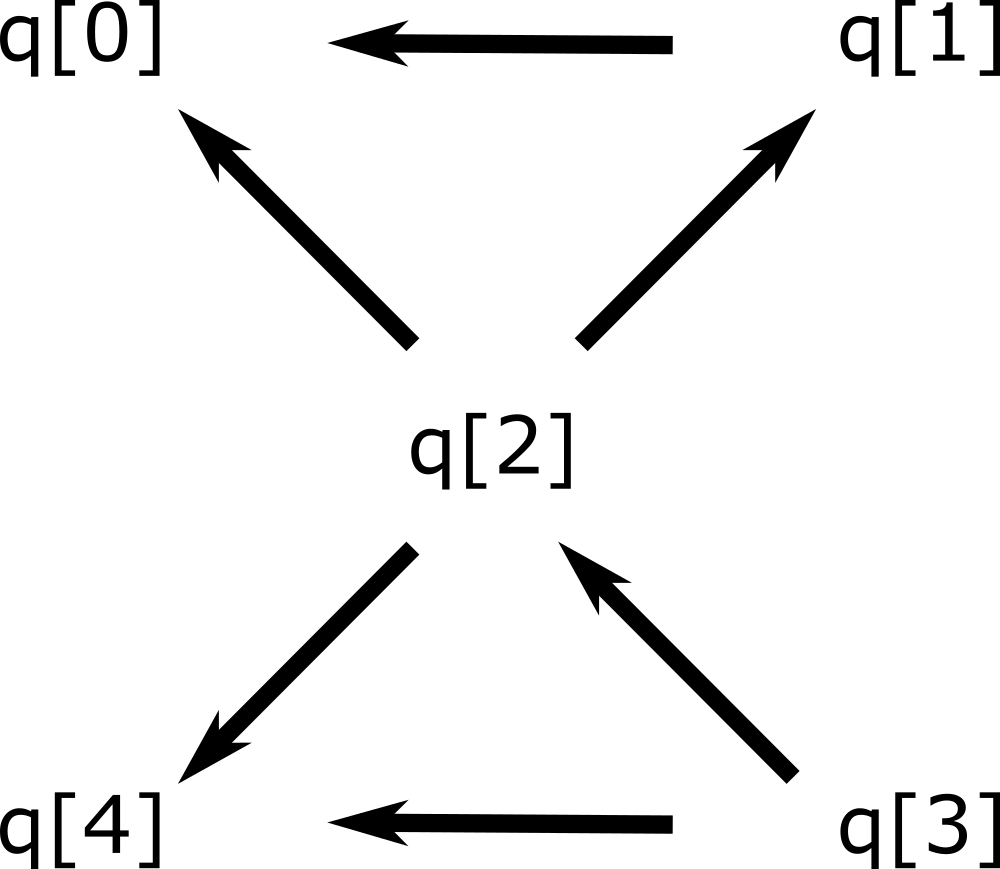}}\caption{Topology 2}\label{fig1_64}
\end{figure}

The following table shows all the allowed combinations for applying a CNOT-gate in QASM using the ibmqx4 backend.

\begin{table}[H]
    \centering
    \caption{CNOT-gate in QASM}
    \label{tab:1_1:Table 10}
    \begin{tabular}{|c|c|c|}  \hline
        Control qubit & Target qubit   & 
        QASM line  \\ \hline
        q[1] & q[0] & cx q[1],q[0] \\ \hline
        
        q[2] & q[1] & cx q[2],q[1] \\ \hline
        
        q[2] & q[0] & cx q[2],q[0] \\ \hline
        
        q[2] & q[4] & cx q[2],q[4] \\ \hline
        
        q[3] & q[2] & cx q[3],q[2] \\ \hline
        
        q[3] & q[4] & cx q[3],q[4] \\ \hline
    \end{tabular}
\end{table}

\item  Which of the following QASM programs will work properly with the ibmqx4 backend, meaning that the gates use allowed configurations of control and target qubits? (select all that are valid)\\
\begin{lstlisting}
include ``qelib1.inc'';

qreg q[5];
creg c[5];

x q[0];
h q[1];
cx q[0],q[1];
       
measure q[0] -> c[0];
measure q[1] -> c[1];
\end{lstlisting}
\begin{lstlisting}
include ``qelib1.inc'';

qreg q[5];
creg c[5];

x q[1];
h q[2];
cx q[2],q[1];

measure q[1] -> c[1];
measure q[2] -> c[2];
\end{lstlisting}

\begin{lstlisting}
include ``qelib1.inc'';

qreg q[5];
creg c[5];

x q[2];
h q[3];
cx q[2],q[3];

measure q[2] -> c[2];
measure q[3] -> c[3];
\end{lstlisting}

\begin{lstlisting}
include ``qelib1.inc'';

qreg q[5];
creg c[5];

x q[3];
h q[4];
cx q[3],q[4];
       
measure q[3] -> c[3];
measure q[4] -> c[4];
\end{lstlisting}

\section{N=1 Data Qubit + 1 Ancilla Qubit}

The following quantum circuit is an example of the Deutsch-Jozsa (DJ) algorithm implemented on a single (N=1) data qubit. In this case, the function being evaluated is $ f(x)=x_0 $, and it is balanced. In addition to the data qubit, the DJ algorithm generally needs a second qubit, an ancilla qubit, to facilitate quantum interference during the algorithm evolution. 

In the input register, qubit q[0] serves as the data qubit, and q[1] the ancilla qubit. The Oracle circuit in this case is a CNOT gate, as seen in the composer image:

\begin{figure}[H] \centering{\includegraphics[scale=.8]{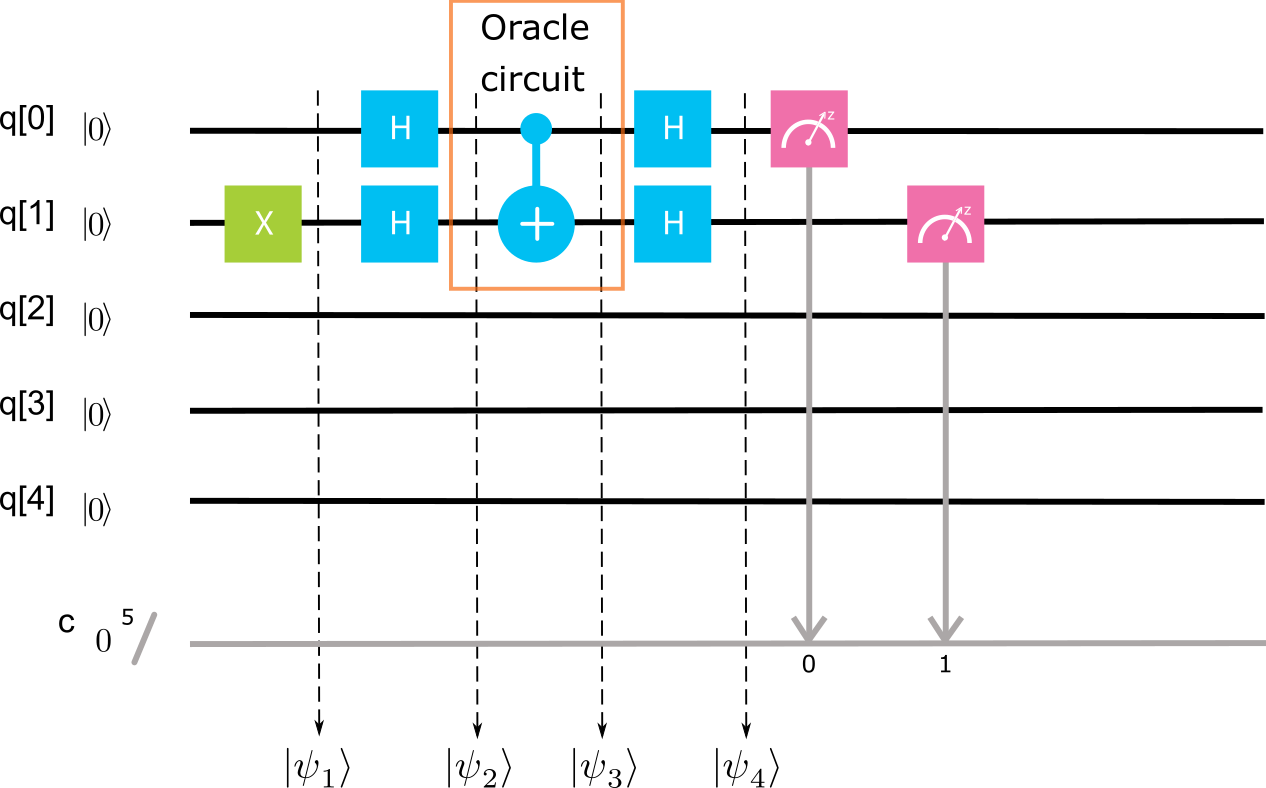}}\caption{N2 balanced circuit}\label{fig1_65}
\end{figure}

Let us now walk through the algorithm step-by-step. All qubits in the IBM Quantum Experience are initialized to state $ \left\vert 0 \right\rangle $,
\begin{equation}\label{eq1_53}
\left\vert \psi _{0}\right\rangle =\left\vert 0\right\rangle \left\vert 0\right\rangle .
\end{equation}

Since the ancilla qubit needs to be prepared in state $ \left\vert1\right\rangle $ as prescribed by the DJ algorithm, an X gate is applied to q[1], 
\begin{equation}\label{eq1_54}
\begin{split}
\displaystyle \left\vert \psi _{1}\right\rangle    \displaystyle & =    \displaystyle \left\vert 0\right\rangle X\left\vert 0\right\rangle ,    \\      
\displaystyle & =    \displaystyle \left\vert 0\right\rangle \left\vert 1\right\rangle .
\end{split}
\end{equation}

Next, a hadamard gate is applied to each qubit to create an equal superposition state of the two qubits.
\begin{equation}\label{eq1_55}
\begin{split}
\displaystyle \left\vert \psi _{2}\right\rangle    \displaystyle & =    \displaystyle H\left\vert 0\right\rangle H\left\vert 1\right\rangle ,\\          
\displaystyle & =    \displaystyle \left( \frac{\left\vert 0\right\rangle +\left\vert 1\right\rangle }{\sqrt{2}}\right) \left( \frac{\left\vert 0\right\rangle -\left\vert 1\right\rangle }{\sqrt{2}}\right) ,     \\     
\displaystyle & =    \displaystyle \frac{1}{2}\left( \left\vert 0\right\rangle \left( \left\vert 0\right\rangle -\left\vert 1\right\rangle \right) +\left\vert 1\right\rangle \left( \left\vert 0\right\rangle -\left\vert 1\right\rangle \right) \right).
\end{split}
\end{equation}

The Oracle then performs a CNOT gate, with q[0] the control qubit and q[1] the target qubit. 
\begin{equation}\label{eq1_56}
\begin{split}
\displaystyle \left\vert \psi _{3}\right\rangle    \displaystyle & =    \displaystyle \frac{1}{2}CNOT_{0,1}\left( \left\vert 0\right\rangle \left( \left\vert 0\right\rangle -\left\vert 1\right\rangle \right) +\left\vert 1\right\rangle \left( \left\vert 0\right\rangle -\left\vert 1\right\rangle \right) \right)         \\ 
\displaystyle & =    \displaystyle \frac{1}{2}\left( \left\vert 0\right\rangle \left( \left\vert 0\right\rangle -\left\vert 1\right\rangle \right) +\left\vert 1\right\rangle \left( \left\vert 1\right\rangle -\left\vert 0\right\rangle \right) \right)     \\     
\displaystyle & =    \displaystyle \frac{1}{\sqrt{2}}\left( \left\vert 0\right\rangle \left( \frac{\left\vert 0\right\rangle -\left\vert 1\right\rangle }{\sqrt{2}}\right) -\left\vert 1\right\rangle \left( \frac{\left\vert 0\right\rangle -\left\vert 1\right\rangle }{\sqrt{2}}\right) \right)          \\
\displaystyle & =    \displaystyle \left( \frac{\left\vert 0\right\rangle -\left\vert 1\right\rangle }{\sqrt{2}}\right) \left( \frac{\left\vert 0\right\rangle -\left\vert 1\right\rangle }{\sqrt{2}}\right).    
\end{split} 
\end{equation}

Two additional Hadamard gates are then applied to the output state of the oracle, facilitating quantum interference and leading to the output state  $ \left\vert f(0) \oplus f(1) \right\rangle $ for q[0] and state $ \left\vert1\right\rangle $ for q[1]. 
\begin{equation}\label{eq1_57}
\begin{split}
\displaystyle \left\vert \psi _{4}\right\rangle    \displaystyle & =    \displaystyle H\frac{\left\vert 0\right\rangle -\left\vert 1\right\rangle }{\sqrt{2}}H\frac{\left\vert 0\right\rangle -\left\vert 1\right\rangle }{\sqrt{2}}\\          
\displaystyle & =    \displaystyle \left\vert 1\right\rangle \left\vert 1\right\rangle
\end{split} 
\end{equation}

In a DJ algorithm, the function is constant if the values of the data qubits in the output register are all 0's. Here, they are not, as$  f(0) \oplus f(1) = 1 $, and so the function is confirmed to be balanced.

Although the DJ algorithm is generally prescribed as N data qubits plus one ancilla qubit, certain specific functions can design a ``simplified'' or ``compiled'' version of the oracle that reduces the overhead in implementing the algorithm. In the above graphic, the oracle shown also performs the DJ algorithm on N=2 data qubits for the function $ f(x)=x_0 \oplus x_1 $. It works, in part, because the function being balanced is determined by $ x_0 $ independent of $ x_1 $ and visa versa.  In this ``compiled'' version of the code, both q[0] and q[1] are data qubits, and q[1] is effectively playing the role of both data qubit and ancilla qubit. At the output, both data qubits are measured and, since they are not both 0's, the function is confirmed balanced.

\item  N=1 Data Qubit, Balanced Function, In the console below write the QASM code that generates the following circuit:

\begin{figure}[H] \centering{\includegraphics[scale=.8]{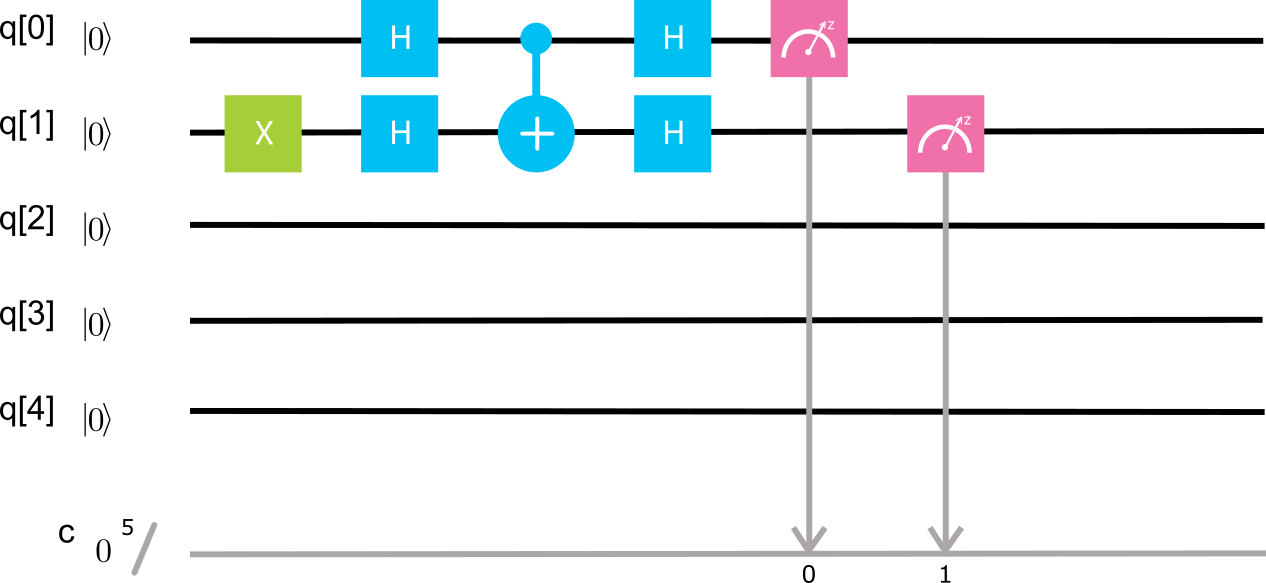}}\caption{N2 balanced circuit NE}\label{fig1_66}
\end{figure}

Submit our response, which will be evaluated with the grader, employing a numerically simulated quantum computer. Once we have a correct submission, our QASM code will automatically be queued to run on a real quantum computer at IBM.\\
\begin{lstlisting}
include ``qelib1.inc'';
qreg q[5];
creg c[5];

x q[1];
h q[0];
h q[1];
cx q[0],q[1];
h q[0];
h q[1];
measure q[0] -> c[0];
measure q[1] -> c[1];
\end{lstlisting}
Solution:\\
we just tested the DJ algorithm for N=2 qubits. This circuit implements in above section. Where the top qubit gives $ f(0)\oplus f(1) $, and the bottom qubit remains unchanged (so it is 1). Thus, since the function is balanced, $ f(0)\oplus f(1)=1 $, and we expect the top quit to be 1, and thus both qubits are measured to be 1. The result should thus be a histogram showing $ |11\rangle $ with high probability, e.g.:

\begin{figure}[H] \centering{\includegraphics[scale=.45]{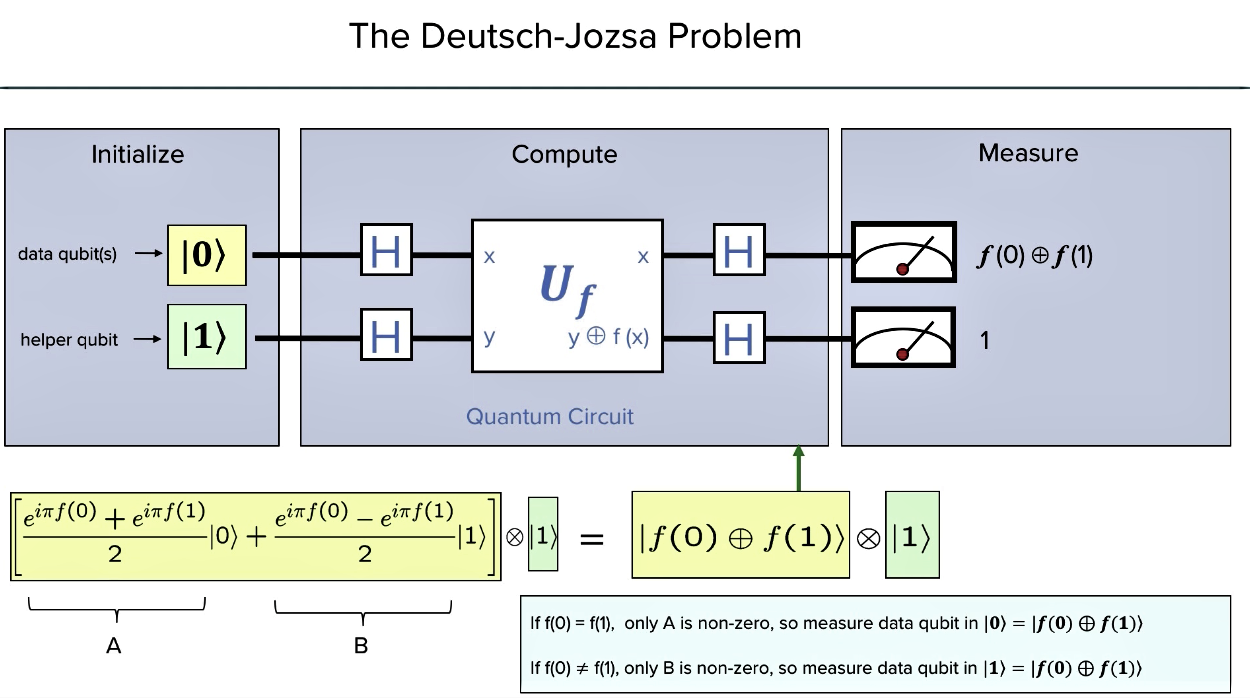}}\caption{DJ circuit-output}\label{fig1_89}
\end{figure}

\begin{figure}[H] \centering{\includegraphics[scale=.6]{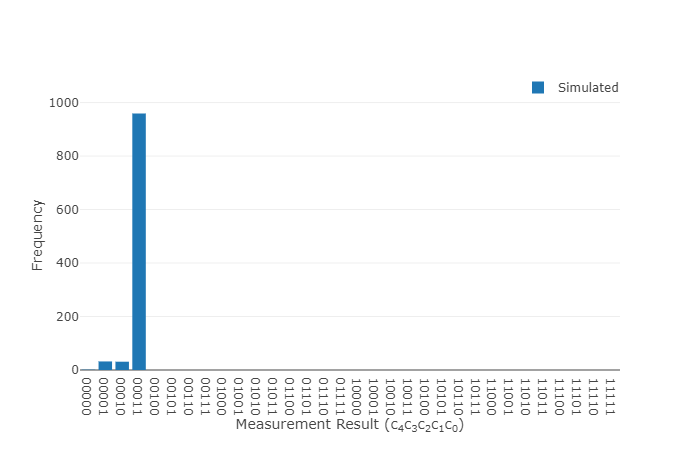}}\caption{Plot}\label{fig1_67}
\end{figure}

\item  N=1 Data Qubit, Constant Function, In the console below write the QASM code that generates the following circuit:

\begin{figure}[H] \centering{\includegraphics[scale=.7]{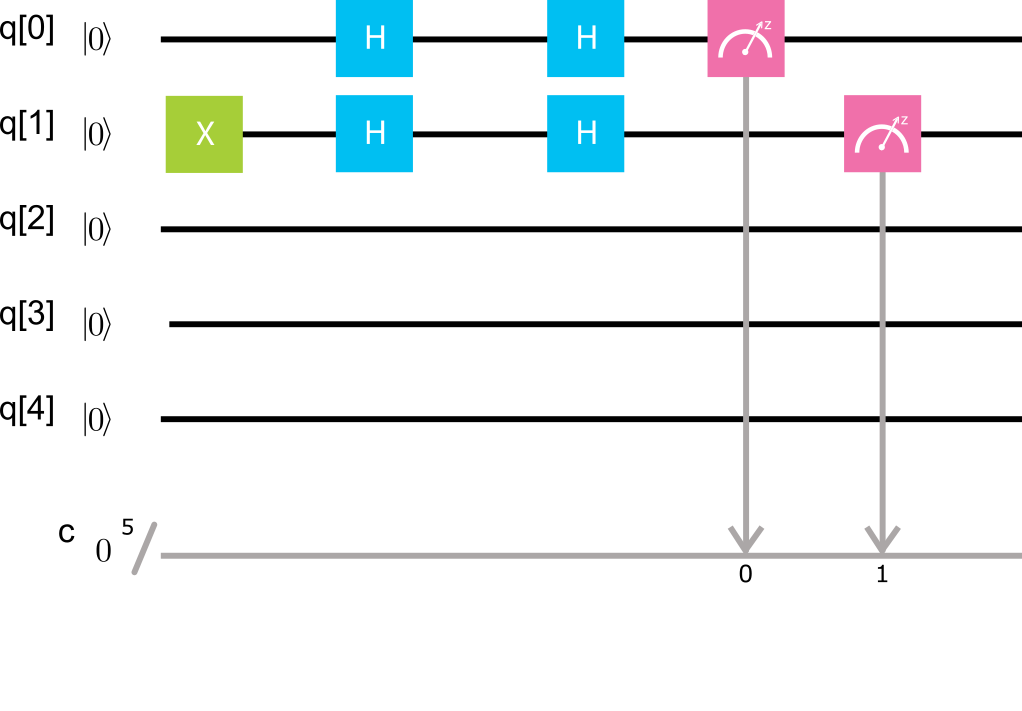}}\caption{N2 constant circuit}\label{fig1_68}
\end{figure}

Submit our response, which will be evaluated with the grader, employing a numerically simulated quantum computer. Once we have a correct submission, our QASM code will automatically be queued to run on a real quantum computer at IBM.\\
\begin{lstlisting}
include ``qelib1.inc'';
qreg q[5];
creg c[5];

x q[1];
h q[0];
h q[1];
h q[0];
h q[1];
measure q[0] -> c[0];
measure q[1] -> c[1];
\end{lstlisting}
Solution:\\
we just tested the DJ algorithm for N=2 qubits for a constant function. Where the top qubit gives $ f(0)\oplus f(1) $, and the bottom qubit remains unchanged (so it is 1). Thus, since the function is constant, $ f(0)\oplus f(1)=0 $, and we expect the top quit to be 0, and thus the output state should be $ |01\rangle $. The result should thus be a histogram showing $ |11\rangle $ with high probability, e.g.:

\begin{figure}[H] \centering{\includegraphics[scale=.5]{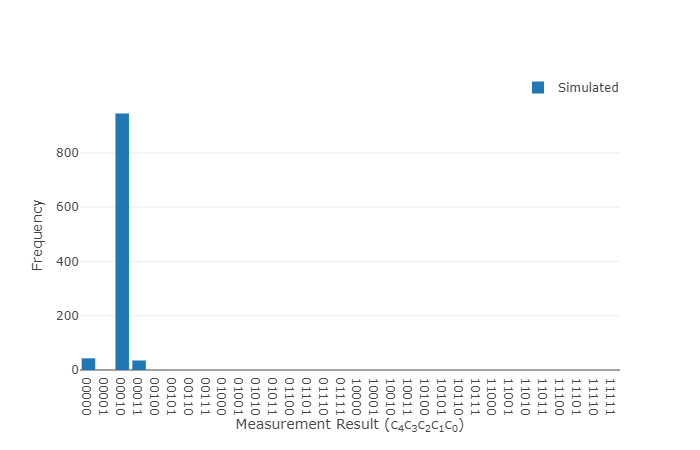}}\caption{Plot}\label{fig1_69}
\end{figure}

\section{N=3 Data Qubits}

The following quantum circuit is an N=3 data qubit example of the Deustch-Jozsa (DJ) algorithm for a balanced function $ f(x) = x_0 \oplus x_1x_2 $. This function can be understood to be balanced, as toggling the first bit  $ f(x) = x_0 $ will toggle the value of f(x) independent of the values of bits $ x_1 $ and $ x_2 $.

Generally, the DJ algorithm prescribes three data qubits plus one ancilla qubit for this function. However, the oracle circuit shown below is an example of a ``compiled'' version of the DJ, that is, an implementation that is simplified by design by taking advantage of a particular (but fairly generic) structure of the problem (note the compilation can be done without knowing anything about the answer). In this case, the input register needs only three qubits, all of which are data qubits to test this particular three-bit function f(x). The quantum interference facilitated by the ancilla qubit in the ``general'' algorithm is built into the oracle circuit in this design.  

\begin{figure}[H] \centering{\includegraphics[scale=.7]{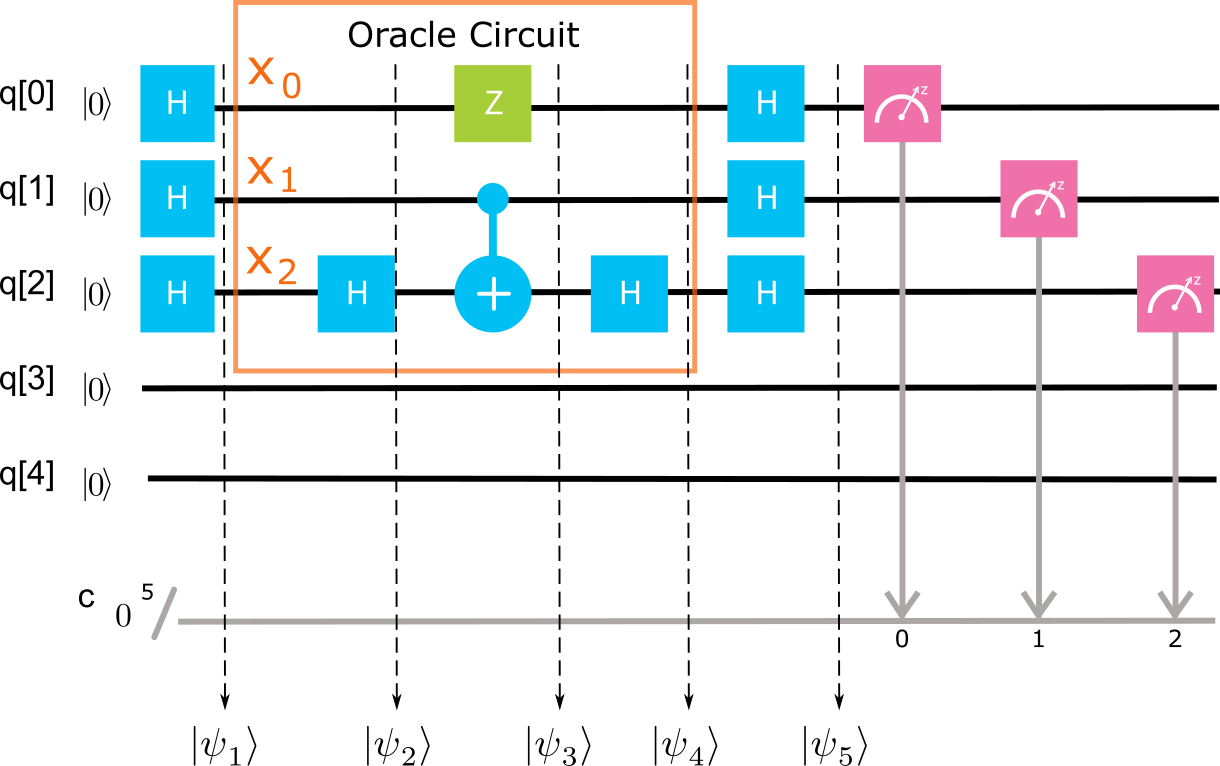}}\caption{Circuit N3 Balanced}\label{fig1_70}
\end{figure}

The three data qubits  q[0], q[1], and q[2]  are all initialized in state $ \left\vert 0 \right\rangle $. Note that none of the qubits are initialized in state $ \left\vert 1 \right\rangle $ as is generally prescribed for the ancilla qubit by the general version of the DJ algorithm.
\begin{equation}\label{eq1_58}
\left\vert \psi _{0}\right\rangle =\left\vert 0\right\rangle \left\vert 0\right\rangle \left\vert 0\right\rangle
(1)
\end{equation}

A three-qubit equal superposition state is then created by applying a Hadamard gate to each qubit.
\begin{equation}\label{eq1_59}
\begin{split}
\displaystyle \left\vert \psi _{1}\right\rangle    \displaystyle &=    \displaystyle H\left\vert 0\right\rangle H\left\vert 0\right\rangle H\left\vert 0\right\rangle \\          
\displaystyle & =    \displaystyle \left( \frac{\left\vert 0\right\rangle +\left\vert 1\right\rangle }{\sqrt{2}}\right) \left( \frac{\left\vert 0\right\rangle +\left\vert 1\right\rangle }{\sqrt{2}}\right) \left( \frac{\left\vert 0\right\rangle +\left\vert 1\right\rangle }{\sqrt{2}}\right)
\end{split}
\end{equation}
     
The oracle circuit $ U_ f $ consists a Z-gate on q[0], and a controlled-Z gate (CZ gate), realized by a CNOT gate  with q[1] the control qubit and q[2] the target qubit  between two Hadamards H applied to q[2]. Starting with the first Hadamard,
\begin{equation}\label{eq1_60}
\begin{split}
\displaystyle \left\vert \psi _{2}\right\rangle    \displaystyle & =    \displaystyle \left( \frac{\left\vert 0\right\rangle +\left\vert 1\right\rangle }{\sqrt{2}}\right) \left( \frac{\left\vert 0\right\rangle +\left\vert 1\right\rangle }{\sqrt{2}}\right) \left( H\frac{\left\vert 0\right\rangle +\left\vert 1\right\rangle }{\sqrt{2}}\right)     \\     
\displaystyle & =    \displaystyle \left( \frac{\left\vert 0\right\rangle +\left\vert 1\right\rangle }{\sqrt{2}}\right) \left( \frac{\left\vert 0\right\rangle +\left\vert 1\right\rangle }{\sqrt{2}}\right) \left\vert 0\right\rangle
\end{split}
\end{equation}

Next, the Z-gate is applied on q[0], and the CNOT gate to q[2] (target) and q[1] (control).
\begin{equation}\label{eq1_61}
\begin{split}
\displaystyle \left\vert \psi _{3}\right\rangle    \displaystyle & =    \displaystyle \left( Z\frac{\left\vert 0\right\rangle +\left\vert 1\right\rangle }{\sqrt{2}}\right) CNOT_{1,2}\left( \frac{\left\vert 0\right\rangle +\left\vert 1\right\rangle }{\sqrt{2}}\right) \left\vert 0\right\rangle     \\     
\displaystyle & =    \displaystyle \left( \frac{\left\vert 0\right\rangle -\left\vert 1\right\rangle }{\sqrt{2}}\right) \left( \frac{\left\vert 0\right\rangle \left\vert 0\right\rangle +\left\vert 1\right\rangle \left\vert 1\right\rangle }{\sqrt{2}}\right)
\end{split}
\end{equation}

And, then the second Hadamard gate is applied to q[2].
\begin{equation}\label{eq1_62}
\begin{split}
\displaystyle \left\vert \psi _{4}\right\rangle    \displaystyle & =    \displaystyle \left( \frac{\left\vert 0\right\rangle -\left\vert 1\right\rangle }{\sqrt{2}}\right) \left( \frac{\left\vert 0\right\rangle H\left\vert 0\right\rangle +\left\vert 1\right\rangle H\left\vert 1\right\rangle }{\sqrt{2}}\right)     \\     
\displaystyle & =    \displaystyle \left( \frac{\left\vert 0\right\rangle -\left\vert 1\right\rangle }{\sqrt{2}}\right) \left( \frac{\left\vert 0\right\rangle \left\vert 0\right\rangle +\left\vert 0\right\rangle \left\vert 1\right\rangle +\left\vert 1\right\rangle \left\vert 0\right\rangle -\left\vert 1\right\rangle \left\vert 1\right\rangle }{2}\right)
\end{split}
\end{equation}
     
After the oracle circuit, each qubit is again operated on by a Hadamard gate, leaving the final state of the 3 qubits as
\begin{equation}\label{eq1_63}
\begin{split}
\displaystyle \left\vert \psi _{5}\right\rangle    \displaystyle & =    \displaystyle H\left( \frac{\left\vert 0\right\rangle -\left\vert 1\right\rangle }{\sqrt{2}}\right) \left( \frac{H\left\vert 0\right\rangle H\left\vert 0\right\rangle +H\left\vert 0\right\rangle H\left\vert 1\right\rangle +H\left\vert 1\right\rangle H\left\vert 0\right\rangle -H\left\vert 1\right\rangle H\left\vert 1\right\rangle }{2}\right)          \\
\displaystyle & =    \displaystyle \left\vert 1\right\rangle \left( \frac{H\left\vert 0\right\rangle H\left\vert 0\right\rangle +H\left\vert 0\right\rangle H\left\vert 1\right\rangle +H\left\vert 1\right\rangle H\left\vert 0\right\rangle -H\left\vert 1\right\rangle H\left\vert 1\right\rangle }{2}\right)    \\      
\displaystyle & =    \displaystyle \frac{\left\vert 1\right\rangle \left\vert 0\right\rangle \left\vert 0\right\rangle +\left\vert 1\right\rangle \left\vert 1\right\rangle \left\vert 0\right\rangle +\left\vert 1\right\rangle \left\vert 0\right\rangle \left\vert 1\right\rangle -\left\vert 1\right\rangle \left\vert 1\right\rangle \left\vert 1\right\rangle }{2}
\end{split}
\end{equation}

In each Deutsch-Jozsa algorithm realization, if the data qubits at the output are all 0's, then the function is constant. Otherwise, it is balanced. In this case, there are no states $ \left\vert 0 \right\rangle \left\vert 0  \right\rangle \left\vert 0 \right\rangle $ in the output superposition state with non-zero probability amplitude, and thus any measurement will have at least one of the qubits in-state $ \left\vert 1 \right\rangle $. Consequently, the function is confirmed to be balanced in just one query.

\item Balanced DJ Algorithm, In the console below write the QASM code that generates the following circuit:

\begin{figure}[H] \centering{\includegraphics[scale=.7]{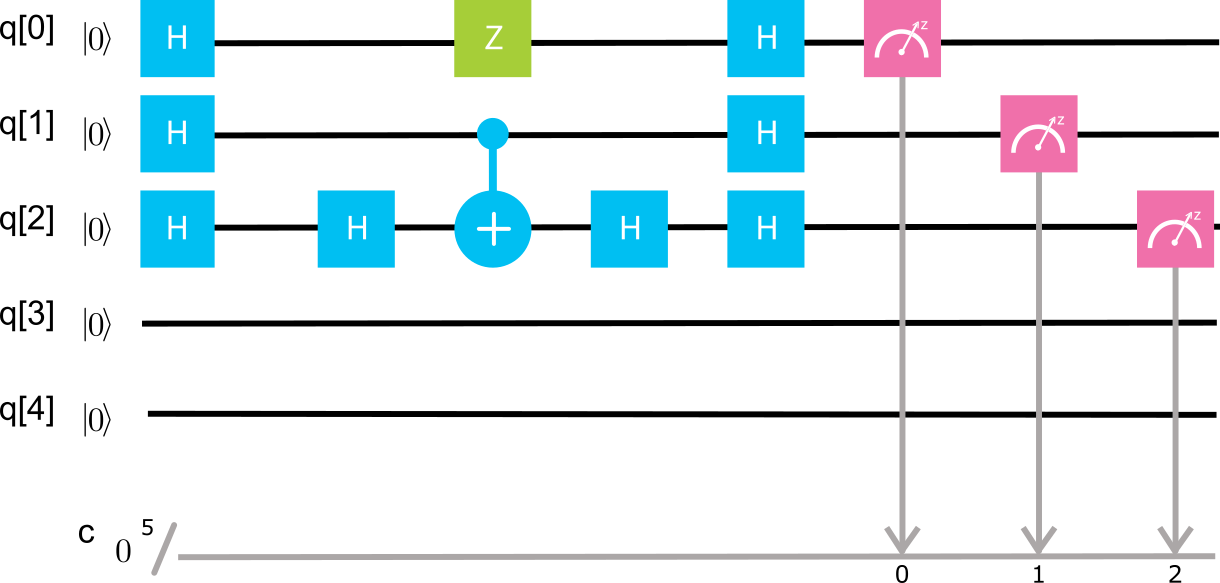}}\caption{Circuit N3 Balanced NE}\label{fig1_71}
\end{figure}

Submit our response, which will be evaluated with the grader, employing a numerically simulated quantum computer. Once we have a correct submission, our QASM code will automatically be queued to run on a real quantum computer at IBM.\\
\begin{lstlisting}
include ``qelib1.inc'';
qreg q[5];
creg c[5];

h q[0];
h q[1];
h q[2];
h q[2];
z q[0];
cx q[1],q[2];
h q[2];
h q[0];
h q[1];
h q[2];
measure q[0] -> c[0];
measure q[1] -> c[1];
measure q[2] -> c[2];
\end{lstlisting}
Solution: we just tested the DJ algorithm for N=3 qubits. The code was run 1024 times. The expected histogram is something like this: four states with nonzero probability all have the first qubit being equal to 1 this is the signature expected for the function being balanced.
\begin{figure}[H] \centering{\includegraphics[scale=.5]{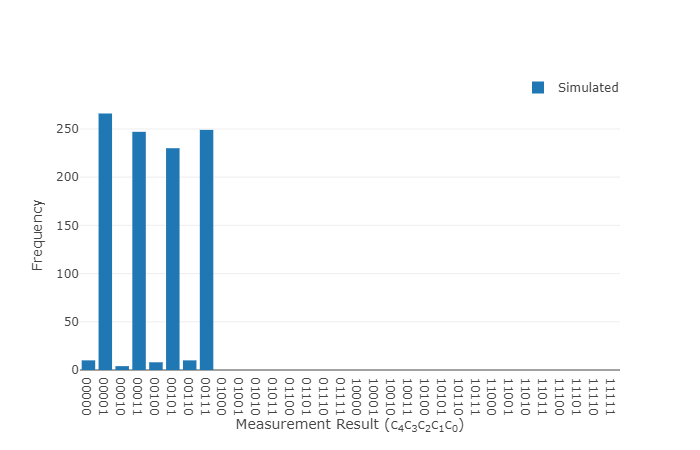}}\caption{Plot}\label{fig1_72}
\end{figure}

\item  Constant DJ Algorithm, In the console below write the QASM code that generates the following circuit:

\begin{figure}[H] \centering{\includegraphics[scale=1]{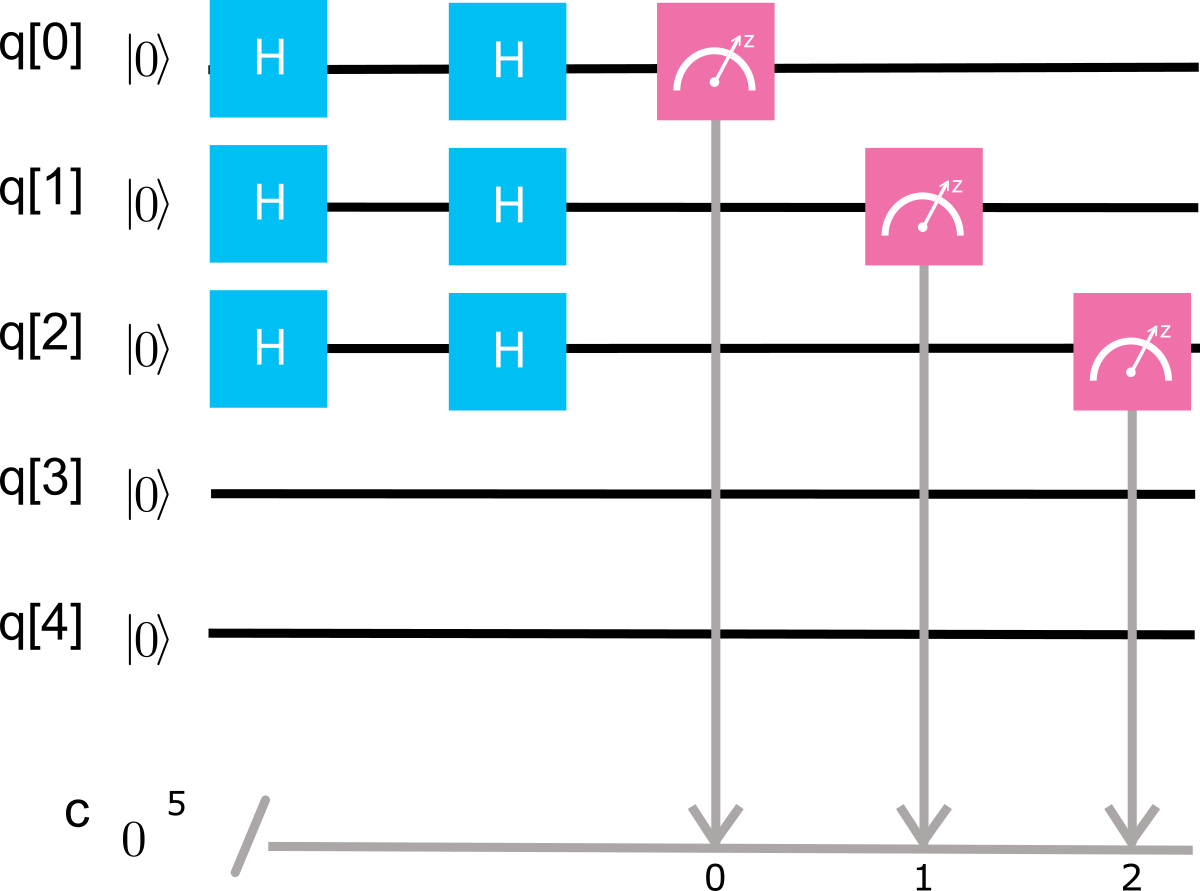}}\caption{Circuit N3 Constant}\label{fig1_73}
\end{figure}

Submit our response, which will be evaluated with the grader, employing a numerically simulated quantum computer. Once we have a correct submission, our QASM code will automatically be queued to run on a real quantum computer at IBM.\\
\begin{lstlisting}
include ``qelib1.inc'';
qreg q[5];
creg c[5];

h q[0];
h q[1];
h q[2];
h q[0];
h q[1];
h q[2];
measure q[0] -> c[0];
measure q[1] -> c[1];
measure q[2] -> c[2];
\end{lstlisting}
Solution: we just tested the DJ algorithm for N=3 qubits. The code was run 1024 times. The expected histogram of measurement results is something like this: fact that the output state is all zeros is expected, because all the Hadamard gates pair up and simply cancel each other.

\begin{figure}[H] \centering{\includegraphics[scale=.6]{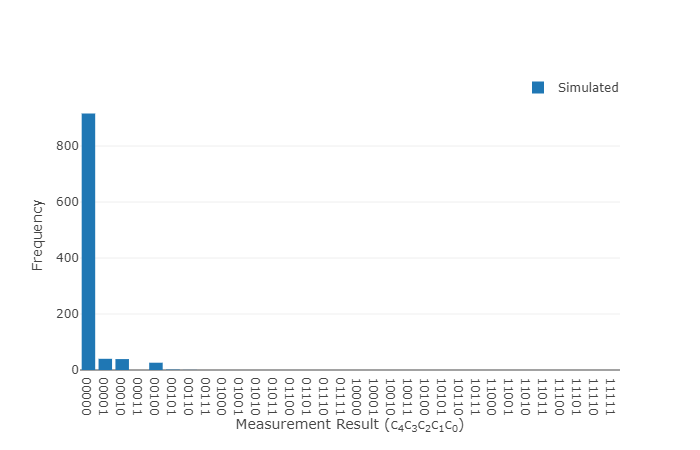}}\caption{Plot}\label{fig1_74}
\end{figure}

\item  QASM Code I
In the console below write the QASM code that generates the following circuit

\begin{figure}[H] \centering{\includegraphics[scale=.5]{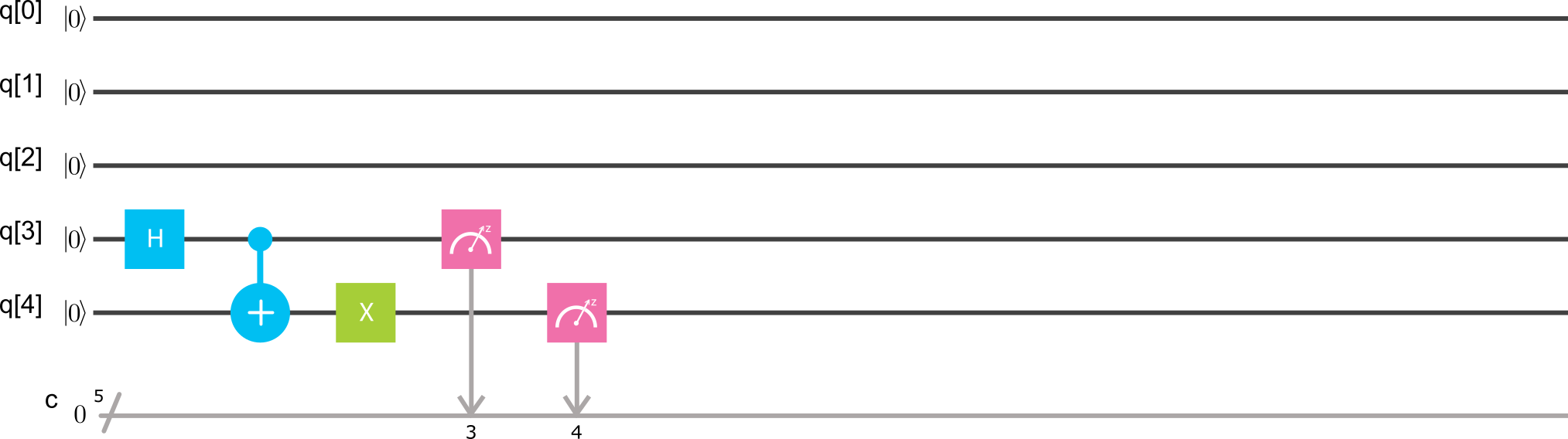}}\caption{GA1}\label{fig1_75}
\end{figure}

\begin{lstlisting}
include ``qelib1.inc'';
qreg q[5];
creg c[5];

h q[3];
cx q[3],q[4];
x q[4];
measure q[3] -> c[3];
measure q[4] -> c[4];
\end{lstlisting}
Solution:\\
The circuit should act on qubits q[3] and q[4], and produce an output histogram like this one: The output is a superposition of $ |00010\rangle and |00001\rangle $; by inspection, this is because the output after the Hadamard and CNOT is (up to normalization) $ |00000\rangle + |00011\rangle $; the last X gate then flips the bottom qubit.

\begin{figure}[H] \centering{\includegraphics[scale=.5]{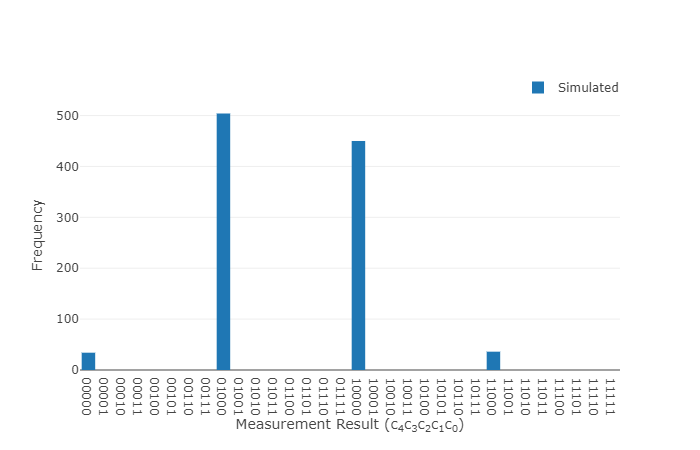}}\caption{Plot}\label{fig1_76}
\end{figure}

\item  QASM Code II
In the console below, write the QASM code that generates the following circuit.

\begin{figure}[H] \centering{\includegraphics[scale=.5]{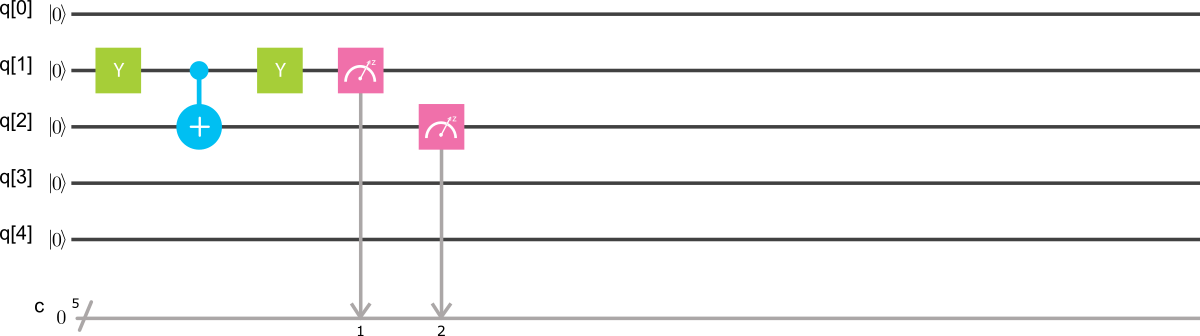}}\caption{GA2}\label{fig1_77}
\end{figure}
\begin{lstlisting}
include ``qelib1.inc'';
qreg q[5];
creg c[5];

y q[1];
cx q[1],q[2];
y q[1];
measure q[1] -> c[1];
measure q[2] -> c[2];
\end{lstlisting}
Solution:\\
The circuit should act on qubits q[1] and q[2], and produce an output histogram like this one: The output is a single state, $ |00100\rangle $; by inspection, this is because the first Y gate flips the control qubit from 0 to 1, causing the CNOT to flip its target qubit; the second Y gate flips the control qubit back to 0.

\begin{figure}[H] \centering{\includegraphics[scale=.5]{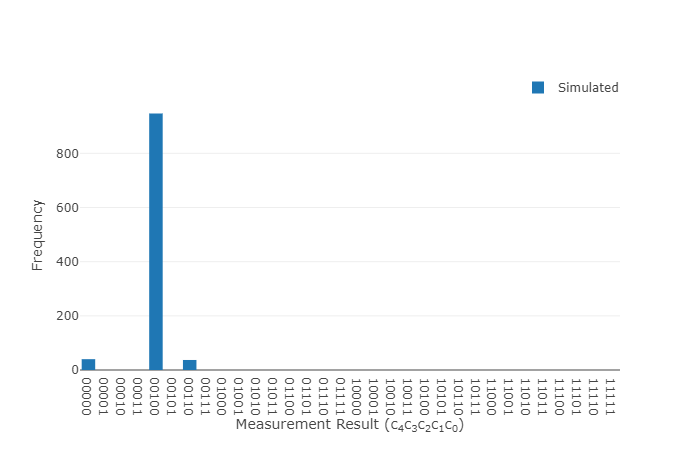}}\caption{Plot}\label{fig1_78}
\end{figure}

\item  QASM Code III
In the console below, write the QASM code that generates the following circuit.

\begin{figure}[H] \centering{\includegraphics[scale=.5]{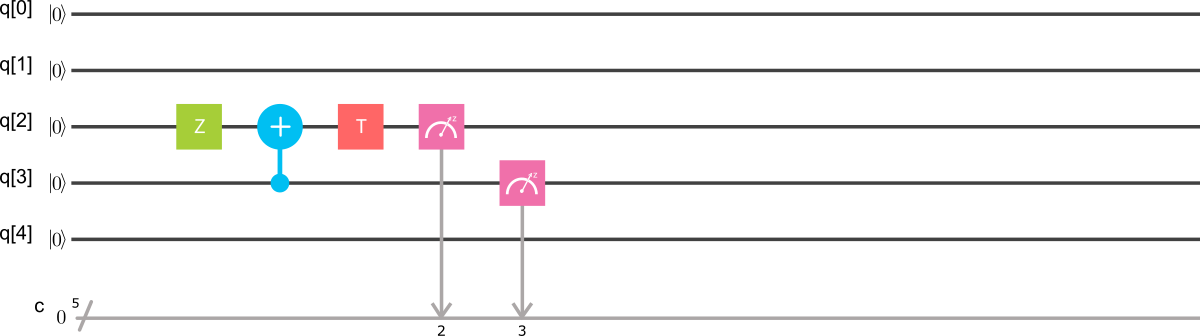}}\caption{GA3}\label{fig1_79}
\end{figure}
\begin{lstlisting}
include ``qelib1.inc'';
qreg q[5];
creg c[5];

z q[2];
cx q[3],q[2];
t q[2];
measure q[2] -> c[2];
measure q[3] -> c[3];
\end{lstlisting}
Solution:\\
The circuit should act on qubits q[2] and q[3], and produce an output histogram like this one: The output is a single state, $ |00000\rangle $; by inspection, this is because the CNOT has a control qubit in the 0 states, so the CNOT does nothing. The Z and T gates also do nothing on their input, the $ |0\rangle  $ state, so the output is all zeros.

\begin{figure}[H] \centering{\includegraphics[scale=.6]{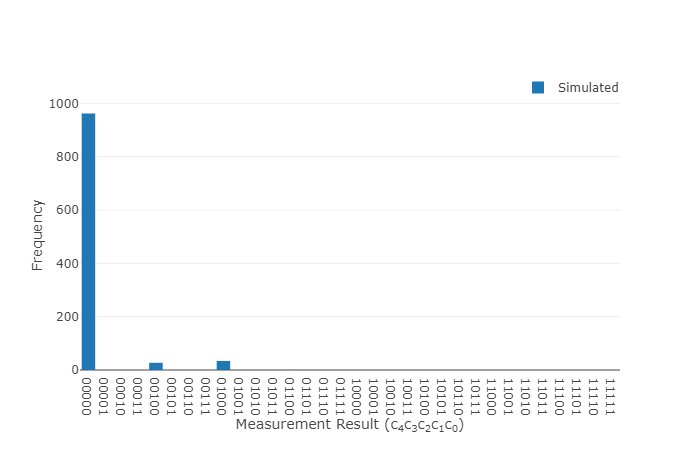}}\caption{Plot}\label{fig1_80}
\end{figure}

\begin{figure}[H] \centering{\includegraphics[scale=.7]{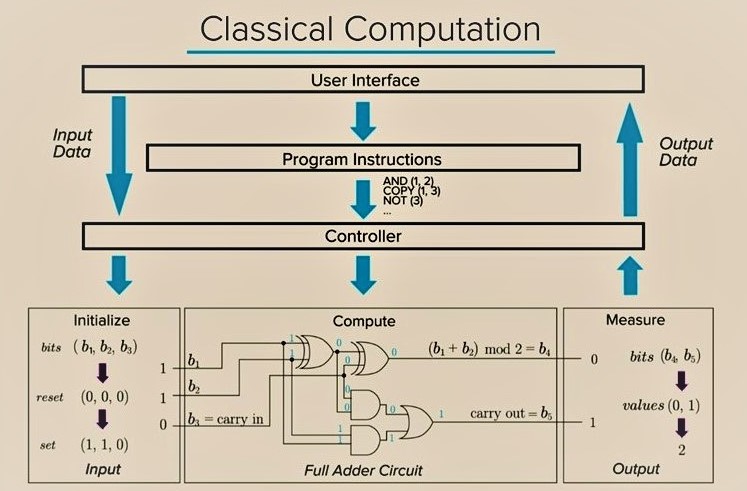}}\caption{Classical Circuit Model}\label{fig1_81}
\end{figure}

\begin{figure}[H] \centering{\includegraphics[scale=.6]{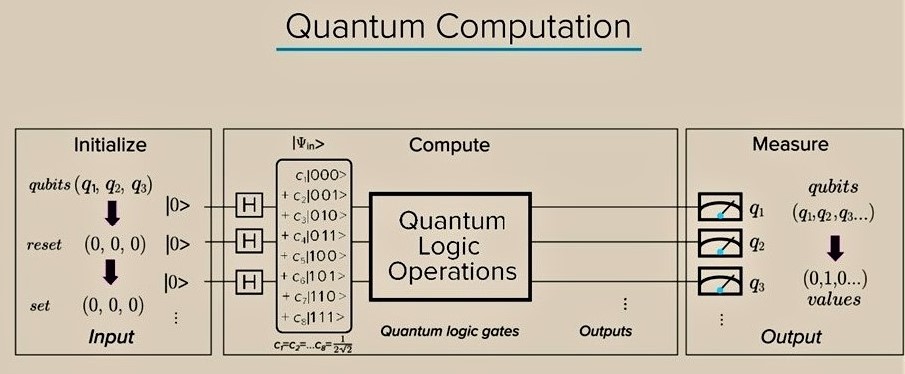}}\caption{Quantum Circuit Model}\label{fig1_82}
\end{figure}

\begin{figure}[H] \centering{\includegraphics[scale=.6]{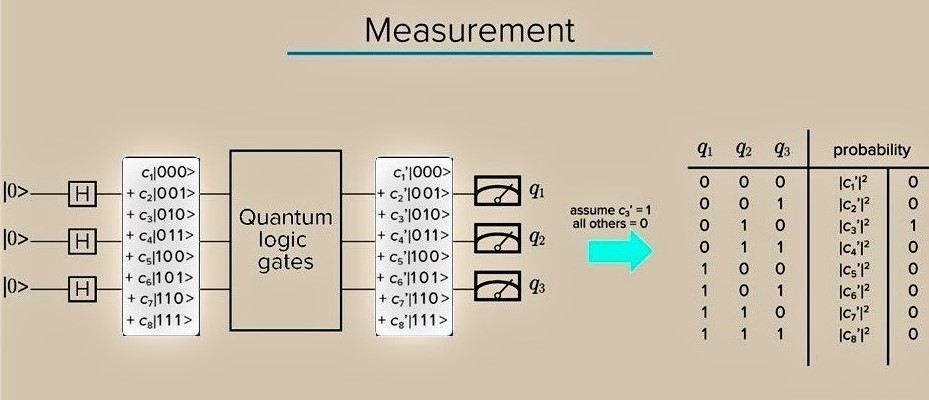}}\caption{Quantum Circuit Model}\label{fig1_83}
\end{figure}

\begin{figure}[H] \centering{\includegraphics[scale=.6]{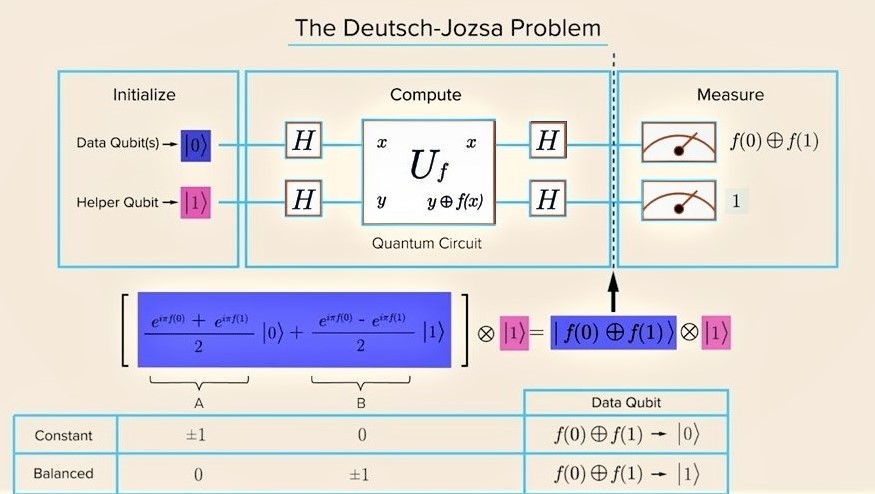}}\caption{The Deutsch-Jozsa Algorithm}\label{fig1_84}
\end{figure}

\begin{figure}[H] \centering{\includegraphics[scale=1]{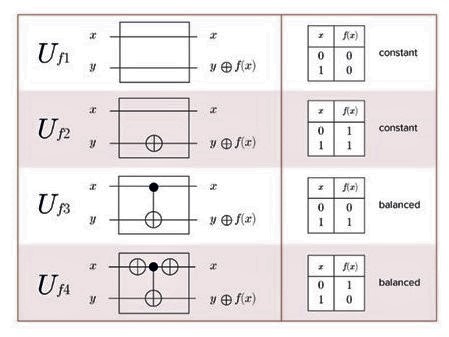}}\caption{The Deutsch-Jozsa Algorithm}\label{fig1_85}
\end{figure}

\begin{figure}[H] \centering{\includegraphics[scale=.6]{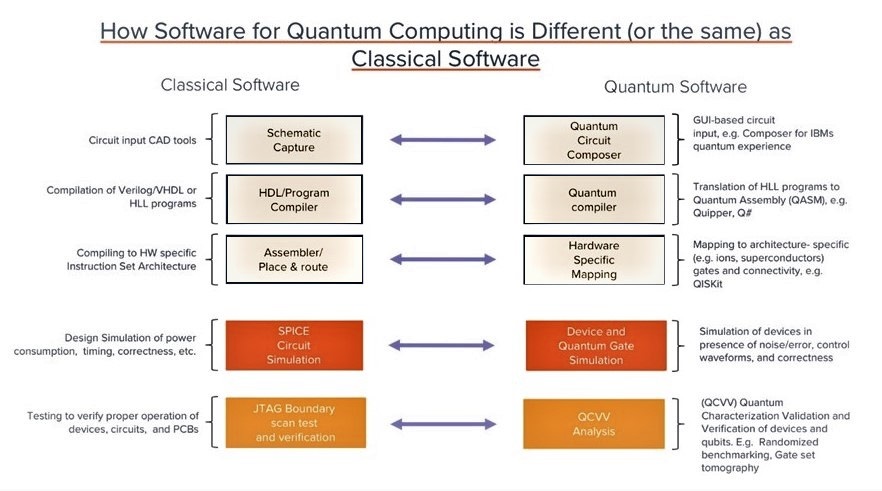}}\caption{Quantum Software}\label{fig1_86}
\end{figure}

\begin{figure}[H] \centering{\includegraphics[scale=.6]{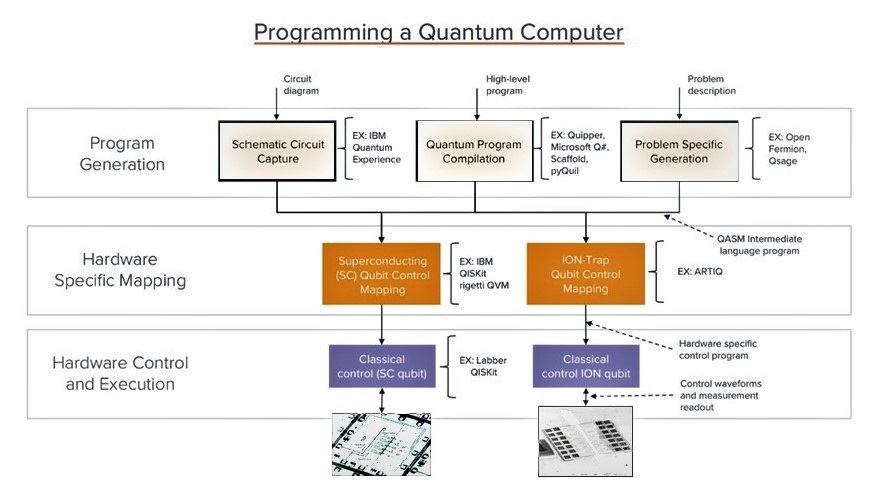}}\caption{Quantum Software}\label{fig1_87}
\end{figure}

\begin{figure}[H] \centering{\includegraphics[scale=.6]{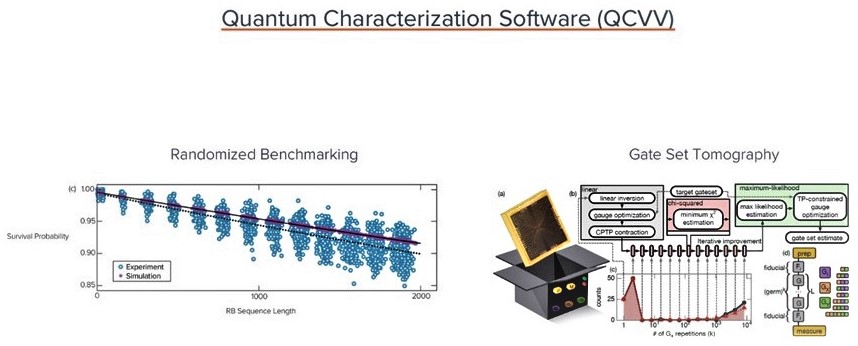}}\caption{Programming Tools}\label{fig1_88}
\end{figure}

\end{enumerate}

\bibliographystyle{IEEEtran}
\bibliography{Bib-Quantum}
\printindex

\part{Quantum-Computation Algorithms}
\chapter{Quantum-Computation Algorithms}

\maketitle

In this research notebook on quantum computation and algorithms for quantum engineers, researchers, and scientists, we will discuss and summarized the application of quantum computing algorithms. The most famous quantum algorithm is for factoring numbers. We begin by describing how this algorithm has an enormous potential impact on the world of public-key cryptography. We began by looking at public-key encryption, and Shor's algorithm, and how Shor's algorithm can efficiently decrypt, for example, the RSA encryption protocol, which is the basis for information security. We discuss the difference between symmetric and asymmetric keys. We look at classical modern cryptography schemes, which are based on one-way functions. How factoring is performed today and why it is so hard on a classical computer. Furthermore, we went into detail about how Shor's algorithm can perform this same task, factoring, or order-finding, more efficiently. We first look at it at an intuitive level, and then went through it in mathematically detail. Next, we describe how quantum key distribution provides a method for regaining secure communications, founded upon the basic principles of quantum physics. Thus, we look at how quantum mechanics can break encryption. However, we are also looking at how quantum mechanics can support the encryption of communication protocols, how we can distribute keys using quantum mechanics. We discuss quantum key distribution (QKD) at some level that distributing these symmetric keys using quantum key distribution is an excellent way to encrypt data. So, we want to have a one-time pad at the disposal to encrypt data, but how do we exchange, or continually refresh that one-time pad? Furthermore, that is the task of QKD. We will discuss two general classes of QKD, one based on single-photon sources and detectors, and the other is based on entangled photons and Bell state measurements. Both rely on quantum concepts, which we will be discussing. For example, entanglement, projective measurement, and the no-cloning theorem are essential, so we are expanding on those concepts in detail. We are also going to discuss the importance of randomness in implementing QKD. Pseudo-random number generators that we have on the computers. They are random, but they are generated by the computers using a seed, and so, they are not truly random, never genuinely random. However, the randomness of projective measurements in quantum mechanics allows us to generate what we believe is true randomness. We will discuss how that is used today and the concept of long-distance quantum repeaters. How can we communicate using QKD over long distances when trying to communicate via photons, and photons can be lost or absorbed on an optical fiber?
Moreover, in a practical sense, we can probably communicate without quantum repeaters over a metropolitan scale, length scale, but if we start to say, communicate over hundreds of miles or thousands of miles, or beyond, then we need to regenerate this quantum information somehow. The no-cloning theorem puts severe restrictions on what we can do. We cannot simplify amplify the quantum information. We need to develop a quantum repeater, and so, we will discuss that as well. Next, we transition to physicist Richard Feynman's original application for quantum computer simulation of the quantum systems and how this is now arising as the key near-term application area for the quantum computation field. Finally, we explore another application area in the optimization, by going in-depth into the quantum search algorithm, and implement the algorithm utilizing a real physical quantum computer by IBM Q Experience.

\section{Introduction to Quantum Information and Cryptography} 
Quantum computing is investigated worldwide, and we have looked at many efforts that are being done in the United States \cite{raymer_us_2019}. Many other efforts are going on worldwide, large efforts in the European Union \cite{riedel_europes_2019}. They have a flagship program that spends about a billion dollars over the next ten years, investigating quantum computing, quantum communication, and quantum sensing \cite{langione_where_nodate}. There are national-level efforts, in Sweden and in Finland, Russia, and Japan \cite{yamamoto_quantum_2019}, Canada \cite{sussman_quantum_2019}, Australia \cite{roberson_charting_2019}. Now many of these are funded by their respective governments, and so different governments have different levels of security. Many of the programs are open, and we can read about them in the public literature, but we should not expect that everything we are seeing is everything that is being done. Many of them are pursuing quantum computing, quantum simulation, universal quantum computation, quantum communication, and quantum sensing. However, in detail, we do not necessarily know what they are doing or what their strategies. In this section, we will manifest quantum mechanical effects that enable quantum computers to process information efficiently \cite{nielsen_quantum_2011,chuang_quantum_2014,preskill_quantum_2019,vazirani_quantum_1997, watrous_theory_2011,vazirani_quantum_2007,aaronson_quantum_2006,shor_quantum_2003,chuang_quantum_2006,aaronson_quantum_2010,harrow_quantum_2018,childs_quantum_2008,cleve_introduction_2007,harrow_why_2012}. we build on the fundamentals studies to probe more deeply the applications of quantum computing \cite{russo_coming_2018,noauthor_next_nodate} and quantum cryptography \cite{barbeau_secure_2019}. We will not worry about practical issues, like noise at this moment, but rather focused on understanding how these algorithms and protocols work \cite{harrow_why_2012}. We will explore the algorithms for cybersecurity, chemistry, and optimization purpose. we want to understand how these algorithms work and their applications. We began by looking at modern cryptography and its use of asymmetric keys, one public, one private, to realize secure public-key cryptography based on one-way functions. We will study in detail one primary example, the RSA cryptosystem \cite{jonsson_pkcs_nodate}. We will discuss that RSA is based on the idea that it is mathematically easy for a classical computer to multiply two large prime numbers, P and Q, and obtain their product N. However. It is a very hard problem for a classical computer to do the inverse, the factor that number N into its constituent primes, P and Q \cite{smolin_pretending_2013}. afterward, we will begin by looking at how Quantum nature can be used to break public-key encryption. We will go into details of Shor's Algorithm \cite{shor_polynomial-time_1997}, which can efficiently defeat certain important encryption protocols if implemented on a quantum computer. For example, RSA encryption, which serves as the basis for information security today, can be compromised by efficiently performing period finding of modular exponentiation, which is related to prime number factorization. We will look at modern cryptography schemes based on one-way functions. We will look at how factoring is performed and why it is so hard on a classical computer. We will then look at how Shor's Algorithm can perform the same task in a much more efficient manner, at an intuitive level, and then deep mathematical level into how his algorithm works. Next, we turn to quantum cryptography and discuss how quantum mechanics could enhance communication security through quantum key distribution. We will discuss single-photon generation and detection, and their use in the BB84 QKD protocol. We then look at entangled photons, Bell States, and Bell State measurements \cite{noauthor_big_nodate,noauthor_big_nodate-1}, and their uses in the Ekert91 protocol, quantum teleportation, and random number generation.Then, we transition to quantum simulation \cite{king_observation_2018}. We first discuss Hamiltonians, describe the dynamics of a quantum system, and the simulation of those dynamics using Trotterization \cite{kivlichan_improved_2019,endo_mitigating_2019}. We then turn to algorithms, such as phase estimation and variational quantum eigensolvers \cite{schuld_circuit-centric_2020,yuan_theory_2019}, that address problem in quantum chemistry \cite{reiher_elucidating_2017,hempel_quantum_2018}, another hybrid quantum-classical algorithm are variational quantum fidelity estimation algorithm\cite{cerezo_variational_2020}, variational quantum factoring (VQF) algorithm \cite{anschuetz_variational_2019}. Next, we will discuss adiabatic quantum computing \cite{farhi_quantum_2000, aharonov_adiabatic_2005}, optimization algorithms, the quantum approximate optimization algorithms (QAOA) \cite{farhi_quantum_2014,venturelli_compiling_2018}. We will also discuss quantum annealing \cite{gabor_assessing_2019,yarkoni_boosting_2019,das_colloquium_2008,vyskocil_embedding_2019,susa_exponential_2018,susa_quantum_2018,kadowaki_quantum_1998,pino_quantum_2018}, and its use of tunneling and relaxation to approach low energy solutions \cite{hegade_experimental_2019}. Although it is not yet known if such processes afford a quantum enhancement, quantum annealers are actively studied in the research, because of optimization problems are ubiquitous. Finally, we will investigate Grover's Search Algorithm \cite{grover_fast_1996}, and how it works within the quantum circuit model, and then do prototype experimental demonstrations and implement it on the IBM quantum experience \cite{kirke_application_2019}. 

\section{Modern Cryptography} 
We generate and communicate tremendous amounts of digital data every day. Nevertheless, how do we protect this data from eavesdroppers? This section will discuss modern cryptography and how it is used to protect data through encryption. In this section, we will look at modern cryptography schemes and their use in information security today. Every day we generate, accumulate, and communicate tremendous amounts of data. Everything from the personal emails, financial transactions, and medical records, to a corporation's intellectual property, business strategy, and online sales, to the daily operation and functioning of the government and military organizations. In each of these cases, we expect, even require, that the information is stored and communicated securely such that it is available to ourselves and the intended recipients, but to nobody else. in general, information is protected using encryption. Encryption takes raw data, such as a message, often referred to as plain text, and translates it or encodes it into an unrecognizable message called a cyphertext. It does this by using an algorithm called a cipher. A cipher takes as inputs the plaintext being encoded and a key, typically a numerical bit string in today's cryptography schemes. The cipher then combines them into an unreadable cyphertext through a prescribed algorithmic protocol based on mathematical operations. Similarly, a decryption key is used with a corresponding decryption protocol to reconstruct the plaintext message. The most common cryptography schemes historically are called symmetric-key cryptography in which the encryption and decryption keys are the same or can be trivially related to one another. Examples include the older data encryption standard, DES, and the more modern advanced encryption standard, AES. The one-time pad that we discussed in section one in the context of quantum key distribution is another example of the symmetric key cryptography scheme. Since symmetric keys are used to both encrypt and decrypt a message, they must remain private. It is not a serious issue, say, a personal computer where we want to encrypt the hard drive, and so, the keys are stored locally. However, it raises a challenge when communicating information. When two parties want to communicate securely, they need to meet in advance to secretly discuss the symmetric keys, not a very practical option, or they need to establish private channels to distribute the key securely. In the future, efficient quantum key distribution protocols may provide a quantum-enhanced security level when transmitting a secret symmetric key. However, today, secure communication links are established using public-key cryptography based on asymmetric keys. Public key cryptography was developed in the 1970s by several groups. Whitfield Diffie and Martin Hellman published the first public-key cryptography scheme, the Diffie-Hellman key exchange protocol, in 1976. Later in 1978, the RSA algorithm was published by Ronald Rivest, Adi Shamir, and Len Adleman. Researchers at the British General Communications Headquarters, GCHQ, had developed similar public-key cryptography schemes in the early 1970s, but only declassified those achievements in 1997. Public key cryptography is based on asymmetric keys. A public key is used for encryption, and a separate independent, private key is used for decryption. The keys are necessarily related to one another mathematically, and the information security relies on it being computationally impractical to determine the private key given the public key. It is possible to use one-way mathematical functions. For example, RSA cryptography \cite{jonsson_pkcs_nodate} is based on the idea that it is mathematically easy for a computer to multiply two large prime numbers, p, and q, and obtain their product, N. However, it is a hard problem for a computer to do the inverse, that is, a factor that number, N, into its constituent prime factors, p, and q\cite{smolin_pretending_2013}. With one-way mathematical functions, ease of use is based on the mathematical simplicity in the forward direction, and security is based on mathematical complexity in the inverse direction. To get an idea for how this works, consider the following cartoon. Bob wants to communicate a message securely to Alice without that message being read by an eavesdropper, Eve. To do this, Bob requests a public encryption key from Alice. Alice then sends a public encryption key but retains her private decryption key. we will represent this by keys with the letter E for encryption, D for decryption, and a notional open lock. During transmission on a public channel, in principle, Eve obtains the open lock and the encryption key. When Bob receives them, he encrypts his plaintext message using the encryption key, creating cyphertext, and effectively locks the padlock. Bob then sends the cyphertext back to Alice on a public channel. Although Eve can obtain the cyphertext in principle, she cannot unlock the padlock since she only has the public encryption key. The security here is based on it is very hard for Eve to figure out the decryption key based on what she has in her possession. Alice, on the other hand, is able to trivially unlock the padlock using her decryption key, thereby converting the cyphertext to plaintext and reading the message. This type of public-key encryption scheme is used ubiquitously today for secure communication. In many cases, it is used to establish a private channel that provides authentication and enables symmetric keys to be exchanged securely. It is because the public key approach based on asymmetric keys is slower to implement than symmetric schemes. Public key encryption is a foundation for information security today. in the next section, we will take a more detailed look at one specific example, RSA cryptography, and how it works. 

The Enigma machine is a legendary cipher machine used by German forces before and during World War II. Enigma is an electro-mechanical rotor cipher machine. A message called "plaintext" written by the operator is encoded by the machine similar in appearance to a typewriter in the following manner: a single character is pseudo-randomly replaced with a different letter according to the wiring of the machine. The subsequent letter is replaced following the same mechanism, but with the rotor shifted to a different position. The shifted rotor position causes the letter to be substituted according to a different electrical pathway and hence character substitution table. Thus, each character of the message is replaced according to a different encoding scheme. The resulting "ciphertext" can subsequently be transmitted via regular public communication channels. It can only be decoded using a machine with the same configuration and capability to reverse the encoding. Finally, a successful decipher trial reveals the true content of the message in the form of a plaintext again.

In 1941, at Bletchley Park, a British team, including Alan Turing, uncovered Enigma's working principles for the Allies. The deciphered information of the intercepted messages provided the Allies with a significant advantage that ultimately became a key element in their victory over the Axis powers.

The Enigma machine is an example of the symmetric encryption scheme, where the same key is used to encode and decode a message. Symmetric encryption schemes are still in use today. However, since World War II, a cryptographic key generation has evolved from electro-mechanical methods to classically constructed digital bit strings serving as encrypting keys. The keys are mathematically challenging to derive and require a significant amount of computational power and time. The computational power and time required to decode are to determine the security of an encoding key.

In 1975, the Data Encryption Standard (DES) was introduced by IBM. The DES algorithm generates a 56-bit symmetric encryption key using a complex mathematical method. Despite this mathematical complexity, DES was known from the beginning to be potentially susceptible to a brute-force key-search attack due to its relatively limited key size. In fact, by the beginning of the 21st century, computational power had sufficiently advanced to the point that one could determine the key in less than a day. For example, Distributed.Net is an organization specializing in the use of idle computer time, utilizing more than 100,000 computers ranging from slow PCs to powerful multiprocessor machines to test 250 billion keys per second and thereby determine a particular 56-bit key in about 23 hours. Given the increasing vulnerability of DES, in 2001, DES was replaced by the Advanced Encryption Standard (AES), a more secure symmetric-key cryptography scheme.

However, there remained a conundrum: how can two parties exchange a symmetric key to initiate secure communication, before they have a secure channel on which to send it? In 1976, a new method for distributing cryptographic keys was introduced by Whitfield Diffie, and Martin Hellman referred to as the Diffie-Hellman key exchange protocol, which heralded the era of public-key cryptography with asymmetric keys. Asymmetric keys comprise a public key used to encrypt messages and a mathematically related (but unique) private key used to decrypt messages. The security of public-key cryptosystems is premised on the use of one-way mathematical functions to create the public and private keys. One-way mathematical functions are straightforward to calculate in the forward direction but prohibitively complex to calculate in the inverse direction. A widely used public-key distribution algorithm was published in 1978 by Ronald Rivest, Adi Shamir, and Len Adleman, referred to as the RSA algorithm. It was later declassified in the late 1990s that asymmetric encryption schemes like RSA were already developed by the British General Communications Headquarters (GCHQ) in the early 1970s. With public-key cryptosystems, one could exchange public keys and establish a secure link, without having symmetric keys in hand at the beginning. Once a secure link was established, one then has the option of continuing to communicate using asymmetric keys, or one can alternatively use this newly established secure link to exchange symmetric keys and use symmetric encryption/decryption protocols.

In practice, public-key encryption techniques are primarily used to establish a secure channel and distribute symmetric keys, which are, in turn, used to encrypt messages. It is because encryption using an asymmetric-key protocol is generally less resource-efficient than using a symmetric-key protocol. For example, a 128-bit symmetric key has $2^{128}$ different possible key combinations. In contrast, a 3000-bit asymmetric key can leverage only the fraction of the $2^{3000}$ possible bit strings that comply with the conditions required by the underlying one-way function. Therefore, whereas the symmetric key of 128 bits and a 3000-bit public key afford a similar level of security, the symmetric-key approach is generally more efficient to implement.

\section{RSA Cryptosystem}

We often communicate securely by encrypting data using cryptographic schemes. However, how do we initiate and establish a secure channel, if we have not yet securely exchanged cryptographic keys? In this section, we will discuss the RSA cryptosystem, a quintessential example of a public-key cryptosystem that leverages asymmetric keys based on one-way mathematical functions to communicate sensitive information. 

In the last section, we introduced modern cryptography schemes and discussed the difference between symmetric and asymmetric keys. we discuss that cryptographic algorithms based on symmetric keys have a conundrum when it comes to key distribution. Namely, how can we securely exchange the keys if we do not yet have them in the possession to encrypt the initial transmission? The answer, as we discuss, is to use public-key cryptography based on asymmetric keys. That is, using a public key for encryption and a different private key for decryption. The keys are related to one another mathematically. Their associated information security relies on it being computationally impractical to determine the private key even when we know the public key. This is achieved through the use of one-way mathematical functions where the ease of use of the cryptosystem is based on the mathematical simplicity of implementing the function in the forward direction, and its security is based on the mathematical complexity of implementing it in the reverse or inverse direction. Thus, now, in this section, we will look at a specific quintessential example of public-key cryptography, the RSA cryptosystem, and discuss the fundamentals of how it works. We will discuss communication in the context of Alice and Bob in the presence of an eavesdropper, Eve. Bob would like to send a secure message to Alice. So, he requests her public key, which we will represent here with the letter E. He uses it to encrypt his plain text message, which we will call M, and generate an unreadable cyphertext called C. Alice retains her private key, D, which she uses to decrypt the cyphertext she receives from Bob, and thereby reconstruct the original plain text message. Although Eve has access to the public key and the cyphertext, she cannot decrypt it to reconstruct the plain text message because she does not have access to the private decryption key. because the two keys are mathematically related to one another through a one-way function, computing the private key from what Eve has in hand is, by design, a very hard problem to solve on a classical computer. The RSA cryptosystem is based on a mathematical concept called modular exponentiation. The basic idea is that one can efficiently define three large positive integers, e, d, and N, such that the following expression is true for all integers to m between 0 and N. Namely, m raised to the power e and then raised again to the power d is equal to itself, modulo N. It is very hard for a computer to calculate the number d even when given the numbers e and N. Now this is a fantastic result because it says that Bob can generate cyphertext by taking a plain text message, converting it to an integer N using an agreed-upon protocol, and then raising N to the power e using the public encryption key. Then Alice can decrypt the cyphertext by raising it to the power d to obtain back the number N, modulo N, from which she can reconstruct the plain text message. The public key comprises the numbers e and N, and the private key comprises the numbers d and N. Now, strictly speaking, the only d needs to be kept private, but Alice will need both d and N to do the decryption. With this background, the RSA cryptosystem can be broken into four steps, key generation, key distribution, encryption, and decryption. Let us take them one at a time. Key generation starts with finding two distinct prime numbers, p, and q, a task which can be performed efficiently on a classical computer using mathematical protocols called primality tests. Prime numbers are those that are divisible by only themselves and unity. importantly, these numbers must be chosen at random. we will discuss later in the section more about how quantum mechanics can, in principle, generate true randomness. For now, though, let us assume that we use a pseudo-random number generator on a classical computer. Next, the prime numbers p and q multiply together to generate the number N. Multiplication, which is also efficient on a classical computer. The number N is used as the modulus for both the public and private keys in the performance of modular exponentiation. The next step is to determine the encryption and decryption keys themselves, e and d. This is done using the periodicity property of modular exponentiation. To see this, let us consider a simple example. we will take the number 3 and its exponents, modulo 10. So, 3 to the ith power, modulo 10, where i is an integer, 1, 2, 3. we see that 3 to the first power modulo 10 is 3. 3 squared modulo 10 is 9. 3 cubed is 27, and so, 3 cubed modulo 10 is 7. The modulo 10 is the remainder when dividing 27 by 10. So, 27 divided by 10 is 2 with a remainder 7. Next, 3 to the fourth power is 81, which modulo 10 is 1. continuing, we see that the numbers repeat, 3, 9, 7, 1. 3, 9, 7, 1. Thus, the periodicity here is 4. we will call this periodicity r, So, r equals 4 in this case. Now there is a relationship between the periodicity r and the prime numbers p and q. Consider a general sequence: x modulo N, x squared modulo N, x cubed modulo N, where N equals p times q and x is a number that is not divisible by p or q. Then the sequence has a period r that evenly divides p minus 1-time q minus 1. Let us see how this plays out in the above example. There, the modulus is N equal 10, which has prime factors p equal 2 and q equals 5. 2 times 5 equals 10. Then p minus 1 is 1. q minus 1 is 4. Thus, the quantity p minus 1 time q minus 1 is 1 time 4, and that equals 4. this evenly divides the period r equals 4. So, that checks. What is important here is that this is the one-way function. If we are given prime numbers p and q, then we can calculate the periodicity r. we will not go into how this is done here. we just note that it can be done efficiently on a classical computer, again, if we know the prime numbers p and q. However, if we are just given N, but we do not know p and q, it is hard to find the periodicity r. typically, the periodicity is very large, unlike the simple example given here. So, a brute force search is not efficient either. So, returning to the key generation, Alice has prime numbers p and q. She used them to determine the period r, and she keeps all these numbers private. She now uses r to determine the keys e and d. First, she chooses an integer e in the range 1 to r that is also co-prime with r. That is, e and r have no common divisor other than 1. Now often, this number is simply chosen so that it is not too big and has mostly 0s to enable efficient encryption. For example, 2 to the 16th power plus 1 equal 65,537, which, in binary, is 1 followed by 15 0s and another 1. This is the public encryption key, e. Then the private decryption key is directly calculated from the expression d times e equals 1 modulo r. In summary, Alice generated the public encryption key e and the private decryption key d from the periodicity r based on randomly chosen prime numbers p and q, which multiply together to yield N. Alice keep private the integers p, q, and r, which she used to calculate her private decryption key d. In contrast, the encryption key e and the number N can be made public. In the next step, key distribution, Alice sends Bob the public encryption key e and the number N. Bob has a plain text message that he wants to transmit to Alice securely. He first turns his message into an integer N using an agreed-upon reversible protocol called padding scheme. He then computes the cyphertext c using modular exponentiation and the public key. c equals N to the eth power modulo N. Bob then sends the ciphertext to Alice. Alice then applies her decryption key d to the cyphertext. Using modular exponentiation once more, c to the d-th power, Alice calculates the result, N modulo N. by reversing the padding scheme, Alice recovers the original plain text message. Again, the security of the RSA cryptosystem is based on the use of a one-way function, in this case, the challenge of period finding in modular exponentiation. As we just discussed, if one knows the prime factors p and q, then one can efficiently find the modular exponentiation period r, and thus generate encryption keys e and d. However, if we are just given the modulus N, there is no known efficient classical algorithm that lets us find its prime factors p and q. Thus, there is no efficient means to determine the period r under these conditions. Thus, even if we reveal the modulus N and the public encryption key e, there are no known efficient means for Eve to calculate the private key d, that is, on a classical computer. In the next section, we will look at Shor's Algorithm and discuss how period finding can be efficiently performed on a quantum computer, thus enabling, in principle, an efficient pathway to breaking the RSA cryptosystem. 

Modern cryptography is a foundation for today's information and economic security, enabling secure communication between individuals, businesses, and governments. There are two basic classes of cryptosystems: symmetric-key and asymmetric-key cryptography. Asymmetric-key cryptography also referred to as public-key cryptography, uses a public key and a private key to encode and decode messages. The two distinct keys are mathematically related via a one-way function, which is computationally efficient to calculate in one direction, but not in the inverse direction.

The workhorse of public-key encryption schemes is the RSA cryptosystem, which can be broken into four steps:

1. key generation (a public key ${\{ e,N\} } $and a private key ${\{ d,N\} }),$

2. public key distribution,

3. message encryption (a message ${m}~ $ $\rightarrow$ a ciphertext ${c})$ and transmission,

4. message decryption.

\textbf{Key Generation and Distribution:}

The encryption and decryption keys are generated using a one-way function related to modular exponentiation. To begin with, the two distinct prime numbers p and q are chosen at random. The key generation then centers on the function.

\begin{equation}\label{eq2_02}
f(x)=a^ x\mod N,
\end{equation}

Where $ a\neq p,q $ is an integer number, the modulus$ N=pq$ is the product of p and q, and ``mod N" means ``modulo N." The term ``modular exponentiation" refers to an exponential $a^{x}$ that is calculated modulo N.

From the number theory of modular exponentiation, it is known that the function $f(x)$ is periodic with period r, and finding the value of r is called order finding. Order finding is a manageable task if one knows p and q, but not if one only knows their product N. This is the one-way function under the security of RSA.

To see this, consider the following. When $x=0$ in the above function, 
\begin{equation}\label{eq2_03}
f(0)=a^0 \mod N = 1.
\end{equation}
  
If f(x) is periodic in r, then it follows that the function f(r), f(2r), f(3r), ... must also equal 1. The period r can be derived using Fermat's little theorem, which uses Euler's totient function
\begin{equation}\label{eq2_04}
r=(p-1)(q-1)
\end{equation}

To find the period r. 

In essence, Euler's totient function gives the number of positive integers smaller than the product $pq=N$, which have no prime factors in common with N, that is, co-prime with p, q, and N and Fermat's little theorem shows that this is the period r. Note that r may be the fundamental period or an integer multiple of the fundamental period. The fundamental period is the least common multiple of $(p-1)$ and $(q-1)$. Both the fundamental period or a multiple of the fundamental period will work.

With p and q are known, the totient can be easily calculated with a classical computer. However, if one is only given N (without knowledge of q and q), then one must factor N to find p and q and thereby find the period r. Factoring is a hard problem for a classical computer\cite{smolin_pretending_2013}.

With the period r in hand, one can generate the public key ${\{ e,N\}}$ and private key ${\{ d,N\}}$. One first chooses a public encryption exponent e that satisfies the following conditions: \\
1.
\begin{equation}\label{eq2_05}
3\leq e \leq (N-1),
\end{equation}

2. e is co-prime with r.

The second condition means that there is no integer (other than 1) that divides both e and r, and so their greatest common divisor (gcd) is 1. This is written as $\gcd (e,r)=1.$

Once a value of e is chosen, then one calculates a private decryption exponent d as the modular multiplicative inverse of e, modulo r:

\begin{equation}\label{eq2_06}
d = e^{-1} \mod r
\end{equation}

Note that when using a modular calculator \cite{noauthor_7-1_nodate}, we will have to use the syntax $A^{B}$mod C and not $\frac{1}{A}$ mod C as the modular multiplicative inverse is only defined for integers. The notation above, while often found, can be misleading in this way.

Intuitively, as we explain in more detail below, the inversion follows that d is used to decrypt a message that is encrypted using e. one can again trivially calculate the private decryption exponent d given the public encryption exponent e, provided one knows the period r.

The steps are summarized in the following table for the specific example of p=17 and q=19.

\begin{table}[H]
\centering
\caption{p=17 and q=19}
\label{tab:2_1:Table 1}
\resizebox{\textwidth}{!}{
\begin{tabular}{|c|c|c|c|}\hline
1 & p,q: & choose two random prime numbers                                      & p= 17, 19                                \\ \hline
2 & N:   & calculate N = pq                                                     & N = 17*19=323                            \\ \hline
3 & r:   & find the period r = ( p-1)(q-1)                                      & r = 16*18=288                            \\ \hline
4 & e:   & choose a public exponent e with greatest common divisor gcd (e,r) =1 & \multicolumn{1}{l|}{e =7, gcd(7,288) =1} \\ \hline
5 & d:   & calculate the private exponent $d = e^{-1} \mod r$                                    & d=247                                    \\ \hline
\end{tabular}}
\end{table}

Having created the keys, Alice sends the public key $\{e, N\}$ to Bob. Importantly, she retains the private key $\{d, N\}$, and does not reveal the values p, q, or r.

\textbf{Message Encryption and Decryption:}

Bob has a message m he wants to send to Alice. Having received the public encryption key \{e, N\}, he encrypts m to create cyphertext c using the public key and modular exponentiation:
\begin{equation}\label{eq2_07}
c = m^e \mod N.
\end{equation}

After the message is sent to Alice, she decrypts it with her private key $\{d,N\}$ using the properties of modular exponentiation,
\begin{equation}\label{eq2_08}
c^d \mod N = (m^e)^d \mod N = m
\end{equation}

provided the original message m is smaller than the modulus N. Eve has access to the public key $\{e, N\}$, but this is no use to her in trying to decrypt the cyphertext, since $ c^e \mod N \neq m.$ 

In summary, the number theory and properties of modular exponentiation underlie the security of the RSA cryptosystem. As we have seen, the challenge of order finding can be related to the problem of factoring. While it is straightforward to multiply two numbers p and q to give the modulus N, the inverse problem is hard on a classical computer. If N is chosen sufficiently large, then a brute-force search to find p and q is also inefficient. Thus, the RSA cryptosystem is believed to be securely provided the prime numbers p and q, and the private exponent d are not made public. However, the encryption exponent e and modulus N can be made public. This type of cryptosystem, where the encryption key is public, and anyone can encrypt a message. In contrast, the decryption key is private, and only the receiver can decrypt the cyphertext, which is the backbone of our modern-day communication systems.

\section{Factoring With a Quantum Computer: Shor's Algorithm} 

The security of the RSA cryptosystem is premised on a one-way function that is difficult to invert using a classical computer. However, using Shor's algorithm, a quantum computer can perform this task much more efficiently. In this section, we will discuss how Shor's algorithm works.

In the last section, we looked at a specific implementation of public-key cryptography in the RSA cryptosystem. The security of RSA is based on the use of a one-way function prime factorization, which is related to the problem of period finding in modular exponentiation. As we discuss, if one picks two prime numbers, p, and q, which multiply together to give a modulus N, then one can efficiently find the modular exponentiation period r, and thus generate a public encryption key e and a private decryption key d. However, if we are just given the modulus N, then we are stuck, because there is no known efficient classical algorithm that lets us find the prime factors p and q, and so, there is no efficient means to determine the period r. Thus, even if we reveal the modulus N and the public encryption key e, there is no known efficient way for an eavesdropper to calculate the private key d on a classical computer. However, if Eve has access to a quantum computer that can run Shor's Algorithm, she can efficiently find the modular exponentiation period r. This is called an order finding. Shor's algorithm finds the order r of a number with respect to the modulus N, where a number is raised to integer powers under modular exponentiation. With a means to efficiently determine, r. Eve can calculate the prime numbers p and q and derive the decryption key d, thus compromising RSA. In this section, we will walk through Shor's Algorithm using the quantum circuit model presented in section 1 \cite{larose_overview_2019}. Then, we will continue with the simple example of modular exponentiation to see the connection between order finding and factorization. Let us begin with the initialization stage. we need two registers of qubits for Shor's Algorithm. Roughly speaking, Register 1 is used to store the results of a period finding protocol, which is implemented using the Quantum Fourier Transform. we will discuss that one in a moment. Register 2 is used to store the values of the modular exponentiation. This is the function that will apply in the computing section. The number of qubits in the first register should generate a state space that is at least large enough to capture the maximum possible period, and do so, at least twice to see it repeat. Let us define N as twice the maximum period. So, we need at least L qubits, such that 2 to the L power is greater than N. Now, in fact, the more qubits we have, the more densely we can sample the solution space and reduce the error in determining the period. Here, we will take 2L qubits for the first register. for simplicity, we will use the same number of qubits in the second register. By choosing 2L qubits, the number of states 2 to the 2L power is greater than N squared. This guarantees that at least N terms are contributing to the probability amplitude when estimating the period. Even as the period r gets exponentially large, approaching N over 2. we prepare the qubits in state 0, which we denote with a superscripted tensor product, 2L. As we did in section 1 for the Deutsch-Jozsa algorithm, we will track the register states by color yellow for Register 1, and green for Register 2. we will track the current position within the algorithm using the vertical line. Next, at the compute stage, we use Hadamard gates to place the Register 1 qubit into an equal superposition state. Again, we use the superscripted tensor product notation to indicate that we apply a Hadamard gate to each of the 2L qubits. This puts each qubit into a superposition state, 0 plus 1. Multiplying out all 2L of these single qubit superposition states results in a single large equal superposition state with 2 to the 2L components. From a component with 2L qubits in state 0 to a component with all qubits in state 1, and with all combinations in between. If we now number these components from 0 to 2 to the 2L minus 1, we can rewrite the superposition state as a sum over x. Each decimal number x corresponds to one of the binary numbers represented by the 2L qubits. Next, the function f is performed. f of x implements modular exponentiation, which outputs a number a to the x power modulo N. we will not present in detail how the modular exponentiation is implemented in a real circuit. There are multiple approaches to doing it based on reversible computing and multiplication algorithms. For the purposes here, we note that this function can be implemented using ancilla qubits and that it is efficient, requiring several gates that scales only polynomials with the number of qubits\cite{bremner_classical_2011}. Now, the number is chosen at random, and it must be co-prime with N. That is, it must have no common factors with N. The resulting output values from the function f are then stored in the Register 2 qubits. Since the values of the power x are taken from the qubits in the first register, implementing f of x makes a conditional connection between the qubits in Register 1 and those in Register 2. Although we will not need to measure the qubits in Register 2, this step sets the stage for quantum interference and quantum parallelism to occur in Register 1 at the next step, which is the Quantum Fourier Transform. The Quantum Fourier Transform is a quantum version of the classical discrete-time Fourier Transform. The classical Fourier Transform takes a vector of numbers, x often a digital sampling of a signal to be analyzed and transforms it into a vector of numbers, y, in the frequency domain. Periodic behavior in the time domain is more easily identified in the transformed frequency domain, which is why we do this. For example, a sine wave that oscillates in time with a specific frequency will transform to a large amplitude spike at plus or minus that frequency in the transform domain. Thus, it is easy to identify the spikes' location and then read off the frequency directly. if we know a signal's frequency, we also know its period, since the frequency is 1 over the period. Now, the Quantum Fourier Transform is essentially the same transformation, which is why we will use it for period finding. It takes a state j and transforms it into a superposition state with specific phase factors. If we start with a superposition state with coefficients Xj, the transformed result is also a superposition state with coefficients Yk, where the probability amplitudes Yk are the classical discrete-time Fourier Transform of the probability amplitudes Xj. due to quantum parallelism and quantum interference, implementing the Quantum Fourier Transform is exponentially faster than implementing the classical fast Fourier Transform algorithm. Now, that sounds great, but there is an important catch related to quantum measurement. As we know, the measurement of a quantum state is probabilistic. Although the output of the Quantum Fourier Transform is a large superposition state with transformed probability amplitudes, we cannot access those amplitudes. the measurement projects out only a single state, and even if we identically prepare the system and measure it many times, we only discuss the magnitude squared of the corresponding coefficients. we lose the phase information. So, we do not have access to the probability amplitudes in the quantum computer, which is quite different from the classical Fourier Transform, where the entire output vector of complex numbers can be directly read. Thus, the Quantum Fourier Transform is most useful in problems with a unique period or phase that can be found with high probability. For now, let us return to Shor's Algorithm. The Quantum Fourier Transform is applied here to the qubits in Register 1, and it imparts specific phase factors that are related to both x and z. That is, they are related to both the coefficients of the qubits on Register 1 and the modular exponentiation stored on Register 2. The phase sampling increment is represented here by omega, and it is the phase 2 $ \pi $ divided by the total number of states 2 to the 2L power. Here we can see that increasing the number of qubits will reduce the step size, and thus the error in estimating the actual period. Essentially, more qubits will more densely sample the phases that represent the period. Lastly, we measure the qubits in Register 1. The projected state that results is a binary value of z, and its decimal value is equal to or an integer multiple of 2 to the 2L-th power divided by r. Thus, it can be used to find the period r using a continued-fraction expansion of z divided by 2 to the 2L. we can discuss more the Quantum Fourier Transform and related topics in the text units following this section. in the next section, we will take a closer look at how the Quantum Fourier Transform mediates the quantum interference terms that constructively converge to estimates for the period r. 

The classical Discrete Time Fourier Transform (DTFT) is a mathematical transformation that takes a vector x of numbers as an input typically a digital sampling of a time-domain signal being analyzed and outputs a vector of numbers y that is its frequency-domain representation:
\begin{equation}\label{eq2_09}
y_ k = \frac{1}{\sqrt {N}}\sum _{j=0}^{N-1} x_ j\exp \left( i\frac{2\pi }{N}jk\right)
\end{equation}

Where N is the total number of points in the vector x, this transform is used ubiquitously in the analysis of signals, because periodic behavior in the time domain is more easily identified in the transformed frequency domain. For example, a sinusoidal signal in time with a frequency f will transform to spikes at $\pm$ f in the frequency domain, find the spikes, and know the frequency of the input signal (or, equivalently, the inverse of its period).

The quantum analog of this transform the Quantum Fourier Transform (QFT) is essentially the same transformation. It takes a single quantum state $\left\vert j\right\rangle$ and transforms it to a superposition of states, each with specific phase factors:
\begin{equation}\label{eq2_10}
\left\vert j\right\rangle \rightarrow \frac{1}{\sqrt {N}}\sum _{k=0}^{N-1}\exp \left( i\frac{2\pi }{N}jk\right) \left\vert k\right\rangle ,
\end{equation}
where N is the total number of states $\left\vert k\right\rangle$; for n qubits, $N=2^ n$. If we start with a superposition of states $\left\vert j\right\rangle$ with coefficients $x_j$, then the QFT yields an output superposition of states $\left\vert k\right\rangle$ with coefficients $y_k$,
\begin{equation}\label{eq2_11}
\sum _{k=0}^{N-1} x_ j \left\vert j\right\rangle \rightarrow \sum _{k=0}^{N-1} y_ k \left\vert k\right\rangle ,
\end{equation}

Where the output amplitudes $y_k$ are the DTFT of the input coefficients $x_j$. Due to quantum parallelism and quantum interference, the QFT can be implemented using only $(n^2)$ gates. This is exponentially faster than the Fast Fourier Transform (FFT) algorithm used to calculate the classical DTFT, which requires $O(n2^n)$ gates.

This sounds terrific, and it is, but there is an important caveat. In general, the QFT cannot be used as a straightforward replacement for the DTFT for real-world data processing applications, and the reason is related to quantum measurement.

In the case of the classical DTFT, even for a real input vector x, the transformed output vector y is complex. We can classically access all of the complex elements of y.

In contrast, in the case of the QFT, a measurement of the output superposition state will project the system into only one of those states with a probability that is the magnitude squared of the corresponding coefficient. We generally have no access to the complex probability amplitudes $y_k$. Even if we identically prepare the system and measure it many times, we can only reconstruct the magnitude squared of those coefficients (the magnitude squared of the Fourier spectrum); we have no access to the phase information, and it will take an exponentially large number of measurements to reconstruct the probability for all $2^n$ states.

Nonetheless, the QFT is a critical element in quantum computing protocols, including phase estimation and period finding. For these types of applications, the desired output from the QFT is highly localized into a few states with high probability. These states encode the desired information, for example, the period r of modular exponentiation used in Shor's algorithm.

\section{Factoring With a Quantum Computer: How Shor's Algorithm Works} 

In the last section, we discuss how Shor's Algorithm is implemented within the circuit model for quantum computing. In this section, we will discuss one more critical step in the algorithm, the application of the quantum Fourier transform. 

In the last section, we presented Shor's Algorithm in the context of the quantum circuit model. we used two registers of qubits. Register 1 was used to store the results of period finding, and Register 2 was used to store the modular exponentiation function's values. we then applied the Quantum Fourier Transform to Register 1 and measured the output. The projected state resulted in a binary value of z, and its decimal value was an integer multiple of 2 to the 2L divided by r. This, in turn, can be used to find the period r using a continued-fraction expansion. In this section, we will take a closer look at how the Quantum Fourier Transform and the resulting quantum interference lead us to this result. To see how this works, we need to rewrite the sum over x as a combination of two sums. we can use the example of an equal 3 and N equal 10, which we discuss has a period r equal 4. we want to sum over all x from 0 to 2 to the 2L power minus 1. The corresponding function f of x, the modular exponentiation is calculated for several values of x, and the period is, indeed, 4. Let us call the values of x in the first period x0, and the resulting values of the function f we will call w. we see that w has only four unique values and repeats for values of x equal to x0 plus integer multiples of the period r. That is, x0 plus y times r, where we use y as the integer. for a given x0, the function f of x0 plus y r is the same value, as indicated in orange for x0 equals 0. we can see the same behavior for x0 equals 1 in green, x0 equals 2 in blue, and x0 equals 3 in gray. we see that in doing this for the four values of x0, we have covered all values of x. Thus, we can rewrite the sum over x as a sum over the four values of x0 and a sum over the integer y, which runs from 0 to 2 to the Lth power divided by r. Together, these two sums run over all x values from 0 to 2 to the L minus 1. Writing in this way enables us to see how quantum interference occurs with these phase vectors. we first replace omega with its explicit value, e to the power i times 2 $ \pi/2 $ over 2 to the 2L. there are two-phase factors here. The first is a fixed phase offset independent of the sum over y. It will rotate the phase of any net result that survives the sum. The second is the quantum interference phase related to the sum over y itself. This sum is performed for every value of x0 and z, and it is essentially a test to see if the value of z is related to the period r when starting from a value x0. The sum over y adds together vectors with specific phases related to x0, z, r, and y. In most cases, these phases are not multiples of 2 $ \pi $, so the corresponding vectors, according to their phases, end up pointing in various directions, uniformly distributed around the unit circle. Since they point in all directions, adding these vectors gives a net result of 0, which is another example of destructive quantum interference. Now, for a few specific cases where z is related to r through x0, the phases are a multiple of 2 $ \pi $. When this is true, the corresponding vectors will always point in the same direction to the right and thus add together constructively. There are 2 to the 2L divided by r unit vectors, and so, the sum is 2 to the 2L divided by r. Mathematically, the sum over y is a geometric sum. we encourage us to calculate this and find the condition which yields non-0 values. we will find the resulting condition is that 2 $ \pi/2 $ to the 2L times z times r is 2 $ \pi $ times an integer, d. Thus, it occurs when z equals d times 2 to the 2 L divided by r. If we plot the probability at the output of the Quantum Fourier Transform, we find a series of spikes that occur when z equals d times 2 to the 2L over r. The probability of measuring any one of these spikes is 1 over r since there are r spikes between 0 and 2 to the 2L over r. This is for a specific value of x0. Whichever one of these spikes we measure, the projected state z with value d times 2 to the 2L over r can be used to find the period r using a continued-fraction expansion procedure. we will not discuss here in detail how that is done. we just note that it is a known algorithm and it can be effectively performed on a classical computer. Now, we just discuss how Shor's Algorithm can find the period r, but how is this connected to the factoring problem in RSA? Again, we will take the example with N equal 10. we want to find the prime factors p and q that, multiplied together, equal 10. To do this, we randomly pick a number, a that is smaller than N and prime. That is, their Greatest Common Divisor-GCD is unity. we will again choose a equal 3. we then use Shor's Algorithm to find the period of a to the x modulo N. Shor's Algorithm, with classical post-processing, will efficiently give us the period r equal 4. Before proceeding, we need to do a couple of checks. First, r must be an even number. second, we need to ensure that a to the r over 2 power plus 1 does not equal 0 modulo N. If we fail either of these tests, we need to go back to step 1 and pick another value of a. However, in the case here, we are fine. For a equal 3 and r equal 4, both conditions check, and we can proceed. we know that a to the r equals 1 modulo N, as this is a property of modular exponentiation. we can rewrite this expression as a to the r-th power minus 1 equals k times N for an integer, k. we then replace N with the unknown prime factors p and q, and we factor the left-hand side to yield a to the r over 2 minus 1 time a to the r over 2 plus 1. Then, p and q can be found from the greatest common divisor of a to the power r over 2 plus or minus 1, and N. Let us take them one at a time. a to the power r over 2 minus 1 is 8. Thus, the greatest common divisor of 8 and 10 is 2, so, p equals 2. Then a to the r over 2 plus 1 is 10, and so, the greatest common divisor of 10 and 10 that is not trivial is 5, so q will equal 5. Therefore, p equals 2, q equals 5, both are prime numbers, and p times q equals 10. Now, this is a very simple example. However, it illustrates how Shor's Algorithm and period finding can be used to calculate prime numbers p and q, and thus calculate the private decryption key and compromise the RSA cryptosystem. More details and background information related to these topics can be found in the following text units.

\section{Quantum Fourier Transform} 
The Quantum Fourier Transform (QFT) is a critical operation used in quantum computing protocols, including period finding and phase estimation. For example, in the period finding protocol, the input state encodes a sequence with an underlying periodic structure, and, following the application of the QFT, the output encodes the period of the sequence.

The QFT is a unitary linear transformation implemented with a series of quantum gates. It maps an n-qubit quantum state $\left\vert j\right\rangle$ as

\begin{equation}\label{eq2_12}
\left\vert j\right\rangle \rightarrow \frac{1}{\sqrt {N}}\sum _{k=0}^{N-1}\exp \left( i\frac{2\pi }{N}jk\right) \left\vert k\right\rangle ,
\end{equation}

N is a constant defined by the number of qubits $N=2^{n}$, and j and k go from 0 to N-1.

Note that the input quantum state can be written as $\left\vert j\right\rangle$ where j is a decimal number spanning 0 to N-1 (as above), or, equivalently, it can be written as a binary number $\left\vert j\right\rangle =\left\vert j_{1}j_{2}...j_{n-1}j_{n}\right\rangle$, where each $j_{i}$ take the value 0 or 1. As short-hand notation, this is sometimes written as the decimal value, e.g. $\left\vert 10\right\rangle = \left\vert 2\right\rangle$ or $\left\vert 11\right\rangle = \left\vert 3\right\rangle.$

In addition to knowing the mapping of the Quantum Fourier Transform, we also need to know how to implement it. For a single-qubit, the mapping of the QFT is:

\begin{equation}\label{eq2_13}
\displaystyle \frac{1}{\sqrt{2^{1}}}\sum _{k=0}^{2^{1}-1}\exp \left( i\frac{2\pi }{2^{1}}0\times k\right) \left\vert k\right\rangle =\frac{\left\vert 0\right\rangle +\left\vert 1\right\rangle }{\sqrt{2}},
\end{equation}

\begin{equation}\label{eq2_14}
\displaystyle \frac{1}{\sqrt{2^{1}}}\sum _{k=0}^{2^{1}-1}\exp \left( i\frac{2\pi }{2^{1}}1\times k\right) \left\vert k\right\rangle =\frac{\left\vert 0\right\rangle -\left\vert 1\right\rangle }{\sqrt{2}},
\end{equation}

This transformation corresponds to a Hadamard gate,
\begin{equation}\label{eq2_15}
H=\frac{1}{\sqrt {2}}\left( \begin{array}{cc} 1 & 1 \\ 1 & -1 \end{array} \right) .
\end{equation}
The figure below shows a graphical representation of the Quantum Fourier Transform for a single-qubit system (the notation $0.j_1$ in the exponent is a binary representation that will be introduced below).

\begin{figure}[H] \centering{\includegraphics[scale=.7]{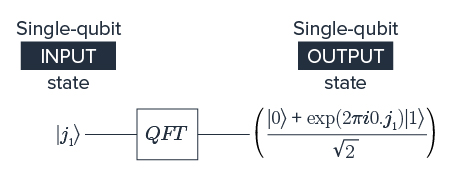}}\caption{Quantum Fourier Transformation}\label{fig2_2}
\end{figure}

To understand how the Quantum Fourier Transformation works, Let us analyze the two-qubit case, n=2. The state $\left\vert j\right\rangle $ transforms to:

\begin{equation}\label{eq2_16}
\displaystyle \left\vert j\right\rangle\displaystyle \rightarrow\displaystyle \frac{1}{\sqrt {2^{2}}}\sum _{k=0}^{2^{2}-1}\exp \left( i\frac{2\pi }{2^{2}}jk\right) \left\vert k\right\rangle ,
\end{equation}

\begin{equation}\label{eq2_17}
\begin{split}
\displaystyle =\displaystyle \frac{1}{\sqrt {2^{2}}}\left( \exp \left( i\frac{2\pi }{2^{2}}j\times 0\right) \left\vert 0\right\rangle +\exp \left( i\frac{2\pi }{2^{2}}j\times 1\right) \left\vert 1\right\rangle \right.\\
\displaystyle \left. +\exp \left( i\frac{2\pi }{2^{2}}j\times 2\right) \left\vert 2\right\rangle +\exp \left( i\frac{2\pi }{2^{2}}j\times 3\right) \left\vert 3\right\rangle \right)
\end{split}
\end{equation}

When the system is initialized in the ground state, j=0, the QFT leaves each qubit in the equal superposition state
\begin{equation}\label{eq2_17_1}
\frac{\left\vert 0\right\rangle +\left\vert 1\right\rangle }{\sqrt {2}}
\end{equation}

This is true for any number of qubits n. The QFT maps the ground state of n qubits into n equal superposition single-qubit states,

\begin{equation}\label{eq2_18}
\lvert 00...0\rangle \to \frac{\left\vert 0\right\rangle +\left\vert 1\right\rangle }{\sqrt{2}}\otimes\frac{\left\vert 0\right\rangle +\left\vert 1\right\rangle }{\sqrt{2}}\otimes...\otimes\frac{\left\vert 0\right\rangle +\left\vert 1\right\rangle }{\sqrt{2}}.
\end{equation}

The table below shows the output state corresponding to each of the values of j.

\begin{table}[H]
    \centering
    \caption{output state}
    \label{tab:2_1:Table 7}
\begin{tabular}{|c|c|c|} 
        \hline
        $j=0$ & $\left\vert 0\right\rangle $ & $\frac{1}{\sqrt {2^{2}}}\left( \left\vert 0\right\rangle +\left\vert 1\right\rangle +\left\vert 2\right\rangle +\left\vert 3\right\rangle \right) $ \\ \hline
        $j=1$ & $\left\vert 1\right\rangle $ &  $ \frac{1}{\sqrt {2^{2}}}\left( \left\vert 0\right\rangle +\exp \left( \frac{1}{2}i\pi \right) \left\vert 1\right\rangle +\exp \left( i\pi \right) \left\vert 2\right\rangle +\exp \left( \frac{3}{2}i\pi \right) \left\vert 3\right\rangle \right)  $\\ \hline
        $j=2$ & $\left\vert 2\right\rangle $ &  $ \frac{1}{\sqrt {2^{2}}}\left( \left\vert 0\right\rangle +\exp \left( i\pi \right) \left\vert 1\right\rangle +\left\vert 2\right\rangle +\exp \left( i\pi \right) \left\vert 3\right\rangle \right)  $ \\\hline
        $j=3$ & $\left\vert 3\right\rangle $ &  $ \frac{1}{\sqrt {2^{2}}}\left( \left\vert 0\right\rangle +\exp \left( \frac{3}{2}i\pi \right) \left\vert 1\right\rangle +\exp \left( i\pi \right) \left\vert 2\right\rangle +\exp \left( \frac{9}{2}i\pi \right) \left\vert 3\right\rangle \right)  $\\
        \hline
    \end{tabular}
\end{table}

When the system consists of more than one qubit and is not initialized in its ground state, the implementation of the QFT is not as simple as implementing Hadamards. To gain insight, Let us write the input and output states as the product of single-qubit states. The input state for an n qubit system is

\begin{equation}\label{eq2_19}
\begin{split}
\displaystyle \left\vert j\right\rangle    \displaystyle & =    \displaystyle \left\vert j_{1}j_{2}...j_{n-1}j_{n}\right\rangle ,\\
\displaystyle & =    \displaystyle \left\vert j_{1}\right\rangle \otimes \left\vert j_{2}\right\rangle \otimes ...\otimes \left\vert j_{n-1}\right\rangle \otimes \left\vert j_{n}\right\rangle
\end{split}
\end{equation}

and the output state is

\begin{equation}\label{eq2_21}
\begin{split}
\displaystyle \left\vert k\right\rangle    \displaystyle & =    \displaystyle \left\vert k_{1}k_{2}...k_{n-1}k_{n}\right\rangle ,\\
\displaystyle & =    \displaystyle \left\vert k_{1}\right\rangle \otimes \left\vert k_{2}\right\rangle \otimes ...\otimes \left\vert k_{n-1}\right\rangle \otimes \left\vert k_{n}\right\rangle
\end{split}
\end{equation}

Next, we should write the indexes j and k in binary using the following procedure:
\begin{equation}\label{eq2_23}
k=k_{1}2^{n-1}+k_{2}2^{n-2}+...+k_{n}2^{0}=\sum _{i=1}^{n}k_{i}2^{n-i},
\end{equation}
and

\begin{equation}\label{eq2_24}
\frac{j}{2^{n}}=j_{1}2^{-1}+j_{2}2^{-2}+...+j_{n}2^{-n}\equiv 0.j_{1}j_{2}...j_{n}.
\end{equation}
Note that the notation $0.j_{1}j_{2}...j_{n}$ is new, and the treatment of the entries is position dependent: that is, the position of the variable $j_n$ to the right of the ``0.'' matters, and not it is subscript. For example, the following equivalences follow:
\begin{equation}\label{eq2_25}
0.j_1 = j_1 2^{-1}
\end{equation}

\begin{equation}\label{eq2_26}
0.j_2 = j_2 2^{-1}
\end{equation}

\begin{equation}\label{eq2_27}
0.j_1j_2 = j_1 2^{-1} + j_2 2^{-2}
\end{equation}

\begin{equation}\label{eq2_28}
0.j_2j_4 = j_2 2^{-1} + j_4 2^{-2}
\end{equation}
and so on.

Replacing the terms   $ k=\sum _{i=1}^{n}k_{i}2^{n-i} $  and $ \frac{j}{2^{n}}=0.j_{1}j_{2}...j_{n}  $ in equation (1), the  n -qubit output state can be written as the product of  n  single-qubit states

\begin{equation}\label{eq2_29}
\displaystyle \nonumber         \displaystyle \left\vert j_{1}\right\rangle \otimes \left\vert j_{2}\right\rangle \otimes ...\otimes \left\vert j_{n-1}\right\rangle \otimes \left\vert j_{n}\right\rangle
\end{equation}

\begin{equation}\label{eq2_30}
\displaystyle \nonumber    \displaystyle \rightarrow    \displaystyle \left( \frac{\left\vert 0\right\rangle +\exp \left( 2\pi i0.j_{n}\right) \left\vert 1\right\rangle }{\sqrt{2}}\right) \otimes \left( \frac{\left\vert 0\right\rangle +\exp \left( 2\pi i0.j_{n-1}j_{n}\right) \left\vert 1\right\rangle }{\sqrt{2}}\right)
\end{equation}

\begin{equation}\label{eq2_31}
\displaystyle \otimes ...\otimes \left( \frac{\left\vert 0\right\rangle +\exp \left( 2\pi i0.j_{2}...j_{n}\right) \left\vert 1\right\rangle }{\sqrt{2}}\right) \otimes \left( \frac{\left\vert 0\right\rangle +\exp \left( 2\pi i0.j_{1}...j_{n}\right) \left\vert 1\right\rangle }{\sqrt{2}}\right) .
\end{equation}

The figure below shows a graphical representation of the Quantum Fourier Transform for an n-qubit system.

\begin{figure}[H] \centering{\includegraphics[scale=1]{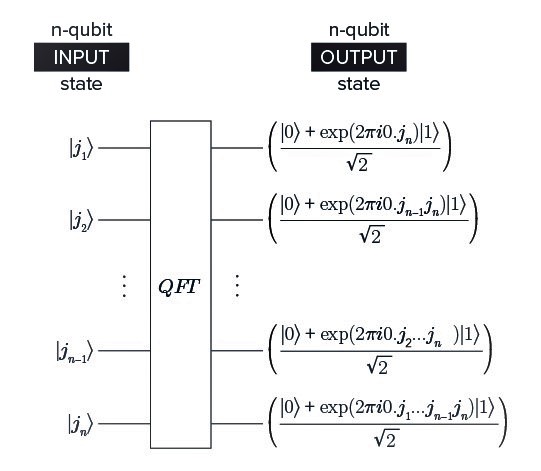}}\caption{Quantum Fourier Transform}\label{fig2_3}
\end{figure}

To understand how this new representation works, Let us go back to the n=2 example. The table below shows the output state corresponding to each of the values of $j_{1}$ and $j_{2}$. Note, that these states are all written in the single-qubit basis $\left\{ \left\vert 00\right\rangle ,\left\vert 01\right\rangle ,\left\vert 10\right\rangle ,\left\vert 11\right\rangle \right\}.$

\begin{table}[H]
    \centering
    \caption{output state}
    \label{tab:2_1:Table 6}
    \resizebox{\textwidth}{!}{
    \begin{tabular}{|c|c|c|c|} \hline
         & binary to decimal & quantum states & \\ \hline
        $j_{1}=0,j_{2}=0$ & $00 \equiv \vert 0\rangle$ & $\left\vert 00\right\rangle =\left\vert 0\right\rangle \otimes \left\vert 0\right\rangle$ & $\frac{\left\vert 0\right\rangle +\left\vert 1\right\rangle }{\sqrt {2}}\otimes \frac{\left\vert 0\right\rangle +\left\vert 1\right\rangle }{\sqrt {2}}$\\ \hline
        $j_{1}=0,j_{2}=1$ & $01 \equiv \vert 1\rangle$ & $\left\vert 01\right\rangle =\left\vert 0\right\rangle \otimes \left\vert 1\right\rangle$ & $\left( \frac{\left\vert 0\right\rangle +\exp \left( \pi i\right) \left\vert 1\right\rangle }{\sqrt {2}}\right) \otimes \left( \frac{\left\vert 0\right\rangle +\exp \left( \pi i\frac{1}{2}\right) \left\vert 1\right\rangle }{\sqrt {2}}\right)$\\\hline
        $j_{1}=1,j_{2}=0$ & $10 \equiv \vert 2\rangle$ & $\left\vert 10\right\rangle =\left\vert 1\right\rangle \otimes \left\vert 0\right\rangle$ & $\left( \frac{\left\vert 0\right\rangle +\left\vert 1\right\rangle }{\sqrt {2}}\right) \otimes \left( \frac{\left\vert 0\right\rangle +\exp \left( \pi i\right) \left\vert 1\right\rangle }{\sqrt {2}}\right)$\\ \hline
        $j_{1}=1,j_{2}=1$ & $11 \equiv \vert 3\rangle$ & $\left\vert 11\right\rangle =\left\vert 1\right\rangle \otimes \left\vert 1\right\rangle$ & $\left( \frac{\left\vert 0\right\rangle +\exp \left( \pi i\right) \left\vert 1\right\rangle }{\sqrt {2}}\right) \otimes \left( \frac{\left\vert 0\right\rangle +\exp \left( \pi i\frac{3}{2}\right) \left\vert \right\rangle }{\sqrt {2}}\right)$\\  \hline
\end{tabular}}
\end{table}

(The binary to decimal column illustrates the connection to the table above, where the decimal short-hand notation was used.)

The single-qubit output states are transformed as:
\begin{equation}\label{eq2_32}
\displaystyle \left\vert 0\right\rangle    \displaystyle \rightarrow    \displaystyle \frac{\left\vert 0\right\rangle +\left\vert 1\right\rangle }{\sqrt{2}},
\end{equation}

\begin{equation}\label{eq2_33}
\displaystyle \left\vert 1\right\rangle    \displaystyle \rightarrow    \displaystyle \frac{\left\vert 0\right\rangle +\exp \left( i\phi _{m}\right) \left\vert 1\right\rangle }{\sqrt{2}}
\end{equation}

where $\phi _{m}$ is a phase between 0 and $2\pi$. If the input state is $\left\vert 0\right\rangle,$ then the output state is always an equal superposition state with zero relative phase. If the input state is $\left\vert 1\right\rangle,$ then the output is an equal superposition state, but with a phase shift $\exp \left( i\phi _{m}\right)$ on state $\left\vert 1\right\rangle.$ Specifically, for n qubits, the phase of the first output qubit is $\phi _{1}=2\pi 0.j_{n},$ the second is $\phi _{2}=2\pi 0.j_{n-1}j_{n},$ the third is $\phi _{3}=2\pi 0.j_{n-2}j_{n-1}j_{n},$ and so on until the last qubit which phase is $\phi _{n}=2\pi 0.j_{1}...j_{n}.$

There are two takeaways:\\
1. When the input state is $\left\vert 0\right\rangle,$ the output state is an equal superposition state with zero relative phases.\\
2. When the input state is$ \left\vert 1\right\rangle,$ the output state is an equal superposition state that undergoes a relative phase rotation.\\

From these two points, it is clear that the implementation of the QFT is related to a controlled operation that modifies a qubit state, as shown in the table below.

From these two points, it is clear that the implementation of the QFT is related to a controlled operation that modifies a qubit state as shown in the table below.
\begin{table}[H]
    \centering
    \caption{QFT}
    \label{tab:2_1:Table 8}
    \begin{tabular}{|c|c|ccc}
        \cline{1-2}
        \multicolumn{2}{|c|}{Controlled Operation} &  &  &  \\ \cline{1-2}
        Input state         & Output state         &  &  &  \\ \cline{1-2}
        $\left\vert 0\right\rangle$                 &   $\frac{\left\vert 0\right\rangle +\left\vert 1\right\rangle }{\sqrt {2}}$                   &  &  &  \\ \cline{1-2}
        $\left\vert 1\right\rangle$                 &        $\frac{\left\vert 0\right\rangle +R_{k}\left\vert 1\right\rangle }{\sqrt {2}}$              &  &  &  \\ \cline{1-2}
    \end{tabular}
\end{table}

The figure below shows the circuit implementation of the Quantum Fourier Transform in terms of multiple Hadamard gates and controlled $R_{k}$ gates. The phase rotation $R_{k}$ is defined as

\begin{equation}\label{eq2_34}
\displaystyle R_{k}\left\vert 0\right\rangle\displaystyle =\displaystyle \left\vert 0\right\rangle ,
\end{equation}

\begin{equation}\label{eq2_35}
\displaystyle R_{k}\left\vert 1\right\rangle\displaystyle =\displaystyle \exp \left(i\frac{2\pi }{2^{k}}\right) \left\vert 1\right\rangle  .
\end{equation}

\begin{figure}[H] \centering{\includegraphics[scale=.5]{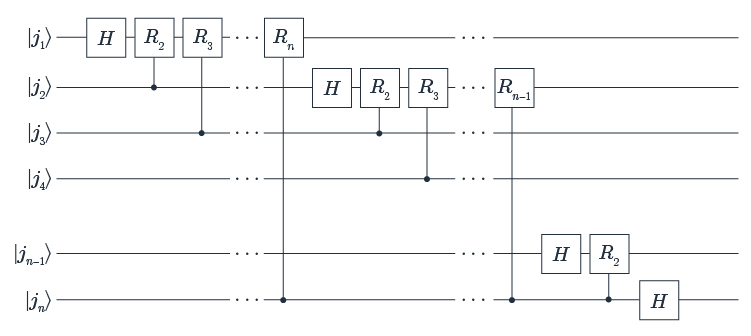}}\caption{Quantum Fourier Transform}\label{fig2_4}
\end{figure}

Let us analyze how the system evolves under this implementation.\\

1. A Hadamard gate is applied on the first qubit

\begin{equation}\label{eq2_36}
\lvert j_1 \rangle \rightarrow \frac{\lvert 0 \rangle + \exp (2\pi i0.j_1)\lvert 1 \rangle }{\sqrt{2}},
\end{equation}

and this puts the system in the state

\begin{equation}\label{eq2_37}
\frac{\left\vert 0\right\rangle +\exp \left( 2\pi i0.j_{1}\right) \left\vert 1\right\rangle }{\sqrt {2}}\otimes \left\vert j_{2}\right\rangle \otimes \left\vert j_{3}\right\rangle \otimes ...\otimes \left\vert j_{n}\right\rangle .
\end{equation}

2. A controlled phase $R_{2}$-gate is applied with the second qubit as control and the first as target,

\begin{equation}\label{eq2_38}
\frac{\left\vert 0 \right \rangle + \exp \left(2\pi i0.j_1 j_2\right) \left\lvert 1 \right\rangle }{\sqrt{2}}\otimes \left\lvert j_2 \right\rangle \otimes \left\lvert j_3 \right \rangle \otimes ... \otimes \left\vert j_n \right\rangle .
\end{equation}

3. A controlled phase $R_{3}$-gate is applied with the third qubit as control and the first as target,

\begin{equation}\label{eq2_39}
\frac{\left\vert 0\right\rangle + \exp \left(2\pi i0.j_1 j_2 j_3 \right) \left \vert 1 \right \rangle}{\sqrt{2}} \otimes \left\vert j_2 \right \rangle \otimes \left\vert j_3 \right\rangle \otimes ... \otimes \left\vert j_n \right\rangle.
\end{equation}

4. This continues until a controlled phase $R_{n}$-the gate is applied with the last qubit as control and the first as a target, leaving the system in the state

\begin{equation}\label{eq2_40}
\frac{\left\vert 0 \right \rangle + \exp \left(2\pi i0.j_1 j_2 j_3 ... j_n\right) \left\vert 1 \right\rangle}{\sqrt{2}}\otimes \left\vert j_2 \right\rangle \otimes \left\vert j_3 \right\rangle \otimes ... \otimes \left\vert j_n \right\rangle.
\end{equation}

5. A Hadamard gate is applied on the second qubit,
\begin{equation}\label{eq2_41}
\left (\frac{\left\vert0\right\rangle+\exp \left (2 \pi i0.j_1j_2j_3...j_n\right) \left\vert 1 \right\rangle}{\sqrt{2}}\right)\otimes \left(\frac{\left\vert 0 \right\rangle + \exp \left(2 \pi i0.j_2\right) \left\vert 1 \right\rangle}{\sqrt{2}}\right) \otimes \left\vert j_3 \right\rangle \otimes ... \otimes \left\vert j_n\right\rangle.
\end{equation}
6. A controlled phase $R_{2}$-gate is applied with the third qubit as control and the second as target,
\begin{equation}\label{eq2_42}
\left(\frac{\left\vert 0 \right\rangle + \exp \left (2 \pi i0.j_1 j_2 j_3 ... j_n\right) \left\vert 1 \right\rangle}{\sqrt{2}}\right) \otimes \left(\frac{\left\vert 0 \right\rangle + \exp \left(2\pi i0.j_2 j_3 \right) \left\vert 1 \right\rangle}{\sqrt{2}}\right) \otimes \left\vert j_3\right\rangle\otimes ... \otimes \left\vert j_n \right\rangle.
\end{equation}
7. This continues until a controlled phase $R_{n-1}$-gate is applied with the last qubit as control and the second as target, leaving the system in the state
\begin{equation}\label{eq2_43}
\frac{\left\vert 0 \right\rangle + \exp \left(2\pi i0.j_1 j_2 j_3 ... j_n \right) \left\vert 1 \right\rangle}{\sqrt{2}}\otimes\frac{\left\vert 0\right\rangle + \exp \left(2 \pi i0.j_2 j_3 ... j_n\right) \left\vert 1 \right\rangle }{\sqrt{2}} \otimes \left\vert j_3 \right\rangle \otimes ... \otimes \left\vert j_n\right\rangle.
\end{equation}
8. This same process is done on each qubit until the last qubit, on which we can only apply a Hadamard gate. This implementation transforms as
\begin{equation}\label{eq2_44}
\displaystyle \nonumber         \displaystyle \left\vert j_{1}\right\rangle \otimes \left\vert j_{2}\right\rangle \otimes ...\otimes \left\vert j_{n-1}\right\rangle \otimes \left\vert j_{n}\right\rangle 
\end{equation}
\begin{equation}\label{eq2_45}
\displaystyle \nonumber    \displaystyle \rightarrow    \displaystyle \left( \frac{\left\vert 0\right\rangle +\exp \left( 2\pi i0.j_{1}...j_{n}\right) \left\vert 1\right\rangle }{\sqrt{2}}\right) \otimes \left( \frac{\left\vert 0\right\rangle +\exp \left( 2\pi i0.j_{2}...j_{n}\right) \left\vert 1\right\rangle }{\sqrt{2}}\right) 
\end{equation}
\begin{equation}\label{eq2_46}
\displaystyle \otimes ...\otimes \left( \frac{\left\vert 0\right\rangle +\exp \left( 2\pi i0.j_{n-1}j_{n}\right) \left\vert 1\right\rangle }{\sqrt{2}}\right) \otimes \left( \frac{\left\vert 0\right\rangle +\exp \left( 2\pi i0.j_{n}\right) \left\vert 1\right\rangle }{\sqrt{2}}\right) .
\end{equation}

However, the output states in equations (\ref{eq2_31}) and (\ref{eq2_46}) are not equal. The output state of the first qubit in (\ref{eq2_31}) is equivalent to the output state of the last qubit in (\ref{eq2_46}). The output state of the second qubit in (\ref{eq2_31}) is equivalent to the output state of the penultimate qubit in (\ref{eq2_46}). The state of qubit n in (\ref{eq2_46}) should be the state of the first qubit in (\ref{eq2_31}), the state of qubit n-1 in (\ref{eq2_46}) should be the state of the second qubit (\ref{eq2_31}), and so on. To return the qubit states to their original ordering, we need to apply swap-gates, which are generally used to exchange qubit states. Thus, the difference between the output states in equations (\ref{eq2_31}) and (\ref{eq2_46}) can be straightforwardly undone by applying a swap gate between qubits ``1'' and ``n'', a swap gate between qubits ``2'' and ``n-1'', a swap gate between qubits ``3'' and ``n-2'', and so on\cite{paler_influence_2019}.

\begin{figure}[H] \centering{\includegraphics[scale=.9]{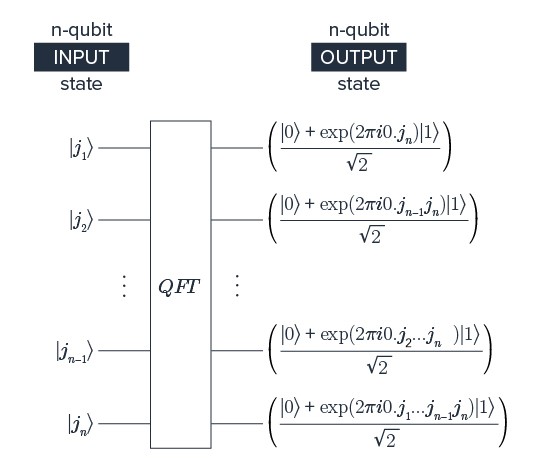}}\caption{RSA Cryptosystem, Factoring, and Shor’s Algorithm Quantum Fourier Transform}\label{fig2_5}
\end{figure}

\section{Period Finding Algorithm} 

The quantum part of Shor's algorithm is called ``period finding.'' In this section, we will discuss this quantum subroutine.

The period finding algorithm determines the period r of a periodic function $f\left( x\right) =f\left( x+r\right)$ from modular exponentiation. For example, to find the period of $f\left( x\right) =2^{x}\mod {15}$, we could test the function for different values of x from its smallest to its greatest value until we find the first x for which $f\left( x\right) =2^{x}\mod {15}=1\mod {15},$ \\
$\displaystyle f\left( 0\right)    \displaystyle =    \displaystyle 2^{0}\mod {15}=1,    $      \\
$\displaystyle f\left( 1\right)    \displaystyle =    \displaystyle 2^{1}\mod {15}=2,$ \\          
$\displaystyle f\left( 2\right)    \displaystyle =    \displaystyle 2^{2}\mod {15}=4,    $      \\
$\displaystyle f\left( 3\right)    \displaystyle =    \displaystyle 2^{3}\mod {15}=8,     $     \\
$\displaystyle f\left( 4\right)    \displaystyle =    \displaystyle 2^{4}\mod {15}=1,    $ \\
$\displaystyle f\left( 5\right)    \displaystyle =    \displaystyle 2^{5}\mod {15}=2,    $      \\
$\displaystyle f\left( 6\right)    \displaystyle =    \displaystyle 2^{6}\mod {15}=4.$\\

From here, we can realize that the function's period is r=4 since the function's values start repeating after x=4. However, for large numbers, finding this period is difficult to implement by hand, but it is also inefficient to find the brute force on a classical computer. However, Shor's algorithm provides an alternative, efficient method.

The quantum period-finding algorithm uses two registers of qubits. The first register is called the source register, and the second one the target register. If the number to be factorized using Shor's algorithm is N, then the number of qubits l in the first register has to satisfy
\begin{equation}\label{eq2_47}
N^{2}\leq 2^{l}\leq 2N^{2}.
\end{equation}

The number of qubits in the first register should generate a state-space that is at least large enough to capture the maximum possible period. The larger the value of $ {l} $, the more densely the solution-space that can be sampled, and thereby, the error in determining the period can be reduced. For simplicity, the number of qubits in the second register will be $ {l} $.

1. The qubits in the both registers start in the $\left\vert 0\right\rangle$ state. The first step is to apply Hadamard gates to the source register to create a large, equal superposition state. This is equivalently sometimes called a QFT, although a QFT is much less efficient to implement in general than a single Hadamard on each qubit. Whether written as Hadamards or as a QFT, the operation transforms the state of the source qubits:
\begin{equation}\label{eq2_48}
\begin{split}
\left\vert 0\right\rangle \otimes \left\vert 0\right\rangle \otimes ...\otimes \left\vert 0\right\rangle \otimes \left\vert 0\right\rangle \rightarrow \left( \frac{\left\vert 0\right\rangle +\left\vert 1\right\rangle }{\sqrt {2}}\right) \otimes \left( \frac{\left\vert 0\right\rangle +\left\vert 1\right\rangle }{\sqrt {2}}\right) \otimes ...\\...\otimes \left( \frac{\left\vert 0\right\rangle +\left\vert 1\right\rangle }{\sqrt {2}}\right) \otimes \left( \frac{\left\vert 0\right\rangle +\left\vert 1\right\rangle }{\sqrt {2}}\right)
\end{split}
\end{equation}

which can also be written as
\begin{equation}\label{eq2_49}
\left\vert 0\right\rangle \rightarrow \frac{1}{\sqrt {2^{l}}}\sum _{x=0}^{2^{l}-1}\left\vert x\right\rangle ,
\end{equation}
where now $\left\vert 0\right\rangle $ is in the l-qubit basis. This leaves the system in the state
\begin{equation}\label{eq2_50}
\frac{1}{\sqrt {2^{l}}}\sum _{x=0}^{2^{l}-1}\left\vert x\right\rangle \left\vert 0\right\rangle
\end{equation}
2. The unitary operation $U_{f} $ is applied. This gate implements the modular exponentiation,$ x\rightarrow f\left( x\right) =a^{x}\mod {N},$ such that the system ends in the state
\begin{equation}\label{eq2_51}
\frac{1}{\sqrt {2^{l}}}\sum _{x=0}^{2^{l}-1}\left\vert x\right\rangle \left\vert a^{x}\mod {N}\right\rangle .
\end{equation}
From here, we can see that the result of the period function $f\left( x\right) =a^{x}\mod {N}$ gets stored on the target register.

3. The second register is measured, and assume that this measurement projects the second register to the state $\left\vert f\left( x_{0}\right) \right\rangle =\left\vert a^{x_{0}}\mod {N}\right\rangle,$ where $x_0 $ are the unique values of x in the first period of f(x) values.

Now, remember that a periodic function with period r only takes r different values. The function $f\left( x\right)$ is unique for distinct periodic values of x,
\begin{equation}\label{eq2_52}
f(x_0)=f(r+x_0)=f(2r+x_0)=f(3r+x_0)=...=a^{x_{0}}\mod {N}
\end{equation}

If the second register is projected to $\left\vert f\left( x_{0}\right) \right\rangle, $the first register will be in a superposition of the states $\left\vert x\right\rangle$ for which$ f\left( x\right) =a^{x_{0}}\mod {N}$. For simplicity, Let us assume that there are $2^{l}/r$ different values of x for which $f\left( x\right) =a^{x_{0}}\mod {N},$ such that
\begin{equation}\label{eq2_53}
f(x_0)=f(r+x_0)=f(2r+x_0)=...=f\left(\frac{2^{l}}{r}r+x_0\right)=a^{x_{0}}\mod {N},
\end{equation}
or
\begin{equation}\label{eq2_54}
\sum _{y=0}^{2^{l}/r-1}f\left(yr+x_0\right)=a^{x_{0}}\mod {N},
\end{equation}
with y an integer from 0 to $ 2^{l}/r-1.$ Once the second register is projected to $\left\vert f\left( x_{0}\right) \right\rangle,$ state of the system can be written as
\begin{equation}\label{eq2_55}
\frac{1}{\sqrt {2^{l}/r}}\sum _{y=0}^{2^{l}/r-1}\left\vert yr+x_{0}\right\rangle \left\vert f\left( x_{0}\right) \right\rangle .
\end{equation}

4. Even though the information of the period r is encoded in the first register, there is still one more step before being able to find its value. A (second) Quantum Fourier Transform is applied on the first register, this leaves the first register on the state
\begin{equation}\label{eq2_56}
\frac{1}{\sqrt {r}}\sum _{z=0}^{r-1}\exp \left( \frac{2\pi i}{r}zx_{0}\right) \left\vert z\frac{2^{l}}{r}\right\rangle .
\end{equation}

5. The first register is measured, projecting the l-qubit system to the state $\left\vert C\right\rangle,$ with $C=z2^{l}/r.$ Knowing the value of C implies knowing the value of the ratio
\begin{equation}\label{eq2_57}
\frac{C}{2^{l}}=\frac{z}{r}.
\end{equation}
If z and r have no common divisors, $\gcd \left( z,r\right) =1,$ the ratio $ C/2^{l} $can be reduced to an irreducible fraction, as
\begin{equation}\label{eq2_58}
\frac{C}{2^{l}}=a_{0}+\frac{1}{a_{1}+\frac{1}{a_{2}+\frac{1}{...}}}.
\end{equation}
The approximations to the ratio z/r can be computed by using the continued fraction expansion relations: 
\begin{center}
$\displaystyle z_{0}    \displaystyle =    \displaystyle a_{0},$     \\     
$\displaystyle r_{0}    \displaystyle =    \displaystyle 1,     $
\end{center}
and       
\begin{center}
$ \displaystyle z_{n}    \displaystyle =    \displaystyle a_{n}z_{n-1}+z_{n-2},          $ \\
$ \displaystyle r_{n}    \displaystyle =    \displaystyle a_{n}r_{n-1}+r_{n-2}.          $ 
\end{center} 

6. Different values for $z_{j}$ and $ r_{j}$ are tested, until one finds values such that $\gcd \left( z_{j},r_{j}\right) =1$. For the valid values of $ r_{j}$, the function $f\left( x\right) =a^{x}\mod {N} $ is evaluated with $x=r_{j}$. If $ f\left( r_{j}\right) =1, $ then the period of f is $ r_{j}$; otherwise, we have to go back to step 1.\\
To understand how this algorithm works, Let us find the period of the function $f\left( x\right) =11^{x}\mod {21}.$

We start by choosing the number of qubits in each register. For the first register, the number $ {l} $ has to satisfy$ 21^{2}\leq 2^{l}\leq 2\times 21^{2} or 441\leq 2^{l}\leq 882.$ Let us consider that each register has $ {l} $ = 9 qubits.

1. All the qubits start in the ground state $\left\vert 0\right\rangle.$ A Hadamard gate is applied on each one of the qubits on the first register. This transforms the state of the first register as
\begin{equation}\label{eq2_59}
\left\vert 0\right\rangle \rightarrow \frac{1}{\sqrt {2^{9}}}\sum _{x=0}^{2^{9}-1}\left\vert x\right\rangle .
\end{equation}

This leaves the system in the state
\begin{equation}\label{eq2_60}
\frac{1}{\sqrt {2^{9}}}\sum _{x=0}^{511}\left\vert x\right\rangle \left\vert 0\right\rangle ,
\end{equation}

Where $\left\vert 0\right\rangle$ represents the ground state of 9 qubits.

2. The unitary operation $U_{f}$ is applied, leaving the system in the state
\begin{equation}\label{eq2_61}
\displaystyle \frac{1}{\sqrt {2^{9}}}\sum _{x=0}^{511}\left\vert x\right\rangle \left\vert 11^{x}\mod {21}\right\rangle    
\end{equation}

\begin{equation}\label{eq2_62}
\displaystyle =    \displaystyle \frac{1}{\sqrt {2^{9}}}\left( \left\vert 0\right\rangle \left\vert 11^{0}\mod {21}\right\rangle +\left\vert 1\right\rangle \left\vert 11^{1}\mod {21}\right\rangle +\left\vert 2\right\rangle \left\vert 11^{2}\mod {21}\right\rangle \right) 
\end{equation}    
\begin{equation}\label{eq2_63}      
\displaystyle +...+\left( \left\vert 509\right\rangle \left\vert 11^{509}\mod {21}\right\rangle +\left\vert 510\right\rangle \left\vert 11^{510}\mod {21}\right\rangle +\left\vert 511\right\rangle \left\vert 11^{511}\mod {21}\right\rangle \right)    
\end{equation}
\begin{equation}\label{eq2_64}      
\displaystyle =    \displaystyle \frac{1}{\sqrt {2^{9}}}\left( \left\vert 0\right\rangle \left\vert 1\right\rangle +\left\vert 1\right\rangle \left\vert 11\right\rangle +\left\vert 2\right\rangle \left\vert 16\right\rangle +...+\left\vert 509\right\rangle \left\vert 2\right\rangle +\left\vert 510\right\rangle \left\vert 1\right\rangle +\left\vert 511\right\rangle \left\vert 11\right\rangle \right)
\end{equation}
     
3. Assume the second register is measured, projecting it into the state $\left\vert f\left( 2\right) \right\rangle =\left\vert 16\right\rangle.$ This leaves the first register in
\begin{equation}\label{eq2_65}
\displaystyle \frac{1}{\sqrt {2^{9}/r}}\sum _{y=0}^{2^{9}/r-1}\left\vert yr+2\right\rangle \left\vert 16\right\rangle    
\end{equation}

\begin{equation}\label{eq2_66}      
\displaystyle =    \displaystyle \frac{1}{\sqrt {2^{9}/r}}\left( \left\vert 2\right\rangle +\left\vert r+2\right\rangle +\left\vert 2r+2\right\rangle +...+\left\vert \left( \frac{2^{9}}{r}-1\right) r+2\right\rangle \right) \left\vert 16\right\rangle         
\end{equation} 

4. A Quantum Fourier Transform is applied on the first register such that
\begin{equation}\label{eq2_67}
\frac{1}{\sqrt {2^{9}/r}}\sum _{y=0}^{2^{9}/r-1}\left\vert yr+2\right\rangle \left\vert 16\right\rangle \rightarrow \frac{1}{\sqrt {r}}\sum _{z=0}^{r-1}\exp \left( 2\frac{2\pi i}{r}z\right) \left\vert z\frac{2^{9}}{r}\right\rangle .
\end{equation} 

5. Assume the first register is measured, projecting it into the state $\left\vert 427\right\rangle$. The irreducible fraction of the ratio $C/2^{9}$ is given by
\begin{equation}\label{eq2_68}
\frac{C}{2^{l}}=\frac{427}{2^{9}}=0+\frac{1}{1+\frac{1}{5+\frac{1}{42+\frac{1}{2+\frac{1}{...}}}}}.
\end{equation} 

Note that we can calculate this irreducible fraction in Wolfram Alpha by typing in the bar ``continued fraction of 427/512.''

6. The values $z_{j}$ and $r_{j}$ are tested for the condition $\gcd \left( z_{j},r_{j}\right) =1$ with j from 0 to 3,\\
$\displaystyle \gcd \left( z_{0},r_{0}\right)    \displaystyle =    \displaystyle \gcd \left( 0,1\right) =1,$ \\          
$\displaystyle \gcd \left( z_{1},r_{1}\right)    \displaystyle =    \displaystyle \gcd \left( 1,1\right) =1,$    \\      
$\displaystyle \gcd \left( z_{2},r_{2}\right)    \displaystyle =    \displaystyle \gcd \left( 5,6\right) =1,$     \\     
$\displaystyle \gcd \left( z_{3},r_{3}\right)    \displaystyle =    \displaystyle \gcd \left( 427,512\right) =1.$    \\      

The function $f\left( x\right) =11^{x}\mod {21}$ is evaluated for different valid values of $r_{j}$,\\
$\displaystyle f\left( 1\right)    \displaystyle =    \displaystyle 11^{1}\mod {21}=11    $      \\
$\displaystyle f\left( 6\right)    \displaystyle =    \displaystyle 11^{6}\mod {21}=1$          \\
$\displaystyle f\left( 512\right)    \displaystyle =    \displaystyle 11^{512}\mod {21}=16$\\
          
With this result, we can conclude that the period of the function f is r=6.

\section{Shor's Algorithm: Reduction of Factoring to Order Finding} 

In the next text units, we will take a deep dive into several details behind Shor's algorithm, including a discussion of the reduction of factoring to order finding, the related Simon's algorithm that helped inspire Shor's algorithm, and the use of Shor's algorithm to perform order finding. 

How do we get factoring from order finding? So, first, let say the second-fastest classical algorithm is the quadratic sieve. So, we would say most classical algorithms\cite{pednault_breaking_2018}. So, suppose we have N equals P times Q. They work in more general cases, but we will just worry about the case of two products. So, the first step, really the only step, the trick is to find x squared congruent to y squared mod N x not congruent to plus or minus y mod N. So, suppose we are trying to factor 21. we get 4 is congruent to 25 mod 21, right? Because 25 minus 4 is a multiple of 21. So, 4, these are both squares.
Moreover, now we have 5 minus 2 times 5 plus 2 is congruent to zero mod 21. whenever this happens, one of these contains P, and the other contains Q. Or rather, whenever neither of these is 0, we have a product of two numbers, which is zero mod 21. Thus, we know that the product has to contain three somewhere, and it has to contain somewhere. If neither of these factors is 0 mod 21, 3 has to be contained in one factor and 7 in the other. Then, what we can do is we can use Euclid's algorithm for GCD to get P and Q. So, we are essentially done whenever we get two squares, which are equal to each other. This is how the quadratic sieve works. That is the second-fastest classical factoring algorithm known. How do we find x squared y squared? Well, let us look at the function f of x equals g to the x mod N. So, this is periodic. If g to the r is congruent to 1, then g to the x plus r is congruent to g to the x mod N. Because if g to the r plus is congruent to 1, then g to the x plus r congruent to g to the x because it is just multiplying both sides of this by g to the x. So, that says that f of x is periodic. We mean, it has to start repeating some time because there are only N possible things here. Whenever it starts repeating, we get a g to the x plus r congruent to g to the x. that means g to the r is congruent to 1. g to the r over 2 squared is congruent to 1 squared if r is even. So, let us do an example. Suppose we are trying to factor 21. So, 2 to the 0, 2 to the first, 2 squared. So, this is 1, 2, 4, 8, 16, 32. 32 minus 21 is 11, so, the next column then is 11. 22 minus 21 is 1. 1. So, 2 to the sixth is congruent to 1 mod 21. That means 2 cubed minus 1 minus time 2 cubed plus 1 is congruent to 0 mod 21. this is 7, and this is 9. 7 contains the factor 7, 9 contains the factor 3. So, the question is, how can this go wrong? Moreover, well, there is a couple of things that could go wrong. We could have started with 4. 4, 16. 16 squared, well, we know 2 to the sixth is congruent to 1 mod 21. So, 4 cubed is congruent to 1 mod 21. So, we get 4 cubed is congruent to 1 mod 21. 4 to the 3/2 does not exist. Well, we cannot compute it. So, this shows that 4 could go wrong. The only thing that could go wrong is we could do this. When we look at the middle element, it is minus 1. that would be a problem because then we have x is equal to minus y mod N. we cannot use that, we cannot use this trick then. So, some things can go wrong, but the great theorem is the probability of success with g to the xth is greater than or equal to 1/2 for a random g between 0 or a random g less than N. So, we choose a random g less than N. we do this repeated squaring. The probability this will work is at least 1/2. If it does not work, we choose another random g. So, eventually, with high probability, we will get a g, which lets us factor.

\section{ Introduction to the Shor Quantum Factoring Algorithm} 
we discovered the factoring algorithm after seeing Simon's algorithm before Simon's algorithm got published. We are going to show we write down a circuit diagram so we can see the similarity. So, Simon's algorithm, remember, we started with n qubits in the plus. Then we went, and here is a black box that computes 0 to the n. From x, it computes f of x. then we did a Hadamard transform to the n, which is just the quantum Fourier transform over the group of two elements to the n-th power. So, n copies of the group of two elements. So, this is just binary strings under addition without carries. After that, we believe we measured this register, and then did classical post-processing. So, what is factoring? So, what we are not going to do directly is a quantum algorithm for factoring. We are going to do a quantum algorithm for an order finding a quantum algorithm for periodicity. So, suppose we have a function over integers, f of x is equal to f of x plus c for some c. here we can see that f of x 2c equals f of x plus 3c. So, it is a periodic function, and a period is well, we call it C right now, but at some point during this section, we are going to start calling it r because that is what we used to call it. Find c. Now we might think this is easy. We just start with f of 0, f of 1, f of 2, f of 3, and keep on going until f starts repeating. So, even if the period is exponentially long. That means that if we want to try to find a period of length r, we can do this length polynomial and log of r. So, we can find the period. On the classical computer, if we have a black-box function, the best thing we can do is to keep on testing values until it starts repeating. So, that means the best we can do is period r. in quantum mechanics, quantum computing, we can do log r. So, it is going to be exponentially long in a period of a number we want to factor. However, if we are just finding a period, find a period in time O of log c. So, we will be able to find the period and time well, that is going to be quicker than log c if we want to do it quicker. Log c squared if we want to do it quickly because we cannot use the exact quantum Fourier transform, but the approximate quantum Fourier transform, which has like log c, log log c gates But here, so, we want to find a period. We have a black box function for f. f has some period of c. So, what is the quantum algorithm for this? Well, we are going to do the same thing here. We are going to start at plus to the c, 0, the n here. Here we have a black-box function, which, if we put in x, computes f of x. Alternatively, actually, it computes well, remember this took x, t to x, t xr f of x. we can assume this does the same thing. x, t to x t xr f of x. nowhere is the only difference, really, between the circuits. We use the quantum for each transform over 2 to the actual, let us do this right. There is 2n. We want 2n qubits, n qubits here. Quantum Fourier transforms over 2 to the n. then we measure. Then we do classical post-processing. So, it looks very similar. So, let us go into detail, in both these cases, what happens to the bottom branch? We do nothing to the bottom branch. We mean that we could measure the bottom branch and throw away the results if we want to. Alternatively, we do not have to. If we measure something and throw away the results, it does not matter whether we measure it. However, we do need to compute it. Because if we do not compute it, what have we done? We have taken the Hadamard transform of x to the n. That is just all 0's. So, if we do not compute f of x, it destroys the interference in the upper half. Again, action in a quantum circuit has back action. 

\section{Simon's Algorithm}
In 1994, Daniel Simon posed the following question: Does a particular function map distinct input elements to unique output elements hence representing a permutation or a one-to-one mapping function or, does it follow a two-to-one mapping scheme? Despite the problem's small practical importance, it is a quantum algorithm that exhibits an exponential speedup. Simon's algorithm served as an inspiration for Peter Shor and his famous quantum algorithm for period finding.

Simon's algorithm addresses a particular black-box function $f:\{ 0,1\} ^ n\rightarrow \{ 0,1\} ^ n$, which is known to map inputs in either a one-to-one or two-to-one manner. Two distinct input bit strings $a_1,a_2 \in \{ 0,1\} ^ n$ are mapped to the same output $f(a_1)=f(a_2)$ if:\\
1. the two bit strings $a_1$ and $a_2$ are equivalent a one-to-one function, or \\
2. there exists a bit string $s \neq 0^ n $such that $a_1\oplus s=a_2 - a$ two-to-one function.

If there exists a single $s\neq 0^ n$, the mapping is ``two-to-one'' and otherwise (if $s=0^ n$ works) it is ``one-to-one''. Simon's algorithm finds the bit string s and the type of function mapping exponentially faster than known classical algorithms.

The algorithm starts with the initialization of 2n qubits in their ground state. Subsequently, half the qubits are rotated with a Hadamard gate to form an equal superposition state.Thereafter, the oracle function $ U_ f$ is applied on all qubits. A second Hadamard gate is applied to the same qubits that received the first Hadamard gate, and then then qubits are read out. The mathematical description of the protocol can be formulated as follows:
\begin{equation}\label{eq2_69}
\begin{split} \lvert 0\rangle \dots \lvert 0\rangle =\lvert 0^ n\rangle \lvert 0^ n\rangle ~ ~ ~  \xrightarrow {H^{\otimes n}I^{\otimes n}}~ & \frac{1}{\sqrt {2^ n}}\sum _{a\in \{ 0,1\} ^ n}\lvert a\rangle \lvert 0^ n\rangle \\  \xrightarrow {~ ~ ~ U_ f~ ~ ~ }& \frac{1}{\sqrt {2^ n}}\sum _{a\in \{ 0,1\} ^ n}\lvert a\rangle \lvert f(a) \rangle \\  \xrightarrow {H^{\otimes n}I^{\otimes n}}&\frac{1}{2^ n}\sum _{b\in \{ 0,1\} ^ n} \sum _{a\in \{ 0,1\} ^ n} (-1)^{a\cdot b} \lvert a\rangle \lvert f(a) \rangle \\ \text {if } f(a_1)=f(a_2) \rightarrow a_1\oplus s=a_2: ~ ~ \\ \frac{1}{2^ n}\sum _{b\in \{ 0,1\} ^ n} \sum _{a_1\in A} ((-1)^{a_1\cdot b}+(-1)^{a_1\oplus s\cdot b}) \lvert a_1\rangle \lvert f(a_1) \rangle \\  \frac{1}{2^ n}\sum _{b\in \{ 0,1\} ^ n} \sum _{a_1\in A}(-1)^{a_1\cdot b} (1+(-1)^{s\cdot b}) \lvert a_1\rangle \lvert f(a_1) \rangle \\ \text {summands~ }\neq 0 \text {~ ~ only if~ } s\cdot b=0: ~ ~ \\  \frac{1}{2^{n-1}}\sum _{b\in \{ 0,1\} ^ n~ \& ~ s\cdot b =0} \sum _{a_1\in A}(-1)^{a_1\cdot b}\lvert a_1\rangle \lvert f(a_1) \rangle \end{split}
\end{equation}

The detectable states non-zero coefficients fulfill the condition of $s\cdot b=s_1b_1\oplus \cdot \cdot \cdot \oplus s_{n-1}b_{n-1}=0$ a dot product modulo 2. To determine s with high accuracy it is necessary to repeat the algorithm to acquire n-1 linearly independent $b_1, \dots , b_{n-1}$. Upon repeated runs of the algorithm, equations $s\cdot b_1=0, \dots , s \cdot b_{n-1}=0 $enable a classical calculation of the solution $s'$. If$ f(s')=f(0^ n)$ holds, the black box function maps its input states in a two-to-one manner. In contrast, if $f(s')\neq f(0^ n)$ with $s'\neq 0^ n$, the derived $s'$ is not a valid solution. Therefore, only $s=0^ n$ satisfies $f(s') = f(0^ n)$ and the function is one-to-one. Simon's algorithm is optimal and yields an answer after $\mathcal{O}(n$) queries. In contrast, a classical algorithm requires at least$ \mathcal{O}(2^{n/2})$ queries.

Both Simon and Shor's algorithms are related to finding the order or the period of a function of interest. In Simon's algorithm, this is accomplished through quantum interference induced by Hadamard gates and multiple measurements. In Shor's algorithm, the period finding is accomplished using the quantum Fourier transform.

\begin{figure}[H] \centering{\includegraphics[scale=.6]{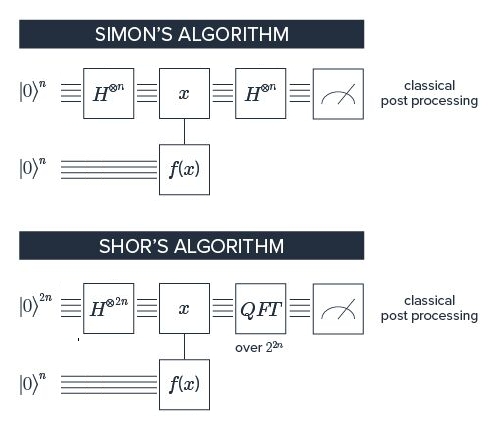}}\caption{SS Algorithm}\label{fig2_6}
\end{figure}

\section{Shor's Algorithm: Detailed Steps of the Quantum Order Finding Algorithm} 
What is the algorithm in detail? So, suppose we have a bound-on r we should call it r-max and choose a power of 2 bigger than 2 r-max. Well, we will write down the algorithm. We start with plus to the 2l 0. That is here. Next, we take the compute f with x. We know that whatever ends in the first register, we compute it in the second register. However, this is just sum i equals 0 to 2 to the 2l. i, where i is just the binary representation of the number. We need to divide by 2 to the l because we want this to be a unit vector. So, we compute f. So, this becomes sum to the 2l i f of i. i equals 0 to 2 to the 2l. Do the quantum Fourier transform. So, that is 1 over to 2 to the 2l summation. We know i is a really bad letter to use because it is also the square root of minus 1. So, let us use x. So, the quantum Fourier transform takes x to, let us call it, z f of x. z equals 0 to 2 to the 2l minus 1. e to the 2 $ \pi $ i xz over 2 to the 2l. So, we have taken x and taken the Fourier transform of it, which gives us this. Now, what happens if we measure f of x? well, we are going to get some value for f of x. we are going to leave that out the normalization constant for the time being because figuring out exactly What is going on with a normalization constant is possible, but it is much tedious algebra. We will come back to it later. So, without a normalized, z equals 0 through 2 to the 2l. Let x0 be the smallest x giving f of x. So, this is going to be z of the form x0 plus y times $ \pi $. Because when we measure f of x, we only get z no, not z. When we measure f of x, we only get x equals x0 plus yr f of x0 z e to the 2 minus 2 $ \pi $ i x well, xz, but x is of the form x0 plus yrz over 2 to the 2l. So, what did we do? We measured f of x here. We are just looking at the ones who give us f of x0. We know all of these x's are the form x0 plus yr. The phase factor does not change, and z does not change. So, now we have the sum. We have to figure out how big it is. Well, we can write sum z equals 0 through 2 to the 2l. Sum y equals 0 through 2 to the 2l over r, plus or minus 1 here, but that does not matter. f of x0 z e to the 2 $ \pi $ i x0 plus yrz over 2 to the 2l. What is this? This is a geometric sum. 
\section{Shor's Algorithm: Order Finding Algorithm Measurement Result as a Geometric Sum} 
So, first, we will do the intuition, and then we will do the math. So, it is a geometric sum, and it is a whole bunch of points around the unit circle. So, we start with e to the minus 2 $ \pi $ i x0. Let us plot these points minus 2 $ \pi $ i x0. now we go r, z over to the 2 to the 2l. So, the next one we want to plot is y equals 1. So, we go some distance around the circle. Let us plot another point. Then we plot another point, another point, another point. We keep ongoing. What we get is quite possibly points that are exactly uniformly distributed over the unit circle, sum close 0. well, and next, what happens is, here was e to the 2 $ \pi $ i x0 2 to the 2l. Here is e to the minus 2 $ \pi $ i x0 plus rz zero over 2 to the 2l. So, unless what happens is when we plot the point, we move around by the angle 2 $ \pi $ i rz over 2 to the 2l. Suppose this angle is tiny. Well, what happens then is we get points uniformly distributed. Suppose it is tiny enough so, that when we add all of these 2 to the l over r points, we get points that are we know, some kind of an arc, which does not even go all the way around the unit circle sum is close to well, how many points were there? There were 2 to the 2l over r. they were all pointing in more or less the same direction. So, an absolute value. So, there are two cases here. 2 $ \pi $ i yr If 2 $ \pi $ i rz over 2 to the 2l is close to 1, which means y rz over 2 to the 2l is close to an integer. Let us call that integer d. So, if this is close to an integer, then we get a large probability of seeing this. If it is not close to an integer, we get a tiny probability of seeing it. So, the geometric sum is 1 minus e to the 2-$ \pi $ i. we want the last term of the sum. So, that is 2 to the 2l rz over 2 to the 2l. No, it is y goes from 1 to 2 to the 2l over r. So, the last term we substitute in 2 to the 2l r over z. So, something like 2 to the 2l over r. we will have to take the floor of that time r times z over 2 to the 2l. Which really, because of this floor could be anywhere on the unit circle. However, it is always less than 2 in absolute value. We are dividing that by 1 minus e to the 2 $ \pi $ i rz over 2 to the 2l. Remember, the sum is, so we are dividing something by something else. It is going to be relatively small unless something else we are dividing is close to 0. So, this is roughly something O of 1 divided by oh, what is this? This is rough rz over 2 to the 2l. The probability of seeing a number z is rough rz over 2 to the 2l squared minus d because d was the integer. So, we have d over r is roughly z over 2 to the 2l. So, what the algorithm does is it output z. we know 2 to the 2l, and we need to round it off to something like d over r. there is a procedure for rounding real numbers of two smaller fractions that is efficient. It is called continuous fractions. However, it is a classical algorithm. So, once we get this, we can round it to d over r, and then we are found r, which was the period we were looking for. 
\section{Demonstrations of Shor's Algorithm: Introduction} 

In the next several sections, we will explore two demonstrations of Shor's algorithm on small-scale quantum-computers realized using liquid-state NMR and trapped ions.  Although the numbers of qubits and the number being factored is rather small (15 = 3 x 5), the demonstrations illustrate that Shor's algorithm works in principle. We will also discuss how many demonstrations have used ``compiled'' versions of Shor's algorithm that are not generalizable. 

To date, there have been a handful of prototype demonstrations of Shor's Algorithm using different qubit modalities, including nuclear spins, photonic qubits, superconducting qubits\cite{kapit_very_2016}, and trapped ions. In this section, we will look at two of these demonstrations in detail. The first is a demonstration of Shor's Algorithm using liquid-state nuclear magnetic resonance, NMR. This was, in fact, the first such experimental demonstration, and it showed the factorization of 15 into the primes 3 times 5. It used molecules with seven individually addressable nuclear spins in a liquid solution at room temperature, and the algorithm comprised a sequence of around 300 pulses. we will discuss this work. The second demonstration is recent, and it was performed using trapped ions \cite{monz_realization_2016}. In this case, seven ions were used to again factor the number 15 into primes 3 and 5. In this example, modular multiplication was implemented, and no simplifications presumed prior knowledge of the result were employed. Instead, a concept called qubit recycling was used to make up for the fact that seven qubits are insufficient to implement the general algorithm. Still, to date, this demonstration constitutes the most realistic implementation of Shor's Algorithm. MIT Professor Isaac Chuang participated in this demonstration with Professor Rainer Blatt's team at Innsbruck. To factor 15 in a straightforward manner, it is estimated that one needs about 12 qubits, eight qubits in the period register, and four qubits in the modular multiplication register. The demonstrations we will discuss in this study had fewer than 12 qubits, and so, certain simplifications had to be made. Now in many cases, the simplifications were quite reasonable. For example, the use of qubit recycling. Nonetheless, it is important to understand the impact of various types of simplifying assumptions. Finally, from a business perspective, these prototype demonstrations of Shor's Algorithm illustrate that quantum computing is real. Although these implementations are not yet at scale, they are concrete indications of and steps towards the promise of quantum computing. From an engineering perspective, these types of demonstrations force groups to confront practical realities of implementation, from qubit control engineering to algorithmic compilation. Although not yet at scale, these algorithms motivate experimental groups to go beyond scientific demonstrations and to meet face-to-face with the practical realities that will need to be accommodated to enable scalable quantum computing. scientifically, these demonstrations test the knowledge of system noise \cite{mavadia_experimental_2018,wallman_noise_2016} and decoherence \cite{herbschleb_ultra-long_2019}. Lastly, from a technology standpoint, these demonstrations serve as a benchmark of progress in the field, indicating where the technology currently stands and clarifying what needs to be addressed in order to reach future milestones. With that brief introduction, let us discuss. 
\section{Demonstrations of Shor's Algorithm: Shor's Algorithm with NMR} 
In 1998, after implementation of the first two-qubit algorithms on the chloroform molecule and using NMR, Isaac Chuang, who supervised the research, gave a great challenge to factor the number 15 within the time span of Ph.D. at Stanford. At that time, Isaac Chuang was transitioning from Los Alamos National Laboratory to the IBM Almaden Research Center outside San Jose. He was finishing the first year of  Ph.D. at Stanford. At that time, Isaac set up his new lab at Almaden, some 30 minutes away from Stanford. Researcher did not know what it would take to factor the number 15, but it did seem like a great challenge. certainly, it was clear that there was an enormous interest in Shor's Algorithm and that it was a major driver for the field. So, we thought it would be very awesome to demonstrate Shor's Algorithm experimentally. So, what would it take to run Shor's Algorithm experimentally? The simplest instance of quantum factoring is to factor the number 15, to pull it apart into 3 times 5. However, how many qubits does that take, and how many operations? we got great insight from a pedagogical article by Dave Beckman from the Preskill group at Caltech. In particular, he analyzed the steps for factoring the number 15. Shor's algorithm, in general, consists of two parts, a classical part and the quantum mechanical part. The quantum part that we run on the quantum computer is designed to find the period of a function, specifically the function a to the power x modulo N, that is to say, the remainder after division of a to the x by N. N here is the number we are trying to factor. A number can be anything as long as it is co-prime with N, and x is the function's argument. For period finding on a quantum computer, we divide the number of qubits into two registers. One register needs to be large enough that it can store the number N. For 15, we need four bits, so, four qubits are needed for that register. The other register, in principle, needs to be twice as large or eight qubits, bringing the total to 12 qubits to factor the number 15. The algorithm consists of, first, a Hadamard operation that prepares an equal superposition of all the states of the first register then a function call of the function a to the x modulo N where x is the states of the first register. the result is mapped into the state of the second register. the third step is the Quantum Fourier Transform applied to the first register. At the end of the computation, we read out this first register. Now, if we begin to analyze how to implement this complex function a to the x modulo N, we realize that it can be broken apart into multiple simpler operations. Each of them is an operation controlled by just one bit out of all the bits of the first register. then we find for the number 15 that we are lucky. Only two of the control qubits play a role in factoring 15. in some cases, some choices of a, even just one of the control qubits, plays a role. So, that allows us to eliminate many qubits of the first register. In principle, we could eliminate all of them except two. We decided that for a meaningful Quantum Fourier Transform, we would like to have not just two qubits but three qubits. With two qubits, the Quantum Fourier Transform comes down to a single two-qubit gate. we thought that was too trivial. So, in the end, we went with an experiment with a total of seven qubits to factor the number 15. So, in NMR, the qubits are represented by the nuclear spins in the molecule, the spins of the atomic nuclei in a molecule. For a successful molecule for NMR qubit experiments, the molecule needs to have rather unusual properties. we need what we call large chemical shifts. That is to say, large separations in frequency between all the spins we want to get involved in the computation for selective addressing of the qubits. nd we want a coupling network that allows sufficient pairs of qubits, pairs of spins, to be coupled to each other so that all the required two-qubit gates can be implemented on the molecule. So, a staff member at IBM Almaden did an extensive search looking through fast tables and documents with tabulated NMR properties of known molecules. after spending days in this process, he concluded that none of the tabulated molecules were good enough for the experiment that were attempting. However, there was one that came close. It consisted of a backbone of four carbon atoms. surrounding it, there were five fluorine atoms that all could serve as a spin one-half and provide a qubit. besides, there was an iron complex. The iron complex created asymmetry in the molecule so that the resonance frequencies of the five fluorine atoms would be separated. To go from five fluorine spins to seven, what we needed to do was to isotopically label two of the carbon spins so that they became carbon-13 and also acquired a spin one-half. Synthesizing this molecule was a real tour de force. Greg Breyta, IBM staff member who usually worked on chemical synthesis in the Resist program, spent over one year trying to synthesize this molecule. at any point in time in this multi-step synthesis, they did not know if this next step would be successful. However, finally, after more than a year, actually just in time, when they were ready to try the experiment, he delivered the molecule. So, the next challenge was to develop the control techniques to take the seven spins through the steps of Shor's Algorithm, to make them do the spin dance with a great degree of control. Here we could benefit from decades of technology development in NMR spectroscopy, where people were interested in revealing the structure of molecules they were synthesizing. Mark Sherwood done great work on NMR spectroscopy at IBM. In the years that followed in 1999 and 2000, we did experiments with the first two, then three, and five spins to discuss how to control the time evolution of increasing spins. They were all coupled together. Nevertheless, the factoring experiment became clear beyond anything that anyone had ever done in terms of controlling many coupled quantum spins. It required more than 300 radio frequency pulses, specially tailored in amplitude and phase, to allow selective addressing of each of the seven spins. Also, we had to account for unintended phase shifts that occurred on any of the qubits when we drove any other qubits. In addition to that, since we were competing against the decoherence timescale, we had to apply simultaneous rotations where crosstalk effects were even more severe. we had to account for those as well. then finally, in NMR in these molecules, the couplings are always on. So, we had to use some sophisticated so, called decoupling sequences that effectively remove the effect of all the couplings except the one coupling that we needed to run a two-qubit gate between two spins. So, together with Matthias Stephan, fellow Ph.D. student at Stanford, who started one year after, they went out to try and control the seven spins, starting in September of 2000. throughout that fall and the winter of 2001, they made much progress. they used a variant NMR four-channel spectrometer to control the seven spins. after some time, they were far enough that the output spectra revealed the information about the period of a to the x modulo 15. from that, they could say that they had factored 15. At the same time, they were disappointed because the output spectra, even though they did give us the right information, they did not look like what ideally expected. even though they added more and more sophistication in the control techniques, it seemed like they were hitting a wall. At some point, they set out to write a model to capture the main effects of decoherence simply. there it was, beautifully reproducing the experimental spectra, the simulated spectra accounting for decoherence. It was interesting because, in the four years of experience programming NMR sequences and operating on molecules, this was the first time that decoherence had become the limiter.
The relationship between duration, strength, and phase of our pulses and their ultimate effect on rotations? so, we can think of operations on a single qubit in terms of a Bloch sphere, or the planet earth, and so we call the North Pole the ground state, state zero. We call the South Pole state one, and those are just classical states, zero, one. But anywhere else on the Bloch sphere, or think of it as planet earth, anywhere else on the earth, we are in a superposition of zero and one. so, we can perform rotations by applying resonant pulses to the qubit, and that will effectively rotate this Bloch vector around the planet earth. So, Let us say we start at the North Pole, and if we, so first we will discuss about duration and strength of the RF pulses. Basically, the amount of rotation we going to get is the area under this pulse. So, it is both the duration, as well as the amplitude of the pulse that matters. Now, we often apply a Gaussian-shaped pulse, or some kind of smoothly-shaped pulse. But if we think of it as a rectangle, then it is very easy to just say, look, it is the duration of the pulse, times the amplitude of the pulse, and we can rotate a given amount of degrees around a certain axis, and we can do that for a short, poll pulse, or we can do that for a very long pulse that is got a low amplitude. As long as they have the same area, we going to rotate the same amount. If we want to rotate further, similarly, we can increase the amplitude more, or we can keep the amplitude fixed, and increase the duration more. So again, it is the area under the pulse that matters, that is going to determine how far we rotate around the Bloch sphere. so, one of the first things that an experimentalist will do is calibrate their pulses, and part of what that means is understanding what duration and amplitude do we need to rotate 90 degrees? And what do we need to rotate 180 degrees? so that has to do with the amount of rotation. Now the second question is, what axis am we rotating around? And so, the, if we think of the, equator around the Bloch sphere, or around the earth, we can put an x and a y axis on here. so x is going to point in one direction, and y will be 90 degrees away from that, pointing in a perpendicular direction. there is plus and minus x, and there is plus and minus y. Now, when we apply this microwave pulse, we can generate that pulse using IQ mixing. so we means in-phase, Q means quadrature, and basically, what this allows us to do is output a pulse where the microwave tone that is resonant with our qubit transition is\cite{randall_efficient_2015,paraoanu_microwave-induced_2006}, has zero phase shift with respect to the source, that would be the end phase channel. Or, we can use the quadrature channel, and output same pulse shape, but with the microwave shifted within that envelope by 90 degrees, And when we use the in phase, the in phase port, then what we will do is we will rotate around the x-axis. when we use the quadrature port, which shifts the microwaves by 90 degrees, we will rotate around the y-axis. so that is basically how the phase of the RF pulses will effect the rotation. It determines what axis these rotations are going to occur around. Now we might say, how do we know what is the x-axis to begin with? It turns out that there is always this relative phase offset that people discuss about. it is a little bit of jargon but, in quantum mechanics, and so, what sets the x-axis, basically, is our first pulse, right? So if our first pulse, Let us say it is got zero phase, we will rotate around a given axis. That axis becomes our x-axis, right? And everything relative to that, now, is either in-phase with that, or in quadrature with that, and if it is in-phase, we will rotate around that axis, and we call it x, and if we are in quadrature with that, we will rotate around the y-axis. 

\section{Demonstrations of Shor's Algorithm: Perspective on NMR}
So, let us reflect on what made this experiment successful. The first is that the coherence times, the t2's in this specially designed molecule, were exceptionally long all more than 1 second. We have to compare, the coherence time to the time scale of the operations that form a single qubit-shaped pulse here varied between 0.2 and 2 milliseconds. We also have to compare to the value of the coupling strings between the spins in the molecule. They varied from about a hertz, which was too low to be usable, to about 200 hertz. In total, the longest sequences we had to run to implement Shor's Algorithm lasted about 700 milliseconds. The second success factor is that in these experiments, we used liquid state NMR based on ensembles of many copies of a molecule, making the detection trivially easy, almost, compared to the great challenge of trying to work with individual spins as is done today.
Furthermore, the liquid state NMR Hamiltonian is relatively transparent and simple. All the terms in the Hamiltonian commute with each other. Single qubit terms have to form omega times $ \sigma $ z. the two-qubit interactions have to form J times $ \sigma $ z on one qubit and $ \sigma $ z on another qubit. Scaling this approach, however, is hard. First, we have already commented on the difficulty of finding a suitable molecule with seven spins.
Moreover, finding molecules with larger numbers of qubits, with larger numbers of suitable spins, to solve more complex problems quickly gets extremely difficult. However, there is a more fundamental and severe limitation to the liquid state NMR approach, and it is that even though we work in large magnetic fields, the fact that we use liquid states pushes us to work at large temperatures. We operated at room temperature. Nuclear spins in a magnetic field at room temperature are in a highly mixed state, that is to say, the probability for a spin to start in the upstate or the downstate are almost equal. We used techniques developed in the community to prepare so-called, effectively pure states. These are states that produce a signal that is proportional to that of a pure state, but the cost of this technique is that the signal is halved for every qubit that we add. Thus, this induces exponential cost that offsets any exponential gain that we hope to get from the quantum algorithms \cite{moussa_function_2019}. The NMR experiments, and in particular the short experiment, had been great fun.
Furthermore, we think we discussed a great deal from it that also benefit and still benefits, today, the community. However, certainly, we wanted to move to technology, where the potential for scaling was far better. We decided to move from nuclear spins to electron spins and from molecules to artificial molecules. The artificial molecules we make use of are built from Silicon quantum dots in which we can find individual electrons and use their spins as qubits\cite{broome_two-electron_2018}. Similar to the NMR molecules, these electron spins in silicon offer long coherence times\cite{yoneda_quantum_2020}, but the devices are much more scalable, and the control, in the end, can take us much further. Matthias Stephan, instead, chose to work on superconducting qubits, and he was one of the early members of the IBM team working on superconducting qubits, a very large and successful program today. Isaac Chuang finally decided to move to trapped ions. He set up a lab at MIT, and in collaboration with the group of Reiner Blatt Some years later. They redid the factoring experiments a beautiful experiment implementing the factoring of 15 on trapped ions. As for the silicon spins that we work, we set state of the art earlier this year in implementing for the first time, in this type of system, again, a two-qubit algorithm. It brought back great memories from '98. However, different from the NMR approach and these early memories, the expectation is that with the group at QuTech in Delft and close collaboration with Intel, with the silicon spin qubits, we can go much beyond factoring 15.

\section{Demonstrations of Shor's Algorithm: Quantum Factoring With Trapped Ions - Ion Trapping System} 

Shor's quantum factoring algorithm has been implemented by several physical systems, including trapped ions. Let us consider what was done and why. Perhaps the most important starting point is to consider the question, what is the smallest meaningful instance of the Shor quantum factoring algorithm? Recall that the intent of Shor's factoring algorithm is to compute the period of this function, f of x equals a to the x mod N. In this expression, N is a number we wish to factor. x is the argument to the function. a is a random number, which is chosen that co-prime to N. It shares no prime factors in common with N. The period of this function, r, allows us to find with some probability, a high probability, a factor of N, which is non-trivial. Quantum factoring, thus, has as a goal finding this period, r. Solving this problem takes an amount of time, or several operations roughly proportional to L cubed where L is the number of digits of the number N. Classically, using the best-known algorithm, the time required, or several operations required roughly go as e to the L/3. The difficulty of this factoring problem is at the root of what secures widely used cryptosystems, such as RSA. To appreciate the smallest meaningful instance of Shor's quantum factoring algorithm, it is helpful to review the quantum circuit for the algorithm. Recall that there are two registers, the top one here, which holds the value x in a superposition of all possible values, and a bottom register, where the result of the modular exponentiation is held. The top register is then fed three inverse Quantum Fourier Transform, which returns the result. There are three main steps in this procedure. First, the input superposition of the top register must be created. Second, the modular exponentiation, a to the x mod N, must be computed. It is the most difficult part of Shor's quantum factoring algorithm because multi-qubit gates are required. The third step is a Quantum Fourier Transform. because the result of the Fourier transform is measured, only single-qubit gates, some of them controlled by classical results, are required in principle. After the result is measured, there are some classical post-processing. this can be done by essentially any classical computer. Now, the number we are factoring is N., And it is clear that the algorithm is too trivial when N is an even number. That is, it has a factor, which is 2. Thus, the smallest meaningful factors are 3 and 5. Here is a generic quantum factoring circuit for the case of N equal 15, namely 3 times 5. What makes this specific for N equal 15 is the choices available for the random base to be chosen, a. a must be co-prime with N. In other words, it must not discuss any factors with N. it must be less than N. Therefore, the only two meaningful cases are a being 7 and 11. Thus, the modular exponentiation circuit in the middle reduces to looking like this, where we have used repeated squaring as it is common to realize the modular exponentiation. Note that each one of these exponentiations is calculated mod 15. So, we began with a controlled multiplication by a, then a controlled multiplication by a square, a to the fourth, a to the eighth, a the 16th, and so, forth, all the way up to a to the 128th power. the result of each of these multiplications is still a number between 0 and 14 because of the modular multiplication being performed. So, two non-trivial values of a may be used in factor 15. we claim that one of these is easy to handle, and the other is hard. Why is this the case? well, let us see why by working out the two cases explicitly. Consider the case of a being equal to 11. Now, what we want is to take powers of a modulo 15. a to the 1 is equal to 11. Now, if we multiply 11 by 11, we get 121. That value, modulo 15, is simply equal to 1. Thus, the pattern continues to repeat. If we multiply again by 11, we get 11 and then 1, and then 11 and 1. So, computing 11 to the x modulo 15 for any value of x is really easy because it only requires one bit of x. thus, the quantum circuit looks simply like this. Because the period is 2 in this case, we find that the quantum circuit required is very simple. All it is a single controlled a multiplication circuit with 1 bit of x involved, the least significant bit. since only two possible values of multiplication by a matter, 11 and 1, it is tempting to replace this controlled by just a single controlled-NOT gate. thus, this case is the easy one. Now, consider the other case of a being equal to 7. Here, we find 7 to the 1 is equal to 7. 7 squared is equal to 49. Modulo 15 is equal to 4. 4 times 7 is 28. 28 modulo 15 is equal to 13. 13 times 7 modulo 15 gives us 1. thus, the period here is 4. So, the quantum circuit required needs 2 bits of x. thus, we find that the quantum circuit is much less trivial. It involves a controlled multiplication by 7, as well as a controlled multiplication by 49 modulo 15. In contrast to the a equal 11 case, this circuit is much more complicated. thus, this is known as the hard case. From this analysis, it is clear that only 7 qubits are needed to realize Shor's Algorithm for factoring 15. Here is the quantum circuit. This circuit includes both easy and hard cases. If we further consider what is needed for the controlled multiplication by a as well as the controlled multiplication by a squared modulo 15, we find that this sequence of gates suffices. It includes two controlled-NOT gates for the easy case and four additional controlled-NOT gates and two Toffoli gates for the hard case. Altogether, including the Quantum Fourier Transform and the superposition creation Hadamard gates, around 20 gates are required. some of these are not elementary gates. The Toffoli gates are three-qubit gates, controlled-controlled NOT gates. they can be very highly non-trivial to realize in a physical system. The first realization of Shor's quantum factoring algorithm was done with a nuclear magnetic resonance system, \cite{vandersypen_experimental_2001}. The NMR experiment did both the easy and hard cases and required around 200 pulses. Each of these pulses realizing some elementary operation, a sequence of which, for example, realized the Toffoli gate. It was done in 2001 at IBM. More recently, in 2007, Shor's Algorithm was also realized using a linear optics setup. However, only the easy case was implemented in that experiment. Thus, only 3 qubits were required, and 13 elementary gates were performed. this experiment required post-selection, meaning that the results were only correct when conditioned on a specific output being in a certain state. The first solid-state qubit implementation of Shor's Algorithm was done with a superconducting qubit system in 2012. Again, this was for factoring the number 15 using 3 qubits, and two controlled-NOT gates. A more complicated case factoring the number 21 was also done using linear optics quantum computing. However, again, only an easy case was performed with a 4-qubit system and 26 gates in 2011. One of the lessons from these experiments is that it is incredibly tempting to reduce this quantum factoring algorithm into something increasingly simple to realize with an experiment. factoring almost any number using a quantum algorithm can be done trivially by choosing the period r is equal to 2. In this case, the quantum circuit for Shor's Algorithm reduces to a very simple 2-qubit circuit shown here. It illustrates the ridiculousness of what happens when one compiles the quantum factoring algorithm down too far. This point is driven home by a beautiful article by John Smolin, Graham Smith, and Alex Vargo, which appeared in Nature in 2013, "Pretending to factor large numbers on a quantum computer." They observed that running algorithms on such tiny experiments is a somewhat frivolous endeavor when such extreme compilation is performed\cite{booth_comparing_2018}. So, one must be very careful in choosing a meaningful instance of Shor's Algorithm as well as a meaningful rendition of that instance in a physical system. If this is done, quantum factoring can be a useful test of scalability. One of the main points of the trapped ion quantum computing realization was how Shor's Algorithm could provide a meaningful and stringent test of the scalability, integration, classical control, and quantum control of quantum computers\cite{monz_realization_2016}. In the trapped ion realization of Shor's quantum factoring algorithm, there were four main goals. First, when doing both the easy and hard cases, it is clear that a cascade of multi-qubit gates is required. if each one of these gates fails with a small probability, the cascade rapidly fails because of the product of those probabilities decreasing very fast. So, to achieve high fidelity, the non-pole selected deterministic output is a challenging goal to realize. The trapped ion demonstration's second goal was to demonstrate the feed-forward control needed for the recipes used in fault-tolerant quantum computation\cite{chow_implementing_2014,corcoles_demonstration_2015,kandala_error_2019,colless_computation_2018}. A third goal was to perform not just quantum computation but also to integrate quantum memory with quantum computation. finally, one of the most important goals of building a process for building quantum computers is to be able to predict their performance in advance of realizing the hardware\cite{mcgeoch_principles_2019,mcgeoch_practical_2019}. thus, one of the goals of the trapped ion demonstration of Shor's quantum factoring algorithm was to demonstrate such predictive modeling of expected system behavior. Altogether, this meant that the goal of the trapped ion quantum computing realization of Shor's quantum factoring algorithm was to use it as a strategic test of scalability elements.

\section{Demonstrations of Shor's Algorithm: Quantum Factoring With Trapped Ions - Ion Trapping System} 
Let now discuss with us a description of the trapped ion quantum computing system employed for the demonstration of Shor's quantum factoring algorithm. Two energy levels of an atom can be employed as a qubit. However, it is ordinarily very hard to hold onto a single atom and not lose it. On the other hand, if an electron is removed from the atom, it becomes a positively charged ion that is much easier to trap and hold. Calcium 40 is a common ion which can be trapped in this way and employed for quantum computation because of its internal level structure. The so-called D5/2 and S1/2 levels of calcium 40 make excellent two-level systems used as a qubit because of its approximately one second lifetime, and the transition can be excited by photons, which is accessible to the solid-state lasers. In reality, however, the atom has other energy levels also involved. For example, the higher energy P3/2 level can connect with the D5/2 level via an 854-nanometer optical transition. The P3/2 level can also decay to the S1/2 ground state via a short wavelength 393-nanometer optical transition. It produces a closed cycle, which is useful for controlling the ion's motion. This cycle and the short wavelength photon emitted are also employed to measure the qubit state since this emission occurs if the qubit is in the S1/2 state but not in the D5/2 state. Two additional levels, the P1/2, and D3/2 levels, also form a closed cycle and provide the primary transitions used for laser cooling of the ion. How the ion itself is trapped, however, only involves its charge. Consider a single ion\cite{delehaye_single-ion_2018,huntemann_single-ion_2016}. Trapping is accomplished based on the following idea. Surround the ion with four electrodes shown here in cross-section. Two additional end cap electrodes in front and the back are not pictured. These metal electrodes hold positive and negative charges via electrical voltage is applied. Opposite charges attract and equal charges repel. However, if the ion were exactly in the middle, all forces would balance. More likely than not, an ion would initially be off-center, and thus, it would move towards the nearest oppositely charged electrode, such as the negatively charged one on the upper left. As the ion moves, we change voltages such that the close electrode becomes positive, repelling the ion before it reaches the electrode. The ion then moves onward, for example, to the next closest negative electrode on the bottom left. Again, we change voltages to keep the ion dancing around, never reaching one of the electrodes, always confined within the region defined by the electrodes. This idea is at the heart of the physical process for which Wolfgang Paul was awarded the 1989 Nobel Prize in physics. The left is a picture of an actual four-electrode quadripolar linear ion guide, which can be employed as an ion trap \cite{mueller_simulating_2011} when the end cap electrodes are added. By applying lasers, which cool the motion of the ion using its internal level structure, the trapped ions can be made to come to rest at the very center of the electrical potential. Such trapping and cooling have been demonstrated with a wide variety of atomic and molecular ions, including strontium, beryllium, ytterbium, barium, magnesium, and calcium. Physical parameters determined by the lasers and by the atomic structure determine how cold the ions get and how cooling competes with heating processes. For scalable quantum computation, the quadripolar ion structure can be simplified as follows. First, imagine rotating the electrode configuration by 90 degrees. Next, envision replacing the three lower electrodes with three coplanar electrodes beneath the ion. The top electrode becomes a top metallic conductor that may be far away from or part of the confining vacuum chamber. This configuration still traps ions. Lasers can be applied from the side just as before for cooling and also to perform quantum gates. The substrate can be cooled, for example, to liquid helium temperatures, to allow integration with devices such as superconducting electronics. This substrate can be micro-fabricated with modern semiconductor lithography techniques with geometries that include sophisticated patterning in the third dimension, such as the segmented structures shown here, comoving ions. A wide variety of such surface electrode ion traps have now been realized in laboratories around the world, including six pictured here from research groups at the University of Mainz, MIT, Maryland, Sussex, Sandia National Lab, and NIST. Designs for trapped-ion chips \cite{stuart_chip-integrated_2019,pino_demonstration_2020,harty_high-fidelity_2014,ballance_high-fidelity_2016,wang_high-fidelity_2020,gaebler_high-fidelity_2016,bruzewicz_trapped-ion_2019} include ones fabricated using sophisticated modern semiconductor line processes such as this one fabricated by Lucent Technologies and pictured installed in a vacuum chamber at MIT. The ion trapped chips have also been operated at cryogenic temperatures, including this one made of silver on quartz used at MIT to trap crystals of strontium ions, shown here with one, two, three, four, and over 20 individual atomic ions visible in the CCD image. The quantum factoring demonstration with trapped ions was carried out in this laboratory at the University of Innsbruck. Pictured here from photos we took during the opening collaboration in 2001, the ion trap here is a macroscopic hand machined quadripolar trap at the center of these magnetic field coils. It shows the same setup from another angle showing the optical components needed to guide all the laser beams to the trapped ions. It is the control computer and electronics used to generate the frequency and amplitude-controlled laser pulses, which perform quantum operations on the ion state. Professor Rainer Blatt, at the University of Innsbruck, his group and David Wineland at NIST set the standard for the field, as well as the character of scientific openness and collaboration that shaped the field of trapped ion quantum information science today\cite{nielsen_quantum_2011,preskill_quantum_2012}. The trapped ion quantum factoring demonstration was done with a linear ion chain of five calcium ions, each of which has the internal level structure shown earlier and reproduced. Recall that the qubit transition is between the D5/2 level and the ground state S1/2 level. Laser pulses focused on individual ions realize single-qubit rotations addressed at individual qubits. The laser is typically tuned to affect rotations about the qubit is z axes. Multi-qubit collective gates are realized either by applying a monochromatic laser beam to multiple ions, which implements in parallel many single-qubit rotations or by applying two simultaneous laser frequencies tuned such that a multi-qubit gate operation is realized, known as the Molmer-Sorensen gate, after its inventors. It is the mathematical form of the gate, which we can discuss more in atomic physics. The basic set of Molmer-Sorensen collective spin-flip and individual light shift operations available can be combined to realize arbitrary multi-qubit quantum gates. For example, the quantum Toffoli gate can be realized as this sequence of five collective three Molmer-Sorensen and three single qubit light shift gates. How well do the traps and gates work in practice? Here are performance numbers for the system employed for the quantum factoring demonstration. The system was demonstrated to reliably trap up to 14 ions with state preparation of 99.5\% fidelity, taking a few milliseconds for each preparation, routinely realizing T1 and T2 coherence times of one second and five seconds respectively. Requiring about 400 microseconds for a quantum state measurement when using a photomultiplier tube, and between 20 and 45 microseconds to realize single and multi-qubit gates, which have gate fidelities of over 99\% for single-qubit operations, and between 99\% and 94\% for two and five ion gates. Additional energy levels within the ions can also be employed as quantum memories with storage fidelity of 99.5\% and write times of 45 microseconds. 
\section{Demonstrations of Shor's Algorithm: Quantum Factoring With Trapped Ions - Experimental Results} 
Let us now go in-depth into the actual experimental demonstration of quantum factoring with trapped ions. Recall that this demonstration has four strategic goals. First, to demonstrate fast feed-forward control, the capability needed for quantum error correction used in fault-tolerant quantum computation\cite{steane_error_1996, jones_layered_2012}. Second, to integrate quantum memory with quantum computation, noting that data memory is a vital component of modern classical computation. Third, to cascade a sizable number of multi-qubit quantum gates deterministically and with high fidelity. fourth, to demonstrate predictive modeling of the system behavior allowing extrapolation to large scale system performance. The following four issues challenge these four goals. First, the Quantum Fourier Transform and the modular exponentiation circuit for the hard case naively requires over 200 gates. even if the fidelity of each is around 99\%, the product of these is expected to have a fidelity lower than 10\%. Second, multi-qubit gates, measurement, and state initialization can be slow, meaning that the total time required can exceed a millisecond, which approaches the available quantum coherence time. Third, multi-qubit gate fidelities typically decrease inversely with the number of qubits. Fourth, measurement readout causes ions to fluoresce, and this scattered light can decohere near ion qubits. This experiment brings two main new ideas to address these challenges. These are a new way to use additional magnetic sub-levels of electronic states as quantum memory and an algorithmic structure for the Quantum Fourier Transform originally developed by Alexei Kitaev, which reduces the number of qubits required. we now go through the challenge for each of the four goals starting with the first to demonstrate feed-forward control. For this, we employ Kitaev's version of the Quantum Fourier Transform, which replaces the multi-qubit circuit by a serial single-qubit circuit based on the fact that the outputs are measured. This example is for the 3-qubit QFT, which is what we need for the factoring demonstration. The trapped ion system is suitable for demonstrating Kitaev's QFT because of the ability to read out single qubits and to re-initialize and reset qubit states, all within a given computation which continues onward. we test Kitaev's QFT using this 3-qubit circuit with three different inputs. The experimental results in red and the expected theoretical results in blue for the three input states are shown here. The expected outputs match the experimental data with classical squared statistical output fidelity of 95\%. It is much better than the full three ion qubit QFT fidelity due to the quantum memory performance being higher than that of multi-qubit gates. The second challenge goal to integrate quantum memory with quantum computation is tested by implementing the quantum order-finding algorithm. Recall, this takes a permutation of N objects as input and finds the power r such that the permutation returns to identity. The quantum phase estimation circuit shown here solves the problem with an exponential speed-up over the classical algorithm compared by counting the number of uses of the permutation operation. For permutations of N equal 4 objects, this is the required quantum circuit. it simplifies to this 3-qubit quantum circuit using Kitaev's Quantum Fourier Transform. The experimental implementation of quantum order-finding tested these four permutations and produced these results shown in red, which matched theoretical expectations shown in blue, with fidelities of the 80\% level. The third challenge goal is to implement both the easy and hard cases of Shor's quantum factoring algorithm. Recall that this algorithm is essentially a special case of order-finding, where this modular multiplication function defines the permutation. Shor's algorithm begins with a random choice of a number a, and the algorithm finds the order of a to the x mod N. For factoring 15, and there are two non-trivial choices of a, namely 11 and 7, as we have seen. The a equal 11 case is easy because it has a power of 2. In contrast, the a equals 7 cases hard because it has a longer period of 4. we employ the Kitaev version of the QFT for the factoring algorithm with 5 qubits, an extra quantum memory qubit state, and feed-forward control. observe that for the a equal 11 easy case, only the controlled a multiplication need be implemented, and that requires only two CNOT gates realized using 16 laser pulses. For the a equals 7 hard cases, controlled a and controlled a squared multiplication must be implemented. after the peephole compilation, which only uses local information and is blind to the final result, the circuits required involve two 3-qubit quantum Fredkin gates and two CNOT gates, altogether needing 52 laser pulses. Such complex 3-qubit gates had not been realized with trapped ion systems before this experiment. Here are the experimental results for the easy and hard cases of Shor's Algorithm. The data are shown in blue, and the ideal noiseless theoretical expectations shown in red agree to over 90\% for both cases. This is an improvement over the NMR implementation from 15 years ago, which had a much worse performance for the hard case than the easy case. The fourth challenge is to predict the actual results, taking into consideration all known imperfections and noise sources without fine-tuning. we developed a tool for this, known as TIQC-SPICE, inspired by the similarly named SPICE tool important in simulating integrated electronic circuits. TIQC-SPICE compiles quantum algorithms into laser pulses, simulates the resulting ion internal and motional state from the pulses, and analyzes performance to produce the same metrics as obtained from the experiment. Shown here, the experimental in red, ideal theoretical in blue, and TIQC-SPICE simulated results in green for the easy and hard cases of the quantum factoring demonstration. The TIQC-SPICE results accurately predict the experimental results to within error bars, including four sources of major errors, namely addressing of individual ions by laser beams, dephasing of ion states due to fluctuations in the laser or magnetic field \cite{klimov_fluctuations_2018}, the spontaneous emission lifetime of the ion qubits, and measurement and gate times. In conclusion, the trapped ion demonstration of quantum factoring provides a meaningful test of scalability by demonstrating fast feed-forward control, quantum memory interleaved with quantum computation, multiple cascaded and deterministic multi-qubit gates, and predictive simulation of outputs using TIQC-SPICE. These results indicate that technical and not fundamental issues limit scalability. For the future, we now understand how modular multiplication circuits can be realized using an algorithm known as the Quantum Fourier Montgomery Multiplier as an elementary building block, such that the factoring of an N bit number would require 2n plus 2 qubits. on the order of N squared gates, of which less than 3N and are 3-qubit Toffoli gates, the hard ones to implement. A feasible future experiment could thus demonstrate the factoring of N equal 21 for which the hard case has a period of 6. Using the Quantum Fourier Montgomery Multiplier circuit, this would require 7 ions and 52 Molmer-Sorensen laser pulses. Altogether, this experiment, while tiny compared with those needed to break cryptographic codes, shows how meaningful conclusions and serious extrapolations can be drawn from this scientific endeavor.

\section{Demonstrations of Shor's Algorithm: Pretending to Factor} 
Peter Shor's 1994 polynomial-time quantum factoring algorithm is what started the quantum computing revolution. It was discovered at IBM Research in 2001 that 15 equals 3 times 5 using seven NMR qubits. Then it was discovered again in 2007 using a different technology, photonics, with just four qubits, and in 2009 using five, finally, in 2012, using only three superconducting qubits. Also, 21 was factored using a photonic qubit and qutrit, which is a three-level system. We have plotted these results versus time, and we may notice something strange. The number of qubits is going down.
Furthermore, one can extrapolate that by 2020, 15 will be factored without any qubits at all. So,  we started to wonder just What is going on here. How many qubits do we need to factor a number? First, a quick review of Shor's Algorithm. The core of the algorithm is as follows. To Factor a number N, we first pick a base a which is some integer. Then we find the period r of a to the power x modulo N. This period finding is the only quantum part of the algorithm. If we find that r is odd, or that a to the r over 2 is minus 1 mod N, we try again.
Otherwise, the greatest common divisor of a to the r over 2 plus or minus 1 and N is a factor. Euclid's algorithm, which is rather old technology circa 300 BCE, is efficient for finding the GCD. The quantum period finding step requires a register big enough to hold N and one more bit, which is used to pull out bits of the Quantum Fourier Transform using a technique known as qubit recycling. So, to Factor N, we need log N plus 1 qubit. For 15, this works out to five bits as the minimum and even more if qubit recycling is not used. So, how did people factor 15 with so, few qubits? It turns out that even the original 2001 experiment with seven qubits used an additional trick called in an abuse of the term compiling. What was noticed was that for some bases a, the period r is small. Then the quantum register used to find the period only needs to be big enough to hold r instead of N. Can we always find a base that results in a small r? It turns out that the answer is yes. We employ a much more modern result than Euclid's algorithm, the Chinese remainder theorem from 400 CE. It says that given two primes p and q, then a squared is 1 mod PQ if and only if a squared is 1 mod p and also mod q. If we then choose a equal plus or minus p PQ plus or minus q QP where PQ is the inverse of p mod q, and QP is the inverse of q mod p, then by construction, a squared equals 1 mod p and mod q.
Furthermore, we can find PQ and QP efficiently using what else Euclid's algorithm. In other words, for any N equals p times q, we have an efficient algorithm for finding a base a with a squared equals 1 mod N, or period r equals 2. So, we can now implement Shor's Algorithm using exactly 2 qubits and Factor an arbitrary N. The circuit looks like this. We can further simplify if we notice that the bottom qubit is never measured and realize that the circuit makes a maximally entangled state, and it measures just one of the qubits. So, all it does is produce random bits. We implemented this using the device shown. Here is the back of the device with a quarter next to it for scale. We factored 15 RSA 768, which we believe is the largest number ever factored on classical computers, and a 20,000-digit number of the own creation called N 20,000. Now, this devolved somewhere into nonsense. The trick the cheating part is that finding a base with a small period is easy only if we already know the factors p and q. Naturally, factoring is easy provided only that we know the factors. As it is the 100th birthday of Richard Feynman, so, let us leave we with this quote from his commencement address to the 1974 class at Caltech. "The first principle is that we must not fool yourself. We are the easiest person to fool. After we have not fooled yourself, it is easy not to fool other scientists." \cite{feynman_simulating_1982,hey_feynman_2018}

\section{Quantum Cryptography - Making Codes}
In the last section, we introduced the basics of modern cryptography schemes with a focus on public-key encryption based on one-way mathematical functions. We discussed the RSA cryptosystem in detail and showed that its security is based on prime number factorization and the related problem of period finding. The period finds it is a hard problem for a classical computer. We showed how it could be efficiently performed on a universal quantum computer running Shor's algorithm. We also discussed the topic of key distribution, namely the conundrum that arises when using symmetric keys for encryption and decryption. How can we securely exchange symmetric keys when we do not yet have the encryption keys in the possession for the first transmission? The answer was to use public-key encryption based on asymmetric keys, like the RSA cryptosystem. However, as we just noted, RSA is vulnerable to attack by a quantum computer running Shor's algorithm. However, where quantum figuratively closes one door, it can also open another. In this section, we will show quantum mechanics not only gives us the tools to compromise secure communication protocols but also provides the resources to enhance secure communication through quantum key distribution or QKD \cite{behnia_tachyon_2018}. In section one, we briefly discussed QKD in the context of a one-time pad, that is, distributing symmetric keys continuously to provide fresh secure keys that can then be used continually to encrypt data.
Here, we will introduce two general classes of QKD \cite{zhuang_floodlight_2016}. One is based on single-photon sources and detectors. The other is based on entangled photons and Bell state measurements. Both rely on many of the quantum concepts we discussed in section one, such as entanglement, projective measurement, and the no-cloning theorem. In this section, we will expand on those concepts. We will also discuss the importance of randomness in implementing QKD. Pseudorandom number generators output sequences of numbers with properties that are approximately random \cite{chattopadhyay_pseudorandom_2019}. However, because it is based on an algorithm that uses an initial value, called a seed, the sequences are never truly random. It has implications for cryptography schemes since, ideally, a current bitstream should not be predictable from the previous bitstream. On the other hand, the randomness of projective measurement is believed to be truly random. This concept, in conjunction with distributed entanglement\cite{wong-campos_demonstration_2017}, is a foundation for quantum random number generators.

\section{Post Quantum Cryptography} 
Cryptography is the art of using mathematical tools to achieve information security while using untrusted channels like the internet or wireless communication. We use this to confirm that the software we download every day by auto-updates came from a legitimate source and did not tamper\cite{williams_tamper-indicating_2016}. It is what we use to know an online shopping or banking web page is authentic. Public key signatures are one of the most commonly used methods to provide authentication and data integrity guarantees. Most of the encryption we use today to protect the confidentiality of information is symmetric-key encryption. However, how are these symmetric keys established? In a typical transaction, we might use a public key signature to authenticate our bank, then public key encryption to establish a symmetric key, and then AES to encrypt our confidential information. However, quantum computers will break the mathematical problems at the core of virtually all public-key cryptographic schemes in use today, namely integer factorization and the discrete logarithm problem. It means RSA and Diffie-Hellman schemes are decimated. Going to larger keys adds very little extra security against quantum attacks. Quantum computers also weaken symmetric cryptography because they can brute force search quadratically faster. That is, a quantum computer can effectively search space of N keys with only square root N quantum steps. Doubling key lengths can effectively mitigate the known quantum attacks against symmetric cryptography. Now, while acronyms like RSA and ECC might not mean much to most users of technology, these tools underpin the security of tools such as secure web browsing, VPN, secure email, auto-updates, blockchains \cite{kiktenko_quantum-secured_2018}, and so, on, which are the basis of technological infrastructure, including the internet, cloud computing, IoT, and so, on. So, to reap the benefits of quantum computing, it is imperative that we first deploy the new cryptographic tools designed to be safe against quantum attacks. It is sometimes called quantum-safe cryptography or post-quantum cryptography\cite{barbeau_secure_2019}. We do not believe that quantum computers capable of breaking RSA and Diffie-Hellman schemes in use today exist, but this does not mean we can just wait and see before doing something about it. Whether we need to take action now depends on three parameters. Firstly, what is the security shelf-life of the information our system is supposed to protect? Call this x year. Secondly, how long will it take to migrate our systems to ones designed to be secure against quantum attacks? Let y be the migration time. Lastly, what is the collapse time, the time until our threat actors can use quantum computers to attack our systems and compromise our information? Now, note that for the next y years, we are stuck with the vulnerable quantum tools. So, information is recorded during this period may be exploited by quantum computers in the future. This information is supposed to be protected for another x years. However, if x plus y is bigger than zed or z, then we will not be providing the x years of confidentiality that we are supposed to. For example, information may be for sale on the dark web while it is still supposed to be confidential. Suppose our information is supposed to remain secure for ten years after transmission, and it takes ten years before the systems protect those transmissions against quantum attacks. If quantum computers capable of decrypting that information are available in 15 years, then there will be a five-year period where those transmissions will be vulnerable to compromise before their 10-year shelf life is up. Further, if y is bigger than z, then systems collapse with no reliable way to rebuild them. That is, there is no effective remediation because the tools to protect against new attacks are simply not available. Also, it is important to note that rushing y will not only be disruptive and expensive but will lead to flawed and vulnerable implementations. While viable candidates exist, there is a long road to wide-scale, practical, robust deployment, including many challenges. we need to identify candidates. we need to scrutinize them against quantum and classical attacks. Protocols using the tools must also be secure. Software and hardware implementations need to be secure. we need to analyze the side channels and design countermeasures. we need to look at the performance requirements. we have to integrate them into tools and applications. we have to worry about migration and compatibility, standardization so, systems will interoperate and validation and certification, who can easily take 10 to 20 years to migrate existing systems properly. In August 2015, the National Security Agency of the United States announced that it would initiate a transition to quantum-resistant algorithms in the not-too-distant future. This sent a very clear signal to the world that quantum computing was something that needs to be taken seriously now. Shortly after, the National Institute of Standards and Technology, NIST, started outlining its plans to move forward and produce standardized post-quantum algorithms \cite{giri_review_2017}. It launched its post-quantum cryptography project, which is a process to solicit, evaluate, and standardize one or more quantum-resistant public-key cryptographic algorithms. The submission deadline for proposed post-quantum algorithms was November 30, 2017, and the first workshop to review the submissions was held in April 2018. Most submissions are based on the hardness of lattice problems or of solving multi-linear equations or of decoding error-correcting codes or computing kernels of elliptic curve isogenies. For signatures, there are also schemes based on the security of hash functions. we expect to see some NIST standards for public-key signatures and key agreement in the early to mid-2020s. Other standard bodies around the world are also studying this challenge and taking steps to repair the wide range of standards needed to protect the information technologies from quantum attacks. For example, standards for how the financial industry will use new algorithms. This need to upgrade the cryptographic systems might seem like a nuisance. However, there may be an additional positive impact. This very advanced warning of the systemic threat to the cyber systems not only gives us a chance to prepare the defenses against this threat, but, if done properly as part of lifecycle management and not as crisis management, we can rebuild the foundations of cybersecurity to be stronger than ever before. we can have a stronger and more agile cyber immune system capable of protecting, detecting, and responding to new, unexpected threats. 

The long-term security of modern cryptographic systems has been brought into question by developments in quantum computing. The existence of Shor's factoring algorithm shows that the inversion of the one-way functions needed for asymmetric cryptography is an instance where quantum computers have a quantum advantage over classical algorithms. It is worth noting that the development of quantum computers is not catastrophic for symmetric-key cryptography schemes. 

Based on the pace of quantum computing research and the effort required to update cryptographic standards, researchers have argued that the process of finding and implementing algorithms that are suitably robust to quantum computers should begin immediately.  Michele Mosca, a leading expert in quantum computing research, has estimated the odds of a quantum computer capable of breaking RSA-2048 an RSA cryptosystem using 2048-bit private and public keys being constructed by 2026 and 2031 at 1/7 and 1/2 respectively. These probabilities are informed estimates based on several parameters such as when a fault-tolerant and scalable qubit will be designed, how many qubits are needed to break RSA-2048, and how long it will take to scale quantum computers to this size.

Professor Mosca argues that, based on these projections, quantum computers capable of breaking RSA-2048 are reasonably likely to be available before the shelf-life of information encrypted with RSA-2048 has expired. Both NIST and NSA have advocated that action should begin now to safeguard cryptographic standards from quantum computers. In 2016, NIST began requesting nominations for post-quantum public-key cryptography algorithms, with the eventual goal of standardizing one or more quantum-resistant algorithms. Additionally, in 2015 the Information Assurance Directorate of the NSA announced it would begin transitioning to quantum-resistant algorithms.

The search for cryptographic algorithms and hardware which are secure against both classical and quantum computers is called post-quantum, or quantum-resistant, cryptography. At present, all post-quantum cryptography algorithms come from either quantum cryptography or by exploiting math problems, similar to prime factorization, which have no known efficient solution on either classical or quantum computers.

Quantum key distribution (QKD) uses fundamental quantum-mechanical properties of systems, usually implemented with light, to securely exchange keys over an insecure public channel. Unlike other solutions, the security of QKD does not rely on assumptions about an adversary's computational resources and can be considered provably secure. Realistically, many QKD implementations are open to attacks on the hardware or software which perform the protocol. At present, QKD is a mature and active area of research with commercial systems available.

Alternatively, two of the most promising post-quantum cryptographic schemes are lattice-based cryptography and code-based cryptography. Lattice-based cryptography makes use of problems on a set of points in n-dimensional space, which have a periodic structure; these sets are called lattices. An example of a lattice problem used in cryptography is the "shortest vector problem," which is based on the shortest non-zero vector for a given lattice. Code-based cryptography uses the properties of error-correcting codes to secure information. Intuitively, these systems consist of a public key that adds random noise to a signal and a private key, an error-correcting code capable of removing the noise. While it is still unclear which of these methods will become a post-quantum cryptographic standard, it is unlikely that the new approach will be compatible with existing systems. Rather, new hardware will likely be required, at significant expense, and therefore careful planning will be required before the current standards are updated.

\section{Single Photon Sources and Detectors} 
Quantum mechanics provides a means to enhance information security by encoding information in quantum mechanical systems, such as qubits. then distributing those qubits from point A to point B. However, how is this done in practice? This section will introduce quantum communication protocols and how they are implemented using single and entangled photons. To get started, before discussing photons, Let us first discuss light in general. Light is an example of an electromagnetic field, and electromagnetic fields are responsible for the many phenomena we commonly associate with electricity and magnetism. For example, it is what causes electricity to flow through a wire, making bar magnets attract or repel one another. when whether propagating in a material or an empty space, the electromagnetic field is a wave that oscillates at a given frequency. these oscillations are what we call light. Now fundamentally, the electromagnetic field is quantum mechanical and so, light, as with all quantum mechanical systems, is quantized into discrete quanta called photons. The fact that light could be described as both a particle and a wave is What is known in quantum mechanics as wave-particle duality. this is why we can discuss discrete particle-like photons having a particular frequency or a wavelength, which are both wave-like properties. Now, like any wave, a photon will oscillate with a given frequency and in a particular direction. So, let us take those one at a time. In the vernacular use of the word, light is generally part of the visible electromagnetic spectrum, the part which we can see, with different colors being different frequencies of light. However, more generally, light as we are using the term and its associated photons can exist at any frequency. From the radio frequencies used in radio communication bands to microwaves and terahertz, infrared, visible, ultraviolet light, and on up through x-rays and even $ \gamma $ rays. Photons are the quanta of light across this entire electromagnetic spectrum. Now, in addition to frequency, an electromagnetic wave oscillates in a particular direction. It is called polarization, and it generally refers to the direction that the electric field is oscillating. Once that is established, Maxwell's equations determine the direction of the magnetic field as well. So, the direction a photon's electric field oscillates is called its polarization \cite{noauthor_classification_nodate}. when a photon travels through space, its electric field can oscillate up and down, it can oscillate side to side, and it can even oscillate in an arbitrary, complex superposition of these two directions. we are saying superposition here deliberately. Horizontal and vertical photon polarizations are orthogonal quantum states. So, we can put these polarization states, H and V, on the poles of a block sphere and treat them as qubit states, zero and one. Similarly, on the equator of the Bloch sphere, the X-axis corresponds to diagonal and anti-diagonal linear polarized light, H plus or minus V. as one goes from the north to the south pole along the X-Z plane, the polarization remains linear and rotates from horizontal at the north pole to diagonal at plus X, to vertical at the south pole, anti-diagonal at minus X, and back to horizontal again at the north pole. Now, the Y-axis corresponds to right and left circularly polarized light. Circular polarization results from having a complex coefficient, such as H plus or minus we times V. so, the polarization rotates in time as the photon propagates. Elsewhere on the surface of the block sphere, the polarization is neither purely linear nor purely circular, but instead is generally elliptical. one can think of the linearly and circularly polarized light as special cases of a distorted ellipse. note that in optics, they give the sphere a different name. They call this a Poincare Sphere. their convention is to swap the Y and Z axes so that circularly polarized light is at the poles, and linearly polarized light is on the equator. In any case, we will use the Bloch convention here. The electromagnetic field is quantized into photons. The polarization of a single photon can be used as a qubit with horizontal and vertical polarization at the north and south poles. Now, we generally do not encounter pure, single-photon states in everyday life. First, there are typical, on average many more than a single photon from a given source. For example, the light coming from our computer screen right now contains many many billions of photons, each with some arbitrary polarization. second, such classical light sources are generally statistical combinations or mixtures of many different photon number states. To implement QKD protocols based on single-photon polarization, we need to find a way to generate single photons with a specific polarization state. To begin with, a pure quantum state of n photons is called a Fock State. it is stated with exactly n photons. what we would like to do here is generate a one-photon Fock State. The Fock States are called non-classical states of light because they generally do not exist in the classical world around us. At least, not on their own. However, there is a form of classical light called a Coherent State, which comprises an infinite superposition of Fock States. Lasers and microwave oscillators \cite{goto_bifurcation-based_2016,goto_quantum_2019}, for example, create this type of light. While a photon Fock State contains a specific number of photons, a coherent state has a probabilistic distribution of photon numbers. For a given coherent state $ \alpha $, the probability of measuring n photons is given by a Poisson Distribution centered around some average number of photons, the magnitude of $ \alpha $ squared. It means that a Coherent State will always have some probability of being in a one-photon Fock State. However, its presence is stochastic and very unlikely when the average photon number is larger than one. Now, a quick and dirty way to produce a single photon is to attenuate a coherent source simply. If we pass laser light through an extremely opaque filter, the average photon number decreases, and the beam gets dimmer. As the average photon number decreases, the Poisson Distribution shifts, and the probability of measuring a low photon number Fock State increases. To make this concrete, imagine we attenuate a laser light until the average photon number is only 0.1. Plugging this into the formula for the probability of measuring n photons, we find that 90\% of the time, we will measure zero photons. Nine percent of the time, we will measure one photon. only one percent of the time will we measure more than one photon. So, with light attenuated to this degree, we can say with reasonably high confidence that if we have a photon, it is likely to be in a single photon state that we need. there is only a one percent chance that it is a multi-photon Fock State. However, while attenuation is by far the simplest way to produce a single photon, this procedure is far from ideal. After attenuation, the light is still in a Coherent State, not the pure single-photon Fock State that we want. while we exceeded increasing the probability of measuring only one photon, this came at the expense of lowering the probability of measuring any photons at all. In the above example, nine out of ten times, there was no photon. as we further lower the average photon number, we will certainly decrease the chance of getting any multi-photon states. However, at the same time, we dramatically increase the likelihood that we obtained no photons. Thus, the rate at which we obtain single photons goes down, which translates to a lower communication rate. A far better scheme for single-photon generation relies on an intermediary quantum system, such as an optically active quantum dot. As we discuss in section one, a quantum dot has a ladder of quantized energy levels. A unique energy splitting separates each pair. transitions between energy levels are generally associated with the absorption or emission of a single photon. The idea is to drive the quantum dot to a higher excited state and then let natural, spontaneous emission occur. The quantum dot relaxes back into its ground state, emitting photons at frequencies corresponding to the energy level spacings of its transitions. Importantly, the final transition from the first excited state to the ground state is generally unique and emits a photon at a specific frequency. We can block all frequencies of light except the frequency of this final single-photon if we filter the photon emission from the quantum dot. This final single photon passes through the filter, creating exactly the single-photon source we need. Although the relaxation process is also stochastic in time, the timing jitter of the single-photon emission can be minimized to the extent to the dot relaxes quickly. in more sophisticated sources, the emission process may even be mediated by a clocked pulse to obtain a heralded stream of single photons. we can then pass these photons through a polarizing filter and prepare the polarization corresponding to the qubit state we need for the QKD protocol. Now, in addition to a single-photon source, we also need a device that can detect single photons at the other end of the communication channel. The challenge here is to convert the small amount of energy stored in a single photon into a measurable electrical signal. A common example of a single-photon detector is a Geiger mode avalanche photodiode. Essentially, a semiconducting material such as silicon with a large external voltage applied across it. When a photon hits the detector, its energy is sufficient to dislodge an electron from the semiconducting material. The external voltage then accelerates this electron through the bulk semiconductor colliding and dislodging more and more electrons, creating an avalanche effect. This cascade of dislodged charges generates a large measurable electrical current, which indicates that a photon was absorbed. Other examples of single-photon detectors include photomultiplier tubes, superconducting nanowires, and superconducting transition-edge sensors. In generating or detecting single photons, these processes should be performed with high quantum efficiency. For example, a photon is generated on each clock cycle, or every photon that is generated is detected without loss. In the coming section, we will see how the source and detector quantum efficiencies come into play when implementing quantum key distribution protocols.

So, the readout for these quantum computers with few qubits are based on bulky microwave components \cite{randall_efficient_2015}. How should these readout chains develop to actually make use of quantum computers with hundreds of qubits? So, that is particularly true for the solid-state implementations. So for example, superconducting qubits, or semiconducting qubits, the ones that obviously are operated in the microwave regime are going to have these microwave components. in fact, we think that, up to about 1,000 qubits or so, we will be able to do what we are doing today, and just bruteforce it. So up to around 1,000, that is not a hard number. But going beyond 1,000, we start thinking about 10,000, 100,000, a million qubits, then, we are not going to be able to bruteforce it, and we have to come up with other ways to readout these qubits. Now, whether that means we are going to take the current, bulky microwave components that we have got and integrate them into smaller form factors, that is one approach. So for example, today we use microwave isolators, or circulators, so the signal goes away from the quantum computer towards some subsequent amplifiers, but noise is coming back down towards the qubit that is either blocked or diverted to another direction. those today are bulky, as we say. But they are ideas, and some early demonstrations using topological materials. A class of materials called topological materials, or two-dimensional materials, that can show something called chirality, or basically quantum Hall-like effect, but without magnetic field. so, these still need to be improved, and proven out, but there are ideas out there of how we can actually integrate some of these bulky components into a smaller form factor, and in particular, in an integrated manufacturing sense. so, these are the kinds of things that have to happen, right, for these, in particular, the quantum computers that use these bulky, microwave components to scale, as we said, beyond the thousands of qubits.\\ 
\textbf{Single-Photon Generation:}\\
The concept of a photon, a quanta of light, was first introduced by Albert Einstein in 1905. A photon embodies a discrete amount of energy proportional to its frequency. Depending on the characteristics and statistical properties of the photons' source, the light may be adequately represented classically called "classical light," or it may necessitate a quantum description called "non-classical light."

The conventional light sources we encounter daily, for example, from a light bulb, are generally classical sources emit so-called "chaotic light." Chaotic light originates from a large ensemble of independent and incoherent photon emitters. The incoherent and uncorrelated behavior of the ensemble can be perceived as being chaotic, having large intensity fluctuations, and thus the term "chaotic light." A special case the laser is an ensemble of photon emitters acting in concert to create a coherent light source that essentially behaves as a classical, coherent sinusoidal wave.

In contrast, a typical example of non-classical light is a single-photon source, a system that emits one (and only one) photon at a time. There are two general approaches to realizing single photons, but only one of them is non-classical. The classical approach is to attenuate a classical multi-photon field such as a laser field to ensure that when a photon arrives, it is likely to be only a single photon. It is just attenuated classical light, and, the attenuation must be large enough to reduce the probability of multi-photon arrivals. While large attenuation ensures that most arrival events are single photons, it comes with the trade-off that those single-photon arrival events are relatively rare, and, most of the time, there are no photons.

The second approach is truly a non-classical light source based on the spontaneous emission of a single photon from a two-level quantum system. A two-level system in its excited state will spontaneously relax to the ground state and, in doing so, emit a single photon with energy $\Delta E $(the energy separation between the two levels). The frequency f of this photon is $\Delta E / h$, where h is Planck's constant. The precise time at which the two-level system emits the photon is generally a random variable described by an exponential probability distribution $(1/\tau ) \exp (-t/\tau ), $where $\tau $is the characteristic excited-state lifetime. In this case, the single-photon emission time is on average the excited state lifetime $\tau.$ Examples of single-photon emitters include quantum dots\cite{hardy_single_2019}, quantum-well heterostructures, and certain atoms or molecules with energy-level transitions at the desired emission frequency.\\

\textbf{Single-Photon Detection:}\\
Single-photon detectors are devices that respond (or figuratively "click") when a photon is absorbed. Importantly, the detector should ideally distinguish between photon absorption events and background noise. It becomes particularly challenging when detecting low-energy photons, such as microwaves.

A detector's "internal quantum efficiency" is a measure of how well a detector works, and it is essentially the probability that a photon is accurately detected, given that it enters the detector. Increasing the photon absorption probability and the detector sensitivity will increase the likelihood of detecting a single photon and generally improve quantum efficiency. However, doing so may also make the detector more sensitive to background noise, generating responses even when no signal photons are present. Such errant photon detection events are referred to as ``dark counts" and reduce the signal to noise ratio (SNR), negatively affecting the overall detection efficiency.

Avalanche photodiodes (APDs) are one type of a single-photon detector. These detectors are generally semiconductor p-i-n diodes, where p, i, n refer to a hole-doped region, an intrinsic region (no doping), and an electron-doped region of the semiconductor. The diode is operated under a strong voltage bias, which imparts a large electric field across the intrinsic region of the device. When a photon is absorbed, it knocks out an electron. This electron is then accelerated by the electric field and proceeds to knock free other electrons. This process cascades, creating an "avalanche" effect that results in a large, measurable electrical current pulse. The presence of an electrical pulse thus indicates that a photon was absorbed.\\

\textbf{Beamsplitters:}\\
A beamsplitter is an optical device that partitions an incoming stream of photons and either transmits or reflects those photons to the output ports. The figure below shows a four-port beamsplitter with input ports (1 and 2) and output ports (3 and 4). As shown, photons impinge on the beamsplitter at port 1 and are transmitted to port 3, and reflected port 4. Photons that arrive at port 2 would similarly be transmitted to port 4 and reflected port 3. There are two basic categories of beamsplitters of interest here: spatial beamsplitters and polarizing beamsplitters.

\begin{figure}[H] \centering{\includegraphics[scale=.5]{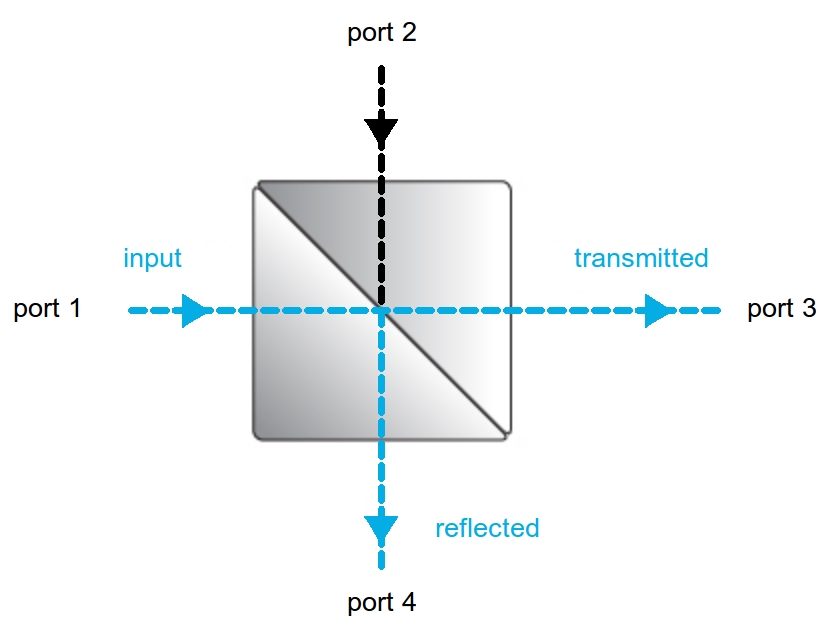}}\caption{Beam Splitter}\label{fig2_7}
\end{figure}
\textbf{Polarizing Beamsplitter:}\\
A polarizing beamsplitter is a device that selectively reflects or transmits photons depending on their polarization (the orientation of its electric field). For example, light with polarization V being reflected and H being transmitted. This type of beamsplitter is generally constructed from anisotropic crystals. Light passing through anisotropic (non-uniform) crystals will refract at different angles and spatially separate depending on the light polarization. By choosing the appropriate crystals and angles of incidence, light with one polarization (e.g., vertical) can be completely reflected, while opposing polarization (e.g., horizontal) is completely transmitted. Polarizing beamsplitters are thus a key component in performing a photon polarization measurement. One can infer the photon polarization with a photon detector at each output by measuring a detection event in either the transmission arm or the reflection arm.

\begin{figure}[H] \centering{\includegraphics[scale=.9]{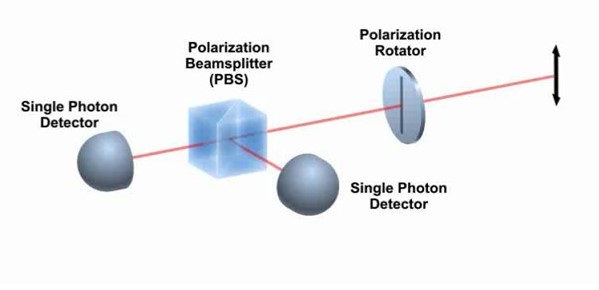}}\caption{Polarized Beam Splitter}\label{fig2_8}
\end{figure}
\textbf{Spatial Beamsplitter:}\\
A spatial beamsplitter takes an incoming photon field and transmits it to the output port with a probability T, and reflects it with a probability R. When T=R=50\%, we call this a ``50:50 beamsplitter.'' A spatial beamsplitter is generally described by a 2x2 matrix $U_{\textrm{BS}}$ that takes incoming photon-field amplitudes (we will call them $a_1$ and$ a_2 $for short) and outputs transmitted and reflected fields ($a_3$ and $a_4$).
\begin{equation}\label{eq2_70}
\displaystyle \left( \begin{array}{cc} a_3 \\ a_4 \end{array} \right)    \displaystyle =    \displaystyle U_{\textrm{BS}}\left( \begin{array}{cc} a_1 \\ a_2 \end{array} \right)     
\end{equation}

\begin{equation}\label{eq2_71}
\displaystyle =    \displaystyle \left( \begin{array}{cc} t & r \\ r & t \end{array} \right) \left( \begin{array}{cc} a_1 \\ a_2 \end{array} \right)
\end{equation}

\begin{equation}\label{eq2_72}
\rightarrow a_3=ta_1+ra_2 ~ ~ \& ~ ~ a_4=ra_1+ta_2
\end{equation}     

Since the fields $a_ i $are amplitudes, it follows that obtaining the photon field energy or photon number requires taking the magnitude squared $|a_ i|^2 = a_ i^* a_ i.$ In cases where these are quantum-mechanical photon-field operators, one similarly calculates an ensemble average $ \langle a_ i^{\dagger } a_ i \rangle$. The matrix unitarity condition $U_{\textrm{BS}}^{\dagger }U_{\textrm{BS}}=I$ requires that $|r|^2+|t|^2=1$ and $ r^*t + t^*r=0$, and this ensures that photons are neither lost nor gained in the beamsplitting process. In other words, it conserves probability T+R=1 where $ T=|t|^2 $is the transmission probability and $R=|r|^2$ is the reflection probability. For a 50:50 beamsplitter, one can take $r=i/\sqrt{2} $and $t=1/\sqrt{2}$. With this matrix relation in hand, one can calculate the average output field intensity and the intensity fluctuations induced due to the beamsplitter partitioning process.

\begin{enumerate}[wide, labelwidth=!, labelindent=0pt]
\item Beam Splitters; When a single photon enters an ideal, lossless 50:50 spatial beam splitter at port 1 shown in Fig. 1, it is partially transmitted to port 3 and partially reflected port 4. Therefore, a single photon leaves the beamsplitter in a superposition state of being in both the transmission and reflection arms. A detection event then randomly collapses the photon into one of those arms, with a 50\% chance of detecting the photon in the transmission arm, and a 50\% chance of detecting it in the reflection arm. On the other hand, if the beam splitter is a polarizing beam splitter illustrated in Fig. 2, then the horizontal component of the photon is transmitted and detected at port 3, whereas the vertical component gets reflected and detected at port 4. Which of the following phrases is valid?
\begin{figure}[H] \centering{\includegraphics[scale=.15]{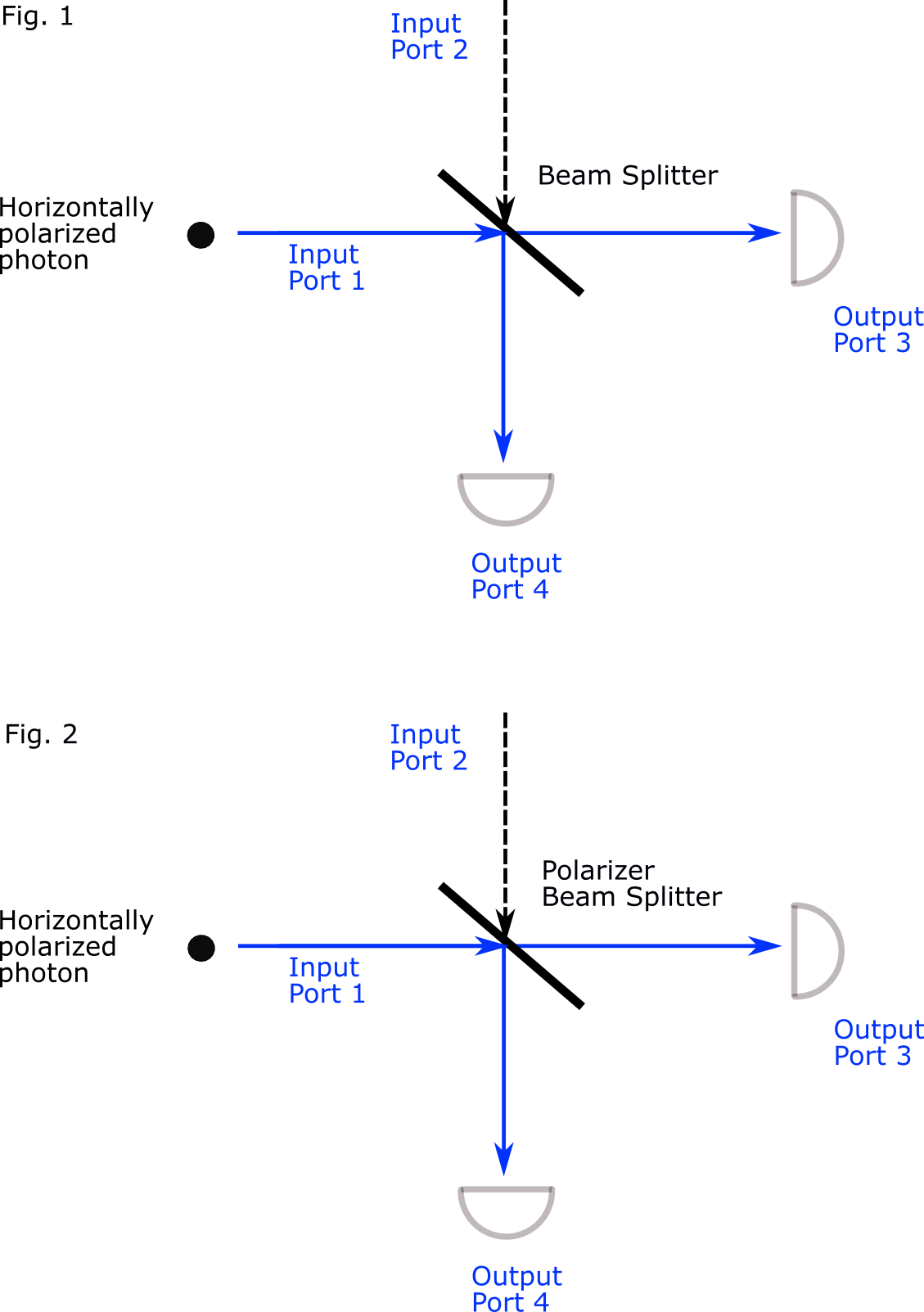}}\caption{Beam Splitters}\label{fig2_33}
\end{figure}

\section{Quantum Key Distribution - BB84} 
Ordinary classical information everybody understands that because it surrounds us. It is like the information in a book. we can copy it at will, and it is not disturbed by reading it. However, what is quantum information\cite{nielsen_quantum_2011,mermin_quantum_2007,national_academies_of_sciences_quantum_2018}? Rather than give a mathematical description, we try to find a metaphor for it. It is rather like the information in a dream. If we try to describe our dream to someone, our memory of it changes, and eventually, we forget the dream, and we only remember what we said about it. we cannot prove to someone else what we dreamed, and we can lie about our dream without getting caught. However, there is no really hard science of dreams, despite the effort of Sigmund Freud. unlike dreams, there is a good mathematical theory of quantum information. Indeed, when we develop it, we will see where classical information comes from, that we used to think was just simple information. well, here are some examples of elementary quantum behavior. we can use single photons of light to carry one bit of information because a horizontal photon, as at the top of the figure, can be reliably distinguished from the vertical photon, as in the middle. if we pass them through a calcite crystal like that, or crystal of any other sufficiently un-symmetric material, the horizontal photons will get passed straight through, and the vertical ones will get deviated by some amount. therefore, we can count them in two separate detectors. that means each photon can carry one reliable bit of classical information. However, if we put in photons at some other angle, instead of being deviated by an intermediate amount, as we might expect, they behave probabilistically. Some go into the upper beam and become horizontal, and the others go into the lower beam and become vertical. It is, we might say, one of the central mysteries of quantum mechanics, the random behavior of things that are individually identically prepared. We can use vertical and horizontal photons to send one bit per photon of the message. However, if we put in diagonal photons, well, simply by rotating the apparatus, we can distinguish a left diagonal from a right diagonal, like a slash from a backslash. However, once that rotated apparatus is used, that thing cannot distinguish the vertical from the horizontal. So, we have four separate states of the photons that are pairwise, perfectly distinguishable. Nevertheless, we cannot distinguish all four kinds. Now we want to get us to worry a little bit about this strange behavior. Bill Wootters, a physics professor, has a pedagogic analog for quantum systems' behavior when measured. He says it is really like one of those old fashioned 19th-century British schools where the students, who are not supposed to ask any questions they were just supposed to respond to the questions the teacher posed, and where the student is the photon, and the teacher is the measuring apparatus. So, the strict teacher is asking the student a question, saying, is our polarization vertical or horizontal? Furthermore, the students start saying, and we polarized about 56, 55 degrees. we believe we asked a question. Are we vertical or horizontal? Oh, horizontal. Did we ever have any other polarization? No, we were always horizontal. So, that is how quantum systems behave when measured, and that is what lies behind many of the interesting things we can do with it. ultimately, when we come to entanglement, we can understand where that behavior comes from. well, now, we will discuss how we put it to use. This is the quantum money that was proposed by Wiesner in 1968. He says a piece of quantum money will have a serial number on it like ordinary money, but it also will have 20 perfectly reflective boxes, each containing a single photon of one of those four states. the bank will have a record based on the serial number of which photon is in which box, which kind of photon, and when we bring the money back to the bank, the bank will open the boxes, one by one, and measure each photon in the way that it would behave reliably, and therefore check that they all agree with what was supposed to be stored there. However, if we do not know What is stored there, we will disturb some of them by measuring them the wrong way, and the chance of passing a fake bill or making a copy that passes is 3/4 to the 20th power, which is small. It is the idea that Gilles Brassard and Peter Shor used in building the quantum cryptography apparatus where we do not try to store the photons, but we send them over a short distance, which in this case was only about 30 centimeters. the idea is that any eavesdropper in that empty middle section of the apparatus would spoil the sum of the transmitted photons by measuring the wrong way, and therefore, would betray their presence. So, in more detail, the way quantum key distribution works are that we have two parties. we call them Alice and Bob. Alice sends a stream of polar ice photons of those four kinds to Bob. Eve is the nickname for the eavesdropper. She can try to measure those photons if she wants, but if she does, she will introduce errors with high probability. the other thing we allow her to do is to listen to the classical discussion between Alice and Bob. that takes place through a series of ordinary classical messages exchanged over a public channel. when they are done with all this, they end up each with a shared secret key, in other words, a random string that they both know, and they have high confidence that no one else knows. well, that is the case if there is not too much eavesdropping, or there is no eavesdropping, but if there is much eavesdropping, they will detect a lot of errors and effort to distill this key will have failed. However, they will know that it has failed with high probability. they just try again tomorrow. So, in more detail, here is how the protocol works, which is called BB84, after the names of the author and the year that it came out. Alice, in the first line, sends a random sequence of these four kinds of photons. Bob, on the other side, makes random measurements uncorrelated. He does not know what Alice would send, and he randomly chooses to make a rectilinear measurement that is vertical versus horizontal or diagonal. So, he does, and then in the third line, we get his measurement results. we including the realistic effect that some of the photons may get lost or his detectors may not be perfectly efficient. So, in some cases, he does not get anything. For example, in the first case, Alice sent a vertical photon. He chose to make a rectilinear measurement, and he gets a vertical result, which is correct. In the second case, Alice sent a horizontal photon, but he decided to make a diagonal measurement, and he gets a random one of the two diagonal results. However, he does not know that it is bad. It is a 45-degree photon, and so, on for the others. The next stage of the protocol now is, after he is received them, Bob tells Alice which kind of measurement he did. He said we measured the first one rectilinearly, the second one diagonally, the third and fourth one rectilinearly, the fifth one we did not get, and so on. That is indicated by the classical message in quotation marks. then Alice tells him in which cases he did the right thing. That is, he made a measurement that did not spoil the polarization that she sent. She very carefully does not say whether she said vertical or horizontal, but she says in the first case, we made a rectilinear measurement, which was good. Keep that photon. Do not tell what it was, because somebody might be listening. In the second case, too bad, we made a diagonal measurement. we sent a rectilinear photon, and so, on. So, they cull from the photons that Bob received, the ones that he spoiled. we can identify, say, a vertical photon is a 1, a horizontal is 0, or for the diagonal photons, they choose to identify the 45 degrees as a 0, and the 135 is a 1. So, what they have, essentially, at this stage is a bunch of bit values that they ought to agree about if there has been no eavesdropping. So, the next question is to find out whether there has been eavesdropping, and we can do those various ways, but the most economical way that does not involve giving away a lot of the data is for Alice to choose a random subset of the bit positions, and announce the parity of that random subset. in the first case, she chose this collection outlined in red, and it has odd parity. She says that it is odd. then the second one, she chooses a different random subset and says that is even, and they go on for  20 of these random subsets. if they agree about all of them, then the chance that there is even one mistake in their sequence is about one in a million. So, they are pretty happy with that. they have to throw away some of the photons to compensate for the fact that they have leaked 20 bits of information to the eavesdropper. So, that is the elementary version of the BB84 protocol. 

Quantum key distribution (QKD) is a protocol used to distribute shared secret keys with security guarantees that are ideally based on the properties of quantum mechanics, rather than on the presumed computational resources of an eavesdropper. This may be contrasted with classical methods of key exchange, such as the RSA cryptosystem, where security is based on the presumed classical computing complexity of the order-finding problem.

The first QKD protocol, now commonly referred to as BB84, was introduced in 1984 by Charles H. Bennett from IBM and Gilles Brassard from the University of Montreal. Since then, there have been several proposed QKD protocols, many of which have been demonstrated, including long-range experimental demonstrations via satellite and even commercially available QKD systems.

The BB84 protocol can be used by a sender (Alice) and a receiver (Bob) who wish to create a shared secret key. It requires that Alice and Bob have access to a quantum channel \cite{shor_capacities_2003}over which Alice will send polarized photons to Bob and a classical, broadcast channel for coordination purposes. By a broadcast channel, we mean that an eavesdropper (Eve) can listen to and record all information sent over this channel, but she cannot alter it.

To begin the protocol, Alice produces, records, and sends Bob single photons with polarization encodings each chosen at random from one of the four polarization states:\\
\begin{center}

$\displaystyle \vert \psi _ H\rangle    \displaystyle =\vert H\rangle ,$    \\     
$\displaystyle \vert \psi _ V\rangle    \displaystyle =\vert V\rangle ,    $     \\
$\displaystyle \vert \psi _+\rangle    \displaystyle =\frac{1}{\sqrt{2}}(\vert H\rangle +\vert V\rangle ),    $ \\     
$\displaystyle \vert \psi _{-}\rangle    \displaystyle =\frac{1}{\sqrt{2}}(\vert H\rangle -\vert V\rangle ).    $ 
\end{center}

The first pair and the second pair of states are orthogonal $\langle \psi _{H}\vert \psi _{V}\rangle =0 $ and $ \langle \psi _{+}\vert \psi _{-}\rangle =0 $ and, therefore, each pair forms a polarization measurement basis. As such, we can think of photon polarization states as qubit states, with the photon polarization being either in states H and V or in states + and -. Importantly, note that while the states within a basis are orthogonal, states between the two bases are not orthogonal with one another, for example, $\langle \psi _{H} \vert \psi _+ \rangle \neq 0.$

Next, Bob measures each photon in one of the two bases either \{ H, V\} or \{ +,-\} which he chooses at random for each photon, and records the measurement results. Once complete, Bob announces the basis he used for each measurement, but not the result, over the broadcast channel. Because Alice and Bob independently chose the polarization basis at random for each photon, there is a 50-50 chance that they chose the same one for any given photon. For each photon where they did choose the same basis, Alice and Bob will have recorded the same polarization value, because the states within each basis are orthogonal, and we assume an ideal measurement.

For example, if Alice sends $\vert \psi _{H}\rangle$ and Bob measures in the $\{ H,V\}$ basis, then Bob will also measure and record an H, assuming perfect measurement, since $ \langle \psi _{H}\vert \psi _{H}\rangle =1$ and $ \langle \psi _{H}\vert \psi _{V}\rangle =0$. It is considered a successful trial because both Alice and Bob recorded an H for this photon. However, for the other half of the photons, where Alice and Bob did not select the same basis, the results are uncorrelated, because the two bases are maximally conjugate. It means that a measurement in the wrong basis is equally likely to result in either of the two incorrect outcomes, for example, $\vert \langle \psi _{H, V}\vert \psi _{+,-}\rangle \vert ^{2}=\frac{1}{2}.$ It would be considered an unsuccessful trial.

The security of BB84 starts with the fact that Eve gains no information from listening to the broadcast channel since only information about bases is available and not the measurement results. Thus, she must try to gain information from the photons in the quantum channel \cite{gyongyosi_survey_2018}. However, any attempt by Eve to extract information reveals her presence through the introduction of abnormalities in the statistics of Bob's measured results. As a result, Eve is unable to extract information encoded by Alice without alerting Alice and Bob that she is trying to do so.

A straightforward strategy Eve could use is to intercept each photon and measure it on some basis, record the result, and then send a photon to Bob based on her measurement outcome. In doing this, Eve destroys Alice's original photon, and due to the no-cloning theorem, Eve cannot precisely reproduce each of Alice's photons. If Eve got lucky and selected the same polarization basis as Alice did, then she, by chance, will send Bob a photon with the same polarization. However, when Eve does not choose the same basis as Alice, Eve obtains the wrong measurement result, infers a different polarization (without knowing it), and thus sends Bob a photon with a polarization different than Alice's original photon. Consequently, the statistics of Alice and Bob's measured results successful trials versus unsuccessful trials and their statistics will be altered by Eve's activities and reveal her presence.

Realistically Alice and Bob do not want to throw out the entire key every time an anomaly occurs since even without an eavesdropper, noise and imperfect measurement will always lead to some level of errors and statistical variation. Alice and Bob adopt a strategy called privacy amplification to deal with this problem, which produces a new, shorter, shared key from the original one, which can, in principle, reduce Eve's inferred information to zero.

Before implementing privacy amplification, Alice and Bob must first ensure their keys match. They can do this using classical error correction methods: having Alice randomly choose two bits from her key, adding them together modulo 2, and announcing the result to Bob. He, in turn, tries the same and replies with his result. If they match, the two bits are kept. If they differ, the two bits are discarded. Just sending the result of modular addition over the classical channel does not reveal any information to Eve since it only indicates the parity of the bits, that is, whether the two bits have the same or different values, but not their specific values. Using this technique, Alice and Bob will arrive at two matching keys. However, Eve may still have knowledge of a few of these bits (but, certainly not all of them).

Next, the privacy amplification protocol is applied. Privacy amplification comprises a series of transformations on Alice and Bob's matching key to create a shorter key, one which Eve cannot replicate without knowing the entire initial key. For example, Alice and Bob jointly choose pairs of bits at random from their keys and replace them with their binary addition. Intuitively, this works because of parity; Eve cannot know the value of the binary addition without knowing the value of both initial bits. Based on the error rates found in Alice and Bob's initial keys, they can estimate how much information Eve has and then reduce it to the desired level (approaching zero) using the privacy amplification protocol.

So far, we have only dealt with eavesdropping in the ideal case; however, real-world implementations of BB84 and other QKD protocols are potentially open to a much wider range of system-level vulnerabilities. For example, if attenuated classical light is used to generate single photons rather than a true single-photon source, photon number splitting attacks become possible on residual states with more than one photon. In this case, Eve can split off and measure one of the photons in the multi-photon state without impacting the polarization that reaches Bob. Besides, access to the hardware at Alice and Bob's locations must be restricted to trusted personnel only. Otherwise, an Eavesdropper can gain access to the random number generator used to select bases or even the encoded keys themselves by other means. There are just a few examples of the types of attacks, many of which are independent of the quantum nature of QKD that may compromise the quantum-enhanced security afforded by an ideally implemented QKD system.

So, the Chinese satellite perform QKD  \cite{liao_satellite-relayed_2018}, used to perform quantum cryptography, and the secure communication, and, how does this work, That satellite is independently communicating with two points on the ground. so, it is the satellite that is doing the exclusive or of the message with the pads, then independently, when it is over one city, it is communicating with that city, and then it carries the information when it is on the other side of the earth, communicating with another city. it is then also communicating with that side. so, the satellite is acting basically as a relay, and performing the quantum cryptography during that transit.

\section{Entangled Photon Sources and Detectors} 
So, for this section, we have introduced single-photon sources and detectors and seen how single photons can be used to encode and transfer quantum information across a communication channel, for example, to distribute a cryptographic key. In this section, we will discuss the generation in the detection of entangled pairs of photons. later this section, we will discuss how entangled photons may also be used for quantum key distribution, as well as for quantum teleportation and random number generation\cite{podoshvedov_efficient_2019}. As discussed in section 1, an entangled state is a manifestly quantum mechanical multi-qubit state that cannot be written as a product of single-qubit states. Let us start with a counter-example. we have seen through section 1 and section 2 that if we apply Hadamard gates to 2 qubits, each prepared in state zero, then we obtain the 2 qubits equal superposition state, 0, 0 plus 0, 1 plus 1, 0, plus 1, 1. While it might appear that this state cannot be factored into single-qubit states, we know it is possible because we generated it by applying Hadamards to single qubits. as we have seen, this state is the tensor product \cite{orus_practical_2014,orus_tensor_2019} of the single-qubit superposition state 0 plus 1. Thus, it is not an entangled state, because we can write it in terms of the single-qubit states ; 0 plus 1 for qubit A, tensor product 0 plus 1 for qubit B. In fact, 0 plus or minus 1 are the excited and ground states along the x-axis of the block sphere. Thus, we can write this as the plus and minus state in the x spaces. In contrast, we can see that state 0, 0 plus 1, 1 cannot be written as a tensor product of single-qubit states. it is an entangled state of 2 qubits called a Bell state. even when we perform a transformation to the x basis, we still get a state that cannot be factored into single-qubit states A, tensor product B. This is a property of entangled states. They remain untangled even on different bases. looking at the circuit model, we see that we needed a conditional 2 qubit gate to generate the entanglement. Single qubit gates alone were not enough. we will return to Bell states in a moment. Recall from earlier this section that single photons have a polarization, for example, horizontal and vertical. since vertical and horizontal polarizations are orthogonal quantum states, we choose to put them at the poles of the block sphere and call this the qubit. For single-photon QKD protocols, like BB84, this is sufficient. Bob preparers a single photon in a particular qubit state that is a particular polarization state and then sends it along the communication channel to Alice. However, as we will see in the next couple of sections, using entangled photons enables certain alternative types of QKD protocols, as well as a means to generate random numbers in teleport quantum information. Thus, the question is, how do we generate and detect entangled photons? The most common method for producing entangled photons is through a process called spontaneous parametric downconversion. Parametric downconversion is essentially a frequency conversion process, where a photon at frequency 2f is converted into two individual photons whose frequencies together add to 2f and conserve energy, for example, each photon having a frequency, f. Now, linear time-invariant systems, by definition, do not convert frequencies. If we take a sine wave and pass it through a linear system, its amplitude and phase may change, but its frequency remains the same. It is essentially a statement that a plane wave e to the i omega t is an eigenfunction of a linear system. Thus, to take a photon at one frequency and convert into two photons at a different frequency, we need a material with strong nonlinear optical properties. a common example of such a crystal is beta barium borate or BBO. When a strong pump beam passes through a BBO crystal, there is a small probability that one of its photons will convert into two photons, one called a signal, and the other called an idler. It is just convention. Now, due to energy conservation, the signal and idler frequencies are each half of the pump frequency. due to momentum conservation, the signal and idler photons will leave the crystal in correlated spatial directions. Additionally, these two photons will also have correlated polarizations. Depending on the type of crystal we use. The two downconverted photons will either have the same polarization, which we call a Type 1 Crystal, or they have opposite polarizations from a type 2 Crystal. BBO is a Type 2 Crystal. We see that the two downconverted photons exit the crystal in directions that form a cone, one having horizontal polarization and one having vertical polarization. The locations of the photons on the cones are correlated, equidistant from the original beam trajectory, again, due to momentum conservation. For example, the colored dots on the polarization cones indicate the correlated spatial positions of the emitted photons. Now, crucially for the two photons to be polarization entangled, their direction of emission cannot indicate which photon is vertically polarized and which is horizontally polarized. Therefore, we use an iris to selectively keep only photons that arise from positions where the two polarization cones cross. Thus, whether the signal and idler photons travel left or right once they exit the crystal, one will have polarization h and the other polarization v. Nevertheless, which is not known. In other words, the direction of propagation does not label the polarization. Under these conditions, the state of the two downconverted photons is polarization-entangled, and it is all of the form H, V plus e to i $ \phi $ V, H. The phase $ \phi $ can be made 0 degrees by choosing the correct phase-matching conditions, crystal thickness, or using a phase shifter. Thus, one has the state H, V plus V, H. This is one of the four Bell states. from this, the other Bell states can be generated by changing the phase $ \phi $ and swapping the polarizations H and V, using a half-wave plate. Now that we are able to generate all four entangled Bell states Let us see how to detect these states. we first need to introduce an optical device called a polarization beam splitter \cite{bouland_generation_2014}. A polarization beam splitter will transmit or reflect a photon depending on its polarization, H or V. at each output of the beam splitter is a single-photon detector. By seeing which detector clicks, we can determine whether the photon had horizontal or vertical polarization. To see how this works, let us try a test case. we begin with a photon polarized in the vertical direction. we also show here for completeness a polarization rotator, a device that changes the polarization of incoming photons. For now, however, we just set it so that it passes photons without making any changes. If we send in a vertically polarized photon, it reflects off the polarization beam splitter, which is detected by the single-photon detector in the reflection arm. On the other hand, if we send in a horizontally polarized photon, it is transmitted by the polarization beam splitter and detected using the single-photon detector in the transmission arm. Thus, we can determine the photon polarization by seeing which detector clicks. So, what happens if we have entangled photons? Let us take the H, V plus V, H entangled state. Photon A travels to the left polarization beam splitter, and Photon B travels to the right polarization being splitter. what we will observe are correlated detection events. That is, if Photon A's polarization analyzer indicates horizontal polarization, then the Photon B analyzer indicates vertical polarization. Similarly, if Photon A is measured to have a vertical polarization, then Photon B must have horizontal polarization. These are correlated detection events. Now, the presence of correlation suggests polarization entanglement, but it is not proof of entanglement. After all, the source may randomly send out classical states of either H, V, or V, H, which would give the same results. However, it is not an entangled quantum state H, V plus V, H. To test for entanglement, one needs to measure the entanglement in different polarization bases\cite{monz_14-qubit_2011}. As we discussed earlier, changing the bases does not change the entanglement. Thus, the measured correlations will be retained in any measurement basis. In practice, rather than changing the measurement hardware, one may instead rotate the input state's polarization. This is experimentally more straightforward, and this is the role of the polarization rotator. With correlation measurements made on different polarization bases, one can compare the measured results with the classical probability model, assuming there is no entanglement. This type of test is called Bell's inequality test. it is used to determine the degree of entanglement of a quantum state within certain assumptions, called loopholes. we will discuss more Bell state measurements, test the Bell's inequality and loopholes later in the section, as well as the application of Bell state measurements to the problem of random number generation. 

Entanglement is a manifestly quantum mechanical state of two or more quantum systems whereby the state as a whole cannot be reduced to a representation of independent, single-system states\cite{song_10-qubit_2017}. For example, the two-qubit state $\vert \Psi \rangle _{12}=\frac{1}{\sqrt{2}}(\vert 0\rangle _{1}\vert 1\rangle _{2}+\vert 1\rangle _{1}\vert 0\rangle _{2}),$ cannot be written as a tensor product of two single-qubit states $\vert \phi \rangle _{1}\otimes \vert \psi \rangle _{2}.$ Moreover, we will find that we always get terms that do not appear in $\vert \Psi \rangle $. The entanglement of two quantum mechanical two-level systems results in what is referred to as Bell states, and there are various means to generate and detect the kinds of quantum states. However, they all share certain features, as we describe below, including the need for some type of two-body ``nonlinearity'' to generate and detect these states, whether a nonlinear crystal, a controlled-NOT gate, photon-number resolving detectors that distinguish 0, 1, and 2 photons.

In the case of photons, a common approach to generated Bell states is spontaneous parametric downconversion (SPDC), as described in the section. In this case, a nonlinear crystal, such as one made from beta-barium borate (BBO), converts incident ``pump'' photons at frequency $\omega _ p $ into two photons: a signal photon at frequency $\omega_ s$ and an idler photon at frequency $\omega_ i$. The process is energy conserving, and hence the frequencies of the two newly created photons will sum to that of the original pump photon: $\omega_ p = \omega_ s+\omega_ i$ (Planck-Einstein relation). Similarly, the process is momentum conserving, and the vector momenta similar sum together as $\vec{k_ p} = \vec{k_ s}+\vec{k_ i}.$ Since the momenta $\vec{k_ s}$ and $\vec{k_ i}$ are vector quantities, momentum conservation correlates both the magnitude of the momentum and the directionality of the emitted photons.

\begin{figure}[H] \centering{\includegraphics[scale=.65]{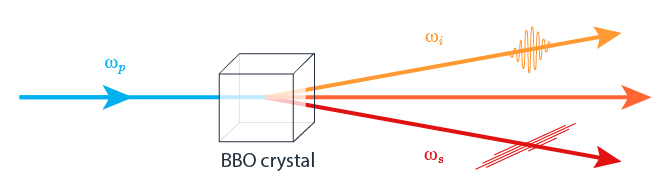}}\caption{Spontaneous Parametric Down conversion}\label{fig2_9}
\end{figure}

There are two general types of SPDC. Type-I SPDC generates two photons with the same polarization, and Type-II SPDC will generate two photons with orthogonal polarization. In both cases, the photons are correlated and, in certain instances, polarization-entangled. As we discuss in the section, BBO is a Type-II SPDC crystal, and so its output comprises a horizontally (H) and a vertically (V) polarized photon. The photon emission forms two cones, one for each polarization. The locations of any two down-converted photons on their respective cones are correlated due to momentum conservation. Thus, in general, the direction of the emission ``labels'' the photon polarization, such that the resulting two-photon state is not polarization-entangled. However, by selecting photon emission only from the two locations where the two cones cross, the emission direction no longer uniquely specifies the polarization and, in this case, a polarization-entangled state is formed of the form $\vert \Psi \rangle =\frac{1}{\sqrt{2}}(\vert H\rangle _{s}\vert V\rangle _{i}+e^{i\phi }\vert V\rangle _{s}\vert H\rangle _{i})$. By choosing the correct phase-matching conditions (e.g., size of the crystal, etc.), one can set the value of $\phi $ and generate the state $\vert \Psi ^-\rangle =\frac{1}{\sqrt{2}}(\vert H\rangle _{s}\vert V\rangle _{i}-\vert V\rangle _{s}\vert H\rangle _{i})$. In the following section, we will look at how to generate the three other Bell states starting from $\vert \Psi ^-\rangle.$

\begin{figure}[H] \centering{\includegraphics[scale=.6]{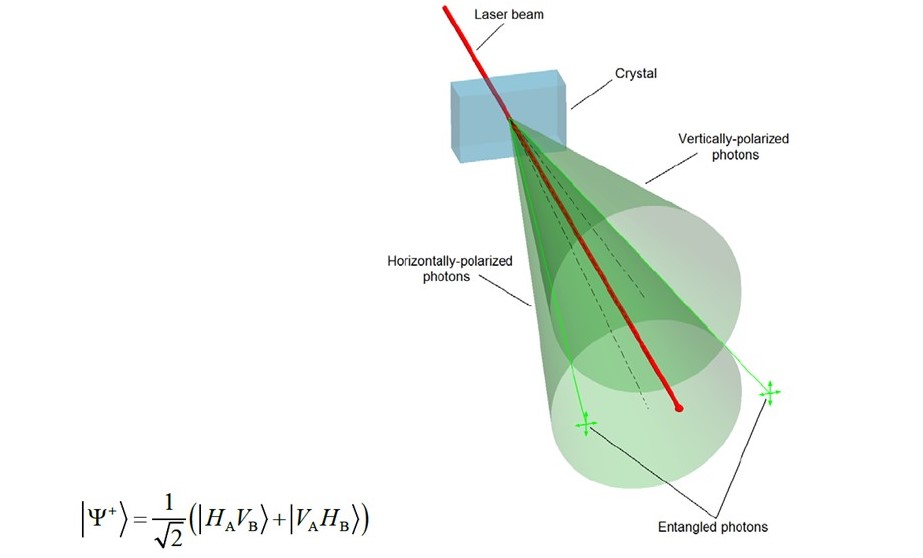}}\caption{Spontaneous Parametric Down conversion}\label{fig2_10}
\end{figure}
\textbf{Generating the Four Bell States:}\\
In general, it is possible to create all four Bell states, given just one of them. For a polarization-entangled Bell state with horizontal and vertical polarization, this can be accomplished using optical elements called ``waveplates.'' Waveplates rotate the polarization axis around the direction of propagation by a specific angle chosen by the design of the waveplate. The waveplate material a birefringent crystal such as quartz has polarization-dependent refractive indices which enable a particular polarization to propagate faster through the material the ``fast axis'' than the polarization orthogonal to it the ``slow axis.'' These two axes form a two-dimensional coordinate system.

Suppose a linearly polarized photon is incident on a waveplate. The overall phase accumulated while the photon travels through the waveplate depends on the waveplate thickness corresponding to the distance the photon has to propagate to pass through the optical element. However, in addition, the photon polarization is projected on to the waveplate's coordinate system, and the portion along the slow axis accumulates an additional phase with respect to the component along the fast axis.

If the ``extra'' phase is equivalent to one-half the incident photon wavelength, the waveplate is referred to as a ``half-wave plate.'' Moreover, the angle between the half-wave plate coordinate system relative to the photon's polarization axis determines the projections onto the fast and slow axes and, thus, the rotation of the polarization.

With two half-wave plates, it is possible to generate all Bell states starting from one of them. Two transformations need to occur: 1) a polarization change that swaps $\vert H\rangle$ and$ \vert V\rangle,$ and 2) a phase shift $\phi =\pi$ that swaps + and -. To swap polarization, the half-wave plate's coordinate system is oriented $45^\circ$ with respect to the photon polarization axis. To impart a phase shift, another half-wave plate's fast axis can be oriented $-45^\circ$ with respect to the horizontal axis to induce a phase inversion.

The following table summarizes the required optical elements a polarization swapping half-wave plate, $HWP_{45^\circ }, $and a phase inversion half-wave plate, $HWP_{0^\circ } -$ to alter the polarization of the first photon $p_1$ or the second photon $p_2$ and what general quantum gate that corresponds to to generate all four Bell states starting with $\vert \Psi ^+\rangle =\frac{1}{\sqrt{2}}(\vert H\rangle _{1}\vert V\rangle _{2}+\vert V\rangle _{1}\vert H\rangle _{2})=\frac{1}{\sqrt{2}}(\vert 0\rangle _{1}\vert 1\rangle _{2}+\vert 1\rangle _{1}\vert 0\rangle _{2}).$

\begin{table}[H]
\centering
\caption{optical elements - a polarization swapping half-wave plate}
\label{tab:2_1:Table 2}
\resizebox{\textwidth}{!}{
\begin{tabular}{|c|c|c|c|c|}  \hline
Target  & p\_2: Optical Element & Gate & Polarization & General \\ \hline
$\vert \Psi ^+\rangle$ & No elements &  $I\otimes I$ &  $\frac{1}{\sqrt{2}}(\vert H\rangle _{1}\vert V\rangle _{2}+\vert V\rangle _{1}\vert H\rangle _{2})$  &    $\frac{1}{\sqrt{2}}(\vert 0\rangle _{1}\vert 1\rangle _{2}+\vert 1\rangle _{1}\vert 0\rangle _{2})$\\ \hline
$\vert \Psi ^-\rangle$    & $HWP_{0^\circ }$ & $I\otimes Z$ & $\frac{1}{\sqrt{2}}(\vert H\rangle _{1}\vert V\rangle _{2}-\vert V\rangle _{1}\vert H\rangle _{2})$ & $\frac{1}{\sqrt{2}}(\vert 0\rangle _{1}\vert 1\rangle _{2}-\vert 1\rangle _{1}\vert 0\rangle _{2})$\\ \hline
$\vert \Phi ^+\rangle$    &  $HWP_{45^\circ }$ & $I\otimes X $ & $\frac{1}{\sqrt{2}}(\vert H\rangle _{1}\vert H\rangle _{2}+\vert V\rangle _{1}\vert V\rangle _{2})$ & $\frac{1}{\sqrt{2}}(\vert 0\rangle _{1}\vert 0\rangle _{2}+\vert 1\rangle _{1}\vert 1\rangle _{2})$ \\ \hline
$\vert \Phi ^-\rangle$    &    $HWP_{0^\circ }, HWP_{45^\circ }$ & $I\otimes ZX $ &  $\frac{1}{\sqrt{2}}(\vert H\rangle _{1}\vert H\rangle _{2}-\vert V\rangle _{1}\vert V\rangle _{2})$ & $\frac{1}{\sqrt{2}}(\vert 0\rangle _{1}\vert 0\rangle _{2}-\vert 1\rangle _{1}\vert 1\rangle _{2})$   \\ \hline
\end{tabular}}
\end{table}
\textbf{Detection of Bell States:}\\
To determine the success of the generation of a particular type of Bell state, detection schemes that distinguish all four Bell states are necessary. A deterministic measurement of Bell states generally requires nonlinear elements and two-particle interactions capable of performing conditional two-qubit rotations, two-particle interferometry (e.g., Hong-Ou-Mandel experiment sensitive to quantum statistics), number-resolving photodetectors, and the like. The details and trade-space of the various types of Bell state analyzers are beyond the scope of this text unit. For our purposes here, we will consider one example of a Bell-state analysis which uses the CNOT gate as the conditional\cite{chen_demonstration_2008}, two-qubit gate and apply it to our photon polarization-entangled Bell-state example.

Since photons interact only weakly (if at all), it is challenging to make reliable, nonlinear optical elements that implement deterministic, nonlinear operations with high efficiency. Linear optical elements are reliable, but they cannot be used to distinguish all four polarization-entangled Bell states deterministically. However, there are probabilistic schemes that can distinguish the Bell states using linear optics. In this case, one must identically prepare the same Bell state many times and run protocols using linear optical elements followed by measurement. Through these schemes, effective nonlinear interactions can be realized probabilistically and detected through an appropriate measurement scheme. When a spectator measurement indicates that an effective interaction has occurred, the results from that specific experimental trial are kept and aggregated to perform the Bell-state analysis; this is called post-selection. For example, although there is no deterministic CNOT gate with solely linear optics, there is a scheme for implementing a probabilistic, CNOT gate with linear optics that can be leveraged with measurement post-selection.

The following general protocol is able to distinguish all four Bell states. It can be implemented in a probabilistic manner with a probabilistic CNOT gate following Emanuel Knill's, Raymond Laflamme's, and Gerard Milburn's proposal and a beamsplitter as a Hadamard gate. Note that in the notation below, we use $\vert 0\rangle$ and $ \vert 1\rangle$ to represent the polarization states $\vert H\rangle $and $ \vert V\rangle, $ so as not to confound the polarization state $\vert H\rangle $ with the Hadamard gate H .\\
$\displaystyle \vert \Psi ^{+}\rangle =\frac{1}{\sqrt{2}}(\vert 0\rangle _{1}\vert 1\rangle _{2}+\vert 1\rangle _{1}\vert 0\rangle _{2}) \xrightarrow {CNOT}$     $\displaystyle \frac{1}{\sqrt{2}}(\vert 0\rangle _{1}\vert 1\rangle _{2}+\vert 1\rangle _{1}\vert 1\rangle _{2}) \xrightarrow {~ H\otimes I~ }    \displaystyle \vert 0\rangle _{1}\vert 1\rangle _{2}     $\\     
$\displaystyle \vert \Psi ^{-}\rangle =\frac{1}{\sqrt{2}}(\vert 0\rangle _{1}\vert 1\rangle _{2}-\vert 1\rangle _{1}\vert 0\rangle _{2}) \xrightarrow {CNOT}$     $\displaystyle \frac{1}{\sqrt{2}}(\vert 0\rangle _{1}\vert 1\rangle _{2}-\vert 1\rangle _{1}\vert 1\rangle _{2}) \xrightarrow {~ H\otimes I~ }    \displaystyle \vert 1\rangle _{1}\vert 1\rangle _{2} $\\
          
$\displaystyle \vert \Phi ^{+}\rangle =\frac{1}{\sqrt{2}}(\vert 0\rangle _{1}\vert 0\rangle _{2}+\vert 1\rangle _{1}\vert 1\rangle _{2}) \xrightarrow {CNOT}    \displaystyle \frac{1}{\sqrt{2}}(\vert 0\rangle _{1}\vert 0\rangle _{2}+\vert 1\rangle _{1}\vert 0\rangle _{2}) \xrightarrow {~ H\otimes I~ }    \displaystyle \vert 0\rangle _{1}\vert 0\rangle _{2}$\\    
      
$\displaystyle \vert \Phi ^{-}\rangle =\frac{1}{\sqrt{2}}(\vert 0\rangle _{1}\vert 0\rangle _{2}-\vert 1\rangle _{1}\vert 1\rangle _{2})\xrightarrow {CNOT}    \displaystyle \frac{1}{\sqrt{2}}(\vert 0\rangle _{1}\vert 0\rangle _{2}-\vert 1\rangle _{1}\vert 0\rangle _{2}) \xrightarrow {~ H\otimes I~ }    \displaystyle \vert 1\rangle _{1}\vert 0\rangle _{2}$ \\

The two-qubit states are measured after a CNOT gate, and the Hadamard gate on the first qubit is applied. The measurement reveals the underlying type of the Bell state.

\section{Quantum Key Distribution - Ekert91 and BBM92} 
There are various forms of quantum key distribution or QKD described by different protocols. They all make use of quantum mechanical properties to establish shared lists of random numbers that can be used for encryption. we have discussed the BB84 QKD protocol where one user, Alice, prepares a specific state that she sends to another user, Bob. Bob then measures the state. Alice and Bob then perform sifting by comparing the preparation basis and the measurement basis and then following several classical postprocessing steps to establish secure encryption keys. Protocols of this sort, which make use of quantum superposition where one party prepares a state and the other party measures it, are called to prepare and measure QKD protocols. An alternative set of protocols is called entanglement-based protocols, and they make use of entanglement. Two common and related entanglement-based protocols are the BBM92 protocol, named for its inventors Bennett, Brassard, and Merman. Its invention, the year of 1992, and E91 or Ekert protocol, was invented by Artur Ekert in 1991. In both of these protocols, a transmitter consisting of an entanglement source generates a two-photon entangled state known as a maximally entangled Bell state, with one photon being sent to Alice and the other photon being sent to Bob. The photons are sent across quantum communication channels that can reliably transmit the single-photon states and preserve their entanglement. The relevant property of Bell states is that the photons are correlated in multiple measurement bases, including the basis consisting of the horizontal and vertical photon polarizations and the 45-degree rotated basis consisting of the diagonal and the antidiagonal polarizations. For an ideal transmission across the channel, if Alice and Bob choose the same measurement basis, their measurement results will be perfectly correlated between the measurement devices. However, individual measurements at each side will look random. Alternately, if Alice chooses the horizontal-vertical basis and Bob chooses the diagonal-antidiagonal measurement basis, the measurements will not be correlated at all. for intermediate basis choices, their measurements will be partly correlated. In the BBM92 protocol, Alice and Bob received their entangled photon, and both choose either the horizontal-vertical or the diagonal-antidiagonal measurement basis. They both record their basis choice and what the actual measurement result is. They repeat this many times, and they now have long lists of chosen measurement bases and measurement results. The remaining protocol steps are identical to the previously described BB84 protocol; namely, Alice and Bob use a classical communication channel to announce their chosen measurement bases to sift out only the instances of them choosing the same basis. They then compare a random subset of measurement results to determine how well the quantum channel performed. if the determined error rate is below 20\%, they perform classical postprocessing to erase any information that a potential eavesdropper could have discovered and correct for remaining errors between Alice and Bob. The benefit of the BBM92 protocol is that the protocol does not require that we trust the source that prepares and sends the state. The protocol ensures that there is no way for a malicious actor to take control of the source and spoof Alice and Bob without them detecting it by observing lower correlation strengths and thus higher error rates than are required to form a secure key. The other entanglement-based protocol, the Ekert protocol, is similar but adds an interesting twist. In this protocol, the same sort of state is generated and distributed between Alice and Bob. Alice randomly chooses between three measurement bases: the horizontal-vertical basis, the 45-degree rotated diagonal-antidiagonal basis, and an intermediate basis rotated by 22 and 1/2 degrees. Bob also randomly chooses between three bases the horizontal-vertical basis, the intermediate 22 and 1/2-degree rotated basis, and a final basis rotated by 67 and 1/2 degrees. The measurement bases and the measurement results are recorded. Alice and Bob repeat this many times and again announce their chosen measurement bases over a public classical communication channel. The instances where they have chosen the same basis are used for the key-generation process. The instances where they have chosen different bases are used to determine the performance of the channel. For these instances, the measurement results are announced publicly and are used in a Bell inequality to determine how well the channel transmission has preserved entanglement. This allows Alice and Bob to have an entanglement-based measurement of the maximum amount of information a potential eavesdropper could have obtained. Alice and Bob then perform classical postprocessing to erase any information an eavesdropper could have discovered and to correct for remaining errors between Alice and Bob. That is how entanglement can establish encryption keys, eliminate the need to trust the system transmitter, and check the security of the encryption keys. There are various other QKD protocols with differing benefits and tradeoffs that may be of interest to specific scenarios. However, the important point is that they all make use of the same basic ideas and concepts that we have described here. 

The first quantum key distribution (QKD) protocol BB84 encoded information in the polarization of single photons. BB84 uses a quantum channel to communicate the photons, and a classical channel to broadcast the chosen polarization bases used for preparation and measurement.

In this section, we will present an alternative method proposed by Artur Ekert in 1991, known as Ekert91, which encodes information using polarization-entangled Bell states. Ekert91 also uses a quantum and classical channel. However, the advantage of the Ekert91 protocol is that its security can be traced directly to fundamental properties of quantum mechanics entanglement and measurement through a test called Bell's inequality. In this section, we will discuss Bell's inequality and its use in Ekert91 to communicate information securely and detect the presence of an eavesdropper.\\
\textbf{Entanglement and Bell's Inequality:}\\
In 1935, Albert Einstein, Boris Podolsky, and Nathan Rosen (EPR) argued that quantum mechanics is an incomplete theory in a scientific article titled ``Can Quantum-Mechanical Description of Physical Reality Be Considered Complete?.'' Their argument was a thought experiment that can be related to two-particle entanglement. Essentially, imagine that we take a polarization-entangled state of two photons, and we then send one photon each to Alice and Bob. A very large distance separates them. If Bob and Alice measure on the same basis, then their results should be correlated, for example, if Bob measures H, then Alice measures V., if they measure on different bases, then their measurements are uncorrelated.

However, what happens if Alice and Bob are separated by a distance so large that they can make their independent measurements faster than it would take light (and thus information) to travel between their locations? Then, we are left with the conundrum called the EPR paradox. Essentially, due to entanglement, there is a quantum-mechanical correlation in the measurements made by Alice and Bob. It is more than simply ``if Alice measures H then Bob measures V.'' Bob's results correlated or uncorrelated depend on Alice's choice of basis; but, Bob makes his measurement long before any knowledge of Alice's choice could have reached him or his measurement apparatus. Einstein did not like such ``spooky action at a distance,'' and thus hypothesized that quantum mechanics theory must be incomplete, that there must be ``hidden variables'' that quantum theory does not account for to connect these measurements. So the perceived randomness of quantum mechanical measurement is a mirage.

Niels Bohr, a contemporary of Einstien, responded the same year in a scientific article with the same title and argued that the theory of quantum mechanics only captures the interaction of a quantum system with a measurement apparatus and not the intrinsic character of it. Therefore, any conclusion based on the assumption that measurement does not disturb the measured system in question such as the EPR conjecture that quantum theory is incomplete without hidden variables must be invalid\cite{bennett_concentrating_1996}.

The debate continued for decades. In 1964, John Bell proposed a theorem later referred to as Bell's inequality, which can be used to test if hidden variables do exist experimentally. Essentially, Bell's inequality is a statement based on purely classical arguments. If the inequality is violated in experimental measurements, quantum mechanics must be complete, and there are no hidden variables. In 1969, John Clauser, Michael Horne, Abner Shimony, and Richard Holt presented a method to quantify Bell's inequality theorem, abbreviated as the CHSH inequality. If the measurement results of a probed quantum system do not exceed a certain threshold, the CHSH inequality is true and hidden variables do indeed exist. In contrast, if the theorem is violated, thus a value exceeding this particular threshold results, true randomness exists.

Suppose Alice and Bob have each a photon from the spin-singlet state $\vert \Psi ^-\rangle =\frac{1}{\sqrt{2}}(\vert 0\rangle _{A}\vert 1\rangle _{B}-\vert 1\rangle _{A}\vert 0\rangle _{B}).$ Alice randomly chooses between two different options to measure her photon, Q or R, once her photon arrives at her laboratory. Similarly, upon arrival of his photon, Bob randomly decides between two other measurements, S or T. Since Alice and Bob decide on and execute the type of measurement of their photon at the same time, any disturbance of each other's process can be ruled out, and information can not propagate faster than light. Their measurements yield either 1 or -1. Considering all different measurement outcomes and some potential noise in the process, the following Bell's inequality with the expectation value $\mathrm{E}(x)$ being the probability-weighted average of all possible measurement outcomes $x\in \{ -1,1\} $ is always true if there exist a set of hidden variables.
\begin{equation}\label{eq2_73}
\vert\mathrm{E}(QS)+\mathrm{E}(RS)+\mathrm{E}(RT)-\mathrm{E}(QT)\vert\leq 2
\end{equation}
Incorporating the rules of quantum mechanics and the definition of Alice's and Bob's measurements as single or combinations of quantum Z-gates and X-gates, Q,R,S, and T can for example be expressed the following way:
\begin{equation}\label{eq2_74}
\hat{Q}=Z_1 ~ ,~ ~ ~ ~ ~ ~ ~ \hat{R}=X_1~ ,~ ~ ~ ~ ~ ~ ~ \hat{S}=\frac{-Z_2-X_2}{\sqrt{2}}~ ,~ ~ ~ ~ ~ ~ ~ \hat{T}=\frac{Z_2-X_2}{\sqrt{2}}
\end{equation}

The expectation value for $\langle \hat{Q}\hat{S} \rangle$ results to be:
\begin{equation}\label{eq2_75}
\displaystyle \langle \Psi ^-\vert \hat{Q}\hat{S} \vert \Psi ^-\rangle    \displaystyle =    \displaystyle \frac{\langle 0\vert _{A}\langle 1\vert _{B}-\langle 1\vert _{A}\langle 0\vert _{B}}{\sqrt{2}} \left(Z_1\otimes \frac{-Z_2-X_2}{\sqrt{2}}\right) \frac{\vert 0\rangle _{A}\vert 1\rangle _{B}-\vert 1\rangle _{A}\vert 0\rangle _{B}}{\sqrt{2}}    
\end{equation}     
\begin{equation}\label{eq2_76}
\displaystyle =    \displaystyle \frac{\langle 0\vert _{A}\langle 1\vert _{B}-\langle 1\vert _{A}\langle 0\vert _{B}}{\sqrt{2}}\frac{-\vert 1\rangle _{A}\vert 0\rangle _{B}-\vert 1\rangle _{A}\vert 1\rangle _{B}+\vert 0\rangle _{A}\vert 1\rangle _{B}-\vert 0\rangle _{A}\vert 0\rangle _{B}}{2})
\end{equation}     
\begin{equation}\label{eq2_77}         
\displaystyle =    \displaystyle \frac{1}{\sqrt{2}}\frac{1+1}{2}={\frac{1}{\sqrt{2}}}={\langle \hat{Q}\hat{S} \rangle }
\end{equation}
    
Similarly, $\langle \hat{R}\hat{S} \rangle =\frac{1}{\sqrt{2}}, \langle \hat{R}\hat{T} \rangle =\frac{1}{\sqrt{2}},$ and $\langle \hat{Q}\hat{T} \rangle =-\frac{1}{\sqrt{2}}$ can be derived. The resulting expectation values following the rules of quantum mechanics violate Bell's inequality.

\begin{equation}\label{eq2_78}
\langle \hat{Q}\hat{S} \rangle +\langle \hat{R}\hat{S} \rangle +\langle \hat{R}\hat{T} \rangle -\langle \hat{Q}\hat{T}\rangle =2\sqrt{2}{>2}
\end{equation}
The violation of Bell's inequality is experimentally confirmed ruling out the existence of hidden variables and thus quantum mechanics is complete.\\
\textbf{Ekert91 and BBM92:}\\
The entangled states can be generated by a trusted source such as Alice or Bob. The required entangled state is the spin singlet state $\vert \Psi ^-\rangle =\frac{1}{\sqrt{2}}(\vert H\rangle _{1}\vert V\rangle _{2}-\vert V\rangle _{1}\vert H\rangle _{2})=\frac{1}{\sqrt{2}}(\vert 0\rangle _{1}\vert 1\rangle _{2}-\vert 1\rangle _{1}\vert 0\rangle _{2}).$ The particles are spatially separated, one is sent to Alice and one to Bob. Both perform measurements on their photon's polarization along three different unit vectors $\vec{a}_ i$ and $\vec{b}_ j$ respectively. Alice measures randomly relative to the horizontal axis along a unit vector $\vec{a}$ with angle $\phi _1^ A=0^\circ, \phi _2^ A=45^\circ, $ or $ \phi _3^ A=90^\circ.$ In contrast, Bob measures randomly along a unit vector $\vec{b}$ with angle $\phi _1^ B=45^\circ, \phi _2^ B=90^\circ, or \phi _3^ B=135^\circ.$

Similarly to BB84, upon completion of all measurements, Alice and Bob publicly share along which orientation each photon was measured. The group for which they picked the same measurement procedure $(\vec{a}_2,\vec{b}_1)$ $ and $ $(\vec{a}_3,\vec{b}_2)$ serve as the group which the key can be generated from. The group of measurements with differing measurement orientations $ (\vec{a}_1,\vec{b}_1),$ $(\vec{a}_1,\vec{b}_2), (\vec{a}_1,\vec{b}_3), (\vec{a}_2,\vec{b}_2), (\vec{a}_2,\vec{b}_3), (\vec{a}_3,\vec{b}_1), $ and $ (\vec{a}_3,\vec{b}_3)$ are used to evaluate the existence of potential eavesdroppers. The measurement results of the second group can be revealed publicly to evaluate via the CHSH inequality if the channel is corrupted or not. The CHSH inequality requires the subset $(\vec{a}_1,\vec{b}_1), (\vec{a}_1,\vec{b}_3), (\vec{a}_3,\vec{b}_1), $ and $ (\vec{a}_3,\vec{b}_3)$ to work and yields $-2\sqrt{2} $if the quantum channel is not corrupted and hence secure.
\begin{equation}\label{eq2_79}
S=\mathrm{E}(\vec{a}_1,\vec{b}_1)-\mathrm{E}(\vec{a}_1,\vec{b}_3)+\mathrm{E}(\vec{a}_3,\vec{b}_1)+\mathrm{E}(\vec{a}_3,\vec{b}_3)=-2\sqrt{2}
\end{equation}

The CHSH inequality is true, meaning the absolute value of the result is $\vert S \vert \leq 2$ if the quantum channel is intercepted and one or both photons are disturbed and hence lost some aspects of their ``quantumness.'' Unless the CHSH inequality is true, Alice and Bob can proceed and define an arbitrary subset of the group with the same measurement procedure to serve as a secret key. Ekert91 utilizes the completeness of quantum mechanics to generate a secret key and distinguishes itself from BB84 by the information carrier of choice, entangled spin-singlet states instead of single photons.

Shortly after, Charles Bennett, Gilles Brassard, and N. David Mermin generalized BB84 and Ekert91, referred to as BBM92. The BBM92 protocol is equivalent to BB84 and a simplified version of Ekert91 that does not require entangled states and Bell's inequality to ensure its security. It is able to withstand all known attacks allowed by the properties of quantum mechanics.

\section{Random Numbers: An Introduction} 
we are going to discuss random numbers. This is a fascinating topic because random numbers are not only really useful in everyday life but thinking about random number generators also brings up some interesting questions about reality and free will. Let us start out discussing why random numbers are so important. If we were a kid in the 80s, we might remember a game show called Press our Luck. Part of the game show had a box moving around different squares on a board that was also changing, and the player would press a button to stop the motion of the box. Whatever square the box was on would be what the player would win, or they could land on a whammy and lose everything. In 1984, Michael Larson went on the show, and he did well. We notice that he pays much attention, almost like he knows when to press the button. he did because the people that made this game did a really bad job randomizing the motion of the box and the squares. They were not random at all, but just randomly picked from a small number of possible sequences. So, Michael Larson spent much time with a VCR watching and memorizing the various, quote-unquote, random sequences that would come up. He ended up setting a record by winning over \$100,000, which we sure had consequences for the people who ran the game show. So, that is one-way random numbers can influence someone's life. here is another more serious one. As we know, the security of many forms of encryption relies on the computational difficulty of solving certain problems. For example, factoring is difficult with a classical computer, so some types of encryption are based on factoring's difficulty, and we can have an encryption key that is the product of two large prime numbers. If one of the prime numbers is reused in another encryption key, that breaks the algorithm's security. Now, this should not be a problem because if prime numbers are selected randomly, the probability of two keys using the same prime should be zero. However, in 2012, researchers found that if they looked at all the publicly accessible keys for RSA, a commonly used public-key cryptosystem, about a quarter of a percent of the time, two unrelated keys shared a prime factor. The authors speculate that the lack of entropy or randomness in one of the inputs to the algorithm resulted in this repeat of factors, which essentially made those systems completely insecure. Beyond public key encryption, many other cryptographic tasks require randomnesses, such as challenge-response protocols, like what we use at an ATM or digital signatures. Modern-day cryptography relies on random numbers, which are also important for scientific applications, like Monte Carlo simulations and lotteries. Now let us look more closely at what a Random Number Generator or RNG \cite{thornton_quantum_2019}. For now, we are going to keep the discussion abstract and just imagine that an RNG is a black box. when we press a button, a one or a zero comes out. to keep the discussion simple, we will assume that they are identically distributed. So, half the time, it is a zero, half the time it is a one. If the black box is truly an RNG, there is no way of knowing what will come out when we press the button, and we can say that each bit coming out has one bit of entropy, which is a measure of randomness. Now it is not just that we or we with limited knowledge have no way of knowing what comes out. It is that nobody, not even somebody with knowledge of everything in the universe, an infinite computational power, can predict it. One problem is that it is hard to tell when something is random or when it is just that we cannot predict it. For example, pseudorandom number generators are often used in practice. With a pseudorandom number generator, we put in a random seed, and what we get out is a lot of numbers that look random. However, a pseudorandom number generator is a classical deterministic algorithm, so it is not random. This quote by John Von Neumann puts it nicely. ``Anyone who considers arithmetical methods of producing random digits is, in a state of sin.'' Because they are not random, pseudorandom number generators are susceptible to algorithmic and computational advances, and they require that the input seed is random and secret. If we cannot use a classical computer to make random numbers, how about it if we take a physical process that we know is random and use that to make the random numbers. So, flip a coin and say heads are zero, and tails are one. That is kind of a slow random number generator. However, people have built much faster generators using ring oscillators, a single photon hitting a beam splitter, and even images of a lava lamp. Many of these things should be random. For example, with a single photon hitting a perfect 50/50 beam splitter, which arm we detect the photon in should be random, but it is hard to verify that everything is working the way we think it is. People have demonstrated the ability to blind detectors or force them to click, for example. here is an example of adversarial control of an RNG demonstrated by two researchers in 2009. They showed that by injecting noise onto a power supply, they could compromise the randomness of a ring oscillator based RNG of the same type used in challenge-response protocols for chip and pin credit card authorization. here is a visual representation of the RNG output from their paper. Before they launched their attack, it looks random, and after, there are patterns. Looking at this, we might say that just means that we should look for patterns. Many statistical tests will look for patterns and quantify the likelihood that a sequence of long numbers is consistent with being generated by a random process\cite{leymann_towards_2019}. The problem is that it only looks for specific patterns, and there are plenty of things that do not have any patterns but are not random the digits of $ \pi $, for instance. So, we cannot just take the output of the RNG and run statistical tests on it to say whether it is an RNG. We can try instead to open up the box, look at What is inside, characterize it, and make sure everything is working. The problem with that is that systems can be complicated and messy. There are many ways that components can fail or compromise, so there is no way of knowing whether or not we are testing for the right things. In 2010, some researchers realized that we could build something close to a black box random number generator. It turns out that we can make an RNG based on a loophole-free test of Bell's inequality. We can quantify the entropy of our RNG without knowing the details of how everything inside the box is working under some general assumptions. 

Random numbers are needed in several applications and areas of research, including cryptography, tests of fundamental physical principles, and simulations based on Monte Carlo methods. True randomness is difficult to realize. Most realistic applications use pseudorandom number generation (PRNG) algorithms, which produce a string of numbers mimics randomness but are based on an algorithm using a private ``seed'' and therefore is ultimately deterministic. Thus, the use of numbers produced by PRNG in principle limits cryptographic security introduces loopholes in fundamental physical tests and may lead to unexpected errors in Monte Carlo simulations.

The necessity of random numbers in cryptography is exemplified by one-time pad encryption, where each user shares an identical secret key as long as the plaintext. To perform encryption, the sender will add each character of the plaintext to the corresponding character of the key using modular arithmetic. It is important to note that for every plaintext and ciphertext pair of equal length, there exists a key of the same length, which will translate between them. Therefore, if a key is chosen at random, every ciphertext is equally likely given any plaintext, and an eavesdropper gains no information by intercepting the ciphertext. This level of security is referred to as ``information-theoretic'' security. It is, in principle, unbreakable by an eavesdropper with infinite resources, as long as she does not possess the key.

Consider a PRNG algorithm called the middle-square method to gain an intuition for the types of problems associated with using PRNG numbers (rather than truly random numbers). In this method, a pseudorandom number of length n is calculated by starting with a seed of the same length, squaring it, and taking the middle n digits of the resulting number. For example, if we want a four-digit pseudorandom number, we could start with the seed 3452, which, when squared, is 11916304, and our new number is the middle four digits given by 9163. This process can then be repeated as many times as needed. One obvious problem with this procedure is that, if at any point, this key is compromised, then an eavesdropper could determine all past and future keys; this property is known as state compromise extension. Another issue with this technique is that it will eventually repeat, which reveals information to an eavesdropper. For example, when starting with the seed 0540, it cycles back to itself after the sequence $0540\rightarrow 2916\rightarrow 5030 \rightarrow 3009 \rightarrow 0540 $. Lastly, some pseudorandom number-generation algorithms will use a shorter seed in length than the message being encrypted, which reduces the number of keys that an adversary needs to search over. Explicitly, a 16-bit plaintext could be encrypted using a 4-bit seed, where the first four rounds of the middle-square method are concatenated to create a 16-bit key. The problem with this is that, even if the eavesdropper does not know the seed, they only need to search over$  2^4$  bits to cover every possible key, rather than the $ 2^{16}$  when each bit of the key is random.

The inherent weaknesses of using PRNG algorithms have motivated significant research to create robust sources of truly random numbers. One of the most promising areas of research in this area is in quantum random number generators (QRNGs), which rely on either superposition of quantum states or quantum entanglement to create randomness\cite{friis_observation_2018}. A simple example of how this works on a single system is to prepare a single qubit in the superposition state given by$  \vert \psi \rangle =\frac{1}{\sqrt{2}}(\vert 0 \rangle + \vert 1 \rangle )$  and then making a measurement in the $  \{ \vert 0\rangle ,\vert 1\rangle \}$  basis. Since $ \vert \psi \rangle $  is an even superposition of both basis states, the outcome of the measurement cannot be predicted, even in principle, and is, therefore, a source of random numbers.

Unfortunately, it can be difficult to verify that a QRNG based on a single system is properly functioning, without intimate knowledge of the system construction. For example, a QRNG could be manipulated to output states which are statistically random but are supplied by a third party, or less maliciously, the superposition of $ \vert \psi \rangle$  could have collapsed before measurement, meaning that the measurement outcomes are actually classical noise and hence predictable in principle given enough knowledge about the system.

To contend with these problems and to guarantee the randomness of numbers, techniques known as ``self-checking'' QRNGs have been developed. Using a Bell state such as $  \vert \Phi \rangle =\frac{1}{\sqrt{2}}(\vert 00\rangle +\vert 11\rangle )$  as a resource, these QRNGs perform what is called a Bell test on $  \vert \Phi \rangle.$  A Bell test is a series of repeated measurements on each qubit of an ensemble of identical entangled pairs \cite{noauthor_big_nodate,noauthor_big_nodate-1}, such as $ \vert \Phi \rangle,$  wherein an instance of the process each qubit can be measured in two different ways, and each of these choices has two different outcomes. Once enough statistics are assembled, it can be determined whether the state being tested violates a Bell inequality. All states which satisfy a Bell inequality can, in principle, be described by a deterministic local hidden variable model (DLHVM), which means that all measurement outcomes on particles described by that state are determined ahead of time, even if they are not available to the experimenter\cite{noauthor_big_nodate,noauthor_big_nodate-1}. Alternatively, if a state violates this inequality, it cannot be described by a DLHVM, and joint measurement outcomes on these states are determined at the time of measurement, regardless of the physical distance between them. While there are many ways to interpret the meaning of a Bell violation\cite{noauthor_big_nodate,noauthor_big_nodate-1}, it suffices to say that the extent of a violation is quantitatively related to the amount of randomness present in the measurement results.

\section{Bells Inequality}
Let us now discuss entanglement, Bell's inequality, and experiments that can tell us something about the nature of reality\cite{noauthor_big_nodate,noauthor_big_nodate-1}. In the particular example we use here, we will be discussing polarization-entangled photons. However, these concepts have been demonstrated in many different types of systems. we will start by discussing polarized single photons. Say we have a vertically polarized photon, which just means that the photon's electric field is oriented vertically. We can send it through some polarization rotating element, and then to a polarizing beam splitter, which has single-photon detectors on each output. Right now, we are not rotating the polarization. the way this polarizing beam splitter works is that if the photon is vertically polarized, it will always reflect off the beam splitter and be detected in this port. However, if the photon is horizontally polarized, it will pass through the beam splitter and be detected in that port. So, this idealized system makes a perfect measurement of the photon polarization as long as it is polarized either horizontally or vertically. we call this measuring on the HV basis. The photons can be polarized in other directions too. For example, diagonally. To measure the diagonal basis, we can rotate the polarization before it gets to the beam splitter. Then this detector will be associated with the diagonally polarized photons. this one is associated with the anti-diagonally polarized photons. This is a measurement in the diagonal, or AD basis. What happens if we take a vertically polarized photon and measure it on the AD basis? In that case, quantum mechanics predicts that it will randomly go to one detector or the other, and we cannot know what will happen in advance. we have been discussing single photons, but quantum theory predicts the existence of entanglement. here is a cartoon of an entangled photon source. It puts out two photons. we will call them A and B, and each goes in a different direction two researchers named Alice and Bob. The quantum state is shown here. what it means is that the source puts out a state that is an equal superposition, a photon A being horizontally polarized and B being vertically polarized, and photon A being vertically polarized and B horizontally polarized. When we measure on the HV basis, we find that if Alice measured an H, Bob would always measure a V and vise versa. So, far this does not seem too weird. if Alice and Bob both make their measurements in the same basis, what they measure is consistent with a source just putting out a classically correlated state, where half the time it sends an H photon to Alice and a V photon to Bob and half the time a V photon to Alice and an H photon to Bob. Now, let us rotate the measurement basis so that both Alice and Bob measure the AD basis. Remember, if we have an H photon and measure it on the AD basis, it will randomly go to one detector or the other with a 50\% probability. Same with a V photon. If we did start with a classical state, we would expect that Alice and Bob's detector clicks would be uncorrelated when they measure on the diagonal basis. On the contrary, countless experiments have verified that the detector clicks are always anti-correlated if Alice and Bob measure on the same basis. These are the stronger than classical correlations that Einstein called a spooky action at a distance. However, another interpretation is possible. It could be that the photons have some hidden variable telling them to go to one or the other detector when they are measured on this or that basis. we can loosely phrase the question of which of these alternatives is correct. Is the world quantum, or is it classical and deterministic? Amazingly, there is an experiment that we can do to rule out one of these possibilities. More precisely, we can rule out something called local realism \cite{collaboration_challenging_2018}. The local part just says that we cannot send out information faster than the speed of light. realism means that things have definite properties independent of whether or not we choose to measure them. This is, we think, how most of us see the world. Hidden variables are an example of the local realistic theory. The experiment involves doing the same measurement of polarization but changing the measurement basis Alice and Bob used. In the simplest form of the experiment, Alice and Bob each measure on two bases. It turns out that for some selection of their measurement basis, there is a correlation function we can define over the inputs, or basis selection, and the outputs, or which detector clicks, where the value of the correlation function is different if the photons have these hidden variables versus what quantum mechanics predicts. Now, there are some loopholes associated with these tests of local realism. The loopholes are ways in which nature could trick us into thinking that local realism is incorrect when, in fact, it is not. First, we need to make sure that Alice does not know in advance what measurement basis Bob will pick and vice versa. If one of them knew the basis that the other chose, they could trivially reproduce the statistics. The only way to guarantee this is to make sure that their basis selections are space-like separated, so there is no time for the information about which basis Alice chose to make it over to Bob before he needs to choose a basis. There is also the free choice loophole, which says the basis selections on either side need to be uncorrelated with any classical variable. In other words, Alice and Bob's basis selections need to be made of their own free will without being affected by anything else. Finally, there is the detection loophole, which says we need to measure enough of the photon pairs to make sure we are not only observing a subset, which appears to violate Bell's inequality. The first experimental test of Bell's inequality was done in the '80s. for a long time after that, people tried to close these loopholes. People managed to close all the loopholes separately, but it was not until 2015 that experimentalists could close the loopholes in a single experiment conclusively. This was a tremendous accomplishment. it implies that the world we live in is inherently random. By the way, if we are the type of person who does not like the idea of the world is random, we should note that no experiment can ever rule out the concept of super determinism, which is the idea that everything in the universe, including Alice and Bob's basis selection, is predetermined. So, we can say that these experiments are consistent with super determinism, but we might also have to accept that we do not have any free will in that case.

\section{Bell's Inequality and Certifiable Random Number Generation} 
Bell's inequality may seem just like an interesting experiment we can do and something to discuss in our first quantum section, but not something that has any real-world applications. However, in 2010, researchers show that we can take the base machinery for demonstrating Bell's inequality and use it to make a remarkable type of quantum random number generator one that is certified by the violation of Bell's inequality under some basic assumptions. They showed that they could use the violation of Bell's inequality to bound the entropy of the outputs of their RNG. Here is how it works. Alice and Bob start with a string of zeros and ones that is used to select one measurement basis or the other. Each time they run the experiment choosing one basis or the other, each gets out a 0 or 1, depending on which detector clicked. They repeat the experiment many times and use the input and output strings to compute the Bell inequality function. This tells them how much entropy is in the output string and uses classical techniques to extract randomness. The amazing thing here is that all we need to know is how much entropy is in the system is the violation of Bell's inequality. So, there is no need to continuously characterize every component in our system to make sure it is working properly. We have one number that we get out of the system that tells us if it is working properly. Again, something can always break, and this does not guarantee that nothing will go wrong in our black box. However, if something goes wrong, either because something broke or because an adversary is hacking the system, we see it reflected in the violation of Bell's inequality. This concept is called device independence or black box security. We have discussed here using this for random number generation, but people have demonstrated black-box security proofs for other applications, such as quantum key distribution. In these protocols, it is important to close the loopholes in Bell's inequality, because those could provide a way for an adversary to compromise the system, for example, by blinding detectors. The verifiable or certifiable random number generator was first demonstrated using trapped ions in 2010. The locality loophole was not closed, but the two halves of the experiment were spatially separated and thought to be not interacting. The only way to know for sure is to close the locality loophole. In 2018, some of the same people that showed they could close the loopholes in Bell's inequality tests then used their system to make a certifiable random number generator. We would like to close by pointing out the timeline for device-independent applications. It all started back in the '30s when Einstein, Podolsky, and Rosen wrote a paper on what they viewed as a paradox and a reason that quantum mechanics was incomplete this concept that they rejected called a spooky action at a distance. Nearly 30 years later, John Bell realized that there is an experiment we could do to shed light on this paradox. People did the first experiments demonstrating the violation of Bell's inequality in the '80s. More than 20 years later, they came up with an idea of device independence and demonstrated it in a real system. Perhaps, someday we will have black box applications on a chip. If so, this would be an amazing story about how something that started as a scientific curiosity, which addressed a question mainly of philosophical interest, had profound effects on security. 

So, can single entanglement exist between two distant locations, or does the entanglement always need to exist in the same locale for entangled QKD methods?  the key property of entanglement, and the one that is so non-intuitive, is the fact that it can be non-local. non-local means, not local, but it means more than that. It means that these two particles, which have been entangled, Let us say two photons in their polarization are separated by such a large distance that even at the speed of light, we cannot communicate a choice to measure, say, particle one, we cannot communicate what we did to make that measurement over to another person over here measuring particle two. so, that is a space-length separation. so these two people could make their measurements on these entangled photons, without even in principle knowing what the other person did and yet the correlations still exist. Meaning that what this person measures is going to depend on the choice of basis that this person over here made. So, that is the spooky action at a distance that Einstein did not like, but as Bell and subsequent Bell test measurements have repeatedly shown, with various loopholes being introduced and subsequently being closed, that this is the way quantum mechanics in nature works, and we think the last loophole, as  mentioned in section, the last loophole that still has to be closed is called the free will loophole. so, if everything is predetermined, right, then that is a different story, but if we believe that these two people, person one and person two, have the free will to independently decide, at any given time, which basis they are going to use, then quantum mechanics will not. Now, the question, if we read into it one bit further, is does the entanglement have to be generated locally? And most of the generations, if not all of them, the schemes start locally? That we do say, for example, parametric down conversion. One photon is generated, or one photon is down converted into two photons which then fly away to distant locales. But we think that that is a generation question. it is not a question about can entanglement exist between two distant locations? Absolutely, and that is the point.

\section{Quantum Teleportation and Entanglement} 
If somebody gives us a single photon of unknown polarization and wants us to transmit it to somewhere else. It is a photon. It naturally likes to travel. we can send it to them. However, suppose it is daytime. we want to send it somewhere else. there are too many photons around. How can we get this photon to another destination intact? well, we could try to measure its polarization. However, if we did, we might measure along the wrong axis. we might spoil it. If we measured it, then We could tell the receiver our result. However, it would be a bad copy. It seems like the uncertainty principle prevents us from getting all the information out about the photon's state. if we cannot get the information out of it, how can we get a copy of this photon or get it to another destination intact? well, quantum teleportation is a way to make it end-run around this logic. it uses entanglement. here is how we do it. First, we will say why it does not work if we do not use entangle. So, we have got this photon of an unknown polarization. The best we can do is measure it. that uses the photon up and gives incomplete information. So, we would then send this partial information. we would make an approximate copy of a copy that might be right or might be wrong or might have a polarization that is not quite right. So, that is the trouble. it would seem that the uncertainty principle prevents this job from being done. well, nowhere is what we can do to get around that problem. we use more photons. we have the original photon A, and we want to get it to another destination. So, we now take two entangled particles, which we will call B and C. they have never been near photon A., And they do not know anything about the state of A. what we do is we make a measurement, not on A alone, but A and B together. since those are two photons, we can get four possible outcomes of that measurement, so we do that. Essentially we measure the relation between their polarizations without measuring the polarization of either one. that spoils both of them. So, this looks like we are worse off than. we do not even have one copy of the information that we wanted to get to the receiver over on the right side of the figure. However, we take this information that we got the relation. that is a two-bit message. we send it to the receiving location. Then, at the receiving location, the receiver takes the other entangled particle C, who is never anywhere near A and does one of four different rotations. we might say applies a corrective treatment. when we do that, it puts it into an exact copy of the state of photon A that we spoiled by measuring it in the first part of the protocol. So, at no point do we exactly discuss the state of this photon. However, we protected it from being damaged. then we have destroyed one copy on the left and produced a perfect replica on the right. that is what we call quantum teleportation. we could draw some equations for this, showing how this works. However, can we do it, metaphorically? Furthermore, the metaphor we have is to say, suppose Alice suppose she has witnessed a complicated crime with possible terrorist implications here in Boston? Now, the FBI in Washington knows that her memory is this fragile dreamlike form. they want to ask her just the right questions in the right order. not to spoil her memory by asking her wrong questions. they certainly do not want to leave the investigation to the Boston Police, who will ask her stupid questions and confuse her. Thus, they ask her to come to Washington and be interviewed by a panel of experts who will ask her just the right questions in just the right order. probably these questions even depend on confidential information they would not want to discuss with the Boston Police. However, Alice does not like travel, and she refuses to go. So, they decide to send one of their agents to Boston. Unfortunately, their agents are all very opinionated about the case. they do not trust each other to ask the right questions. well, two agents seem to be pretty useless. They are around the FBI, but they do not seem to understand anything. They are twin brothers, Remus and Romulus. they do not have an opinion about anything. If we ask either of them a question about any subject at all, they give a completely random answer. However, for some reason, they would always give the same answer as each other. So, discussing this problem of who is going to go to Boston and talk to Alice, Remus says, well, we do not know anything about the case. So, we less likely to influence her than any of us. Besides, we like to travel, and Romulus agrees. So, they send Remus to Boston to meet Alice, And they explain that this is a speed date. They are not supposed to discuss anything substantive. They just have to figure out whether they click in their relationship. well, it goes badly, and Alice leaves in a few minutes, saying we cannot stand this guy. nd for some reason, it has been so stressful that we forget everything about this crime. the Boston Police thank her and say she can go home. Then they get on the phone to Washington and say, well, Alice and Remus do not get along. So, the FBI, the x, then go to Romulus and say, well it looks like Alice and our brother do not get on. So, any question that we would have asked Alice, we can ask us. we know that whenever we say yes, she would have said no. So, they proceed with their careful questioning, reversing every one of Romulus's answers to get what Alice would have answered. In other words, the metaphor is that by the speed date, Alice has transplanted her state, not into Remus, but his twin brother but with a random rotation. In this case, a reversal of all the answers. So, when we work for the math of teleportation, essentially, that is what it does. 

Quantum teleportation is a method to transfer the state of a local quantum system from one location to another location through the use of a pre-shared Bell state and the transmission of two bits of classical information. This protocol, first discovered in 1993, represents a fundamental building block for communicating quantum information. For this reason, it has been studied intensely and has been demonstrated numerous times experimentally. Despite the colloquial definition of the word teleportation as popularized by science fiction, no physical object is transferred by the process; instead, just the quantum information describing the state of a qubit is transferred between two local quantum systems.

To be precise, we can assume that a user Alice is in possession of a qubit in the state $\vert \psi _0 \rangle_{C} =a \vert 0 \rangle_{C} + b\vert 1 \rangle_{C}$ and wishes to transfer this state to another user, Bob. Alice and Bob have pre-shared one of the four Bell states with subscripts A and B which are given by:
\begin{equation}\label{eq2_80}
\vert \Psi ^{\pm }\rangle =\frac{1}{\sqrt{2}}\left(\vert 0 \rangle _{A} \vert 1 \rangle _{B} \pm \vert 1 \rangle _{A}\vert 0 \rangle _{B}\right)~ ~ ~ ~ \& ~ ~ ~ ~ \vert \Phi ^{\pm }\rangle =\frac{1}{\sqrt{2}}\left(\vert 0 \rangle _{A} \vert 0 \rangle _{B} \pm \vert 1 \rangle _{A}\vert 1 \rangle _{B}\right)
\end{equation}

Suppose Alice and Bob have chosen to pre-share the $\vert \Psi ^{-}\rangle$ state (similar results would be found for the other choices). The initial state of the system can now be written as
\begin{equation}\label{eq2_81}
\vert \psi _{0}\rangle _{C}\otimes \vert \Psi ^{-}\rangle _{AB}=\frac{1}{\sqrt{2}}\left(a\vert 0 \rangle _{C}+ b \vert 1 \rangle _{C}\right)\left(\vert 0 \rangle _{A} \vert 1 \rangle _{B} - \vert 1 \rangle _{A}\vert 0 \rangle _{B}\right).
\end{equation}

By expanding equation (\ref*{eq2_81}) and grouping Alice's qubits together, the same state can be equivalently expressed as

\begin{equation}\label{eq2_82}
\begin{split}
\displaystyle & \vert \psi _{0}\rangle _{C}\otimes \vert \Psi ^{-}\rangle _{AB} \displaystyle =\\ & \displaystyle \frac{1}{\sqrt{2}}\left[a\vert 0 \rangle _{C}\vert 0 \rangle _{A} \vert 1 \rangle _{B}+ b \vert 1 \rangle _{C}\vert 0 \rangle _{A} \vert 1 \rangle _{B} -a\vert 0 \rangle _{C}\vert 1 \rangle _{A} \vert 0 \rangle _{B}- b \vert 1 \rangle _{C}\vert 1 \rangle _{A} \vert 0 \rangle _{B}\right]
\end{split}        
\end{equation}

\begin{equation}\label{eq2_83}     
\displaystyle =    \displaystyle \frac{1}{2}\big [a\left(\vert \Phi ^{+}\rangle _{CA}+\vert \Phi ^{-}\rangle _{CA}\right) \vert 1 \rangle _{B}+ b \left(\vert \Psi ^{+}\rangle _{CA}-\vert \Psi ^{-}\rangle _{CA}\right) \vert 1 \rangle _{B}\\
\end{equation}
\begin{equation}\label{eq2_84} 
-a\left(\vert \Psi ^{+}\rangle _{CA}+\vert \Psi ^{-}\rangle _{CA}\right) \vert 0 \rangle _{B}         
\displaystyle - b \left(\vert \Phi ^{+}\rangle _{CA}-\vert \Phi ^{-}\rangle _{CA}\right) \vert 0 \rangle _{B}\big ]     
\end{equation}

\begin{equation}\label{eq2_85}    
\displaystyle =    \displaystyle \frac{1}{2}\big [\vert \Psi ^{-}\rangle _{CA} (-a\vert 0\rangle _{B}-b\vert 1\rangle _{B})+\vert \Psi ^{+}\rangle _{CA} (-a\vert 0\rangle _{B}+b\vert 1\rangle _{B})
\end{equation}

\begin{equation}\label{eq2_86}
\displaystyle +\vert \Phi ^{-}\rangle _{CA} (a\vert 1\rangle _{B}+b\vert 0\rangle _{B})+\vert \Phi ^{+}\rangle _{CA} (a\vert 1\rangle _{B}-b\vert 0\rangle _{B})\big ]
\end{equation}

where we have used the following identities
\begin{equation}\label{eq2_87}
\displaystyle \vert 0\rangle \vert 0 \rangle    \displaystyle =    \displaystyle \frac{1}{\sqrt{2}}\left(\vert \Phi ^{+}\rangle +\vert \Phi ^{-}\rangle \right)     \end{equation}
\begin{equation}\label{eq2_88}
\displaystyle \vert 0\rangle \vert 1 \rangle    \displaystyle =    \displaystyle \frac{1}{\sqrt{2}}\left(\vert \Psi ^{+}\rangle +\vert \Psi ^{-}\rangle \right)          
\end{equation}
\begin{equation}\label{eq2_89}
\displaystyle \vert 1\rangle \vert 0 \rangle    \displaystyle =    \displaystyle \frac{1}{\sqrt{2}}\left(\vert \Psi ^{+}\rangle -\vert \Psi ^{-}\rangle \right)          
\end{equation}
\begin{equation}\label{eq2_90}
\displaystyle \vert 1\rangle \vert 1 \rangle    \displaystyle =    \displaystyle \frac{1}{\sqrt{2}}\left(\vert \Phi ^{+}\rangle -\vert \Phi ^{-}\rangle \right)
\end{equation}
      
The protocol then begins with Alice projecting the state of her two qubits onto one of the four Bell states in a process known as Bell state measurement (BSM). It leaves Bob with one of four qubit states, which for the case of $\vert \Psi ^{-}\rangle_{CA}$ is already equal to $ \vert \psi_{0} \rangle$ up to a meaningless global phase and for the other states can be transformed into $\vert \psi _{0}\rangle $ using either X, Y, or a Z-gate. However, it should be noted that if Alice does not send Bob the results of her BSM, then Bob does not know which transformation to apply and is unable to recover $\vert \psi _{0}\rangle$.

As an example, if the result of Alice's BSM is $ \vert \Phi ^{-}\rangle_{CA}, $ then Bob would be left with the state $a\vert 1\rangle _{B} + b \vert 0\rangle _{B},$ which can be rotated into the desired state with an application of the X-gate.

Intuitively, we can think of teleportation as first comparing the states of Alice's two qubits through a BSM, and then exploiting the fact that the qubits in the pre-shared Bell state are correlated and our knowledge of the BSM outcome to recover the state of the original qubit.

For example, if the BSM between Alice's two qubits results in $\vert \Psi ^{-}\rangle,$ we know that Alice's qubits are anti-correlated. Then, since Alice and Bob's initial qubits are also anti-correlated, the two anti-correlations cancel, and Bob is left with the same qubit up to a meaningless global phase as the one Alice wanted to send.

Quantum teleportation is consistent with the physical principles that information cannot be sent faster than light and that quantum information cannot be cloned. The first is apparent because Bob is unable to recover $\vert \psi _{0}\rangle $ until he is told the result of Alice's BSM, which must be sent to Bob using some sort of classical method, and therefore cannot be transferred faster than light. The principle of no-cloning is satisfied because Alice's BSM leaves her two qubits in a Bell state, which is now completely independent of the state $\vert \psi _{0}\rangle$. Explicitly, if we were to look at the state of either of Alice's qubits individually after the BSM, we would find an even statistical mixture of the states $ \vert 0\rangle $ and $ \vert 1\rangle $ which contains no information related to $ \vert \psi _{0}\rangle$.

\section{Quantum Repeaters}
Long-distance quantum communication is problematic due to photon loss inherent in realistic optical channels. This problem is dealt with classically using amplifiers, known as ``repeater stations,'' which are interspersed along a transmission channel to compensate for the reduction in signal power due to loss. Unfortunately, the no-cloning theorem prohibits the amplification of an unknown quantum state. It would allow for a single photon to be copied into more than one photon carrying the same quantum information. Therefore, classical amplification techniques cannot be used to extend the range over which quantum information can be transmitted.

Optical fibers are made of silica $(SiO_{2})$, which has two leading loss mechanisms, each of which becomes dominant at different wavelengths. At short wavelengths, the primary loss mechanism is elastic scattering, in the form of Rayleigh scattering, which can change the propagation direction of the light enough to allow it to escape the fiber. Alternatively, at long wavelengths, the dominant loss mechanism becomes absorption by the material itself. At approximately 1550~ nm, the combined loss due to both of these effects are at a minimum of about 0.2 dB/km, a value which is referred to as the attenuation coefficient. Here, the decibel (dB) scale quantifies the ratio of output to input power as $10\cdot\log_{10}\left(\frac{\textrm{output}}{\textrm{input}}\right)$. Outside of this optimal wavelength, the attenuation coefficient increases. It defines a relatively small range of wavelengths useful for telecommunications in optical fiber, limiting the prospect of multiplexing signals to overcome the loss. It should be noted that the optimal attenuation coefficient for silica of 0.2 dB/km should be thought of as the lower bound for attenuation coefficients found in real fibers where other loss mechanisms such as absorption due to material impurities or fiber bending also contribute to the overall attenuation.

Fiber loss is approximately constant, so the total attenuation is proportional to the distance of transmission. For context, assuming the optimal attenuation coefficient for silica of 0.2 dB/km, $95\% $ and $ 1\% $ of the initial signal power remains after a transmission distance of 1km and 100 km, respectively. Unfortunately, the transmitted power scales very poorly with distance, and at 500km (100dB loss), only one in every $10^{10}$ photons is transmitted through the channel.

A sophisticated approach to overcoming the limits imposed by attenuation and the no-cloning theorem is the concept of a quantum repeater. Quantum repeaters operate on two entangled states that span consecutive distances of L/2 such that the final state of the system is a single entangled state which spans the entire distance L. To be more specific, if we have two Bell states which span the L/2 distances then a Bell state measurement on the two collocated qubits in the center along with classical communication of the measurement result to the end stations will result in the entanglement between the two outermost qubits which span L. This process can be thought of as the quantum teleportation of a quantum state from the central station to one of the end stations where the initial entanglement of the state is teleported along with it. In general, this process can be nested as many times as needed to span a given initial distance L.

The final entangled state distributed by quantum repeaters can be used as a resource for any quantum communication protocol. For example, this state can now be used to teleport an unknown quantum state over the total distance L in a deterministic fashion as opposed to direct transmission, which would have only been successful with very low probability due to the attenuation. It is important to note that in this example the quantum repeaters have not only allowed for the transmission of a quantum system over a large distance, the original aim but have also shifted the problem of probabilistic losses to a part of the protocol which can be repeated until successful without risking the loss of the unknown state.

\begin{figure}[H] \centering{\includegraphics[scale=.8]{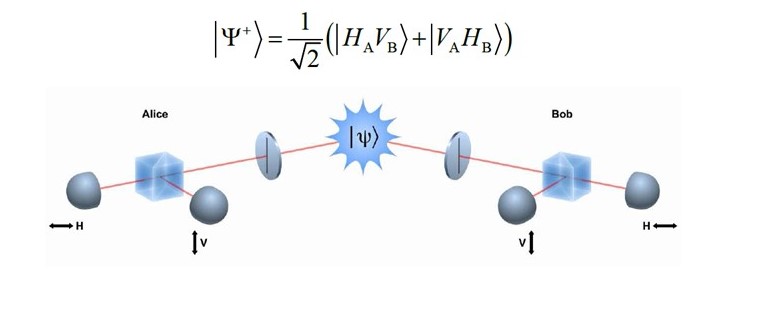}}\caption{Quantum Makes New Cryptography Entangled Photon Schemes-Entangled Photon Sources and Detectors}\label{fig2_11}
\end{figure}

\section{Quantum Simulation: Ideas and Issues} 

Over the last two sections, we looked at how quantum computers running Shor's algorithm enable us to compromise conventional encryption schemes, such as the RSA cryptosystem. we also discuss how quantum mechanics provides a workaround via quantum cryptography and quantum key distribution. In this section, we will turn the attention to the subject of quantum simulation. It was Feynman's original insight that, if we want to simulate a quantum system, we would better use a quantum system to do it, because it is a really hard problem for a classical computer to solve. Itis due to the exponentially large number of quantum states that we need to keep track of. However, what does it mean to simulate a quantum system, and how do we represent the quantum system that we want to simulate? Moreover, what does quantum mechanics do to make these simulations tractable? Further, what can go wrong? we will address these questions this section, and review several applications based on quantum simulation. However, to get started, let us first discuss about ideas and issues associated with quantum simulation. Some of the issues and the challenges when working through quantum simulation, first, we would like to have, in principle, the knowledge that we can do the simulation. the way one approaches this is to define the problem by saying the goal is to simulate Hamiltonian; we will call the Hsys. However, we are given a different set of Hamiltonians, not just one Hamiltonian, but a whole family of them, which will be the Hamiltonian of the quantum computing units. Like each of the gates or each of the subsystems. we can turn them on and off at will. Then the question is, in a mathematical way, whether or not HkQC can generate Hsys. Mathematically, this is a beautiful group theory question about whether the algebra of H quantum computer spans the algebra of Hsys. However, this is the starting point for quantum computation or quantum simulation. A second idea and issue are how does one transform this kind of a set of Hamiltonians into the system that we would like to simulate. one of the earliest approaches of this is to use analog techniques, which go under the name, for example, of analog deformation. that is that we might have a Hamiltonian which has a number of terms, and our strategy would be then to manipulate these, deform them to make this Hamiltonian look like this one, for example, in a time-averaged way or in a stroboscopic way. So, for example, at every time, capital T, 2T, 3T, 4T, and so, we would like this Hamiltonian to equal that Hamiltonian, not at all times. this is the strategy and approach, for example, employed by ultracold atom gases\cite{bloch_quantum_2012}, Bose-Einstein condensates, when they are being utilized to simulate, for example, a superconducting system. that naturally works out, in many ways, because they can be controlled to have similar Hamiltonians. the Bose-Einstein condensate is very controllable, and we can get an equivalent lattice spacing that is much larger than the physical lattice spacing of atoms inside a superconducting solid. Thus, we can engineer the kind of Hamiltonian we would like our quantum computer to simulate in many rich ways. So, this has been a very effective and useful approach. Here are another idea and challenge, however. Even if we can simulate so, this is, in principle, this is one method by which we can simulate we are still faced by a rather annoying and pressing challenge and question, which is what to measure. Even if we just tell us, we can simulate. If we say, we are given some state which is e to the minus iht times some initial state, if we measure this, it collapses. then we get only one statistical sample of that wave function. Does that tell us what we want? Does that give us an actual output? Furthermore, what are its statistical errors? Because after all, nothing is real unless it is reproducible. in Junior Physics Lab here, those of us who have taken Junior Lab to know nothing is real unless it has an error bar. we do not get an error bar from a single instance of something. However, if we have to repeat this many times, our errors now, normally, if there is just a lot of large numbers distributed, going to decrease as the square root of the number of tries. that often is not good enough, because we do not want to try an exponentially large number of times. However, naively, this is what we would have to do here. So,  what we want to do is not to measure the state.  what we would like to get is some expectation value of some operator.  it is going to be easier to measure that operator than it is to, say, get all the coefficients of a wave function. However, even this is fraught with difficulty because it still has the statistical randomness that we just described. It is an issue of precision. we will try to illustrate this issue with the Grover algorithm simulation topic that we will come to in two steps. then let us leave we with one more issue and idea, which is, and this is kind of outside of the scope of quantum simulation of quantum, but it is a very real one for quantum simulations \cite{bloch_quantum_2012,bauman_quantum_2019}. Instead of building a quantum simulator, why not just improve our classical simulation of the quantum system by using a better approximation?\cite{chen_classical_2018} Moreover, this approximation might be quantum-inspired. this has happened over the last decade or so. A whole variety of new approximations have come because of quantum computation that allows us more efficient classical algorithms \cite{tang_quantum-inspired_2019}. these have swept through condensed matter theory, and are now entering into the realm of quantum chemistry\cite{hempel_quantum_2018}. they go by a variety of names. The technique that started all of this in physics was the density matrix renormalization group idea\cite{de_chiara_density_2008,brabec_massively_2020}, which was awarded the Nobel Prize, Ken Wilson. However, now, we have many more advanced techniques. So, not just DMRG and we will toss out a bunch of acronyms here, for those of we interested in this but also things like PEPS \cite{lubasch_algorithms_2014}, MERA, and all sorts of things. we know we are entering chemistry when all we see are acronyms. These are real ways of getting very useful, highly accurate approximations to answers that we want to find in quantum chemistry\cite{reiher_elucidating_2017}.

There is active research going on relating to quantum computing \cite{bo_ewald_introduction_2019,maslov_outlook_2019} in support of improved classical quantum chemistry computing. So, can we comment on how we see using quantum computing to perform quantum simulations in model systems to improve approximate descriptions \cite{paini_approximate_2019} using classical quantum chemistry and material science calculations? For example, bond-breaking and electron correlation effects. indeed, those are very challenging, in particular, the correlation effects are very challenging for mean-field theories, and density functional theory. So when we try to come up with these simulations on classical computers, a lot of times, we lose fidelity in the answers, because at various places, we have to make approximations, or truncation, and as a result, unless we start out with something that is very close to the final answer, the results we get are quite fuzzy. this, is exacerbated by, we know, when we are trying to simulate larger and larger molecules, with many atoms and electrons and degrees of freedom. Now, quantum computers, are built upon having entanglement. Long range, large scale correlations built-in. it is part of their instruction set, if we want to think of it that way. so, it seems natural that there will be an advantage using a quantum computer to solve these types of simulation problems. in fact, this really was the genesis of quantum computing with Richard Feynman, saying, we know, if we want to simulate a quantum system, we should use a quantum system to do it. Because, this is where we can access a large number of degrees of freedom. Basically, the Hilbert space gets very large in a quantum computer. so, in principle, we can do that. Now, in practice, we have to build a quantum computer that is large enough to capture these degrees of freedom, and the quantum computer has to operate for a long enough period of time that we can complete that calculation. Now, a fault-tolerant universal quantum computer will be able to do this, and when we can build that computer, we will be able to perform these algorithms. We know that, if we can build, and when we can build that computer, we will be able to have quantum advantage for these types of problems. Now, the outstanding question today in a very active area of research, is what can we do with the qubits that we have today? And so there is this acronym NISQ, Noisy Intermediate Scale Quantum Computing \cite{preskill_quantum_2018}, or quantum algorithms \cite{giri_review_2017}. this is a class of algorithms that is being proposed, and investigated, where typically nodes in a quantum computer will work with, in conjunction, or subservient to a classical computer, where the calculations being performed, and the classical computer will periodically poll the smaller-scale quantum computer with these noisy qubits \cite{bremner_achieving_2017}, and ask it specific questions, and get specific answers back. the question that researchers are focusing on is, can we develop such algorithms where, using the qubits we have today, with the level of coherence we have today, and the fidelities we have today, can we achieve quantum advantage in this coprocessor architecture configuration? And so, there is a lot of work in this area, and we think that what many people hope is that with these NISQ-era algorithms, we will really be able to start to demonstrate quantum enhancement for important problems, which are then commercialization, and this will kick in this virtuous cycle of development. 

In 1980 Yuri Manin and 1982 Richard Feynman noticed that calculating the evolution of a quantum system using classical computational devices is inefficient and quickly becomes impossible, as the computational time scales exponentially with the number of particles involved. Their solution to this problem: use a quantum device for the calculation as it will scale much more favorably. Today, only a hand full of quantum systems can be efficiently simulated on classical computers, and exact solutions can only be determined for low-dimensional Hilbert spaces, roughly speaking the space of solutions that corresponds to the degrees of freedom involved in a problem. Larger scale simulations of materials \cite{bassman_towards_2020}, which a large part of the world-wide supercomputing time is devoted to, require approximations that unavoidably lead to errors and make, say industrially-relevant, predictions impractical. In contrast, a large-scale quantum computational device can naturally simulate higher-dimensional Hilbert spaces and hence more complex quantum problems with many particles and degrees of freedom. 

Quantum simulation has applications in several fields in physics and chemistry\cite{hempel_quantum_2018,mccaskey_quantum_2019,lanyon_towards_2010}. In physics, the dynamics of solid-state systems involving many atoms is a vast field of a study investigating, e.g., magnetic materials used in hard drives or the electrons in a superconducting material, where collective quantum effects allow for the loss-less transmission of electricity. Similarly, in chemistry, it is a challenge to describe the complicated energy level structure of atoms and molecules and the processes involved in a chemical reaction. This was famously stated by Paul Dirac in 1929 when he remarked that ``The underlying physical laws necessary for the mathematical theory of a large part of physics and the whole of chemistry are thus completely known, and the difficulty is only that the exact application of these laws leads to equations much too complicated to be soluble.'' \cite{dirac_fundamental_1925, dirac_quantum_1926,dirac_mathematical_1978,dirac_basis_1929,dirac_quantum_1929,dirac_theory_1926}. For example, nitrogen fixation is a process in which nitrogen from the air is bound in the form of ammonia $ (NH_{3}) $, which makes it accessible to plants. It is common in a range of bacteria, yet it has been elusive to reproduce efficiently on an industrial scale. Today, much of the fertilizer production still relies on the very energy-intensive Haber-Bosch process, which requires a pressure of around 200 atmospheres and temperatures of around  400\textdegree{}C (842 \textdegree{}F). Quantum simulation is a promising tool to describe and understand this problem already with a medium scale quantum computer, which was worked out as an example in 2017 \cite{reiher_elucidating_2017}.

So, about quantum chemistry, and it says, it is often-cited that, this is likely the near term, killer application, what does this mean, and when do we think a meaningful quantum advantage might be demonstrated? we are going to look at this in more detail, so next section, we will be looking at Schrodinger's equation, and how to simulate that, how to simulate Hamiltonians and dynamics. we think it is true that this is the killer application. The governments care a lot about security but in terms of breaking cryptographic protocols, or encoding information \cite{setia_superfast_2019}, in particular, the breaking the cryptographic protocols, that will not be a huge, there is not going to generate a huge business around it, because there are only a few people who really want to use that to break other people's codes. But quantum chemistry, simulating say quantum chemistry, or new materials, pharmaceuticals, drug discovery, things like this \cite{bassman_towards_2020}. This is where we can really imagine large businesses being developed. so, it is intriguing because, many of these quantum chemistry problems actually do, the ones that are hard, actually do rely on, say, entanglement, and we can map them over on to the working principles of a quantum computer, we know, quantum mechanics. so, we will discuss much more about this next section. as to when we will have them, there are two pathways here, and one is we build a fault-tolerant, error corrected, universal quantum computer, and when we have that one, then we will be able to get quantum chemistry. But in the near term, it is developing these NISQ algorithms, very likely in a coprocessor configuration where a classical computer is, as we have discussed about earlier, asking questions to a smaller quantum computer, or set of quantum computing nodes, and periodically polls it for answers. if these nodes can return an answer with quantum advantage, then fine, overall, there is a quantum advantage to doing this.

Many challenges remain, both in the size of the calculable problem (related to the number of qubits) and length of simulation (related to the coherence time and accumulation of errors). Hence, various approaches to quantum simulation are being pursued, divided coarsely along the lines of digital simulations, operating a quantum computer as a universal, gate-based machine, and analog simulations in which the available Hamiltonian on a quantum device is mapped and/or deformed to mimic the behavior of another quantum system. A fairly recent step is the introduction of hybrid simulations \cite{chen_hybrid_2019,mcclean_theory_2016}, which combines analog elements with digital gates and generally run in the form of optimization algorithms, where a classical computer guides the computation and the quantum devices sequentially scan the parameter space to find, e.g., an energy minimum. The Variational Quantum Eigensolver discussed later during this section, is an example of one such algorithm. Another hybrid quantum-classical algorithm is Variational Quantum Fidelity Estimation\cite{cerezo_variational_2020},variational quantum factoring (VQF) algorithm\cite{anschuetz_variational_2019,peruzzo_variational_2014,li_efficient_2017}.

\section{What is a Hamiltonian}
We introduced several ideas and issues at play in the previous section when implementing quantum simulations on a quantum computer. The simulation begins with a representation of the system being simulated, and once we have this representation, we can then choose how to simulate it. Quantum systems are generally defined in terms of a mathematical representation called a Hamiltonian\cite{rue_mathematical_nodate}. with a Hamiltonian in hand, as discussed in section one, one can simulate the Hamiltonian dynamics using universal gates. It is called a digital simulator. Alternatively, one can use the qubits and the couplings to mimic the system and its Hamiltonian directly. It is called an analog simulator or an emulator\cite{villalonga_flexible_2019}. In either case, we need to define the simulation problem in terms of a Hamiltonian, which characterizes the system of interest. Thus, that leads us to the subject of this section, what is a Hamiltonian? A Hamiltonian is a mathematical construct that can be used to describe the time evolution of a system. Now, these systems can be classical, like a ball rolling off a table, or they can be quantum mechanical, like the qubits we have been looked at throughout the section. It sounds pretty abstract, so Let us take a look at how to use a Hamiltonian more concretely. we might remember from physics how to predict an object's motion using Newton's laws. First, we write down all the forces acting on our system, taking note of both the strength and direction of each force. we can keep track of all these forces using a free body diagram, like the one shown here. Once we have written down all the forces, we can invoke one of the fundamental laws of physics. Newton's second law of motion tells us that the sum of all the forces acting on a system is equal to the mass of the system multiplied by its acceleration. Now, this equation is extremely powerful. Recall that acceleration is the rate of change of velocity with respect to time, like our car going from 0 to 60 miles an hour on the highway. Meanwhile, velocity is simply the rate of change in an object's position with respect to time. We can rewrite the acceleration as the second derivative of the system's position with respect to time. When we make the substitution to Newton's second law, we see that we now have a differential equation, which relates the second time derivative of position to the sum of the forces acting on the system. These forces can be constant, or they can be some function of the position of the system. When we solve this differential equation \cite{richardson_ix_1911}, we get an equation x of t, which tells us the system's position at any arbitrary time. So, we have used the forces acting on the system to figure out how it moves in time. As it turns out, this is not the only method we can use to arrive at the exact same result. Instead of thinking of the forces acting on the system, Let us focus on the system's energy. Generally, a system has two types of energy, kinetic energy, the energy that comes from a system's motion, and potential energy, the latent energy stored in the system waiting to be converted into motion. Let us write down each of these energies as equations. A systems kinetic energy is equal to its momentum squared divided by 2 times its mass. Momentum, we might remember, is related to an object's speed. So, the faster an object moves, the more kinetic energy it has. A system's potential energy depends on the forces that it is experiencing. So, we need to know exactly what forces we are dealing with and order to write down an exact equation. However, in general, we can say that the potential energy is some function of the system's position, just like the forces were. So, we have two types of energy, one which depends on the system's momentum, and one which depends on its position. Let us add these two energies together to get an expression for the total energy stored in the system. This expression for the total energy in the system is called it is Hamiltonian. as we can see, the Hamiltonian depends on two parameters, position, and momentum. These two parameters have a special role in physics, and we call them the canonical coordinates of the system. So, why is the Hamiltonian important? As it turns out, just as we previously used Newton's laws to find the system's motion in time from a set of forces, we can use the Hamiltonian to arrive at the same result. To do this, we invoke two extremely elegant equations known as Hamilton's equations. The first equation tells us that the rate of change of the systems' position in time will be equal to the derivative of its Hamiltonian with respect to momentum. Symmetrically, the second equation then tells us that the rate of change of the momentum of the system in time is equal to the derivative of its Hamiltonian with respect to position. Just like Newton's laws, these are differential equations, and if we solve these equations, we arrive at expressions that tell us how the systems position and momentum change in time. With these expressions in hand, we can see where the system is and how fast it is moving at any arbitrary time. To summarize, as long as we have the Hamiltonian of the system and an initial set of conditions, we can calculate exactly how the system will evolve in time. So, far everything we just discussed has referred to classical systems. What about quantum systems? The formalism of Hamiltonian physics is the architecture of quantum mechanics \cite{smith_practical_2017,linke_experimental_2017}. Like in classical mechanics, we can write down the total energy of the quantum system as a Hamiltonian. This Hamiltonian is simply the sum of the systems kinetic and potential energies, as in the classical case. However, note one minor difference. Whereas in the classical case, the Hamiltonian was a scalar function of the system's momentum and position, in quantum mechanics, the Hamiltonian takes on the form of a quantum operator, which acts on a given quantum state. we now invoke a fundamental relationship of quantum mechanics. The operation of a Hamiltonian on a given quantum state is equal to the rate of state change in time. This relationship is called Schrodinger's equation, and it tells us that if we know the operator responsible for the energy of the system, that is the Hamiltonian, we can solve for the evolution of that state in time. For this reason, we often say that the Hamiltonian generates the dynamics of a quantum system. Let us look at how to use the Hamiltonian more concretely in quantum mechanics. Imagine that the quantum system is a single qubit. we can write the qubit's state as a vector with two entries, A and B, which correspond to the probability of measuring the system, either at a 0 or 1 state at a given time. Just like in the classic example, the exact form of the Hamiltonian depends entirely on what forces that are acting on the qubit. However, in general, we can always write the Hamiltonian of a qubit as a two by two Hermitian matrix. A Hermitian matrix is a special type of square matrix that is equal to its conjugate transpose. When we plug the state vector and Hamiltonian matrix into Schrodinger's equation, we get two linearly independent differential equations that can be solved to find the values of A and B at any arbitrary time. When we solve these differential equations, we find that for an arbitrary Hamiltonian, an initial state, psi 0, the quantum state at the time, t, will be equal to the complex exponential, e to the iHt time the initial state, psi 0. Recall that we just said the Hamiltonian is a Hermitian matrix. When we work out the math, we can show that e to the iHt is a new matrix, which is unitary. By this point in the section, we are already intimately familiar with unitary operators, like x, y, and z rotations or Hadamard gates. we can now see where these gates physically come from. To implement a particular qubit gate, we must engineer a qubit with the correct Hamiltonian, and apply it for the right amount of time, such that the evolution of the state produces the unitary operator that we need. So, we just showed how to solve for the dynamics of a single qubit. Let us see how this problem scales when we try solving the dynamics of many qubits. To represent a system of n qubits, we need a state vector with 2 to the n entries, corresponding to each of the possible states we might measure the system in. To generate the dynamics of this quantum state, we need a Hamiltonian matrix with 2 to the n by 2 to the n entries. Indeed, as we add more qubits to the system, the number of matrix elements that we need to keep track of in the Hamiltonian scales exponentially. To make matters worse, if we want to solve the dynamics of the system, we need to exponentiate this massive matrix, which is an extremely computationally demanding operation. As we can see, solving for the dynamics of an arbitrary quantum state is an exponentially hard problem classically\cite{calude_-quantizing_2007}. This observation has profound consequences, and as we will see in this section, it provides the primary motivation for quantum simulation on a quantum computer. 

The Hamiltonian is a physical quantity that represents the total energy of a system. For a closed system, one that is isolated from its environment, so we can focus solely on the system itself, the Hamiltonian is the sum of the kinetic and potential energy. For example, for a one-particle quantum system, the Hamiltonian is
\begin{center}
 $\hat{H}=\frac{\hat{p}^{2}}{2m}+\hat{V}$ 
\end{center}
where $ \frac{\hat{p}^{2}}{2m}$ corresponds to the kinetic energy with momentum operator $\hat{p},$ and $\hat{V}$ is the potential energy of the particle.

The Hamiltonian enables one to find the time evolution of a quantum system. In the Schrödinger formulation of quantum mechanics, the state of a particle is represented by a time-dependent wave function $\psi (\vec{x},t)$. The time evolution of a quantum system is then described by the Schrödinger equation:
\begin{equation}\label{eq2_91}
\hat{H}\psi (\vec{x},t)=i\hbar \frac{d}{dt}\psi (\vec{x},t),
\end{equation}

where $\hbar \equiv h/2 \pi $ is Planck's constant h divided by $2 \pi$. Erwin Schrödinger first developed this equation (\ref{eq2_91}) in 1925, and it was the basis for his 1933 Nobel Prize in Physics.

The Schrödinger equation has a formal solution,
\begin{equation}\label{eq2_92}
\psi (\vec{x},t) = e^{-i \hat{H}t/\hbar } \psi (\vec{x},0),
\end{equation}

Which one can use to find the time-evolution of the quantum state given its initial condition? However, because the size of the Hamiltonian increases exponentially with the size of the system, for example, for n qubits, the size of the matrix representing the Hamiltonian operator is $2^{n}\times 2^{n}$  solving the differential equation becomes intractable on a classical computer for large n. Thus, another approach is warranted and, throughout this section, we will introduce several techniques that have been developed for quantum computers to simulate the static and dynamical properties of Hamiltonian systems.

\section{Simulation Example Particle in Box} 
Let us work through something concrete. First, we would like to show we the example of a particle in a one-dimensional well. It is entry-level quantum mechanics. We start at this point because we are very familiar with it, and it will illustrate the first two steps in a quantum simulation process. Imagine that we have a particle located on a one-dimensional axis, and there is some potential V of x—the somewhere delocalized in here. Let the scale of this heat from minus d to plus d or so.
Furthermore, the particle is governed by a Hamiltonian, which is of the form p squared over 2m plus V of x. So, this is the energy of a free particle, the kinetic energy. It is the potential energy. This is the set of T terms, and this is the set of V terms that are canonical in a molecule as well. This is a very simplistic scenario, but it is not all that far away from the more complicated molecular Hamiltonian\cite{omalley_scalable_2016}.
Moreover, the first step that one employs in simulating the dynamics of this physical system is to discretize. So, normally, we will have some continuous set over the infinite one-dimensional axis. So, we can imagine that these are some coefficients that we have. So, this has now expanded into the eigenbasis of position x. the natural thing for us to do is to replace this with something discrete, like k delta x, so, that this is in some way now going to be a sum over some coefficients, C sub k of k delta x where we choose the delta x to be sufficiently small and defined over this discrete interval from minus d to d. So, many of we have done this, especially if we are doing a computer simulation of such a particle in a potential. It should be very comfortable. The second thing that we do is to be able to simulate the Hamiltonian. So, we evolve the system. Now, how does one evolve a system under a Hamiltonian? The problem is that this Hamiltonian here is comprised of two parts, H1 and H2, and they do not commute. Just let evolution, e to the minus i h t we cannot simply let this be e to the minus i h1 t e to the minus i h2 t. we cannot evolve with one Hamiltonian then evolve with a second Hamiltonian, even though they are added here because they do not commute with each other.
Furthermore, this should be something we are very familiar with. Therefore, we have to apply an approximation. The approximation we apply in order to simulate this some of the Hamiltonians is the Lie Product formula, which says that in the limit as we take an infinite number of partitions of e to the i h1 t divided by n, e to the minus i h2 t divided by n, where we evolve by the first Hamiltonian for a small sliver of time, t divided by n, then the second one for another small sliver of time. Repeat this many times, taking the slivers to become smaller and smaller. It gives us equality, which is the sum of the two Hamiltonians. Now, this is a beautiful mathematical fact. Unfortunately, when we do the actual simulation, we have to truncate it at some point, and errors start to pop in. When we truncate it at a certain point, this method is called Trotterization \cite{kivlichan_improved_2019}. 

The system of a particle in a box is a useful example to study, as it is one of the few systems that can be solved analytically.

A particle with mass m in one dimension that can move freely over a length $a_{x}$ between $x=0 $ and $ x=a,$ but is restricted from entering positions at $ x<0 $ or $ x>a,$ is referred to as a particle in a one-dimensional box or a one-dimensional rectangular potential well. The particle is subject to a potential energy $\hat{V}(x)=\infty$ outside the box $(x<0 $ and $ x>a) $ and $\hat{V}(x)=0 $ inside the box.

The time-independent Schrödinger equation for a particle in such a one-dimensional rectangular potential well is:
\begin{equation}\label{eq2_93}
\displaystyle \hat{H}\psi (x)    \displaystyle =    \displaystyle \left( \frac{\hat{p}^{2}}{2m}+\hat{V}(x)\right) \psi (x)=-\frac{\hbar ^{2}}{2m}\frac{\partial ^{2}}{\partial x^{2}}\psi (x)
\end{equation}
    
$\textrm{for}~~0\leq x\leq a.$

where $\hat{p}=-i\hbar \partial /\partial x $ is the momentum operator. In contrast, outside of the potential well $(x<0 $ and $ x>a)$ the potential energy is $\hat{V}(x)=\infty$, and the wavefunction $\psi(x)$ does not enter this region.

The most general solution is
\begin{equation}\label{eq2_94}
\psi (x)=A\sin (kx)+B\cos (kx),
\end{equation}

with $ k^{2}=2mE/\hbar ^{2}.$ The constant coefficients A and B are determined by the boundary conditions
\begin{equation}\label{eq2_95}
\psi (0)=\psi (a_{x})=0,
\end{equation}

due to the fact that the particle is not able to enter the infinitely high potential well. The first condition at x=0 yields B=0, and the second condition at $x=a_{x}$ requires $ A\sin (kx)=0.$ Therefore, $ ka_{x}=n\pi,$ with $n=1,2,\cdots. $ With this solution, the Schrödinger equation gives energy eigenvalues
\begin{equation}\label{eq2_96}
E_{n}=\frac{h^{2}n^{2}}{8ma_{x}^{2}}.
\end{equation}
The integer n is called a quantum number, and it labels the energy levels of the system.

In conclusion, the wave function of a one-dimensional rectangular potential well can be written as
\begin{equation}\label{eq2_97}
\psi _{n}(x)=A\sin (n\pi x/a_{x}).
\end{equation}
The parameter A normalizes the wave function to ensure 
\begin{center}
$\int _{-\infty }^{\infty }|\psi _{n}(x)|^{2}dx=1, $
\end{center}
that is, the integral of the probability density function over the entire space is unity or, in other words, the particle is somewhere in the box with absolute certainty. Since the wave function $\psi _{n}(x)$ is only non-zero inside the potential well,
\begin{center}
$ \int _{0}^{a}|\psi _{n}(x)|^{2}dx=1$ 
\end{center}
yields normalization parameter $A=\sqrt{2/a}.$

The problem of a particle in a one-dimensional rectangular potential well can be generalized to two and three dimensions. For a particle in a two-dimensional box of length $a_{x}$ and $a_{y}$ the Schrödinger equation inside the potential well is given as follows:
\begin{equation}\label{eq2_98}
\left[ -\frac{\hbar ^{2}}{2m}\frac{\partial ^{2}}{\partial x^{2}}-\frac{\hbar ^{2}}{2m}\frac{\partial ^{2}}{\partial y^{2}}\right] \psi (x,y) = E \psi (x,y)
\end{equation}

The potential $V(x,y)$ is again zero inside the potential well, and infinite elsewhere. The solutions can be derived using the method of separation of variables. Separation of variables expresses the total wave function as a multiplication of two different functions, one for each dimension:
\begin{equation}\label{eq2_99}
\psi (x,y)=\Psi _{x}(x)\Psi _{y}(y).
\end{equation}

The two-dimensional Schrödinger equation subsequently results as:
\begin{equation}\label{eq2_100}
-\Psi _{y}(y)\frac{\hbar ^{2}}{2m}\frac{\partial ^{2}}{\partial x^{2}}\Psi _{x}(x)-\Psi _{x}(x)\frac{\hbar ^{2}}{2m}\frac{\partial ^{2}}{\partial y^{2}}\Psi _{y}(y)=E\Psi _{x}(x)\Psi _{y}(y),
\end{equation}

wich can be rewritten as
\begin{equation}\label{eq2_101}
-\frac{1}{\Psi _{x}(x)}\frac{\hbar ^{2}}{2m}\frac{\partial ^{2}}{\partial x^{2}}\Psi _{x}(x)-\frac{1}{\Psi _{y}(y)}\frac{\hbar ^{2}}{2m}\frac{\partial ^{2}}{\partial y^{2}}\Psi _{y}(y)=E.
\end{equation}

If the sum of two independent functions is constant, both functions have to be constant as well:
\begin{equation}\label{eq2_102}
\begin{split}
-\frac{1}{\psi _{x}(x)}\frac{\hbar ^{2}}{2m}\frac{\partial ^{2}}{\partial x^{2}}\psi _{x}(x)=E_{x}, &\\
-\frac{1}{\psi _{y}(y)}\frac{\hbar ^{2}}{2m}\frac{\partial ^{2}}{\partial y^{2}}\psi _{y}(y)=E_{y}, &\\
E_{x}+E_{y}=E
\end{split}
\end{equation}

These two equations are two, independent one-dimensional problems, one in the x- and one in the y-direction. Thus, the solutions for the energies in each dimension are
\begin{equation}\label{eq2_103}
E_{x}=\frac{h^{2}n_{x}^{2}}{8ma_{x}^{2}}
\end{equation}

and
\begin{equation}\label{eq2_104}
E_{y}=\frac{h^{2}n_{y}^{2}}{8ma_{y}^{2}}
\end{equation}
with quantum numbers $n_{x}$ and $ n_{y}$. The wave functions $\psi _{x}(x)$ and $ \psi _{y}(y)$ are normalized by $A_{x} $ and $ A_{y}$ respectively. Finally, the total normalized wave function can be expressed as
\begin{equation}\label{eq2_105}
\psi (x,y)=\sqrt{4/(a_{x}a_{y})}\sin (n_{x}\pi x/a_{x})\sin (n_{y}\pi y/a_{y}).
\end{equation}

The generalization to three dimensions follows the same mathematical procedure with a separation of variables and three dimensional boundary conditions. The normalized wave function results as
\begin{equation}\label{eq2_106}
\psi (x,y,z)=\sqrt{8/(a_{x}a_{y}a_{z})}\sin (n_{x}\pi x/a_{x})\sin (n_{y}\pi y/a_{y})\sin (n_{z}\pi z/a_{z}).
\end{equation}

\section{Hamiltonian Simulation: Trotterization} 
In the previous section, we discussed the quantum simulation of a particle in a box. Towards the end of the section, we reviewed an important technique for quantum simulation called Trotterization. In this section, we will take a closer look at Trotterization and see why it is such an important tool for quantum simulations. As we have already seen, the first step in quantum simulation is to write down the Hamiltonian, which governs the dynamics of the system we want to simulate. the Hamiltonians have generally separated into different terms corresponding to the system's different types of energy, for example, the kinetic and potential energies. Alternatively, for systems of many qubits, we might instead use a sum of single-qubit energies plus terms which correspond to qubit-qubit interactions. However, we define the Hamiltonian, the goal of the quantum simulation is to take an initial quantum state psi at time t0 and find out where it ends up at a later time t. as we introduced in the ``What is a Hamiltonian?'' section, the state psi at time t results from the initial state psi 0 acted upon by the operator e to the iHt where H is the Hamiltonian. Now, e to the iHt is a unitary operator, and, in principle, it can be decomposed into a series of qubit gates and placed into a block in the circuit diagram. However, these operations may be hard to realize in practice. Alternatively, we can break the Hamiltonian into pieces that are more tractable to implement, such as single-qubit rotations and elementary two-qubit gates. To get started, let us say we have a Hamiltonian H that can be broken into two terms, H0 and H1. while H is very difficult to implement, H0 and H1 are much easier. So, how can we use these simpler terms to implement the total Hamiltonian H? Now, we might think that we can simply implement these two terms as two sequential time evolutions, each acting on the system for the total time t, and then we are done. However, unfortunately, this does not work. While it is certainly true that H equals H0 plus H1, it is not necessarily true that e to the iHt equals e to the iH0t times e to the iH1t. It is because, in general, the operators H0 and H1 do not commute with one another. Commutation is a mathematical concept that we take for granted in basic arithmetic. We can multiply numbers in any order and still get the same answer. For example, 3 times 6 is equal to 6 times 3. The order in which we multiply does not matter. Because we can rearrange the numbers on either side of the multiplication sign, we say that multiplication of scalar numbers is commutative and that these numbers commute. Now, it turns out, and this statement is not always true for quantum operators. As we discuss in section 1, we can represent quantum operators as unitary matrices. To check if multiplication still commutes, Let us consider two elementary quantum operators, the z and x operators. When we multiply these matrices, it is clear that z times x does not equal x times z. Therefore, in general, operator matrix multiplication does not obey the commutative property, and we say that the operators z and x do not commute with one another. This means that order matters when we apply these operators to qubits. Change the order, and we get a different outcome. Now, returning to the Hamiltonian problem, we see that implementing H0 first and then H1 will give us a different result than if we first implemented H1 and then H0. How can we take advantage of the simpler terms, H0 and H1, and simulate the full Hamiltonian? This is where the concept called Trotterization comes in. The Suzuki-Trotter method, or Trotterization for short, takes advantage of a clever decomposition of matrices, which allows us to simulate the action of a Hamiltonian using only its constituent pieces. Instead of just breaking the Hamiltonian into two terms and evolving each for the full-time t, we will also break the time evolution into n small pieces and run each evolution for a small fraction t over n of the total time t. we will then repeatedly implement this smaller interleaved evolution a total of n times such that, by the end of the protocol, the system is evolved for the full-time t. as we make n larger and larger that is, as we make the stepwise evolution smaller and smaller, the Trotterization procedure becomes equivalent to evolving the original Hamiltonian H for the full-time t. This remarkable relationship is known as the Trotter equation. By breaking the Hamiltonian time evolution into small pieces and toggling between each term of the Hamiltonian, Trotterization enables us to simulate the full Hamiltonian evolution using only its constituent terms, even if those terms do not commute with one another. as a result, this concept is a powerful, almost universally applicable approach to the simulation of quantum systems and their dynamical evolution. 

The time evolution of a quantum system with a time-independent Hamiltonian $\hat{H}$ is described by $\vert \psi (t)\rangle =e^{-i\hat{H} t/\hbar }\vert \psi (0)\rangle$. In this expression, the operator $\hat{U}(t)=e^{-i\hat{H} t/\hbar }$ is a unitary operator \cite{franson_limitations_2018} that describes the evolution of the state.

In cases where determining or implementing a single Hamiltonian $\hat{H}$ is challenging, it can often be expressed as $\hat{H}=\sum _ n \hat{H}_ n,$ a sum of simpler Hamiltonians   $\hat{H}_i$ that are chosen to be more straightforward to implement. However, the sum of Hamiltonians does not imply $\hat{U}(t)\neq \Pi _ n e^{-i H_ n t/\hbar }$, because Hamiltonians do not commute with each other in general.

If two operators $\hat{A}$ and $\hat{B}$ do not commute that is, $[\hat{A},\hat{B}]=\hat{A}\hat{B}-\hat{B}\hat{A} \equiv c,$ where c is a scalar number not equal to zero then the exponentiation of a sum of $\hat{A}$ and $\hat{B}$ differs from the multiplication of each exponentiated operator:
\begin{equation}\label{eq2_107}
e^{\hat{A}+\hat{B}} \neq e^{\hat{A}}e^{\hat{B}}.
\end{equation}

Rather, the correct expression is given by the Baker-Campbell-Hausdorff formula:
\begin{equation}\label{eq2_108}
e^{\hat{A}+\hat{B}} =e^{\hat{A}}e^{\hat{B}}e^{\frac{1}{2}[\hat{A},\hat{B}]}.
\end{equation}

The Lie product formula provides a means to simplify the simulation of a sum of Hamiltonians.
\begin{equation}\label{eq2_109}
\hat{U}(t)=e^{-i\hat{H} t/\hbar }=\lim _{\tau \rightarrow \infty } \left(\Pi _ n e^{-i\hat{H}_ n t/\tau \hbar }\right)^\tau
\end{equation}

Essentially, the Lie product formula divides the system time evolution in small time-steps $t/\tau$, each of which implements one factor in the product form of the full Hamiltonian, and then repeats it $\tau$ times. In practice, if the time steps are small enough, this approach called ``Trotterization'' is a good approximation to the original time evolution. In the limit $\tau\rightarrow\infty$, corresponding to infinitesimal time slices repeated an infinite number of times, this relation is exact.

Trotterization: The total Hamiltonian of a quantum system can be defined by more than one term, such as  $H=H_1+H_2.$, When $H_1H_2\neq H_2H_1 $it is said that $H_1$ and $H_2 $ do not commute. Evolving the system under H is not equivalent to evolving it under $H_1 $and then$ H_2,$, or vice versa.

If a quantum system in an initial state $\lvert \psi _0\rangle$ is allowed to evolve for a time t under the Hamiltonian H, then the state of the system after evolution is given by
\begin{equation}\label{eq2_110}
\lvert \psi (t) \rangle = \exp {\left(-iHt\right)}\lvert \psi _0 \rangle .
\end{equation}

Since $H_1$ and $H_2$ do not commute, finding an explicit expression for $\lvert \psi (t) \rangle$ is not as simple as computing
\begin{equation}\label{eq2_111}
\lvert \psi (t) \rangle = \exp {\left(-iH_1t\right)}\exp {\left(-iH_2t\right)}\lvert \psi _0 \rangle \textrm{     (wrong)}.
\end{equation}
Which of the following phrases is not valid?

Trotterization is a method that determines the evolution of a quantum system under non-commuting Hamiltonians
The Trotter-Suzuki expansion comes from truncating the Lie product formula
The Lie product formula given by,
\begin{equation}\label{eq2_112}
\exp {\left(-i(H_1+H_2)t\right)}=\lim _{n\to \infty }\left(\exp {\left(-iH_1\frac{t}{n}\right)}\exp {\left(-iH_2\frac{t}{n}\right)}\right)^ n,
\end{equation}
can be used to determine the time evolution of$ \exp{\left(-iHt\right)\lvert \psi _0 \rangle} $ and is exact in the limit $n\to\infty$
The Trotter-Suzuki expansion can only be used for Hamiltonians of the form $H=H_1+H_2$

\section{Simulation of Chemistry Problems} 
we want to illustrate the simulation challenge to we today with one very specific thing, and this specific thing has to do with a well-known task known as nitrogen fixation. nitrogen fixation is a process that happens in nature that involves taking gaseous nitrogen from the atmosphere and then combining that with protons, typically from water, and then producing as a result of these two molecules of ammonia. In natural systems, in biological systems, this is not usually ammonia but probably ammonium. However, this is one of the most important processes for biological systems, including our life. It is the rate of this nitrogen fixation process that limits the production, for example, of food. It is an important and interesting reaction can be illustrated by looking at the reaction diagram of the process. So, let this be a reaction coordinate so, progress as the reaction proceeds from the left-hand side to the right-hand side of this equation. the fact is that the reactants coming in have a certain kind of energy and not be too specific about it. However, they have to overcome an enormous activation energy barrier before being able to exit the reaction as a product, the ammonia that we have here. This energy difference here is on the order of 50 kilojoules per mole of the reactants. This energy activation barrier here is much higher, on the order of 400 kilojoules per mole. Thus, although it is desirable to be able to fixate the nitrogen and produce ammonia, we have to at least, if we have no other recourse, supply a great deal of energy before we can get this recovery of energy since this is an endothermic reaction. Many of we may not know, but we have a lot of nitrogens in the body 50\% of the nitrogens in our body, according to an article in nature in the early 2000s, comes from a single chemical process. it is remarkable to realize and appreciate that so much of ourselves, the bodies, and the nitrogen are dependent on this process known as the Haber-Bosch process circa 1911. Fritz Haber and Bosch won Nobel prizes for their work in this area. It is a process, which creates high ammonia from nitrogen and steam by cracking it. Thus, it utilizes metal catalysts at a very high temperature and pressure, 200 megapascals, and something like 450 degrees Celsius in a very large-scale industrial process to generate ammonia. what is the ammonia then used for? It turns out that it is used for fertilizer, and that fertilizer goes into us, the human body, and we can estimate that it probably is responsible for something on the order of the sustenance of  40\% of the world's population. Enormous industrial-scale production. at the same time, the cost is between 1\% and 2\% of the world's energy. Now, this is fascinating. It is a strange place, perhaps to be discussing starting a section on quantum computation. However, the idea that nitrogen fixation is not done by nature using the Haber-Bosch process, using this very high temperature, a very high-pressure industrial process. Instead, nature does it with catalysts, not metal catalysts, but biological catalysts, which are enzymes, such as in bacteria, at room temperature, and room pressure. it does it a remarkably different way, which we do not understand today. the example of how this is done is with a special enzyme known as nitrogenase. this is a bacteria enzyme, one of many different kinds of enzymes, which are able to fixate nitrogen from the atmosphere and turn it into ammonium and ammonia. However, what is remarkable about this enzyme is that we do not understand how and why it works today.  on the order of a dozen other kinds of varieties of biological enzymes have been manufactured or extracted to accomplish the same task that the Haber-Bosch process does. However, none of them is quite stable and efficient and long-running as what happens in biology today. Now, the reason we do not understand this is because this is all happening in a complex molecule. The actual reaction, which we wrote over there, involves much more than just nitrogen and two molecules of water or hydrogen coming in. Let us balance the reaction over here. Many other things are happening inside these processes. For example, there are 8 protons and 8 electrons that come in typically. It may also involve a proton pump with adenosine triphosphate, losing phosphorus to become on the other side, 16 ADP, and 16 phosphorus like this. So, nature uses a very small amount of energy to be able to catalyze this and take this nitrogen fixation reaction over the activation energy barrier. Now, why is this possible? It turns out there is this amazing center inside this nitrogenase called the iron-molybdenum complex. It has on the order of 100 kilodaltons of atoms inside this, so we are discussing it on the order of  1,000 carbon atoms. they are all shaped in various ways with alpha helices and beta sheets. these shapes then allow of the nitrogen and the hydrogen coming in to dock in a certain way. Then because of where they are docked, they can react and do so much more efficiently than they are able to do in this heterogeneous type of catalytic reaction that is industrially done. So, we would love to, in quantum chemistry, synthesize molecules that have such functionality because we could catalyze all sorts of other reactions. we cannot understand why this is happening from a fundamental standpoint today because we do not know what is happening in a quantum mechanical way with this molecule. Only very recently was even the structure of the molecule understood as with X-ray crystallography and the like. So, we, in principle, could do so. In principle, we can write down all we need in order to figure out What is happening in this because the system is a very simple Hamiltonian in principle. There is the kinetic energy of nucleons, there is the kinetic energy of the electrons, and then there are interaction terms between the nucleons and nucleons, between the nucleons and the electrons, and electrons and electrons. we will expand into this Hamiltonian in a little bit to show we concretely what this means. However, suffice to say that we can write down a very accurate description of the system. We cannot do what we do not know how to do, but to solve for things like the reaction rate and even more simply, just the ground state of the system.

\section{Phase Estimation} 
we have taken H. we have modeled it with some simplified h prime. Then the two pathways the community has taken, which are rather distinct, are first to utilize what we have already discussed in this section, the phase estimation algorithm. This will give us the ground state and the energy of the ground state approximately. The second approach does something different and is known as a variational quantum eigensolver. We will now find the literature replete with people discussing the VQE algorithm. We want us to know what it is because there is a lot of discussion about that. This will give we a state psi, which might be if we are lucky to close to the ground state, and energy, which might be if we are lucky, the ground state but has a certain guarantee that this energy will be no smaller than the ground state energy that we are looking for. So, let us look at these two different routes, one at a time. The first approach, phase estimation, is very simple to understand, as we described because it is no more than the application of phase estimation. We start with registered qubits, where we are going to store the eigenenergy. So, this is determined by the imprecision that we want the number of qubits. We also start with a state psi, which is an approximation of the ground state, but we do not know how good it is. So, it is the ground state plus epsilon times some other component, which is orthogonal to the ground state. Allow that we have prepared this somehow. We may prepare it badly. That is still fine. we perform It is going to be some sum over all values k. we perform a controlled h to the k operation over here. It is feasible because we have shown we how to trotterize. However, this is painful because we now have to trod right and also trod right to the kth power. So, now we have to rerun that [troderization] an exponentially large number of a span of energy, for example. But allow this. We know from the phase estimation algorithm that what we get as the output of this is going to be a sum over energy eigenstates and then phases, which are on the order of some constant over e sub k. So, these are going to give us the energies of the system. It is the top register, and this is the bottom register. There are some coefficients. When we measure this top register after the quantum Fourier transform, we collapse into an eigenstate down here. If our initial overlap is sufficiently large, so, epsilon is small, then we get the ground state that we desire. We will have a measure, which is the energy eigenstate. However, by the way, this is exactly Shor's algorithm. It is a quantum factoring algorithm also being used for quantum simulation. It is not accidental that all known exponentially fast algorithms today, quantum algorithms are essentially the structure of Shor's factoring algorithm. We put something in superposition. We do control something to a higher power based on the superposition. We do the inverse quantum Fourier transform. So, if we know Shor's factoring algorithm, we essentially know this whole route to quantum simulation. Why is this challenging? Well, first, it is good because we get the eigenstate and eigenenergy, the ground state, right away. However, the challenge is that we need a perfect quantum computer with long coherence times. There is a beautiful analysis of the requirements by Matthias Troyer and others from now at Microsoft \cite{svore_q_2018,soeken_programming_2018}, where he shows that if one uses this phase estimation approach, one can simulate nitrogenase exactly what we started in section. If we can run the quantum computer for about a day with a relatively fast clock cycle as well, the longest coherence time we have for many of the quantum bits like in superconductors is on the order of a millisecond \cite{soloviev_beyond_2017}. For ions, it is perhaps tens of seconds. So, a day is a bit of a stretch right now. This might need something on the order of a million qubits as well, whereas, in the actual deliverable common computers, companies are discussing 50 qubits today. So, that gives us an idea of the gap between the ideal system needed and that which is available. 

Phase estimation is an important module of several quantum algorithms, such as Shor's factorization algorithm. Suppose there is a unitary operator $\hat{U}$ with eigenvector $\lvert u \rangle $ and eigenvalue $e^{2\pi i \phi } $ with an unknown value $\phi$. As the name ``phase estimation'' suggests, the module estimates the phase $\phi$ of an eigenvalue of a system with eigenvector $\lvert u \rangle$ and operator $\hat{U}$.

The quantum phase estimation protocol requires a register with n qubits initialized in the ground-state and a second register representing state $\lvert u \rangle $. The number of qubits n can be chosen arbitrarily and ultimately defines the digital accuracy of the estimated phase $\phi$ as well as the probability of success.

The first step of the protocol creates a superposition state composed of the n qubits of the first register. The second step involves controlled $U^{2^ j} $ operation on state $\lvert u \rangle$. The non-negative integer j in $U^{2^ j}$ depends on the individual controlling qubit, $j=0,\cdots , n-1.$
\begin{figure}[H] \centering{\includegraphics[scale=.8]{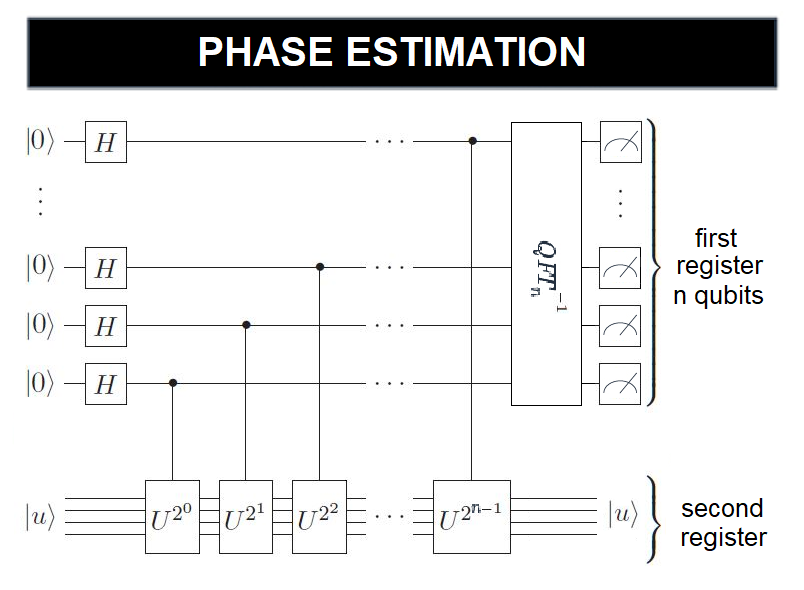}}\caption{Phase Estimation}\label{fig2_12}
\end{figure}

The first register prior to implementing the inverse quantum Fourier transformation can be expressed as:
\begin{equation}\label{eq2_113}
\displaystyle \frac{1}{2^{n/2}}\left(\lvert 0 \rangle + e^{2\pi i 2^{n-1}\phi }\lvert 1 \rangle \right)\left(\lvert 0 \rangle + e^{2\pi i 2^{n-2}\phi }\lvert 1 \rangle \right)\cdots \left(\lvert 0 \rangle + e^{2\pi i 2^{0}\phi }\lvert 1 \rangle \right)
\end{equation}
\begin{equation}\label{eq2_114}
=\frac{1}{2^{n/2}}\sum _{k=0}^{2^ n-1}e^{2\pi i k\phi }\lvert k \rangle    
\end{equation}
      
Subsequently, an inverse quantum Fourier transformation is applied on the first register, yielding:
\begin{equation}\label{eq2_115}
\displaystyle \frac{1}{2^{n/2}}\sum _{k=0}^{2^{n}-1}e^{2\pi i k\phi }\lvert k \rangle \xrightarrow {QFT_ n^{-1}}\frac{1}{2^{n}}\sum _{k,l=0}^{2^ n-1}e^{\frac{-2\pi i k l}{2^ n}} e^{2\pi i k\phi }\lvert l\rangle    \displaystyle =    \displaystyle \frac{1}{2^{n}}\sum _{l=0}^{2^{n}-1}\left[\sum _{k=0}^{2^{n}-1}\left(e^{\frac{-2\pi i l}{2^ n}} e^{2\pi i \phi }\right)^ k\right]\lvert l \rangle     
\end{equation}
\begin{equation}\label{eq2_116} 
\displaystyle =    \displaystyle \frac{1}{2^{n}}\sum _{l=0}^{2^{n}-1}\frac{1-e^{2 \pi i (2^{n}\phi -l)}}{1-e^{2 \pi i (\phi -l/2^{n})}}\lvert l \rangle
\end{equation}

The sum in the square brackets is a geometric sum and can be simplified. The resultant state contains the estimation of the phase $\phi$. Suppose $\phi$ can be expressed with n bits as $\phi =0.\phi _1\dots \phi _ n$. Then, the exponential term $(2^ n\phi -l)$ can be expressed as $(\phi _1\dots \phi _ n-l) $ with l being an integer between 0 and $2^ n-1$. Note that exponential terms are periodic in $2 \pi i $ thus $ e^{2\pi i 1.\phi _ n}=e^{2\pi i (1+0.\phi _ n)}=e^{2\pi i 0.\phi _ n}. $ The last term $l=2^{n}-1$ contains all digits of $\phi$ up to the $n^{th} $ digit, which determines the precision of the phase estimation protocol.

The phase estimation protocol enables the estimation of the eigenvalue's phase $\phi$ of a quantum system with an operator $ \hat{U}$ and eigenvector $ \lvert u \rangle. $ The output of the measurement of n qubits estimates the phase $\phi $ up to a precision $\propto n$.

\section{Variational Quantum Eigensolver} 
So, now we come to the second approach. The point of this second approach, VQE, is to get by with a much poorer, worse quantum computer by not asking for quite as much, and we will see how that works by describing the algorithm. The point of this argument is to rely upon the variational principle, which in this context says that if we were given some state, psi, which is not the ground state. We measure the Hamiltonian's expectation value, which is full total Hamiltonian with respect to that state. We will get some expectation value average energy, which is no smaller than the energy of the ground state. So, therefore, we might do something like introducing a bunch of parameters here. Let us call these $ \theta $. So, $ \theta 1 $, $ \theta 2 $, $ \theta 3$, and say we use the $ \theta $ to manipulate the degrees of freedom we have in the input state to this expectation value. Then again, it will also be no smaller than that. So, we might define this thing as an expectation value of H parameterized by some $ \theta $. we can do this on a classical computer, but the hardest part about the classical computation is preparing this state and then computing this matrix element of H with that state. the VQE algorithm proposes to do those two steps on a quantum computer. the rest, it does with a classical computer. So, here are the steps. Prepare psi of $ \theta $, where $ \theta $ it is controlled by a classical algorithm, and this is done on a quantum computer. ensure this H of $ \theta $, and there are many ways of doing this measurement. One of which we will describe in a moment if we have time on a quantum computer. then, third, take this classical measurement result that we now have and run an optimization algorithm. For example, the Nelder-Mead simplex algorithm, or stochastic gradient descent, or we choose our favorite optimization algorithm, changing $ \theta $ to minimize each of $ \theta $. go back to the beginning, and this is done on a classical computer. So, now we have a hybrid, where we have a quantum computer executing some steps, and a classical computer executing some other steps. we can show that this kind of a variational expectation value is true even if we have a natty density matrix here, instead of a pure state. So, if our quantum computer starts to devolve and go not so, quantum on we halfway through, it still cannot give us an expectation value that is better than the ground state. there are various versions of these sets of theorems, which apply to variances in other quantities that we might want to measure. Thus, one can try to assert, although it is largely done without proof, that this argument will still work reasonably fine on a noisy quantum computer with not so good quantum bits. therefore, companies can deliver to we a product which is a quantum computer with poor quantum bits, and still expect we to pay a lot of money for it because it could run this algorithm and give we some result. There are not many proofs that it will do so, but it is a very compelling argument and a very interesting direction. there are now experimental results showing small one and two-qubit quantum computers doing exactly these steps, the variational quantum eigensolver. So, the challenge of this, there are two, we would point out most. First, does this expectation value converge? we have to rely upon this classical optimization algorithm to get we better and better values. However, we do not have proof that we got the lowest possible value or the desired ground state. Second, this state preparation. It is unclear how this state is prepared. There are some ideas on how to prepare this. For example, with adiabatic quantum computation. we start with a beginning Hamiltonian, which we can create, and try to construct an end Hamiltonian whose ground state is this state that we want to create. since we have this parametrized in principle, we have some means by which we can do that. This parameter, by the way, could be things, for example, like the adiabatic path that we use in the quantum computer. related to this first challenge is how we measure H., And one way people have done this is to measure the terms of H. So, we can say this is equal to the average expectation value of H1 plus H2, plus all the way up to however many terms there are in here. So, we can measure, in principle, the expectation value turn by turn in the Hamiltonian. if the Hamiltonian terms are simple, like Pauli matrices\cite{smart_experimental_2019}, we have a feasible way of doing it. Unfortunately, that means that we also now added the error from each one of these expectation values. So, the error can grow very rapidly as we are trying to estimate the expectation value this way. Still, the variational quantum eigensolver is different. we have not looked at any kind of algorithm like that in this section. How do we utilize a system with a quantum computer and a classical computer together so that we can get something of the best of both systems? How can we make them work in concert so that one handoff information to the other one so that now we can feedback and try to speed up, while only have a limited amount of quantum coherence and a limited number of qubits at our disposal? we think this is a fascinating topic. we want to leave us with that challenge, and we want to also leave us with the challenge of thinking about how we can relate Hamiltonians and quantum algorithms \cite{de_ridder_quantum_2019}. 

Phase estimation determines the eigenenergies of a quantum system up to a specified accuracy. However, stringent requirements on the quantum processor's performance, such as the available coherence time, limits the current applicability of this method as high accuracy requires long quantum circuits with many gates (high gate depth)\cite{terhal_adaptive_2004,pednault_breaking_2018}. In contrast, quantum optimizers such as the variational quantum eigensolver (VQE) have more relaxed requirements on the quantum processor performance and work with shorter quantum circuits (``shallow depth'') contain parameterized, analog elements\cite{verdon_quantum_2019,thornton_quantum_2019}.

A VQE uses an iterative hybrid approach \cite{shaydulin_hybrid_2019} to calculate the ground-state energy of a problem-specific Hamiltonian by creating a trial wave function on a quantum processor and measuring its energy with respect to the desired Hamiltonian. The trial wave function is made using a quantum circuit based on multi-qubit and single-qubit gates containing tunable parameters and optimized using a classical computer working in tandem with the quantum processor. An optimization loop repeatedly alters between quantum and classical processing until the optimal parameter set has been found, which generates the minimum energy corresponding to the ground state wave function. 

As explained in the section, the VQE protocol is initiated with a trial wave function $\Psi (\alpha )$ on a quantum computer. Here, $\alpha$ is the tuning parameter that will successively get optimized to yield the minimum energy. The quantum processor evaluates the expectation value of the trial wave function, $\langle \Psi (\alpha )\vert H \vert \Psi (\alpha )\rangle,$ where $H$ represents the problem-specific Hamiltonian, and $\langle H \rangle $ correspond to the energy of the prepared wave function with respect to the Hamiltonian. Generally, this requires repeated runs for the same parameter set, not just to obtain enough statistics but also to measure in all the different measurement bases specified in the Hamiltonian (a process often referred to as Hamiltonian averaging). A classical optimizer is then used to minimize the expectation value (energy) by varying the trial wave function through the parameter $\alpha $. The process repeats until E converges to a steady-state close or equal to the ground state energy $ E_0.$

Early demonstrations of VQEs using superconducting qubits simulated the ground-state energies for a hydrogen molecule $H_2$ \cite{bauman_downfolding_2019}, lithium hydride LiH and beryllium hydride $BeH_2$ molecule. A demonstration using trapped-ion qubits \cite{schafer_fast_2018}also used VQE to simulate a hydrogen molecule $ H_2 $ and lithium hydride LiH., photonic qubits have been used to find the ground-state energy of a $He-H^+$ molecule\cite{zhang_observation_2017,kandala_hardware-efficient_2017,omalley_scalable_2016}.

\section{Variational Quantum Eigensolver (VQE)} 
Let us start with the energy graph of a simple diatomic molecule. When the atoms are close to each other, the repulsion between the individual nuclei dominates the energy of the system and leads to an unstable configuration. At the other extreme is the case when the two atoms are dissociated. The point of minimum energy is the equilibrium configuration when the bond distance separates the atoms. Evaluating these energies to high accuracy is extremely crucial for accurately predicting chemical reaction rates and pathways and developing insights into molecular structure\cite{omalley_scalable_2016}. The evaluation of these energies essentially comes down to solving the Schrodinger equation, with the following electronic Hamiltonian, which captures the electronic kinetic energies and the Coulomb interaction between pairs of electrons and nuclei, as well as pairs of electrons. However, this eigenvalue problem is one with the exponential cost when addressed by classical computation. this was indeed one of the initial motivations for quantum computers, the idea that one could use a highly controllable programmable quantum processor to efficiently study and predict the properties of quantum systems of interest. The first step of such simulations, then, involves encoding the problem to the quantum processor, in the specific case, mapping the electronic orbitals onto qubits. One then needs an algorithm that solves the eigenvalue problem to evaluate the ground state energy. In this context, hybrid quantum-classical algorithms with short-depth circuits\cite{bauman_downfolding_2019}, such as the variational quantum eigensolver\cite{kandala_hardware-efficient_2017,thornton_quantum_2019,yuan_theory_2019}, are well-suited to the capabilities of existing hardware. Here, the trials, or guesses, to the ground state are prepared on the quantum processor, parametrized by some experimental controls $ \theta $. The energy of these states is measured and fed to an optimization routine run on a classical computer that updates these parameters to attempt a new trial state. this is now run its iteratively until the energy converges towards its minimum value. It raises three important questions for implementing this algorithm. Are there encoding schemes that are efficient towards the number of qubits required? Are there trial states that are efficient towards the number of gates required while faithfully representing the ground state of the molecule? It is important since this can help reduce the effect of decoherence during state preparation. Finally, are classical optimization routines that can effectively deal with a large number of variational parameters and experimental lines? These were all important components of a recent experiment at IBM. Taking advantage of natural symmetries in the molecular Hamiltonian\cite{omalley_scalable_2016}, we were able to encode eight orbitals of beryllium hydride onto a reduced number of 6 qubits at no loss of information. Also, we constructed trial states that used layers of untangling gates that are naturally accessible to the experimental hardware, which is then interleaved with arbitrary Euler rotations. The circuit dept or the number of entangling layers is then set based on the available quantum coherence of the hardware. Making efficient use of coherence time and qubit resources, these states are dubbed hardware-efficient trial states. Using such trial states, we were able to perform energy minimizations of up to 6-qubit Hamiltonians. Here is one such example. In the classical optimization routine, 13 Euler angles, which serve as the variation of parameters, are simultaneously updated as the local reading is approximated using solely two energy measurements by iteration, shown here in red and blue. The iterations are stopped once the energy flattens out towards its minimum. Using these techniques, we were able to address different electronic structure problems \cite{motta_low_2018,babbush_low-depth_2018,mcclean_openfermion_2019,higgott_variational_2019,endo_variational_2019,mcardle_variational_2019} of fairly elementary molecules using 2 qubits for hydrogen, 4 qubits for lithium hydride, and 6 qubits for beryllium hydride. These are problems that are small enough to solve even on a laptop. However, a big motivation for this work is to be able to understand the sources of error and when addresses these problems using the normal paradigm of computing, using real quantum hardware. by comparing the experimental results, shown in these black dots with numerical simulations of the noisy device, we see the decoherence at insufficient circuit depth is some of the important sources of error. So, now this raises the question, in the absence of quantum error correction\cite{edmunds_measuring_2017}, is there a way to do something meaningful with near-term hardware? Is it possible to obtain estimates for, say, these energies in the absence of incoherent noise, despite using a noisy quantum processor? It turns out there are methods that we call error mitigation\cite{endo_practical_2018}. If there is an easy way to amplify the strength of the noise and measure the expectation values of interest, we can then use a powerful numerical technique Richardson extrapolation to obtain a zero-noise estimate. Here is a case of two wrongs making a right. In contrast to quantum error correction that has a resource requirement that is perhaps beyond near-term hardware, this error mitigation technique is fairly accessible since it does not require any additional quantum resources\cite{kandala_error_2019}. Now, is there a way for us to controllability amplify the strength of the noise? For instance, warming up the dilution refrigerator, which houses the superconducting processor\cite{gambetta_building_2017,xu_coherent_2016}, certainly increases decoherence. However, there is not a precise way of scaling the noise strength. However, an alternate, more accessible approach involves rescaling the dynamics of the state preparation. It can be shown to be equivalent to amplifying the strength of the noise if the noise is constant over the duration of the rescale measurements. Let us now revisit the molecular simulation of lithium hydride with 4 qubits. The original experiments used just depth-one circuits to reduce the effect of decoherence. More recently, the mitigation of errors from decoherence enables the benefit of longer computations with increased circuit depth, to achieve accuracies that were otherwise believed to be beyond the scope of the noisy superconducting processor. Although discussed with specific examples of quantum chemistry, the variational algorithms \cite{kshetrimayum_simple_2017} and error mitigation techniques discussed here can be extended to a host of different problems that rely on computations with expectation values problems in quantum simulation addressing fundamental particle physics and condensed matter physics, and even problems in optimization and machine learning \cite{shiba_convolution_2019,riste_demonstration_2017,schuld_machine_2019,havlicek_supervised_2019}. These techniques will likely be crucial to enhance the quality of quantum computation and simulation in this era of noisy intermediate-scale quantum computers.

\begin{figure}[H] \centering{\includegraphics[scale=.4]{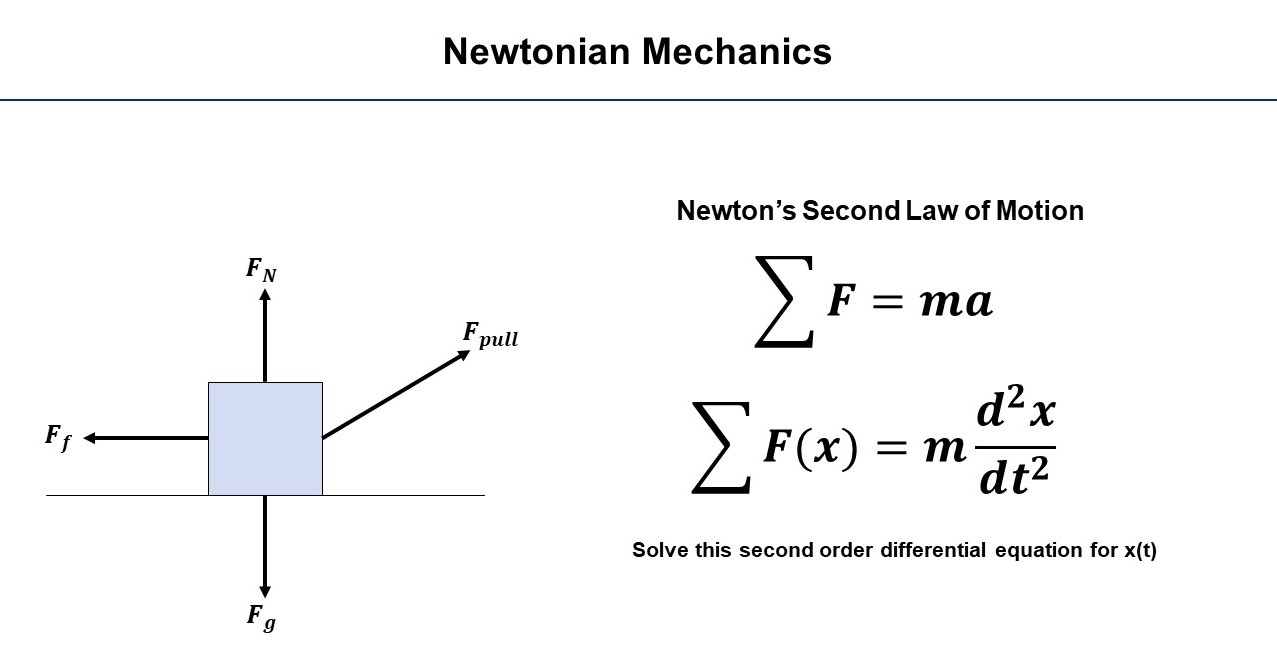}}\caption{Newtonian Mechanics}\label{fig2_13}
\end{figure}

\begin{figure}[H] \centering{\includegraphics[scale=.7]{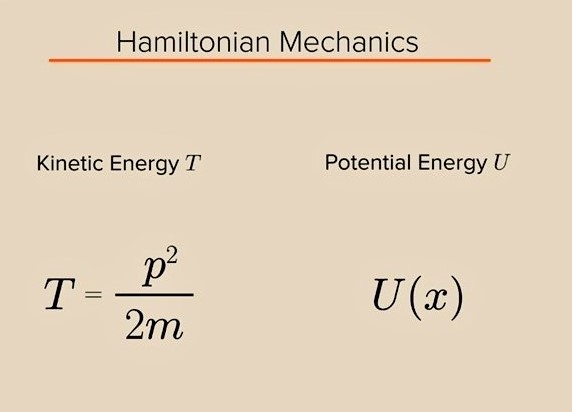}}\caption{Hamiltonian Mechanics}\label{fig2_14}
\end{figure}

\begin{figure}[H] \centering{\includegraphics[scale=.7]{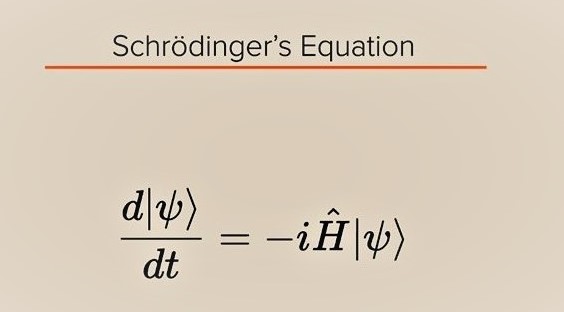}}\caption{Schrodingers Equation}\label{fig2_15}
\end{figure}

\begin{figure}[H] \centering{\includegraphics[scale=.6]{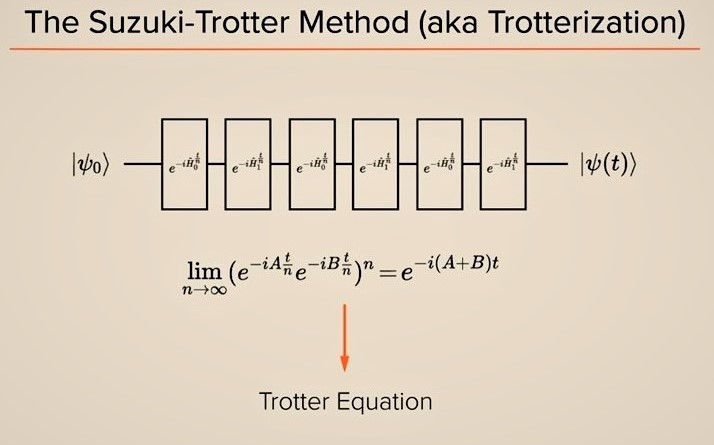}}\caption{Trotterization}\label{fig2_16}
\end{figure}

\section{Introduction to Quantum Optimization}
In this section, we turn our attention to optimization problems and their implementation on a quantum computer. Optimization problems are ubiquitous in industry and government, for example, routing signals in electronics to minimize cross-talk, distributing financial portfolios to maximize profit while minimizing risk, or identifying the most likely nexus within a dense network of telephone calls. Such optimization problems with large numbers of variables and constraints are challenging to implement on classical computers. There are significant research efforts aimed at developing quantum optimization algorithms that may yield quantum enhancement.

There are several approaches to quantum optimization. The first is adiabatic quantum computing and its related\cite{aharonov_adiabatic_2005}, restricted cousin, quantum annealing\cite{yarkoni_boosting_2019,das_colloquium_2008}. In this case, a cost function is encoded into a Hamiltonian using a prescribed approach. The quantum device then seeks to minimize the cost function by finding the ground state of this Hamiltonian. A second approach is to use a universal, digital quantum computer to find a solution that satisfies (to the extent possible) a set of boolean clauses. We will discuss this section more about the advantages and challenges of these differing approaches.

We do not always need exact solutions to optimization problems; approximate solutions are often ``good enough.'' For example, in establishing an optimal route that minimizes the distance one needs to travel between several cities separated by 100's of kilometers, one probably would not distinguish between several high-quality solutions that differ by only a few meters or even a few hundred meters. Thus, there is intense interest in identifying quantum approximate optimization algorithms (QAOA) \cite{farhi_quantum_2014,venturelli_compiling_2018} that feature quantum advantage in providing high-quality solutions, without necessarily obtaining the optimal answer. Such algorithms may be realizable on nearer-term, smaller-scale quantum computers, whether acting alone or as the co-processor working in tandem with a classical computer.

Next, we discuss that the output of quantum computer was one value with a higher probability, and the question is can a simulation algorithm result in a distribution of probabilities? and the answer is yes it can. if there really is only one value, then that value has probability one. because we always normalize the probability to one. so, if any given answer is not obtained with unity probability. it means is that there are other answers floating around with lower probability. such that when we add up all the possible answers then the probabilities of all those answers is that we get one. So, in quantum simulations it could be depending on what we are simulating that we would get a distribution of many different answers with different weightings. the problem with that is we have to at any given instance or run the algorithm we are going to make a measurement projectively and get one of those answers out.
so to estimate the probability of each one and sampling all different answers. we have to, identically prepare and repeat the algorithm many times. and if this solution space is exponentially large then that is going to take an exponentially long time. So that is not really the right way to ask a quantum computer a question. Rather, the simulations, or the optimizations, need to ask the questions in such a way that the solution space is not exponentially large, but reduced to either one answer, with high probability, or a set of answers, that all of which are high quality, so that it does not really matter which one we get. so, the example that we have given, we think, before, is that if we are trying to optimize, which route to take home from work, and it is 10 miles away, there may be a whole set of answers that are accurate enough that we do not care which one it is. Let us say that there are a hundred different ways we could get home, and they differ by 100 meters? and it is 10 kilometers to get from home, we know, from work to home. we do not care? it is only 100 meters difference. What matters is not off by a kilometer. So, in these types of cases, optimization problems, for example, there need not be, there may be one truly optimal answer, but in all likelihood, there may be many answers which are equally good from a practical sense. so as long as the algorithm is set up to find those answers, or that set of answers, samples any one with high probability, then it is fine. we can run the algorithm a few times, not an exponentially large number of times, if we can ask it a few times, and we start to sample some of these potential good solutions, and eventually, we gain some confidential cases, it is probably a good solution.

\section{Adiabatic Quantum Computing}
Optimization problems are ubiquitous in business, government, and our daily lives. They also tend to be hard problems to solve, because of the large number of variables and connectivity within the problem. In this section, we will discuss adiabatic quantum computing, one approach to addressing optimization problems. We will also discuss how adiabatic quantum computers are different from gate-model quantum computers. 

In the last section, we discussed several topics related to quantum Hamiltonian simulation, including methods like Trotterization \cite{endo_mitigating_2019}, phase estimation, and variational quantum eigensolvers. In this section, we will turn the attention to quantum optimization algorithms. Optimization problems are ubiquitous, from supply transport optimization to sensor and satellite tasking, pattern recognition. Many important practical problems can be reduced to satisfy, in an optimal manner, as many constraints as possible. we will first discuss optimization problems and their implementation on an adiabatic quantum computer. we will then turn the attention to quantum annealers, machines that target classical optimization problems. lastly, we will look at Grover's algorithm and how it works. at the end of the section, we will implement Grover's algorithm on the IBM Quantum Experience \cite{lesovik_arrow_2019}. To get started, let us first study adiabatic quantum computers and quantum optimization algorithms. Adiabatic quantum computing is an all-purpose approach to solving combinatorial optimization problems\cite{goto_combinatorial_2019}, or generally for finding the minimum of an objective function, using a quantum computer. we know that all systems in nature obey the Schrodinger equation. It is a first-order differential equation, which says that the wave function's rate of change is proportional to the Hamiltonian operator times the wave function. So, adiabatic quantum computing uses the Schrodinger equation as the basis of the algorithm. It is distinct from the gate model in that sense because we are going to have continuous-time evolution. The quantum state is going to change continuously, as opposed to the gate model, where sequences of discrete transformations update the state. They are very closely related. However, in this case, we think it is a lot easier to think about how the algorithm works if we think in the context of the continuous-time evolution. Let us try to give us a feel for how the adiabatic algorithm works. we want to find the minimum of a complicated cost function. in general, there is no all-purpose algorithm that will guarantee to find the minimum of any cost function in all circumstances. We are trying to find something which we hope will work well in practice in situations we care about. So, the idea is this. We are going to attempt to find the minimum cost function by encoding the cost function in a quantum Hamiltonian. we are going to take a quantum Hamiltonian, whose ground state that is its lowest energy state we know, and we are going to prepare the quantum system in the ground state of this known Hamiltonian. Sometimes we call that the beginning Hamiltonian. If we look at this equation here, we see that the Hamiltonian that controls the evolution of the system has two parts, the beginning Hamiltonian and the problem Hamiltonian. Now, we are going to start the system off in the ground state of the beginning Hamiltonian, because the beginning Hamiltonian is something we know how to construct, and we know its ground state. then what we are going to do is slowly turn off the beginning Hamiltonian and slowly turn on the problem Hamiltonian. So, we are going to go from one to the other slowly. However, the problem Hamiltonian has been constructed so that its ground state encodes the solution to the minimization problem, which is the goal. we are going to take the system, and we are going to slowly change the Hamiltonian as the state evolves, according to the Schrodinger equation, the first equation we showed us. Let us give us a little example of how this might work. Suppose that we put a little ball in hand. it is just sitting there. Imagine it in our mind's eye. it is sitting in it is in the minimum of a potential well formed by hand. Now we slowly move the hand. the little ball, as we can see in our mind's eye, stays right there in hand. What is happened in that case is that we are changing the physical system, and we are dragging along the minimum energy configuration. Now, suppose that we put that little ball there, and we jerked hand. The ball would fall away. So, if we let the hand go slowly enough, we will drag along the little ball. the quantum system works in a very similar fashion. If we put a quantum state in Hamiltonian's ground state, then we slowly change the Hamiltonian. If we go slowly enough, the Hamiltonian ground state will correspond to the evolving quantum state. In other words, we can shlep the ground state from the ground state of the beginning Hamiltonian to the ground state of the problem Hamiltonian, if we go slowly enough. that means that we have a way of finding the ground state, which corresponds to the minimum energy configuration, or the lowest value or the cost function that we sought. It will work, for sure, if we go infinitely slowly. Now, infinitely slowly requires an infinite amount of time. Thus, that is not a practical algorithm. So, the real question becomes, can we go slowly enough but not so, slowly that it is impractical, that we can bring the system to the place we want? 

Adiabatic quantum computing (AQC) is a quantum computing model that casts the problem of finding the solution of a computational problem into the problem of obtaining the lowest energy eigenstate of a specified Hamiltonian. Edward Farhi theoretically proposed it, Jeffrey Goldstone, Sam Gutmann, and Michael Sipser in 2000 \cite{farhi_quantum_2000,farhi_limit_1998}, as a general computational model, and it was later shown to be polynomially equivalent to universal, circuit-model quantum computation by Dorit Aharonov and colleagues in 2004 \cite{aharonov_adiabatic_2005}.

The physical principle governing AQC is the adiabatic theorem. Essentially, this theorem states that a physical system will remain in its lowest-energy eigenstate under sufficiently slow changes to the parameters of its Hamiltonian. This fact can be put to computational use by initializing a computer in the ground state of a known Hamiltonian, called the ``initial Hamiltonian'', and then adiabatically evolving its parameters into the ``problem Hamiltonian'' with a ground state that encodes the solution to the computational problem. The slow evolution of the system is intended to guarantee that the system remains in the ground-state throughout this process. However, as indicated below, this is generally not achievable for problems of practical size due to the ``minimum gap'' problem.

An adiabatic quantum algorithm \cite{farhi_quantum_2000} for a given computational problem is comprised of choice for the initial Hamiltonian, the evolution path, and Hamiltonian. None of these choices are unique (except, possibly, the problem Hamiltonian). The initial Hamiltonian is generally chosen to have a known ground state, and simple implementation, e.g., all spins aligned along the X-axis of the Bloch sphere (state $\vert+\rangle = \frac{1}{\sqrt{2}}\left(\vert0\rangle+\vert 1\rangle\right)$ due to an interaction with a large control field. The problem of interest is then embedded onto the hardware to create the problem Hamiltonian, within the constraints imposed by the available qubits and their interactions. Lastly, an evolution path from the initial Hamiltonian to the Hamiltonian's target is chosen, and ideally, this choice respects the conditions imposed by the adiabatic theorem. Using this approach, small-scale adiabatic quantum algorithms for search, factorization, and optimization have been experimentally demonstrated with nuclear spins and magnetic resonance, superconducting qubits, and photonic qubits \cite{thornton_quantum_2019}.

The choice of initial Hamiltonian, evolution path, and problem Hamiltonian together determine how rapidly the system can be evolved and, thus, determine the runtime complexity of the adiabatic algorithm. In principle, the shorter the runtime, the more efficient the algorithm. However, running the algorithm faster than is allowed by the adiabatic theorem decreases the chances of obtaining a correct solution at the end of the evolution process.

In fact, for larger problems of practical significance, adiabatic quantum algorithms generally cannot be implemented without the system leaving the ground state. It is because of the ``minimum gap'' problem. As the system evolves from initial to problem Hamiltonian, the energy levels eventually come close and form an avoided crossing, i.e., come close but do not quite touch. One can intuitively think of this as a ``crossing point'' where the system leaves its initial-Hamiltonian-like nature behind and starts adopting problem-Hamiltonian-like characteristics. In physics jargon, this is essentially a phase transition \cite{zhang_observation_2017}.

The problem is that the gap becomes smaller and smaller as the problem size increases. This results in two key issues. First of all, even in the absence of noise, the system must evolve more slowly through the gap region as the gap gets smaller in order to maintain adiabaticity. If the system evolves too quickly, the system will effectively ``jump'' to the excited state via a nonadiabatic process called a Landau-Zener transition. Avoiding Landau-Zener transitions requires longer evolution times, thus increasing the time-to-solution and slowing the computation.

Second, and more insidious is that there is always some level of noise. As the gap size decreases, noise from the thermal environment and noise from the control fields implementing the evolution will drive transitions from the ground state to the excited state. For example, once the gap energy becomes smaller than the thermal energy $k_{\textrm{B}}T$ at temperature T, the environment will drive transitions. Thus, for large enough problems, it will be practically prohibitive to evolve the system while remaining in the ground state.

Given that an adiabatic quantum computer will leave the ground state at some point during its evolution (for a sufficiently large problem), one may wonder if that destroys any potential for quantum enhancement. After all, what if the computer could return to the ground state after passing through the minimum gap region?

In 1998, Tadashi Kadowaki and Hidetoshi Nishimori proposed such a quantum annealer to solve classical optimization problems. A quantum annealer leverages quantum tunneling \cite{hegade_experimental_2019}and dissipation (relaxation processes) to return a system to its ground state (or a lower-energy state) after passing a small avoided crossing. While such quantum annealers seemingly bypass many of the constraints imposed by an adiabatic quantum computer, there is no evidence that quantum annealing computers feature quantum enhancement to date\cite{vyskocil_embedding_2019,susa_exponential_2018,susa_quantum_2018,kadowaki_quantum_1998,pino_quantum_2018}.

\section{Adiabatic QC}

How do adiabatic algorithms solve optimization problems? In this section, we will discuss the adiabatic algorithm \cite{farhi_limit_1998}, how it is used to minimize ``cost functions,'' and how it is implemented using spin Hamiltonians.

Let us go back to the concept of adiabatic evolution, but let us be a little bit more specific now. So, we want to imagine that we want to find the minimum of a cost function, which is defined over bit strings. that cost function is the sum of terms, each of which depends on a subset of the bits and has a value of 1. Let us say, on certain assignments of those bits in the subset, and has the value 0 on other assignments. For example, if an individual term might look at 2 bits and it will say, if the 2 bits agree in their value, we are going to give it the value 1, but if they disagree, the function has the value of 0. So, now, we want to imagine that the optimization problem is to find the minimum of a function, which is a sum of terms, each of which acts on a few bits. finding the minimum of such a function, in general, is what we call NP-hard. It is not believed that there is any kind of classical time algorithm that can, in polynomial time, solve that problem in general. Nonetheless, we are going to attempt the quantum approach to it. So, the first thing we are going to do is to find what we told us before was the problem Hamiltonian. the problem Hamiltonian we are going to take to be diagonal in the computational basis, so, when the problem Hamiltonian acts on a ket with the string z in it, it will give the value of the cost function times the cat. So, it is diagonal in the computational basis. Note that the minimum string that achieves the minimum value of the cost function can be viewed as the ground state of this particular quantum Hamiltonian because it is diagonal on a computational basis. Now, we need to take a beginning Hamiltonian too. from that, we typically use a magnetic field in the x-direction. If we have a magnetic field on a single spin a half particle and it is pointing in the x-direction, it is represented by the operator. we will take it to be minus $ \sigma $ x so that the state in the x-direction is the minimum with the value minus 1. So, the Hamiltonian we are going to take for the beginning Hamiltonian is the sum, a minus in front, the sum of the $ \sigma $ x is one for each qubit. That is a very simple Hamiltonian. it is very easy to initialize a quantum state as the ground state of that Hamiltonian because all we have to do is take the spins and line them up in the x-direction. Then we are automatically in the ground state of the beginning Hamiltonian. So, now for all full evolution, we are going to take a Hamiltonian that it interpolates between the beginning Hamiltonian we prescribed and the problem Hamiltonian. So, that is all very simple to write down. Not only is it simple to write down, but it is also easy to imagine implementing such a thing in the lab or on a quantum computer because the terms in the problem Hamiltonian are a sum of local terms diagonal in the computational basis. the beginning Hamiltonian is just a magnetic field in the x-direction. So, we can imagine building this thing. what the D Wave company did is build a machine that runs this algorithm. They chose to implement the quantum adiabatic algorithm because it was simple to imagine building a device that could do that. the way we have discussed it, it is kind of an all-purpose minimizer. So, we now have gone as far as to see that we can have a specific implementation of the adiabatic algorithm. So, What is the killer here? What slows it down? What is the thing that makes it hard to implement this in reality and means that it is hard to show the algorithm works all the time? Furthermore, that has to do with how slowly we have to go to guarantee that we are going to go to the place we want to end up at. that slowness is determined by something we call the minimum gap. At the beginning of Hamiltonian, we have a magnetic field in the x-direction. If we flip a single spin, the energy of that single spin goes from Let us say minus 1 to plus 1. So, the energy difference is 2. That is a big energy difference between the ground state and the first excited state. In the problem Hamiltonian, we told we that all terms were zeros and ones, because the cost function, the way we happened to define it, was a sum of terms, which were all 0 or 1. So, that means that the energy difference between the ground state and the first excited state is Let us say, 1 or 2, but it is going to be an integer. However, as the system evolves, we can track the ground state energy and the energy of the first excited state. we find in these situations that as we move along, the two approach each other, coming very, very close. then they are going to spread out back to that integer at the end. the question is, What is that value, the smallest energy difference between the first excited state and the ground state? we call that the gap. the runtime of the algorithm is determined by the gap. The minimum runtime of the algorithm scales like 1 over the minimum gap squared. That is just a little math fact, which we could establish. The whole issue from a physics point of view in the adiabatic algorithm is, does the system, as it evolves, make the energy difference between the ground state and the first excited state get too small? Because if it is too small, the run time required to stay in the ground state is too big. So, one of the things we have to study when we think about adiabatic algorithms is the gap between the first excited state and the ground state as the system evolves.

\section{Quantum Approximate Optimization Algorithm}
In this section, we continues discussion of optimization algorithms \cite{farhi_limit_1998}. Here, we will discuss combinatorial search problems and the quantum approximate optimization algorithm (QAOA)\cite{farhi_quantum_2014,hadfield_quantum_2019}.

we have already established what we think throughout this section, and most of us know that optimization is a key problem in all of computer science. Now, let us get a little more specific and think about what optimization means in combinatorial search. There we have a problem defined on bit strings. we have clauses or constraints, which look at subsets of the bits. for certain assignments of the bits, those clauses are satisfied. for others, they are not. What we want to do in combinatorial optimization, is find a string that can satisfy as many clauses as possible. Often we cannot satisfy all the clauses. Because the clauses have intrinsic contradictions, so that assignments for one clause that satisfy it, force other clauses not to be satisfied. The question then becomes, what is the string that satisfies, Let us say, as most clauses as possible? Now that is very ambitious. Because satisfying the most clauses as possible, that is NP-hard. if we could do that in general, we would have a huge computer science breakthrough. Most people believe that there could be no polynomial-time algorithm for that. So, instead, we are going to come back a little bit and look for a string that satisfies many of the clauses, or as many as the algorithm can produce. So, what we are looking for is a good solution, not the best solution. we are looking for a good string, not the best. In that case, if our good solution is very, very close to the best,  it is as good as the best; but if it is a little off, it could still be of great value to us. Thus, what we are going to be discussing is an algorithm that is an approximate optimizer. It is looking for approximate solutions in the sense we just said. we are going to try this with a quantum algorithm. However, the quantum algorithm is now not like in the adiabatic, and it is going to be based on a continuous-time evolution. However, rather it is going to be in the gate model, which means we are going to apply a sequence of discrete unitary transformations to an initial state, and then make a measurement, and hope that that measurement gives a string that satisfies a high fraction of the clauses that could be satisfied. So, in order to build this up, we need a few ingredients. we need the initial state and the unitaries that we are going to apply. So, let us think about that. well, for the initial state, what we are going to take is all the spins lined up in the x-direction. the reason we like that is we probably already know that if our spin is in the x-direction, it is a uniform superposition of up and down in the z. In other words, it is an equal combination of 0 and 1. So, if all our spins are lined up in the x-direction, our state is a uniform superposition of all possible bit strings. that is kind of a good place to start. Because we have everything in there when we begin, in the computation basis. Now we are going to apply two operators. The first operator is going to depend on the actual operator-valued cost function. The cost function acting on a string gives us the number of clauses satisfied by the total that string. That is an easy-to-compute function. However, we could promote that function to a quantum operator, and that quantum operator acting on the ket labeled by the bit string z will be the function, which is the number of satisfied clauses times the ket. So, this operator-valued cost function is diagonal in the computational basis, and it is easy to construct. However, it is a Hermitian operator. we would like to apply a unitary transformation that depends on it. So, we are going to introduce a parameter, $ \gamma $. we are going to write down the unitary here, e to the minus i $ \gamma $ times that cost function. So, this object is a unitary transformation, which we can easily apply to any state. it depends on a parameter, $ \gamma $, which we will discuss more later. Now the next operator, which we would like to apply again, there is much room for choice. However, we make a choice that will depend on the operator, which we call B, which is the sum of the $ \sigma $ x's, the sum of the Pauli operators in the x-direction. That is an operator, which we discuss in the adiabatic case, we call the beginning Hamiltonian. This operator B, which is the sum of the $ \sigma $ x's, is easy to construct. However, now what we want to do, though, is make a unitary operator, which depends on that. we are going to do that by writing e to the minus i $ \beta $, which is the parameter we have just introduced, times the operator B. So, this operator, if we look at it, e to the minus i $ \beta $ B, since each term in B is a $ \sigma $ x, what it is doing is rotating the spins around the x-direction by an angle, $ \beta $. So, now we have the two building block unitary operators, each of which depends on one parameter. So, the algorithm consists of the following. Start with the initial state, which we already described. Choose $ \gamma $. Apply the unitary e to the minus i $ \gamma $ C to that state. Then choose the $ \beta $. Apply either the minus i $ \beta $ B to the state, and now measure in the computational basis. we get out a string. now, What is interesting is that we can show that for certain problems, we can find values of $ \gamma $ and $ \beta $ such that the string that is produced will satisfy a high fraction of the clauses. we are not going to show we now how we find the $ \gamma $ and $ \beta $ to do that. However, that can be done for certain problems. in fact, for one problem, called MAX-CUT, when we first introduced this algorithm in 2014, we were able to show that we could achieve a restricted version of MAX-CUT \cite{guerreschi_qaoa_2019}. This certain approximation ratio was better than we would get if we just picked a string at random. Now we could say, so, what? we did a little better than random guessing. However, we managed to show that this worked in every case, no matter how big the instance was, no matter how many bits or qubits were involved. With the right choice of $ \gamma $ and $ \beta $, we were guaranteed to get a good answer. So, in this case, we were able to show that the quantum computer would perform at a certain level, even on instances that we had not yet seen. The factoring algorithm has that great virtue. we know that if we had a functioning quantum computer, no matter what integer we give, it will be able to find its factors. Thus, we were able to show that we could get an approximation ratio that beat random guessing, no matter what we gave it. Nevertheless, we were not beating the best classical algorithm. then we looked at another problem called E3LIN2. We could be much more specific. However, for now, just go with the story. we looked at this problem, E3LIN2. We showed that with just this shallow depth circuit, we could achieve an approximation ratio, which is 2014 beat the best classical algorithm for this problem. Now at that moment, We had a quantum algorithm that beat the best classical algorithm for a certain task in all cases. The word of this spread. The way it spread was basically because Scott Aaronson blogged about it and said there was this neat new quantum algorithm, which got the classical computer scientists and the non-quantum people together. 10 of them ganged upon, and they devised a classical algorithm that beat the quantum algorithm for this task. Nevertheless, what was interesting there is we got a competition going between the classical and the quantum. at this moment, in terms of provable worst-case performance guarantees, the classical algorithm is outperforming the QAOA. The QAOA is the Quantum Approximate Optimization Algorithm. Now we can go further with the QAOA than we have already outlined. Because of the way we have outlined it, it is a very simple algorithm with just two unitaries. we apply the first one, then the next one, then we measure. Nevertheless, we can go further. we can apply the first one and then the second one. Then we can apply the first one again with a different parameter. then, we can apply the $ \beta $ dependent operator with a different parameter. Then if we keep going, by alternations of those two operators, it turns out that we can only improve the performance of the algorithm. Now we have had trouble from a technical point of view showing how much that improvement consists of. It has been too hard for us to analyze it beyond the very shallowest depth circuits that we did. However, we optimistic that when we run a near term quantum computer, if we apply this algorithm which should be fairly simple to do because it is a shallow depth gate model with simple-to-construct operators; that we might discover that there are parameter choices for which the solutions we get are quite good,  even as good as what we get from classical algorithms,  even better. So, we think that applying the Quantum Approximate Optimization Algorithm to near-term quantum devices is a definite avenue for this community to go down. we look forward to seeing the algorithm run on any platform capable of running it. 

A combinatorial optimization algorithm finds solutions to a problem subject to a set of constraints, and the optimal solution is the one that satisfies the largest number of constraints.

The problem is defined by an object function C(b), which is a sum of m constraint clauses $C_\alpha$, where $\alpha =1...m.$ Each clause is a Boolean function that captures one specific constraint of the optimization problem. The Boolean functions are constructed from n Boolean variables $b_1 \dots b_ n$, which together form an n-bit string b, and there are $2^ n$ such strings. Note that not all Boolean variables need to be used in every clause, and in general, they are not. If a given b satisfies a clause, then $C_{\alpha }(b)=1$ according to the Boolean function it represents; otherwise, $C_{\alpha }(b)=0$.

The $b=b_{\textrm{opt}} $ that maximizes the object function $C(b)$,
\begin{equation}\label{eq2_117}
C(b_{\textrm{opt}})=\sum _{\alpha =1}^ m C_{\alpha }(b_{\textrm{opt}}) \geq \sum _{\alpha =1}^ m C_{\alpha }(b \neq b_{\textrm{opt}})
\end{equation}

Is the optimal solution (or, an optimal solution if more than one solution exists). If all clauses are satisfied, $C(b_{\textrm{opt}})=m,$ but this is not required.

In general, evaluating the optimal solution is a hard problem. An approximate solution that is not necessarily optimal but is still of reasonably high quality is sufficient for many optimization problems and maybe computed more efficiently. The quantum approximate optimization algorithm (QAOA) is proposed to find approximate solutions to a combinatorial optimization problem.

QAOA is a hybrid approach that utilizes a digital quantum computer (QC) to evaluate the object function and a classical optimizer (CO) to update trial solutions to optimize the object function. The iterative process on n qubits approximated with p optimization steps can be summarized as follows:\\
$\displaystyle 1.~ ~ ~$    $\displaystyle \text {CO: generate initial set of angles $\gamma _1$ and $\beta _1$ to optimize the object function}    $     \\
$\displaystyle 2.~ ~ ~    $ $\displaystyle \text {QC: create superposition of $n$ qubits in computational basis ground-state}     $     \\
$\displaystyle 3.~ ~ ~$         $\displaystyle \text {QC: rotate qubits on Bloch sphere by angles $\gamma _1$ (z-axis) and $\beta _1$ (x-axis)}$    \\      
$\displaystyle 4.~ ~ ~$     $    \displaystyle \text {QC: measure qubits in computational basis and evaluate the object function}    $      \\
$\displaystyle 5.~ ~ ~    $ $    \displaystyle \text {CO: generate new set of angles $\gamma _2$ and $\beta _2$ to optimize the object function}$     \\     
$\displaystyle 6.~ ~ ~ $   $\displaystyle \text {QC: create superposition of $n$ qubits in computational basis ground-state}    $      \\
$\displaystyle 7.~ ~ $    $\displaystyle \text {QC: rotate qubits on Bloch sphere by angles $\gamma _1,\gamma _2$ and $\beta _1,\beta _2$}    $     \\ 
$\displaystyle 8.~ ~ ~$        $ \displaystyle \text {QC: measure qubits in computational basis and evaluate the object function}$     \\     
$\displaystyle \vdots ~ ~ ~ ~    $      \\          
$\displaystyle 4p-3.         \displaystyle \text {CO: generate new set of angles $\gamma _ p$ and $\beta _ p$ to optimize the object function}$    \\      
$\displaystyle 4p-2.         \displaystyle \text {QC: create superposition of $n$ qubits in computational basis ground-state}     $     \\
$\displaystyle 4p-1.         \displaystyle \text {QC: rotate qubits on Bloch sphere by angles $\gamma _0,\gamma _1,\dots ,\gamma _ p$ and $\beta _0,\beta _1,\dots ,\beta _ p$}          $
$\displaystyle 4p.~ ~         \displaystyle \text {QC: measure qubits in computational basis and evaluate the object function}$

The algorithm is composed of two types of operators constructed with unitary quantum gates to rotate the qubits by angles $\gamma $ and $\beta$.
\begin{equation}\label{eq2_118}
\displaystyle C(b)=\sum _{\alpha =1}^ m C_\alpha (b)    \displaystyle \rightarrow    \displaystyle U(C,\gamma )=e^{-i\gamma C}=\Pi _{\alpha =1}^ m e^{-i\gamma C_\alpha }
\end{equation}
         
\begin{equation}\label{eq2_119}
\displaystyle B=\sum _{j=1}^ n X_ j    \displaystyle \rightarrow    \displaystyle U(B,\beta )=e^{-i\beta B}=\Pi _{j=1}^ n e^{-i\beta X_ j}
\end{equation}

The operator C is based on the object function C(b), and it conditionally rotates qubits by an angle $\gamma$. For example, a clause $C_{\alpha }=\frac{1}{2}(-Z_ jZ_ k+1)$ rotates qubits j and k by an angle $\gamma$ around the z-axis if they are in opposite states. Therefore, only qubits complying with the clauses of the objective function are affected by the rotation.

The second operator, B, applies a rotation of angle $\beta $ around the x-axis to all qubits. Note that states co-aligned with the x-axis remain unchanged for an X-rotation.

An initial state $ \lvert \psi \rangle =\frac{1}{\sqrt{2^ n}}\sum _{s=1}^{2^ n} \lvert s\rangle $ is iteratively rotated conditionally around the z-axis, and then around the x-axis, to reach a particular location on the two-dimensional Bloch sphere. The measured resulting state after p optimization steps is:
\begin{equation}\label{eq2_120}
\lvert \vec {\gamma }, \vec {\beta } \rangle = U(B,\beta _ p)U(C,\gamma _ p)\cdots U(B,\beta _1)U(C,\gamma _1)\lvert s \rangle
\end{equation}

It can be shown that the expectation value of the object function improves with additional iterations and ultimately maximizes it. The performance of the underlying quantum processor usually determines the approximation accuracy.
\begin{equation}\label{eq2_121}
\max _{\vec {\gamma }, \vec {\beta }} \langle \vec {\gamma }, \vec {\beta } \vert C \vert \vec {\gamma }, \vec {\beta } \rangle =M_ p \geq M_{p-1}~ ~ \Rightarrow ~ ~ \lim _{p\rightarrow \infty } M_ p=\max _ z C(b)
\end{equation}

The algorithm's key element is the ability to conditionally project qubits on arbitrary locations on the Bloch sphere via iterative rotations induced by two distinct unitary operators. Each iterative step starts with a new set of classically optimized angles based on the qubit measurement of the prior step. A sufficient number of repetitions approximates the object function C(b), and the measured qubit state reflects the approximately optimal bit string b.

There was a question about quantum algorithms, Is it always possible to transform any classical equation or algorithm into a quantum equivalent, or vice-versa, a quantum equation into a graphical tool, or back into a classical equivalent, The answer is yes, if we have a universal, fault-tolerant quantum computer, it is universal, and in principle, it can perform any algorithm, quantum or classical. Now, the key here though is that, the quantum computer may do no better than a classical computer, and oftentimes, it may do far worse, because it runs at a much slower rate, it does not have quantum advantage for that particular type of problem. so, although in principle it can do it, it may not be any better than a classical computer, and in fact, may be slower. Now similarly, we can take a quantum equation, or quantum algorithm, and we can simulate that, or implement that, on a classical computer, but to do that, we need to discretize the problem. We often need to make some assumptions, truncate the problem, reduce it down in some way so that it is tractable to implement on a classical computer and in doing that, we lose information, we lose fidelity. that is why it is generally hard to implement quantum algorithms, or quantum systems, to simulate quantum systems on a classical computer. Now, we could do it in full glory, but then that takes the age of the universe? it is this type of problem if we want to do it and simulate it, with the highest possible fidelities on a classical computer, it is going to take way too long, or it takes way too much memory to implement or to be practicable. If we cut down the problem so that we can do it on our classical computers, then we do not get accurate answers? And so, this motivates why we need to have a quantum computer. Now, is it always possible to transform quantum equations into its equivalent in a graphical tool set, like we have  experienced with the IBMQ Quantum Experience. In principle, yes, but, a graphical user interface is an abstraction, so it is a convenience that we build in to, our human-computer interfaces, it is a tool. So, the graphical interface is limited to the extent that some of these programs has capability into it. so, perhaps there are gate operations that are not represented yet in their full generality in that graphical user interface. So it is in principle possible, but this is one reason why we do programming, using programming languages, rather than using a GUI interface, is because we can do operations with more generality when we do a lower level of programming rather than a higher level of graphical user interface. But as an in-principle, sure, there is nothing that prevents us from implementing everything that we would possibly calculate or program into a graphical user interface. It just may become bulky, and it needs to be designed very well, so that human-computer interface is still efficient with all that information. 

\section{Quantum Annealing: Concept} 

How are quantum annealers different from adiabatic quantum computers? Furthermore, how are they used to solve optimization problems? In this section, we will discuss quantum annealers.

In the previous section, we discussed a paradigm of quantum computing known as adiabatic quantum computation. In this section, we are going to introduce another related quantum computing paradigm called quantum annealing. As we will see, quantum annealing bears many similarities to adiabatic quantum computation, though with a few key differences. First, recall that adiabatic quantum computation is a form of analog quantum computing. That is, rather than digitizing the computation into discrete gates, AQC relies on the continuous transition of a quantum system from the ground state of one Hamiltonian to the ground state of another. While it is true that adiabatic quantum computation is a provably viable method of universal quantum computation\cite{kitaev_classical_2002,goto_universal_2016}, ACQ presents a serious practical challenge. In order for the computation to succeed, the system must remain in the ground state for the duration of the computation. If the evolution is not truly adiabatic, and the system ends up in an excited state at the end of the protocol, the computation fails. Unfortunately, this is an extremely hard condition to avoid in any realistic quantum computer. To see this, let us consider what the system's energy levels might look like as we perform the computation. In particular, Let us focus on the energy gap between the ground state of the system and its first excited state. As the Hamiltonian of the system changes throughout the computation, the gap between the ground and excited states also changes, and in general, it can become very small. It is where the trouble starts. For any physical system at finite temperature, the environment around the qubits generates thermal photons with an energy proportional to the environment's temperature. When these photons encounter the qubits, their energy is absorbed, and the system has some probability of being excited to a higher state. The smaller the gap with respect to the photon energy, the higher the probability that excitation will occur. If we want to reduce the probability of an excitation occurring for a fixed gap size, we need to cool the system to a temperature much lower than the energy gap between the qubit states. It is precisely why we operate superconducting qubits in a dilution refrigerator cooled to a temperature of around 10 millikelvins. For typical superconducting qubits, the gap between the bare qubits' states is on the order of 5 gigahertz, which corresponds to a temperature of around 250 millikelvins. By operating the qubits in an environment order of magnitude colder, we dramatically reduce the probability of an unintentional excitation occurring during the computation. Unfortunately, as we change the Hamiltonian during an adiabatic quantum computation, new avoided crossings will open up between the states, which can be much smaller than the energy gaps between the barer qubit states. If we want to avoid transitioning across these small gaps in the case, say, of a superconducting annealer, we would need to engineer a refrigerator that operates much colder than 10 millikelvin, which is not practical given current cryogenic technology \cite{bardin_28nm_2019}. this type of problem generalizes other qubit technologies as well. there is a second problem. As we evolve the computer through a small gap region, if we sweep through the gap too quickly, we can effectively jump to the excited state. It is called a Landau-Zener transition. So, even if we could operate the computer at zero temperature, we would need to evolve the system sufficiently slowly to accommodate smaller and smaller gaps and thereby increase the time to solution. So, although adiabatic quantum computation can, in principle, be universal, it has major drawbacks related to the small gap sizes that occur at larger problem sizes. Given this reality, it is natural to ask, what happens if we allow the computer to leave the ground state during the evolution. Can it return to the ground states, or even a lower excited state, through quantum tunneling and relaxation processes? Furthermore, more importantly, is leaving the ground state consistent with the quantum-enhanced operation? Quantum annealing is an adiabatic quantum computer that has left its ground state at some point during the evolution. conceptually, a combination of quantum tunneling and relaxation processes are leveraged to allow the computer to return to its ground state, or at the very least, a lower energy state. However, to date, there was no proof or evidence that the quantum and dissipative dynamics in quantum annealing could then forward a quantum enhancement for a general class of problems. It remains an open research question. Much like AQC, the principle of quantum annealing involves evolving a quantum state according to some time-dependent Hamiltonian such that the state of the system at the end of the computation encodes the result of the problem of interest. However, unlike ACQ, which is universal and can, in principle, implement any quantum algorithm, quantum annealing is designed only to solve a particular class of computations known as classical optimization problems. we deal with classical optimization problems in every industry and throughout the daily lives. For example, What is the best portfolio of stocks to maximize return on investment? What is the best fighter jet designed to minimize drag? Or to use a more mundane example, What is the best path home from work? The key to an optimization problem is that there are many potentially viable answers to a given problem. For example, many paths will take us home from our office. The goal of an optimization problem is to find the most efficient solution from amongst all the possible solutions. we quantify this by introducing a cost function, which tells us the relative efficiency of a solution. For the commute home, the cost function could be the time it takes to get home if we followed a particular route. Alternatively, it could be the amount of gas we use. Whatever we are optimizing against, the optimal solution tells us the one that minimizes the cost function. although there may be a truly optimal solution, it is very often the case that there are several nearly optimal solutions that are all, in practice, very good. Commutes that differ by a few meters on 10 kilometers or differ by a few drops of gas on a liter are essentially interchangeable. the hope for these kinds of problems is that leaving the ground state may be fine if we end up in a reasonably low energy state. To map an optimization problem onto a quantum annealing problem, we make use of a particular Hamiltonian known as an Ising Hamiltonian. The Ising Hamiltonian was first formulated in the context of quantum magnetism between atomic spins. For a system of physical spin 1/2 particles, the Ising Hamiltonian tells us the energy of the spins as they interact with external magnetic fields plus the energy of the magnetic interactions between pairs of spins. So, how can this physics problem be used for optimization? As it turns out, we can use the Ising Hamiltonian to encode so,-called classical optimization problems\cite{cervera-lierta_exact_2018}. By choosing the right combination of interaction strings between spins, we can mimic the cost function of the problem we are interested in. When we encode this optimization problem into the Ising Hamiltonian, we then find that the Hamiltonian's energy levels corresponding to the possible solutions to the optimization problem with the energy at each level corresponding to the value of the cost function for that particular solution. Higher energies incur higher penalties. The goal of quantum annealing is to engineer a qubit system with an Ising Hamiltonian corresponding to the optimization problem of interest \cite{inagaki_coherent_2016}. we then want to evolve the system such that it ends up in a low-energy eigenstate of this Hamiltonian at the end of the computation. The lower the energy of this final stage, the more optimal the solution. To perform this evolution, we start by initializing the system in the ground state of a much simpler Hamiltonian. Conventionally, this Hamiltonian is the sum of the x operator acting on every qubit, which is analogous to having an array of spin 1/2 particles in a global magnetic field pointing transversely to the spins. When we work out the math, we find that the ground state of this Hamiltonian is straightforward. It is the state where all the qubits are in the plus x state on the equator of the block sphere, which is a relatively easy state to the initialize system in. Once we prepare the system in the ground state of this initial Hamiltonian, we gradually reduce the strength of this initial Hamiltonian and turn on the Hamiltonian, which encodes the optimization problem. In the ideal case, the system evolves adiabatically and remains in the ground state throughout the evolution. It would be an adiabatic quantum computer simulating an Ising Hamiltonian. Ideally, at the end of the evolution, the system would still be in the ground state of the problem Hamiltonian corresponding to the solution, which minimizes the cost function. In other words, we would find the absolute shortest route home. However, as discussed at the beginning of this section, it is very unlikely that we will remain in the ground state throughout the computation. It does not necessarily ruin the computation. Because every energy state of the problem Hamiltonian corresponds to a possible solution, even if we exit the ground state and end up in an excited state at the end of the computation, this excited state may still give us a useful answer to the optimization problem. The question is, how good is this answer. Unfortunately, while quantum annealing relaxes the constraint that we stay in the ground state, it makes no guarantees that the result we arrive at is any more optimal, or any more efficiently found, than the result we might calculate on a classical computer. As such, unlike AQC, quantum annealing can make no provable claim to either universality or a quantum enhancement, and there is no demonstration to date of quantum enhancement for a class of optimization problems. That said, even if a quantum annealer ends upscaling with problem size like the classical computer, it may still be useful if it offers a significant advantage over existing classical computing alternatives. However, this also has yet to be demonstrated. Nonetheless, there is tremendous interest in quantum annealers because optimization problems are ubiquitous. while large-scale universal quantum computers are still a long way from commercial reality, the first generation of commercial quantum annealers designed by D-Wave currently boasts up to 2,000 qubits\cite{mcgeoch_practical_2019,pelofske_solving_2019}. while the D-Wave system has yet to demonstrate quantum enhancement for a general class of optimization problems \cite{stollenwerk_flight_2019}, these machines, as currently constructed, only explore a small part of the quantum annealing design space\cite{streif_solving_2019}. As work gets underway to design the next generation of quantum annealers, it remains an open question if systems with higher quantum coherence and more sophisticated qubit connectivity might demonstrate a truly conclusive quantum enhancement.

Quantum annealing (QA) is a method for finding solutions to classical optimization problems. It is essentially an adiabatic quantum computer that is allowed to leave its ground state during its evolution. Then, through a combination of quantum tunneling and relaxation, the system finds its way back to the ground state (or a lower-energy state), which encodes an (approximate) solution to the optimization problem.

Quantum annealing borrows its name from thermal annealing. For example, in metallurgy, annealing is a process by which a metal is heated such that it is physical shape and composition can be modified. Then, upon cooling, the metal relaxes into this new shape and composition, from which it will not change unless heated. Loosely speaking, the thermal fluctuations present during the heating process enable the metal to reconfigure into an alternate, advantageous configuration.

To gain insight, it is instructive to consider a classical algorithm called simulated annealing (SA)\cite{isakov_optimised_2015}, which uses a thermal activation model to find approximate solutions to optimization problems. Intuitively, SA algorithms seek to replicate thermodynamics, where the cost function is encoded into the total energy of the system. The algorithm seeks to find a low-energy configuration as a parameter analogous to temperature T ``reduced'' to zero. Operationally, each iteration of an SA algorithm begins with a valid solution with energy E and randomly selects another solution near the current one. The new solution can either have a higher $(E'>E) $or lower $ (E'<E)$ energy, where E' is the energy of the newly selected solution. In one of the most basic implementations of SA, if $ (E'<E)$, then the new solution is chosen as the starting point for the next iteration. However, a key feature of SA is that even when $E'>E$, although representing a less optimal solution, there is a non-zero probability that the algorithm will use it in the next round. It is intended to mimic thermal fluctuations transiently exciting a physical system, and it helps the algorithm escape from local minima in the potential landscape. The algorithm starts at ``high temperature'', and the probability of keeping these $E'>E$ solutions is relatively large. As the algorithm proceeds, the temperature is ``reduced,'' and the probability of retaining such solutions is reduced until the temperature reaches zero. The system is ``frozen'' into a local or global minimum.

Quantum annealing is an analog to simulated annealing\cite{isakov_optimised_2015}. Rather than thermal fluctuations, however, quantum annealing relies on ``quantum fluctuations.'' Quantum fluctuations are generally manifest in the quantum wavefunctions that represent quantum states. In general, a wavefunction is not localized to a single point, but, rather, is distributed or ``extended'' over some distance in parameter space. Thus, the wavefunction has both mean value and a standard deviation, which intuitively gives some notion of quantum fluctuation. Repeated measurements of an identically prepared system would project out different measurement results. The wavefunction is extended, and the variation in measurement results is a manifestation of the underlying quantum fluctuations.

In quantum annealing, when the system wavefunction is extended between different local minima in the potential landscape that describes the problem, the corresponding ``quantum fluctuations'' are manifest as quantum tunneling between potential wells. Thus, quantum annealing relies on quantum fluctuations that enable quantum tunneling through barriers rather than thermal fluctuations, activating the system over a potential barrier. The probability of this tunneling is controlled by the strength of a transverse field in the annealing Hamiltonian a field along the x- or y-axis of the Bloch sphere instead of the temperature.

As with adiabatic quantum computing, the cost function for QA is encoded into a Hamiltonian $H_{\textrm{problem}}$ such that the system ground state represents the optimal solution. The system is initialized in the known ground state of a starting Hamiltonian $H_{\textrm{initial}}$, which typically has all qubits co-aligned with a large, transverse X-field applied to all qubits. Then, the system is adiabatically evolved from the initial Hamiltonian to the final problem Hamiltonian:
\begin{equation}\label{eq2_122}
H(t) = A(t) H_{\textrm{initial}} + B(t) H_{\textrm{problem}},
\end{equation}

where $A(0)=1 $and $B(0)=0$ at the beginning of the evolution, and $A(t_{\textrm{final}})=0$ and $B(t_{\textrm{final}})=1$ at the end of the evolution.

Moreover, as with an adiabatic quantum computer, the system will encounter a minimum gap at some point during the evolution, and, for practical problems of interesting size, almost certainly leave the ground state due to noise and Landau-Zener transitions. The hope for quantum annealing is that through a combination of quantum tunneling and dissipation, the system can return to lower-energy states or even the ground state.

The most advanced quantum annealing system is the commercially available D-Wave computer, which features more than 2000 superconducting qubits with fixed connectivity called the ``chimera graph''. The D-Wave computer is certainly an engineering achievement, having integrated classical superconducting electronics to calibrate the more than 2000 qubits and implement an annealing schedule. However, despite many years of testing, there is no theoretical or experimental evidence that quantum annealing affords a quantum enhancement for a general class of optimization problems.

Whether quantum annealing will ever be able to provide quantum enhancement remains an open question. In rapidly scaling to 2000+ qubits, D-wave engineers had to make system choices that constrained them to use relatively low-coherence qubits, fixed connectivity, and solely ZZ interactions. It represents only a small part of the possible annealing ``phase space.'' Using higher-coherence qubits, other types of interactions, e.g., XX or YY, multi-body interactions, e.g., XXX, and enhanced connectivity graphs, may improve performance. Even if quantum annealers are ultimate ``just classical computers'' that scale classically with problem size, they may still be valuable if they can outperform existing classical computers for certain problems.

\section{Quantum Annealing} 
In this section, we will continue our exploration of quantum annealing and on the history and challenges of quantum annealing. we think we should say something about D-wave. It certainly is an interesting story, so that we would put a timeline. In 1999, D-wave was founded. presumably, they came out of nowhere. The founders must have been doing some work before in 1999. In 2000, Fahri, Goldstone, Gutmann, and Sipser wrote a paper that proposes adiabatic quantum computation. a couple of years later, Aharanov, unfortunately, we did not look up the exact date before this. A couple of years later, Aharanov et. al. Show it is universal. in 2007, D-wave announces first a 16-bit prototype adiabatic quantum computer. there is a lot of skepticism about this. So Scott Aaronson goes on a tirade in his blog and launches a crusade against them, saying they could not possibly say the things are saying could not possibly be true. that is well, and there is a couple of questions here. When they were founded, adiabatic quantum computation did not exist. So presumably, they were founded on some other kind of quantum computation. at some point, we decide adiabatic quantum computation was way to go. sometime between 2007 and today, we could not figure out when D-wave now says they do quantum annealing. So what is the difference? Well, adiabatic quantum computation, we are in the ground state, and we would like to operate at a temperature low enough to stay in the ground state. So what is this temperature? We call quantum adiabatic computing ground state. Let us have a first excited state. this is what is known as an avoided crossing, where this line and this line have roughly the same structure of the quantum state, as do this line and this line. we get a superposition of these two structures right here. the more these two structures differ, the smaller the energy gap is between them. So that is an oversimplification and waving hands, but it is something like that.  Can we evolve our Hamiltonian quite quickly most of the time, and then just slow down that evolution, when we are going past we can, we know where the gaps are, or rather, if we know where these things are, we can evolve it quickly most of the time. in fact, in Grover's algorithm, we can get a square root speed up by adiabatic computing, or a square root speed up over the classical search by adiabatic computing. However, the way to do that is in Grover's algorithm, and there is one spot where we have an avoided crossing, which is we think 1 over square root of n. to get the speed up, we have to go quickly most of the time, then slow down right before the avoided crossing. then speed up right after it. If we do not know where these points where they come close are, it is going to be hard to speed up and slow down our evolution to account for them. the temperature means that one of our qubits might get excited, but if we excite this state, it is not going to go to this state, because the state and this state are structurally quite different with exciting qubit in this state, and we go up to this state. So the temperature is bigger than the smallest gap. So gaps account for energy, and energy converts to temperature by thermodynamics. So the right temperature is probably going to be bigger than the smallest gap. However, it is still small. D-wave does not operate at a temperature small enough to avoid thermal excitations. that is not because they do not operate at a small temperature. They operated at extremely low temperatures. it is just that the structure of their ships allows for thermal excitations, even at that temperature. So people were arguing with them about whether they could call this quantum adiabatic computation. So they decided to call it quantum annealing instead. So what does that mean? If quantum adiabatic computation were quantum, there would be proof that the quantum computer was universal, which means that theoretically, it could solve any problem that any quantum computer can. However, the problem of blowing up the size is due to a polynomial factor converting from the gate model to the adiabatic computation model and back. So if it were a quantum adiabatic computer, they would have a good reason to claim that it could do things that classical computers cannot. Well, they have a quantum annealing computer. it is certainly using quantum processes and operating in the quantum, but is it more powerful? Moreover, we want to say no theoretical results. There are chips that we can program the Hamiltonian, and we can program a Hamiltonian to change from one Hamiltonian to another Hamiltonian slowly. we can put the starting Hamiltonian into the ground state. we are operating at a temperature too large not to have thermal excitations. So presumably, we could take this process and construct some mathematical model of it. Although we might have to worry about what the right temperature for this mathematical model is. then we can ask, does this mathematical model let us do anything that we cannot do on classical computers? we do not know. Nobody knows. the other problem is that they have experiments that they claim show that the computer does things better than classical computers, but is it an exponential speed up? Or is it just that their computer has electrical? It has a Hamiltonian trying to find its ground state, and it does this reasonably fast. there is a constant factor better than classical computers that are letting it find the ground state better than classical computers can because physics processes can be really fast. So we do not know whether D-wave is more powerful.

    \item Quantum Annealing II; In quantum annealing, what is the most relevant consequence of the system temperature not being low enough?
    \begin{itemize}
        \item The minimum energy gap between the ground state and the first excited state is reduced due to the temperature, inducing Landau-Zener transitions.
        \item energy from the environment excites the system, bringing it out of the ground state.
        \item The coherence time of the spins is reduced, and the gate fidelity is reduced.
        \item  The minimum computational time necessary to keep the system in its ground state decreases.
    \end{itemize}
    Solution:\\
    The minimum energy gap does not generally depend on temperature.
    Although the coherence of a single spin might be affected by temperature, quantum annealing is not based on digital gate operations.
    The minimum computational time refers to the minimum time required to evolve the system to the ground state of the desired Hamiltonian without excitation to an excited state. The minimum gap energy sets it, and it generally does not depend on the temperature of the system.

\section{Grover’s Algorithm: Quantum Search} 
In this section, we will discuss Grover's algorithm and how it works. At the end of this section, we will implement Grover's algorithm on the IBM quantum computer. 

In this section, we will take a look at Grover's algorithm. as we have done previously with Deutsch-Jozsa and Shor's algorithm \cite{shor_polynomial-time_1997}, we will do it in the quantum circuit model. by the end of the section, we will implement it for yourselves on a real cloud-based quantum computer\cite{dumitrescu_cloud_2018}. Grover's algorithm is often described in the context of searching an unstructured database. Here though, Let us generalize the search problem to the related problem of function inversion. Consider a function, f of x, which can take on the value 0 or 1, depending on the argument x. For example, for x equals 10, the function f of 10 equals 1, and 0 for all other x values. Now, the function inversion problem is the following. Given f of x equals 1, can we find the value of x or a value of x if there is more than one solution. Classically, one has little choice but to try different x-values and test the function. So, for n total x-values of which M satisfy f of x equals 1, it will take on the order of N divided by M tries to succeed with high probability. However, on a quantum computer, the same task succeeds with high probability in the only square root of N over M tries, a quadratic improvement over classical computing. while a quadratic improvement may not seem like a lot, it does become practically significant for problems with large N. To implement Grover's algorithm, and we need an operational block that can recognize if a test value of x satisfies the function or not. This operation is called an oracle. it is a key part of each trial, which we will call a Grover iteration. Now, it may seem challenging or even cheating to define an oracle that represents, for example, the database. After all, in loading such a database into the quantum computer, have not we already searched it? Now, that is certainly true, but this is not how an oracle works. Again, it is worthwhile to think in terms of a mathematical function. The oracle does not need to know every answer to every function call, nor does it need to have a ledger of all the input and output values. Rather, it just needs to be able to calculate the function and recognize when a particular condition is satisfied. we can conceptually program different functions into the oracle while keeping the underlying machinery in place. we will not discuss in detail how this is done, as it is problem-dependent. However, we just note that this type of oracle can be designed using reversible logic approaches. The idea behind Grover's algorithm is the following. we start with an index register of values x running from 0 to n minus 1. The oracle takes a test value x, calculates the function f of x, and performs modulo-2 addition of f of x with a helper qubit, which we will call the oracle qubit. The oracle qubit is prepared in a superposition state 0 minus 1, and it plays a similar role as the helper qubit did in the Deutsch-Jozsa algorithm studied in section 1. If f of x equals 0, then the modulo-2 addition leaves the superposition state unchanged. However, if f of x equals 1, then there is a sign change on the oracle qubit. 0 minus 1 becomes 1 minus 0. by pulling out this minus sign, we return to the original superposition state, 0 minus 1, but with a minus sign in front. as we discuss with the Deutsch-Jozsa algorithm, this minus sign can be written as minus 1 to the f of x power. Thus, the minus 1 is realized when f of x equals 1, that is when a particular value of x is a solution to the search problem. It is how the oracle marks a quantum state when it recognizes a satisfied condition. To implement Grover's algorithm, we start with the initialization stage of the quantum circuit model. Since there are n possible values of x, we will choose q qubits for register 1. register 2 will hold a single qubit, which is the oracle qubit. Register 1 qubits are prepared in state 0 and the oracle qubit in state 1. as before, we will associate register one with yellow and register 2 with green to help keep track as the algorithm unfolds. In the compute stage, the first step is to create a large equal superposition state of register 1 qubits using q Hadamard gates. This state has n equals 2 the q-th power components. as we did before, we will replace the 2 to the q-th power binary combinations of q qubits with decimal equivalence x equals 0 to n minus 1. The oracle qubit is placed into the superposition state 0 minus 1 using a single Hadamard gate. The next step is to implement a Grover iteration, which we will call G, and we will do this on the order of square root of n times to obtain a solution with order unit probability. Each G contains the oracle operator O, which with the help of the oracle qubit, will leave a minus 1 on the register 1 states for values x, which satisfy f of x equals 1. as we will see in a moment, the oracle operation has a geometric interpretation. It is a reflection operation. we will come back to that in a moment. Now, the remainder of the Grover iteration is a conditional phase shift nestled between Hadamard gates. The conditional phase shift leaves state 0 alone and, otherwise, places a minus 1 phase shift in front of all other states x. It can be written in terms of an equal superposition state, just like the one we started with. we will call this state CI. we encourage us to work this out for yourselves. as we will see in a moment, this set of three operations can also be viewed as a reflection across the vector psi. Together with the oracle, these two reflections in the Grover iteration serve to rotate the quantum state by an angle $ \theta $. we will look at the geometric picture in a moment. However, for now, let us continue with the circuit model. To see how the Grover iteration acts on register 1, Let us rewrite the sum over x as two terms. One term will be the sum over all x-values that are not solutions, that is, give f of x equals 0. The second term will be the remaining x-values that are a solution and give f of x equal 1. This, in turn, can be written in terms of normalized states, $ \alpha $, and $ \beta $. we then run the Grover iteration on the order of square root of n times. Let us assume there are k iterations, with each iteration imparting a phase shift $ \theta $. The net result is the cosine of 2k plus 1 over 2 in front of state $ \alpha $, which is not solutions, and sine 2k plus 1 over 2 in front of state $ \beta $, which is the solutions. the oracle qubit remains unchanged throughout. Starting from a small angle $ \theta $, much less than $ \pi/2 $, we see that as k increases, the cosine term goes to 0 suppressing the amplitude of the wrong answers and the sine term towards 1, enhancing the amplitude of the correct answers. Then a measurement is made, and one of the x-value solutions in the state $ \beta $ is measured with high probability. Let us now look at the geometric interpretation to see how repeated Grover iterations lead to the correct solution. The state CI is in the register 1 equals superposition state with end components. we write it in terms of the states $ \alpha $ and $ \beta $, where $ \alpha $ is all the x-values that are not solutions, and $ \beta $ represents the single x-value or multiple end values that are solutions yielding f of x equals 1. The state's $ \alpha $ and $ \beta $ are orthogonal, and we can represent them as orthogonal vectors. Since the state CI is normalized to 1, we can also write state CI in terms of the cosine and sine $ \theta/2 $. the state CI makes an angle $ \theta/2 $ with the $ \alpha $ axis., and Now since we start with all x-values equally likely, and we assume there are only one or a few solutions, the vector CI is close to the $ \alpha $ axis. The oracle acts first, and it places a minus sign in front of the state $ \beta $, which represents the valid solutions. It is a reflection over the $ \alpha $ axis. The remainder of the Grover iteration, the conditional phase shifts nestled between Hadamard gates, is a reflection of the state CI. The result from a single Grover iteration is then to move the state CI and angle $ \theta $ towards the $ \beta $ axis. Essentially, the amplitudes of the solutions are enhanced, while the amplitudes of the remaining states are being suppressed. The Grover iteration is then repeated k times, where k is on the order of square root of N. The iterative reflections continue to push the vector toward the $ \beta $ axis, increasing the likelihood that when a measurement is made, the state will be projected into a value of x that satisfies f of x equal 1. Lastly, we can show that the number of iterations. R is of the order square root of n. The state CI is an angle $ \gamma $ away from the $ \beta $ axis, where $ \gamma $ equals arccosine square root of M over N., and we know that each iteration pushes us an angle $ \theta $ toward the $ \beta $ axis. So, the number of repetitions we need is the closest integer to the ratio of $ \gamma $ over $ \theta $. we want to rotate the state CI at least pass $ \pi/4 $ from the $ \alpha $ axis so that the probability of measuring a $ \beta $ solution is greater than a 1/2, and In fact, we can place an upper bound on the number R. It is$ \pi/4 $ times the square root of M divided by N. This shows that Grover's algorithm will find a solution with a high probability for only order square root of n iterations. In contrast, it will take a classical computer on the order of n queries. 

what is the current status or capability for the most capable quantum computers today? For running Shor's algorithm, or Grover's algorithm \cite{grover_fast_1996}. So, the status today is that, all of these algorithms, like Shor's algorithm and Grover's algorithm, they are done in prototype fashion, or demonstration fashion, which means that they are not exceeding the capabilities of classical computer today \cite{kirke_application_2019}. we are demonstrating on smaller scale quantum computers that these algorithms, in fact, work. That they in fact return the correct answer. that gives us confidence that when we build larger quantum computers that those algorithms will still work. then the question would be, when are we going to get to this threshold or breaking point where a quantum computer can do something that classical computers, in principle could do, but cannot do in any kind of reasonable timeframe? And so that is called quantum supremacy, that turning point. there are many groups now, and industry, in particular, Google is pursuing a quantum supremacy \cite{harrow_quantum_2017, arute_quantum_2019,markov_quantum_2018} experiment with their Bristlecone chip.\cite{mcclean_openfermion_2019}, we know, it is still not close to being able to implement, say, Shor's algorithm, or Grover's algorithm, yet, in a way that cannot be done on a classical computer. Very likely what they are going to do is demonstrate a particular physics problem. It will be a well-posed problem. what they will do is they will demonstrate that they can solve that problem on a quantum computer and get the answer out, whereas trying to solve that exact same problem on a classical computer just does not work in a reasonable amount of time. that will be the first demonstration of quantum supremacy \cite{harrow_quantum_2017, boixo_characterizing_2018,arute_quantum_2019}. Now, in some sense, one might say, we have not already done that? we know, neutral atoms, or in other physical systems, we basically set it up, it is a complicated physical system, we set it up, and we look at the answer, and basically, we let our quantum computer emulate another system? And then we say, we could not calculate what was going to happen, so that is quantum supremacy. we think that there is a distinction to be made here. Those are very important experiments, by the way, we not saying they are not. But we think that when we discuss about quantum supremacy, what we mean here is that, not that no classical computer could ever simulate? that is not what we are discussing about. What we are discussing about is using single and two qubit gates, a universal logic set of gates applied to a problem, and that that problem exhibits quantum advantage beyond what a classical computer could do. the reason that distinction is important is because with a universal set of gates, in principle, we could do any algorithm? And we could solve any problem? And so, a demonstration of quantum supremacy should be a demonstration that is abstracted into quantum gates which in principle could do any algorithm? Now, it is only going to be applied to this one, early on, this one particular problem, or set of problems that shows quantum supremacy, and to apply it to practical problems that we care about that may still require yet a larger quantum computer. that is almost certainly going to be true. But that distinction is important. we think emulation in general is not the same as quantum supremacy. we are going to discuss about NISQ algorithms, algorithms that people are trying to develop today with the qubits that we have today. we are going to look at different types of noise, what do we mean by noise? What does noise do to cause decoherence? What types of noise exist? For example, there can be systematic noise. So, for example, we want to rotate from the North Pole to the South Pole, but every time we do that, we go a little bit too far, or we do not go quite far enough. every time we do it, we get the same error, it is just off by the same amount, that is a systematic error. There can also be errors which are stochastic. That, each time there is some noise on my control lines which either over-rotate or under-rotate a little bit, but how much it over or under-rotates is different in time, it changes in time, and we will discuss about these different types of noise, and we will also discuss about, ways that we can mitigate noise. error correction is one way that we can mitigate noise, and that involves adding redundancy, so we add more qubits, but by adding more qubits, and in particular qubits with a high enough fidelity, it exceeds some threshold. When they exceed that threshold, adding more improves the overall situation of all the qubits? And the way that this is implemented is through the use of syndromes, and we will go into detail about what that means. Now, there are other types of mitigation strategies that we can implement, and those are related, primarily to dephasing. Now, dephasing, it turns out in principle, it is reversible. it is a coherent error, as we call it. so, by doing certain pulse sequences, we can dynamically mitigate the error, and we will discuss through that in some detail. we think if these are topics that we find interesting, and in particular, we want to know more than just we have, that we want to take it one level deeper, look at how quantum computing works, and in particular how we are going to fight against noise that we have, either systematic or stochastic noise, and in principle, how error correction works. If we want to understand how error correction works, what fault tolerance means, what these syndromes are, how they are implemented. we will give the picture of how these protocols work, how error correction works. We are trying to introduce terminology, what it means, give some physical intuition.

So, can quantum computers create the ability to manipulate time, and time travel? What does this really mean?  we know, quantum computing is what is called reversible, and we think we may have discussed about that when we looked at single and two qubit gates. But with classical gates, and quantum gates. in classical boolean logic, many of the gates are irreversible. So for example, the AND gate. we have two inputs, we AND them, and we get one number out. But if we look at the output number, we cannot generally go backwards and tell we what the inputs were. Now, for an AND gate, if we get a one out, then we know that the two inputs were one and one. Fine, but if we get a zero out, we cannot tell whether it was zero zero, zero one, or one zero. All three of those combinations of input bits will give a zero out. So in that sense, that is not a reversible gate. so, one might think that we can flow forwards in time and get the output, but we cannot flow backwards in time and derive the inputs from the output. Now, there are a few reversible, there is a concept of reversible logic in classical computing, and in fact, in quantum systems, quantum computers, and quantum qubits, and these operations, these are unitary, and certainly in the absence of noise, they are unitary, and so, ideally, we go forwards, we can also run an algorithm backwards and go from the output to the input. we can run it forwards and go from the input to the output. so, in this sense, if we did those experiments, and many people have, but if we wanted to sell it as, we reverse time, we could think of it that way, but we are not really reversing time. we are just implementing an algorithm backwards. 

\section{Introduction to Grover's Algorithm on the Quantum Experience}

In this section, we will continue our exploration of the Grover search algorithm, and discuss how it is applied to a specific problem called the ``exactly-1 3SAT'' problem\cite{gabor_assessing_2019}. 

In this section, we will discuss how to implement and run Grover's algorithm. First, let us briefly recall What the problem solved by this algorithm is, and what are its benefits compared to classical algorithms. Grover's algorithm is designed to solve a completely unstructured search problem. Imagine that we have a list of two to the n numbers, and we want to find the position of a given number that appears once in this list. If we know that the list is sorted, we can use binary search, and find the position of the desired number looking at just an element out of the two to the n in the list. However, if we do not have any information whatsoever on the list, we may have to look at all of its elements to find the position of the desired number. In fact, given any deterministic classical algorithm to solve this search problem, we can construct a list on which the algorithm has to look at all of the two to the n elements before it finds the correct answer. Even a randomized classical algorithm would have to look at approximately two to the n elements on average. Using a quantum algorithm, we can do better than this. Let us properly state our assumptions first. we have a function, f, that maps any n digit binary string into zero or one, with the property that there exists a unique binary string for which the function has value one. This is the magic binary string that we want to find. Let us call l the integer that corresponds to it. So, that f of l is equal to one. we want to determine l while evaluating f as few times as possible. Since there is no property of f that can be exploited, we know that we can do it in two to the n evaluations of f classically, simply by applying f to all binary strings. However, can we do better? Grover's algorithm can do this with only the square root of two to the n evaluations of f, a quadratic improvement. The function f is assumed to be encoded by a quantum circuit that acts on n plus one qubit. The first n qubits are the input register. The last qubit is the output register. The circuit that implements f does not modify the input register. It performs the modulo 2 addition between the output register and the result of f applied to the input register. Separating the input and the output registers are standard in the quantum world because it allows reversibility of the function. we are given a three-set formula that is the logical end of a series of clauses and Boolean variables, where each clause is the logical, or of three Boolean variables or their negation. we want to find an assignment of the Boolean variables that makes the formula evaluate to true, ensuring that there is exactly one true literal per clause. It is a variation of the well-know three-satisfiability problem, and it is NP-hard. In the implementation, when we use the instance of exactly one three-satisfiability, specified by the formula that we can see on the screen. In this section, we will only work with the formula shown here. To construct the function f, we need a quantum circuit that acts on qubits encoding the Boolean variables and determines if this formula is satisfied. Notice that the circuit that implements f should simply be able to decide if the Boolean assignment corresponding to the binary string it is given as input satisfies the formula. Determining and returning the satisfying assignment, we are the goal of Grover's algorithm, not of the function f. The example formula has three Boolean variables, x1, x2, and x3. There are eight possible assignments. we can see in this table that only one possible assignment evaluates to true. we can verify this. As usual, one means true, and zero means false. Let us call Uf the unitary matrix that implements f. we can implement Uf in several ways. For simplicity, we have composed the problem of computing Uf by introducing three auxiliary qubits, one for each clause. For each clause, we construct a circuit that sets the corresponding zero qubits to one if and only if the clause has exactly one true term. Finally, the output register of Uf is set to one if and only if all three auxiliary qubits are one. For example, the circuit sets the bottom qubit, y1 to one for the close x1, or x2, or not x3. The x-gate flips the qubit corresponding to x3 because x3 appears negated in the clause. Using three CNOT gates, we set y1 equal x1, xor, x2, xor, not x3, implying that y1 is equal to one if an odd number of literals is satisfied. Since we want y1 equals to one if and only if exactly one literal is satisfied, we use a triply-controlled not gate to implement the desired formula finally. The last x-gate simply resets the state of the qubit x3. Similarly, we can implement the circuit that checks whether the second clause, not x1, or not x2, or not x3 is satisfied, and one for the third clause, not x1, or x2, or x3. To implement this circuit, there is a small obstacle; the triply-controlled not gate is not part of the basic gate set. Such a gate can be implemented in several ways. For simplicity, we choose to do it using 3 doubly-controlled, not gates, and one auxiliary qubit, as we show on the screen. In the circuit, we can quickly verify that q4 is set to one if and only if q0, q1, q2 are one. The final doubly-con trolled not resets the state of the auxiliary qubit q3. Remember that if we plan to re-use the auxiliary qubits, we should leave them in the same state as they begin. With these blocks, we can construct the full circuit that implements Uf using four auxiliary qubits, one for each of the three clauses, and one for the triply-controlled not. For each of the three clauses, we set the corresponding auxiliary qubit to one if the clause is satisfied. Then we apply a logical end between these three auxiliary qubits. Finally, we run the same circuit in reverse to reset the state of the auxiliary qubits. we have seen how to implement the function Uf used in one of the building blocks of Grover's algorithm. Remember that Grover's algorithms have three steps, initialization, sign flip, and inversion about the average. Let us see how to implement these steps. we will use the first three qubits lines for the input, the fourth as output, and the remaining qubits will be auxiliary qubits. For the initialization, we apply Hadamard gates on the input lines. on the output line, we apply an x-gate and then a Hadamard. It will be important for the science lab. At this point, the state of the input qubits is in the uniform superposition. All amplitudes are equal. For the sign flip, we simply applied the circuit that we constructed for Uf. Because we prepared the state of the output qubit in the way we described, this has the effect of flipping the sign of the amplitude corresponding to the basis state for which Uf outputs one. The last step is the inversion of the average. It is implemented by this simple circuit, which we can easily verify applies the matrix shown at the bottom. The matrix has the effect of mapping each amplitude to twice the average coefficient minus the amplitude itself. Because the average is smaller than most of the coefficients, all amplitudes except the negative one get reduced. the negative one gets amplified. It is one iteration of Grover's algorithm. The optimal number of iterations, k, can be found using the formula shown on the screen from which we derive that we must perform two iterations. It simply means that we should append to the overall circuit another copy of the circuit that performs a sign flip, and another copy of the circuit that performs the inversion about the average. that is it. we can now run this circuit on the simulator and look at the results. This circuit is, unfortunately, a bit too large to run on a real device, so we will have to rely on the simulator. This is the histogram of the outcome probabilities after 2,048 samples. After two iterations and only two calls to the black box function Uf, we sample the correct string, 101, with a probability of approximately 95\%. In comparison, any deterministic classical algorithm would have needed eight calls in the worst case. Notice that as we increase the number of iterations of Grover's algorithm, the probability of sampling the correct string goes down. This is the histogram for three iterations, and the probability dropped down to 32\%. The number of iterations must be chosen optimally.

The Exactly-1 3-SAT problem requires us to find an assignment such that f = 1 (or, True), and there is exactly one literal that evaluates to True in every clause of f.

Regarding the Exactly-1 3-SAT problem, which of the following statements are valid for the Boolean function:
\begin{equation}\label{eq2_123}
f(x_1, x_2, x_3) = (x_1 \vee x_2 \vee \neg x_3) \wedge (\neg x_1 \vee \neg x_2 \vee \neg x_3) \wedge (\neg x_1 \vee x_2 \vee x_3)?
\end{equation}

Recall that $x_1 \vee x_2  $ means  $x_1 $ OR $ x_2; x_1 \wedge x_2 $ means  $ x_1 $ AND $ x_2; $ and  $\neg x_1 $ means  NOT $ x_1 $ \\
1. The second clause has  $x_1, x_2, $ and $ x_3 $ as its literals\\
2. The input $ (x_1,x_2,x_3)=(0,1,0) $ has three True literals in the third clause\\
3. Each clause of the Exactly-1 3-SAT problem can be represented by a sub-circuit and evaluated independently.\\
4. The oracle operation is obtained by combining all sub-circuits for all clauses, such that it outputs True if and only if all clauses are satisfied with exactly one True literal each.\\

\section{Grover's Algorithm: Introduction}
Grover's algorithm can be decomposed into three steps: Initialization of the system, and Oracle operator O, and a (second) reflection operator W. For a list of $N=2^n$ items,  the system is initialized in the n-qubit equal superposition state. The Oracle operator encodes and marks the information in the list onto the quantum system; this can be viewed as the first reflection. Once the list is encoded, the operator W applies a second reflection. The resulting state after these two operations is closer to the solution state corresponding to the marked item. The product of the two reflections G=WO  the Grover iteration   is applied several times, and the number of repetitions necessary to find the marked item is upper bounded by  $\frac{\pi}{4}\sqrt{N}.$

we will implement Grover's algorithm for N=4. To make the problem easier to program, we will use a compiled version of the algorithm, that is, n=2 qubits instead of n=2+1 qubits. The implementation is divided into five sections: Introduction, Initialization, Oracle O,  Grover Operator, and Putting it All Together. 

Introduction
Grover's algorithm allows we to find marked items from an unstructured list. We will assume that there is only one marked item, which we will call $\beta$. We start by defining a function f(x) such that the function is true, $f(x)=1$, only if the input x corresponds to the marked item $\beta$, and zero otherwise. Mathematically, this can be  written as: 
\begin{equation}\label{eq2_124}
f\left( x\right) =\left\{ \begin{array}{ccc} 0 & \text {If} & x\neq \beta ,\\ 1 & \text {if} & x=\beta.\end{array}\right.
\end{equation}

For a list of $N=2^ n$ items, Grover's algorithm requires an n-qubit quantum system. We can associate each one of the N items of the list,  x, with one of the states of the n-qubit computational basis $x\to\lvert x\rangle.$
In this IBM Q experience \cite{sisodia_circuit_2018}, we will consider the following associations:

For a list of N=4 items and a system of  two-qubits 
\begin{table}[H]
\centering
\caption{N=4 items and a system of two-qubits}
\label{tab:2_1:Table 3}
\begin{tabular}{|c|c|c|}\hline
Item in decimal & Item in binary & Basis state \\\hline
    0&   00              &  $\lvert 00\rangle$           \\\hline
    1&   01              &  $\lvert 01\rangle$           \\\hline
    2&   10              &  $\lvert 10\rangle$           \\\hline
    3&   11              &  $\lvert 11\rangle $         \\\hline
\end{tabular}
\end{table}

For a list of N=8 items and a system of three-qubits

\begin{table}[H]
\centering
\caption{N=8 items and a system of three-qubits}
\label{tab:2_1:Table 4}
\begin{tabular}{|c|c|c|} \hline
Item in decimal & Item in binary & Basis state \\\hline
0&   000              &  $\lvert 000\rangle$           \\\hline
1&   001              &  $\lvert 001\rangle$           \\\hline
2&   010              &  $\lvert 010\rangle$           \\\hline
3&   011              &  $\lvert 011\rangle $          \\\hline
4&   100              &  $\lvert 100\rangle$           \\\hline
5&   101              &  $\lvert 101\rangle$           \\\hline
6&   110              &  $\lvert 110\rangle$           \\\hline
7&   111              &  $\lvert 111\rangle $          \\\hline
\end{tabular}
\end{table}

To encode the function f(x) into a quantum computer, the Oracle operator O is applied onto the equal n-qubit superposition state. This operation acts as an identity for each basis state that does not correspond to the marked item, and adds a negative sign to the solution state $\left\vert\beta\right\rangle,$
\begin{equation}\label{eq2_125}
\left\vert x\right\rangle \rightarrow \left( -1\right) ^{f\left( x\right) }\left\vert x\right\rangle =\left\{ \begin{array}{ccc} \left\vert x\right\rangle & \text {if} & x\neq \beta, \\ -\left\vert x\right\rangle & \text {if} & x=\beta.\end{array}\right.
\end{equation}

N=4 example: If the marked item is $\beta=2$ in decimal, then the function f is given by
\begin{equation}\label{eq2_126}
f\left( x\right) =\left\{ \begin{array}{lll} 0 & \text {if} & x=00,01,11 ,\\ 1 & \text {if} & x=10.\end{array}\right.
\end{equation}
and therefore the oracle O must transform the state according to:

\begin{equation}\label{eq2_127}
\displaystyle \left\vert 00\right\rangle    \displaystyle \rightarrow    \displaystyle \left( -1\right) ^{f\left( 00\right) }\left\vert 00\right\rangle =\left\vert 00\right\rangle ,
\end{equation}

\begin{equation}\label{eq2_128}     
\displaystyle \left\vert 01\right\rangle    \displaystyle \rightarrow    \displaystyle \left( -1\right) ^{f\left( 01\right) }\left\vert 01\right\rangle =\left\vert 01\right\rangle ,
\end{equation}

\begin{equation}\label{eq2_129}          
\displaystyle \left\vert 10\right\rangle    \displaystyle \rightarrow    \displaystyle \left( -1\right) ^{f\left( 10\right) }\left\vert 10\right\rangle =-\left\vert 10\right\rangle ,
\end{equation}

\begin{equation}\label{eq2_130}          
\displaystyle \left\vert 11\right\rangle    \displaystyle \rightarrow    \displaystyle \left( -1\right) ^{f\left( 11\right) }\left\vert 11\right\rangle =\left\vert 11\right\rangle .
\end{equation}
     
N=8 example: If marked item is $\beta=2$ in decimal, then the function f is given by
\begin{equation}\label{eq2_131}
f\left( x\right) =\left\{ \begin{array}{ll} 1 & \text {if }  x=010 ,\\ 0 & \text {otherwise,}  \end{array}\right.
\end{equation}

and therefore the oracle $O$ must transform according to:
\begin{equation}\label{eq2_132}
\left\vert 010\right\rangle \rightarrow \left( -1\right) ^{f\left( 010\right) }\left\vert 010\right\rangle =-\left\vert 010\right\rangle
\end{equation}

and
\begin{equation}\label{eq2_133}
\left\vert x\right\rangle \rightarrow \left( -1\right) ^{f\left( x\right) }\left\vert x\right\rangle =\left\vert x\right\rangle \text {, for }x\neq 010.
\end{equation}

How Many Qubits? N=512; For a list of N=512 items, how many qubits would we need to implement Grover's algorithm?
    
Solution:For a list of $ N=2^n $ elements, the Grover's algorithm needs n qubits. For 512 items: $ 512=2^n\to n=9 $

\section{Grover's Algorithm: Initialization}

State preparation:\\
At the beginning of the protocol, there is no information about where the marked item is. Thus, it is convenient to start with an equal superposition state
\begin{equation}\label{eq2_134}
\left\vert \psi\right\rangle =\frac{1}{\sqrt{N}}\sum _{x=0}^{N-1}\left\vert x\right\rangle ,
\end{equation}

such that each basis state $\left\vert x\right\rangle$ will have the same probability of being measured:
\begin{equation}\label{eq2_135}
\frac{1}{N}=\frac{1}{2^{n}}.
\end{equation}

Even though we do not know which item is the marked item at the beginning of the algorithm, we can break the equal superposition state into a sum over the marked items and a sum over all other items,
\begin{equation}\label{eq2_136}
\begin{split}
\displaystyle\left\vert\psi\right\rangle    \displaystyle & =    \displaystyle \frac{1}{\sqrt{N}}\sum _{x=0}^{N-1}\left\vert x\right\rangle ,    \\     
\displaystyle & =    \displaystyle \frac{1}{\sqrt{N}}\sum _{x:f(x)=0}\left\vert x\right\rangle +\frac{1}{\sqrt{N}}\sum _{x:f(x)=1}\left\vert x\right\rangle ,    \\     
\displaystyle & =    \displaystyle \underbrace{\frac{1}{\sqrt{N}}\sum _{x:f(x)=0}\left\vert x\right\rangle }_{\text {Non-solution state}}+\underbrace{\frac{1}{\sqrt{N}}\left\vert \beta\right\rangle }_{\text {Solution state}}
\end{split}
\end{equation}
         
To understand the geometrical interpretation of the algorithm, it is convenient to define the normalized quantum state
\begin{equation}\label{eq2_137}
\left\vert \alpha\right\rangle \equiv \frac{1}{\sqrt{\frac{N-1}{N}}}\left( \frac{1}{\sqrt{N}}\sum _{x:f\left( x\right) =0}\left\vert x\right\rangle \right) ,
\end{equation}

such that the equal superposition state can be written as the sum
\begin{equation}\label{eq2_138}
\left\vert \psi\right\rangle =\underbrace{\sqrt{\frac{N-1}{N}}\left\vert \alpha\right\rangle }_{\text {Non-solution state}} +\underbrace{\frac{1}{\sqrt{N}}\left\vert \beta\right\rangle }_{\text {Solution state}},
\end{equation}

or, equivalently,
\begin{equation}\label{eq2_139}
\left\vert \psi\right\rangle =\cos \left(\frac{\theta }{2}\right)\left\vert \alpha\right\rangle +\sin \left(\frac{\theta }{2}\right)\left\vert \beta\right\rangle ,
\end{equation}

where we have defined the angle $\theta$ such that
\begin{equation}\label{eq2_140}
\begin{split}
\displaystyle \cos \left(\frac{\theta }{2}\right)    \displaystyle & =    \displaystyle \sqrt{\frac{N-1}{N}}, \\         
\displaystyle \sin \left(\frac{\theta }{2}\right)    \displaystyle & =    \displaystyle \frac{1}{\sqrt{N}}.
\end{split}
\end{equation}
         
The figure below shows a graphical representation of quantum states $\left\vert \psi\right\rangle, \left\vert \beta\right\rangle, $ and $\left\vert\alpha\right\rangle $ as vectors in a two-dimensional space. The projection of the superposition state $\left\vert\psi\right\rangle$ onto the solution state $\left\vert \beta\right\rangle$ is equal to $1/\sqrt{N}$, and the projection of $\left\vert\psi\right\rangle$ on $\left\vert\alpha\right\rangle$ is equal to $\sqrt{(N-1)/N}$.

\begin{figure}[H] \centering{\includegraphics[scale=.1]{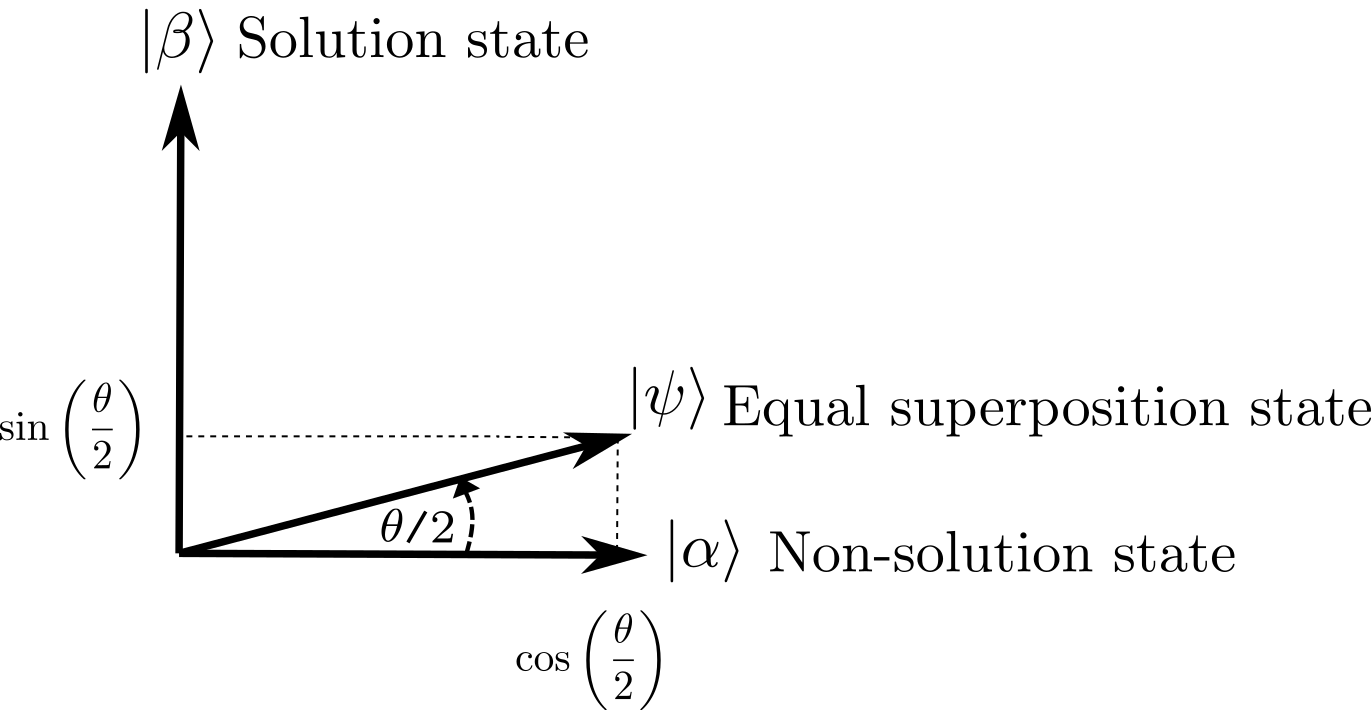}}\caption{Initialization N3}\label{fig2_17}
\end{figure}

Two-qubit initialization:\\

To implement Grover's algorithm for a list of $N=2^ n=4$ elements, the compiled algorithm requires two qubits, $n=2$. To prepare the two qubits in an equal superposition state $\left\vert \psi\right\rangle,$ a Hadamard gate is applied to each qubit,

\begin{equation}\label{eq2_141}
\begin{split}
\displaystyle \left\vert 0\right\rangle \left\vert 0\right\rangle & \displaystyle \to    \displaystyle \frac{\left\vert 0\right\rangle +\left\vert 1\right\rangle }{\sqrt{2}}\frac{\left\vert 0\right\rangle +\left\vert 1\right\rangle }{\sqrt{2}},     \\     
\displaystyle & =    \displaystyle \frac{\left\vert 0\right\rangle \left\vert 0\right\rangle +\left\vert 0\right\rangle \left\vert 1\right\rangle +\left\vert 1\right\rangle \left\vert 0\right\rangle +\left\vert 1\right\rangle \left\vert 1\right\rangle }{2}.
\end{split}
\end{equation}
          
If we assume that the marked item is given by $\beta=3$ (decimal representation), then it is convenient to write the two-qubit equal superposition state as
\begin{equation}\label{eq2_142}
\left\vert \psi\right\rangle =\underbrace{\frac{\left\vert 0\right\rangle \left\vert 0\right\rangle +\left\vert 0\right\rangle \left\vert 1\right\rangle +\left\vert 1\right\rangle \left\vert 0\right\rangle }{2}}_{\text {Non-solution state }}+\underbrace{\frac{\left\vert 1\right\rangle \left\vert 1\right\rangle }{2}}_{\text {Solution state}},
\end{equation}

or as a superposition of the non-solution and solution states
\begin{equation}\label{eq2_143}
\displaystyle \left\vert \psi\right\rangle    \displaystyle =    \displaystyle \frac{\sqrt{3}}{2}\left\vert \alpha\right\rangle +\frac{1}{2}\left\vert \beta\right\rangle      
\displaystyle =    \displaystyle \cos (30^{\circ})\left\vert \alpha\right\rangle +\sin (30^{\circ})\left\vert \beta\right\rangle ,
\end{equation}
         
where the non-solution state is given by
\begin{equation}\label{eq2_144}
\left\vert \alpha\right\rangle =\frac{1}{\sqrt{3}}\left( \left\vert 0\right\rangle \left\vert 0\right\rangle +\left\vert 0\right\rangle \left\vert 1\right\rangle +\left\vert 1\right\rangle \left\vert 0\right\rangle \right) ,
\end{equation}

and the solution state by
\begin{equation}\label{eq2_145}
\left\vert \beta\right\rangle =\left\vert 1\right\rangle \left\vert 1\right\rangle .
\end{equation}

The figure below shows the graphical representation of the two-qubit equal superposition state $\left\vert \psi\right\rangle$. The projection of $\left\vert \psi\right\rangle$ on $\left\vert \alpha\right\rangle$ and $\left\vert \beta\right\rangle$ are $\sqrt{3}/2 $and $1/2$ respectively.

\begin{figure}[H] \centering{\includegraphics[scale=.1]{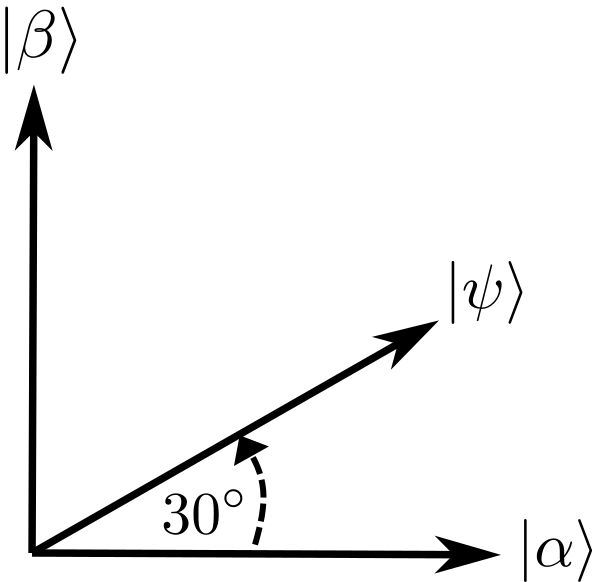}}\caption{Initialization N2}\label{fig2_18}
\end{figure}

\item Two-qubit Equal Superposition State; Write a QASM code \cite{noauthor_qiskitopenqasm_2020}that generates the equal superposition state for two-qubits. To do this:
    \begin{itemize}
        \item Apply a Hadamard gate in the first qubit
        \item Apply a Hadamard gate in the second qubit
        \item Save the measurement's result of the first qubit in the first bit
        \item Save the measurement's result of the second qubit in the second bit
    \end{itemize}
\begin{figure}[H] \centering{\includegraphics[scale=.8]{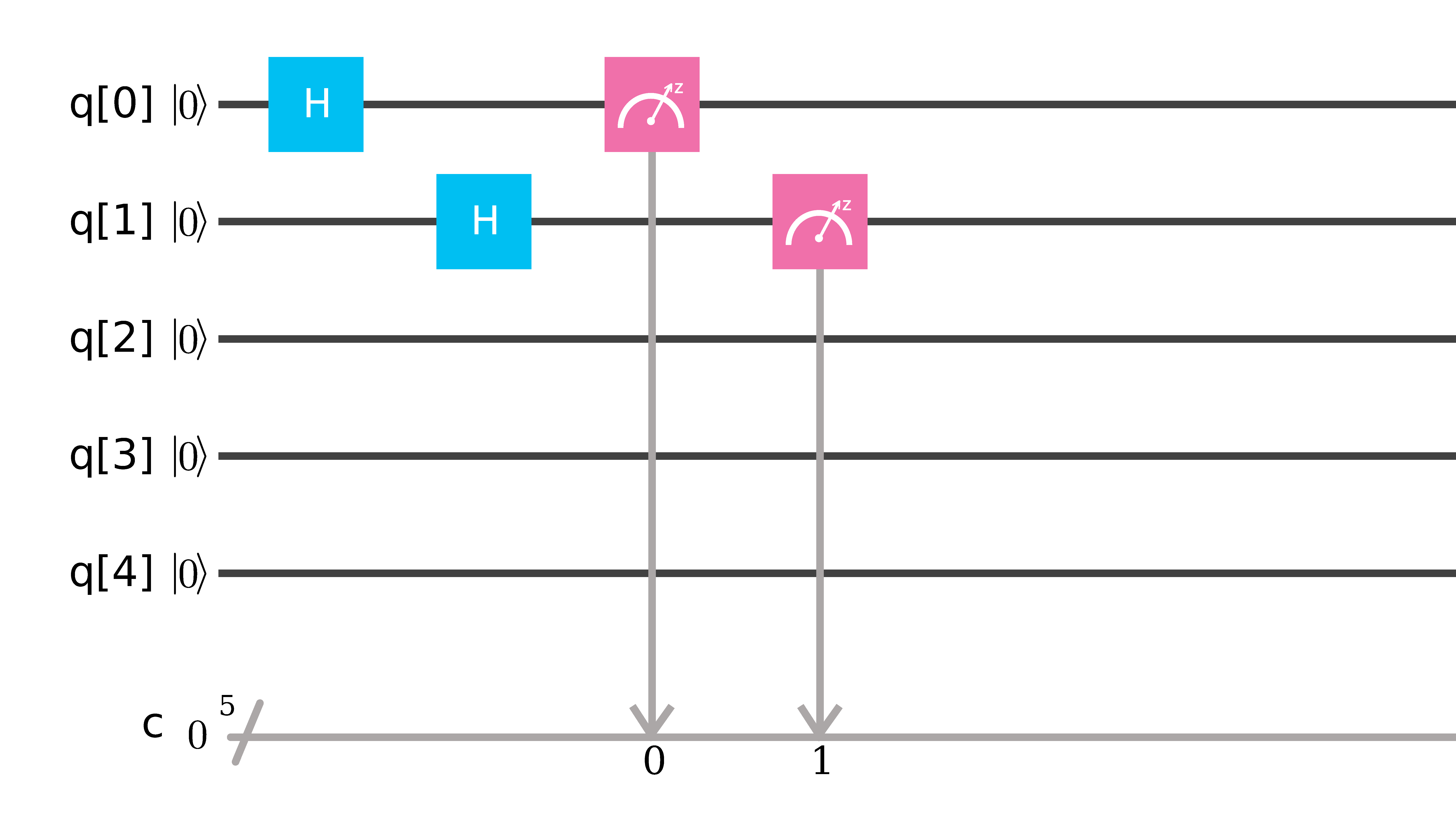}}\caption{Two-qubit Equal Superposition State}\label{fig2_44}
\end{figure}
Solution:\\
\begin{lstlisting}
include ``qelib1.inc'';
qreg q[5];
creg c[5];
h q[0];
h q[1];
measure q[0] -> c[0];
measure q[1] -> c[1];
\end{lstlisting}
A Hadamard gate transforms the states $ \lvert 0 \rangle $ and $ \lvert 1 \rangle $ into $ \frac{\lvert 0 \rangle+\lvert 1 \rangle}{\sqrt{2}} $ and $ \frac{\lvert 0 \rangle-\lvert 1 \rangle}{\sqrt{2}} $ respectively. By applying a Hadamard gate on each qubit, we have just created the equal superposition two-qubit state
$ \frac{\lvert 0\rangle \lvert 0\rangle+ \lvert 0\rangle \lvert 1\rangle+\lvert 1\rangle \lvert 0\rangle+\lvert 1\rangle \lvert 1\rangle}{2} $.
\item Three-qubit Equal Superposition State; Write a QASM code that generates the equal superposition state for three-qubits. To do this:
\begin{itemize}
    \item Apply a Hadamard gate in the first qubit
    \item Apply a Hadamard gate in the second qubit
    \item Apply a Hadamard gate in the third qubit
    \item Save the measurement's result of the first qubit in the first bit
    \item Save the measurement's result of the second qubit in the second bit
    \item Save the measurement's result of the third qubit in the third bit
\end{itemize}

\begin{figure}[H] \centering{\includegraphics[scale=.8]{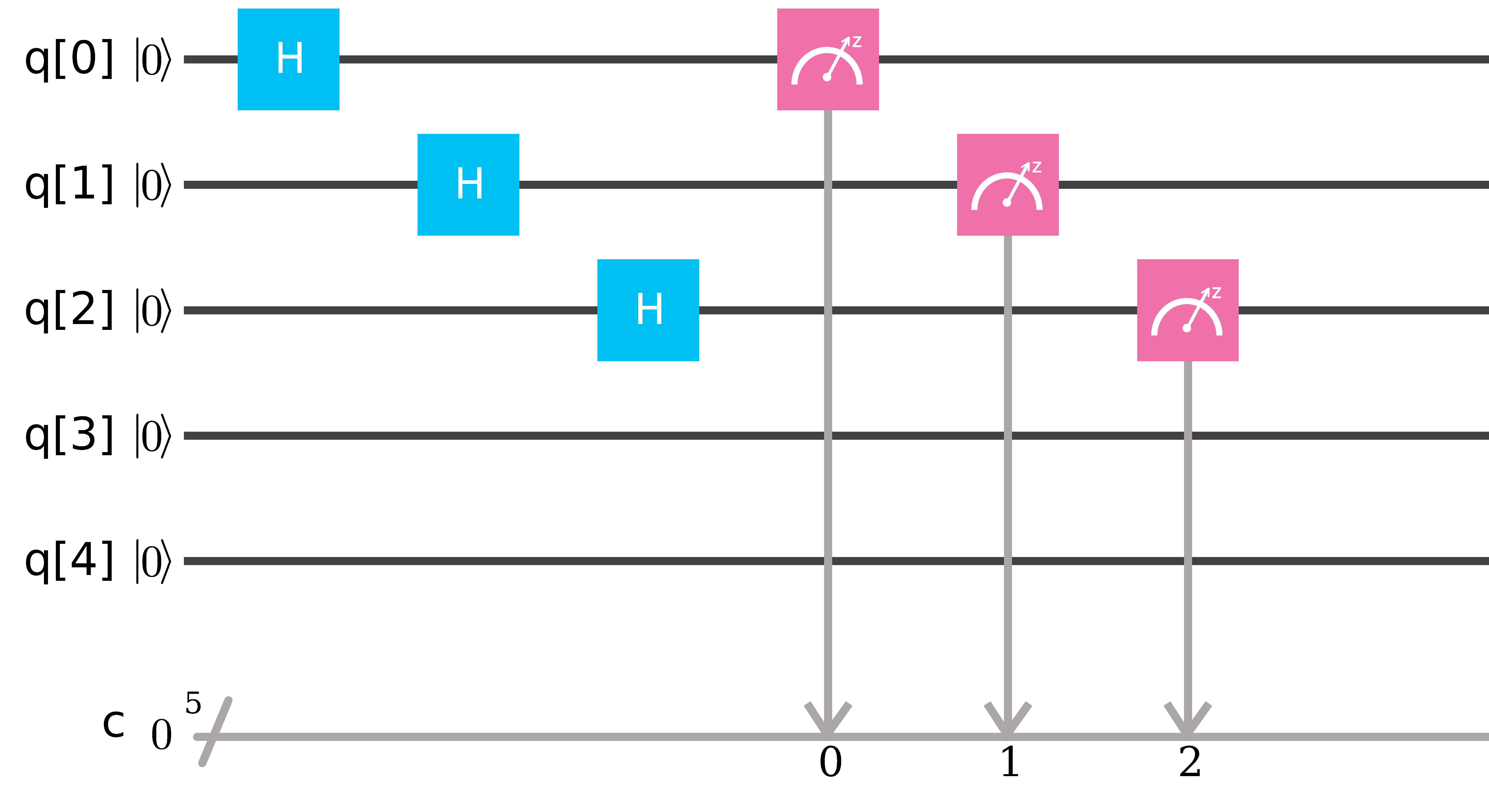}}\caption{Three-qubit Equal Superposition State}\label{fig2_45}
\end{figure}
Solution:\\
\begin{lstlisting}    
include ``qelib1.inc'';
qreg q[5];
creg c[5];
h q[0];
h q[1];
h q[2];
measure q[0] -> c[0];
measure q[1] -> c[1];
measure q[2] -> c[2];
\end{lstlisting}

\section{Grover's Algorithm: Oracle O }

Oracle operation:\\
The role of the oracle, O, is to encode the function f(x) in the quantum system and mark the solution. The operation transforms each state of the computational basis as
\begin{equation}\label{eq2_146}
\left\vert x\right\rangle \to \left( -1\right)^{f\left( x\right) }\left\vert x\right\rangle .
\end{equation}

In the first iteration of the protocol, this operation transforms the equal superposition state according to:

\begin{equation}\label{eq2_147}
\displaystyle O\left\vert \psi\right\rangle    \displaystyle =    \displaystyle \cos \left(\frac{\theta }{2}\right)O\left\vert \alpha \right\rangle +\sin \left(\frac{\theta }{2}\right)O\left\vert \beta\right\rangle           
\displaystyle =    \displaystyle \cos \left(\frac{\theta }{2}\right)\left\vert \alpha\right\rangle -\sin \left(\frac{\theta }{2}\right)\left\vert \beta\right\rangle .
\end{equation}
          
The figure below shows the graphical representation of the equal superposition state after the Oracle operator, $O\left\vert \psi \right\rangle.$ The amplitude of $\left\vert \psi\right\rangle$ in front of $\left\vert \alpha\right\rangle$ remains the same as before. However, the amplitude in front of $\left\vert \beta\right\rangle $ acquires a negative sign. The oracle operation serves to reflect the state $ \left\vert \psi\right\rangle $ about the non-solution state $ \left\vert \alpha\right\rangle$.

\begin{figure}[H] \centering{\includegraphics[scale=.15]{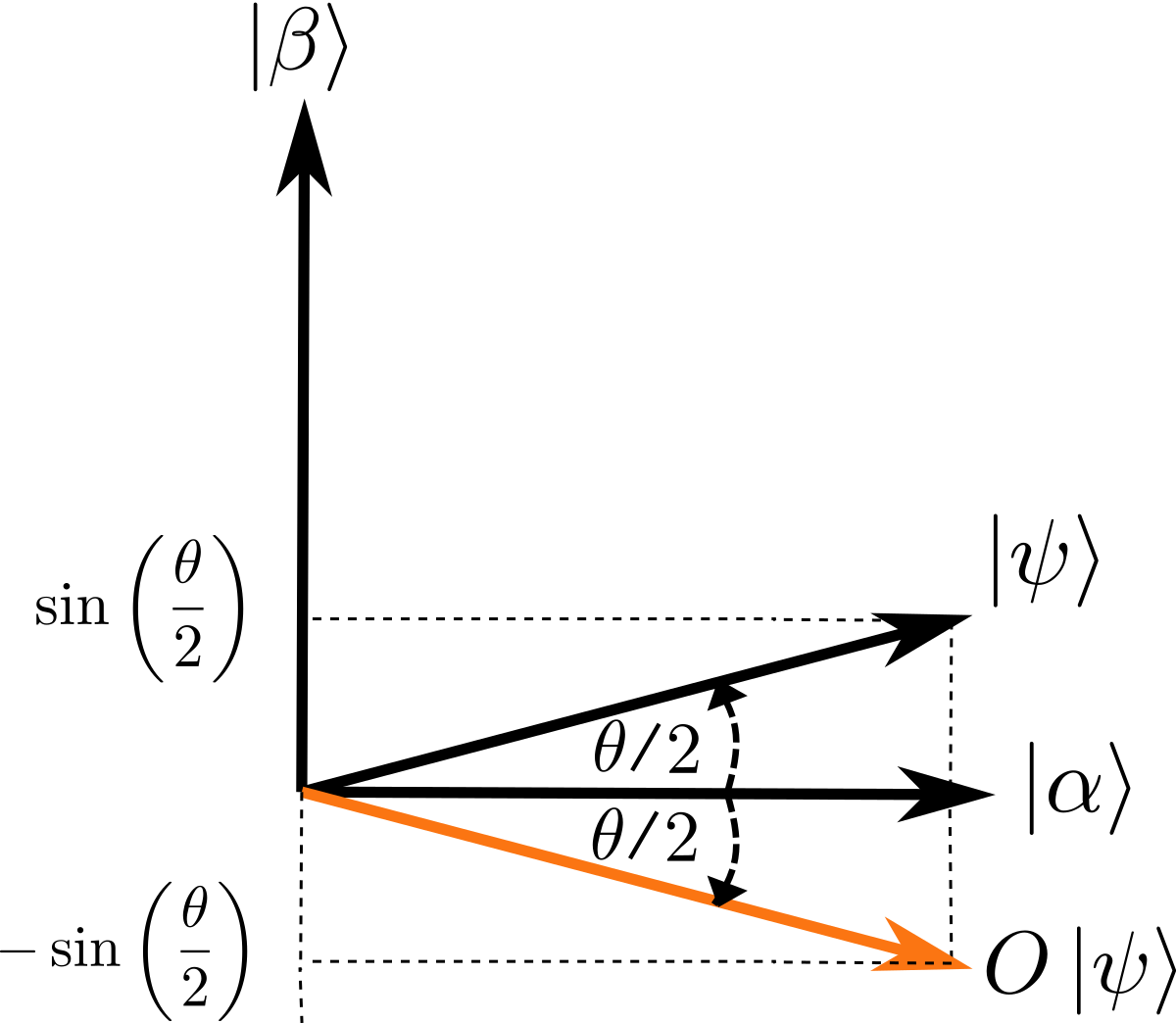}}\caption{Oracle N}\label{fig2_19}
\end{figure}

N=4 item list \\
Example 1: If the marked item is $\beta=3 $(decimal representation), then the function f is given by
$f\left( x\right) =\left\{ \begin{array}{cc} 1, & \text {if }x=11, \\ 0, & \text {otherwise,}\end{array}\right.$

and therefore the oracle O must transform according to:
\begin{equation}\label{eq2_148}
\left\vert \psi \right\rangle \rightarrow \cos \left( 30^{\circ}\right) \left\vert \alpha \right\rangle -\sin \left( 30^{\circ}\right) \left\vert \beta\right\rangle ,
\end{equation}

where the non-solution state is 
\begin{equation}\label{eq2_149}
\left\vert \alpha \right\rangle =\frac{1}{\sqrt{3}}\left( \left\vert 0\right\rangle \left\vert 0\right\rangle +\left\vert 0\right\rangle \left\vert 1\right\rangle +\left\vert 1\right\rangle \left\vert 0\right\rangle \right) ,
\end{equation}

and the solution state is
\begin{equation}\label{eq2_150}
\left\vert \beta\right\rangle =\left\vert 1\right\rangle \left\vert 1\right\rangle .
\end{equation}

In particular, the Oracle must transform each element of the two-qubit basis according to:
\begin{equation}\label{eq2_151}
\displaystyle \left\vert 0\right\rangle \left\vert 0\right\rangle    \displaystyle \rightarrow    \displaystyle \left( -1\right) ^{f\left( 00\right) }\left\vert 0\right\rangle \left\vert 0\right\rangle =\left\vert 0\right\rangle \left\vert 0\right\rangle ,
\end{equation}
\begin{equation}\label{eq2_152}          
\displaystyle \left\vert 0\right\rangle \left\vert 1\right\rangle    \displaystyle \rightarrow    \displaystyle \left( -1\right) ^{f\left( 01\right) }\left\vert 0\right\rangle \left\vert 1\right\rangle =\left\vert 0\right\rangle \left\vert 1\right\rangle ,
\end{equation}    
\begin{equation}\label{eq2_153}      
\displaystyle \left\vert 1\right\rangle \left\vert 0\right\rangle    \displaystyle \rightarrow    \displaystyle \left( -1\right) ^{f\left( 10\right) }\left\vert 1\right\rangle \left\vert 0\right\rangle =\left\vert 1\right\rangle \left\vert 0\right\rangle ,
\end{equation}
\begin{equation}\label{eq2_154}          
\displaystyle \left\vert 1\right\rangle \left\vert 1\right\rangle    \displaystyle \rightarrow    \displaystyle \left( -1\right) ^{f\left( 11\right) }\left\vert 1\right\rangle \left\vert 1\right\rangle =-\left\vert 1\right\rangle \left\vert 1\right\rangle .
\end{equation}
          
The figure below shows a graphical representation of the non-solution, solution, and reflected states. 

\begin{figure}[H] \centering{\includegraphics[scale=.15]{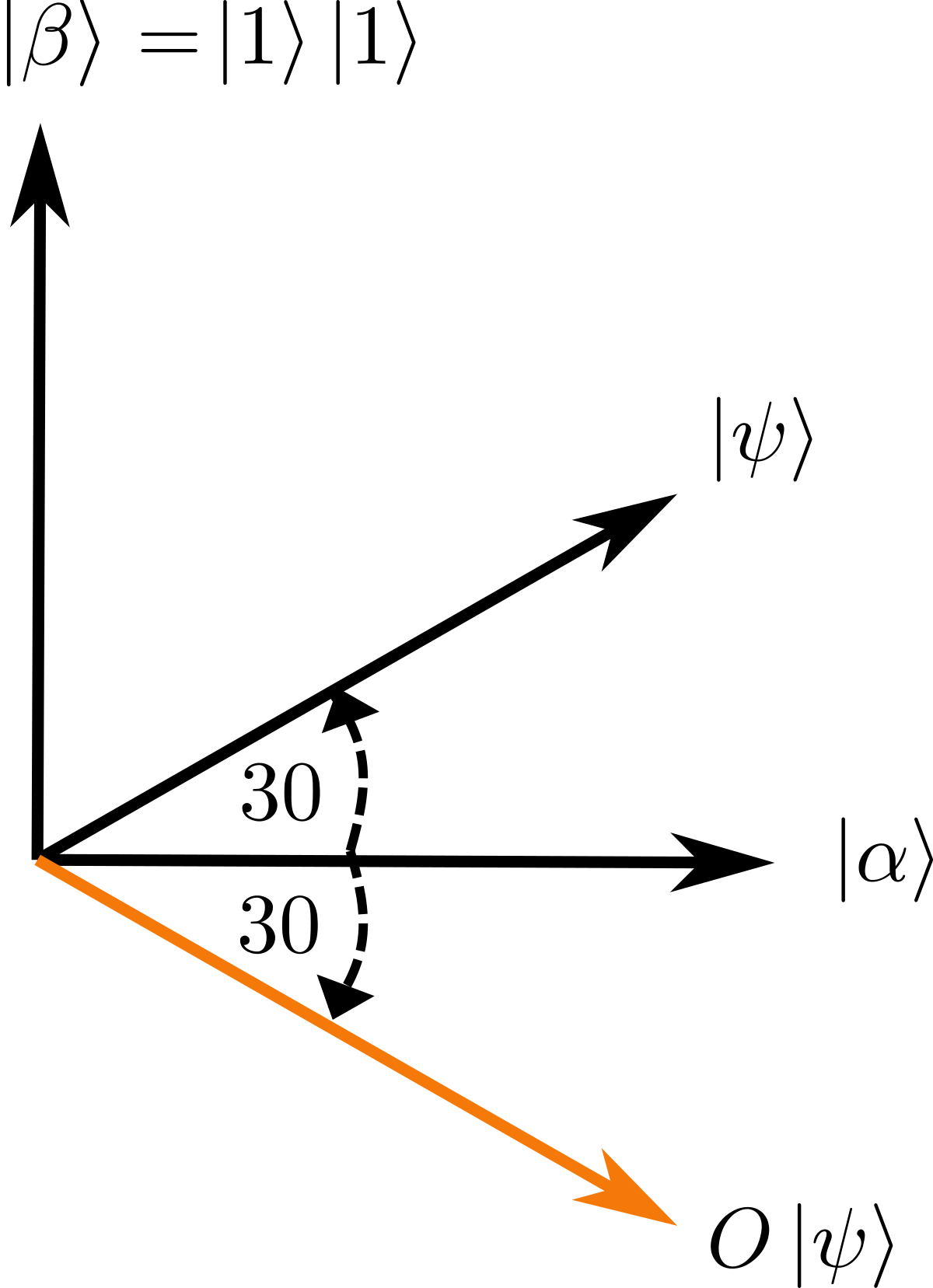}}\caption{Oracle 2}\label{fig2_20}
\end{figure}

The quantum circuit below implements the Oracle for this marked item. We can verify this by computing the state of each element of the basis after the operation. In this case, the Oracle is composed of two Hadamard gates on the second qubit, and one CNOT gate with the first qubit as control and the second as the target. Remember that the Hadamard gate maps a single qubit state according to:
\begin{equation}\label{eq2_155}
\begin{split}
\displaystyle \left\vert 0\right\rangle    & \displaystyle \rightarrow    \displaystyle \frac{\left\vert 0\right\rangle +\left\vert 1\right\rangle }{\sqrt{2}},     \\     
\displaystyle \left\vert 1\right\rangle    & \displaystyle \rightarrow    \displaystyle \frac{\left\vert 0\right\rangle -\left\vert 1\right\rangle }{\sqrt{2}}.
\end{split}    
\end{equation}
      
And, the CNOT gate maps a two-qubit state according to:
\begin{equation}\label{eq2_156}
\displaystyle \left\vert 0\right\rangle \left\vert 0\right\rangle    \displaystyle \rightarrow    \displaystyle \left\vert 0\right\rangle \left\vert 0\right\rangle ,
\end{equation}
\begin{equation}\label{eq2_157}          
\displaystyle \left\vert 0\right\rangle \left\vert 1\right\rangle    \displaystyle \rightarrow    \displaystyle \left\vert 0\right\rangle \left\vert 1\right\rangle ,    
\end{equation}
\begin{equation}\label{eq2_158}      
\displaystyle \left\vert 1\right\rangle \left\vert 0\right\rangle    \displaystyle \rightarrow    \displaystyle \left\vert 1\right\rangle \left\vert 1\right\rangle ,    
\end{equation}
\begin{equation}\label{eq2_159}      
\displaystyle \left\vert 1\right\rangle \left\vert 1\right\rangle    \displaystyle \rightarrow    \displaystyle \left\vert 1\right\rangle \left\vert 0\right\rangle .
\end{equation}
     
\begin{figure}[H] \centering{\includegraphics[scale=.9]{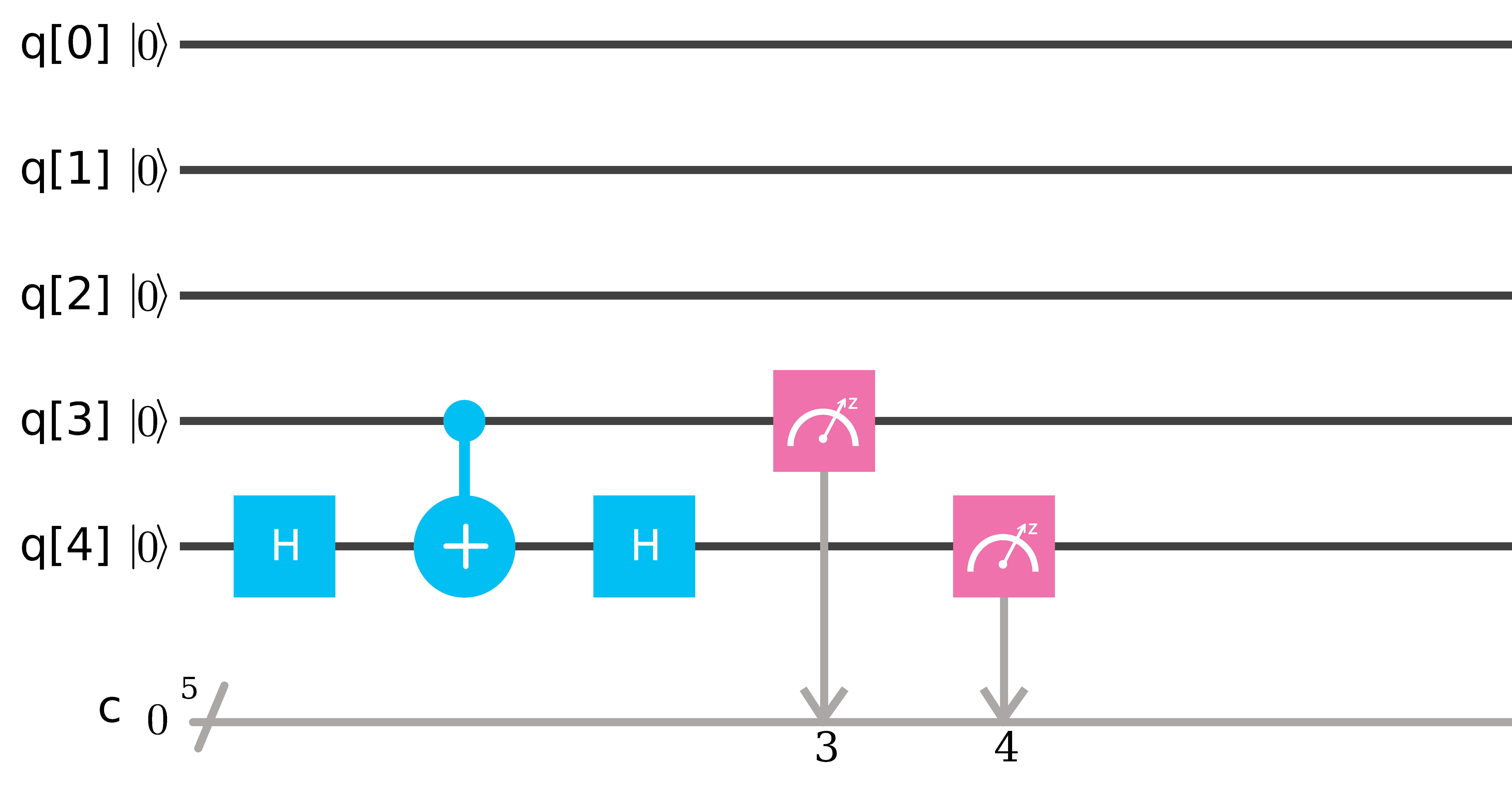}}\caption{Oracle 11}\label{fig2_21}
\end{figure}

The table below shows the evolution of the two-qubit basis states. we can see that the only basis state that acquires a negative sign is the solution state. 
\begin{table}[H]
\centering
\caption{two-qubit basis states}
\label{tab:2_1:Table 5}
\begin{tabular}{|c|c|c|c|}\hline
\begin{tabular}[c]{@{}l@{}}Basis\\ State\end{tabular} & \begin{tabular}[c]{@{}l@{}}After first\\ Hadamard H\end{tabular} & \begin{tabular}[c]{@{}l@{}}After \\ CNOT gate\end{tabular} & \begin{tabular}[c]{@{}l@{}}After second \\ Hadamard H\end{tabular} \\ \hline
$\left\vert 0\right\rangle \left\vert 0\right\rangle$ &$\left\vert 0\right\rangle \frac{\left\vert 0\right\rangle +\left\vert 1\right\rangle }{\sqrt{2}}$                                                                  &$\left\vert 0\right\rangle \frac{\left\vert 0\right\rangle +\left\vert 1\right\rangle }{\sqrt{2}}$                                                            &$\left\vert 0\right\rangle \left\vert 0\right\rangle$                                                                    \\ \hline
$\left\vert 0\right\rangle \left\vert 1\right\rangle$& $\left\vert 0\right\rangle \frac{\left\vert 0\right\rangle -\left\vert 1\right\rangle }{\sqrt{2}}$                                                                 &$\left\vert 0\right\rangle \frac{\left\vert 0\right\rangle -\left\vert 1\right\rangle }{\sqrt{2}}$                                                            &$\left\vert 0\right\rangle \left\vert 1\right\rangle$                                                                    \\ \hline
$\left\vert 1\right\rangle \left\vert 0\right\rangle$& $\left\vert 1\right\rangle \frac{\left\vert 0\right\rangle +\left\vert 1\right\rangle }{\sqrt{2}}$                                                                 &$\left\vert 1\right\rangle \frac{\left\vert 0\right\rangle +\left\vert 1\right\rangle }{\sqrt{2}}$                                                            &$\left\vert 1\right\rangle \left\vert 0\right\rangle$                                                                    \\ \hline
$\left\vert 1\right\rangle \left\vert 1\right\rangle$&$\left\vert 1\right\rangle \frac{\left\vert 0\right\rangle -\left\vert 1\right\rangle }{\sqrt{2}}$                                                                  &$-\left\vert 1\right\rangle \frac{\left\vert 0\right\rangle -\left\vert 1\right\rangle }{\sqrt{2}}$                                                            &$-\left\vert 1\right\rangle \left\vert 1\right\rangle$ \\ \hline
\end{tabular}
\end{table}

    \item N=4: Oracle Operator $ \beta=3 $; For a list of N=4 items, write the QASM code that generates the Oracle operator when the marked item is $ \beta=3  $(decimal representation). Use qubits q[3] and q[4] as the first and second qubit respectively. Remember that we are using the ibmqx4 backend \cite{shukla_complete_2018}, and do not forget to measure our two qubits. Compare our results with the table above using $ q[3], q[4] = \lvert 0 \rangle \lvert 0 \rangle. $
    \begin{figure}[H] \centering{\includegraphics[scale=.9]{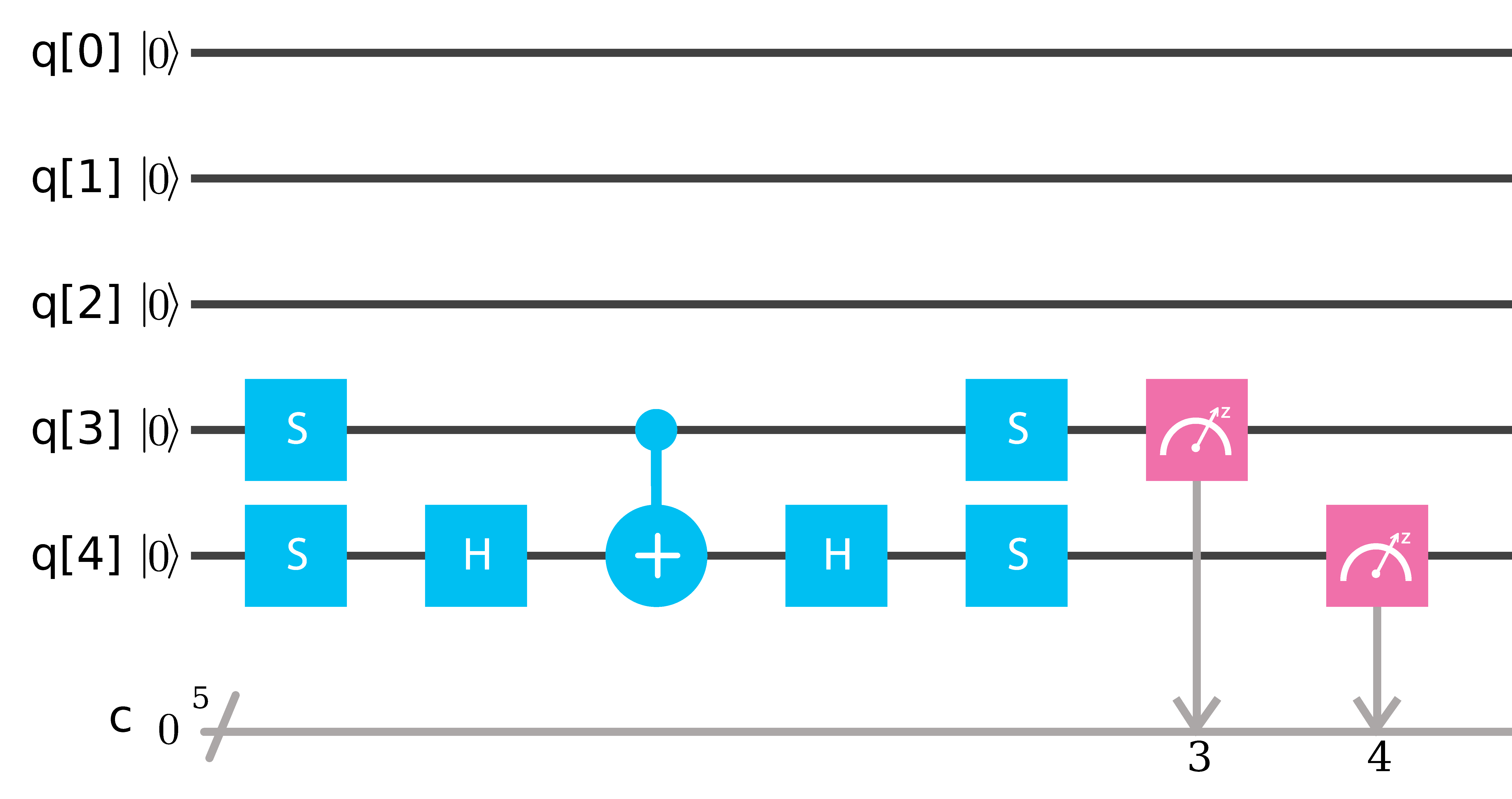}}\caption{N=4: Oracle Operator $ \beta=3 $}\label{fig2_46}
    \end{figure}
    Solution:\\
    \begin{lstlisting}
    include ``qelib1.inc'';
    qreg q[5];
    creg c[5];
    h q[4];
    cx q[3],q[4];
    h q[4];
    measure q[3] -> c[3];
    measure q[4] -> c[4];
    \end{lstlisting}

Example 2: If the marked item is $ \beta=2 $ (decimal representation), then the function f is given by
\begin{equation}\label{eq2_159_1}
f\left( x\right) =\left\{ \begin{array}{cc} 1, & \text {if }x=10, \\ 0, & \text {otherwise.}\end{array}\right.
\end{equation}

The Oracle must transform each element of the two-qubit basis according to:
\begin{center}
$ \displaystyle \left\vert 0\right\rangle \left\vert 0\right\rangle    \displaystyle \rightarrow    \displaystyle \left( -1\right) ^{f\left( 00\right) }\left\vert 0\right\rangle \left\vert 0\right\rangle =\left\vert 0\right\rangle \left\vert 0\right\rangle $          \\ 
$ \displaystyle \left\vert 0\right\rangle \left\vert 1\right\rangle    \displaystyle \rightarrow    \displaystyle \left( -1\right) ^{f\left( 01\right) }\left\vert 0\right\rangle \left\vert 1\right\rangle =\left\vert 0\right\rangle \left\vert 1\right\rangle  $         \\ 
$ \displaystyle \left\vert 1\right\rangle \left\vert 0\right\rangle    \displaystyle \rightarrow    \displaystyle \left( -1\right) ^{f\left( 10\right) }\left\vert 1\right\rangle \left\vert 0\right\rangle =-\left\vert 1\right\rangle \left\vert 0\right\rangle $           \\
$ \displaystyle \left\vert 1\right\rangle \left\vert 1\right\rangle    \displaystyle \rightarrow    \displaystyle \left( -1\right) ^{f\left( 11\right) }\left\vert 1\right\rangle \left\vert 1\right\rangle =\left\vert 1\right\rangle \left\vert 1\right\rangle $           
\end{center}

\begin{figure}[H] \centering{\includegraphics[scale=.9]{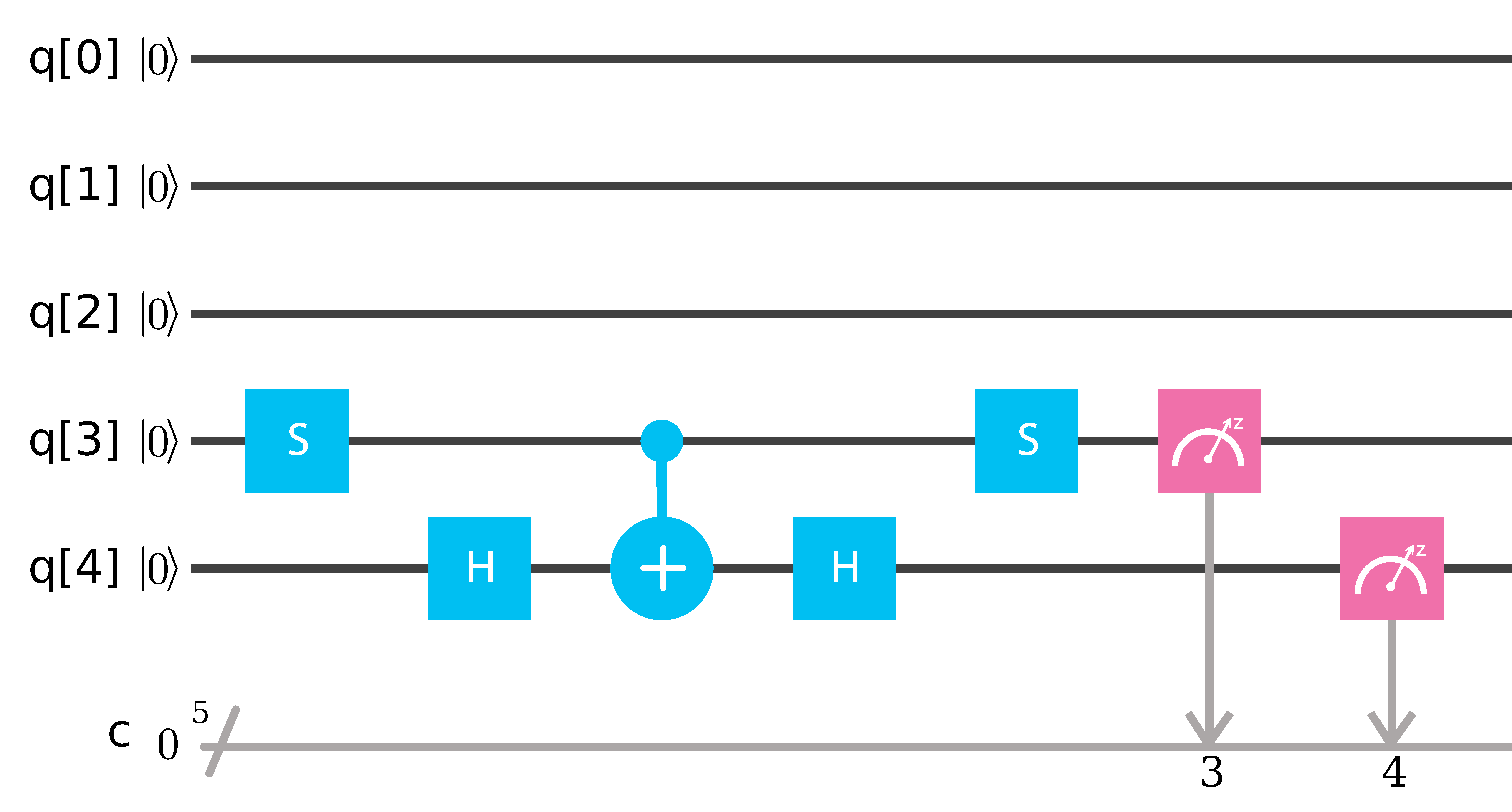}}\caption{Oracle 10}\label{fig2_47}
\end{figure}

The circuit above implements the Oracle for the item $ \beta=10 $, where the S gate transforms according to:
\begin{center}
$ \displaystyle \left\vert 0\right\rangle    \displaystyle \rightarrow    \displaystyle \left\vert 0\right\rangle $      \\     
$ \displaystyle \left\vert 1\right\rangle    \displaystyle \rightarrow    \displaystyle i\left\vert 1\right\rangle $    
\end{center}      
The table below shows the evolution of the basis state under the Oracle. we can see that the only basis state that acquires a negative sign is the solution state $ \left\vert 1\right\rangle \left\vert 0\right\rangle $

\begin
{table}[H]
    \centering
    \caption{basis state under the Oracle}
    \label{tab:2_1:Table 9}
    \begin{tabular}{|c|c|c|c|c|c|}\hline
        \begin{tabular}[c]{@{}l@{}}Basis\\ State\end{tabular} &
        \begin{tabular}[c]{@{}l@{}}After first\\ S\end{tabular} &
        \begin{tabular}[c]{@{}l@{}}After first\\ H\end{tabular} &
        \begin{tabular}[c]{@{}l@{}}After \\ CNOT \end{tabular} & \begin{tabular}[c]{@{}l@{}}After second\\ H\end{tabular} &  \begin{tabular}[c]{@{}l@{}}After second \\ S\end{tabular} \\ \hline

$ \left\vert 0\right\rangle \left\vert 0\right\rangle $    & $ \left\vert 0\right\rangle \left\vert 0\right\rangle $    & $ \left\vert 0\right\rangle \frac{\left\vert 0\right\rangle +\left\vert 1\right\rangle }{\sqrt{2}} $ &    $ \left\vert 0\right\rangle \frac{\left\vert 0\right\rangle +\left\vert 1\right\rangle }{\sqrt{2}} $ &    $ \left\vert 0\right\rangle \left\vert 0\right\rangle $ &    $ \left\vert 0\right\rangle \left\vert 0\right\rangle $ \\ \hline
$ \left\vert 0\right\rangle \left\vert 1\right\rangle $    & $ \left\vert 0\right\rangle \left\vert 1\right\rangle $ &    $ \left\vert 0\right\rangle \frac{\left\vert 0\right\rangle -\left\vert 1\right\rangle }{\sqrt{2}} $ &    $ \left\vert 0\right\rangle \frac{\left\vert 0\right\rangle -\left\vert 1\right\rangle }{\sqrt{2}} $ &    $ \left\vert 0\right\rangle \left\vert 1\right\rangle $ &    $ \left\vert 0\right\rangle \left\vert 1\right\rangle $ \\ \hline

$ \left\vert 1\right\rangle \left\vert 0\right\rangle $ &    $ i\left\vert 1\right\rangle \left\vert 0\right\rangle $ &    $ i\left\vert 1\right\rangle \frac{\left\vert 0\right\rangle +\left\vert 1\right\rangle }{\sqrt{2}} $ &    $ i\left\vert 1\right\rangle \frac{\left\vert 0\right\rangle +\left\vert 1\right\rangle }{\sqrt{2}} $ &    $ i\left\vert 1\right\rangle \left\vert 0\right\rangle $ &    $ -\left\vert 1\right\rangle \left\vert 0\right\rangle $ \\ \hline
$ \left\vert 1\right\rangle \left\vert 1\right\rangle $ &    $ i\left\vert 1\right\rangle \left\vert 1\right\rangle $ &    $ i\left\vert 1\right\rangle \frac{\left\vert 0\right\rangle -\left\vert 1\right\rangle }{\sqrt{2}} $ &    $ -i\left\vert 1\right\rangle \frac{\left\vert 0\right\rangle -\left\vert 1\right\rangle }{\sqrt{2}} $ &    $ -i\left\vert 1\right\rangle \left\vert 1\right\rangle $ &    $ \left\vert 1\right\rangle \left\vert 1\right\rangle $ \\ \hline
\end{tabular}
\end{table}

    \item N=4: Oracle Operator $ \beta=2 $; For a list of N=4 items, write the QASM code \cite{cross_open_2017} that generates the Oracle operator when the marked item is $ \beta=2 $(decimal representation). Use qubits q[3] and q[4] as the first and second qubit respectively.
    \begin{figure}[H] \centering{\includegraphics[scale=.9]{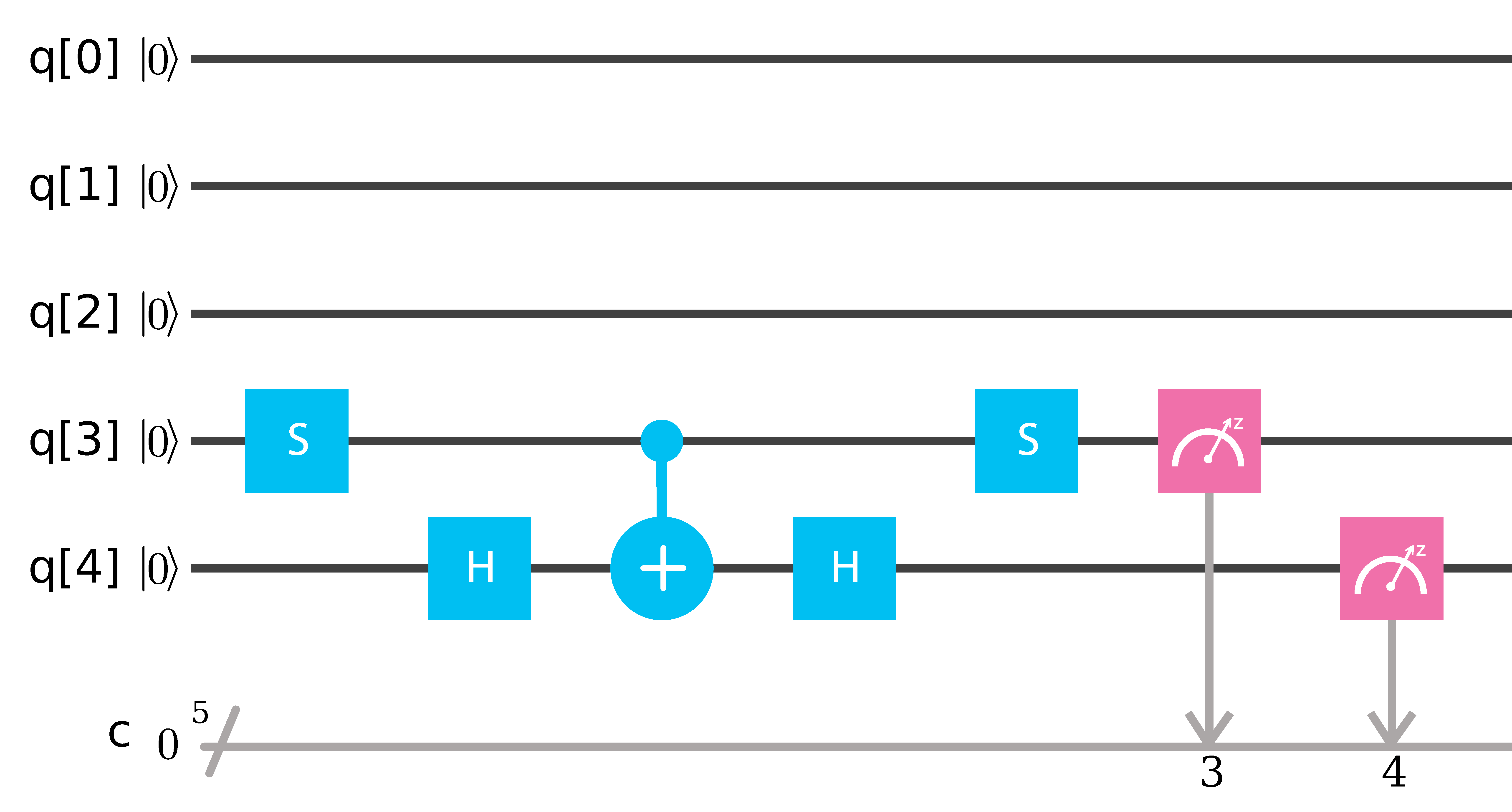}}\caption{N=4: Oracle Operator $ \beta=2 $}\label{fig2_48}
    \end{figure}
    Solution:\\
    \begin{lstlisting}
    include ``qelib1.inc'';
    qreg q[5];
    creg c[5];
    s q[3];
    h q[4];
    cx q[3],q[4];
    h q[4];
    s q[3];
    measure q[3] -> c[3];
    measure q[4] -> c[4];
    \end{lstlisting}

\section{Grover's Algorithm: Grover Operator}
Reflection W:\\
The third step in Grover's algorithm is given by the implementation of the second reflection, which we will call W. Let us recall the state after the first implementation of the Oracle Operator,

\begin{equation}\label{eq2_160}
O\left\vert \psi \right\rangle =\cos \left( \frac{\theta }{2}\right) \left\vert \alpha \right\rangle -\sin \left( \frac{\theta }{2}\right) \left\vert \beta\right\rangle.
\end{equation}

The $\beta $operator then corresponds to a reflection about the equal superposition state $\left\vert \psi \right\rangle$. Since this operator does not depend on the function f(x) and, rather, depends only on the superposition state, W is the same for any marked item $\beta$. 

Grover Operator\\
We define the Grover operator as 
$G=WO,$

that is, the application of the Oracle operator followed by the W reflection.
\begin{equation}\label{eq2_161}
G\left\vert \psi \right\rangle =\cos \left( \frac{3\theta }{2}\right) \left\vert \alpha \right\rangle +\sin \left( \frac{3\theta }{2}\right) \left\vert \beta\right\rangle
\end{equation}

\begin{figure}[H] \centering{\includegraphics[scale=.15]{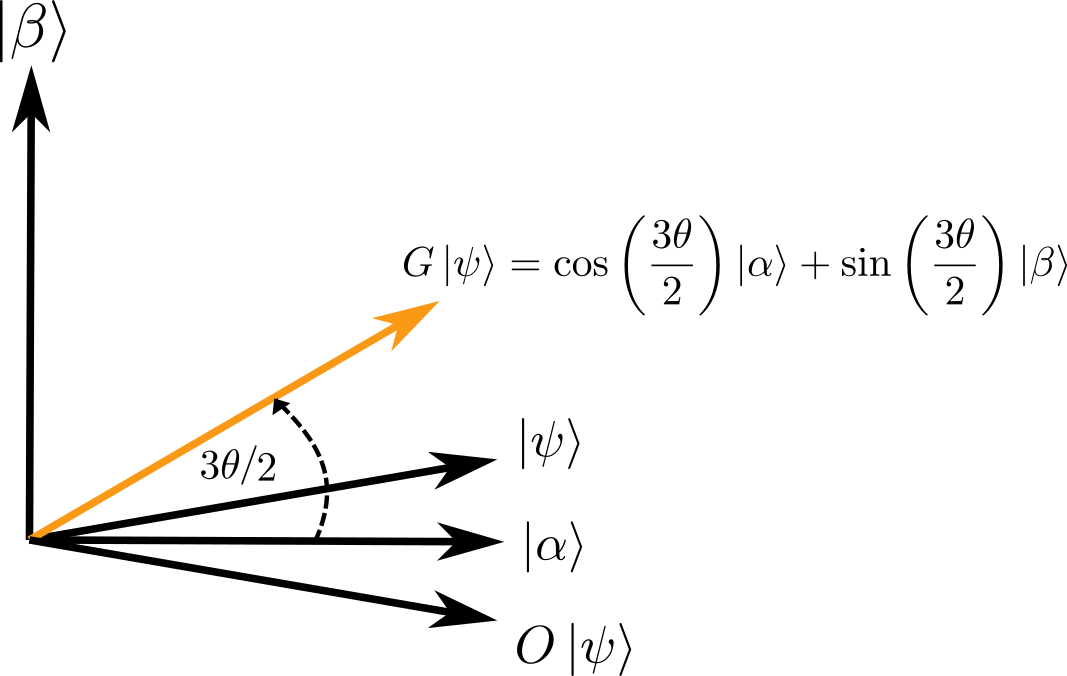}}\caption{N Grover}\label{fig2_22}
\end{figure}

The state after the second Grover iteration is:
\begin{equation}\label{eq2_162}
G^{2}\left\vert \psi \right\rangle =\cos \left( \frac{5\theta }{2}\right) \left\vert \alpha \right\rangle +\sin \left( \frac{5\theta }{2}\right) \left\vert \beta\right\rangle
\end{equation}

\begin{figure}[H] \centering{\includegraphics[scale=.15]{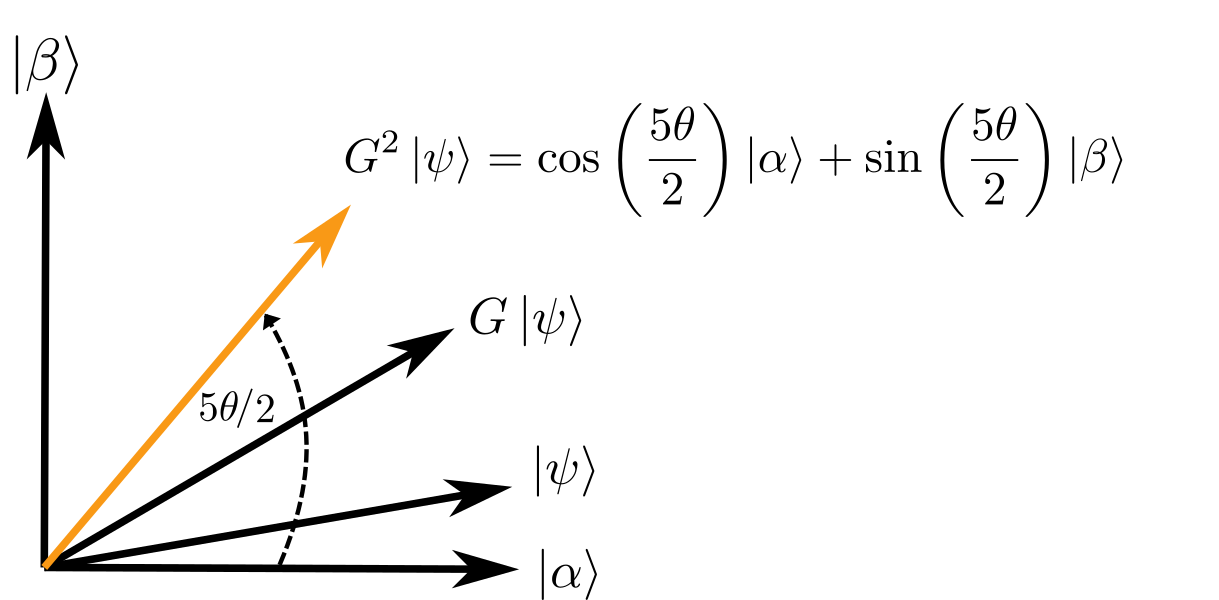}}\caption{N Grover}\label{fig2_23}
\end{figure}

In general, the state after the k-th Grover iteration will be given by
\begin{equation}\label{eq2_163}
G^{k}\left\vert \psi \right\rangle =\cos \left( \frac{2k+1}{2}\theta \right) \left\vert \alpha \right\rangle +\sin \left( \frac{2k+1}{2}\theta \right) \left\vert \beta\right\rangle\end{equation}

$N=4$ items: Marked item $\beta=11$\\

As we discussed in the previous section, the equal superposition state for a list of 4 items is given by:
\begin{equation}\label{eq2_164}
\left\vert \psi \right\rangle =\cos \left( 30^{\circ}\right) \left\vert \alpha \right\rangle +\sin \left( 30^{\circ}\right) \left\vert \beta\right\rangle ,
\end{equation}

where the non-solution and solution states are respectively given by
\begin{equation}\label{eq2_165}
\begin{split}
\displaystyle \left\vert \alpha \right\rangle    \displaystyle & =    \displaystyle \frac{\left\vert 0\right\rangle \left\vert 0\right\rangle +\left\vert 0\right\rangle \left\vert 1\right\rangle +\left\vert 1\right\rangle \left\vert 0\right\rangle }{\sqrt{3}},    \\      
\displaystyle \left\vert \beta\right\rangle    \displaystyle & =    \displaystyle \left\vert 1\right\rangle \left\vert 1\right\rangle
\end{split}
\end{equation}
          
After the Oracle operator for the item $\beta=11$ is applied, the equal superposition state $\lvert \psi\rangle $is reflected about the non-solution state $\lvert \alpha\rangle$

\begin{equation}\label{eq2_166}
O\left\vert \psi \right\rangle =\cos \left( 30^{\circ}\right) \left\vert \alpha \right\rangle -\sin \left( 30^{\circ}\right) \left\vert \beta\right\rangle .
\end{equation}
The second reflection is applied, and this reflects the state $O\left\vert \psi \right\rangle$ about the equal superposition state $\left\vert \psi \right\rangle$,
\begin{equation}\label{eq2_167}
WO\left\vert \psi \right\rangle =\cos \left( 90^{\circ}\right) \left\vert \alpha \right\rangle +\sin \left( 90^{\circ}\right) \left\vert \beta\right\rangle .
\end{equation}

How many times should we apply the Grover Operator? For a list of $N=4$ items and one marked item $\beta$, the number of necessary repetitions R required to find w is bounded above by:
\begin{equation}\label{eq2_168}
\begin{split}
\displaystyle R    &\displaystyle \leq    \displaystyle \frac{\pi }{4}\sqrt{4},          \\
& \displaystyle \leq    \displaystyle \frac{\pi }{4}\sqrt{4},          \\
& \displaystyle \leq    \displaystyle 1.5
\end{split}
\end{equation}
          
Since the upper bound is $\approx 1.5$, we know that the optimal number of iterations to find the marked item is equal to one. The state after the first Grover iteration, $G=WO$, is given by
\begin{equation}\label{eq2_169}
G\left\vert \psi \right\rangle =\cos \left( 90^{\circ}\right) \left\vert \alpha \right\rangle +\sin \left( 90^{\circ}\right) \left\vert \beta\right\rangle .
\end{equation}

The figure below illustrates the final state after the first Grover iteration. we can see that the final state matches the solution state. 
\begin{figure}[H] \centering{\includegraphics[scale=.15]{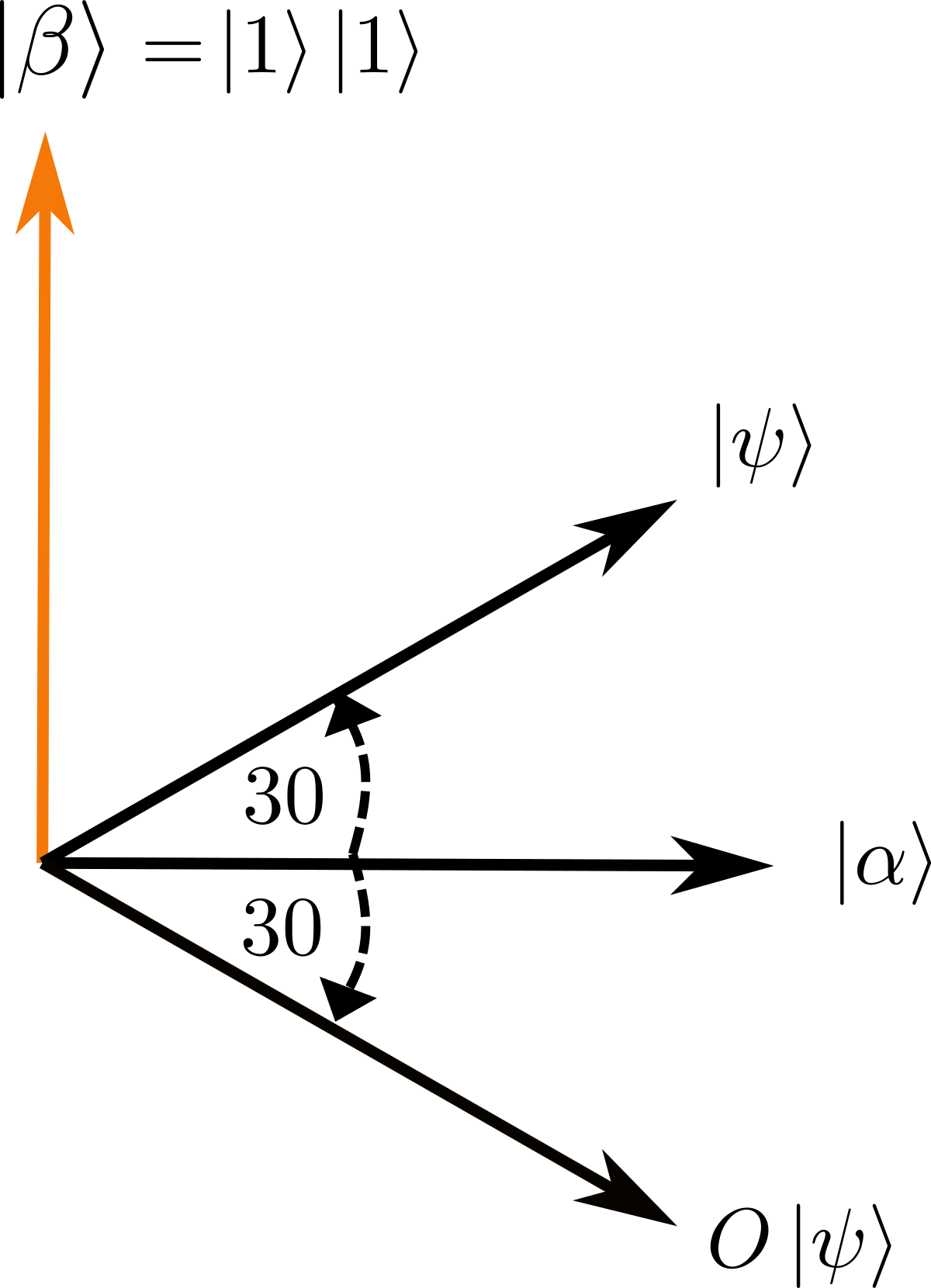}}\caption{N4 Grover}\label{fig2_24}
\end{figure}

From the figure below, we can see that applying the Grover operator a second time, $G^{2}=\left( WO\right) ^{2}$, does not improve the performance of the algorithm. After a second iteration, the resulting state moves away from the solution state,

\begin{equation}\label{eq2_170}
G^{2}\left\vert \psi \right\rangle =\cos \left( 150\right) \left\vert \alpha \right\rangle +\sin \left( 150\right) \left\vert \beta\right\rangle 
\end{equation}

\begin{figure}[H] \centering{\includegraphics[scale=.09]{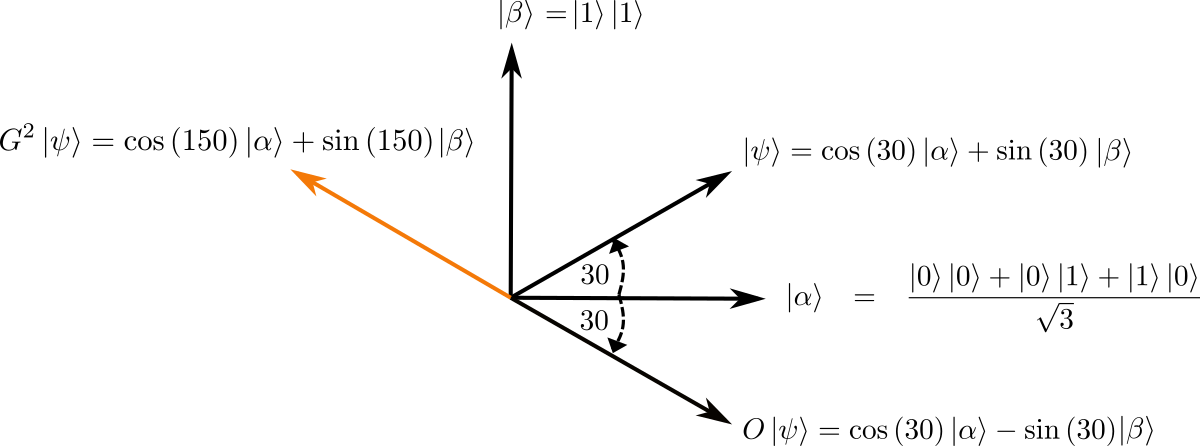}}\caption{N4 Grover}\label{fig2_25}
\end{figure}

    \item W Reflection for N=4 items; The figure below shows the quantum circuit used to implement the reflection W for a list of N=4 items. Write the QASM code that generates this quantum circuit.
    \begin{figure}[H] \centering{\includegraphics[scale=.8]{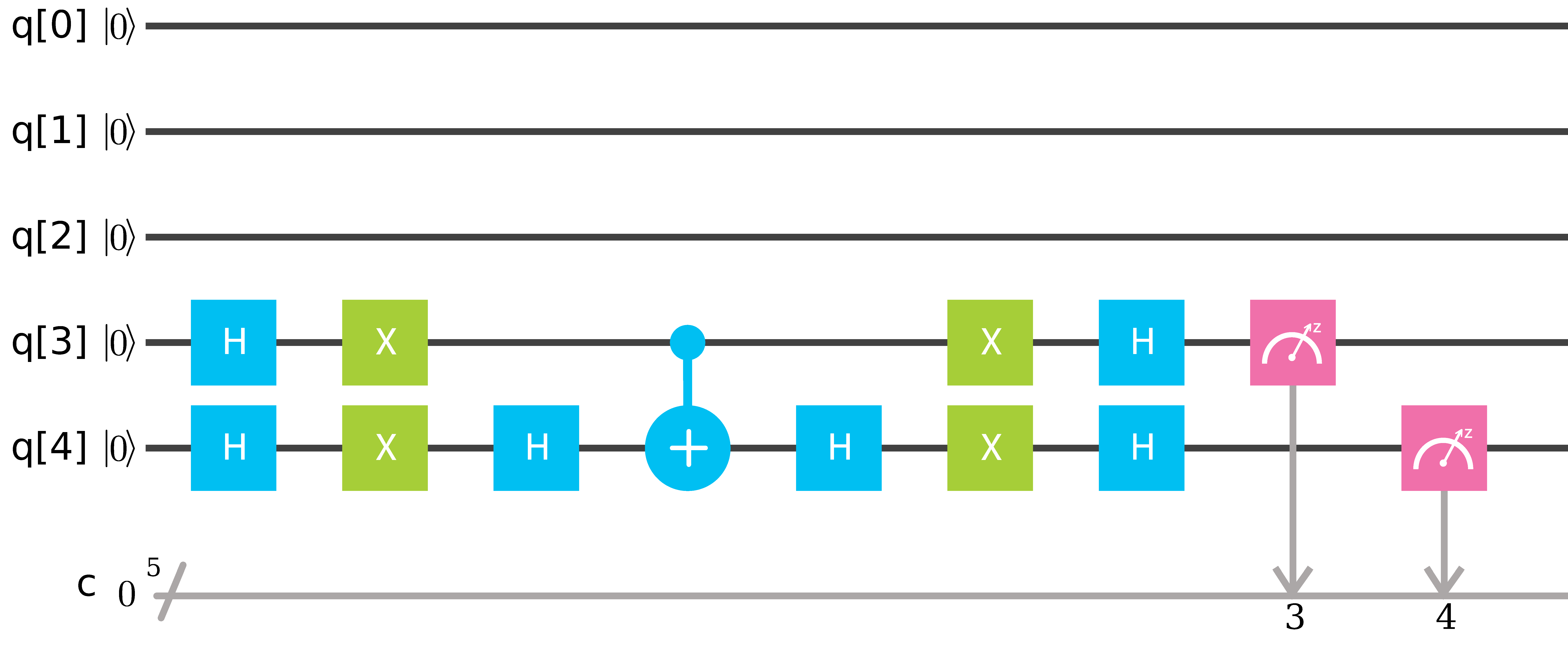}}\caption{Reflection N4}\label{fig2_49}
    \end{figure}
    Solution:\\
    \begin{lstlisting}
    include ``qelib1.inc'';
    qreg q[5];
    creg c[5];
    h q[3];
    h q[4];
    x q[3];
    x q[4];
    h q[4];
    cx q[3],q[4];
    h q[4];
    x q[3];
    x q[4];
    h q[3];
    h q[4];
    measure q[3] -> c[3];
    measure q[4] -> c[4];
    \end{lstlisting}

\section{Grover's Algorithm: Putting it All Together}
By now, we have seen how to implement each step of Grover's algorithm: Initialization, Oracle $O$, and Reflection W, where the last two correspond to the implementation of the Grover operator $G=WO$. In particular, we discussed the compiled quantum circuit to implement the Oracles for two marked items in a list of $N=4$ items. In this section of the IBM Q experience, we will program Grover's algorithm for all the possible marked items when $N=4$.

    \item Grover's Algorithm: N=4 Items with $ \beta=00 $; The figure below shows the quantum circuit to implement Grover's algorithm for a list of N=4 items with $ \beta=00 $. Write the QASM code that generates this quantum circuit.
    \begin{figure}[H] \centering{\includegraphics[scale=.55]{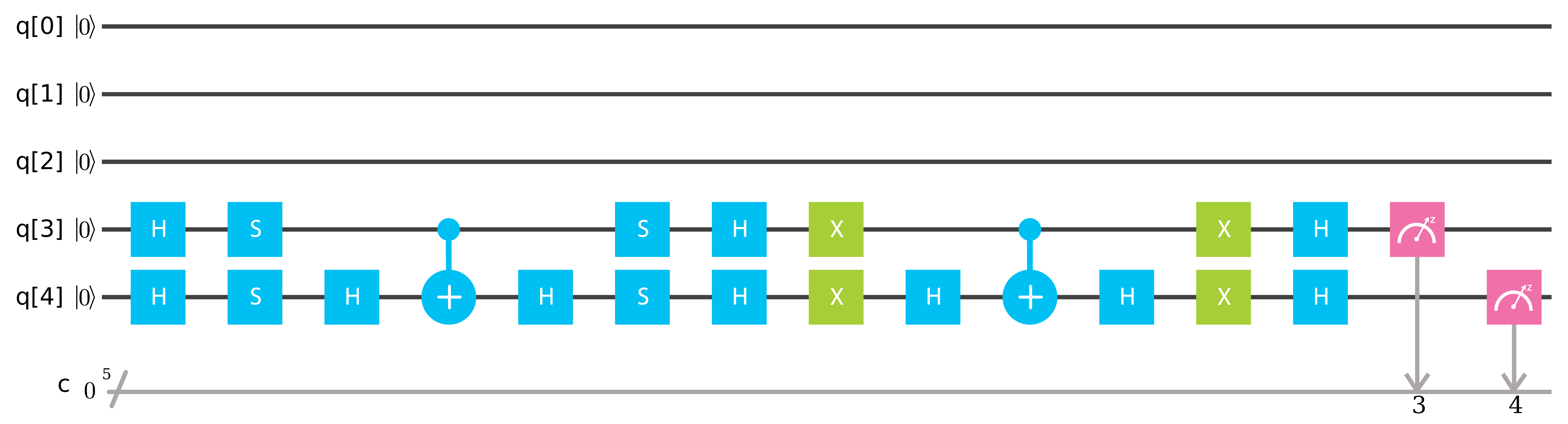}}\caption{Grover final}\label{fig2_50}
    \end{figure}
    Solution:\\
    \begin{lstlisting}
    include ``qelib1.inc'';
    qreg q[5];
    creg c[5];
    //INITIALIZATION
    h q[3];
    h q[4];
    //ORACLE 00
    s q[3];
    s q[4];
    h q[4];
    cx q[3],q[4];
    h q[4];
    s q[3];
    s q[4];
    //REFLECTION
    h q[3];
    h q[4];
    x q[3];
    x q[4];
    h q[4];
    cx q[3],q[4];
    h q[4];
    x q[3];
    x q[4];
    h q[3];
    h q[4];
    measure q[3] -> c[3];
    measure q[4] -> c[4];
    \end{lstlisting}
    In this circuit, the first two Hadamard gates, H, create the equal superposition two-qubit state
    \begin{center}
    $ \frac{\left\vert 0\right\rangle \left\vert 0\right\rangle +\left\vert 0\right\rangle \left\vert 1\right\rangle +\left\vert 1\right\rangle \left\vert 0\right\rangle +\left\vert 1\right\rangle \left\vert 1\right\rangle }{2} $
    \end{center}
    The following combination of four S gates, two H gates, and one CNOT gate implements the oracle O for the marked item $ \beta =00 $ in binary notation. This oracle leaves the two-qubit system in
    \begin{center}
    $ \frac{-\left\vert 0\right\rangle \left\vert 0\right\rangle +\left\vert 0\right\rangle \left\vert 1\right\rangle +\left\vert 1\right\rangle \left\vert 0\right\rangle +\left\vert 1\right\rangle \left\vert 1\right\rangle }{2} $
    \end{center}
    The remaining section of the circuit implements the reflection W for a two-qubit system. The Grover operator $ G=WO $ maps the equal superpostition state as
    \begin{center}
    $ \frac{\left\vert 0\right\rangle \left\vert 0\right\rangle +\left\vert 0\right\rangle \left\vert 1\right\rangle +\left\vert 1\right\rangle \left\vert 0\right\rangle +\left\vert 1\right\rangle \left\vert 1\right\rangle }{2}\rightarrow \left\vert 0\right\rangle \left\vert 0\right\rangle  $
    \end{center}
    we can see that the analytical result agrees with the simulation. The higher frequency is given by the measurement result $ (c_4,c_3,c_2,c_1,c_0)=(0,0,0,0,0) $, where$  c_0, c_1, c_2, c_3 $, and $ c_4 $ refers to the measurement result of qubit $ q_0, q_1, q_2, q_3, $ and $ q_4 $ respectively.
    \item Grover's Algorithm: N=4 Items with $ \beta=01 $; The figure below shows the quantum circuit to implement Grover's algorithm for a list of N=4 items with $ \beta=01 $. Write the QASM code that generates this quantum circuit.
    \begin{figure}[H] \centering{\includegraphics[scale=.55]{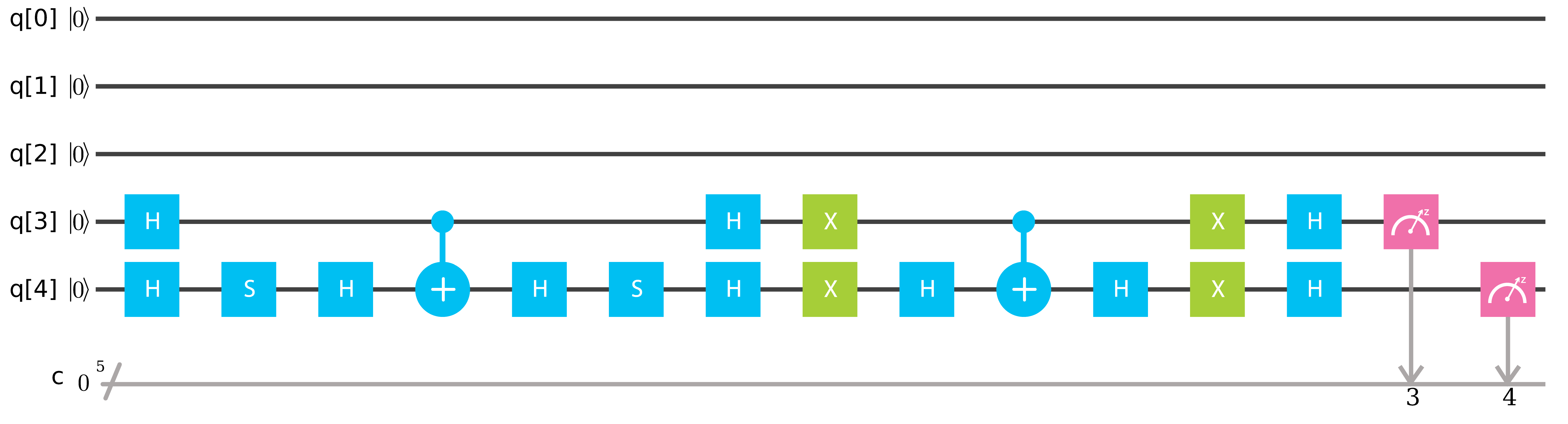}}\caption{Grover final}\label{fig2_51}
    \end{figure}
    Solution:\\
    \begin{lstlisting}
    include ``qelib1.inc'';
    qreg q[5];
    creg c[5];
    //INITIALIZATION
    h q[4];
    //ORACLE 01
    s q[4];
    h q[4];
    cx q[3],q[4];
    h q[4];
    s q[4];
    //REFLECTION
    h q[3];
    h q[4];
    x q[3];
    x q[4];
    h q[4];
    cx q[3],q[4];
    h q[4];
    x q[3];
    x q[4];
    h q[3];
    h q[4];
    measure q[3] -> c[3];
    measure q[4] -> c[4];
    \end{lstlisting}
    In this circuit, the first two Hadamard gates, H, create the equal superposition two-qubit state
    \begin{center}
    $ \frac{\left\vert 0\right\rangle \left\vert 0\right\rangle +\left\vert 0\right\rangle \left\vert 1\right\rangle +\left\vert 1\right\rangle \left\vert 0\right\rangle +\left\vert 1\right\rangle \left\vert 1\right\rangle }{2} $
    \end{center}
    The following combination of two S gates, two H gates, and one CNOT gate implements the oracle O for the marked item $ \beta =01 $ in binary notation. This oracle leaves the two-qubit system in
    \begin{center}
    $ \frac{\left\vert 0\right\rangle \left\vert 0\right\rangle -\left\vert 0\right\rangle \left\vert 1\right\rangle +\left\vert 1\right\rangle \left\vert 0\right\rangle +\left\vert 1\right\rangle \left\vert 1\right\rangle }{2} $
    \end{center}
    The remaining section of the circuit implements the reflection W for a two-qubit system. The Grover operator $ G=WO $ maps the equal superpostition state as
    \begin{center}
    $ \frac{\left\vert 0\right\rangle \left\vert 0\right\rangle +\left\vert 0\right\rangle \left\vert 1\right\rangle +\left\vert 1\right\rangle \left\vert 0\right\rangle +\left\vert 1\right\rangle \left\vert 1\right\rangle }{2}\rightarrow \left\vert 0\right\rangle \left\vert 1\right\rangle  $
    \end{center}
    we can see that the analytical result agrees with the simulation. The higher frequency is given by the measurement result $ (c_4,c_3,c_2,c_1,c_0)=(1,0,0,0,0) $, where$  c_0, c_1, c_2, c_3 $, and $ c_4 $ refers to the measurement result of qubit $ q_0, q_1, q_2, q_3, $ and $ q_4 $ respectively.
    \item The figure below shows the quantum circuit to implement Grover's algorithm for a list of N=4 items with $ \beta=10 $. Write the QASM code that generates this quantum circuit.
    \begin{figure}[H] \centering{\includegraphics[scale=.55]{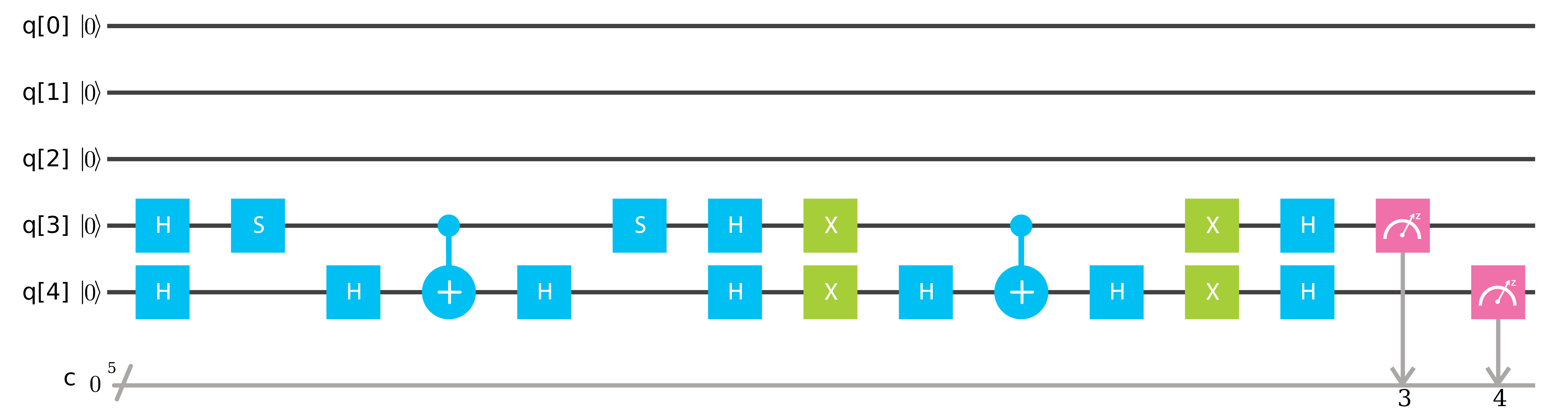}}\caption{Grover final}\label{fig2_52}
    \end{figure}
    Solution:\\
    \begin{lstlisting}
    include ``qelib1.inc'';
    qreg q[5];
    creg c[5];
    //INITIALIZATION
    h q[3];
    h q[4];
    //ORACLE10
    s q[3];
    h q[4];
    cx q[3],q[4];
    h q[4];
    s q[3];
    //REFLECTION
    h q[3];
    h q[4];
    x q[3];
    x q[4];
    h q[4];
    cx q[3],q[4];
    h q[4];
    x q[3];
    x q[4];
    h q[3];
    h q[4];
    measure q[3] -> c[3];
    measure q[4] -> c[4];
    \end{lstlisting}
    In this circuit, the first two Hadamard gates, H, create the equal superposition two-qubit state
    \begin{center}
    $ \frac{\left\vert 0\right\rangle \left\vert 0\right\rangle +\left\vert 0\right\rangle \left\vert 1\right\rangle +\left\vert 1\right\rangle \left\vert 0\right\rangle +\left\vert 1\right\rangle \left\vert 1\right\rangle }{2} $
    \end{center}
    The following combination of two S gates, two H gates, and one CNOT gate implements the oracle O for the marked item $ \beta =10 $ in binary notation. This oracle leaves the two-qubit system in
    \begin{center}
    $ \frac{\left\vert 0\right\rangle \left\vert 0\right\rangle +\left\vert 0\right\rangle \left\vert 1\right\rangle -\left\vert 1\right\rangle \left\vert 0\right\rangle +\left\vert 1\right\rangle \left\vert 1\right\rangle }{2} $
    \end{center}
    The remaining section of the circuit implements the reflection W for a two-qubit system. The Grover operator $ G=WO $ maps the equal superpostition state as
    \begin{center}
    $ \frac{\left\vert 0\right\rangle \left\vert 0\right\rangle +\left\vert 0\right\rangle \left\vert 1\right\rangle +\left\vert 1\right\rangle \left\vert 0\right\rangle +\left\vert 1\right\rangle \left\vert 1\right\rangle }{2}\rightarrow \left\vert 1\right\rangle \left\vert 0\right\rangle $
    \end{center}
    we can see that the analytical result agrees with the simulation. The higher frequency is given by the measurement result $ (c_4,c_3,c_2,c_1,c_0)=(0,1,0,0,0), $ where $ c_0, c_1, c_2, c_3, $ and $ c_4 $ refers to the measurement result of qubit $ q_0, q_1, q_2, q_3, $ and $ q_4 $ respectively.
    \item The figure below shows the quantum circuit to implement Grover's algorithm for a list of N=4 items with $ \beta=11 $. Write the QASM code that generates this quantum circuit.
    \begin{figure}[H] \centering{\includegraphics[scale=.55]{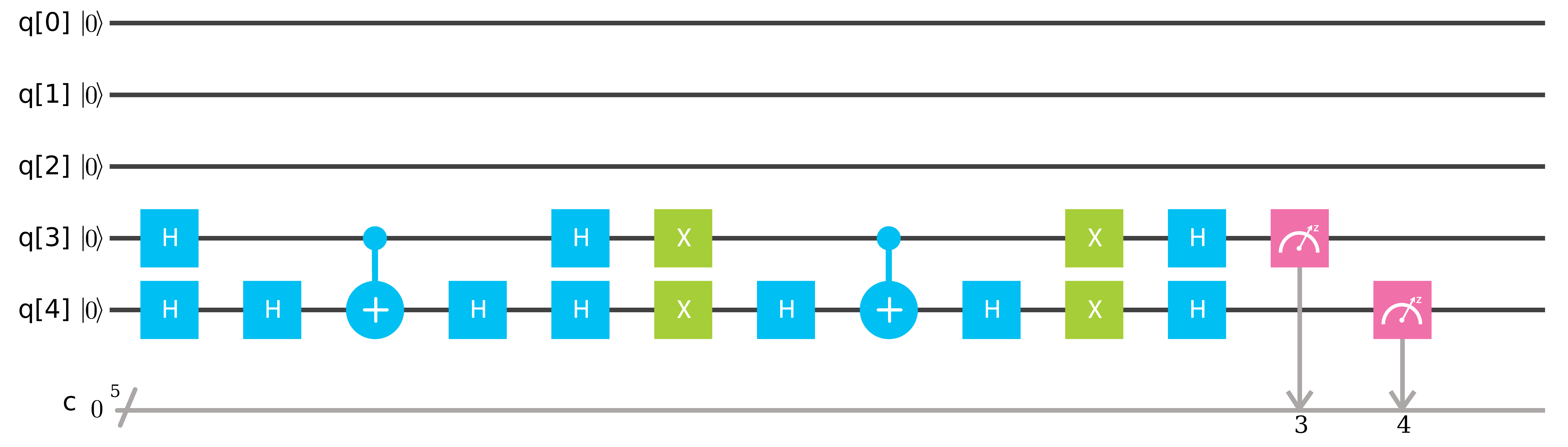}}\caption{Grover final}\label{fig2_53}
    \end{figure}
    Solution:\\
    \begin{lstlisting}
    include ``qelib1.inc'';
    qreg q[5];
    creg c[5];
    //INITIALIZATION
    h q[3];
    h q[4];
    //ORACLE11
    h q[4];
    cx q[3],q[4];
    h q[4];
    //REFLECTION
    h q[3];
    h q[4];
    x q[3];
    x q[4];
    h q[4];
    cx q[3],q[4];
    h q[4];
    x q[3];
    x q[4];
    h q[3];
    h q[4];
    measure q[3] -> c[3];
    measure q[4] -> c[4];
    \end{lstlisting}
    In this circuit, the first two Hadamard gates, H, create the equal superposition two-qubit state
    \begin{center}
    $ \frac{\left\vert 0\right\rangle \left\vert 0\right\rangle +\left\vert 0\right\rangle \left\vert 1\right\rangle +\left\vert 1\right\rangle \left\vert 0\right\rangle +\left\vert 1\right\rangle \left\vert 1\right\rangle }{2} $
    \end{center}
    The following combination of two H gates, and one CNOT gate implements the oracle O for the marked item $ \beta =11 $ in binary notation. This oracle leaves the two-qubit system in
    \begin{center}
    $ \frac{\left\vert 0\right\rangle \left\vert 0\right\rangle +\left\vert 0\right\rangle \left\vert 1\right\rangle +\left\vert 1\right\rangle \left\vert 0\right\rangle -\left\vert 1\right\rangle \left\vert 1\right\rangle }{2} $
    \end{center}
    The remaining section of the circuit implements the reflection W for a two-qubit system. The Grover operator $ G=WO $ maps the equal superpostition state as
    \begin{center}
    $ \frac{\left\vert 0\right\rangle \left\vert 0\right\rangle +\left\vert 0\right\rangle \left\vert 1\right\rangle +\left\vert 1\right\rangle \left\vert 0\right\rangle +\left\vert 1\right\rangle \left\vert 1\right\rangle }{2}\rightarrow \left\vert 1\right\rangle \left\vert 1\right\rangle  $
    \end{center}
    we can see that the analytical result agrees with the simulation. The higher frequency is given by the measurement result $ (c_4,c_3,c_2,c_1,c_0)=(1,1,0,0,0), $ where $ c_0, c_1, c_2, c_3, $ and $ c_4 $ refers to the measurement result of qubit $ q_0, q_1, q_2, q_3, $ and $ q_4 $ respectively.

\end{enumerate}

\begin{figure}[H] \centering{\includegraphics[scale=.8]{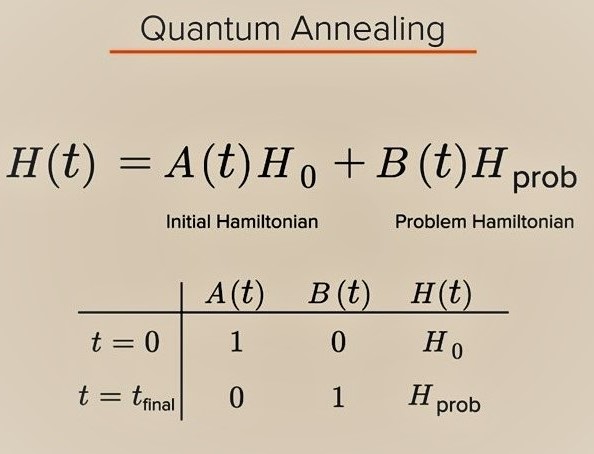}}\caption{Quantum Annealing}\label{fig2_26}
\end{figure}

\begin{figure}[H] \centering{\includegraphics[scale=.5]{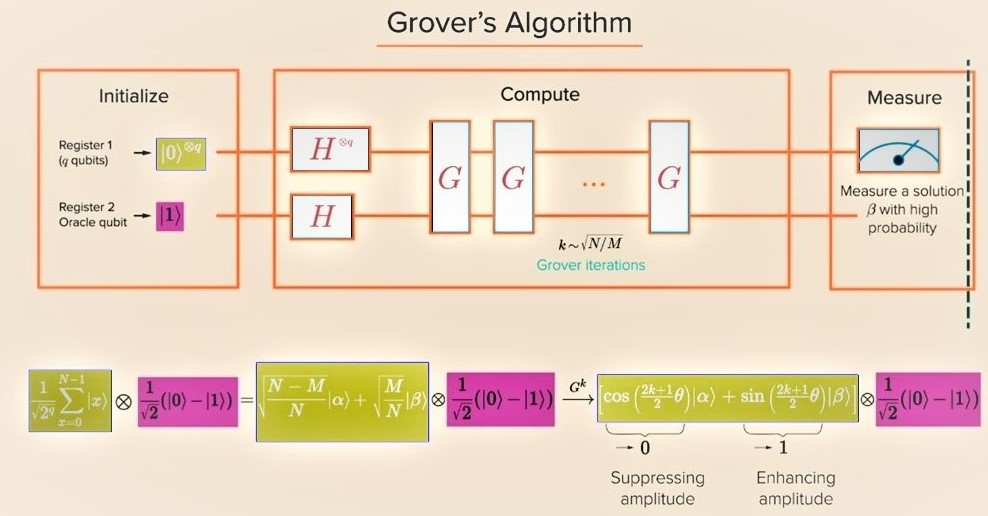}}\caption{Grovers Algorithm}\label{fig2_27}
\end{figure}

\bibliographystyle{IEEEtran}
\bibliography{Bib-Quantum}
\printindex

\part{Quantum-Computation and Quantum-Communication}
\chapter{Quantum-Computation and Quantum-Communication}

\maketitle

In this research notebook on quantum computation and quantum communication protocol for quantum engineers, researchers, and scientists, we will discuss and summarize the practical challenges of implementing quantum computing algorithms and quantum communication protocols in today's real-world hardware systems. We will begin the discussion about the noise and its effect on quantum coherence and the mathematical models to represent the impact of noise on the physical systems. Furthermore, we will discuss the basic concepts of noise that practically limit quantum computation and quantum communication. We will discuss several mathematical methods to characterize, model, and represent the practical impact of noise. Next, we will discuss how photon loss impacts quantum communication protocols. We will summarize the current physical limitations and challenges facing quantum computation and quantum communication over optical fiber. We will also discuss technology options that can mitigate or eliminate the physical limitations for realizing quantum computation and quantum communication protocol. Later we will consider the methods being developed to mitigate such losses in larger-scale communication systems. We will investigate the current limitations of quantum computing algorithms and the need to develop near-term practical applications that can be successfully performed on noisy qubits. We will also review the challenges associated with implementing algorithms on the relatively noisy, intermediate-scale qubits used in today's technology domain. Finally, we will investigate how to benchmark qubits and quantum gates in the presence of noise. We will discuss several methods to benchmark quantum states and quantum gates and finally implement benchmarking of qubits in a real physical quantum computer on the IBM Q Experience.

\section{Introduction}
We will manifest quantum mechanical effects that enable quantum computers to process information efficiently, \cite{nielsen_quantum_2011,chuang_quantum_2014,preskill_quantum_2019,vazirani_quantum_1997, watrous_theory_2011,vazirani_quantum_2007,aaronson_quantum_2006,shor_quantum_2003,chuang_quantum_2006,aaronson_quantum_2010,harrow_quantum_2018,childs_quantum_2008,cleve_introduction_2007,harrow_why_2012} facing the practical challenges in implementing quantum algorithms in quantum communication. We will look at noise, its impact on quantum coherence, and how it impacts gate operations. We will also look at how we can mitigate certain types of noise to retain some of that coherence. We begin by discussing the ubiquity and challenges of noise and its impact on quantum information. We will discuss how to characterize and quantify noise processes and how to represent noisy qubits using the density matrix formalism \cite{song_10-qubit_2017,de_chiara_density_2008}. We then study the difference between classical and quantum noise, and we will visualize the resulting qubit dephasing and energy decay on the Bloch sphere. Then, We will look at practical algorithms being implemented at a small scale, and discuss the challenges associated with running those algorithms in the presence of noise \cite{lloyd_ultimate_2000,wallman_noise_2016}. We will discuss the practical realities of implementing the quantum algorithms and quantum cryptography \cite{barbeau_secure_2019}, using the smaller error-prone machines we have today. Next, we will look at the practical aspects of quantum communication, including photon loss and its impact on long-distance quantum communication. We turn to the practical issues and challenges of implementing quantum communication schemes. We start by revisiting quantum entanglement \cite{monz_14-qubit_2011}, the generation and detection of Bell states, and the challenges associated with implementing Bell's inequality experiments \cite{noauthor_big_nodate,noauthor_big_nodate-1}.
We can use entanglement to build a quantum version of an optical repeater to extend the communication distance \cite{wong-campos_demonstration_2017}. We then will discuss the practical issues associated with distributing such entanglement in the presence of photon loss in optical fibers \cite{wong-campos_demonstration_2017,friis_observation_2018}. It motivates the need for quantum repeaters. We will study several challenges associated with quantum key distribution, from having bright photon sources to high fidelity, fast photon detectors, all of which enable high key generation rates. Next, we will transition to the challenges and opportunities of contemporary quantum computing \cite{russo_coming_2018,noauthor_next_nodate}. That is, what algorithms can implement now or in the near future, given the noisy short-depth circuits we have available today. We will discuss at contemporary demonstrations of noisy intermediate-scale quantum algorithms, the so-called NISQ algorithms \cite{preskill_quantum_2018,villalonga_flexible_2019}, targeting machine learning, support vector machines \cite{havlicek_supervised_2019}, and variational quantum eigensolvers \cite{schuld_circuit-centric_2020,kandala_hardware-efficient_2017,thornton_quantum_2019,yuan_theory_2019}, as well as the HHL algorithm for linear systems \cite{harrow_quantum_2009}. Another hybrid quantum-classical algorithm is Variational Quantum Fidelity Estimation \cite{cerezo_variational_2020}, Variational Quantum Factoring (VQF) algorithm \cite{anschuetz_variational_2019}. Finally, we will investigate how to benchmark qubits and benchmark quantum gates in the presence of noise. We will discuss several techniques used to quantify the quality of realistic quantum machines. We study state and process tomography techniques \cite{mohseni_quantum-process_2008} and how they are applied in practice to determine the fidelity of qubit states and quantum operations. 
In the end, we will characterize noise in benchmark qubits on the IBM quantum experience \cite{lesovik_arrow_2019,sisodia_circuit_2018} and implement quantum state tomography \cite{thew_qudit_2002} using the qubits on the IBM quantum experience \cite{lesovik_arrow_2019}. 

\section{Density Matrices Introduction} 
The definitions of the Dirac notation used in this discussion\cite{dirac_fundamental_1925,dirac_quantum_1926,dirac_mathematical_1978,dirac_basis_1929,dirac_quantum_1929,dirac_theory_1926}. The Hadamard transform (or Hadamard gate) is an operation that maps state
\begin{center}
$|0\rangle \rightarrow \frac{\left(|0\rangle + |1\rangle\right)}{\sqrt{2}} $ and $ |1\rangle \rightarrow \frac{\left(|0\rangle - |1\rangle\right)}{\sqrt{2}}$ 
\end{center}
In this text, we will find the basics on Dirac notation and matrix representation of quantum states and operators.\\
\textbf{Single-qubit system:}\\
The computational basis for a single-qubit is expanded by the eigenvectors of the Pauli matrix Z-gate or $\sigma _{Z},$
\begin{equation}\label{eq3_01}
Z=\left( \begin{array}{cc} 1 & 0 \\ 0 & -1\end{array}\right) .
\end{equation}

This eigenvectors are written in Dirac notation as $\left\vert 0\right\rangle and \left\vert 1\right\rangle$, and in matrix form as
\begin{equation}\label{eq3_02}
\left\vert 0\right\rangle =\left( \begin{array}{c} 1 \\ 0\end{array}\right) \text {, }\left\vert 1\right\rangle =\left( \begin{array}{c} 0 \\ 1\end{array}\right) .
\end{equation}

In Dirac notation, their conjugate transponse are $\left\langle 0\right\vert and \left\langle 1\right\vert,$ and in matrix form
\begin{equation}\label{eq3_03}
\left\langle 0\right\vert =\left( \begin{array}{cc} 1 & 0\end{array}\right) \text {, }\left\langle 1\right\vert =\left( \begin{array}{cc} 0 & 1\end{array}\right) .
\end{equation}

The computational basis is defined as orthonormal, since its elements are perpendicular and have norm one,
\begin{equation}\label{eq3_04}
\begin{split}
\displaystyle \left\langle 1\right. \left\vert 0\right\rangle    \displaystyle =    \displaystyle  \left( \begin{array}{cc} 0 & 1\end{array}\right)\left( \begin{array}{c} 1 \\ 0\end{array}\right) =0,    \\      
\displaystyle \left\langle 0\right. \left\vert 1\right\rangle    \displaystyle =    \displaystyle \left( \begin{array}{cc} 1 & 0\end{array}\right)\left( \begin{array}{c} 0 \\ 1\end{array}\right) =0,
\end{split}
\end{equation}          

and

\begin{equation}\label{eq3_05}
\sqrt{\left\langle 0\right. \left\vert 0\right\rangle }=1, \sqrt{\left\langle 1\right. \left\vert 1\right\rangle }=1.
\end{equation}
While the single-qubit identity matrix is given by
\begin{equation}\label{eq3_06}
I=\text { }\left( \begin{array}{cc} 1 & 0 \\ 0 & 1\end{array}\right) 
\end{equation}
the projectors of $\left\vert 0\right\rangle $and $\left\vert 1\right\rangle$ are
\begin{equation}\label{eq3_07}
\Pi _{0}=\left\vert 0\right\rangle \left\langle 0\right\vert =\text { }\left( \begin{array}{cc} 1 & 0 \\ 0 & 0\end{array}\right) 
\end{equation}
and
\begin{equation}\label{eq3_08}
\Pi _{1}=\left\vert 1\right\rangle \left\langle 1\right\vert =\text { }\left( \begin{array}{cc} 0 & 0 \\ 0 & 1\end{array}\right) 
\end{equation}
An arbitrary single-qubit state, $\left\vert \psi \right\rangle =\alpha \left\vert 0\right\rangle +\beta \left\vert 1\right\rangle$, can be written in matrix form as
\begin{equation}\label{eq3_09}
\left\vert \psi \right\rangle =\left( \begin{array}{c} \alpha \\ \beta \end{array}\right) 
\end{equation}
and its norm can be computed by finding $ \left\vert \left\vert \psi \right\rangle \right\vert =\sqrt{\left\langle \psi \right. \left\vert \psi \right\rangle }$ or
\begin{equation}\label{eq3_10}
\displaystyle \left\vert \left\vert \psi \right\rangle \right\vert    \displaystyle =    \displaystyle \sqrt{\left( \begin{array}{cc} \alpha ^{\ast } & \beta ^{\ast }\end{array}\right) \left( \begin{array}{c} \alpha \\ \beta \end{array}\right) }          
\displaystyle =    \displaystyle \sqrt{\left\vert \alpha \right\vert ^{2}+\left\vert \beta \right\vert ^{2}},
\end{equation}          
where $\alpha ^{\ast }  $and  $\beta ^{\ast } $ are the complex conjugate of  $\alpha $ and $ \beta $.

The projector of any pure state  $\left\vert \psi \right\rangle $ can be written as
\begin{equation}\label{eq3_11}
\displaystyle \left\vert \psi \right\rangle \left\langle \psi \right\vert    \displaystyle =    \displaystyle \left( \begin{array}{c} \alpha \\ \beta \end{array}\right) \left( \begin{array}{cc} \alpha ^{\ast } & \beta ^{\ast }\end{array}\right)      
\displaystyle =    \displaystyle \left( \begin{array}{cc} \left\vert \alpha \right\vert ^{2} & \alpha \beta ^{\ast } \\ \beta \alpha ^{\ast } & \left\vert \beta \right\vert ^{2}\end{array}\right) 
\end{equation}          
The tensor product between two single-qubit states, $\left\vert \psi \right\rangle and \left\vert \phi \right\rangle,$ can be computed in matrix form as
\begin{equation}\label{eq3_12}
\displaystyle \left\vert \psi \right\rangle \otimes \left\vert \phi \right\rangle    \displaystyle =    \displaystyle \left( \begin{array}{c} \alpha \\ \beta \end{array}\right) \otimes \left( \begin{array}{c} \gamma \\ \delta \end{array}\right) 
\displaystyle =    \displaystyle \left( \begin{array}{c} \alpha \left( \begin{array}{c} \gamma \\ \delta \end{array}\right) \\ \beta \left( \begin{array}{c} \gamma \\ \delta \end{array}\right) \end{array}\right)           
\displaystyle =    \displaystyle \left( \begin{array}{c} \alpha \gamma \\ \alpha \delta \\ \beta \gamma \\ \beta \delta \end{array}\right) 
\end{equation}      
\textbf{Two-qubit system:}\\

The computational basis is spanned by
\begin{equation}\label{eq3_13}
\left\vert 0\right\rangle \left\vert 0\right\rangle =\left( \begin{array}{c} 1 \\ 0 \\ 0 \\ 0\end{array}\right) \text {, }\left\vert 0\right\rangle \left\vert 1\right\rangle =\left( \begin{array}{c} 0 \\ 1 \\ 0 \\ 0\end{array}\right) \text {, }\left\vert 1\right\rangle \left\vert 0\right\rangle =\left( \begin{array}{c} 0 \\ 0 \\ 1 \\ 0\end{array}\right) \text {, }\left\vert 1\right\rangle \left\vert 1\right\rangle =\left( \begin{array}{c} 0 \\ 0 \\ 0 \\ 1\end{array}\right) ,
\end{equation} 
which comes from calculating the tensor product between $\left\vert 0\right\rangle$ and $\left\vert 1\right\rangle,$ i.e
\begin{equation}\label{eq3_14}
\displaystyle \left\vert 0\right\rangle \left\vert 0\right\rangle    \displaystyle =    \displaystyle \left( \begin{array}{c} 1 \\ 0\end{array}\right) \otimes \left( \begin{array}{c} 1 \\ 0\end{array}\right)           
\displaystyle =    \displaystyle \left( \begin{array}{c} 1\left( \begin{array}{c} 1 \\ 0\end{array}\right) \\ 0\left( \begin{array}{c} 1 \\ 0\end{array}\right) \end{array}\right) 
\displaystyle =    \displaystyle \left( \begin{array}{c} 1 \\ 0 \\ 0 \\ 0\end{array}\right) 
\end{equation}           
For simplicity, most people usually write $\left\vert 00\right\rangle$ instead of $\left\vert 0\right\rangle \left\vert 0\right\rangle$, and the same for the other elements of the basis.
\begin{equation}\label{eq3_15}
\displaystyle \left\vert 01\right\rangle \equiv \left\vert 0\right\rangle \left\vert 1\right\rangle ,          
\displaystyle \left\vert 10\right\rangle \equiv \left\vert 1\right\rangle \left\vert 0\right\rangle ,          
\displaystyle \left\vert 11\right\rangle \equiv \left\vert 1\right\rangle \left\vert 1\right\rangle .    
\end{equation}       
To keep track of which qubit the state belongs, we can use the subscripts in the labels of the state, such as $\left\vert 0_{A}\right\rangle$ to indicate that this is the state $\left\vert 0\right\rangle$ for qubit A.

The two-qubit identity matrix is given by
\begin{equation}\label{eq3_16}
I=\text { }\left( \begin{array}{cccc} 1 & 0 & 0 & 0 \\ 0 & 1 & 0 & 0 \\ 0 & 0 & 1 & 0 \\ 0 & 0 & 0 & 1\end{array}\right)
\end{equation}
and the projectors of the computational basis by
\begin{equation}\label{eq3_17}
\Pi _{00}=\left\vert 00\right\rangle \left\langle 00\right\vert, \Pi _{01}=\left\vert 01\right\rangle \left\langle 01\right\vert, \Pi _{10}=\left\vert 10\right\rangle \left\langle 10\right\vert,
\end{equation}

and $\Pi _{11}=\left\vert 11\right\rangle \left\langle 11\right\vert.$
 
An arbitrary two-qubit state with qubit A and B,
\begin{equation}\label{eq3_18}
\left\vert \psi _{AB}\right\rangle =\alpha \left\vert 0_{A}0_{B}\right\rangle +\beta \left\vert 0_{A}1_{B}\right\rangle +\gamma \left\vert 1_{A}0_{B}\right\rangle +\delta \left\vert 1_{A}1_{B}\right\rangle,
\end{equation}
can be written in matrix form as
\begin{equation}\label{eq3_19}
\left\vert \psi _{AB}\right\rangle =\left( \begin{array}{c} \alpha \\ \beta \\ \gamma \\ \delta \end{array}\right)
\end{equation} 
and its norm can be computed by finding $\left\vert \left\vert \psi _{AB}\right\rangle \right\vert =\sqrt{\left\langle \psi _{AB}\right. \left\vert \psi _{AB}\right\rangle }$ or
\begin{equation}\label{eq3_20}
\displaystyle \left\vert \left\vert \psi _{AB}\right\rangle \right\vert    \displaystyle =    \displaystyle \sqrt{\left( \begin{array}{cccc} \alpha ^{\ast } & \beta ^{\ast } & \gamma ^{\ast } & \delta ^{\ast }\end{array}\right) \left( \begin{array}{c} \alpha \\ \beta \\ \gamma \\ \delta \end{array}\right) }          
\displaystyle =    \displaystyle \sqrt{\left\vert \alpha \right\vert ^{2}+\left\vert \beta \right\vert ^{2}+\left\vert \gamma \right\vert ^{2}+\left\vert \delta \right\vert ^{2}},
\end{equation}          
where $\alpha ^{\ast },\beta ^{\ast }, \gamma ^{\ast }, \delta ^{\ast }$ are the complex conjugate of $\alpha, \beta, \gamma, \delta$.

The projector of the two-qubit state $\left\vert \psi _{AB}\right\rangle$ can be written as
\begin{equation}\label{eq3_21}
\displaystyle \left\vert \psi _{AB}\right\rangle \left\langle \psi _{AB}\right\vert    \displaystyle =    \displaystyle \left( \begin{array}{c} \alpha \\ \beta \\ \gamma \\ \delta \end{array}\right) \left( \begin{array}{cccc} \alpha ^{\ast } & \beta ^{\ast } & \gamma ^{\ast } & \delta ^{\ast }\end{array}\right) ,          
\displaystyle =    \displaystyle \left( \begin{array}{cccc} \left\vert \alpha \right\vert ^{2} & \alpha \beta ^{\ast } & \alpha \gamma ^{\ast } & \alpha \delta ^{\ast } \\ \beta \alpha ^{\ast } & \left\vert \beta \right\vert ^{2} & \beta \gamma ^{\ast } & \beta \delta ^{\ast } \\ \gamma \alpha ^{\ast } & \gamma \beta ^{\ast } & \left\vert \gamma \right\vert ^{2} & \gamma \delta ^{\ast } \\ \delta \alpha ^{\ast } & \delta \beta ^{\ast } & \delta \gamma ^{\ast } & \left\vert \delta \right\vert ^{2}\end{array}\right) .    
\end{equation}      

Further tensor product examples\\

The tensor product between an operator A and the identity I is
\begin{equation}\label{eq3_22}
\displaystyle A\otimes I    \displaystyle =    \displaystyle \left( \begin{array}{cc} a_{A} & c_{A} \\ b_{A} & d_{A}\end{array}\right) \otimes \text { }\left( \begin{array}{cc} 1 & 0 \\ 0 & 1\end{array}\right)           
\displaystyle =    \displaystyle \left( \begin{array}{cc} a_{A}\left( \begin{array}{cc} 1 & 0 \\ 0 & 1\end{array}\right) & c_{A}\left( \begin{array}{cc} 1 & 0 \\ 0 & 1\end{array}\right) \\ b_{A}\left( \begin{array}{cc} 1 & 0 \\ 0 & 1\end{array}\right) & d_{A}\left( \begin{array}{cc} 1 & 0 \\ 0 & 1\end{array}\right) \end{array}\right)
\end{equation}
    
\begin{equation}\label{eq3_23}
\displaystyle =    \displaystyle \left( \text { }\begin{array}{cccc} a_{A} & 0 & c_{A} & 0 \\ 0 & a_{A} & 0 & c_{A} \\ b_{A} & 0 & d_{A} & 0 \\ 0 & b_{A} & 0 & d_{A}\end{array}\right)
\end{equation}    
      
and

\begin{equation}\label{eq3_24}
\displaystyle I\otimes A    \displaystyle =    \displaystyle \left( \begin{array}{cc} 1 & 0 \\ 0 & 1\end{array}\right) \otimes \left( \begin{array}{cc} a_{A} & c_{A} \\ b_{A} & d_{A}\end{array}\right) \text { }          
\displaystyle =    \displaystyle \left( \begin{array}{cc} 1\left( \begin{array}{cc} a_{A} & c_{A} \\ b_{A} & d_{A}\end{array}\right) & 0\left( \begin{array}{cc} a_{A} & c_{A} \\ b_{A} & d_{A}\end{array}\right) \\ 0\left( \begin{array}{cc} a_{A} & c_{A} \\ b_{A} & d_{A}\end{array}\right) & 1\left( \begin{array}{cc} a_{A} & c_{A} \\ b_{A} & d_{A}\end{array}\right) \end{array}\right) 
\end{equation}
\begin{equation}\label{eq3_25}          
\displaystyle =    \displaystyle \left( \begin{array}{cccc} a_{A} & c_{A} & 0 & 0 \\ b_{A} & d_{A} & 0 & 0 \\ 0 & 0 & a_{A} & c_{A} \\ 0 & 0 & b_{A} & d_{A}\end{array}\right)    
\end{equation}
      
Given two 2 by 2 matrices, A and B,
\begin{equation}\label{eq3_26}
A=\left( \begin{array}{cc} a_{A} & c_{A} \\ b_{A} & d_{A}\end{array}\right) \text {, }B=\left( \begin{array}{cc} a_{B} & c_{B} \\ b_{B} & d_{B}\end{array}\right) \text {,}
\end{equation}

The tensor product between them is given by

\begin{equation}\label{eq3_27}
\displaystyle A\otimes B    \displaystyle =    \displaystyle \left( \begin{array}{cc} a_{A} & c_{A} \\ b_{A} & d_{A}\end{array}\right) \otimes \left( \begin{array}{cc} a_{B} & c_{B} \\ b_{B} & d_{B}\end{array}\right)
\end{equation}
\begin{equation}\label{eq3_28}          
\displaystyle =    \displaystyle \left( \begin{array}{cc} a_{A}\left( \begin{array}{cc} a_{B} & c_{B} \\ b_{B} & d_{B}\end{array}\right) & c_{A}\left( \begin{array}{cc} a_{B} & c_{B} \\ b_{B} & d_{B}\end{array}\right) \\ b_{A}\left( \begin{array}{cc} a_{B} & c_{B} \\ b_{B} & d_{B}\end{array}\right) & d_{A}\left( \begin{array}{cc} a_{B} & c_{B} \\ b_{B} & d_{B}\end{array}\right) \end{array}\right) 
\end{equation}
\begin{equation}\label{eq3_29}          
\displaystyle =    \displaystyle \left( \begin{array}{cccc} a_{A}a_{B} & a_{A}c_{B} & c_{A}a_{B} & c_{A}c_{B} \\ a_{A}b_{B} & a_{A}d_{B} & c_{A}b_{B} & c_{A}d_{B} \\ b_{A}a_{B} & b_{A}c_{B} & d_{A}a_{B} & d_{A}c_{B} \\ b_{A}b_{B} & b_{A}d_{B} & d_{A}b_{B} & d_{A}d_{B}\end{array}\right) 
\end{equation}          

We have described quantum mechanics using four postulates. However, for quantum error correction\cite{jones_layered_2012,steane_error_1996}, we need more. We need classical statistics. Consider this question with probability 1/2, and we were given either psi or phi. How do we describe this state? It is not a single quantum state, but rather a statistical combination of two possible states with equal probability. We may recall that in quantum mechanics, this is described by a density matrix\cite{de_chiara_density_2008}. Here, it is the sum of the two outer product vectors of the quantum state. we call this $ \rho $. To further appreciate the need for density matrices in describing stochastic combinations of quantum states, consider the following scenario, we have a two-part state psi AB. This state may draw as this quantum circuit, which originates from a single place. The two-qubit state will be this particular position and have parts a and b. A will be the first label, and B will be the second label in each of the kets. Is the question supposing B measures B's part of the state? What describes A's part of this bipartite quantum state? This state will be a statistical mixture, because what A obtains differs, depending on what B's measurement result. If B measures a 0, then A has a 0. If B measures a 1, A has a 1. each one of these happens with a different amplitude. So, we get one or the other of these two states with the two different amplitudes. Let us call this statistical mixture, Mixture 1. now, imagine another separate scenario. In this scenario, B measures but on a different basis, rather than in this computational 0 and 1 basis, as shown above. We have the same bipartite state as an input to this quantum circuit. B before a measurement, let us say, performs a Hadamard transform. The question again is, what is A's state? A will see something that we may interpret as being different, depending on what B's measurement result is. The state before B's measurement is written here. It is a superposition where the second ket has had a Hadamard operation applied to it. Therefore, we can read off from this what A state is going to be. A will see when B measures a 0, a superposition of root 3 or 4, 0, and route 1 over 4, 1. If B measures a 1 instead, A state will be different with a minus sign instead of a plus sign. Now, this is another statistical mixture. The mixture is of these two states, one with a plus, and one with a minus. They are superpositions in contrast to Mixture 1. Let us call this Mixture 2. certainly, and it looks very different from Mixture 1. The question is, how are these two statistical mixtures different? Or might they be the same mixture in some sense? To answer this question, we will need to turn to the definition of a density matrix. For these statistical mixtures, we have written here using this circle-plus as an or sign. The density matrix is defined as the sum over the outer product of each of the elements in this statistical mixture as given here. We call $ \rho $ the density matrix. Let us explore some sample density matrices. Recall that the state 0 is a two-component vector, 1, 0, and the state 1 is 0, 1. The density matrices corresponding to these are given by the outer products. So, 0 is 1, 0, 0, 0. And 1 is 0, 0, 0, 1. The cross term, 0, 1, has the 1 in the upper right-hand corner.
Furthermore, this is easy to understand by labeling the rows and columns appropriately with 0 on the left and the top. We may also explicitly multiply the column and row vector to obtain this matrix form. Using these matrices, we now write off the density matrices, which correspond to the mixtures that we have studied. The first statistical mixture let us call it row 1, turns out to be the density matrix given by the sum 3/4 of 1 and 1/4 of 1 gives we 1/4 times 3001. The second statistical mixture is where we need the cross-terms because the states are superpositions. These cross-terms give rise to off-diagonal elements in the density matrix, summing as before, we find, though, something very interesting. The result is the same density matrix as for Mixture 1. we conclude from this that the two are the same state. Note: The section above currently contains an error in the treatment of mixture 2. Firstly, the normalization factor $\frac{1}{\sqrt{2}}$ is missing and, secondly, the final calculation of the density matrix should read
\begin{equation}\label{eq3_30}
\rho_2 = \frac{1}{8} \begin{bmatrix} 3 & \sqrt{3}\\ \sqrt{3} & 1 \end{bmatrix} + \frac{1}{8} \begin{bmatrix} 3 & -\sqrt{3}\\ -\sqrt{3} & 1 \end{bmatrix} = \frac{1}{8} \begin{bmatrix} 6 & 0\\ 0 & 2 \end{bmatrix} = \frac{1}{4} \begin{bmatrix} 3 & 0\\ 0 & 1\end{bmatrix}.
\end{equation}

\section{Density Matrices Properties} 
More about this can be understood by returning to the formalism of density matrices. Let us begin by defining a density matrix \cite{song_10-qubit_2017}. A density matrix is a matrix $ \rho $, which has two particular properties. First, the trace of $ \rho $ must equal 1. Second, the density matrix $ \rho $ must be positive. In other words, for all possible states that one might choose, the expectation value of $ \rho $ must be real and greater than or equal to 0. How do we construct a density matrix from states? First, let us make this claim, that a density matrix may be constructed from a probabilistic combination of pure states. P here is a probability. Psi is stated, which for now let us take to be orthogonal. We may prove that this statistical combination is a density matrix by checking the two constraints. First, we know that its trace is equal to 1 since p's are probabilities.
Moreover, second, it is clear that the matrix is positive. So, we can construct density matrices this way. In the reverse direction, given a density matrix, how can we interpret it? We claim that any density matrix $ \rho $ can be expressed as a stochastic combination of pure states. So, that is exactly in the form we have just seen using a stochastic combination of pure states. Let us prove this claim as follows. A density matrix is positive, and thus it has a spectral decomposition into eigenvalues and eigenvectors.
Furthermore, this spectral decomposition gives us exactly what we want. We may see this straightforwardly by writing out the spectral decomposition with $ \lambda $ sub k being the eigenvalue for state k. Note that the trace is 1, and thus the eigenvalues must sum to 1, meaning they are a probability distribution. They are also real-valued. $ \rho $ is positive by construction. This decomposition, known as an unraveling, allows us to consider some useful definitions. $ \rho $ is known as being pure when its decomposition results in a single pure state alone. Otherwise, $ \rho $ is known as being a mixed state, meaning that it is a stochastic combination of multiple pure states. Given these two constructions, let us consider a quick question. How about a stochastic combination of density matrices? Given p is as probabilities and $ \rho $s as density matrices, is this combination a legal density matrix? It is a simple thing to work out. The answer is, yes. So, try it ourselves and see if we can prove this to be true. The concept of unraveling a density matrix into a stochastic combination of pure states was considered by Von Neumann and is very interesting. It is an important lemma about unravelings. Let $ \rho $ be unraveled as a stochastic combination with probabilities p and states psi. It turns out that this unraveling is not unique in general for mixed states. For a mixed state, the density matrix may also be unraveled with probabilities p and states $ \phi $ under the following condition. That is, that the square root in the sense of this decomposition. The square root of p and state psi is related by a unitary transformation to the other decomposition with square root q and states $ \phi $. The physical origin of this infinity of possibilities of unravelings may be interpreted as arising from the fact that it does not matter which basis b measures in the original scenario. We started the discussion today. Let us now turn to another definition. We will say that purification of a density matrix, $ \rho $s of a, is a state size of a b, a pure state, such that $ \rho $s of a is obtained by taking the partial trace over a b of the density matrix given by the pure state. This partial trace operation removes the degrees of freedom associated with b and can be viewed in terms of the quantum circuit that we started today. Namely, that we have a bipartite system, and the b part of the system is measured, leaving $ \rho $ of a as the final state. This concept of having infinite unravelings and the interpretation of density matrices in purified forms will be key ideas in understanding the intuition behind quantum error correction\cite{edmunds_measuring_2017}. 

A known and well-defined quantum state that can be written as a single state $\lvert \phi \rangle $ is referred to as a pure state. In contrast, the system states that it can only be expressed as statistical mixtures of several pure states called mixed states. A general approach that accommodates both pure and mixed states uses density operators and matrix form density matrices $\rho$.

A pure state $\lvert \phi \rangle$ is mathematically represented using a single state vector, for example\cite{rue_mathematical_nodate},
\begin{equation}\label{eq3_31}
\lvert \phi \rangle =\lvert 0 \rangle =\begin{pmatrix}1\\0 \end{pmatrix}
\end{equation}

Note, a coherent superposition of pure states, such as $\alpha \lvert 0 \rangle +\beta \lvert 1 \rangle,$ is also a pure state.

The density matrix representation of a pure state is the outer product of the pure state with itself, for example:
\begin{equation}\label{eq3_32}
\rho=\lvert \phi \rangle \langle \phi \lvert=\begin{pmatrix}1\\0\end{pmatrix}\begin{pmatrix}1&0\end{pmatrix}=\begin{pmatrix}1&0\\0&0\end{pmatrix}.
\end{equation}

By extension, a mixed state is a statistical mixture of pure states $\lvert \phi _ i \rangle \langle \phi _ i \lvert $ with individual weights or probabilities $0 \leq p_ i \leq 1,$ and it is expressed as sum over all possible states
\begin{equation}\label{eq3_33}
\rho =\sum _ i p_ i \lvert \phi _ i \rangle \langle \phi _ i \lvert .
\end{equation}

Note that some states in the sum may have $p_ i=0.$

A density matrix that describes pure and mixed states has the following properties:
\begin{equation}\label{eq3_34}
\displaystyle Tr(\rho ) = \sum _ i \rho _{ii}=1~~~~~~\text {normalization}
\end{equation}
\begin{equation}\label{eq3_35}                  
\displaystyle \langle \psi \vert \rho \vert \psi \rangle =\sum _ i p_ i \lvert \langle \psi \lvert \phi _ i\rangle \vert ^2 \geq 0~~~~~~\text {positivity}    
\end{equation}
              
First, a density matrix is normalized to unity. In general, the trace of a matrix equals the sum of its eigenvalues. In the case of a density matrix, the basis of the density matrix spans a complete set of states, and the diagonal elements represent the probabilities of the system being in these various basis states. Therefore, the trace of the density matrix equals one.

Second, the expectation value that a system is found in state $\vert \psi \rangle$ is calculated using the density matrix, as shown above. The result is manifestly positive because the density matrix itself is expanded into a basis of projectors with manifestly positive weighting coefficients. This property is called positivity.

The density matrix $\rho_p$ of a pure state is a projection matrix a matrix that projects the system onto the basis state (or states) that comprise the pure state. For a pure state, a second projection leaves the result unchanged, and hence $Tr(\rho_p)=Tr(\rho _ p^2)=1$ for a pure state. In contrast, a mixed state represented by a density matrix $\rho_m $ is not a projection matrix, and consequently $Tr(\rho _ m^2)<1$. This follows, because the weightings $0<p_ i<1$ of the mixed state become smaller when squared, $p_ i^2<p_ i$. In the case of the pure state discussed before, only one particular $p_ i$ equals one, and so it remains one when squared.
\begin{equation}\label{eq3_36}
\displaystyle \text {pure state:}~ ~ ~ ~ Tr(\rho _ p^2)=Tr(\vert \phi \rangle \langle \phi \vert \phi \rangle \langle \phi \vert )=Tr(\vert \phi \rangle \langle \phi \vert )=Tr(\rho _ p)    \displaystyle =    \displaystyle {1}
\end{equation}
\begin{equation}\label{eq3_37}          
\displaystyle \text {mixed state:}~ ~ ~ Tr(\rho _ m^2)=Tr\Big(\sum _{i,j} p_ i p_ j \vert \phi _ i \rangle \langle \phi _ i \vert \phi _ j \rangle \langle \phi _ j \vert \Big)=\sum _ i p_ i^2    \displaystyle <    \displaystyle {1}     
\end{equation}
     
In conclusion, the trace of a density matrix is required to be unity, whether a pure state or a mixed state. However, the trace of the density matrix squared depends on the degree of purity, and it is thus used to assess the purity of an arbitrary system state.

A common visualization of density matrices is illustrated in the following figure for a pure state, a mixture of pure states, and Bell-state entanglement of pure states $\vert \Phi ^- \rangle = \frac{1}{\sqrt {2}}\left(\vert 0\rangle \vert 0 \rangle -\vert 1\rangle \vert 1 \rangle \right)$.

\begin{figure}[H] \centering{\includegraphics[scale=.25]{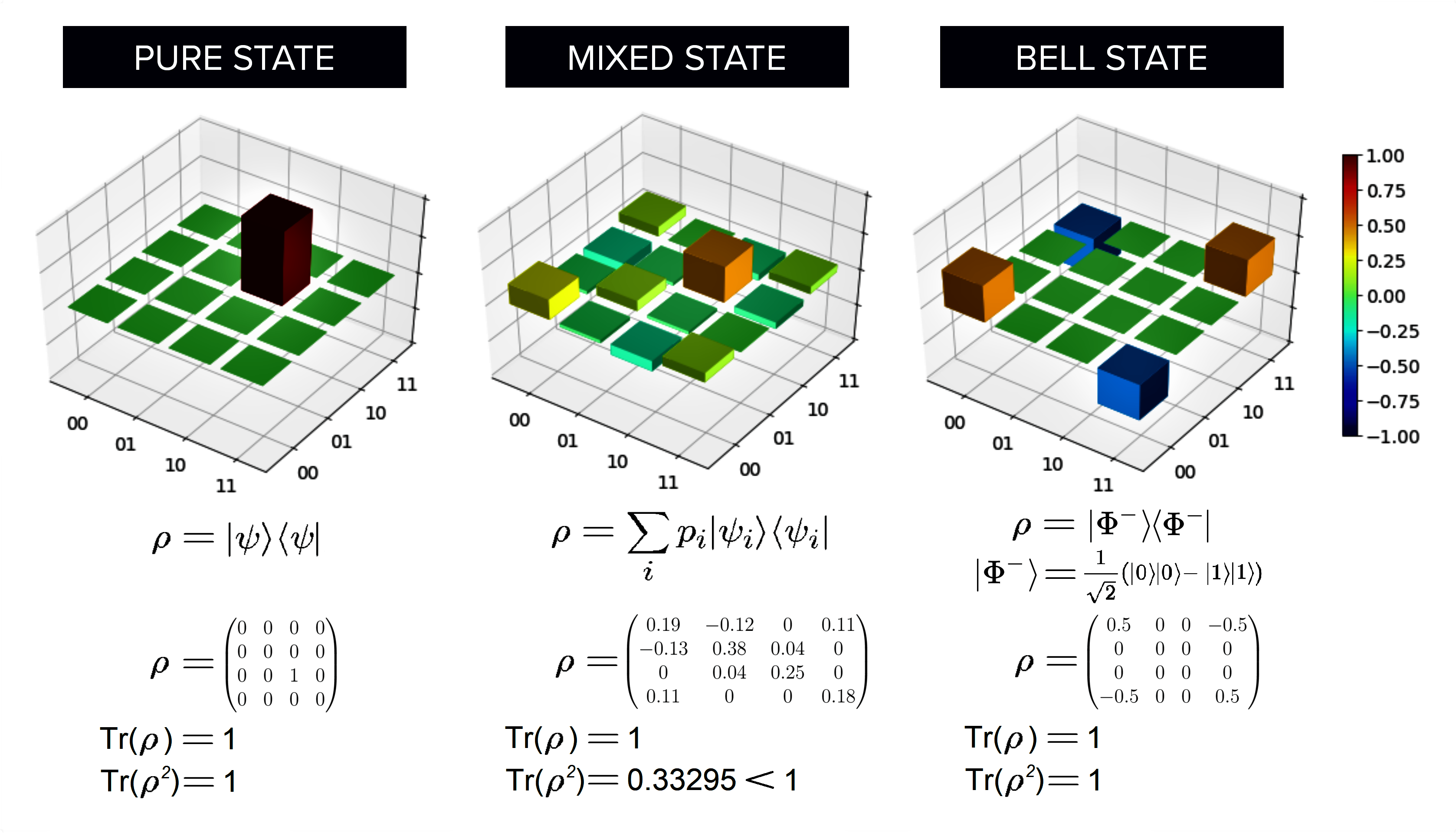}}\caption{DVS}\label{fig3_1}
\end{figure}

Unitary operations alter quantum states and, thus, they also modify density matrices\cite{franson_limitations_2018}. For example, a pure state $\vert \phi \rangle$ altered by a unitary operator U is transformed into $U\lvert \phi \rangle =\lvert \tilde {\phi }\rangle.$ The density matrix is modified according to:$ \tilde {\rho }=\vert \tilde {\phi }\rangle \langle \tilde {\phi }\vert =U\lvert \phi \rangle \langle \phi \vert U^\dagger =U\rho U^\dagger.$ This transformation generalizes for any density matrix: $ U$ alters $\rho$ according to $ U\rho U^\dagger.$

\begin{enumerate}[wide, labelwidth=!, labelindent=0pt]

\item A pure state $ \left\vert \psi \right\rangle $ can be written as a density matrix $ \rho =\left\vert \psi \right\rangle \left\langle \psi \right\vert $.
       
Solution:\\
    Any pure state, such as\\
        $ \left\vert \psi \right\rangle _{AB}=\alpha \left\vert 0\right\rangle _{A}\left\vert 0\right\rangle _{B}+\beta \left\vert 1\right\rangle _{A}\left\vert 1\right\rangle _{B}=\left( \begin{array}{c} \alpha \\ 0 \\ 0 \\ \beta \end{array}\right)  $,\\
    can be written as a density matrix $ \rho $,\\
    $ \displaystyle \rho    \displaystyle =    \displaystyle \left\vert \psi \right\rangle \left\langle \psi \right\vert ,\\          
    \displaystyle =    \displaystyle \left( \alpha \left\vert 0\right\rangle _{A}\left\vert 0\right\rangle _{B}+\beta \left\vert 1\right\rangle _{A}\left\vert 1\right\rangle _{B}\right) \left( \alpha ^{\ast }\left\langle 0\right\vert _{A}\left\langle 0\right\vert _{B}+\beta ^{\ast }\left\langle 1\right\vert _{A}\left\langle 1\right\vert _{B}\right) ,    \\      
    \displaystyle =    \displaystyle \left( \begin{array}{c} \alpha \\ 0 \\ 0 \\ \beta \end{array}\right) \left( \begin{array}{cccc} \alpha ^{\ast } & 0 & 0 & \beta ^{\ast }\end{array}\right) ,\\          
    \displaystyle =    \displaystyle \left( \begin{array}{cccc} \left\vert \alpha \right\vert ^{2} & 0 & 0 & \alpha \beta ^{\ast } \\ 0 & 0 & 0 & 0 \\ 0 & 0 & 0 & 0 \\ \beta \alpha ^{\ast } & 0 & 0 & \left\vert \beta \right\vert ^{2}\end{array}\right). $    \\ 
    The diagonal elements of a density matrix are the probabilities associated with each projector on to the elements' corresponding basis states. In the previous example, the first diagonal element, $ \left\vert \alpha \right\vert ^{2} $, is the probability associated with the projector $ \left\vert 0\right\rangle _{A}\left\vert 0\right\rangle _{B}\left\langle 0\right\vert _{A}\left\langle 0\right\vert _{B} $. Since the diagonal elements are measurement probabilitites across a complete basis set, they must sum to unity, $ \left\vert \alpha \right\vert ^{2}+\left\vert \beta \right\vert ^{2}=1 $. Density matrices not only can be constructed as probabilistic combinations of pure states, they can also be constructed as combinations of mixed states.

\section{Density Matrices}

Consider the bipartite two-qubit state discussed in the first section
\begin{equation}\label{eq3_38}
\left\vert \psi _{AB}\right\rangle =\sqrt{\frac{3}{4}}\left\vert 0_{A}\right\rangle \left\vert 0_{B}\right\rangle +\sqrt{\frac{1}{4}}\left\vert 1_{A}\right\rangle \left\vert 1_{B}\right\rangle =\left( \begin{array}{c} \sqrt{\frac{3}{4}} \\ 0 \\ 0 \\ \sqrt{\frac{1}{4}}\end{array}\right)
\end{equation}

where A and B are the labels assigned to each qubit. State $ \left\vert 0_{A}\right\rangle $ corresponds to the $ \left\vert 0\right\rangle $ state for qubit A, and$  \left\vert 0_{B}\right\rangle $ corresponds to the $ \left\vert 0\right\rangle $ state for qubit B. 

In the first part of this exercise, we will work through the steps to find the probability of projecting qubit B on to state $ \left\vert 0_{B}\right\rangle $, and then determine the resulting density matrix for qubit A following this specific projection. In the second part of this exercise, we will apply the same steps to find the probability of projecting qubit B on to state $ \left\vert 1_{B}\right\rangle $, and then find the resulting density matrix for qubit A. In the last part of this exercise, we will use the above results to write the density matrix for qubit A.\\
Section I\\

The conjugate transpose of the two-qubit state $ \left\vert \psi _{AB}\right\rangle $ is
\begin{equation}\label{eq3_39}
\left\langle \psi _{AB}\right\vert =\sqrt{\frac{3}{4}}\left\langle 0_{A}\right\vert \left\langle 0_{B}\right\vert +\sqrt{\frac{1}{4}}\left\langle 1_{A}\right\vert \left\langle 1_{B}\right\vert ,
\end{equation}

or, alternatively in matrix form,
\begin{equation}\label{eq3_40}
\left\langle \psi _{AB}\right\vert =\left( \begin{array}{cccc} \sqrt{\frac{3}{4}} & 0 & 0 & \sqrt{\frac{1}{4}}\end{array}\right)
\end{equation}

If qubit B is projected on to state $ \left\vert 0_{B}\right\rangle $, then the two-qubit state after the measurement becomes

\begin{equation}\label{eq3_41}
\begin{split}
\displaystyle\frac{ \left( I_{A}\otimes \Pi _{0}\right) \left\vert \psi _{AB}\right\rangle}{\sqrt{N}}    \displaystyle & =    \displaystyle \frac{\left( I_{A}\otimes \left\vert 0_{B}\right\rangle \left\langle 0_{B}\right\vert \right)}{\sqrt{N}} \left( \sqrt{\frac{3}{4}}\left\vert 0_{A}\right\rangle \left\vert 0_{B}\right\rangle +\sqrt{\frac{1}{4}}\left\vert 1_{A}\right\rangle \left\vert 1_{B}\right\rangle \right) \\          
\displaystyle = &    \displaystyle\frac{1}{\sqrt{N}}\sqrt{\frac{3}{4}}\left( I_{A}\otimes \left\vert 0_{B}\right\rangle \left\langle 0_{B}\right\vert \right) \left\vert 0_{A}\right\rangle \left\vert 0_{B}\right\rangle \\ & +\frac{1}{\sqrt{N}}\sqrt{\frac{1}{4}}\left( I_{A}\otimes \left\vert 0_{B}\right\rangle \left\langle 0_{B}\right\vert \right) \left\vert 1_{A}\right\rangle \left\vert 1_{B}\right\rangle 
\end{split}
\end{equation}

where I is the identity operator, $ \Pi _{0} $ is the single-qubit projector  $ \left\vert 0\right\rangle \left\langle 0\right\vert $, and N is a normalization constant. We can simplify equation (\ref{eq3_41}) by replacing with the values of the inner products $ \left\langle 0_{B}\right. \left\vert 0_{B}\right\rangle =1, \left\langle 0_{B}\right. \left\vert 1_{B}\right\rangle =0 $, and noting that the identify operator will leave a state unchanged,
\begin{equation}\label{eq3_42}
\displaystyle\frac{ \left( I_{A}\otimes \Pi _{0}\right) \left\vert \psi _{AB}\right\rangle}{\sqrt{N}}    \displaystyle =    \displaystyle \frac{1}{\sqrt{N}}\sqrt{\frac{3}{4}}\left\vert 0_{A}\right\rangle \left\vert 0_{B}\right\rangle     \end{equation}
    
From equation (\ref{eq3_42}) we determine that the normalization constant is $ N=\frac{3}{4} $, such that the norm of $ \displaystyle \left( I_{A}\otimes \Pi _{0}\right) \left\vert \psi _{AB}\right\rangle /\sqrt{N}  $is equal to one,

\begin{equation}\label{eq3_43}
\left( \frac{1}{\sqrt{N}}\sqrt{\frac{3}{4}}\left\langle 0_{A}\right\vert \left\langle 0_{B}\right\vert \right) \left( \frac{1}{\sqrt{N}}\sqrt{\frac{3}{4}}\left\vert 0_{A}\right\rangle \left\vert 0_{B}\right\rangle \right) =1.
\end{equation}

The state of qubit A after the measurement is 
\begin{equation}\label{eq3_44}
\displaystyle \left\vert \psi _{A,0}\right\rangle    \displaystyle =    \displaystyle \sqrt{\frac{4}{3}}\sqrt{\frac{3}{4}}\left\vert 0_{A}\right\rangle      
\displaystyle =    \displaystyle \left\vert 0_{A}\right\rangle 
\end{equation}
     
The probability of projecting qubit B on to state $ \left\vert 0_{B}\right\rangle $ is given by

\begin{equation}\label{eq3_45}
\begin{split}
\displaystyle p_{B,0}    \displaystyle & =    \displaystyle \left\langle \psi _{AB}\right\vert \left( I_{A}\otimes \Pi _{0}\right) \left\vert \psi _{AB}\right\rangle \\          
\displaystyle & =    \displaystyle \left\langle \psi _{AB}\right\vert \left( I_{A}\otimes \left\vert 0_{B}\right\rangle \left\langle 0_{B}\right\vert \right) \left\vert \psi _{AB}\right\rangle \\          
\displaystyle & =    \displaystyle \left\langle \psi _{AB}\right\vert \sqrt{\frac{3}{4}}\left\vert 0_{A}\right\rangle \left\vert 0_{B}\right\rangle \\     
\displaystyle & =    \displaystyle \sqrt{\frac{3}{4}}\left[ \sqrt{\frac{3}{4}}\left\langle 0_{A}\right\vert \left\langle 0_{B}\right\vert +\sqrt{\frac{1}{4}}\left\langle 1_{A}\right\vert \left\langle 1_{B}\right\vert \right] \left\vert 0_{A}\right\rangle \left\vert 0_{B}\right\rangle \\          
\displaystyle & =    \displaystyle \sqrt{\frac{3}{4}}\left[ \sqrt{\frac{3}{4}}\left\langle 0_{A}\right. \left\vert 0_{A}\right\rangle \left\langle 0_{B}\right. \left\vert 0_{B}\right\rangle +\sqrt{\frac{1}{4}}\left\langle 1_{A}\right. \left\vert 0_{A}\right\rangle \left\langle 1_{B}\right. \left\vert 0_{B}\right\rangle \right] \\          
\displaystyle & =    \displaystyle \frac{3}{4}
\end{split}     
\end{equation}

 \item Writing a Density Matrix I; if qubit B is projected to state $ \left\vert 1_{B}\right\rangle $, what is the state of qubit A after the measurement?
    \begin{itemize}
         \item $ \left\vert \psi _{A,1}\right\rangle  $ = $ 0\left\vert 0_{A}\right\rangle$ + $ 1\left\vert 1_{A}\right\rangle  $ 
        \item What is the probability of projecting qubit B to state $ \left\vert 1_{B}\right\rangle $?
        \item $ p_{B,1}= 0.25 $    
        \end{itemize}
    Solution:\\
    If qubit B is projected to state $ \left\vert 1_{B}\right\rangle $, then the two-qubit state in Dirac notation becomes
    \begin{center}
    $ \displaystyle \frac{ \left( I_{A}\otimes \Pi _{1}\right) \left\vert \psi _{AB}\right\rangle}{\sqrt{N}}    \displaystyle =    \displaystyle \frac{1}{\sqrt{N}}\left(\left( I_{A}\otimes \left\vert 1_{B}\right\rangle \left\langle 1_{B}\right\vert \right) \left( \sqrt{\frac{3}{4}}\left\vert 0_{A}\right\rangle \left\vert 0_{B}\right\rangle +\sqrt{\frac{1}{4}}\left\vert 1_{A}\right\rangle \left\vert 1_{B}\right\rangle \right) \right),          
    \displaystyle =    \displaystyle\frac{1}{\sqrt{N}} \sqrt{\frac{1}{4}}\left\vert 1_{A}\right\rangle \left\vert 1_{B}\right\rangle  $,
   \end{center}          
    From here, we notice that the normalization constant is equal to $ N=1/4 $, such that the state of qubit A after the measurement is
    
    $ \displaystyle \left\vert \psi _{A,1}\right\rangle    \displaystyle =    \displaystyle \sqrt{\frac{4}{1}}\sqrt{\frac{1}{4}}\left\vert 1_{A}\right\rangle,          
    \displaystyle =    \displaystyle \left\vert 1_{A}\right\rangle  $.

    The probability of projecting qubit B to state $ \left\vert 1_{B}\right\rangle $ is given by
    
    $ \displaystyle p_{B,1}    \displaystyle =    \displaystyle \left\langle \psi _{AB}\right\vert \left( I_{A}\otimes \Pi _{1}\right) \left\vert \psi _{AB}\right\rangle ,          
    \displaystyle =    \displaystyle \left\langle \psi _{AB}\right\vert \sqrt{\frac{1}{4}}\left\vert 1_{A}\right\rangle \left\vert 1_{B}\right\rangle ,          
    \displaystyle =    \displaystyle \frac{1}{4} $.
    
    \item  Writing a Density Matrix II; If there is no information regarding on to which state qubit B was projected, then the state of qubit A has to be described by a density matrix $ \rho _{A} $ given by: the probability of projecting qubit B on to $ \left\vert 0_{B}\right\rangle $ times the projector for the qubit-A state after the measurement, $ \left\vert \psi _{A,0}\right\rangle \langle \psi _{A,0}\rvert $; plus the probability of projecting qubit B on to$  \left\vert 1_{B}\right\rangle $ times the projector for the qubit-A state after the measurement, $ \left\vert \psi _{A,1}\right\rangle \langle \psi _{A,1}\rvert $.
    \begin{itemize}
        \item Write the density matrix $ \rho _ A $ using Dirac notation
        
        $ \rho _ A $    =    $ 0.75 \left\vert \psi _{A,0}\right\rangle \left\langle \psi _{A,0}\right\vert $    +    $ 0.25    \left\vert \psi _{A,1}\right\rangle \left\langle \psi _{A,1}\right\vert $
        
        \item Write the density matrix $ \rho _ A $ using matrices
        
        $ \rho _ A $    =    $ 0.75    \left(\begin{array}{cc} 1 & 0 \\ 0 & 0\end{array}\right) $    + $ 0.25    \left(\begin{array}{cc} 0 & 0 \\ 0 & 1\end{array}\right) $
    \end{itemize}

\section{Dephasing Channel on the Bloch Sphere}
In order to be useful machines, quantum computers like classical computers must be built from robust elements. Gate fidelity is one metric that quantifies the robustness of a qubit. Intuitively, it is related to the number of gates one can apply to the qubit, on average, before an error occurs, and the qubit state is lost. As we in section one, there are two fundamental ways in which a qubit loses quantum information. The first is energy exchange with the environment related to the coherence time, t1. The second is dephasing, a loss of coherence in a superposition state. The coherence time t2 is then related to both the dephasing and energy exchange processes. In this section, we will look at a method to treat qubit errors on the Bloch sphere using probabilistic error channels based on qubit gates. While this simple model is not entirely general, it does provide a simple means to visualize concepts like depolarization and dephasing as represented by a density matrix. Thus, with that introduction, let us discuss about error channels, depolarization, dephasing, and the Bloch sphere. If we want to discuss quantum channels in the most generality \cite{shor_capacities_2003}, there are completely positive trace-preserving maps, and they are complicated, even to write down the definition of. Well, maybe not to write down the definition, but intuitively, they are quite difficult to understand. Today, we will tell we the simplest kind of quantum channel\cite{gyongyosi_survey_2018}, which is we mean, we can do everything we want to in this section with that kind of channel. So, what is a quantum channel? So, what we are going to do is we are going to expressive a quantum channel as a mixture of simple channels. So, that is not always true. Let us do the easiest one first, what we will call the dephasing channel. So, let us say with probability one minus p, we do nothing. Probability, p, apply $ \sigma $ z. So, let us put a quantum state into the depolarizing channel. So, suppose we put in $ \alpha $ zero plus $ \beta $ one. we get out a mixture. we get out one minus p, probability, $ \alpha $ zero plus $ \beta $ one, and with p probability, we get out $ \alpha $ zero minus $ \beta $ one. from the last section, we know this can be expressed as a density matrix. So, the density matrix is one minus p, $ \alpha $ zero plus $ \beta $ one, $ \alpha $ zero plus $ \beta $ one, plus p. Same thing with minus signs. That is one minus p, $ \alpha $ squared. Let us assume $ \alpha $ and $ \beta $ are real, because $ \alpha $ $ \beta $, $ \alpha $ $ \beta $, $ \beta $ squared because otherwise, we get a whole bunch of complex conjugates, which we do not want to mess up. Plus, now down here, we get $ \alpha $ zero minus $ \beta $ one. That trick gives we $ \alpha $ squared, minus $ \alpha $ $ \beta $, minus $ \alpha $ $ \beta $, $ \beta $ squared is equal to well, one minus p $ \alpha $ squared plus $ \alpha $ squared. P $ \alpha $ squared is just $ \alpha $ squared. the same thing for $ \beta $ squared. $ \alpha $ squared, one minus 2p, $ \alpha $ $ \beta $, one minus 2p $ \alpha $ $ \beta $, $ \beta $ squared. the density matrix we put in was $ \alpha $ squared, $ \alpha $ $ \beta $, $ \alpha $ $ \beta $, $ \beta $ squared. So, the depolarizing channel we can see from this example.
Moreover, this example is, well, this is as general as possible, really, except we also should check that it works with $ \alpha $ and $ \beta $ complex. So, what the depolarizing channel does is it multiplies the off-diagonal elements of the density matrix. Sorry, the dephasing channel. The dephasing channel multiplies the off-diagonal elements of the density matrix by some constant. Next, we want to tell us what happens to the dephasing channel when we look at it in the Bloch sphere. Let us go back and draw the Bloch sphere. So, the Bloch sphere looked like zero one, plus-minus. When we multiply the off-diagonal elements by q, we get one, one, one, one. That is one half goes to one half, one, one minus q, one minus q, one. Are we are just moving the Bloch sphere into the middle, because we are taking this. We are multiplying it by one minus q, and we are adding q amount of diagonal, or q amount of the identity. So, that goes here.
Furthermore, in general, this is going to go to an ellipsoid. It is still zero, this is still one, and this is one, one minus q, one minus q, one. So, we shrink the Bloch sphere into an ellipsoid with the vertical axis stays the same, and everything else shrinks by the same factor.

\section{Depolarizing Channel on the Bloch Sphere} 
There is another channel we would like to discuss the depolarizing channel. $ \rho $ goes to 1 minus p $ \rho $ plus p/3 $ \sigma $ x $ \rho $ $ \sigma $ x plus 3/p $ \sigma $ y $ \rho $ $ \sigma $ y plus p/3 $ \sigma $ z $ \rho $ $ \sigma $ z. So, with 1 minus p, we do nothing. Probability of the p/3 we will apply each of the three Pauli matrices. So, is there an easier formula for this? Yes, $ \rho $ goes to, we guess, 1 minus 4/3 p $ \rho $ plus identity. the reason for this is for any density matrix, $ \rho $ plus $ \sigma $ x $ \rho $ $ \sigma $ x plus $ \sigma $ y $ \rho $ $ \sigma $ y plus $ \sigma $ z $ \rho $ $ \sigma $ z equals. So, this is completely unnormalized. So, we want this to be we/2. we want this to be 1/4. probability 1/4, we apply $ \sigma $ x. 1/4, we apply $ \sigma $ y. 1/4 we apply $ \sigma $ z. we get we/2. Proof that is tau. $ \sigma $ x tau $ \sigma $ x equals $ \sigma $ y tau. $ \sigma $ y equals $ \sigma $ z tau. $ \sigma $ z equals tau. Why is that? If we multiply this by $ \sigma $ y in front and $ \sigma $ y behind, this gets taken to this $ \sigma $ y. This gets taken to $ \rho $. $ \sigma $ x $ \sigma $ $ \rho $ $ \sigma $ x gets taken to $ \sigma $ z, $ \rho $ $ \sigma $ z. this gets taken to this. So, it is invariant. So, this is true of tau. Now the only density matrix tau, which has this property, is we/2. Let us say tau equals a b c d. $ \sigma $ x tau $ \sigma $ x equals b d a c. So, a has to be equal to b. $ \sigma $ z tau $ \sigma $ z equals a minus z minus d b. So, c has to be 0, and d has to be 0. So, that means tau has to be the identity. Alternatively, the identity over 2, because we know it is the trace, is 1 because it is the density matrix. So, what does the depolarizing channel look like in the Bloch picture? Furthermore, everything moves to the center at the same rate. So, it goes to that. We probably should draw the axes, so; we know it is not misplaced. So, these are the two channels we will be discussing. There are many, many other ways to get the depolarizing channel. We should also say that the binary symmetric channel corresponds to the depolarizing channel because the depolarizing channel, there is no preferred basis. The binary symmetric channel, remember, took the 0 to 1 with probability p and vice versa. There is no preferred input. So, we take a density matrix $ \rho $, and with probability 1/4, we do nothing. We apply the identity. Probability 1/4 we apply $ \sigma $ x. Probability 1/4 we apply $ \sigma $ y. Probability 1/4 we apply $ \sigma $ z. it is now easy to see that $ \sigma $ x tau $ \sigma $ x is equal to tau because what happens when we apply $ \sigma $ x to tau? Well, $ \rho $ goes to $ \sigma $ x $ \rho $ $ \sigma $ x. And $ \sigma $ x $ \rho $ $ \sigma $ x goes to $ \rho $. Since $ \sigma $ x $ \sigma $ y is equal to, we $ \sigma $ z, $ \sigma $ y $ \rho $ $ \sigma $ y goes to $ \sigma $ z $ \rho $ $ \sigma $ z and vice versa. we should put a conjugate here because $ \sigma $ y is not self-adjoint, is not self-adjoint. So, we really should put all of these conjugates on all of these and conjugate here and a conjugate here. 

The robustness of quantum systems is limited by noise. The effect of noise on a system of qubits can be described within the density matrix formalism by using a mapping $ \rho ~ \rightarrow ~ \mathcal{E}(\rho ) $. The different noise mechanisms and the means by which they affect the system are called noise channels. In this text unit, we will look at three simple models for noise channels: depolarization, amplitude damping, and dephasing.

Consider a quantum system that is initialized in a well-defined pure state. Over time, environment noise ``mixes'' the quantum system with a large ensemble of unknown quantum systems in the environment. The initialized quantum system loses its ``purity,'' and this results in a probabilistic manifestation of errors.

In the case of the depolarizing channel, we assume that a quantum system interacting with its environment will have an error with probability p. We model these errors through the use of quantum gates, an X-gate, a Y-gate, and a Z-gate, and we assume that each is equally likely with probability p/3. In this sense, the depolarization channel models errors as bit flips and phase flip.

The mapping of a density matrix assuming a depolarizing channel is:
\begin{equation}\label{eq3_46}
\rho ~ \rightarrow ~ \mathcal{E}(\rho )=(1-p)\rho +\frac{p}{3}(X\rho X+Z\rho Z+Y\rho Y),
\end{equation}

Where the density matrix remains unaffected with probability (1-p), it has an error with probability p. It is modelled as an X-gate, Y-gate, and Z-gate rotation with probability p/3 each. The net effect of many such errors is to stochastically depolarize the qubit state to the center of the Bloch sphere, and thus the origin of the name.
\begin{equation}\label{eq3_47}
\mathcal{E}(\rho (\vec{q}))=\frac{1}{2}\left(I+\left(1-\frac{4}{3}p\right)\left(q_1X+q_2Z+q_3Y\right)\right).
\end{equation}

A second noise model is amplitude damping, which is a model of energy loss from the quantum system to the environment. Amplitude damping channel maps a density matrix $ \rho $ according to,
\begin{equation}\label{eq3_48}
\rho ~ \rightarrow ~ \mathcal{E}(\rho )=\begin{pmatrix}1& 0\\ 0& \sqrt {1-p}\end{pmatrix}\rho \begin{pmatrix}1& 0\\ 0& \sqrt {1-p}\end{pmatrix}+\begin{pmatrix} 0& \sqrt {p}\\ 0& 0\end{pmatrix}\rho \begin{pmatrix}0& 0\\ \sqrt {p}& 0\end{pmatrix}.
\end{equation}

Where an amplitude damping event occurs with probability p. If p=0, the density matrix remains unchanged. With amplitude damping, a qubit represented on the Bloch sphere contracts to a scaled spheroid with a center that moves toward the ground state (north pole).

Lastly, the dephasing channel acts to dephase the qubit through the application of the phase gate Z according to the mapping:
\begin{equation}\label{eq3_49}
\rho ~ \rightarrow ~ \mathcal{E}(\rho )=(1-p)\rho +p(Z\rho Z)=\begin{pmatrix} \rho _{11}& (1-2p)\rho _{12}\\ (1-2p)\rho _{21} & \rho _{22} \end{pmatrix},
\end{equation}

The probability of an error is p, and the density matrix remains unchanged with probability (1-p). The phase damping channel transforms the Bloch sphere to a prolate spheroid along the z-axis with scaled x and y axes.

\begin{figure}[H] \centering{\includegraphics[scale=.17]{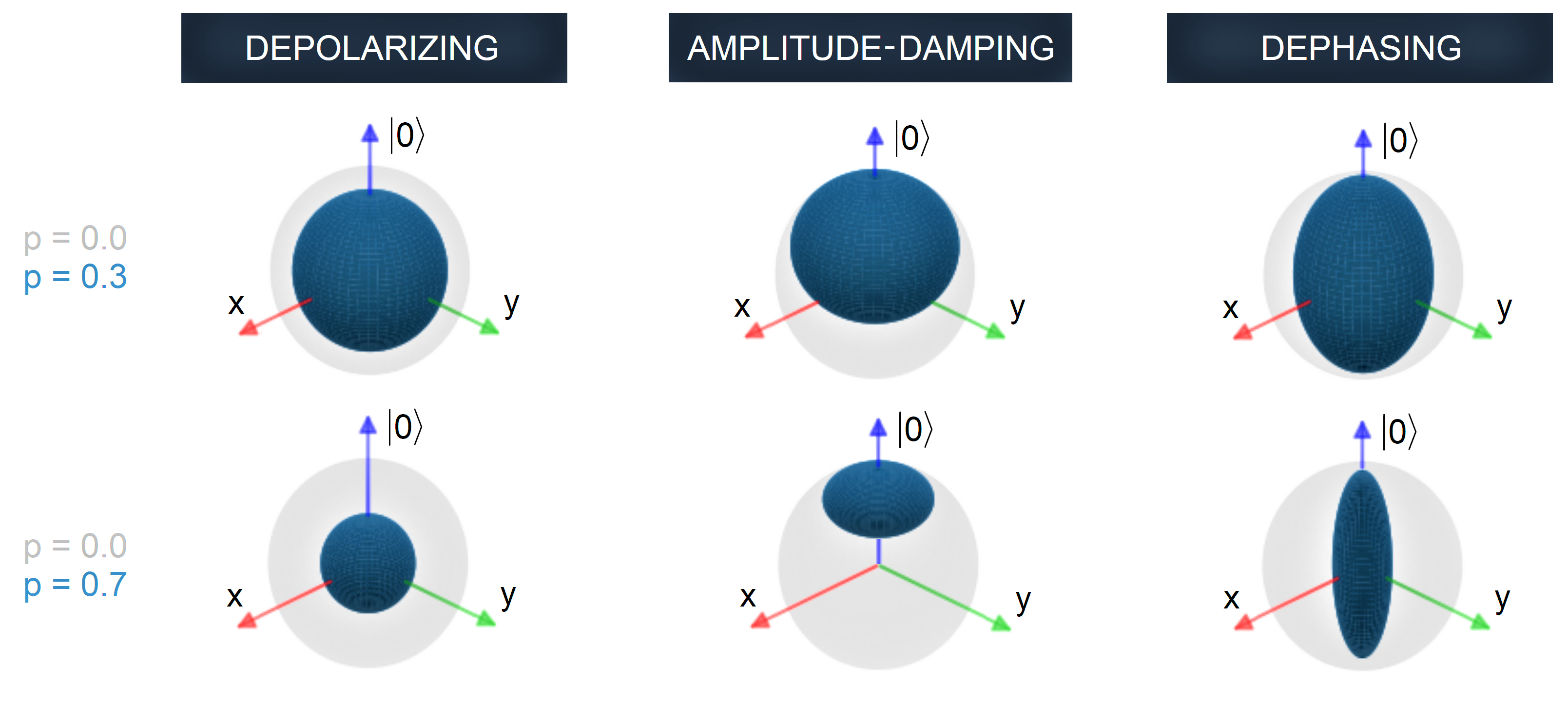}}\caption{shrinking of the Bloch sphere}\label{fig3_2}
\end{figure}

The loss in purity represented by the shrinking of the Bloch sphere is detrimental to any computation. In order to reliably trace states and exploit quantum interference with maximal efficiency, pure states are needed, and these are only found on the surface of the sphere. Points inside the sphere correspond to mixed states and are indicative of a loss of quantum coherence.

\section{Noise Processes} 
We looked at how quantum computing algorithms and quantum communication protocols work in an ideal sense in section two. We assume perfect qubits on which to apply our algorithms or perfect control electronics and optics to implement these protocols \cite{bardin_28nm_2019}. In this section, we are now turning our attention to the practical realities of quantum technologies. one of the main challenges to building robust systems is the presence of noise. Noise has an impact on nearly every aspect of quantum information. For example, in quantum communication, it can be manifest as photon loss in a long optical fiber or a measurement error in a photodetector. quantum computing causes qubit decoherence or control errors when implementing a qubit gate, both of which reduce gate fidelity. Indeed, noise is ubiquitous. it is a problem because it causes errors. practically, these errors translate to reduced communication rates, limits on the length or circuit depth of a computation, an increase in the amount of overhead needed to combat those errors, and, in the worst case, even system failure. we will focus here on examples related to quantum computing. However, the concepts are relevant for quantum communication as well. Noise can be categorized into two primary types systematic noise and stochastic noise. Systematic noise arises from a process that ultimately can be traced back to a systematic error. For example, we want to apply an X-gate to a qubit. However, the control field is not tuned properly. So, rather than rotating the qubit 180 degrees, the pulse slightly over-rotates or under-rotates by a fixed amount. Now, in this case, the underlying error is always the same. it is, in this sense, systematic. However, the symptoms we observe in measuring the output of our qubit may look like noise. They are not always the same. It is because we do not always apply the pulse in the same way. It could be applied to a different number of times, or it could be interspersed with different pulses in different ways. Thus, in general, it differs from experiment to experiment. However, such systematic errors can generally be corrected once They are identified through proper calibration or improved hardware. The second type of noise is stochastic noise. this is the random fluctuation of a parameter that is coupled to our qubit. For example, the thermal noise of a 50-ohm resistor will have voltage, and current fluctuations called Johnson noise. They are proportional to temperature. Alternatively, the oscillator we using to create an X-pulse has amplitude fluctuations due to noise in the internal electronics \cite{goto_bifurcation-based_2016,goto_quantum_2019}. In principle, any randomly fluctuating parameter that couples to the qubit can cause decoherence. In the case of superconducting qubits \cite{motzoi_simple_2009,muller_interacting_2015}, this could be a fluctuating magnetic field through the superconducting qubit loop or trap magnetic field vortices that move around \cite{xu_coherent_2016}. It could also be fluctuating charges in the substrate or charges we call quasi-particles that tunnel across the Josephson junction \cite{abdo_active_2019,bergeal_analog_2010,abdo_josephson_2011,}. It can be the loss of energy due to phonon or photon emission. it could also be noise that comes in and out of our control line. there are many, many others. To see how noise affects qubit operations, let us first examine an example of qubit control in the absence of noise and compare those ideal results to the same experiment in the presence of noise. In this case, we apply a control field resonant with the qubit along the x-axis of the block sphere. we leave this control field on. The block vector then rotates around this field from north pole to south pole and back again and continues to do this. The block vector's projection along the z-axis is plotted on the right, and it oscillates up and down between the north and south poles. It is called a Rabi oscillation. the frequency of this oscillation is proportional to the amplitude of the control field. Now, we notice that this looks like an ideal cosine. It oscillates with full contrast between plus and minus one, and it has a fixed frequency. This is an ideal result because we have assumed that there is no control or environmental noise. remember that when we measure the z-axis, we get either state zero, a plus one on the graph, or state one, a minus one on the graph, and that is it. So, we have two identically prepare, control, and measure the qubit a large number of times at each time step and average these results to generate the cosine experimentally. For example, when the qubit is at the north pole, all measurements will yield a plus one, and the average is obviously plus one. when the qubit is at the south pole, all measurements yield a minus one. Thus, the average there is minus one. Now, on the equator, for example, the qubit is in an equal superposition state. Thus, half of the time, we measure one, and half of the time, we measure minus one. Thus, the average there is zero. The cosine trace represents the average of many, many measurements, and, in fact, so many in this case that we cannot even resolve the statistical sampling error. Now, as an aside, if we had stopped the pulse after the first quarter period of the cosine wave, then the block vector stops on the equator, and this is $ \pi/2 $ pulse around the x-axis. If we had instead stopped the pulse after half a period, the block vector rotates 180 degrees around the x-axis, and this is an X-gate. so, what happens in the presence of noise? we will assume that we have one specific type of noise, amplitude noise on the control field. since the Rabi frequency is proportional to amplitude, fluctuations in the control field amplitude cause fluctuations in the Rabi frequency. To help visualize what is happening, we will also assume that the noise is quasi-static, which means that the amplitude is fixed for one whole Rabi trace, but fluctuates to a different value according to some probability distribution for each new Rabi trace. Essentially, we are saying that the amplitude changes slowly in time, or equivalently, that this is low-frequency amplitude noise. Again, we simulate multiple nominally identical trials. However, due to the noise, different trials have different Rabi frequencies. The Bloch vector begins to diffuse. Some instances are rotating faster, and others are rotating slower. on the right, individual instances are shown in gray while the average value is green. The net result is that the Rabi oscillations decay in time due to the frequency fluctuations caused by the amplitude noise of the control field. The actual form of the decay function turns out to be Gaussian because we assumed low-frequency quasi-static noise. we can see how this type of noise then causes small errors in the $ \pi/2 $ in $ \pi $ pulse gates. Although this example considered only one type of noise, control field amplitude fluctuations, it gives an intuitive picture of how noise can cause gate errors in general. In the next section, we will examine how noise is measured and characterized through a two-time correlation function and a corresponding noise power spectral density.

\section{Noise Characterization and the Noise Power Spectral Density} 

In the last section, we an example of how a specific type of noise amplitude noise on a control field can cause Rabi oscillations to decay due to the resulting fluctuation of the Rabi frequency. However, how do we characterize and quantify these types of noise processes in a real physical system? Moreover, how do we relate the temporal behavior of noise to its frequency spectrum? In this section, we will overview the standard approach to characterizing and representing noise processes via the autocorrelation function and the noise power spectral density. To get started, let us consider a system with a noisy parameter, x. As an experimentalist, we set up a detector in our system to measure x and record how it varies in time\cite{kandala_error_2019}. ideally, we would like to calculate the statistics of x from the single time trace. After all, it is a straightforward measurement. we only needed one system to do it. we do not necessarily know the explicit probability density function of the noise processes that underlie in the fluctuations in x. So, instead, we just measure them. in the limit, we measure for a long time. we get a reasonable estimate for the statistics of x. as long as it fluctuates the same way tomorrow as it did today, this approach works. On the other hand, we could, in principle, set up an ensemble of n identical systems and record one sample of x from each of them at the same time. Then, we could calculate the ensemble average statistics across all end samples. in the limit of large n, we would, again, be able to obtain the statistics of x. It is notionally how theorists would calculate the properties of a system. They ideally have a model in mind that yields a probability distribution function that describes the fluctuations in x. It also seems reasonable as long as it describes the system accurately at any time. For example, the first-order time average of x is the time integral of x over a duration, T, that we take to be large and then normalized by T to obtain the average value of x. we will denote time averages with a bar over the variable x. In comparison, the first order ensemble average, which we denote with brackets, is defined as a sum over all members of an ensemble, at a particular time, T1, and then divided by n to obtain the average value. We generally do not have identical copies of an experiment. Thus, the calculation is performed using a presumed known probability density function, p. P tells us the probability of obtaining any value x at a time, T. the ensemble average follows by integrating, overall, x. The variable x times the probability distribution function, p. Similarly, one can define an autocorrelation function between two points of a measured time trace separated by a time tau. Alternatively, one can find the average ensemble version of this, the covariance, by using a joint probability distribution function, as shown here. Now, ideally, we would like to work with systems where these two approaches give the same results that time averaging is equivalent to ensemble averaging. Such systems are called ergodic ensembles, and their statistics are called ergodic. if the statistics calculated from a time trace are independent of time, then the noise is statistically stationary. For example, the autocorrelation function of a statistically stationary process would depend only on the time difference tau between the two samples independent of when it is measured. while ergodicity implies stationarity, the opposite is not always true. Now, ergodicity and stationarity, in the strictest sense, must be true to all orders of the statistics. However, for what we will look at next, we only require what is called weak stationarity or wide sense stationarity, which means that the mean value is constant in time. the autocorrelation function depends only on the time difference tau. We will see, if we know or can measure the autocorrelation function of a wide sense stationary process, we can find the power spectral density using the Wiener-Khinchine theorem. Wiener-Khinchine tells us that the autocorrelation function and the power spectral density are Fourier transform pairs. Let us assume we have a fluctuating parameter $ \lambda $, for example, magnetic flux. Then if we take the Fourier transform of an autocorrelation function over the time difference, tau, we will obtain the power spectral density in units of $ \lambda $ squared per Hertz. In turn, if we take the inverse Fourier transform of the power spectral density, we will get back the autocorrelation function with units, $ \lambda $ square. S of omega is called the noise power spectral density. we plot S of omega for both positive and negative frequencies. Now, for noise at the qubit frequency, which can exchange energy with the environment, positive frequencies correspond to the qubit emitting energy to the environment. negative frequencies correspond to the qubit absorbing energy from the environment. we will return to this in a moment. First, though, let us look at an example of low-frequency noise, such as 1 over f noise. In this case, S of omega's peaked at zero frequency. it decreases with frequency as 1 over f. we have taken the noise to be symmetric about zero frequency. as we may remember for Fourier transform pairs, symmetric infrequency implies that a signal is manifestly real in the time domain. Thus, this represents a real fluctuating signal. we can represent it using a classical variable of $ \lambda $. Thus, these symmetric power spectra are often called classical noise. Next, we can include thermal noise or a Johnson noise. Johnson noise is proportional to temperature. it is called white noise because it is constant over all frequencies. In electrical systems, Johnson noise is the voltage or current noise present in a resistor at a given temperature, T. This is also an example of classical noise because it is symmetric in frequency. Next, let us consider an example of quantum noise or Nyquist noise, which is not symmetric in frequency. Nyquist noise is proportional to frequency omega. it is only present at positive frequencies. This asymmetry implies that the time domain signal is not manifestly real. this arises from the fact that Nyquist noise must be described using quantum operators. we promote the $ \lambda $ variable to an operator. the operators, $ \lambda $ T, do not commute with one another at different times. The computation relation can include imaginary terms. Thus, the power spectrum needs no longer be symmetric. Nyquist noise is responsible for spontaneous emission of energy from a qubit to its environment. That is, if we put the qubit in its excited state, it will relax back to its ground state by emitting a photon at a frequency omega q. It happens, even if the environments at zero temperature, and is a complete absence of classical noise. Now, a zero-temperature environment cannot drive the qubit from its ground state to its excited state. It has no energy to do this. if the environments at non-zero temperature, then we can have classical Johnson noise, which drives transitions in both directions, 0 to 1 and 1 is 0, because of its symmetric infrequency. However, Nyquist noise is different. It represents an additional spontaneous emission component. Together, these two processes are called Johnson-Nyquist noise. Now, note that generally speaking, only noise at the qubit frequency, at plus or minus omega q, causes transitions and energy exchange with the environment. These are T1 processes. In contrast, classical noise that is off-resonance with the qubit contributes to its dephasing. With knowledge of the noise power spectral density, as seen by the qubit, one then understands the frequency content of the various noise sources that cause decoherence.

In this section, we will review the distinction between time averaging and ensemble averaging. We will also review the Wiener-Khinchine theorem, which relates the two-time correlation function to the noise power spectral density.

There were two types of noise discussed in the section: systematic noise and stochastic noise. Systematic noise arises from a systematic source, for example, a fixed offset in driving field amplitude that causes a fixed amount of over-rotation or under-rotation when performing a particular quantum gate. Despite being systematic errors, measurement results may still differ in the presence of such errors because the experiments may differ. For example, when applying different numbers of gates or interspersing the gates with different assortments of other gates, measurement results may give an appearance of noise. This type of error can generally be corrected with better calibration or hardware improvements.

Here, we will focus on stochastic noise processes. Noise is a random fluctuation of a parameter in space and time. It is characterized statistically, either via temporal averaging or through ensemble averaging. The distinction is related to how averaged quantities are determined:

1. Temporal (spatial) averaging: averaging the time (or spatial) record of a single system

2. Ensemble averaging: averaging over many identical systems at a fixed time (or spatial position)\\

Time averaging and ensemble averaging\\

\begin{equation}\label{eq3_50}
\displaystyle \overline{x^{(i)}(t)}    \displaystyle =    \displaystyle \lim _{T\rightarrow \infty } \frac{1}{T} \int _{-T/2}^{T/2} x^{(i)}(t) \; dt    
\end{equation}

\begin{equation}\label{eq3_51}
\overline{\left[ x^{(i)}(t) \right]^2}    \displaystyle =    \displaystyle \lim _{T\rightarrow \infty } \frac{1}{T} \int _{-T/2}^{T/2} \left[ x^{(i)}(t) \right]^2 dt         
\end{equation}

\begin{equation}\label{eq3_52}
\displaystyle \overline{x^{(i)}(t)x^{(i)}(t+\tau )}    \displaystyle =    \displaystyle \lim _{T\rightarrow \infty } \frac{1}{T} \int _{-T/2}^{T/2} x^{(i)}(t)x^{(i)}(t+\tau ) \; dt \equiv \phi _ x^{(i)}(\tau )         
\end{equation}

In the above definitions, the time T is the length of the time window being used to estimate the statistics of $ x^{(i)}(t) $, and the estimate becomes ``sharp'' in the limit $ T \rightarrow \infty $. We have also defined the autocorrelation function  $ \phi _ x^{(i)}(\tau ) $ in the last line. Note that the ``overbar'' is used to represent time-averaged quantities.

Similarly, one can respectively define ensemble-averaged quantities using N identical systems, as defined at a certain point in time $ t_1 $ for a parameter $ x(t_1) $ as:
\begin{equation}\label{eq3_53}
\displaystyle \langle x(t_1)\rangle    \displaystyle =    \displaystyle \lim _{N\rightarrow \infty } \frac{1}{N} \sum _{i=1}^ N x^{(i)}(t_1) = \int _{-\infty }^{\infty } x_1 \; p_1(x_1;t_1) \; dx_1         
\end{equation}

\begin{equation}\label{eq3_54}
\displaystyle \langle \left[ x(t_1) \right]^2 \rangle    \displaystyle =    \displaystyle \lim _{N\rightarrow \infty } \frac{1}{N} \sum _{i=1}^ N \left[ x^{(i)}(t_1) \right]^2 = \int _{-\infty }^{\infty } x_1^2 \; p_1(x_1;t_1) \; dx_1         
\end{equation}    

\begin{equation}\label{eq3_55}
\displaystyle \langle x(t_1) x(t_2) \rangle    \displaystyle =    \displaystyle \lim _{N\rightarrow \infty } \frac{1}{N} \sum _{i=1}^ N x^{(i)}(t_1) x^{(i)}(t_2) = \int _{-\infty }^{\infty } x_1 x_2 \; p_2(x_1,x_2;t_1,t_2) \; dx_1 dx_2
\end{equation}

In these definitions, N is the number of ensemble members used to make an estimate for the statistics of $ x(t_1), $ and the estimate becomes ``sharp'' in the limit $ N \rightarrow \infty $; we have implicitly defined $ x_1=x(t_1) $ and $ x_2=x(t_2) $; and the quantities $ p_1(x_1;t_1) $ and $ p_2(x_1,x_2;t_1,t_2) $ are respectively the first-order and second-order joint probability density functions.\\

Ergodicity and Stationarity\\

The ensemble average is generally a theoretical concept that describes a system's statistics by a probability density function derived from a particular theoretical model. Since it is generally impractical to make N identical systems and measure them all simultaneously, one generally characterizes the statistics of a single system through time averaging of a measured time trace. These theoretical and experimental approaches are equivalent to an ergodic ensemble, that is when the system is said to be ergodic.

A system is stationary if quantities such as $ \overline{x^{(i)}(t)} $ and $ \overline{\left[ x^{(i)}(t) \right]^2} $ are independent of time (when they are measured), and quantities such as $ \phi _ x^{(i)}(\tau ) $ depend only on the time difference $ \tau $. In other words, the statistics do not change with time. This can be further quantified by the order of stationarity: a stochastic process is stationary to order k if its $ k^{\textrm{th}} $-order probability density function is time-independent. Similarly, there are degrees of ergodicity (e.g., ergodicity in the mean, etc.).

Ergodicity implies stationarity since one sample function represents the entire process. However, although stationarity implies ergodicity (that time averaging and ensemble averaging is equivalent), this is not always the case. For example, consider a bucket of batteries of varying types and select one at random. Let us say we picked a 9V battery. We find that the mean battery voltage $ \overline{v(t)} $ is independent of time, so we infer that it is the first-order stationery. However, this battery is but one type of battery. What if the bucket also had other types of batteries, such as 1.5V batteries? In that case, the statistical ensemble average would be different. Thus, one, in general, cannot infer ergodicity from stationarity, without additional (often reasonable) assumptions. Here, we would need to assume that the bucket only holds nominally identical 9V batteries, for instance.\\

Noise power spectral density and the Wiener-Khintchine theorem\\

A system is called wide-sense stationary or weakly stationary if its mean and auto-correlaton function are stationary. For wide-sense stationary processes, the Wiener-Khintchine theorem relates the autocorrelation function $  \phi _ x^{(i)}(\tau ) $ to the noise power spectral density $ S_ x(\omega ) $:
\begin{equation}\label{eq3_56}
\displaystyle S_ x(\omega )    \displaystyle =    \displaystyle \int _{-\infty }^{\infty } \phi _ x(\tau ) \; e^{-i \omega \tau } \; d\tau ,         (3.7)
\end{equation}

\begin{equation}\label{eq3_57}
\displaystyle \phi _ x(\tau )    \displaystyle =    \displaystyle \frac{1}{2 \pi } \int _{-\infty }^{\infty } S_ x(\omega ) \; e^{i \omega \tau } \; d\omega ,         (3.8)
\end{equation}

Where we have defined $  S_ x(\omega ) $ as a bilateral spectral density defined overall negative and positive frequencies to accommodate the potential for asymmetric spectra, negative frequencies correspond to noise processes associated with a qubit absorbing photons from the environment, and positive frequencies correspond to the qubit emitting photons to the environment. The asymmetry between photon absorption and emission arises due to spontaneous emission: a qubit in its excited state can emit a photon to its environment even if that environment is at zero temperature whereas it cannot absorb energy from a zero-temperature environment (the environment has no energy to give). The origin of asymmetric spectra is traceable to the fact that quantum operators generally do not commute at different times. Instead, their commutation relation is in general imaginary, and Fourier spectra are only symmetric in frequency when their time-domain counterpart is manifestly real.

Characterizing the power spectral density of noise seen by the qubit is a first step to mitigate that noise, whether by reducing the source of the noise or by designing the qubit to be insensitive to it. With knowledge of the frequency spectrum, one can design noise filter functions that decouple the qubit from certain types of low-frequency noise. This type of noise mitigation is achieved by applying gate operations (for instance, X-gates, Y-gates, ) to the qubit in a specified manner. The pulse sequence in the time domain corresponds to a filter function in the frequency domain. These types of noise mitigation techniques are essentially a filter-design problem, and they are enabled by knowing the noise power spectral density.

Lastly, its worth noting that the auto-correlation function and noise power spectral density are second-order statistics. While this generally captures the predominant noise sources affecting qubits today, environmental noise may also exhibit higher-order correlations and correspondingly higher-order noise spectra.

The image below represents the ``static noise'' that we may remember seeing when we had poor analog TV reception. Let us take the reception here to be very poor, such that there is no underlying image and only static noise.

For this problem, let us assume that each pixel has a ``brightness'' that varies continuously between white (bright) and black (dark), with all shades of grey in between. For a given pixel, we will assume the brightness is fluctuating randomly in time according to a uniform probability distribution. That is, all shades of grey (including black and white) are equally likely to appear at any given time. We will also assume that the correlation time is infinitesimally short. Finally, we will assume that all pixels behave this way, but are completely independent of one another. To characterize this noise, let us assign the value zero to black, one to white, with all grey shades being the values in between zero and one.

\begin{figure}[H] \centering{\includegraphics[scale=2]{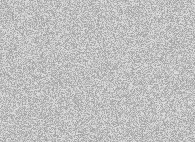}}\caption{static noise}\label{fig3_30}
\end{figure}

There are two ways to calculate statistical averages of this noise:\\
1. By looking at one pixel and averaging its fluctuating brightness over time.\\
2. By taking a screenshot at a fixed time, and averaging the brightness across all pixels.\\
These two alternatives correspond to the temporal average and the ensemble average, respectively. By assumption, the temporal auto-correlation of a given pixel is how the brightness of a pixel at one time is related to its brightness at a later time to be very small. Since the auto-correlation is very short in time, by Fourier transform properties, the noise power spectral density is essentially uniformly distributed in frequency (``white noise''). After carefully staring at the screen for a very long time, we determine that both the mean and the autocorrelation function in time are equivalent to the mean and covariance of the pixel ensemble.

\item  Density Matrices; In Density Matrix, we discussed how to write the density matrix of a single-qubit. To do so, we considered the two-qubit state $ \left\vert \psi _{AB}\right\rangle =\sqrt{\frac{3}{4}}\left\vert 0_{A}\right\rangle \left\vert 0_{B}\right\rangle +\sqrt{\frac{1}{4}}\left\vert 1_{A}\right\rangle \left\vert 1_{B}\right\rangle $ and found the probability of projecting qubit B onto the states $ \left\vert 0_{B}\right\rangle and \left\vert 1_{B}\right\rangle $. we used these results to write the density matrix of qubit A after the measurement of B,
\begin{equation}\label{eq3_58}
\displaystyle \rho _{A}    \displaystyle =    \displaystyle p_{B,0}\left\vert \psi _{A,0}\right\rangle \left\langle \psi _{A,0}\right\vert +p_{B,1}\left\vert \psi _{A,1}\right\rangle \left\langle \psi _{A,1}\right\vert ,          
\displaystyle =    \displaystyle \frac{3}{4}\left( \begin{array}{cc} 1 & 0 \\ 0 & 0\end{array}\right) +\frac{1}{4}\left( \begin{array}{cc} 0 & 0 \\ 0 & 1\end{array}\right) .
\end{equation}

In this assessment, we should use the same steps used to compute the density matrix of qubit A after measuring qubit B if a Hadmard gate is applied on B before the measurement.

Start by considering the two-qubit state after the Hadamard gate and its complex conjugate
\begin{equation}\label{eq3_59}
\begin{split}
\displaystyle \left\vert \psi _{AB}^{\prime }\right\rangle    \displaystyle & =    \displaystyle \sqrt{\frac{3}{4}}\left\vert 0_{A}\right\rangle \frac{\left\vert 0_{B}\right\rangle +\left\vert 1_{B}\right\rangle }{\sqrt{2}}+\sqrt{\frac{1}{4}}\left\vert 1_{A}\right\rangle \frac{\left\vert 0_{B}\right\rangle -\left\vert 1_{B}\right\rangle }{\sqrt{2}} \\     
\displaystyle \left\langle \psi _{AB}^{\prime }\right\vert    \displaystyle & =    \displaystyle \sqrt{\frac{3}{4}}\left\langle 0_{A}\right\vert \frac{\left\langle 0_{B}\right\vert +\left\langle 1_{B}\right\vert }{\sqrt{2}}+\sqrt{\frac{1}{4}}\left\langle 1_{A}\right\vert \frac{\left\langle 0_{B}\right\vert -\left\langle 1_{B}\right\vert }{\sqrt{2}}
\end{split}
\end{equation}
      
The probability of projecting qubit B on to state $ \left\vert 0_{B}\right\rangle $ and $ \left\vert 1_{B}\right\rangle $ is given by $ p_{B,0}^{\prime }=\left\langle \psi _{AB}^{\prime }\right\vert \left( I_{A}\otimes \Pi _{0}\right) \left\vert \psi _{AB}^{\prime }\right\rangle $ and $ p_{B,1}^{\prime }=\left\langle \psi _{AB}^{\prime }\right\vert \left( I_{A}\otimes \Pi _{1}\right) \left\vert \psi _{AB}^{\prime }\right\rangle $ respectively. Here, I is the identity operator, $ \Pi _{0} $ is the single-qubit projector $ \left\vert 0\right\rangle \left\langle 0\right\vert, $ and $ \Pi _{1} $ is the single-qubit projector $ \left\vert 1\right\rangle \left\langle 1\right\vert $. To find the probabilities $ p_{B,0}^{\prime } $ and $ p_{B,1}^{\prime } $, it is easier to first calculate

\begin{equation}\label{eq3_60}
\left( I_{A}\otimes \Pi _{0}\right) \left\vert \psi _{AB}^{\prime }\right\rangle =\frac{1}{\sqrt{2}}\left( \sqrt{\frac{3}{4}}\left\vert 0_{A}\right\rangle \left\vert 0_{B}\right\rangle +\sqrt{\frac{1}{4}}\left\vert 1_{A}\right\rangle \left\vert 0_{B}\right\rangle \right) \end{equation}

and
\begin{equation}\label{eq3_61}
\left( I_{A}\otimes \Pi _{1}\right) \left\vert \psi _{AB}^{\prime }\right\rangle =\frac{1}{\sqrt{2}}\left( \sqrt{\frac{3}{4}}\left\vert 0_{A}\right\rangle \left\vert 1_{B}\right\rangle -\sqrt{\frac{1}{4}}\left\vert 1_{A}\right\rangle \left\vert 1_{B}\right\rangle \right) \end{equation}

The probabilitites are then given by:

\begin{equation}\label{eq3_62}
\begin{split}
\displaystyle p_{B,0}^{\prime }    \displaystyle & =    \displaystyle \left\langle \psi _{AB}^{\prime }\right\vert \left( I_{A}\otimes \Pi _{0}\right) \left\vert \psi _{AB}^{\prime }\right\rangle \\          
\displaystyle & =    \displaystyle \frac{1}{\sqrt{2}}\left\langle \psi _{AB}^{\prime }\right\vert \left( \sqrt{\frac{3}{4}}\left\vert 0_{A}\right\rangle \left\vert 0_{B}\right\rangle +\sqrt{\frac{1}{4}}\left\vert 1_{A}\right\rangle \left\vert 0_{B}\right\rangle \right) \\
\displaystyle &=    \displaystyle \frac{1}{2}\\
\end{split}
\end{equation}
          
and

\begin{equation}\label{eq3_63}
\begin{split}
\displaystyle p_{B,1}^{\prime }    \displaystyle & =    \displaystyle \left\langle \psi _{AB}^{\prime }\right\vert \left( I_{A}\otimes \Pi _{1}\right) \left\vert \psi _{AB}^{\prime }\right\rangle \\          
\displaystyle & =    \displaystyle \frac{1}{\sqrt{2}}\left\langle \psi _{AB}^{\prime }\right\vert \left( \sqrt{\frac{3}{4}}\left\vert 0_{A}\right\rangle\left\vert 1_{B}\right\rangle- \sqrt{\frac{1}{4}}\left\vert 1_{A}\right\rangle \left\vert 1_{B}\right\rangle \right) \\          
\displaystyle & =    \displaystyle \frac{1}{2}\\
\end{split}
\end{equation}
          
We may wish to use Wolfram Alpha online to calculate square roots. Round results to two decimal places, if necessary.

\item  Density Matrices Part A; 
\begin{itemize}
\item What is the state of qubit A if B is projected to$  \left\vert 0_{B}\right\rangle $?
\item $ \left\vert \psi _{A,0}^{\prime }\right\rangle $ = 0.86
$ \left\vert 0_{A}\right\rangle $ + 0.5 $ \left\vert 1_{A}\right\rangle $ \\
\item What is the state of qubit A if B is projected to $ \left\vert 1_{B}\right\rangle $?
\item $ \left\vert \psi _{A,1}^{\prime }\right\rangle = $  0.86 $ \left\vert 0_{A}\right\rangle - 0.5 \left\vert 1_{A}\right\rangle $ \\
Note that we ask about the amplitudes, not the probabilities, in the fields above (round to two decimal places).
\end{itemize}
Solution:\\
The probability of projecting qubit B to state $ \left\vert 0_{B}\right\rangle $ is given by $ p_{B,0}^{\prime }=\left\langle \psi _{AB}^{\prime }\right\vert \left( I_{A}\otimes \Pi _{0}\right) \left\vert \psi _{AB}^{\prime }\right\rangle $, where I is the identity operator and $ \Pi _{0} $ is the single-qubit projector $ \left\vert 0\right\rangle \left\langle 0\right\vert $. To find the probability $ p_{B,0}^{\prime } $ we first calculate

$ \displaystyle \left( I_{A}\otimes \Pi _{0}\right) \left\vert \psi _{AB}^{\prime }\right\rangle    \\ \displaystyle =    \displaystyle \left( I_{A}\otimes \left\vert 0_{B}\right\rangle \left\langle 0_{B}\right\vert \right) \left( \sqrt{\frac{3}{4}}\left\vert 0_{A}\right\rangle \frac{\left\vert 0_{B}\right\rangle +\left\vert 1_{B}\right\rangle }{\sqrt{2}}+\sqrt{\frac{1}{4}}\left\vert 1_{A}\right\rangle \frac{\left\vert 0_{B}\right\rangle -\left\vert 1_{B}\right\rangle }{\sqrt{2}}\right)          
\\\displaystyle =    \displaystyle \frac{1}{\sqrt{2}}\left( \sqrt{\frac{3}{4}}\left\vert 0_{A}\right\rangle +\sqrt{\frac{1}{4}}\left\vert 1_{A}\right\rangle \right) \left\vert 0_{B}\right\rangle , $    \\      
and then

$ \displaystyle p_{B,0}^{\prime }    \displaystyle =    \displaystyle \left\langle \psi _{AB}^{\prime }\right\vert \left( I_{A}\otimes \Pi _{0}\right) \left\vert \psi _{AB}^{\prime }\right\rangle ,          
\displaystyle =    \displaystyle \frac{1}{2}. $\\     
     
In the same way, to find the probability of projecting qubit B to state $ \left\vert 1_{B}\right\rangle $, we first calculate

$ \displaystyle \left( I_{A}\otimes \Pi _{1}\right) \left\vert \psi _{AB}^{\prime }\right\rangle    \\\displaystyle =    \displaystyle \left( I_{A}\otimes \left\vert 1_{B}\right\rangle \left\langle 1_{B}\right\vert \right) \left( \sqrt{\frac{3}{4}}\left\vert 0_{A}\right\rangle \frac{\left\vert 0_{B}\right\rangle +\left\vert 1_{B}\right\rangle }{\sqrt{2}}+\sqrt{\frac{1}{4}}\left\vert 1_{A}\right\rangle \frac{\left\vert 0_{B}\right\rangle -\left\vert 1_{B}\right\rangle }{\sqrt{2}}\right) ,          
\\\displaystyle =    \displaystyle \frac{1}{\sqrt{2}}\left( \sqrt{\frac{3}{4}}\left\vert 0_{A}\right\rangle -\sqrt{\frac{1}{4}}\left\vert 1_{A}\right\rangle \right) \left\vert 1_{B}\right\rangle ,     $      \\
and then

$ \displaystyle p_{B,1}^{\prime }    \displaystyle =    \displaystyle \left\langle \psi _{AB}^{\prime }\right\vert \left( I_{A}\otimes \Pi _{1}\right) \left\vert \psi _{AB}^{\prime }\right\rangle ,          
\displaystyle =    \displaystyle \frac{1}{2}. $\\
          
The two-qubit state after qubit B is projected to $ \left\vert 0_{B}\right\rangle $ and $ \left\vert 1_{B}\right\rangle $ is given by\\

$ \frac{\left( I_{A}\otimes \Pi _{0}\right) \left\vert \psi _{AB}^{\prime }\right\rangle }{\sqrt{\left\langle \psi _{AB}^{\prime }\right\vert \left( I_{A}\otimes \Pi _{0}\right) \left\vert \psi _{AB}^{\prime }\right\rangle }}=\left( \sqrt{\frac{3}{4}}\left\vert 0_{A}\right\rangle +\sqrt{\frac{1}{4}}\left\vert 1_{A}\right\rangle \right) \left\vert 0_{B}\right\rangle , $\\

and\\

$ \frac{\left( I_{A}\otimes \Pi _{1}\right) \left\vert \psi _{AB}^{\prime }\right\rangle }{\sqrt{\left\langle \psi _{AB}^{\prime }\right\vert \left( I_{A}\otimes \Pi _{1}\right) \left\vert \psi _{AB}^{\prime }\right\rangle }}=\left( \sqrt{\frac{3}{4}}\left\vert 0_{A}\right\rangle -\sqrt{\frac{1}{4}}\left\vert 1_{A}\right\rangle \right) \left\vert 1_{B}\right\rangle . $\\

Thus the state of qubit A after qubit B is projected to $ \left\vert 0_{B}\right\rangle $ and $ \left\vert 1_{B}\right\rangle $ is

$ \displaystyle \left\vert \psi _{A,0}^{\prime }\right\rangle    \displaystyle =    \displaystyle \sqrt{\frac{3}{4}}\left\vert 0_{A}\right\rangle +\sqrt{\frac{1}{4}}\left\vert 1_{A}\right\rangle ,     \\     
\displaystyle \left\vert \psi _{A,1}^{\prime }\right\rangle    \displaystyle =    \displaystyle \sqrt{\frac{3}{4}}\left\vert 0_{A}\right\rangle -\sqrt{\frac{1}{4}}\left\vert 1_{A}\right\rangle ,     $      
respectively.

\item  Density Matrices Part B; 
\begin{itemize}
\item  If there is no information about to which state qubit B was projected, what is the density matrix that describes the state of qubit A?
\item In Dirac notation,
\item $ \rho _{A}=  $ 0.5 $ \left\vert \psi _{A,0}^{\prime }\right\rangle \left\langle \psi _{A,0}^{\prime }\right\vert +  $ 0.5 $ \left\vert \psi _{A,1}^{\prime }\right\rangle \left\langle \psi _{A,1}^{\prime }\right\vert $
\item In matrix form,
\item $ \rho _{A} $= 0.75 $ \left( \begin{array}{cc} 1 & 0 \\ 0 & 0\end{array}\right)  $+  0.25 $ \left( \begin{array}{cc} 0 & 0 \\ 0 & 1\end{array}\right) $
\end{itemize}
Solution:\\
If there is no information about to which state qubit B was projected, the state of qubit A can be described by the density matrix

$ \displaystyle \rho _{A}    \displaystyle =    \displaystyle p_{B,0}\left\vert \psi _{A,0}^{\prime }\right\rangle \left\langle \psi _{A,0}^{\prime }\right\vert +p_{B,1}\left\vert \psi _{A,1}^{\prime }\right\rangle \left\langle \psi _{A,1}^{\prime }\right\vert ,     \\     \\
\displaystyle =    \displaystyle \frac{1}{2}\left( \sqrt{\frac{3}{4}}\left\vert 0_{A}\right\rangle +\sqrt{\frac{1}{4}}\left\vert 1_{A}\right\rangle \right) \left( \sqrt{\frac{3}{4}}\left\langle 0_{A}\right\vert +\sqrt{\frac{1}{4}}\left\langle 1_{A}\right\vert \right) \\+\frac{1}{2}\left( \sqrt{\frac{3}{4}}\left\vert 0_{A}\right\rangle -\sqrt{\frac{1}{4}}\left\vert 1_{A}\right\rangle \right) \left( \sqrt{\frac{3}{4}}\left\langle 0_{A}\right\vert -\sqrt{\frac{1}{4}}\left\langle 1_{A}\right\vert \right) ,          \\\\
\displaystyle =    \displaystyle \frac{3}{4}\left\vert 0_{A}\right\rangle \left\langle 0_{A}\right\vert +\frac{1}{4}\left\vert 1_{A}\right\rangle \left\langle 1_{A}\right\vert ,         \\ \\
\displaystyle =    \displaystyle \frac{3}{4}\left( \begin{array}{cc} 1 & 0 \\ 0 & 0\end{array}\right) +\frac{1}{4}\left( \begin{array}{cc} 0 & 0 \\ 0 & 1\end{array}\right) . $

\begin{figure}[H] \centering{\includegraphics[scale=.8]{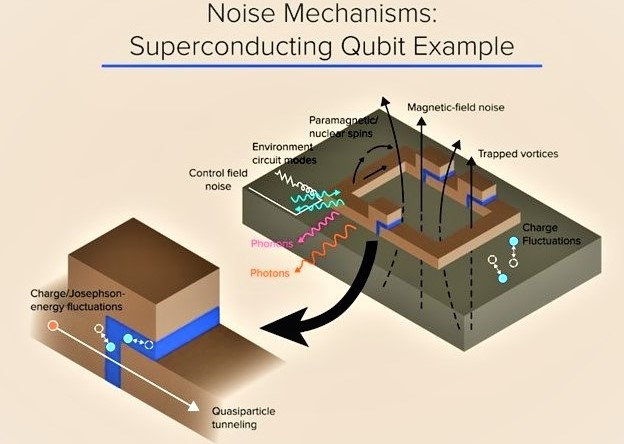}}\caption{Noise Mechanisms}\label{fig3_3}
\end{figure}

\begin{figure}[H] \centering{\includegraphics[scale=.7]{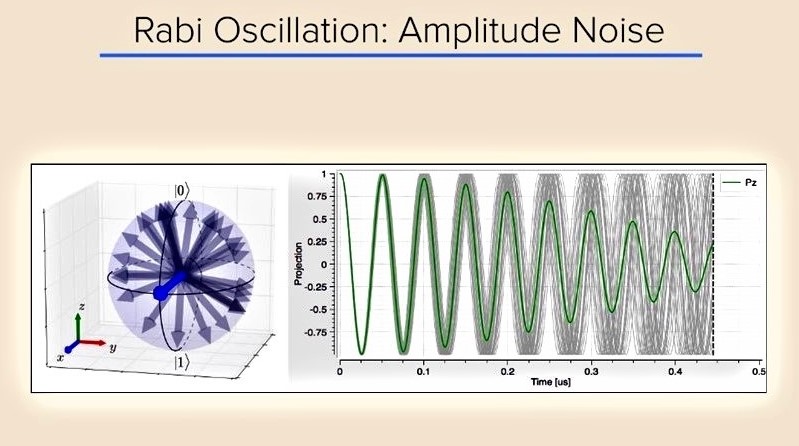}}\caption{Rabi Oscillation}\label{fig3_4}
\end{figure}

\begin{figure}[H] \centering{\includegraphics[scale=.6]{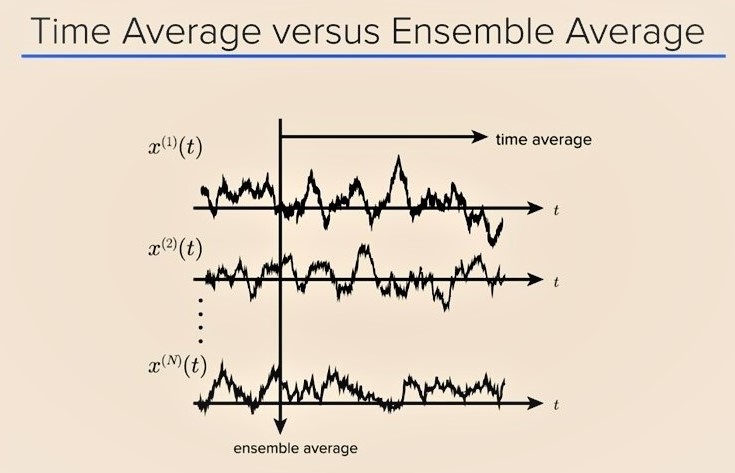}}\caption{Noise Characterization and the Noise Power Spectral Density}\label{fig3_5}
\end{figure}

\section{Quantum Communication Perspectives} 
In this section, we will look at the practical challenges faced by realistic quantum communication today. As we have seen throughout the section, the promise of quantum communication centers on achieving quantum advantage for applications such as secret sharing, networking, authentication, and key distribution. For many of these schemes, entanglement is a resource that enables quantum enhancement when shared between two or more parties\cite{monz_14-qubit_2011}. However, we face several practical issues when implementing quantum communication protocols, and we will touch on several of those in this section. We will revisit Bell states and discuss a game related to Bell's classical argument that forms a well-defined inequality that classical probability arguments must respect, but one that quantum mechanics consistently violates. We will also look at the challenges of long-distance entanglement distribution using repeaters to mitigate photon loss and other schemes being developed to enable long-distance QKD.
Moreover, finally, we will take a deep dive into quantum entanglement, exploring its mathematical representation and measures\cite{laforest_mathematics_nodate}, as well as its fungibility. However, first, to get started, let us discuss about entropy and channel capacity, fundamental concepts of information theory, and their representations in noisy classical and quantum communication. An insightful perspective for quantum communication is provided by starting with the scenario of classical communication. Here, the goal is to transmit a message, say, x, described by its probability distribution function. What is received is a noisy message from the output of the channel, call that y. The input to the channel is x. Two principal ideas describe this scenario mathematically. First, H of x, known as the entropy of x, is the number of bits needed to represent the message, x, on average faithfully. Second, the capacity, c, is the max overall probability distributions of something called the mutual information. We are x, y, which is the maximum error-free communication rate achievable over the noisy channel. The two key concepts here are the entropy, H of x, and the mutual information, we of x, y. There are parallels to this in quantum communication. In this scenario, the source messages are described by a density matrix. It is a distribution, which may be viewed as a distribution over pure states, for example. These may be non-orthogonal states, in contrast to the classical case. A density matrix also describes the received message.
Furthermore, in contrast to the classical case, this is possibly decoded quantum mechanically, say by a quantum computer. Analogous to the classical case, the Von Neumann entropy describes the number of qubits needed to represent the density matrix $ \rho $ faithfully. This is asymptotical. This entropy is similar to but different from the Shannon entropy, as we will see. Also, there are several distinct scenarios for noisy communication. After all, classical data may be the source as well as the received data. Alternatively, this source may be quantum data. Alternatively, the receiver may receive and decode quantum mechanically. Alternatively, both the sender and receiver may be quantum. It is known that these channel capacities are ordered. CCC is less than CQC and CCQ and is less than CQQ. There are even richer scenarios beyond these four. Entanglement may also be used to assist, for example, classical communication. When the sender and receiver have pre-shared entangled qubits, this may change the time of capacity. Another set of capacity measure results when quantum information is allowed to be sent, but only one way or two ways, and possibly, in addition to having a classical side channel. Beyond this, quantum communication may also involve multiple parties. Quantum communication is thus a fascinating field of study. 

Information exchange between a sender and receiver is conducted over communication channels. Wherever previously, we considered ideal communication channels, we will address the practical issues and challenges associated with communicating on realistic, physical channels. A primary consideration is the reduction of signal integrity due to information loss in the channel and/or the addition of noise to the channel.

Its classical counterpart inspires the characterization of quantum communication. The characterization of classical communication comprises two primary figures of merit. The first figure of merit is the entropy H, which is related to the average number of bits m required to represent a message M. The averaged Shannon entropy is defined as $  H(M)=-\sum _ i^{2m} p(M_ i) \log _2 p(M_ i)=m $, with each bit being equally likely $ p(M_ i)=\frac{1}{2} $. Within this definition, the entropy H(M) is the same for all messages M with m bits. The second figure of merit is the maximal achievable error-free communication rate, referred to as the channel capacity, which can be deduced by comparing the shared mutual information of the received message $ \tilde{M} $ with the original message M.

The characterization of quantum information on quantum channels follows an analogous definition for entropy. The von Neumann entropy $ S(\rho )=-Tr(\rho \ln \rho ) $ describes the number of qubits required to represent a density matrix $ \rho $ corresponding to the quantum information being transmitted. Similarly, the channel capacity can be inferred by comparing the received density matrix $ \tilde{\rho } $ to the initial density matrix$  \rho $, and there are several definitions for channel capacity, depending on how the information is sent and received. 

While a quantum communication channel generally transmits information encoded in quantum states, in practice, many schemes also require parallel classical channels to implement fully a particular protocol. Upon receipt, the information may either remain quantum-mechanical, for example, be stored in a quantum system for further processing, or it may also be measured to retrieve an underlying classical output.

\section{Quantum Weirdness} 
we will be discussing quantum weirdness. We think of discussing this, and we first want to go back to a paper by Einstein, Podolsky, and Rosen. That is EPR 1935. What Einstein argued was that quantum mechanics is incomplete. So, what Einstein used in this paper were position and momentum. However, we are working with discrete quantum mechanics, so we will just use, well, up and down or 0 and 1. So, we want to give Einstein's argument. So, suppose Alice and Bob share a pair of entangled particles like this. Alice measures in the 0, 1 basis. Bob will have the opposite particle, no matter what Alice measures. So, if she measures a 0, he gets a 1, and if she measures a 1, he gets a 0. if she measures in the plus or minus basis, Alice gets a plus, and Bob will always get a minus. If Alice gets a minus, Bob will always get a plus. Because Alice and Bob can be reasonably far apart and can measure at the same time with relativistic speeds, Bob's particle cannot know whether Alice measured a 0 or 1 when Bob measures it. If Alice can predict what Bob gets, which she can because she can predict in the plus, minus basis if she measures in the plus, minus basis. If she can predict on the 1, 0 bases if Bob measures on the 1, 0 bases. If Alice can always predict 100\% what Bob will do, then there must be something in the quantum state that tells what Bob will get when he measures it. So, we would think that there is something in this quantum state that tells whether it is going to be a 0 or 1 when Bob measures it. However, we know, according to quantum mechanics, that there is nothing in this state that tells whether it is going to be a 0 or a 1. So, what that means is that there must be a more complete theory of reality than quantum mechanics. So, this is what Einstein argued. Schrodinger wrote a paper that answered that. Schrodinger's paper said that qubits particles are verschrankten, which is a German word that means crossed arms or crossed legs, or intertwined fingers. That is why this happens, which is not that great of an explanation. Then Schrodinger, we know, later translated verschrankten into English as entangled, because crossed would not have worked. Nobody paid much attention to Einstein's argument for the next 25 or 30 years or so. Einstein was never satisfied with quantum mechanics completely. However, quantum mechanics worked. So, they ignored this objection to it. Then, in 1964, Bell published a paper that said Einstein was both rights in that there is something very strange going on here. However, wrong in that, there is no way to complete this theory to a real theory because the arguments of quantum mechanics contradict classical probability theory.
 
\section{Bell's Argument as a Classical Game} 
we are going to present Bell's argument first. Originally, it is a statement about the marginal probabilities of a joint probability distribution. Nevertheless, we find it makes much more intuitive sense if we discuss it as a game that two players are cooperatively trying to play. It is called the CHSH game with Alice and Bob. So, Alice and Bob are two of the prototypical participants in cryptosystems. So, they are playing a game. They want to cooperate to win with as high a probability as possible. So, we have a referee who does not make any decisions. He just flips coins and does what the instructions tell him so, and he sends a bit to Alice and Bob. So, we will have him send A to Alice and B to Bob. Then Alice and Bob, who are in separate rooms and cannot communicate, get these two bits and output two more bits x, y, two bits. We should say the referee chooses his bits completely at random. So, now, we need to tell us when people win and lose to determine the game. So, Alice and Bob win if x plus y equals a times b. we should say x plus y mod 2. So, that is an x exclusive bar y. So, that should be pretty clear, but we want to put up a little chart. So, a can either be 0 or 1 b can either be 0 or 1. Alice and Bob win if x plus y equals 0 x plus y equals 0, x plus y equals 0, and x plus y equals 1. Now let us pretend that Alice and Bob have to do this deterministically. So, Alice has a little chart. If x equals 0, output 0. If x equals 1, output 1 or something like that. There is a general theory in games like this, Alice and Bob cannot do any better with a probabilistic strategy than a deterministic strategy. We will get to that later. So, this is x0 plus y0 x0 being the bit that Alice outputs if a is 0. It is x0 plus y1. It is x1 plus y0, and this is x1 plus y1. These are all mod 2. So, this is what Alice and Bob want. Now, can anyone tell why they cannot win with probability 100\%? That is right. So, we sum everything up. That is 2x0 plus 2x1, plus 2y0, plus 2y1 is equal to 1 mod 2. That does not work, because there are 2 x1's in this chart, 2 x0's, 2 y1's, and 2 y0's. Alice and Bob can only win with a probability of 75\%. Why is this? So, there is no choice of x0 and x1 for Alice and Bob x0, x1, y0, and y1 that makes this perfect. So, let us suppose that Alice and Bob have agreed in advance to some combination of x0, x1, y0, and y1. The referee comes along and sends them two bits, which randomly select one of these four boxes. We know one of these four boxes has to disagree with what Alice and Bob want because they cannot all work. So, at least one has to be wrong. If one is wrong, and the referee picks that one, which he does 1/4 of the time, then Alice and Bob lose. So, that is 75\%. Now, we want to argue why a random strategy cannot do any better. So, for a random strategy, we have a probability that they win equals the sum probability of So, random strategy, we can always think of this as Alice, and Bob gets some set of random bits. We do not care whether They are correlated or not. Then, for these random bits, they pick a deterministic outcome of x0, x1, y0, and y1. So, the probability that Alice and Bob win with a random strategy is equal to sum times the probability of string r, random bits, times the probability that Alice and Bob win given r. So, this is essentially an average of how often they win given any particular string of random bits. So, if this was greater than 3/4, there must be some string of random bits, which gives Alice and Bob a deterministic strategy of winning with probability at least 3/4. We just argued that it was impossible. So, that is the classical part. So, now we want to say that if Alice and Bob have an EPR pair of qubits 0,1 minus 1,0, then they can win with probability larger than 75\%. The problem with 1 over 2, 0,1 minus 1,0 is that Alice and Bob get opposite answers every time. We find it much harder to think about getting opposite answers than getting the same answer. So, we want to find a way that Alice and Bob can arrange to get the same answer. In the CHSH game, Alice and Bob have spatially separated from each other from the time that the game starts until it is over, and they cannot communicate. The game starts when the referee randomly selects two bits, a and b, and sends a to Alice and b to Bob. Alice outputs the bit x, Bob outputs bit y, and both send their bits to the referee. After receiving the output bits x and y, the referee determines if the input and output bits satisfy the condition $ x\oplus y=ab $
If the condition is satisfied, then Alice and Bob win; otherwise, they lose.

 \section{EPR Pair Violation: Introduction} 
Alice and Bob share 1 over root 2 0 0 plus 1 1. Fact, if Alice and Bob measure the form $ \alpha $ 0 plus $ \sigma $ x and $ \sigma $ z so, they both make the same measurement, which is on this great circle of the Bloch sphere they get the same outcome. if they measure with the observable $ \sigma $ y, they get opposite outcomes. So, we can easily check that proof for $ \alpha $ equals 0 and $ \alpha $ equals 1. here, we probably should have said $ \alpha $ squared plus $ \beta $ squared is one, $ \alpha $, $ \beta $ real. So, proof for $ \alpha $ equals 0 and $ \alpha $ equals 1 $ \alpha $ equals 1, that is $ \sigma $ z. they just measure in the 0 1 basis, which means they either get both 0 or they both get 1. $ \alpha $ equals 0 is easy. $ \alpha $ equals 1 that is 1 over root 2 0 1, 0 0 plus 1 1. we are going to put another one of root 2 here. So, this is Alice's state, and this Alice Bob, Alice Bob. When we multiply this by this, we get 0 Bob. we multiply this by this, and we get 0 1 Bob. there is a 1/2 in front of here. similarly, if we had a minus here and a plus here, we would get a minus there. So, when we take the projection onto 0 plus 1, we get 0 plus 1. When we take the projection onto 0 minus 1, we get 0 minus 1. the proof for arbitrary $ \alpha $ is not so hard. we are not going to go through it right now for lack of time. However, we want to use this. What is our strategy? Well, so, 0 1 plus-minus and these are the points that are 45 degrees. So, this is the slice, vertical slice through the Bloch sphere. So, Alice chooses 0 1 or plus-minus. So, Alice and Bob are going to measure each going to measure their half of this state. we want to say, Alice chooses 0 1 or plus-minus with a probability depending on what bit a was. So, a equal 0 and a equal 1. we need variables. So, let us call this s and this axis t. So, Bob chooses s and the opposite of s, if b equals 0. he chooses t an opposite state of t if b equals 1. 
\section{EPR Pair Violation: Protocol} 
We want to say the first column here, the output is going to be 0, and the right column here will be the output is going to be 1. So, let us label this diagram. So, this is Alice, Alice, Alice, Alice, Bob, Bob, Bob, Bob. here, this is 0. we want a equal 0, a equal 1, b equals 0, and b equals 1. the outputs are going to be 0, 0, 0, 0. So, this is x. This is y. This is x. This is y. x equals 1, y equals 1, x equals 1, y equals 1. So, Alice gets a bit from the referee, and if the bit is 0, she chooses this basis, and if the bit is 1, she chooses that basis. Ditto from Bob. if Bob gets this state, he outputs 0, and if he gets this state, his outputs 1. So, suppose Alice gets a equal 0, and Bob gets b equals 0. Alice gets observable $ \sigma $ z, and now we see that we should never have used x and y as the bits, because we have $ \sigma $ z, which is and Bob measures the observable 1 over 2, $ \sigma $ x plus $ \sigma $ z. So, suppose Alice measures 0, and Bob has qubit 0. He measures it with $ \sigma $ x plus $ \sigma $ z over 2. So, what is the probability that he gets the eigenvector plus 1 for this? One way to do this is to compute the eigenvector plus 1 and take the inner product of it with 0 and square it. However, there is an easier way. It should be a root 2. The expectation of Bob's value is $ \sigma $ x plus $ \sigma $ z over root 2 times 0 is 0. So, how do we compute that? Well, it is just 1, 0. That is 0. 1, 1, 1, minus 1. So, that is $ \sigma $ x plus $ \sigma $ z. Because $ \sigma $ x was 1. 1 here. And $ \sigma $ z was 1 minus 1. Must put a 1 over root 2 in front of that, times. It should have been a row vector, and this should be a column vector. These times that is 1, so this is 1 over root 2. Nevertheless, this is equal to the probability that Bob gets 0 minus the probability that Bob gets 1 because the expectation of this observable, and this observable has two eigenvalues, 1 and minus 1. The probability and the one eigenvalue corresponds to Bob getting a 0 are getting y equals 0, and the minus eigenvalue corresponds to an ability that Bob gets y equals 1. Thus, if we want two numbers whose difference is 1 over root 2 and whose sum is 1 because of the probability sum to 1. So, this is 1/2 half plus 1 over root 2, and this is 1/2 minus 1 over root 2. So, this is the probability that Bob wins if Alice measures a 0. So, what is that probability? It is 0.854 roughly. let us go back and look. it is also equal to the cosine of $ \pi/8 $ squared. So, let us go back and look at this picture again. That was the probability that Alice and Bob won if they, we know if they got a equal 0, b equals 0. Well, the probability that we win if they get b equals 0 and a equal 1 is also Yes? 1 over 2 roots 2 and 1 over 1/2 minus 1 over 2 roots 2. Thank we very much. However, this is 0.854. 
\section{EPR Pair Violation: Violation} 
So, let us go back and look at this diagram some more. That was the probability Alice and Bob win if they use these two axes, or these two yes, these two points on the Bloch sphere, which have, differ by an angle of $ \pi/4 $. But we know. If they use these two points on the Bloch sphere, which also differ by $ \pi/4 $, they also win without probability, because the only thing that matters when telling whether two axes on the Bloch sphere agree is the angle between the points. For b equals 1 and a equal 1. Well, how about b equals 1 and a equal 0? we know that a equal 1 and b equals 1, Bob gets 0, and Alice gets 1. So, this point and this point, and we want to know how often does occur together. Well, they occur together with 1 minus the probability that these two give the same answer, because y is 0, and here y is 1. So, x and y agree for these exactly when x and y disagree for these. Nevertheless, x and y disagree exactly when, but we know, if Alice will get this answer and Bob will get this answer, the probability is also 0.854. So, that means that Bob and Alice got (0,1), (0,0) and (1,0), they wanted their bits to agree. If they chose a equal 1 and b equals 1, they want their bits to disagree, and they disagree with exactly the same probability that these two things would agree. We do not think we said that very well, but hopefully, we understand what we mean. X equals 0, and b equals 1 are the vice versa, x plus y equals one with a probability of 0.854. So, this is what we wanted to happen. So, Alice and Bob can win this game with a probability of 0.854, which is bigger than 0.75. There are two more things we want to say. Let us draw another thing. If this is $ \theta $, these two observables agree with probability cos squared $ \theta/2 $. So, that is cos squared $ \pi/8 $. However, we know, if Bob, we know, if we have measured these two observables simultaneously, the probability that they would be agreed would be called cos squared $ \pi/4 $, which is 1/2. We know that these two observables, we know, that Alice measures 0 and Bob measures plus, they agree with the probability of 1/2. These two observables never agree, and that is cosine squared $ \pi/2 $. So, this is the general formula for, we know, if Alice and Bob have an entangled state, and they measure something, two observables $ \theta $ apart, its cosine squared $ \theta $ over 2. we are not going to prove that. Yes. So, this means that this particular strategy is the optimal one? That is the next thing we were going to say. We can prove that this particular strategy is the optimal one. They cannot do better than 0.854 at this game, and that is well. We are not going to prove it in this section because it is a little more complicated than everything we have done so far. However, we think it is the one who proved it. So, now this gives a game that Alice and Bob can win with higher probability if they have an entangled state than if not.

We generally observe and accept that quantum systems behave differently than classical systems. However, in practice, how are they different? Furthermore, are there means to quantify or even prove this distinction? For example:

1. Is the seemingly ``weird behavior'' (i.e., non-classical behavior) associated with quantum systems simply a result of quantum mechanics being somehow an ``incomplete description'' of the system behavior implying that, if we had a complete description, then such quantum ``weirdness'' would be understandable and consistent with our classical notion of proper behavior and no longer strange?

2, Or, is quantum mechanics fundamentally different, implying that the quantum description is already complete, and the observed ``weirdness'' is fundamental?

The weirdness of quantum mechanics and quantum entanglement, in particular, has bothered many great minds since the advent of quantum theory. Most notably, in 1935, Albert Einstein, Boris Podolsky, and Nathan Rosen (EPR) presented a thought experiment on quantum weirdness, which initiated a several-decades-long debate on the theory of quantum mechanics itself.

The thought experiment called the EPR paradox centers around two, distant entangled particles; for example, two polarization-entangled photons, one in Alice's lab and one in Bob's lab, with the labs separated by a very large distance. Let us take as an example the entangled state $ \vert \Psi ^- \rangle =(1/\sqrt {2}) \left(\vert H\rangle _ A \vert V\rangle _ B-\vert V \rangle _ A \vert H \rangle _ B\right) $, where $ \vert H\rangle $ and $ \vert V\rangle $ are the two basis states for the polarization of the photon, which stand for horizontal and vertical, respectively. As usual, $ \langle V \vert H\rangle = 0 $.

We know that when Alice and Bob measure their photons using the same measurement basis, they will have correlated measurement results. For example, if Alice measures H polarization, then Bob will measure V polarization and visa versa. It is not yet the weird part, as we know that if one photon is H, then the other must be V.

Now, if Alice and Bob choose to measure in different measurement bases, then the result may change. For example, let us say Alice again measures in the H-V basis, and Bob decides to measure in the A-D basis, where $ \vert A/D \rangle \equiv (1/\sqrt{2}) \left(\vert H \rangle \pm \vert V \rangle \right) $. In this case, no matter what Alice measures whether an H or a V Bob will always have a 50/50 chance of measuring A or D. In other words, there is no longer any correlation between Alice's result and Bob's result.

Here is where it gets weird. Assume that Alice and Bob are separated by a very large distance much larger than the distance it would take light to travel in a reasonable amount of time such that choice of Alice's measurement basis cannot reach Bob in time to influence his choice of basis or measurement outcome. We will assume that Alice and Bob each independently, randomly, and at the same time, decide a basis to measure in, either the H-V basis or the A-D basis. In this situation, the choice of Alice and Bob's measurement bases whether the same or different bases will affect the outcome of their measurement, but there is no way (even traveling at the speed of light) that any information about one's choice of basis could influence the other's measurement! Nevertheless, it does: the presence or absence of observed correlations depended on their independent choices.

Einstein, Podolsky, and Rosen did not like this seemingly ``spooky action at a distance,'' asserting that Alice's measurement cannot possibly influence Bob's measurement due to the large distance. The only explanation for the observed correlations is the underlying ``hidden variables'' that are not captured in the quantum theory that predetermines the measurement result. In other words, quantum mechanics is incomplete.

In response, Niels Bohr, a contemporary of EPR, argued that the theory of quantum mechanics merely describes the interaction of the measured quantum system with the measurement apparatus and not the intrinsic character of the quantum system itself. Therefore, one cannot assume that measurements do not disturb a quantum system. Hence, EPR's conjecture that quantum theory must be incomplete without hidden variables does not warrant a rather uninspiring explanation.

In 1964, John Bell developed a theorem Bell's inequality based on classical probability arguments that could distinguish whether quantum mechanics is incomplete requiring hidden variables or is indeed fundamentally correct ``as is''. Five years later, John Clauser, Michael Horne, Abner Shimony, and Richard Holt presented a physical method to test a specific case of Bell's inequality theorem, referred to today as the CHSH inequality. Essentially, if observed experimental results respect the CHSH inequality based on classical probability arguments, then a ``hidden variables''-type argument must be invoked. In contrast, if experimental results do not satisfy the inequality referred to as a ``violation'' of the inequality, quantum mechanics is a complete theory.

\begin{figure}[H] \centering{\includegraphics[scale=.3]{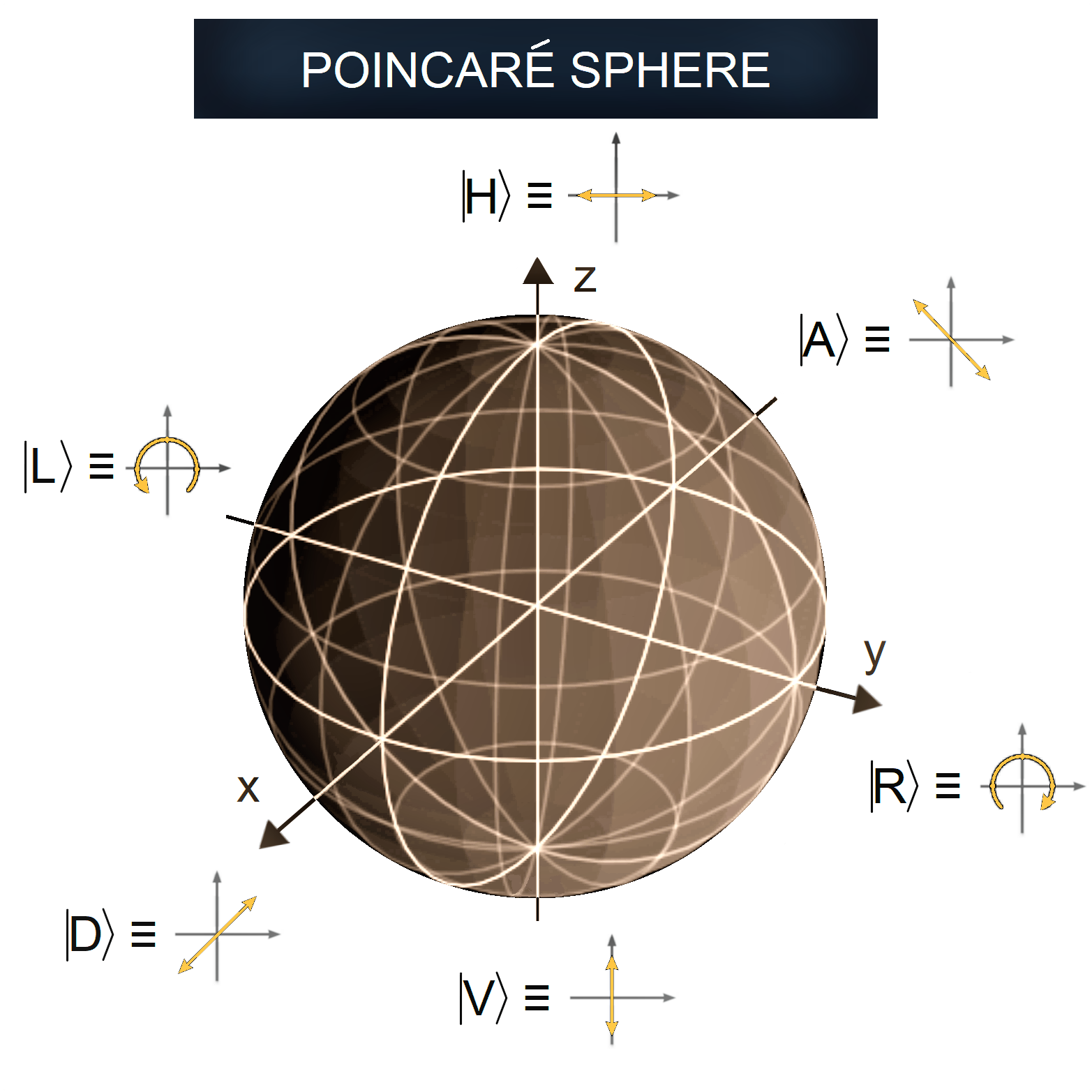}}\caption{Poincare}\label{fig3_6}
\end{figure}

Let us consider an example of the CHSH inequality. Suppose there is a source creating pairs of polarization-entangled photons such as $ \vert \Psi ^- \rangle =1/\sqrt {2}\left(\vert H\rangle _ A \vert V\rangle _ B-\vert V \rangle _ A \vert H \rangle _ B\right) $. One photon of the entangled pair is routed to Alice and one to Bob. Alice and Bob simultaneously measure their photons in their distant labs upon arrival of their share of the entangled photon pair. The measurement scenario is a bit different than the one described above. Alice randomly measures her photon either along the Poincaré sphere's z-axis horizontal $ \vert H \rangle $ or vertical $ \vert V \rangle $ polarization or along the x-axis diagonal $ \vert D \rangle $ or antidiagonal $ \vert A \rangle $ polarization denoted as$  M_ A^ z $ and $ M_ A^ x $, respectively. Bob, on the other hand, randomly measures his photon along with a different pair of axes either along the axis halfway between the z- and x-axes or the axis halfway between the z-axis and the negative x-axis here referred to as $ M_ B^{z+x}  $and $ M_ B^{z-x} $, respectively. Using any of these four bases, a measurement will project out either ``up'' or ``down'' along that axis, corresponding to 1 or -1, respectively.

According to the EPR argument, measurement results depend on underlying ``hidden variables'' such that a measurement of one party has no ``non-classical'' influence on the other party's measurement. To see this, let us define a quantity: $ Q=M_ A^ z(M_ B^{z+x}+M_ B^{z-x})+M_ A^ x(M_ B^{z+x}-M_ B^{z-x}) $ and make a table of all possible measurement outcomes with the EPR assumptions.

\begin{table}[H]
\centering
\caption{All possible measurement outcomes with the EPR assumptions}
\label{tab:3_1:Table 1}
\resizebox{\textwidth}{!}{
\begin{tabular}{|c|c|c|c|c|c|}\hline
\multicolumn{2}{c}{\multirow{2}{*}{Measurement Results and Expectation Value for Q}} & $ M_\textrm{A}^z\Rightarrow 1 $  & $    M_\textrm{A}^z\Rightarrow1 $ & $ M_\textrm{A}^z\Rightarrow-1 $ & $M_\textrm{A}^z\Rightarrow-1 $  \\ 
\multicolumn{2}{c}{} & $ M_\textrm{A}^x\Rightarrow1 $ & $ M_\textrm{A}^x\Rightarrow-1 $     & $ M_\textrm{A}^x\Rightarrow1 $ & $ M_\textrm{A}^x\Rightarrow-1 $ \\\hline
$ M_\textrm{B}^{z+x}\Rightarrow1 $        &                                             $ M_\textrm{B}^{z-x}\Rightarrow1 $                        & 2 & 2 & -2 & -2 \\\hline
$ M_\textrm{B}^{z+x}\Rightarrow1 $    &                                         $ M_\textrm{B}^{z-x}\Rightarrow -1 $                        & 2 & -2 & 2 & -2 \\\hline
$ M_\textrm{B}^{z+x}\Rightarrow -1 $    &                                          $ M_\textrm{B}^{z-x}\Rightarrow1 $                               & -2 & 2 & -2 & 2 \\\hline
$ M_\textrm{B}^{z+x}\Rightarrow -1 $    &                             $ M_\textrm{B}^{z-x}\Rightarrow -1 $                                    & -2 & -2 & 2 & 2 \\ \hline
\end{tabular}}
\end{table}

As seen from the table, over many measurements, the expectation value E(Q) must satisfy:
\begin{equation}\label{eq3_64}
-2\leq E(Q)\leq 2
\end{equation}

\begin{equation}\label{eq3_65}
-2\leq E(M_ A^ z M_ B^{z+x})+E(M_ A^ z M_ B^{z-x})+E(M_ A^ x M_ B^{z+x})-E(M_ A^ x M_ B^{z-x})\leq 2
\end{equation}

The expanded version of E(Q) is called the CHSH inequality.

As intended, results which satisfy the inequality are consistent with classical probability arguments; the previous paragraph ignored the impact of quantum mechanics on measurement outcomes. If we instead now assume that measuring the polarization of a photon is a quantum measurement and therefore requires a quantum mechanical description, the expectation value of $ \langle Q \rangle $ yields a different result. For example, a calulation of $ \langle \hat{M}_ A^ z\hat{M}_ B^{z+x} \rangle $ yields:

\begin{equation}\label{eq3_66}
\begin{split}
\displaystyle \langle \Psi ^-\vert \hat{M}_ A^ z\hat{M}_ B^{z+x} \vert \Psi ^-\rangle \displaystyle  =    \displaystyle \frac{\langle H\vert _{A}\langle V\vert _{B}-\langle V\vert _{A}\langle H\vert _{B}}{\sqrt {2}} \left(\hat{z}\otimes \frac{\hat{z}+\hat{x}}{\sqrt {2}}\right) \frac{\vert H\rangle _{A}\vert V\rangle _{B}-\vert V\rangle _{A}\vert H\rangle _{B}}{\sqrt {2}}    \\      
\displaystyle  =    \displaystyle \frac{\langle H\vert _{A}\langle V\vert _{B}-\langle V\vert _{A}\langle H\vert _{B}}{\sqrt {2}} \frac{\vert H\rangle _{A}(\vert H\rangle _{B}-\vert V\rangle _{B})+\vert V\rangle _{A}(\vert H\rangle _{B}+\vert V\rangle _{B})}{2}          \\ 
\displaystyle =    \displaystyle \frac{1}{\sqrt {2}}\frac{-1-1}{2}={-\frac{1}{\sqrt {2}}} = {\langle \hat{M}_ A^ z\hat{M}_ B^{z+x} \rangle }
\end{split}
\end{equation}

And, one can similarly derive: $  \langle \hat{M}_ A^ x\hat{M}_ B^{z+x} \rangle =-1 / \sqrt {2}, \langle \hat{M}_ A^ z\hat{M}_ B^{z-x} \rangle =-1 / \sqrt {2} $, and $ \langle \hat{M}_ A^ x\hat{M}_ B^{z-x} \rangle =1 / \sqrt {2} $. The resulting expectation values following the rules of quantum mechanics violate the CHSH inequality.
\begin{equation}\label{eq3_67}
\langle Q \rangle = \langle \hat{M}_ A^ z\hat{M}_ B^{z+x} \rangle +\langle \hat{M}_ A^ z\hat{M}_ B^{z-x} \rangle +\langle \hat{M}_ A^ x\hat{M}_ B^{z+x} \rangle -\langle \hat{M}_ A^ x\hat{M}_ B^{z-x}\rangle ={-2\sqrt {2}<-2}
\end{equation}

With such a ``yes/no'' inequality in hand, one can experimentally test whether the calculated inequality is satisfied or not. To date, numerous examples of experiments have violated Bell's inequality using a number of physical systems.

However, as we might expect, scientists over the years have pointed out ``loopholes'' in the experimental realizations of the inequality test, such as the measurement loophole, or the communication loophole, and related to the fact that our detectors are not perfect, or the distance may not be large enough to avoid speed-of-light communication. Recently, all of these loopholes have been closed except one: the ``free-will'' loophole, which essentially questions whether or not Alice and Bob can truly pick their measurement bases independently and at random in other words, do we have ``free will''. Nonetheless, with this one exception, all experimental results are consistent with a quantum description of the world, and hidden variables are not required to describe physical reality.

\section{Quantum Repeaters: Introduction} 
In the section, we have looked at quantum communication and the promise for quantum advantage in applications such as secret sharing, networking, and authentication. we have also looked in detail at how to implement quantum key distribution using single and entangled photon polarization states. For many of these schemes, we have seen that entanglement is a resource that needs to be distributed between two or more parties separated by some distance in order to implement the protocols. In this section, we will address a practical concern that we will face in any realistic quantum communication implementation. Namely, over long distances, how do we ensure that our photons and quantum information make it through the communication channel without getting lost along the way? While this might seem like a trivial concern at first glance, we know from the no-cloning theorem that quantum states cannot simply be amplified in the way that we amplify classical signals, and so, the nature of quantum mechanics leads to a fundamental complication which demands a more clever solution. as we will see, here again, quantum mechanics takes away with one hand and gets back with another. Consider the following. Alice and Bob want to establish a secure communication link using a QKD protocol. Both Alice and Bob have access to the required photon sources, detectors, and other tools that we have discussed previously. There is just one problem. Alice and Bob live in cities that are 500 kilometers away from one another. To complicate matters, there is a large mountain range in between these two cities which prevents any direct free space communication between the two. There are other issues with free-space communication, but let us go with the scenario. Free space communication is just not possible here. Thankfully, the two cities are connected by a run of optical fiber. just like the fiber optic cable that an internet provider uses to deliver high-speed web access, this cable is simply a flexible transparent glass fiber designed to guide and transmit light over long distances. So, in principle, all Alice and Bob need to do is plug their photon sources and detectors into each end of the fiber, and it will guide their photons from one city to the other, winding over and between the mountains in between. The practical issue is that, as we may already know or may have surmised, optical fiber does not transmit light perfectly. After a photon enters one end of the fiber, as it is guided along, there is a small but non-negligible probability that the photon will be lost along the way and not make it out the other side. It is usually due to a combination of incoherent scattering and photon absorption due to the various types of inhomogeneity in the glass. For typical optical fibers, the losses are reably low, only about 0.2 dB per kilometer. There is only a 5\% chance of a photon getting lost for a one-kilometer length of the fiber. so, that is not too bad. However, the problem is that this loss scales exponentially with the length that the cable. If we assume a constant loss of 0.2 dB kilometer and calculate the loss, we would expect over 500 kilometers connecting Alice and Bob, and we find that the probability of a single photon successfully making it through is only about one part in 10 billion, and continues to plummet as we add each additional kilometer. Now, we use optical fiber for intercity and classical transoceanic communication all the time. So, we have solved this problem for classical optical signals. For these signals, we overcome the intrinsic loss by installing amplifiers periodically throughout the fiber. When a given photon successfully travels through a length of the fiber and reaches an amplifier, it is amplified, or duplicated into multiple copies, which then continue through the fiber. Now, many of these copies will subsequently be lost in the next section of fiber. However, as long as the amplifiers appear often enough, some fraction of the photons will succeed in making it to the next amplifier. in this way, the loss is overcome, and the signals faithfully transmitted from one end to the fiber to the other. That is classical optical communication. We know from this section that this type of amplification will not work for quantum communication due to the no-cloning theorem. Let us say Alice preparers a single photon in a particular but arbitrary polarization stateside. The photon travels a length of the fiber and encounters some amplifier, which we hope would somehow duplicate the photon and its quantum state. However, this sort of amplifier violates the no-cloning theorem. There is simply no physical device that can take an arbitrary quantum state and generate multiple copies. So, classical amplification is not going to help overcome photon loss in our fiber. Fortunately, all is not lost. As it turns out, we can use quantum entanglement to our advantage here. As we have seen previously, entanglement can be swapped between pairs of qubits\cite{paler_influence_2019}. we can use a variant of entanglement swapping to teleport the state of one qubit onto another qubit without the two qubits ever directly interacting. To see how this helps, we will take a step back and review how quantum teleportation works. we will start with Alice and her polarized photon at the stateside, which she is trying to send to Bob. Imagine that halfway down the fiber line connecting Alice and Bob, we install an entangled photon source. This source generates two photons in a maximally entangled bell state. the two photons are sent in opposite directions through the fiber, one to Alice and one to Bob. So, Alice now has two photons. The photon whose stateside she wants to send to Bob, and one photon of four entangled pair that she shares with Bob. Alice now takes her two photons and performs a Bell state measurement. Regardless of the initial state of the two photons, this measurement process untangles the two photons and projects them onto one of the foreign Bell state measurement outcomes. It yields two classical bits of information. So, what happens to Bob's photon? Alice's measurement has the remarkable effect of projecting Bob's photon onto one of four disentangled states, depending on the outcome of Alice's Bell measurement. as we can see, the state of Bob's photon is now essentially the same state as Alice's original photon, say, for one of four different rotations corresponding to Alice's for measurement outcomes. To undo these rotations, Alice simply needs to call Bob on the phone and tell him the outcome of her measurement, the two classical bits of information that she gleaned. Bob then performs the necessary inverse rotation and recovers the state of Alice's original photon. So, what did all this buy us? Notice that by implementing this protocol, Alice successfully transferred the quantum state of her photon to Bob 500 kilometers away. However, none of these photons ever had to travel more than 250 kilometers, half the fiber length. since photon loss scales exponentially with length, even cutting the distance in half has dramatically increased the likelihood that the photons will make it through the fiber. we do not need to stop there. we can imagine placing many such sources periodically along the length of the fiber, swapping entanglement pairwise along all of the nodes, and thereby establishing a quantum link between Alice and Bob along which a quantum state can then be teleported. Each node in this network of entangled photon sources and detectors is called a quantum repeater. with each additional quantum repeater, the average distance that each individual photon needs to travel decreases linearly. It, in turn, exponentially decreases the probability of a photon getting lost in the fiber. Thus, although we never amplify the quantum signal, we do prevent it from being lost. Now, there are other challenges associated with building quantum repeaters. As we discuss, the generation of entangled photons is stochastic, and so, entangled photons generally will not arrive at each node simultaneously. Thus, we need a means to store quantum information at each node, effectively a quantum memory, until all entangled pairs are present, and the entire link is ready to be established. we may also want to error correct that memory in order to increase the overall likelihood of success. Although there is still much work to do, quantum repeaters will enable long-distance entanglement distribution in quantum communication. As we just discussed, with only a linear increase in resources, we could exponentially suppress the loss of quantum information over long distances. In the next section, we will discuss more quantum repeaters and their application to quantum networks \cite{dahlberg_link_2019,shang_continuous-variable_2019,liao_satellite-relayed_2018}. 

Long-distance quantum communication is challenging due to photon loss inherent in realistic optical channels. This problem is dealt with classically using amplifiers, known as ``repeater stations,'' which are interspersed along the transmission channel to compensate for the reduction in signal power due to loss. Unfortunately, the no-cloning theorem prohibits the amplification of an unknown quantum state, as it would allow for a single photon to be copied into more than one photon carrying the same quantum information. Therefore, classical amplification techniques cannot be used to extend the range over which quantum information can be transmitted.

Optical fibers are made of silica (Si$\text{O}_{2}$), and it has two leading loss mechanisms which dominate at different wavelengths. At shorter wavelengths, the primary loss mechanism is elastic scattering, in the form of Rayleigh scattering, which can change the propagation direction of the light enough that it escapes the fiber. Alternatively, at longer wavelengths, the dominant loss mechanism is absorption by the material itself. At approximately 1550 nm, the combined loss due to both of these effects is at a minimum of about 0.2 dB/km, a value which is referred to as the attenuation coefficient. Away from this optimal wavelength, the attenuation coefficient increases, defining a ``sweet spot'' around 1550 nm that is widely used for telecommunications today. It should be noted that the optimal attenuation coefficient for silica of 0.2 dB/km should be thought of as the lower bound for attenuation coefficients found in real fibers where other loss mechanisms such as absorption due to material impurities or fiber bending also contribute to the overall attenuation.

Fiber loss is approximately constant throughout a fiber, so the total attenuation is related to the transmission distance. For context, assuming the optimal attenuation coefficient for silica of 0.2 dB/km, 95\% and 1\% of the initial signal power remains after a transmission distance of 1km and 100 km, respectively. Unfortunately, the transmitted power scales very poorly with distance, and at 500km (100dB loss), only one in every $ 10^{10} $ photons is transmitted through the channel.

A sophisticated approach to overcoming the limits imposed by attenuation and the no-cloning theorem is the ``quantum repeater.'' Quantum repeaters operate on two entangled states that span consecutive distances of L/2, such that, through a process of measurement and entanglement swapping, the final state of the system is a single entangled state which spans the entire distance L. To be more specific, if we have two Bell states which span the L/2 distances, then a Bell state measurement on the two collocated qubits in the center along with classical communication of the measurement results to the end stations will result in the entanglement between the two outermost photons which span L. This process can be thought of as the quantum teleportation of a quantum state from the central station to one of the end stations where the initial entanglement of the state is teleported along with it. In general, this process can be nested as many times as needed to span a given initial distance L.

The final entangled state distributed by quantum repeaters can be used as a resource for many quantum communication protocols. For example, this state can now be used to teleport an unknown quantum state over the total distance L in a deterministic fashion. It could be contrasted with direct transmission, which would have only been successful if the photon had not been lost in transit over this entire distance. It is important to note that, in this example, the quantum repeaters have not only allowed for the transmission of a quantum system over a large distance the original aim but have also shifted the problem of probabilistic losses to a part of the protocol, e.g., the entangled-state generation which can be repeated until successful without risking the loss of the unknown state. In realistic quantum repeaters, this will be facilitated by local quantum memories that can store entangled states as they are generated until each node of the quantum link is established. 

\section{Quantum Repeaters: Applications} 
In classical communication systems, amplifiers, and the ability to repeat or regenerate, a signal is taken for granted to extend the transmission distance between the sender and receiver. Fundamentally, quantum communication is complicated by the no-cloning theorem for quantum information. It has been leveraged for very useful secure communication in quantum key distribution, or QKD, systems, but it presents a fundamental complication to long-distance quantum communication. QKD systems to date have therefore been limited to a maximum practical distance on the scale of 100 kilometers in optical fiber, far below the global reach of today's classical internet. The goals for communicating quantum information extend far beyond the challenge of increasing the maximum distance of QKD systems. As quantum computers increase in capability, an increasing need to interface and connect these systems is merging. The vision of a quantum internet has been an exciting prospect for the community for a long time, most famously spelled out in a 2008 Nature Review article by Jeff Kimble. On a practical level, protocols from blind quantum computing, which gives the true user and privacy guarantees in a cloud computing model, and distributed quantum computing, which may enable new computational scaling flexibility, require the exchange of quantum information. Within the space of quantum metrology and sensing \cite{takeuchi_quantum_2019}, long-range entanglement can be used as a resource to increase the precision of measurements. The situation is even further complicated because quantum information is often stored in a form that is very hard to transmit in the first place. The superconducting circuit qubit\cite{devoret_superconducting_2013} platforms that IBM, among many others, is seeking to develop store quantum information in a microwave state \cite{randall_efficient_2015,paraoanu_microwave-induced_2006} that is only free of a thermal background at extremely low temperatures. Other platforms store such quantum information as electronic or nuclear spins. In all of these cases, the information must first be transduced from the initial state to an optical photon through a quantum state preserving coherent process. Even in systems where direct optical interrogation of the quantum state is possible, such as in trapped ion qubits\cite{pino_demonstration_2020,schafer_fast_2018,harty_high-fidelity_2014,ballance_high-fidelity_2016,wang_high-fidelity_2020,gaebler_high-fidelity_2016,bruzewicz_trapped-ion_2019}, the optical frequency of the atomic transitions rarely matches with the required frequencies for free space or fiber communication and must be converted for transmission. The high state-energy, weak light-matter interaction in existing telecommunication infrastructure makes optical photons ideal carriers of quantum information. However, optical photons still suffer exponential loss in transmission, as described by the Beer-Lambert Law. In the absence of amplification, we can expect the error rate of transmission to increase exponentially with the distance of transmission. Technological limits, such as the dark count rate of the receiving single-photon detector, will set a maximum transmission distance for quantum information. To avoid these limitations, Knill and Laflamme introduced the first quantum computer proposal in 1996, based on the concepts of quantum error correction. Here, an intermediate node, analogous to classical communication amplifier, could be used to restore the received quantum information for retransmission. However, the constraint of the no-cloning theorem places a hard boundary at 50\% maximum link loss between repeater nodes. If we revisit the problem of attenuation and transmission of optical photons, there is a very inconvenient case. Even with that 0.2 dB per kilometer loss of modern optical fiber, the maximum distance between repeaters would be limited 15 kilometers in the most optimal case. In 1998, the authors of the BDCZ Proposal introduced a landmark quantum repeater protocol with three key elements. First, it used two-way communication between nodes to achieve entanglement purification, even when using imperfect components. Second, the protocol supported swapping the intermediate nodes and entanglement in a nested protocol to entangle the endpoints. Third, the computational resources required that each node had a modest scaling with the transmission distance. In this scheme, the goal is to establish entanglement between two remote nodes to be used as a resource for any communication, computation, or measurement application. The challenge in all of this is the requirement that the quantum computer nodes contain a quantum memory that remains coherent over the total communication latency needed to establish the entanglement. The magnitude of this quantum memory challenge resulted in an alternative proposal in 2001 known as the DLCZ Protocol. In this scheme, atomic ensembles are used as the long-coherence memory and entangled by measuring the emitted photons from an optical ray pulse through a beam splitter\cite{bouland_generation_2014}. These neighboring ensembles can then be entangled by sending a read pulse and measuring the emitted photons through a beam splitter, as before. At that time, ultra-cold atom systems were the only known implementation. Nevertheless, recent progress has extended this concept to rare-earth ions and crystals to form a solid-state memory that can be leveraged as a resource for a quantum repeater. Still, many technical challenges remain for a DLCZ-based link to offer an improved endpoint entanglement rate relative to direct transmission. As the field of quantum information science has matured over the past two decades, the theoretical requirements for a BDCZ-based quantum repeater protocol have been reduced. At the same time, the capabilities of existing hardware systems have increased. Today, robust quantum repeater protocols can support a fixed resource per node on the order of 100 qubits, with memory coherence times that are on the order of 100 milliseconds. Systems of this order are not far from what can be envisioned with a foreseeable roadmap of quantum computing systems. As such, a practical quantum repeater demonstration is now being discussed as the next achievable milestone in quantum information science\cite{preskill_quantum_2012}.

The secure communication of information on a global scale is fundamental to our digital world. As we have seen throughout contemporary encryption methods used to communicate information generally rely on an assortment of mathematical constructs that are presumed to be hard to compromise using available computational resources. As such, these conventional protocols are generally not proven to be perfectly secure. Rather, they are constructed to offer acceptable levels of security based on reasonable assumptions surrounding the mathematical problems and computing power available today\cite{williams_tamper-indicating_2016}. Alternatively, the various quantum key distribution (QKD) protocols offer the promise of communication with quantum-enhanced levels of randomness and security.

In QKD, two distant users generate a shared random bit string, referred to as a secret key, which can subsequently be used to encrypt a message using a symmetric-key protocol. The key generation process generally involves one user preparing quantum states of light and transmitting them to the other user, who then measures them. Subsequently, they compare what was sent versus what was measured via classical communication channels. Attempts to eavesdrop on the transmitted states will disturb the measured results in a detectable manner, enhancing the protocol's security.

A major drawback of current QKD implementations is its relatively limited operational distance compared with conventional classical-communication methods. For example, QKD realizations based on single or few-photon quantum states are limited to operational distances of at most a few hundred kilometers to retain any notion of a practical communication rate. Due to the exponential scaling of photon loss with distance relevant for both free-space atmospheric links and optical fiber links, a problem is generally addressed using distributed amplification in conventional classical communication schemes. In the quantum setting, however, conventional amplification is not an option due to the no-cloning theorem.

One possibility for overcoming photon loss in QKD systems is to use quantum repeaters. Quantum repeaters improve the rate-loss scaling of long-distance quantum communication by converting the problem of distributing a single quantum state over a long distance into one of distributing several quantum states over shorter distances and then coordinating them to create a long-distance quantum link. While quantum repeater research is progressing, many engineering challenges remain to be addressed before quantum repeaters will be ready for practical deployment.

In the context of free-space communication, the use of a satellite or a constellation of satellites as an intermediary between ground stations affords an improvement over solely terrestrial free-space communication. In these schemes, a ground-based station communicates first with a satellite orbiting the Earth. Although the overall distance from the ground station to the satellite remains considerable (several hundred to thousands of kilometers), the thinning of the atmosphere with altitude means that loss due to atmospheric turbulence and absorption decreases rapidly with distance. It significantly limits the overall loss compared with manifestly terrestrial free-space links of the same distance. Once the quantum information is at a satellite, it can then be passed between other orbiting satellites (e.g., as in a repeater scheme) with much lower loss before being transmitted back to Earth. Further, the use of satellites enables the option of having the satellite implement the QKD protocol based on keys shared independently with each ground station. For example, once a satellite establishes independent keys with each ground station, it can encrypt and transmit a shared key to both locations.

In 2016, China launched the satellite 'Micius' with the primary purpose of performing quantum information science demonstrations\cite{liao_satellite-relayed_2018}. Micius orbits the Earth at an altitude of 500 km and a speed of 7.6 km/s. Since its launch, Micius has been used to demonstrate QKD at kHz rates over satellite-to-ground distances of 1200 km. Originating the QKD protocol at the satellite, as opposed to the ground stations, is because of atmospheric turbulence. Turbulence causes ``beam wandering,'' the redirection of an optical beam due to passage through a turbulent atmosphere, and such atmospheric turbulence is largest near the surface of the Earth. Since a beam naturally broadens as it propagates, the wandering is a proportionately smaller effect if it occurs once the beam has already broadened (rather than while it is tightly confined at the source with a long distance still to go). For context, the beam used to communicate with Micius broadens to a diameter of 12 m after traveling 1200 km.

Micius has also been used to demonstrate QKD between two different ground stations simultaneously, one in Europe and one in China, separated by 7600 km on the Earth. Then, as described above, the satellite could use these two independent QKD sessions to create a shared secret key between the two ground stations, thereby performing intercontinental quantum communication. Finally, the shared key was used to seed a conventional AES (advanced encryption standard) protocol with 128 bits in order to perform 75 minutes of encrypted video conferencing between Europe and China, consuming a total of 70 kB of the shared key.

The Micius experiments definitively demonstrate that a global communications network based on QKD-enhanced key exchange is possible. However, many challenges remain before satellite QKD becomes practical; in particular, the need to increase communication rates from the kiloHertz range to more practically useful levels.

\section{QKD: Practical Issues and Challenges} 
In section two, we discussed quantum key distribution, or QKD, in which the BB84, Ekert 91, and BBN92 protocols were explained. All of these protocols enable two remote users, Alice and Bob, to accumulate a shared string of random bits while precluding an eavesdropper, known as Eve, from learning anything about those bits. Randomness, as we discussed, is a valuable resource. Shared randomness, in particular, is of great value for secure communication because, as we have also seen, it enables Alice and Bob to communicate in complete secrecy employing one-time pad encryption and decryption. It is especially important because the advent of quantum computers with their possibility of running Shor's factoring algorithm will break the RSA public key infrastructure on which internet commerce relies. Thus, it should come as no surprise that QKD systems are now commercially available and that QKD networks have been demonstrated in the United States\cite{raymer_us_2019}, Canada\cite{sussman_quantum_2019}, Europe\cite{riedel_europes_2019}, and Japan\cite{yamamoto_quantum_2019}. one is entering a large scale deployment in China. Nevertheless, there is a host of issues related to QKD day that awaits a solution before one-time pad communication becomes the norm. These include providing security against quantum hacking, increasing the distance over which QKD can be accomplished, and increasing the rate at which the secret key can be accumulated. QKD security proofs generally guarantee the security of the protocol, not its hardware implementation. It leaves open the possibility of quantum hacking, in which Eve exploits equipment vulnerabilities to break the protocol by gaining partial or complete information about Alice and Bob's key. In BB84, for example, Alice sends bits at the single-photon level to Bob, who performs polarization analysis on them using single-photon detectors. In a blinding attack, Eve illuminates Bob with laser pulses that enable her to control which detector will click, hence giving her what she needs to obtain information about the key. Measurement device-independent QKD eliminates Alice and Bob's vulnerability to photodetector hacking but at the expense of some implementation complications not present in BB84. More generally, anticipating all possible hacking attacks on a particular QKD implementation can be a daunting task. The exception is device-independent QKD, a protocol based on the Bell inequality, which makes no assumptions about the functioning of the equipment used in its implementation. The price paid for this immunity, however, is severe. Device-independent QKD's predicted secret key rate is far below what BB84 affords. its maximum distance of operation is much shorter than what BB84 offers. The distance over which a QKD system operates, and the secret key rate that it provides at that distance, are quintessential figures of merit. If an optical fiber links Alice and Bob with no intervening equipment, then a BB84 system in which Alice sends single photons at a rate of R per second, and Bob has a perfect single-photon detector, gives a secret key rate in bits per second that is proportional to R times the fraction of Alice's photons that reach Bob. For standard low loss fiber, that fraction is 10\% at a 50-kilometer distance. 1\% at 100 kilometers. 10-billionths of a percent at 500 kilometers. A direct fiber connection between Alice and Bob will not do for transcontinental or intercontinental QKD. One approach to reaching such distances is to employ quantum repeaters. Let us see how this can be done. Suppose Alice and Bob each have entanglement sources that produce a pair of polarization-entangled signal and idler photons. that Alice and Bob each send their signal photons to a repeater station while retaining their idler photons in quantum memories. The repeater station performs a Bell state measurement on the signal photons it receives and then uses a classical link to communicate that measurement outcome either 00, 01, 10, or 11 to Alice and Bob. They then will know that their stored idler photons are in a particular polarization-entangled state. This operation, called entanglement swapping, performed over a chain of such repeater stations can distribute entanglement and hence enable QKD between much longer distances than is practical with a direct fiber connection. Making such a repeater chain practical requires substantial technology developments in quantum memories and quantum processors. Hence, an alternative approach to long-distance QKD, the use of quantum satellites recently demonstrated by China, is worth considering. Putting aside for now extending the reach of QKD beyond metropolitan area distances, let us address the secret key rates that can be achieved within that more limited operational range. The state of the art is a BB84 system that provides a megabit per second class secret key rate over 50 kilometers of optical fiber. However, is this enough? For one-time pad encryption of large files to be transmitted at internet speeds, gigabit per second secret key rates should be available. To push the record-setting BB84 system to such rates at 50-kilometer range would require increasing its clock rate by a factor of 10 and multiplexing 100 wavelength channels, all operating at that increased clock rate. Next, we will look at a new QKD protocol for reaching gigabit per second secret key rates in a metropolitan area setting with a substantially lower implementation burden than 100 channel BB84. 
\section{Floodlight QKD}
In this section, we will continue the discussion of QKD. specifically, why BB84 and the other QKD protocols we have seen are incapable of a gigabit per second secret key rates at metropolitan area distances unless they use massive amounts of wavelength division multiplexing. we will also look at a new protocol that offers such performance without any multiplexing. To begin, however, let us explain how classical fiber optic communication easily achieves multi-gigabyte per second rates over such distances in single wavelength operation, and why those approaches are unavailable for BB84. Remember that without quantum repeaters, only 10\% of the BB84 light is entering 50 kilometers long, and low-loss optical fiber will emerge from that far end. for 100 kilometers long fiber, that fraction drops to 1\%. Classical communication systems overcome these losses by transmitting many photons per bit and using an optical amplifier when necessary to boost signal strength. BB84, like all QKD protocols, must be secure against an all-powerful Eve, which includes an Eve who collects all the light lost in propagation from Alice to Bob but does nothing that affects the light reaching Bob. Alice's transmitting her BB84 bits at the single-photon level offers protection against this passive eavesdropper, because quantum mechanics No-Cloning Theorem precludes Eve from perfectly duplicating Alice's photons. No Cloning, unfortunately, also impacts Bob, who is unable to noiselessly boost the signals he receives from Alice, because amplified, spontaneous emission noise will unavoidably be present in his amplifier's output. Floodlight QKD is a new protocol, in which Alice transmits many photons per bit, and Bob uses an optical amplifier\cite{zhuang_floodlight_2016}. However, it is immune to passive eavesdropping and is capable of a gigabit per second, secret key rates at metropolitan area distances in single wavelength operation. Floodlight QKD, unlike BB84, is a two-way protocol. Alice first sends unmodulated light to Bob. Bob encodes a random bit string on the light he receives, then amplifies that light, and sends it back to Alice for decoding. In particular, Alice uses an optical amplifier to produce terahertz bandwidth, high brightness, Amplified Spontaneous Emission light, containing tens of thousands of photons per second per hertz of optical bandwidth. She splits off a tiny fraction of that ASE light containing, on average, less than one photon per second per hertz of optical bandwidth. She transmits it to Bob while retaining the high brightness remainder in optical fiber for decoding the light she will receive from Bob. Bob imposes his random bit string on the light he receives at a multi-gigabit per second rate employing binary phase-shift keying, in which 0’s and 1s are represented by 0 and $ \pi $ radian phase shifts. Bob amplifies his modulated light to overcome propagation loss on the return path to Alice, which has the added benefit of bearing that amplified, modulated light in his amplifier’s high brightness, ASE noise. Nevertheless, Alice can successfully recover Bob's bit string by a coherent detection process in which he beats the light she receives from him against the ASE light she has stored on a conventional linear mode photodiode. That coherent detection affords Alice a processing gains equal to the product of our ASE light's optical bandwidth and Bob's bit time. Eve, operating as a passive eavesdropper, cannot decode Bob's bit string. The light she collected from the Alice-to-Bob channel is too low brightness for coherent detection to work, and the No Cloning Theorem forbids her from faithfully converting it to a high-brightness replica. Let us put some numbers together. Suppose that a pair of 50-kilometer-long optical fibers connect Alice and Bob. Alice sends 2 terahertz bandwidth light to Bob, and Bob uses 10 gigabits per second modulation, and a 40 decibel gain optical amplifier. Then, Alice is sending 20 photons for each of Bob's bit times at of brightness of 1/10th of a photon per second per hertz of optical bandwidth. Although Alice receives 2,000 photons of modulated light per bit embedded in 20,000 ASE photons per bit, her 200-fold processing gain, 1/10th nanosecond times 2 terahertz of bandwidth, makes that effectively 2,000 signal photons embedded in 100 ASE photons. There is, unfortunately, a very definite fly in the preceding QKD ointment. Because Floodlight QKD is a two-way protocol, Eve can injector her own broadband, low-brightness light into Bob's terminal, while retaining a suitable high-brightness reference for decoding Bob's modulation of her light, despite its being buried in the ASE from Bob's amplifier. To defeat this active eavesdropping attack, Alice and Bob perform a channel monitoring operation, in which they rely on the photon pairing mission from a spontaneous parametric down converter. Alice has that source and multiplexes its signal beam output in with the ASE she is transmitting to Bob while recording time-tagged values of idle or photon detections. Bob taps a small portion of the light he receives from Alice and records the time tag values of his photon detections. By sharing their photon detection times, Alice and Bob can calculate the fraction of light entering Bob's terminal that was due to Eve, and thus determine how much secret information they can safely distill. Where is floodlight QKD now? Tabletop demonstration of Floodlight QKD has demonstrated 1.3 gigabits per second secret key rate through a channel with attenuation that was equivalent to 50 kilometers of optical fiber. Work is continuing on both the theory and experiment of Floodlight QKD, with major tasks on the agenda being bringing the current security level up from a collective attack to the ultimate coherent attack and implementing the protocol on a deployed metropolitan area scale fiber. FL-QKD is immune to a passive-eavesdropping attack. This immunity arises from the high number of photons transmitted per bit over a more significant number of optical modes as compared to other QKD protocols, which transmit no more than one photon per bit. If we want to know more about FL-QKD, we encourage to go through this paper \cite{zhuang_floodlight_2016}.

\section{QKD: Implementation Challenges and Potential Solutions} 
The end goal of quantum key distribution, or QKD, is to establish a shared list of random numbers that can be used for encryption. This list of random numbers is called a key. After a successful QKD session, the transmitter, Alice, and the receiver, Bob, should end up with identical copies of the key. At the same time, an eavesdropper, Eve, should have no information about the random numbers in the key. One way to assess the performance of a QKD system is to look at its secret key rate, that is, how quickly the secret random numbers are shared in units of bits per second. The required secret key rate depends on the encryption algorithm that will use the keys and the desired rate of secure communication\cite{behnia_tachyon_2018}. The holy grail of secure communication is the one-time pad, an encryption algorithm that is mathematically proven secure. Even if Eve had the most powerful supercomputer in the world, or even if she had a quantum computer, she would not be able to crack the one-time pad. However, the one-time pad is a challenging encryption scheme to use because it consumes one bit of the secret key for each bit of the message to be encrypted. the secret keys cannot be reused. There are many different implementations of QKD transmitters and receivers. In general, the hardware limits how quickly the transmitter can produce quantum states and how quickly the receiver can detect them. For example, it might take a finite amount of time to prepare a quantum state. Alternatively, after a receiver detects a photon, it might take a finite amount of time to reset the receiver and prepare it to detect another photon. As a result, both the transmitter and receiver place limits on the achievable secret key rate. The secret key rate is also strongly affected by the quantum channel. A crucial issue that affects every real-world QKD system is the loss in the quantum channel. Because of loss, only a fraction of Alice's transmitted photons will reach Bob's receiver. The value of the fraction depends on the properties of the quantum channel. One of the most common channels is an optical fiber. For example, an illustration of an optical fiber link between Lincoln Laboratory and MIT is used for quantum communication experiments. Over an optical fiber, the probability of successful photon transmission decreases exponentially as the fiber gets longer. For example, if the quantum channel is 50 kilometers of standard optical fiber, we would expect only 10\% of the transmitted photons to reach the receiver. If the fiber length is doubled from 50 kilometers to 100 kilometers, then the transmission is reduced from 10\% to 1\%. These are the ideal transmission values for those fiber lengths, but real-world fiber optic links tend to have lower than ideal transmission. Our 43-kilometer fiber has between 2 and 1/2\% and 5\% transmission. In classical optical fiber communication, a decrease in transmission can be overcome by optically amplifying the signal. However, optical amplifiers cannot be used in quantum communication and QKD. Because according to the laws of quantum mechanics, it is impossible to make a perfect copy of a quantum state. It means that the transmission of the quantum channel fundamentally limits the achievable secret key rate. Many researchers are working on new technologies that can overcome this limitation, but a practical solution is still a few years away. In the meantime, QKD researchers are investigating different strategies to increase secret key rates using currently available technology. One strategy is to take advantage of the properties of optical fiber. A single optical fiber can guide many different wavelengths, or colors, of light at the same time. Alice and Bob could raise their secret key rate by running several QKD systems simultaneously, with each system operating at a unique wavelength. Using a technique called wavelength division multiplexing, Alice can combine the quantum outputs of all of her transmitters onto the same optical fiber, and Bob can separate the quantum signals so that a separate receiver detects each wavelength. The resulting secret key rate scales linearly with the number of multiplex systems. Two systems would result in double the secret key rate of a single system, and four systems would quadruple the rate of the single system, and so on. The drawback of this strategy is that we need multiple transmitters and multiple receivers. It could be challenging if we are operating under constraints on the available size, weight, power consumption, or cost of the system. Another strategy involves changing the way the photons are used to carry information. we have discussed encoding information in a photon's polarization state. For example, horizontal polarization could correspond to bit value zero, and vertical polarization could correspond to bit value one. Because there are two possible bit values, 0 and 1, the polarization encoded photon could contribute at most one bit of information to the secret key. It would be useful if there were a way to encode information in photonic states and get more than one bit of information from a single photon. It is called high dimensional encoding. As an example, if a photon could be in one of four possible states with bit values 0 0, 0 1, 1 0, and 1 1, then that photon could carry up to two bits of information. However, on a given basis, there are only two possible polarization states. So, instead of using polarization, Alice and Bob need to encode information in the photon's different properties. Several properties can be used, but not all of them are compatible with optical fiber channels. For example, a popular research area is encoding more than one bit of information using what is called the orbital angular momentum modes of a photon. However, those modes cannot be easily transmitted over optical fiber. Instead, for a fiber-compatible, high-dimensional encoding scheme\cite{setia_superfast_2019}, the transmitter can prepare the photon by putting it in one of many different time slots. The receiver can detect which time slot had the photon in it. Each time slot corresponds to a different bit value, and the total number of time slots determines the number of bits of information that each photon could carry. Four-time slots correspond to two bits per photon, eight times slots correspond to three bits, 16-time slots correspond to four bits, and so on. The more possible time slots there are per photon, the fewer possible photons can be transmitted per second. That is, there is a trade-off between how much information a photon can carry and how many photons can be transmitted per second. It is useful because, as discussed earlier, there is often a mismatch between the transmitter and receiver. So, the transmitter can produce quantum states much more quickly than the receiver can detect them. The photons that are not detected and the information that they carry are wasted and cannot contribute to this shared key. This high dimensional time slot scheme encodes the same amount of information into a smaller number of photons, which makes this scheme especially useful when the receiver is saturated. Practically speaking, this often happens over metropolitan area distances of a few tens of kilometers. There is a potential drawback to using high dimensional encoding for QKD. The transmitter and receiver become more complex compared to a binary encoded system. Depending on the situation, the potential gain in the secret key rate could be worth the additional complexity. we have discussed the constraints on the secret key rate achievable by a QKD system and about strategies to increase the secret key rate under those constraints. There is no single best QKD system or strategy that covers all operation scenarios. The most important point is that the optimal system and strategy depend heavily on the available hardware and the channel properties.

\section{Entanglement as a Physical Resource} 
In this section and previously, we have seen how quantum entanglement is used as a resource for various aspects of quantum communication, from key distribution to random number generation to quantum repeaters. In this deep dive, we will explore the resource model of quantum entanglement. We will start by asking what it means for entanglement to be a resource, both at an intuitive level, as well as in mathematical terms. We will then provide a general definition of entanglement and introduce mathematical measures, the entropy, and the Schmidt number that quantify the degree of entanglement in a given system. Lastly, we will show that entanglement is fungible. That is, entangled bipartite pure states are asymptotically equivalent and, therefore, constitute a physical resource. So, with that brief introduction, let us discuss about the resource model of quantum information theory and quantum entanglement. Space, time, and energy are fundamental physical resources. Does quantum information theory introduce additional such resources? For example, we have seen how noisy classical channels may provide a correlation between the two signals, x, and y. Shared classical randomness is known to be useful for cryptography. Might noisy quantum channels also be resources? Could entangled quantum states be resources? Well, let us see what entanglement may be useful. We have seen that teleportation \cite{podoshvedov_efficient_2019}may be utilized to turn one entangled pair, called an ebit, plus two classical bits into the ability to communicate one quantum bit. Such conversion is not possible without entanglement. Another well-known protocol is superdense coding by which an ebit and one qubit are used to communicate two classical bits. Entangled quantum states are also known to be useful for clock synchronization, distributed quantum computation, and cryptography, such as quantum key distribution. Entanglement may also sometimes replace classical shared randomness. So, given that entanglement is useful, the problem is, is entanglement a resource, and by this, we mean a formal mathematical definition of resource? For example, here are two entangled states, 0 0 plus 1 1 versus 0.9 square root 0 0 plus 0.1 square root 1 1. How are these the same or different? After all, one cannot have different kinds of resources for different states. For entanglement to be a resource, there must be a way to convert between states of different qualities of entanglement.
One strict criterion for a definition of a resource is to say that A and B are equivalent if there is a mechanism to convert from A to B and from B to A. Does this apply to entangled states? In fact, yes, to some degree. A state psi and another $ \phi $, which are pure bipartite states, are equivalent under local operations and classical communication\cite{bennett_concentrating_1996}, that is, operations which do not create entanglement, if and only if psi majorizes $ \phi $ and $ \phi $ also majorizes psi. Here, $ \psi $ and $ \phi $ are the Schmidt decomposition coefficients of the pure bipartite states. Equivalently, the eigenvalues of the reduced density matrices of psi and $ \phi $ for one side or another must be the same. It is a fine criterion, but it is too strict because it means that we would have many different categories of entanglement. A better approach utilizes a broader concept for what a resource is. It is an asymptotic equivalence.
For example, currency, such as dollars and the English pound, are considered equivalent, because they may be converted to each other at a constant rate with a fixed fee. We capture this idea mathematically with the following definition, saying that A and B are asymptotically equivalent, if there exists a conversion rate that is a ratio, which we will call R, such that for some epsilon and delta greater than 0, there exists a size n such that when the amount being converted is larger than capital N, we have a procedure which takes n times R plus delta copies of A and returns n copies of B and n copies of B may be converted to n times R minus delta copies of A. These conversions should be possible with an error less than epsilon. Think of R as being the exchange rate and delta as being the fee. This model works for many resources. Does it work for entanglement? Do entangled states have a well-defined notion of asymptotic equivalence? Let us find out.
 
\section{Entanglement - Definition and Measures} 
Let us define quantum entanglement and how entanglement may be quantitatively measured. For a pure state psi AB, a composite state of two systems A and B, we say that psi is entangled if and only if there does not exist a pure state $ \phi $ of A and chi of B, such that $ \psi $ of A and B is the tensor product of $ \phi $ A and chi B. Entanglement is more than just a binary state, however. There are measures that quantify how entangled a given state is. The most important of these is the entropy. Let $ \rho $ sub-A be the trace over B of the bipartite AB system. The entanglement of psi AB is defined as the Von Neumann entropy of this reduced density matrix on A or the reduced density matrix on B. This is called The Entanglement, with a capital E because of how important this measure is. For example, consider the bipartite state 0-0 plus 1-1. The density matrix of this state is a familiar quantity, which has four 1/2 values in the corners. The reduced density matrix is the identity matrix divided by 2, and thus the entropy is one bit, and the entanglement is, therefore, one ebit. We will see this defines an ebit. Another useful measure is the Schmidt number. It is the number of non-zero coefficients that appear in the Schmidt decomposition of a bipartite pure state. More explicitly for psi AB, this means we take the Schmidt decomposition, which is a sum over square root $ \lambda $ times K sub-A, K sub B, where the square root of $ \lambda $ sub K is the Schmidt coefficient. For example, consider two ebits of the form 0-0 plus 1-1.
Let us expand this, keeping A in blue and B in orange. When we expand this product, we end up with the A and B labels all mixed up. This is a very inconvenient representation. So, let us collect all the A terms together and all the B terms together. That gives us 0-0 0-0 plus 0-1 0-1 plus 1-0 1-0 plus 1-1 1-1. this is conveniently represented by taking x to be an integer from 0 to 3, and it is a sum of x x. The Schmidt number for this state is 4. this representation, a sum over x of x x, is a general and very useful representation for maximally entangled states.
 
\section{The Fungibility of Entanglement} 

we now show that quantum entanglement can be thought of as a physical resource, in the sense of this claim. we claim that all entangled bipartite pure states are asymptotically equivalent. Different forms of entanglement are fungible. we may prove this claim by showing there is a gold standard for entanglement. we will use the state 00 plus 11. Specifically, there are two parts to the argument. First, is entanglement concentration. There must exist a mechanism to take n copies of a partially entangled state psi and turn them into a smaller number of gold standard states, which we will call $ \phi $, capital $ \phi $. In particular, we will show that concentration produces N times E of psi minus delta of these states’ $ \phi $. The second part of the argument is entanglement dilution. we start with N times E of psi plus delta copies of the maximum entangled state. obtain n copies of the less entangled state. These protocols will work with epsilon-delta parameters greater than zero, and in the asymptotic limit of sufficiently large N. Epsilon, here is the probability of error of the protocol. Let us first look at the concentration. Without loss of generality, we may take the state psi as being a bipartite state with coefficients 1 minus $ \phi $ square root 00, and the square root of P, 11. This is the form provided by the Schmidt Decomposition. Recall that the entanglement of this state psi33 00:01:42,210 $->$ 00:01:45,990 is given by the Von Neumann entropy of the reduced density matrix of half of this state. we find this is nothing more than the binary entropy of P, which we have seen several times. What we need from concentration is to be able to take n copies of the state psi39 00:02:01,410 $->$ 00:02:05,340 into approximately E of psi copies of the maximum entangled state. Delta here will be the $ \phi $ that is paid to be able to create these EPR pairs. Since the protocol uses N copies of psi, let us explicitly write out this tensor product of N copies. we may express this tensor product using a binomial expansion where each of the usual binomial coefficients multiplies the term xx. Here, the absolute value of x denotes the number of ones in the binary string x. Let us rewrite using the number of ones as w, giving us a slightly simpler expression where we may take into account that there are multiple values of x with the same weight w. The superposition of all of the states with the same weight, w, is useful to define. Here we give that the name s sub w and normalize it properly by the number of possibilities that is the square root of n choose w. Each s sub w state is a maximally entangled state in a Hilbert space of dimension n choose w. Meaning equivalently, log n chooses w E bits. What we have, though, is not a single s sub w state, but rather with side psi to the n, a superposition over a variety of s sub w states. Each with an amplitude with a form, which has a binomial coefficient, as shown here. This superposition of maximally entangled states is quite useful, despite the different sizes.  that is because we may obtain one of them simply by following this procedure, Alice and Bob, A and B, both independently measure w, the Hamming weight of the strings that they have. This collapses the superposition into a single s sub w requiring no communication between the two parties. The probability of a given weight is n choose w, p to the w, 1 minus p to the n minus w.  this is approximately Gaussian for large n. The mean value of w obtained is n times p. nd the variance is NP times 1 minus p. The average number of entangled pairs obtained is the log of n choose NP, which is approximately the binary entropy of p.  this is nothing more than n times e, the entanglement. To be a little more precise, we do not get exactly nE, but rather nE minus some overhead, which goes something like the width of the variance that is the square root of n.  this determines our overhead $ \phi $ cost delta. Choosing delta to go as 1 over square root of n, thus allowing us to obtain a final protocol, which succeeds in producing approximately n times e minus delta EPR pairs with high probability. This completes the argument for part one of the proof. The second part of the proof requires that we go in the opposite direction to dilute a given number of EPR pairs. Specifically, n times e plus delta of these EPR pairs into n copies of psi. we want to turn ebits into arbitrating bipartite entangled states. An elegant way to do this uses teleportation. One party, say, Alice. Starts by generating n copies of the state psi locally, in her own apparatus. Remember, this is a bipartite state. She takes half of this state n qubits and compresses it to produce using Schumacher compression. For example, an output that has approximately nE qubits. Here, the delta is the overhead required for the compression. Separately, Bob shown here in purple produces n perfect EPR pairs. Perhaps with a little overhead delta, Bob, who holds this entangled state, then sends half of those qubits that are approximately nE qubits over to Alice. Alice performs a Bell basis measurement between the compressed state and this half of the EPR pairs, sends the classical measurement result to Bob. He fixes up the other half of the EPR pairs using the normal teleportation protocol.  remember, there are nE of these qubits. Decompresses with the reverse of the Schumacher compression, obtaining a state with n qubits. Altogether then, what Bob and Alice have is with high probability n copies of the state psi ab. For this protocol, the fixed parameter is delta. The overhead and epsilon the error is determined by the Schumacher compression and decompression steps. This completes part two of the proof. Thus, now we have the full proof providing a foundation for conceptually understanding how entangled bipartite pure states a physical resource is, which are fungible and may hold quantifiable intrinsic value. 

\item  CHSH-type Game: Classical Approach; In the CHSH-type game, Alice and Bob are spatially separated from each other from the time that the game starts until it is over, and they cannot communicate. The game starts when the referee randomly selects two bits, a and b, and sends a to Alice and b to Bob. Alice and Bob then return values 0 or 1 for the bits x and y. After receiving the values for bits x and y, the referee determines if the input and output bits satisfy the condition
\begin{equation}\label{eq3_68}
x\oplus y=ab.
\end{equation}

If the condition is satisfied, then Alice and Bob win; otherwise, they lose. Their goal is to find the best strategy that maximizes their likelihood of winning the game, and this depends on how they choose their bit values $ x_{0,1} $ and $ y_{0,1} $. Note that we do not specify which of $ x_{0,1} $ and M take on value 0 or 1. That assignment is the strategy that Alice and Bob must find.

In the classical deterministic case, Alice agrees to output the bit value $ x=x_0 $ when M, and M when M. Similarly, Bob agrees to output the bit valueM when M, and M whenM. The table below shows the different possibilities of the game for the possible values of a and b.
\begin{table}[H]
\centering
\caption{Different possibilities of the game for the possible values of a and b}
\label{tab:3_1:Table 2}
    \begin{tabular}{|c|c|c|c|} \hline
    a    & b & ab  &  $ x\oplus y $\\ \hline
    0    & 0 & 0   & $ x_0\oplus y_0 $\\ \hline
    0    & 1 & 0   & $ x_0\oplus y_1 $\\ \hline
    1    & 0 & 0   & $ x_1\oplus y_0 $\\ \hline
    1    & 1 & 1   & $ x_1\oplus y_1 $ \\ \hline
    \end{tabular}
\end{table}
    
When $ a=0 $, Alice outputs $ x_0$ . However, there is no way to infer from the table that Alice has assigned $ x_0=0 $ and $ x_1=1 $, or visa versa. 

\item  CHSH-type Game: Quantum Approach; In the CHSH-type game, Alice and Bob are spatially separated from each other from the time that the game starts until it is over, and they cannot communicate. The game starts when the referee randomly selects two bits, a and b, and sends a to Alice and b to Bob. Alice and Bob then return values 0 or 1 for the bits x and y. After receiving the values for bits x and y, the referee determines if the input and output bits satisfy the condition
\begin{equation}\label{eq3_69}
x\oplus y=ab.
\end{equation}

If the condition is satisfied, then Alice and Bob win. Otherwise, they lose. Their goal is to find the best strategy that maximizes their likelihood of winning the game,\cite{khan_nash_2019} and this depends on how they choose their bit values $ x_{0,1} $ and $ y_{0,1} $. Note that we do not specify which of $ x_{0,1} $ and $ y_{0,1}  $take on value 0 or 1. That assignment is the strategy that Alice and Bob must find.

In the quantum version of the CHSH game, Alice and Bob start with a shared quantum state, for example, a Bell state:
\begin{equation}\label{eq3_70}
\frac{\lvert0_A1_B\rangle+\lvert1_A0_B\rangle}{\sqrt{2}}.
\end{equation}

In order to maximize their likelihood of winning, Alice and Bob must decide how they will perform measurements on this state and how the results of those measurements will determine the bit values $ x_{0,1} $ and $ y_{0,1} $ that they will return to the referee.

In general, if Alice receives the bit $ a=0 $, she measures in the basis expanded by the states
\begin{equation}\label{eq3_71}
\begin{split}
\displaystyle \lvert \psi _{A,0}\rangle    \displaystyle & =    \displaystyle \cos \left(\theta _{A,0}\right)\lvert 0_ A\rangle +\sin \left(\theta _{A,0}\right) e^{i \phi_{A,0}}\lvert 1_ A\rangle \\    
\displaystyle \lvert \psi _{A,0}^{\perp }\rangle    \displaystyle & =    \displaystyle \sin \left(\theta _{A,0}\right)\lvert 0_ A\rangle -\cos \left(\theta _{A,0}\right) e^{-i \phi_{A,0}} \lvert 1_ A\rangle
\end{split}
\end{equation}
         
where $ \theta_{A,0} $ and $ \phi_{A,0} $ are the angles that define the states in the Bloch sphere.

If she receives the bit $ a=1 $, she measures in the basis spanned by the states
\begin{equation}\label{eq3_72}
\begin{split}
\displaystyle \lvert \psi _{A,1}\rangle    \displaystyle & =    \displaystyle \cos \left(\theta _{A,1}\right)\lvert 0_ A\rangle +\sin \left(\theta _{A,1}\right) e^{i \phi_{A,1}} \lvert 1_ A\rangle \\         
\displaystyle \lvert \psi _{A,1}^{\perp }\rangle    \displaystyle & =    \displaystyle \sin \left(\theta _{A,1}\right)\lvert 0_ A\rangle -\cos \left(\theta _{A,1}\right) e^{-i \phi_{A,1}} \lvert 1_ A\rangle
\end{split}
\end{equation}
         
\item When Bob receives the bit $  b=0 $ or $ b=1 $, he measures in one of the two bases spanned by analogous states as Alice's states above.

Solution:\\
Alice and Bob are not constrained to always measure on the same basis. , their best strategy in the quantum case is to measure in different bases.

\item  Entanglement: Definition and Measures; Regarding the quantification of entanglement.
Solution:\\
Quantum entanglement is not a binary quantity, meaning that it is not just zero or one (if we normalize it), but it can also take values in between. For a product state as $ \lvert 0\rangle _ A\lvert 0\rangle _ B $ is zero, and for a Bell state $ \frac{\lvert 0\rangle _ A+\lvert 0\rangle _ B+\lvert 1\rangle _ A\lvert 1\rangle _ B}{\sqrt{2}} $ is maximal. However, for an arbitrary state

$ \alpha \lvert 0\rangle _ A\lvert 0\rangle _ B+\beta \lvert 0\rangle \lvert 1\rangle +\gamma \lvert 1\rangle \lvert 0\rangle +\delta \lvert 1\rangle \lvert 1\rangle $ ,

it can take continuous values in between depending on the parameters $ \alpha, \beta, \gamma, $ and $ \delta $.

\begin{figure}[H] \centering{\includegraphics[scale=.55]{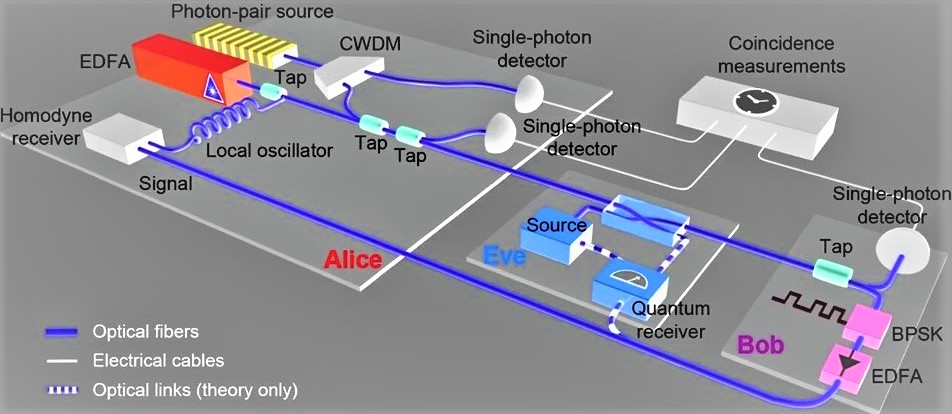}}\caption{QKD Issues and Challenges}\label{fig3_7}
\end{figure}

\section{Introduction to NISQ and Short-depth Quantum Circuits} 
In this section, we will look at the practical challenges and opportunities faced when implementing realistic quantum algorithms on the hardware that is available today\cite{de_ridder_quantum_2019}. As we have seen throughout the section, the promise of quantum computing centers on finding quantum advantage for useful, meaningful problems that do not currently have a known fast  \cite{tang_quantum-inspired_2019}, and one for which a fast quantum algorithm exists. In sections one and two, we looked at several examples, such as Shor's algorithm, the simulation of quantum chemistry and quantum materials \cite{bassman_towards_2020} in optimization problems \cite{inagaki_coherent_2016}. In all of these cases, we assumed that we had perfect qubits. However, as we now know, the physical qubits we have today are far from perfect. They are subject to noise, and the noise reduces the overall gate fidelity, limiting the number of gates that one can perform before quantum information is lost\cite{nielsen_quantum_2011}.
Furthermore, as we will see in section four, the path forward is to implement fault-tolerant quantum error correction \cite{chow_implementing_2014,corcoles_demonstration_2015,kandala_error_2019,colless_computation_2018}. Basically, by adding additional qubits, we can achieve system resilience through hardware redundancy. With error-protected logical qubits, we can envision making fault-tolerant circuits with arbitrarily large complexity. Now, that is fine, but at the current error rates we have today, even for the leading qubit modalities, it would require around 1,000 physical qubits to error-protect a single logical qubit\cite{kapit_very_2016}. This line of work certainly needs to be done, and it will be, but still, several years before we have fault-tolerant error-corrected systems.
Furthermore, that is the subject of section four. For now, and over the next several years, we will have quantum computing testbeds with a few hundred to a few thousand noisy physical qubits that are not fully error corrected. The question is, what useful, commercialize tasks can we perform with these types of computers? This is a very important question because we know from experience, much like with classical digital computers, that the investment required to push the technology development that enables scaling to larger sizes, that this requires a commercial product and revenue stream to supplement government investments. The revenue from this generation of quantum computers would be used to push the next node and enable better performance in the next generation, which in turn generates more revenue.
Furthermore, this virtuous cycle continues. The status today is that we have known useful algorithms that require many more qubits than we currently have in practice. The question is, are there near-term applications that solve real-world problems using noisy, intermediate-scale quantum computers, or NISQ computers \cite{preskill_quantum_2018}. It is currently an area of active research and is growing rapidly due to the need to find such applications. There is certainly hope that smaller-scale quantum computers, working as a co-processor for a classical computer, may provide a quantum advantage for certain types of problems in machine learning \cite{shiba_convolution_2019}, simulation, optimization, or quantum dynamics. However, the reality is, we just do not know yet, because we are just getting started. Getting 50 to 100 qubit testbeds online so that computer scientists can start playing with them is a very important step. Also, there is a strong motivation to continue to improve gate fidelities, reduce error rates to allow for more gates that can be used for larger quantum circuits \cite{terhal_adaptive_2004}, even before fault-tolerant error correction is implemented. Finding a useful killer app that offers some level of an advantage when using NISQ scale quantum computers would be a tremendous advance, one that would potentially kickstart a virtuous cycle that ultimately leads, in the long term, to large-scale fault-tolerant quantum computers.
Furthermore, that is what we need in the end to realize the promise of quantum computation fully\cite{kitaev_classical_2002}. Thus, in the meantime, we need to identify NISQ scale applications, which brings us to this section, realistic quantum computation \cite{nielsen_quantum_2011}today, challenges, and opportunities. This section will discuss what types of algorithms have been demonstrated at a small scale, the challenges they faced, how they benchmarked their systems, what they learned, and its implications as we move into the NISQ scale era of quantum computing \cite{maslov_outlook_2019}. 
There is question about, are quantum computers fundamentally realizable? 
in a quantum computer, there are so many degrees of freedom, Let us say that we have say, 300 qubits, so we have two to the 300 degrees of freedom in our Hilbert state, and based on that, we have to control all two to the 300 degrees of freedom, which basically, if we calculate that out, that is more than all of the atoms in the known universe. so, if that were true, that would be a very difficult prospect. But, we know that is not the case, we have to control all two to the 300 degrees of freedom explicitly. That quantum mechanics provides quantum parallelism in a way that allows us to control large areas of the Hilbert space simultaneously, with far fewer gates than that. So that is one major misconception. Another one is this idea that, we will never be able to control the noise to this level, and that is going to be there, errors in these computers that cannot be corrected, and we think that this gets into the concept of fault-tolerance and error correction. The idea here is that, we know, quantum computing, is indeed analog up to some extent, but measurement allows us to digitize quantum information, and when we do a projective measurement, and in particular the syndrome measurements that we need to perform to do error correction, we can project a quantum system on to a state of error, or no error. that is a very important statement. Because, in doing that, we can do those projections, as we will discuss, we can do those projections in a way that tells us there is an error. It can tell us the type of error. But it never projects out the quantum information. so, once we know the type of error, and where it occurred, we can go back in and correct it. For example, we might discuss that a bit-flip occurred. State zero became state one, or state one became state zero. We can look that occurred, but we never look that the computer was in state one or state zero. we just discussed that a bit-flip occurred, and so we can go back in and fix it, by flipping the bit back. Now, we never lose the quantum information in this process, and it requires specially-designed syndromes, and also, the concept of fault tolerance. That when an error occurs, it does not propagate into more and more errors downstream, and that is an architectural question. we think that is important to realize. Now, it is certainly true that there is a lot of hype in the field these days, probably more than is warranted at this stage. This is a new technology, and developing technologies takes time, it takes engineering. it is natural that with any exciting technology, that we get excited about it, and we want it to occur faster than it is going to occur. But as we discuss, if we look at the time scale over which classical computing developed, we remember that the triode diode was invented in 1906, and it was not until 1946 that we had ENIAC, which was the first very, we know, by today's standards, a very primitive classical electronic computer. the transistor was invented a year later, but it was not until the '70s and '80s that we had chips that we think of in our computers today., whereas we started with vacuum tubes, and we moved to discrete transistors, and eventually we moved on to CMOS where are today. we know, there were many steps along the way. The technologies changed and pivoted. But we think, had we not followed this thread of development, we would not have got to where we are today. So certainly there is a lot of hype today, and this is a hard problem. But if we think back to many of the major technological advances that have been made over the decades, and even centuries, it is littered with people who say that things just cannot happen, right? And, they are almost always proven wrong. so, we think that quantum computing is hard, but we will succeed, and we will have quantum computers, and what they look like, that is to debate, but they will happen.

\section{Quantum Volume} 
How powerful is a quantum system? Furthermore, how do we compare quantum computers based on different technologies? There is much more to the answer than the number of qubits, just as we would not think of measuring a laptop performance by the number of transistors. Meaningful performance metrics are just being formulated. The quantum volume was created to capture key aspects of noisy intermediate quantum systems that operate using quantum circuits constructed from the universal gate set\cite{pednault_breaking_2018}. Note that this metric would not be appropriate for quantum annealers, nor fully fault-tolerant quantum computers. The quantum volume metric incorporates the number of physical qubits and errors due to gates, quantum crosstalk, decoherence, measurements, the topology of qubit connectivity, the gate set implemented, and parallelism. How many gates can be run in parallel? Let us think about each of these issues separately. Consider gate errors when preparing a 3 qubit GHZ state. A first CNOT is used to entangle the top and middle qubits. Then, a second CNOT entangles them with the bottom qubit. If each CNOT has an error of epsilon, then the total gate error is the product of the fidelities, or the quantity 1 minus epsilon squared, for this example. To first order an epsilon, the fidelity is then 1 minus 2 times epsilon. For the quantum volume to represent typical quantum approximate optimization algorithms \cite{farhi_quantum_2014, paini_approximate_2019,hadfield_quantum_2019,guerreschi_qaoa_2019} or variational quantum eigensolvers\cite{kandala_hardware-efficient_2017,thornton_quantum_2019}, we must implement a circuit comprised of d layers of U4 or general two-qubit unitaries, which can be decomposed into three CNOT gates\cite{chen_demonstration_2008}. These errors of each U4 is then 3 times epsilon, to first order in epsilon. Other errors could also be introduced when making the GHZ state. For example, during the first CNOT gate between the top and middle qubits, the state of the bottom qubit could be affected by the control signal. We call this spectator qubit error. Significant errors also accrue if the qubits decohere appreciably while executing a circuit of great depth.
Furthermore, measurement errors directly impact the fidelity of a circuit. We have chosen to implement the quantum volume metric with randomly chosen U4 unitaries between random qubit pairs. Qubit technologies with gates only between neighboring qubits must then teleport or swap the quantum states around an interconnection lattice to cause the desired qubit states to interact. Either of these operations introduces more gates and errors. For example, to implement a U4 unitary between non-neighbor qubits 0 and 2, a 3 CNOT sequence to swap qubits 0 and 1 would first need to be executed. Each quantum system may implement a different set of gates, natively in their control hardware. The Solovay-Kitaev algorithm approximates any two-qubit unitary to within a specified error with multiple gates. Therefore, systems that natively implement more unitaries than the simplest universal gate set accrue fewer gates and less error. Finally, overall execution time, hence, decoherence, can be reduced if the unitaries can be implemented in as parallel a manner as possible. So, how are all these effects incorporated into one expression for the quantum volume metric? The quantum volume expresses accessible quantum states, so, it is exponential in the number of accessible qubits, m. If the error per 2 qubit gates is large, then not all qubits can be entangled with useful fidelity. So, m is less than or equal to the number of physical qubits. The depth of the quantum circuit, d, is defined as the number of layers of permutations, $ \pi $, and U4 unitaries executed on the hardware. The number of accessible qubits is then the minimum of m prime and the circuit depth, d. If the gate error is high, the depth may be small. If too large, an m prime is chosen. We maximize the quantum volume by choosing the largest n prime less than or equal to the number of available qubits that maximizes m. we have implemented this algorithm on our quantum experience hardware. We compiled random circuits offline using a greedy algorithm to determine the sequence of random 2 qubit unitaries combined with added swaps or teleportation to execute on our hardware. After multiple random trials, we arrive at an average quantum volume of 8 for a 5-qubit quantum experience processor. The quantum volume metric is a stringent test of near-term universal quantum hardware capabilities. We expect this and new metrics to evolve as we discuss more benchmarking quantum systems\cite{villalonga_flexible_2019, mavadia_experimental_2018}. 

Quantum volume is a metric that attempts to quantify the computational utility of a quantum computing platform. It is designed to be an architecture-neutral means to compare capabilities across platforms with potentially widely differing characteristics\cite{smith_practical_2017}. Platforms with higher values of quantum volume can, in principle, implement more complex quantum algorithms.

The computational power of a quantum device depends not only on the number of qubits but also on their gate fidelities, the types of qubit interactions, their connectivity, as well as several other characteristics that may influence computing performance. For example, the increased connectivity of an ion-based quantum processor may offer certain advantages over the generally nearest-neighbor connectivity of a superconducting qubit processor \cite{gambetta_building_2017}. This goes beyond a solely fidelity-based comparison between platforms.

Conceptually, while the quantum volume is a metric for the utility of a quantum processor when executing a particular algorithm, it does not provide any insight into specific aspects such as the overall computational time. Rather, it quantifies the ``space-time'' volume occupied by a hypothetical circuit with random two-qubit interactions that can be reliably executed on the processor. The ``space,'' as its represented here, translates to the number of active qubits n' out of all available physical qubits n in the quantum processor. The ``time,'' in turn, translates to the depth of the circuit d. The quantum circuit depth reflects the maximum possible number of sequential quantum operations that can be reliably applied to the processor. Implementing a model algorithm with a depth greater than d runs the risk of erroneous output due to the accumulation of noise and gate errors. The circuit depth depends on the number of active qubits and the per-gate error rate $ \epsilon _\mathrm{eff} $ such that $ d=1/n'\epsilon _\mathrm{eff} $.
The quantum volume QV is an expression for the ``reachability'' of the quantum state space, and it is thus exponential in the number of ``reliable'' qubits m.  The number m is smaller or equal to the actual number of active physical qubits, as it also depends on the per-gate error rate.
\begin{equation}\label{eq3_73}
QV=2^ m=2^{max\{ min[n',d]\} }=2^{max\{ min[n',1/n'\epsilon _\mathrm{eff}]\} }.
\end{equation}

In this way, the QV attempts to quantify the state-space accessibility, a proxy for the utility of a quantum processor, which, in aggregate, effectively incorporates the different strengths and weaknesses of different quantum computing architectures and platforms \cite{smith_practical_2017,linke_experimental_2017}. Further reading, Validating quantum computers using randomized model circuits \cite{cross_validating_2019}, Quantum optimization using variational algorithms on near-term quantum devices\cite{moll_quantum_2018}.

There are several aspects that need to be considered when quantifying the efficiency or power of a quantum computer. In March 2017, IBM published the paper, ``Quantum optimization using variational algorithms on near-term quantum devices.'' In this paper \cite{moll_quantum_2018}, they defined a new architecture-neutral metric based on the question: Can this quantum computer run a given algorithm? Their metric, called Quantum Volume, QV, takes into account the minimum number of qubits, N, required to run a quantum algorithm, and the necessary number of steps, also known as the circuit depth d.
Some of the hardware aspects that Quantum Volume QV incorporates are.
\begin{itemize}
\item The number of gates that can be applied before decoherence affects the results.
\item Connectivity between qubits.
\item Available quantum gates.
\item The number of gates that can be run in parallel.
\end{itemize}

\section{Demonstration of Quantum Advantage in Machine Learning} 
As we have seen so far, currently available quantum processors are mostly limited to the handful of qubits and noisy quantum gates. For this reason, demonstrating a definite advantage of quantum over classical computing is an ongoing challenge\cite{riste_demonstration_2017,schuld_machine_2019}. In this section, we will present an experiment we did at Raytheon BBM Technologies in collaboration with IBM Research, where a clear quantum advantage for a specific algorithm appears with only a few highly noisy qubits. The problem we consider is known in classical computing as Learning Parity with Noise, or LPN for short. At the center of the problem is a device, also called an oracle, with an unknown string, k, encoded on it. Our goal is to discuss what the k is by observing the effect of the oracle on random input bits. The bitstream, k, is related to the action of the oracle on the input, D, by this relation. In words, the oracle takes the bit, Di, for which ki equals 1, calculates their parity, and writes it into the result bit, A. For instance, let us take an oracle of dimension 2 and k equals 1, 1. These are some of the possible outcomes. After a few attempts, it is easy to figure out that it contains the result of both bits. So, we conclude that k equals 1, 1. However, the problem is not trivial in the presence of errors. Imagine that with some probability, the result of either d1, d2, or are randomly flipped. As we can see, it is now much harder to find the relation between the output. It can be shown that the number of queries grows nearly exponentially with the amount of noise. Interestingly, if one uses a quantum oracle, the difficulty of the problem can be drastically reduced. The circuit implements a quantum version of the Oracle. Now, the input and the result bits are quantum bits. nd the function is implemented as a series of Hadamard gates and CNOT gates. The Hadamard gates create an equal superposition of input states. Then, a CNOT is applied between every qubit Di for which ki equals 1 and the target A. The effect of the CNOT gate is to map the party of qubits D1 up to the Dn onto qubit A, reproducing the classical LPN Oracle on a quantum circuit. Our system of choice to implement the circuit is an IBM five-qubit superconducting quantum processor. Each qubit is coupled to a microwave resonator, like the one shown here, to deliver control and readout pulses. The interaction between the middle qubit, A, and the other four are mediated by other resonators, in blue.  this is what allows us to implement the desired CNOT gate. we use this circuit to implement an LPN oracle. we now compare these strategies to find k for a classical and for a quantum learner, both using the results of this oracle. In a classical case, for every call of the oracle, the learner simply measures the output register and collects the results. Let us go back to the example with k equals 1, 1. These are some of the possible results we may obtain for five queries. Remember that, ideally, a should equal the parity of d1, d2. Even in the presence of errors, with enough statistics, one can find that the most consistent solution for these data is k equals 1, 1. This is the experimental result for the four different two-bit oracles. As expected, the higher the number of queries, the lower the error probability. Despite using a quantum oracle, the learning here is completely classical. we take the measurement results, which are classical, and analyze them on a classical computer. The power of quantum processing only appears when we are allowed to manipulate the output state of the oracle. Because the state is quantum, we can apply a Hadamard gate on all qubits. This is the final stage before measurement. This is an entangled state of the ancilla with the data qubits. As we can see, when the ancilla is measured to be 0, the data qubits are also projected into 0. we obtain no information on k. Because of this, we discard all the results were an equal 0. On the other hand, when a equal 1, we project the data qubits onto k, which is exactly the solution we are looking for. Finally, we take the bitwise majority vote for d1 and d2 to find the best guess for k. As we will see later, this makes the protocol particularly robust to errors. This is now how the error probability decreases with the number of queries in the experiment. Let us look again at the classical solver for comparison. Whereas the error probability is comparable in the two cases, for k equals 0, 0, 0, 1, or 1, 0, the behavior starts to differ for k equals 1, 1. This is the most complex of the four circuits, including two CNOT gates, each introducing errors. This suggests that as the problem gets harder, quantum learning is more efficient. Let us indeed increase the problem size in the experiment. we go from two to three data qubits for the oracle.  this is the result we obtain for the error averaged over the eight computational input states. Clearly, not only the quantum learner retains an advantage, but the gap with the classical learner increases with the oracle size. Finally, let us consider how robust this protocol is to additional errors. As an example, consider the case where we introduce additional noise on the result qubit, a. On the x-axis, we denote with eta the probability that we make an error in reading its state. On the y-axis, we indicate the number of queries required to find k with a 1\% error probability. As eta increases, this number only slowly increases for a quantum learner. On the other hand, the number of queries required to a classical learner grows by up to two orders of magnitude. To summarize, the learning parity with noise is an experimental realization of a quantum advantage using current noisy quantum circuits. Not only is quantum learning more efficient in this case than the classical counterpart, but also, the performance gap increases with the difficulty of the problem, either by increasing the oracle size or by adding noise to the system. 

Encoding messages using a parity function and a secret bit string, or ``key,'' is a straightforward encryption scheme. The output is formed by a single-bit parity measurement of an n-bit message that is bitwise combined with an n-bit key using an XOR operation. To unravel the key using classical methods, up to $ 2^n $ different input messages have to be compared with their corresponding encrypted output bit. As we might expect, the power of quantum parallelism and interference may again provide an advantage over classical approaches.

A parity encryption scheme consists of an input bit string $ {D} $ bit-wise combined with the key $ {k} $ such that $ f({D} $,$ {k})={D}\cdot {k}\mod 2 $. This represents the bit-wise multiplication of the elements of vectors $ {D} $ and $ {k} $, followed by the mod-2 addition of the resulting values, which is very similar to how a dot product between vectors works. The parity function $ f({D},{k}) $ yields a 0 for an even parity, or a 1 for an odd parity. The algorithm's task is to identify the underlying key $ {k} $.

Suppose a quantum processor is comprised of 3 qubits, with 2 qubits serving as input states $ {D}= [d_1, d_2] $, and the remaining qubit as an auxiliary qubit a on which the result of $ f({D},{k}) $ is stored. The unknown bit string $ {k} $ is implemented on the quantum processor as an oracle function. For an n=2-bit key $ {k} $, there are $ 2^ n=4 $ different possible keys $ {k}=[k_1,k_2] $, which are $ \{11,10,01,00\} $. The following tables summarize the value of a depending on the input bits$  d_1, d_2, $ and the bit string $  {k} $.

\begin{figure}[H] \centering{\includegraphics[scale=.25]{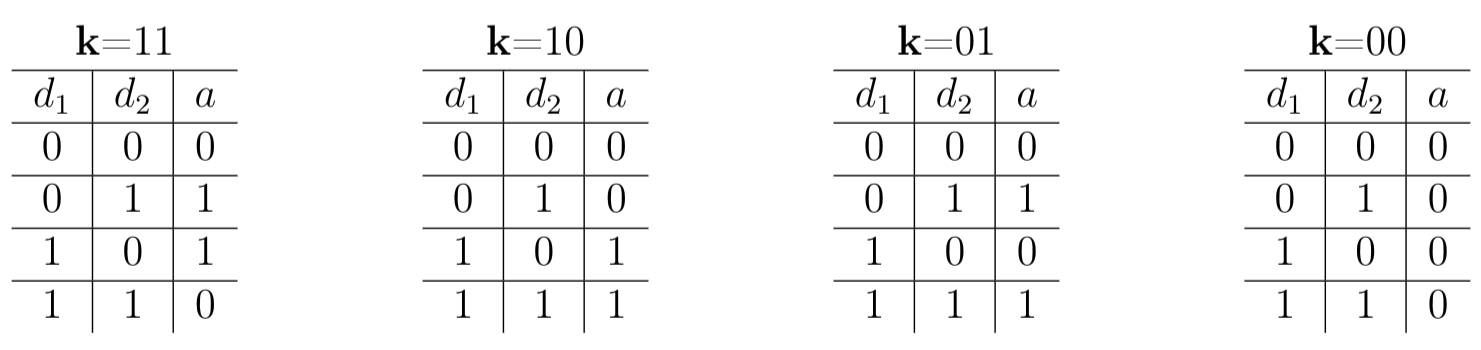}}\caption{Riste}\label{fig3_8}
\end{figure}

A quantum algorithm to determine an encoded 2-bit key $ {k} $ with 2 qubits serving as input states for $ {D}=[d_1, d_2] $ and 1 auxiliary qubit $ \vert 0 \rangle _ a $. At first, all qubits are initialized in the ground state. Then, a Hadamard gate acts on each of the data qubits. Next, the oracle implements the parity operation with the bit string $ {k} $, storing the result on the auxiliary bit. $ k_ i=0 $ is equivalent to an identity gate, and $ k_ i=1 $ can be implemented with a $ CNOT_{i\rightarrow a} $ gate acting on the auxiliary qubit with qubit i as the control \cite{farhi_limit_1998}. After the oracle function, Hadamard gates are again applied to the qubits to leverage quantum parallelism and quantum interference. (If we were to measure the qubits before performing the last Hadamard gates and thus omitting the quantum parallelism and interference step, we would essentially mimic the output of a classical processor.)

An example of this step-by-step can be seen in the following diagram, but note that the second line flips the middle two terms in the superposition state. This does not affect the result, but it might be easier to follow knowing this.

\begin{figure}[H] \centering{\includegraphics[scale=.25]{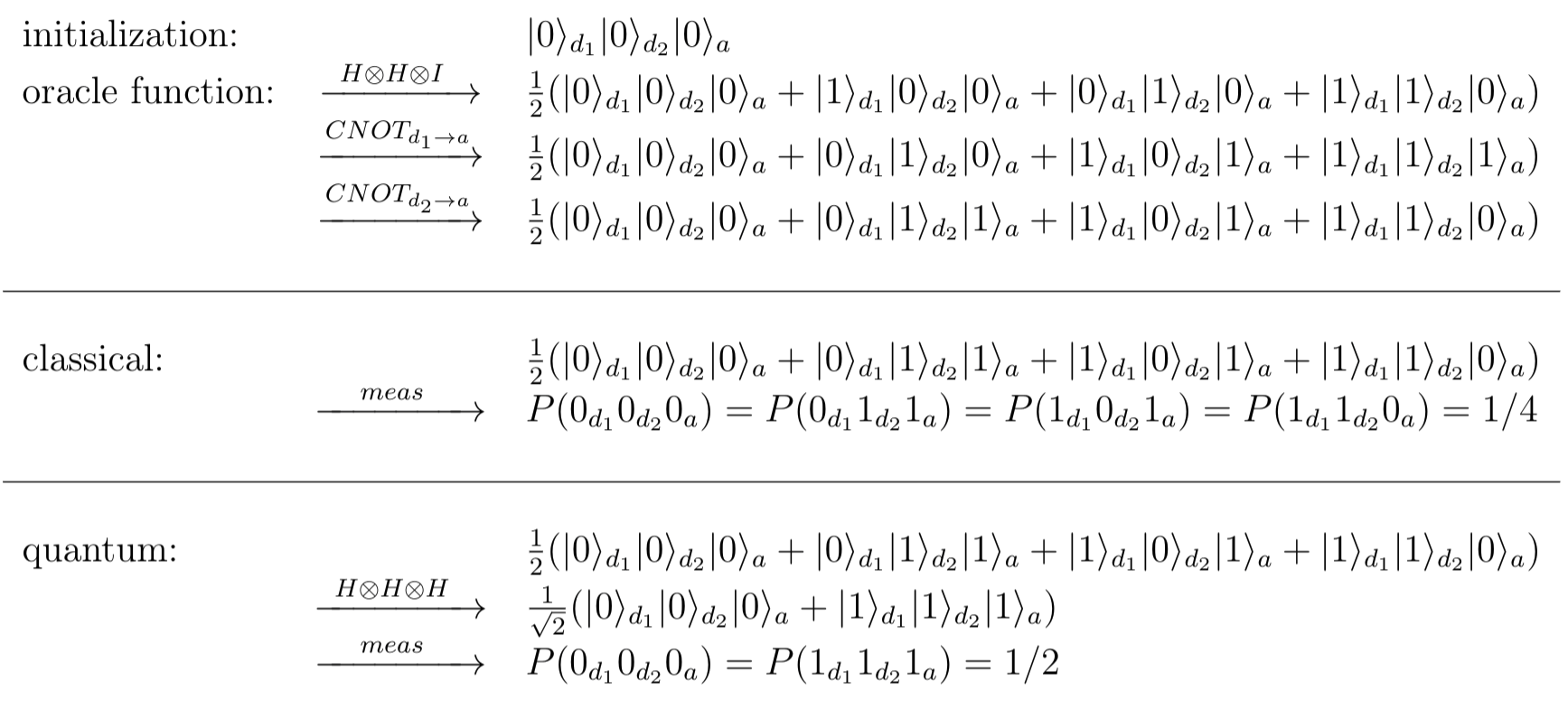}}\caption{Riste}\label{fig3_9}
\end{figure}
An immediate measurement after the oracle function for $ {k}=11  $ generates one of four distinct output states with equal probability. In fact, in a noisy environment, identification of the unknown bit string $ {k} $ may require even more than the ideally expected 4 queries to claim the identification of $ {k} $ with sufficiently high confidence.

In contrast, a quantum algorithm with classical post-processing a hybrid approach \cite{bauman_downfolding_2019} benefits from quantum parallelism and interference through additional Hadamard gates applied to the qubits before the measurement \cite{shaydulin_hybrid_2019}. The measurement projects the state onto the qubit's ground-state or a more complex state, both with probability 1/2. Post-selecting the measurement and only considering the states with an auxiliary qubit in state 1 further improves the performance of the hybrid approach\cite{chen_hybrid_2019}. Similar to the classical approach, interfering noise processes enforce multiple measurements and averaging of the results to determine $ {k} $ with certainty.

Even under noisy conditions, the hybrid approach outperforms the classical with respect to the number of measurements to identify $ {k} $ with sufficiently high certainty. The example can be generalized to other 2-bit or n-bit keys $ {k} $. Note, the quantum advantage can only be achieved if the data qubits are initialized in a superposition state, reflecting all possible data bit inputs. The experiments indicate that under noisy conditions, a hybrid approach using quantum interference and parallelism outperforms its classical counterpart and determines the $ {k} $ faster and hence with fewer queries.

For further reading, the publication describing the experiment discussed above: Demonstration of quantum advantage in machine learning \cite{riste_demonstration_2017}

\section{Quantum Support Vector Machine} 
So, suppose we have a classification problem we have to solve. this classification problem is given to us as n-dimensional vectors that are our data. these vectors, they belong to several classes. for simplicity, we are going to keep this discussion to only two classes. So, each vector either belongs to one class or another, and we need to classify into these two classes the data that we are given. So, in this example, we are going to have a training set that is given to us, of which we know the data and the labels. we are going to use this training set to find a cutting hyperplane that separates this data, and that we can use to classify any future data whose label we do not know and that is given to us. we assume that the data is already linearly separable. with that assumption, we can go ahead and do cutting in our training set that separates the training set we are being given. Now, we can use this cutting hyperplane to classify any future database given to us. we will do that with some degree of success. However, we want to maximize that success. So, we have a choice of cutting hyperplanes we can choose from our training data. Which one is the best one which we could choose? So, there is a method that people use to determine what is the most optimal one, which is called a support vector machine. what this method does is it places the cutting hyperplane as far as possible to any data point in the training set. Now, the closest data points to the cutting hyperplane in the training sets are going to be the data points that are more difficult to classify. once we find them, that is equivalent to finding the cutting hyperplane. These data points are called the support vectors. That is where the name comes from. once we know the support vectors, we know the cutting hyperplane, and we know it is the best we can do for this training set, now, with that cutting hyperplane, we can take new data whose label we do not know and classify it into one of the two labels. Now, this is if the data is linearly separable. If it is not, like this example, we are showing here where we have one-dimensional data. We cannot do a single cut that will separate the two labels, what we can do, the trick that people use is we can map that data into higher dimensions until we find a space in which we can linearly separate the data. This transformation is called a feature map. here, we give an example of where we transfer the one-dimensional data into two-dimensional data. here, we see that we can separate it. Now, as the problems get more complex, we can imagine that the size of our feature space, the target of the feature map, can grow significantly in a number of dimensions. However, it is not only the dimensionality that makes that problem complex. Alternatively, it is not precisely the dimensionality because what we are doing for classifying is, and we want to see how close points are to each other in the new feature space. We do not need to know where each point is, we need to know how close each point is from all the other points. So, there is a magical tool that one can use for finding how close the points are in a given space, which is called the inner product. If we can know the inner product of every pair of data in our training set, we do not need to apply, and we do not need to know the feature map explicitly. we can just try to get the inner product after applying the feature map, without applying the feature map. So, the information that gives us the inner product in the feature space between each two data points in a training set is called a kernel. once we know the kernel, we know the distance between each of the points. then, from there, we can obtain the support vectors. Now, how can quantum computing help with this problem? Well, one way is, if we have a weird data set, like the one we see here, this data set, for simplicity, has only two features, two dimensions, and two classes, which we see here as red and blue. these two classes are separated by a white gap to make them a bit easier to classify. This data set has a property that we know a feature map that will take this space, where the data is, and map it into a two-qubit state space. This is a space of two-qubit density matrices. we know that, if we have a two-qubit quantum processor, we can efficiently do operations in that space. if this data set with this same structure was much, much larger in dimensionality, we know that we could map it into a large enough quantum processor. If we had enough low noise, then we could estimate inner products in that space because it is not only the dimensionality. we are using a data set that we know is going to be difficult to estimate inner products if the set is complex enough. With this data set, we can see here how the feature map works for a single qubit because it is simple to visualize. we can see that a one-dimensional line from 0 to 2 $ \pi $ will map in the single-qubit Bloch sphere in the way we see. This is attained with just basic changes and phase gates. In the case of more qubits, we will add entanglements as CZ interactions, and so on. So, now we can go ahead and take a training set that we see here of 20 points per label. what we do with this training set is we map it to the two-qubit displays. We then use a quantum processor to estimate the inner product of each two points, and we obtain the kernel. here we can see the kernel. This kernel is just a matrix of 40 by 40. It is symmetric and semi-definite positive. this is real data from a quantum computer. So, this matrix is going to tell us all we need to know about this training set for classification purposes. So, from this data, we can estimate the support vectors. we can do that already outside of the quantum processor because it is easy to estimate the support vectors from a kernel if we know the kernel. It is getting the kernel that might be difficult. So, here, we see the support vectors circled in green. So, these are the data points in the training set that is closest to the hyperplane. with these support vectors, we can get new data, and we can compare it or see how close the new data is to each of the support vectors. we forget about any other point in the training data set. Then we can classify any new data. This is a randomly drawn new set of data points that we are going to try to classify. So, we have the black squares belonging to class red, and the white squares belonging to class blue. if we take each of these points and do the inner product in the quantum processor, and then see at what side of the cutting hyperplane they fall, we see that we can classify them with 100\% success, that the black squares will fall on the right-hand side of the hyperplane and the white squares will fall on the left-hand side. we can say we can classify them perfectly well. Well, this was just a toy example, without any connection with real-life data problems, or data sets. However, we think that, given the spectacular progress in quantum hardware over the last decade or so, we are starting to get to a point where these processors are powerful enough to try new ideas. we know that the classical machine learning methods have been around for about half a century, and only reasonably recently, they have been starting to attract attention.  that was because of the development of computer hardware, and the computing ability of our processor became much, much larger.  these ideas that existed for artificial intelligence could be tested, and then the field advanced that way. It is an exciting time, really, for quantum computing because we are at a time where we can get inspiration from the classical methods in artificial intelligence and try to produce new ideas that will be applicable to the field. Support Vector Machines (SVM); The Quantum Support Vector Machine is an implementation of a machine-learning algorithm on a quantum computer\cite{havlicek_supervised_2019}, and it is used to classify data. The following statements describe the steps of the SVM algorithm. A quantum computer is used to efficiently calculate the inner product among points in a data set. In the case of the quantum support vector machine algorithm, the quantum computer is used for two purposes: to obtain the kernel of a known data set, and to efficiently classify any new data sets.

\section{Practical Quantum Simulation: Introduction} 
Quantum mechanics is a remarkable theory in that its principles apply not only to atomic-scale particles but also to macroscopic objects such as electrical circuits. As we have already covered in this section, in the race towards building a quantum computer, resonant circuits constructed from superconducting metals have progressed tremendously over the past 10 years with coherent lifetimes increasing from a few nanoseconds to almost a millisecond in state of the art devices, making nascent quantum processors a reality. These solid-state objects enjoy all the rights and privileges of qubits. They can be placed in a coherent superposition that is collapsed by the act of measurement, just like an atomic spin interacting with photons. So, why cannot we just print thousands of superconducting qubits if we know the basic recipe for each one? Well, a major challenge even at the level of 100 qubits arises from the fact that though there may be plenty of space at the bottom as Feynman said, at the nanoscale, say, to have a high density of quantum bits, there is not much space at the top where wires and electronics have to be carefully arranged to handle and process quantum information efficiently\cite{feynman_simulating_1982,hey_feynman_2018}. Thus, the road to a universal quantum computer \cite{goto_universal_2016} robust to errors is still several orders of magnitude away in qubit density and lifetime. That being said, there are still different types of quantum algorithms that can be executed with technologies at hand and can solve specific problems of great scientific interest. A major focus of research today is to produce those algorithms compatible with noisy intermediate-scale quantum devices, or NISQ technology. One class of problems that can be tackled on NISQ hardware are associated with ideas put forward by Richard Feynman and are called quantum simulations\cite{king_observation_2018,bloch_quantum_2012,bauman_quantum_2019}, where quantum bits are coupled in a specific geometry, allowed to interact with each other, and finally probed after they have reached a final configuration. Such a hardware simulator eliminates the need to describe and track all the possible arrangements that this array of qubits can access\cite{villalonga_flexible_2019}. Doing so would require an exponentially large set of classical numbers and is the fundamental reason problems of this type cannot be performed on a classical computer. In the quantum simulation paradigm, only a few select properties of the system have to be probed at the end of an algorithmic sequence. With progress in next-generation quantum algorithms, there are now many types of scientific calculations that can be efficiently mapped onto a quantum simulator \cite{zhang_observation_2017}ranging from problems involving the energy spectra of molecules and complex solid-state materials such as high-temperature superconductors and quantum magnets to the dynamics present in natural and artificial light-harvesting complexes to models of the cosmos were ideas and quantum field theory and gravity can be tested \cite{soloviev_beyond_2017}. Let us focus for a moment on an example from quantum chemistry. If we start with a modest patch of electronic matter which we can approximate for the moment as distinguishable particles, given an m by m grid of electrons which we describe each with a 100 orbitals, for an 80 electron system, this tiny patch of matter would require 100 to the power 80 numbers to describe. To put this in context, this is much larger than a mole or the number of particles in a tangible piece of matter. in fact, and it is larger than the number of particles in the entire universe. Thus, directly calculating the structure of even a small number of quantum particles is very computationally intensive. The number of atoms that can be treated by various computational chemistry methods is shown in the plot on the left. A direct computation of Schrodinger's equation limits one to less than 10 atoms. Other methods like density functional theory give access to a much larger sample size but are approximate and not accurate for all types of materials \cite{bassman_towards_2020}. For calculating energy levels, classical computers simply do not have enough memory to describe the large combinatorial space quantum systems live in\cite{goto_combinatorial_2019}. even the most powerful classical supercomputers with petabytes of memory can only tackle of order 50 qubits. Certain clever methods using tensor networks \cite{orus_practical_2014,kshetrimayum_simple_2017} and renormalization group theory can boost this number to say, 70 elements, but nonetheless do not approach the realm of complex materials. Quantum simulators provide access to classically intractable problems in chemistry\cite{hempel_quantum_2018,mccaskey_quantum_2019,lanyon_towards_2010}. A grand challenge, for example, in the field, is to study catalysis in particular with respect to the production of ammonia. Ammonia is a key agent used in fertilizer production and other industrial products. the industrial scale Haber process, which combines nitrogen and hydrogen at 400 degrees centigrade and 200 times atmospheric pressure, consumes of water a few percent of all the energy on Earth. Remarkably, bacteria execute nitrogen fixation at ambient conditions, making use of a catalyst, which in principle can be studied with a quantum computer. Solving problems of this type, though still some distance away from the capabilities of current quantum machines, nevertheless promises to have a tremendous impact on human civilization. 

\section{Practical Quantum Simulation: Contemporary VQE}
Let us work with a simpler molecule, hydrogen. we want to describe how one can develop a quantum algorithm to treat electronic structure problems in chemistry\cite{motta_low_2018,babbush_low-depth_2018,mcclean_openfermion_2019,higgott_variational_2019,endo_variational_2019,mcardle_variational_2019}. we first note from quantum mechanics that an electron is a fermion and obeys symmetry rules associated with such particles, and moreover, is indistinguishable from other electrons. Qubits, on the other hand, are certainly indistinguishable, since they can be labeled and addressed individually. In fact, one addresses them by the use of poly operators, and the first step in mapping a chemistry problem onto a quantum computer is to rewrite the Hamiltonian, which describes the total energy of the system from electronic creation and annihilation operators, labeled a and a dagger, into polyspin operators, $ \sigma $.  this can be accomplished, for example, by the famous Jordan-Wigner transformation. This mapping preserves the number of terms in the Hamiltonian. However, the locality is lost in that spin operators from different sites are convolved, and the coefficients, g, shown here have to be computed classically. It turns out that this can be done without too much trouble. The next step is then to evaluate the terms in the Hamiltonian, which involve products of spin operators and are analogous to calculating correlation functions. This is where the quantum simulation magic comes in., we going to describe a clever algorithm, known as the Variational Quantum Eigen solver, or VQE. In this approach, we rewrite the electronic structure Hamiltonian in a power series expansion of poly operators. we are interested in computing the energy of the molecule, expressed as the expectation value of the Hamiltonian, which by linearity, can be expressed as a sum of the energies associated with each poly correlator in the sum. To calculate each term, we can arrange our quantum simulator in the corresponding configuration by way of single and two-qubit logical gate operations. One of the powerful features of this technique is that each term can be evaluated independently of the other, thereby reducing the number of quantum logic operations that have to be run in succession to only a few. This process allows current quantum hardware to execute the routine. Finally, the task of combining these terms is performed using a classical computer. Since we do not make a priori, know the electronic configuration that will yield the lowest energy, a set of parameters is scanned, and the energy landscape is produced. So, in summary, executing the algorithm consists of parameterizing the quantum state in terms of the classical parameter. Setting the knobs, we can turn in the experiment to a particular value, computing the terms in the Hamiltonian using our quantum hardware, and then finally turning our control knobs until a minimum is found. Let us see this in action for hydrogen. we are going to ignore the motion of the heavy nuclei in this calculation and treat only the electronic structure problem. In the first step, we choose a common basis of orbitals to describe the electrons in the system and then transform the Hamiltonian into the poly operator basis that we described earlier. This is called the pre-compilation step\cite{booth_comparing_2018}. Since it is a classical computation step, we can show it as blue in the figure. In this problem, we have six degrees of freedom corresponding to two electrons, and we prepare the system in a target configuration by adjusting six experimental parameters that we have access to in our qubit system. In this case, these parameters correspond to the amplitude and phase of the excitations we can apply to each individual qubit and the length and phase of a pulse used to create quantum entanglement between the two qubits. This set of six parameters is simply expressed as a single vector, $ \theta $. This step prepares the quantum hardware in the calculation and, as shown in tan. Then, to extract the individual terms that go into the energy calculation, we measure the specific correlators that appear in the Hamiltonian and use them to calculate the ground state energy. we can then repeat the process for a variety of different values of $ \theta $ until the minimum is reached. Finally, there is a neat trick to calculate the excited states as well. Once we obtain our best guess for the ground state, we can perform an operation analogous to a Taylor expansion, where operators that cost excitations in the system can be approximated in the vicinity of the energy minimum. This can be done with only a modest number of additional measurements. Putting it all together, voila. we have the full energy spectrum of the hydrogen molecule computed using a specialized quantum computer, running a hybrid algorithm that makes efficient use of both classical and quantum computing resources. The data points in the figure represent the values obtained from the quantum machine, and the solid lines are the predicted exact values obtained from a conventional computer. The accuracy of the result is shown in the upper right inset, which shows that the quantum computations are hovering around a milli-Hartree discrepancy in energy. This level of error is near what is called chemical accuracy and indicates that the results are good enough to predict chemical processes. Though we can compute the classical energy spectrum for a simple molecule like hydrogen, the complexity of the calculation grows exponentially with system size.  we would rapidly exceed the computational power of a supercomputer with say 100 qubit system. The key science questions that must now be tackled gravitate around how VQE will perform for larger numbers of qubits. Will sources of error scale in a way that does not require large resources for error correction? Will the quantum processor be stable over the time needed to make all the correlator measurements? These questions all point to the fact that this is an exciting time in quantum information science as we explore the advantage of emerging quantum hardware\cite{nielsen_quantum_2011,mermin_quantum_2007,national_academies_of_sciences_quantum_2018}. In summary, specialized quantum computers are already here and have an important role to play in many optimization problems in basic science from chemistry to cosmology. we need to study sources of error in hybrid algorithms like VQE and develop efficient routines to mitigate them\cite{chen_hybrid_2019,mcclean_theory_2016,peruzzo_variational_2014,li_efficient_2017}. Relatedly, can the classical information processing needed to execute such routines be efficiently handled using machine learning methods? All of these points put together will help us determine the true power of these types of quantum machines. 

State-of-the-art quantum computers today have 10-100 qubits, and in the not-too-distant future, we can foresee a demonstration of ``quantum supremacy''\cite{harrow_quantum_2017, arute_quantum_2019,markov_quantum_2018}.  Loosely defined, quantum supremacy is the demonstration of a task that can be successfully performed on a quantum computer, and yet cannot be performed on existing classical computers in any reasonable amount of time. For the purposes of demonstrating quantum supremacy, the actual problem implemented need not be useful, commercialize able algorithm that addresses an important problem to humankind\cite{mohseni_commercialize_2017}; rather, its sole purpose could simply be to demonstrate that quantum supremacy is possible.

Thus, while the demonstration of quantum supremacy will stand as an important milestone in the to-be-told history of quantum computing\cite{boixo_characterizing_2018}, it might not represent the turning point when quantum computers suddenly become useful machines. This is because all of the known quantum algorithms that do address important problems in cybersecurity, simulation, and optimization, require very large numbers of essentially error-free logical qubits, and the required technology to build such universal and fault-tolerant quantum processors is likely to be at least a decade or two away\cite{mohseni_commercialize_2017}.

Between the first demonstration of quantum supremacy and the advent of fault-tolerant quantum processors, the question is, what can we do in the meantime? Are there practical, useful algorithms that we can do with the noisy, intermediate-scale quantum (NISQ) computers we have now and will have for the foreseeable future? This is an important question to answer because we know that technology development requires significant investment. Investment by governments worldwide have helped to start the field of quantum computing, and this investment will continue to play an important role. However, government investment alone might not be sufficient to realize the potential of quantum computation.

It is imperative that the community identify quantum algorithms that run on NISQ computers and address important, meaningful problems that can be commercialized. This will kickstart a virtuous cycle of technology development. The revenue from the current generation of NISQ computers will support the development of a next-generation, higher-performance NISQ computer. There is currently significant research efforts aimed at developing such NISQ-era algorithms, starting with the types of prototype algorithms. Already there are blue-chip and start-up companies working to commercialize cloud-based access to quantum computers\cite{dumitrescu_cloud_2018}. With more access to such quantum computers, in conjunction with the development of a software stack needed to facilitate quantum-computer programming\cite{larose_overview_2019}, more and more researchers will be able to develop and test prototype algorithms that target commercialize applications.

\section{Solving Linear Systems of Equations} 

One not so, the obvious application of the Hamiltonian simulation algorithms \cite{bravyi_tapering_2017}. we have seen is a quantum algorithm to solve the linear system of equations, which becomes particularly interesting when these are very large systems of equations. So, we are going to tell us about that algorithm now. First, we are going to tell us about what we mean by solving a linear system of equations, and we will tell us about classical algorithms for the problem\cite{pednault_breaking_2018}. Then, we will discuss a quantum approach to this and discuss its performance, and then we will discuss some issues that arise with the quantum algorithm that is not present for the classical algorithm \cite{giri_review_2017}. So, by a system of linear equations, what we mean is that we have a vector of unknowns that we would like to find. we will call them x1, x2 through xn, which for our purposes will be complex numbers, although reals would also work. then, we are given the coefficients of those equations. So, we would be given an equation of the form 3x1 plus x3 equals 5, x2 minus ix3 equals 0, and so on. we want to find a bunch of x's that satisfy all of those equations simultaneously. It is more convenient to put these equations into the form of a matrix. So, we would have a matrix equation Ax equals b, where A is an N by N matrix that we are given, x is our vector of unknowns, and b is another vector of coefficients that we are given as part of the input. We are going to tell us about the complexity of this problem and how hard it is to solve classically in quantumly. To discuss complexity, this needs to be parameterized by several things. One may be an obvious thing to parameterize it by is the size of these vectors and matrices. we say that the dimension N is the length of the vector x, and the size of the matrix A is N by N. But, there are other parameters as well, which depend on features of the input. So, one of them is called the sparsity of the matrix A. In general, we want to work with matrices where most of the entries are 0, and only a few are non-zero. So, we say that the sparsity is S if there are at most S non-zero entries in A per row or column. S could be as large as n, in which case we would have a dense matrix. However, often, we are interested in matrices where S happens to be small and or our algorithms can make use of this to run more efficiently. Another important parameter is the condition number, kappa, which we will discuss more later. Kappa is defined as the ratio of the largest singular value of A to the smallest singular value of A. When A is a unitary matrix, it is one. This is the smallest that kappa can be. As an approach to a non-invertible matrix, kappa approaches infinity. Thus, we can see that kappa represents, in some sense, the difficulty of inverting the matrix A. One nice feature of it is it is invariant under rescaling in matrix A by a scalar. In some sense, it measures the numerical instability of the matrix inversion process. What we mean by that is if we perturb the input b a little bit, that will cause some change in the vector x. how large that change depends on the condition number. If the condition number is small, a small change in the input will produce a small change in the output. If the condition number is large, then there exist small changes in the input b that will cause very large changes in the output x. So, that is what we mean by numerical instability. we will describe two classical algorithms for this problem or rather families of classical algorithms before we get to the quantum algorithm. One family of classical algorithms is called iterative solvers because they start with a partial solution which they iteratively improve. These are best described in terms of a closely related problem, which is not solving Ax equals b but minimizing the norm of Ax minus b. Because this norm is a convex function, if we use gradient descent, that will work well. we should note that this generalizes the linear systems problem because it is also interesting even when there is no x that exactly solves Ax equals b, we still might want to minimize the norm of Ax minus b. In other words, this minimum might not reach 0, but still, we want to find the minimum. This is useful, for example, in the problem of linear regression \cite{dutta_demonstration_2018}, where we want to find the best linear model to fit a set of data points. So, if we apply gradient descent to this problem, the norm is a very simple function. It just looks like a giant paraboloid if we plot it. What we have shown on the screen are the level sets, which are these nested ellipses. if we follow the path of gradient descent, we will follow these blue arrows. It will sort of move towards the minimum in this zigzags path. we can see that the more distorted the ellipses are, the farther from spherical they are, the more this path will zigzag. this is precisely what the condition number measures. The condition number is the ratio of the longest axis to the smallest axis. Thus, the larger the condition number is, the more stretched this ellipsoid is, and the longer it will take gradient descent to find the middle. we can see that this will converge to error epsilon in time that scales like that condition number kappa times log 1 over epsilon. This log 1 over epsilon dependence is not too bad. However, the linear scaling with condition number is why we say that the condition number reflects the difficulty of solving the new system of equations in this method. So,  the number of iterations is kappa log 1 over epsilon. Each iteration requires multiplying a matrix, in fact, matrix A by a vector. doing this involves iterating over all the entries of the matrix A. the time for that is just the number of non-zero entries, which is order N, the dimension, times s, the number of non-zero entries per row. If we put this all together, we see the time is on the order of Ns kappa log 1 over epsilon. Another class of solvers has better dependence on kappa but much worse dependence on dimension. The dimension dependency goes like order N cubed, which makes them non-competitive for very large matrices. These are called direct solvers. Examples of these are Gaussian elimination and the LU decomposition. If we remember back in high school, learning to solve a linear system of equations, this was probably the method that we learned to use. They work well for small systems but are generally not used for large N because the iterative methods will be performed much better as N becomes large.

\section{Quantum Algorithm for Linear Systems}
Now that we have seen a classical algorithm to solve a linear system of equations, we will describe to us a quantum algorithm. The appeal of this algorithm we can see in the run time, which is similar to the classical one in its dependence on the sparsity, the condition number, and the error, but improves that linear dependence in N to an order log N dependence. So, when N is large, this can mean dramatic savings in run time. Just a reminder that parameters N, S, and kappa refer to dimension, sparsity, and condition number. The way the algorithm is able to achieve this is by not writing down the entire inputs and outputs just as states in memory because by the time we would have written it down, we already would have had to expend effort on the order of N. Instead, the input and output will be represented as quantum states of N dimensions, which means they can be encoded into log N qubits. So, in this way, we never have to work with N actual pieces of memory or expend effort linear in N. The matrix A also cannot afford to fully write down because that would involve N times S pieces of memory. Instead, we will assume that A is computable on the fly. We will need to be able to, given the index of a row of A, some procedure that will output the non-zero entries of that row along with their locations. So, this input-output model is what makes possible the run time scaling with log N. As a result, it means that the quantum algorithm achieves something which is not directly the same as in the classical case. we cannot just take place where we would have used the classical linear system solver and plug in the quantum one. we have to make sure that we can meet the input and output requirements in order to run the algorithm. This is analogous to what we discuss with the quantum Fourier transform. The quantum Fourier transform took something that classically ran in time N log N. It turned it into a quantum run time that was polynomial in log N \cite{bremner_classical_2011,calude_-quantizing_2007}. The only way for this to be possible is to transform the amplitudes of the state rather than just a list of numbers. Similarly, the quantum algorithm will solve a linear system of equations where the input and the output are amplitudes of a quantum state rather than lists of numbers stored in memory. So, lets now tell us how the algorithm works. There are three building blocks that we will put together in order to create the quantum linear system solving algorithm.  for the purpose of this example, we are going to assume that A is a Hermitian matrix. The algorithm applies more generally as well, but the extension to non-Hermitian matrices is something that we will not mention right now, how that works. So, the three building blocks are Hamiltonian simulation, which we discussed earlier. Here, what this means is we want to apply e to the iAt, and the time required to do this should scale like the norm of A as well as the time, the evolution time, t. We discuss this in the context of Hamiltonian's, but the same ideas would apply to any matrix A presented in the form that we described where we have an efficient way, given a row index, to find the non-zero entries of that row. So, in other words, it applies to matrices that are not necessarily Hamiltonian's of physical systems, but may, for example, be the matrix of coefficients for the linear system of equations that we want to solve. The second building block is phase estimation. Here, we assume that we can apply something of form e to the i $ \lambda $ t for some, perhaps, scalar $ \lambda $, and we want to estimate $ \lambda $. Phase estimation is a way of doing this where the accuracy goes like 1 over t, where t is the amount of time that we have evolved. Essentially, this can be viewed as frequency-time uncertainty. If we look at some sample for up to time t, then we have frequency resolution, which goes like 1 over t. The third ingredient we could call filtering. In this case, we will make a measurement and condition on the measurement coming out a particular way. This is something that only succeeds with some probability, not necessarily probability one. the result of doing it is that the state will evolve in a way that is still linear but can be non-unitary. Thus, this is called filtering because it is analogous to the way when the light goes through the filter; it can be transformed. For example, we can remove some colors or attenuate the amount of some frequencies. So, how do we put these together into the algorithm? First is a mathematical step. Nothing happens but will express the input b in terms of the eigenbasis of A. So, we can expand b out into a sum of eigenstates of A, call them A sub i, and coefficients which we will call b sub i. We will then apply the phase estimation to estimate the phase that we get from doing the Hamiltonian simulation to apply e to the iAt. When we do e to the iAt, each eigenstate of A will pick up phase at some rate, depending on what the corresponding eigenvalue is of A. Call those eigenvalues, $ \lambda $ sub i. If we combine the two, then what we get is we get an estimate for each eigenstate of A of the corresponding eigenvalue. we have put an approximate equal sign because these procedures are not perfect, but we can bound their error. we can rigorously say how accurate they are. So, what this does is it attaches to each A sub i an estimate, $ \lambda $ i, of the corresponding eigenvalue. Once we have this estimate, we can perform essentially any function of $ \lambda $ that we would like. For our purposes, we are interested in matrix inversion, but other work has extended this to other functions of $ \lambda $ as well, which in other cases can be interesting. To do matrix inversion, what we want to do is we want to add a factor of 1 over $ \lambda $ i here. Essentially, we want to apply A inverse to the state, which means dividing the corresponding eigenvector, A sub i, by the eigenvalue, $ \lambda $ sub i. To do this, we will do a filtering operation where we, in some cases, accept and, in some cases, reject.  our probability of acceptance should scale like 1 over $ \lambda $ sub i. If we do this and condition on acceptance, then what we will have done is essentially added a 1 over $ \lambda $ i term to the appropriate point in the sum. we then want to un-compute $ \lambda $ sub i. This means reversing step 2, which will erase that register. we will still be left with this coefficient of 1 over $ \lambda $ sub i, but we will not have that extra quantum state, $ \lambda $ i, which had we kept could have caused decoherence. we are left in the last line with this term that essentially looks like A inverse applied to b, which is the state that we wanted to solve, cat x, the solution to the linear system of equations. 

\section{Discussion of Quantum Linear Systems Algorithm} 
Now that we have seen this algorithm, how should we think about its performance? Since we have had such a dramatic improvement in the dependence on n, now it makes sense to look at the other parameters since those dependencies have not been reduced. If the condition number, kappa, is still large, then we will not see such a big speedup. Whereas conversely, the places where we should find a big quantum speedup are where the condition number is very small.  we can show, in fact, that this dependence and condition number cannot be reduced. So, this linear and kappa dependence, that we see in both classical and quantum, we are essentially stuck with. One way to think about the role of condition number, here, is that what we have discussed about quantum circuits, that they always produce unitary evolutions, is more of a guideline than the hard and fast rule. If kappa is one, then we are inverting a unitary matrix. we have no overhead from kappa. Everything runs very efficiently, but we can also take kappa to be a little bit larger, and then our runtime will only be a little bit larger as well. So, we can afford to do something that is slightly non-unitary and just pay a little bit of a price for it in terms of runtime. This idea has been used in places that go far beyond the linear system of equations algorithms we have just described. One example is an improved algorithm for Hamiltonian simulation, which is based on taking the Taylor series, for e to the minus IHT, spanning it out to a ton of terms, many of which might be either non-unitary or proportional to unitaries, and taking linear combinations of those. The intermediate steps of that algorithm involve things that are non-unitary, but only mildly non-unitary. so, the cost of doing this is not too bad. They are able to put together the whole package for a very efficient overall algorithm to do Hamiltonian simulation. Another example is due to Dominic Berry, who describes an algorithm to simulate ODEs, where the cost scales with how non-unitary that time evolution is. The Schrodinger equation can be thought of as an example of an ODE, where the time evolution is exactly unitary. However, Berry extends this to things where the time evolution can be a little bit non-unitary. the cost only goes up with the amount of non-unitary that we experience. Another important issue with the algorithm that distinguishes it from the classical version is the format of the input. As we have seen, the input should be a vector CatB b, where the entries of the vector are encoded as the amplitudes of the input state. Similarly, the output we receive in the form of amplitudes of a quantum state. The matrix a also is not something that is just written down as a list of numbers. However, it should be something that we can compute on the fly, meaning that we are given a row index we and it can efficiently describe all the non-zero entries on that row of a. Similar issues here also apply to recent work on quantum algorithms for machine learning, for which there has been much recent research. What we need to do is find a way to put the algorithm, such as the linear systems algorithm, into a complete package. This is a little bit analogous to what Shor's algorithm does with the quantum Fourier transform. If we remember, the quantum Fourier transform is a purely quantum procedure. It takes inputs that are amplitudes of quantum states and transforms into an output, which is also an amplitude of a quantum state. To make that useful, we need an entire package that begins and ends with classical data. So, how do we do this for linear systems? we need to take a classical input string, let us say if little n parameters, y1 through yn, and transforms it into a high dimensional state, which can be the input to the quantum linear systems algorithm \cite{harrow_quantum_2009}. we also need to do something about the output, but this turns out to be much easier. If we have an output state, then we can measure it and estimate observables. This is often closer to what we already need. However, dealing with input is an important algorithmic problem. This is an ongoing research question. So, we will not be able to present a full answer, but we will mention two solutions that people have considered. One of them is to build a new type of quantum hardware, called a quantum RAM \cite{zeng_quantum_2016}, that would allow quantum queries to a large classical data set. In this case, the dimension capital n would be on the same order as the number of input parameters little n. One model of what this might look like is something like a CD-ROM, where we have a large amount of classical data that could, in principle, be queried by a photon that is in a superposition of different modes that picks up phase information corresponding both to its quantum data about which cell is querying and to the classical information in that cell. This would be a powerful thing to be able to build, but we do not have large examples of this yet. Just like we do not have large quantum computers. The second approach is what we call algorithmic generation. In this case, we have a small classical input. Let us say of little n degrees of freedom. we use this with some algorithm to create a high dimensional quantum state based on this. This might be like a nonlinear embedding of the classical input into a larger space. Another example of where this might come up is if our classical input describes the shape and our state that we are applying the linear systems algorithm to comes from a finite element model, where we have discretized space very finely. Our shape just tells us where There is material or where there is not. This would come up if we were to try to solve things like the wave equation. So, in this case, we could algorithmically generate a high dimensional state from a small input. These are two ways of trying to address the issue of how to construct an input in order to put the linear systems algorithm into an overall package that would start and end with classical data and solve a problem more efficiently than we could solve on a classical computer. These are still stepping that would need to be worked out in order to solve practical problems with this linear systems solver.

\item  Linear System of Equations: Characterization; One application of Hamiltonian simulation is solving a linear system of coupled equations. Let us consider the example from the section : Solving a Linear System of Equations,
\begin{equation}\label{eq3_74}
\displaystyle 3x_1+x_3    \displaystyle =    \displaystyle 5,         
\displaystyle x_2-ix_3    \displaystyle =    \displaystyle 0,         
\displaystyle -2x_1+x_2    \displaystyle =    \displaystyle -1.
\end{equation}

In this system, the variables are $ x_1, x_2 $ and $ x_3 $. All the equations have linear combinations of these variables. The system is considered to be solved when one finds the values of $ x_1, x_2 $ and $ x_3 $ that satisfy the three equations simultaneously.

To solve linear systems of equations, it is convenient to write the system as one equation of the form$  Ax=b $, where x and b are vectors of length N, with N equal to the number of variables, and where A is an $ N\times N $ matrix. The length of the vector x is also called the dimension of the system.

Following the example above, the system of equations can be written as
\begin{equation}\label{eq3_75}
\left( \begin{array}{ccc} 3 &  0 &  1\\ 0 &  1 &  -i\\ -2 &  1 &  0 \end{array} \right) \left( \begin{array}{c} x_1\\ x_2\\ x_3 \end{array} \right) = \left( \begin{array}{c} 5\\ 0\\ -1 \end{array} \right),
\end{equation}
In this example, the length of the vectors x and b is $ N=3 $, and the matrix A is a $ 3-by-3 $ matrix.
The difficulty in finding a solution of a linear system of equations can be parametrized by several parameters. The trace of a matrix representing a system of linear equations does not indicate the difficulty in finding a solution. The matrix trace is equivalent to the sum of the eigenvalues of the matrix.

\item  Quantum Algorithm for Linear Systems I; Given a hermitian N-by-N matrix A, and a unit vector $ \vec{b} $, the quantum algorithm for linear systems finds a sample vector $ \vec{x} $, such that approximately $ A\vec{x}=\vec{b} $.
According to the section ``Quantum Algorithm for Linear Systems,'' following is building block of the quantum linear system solving algorithm. There are three building blocks of the quantum algorithm for solving systems of linear equations: Hamiltonian simulation, phase estimation, and filtering. To solve the eigenvalue equation $ e^{iAt}\lvert b\rangle = e^{i\lambda t}\lvert b\rangle$, it is necessary to find the eigenvalues $ \lambda $ of matrix $ A $. The phase estimation algorithm estimates the eigenvalues $ \lambda $ with accuracy that scales as $ 1/T $, where $ T $ is the amount of time that the system has evolved. To learn more about the HHL algorithm, we encourage to read ``Quantum algorithm for linear systems of equations'' \cite{harrow_quantum_2009}

\item Quantum Linear Systems Algorithm; Given a hermitian $ N-by-N $ matrix A, and a unit vector $  \vec{b} $, the quantum algorithm for linear systems finds a sample vector $ \vec{x} $, such that $ A\vec{x}=\vec{b} $. The condition number $ \kappa $ is one of the parameters used to characterize the efficiency of the quantum linear system algorithm, and it is given by
\begin{equation}\label{eq3_76}
\kappa =\lvert \lvert A\rvert \rvert \cdot \lvert \lvert A^{-1}\rvert \rvert ,    
\end{equation} where $ A^{-1} $ is the inverse of the matrix A.
the condition number measure how far A is from an invertible matrix. The performance of the quantum linear system algorithm depends on the dimension $ N $ of  $ \vec{x} $, and the condition number $ \kappa  $. If the condition number is one, then the matrix is invertible. As $ \kappa  $ becomes larger, the runtime of the algorithm increases, but the algorithm can still provide a quantum enhancement. 

The quantum LPN algorithm is robust against noise, but it cannot identify where an error has occurred. It requires fewer queries than its classical counterpart for the same level of noise. 

\item  Noisy Intermediate-Scale Quantum (NISQ) Computing; John Preskill introduced the concept of noisy intermediate-scale quantum (NISQ) computing. ``Intermediate-scale'' refers to computers sized between today's 50-100 qubit machines and the large-scale, fault-tolerant machines of the future. ``Noisy'' refers to the fact that these machines will not be error-corrected\cite{bremner_achieving_2017}. NISQ hardware can be used to do tasks such as quantum simulations, in which qubits are coupled in some geometry, allowed to interact with each other, and measured. Regarding NISQ hardware, Even the most powerful classical supercomputers with petabytes of memory can only exactly simulate quantum systems comprising on the order of 50-70 qubits.There are many types of scientific calculations that can be efficiently mapped onto a quantum simulator. For calculating energy levels, classical computers do not have enough memory to describe the large combinatorial space quantum systems live in. More about NISQ read ``Quantum Computing in the NISQ era and beyond''\cite{preskill_quantum_2018}.

\item  Variational Quantum Eigensolver (VQE); A variational quantum eigensolver was used in the sections to compute the energy of an atom or molecule, where the energy is given as the expectation value of the Hamiltonian. When using the VQE approach, the Hamiltonian is re-written as a power series expansion of Pauli operators, $ \sigma _ x $, $ \sigma _ y $, and $ \sigma _ z $ (or, X, Y, and Z gates). The expectation value of the total Hamiltonian can be written as a sum of energies associated with each Pauli operator term. VQE is a quantum/classical hybrid algorithm, which uses a classical computer to generate trial wavefunctions, and a quantum computer to find their energy for a given Hamiltonian. The VQE is used for atomic and molecular systems\cite{omalley_scalable_2016}, but it can, in principle, be used to find the ground state energy of any quantum system. 

\item  Quantum Algorithm for Linear Systems II; According to the section ``Quantum Algorithm for Linear Systems,'' Three building blocks of the quantum algorithm for solving systems of linear equations: Hamiltonian simulation, phase estimation, and filtering. Techniques of hamiltonian simulation are used to apply $ e^{iAt} $ to the state $ \lvert b\rangle $. This is done on time $ O\left(\lvert \lvert A\rvert \rvert t\right) $, where $ \lvert \lvert A\rvert \rvert $ is the norm of the matrix $ A $.

\begin{figure}[H] \centering{\includegraphics[scale=.7]{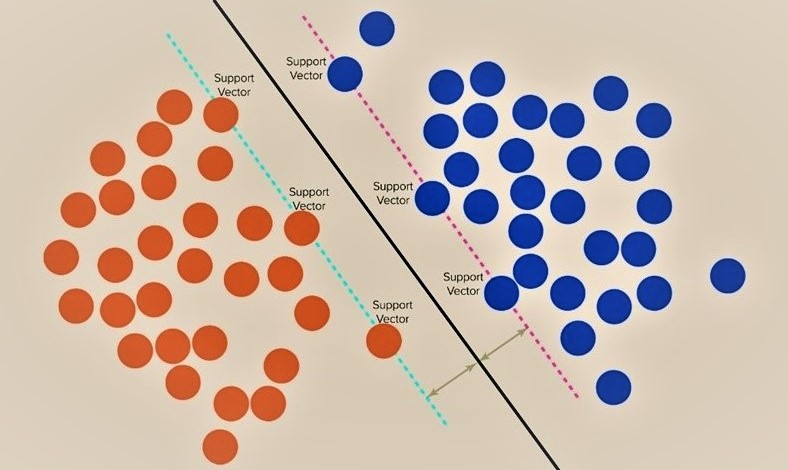}}\caption{Support Vectors}\label{fig3_10}
\end{figure}

\begin{figure}[H] \centering{\includegraphics[scale=.7]{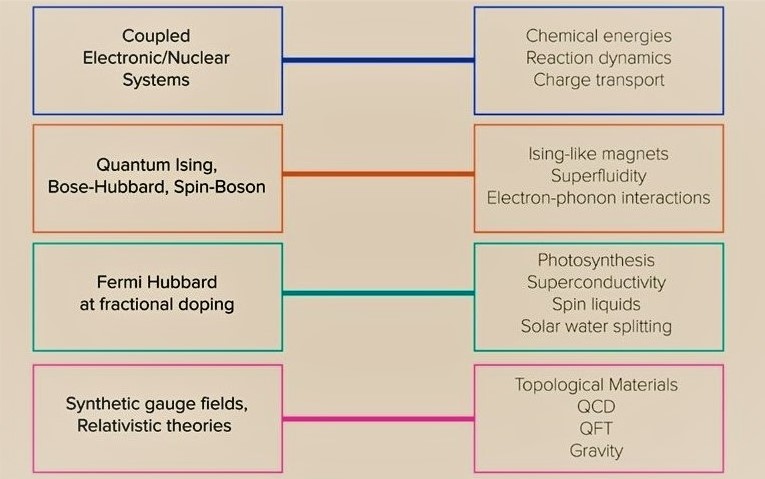}}\caption{Quantum Simulator}\label{fig3_11}
\end{figure}

\section{Introduction to the Benchmarking Quantum States and Quantum Gates} 

In the last section, we introduced noisy intermediate-scale quantum computers and discussed examples of the types of algorithms that one might foresee running on them. NISQ computers are smaller scale quantum computers, say, 50 to 100, or 1,000 or a few thousand qubits that we either currently have, or will become available within the next several years. one of their main features is that They are not yet error corrected. This means that their utility is predominantly limited by their intrinsic gate fidelity, the types of gates enabled by the hardware, and, ultimately, the circuit depth they can support. As discussed in the last section, the quantum volume \cite{cross_validating_2019} is one metric that has been developed to compare different modalities and architectures. Essentially, it is a metric that attempts to capture the aggregate effect of the many characteristics that influence how well a particular quantum computer can implement a given algorithm. NISQ computers will likely implement small application-specific algorithms. For example, implementing an algorithm as a quantum coprocessor to a classical computer. we discuss an example of this last section with the variational quantum eigensolver and its application to determining the ground state energy of atoms and molecules. There is significant research currently in trying to develop useful algorithms where a small-scale quantum computer can implement a meaningful task within the circuit depth afforded by the processor. Finding such a commercialized application would be an important milestone in the development of quantum computing. The promise of universal quantum computation will require error protected qubits to perform computation with large circuit depth. whether for universal error protected computers, or for NISQ computers, it is imperative to understand how well our data operations perform at the physical level in the presence of noise. Thus, this section will take a deeper look at how we benchmark quantum states in quantum gates to understand how well we can prepare, operate, and even measure a quantum system. The types of state and process tomography that we will discuss this section will be referenced throughout section four as we introduce fault-tolerant error correction and the threshold theorem. we will start with how to benchmark quantum states, and then we will use this state tomography to benchmark quantum gates, called process tomography \cite{mohseni_quantum-process_2008}. we will then look at experimental, practical aspects of benchmarking, the overhead required for tomographic methods, and the use of randomized benchmarking protocols \cite{corcoles_process_2013} to reduce that overhead. Finally, at the end of the section, we will have the opportunity to implement state tomography for ourselves on a real quantum computer. 
\section{Quantum State Tomography} 
Quantum state tomography is a vital experimental technique at the heart of debugging the operation of quantum computers, which essentially reads out a quantum state \cite{thew_qudit_2002}. Recall that the quantum state of a single qubit is convenient to represent using the Bloch sphere picture. In this three-dimensional sphere, the north pole represents the 0 states of a qubit, while the south pole represents the 1 state. The positive x-axis point is the equal superposition of 0 and 1. the positive y-axis point is an equal superposition with a complex i phase factor. An arbitrary state, shown here as this green point, can be parameterized with two variables $ \theta $ and $ \phi $. The Bloch sphere is also useful for representing density matrix states, as we have seen. The 0 and 1 states are the same. the surface of the sphere is the location of all density matrices made of a single quantum state that is a pure state. An equal statistical mixture of 0 and 1 is the density matrix represented by a point at the center of the sphere. This is the identity matrix. all other points in the unit ball represent other statistical mixtures. This background allows us a concrete visualization for the question answered by state tomography. Given many copies of a single qubit in-state $ \rho $, how do we experimentally determine $ \rho $? For example, $ \rho $ could be at this point or is this one, and so, forth even possibly a point on the surface of the ball that is a pure state. The idea behind state tomography is that the point representing $ \rho $ can be represented as a vector r, which, for a qubit, is known as the Bloch vector. Mathematically, this is because any valid density matrix can be written as the linear combination of the identity matrix and the three Pauli matrices\cite{smart_experimental_2019}. Recall that the Pauli matrices are denoted as $ \sigma $ x, $ \sigma $ y, and $ \sigma $ z. It is convenient to write $ \rho $ in this way, as 1 plus r dotted with the $ \sigma $ vector which is a vector of matrices all divided by 2. The key idea behind state tomography of a single qubit is to measure the three vector components Rx, ry, and rz. we claim that this can be done by recognizing that each component is the statistical expectation of a Pauli matrix. For example, rz is the expectation value of $ \sigma $ z. This can be seen by recalling the expectation is defined as a trace of the density matrix times the observable here, $ \sigma $ z. And $ \sigma $ z is 0,0 minus 1,1 matrices. So, the expectation value is just the 0,0 entry of $ \rho $ minus the 1,1 entry of $ \rho $ that is the projection along the z-axis, as we might expect. Mathematically, we can formally prove the claim by substituting the Bloch vector expression for $ \rho $ into the expectation value, expanding to see that the products of Pauli matrices then appear, and recognizing that all these matrices are traceless except $ \sigma $ z times $ \sigma $ z, meaning only the last term with r's of z remains. the trace of the identity matrix gives 2, completing the proof. So, the procedure for state tomography of a single qubit reduces to measuring the Bloch vector components. For $ \sigma $ z, this is done by obtaining the probability of a qubit being 1 in the computational basis. For $ \sigma $ x, this is done by rotating minus 90 degrees around the y-axis to move plus x to plus z, then measuring. Moreover, for $ \sigma $ y, we first rotate by 90 degrees about x to move plus y to plus z before measuring. The Bloch vector components are, thus, these three expectation values. The single-qubit picture of state tomography was worked out in the 1950s for nuclear magnetic resonance, but its utility emerges for tomography of multi-qubit states. For two qubits, the density matrix may similarly be represented as a weighted sum of now tensor products of Pauli matrices. the measurement method is, as before, accomplished by rotating each qubit by 90 degrees before measurement, which, for two qubits, requires nine different measurements. State tomography has been experimentally realized in a wide number of physical systems. Shown here is one example from the 2003 experiment by Rainer Blatt's group at the University of Innsbruck, depicting the density matrix for two ions in the entangled state SS plus DD. we see that the real part of the density matrix has the expected for non-zero equal values with the off-diagonals being the signature of the entanglement of the state. we also see that the imaginary part of the density matrix is 0 to within experimental errors. This state tomography measurement was one of the first to be accomplished in a well-controlled multi-qubit system. many telegraphic measurements of even much larger quantum systems have been experimentally accomplished since these early results. State tomography remains an important foundational experimental tool today for diagnosing quantum states in quantum computers.

Quantum state tomography (QST) is a method used to characterize the quantum state of a physical system through a sequence of projective measurements on different bases. Due to the probabilistic nature of quantum states, a single projective measurement does not reveal complete information about an unknown state. Further, the act of measurement generally changes the system itself, limiting the utility of making additional measurements on the same system.

For these reasons, QST can only be successfully performed on a large ensemble of identically prepared quantum states. The availability of multiple identically prepared states enables the expectation values of measurements in different measurement bases to be estimated through repeated projective measurements. These estimated expectation values can then be used to infer the state of the system.

We will now illustrate the QST procedure where the unknown system is a single qubit. Regardless of the encoding, the density matrix of a single qubit, whether pure or mixed, can be represented as a sum of the identity matrix and the three Pauli matrices in the following form
\begin{equation}\label{eq3_77}
\rho=\frac{1}{2}(I+r_{x}\sigma_{x}+r_{y}\sigma_{y}+r_{z}\sigma_{z})=\frac{I+\vec{r}\cdot\vec{\sigma}}{2}
\end{equation}

where I is the identity matrix in two dimensions and the three Pauli matrices being the previously introduced X, Y, and Z gates given by
\begin{equation}\label{eq3_78}
X=\sigma_{x}= \left( \begin{array}{cc}  0 & 1 \\ 1 & 0  \end{array} \right), \quad Y=\sigma_{y}= \left( \begin {array}{cc} 0 & -i \\ i & 0  \end{array} \right), \quad Z=\sigma_{z}= \left( \begin{array}{cc} 1 & 0 \\ 0 & -1  \end{array} \right). \quad    
\end{equation}

The Pauli matrices are combined into a vector $ \hat{\sigma}=(\sigma_{x},\sigma_{y},\sigma_{z}) $, and this vector is acted on by $ \vec{r}=(r_{x},r_{y},r_{z}), $ which is known as the Bloch vector. The Bloch vector determines the state of a single qubit, and it has magnitude $ \vert \vert \vec{r}\vert \vert\leq 1 $, with $ \vert \vert \vec{r}\vert \vert = 1 $ representing pure states.

The aim of single-qubit tomography is to determine $ \vec{r} $ for a given quantum state. This can be accomplished by relating the expectation value of a projective measurement on $ \rho $ along a direction $ \hat{m} $ to the components of $  \vec{r} $ using $ \langle \sigma_{m}\rangle = \hat{m}\cdot\vec{r} $. Experimentally, this means projecting N identically prepared states represented by $ \rho $ along$  \hat{m} $ and calculating $  \langle \sigma_{m}\rangle_{E} = (N_{\uparrow}-N_{\downarrow})/N $ where $ N_{\uparrow} (N_{\downarrow}) $ represents the number of times $ \rho $ was measured to be in the ``up'' state with eigenvalue 1 (``down'' state with eigenvalue -1) along the $ \hat{m} $ direction. The subscript E indicates it is an estimate.

In the simplest scenario, called direct inversion, these expectation values are obtained along the Cartesian coordinates, such that $ \vec{r} $ can be reconstructed from them using $  \vec{r}_{E}\approx\left(\langle \sigma_{x}\rangle_{E},\langle \sigma_{y}\rangle_{E},\langle \sigma_{z}\rangle_{E}\right) $. While this approach works well, there are numerous sources of error that may impact the estimate of the Bloch vector. For example, state preparation errors, errors in measurement, and stochastic sampling errors may change to the length of the Bloch vector, making it smaller or larger than the ideal value. Numerous techniques have been developed to account for these errors from independently measuring the state-preparation and measurement (SPAM) errors and then effectively ``subtracting them off'' to Bloch vector renormalization, more sophisticated forms of maximum likelihood estimation, and Bayesian estimation methods.  

Extensions of the single-qubit QST procedure to multi-qubit states are straight forward, with the primary difference being that expectation values are over joint projective measurements. The density matrix for two-qubit states with qubit A and qubit B results as

\begin{eqnarray*}\label{eq3_79} \rho&=&\frac{1}{4}(I_{A}I_{B}+r_{xx}\sigma_{x}^{A}\sigma_{x}^{B}+r_{xy}\sigma_{x}^{A}\sigma_{y}^{B}+r_{xz}\sigma_{x}^{A}\sigma_{z}^{B} \\ & &~~+r_{yx}\sigma_{y}^{A}\sigma_{x}^{B}+r_{yy}\sigma_{y}^{A}\sigma_{y}^{B}+r_{yz}\sigma_{y}^{A}\sigma_{z}^{B}+r_{zx}\sigma_{z}^{A}\sigma_{x}^{B}+r_{zy}\sigma_{z}^{A}\sigma_{y}^{B}+r_{zz}\sigma_{z}^{A}\sigma_{z}^{B})\\ 
\end{eqnarray*}

The examples for single-qubit and two-qubit QST indicate that there are three and nine measurement settings required, respectively, each estimating a Cartesian component of $  \vec{r} $ or a combination of them.

For n-qubits, a minimum of $ 3^n $ measurement settings are required. However, due to experimental constraints, in practice, the number of measurement settings required to perform QST scales closer to $ 4^n $. This scaling is a consequence of individual evaluations of all entries of the density matrix representing an n-qubit state and hence $  4^n $ entries.

Density Matrices on The Bloch Sphere I; The Bloch Sphere can also be used for representing single-qubit density matrices. The north pole corresponds to the projector $ \lvert 0\rangle \langle 0\rvert $, and south pole to the projector $ \lvert 1\rangle \langle 1\rvert $. The surface of the unit sphere (i.e., a sphere with unit radius) represents single-qubit density matrices that correspond. In the Bloch Sphere representation for density matrices, the matrices located in the surface of the unit sphere are given by the density matrix for $ \rho =\lvert \psi \rangle \langle \psi \rvert $ of a pure quantum state $ \lvert \psi \rangle $. All of the density matrices inside the unit sphere are mixed states that can be written as a probabilistic combination of pure states, such as $ \rho _ M=\frac{1}{4}\lvert 0\rangle \langle 0\rvert +\frac{3}{4}\lvert 1\rangle \langle 1\rvert $. Since we are representing single-qubit density matrices, these cannot correspond to two-qubit states or entangled states. 

\section{Quantum State Fidelity} 

This is a brief note about quantum state fidelity, a measure of practical importance in experimental quantum computation. Recall that for pure quantum states, and the fidelity answers the question, how well does one state, say psi, represent another state, say $ \phi $. The fidelity, F, is given by the absolute value of the overlap between the two states. Geometrically, the two states can be represented as unit vectors, so F is also interpretable as the projection of $ \phi $ onto psi or vice versa. F is thus a number between 0 and 1. Low fidelity means close to 0, while high fidelity means close to 1. Note that in the literature, sometimes an alternate definition is used, which is the square of the expression we have given to the left. Call this the square fidelity. It has the advantage that it is interpretable as a probability, though the original one reduces directly to the longstanding classical definition of fidelities between probability distributions. So, how fidelity measures extend to noisy quantum states? That is density matrices. Specifically, how well does some density matrix $ \rho $ represent some state $ \phi $? This is given by the square root of the overlap of $ \phi $ and $ \rho $, as shown here. It also turns out this is equal to the largest possible overlap between a purification of $ \rho $ and the state $ \phi $. If psi is a purification of $ \rho $, then it lives in a higher-dimensional space, such that forgetting the extra part produces $ \rho $. See standard textbooks for more. Note that the fidelity between a density matrix and a pure state is commonly desired in evaluating experiments. $ \phi $ is the ideal, theoretical state, while $ \rho $ is the actual noisy experimental result. These are just a few of the most useful expressions and properties of quantum state fidelities. 

\item  Noisy Quantum-State Fidelity; For noisy quantum states or density matrices, the quantum state fidelity characterizes how well an experimental density matrix $ \rho $ represents an ideal pure state$  \lvert \phi \rangle $. In this context, the fidelity is given by
\begin{equation}\label{eq3_80}
F(\lvert \phi \rangle ,\rho )=\sqrt{\langle \phi \rvert \rho \lvert \phi \rangle  }.    
\end{equation}
the quantum state fidelity is valid for.
\begin{itemize}
\item Fidelity represents the maximum projection probability of the state represented by $ \rho $ onto state $ \lvert \phi \rangle $.
\item Fidelity equals the maximum overlap between a purification of $ \rho, $ across all states $ \lvert \psi \rangle $, and a pure state 
\begin{equation}\label{eq3_81}
\phi,
F(\lvert \phi \rangle ,\rho )=\max _{\lvert \psi \rangle }\left\vert \langle \psi \right.\lvert \phi \rangle \rvert    
\end{equation}
\item While $ \lvert \phi \rangle $ is a theoretically ideal pure state, $ \rho $ represents an actual state that is experimentally realized in the presence of noise.
\end{itemize}

\section{System-Environment Model} 
For quantum error correction. We need to describe what an error is in a quantum mechanical context. More generally, let us try to describe all possible ways in which an input density matrix may evolve to an output density matrix. We will describe this via a map called E. E. It should be able to capture any physical process happening to this input state, not just unitary evolution, but non-unitary as well. The model which we all employ will be one which has two parts. We will suppose that we have a system, which may start at a pure state, say psi for now. The system evolves in a unitary manner, but not just by itself. Rather, it is coupled to another state, which we will call the environment. Let us say the initial state of the environment is at e, a pure state. However, that need not be a pure state in general. The environment, let us say, is measured after this unitary transform. We will consider the measurement as happening on an orthonormal basis, e0, e1,  up through the dimension of the environment. The question is, what is the output state $ \rho $? We may write the output stage before the measurement by understanding that the unitary transform gives us U times e times psi. It is convenient to invent some notation to describe this output state. We will use the circle-plus symbol to represent or. Using this, then we may write the output state $ \rho $ as a stochastic combination, where when the measurement result is 0, we get e0, U, e. this is an operator which acts on psi. Alternatively, if the measurement result is 1, we get e1, U, e acting on psi. This may be compactly written as a sum over all possible measurement outcomes where Ek, U, e .we emphasize is a matrix, an operator that acts on psi. We call this operator an operation element. Historically, it is also called sometimes a Kraus operator. This operator E sub k will play a principal role in our study of quantum error correction. We may also sometimes call them error operators. Concisely, we may write the output density matrix in this form as the sum over k Ek $ \rho $ Ek dagger. This is the map E acting on $ \rho $. The main constraint is that the sum of the square of these operators Ek dagger Ek must equal identity. We can check that this is true explicitly by substituting in the definition of E sub k, first using Ek dagger, and then Ek. Then noting that this term in the middle, the sum over Ek Ek, is an identity because this is a complete set of states for the basis of the environment. Thus, this sum overall gives the identity operator as desired. This setup with the system environment model giving us an equation that maps the input density matrix to the output density matrix is an important idea for quantum error correction. We will return to this model several times. 

\section{Quantum Operations} 
Let us now turn to a formal definition of what a quantum operation is. Recall that our goal is to describe in general what a legal map is for $ \rho $ going to E of $ \rho $. Not just including unitary operations, we find that a map E is a valid quantum operation if and only if three criteria are met. First, the output density matrix must have a trace of 1. Second, E must be a linear and convex operation that is acting on a statistical mixture of density matrix inputs. It must also produce a statistical mixture and do so linearly. These two constraints are very natural from the standpoint of quantum mechanics being a linear matrix theory and also that the output must be a density matrix. So, therefore the map must be trace-preserving. A third criterion has two parts. That is that the output density matrix must also be positive. It is natural to require that the output of the map be positive. However, there is more. It turns out that one must also allow for the output to be positive, even when the map is only operating on the part of the system. Suppose we have a reference and a quantum system, R and Q, and then if E only acts on Q, then the output still must be positive, even though nothing was done to the reference part of the system. This concept is known as complete positivity, and it is much more general than just positivity. The complete positivity requirement can be understood and appreciated by looking at the quantum circuit representation of this criterion. The initial density matrix has E acting on one part, namely the quantum part, whereas the referenced part has nothing acting on it. Despite E only acting on this subsystem of the density matrix, it must still produce a valid output density matrix. Thus, this output state must be positive. We can further understand this criterion by looking at a specific example. Consider this particular map which acts on a single qubit, say a, b, c, d. it transforms that qubit by transposing the off-diagonal elements. It is straightforward to see that a positive density matrix is transformed into a positive density matrix by this transpose map. However, if we look at the operation of this map on a two-qubit system, where the first qubit has nothing happening to it, we may then, by looking at the four by four matrix, which corresponds to this transform we tensored with E, see that the operation of the trace acting on the second qubit flips these off-diagonal elements in each local quadrant of the four by four matrix. This is known as a partial transpose operation. Now let us apply this partial transpose operation to a specific density matrix, which is this 1, 0, 0, 1, 0, 0, 0, 0, 0, 0, 0, 0, 1, 0, 0, 1 state. That is, the density matrix for the two-qubit states 0,0 plus 1,1 normalized. The partial transpose produces this output density matrix state. The density matrix just has ones and zeros and looks tantalizingly legal and has a trace of one. However, it is not positive. It is an illegal state, not a density matrix. That is because the inner block matrix has an eigenvalue of minus 1. The transpose is thus a nonphysical operation, but it is still mathematically useful as a negative partial transpose test for entanglement. 

Entanglement is a unique feature of quantum mechanics and can cause peculiar correlated quantum phenomena between distant quantum systems\cite{friis_observation_2018}. The information on the ``strength'' of the entanglement between two quantum systems is contained in the density matrix. As discussed in the section, a density matrix is trace-preserving and completely positive. Therefore, a completely positive density matrix must possess exclusively non-negative eigenvalues and hence map all positive elements to positive elements.

To evaluate if two quantum systems A and B composing a quantum state are entangled, it is sufficient to test if the resultant quantum state is not separable. The separability of a quantum state can be evaluated with the following criterion, referred to as the Peres-Horodecki criterion. Transposing one quantum system but not the other, referred to as a partial transpose, can only result in a valid density matrix if the two quantum systems are indeed separable and hence not entangled. Suppose a density matrix $ \rho_{AB} $ representing quantum systems A and B is acted on by a partial transpose operation on quantum system B. We can write the resulting density matrix as $ \rho_{AB}^{T_B} $.

Suppose there are two normalized quantum states $ \vert\psi_A\rangle=\alpha\vert 0 \rangle+\beta \vert 1 \rangle $ and $ \vert\psi_B\rangle=\gamma\vert 0 \rangle+\delta \vert 1 \rangle $ with complex amplitudes.

\begin{eqnarray*}\label{eq3_82}
\rho_{AB}=\left( \begin{array}{c c} \vert \alpha\vert^2 & \alpha \beta^* \\ \alpha^* \beta & \vert \beta \vert^2 \end{array} \right) \otimes \left( \begin{array}{c c} \vert \gamma\vert^2 & \gamma \delta^* \\ \gamma^* \delta & \vert \delta \vert^2 \end{array} \right) &=&\left( \begin{array}{c c c c} \vert\alpha\vert^2\vert \gamma\vert^2 & \vert\alpha\vert^2\gamma \delta^* & \alpha\beta^*\vert \gamma\vert^2 & \alpha\beta^*\gamma \delta^* \\ \vert\alpha\vert^2\gamma^* \delta & \vert\alpha\vert^2\vert \delta \vert^2 & \alpha\beta^* \gamma^* \delta &  \alpha\beta^*\vert \delta \vert^2 \\ \alpha^*\beta\vert\gamma\vert^2 & \alpha^*\beta\gamma \delta^* & \vert\beta\vert^2\vert \gamma\vert^2 & \vert\beta\vert^2\gamma \delta^*\\ \alpha^*\beta\gamma^* \delta & \alpha^*\beta\vert \delta \vert^2 & \vert\beta\vert^2\gamma^* \delta & \vert\beta\vert^2 \vert \delta \vert^2 \end{array} \right) \\ \rho_{AB}^{T_B}=\left( \begin{array}{c c} \vert \alpha\vert^2 & \alpha \beta^* \\ \alpha^* \beta & \vert \beta \vert^2 \end{array} \right) \otimes \left( \begin{array}{c c} \vert \gamma\vert^2 & \gamma^* \delta \\ \gamma\delta^* & \vert \delta \vert^2 \end{array} \right) &=&\left( \begin{array}{c c c c} \vert\alpha\vert^2\vert \gamma\vert^2 & \vert\alpha\vert^2\gamma^* \delta & \alpha\beta^*\vert \gamma\vert^2 & \alpha\beta^*\gamma^* \delta \\ \vert\alpha\vert^2\gamma \delta^* & \vert\alpha\vert^2\vert \delta \vert^2 & \alpha\beta^* \gamma \delta^* & \alpha\beta^*\vert \delta \vert^2 \\ \alpha^*\beta\vert\gamma\vert^2 & \alpha^*\beta\gamma^* \delta & \vert\beta\vert^2\vert \gamma\vert^2 & \vert\beta\vert^2\gamma^* \delta\\ \alpha^*\beta\gamma \delta^* & \alpha^*\beta\vert \delta \vert^2 & \vert\beta\vert^2\gamma \delta^* & \vert\beta\vert^2 \vert \delta \vert^2 \end{array} \right)\\ 
\end{eqnarray*}

The resulting density matrix $ \rho_{AB}^{T_B} $ after transposing the B-qubit density matrix (compare the construction of $ \rho_{AB} $ and $ \rho_{AB}^{T_B} $ above) represents the situation where quantum system A is unchanged, and quantum system B is flipped with respect to the xz-plane of the Bloch sphere. The trace of the density matrix $ \rho_{AB}^{T_B} $ remains unchanged and is completely positive.

Now, suppose instead that the quantum systems are entangled, such as one of the four Bell states:
$ \vert\Psi^+\rangle=(\vert 01\rangle +\vert 10 \rangle)/\sqrt{2} $. We then define the density matrix $ \rho_{AB} $ and the partially transposed version $ \rho_{AB}^{T_B} $ (as before, the transpose of the B-qubit density matrix only).

\begin{equation*} \rho_{AB}=\frac{1}{2}\left( \begin{array}{c c c c} 0 & 0 & 0 & 0 \\ 0 & 1 & 1 & 0 \\0 & 1 & 1 & 0\\0 & 0 & 0 & 0 \end{array} \right) \qquad\qquad\qquad \rho_{AB}^{T_B}=\frac{1}{2}\left( \begin{array}{c c c c} 0 & 0 & 0 & 1 \\ 0 & 1 & 0 & 0 \\0 & 0 & 1 & 0\\1 & 0 & 0 & 0 \end{array} \right)  \end{equation*}
Again, the trace the sum of the eigenvalues of the resulting matrix $ \rho_{AB}^{T_B} $ remains unaffected. However, something has changed: we now find that one of the eigenvalues is negative. 

To see this, we diagonalize $ \rho_{AB}^{T_B} $: 

\begin{equation}\label{eq3_83} 
\rho_{AB}^{T_B}=\frac{1}{2}\left( \begin{array}{c c c c} 0 & 0 & 0 & 1 \\ 0 & 1 & 0 & 0 \\0 & 0 & 1 & 0\\1 & 0 & 0 & 0 \end{array} \right)\quad \xrightarrow{diagonalization} \quad \frac{1}{2}\left( \begin{array}{c c c c} -1 & 0 & 0 & 0 \\ 0 & 1 & 0 & 0 \\0 & 0 & 1 & 0\\0 & 0 & 0 & 1 \end{array}\right) 
\end{equation}
The diagonal elements of a diagonal matrix are its eigenvalues. As we can see, one eigenvalue is negative. Therefore, the partially-transposed matrix  $ \rho_{AB}^{T_B} $ is not a positive matrix. This ``test'' finding negative eigenvalues in $ \rho_{AB}^{T_B} $ indicates that the original density matrix $ \rho_{AB} $ represents an entangled state.

Since we only require the sign of each eigenvalue, we can employ an alternative to diagonalization by using Sylvester's criterion. The criterion states that if all leading principal minors are positive, all eigenvalues are positive as well. However, first, what are leading principal minors? The leading principle minors are the determinants of the product of the eigenvalues of the upper-left 1-by-1 corner, the upper-left 2-by-2 corner. The protocol is shown below for the discussed matrix without the prefactor 1/2.

\begin{figure}[H] \centering{\includegraphics[scale=.25]{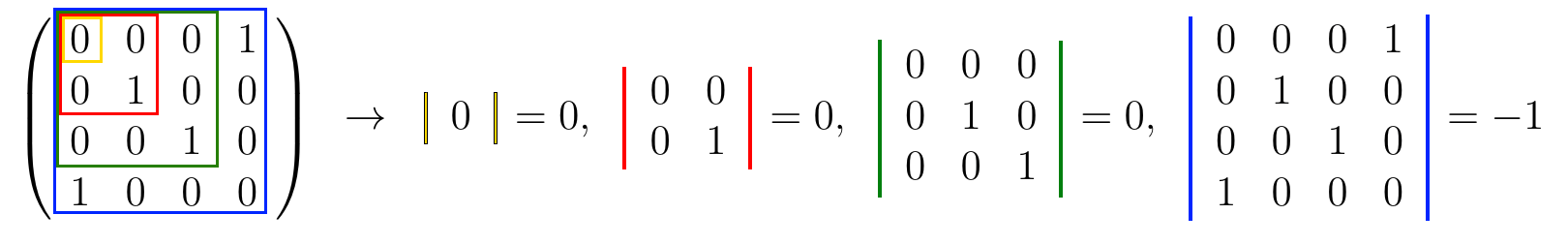}}\caption{Formula}\label{fig3_12}
\end{figure}

Since the last determinant is negative, we know that the eigenvalues of $ \rho_{AB}^{T_B} $ are not all positive, and therefore $ \rho_{AB}  $ represents an entangled state. This is useful because sometimes computing the principal minors is easier than diagonalizing the matrix.

With the ability to confirm if two quantum systems are entangled, one can further ask the question, ``to what degree are they entangled?'' The answer can be determined by probing the density matrix once more. If the partial trace of a density matrix results in a diagonal matrix with equal probabilities as entries, the quantum systems are referred to as maximally entangled.

For example, the partial trace on the entangled Bell state $ \vert\Psi^+\rangle=(\vert 01\rangle +\vert 10 \rangle)/\sqrt{2} $ yields:

\begin{eqnarray*}\label{eq3_84} 
\begin{aligned}
\rho_{A}=Tr_B\left(\frac{1}{2}\left( \begin{array}{c c c c} 0 & 0 & 0 & 0 \\ 0 & 1 & 1 & 0 \\0 & 1 & 1 & 0\\0 & 0 & 0 & 0 \end{array} \right)\right)& \\=&Tr_B\left(\frac{\vert 01\rangle\langle 01 \vert +\vert 10 \rangle\langle 01 \vert+\vert 01\rangle\langle 10 \vert +\vert 10 \rangle\langle 10 \vert}{2}\right)\\ &\\=&\frac{\vert 0\rangle\langle 0\vert\langle 1\vert 1\rangle +\vert 1 \rangle\langle 0 \vert\langle 1\vert 0\rangle+\vert 0\rangle\langle 1 \vert\langle 0\vert 1\rangle +\vert 1 \rangle\langle 1 \vert\langle 0\vert 0\rangle}{2}\\ &\\=&\frac{\vert 0\rangle\langle 0\vert +\vert 1 \rangle\langle 1 \vert}{2}=\frac{1}{2}I.\\ 
\end{aligned}
\end{eqnarray*}

The reduced density matrix indicates that $ \vert \Psi^+\rangle $ is maximally entangled. This observation is true for all Bell states. The reduced density matrix describes a mixed state. As we have shown in a previous section, Bell states are pure states. Partial traces of pure entangled states represent mixed states.

In conclusion, a partial transpose can be used to evaluate if two quantum systems are entangled. If the resultant matrix is a density matrix, the quantum systems can be described separately. The two quantum systems are maximally entangled if the partial trace of their density matrix results in a diagonal matrix and entries with equal probability. A quantum operation is linear, meaning that if the operation is performed over a statistical mixture of states with a given probability, it acts on every state and preserves the probability distribution of the mixture intact.

\section{Quantum Gate Fidelity} 
Given the formalism for quantum noise, we may now see how the fidelity of gates versus states can be defined and measured in practice. The question is, how well does an experimental quantum operation E represent a theoretically desired unitary gate or circuit U. One simple answer to this is given by the minimum gate fidelity, also known just as gate fidelity and defined as the minimum fidelity between we acting on some input psi, and E acting on the same input. This measure varies between 0 and 1. For example, consider a noisy X gate, which with probability p performs as zed instead of the desired X. Plugging E and U into the definition, we find the minimization is over an expectation of the y operator. So, the final result is just the square root of 1 minus p. This agrees with our intuition that the failure probability is p. The minimum gate fidelity, unfortunately, can be hard to compute and may not reflect experimental impact in practice where the minimum is not dominant. Another measure is the entanglement fidelity of a gate, defined here by the fidelity between $ \phi $ and the combination of E and the inverse of U acting on just half of $ \phi $.
Here, $ \phi $ is a maximally entangled state. We can understand the intuition behind this expression by looking at the quantum circuit realization, where it is evident that ideally, U dagger should cancel E and leave the entanglement between the two halves of the state intact. The entanglement fidelity thus measures how much entanglement is left intact by the quantum operation E modulo the ideal gate U. This measure is in many ways the gold standard of quantum gate fidelity because entanglement is the precious resource utilized by quantum computation. Unfortunately, it is largely impractical to measure directly today. The third measure of quantum gate fidelity is the average gate fidelity, given by computing the fidelity between U acting on a state psi and E acting on the same state and averaging evenly over all states. This is evidently much easier to measure. However, how meaningful is it? It turns out there is a practical way to get the desired entanglement fidelity for experimental realizations. This is via a deep relationship between the entanglement fidelity of a gate and its average gate fidelity. The average fidelity is d times the entanglement fidelity plus 1 all divided by d plus 1, where d is the dimension of the system. This means for a large dimension, they are the same. However, for qubits, they may be quite different. The neat thing is that this relationship gives a practical way to measure entanglement fidelity explicitly for a qubit. One can compare the outputs of the experimental and expected inputs $ \sigma $ j, which are just the three Pauli matrices. Combining these three state tomography measurements gives the qubit gate fidelity. Generalizations of this to multi-qubit systems exist, and today, quantum gate fidelity is routinely measured as a natural part of practical quantum computation.
 
\section{Quantum Process Tomography} 
Quantum state tomography is a method that allows noisy quantum states to be experimentally quantified. Quantum process tomography does the same for gates. This problem of knowing what actual quantum operation is being performed by a quantum apparatus is vital, for example, in evaluating the performance of quantum computing hardware \cite{shukla_complete_2018,mcgeoch_principles_2019,mcgeoch_practical_2019}. Mathematically, the problem is framed by recalling the operator sum representation of a quantum operation, shown here in terms of which the goal is to determine the operation elements e sub k. The e sub k are matrices. It is convenient to expand them in terms of a linear combination where $ \sigma $ j are generalized Pauli matrices, and the C sub-KJ is an ordinary real number of coefficients. With this, we may then re-write the quantum operation e in this form, where the numerical coefficients form the so,-called chi matrix. It becomes our goal to find this matrix. How many degrees of freedom are in the chi matrix? If the system $ \rho $ has dimension d, then the chi matrix has d to the fourth minus d squared degrees of freedom.
Specifically, for example, for a single cubit, d equals 2, and thus, Chi has 12 parameters. Of these, three parametrize the possible single keyword unitary gates, and nine capture decoherence and damping. There are many more ways to decohere and damp than to be a perfect gate. For two qubits, the number of parameters grows quickly to become 240. Twenty years ago, when it was first computed this number, researchers are so surprised and shared with Peter Shor. For n qubits, the number is 2 to the 4 times n minus 2 to the 2 times n. Unsurprisingly, this is an exponential increase in the number of parameters. These parameters may be measured using the following four steps. First, prepare a basis set of linearly independent inputs, $ \rho $ sub a. These may be the generalized poly matrices, for example. Second, send each $ \rho $ sub a into the apparatus. Formally, we expect this to depend on the chi matrix. We may expand its matrix coefficients in terms of other basis states. Third, measure each output using the quantum state tomography procedure and express each resulting density matrix again in terms of the basis states giving coefficients c sub ab. Fourth, invert the linear equation relating c sub ab to the chi matrix to obtain the chi matrix. We are done in principle. This baseline method is known as standard quantum process tomography. There are some challenges, however. First, the exponential growth in the number of parameters makes this impractical for large numbers of qubits. The inversion of c sub ab is also difficult in practice when measurement results are noisy, as they inevitably are due to the statistical sampling needed for state tomography. This noise may also lead to chi matrices, which are invalid quantum operations because of violations of strict properties necessary. These operations must be completely positive and trace-preserving. Options are known for meeting these challenges. For example, one may employ statistical methods such as maximum likelihood optimization to find the closest legitimate inverse. One may also constrain the inversion always to produce a valid answer. Adding assumptions given by prior information about the operation may also dramatically reduce the number of parameters needed. As experimentalists build and test small quantum machines, quantum process tomography and its variants have become one of the most important and widely employed diagnostic tools in quantum computing laboratories.

As we have seen, quantum state fidelity quantifies the degree of similarity between an experimentally generated quantum state and its ideal counterpart. In this unit, we will study quantum process tomography, which uses quantum state tomography to characterize the fidelity of a quantum operation, such as a single gate operation or multiple gate operations. These types of operations are referred to as a ``quantum process,'' thus the nomenclature. For quantum computing, we are most often interested in the fidelity of the gates that comprise a universal gate set. However, more generally, quantum process tomography characterizes the quality of any quantum process.

Suppose a quantum state is manipulated with a known set of quantum gates. The ideal output quantum state can be calculated for a particular input state. We can further perform quantum state tomography to see how close our experimental output state matches our theoretical one. In practice, due to noise, we will often see a reduced quantum state fidelity. A method to investigate the underlying reasons for this reduced fidelity for any input state is called quantum process tomography (QPT), also referred to as quantum gate tomography. 

QPT essentially implements quantum state tomography for a set of input states that spans the space of all input states for a given system. We will focus on systems of qubits. Note that quantum state tomography (QST), as we have seen, performs repeated measurements of an identically prepared input state across a basis that spans all output states. As we have seen, spanning this space requires approximately $ 2^{2n}=4^n $ matrix elements to be estimated, where n is the number of qubits. With QPT, this state tomography is now performed for a set of input states that generally spans the same space, that is, another $ 4^n $ elements. Since all output elements must be estimated for each input element, the number of total elements that must be estimated is $ 4^n\times4^n=4^{2n} $. As we discuss, there are a few constraints that may reduce this number by a small amount, because the quantum process must take quantum states to other valid quantum states. In the language of density matrices, there are two constraints on a quantum process: it must be (1) completely positive, which means that the density matrix after the process still has positive eigenvalues, and (2) trace-preserving because density matrices must have a trace of 1. In total, this constrains $ 4^n $ of the elements we are trying to find, so there are $ 4^{2n}-4^n $ elements to determine in total. Since the number of measurements scales exponentially in the number of qubits, QPT becomes intractable on larger systems of qubits. 

The result of QPT is reported as a fidelity of the realized gate with respect to the desired quantum gate. Obtaining this fidelity is related to the construction of a ``superoperator'' and its matrix form. A superoperator is a map of density matrices to density matrices; for example, the density matrices of the input basis states to the density matrices for the output basis states. From this experimentally determined mapping, one can define a gate fidelity that describes how well the experimentally implemented gate matches the desired, ideal operation.

\section{Tomography and Characterization} 
At a high level, running quantum algorithms involves applying a series of gate operations to a set of quantum bits in order to generate quantum states. In other sections, we have seen details on those algorithms, how they work, and what problems they are designed to solve. However, before we start applying algorithms, we need to verify that our dates are operating in the way we expect them to and that we are creating the states we intend to create. Sounds simple, but is it? To understand how to characterize a quantum state, we first need to understand how to represent a quantum state, and what happens when we perform a measurement. A pure quantum state can be expressed as a sum of basis states where the coefficients of the sum are complex, and the magnitude of the coefficients squared must sum to one. For a qubit, this means that there are two parameters that characterize a quantum state. These parameters correspond well to a point on the surface of a sphere. This is known as the Bloch sphere representation of a qubit state. However, any measurement of this quantum state projects the qubit into either the state up, the point at the top of the sphere, or the state down, the point at the bottom of the sphere. The probability of measuring up or down is given by how much of the qubit state was in the top or bottom half of the Bloch sphere. In the state representation, this means that the magnitude of the coefficient squared is the probability of measuring that particular state. In a single quantum measurement, we only obtain a single bit of information, and the original state is lost. So, how do we measure a quantum state given these constraints? The answer is tomography. Tomography means reconstructing something from many pieces of incomplete data. It is a word that we often hear from the medical field, such as X-ray tomography. True to the word, X-ray tomography is the process of building a 3D image of a person's body from 2D X-rays taken from a number of different directions. To measure a quantum state, we perform quantum state tomography, reconstructing a quantum state from multiple incomplete measurements. To see how tomography works, we use as an example, the maximal superposition state, which is a state on the equator of the Bloch sphere. If we measure the qubit, it will find it up or down with equal probability. If we prepare the state, again, and measure, we will, again, measure up or down with equal probability. Repeating this many times, we will get a histogram of measurements that should converge to 50/50. This is the start of a state tomography experiment because we then know that the qubit lies on the equator of the Bloch sphere, but we do not know the angle of the state $ \phi $. Therefore, we apply a $ \pi/2 $ rotation to the state around the x-axis in the Bloch sphere. This mixes $ \phi $ along the measurement axis. If we repeat this procedure many times, we can construct a new histogram and determine $ \phi $. For example, if the state originally lied along the x-axis in the equator, the histogram after applying the pulse will, again, be 50/50. If the state originally lied on the y-axis in the Bloch sphere, the histogram after the pulse will be 100\% in the upstate. This is the essence of quantum state tomography, but note the fuzziness of the process. we needed to take enough data so that we could get a representative histogram, but how much data does that require? What if the measurement itself has a bias? Also, we need to be able to apply a perfect $ \pi/2 $ pulse and prepare the exact same state many times. These are some of the challenges with quantum tomography. The above example is oversimplified because a general quantum state can also include classical probabilities. This is represented by a density matrix, which is a sum of the classical probabilities of being in a particular quantum state. These probabilities must sum to one. For a qubit, the density matrix is a two by two matrix, which is given as an example for our superposition state. If the state was instead a classical probabilistic state, that is a coin flip between up and down, the off-diagonal elements of the density matrix to go to zero. In the Bloch sphere representation, this means that the qubit state can also lie inside the sphere. Because there are more terms in a density matrix, we need to do more measurements than in the example that was given before. The general procedure for quantum state tomography given a completely unknown state that can be prepared many times is shown. we perform measurements on the state after applying no pulse, a $ \pi/2 $ pulse along the x-axis, and a $ \pi/2 $ pulse along the y-axis. From these measurements, we get three numbers. These are fed into a classical optimization routine that obtains the most likely density matrix that corresponds to a physical quantum state. For end cubic quantum state tomography, we need to do three to the power of N measurements, and the density matrix has four to the power of N elements. Clearly, tomography is not a scalable procedure as the number of qubits grows. What about measuring a process, a gate versus a state? If we have a Blackbox classical gate with two inputs and one output, we construct a truth table. we input all possible states and measure the outputs. This is straightforward for the classical gate because, one, we can exactly measure the output state, and two, there are only four input states. The quantum case is similar, except we need to input quantum states and measure the output quantum states. As we know from the previous part of the discussion, measuring quantum states requires quantum state tomography. Therefore, quantum process tomography is just the repeated application of quantum state tomography, three to the power of N times. How do we represent a quantum process? Pure quantum processes or unitary matrices that take an input state to an output state. A general quantum process maps an input density matrix to an output density matrix. There is a matrix form for these general quantum processes, but this matrix has 16 to the power of N elements. therefore, it is very difficult to represent a process involving more than a handful of qubits, but what if we just want to verify the gate is doing what we expected to be doing? Moreover, we do not want to characterize the gateway fully. we just need to know how close it is to the gate that we want. This is analogous to getting in a grade on a test. A grade is a single number that tells us how we performed. However, it does not indicate what we did wrong or how we would need to fix it. What kind of grade can we give a gate? One metric is known as gate fidelity. This fidelity is related to a similar thing called state fidelity. The state fidelity is the overlap between two quantum states. Gate fidelity is the overlap between the output states of our gate versus the output states of the expected gate averaged over all possible input states. Now, this might not sound easier than tomography. However, it can be shown that there is a simple experimental procedure to perform this type of measurement by repeated application of gates to a set of qubits starting in the ground state. This is known as randomized benchmarking, and we will go over the details of this in the next section. Single-Qubit State Tomography; Single-qubit state tomography is a procedure that enables one to reconstruct the density matrix of a given system. When implementing single-qubit state tomography all of the steps one makes in performing single-qubit state tomography.
\begin{itemize}
\item The state of the system is measured in the bases of the three Pauli matrices.
\item The probabilities of projection onto the eigenstates of the three Pauli matrices are calculated based on the measurement results.
\item The elements of the reconstructed density matrix are given by combinations of the expectation values of the three Pauli matrices.
\end{itemize}

\section{Randomized Benchmarking}
In the section on tomography, we discuss that to characterize a quantum process fully requires an exponential number of measurements. Additionally, tomographic methods have many built-in assumptions since we are trying to reconstruct a process from slices of data. Instead, we would like a scalable method to verify that our gates are operating as we expect without needing to do full characterization. One such method is known as randomized benchmarking \cite{villalonga_flexible_2019,corcoles_process_2013,sheldon_characterizing_2016,magesan_characterizing_2012}. Randomized benchmarking starts by constructing a circuit from gates randomly selected from a particular set of gates known as the Clifford group. The Clifford group is, as the name suggests, a mathematical group, and so, a gate formed from composite Clifford gates is itself a Clifford gate. The Clifford group includes commonly used gates, such as the $ \pi $ pulse, the $ \pi/2 $ pulses, and the 2 qubits CNOT gate. The Clifford group is not an esoteric set of gates. It can, for example, be used to create a maximally entangled Bell state. The Clifford group has two important properties for the purposes of randomized benchmarking. First, it is classically easy to compute the gate formed from a sequence of Clifford gates. This is known as the Gottesman-Knill theorem. This theorem means that the Clifford group is not a universal set of gates for quantum computing. However, it is still considered a representative set of gates, as is evident from the list of gates we mentioned. All that is needed to extend the Clifford group to a universal set of gates is a T gate, a $ \pi/4 $ z rotation. The second relevant property of the Clifford group is that randomly sampling from the group ensures that we are doing a fair random sampling of the states around the Bloch sphere. we will elaborate on this point later. The first point is important because the second step of a randomized benchmarking experiment is to compute the inverse gate for the entire random sequence that we have just selected. In other words, we are constructing a large sequence of gates that is the main purpose is to do nothing, and this is precisely the point of randomized benchmarking. If we apply the sequence to a set of qubits starting in the ground state, we measure the population of each qubit in the excited state after applying the sequence of gates. For a single qubit in the Bloch sphere representation, we start in the downstate, and we apply the sequence of gates that takes the state around the Bloch sphere, then ideally, we return back to the downstate. This highlights the importance of the second property of the Clifford group on average, and the random Clifford sequences will fairly sample the entire space of the Bloch sphere. we perform this procedure for different sequence lengths and very clearly observe that as the number of gates increases, the less likely the qubit is to return to the ground state. For extremely long sequences, the qubit state is completely depolarized. The state is equally likely to be up or down. This process is repeated again for a new random sequence, and the results are averaged. Because the Clifford group fairly samples the Bloch sphere, the average curve will fit an exponential decay. Unlike tomography, errors in the measurement or errors in the preparation of the initial state of the qubits do not affect our ability to fit this decay coefficient. How does this decay coefficient allow us to verify the gate? Well, the decay coefficient is simply related to a quantity known as the average gate fidelity. The average gave fidelity is a metric that tells us how close our gate is to the ideal gate we were attempting to implement. Because we sample from the Clifford group, we actually measure the average fidelity of the average Clifford gate. Specifically, gate fidelity is defined as the state fidelity of the output state from the gate with respect to the output state of the ideal gate, averaged over all input states, and state fidelity is defined as the overlap between states. For pure states, this is the square of the inner product of the state vectors. Randomized benchmarking is a versatile procedure that can be used to characterize single or multi-qubit circuits, but it also has variants that can be used for more specialized measurements. If we want to measure the fidelity of an individual Clifford gate, we can perform randomized benchmarking and then repeat the experiment again with the specific gate in between each Clifford of the sequence. This is known as interleaved randomized benchmarking. we can also perform independent benchmarking in parallel on different subsystems, which is known as simultaneous randomized benchmarking. This is a method for a qualitative understanding of crosstalk. Randomized benchmarking does not tell us what the errors are or how to fix them. It only gives us a single number to characterize the average performance of our multi-qubit circuit. Therefore, to improve the performance of quantum circuits, randomized benchmarking must be part of a larger toolbox of error amplification and detection routines. Fidelity from Randomized Benchmarking; There is more than one way to characterize the fidelity of a quantum gate. However, for practical reasons, not all of them are equivalently easy to measure. The gate fidelity can be more experimentally feasible to characterized by the average fidelity is easy to calculate using only state tomography. For example, for $ d=2 $, it can be done using state tomography to determine the expected values of the three Pauli matrices. For $ d $ greater than two, these are generalizations.

\section{Single-Qubit State Tomography: Theory}

In this IBM Q experience, we will implement quantum state tomography for a single-qubit system. In the first section, we will discuss the basic concepts necessary to understand the single-qubit quantum state tomography. 

In Single-Qubit State Tomography: Example, we will follow a step-by-step demonstration of how to implement single-qubit state tomography using QASM \cite{cross_open_2017,noauthor_qiskitopenqasm_2020} and IBM Q for a particular density matrix.

In Single-Qubit State Tomography: Experience, we will implement quantum state tomography for a given single-qubit quantum state. To do this, we will

1. Write the QASM codes to measure the system in each one of the bases of the Pauli matrices X, Y, and Z.

2. Use the simulation results to compute the probabilities of projecting the state to each one of the eigenstates of the Pauli matrices.

3. Calculate the expectation values of the Pauli matrices using the projection probabilities.

4. Reconstruct the density matrix from the expectation values.

Arbitrary single-qubit density matrices\\

An arbitrary single-qubit density matrix $ \rho $ can be written in terms of the parameters, $ S_{0}, S_{1}, S_{2}, $ and $ S_{3}, $ as
\begin{equation}\label{eq3_85}
\rho =\frac{1}{2}S_{0}\sigma _{0}+\frac{1}{2}S_{1}\sigma _{1}+\frac{1}{2}S_{2}\sigma _{2}+\frac{1}{2}S_{3}\sigma _{3},
\end{equation}

where $ \sigma _{0} $ is the 2-by-2 identity matrix, and $ \sigma _{1}, \sigma _{2}, and \sigma _{3} $ are the Pauli matrices X, Y, and Z,
\begin{equation}\label{eq3_86}
\sigma _{0}=\left( \begin{array}{cc} 1 & 0 \\ 0 & 1\end{array}\right) \text {, }\sigma _{1}=\left( \begin{array}{cc} 0 & 1 \\ 1 & 0\end{array}\right) \text {, }\sigma _{2}=\left( \begin{array}{cc} 0 & -i \\ i & 0\end{array}\right) \text {, }\sigma _{3}=\left( \begin{array}{cc} 1 & 0 \\ 0 & -1\end{array}\right) \text {.}    
\end{equation}

The $ S_ i $ parameters are given by the expectation values of the Pauli matrices, $ S_ i=Tr\left(\sigma _ i\rho \right) $.

Let us recall that the probability of projecting$  \rho $ onto a pure state $ \left\vert \psi \right\rangle $ is$  P_{\left\vert \psi \right\rangle }=\left\langle \psi \right\vert \rho \left\vert \psi \right\rangle $. The expectation values $ Tr\left(\sigma _ i\rho \right) $ can be written in terms of the probabilities of projecting $ \rho $ to the eigenstates of the Pauli matrices,
\begin{equation}\label{eq3_87}
\begin{split}
\displaystyle S_{0}    \displaystyle & =    \displaystyle P_{\left\vert 0\right\rangle }+P_{\left\vert 1\right\rangle }=1\\          
\displaystyle S_{1}    \displaystyle & =    \displaystyle P_{\frac{\left\vert 0\right\rangle +\left\vert 1\right\rangle }{\sqrt{2}}}-P_{\frac{\left\vert 0\right\rangle -\left\vert 1\right\rangle }{\sqrt{2}}}\\          
\displaystyle S_{2}    \displaystyle & =    \displaystyle P_{\frac{\left\vert 0\right\rangle +i\left\vert 1\right\rangle }{\sqrt{2}}}-P_{\frac{\left\vert 0\right\rangle -i\left\vert 1\right\rangle }{\sqrt{2}}}\\      
\displaystyle S_{3}    \displaystyle & =    \displaystyle P_{\left\vert 0\right\rangle }-P_{\left\vert 1\right\rangle }
\end{split}    
\end{equation}
         
\textbf{Measurement Basis:}\\
To reconstruct an implemented density matrix $ \rho $, we need to compute the expectation values of the three Pauli matrices X, Y, and Z. To do this, we need to measure the system on the basis of each Pauli matrix. The IBM Q online composer comes with one pre-defined measurement. This measurement projects the state of the system to one of the eigenstates of Z (or $ \sigma _{3} $) $ \left\vert 0\right\rangle $ and $ \left\vert 1\right\rangle $. The IBM Q composer symbol for this measurement is

\begin{figure}[H] \centering{\includegraphics[scale=3]{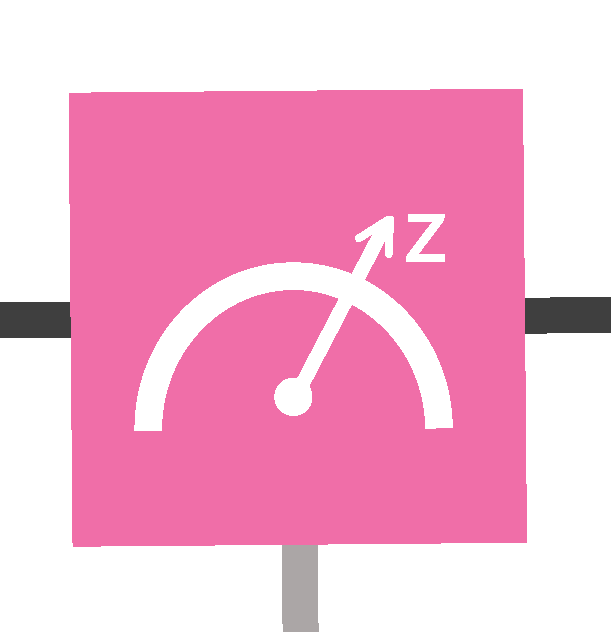}}\caption{Measurement-Sigma Z}\label{fig3_13}
\end{figure}

Once the system is measured, the IBM Q platform outputs the measurement probabilities $ P_{\left\vert 0\right\rangle } $ and $ P_{\left\vert 1\right\rangle } $. For more information on how to interpret the IBM Q measurement results, read IBM Q Tutorial.

For measuring in the X (or $ \sigma _{1} $) basis, and thereby projecting the system to $ \frac{\left\vert 0\right\rangle + \left\vert 1\right\rangle }{\sqrt{2}} $ or$  \frac{\left\vert 0\right\rangle - \left\vert 1\right\rangle }{\sqrt{2}} $, a Hadamard gate H has to be placed just before the pre-defined measurement,

\begin{figure}[H] \centering{\includegraphics[scale=3]{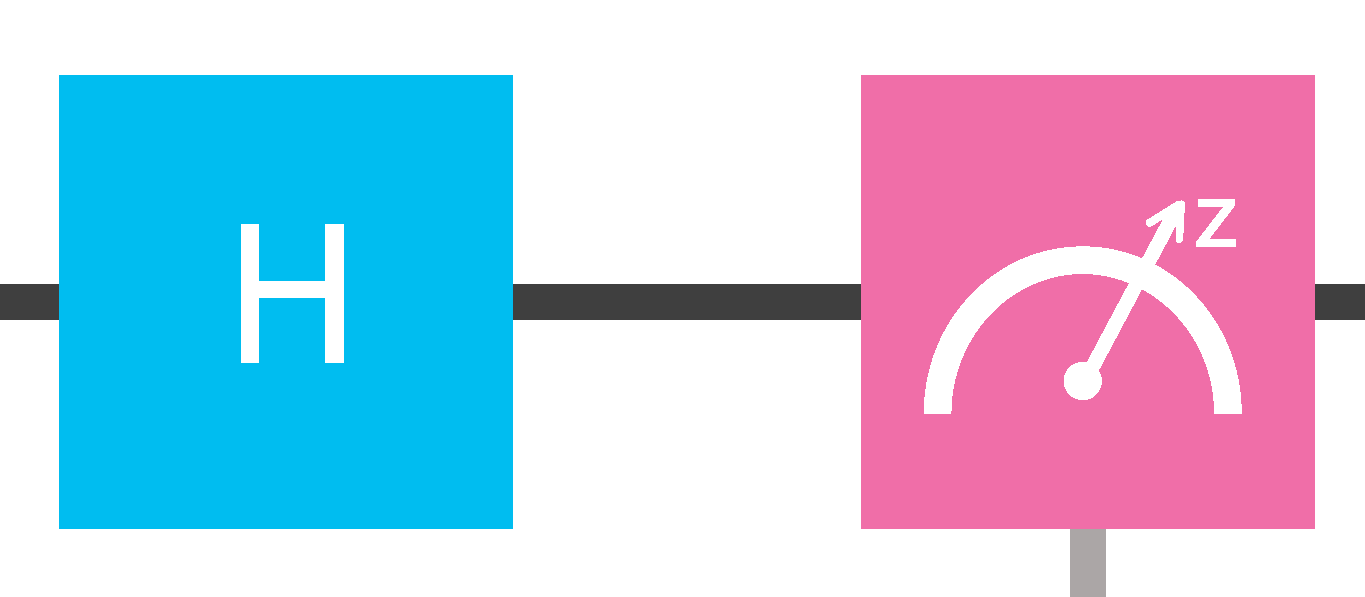}}\caption{Measurement-Sigma X}\label{fig3_14}
\end{figure}

In this way, the IBM Q outputs are now equivalent to $ P_{\frac{\left\vert 0\right\rangle +\left\vert 1\right\rangle }{\sqrt{2}}} $ and $ P_{\frac{\left\vert 0\right\rangle -\left\vert 1\right\rangle }{\sqrt{2}}} $.

To measure in the Y (or $ \sigma _{2} $) basis, and thereby projecting the system to $ \frac{\left\vert 0\right\rangle + i\left\vert 1\right\rangle }{\sqrt{2}} $ or $ \frac{\left\vert 0\right\rangle - i\left\vert 1\right\rangle }{\sqrt{2}} $, a Hadamard gate H and phase gate S have to be placed just before the measurement,

\begin{figure}[H] \centering{\includegraphics[scale=3]{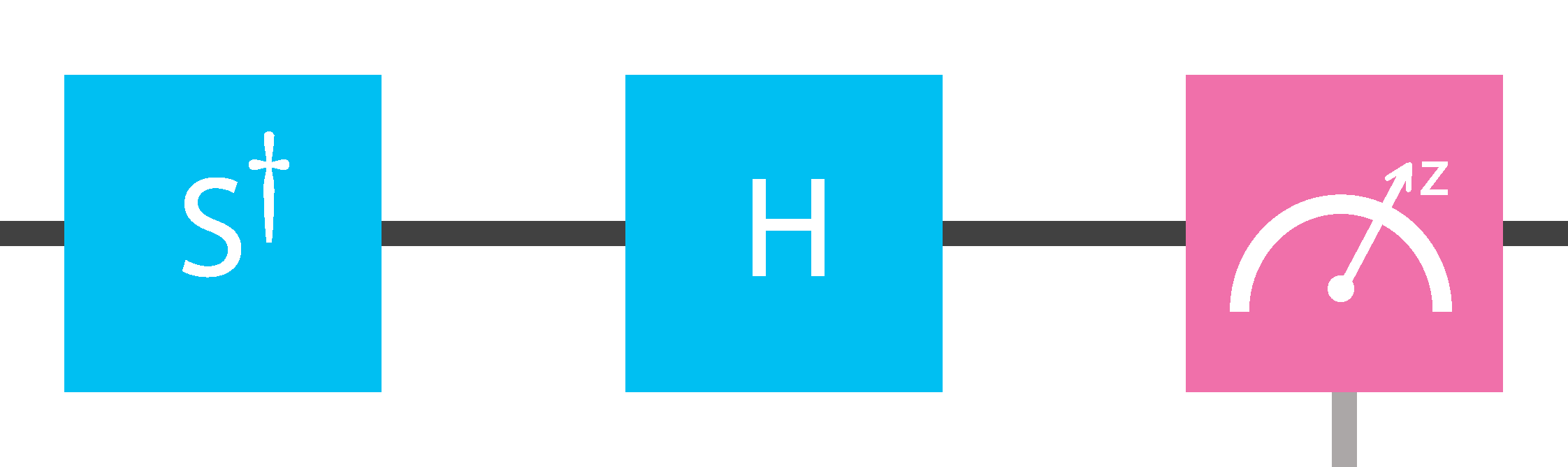}}\caption{Measurement-Sigma Y}\label{fig3_15}
\end{figure}

By adding these two gates, the IBM Q outputs are now equivalent to $ P_{\frac{\left\vert 0\right\rangle +i\left\vert 1\right\rangle }{\sqrt{2}}} $ and $ P_{\frac{\left\vert 0\right\rangle -i\left\vert 1\right\rangle }{\sqrt{2}}} $.\\

\textbf{Density Matrix Reconstruction:}\\
The density matrix $ \rho $ can be reconstructed from the expectation values of the Pauli matrices as
\begin{equation}\label{eq3_88}
\displaystyle \rho ^{E}    \displaystyle =    \displaystyle \frac{1}{2}S_{0}\sigma _{0}+\frac{1}{2}S_{1}\sigma _{1}+\frac{1}{2}S_{2}\sigma _{2}+\frac{1}{2}S_{3}\sigma _{3},          
\displaystyle =    \displaystyle \left( \begin{array}{cc} \frac{1}{2}S_{0}+\frac{1}{2}S_{3} & \frac{1}{2}S_{1}-\frac{1}{2}iS_{2} \\ \frac{1}{2}S_{1}+\frac{1}{2}iS_{2} & \frac{1}{2}S_{0}-\frac{1}{2}S_{3}\end{array}\right)    
\end{equation}
    
or

\begin{equation}\label{eq3_89}
\displaystyle \rho ^{E}    \displaystyle =    \displaystyle \left( \begin{array}{cc} \frac{1}{2}+\frac{1}{2}S_{3} & \frac{1}{2}S_{1}-\frac{1}{2}iS_{2} \\ \frac{1}{2}S_{1}+\frac{1}{2}iS_{2} & \frac{1}{2}-\frac{1}{2}S_{3}\end{array}\right),        
\end{equation}
      
Where we have used that $ S_0=1 $.

\section{Single-Qubit State Tomography: Example}

\textbf{State preparation:}\\
Consider the single-qubit quantum state
\begin{equation}\label{eq3_90}
\displaystyle \left\vert \psi _{1}\right\rangle    \displaystyle =    \displaystyle TH\left\vert 0\right\rangle \\     
\displaystyle =    \displaystyle \left(\begin{array}{c} \frac{1}{\sqrt{2}}\\ \frac{1+i}{2} \end{array}\right)    
\end{equation}
        
The figure below shows the generation of state $ \left\vert \psi _{1}\right\rangle $ using the IBM Q composer.

\begin{figure}[H] \centering{\includegraphics[scale=2.5]{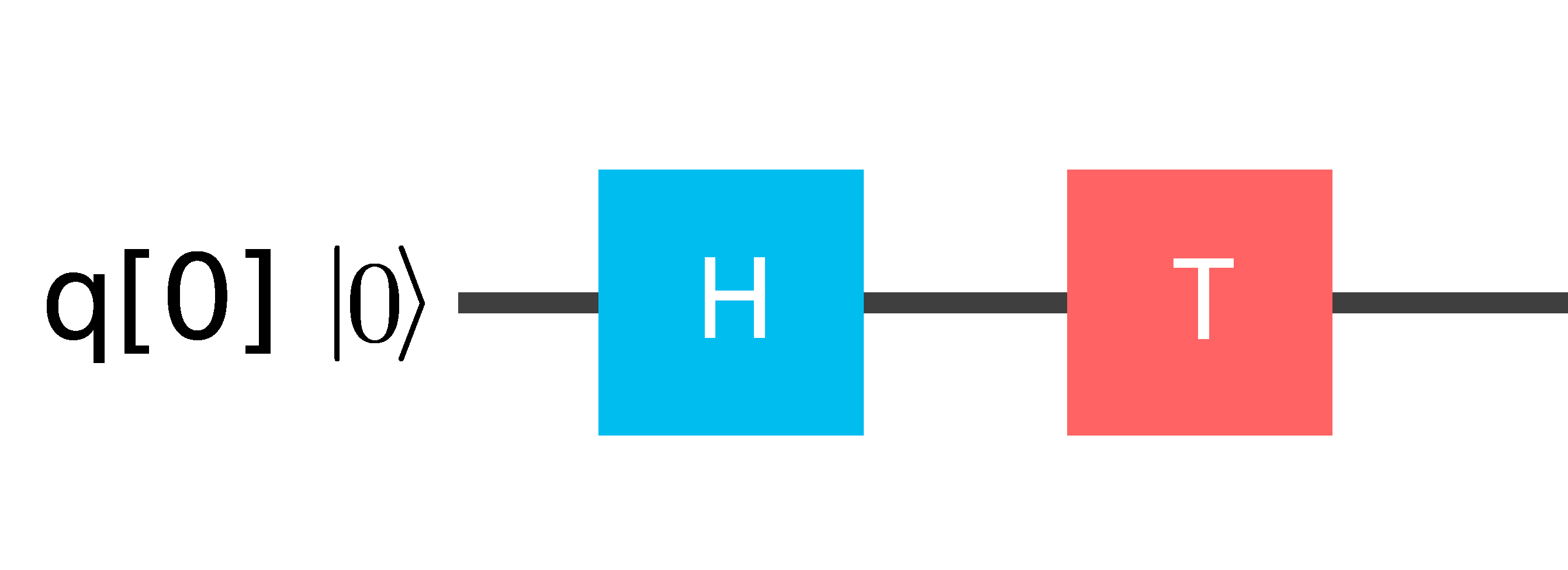}}\caption{Generation of state}\label{fig3_16}
\end{figure}

\begin{lstlisting}
The QASM code that generates this quantum circuit is
include ``qelib1.inc'';
qreg q[5];
creg c[5];
// State Preparation 
h q[0];
t q[0];
\end{lstlisting}

Quantum state tomography allows us to verify that the quantum state was prepared correctly. To do quantum state tomography, we start by writing the density matrix representation of the quantum state,
\begin{equation}\label{eq3_91}
\displaystyle \rho _{1}    \displaystyle =    \displaystyle \left\vert \psi _{1}\right\rangle \left\langle \psi _{1}\right\vert ,
\end{equation}

\begin{equation}\label{eq3_92}
\displaystyle =    \displaystyle \left( \begin{array}{c} \frac{1}{\sqrt{2}} \\ \frac{1+i}{2}\end{array}\right) \left( \begin{array}{cc} \frac{1}{\sqrt{2}} & \frac{1-i}{2}\end{array}\right) ,    
\end{equation}              
          
\begin{equation}\label{eq3_93}
\displaystyle =    \displaystyle \left( \begin{array}{cc} \frac{1}{2} & \frac{1}{4}\sqrt{2}-\frac{1}{4}i\sqrt{2} \\ \frac{1}{4}\sqrt{2}+\frac{1}{4}i\sqrt{2} & \frac{1}{2}\end{array}\right) ,    
\end{equation}

We write $ \rho _1 $ using numerical values to make it easier to compare with the reconstructed density matrix from the simulation results,
\begin{equation}\label{eq3_94}
\rho _1\approx \left( \begin{array}{cc} 0.5 & 0.353\, 55-0.353\, 55i \\ 0.353\, 55+0.353\, 55i & 0.5\end{array}\right) 
\end{equation}
\textbf{Measurement Basis:} \\
The state$  \lvert \psi _1\rangle $ was simulated in the IBM Q composer 1024 times using the IBM Q 5 Tenerife backend. In order to reconstruct the density matrix $ \rho _1=\lvert \psi \rangle \langle \psi _1\rvert $, and verify the state preparation, we measured qubit q[0] in each of the Pauli-matrix bases.

The figure below shows the quantum circuit implemented to measure $ \rho _1 $ in the $  \sigma _{3} $ or Z basis:

\begin{figure}[H] \centering{\includegraphics[scale=.8]{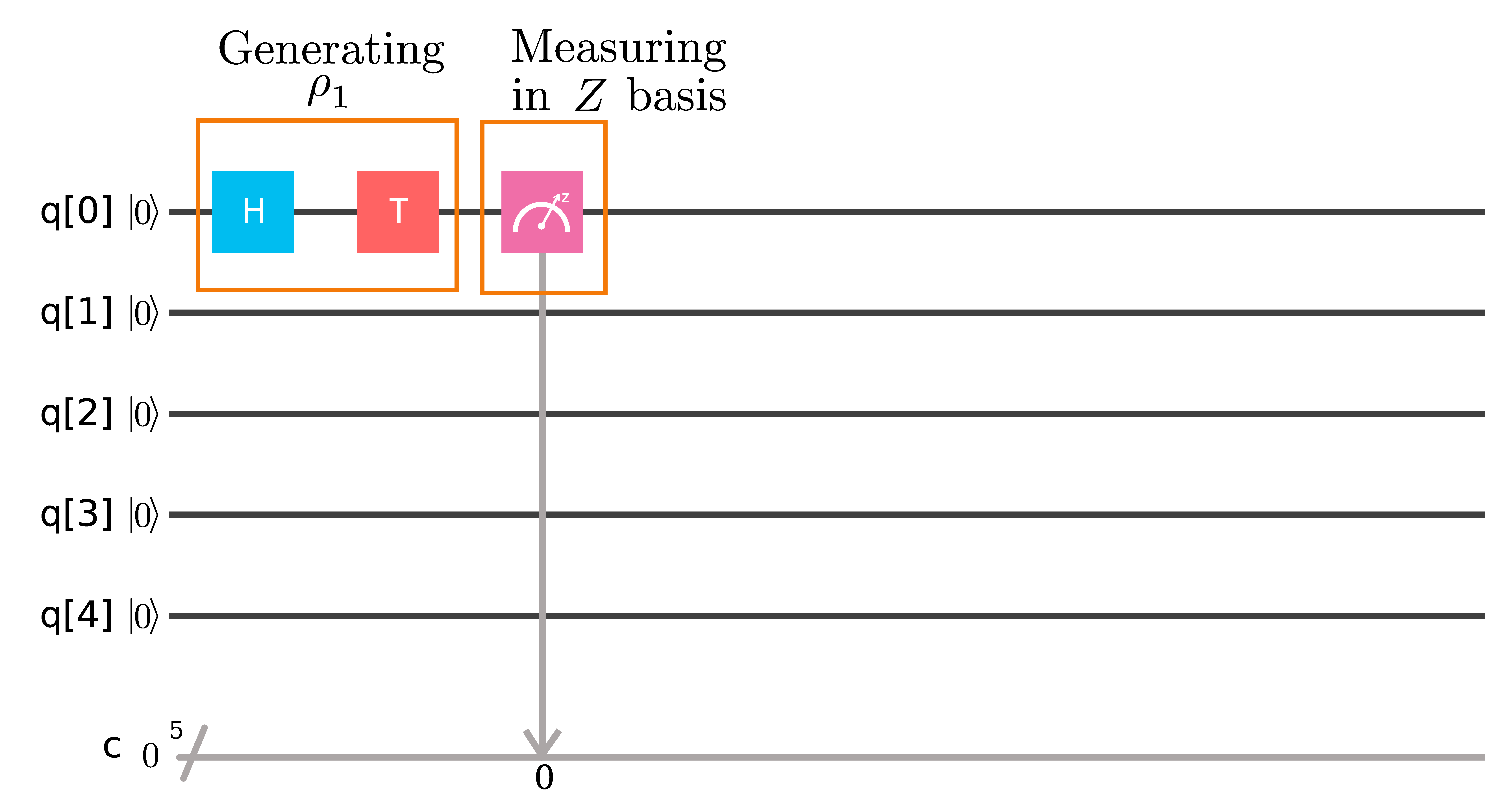}}\caption{Sigma Z}\label{fig3_17}
\end{figure}

\begin{lstlisting}
The QASM code that generates this circuit is:
include ``qelib1.inc'';
qreg q[5];
creg c[5];
// State Preparation
h q[0];
t q[0];
// Z or sigma\_3 basis
measure q[0] $->$ c[0];
\end{lstlisting}

The figure below shows the quantum circuit implemented to measure $ \rho _1 $ in the $ \sigma _{1} $ or X basis:

\begin{figure}[H] \centering{\includegraphics[scale=.8]{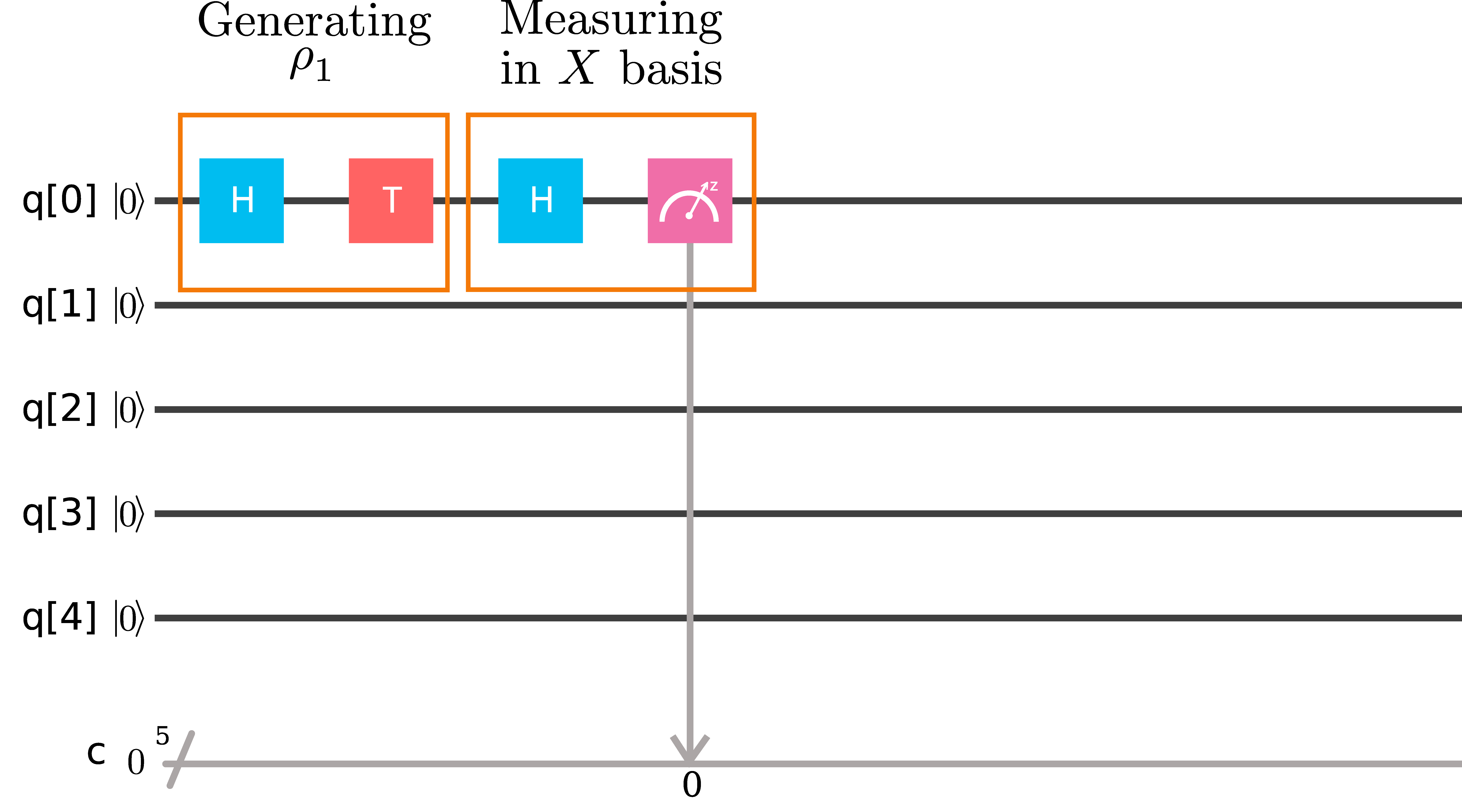}}\caption{Sigma X}\label{fig3_18}
\end{figure}

\begin{lstlisting}
The QASM code that generates this circuit is:
include ``qelib1.inc'';
qreg q[5];
creg c[5];
// State Preparation
h q[0];
t q[0];
// X or sigma\_1 basis
h q[0];
measure q[0] $->$ c[0];
\end{lstlisting}

The figure below shows the quantum circuit implemented to measure $ \rho _1  $in the $ \sigma _{2} $ or Y basis:

\begin{figure}[H] \centering{\includegraphics[scale=.8]{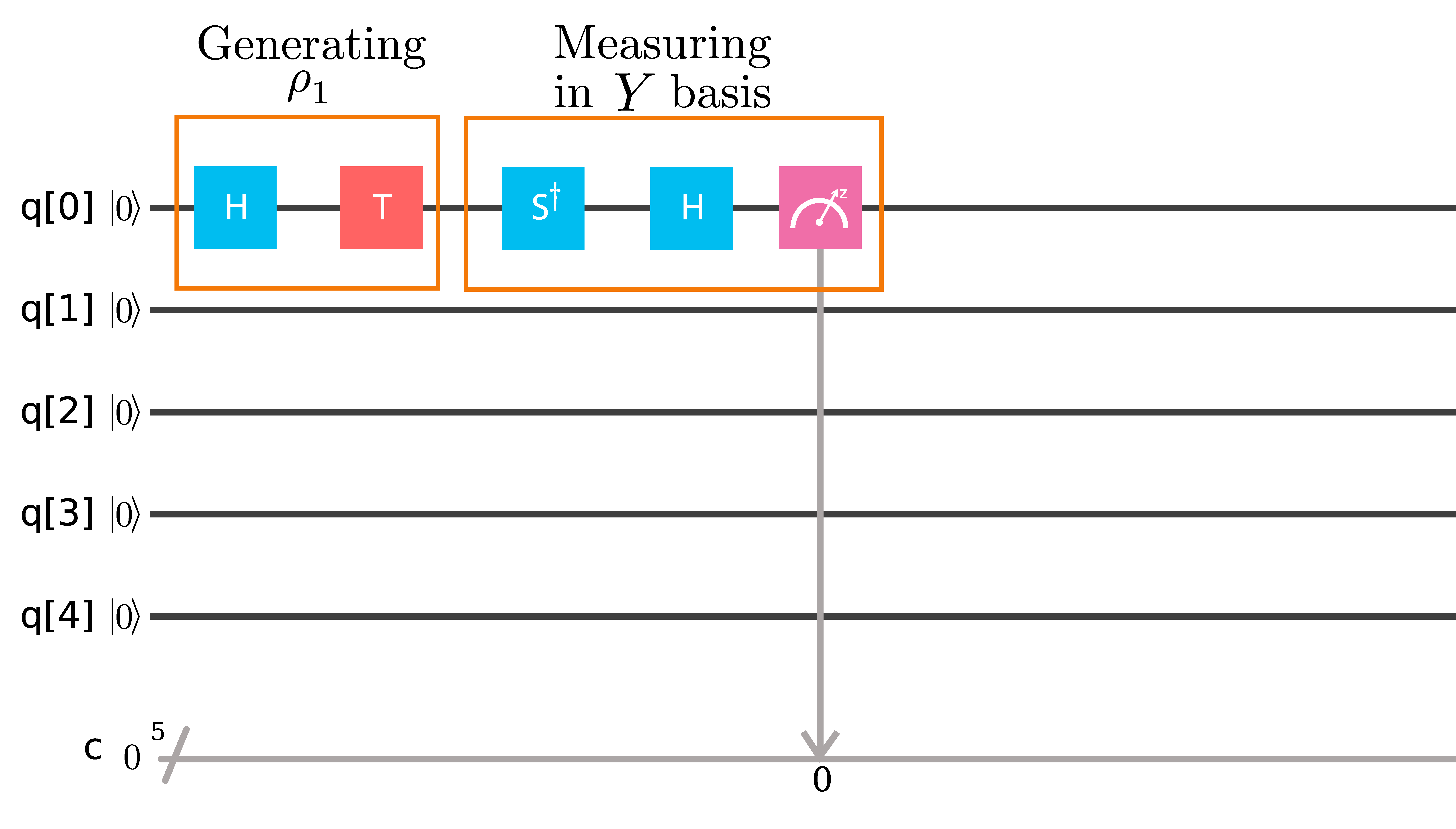}}\caption{Sigma Y}\label{fig3_19}
\end{figure}

\begin{lstlisting}
The QASM code that generates this circuit is:
include ``qelib1.inc'';
qreg q[5];
creg c[5];
// State Preparation
h q[0];
t q[0];
// Y or sigma\_2 basis
sdg q[0];
h q[0];
measure q[0] $->$ c[0];
\end{lstlisting}
\textbf{IBM Q Results:}\\
The figure below shows the simulation results obtained when qubit q[0] was measured in the $ \sigma _{3}  $basis. The system was projected to state $ \lvert 0\rangle $ 489 times, and to state $ \lvert 1\rangle $ 535 times.

\begin{figure}[H] \centering{\includegraphics[scale=.4]{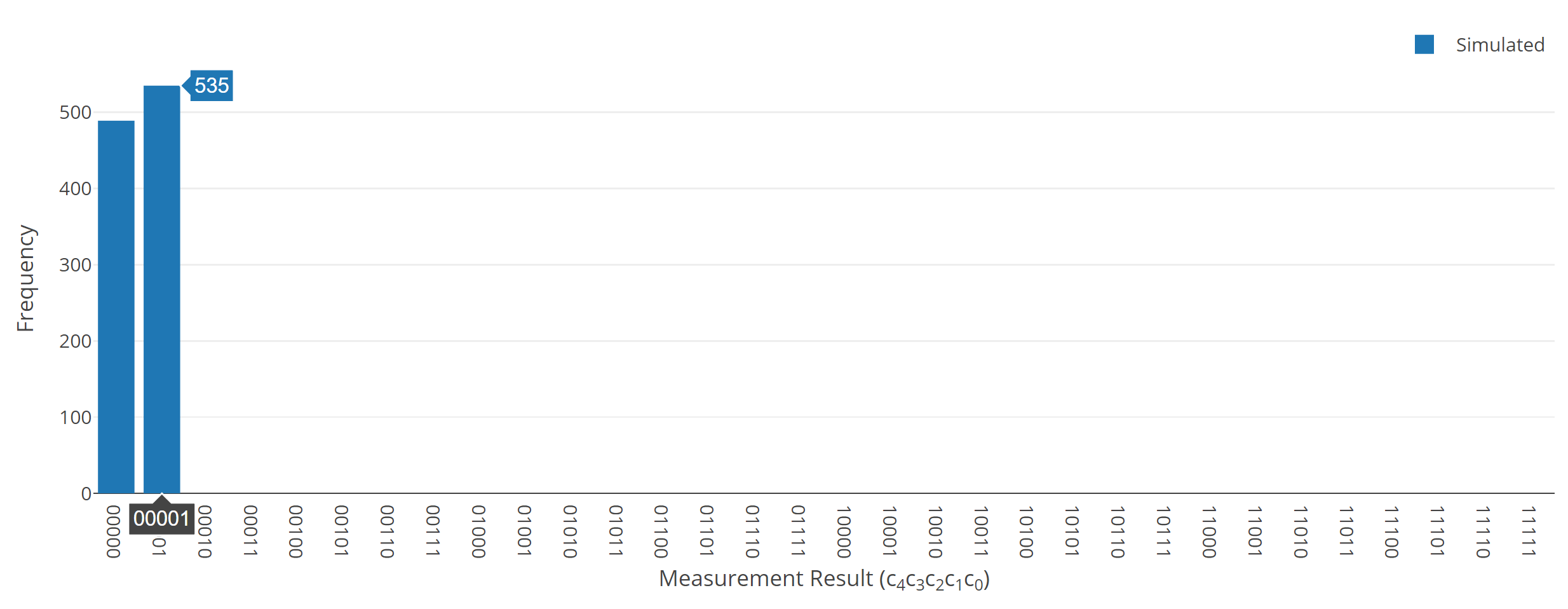}}\caption{BAR-Measurement Sigma Z}\label{fig3_20}
\end{figure}

The figure below shows the simulation results obtained when qubit q[0] was measured in the $ \sigma _{1} $ basis. The system was projected to state $ \frac{\left\vert 0\right\rangle +\left\vert 1\right\rangle }{\sqrt{2}} $ 835 times, and to state $ \frac{\left\vert 0\right\rangle -\left\vert 1\right\rangle }{\sqrt{2}}  $189 times.

\begin{figure}[H] \centering{\includegraphics[scale=.4]{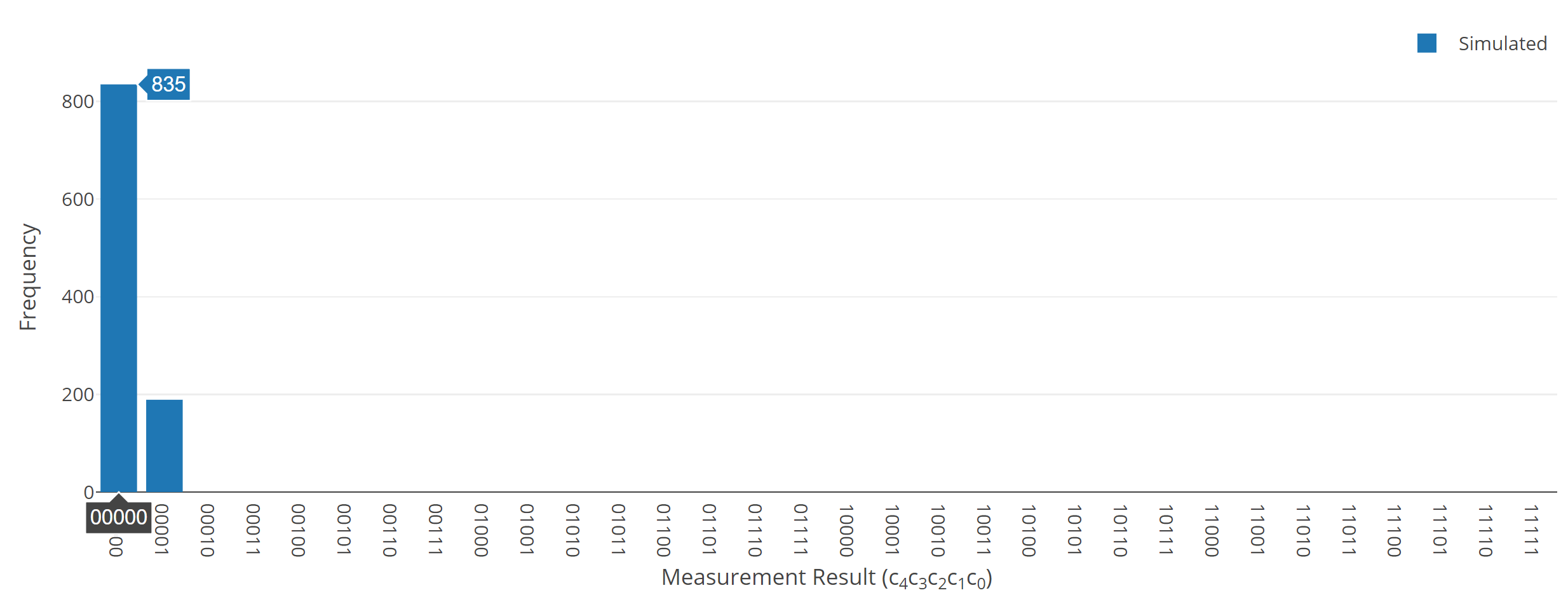}}\caption{BAR-Measurement Sigma X}\label{fig3_21}
\end{figure}

The figure below shows the simulation results obtained when qubit q[0] was measured in the $ \sigma _{2} $ basis. The system was projected to state$  \frac{\left\vert 0\right\rangle +i\left\vert 1\right\rangle }{\sqrt{2}} $ 845 times, and to state $ \frac{\left\vert 0\right\rangle -i\left\vert 1\right\rangle }{\sqrt{2}} $ 179  times.

\begin{figure}[H] \centering{\includegraphics[scale=.3]{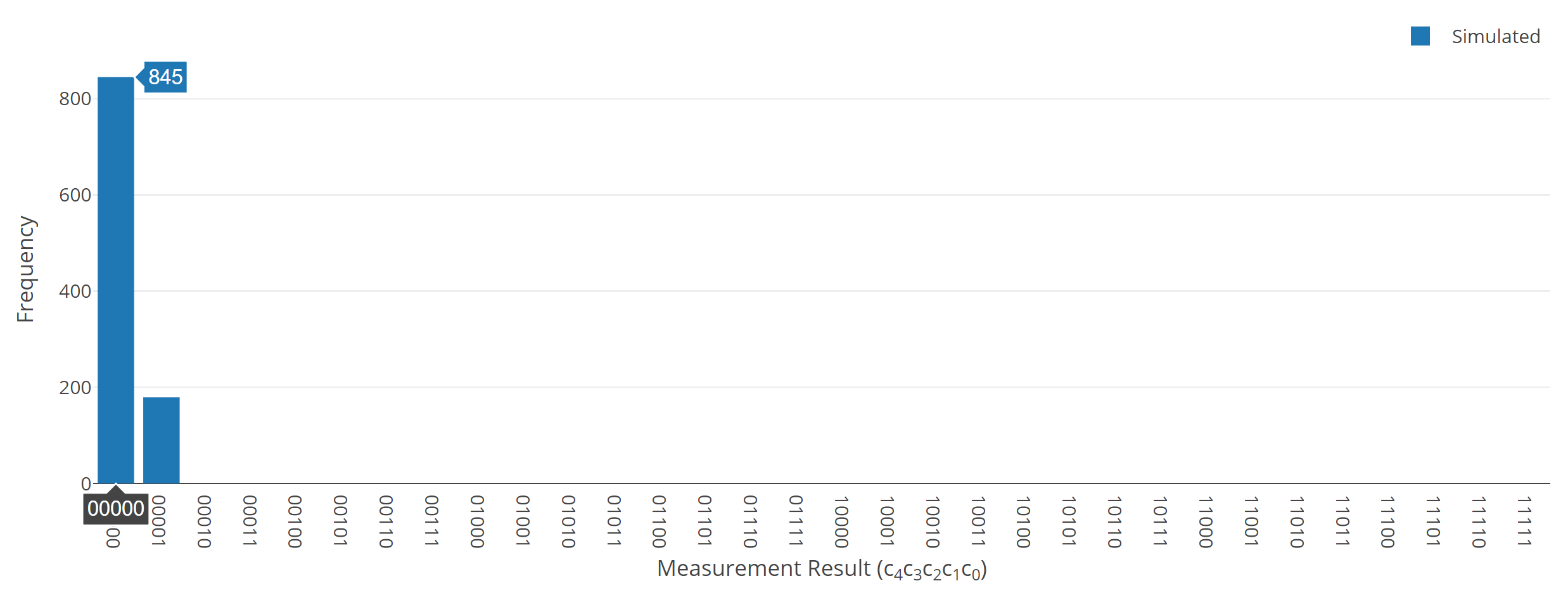}}\caption{BAR-Measurement Sigma Y}\label{fig3_22}
\end{figure}

We can compute the projection probabilities by normalizing the number of projections to a particular state by the total number of simulations 1024.

\begin{table}[H]
\centering
\caption{Projection probabilities}
\label{tab:3_1:Table 3}
\begin{tabular}{|c|c|}  \hline
        Projection probability & IBM Q results     \\  \hline
        $ P_{\left\vert 0\right\rangle } $  & 0.478  \\ \hline
$ P_{\left\vert 1\right\rangle } $    & 0.522 \\ \hline
$ P_{\frac{\left\vert 0\right\rangle +\left\vert 1\right\rangle }{\sqrt{2}}} $    & 0.815 \\ \hline
$ P_{\frac{\left\vert 0\right\rangle -\left\vert 1\right\rangle }{\sqrt{2}}}$ &    0.185 \\ \hline
$ P_{\frac{\left\vert 0\right\rangle +i\left\vert 1\right\rangle }{\sqrt{2}}} $ &    0.825 \\ \hline
$ P_{\frac{\left\vert 0\right\rangle -i\left\vert 1\right\rangle }{\sqrt{2}}}     $ & 0.175 \\ \hline
\end{tabular}
\end{table}
\textbf{Matrix Reconstruction:}\\
We use our results of the projection probabilities to compute the expectation values of the Pauli matrices,
\begin{center}
$ \displaystyle S_{1}    \displaystyle =    \displaystyle  0.815-0.185=0.630, $     \\     
$ \displaystyle S_{2}    \displaystyle =    \displaystyle 0.825-0.175=0.650, $     \\     
$ \displaystyle S_{3}    \displaystyle =    \displaystyle 0.478-0.522=-0.044. $
\end{center}
     
Once we have the experimental values of the $ \displaystyle S_{i} $ parameters, we reconstruct the density matrix
\begin{equation}\label{eq3_95}
\displaystyle \nonumber \rho _{1}^{E}    \displaystyle =    \displaystyle \left( \begin{array}{cc} \frac{1}{2}+\frac{1}{2}S_{3} & \frac{1}{2}S_{1}-\frac{1}{2}iS_{2} \\ \frac{1}{2}S_{1}+\frac{1}{2}iS_{2} & \frac{1}{2}-\frac{1}{2}S_{3}\end{array}\right)      
\displaystyle    \displaystyle =    \displaystyle \left( \begin{array}{cc} 0.478 & 0.315-0.325i \\ 0.315+0.325i & 0.522 \end{array}\right)
\end{equation}

We copy Equation (\ref{eq3_94}) here for convenience,
\begin{equation*}\label{eq3_96}
\rho _1\approx \left( \begin{array}{cc} 0.5 & 0.353\, 55-0.353\, 55i \\ 0.353\, 55+0.353\, 55i & 0.5\end{array}\right)    
\end{equation*}

Comparing Equation (\ref{eq3_94}) with Equation (\ref{eq3_95}), we can see that $ \rho _1 $ is in good agreement with the reconstructed density matrix $ \rho _1^ E $. The results are not identical due to the presence of noise and statistical sampling error. We can conclude that state preparation was successful.

\section{Single-Qubit State Tomography: Experience I}
\textbf{State Preparation:}\\
Consider the single-qubit quantum state
\begin{equation}\label{eq3_97}
\displaystyle \left\vert \psi _{1}\right\rangle    \displaystyle =    \displaystyle S^{\dagger }H\left\vert 0\right\rangle ,         
\displaystyle =    \displaystyle \frac{1}{\sqrt{2}}\left(\begin{array}{c} 1\\ -i \end{array}\right).    
\end{equation}
         
The figure below shows the generation of state $ \left\vert \psi _{2}\right\rangle $ using the IBM Q composer.

\begin{figure}[H] \centering{\includegraphics[scale=2]{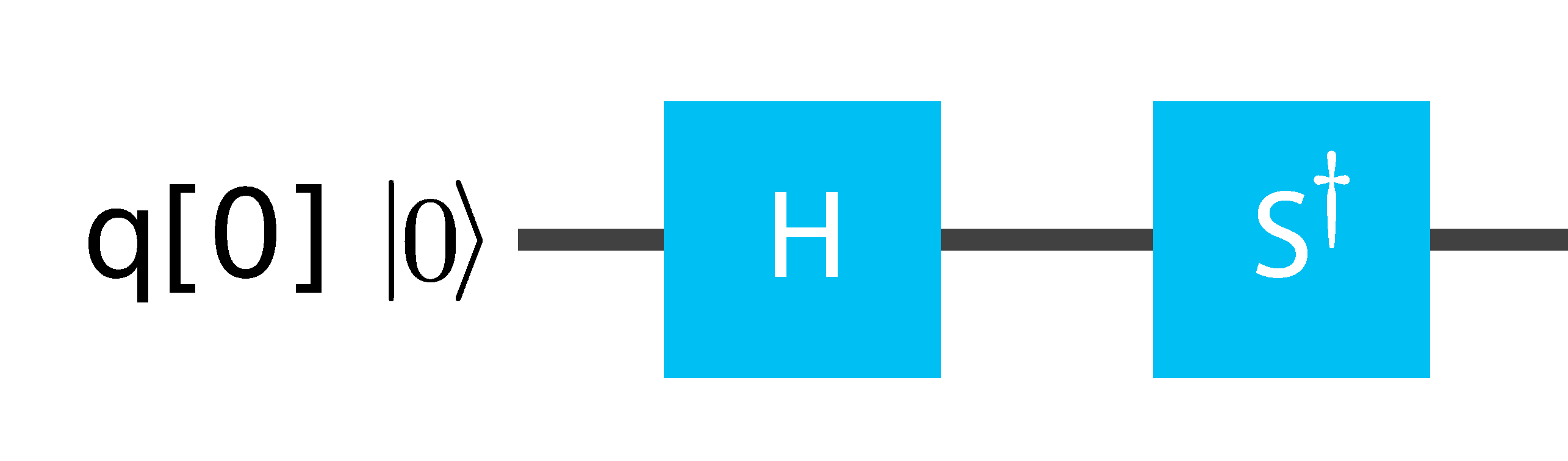}}\caption{Generation of state}\label{fig3_23}
\end{figure}

\begin{lstlisting}
The QASM code that generates this circuit is 
include ``qelib1.inc'';
qreg q[5];
creg c[5];
// State Preparation 
h q[0];
sdg q[0];
\end{lstlisting}

The density matrix representation of the quantum state $ \left\vert \psi _{2}\right\rangle $ is given by
\begin{equation}\label{eq3_98}
\displaystyle \rho _{2}    \displaystyle =    \displaystyle \left\vert \psi _{2}\right\rangle \left\langle \psi _{2}\right\vert           
\displaystyle =    \displaystyle \frac{1}{\sqrt{2}}\left( \begin{array}{c} 1 \\ -i\end{array}\right) \frac{1}{\sqrt{2}}\left( \begin{array}{cc} 1 & -i\end{array}\right)           
\displaystyle =    \displaystyle \left( \begin{array}{cc} 0.5 & 0.5i \\ -0.5i & 0.5 \end{array}\right) 
\end{equation}

In the following three IBM Q assessments, we will measure qubit q[0] in each of the Pauli-matrix bases: the$  \sigma _{1} $ basis, $ \sigma _{2} $ basis, and $ \sigma _{3} $ basis (equivalently, the X basis, Y basis, and Z basis, respectively). In each case, we will simulate and run the circuit 1024 times using the  IBM Q 5 Tenerife backend.

\item  Measuring in the Z Basis; The quantum circuit below shows the state preparation of $  \left\vert \psi _{2}\right\rangle $ and its measurement in the $  \sigma _{3} $ or Z basis.

\begin{figure}[H] \centering{\includegraphics[scale=1]{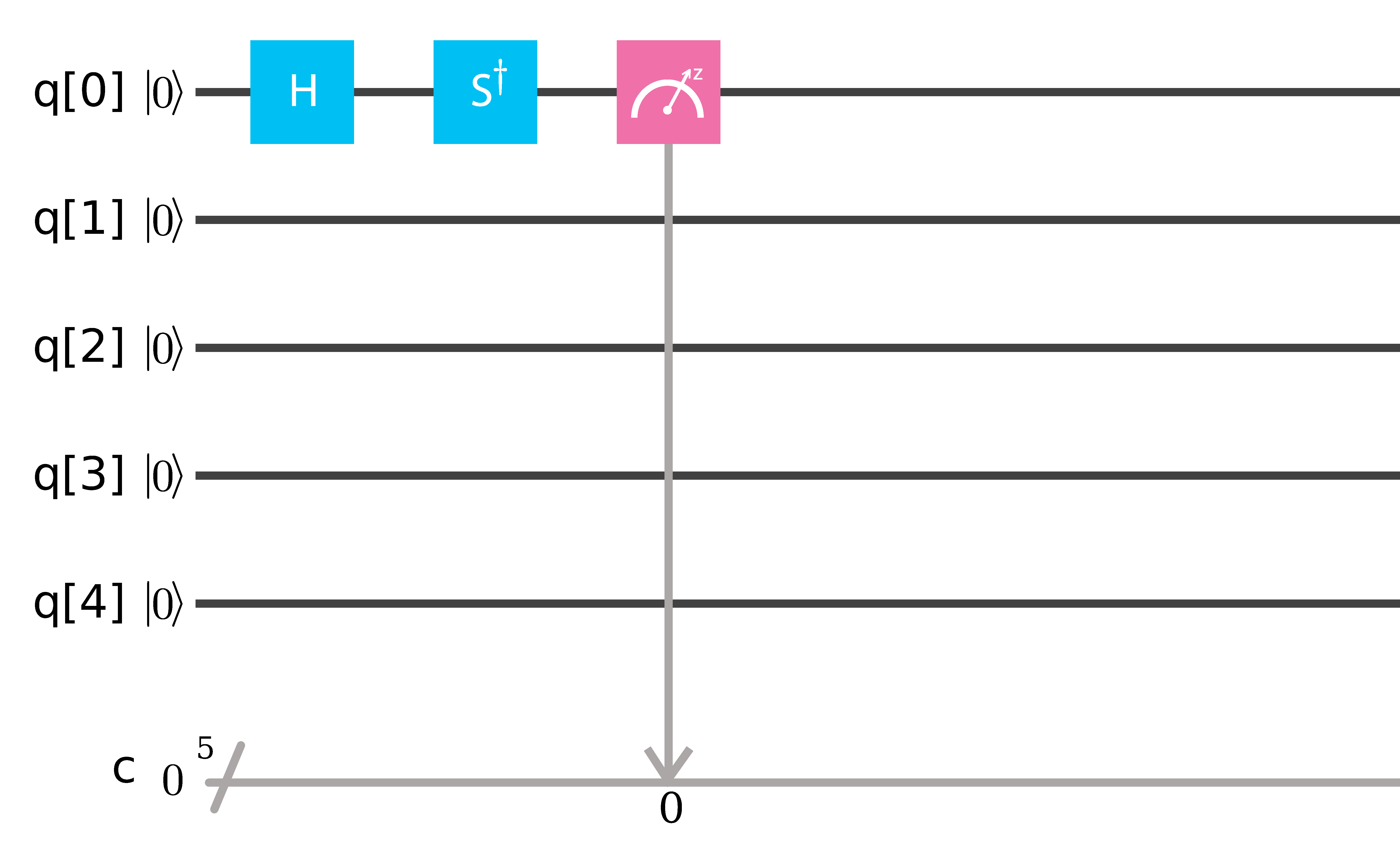}}\caption{Sigma Z}\label{fig3_24}
\end{figure} 

Write the QASM code that implements this quantum circuit.
\begin{lstlisting}
include ``qelib1.inc'';
qreg q[5];
creg c[5];
// State preparation
h q[0];
sdg q[0];
// Measurement in Z  
measure q[0] -> c[0];
\end{lstlisting}
Solution:\\
In the bar diagram, the bar for the measurement result labeled 00000 corresponds to the number of times that q[0] was projected to state $ \left\vert 0\right\rangle, $ and the bar for the measurement result labeled 00001 corresponds to the number of times that q[0] was projected to state $ \left\vert 1\right\rangle $. With these results, we can now compute the projection probabilities $ P_{\lvert 0\rangle } $ and $ P_{\lvert 1\rangle } $. 

\item  Measuring in the X Basis; The quantum circuit below shows the state preparation of $ \left\vert \psi _{2}\right\rangle $ and its measurement in the $ \sigma _{1} $ or X basis.

\begin{figure}[H] \centering{\includegraphics[scale=1]{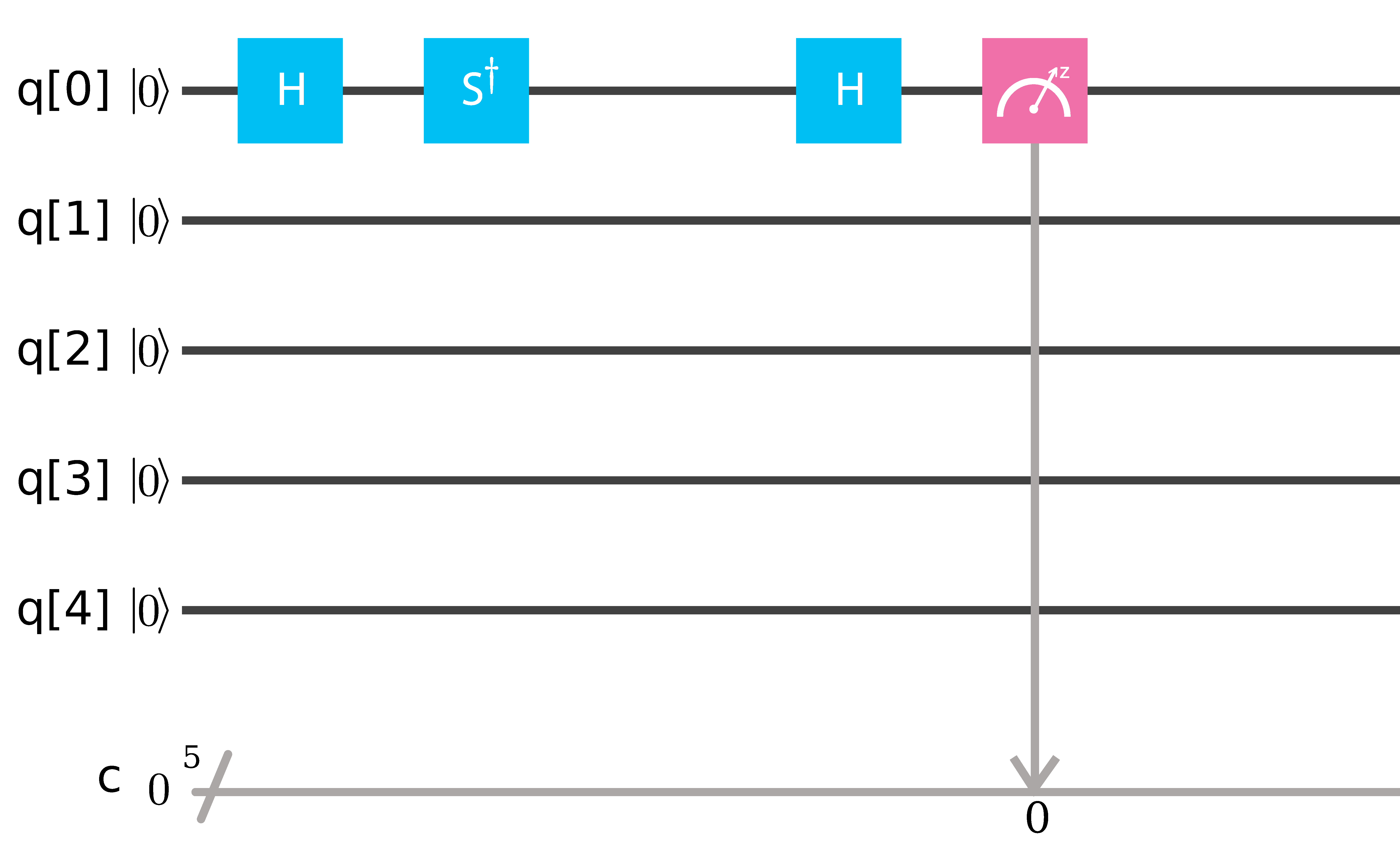}}\caption{Sigma X}\label{fig3_25}
\end{figure} 
Write the QASM code that implements this quantum circuit.
\begin{lstlisting}
include ``qelib1.inc'';
qreg q[5];
creg c[5];
// State preparation
h q[0];
sdg q[0];
// Measurement in X  
h q[0];
measure q[0] -> c[0];
\end{lstlisting}
Solution:\\
In the bar diagram, the bar for the measurement result labeled 00000 corresponds to the number of times that q[0] was projected to state $ \frac{\left\vert 0\right\rangle +\left\vert 1\right\rangle }{\sqrt{2}} $, and the bar for the measurement result labeled 00001 corresponds to the number of times that q[0] was projected to state $ \frac{\left\vert 0\right\rangle -\left\vert 1\right\rangle }{\sqrt{2}} $. With these results, we can now compute the projection probabilities $ P_{\frac{\left\vert 0\right\rangle \pm \left\vert 1\right\rangle }{\sqrt{2}}} $. 

\item  Measuring in the Y Basis; The quantum circuit below shows the state preparation of $ \left\vert \psi _{2}\right\rangle $ and its measurement in the $ \sigma _{2} $ or Y basis.

\begin{figure}[H] \centering{\includegraphics[scale=1]{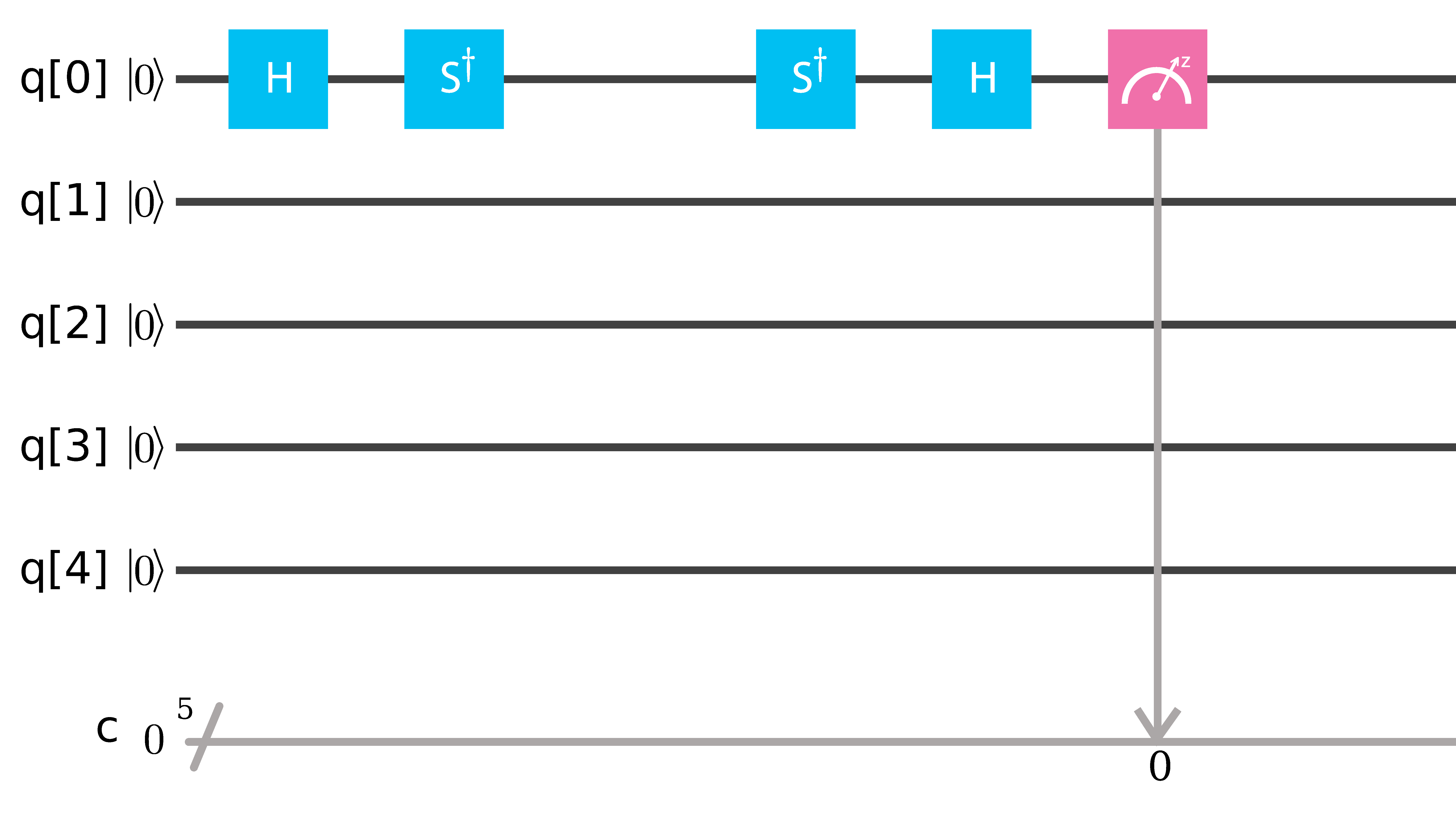}}\caption{Sigma Y}\label{fig3_26}
\end{figure}

Write the QASM code that implements this quantum circuit.
\begin{lstlisting}
include ``qelib1.inc'';
qreg q[5];
creg c[5];
// State preparation
h q[0];
sdg q[0];
// Measurement in Y  
sdg q[0];  
h q[0];
measure q[0] -> c[0];
\end{lstlisting}

Solution:\\
In the bar diagram, the bar for the measurement result labeled 00000 corresponds to the number of times that q[0] was projected to state $ \frac{\left\vert 0\right\rangle +i\left\vert 1\right\rangle }{\sqrt{2}} $, and the bar for the measurement result labeled 00001 corresponds to the number of times that q[0] was projected to state $ \frac{\left\vert 0\right\rangle -i\left\vert 1\right\rangle }{\sqrt{2}} $. With these results, we can now compute the projection probabilities $ P_{\frac{\left\vert 0\right\rangle \pm i\left\vert 1\right\rangle }{\sqrt{2}}} $.

\section{Single-Qubit State Tomography: Experience II}
\textbf{Expectation Values of the Pauli Matrices:}\\
In order to obtain the expectation values of the Pauli matrices, we need to compute the probabilities of projecting the system onto each one of the eigenstates of the Pauli matrices. Each probability will be given by the quotient of the number of times the system was projected onto one of the eigenstates and the total number of times the state was projected. In this example, we will always simulate and run the experiments 1024 times.

As an example, if the system is projected onto state $ \left\vert 0\right\rangle$ $\# 0$  times and to state $ \left\vert 1\right\rangle$ $\# 1$ times, then the corresponding projection probabilitites  are
\begin{equation}\label{eq3_99}
\begin{split}
\displaystyle P_{\left\vert 0\right\rangle } \displaystyle =    \displaystyle \frac{\#{0}}{\text{1024}} \\ 
\displaystyle P_{\left\vert 1\right\rangle } \displaystyle =    \displaystyle \frac{\#{1}}{\text{1024}} 
\end{split}
\end{equation}

\item  Z Basis Projection Probabilitites; Using the simulation results that we obtained measuring $ \left\vert \psi _{2}\right\rangle $ in the Z basis, compute the values of the following projection probabilities $ P_{\left\vert 0\right\rangle } $ and $ P_{\left\vert 1\right\rangle } $. Approximate our values using three significant figures.
\begin{equation}\label{eq3_99_1}
\begin{split}
P_{\left\vert 0\right\rangle }=0.5 \\
P_{\left\vert 1\right\rangle }=0.5     
\end{split}
\end{equation}
Solution:\\
The projection probabilities to states $ \left\vert 0 \right\rangle $ and $ \left\vert 1 \right\rangle $ can be analytically computed as 
\begin{center}
$ P_{\left\vert 0\right\rangle} =\left(\begin{array}{c c}1~~~ 0\end{array}\right)\left(\begin{array}{c c}0.5& 0.5i\\ -0.5i & 0.5\end{array}\right)\left(\begin{array}{ c}1\\ 0\end{array}\right)=0.5 $ \\
$ P_{\left\vert 1\right\rangle} =\left(\begin{array}{c c}0~~~ 1\end{array}\right)\left(\begin{array}{c c}0.5& 0.5i\\ -0.5i & 0.5\end{array}\right)\left(\begin{array}{ c}0\\ 1\end{array}\right)=0.5 $ \\
\end{center}
If in the IBM Q simulation the state was projected to $ \left\vert 0 \right\rangle $ and $ \left\vert 1 \right\rangle $ $ \# 0 $ and $ \#1 $ times respectively, the projection probabilitites are given by 
\begin{equation}\label{eq3_99_2}
\begin{split}
P_{\left\vert 0\right\rangle }    =      \frac{\# 0}{1024}  \\     
P_{\left\vert 1\right\rangle }    =      \frac {\# 1}{1024} 
\end{split}     
\end{equation}    
Note that since we are using our IBM Q simulation results for calculating projection probabilities, the accepted answers have a 0.05 tolerance.

\item  X Basis Projection Probabilitites; Using the simulation results that we obtained measuring $ \left\vert \psi _{2}\right\rangle $ in the X basis, compute the values of the following projection probabilities $ P_{\frac{\left\vert 0\right\rangle \pm \left\vert 1\right\rangle }{\sqrt{2}}} $. Approximate our values using three significant figures.
\begin{equation}\label{eq3_99_3}
\begin{split}
P_{\frac{\left\vert 0\right\rangle + \left\vert 1\right\rangle }{\sqrt{2}}}= 0.5\\
P_{\frac{\left\vert 0\right\rangle - \left\vert 1\right\rangle }{\sqrt{2}}}= 0.5    
\end{split}     
\end{equation}
Solution:\\
The projection probabilities to states $ \frac{\left\vert 0\right\rangle + \left\vert 1\right\rangle}{\sqrt{2}} $ and $ \frac{\left\vert 0\right\rangle - \left\vert 1\right\rangle}{\sqrt{2}} $ can be analytically computed as
\begin{center}
$ P_\frac{\left\vert 0\right\rangle + \left\vert 1\right\rangle}{\sqrt{2}} =\frac{1}{\sqrt{2}}\left(\begin{array}{c c}1~~~ 1\end{array}\right)\left(\begin{array}{c c}0.5 & 0.5i\\ -0.5i & 0.5\end{array}\right)\frac{1}{\sqrt{2}}\left(\begin{array}{ c}1\\ 1\end{array}\right)=0.5 $\\
$ P_\frac{\left\vert 0\right\rangle - \left\vert 1\right\rangle}{\sqrt{2}} =\frac{1}{\sqrt{2}}\left(\begin{array}{c c}1~ -1\end{array}\right)\left(\begin{array}{c c}0.5 & 0.5i\\ -0.5i & 0.5\end{array}\right)\frac{1}{\sqrt{2}}\left(\begin{array}{ c}1\\ -1\end{array}\right)=0.5 $ \\
\end{center}
Because of the Hadamard gate before the pre-defined measurement, the result of the number of times that the state was projected to $ \left\vert 0\right\rangle $ and $ \left\vert 1\right\rangle $, $ \# 0 $ and $ \# 1 $, now recorrespond to the number of times the state was projected to $ \frac{\left\vert 0\right\rangle + \left\vert 1\right\rangle}{\sqrt{2}} $ and $ \frac{\left\vert 0\right\rangle - \left\vert 1\right\rangle}{\sqrt{2}} $ respectively.
\begin{equation}\label{eq3_99_4}
\begin{split}
P_\frac{\left\vert 0\right\rangle + \left\vert 1\right\rangle}{\sqrt{2}}     =      \frac{\# 0}{1024}       \\
P_\frac{\left\vert 0\right\rangle - \left\vert 1\right\rangle}{\sqrt{2}}  =     \frac{\# 1}{1024}       \\    
\end{split}     
\end{equation}
Note that since we are using our IBM Q simulation results for calculating projection probabilities, the accepted answers have a 0.05 tolerance.

\item  Y Basis Projection Probabilitites; Using the simulation results that we obtained measuring $ \left\vert \psi _{2}\right\rangle $ in the Y basis, compute the values of the following projection probabilities$  P_{\frac{\left\vert 0\right\rangle \pm i\left\vert 1\right\rangle }{\sqrt{2}}} $. Approximate our values using three significant figures.
\begin{equation}\label{eq3_99_5}
\begin{split}
P_{\frac{\left\vert 0\right\rangle + i\left\vert 1\right\rangle }{\sqrt{2}}}= 0\\
P_{\frac{\left\vert 0\right\rangle - i\left\vert 1\right\rangle }{\sqrt{2}}}=1    
\end{split}     
\end{equation}

Solution:\\
The projection probabilities to states $ \frac{\left\vert 0\right\rangle + i\left\vert 1\right\rangle}{\sqrt{2}} $ and $ \frac{\left\vert 0\right\rangle - i\left\vert 1\right\rangle}{\sqrt{2}} $ can be analytically computed as 
\begin{center}
$ P_\frac{\left\vert 0\right\rangle + i\left\vert 1\right\rangle}{\sqrt{2}} =\frac{1}{\sqrt{2}}\left(\begin{array}{c c}1~-i\end{array}\right)\left(\begin{array}{c c}0.5& 0.5i\\ -0.5i & 0.5\end{array}\right)\frac{1}{\sqrt{2}}\left(\begin{array}{ c}1\\ i\end{array}\right)=0 $\\
$ P_\frac{\left\vert 0\right\rangle - i\left\vert 1\right\rangle}{\sqrt{2}} =\frac{1}{\sqrt{2}}\left(\begin{array}{c c}1~~~ i\end{array}\right)\left(\begin{array}{c c}0.5& 0.5i\\ -0.5i & 0.5\end{array}\right)\frac{1}{\sqrt{2}}\left(\begin{array}{ c}1\\ -i\end{array}\right)=1 $ \\
\end{center}
Because of the Hadamard gate and $ S^\dagger $ gates before the pre-defined measurement, the result of the number of times that the state was projected to $ \left\vert 0\right\rangle $ and $ \left\vert 1\right\rangle $, $ \# 0 $ and $ \# 1 $, now recorrespond to the number of times the state was projected to $ \frac{\left\vert 0\right\rangle + i\left\vert 1\right\rangle}{\sqrt{2}} $ and $ \frac{\left\vert 0\right\rangle - i\left\vert 1\right\rangle}{\sqrt{2}} $ respectively.
\begin{equation}\label{eq3_99_6}
\begin{split}
P_\frac{\left\vert 0\right\rangle + i\left\vert 1\right\rangle}{\sqrt{2}}     =      \frac{\# 0}{1024} ,      \\
P_\frac{\left\vert 0\right\rangle - i\left\vert 1\right\rangle}{\sqrt{2}}  =      \frac{\# 1}{1024} ,      \\    
\end{split}     
\end{equation}
Note that since we are using our IBM Q simulation results for calculating projection probabilities, the accepted answers have a 0.05 tolerance.\\
\textbf{Expectation Values:}\\
Compute the expectation values of the Pauli matrices $ \sigma_1, \sigma_2, $ and $ \sigma_3 $ using the following equations,

\begin{equation}\label{eq3_103}
\displaystyle S_{1}    \displaystyle =    \displaystyle P_{\frac{\left\vert 0\right\rangle +\left\vert 1\right\rangle }{\sqrt{2}}}-P_{\frac{\left\vert 0\right\rangle -\left\vert 1\right\rangle }{\sqrt{2}}},    
\end{equation}          

\begin{equation}\label{eq3_104}
\displaystyle S_{2}    \displaystyle =    \displaystyle P_{\frac{\left\vert 0\right\rangle +i\left\vert 1\right\rangle }{\sqrt{2}}}-P_{\frac{\left\vert 0\right\rangle -i\left\vert 1\right\rangle }{\sqrt{2}}},    
\end{equation}          
    
\begin{equation}\label{eq3_105}
\displaystyle S_{3}    \displaystyle =    \displaystyle P_{\left\vert 0\right\rangle }-P_{\left\vert 1\right\rangle }.    
\end{equation}

\item  Expectation Values for Pauli Matrices: X; Enter our numerical result for the expectation value of the Pauli matrix X or $ \sigma_1 = 0 $.\\
Solution:\\
The analytical result of the expectation value of X or $ \sigma _{1} $ is given by
\begin{center}
$ \displaystyle S_{1}    \displaystyle =    \displaystyle P_{\frac{\left\vert 0\right\rangle +\left\vert 1\right\rangle }{\sqrt{2}}}-P_{\frac{\left\vert 0\right\rangle -\left\vert 1\right\rangle }{\sqrt{2}}},          
\displaystyle =    \displaystyle 0.5-0.5,          
\displaystyle =    \displaystyle 0. $    
\end{center}
Since we are using our IBM Q simulation results for calculating the expectation value, the accepted correct answers have a 0.07 tolerance, 

\item  Expectation Values for Pauli Matrices: Y; Enter our numerical result for the expectation value of the Pauli matrix Y or $ \sigma_2  = -1 $.\\
Solution:\\
The analytical result of the expectation value of Y or $ \sigma _{2} $ is given by\\
\begin{center}
$ \displaystyle S_{2}    \displaystyle =    \displaystyle P_{\frac{\left\vert 0\right\rangle +i\left\vert 1\right\rangle }{\sqrt{2}}}-P_{\frac{\left\vert 0\right\rangle -i\left\vert 1\right\rangle }{\sqrt{2}}},          
\displaystyle =    \displaystyle 0-1.0,          
\displaystyle =    \displaystyle -1.0. $
\end{center}
\item  Expectation Values for Pauli Matrices: Z; Enter our numerical result for the expectation value of the Pauli matrix Z or $ \sigma_3 = -0 $.\\
Solution:\\
The analytical result of the expectation value of Z or $ \sigma _{3} $ is given by
\begin{center}
$ \displaystyle S_{3}    \displaystyle =    \displaystyle P_{\left\vert 0\right\rangle }-P_{\left\vert 1\right\rangle },          
\displaystyle =    \displaystyle 0.5-0.5,          
\displaystyle =    \displaystyle 0. $
\end{center}
Since we are using our IBM Q simulation results for calculating the expectation value, the accepted correct answers have a 0.07 tolerance.

\section{Single-Qubit State Tomography: Experience III}

Let us recall that the original density matrix implemented on the IBM Q Experience was 
\begin{equation}\label{eq3_106}
\displaystyle \rho _{2}    \displaystyle =    \displaystyle \left\vert \psi _{2}\right\rangle \left\langle \psi _{2}\right\vert           
\displaystyle =    \displaystyle \left( \begin{array}{cc} 0.5 & 0.5i \\ -0.5i & 0.5 \end{array}\right)
\end{equation}
      
By measuring our system on three different bases, we can now reconstruct the original density matrix.

\begin{figure}[H] \centering{\includegraphics[scale=.3]{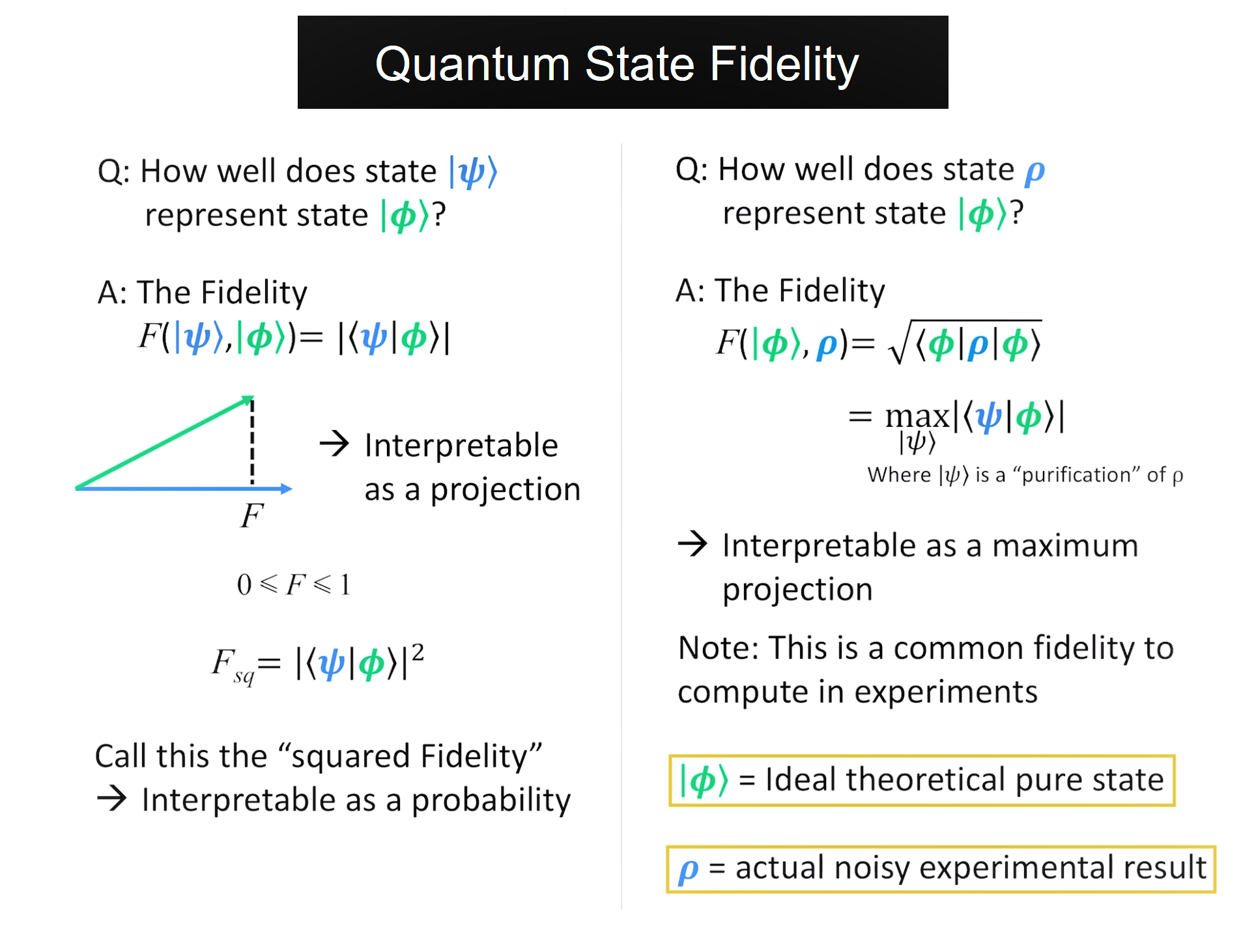}}\caption{Quantum State Fidelity}\label{fig3_28}
\end{figure}

\begin{figure}[H] \centering{\includegraphics[scale=.3]{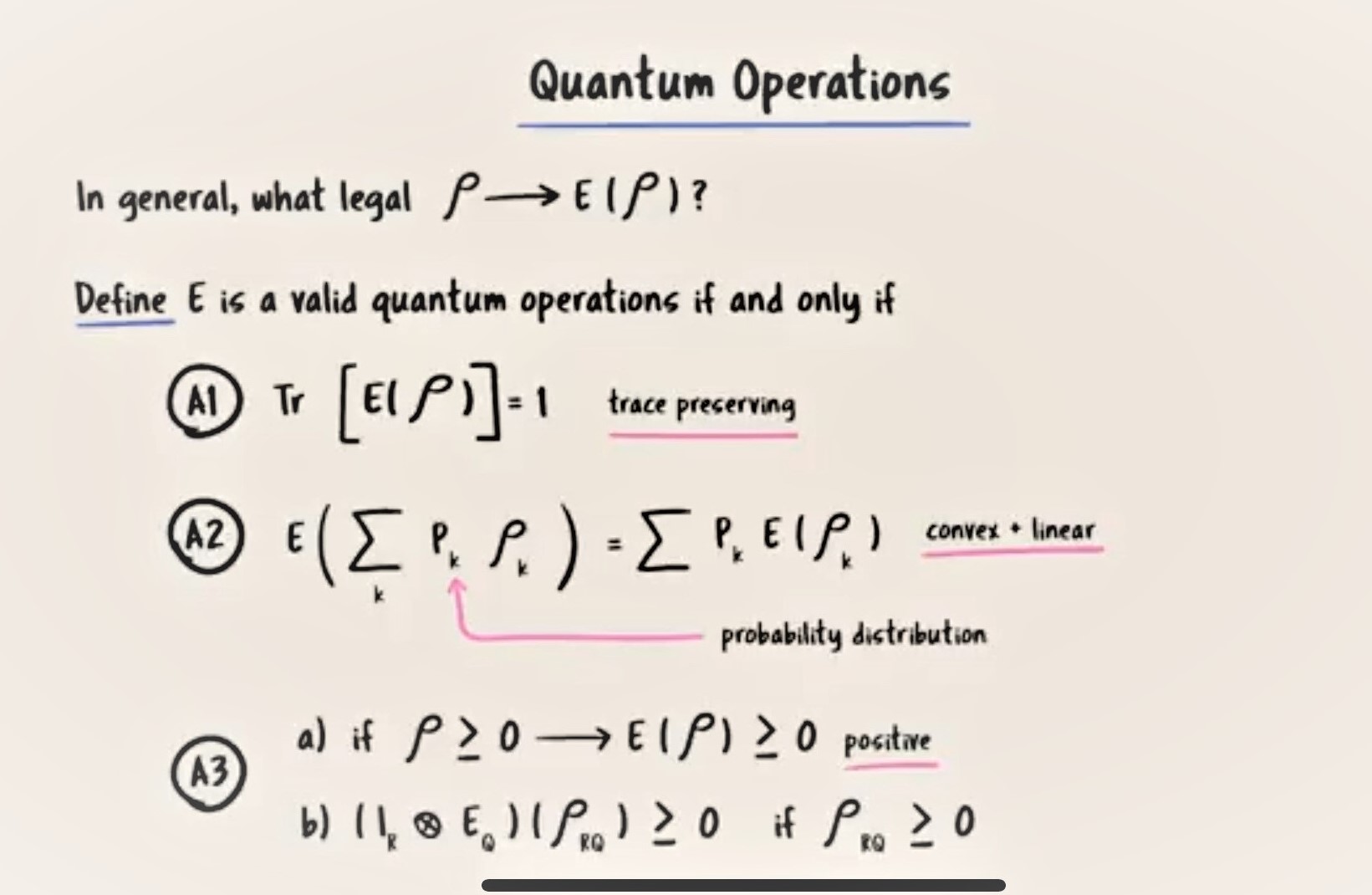}}\caption{Quantum Operations}\label{fig3_29}
\end{figure}

\end{enumerate}

\bibliographystyle{IEEEtran}
\bibliography{Bib-Quantum}
\printindex

\part{Universal Quantum-Computation}
\chapter{Universal Quantum-Computation}

\maketitle

In this research notebook on universal quantum computation for quantum engineers, researchers, and scientists, we will explore quantum advantage, referred to as quantum supremacy. We will put into practice the dynamical decoupling and free evolution. We will also discuss requirements and strategies for the realization of large-scale quantum computation robust and resilient to errors. We begin with an introduction to the concepts of error detection and error correction in classical and quantum systems, and we will introduce several examples to build intuition. We will discuss in detail one quantum error correction code and its code parameters. Next, we will describe how reliable classical and quantum machines can be built from unreliable components, introduce the threshold theorem, and discuss fault tolerance principles.
Furthermore, describe the conceptual role of fault-tolerance in realizing large-scale quantum and classical computation. After that, we will transition to quantum error suppression and error correction in practice, beginning with examples of composite pulses and dynamical decoupling sequences. We will then conclude with a detailed discussion on the surface code. We will discuss the scientific basis of families of quantum error correction codes. We will understand the different strategies for counteracting systematic versus random error in quantum computation. Next, we introduce computational complexity classes for classical and quantum computers and explore the meaning of quantum advantage, often referred to as ``quantum supremacy.'' Furthermore, in the end, we will implement dynamical error suppression protocols on the qubits of a real physical quantum computer, the IBM Quantum Experience.

\section{Strategies for Mitigating and Correcting Errors in Quantum Computers: Introduction} 
We will discuss what is needed to operate larger-scale machines for longer periods of time and in a fault-tolerant manner,\cite{nielsen_quantum_2011,chuang_quantum_2014,preskill_quantum_2019,vazirani_quantum_1997, watrous_theory_2011,vazirani_quantum_2007,aaronson_quantum_2006,shor_quantum_2003,chuang_quantum_2006,aaronson_quantum_2010,harrow_quantum_2018,childs_quantum_2008,cleve_introduction_2007,kivlichan_improved_2019}. We will begin by introducing strategies for detecting and correcting errors in quantum systems. We will discuss several simple examples of quantum error-correcting codes and review the importance of parity measurements in the implementation of quantum error detection. Then, we will consider how reliable classical and quantum machines are built from unreliable components. We will discuss in detail at the threshold theorem and the principles of fault tolerance as applied to both classical and quantum circuits \cite{pednault_breaking_2018}. Then, we will discuss at quantum error mitigation \cite{endo_mitigating_2019,endo_practical_2018} and quantum error correction in practice\cite{kandala_error_2019}. We will begin with examples of composite pulses and dynamical error suppression. We will then feature a study on realizing fault-tolerant quantum computation using the surface code \cite{kitaev_classical_2002}. Finally, we will address computational complexity topics and consider what it means to demonstrate a quantum computational advantage over classical computers. This milestone is often referred to as quantum supremacy\cite{harrow_quantum_2017,boixo_characterizing_2018}. In the end, we will implement dynamical error-mitigation protocols on the IBM quantum experience\cite{sisodia_circuit_2018}.

\section{Introduction to the Bit-Flip Code: Concept} 
In order to be useful, quantum computers like classical computers must be built from robust elements. Transistors are amazingly successful circuit elements, and this is in large part because they rarely fail. The failure rate of a transistor is around one part and say, 10 to the 20. In comparison, today's qubits are far more error-prone, with error rates as we have seen around one part in 10 to the 3 or 10 to the 4. It is a problem because algorithms that solve importantly, large scale problems will generally need to implement far more than 1,000 or 10,000 gates. So, what can be done? One thing that can be done, as we have discussed in section three, is to develop algorithms that provide a quantum advantage with the noisy devices we have today, working as a co-processor with a classical computer. That research is ongoing. However, for large scale, fault-tolerant quantum computing applications, we will need to make quantum devices more resilient to error. It is not just quantum systems. Most classical, electronic, and communication systems are also subject to errors at some level. Whether a classical system or quantum system, one means to improve system reliability is to use error correction protocols\cite{edmunds_measuring_2017}. Conceptually, systems are built from devices, and devices can be faulty. Let us say we have an imperfect device that is somewhat insensitive to small input errors but becomes less resilient with larger errors. If they are individually good enough, such devices can often be made more resilient by adding additional resources. Say, by building in redundancy. So, we improve system reliability, but it is done at the expense of additional overhead. To see how this works, let us discuss at a specific example of an error-correcting code called the bit-flip code, which can be implemented on both classical and quantum bits. We will start with a single qubit—a physical qubit with a probability, p, of a single error. As it is around 10 to the minus 3 or 10 to the minus 4. let us say this qubit is in a superposition state, $ \alpha $(0) + $ \beta $(1). we add redundancy by encoding the qubit state on to 3 qubits, $ \alpha $(0,0,0) + $ \beta $(1,1,1). Using the code words ``0, 0, 0'' to represent the logical state 0, and ``1,1,1'' to represent logical state 1. As we will see, this encoding protects against a single bit flip error on one of the qubits\cite{setia_superfast_2019}. That is, one of the qubits flips from 0 to 1 or from 1 to 0. If that error occurs on at most one of the qubits, we can detect that it happens and fix it. However, the error code will fail if there are two or more errors on the encoded qubits. The probability of getting two or more errors is 3p squared times 1 minus p plus p cubed. We will write this as a constant, c, times p squared to the leading order in the probability, p. Then the constant, c, conceptually, incorporates the number of ways these errors can occur, and the amount of overhead used in the code. So, does this encoding help us? Adding redundancy three qubits instead of one qubit is a net win if the encoded error probability, c times p squared, is less than the single-qubit error probability, p. conceptually, this sets a threshold condition to achieve a net win. The single-qubit error probability, p, must be less than 1/c. It means that if the single-qubit error rate is low enough, then introducing redundancy by adding more such qubits would reduce the overall error rate. The cost we pay to achieve this improvement is the overhead required to implement the error correction. This turning point probability, where single qubit performance becomes good enough for error correction to work, is called the threshold probability, or just the threshold for short\cite{mcgeoch_principles_2019,mcgeoch_practical_2019}. Here, the conceptual threshold is 1/c. We will discuss the formal definition of the threshold and its connection to extensible error correction protocols in future sections.

All realistic communication and computation schemes experience some form of noise. When noise induces errors in our system\cite{wallman_noise_2016}, we need tools that allow us to detect when an error occurred. One means for implementing error detection is based on a parity check. Parity checks are used in classical and quantum error correction schemes to determine the presence and location of an error. With that knowledge, one can go back and correct the error.

What is parity? Parity is the mathematical term which tells us if two things are the same, or if they are different\cite{laforest_mathematics_nodate}. For example, if two data bits $ b_1 $ and$  b_2 $ are the same, we can say that they have a parity equal to 0. However, if the two data bits are different, we can say they have a parity of 1. In other words, we can describe the parity of two data bits using a third ``parity bit'' $ b_ p $, which stores value 0 when the data bits are the same, and it stores 1 when they are different:
\begin{center}
$ b_1 b_2 = 00~ \rightarrow ~ b_ p=0, $ \\
$ b_1 b_2 = 01~ \rightarrow ~ b_ p=1, $ \\
$ b_1 b_2 = 10~ \rightarrow ~ b_ p=1, $ \\
$ b_1 b_2 = 11~ \rightarrow ~ b_ p=0 $ \\
\end{center}

Defined in this way, the parity bit is the exclusive-OR (XOR) of the two data bits, $ b_ p = b_1 \oplus b_2 $. Quantum mechanically, the XOR gate is implemented using a CNOT gate, for example, with qubit $ b_2 $ the target qubit.

Parity measurements can be generalized to larger numbers of bits, defined as the XOR sum of those bits. In general, the parity is 1 if there are an odd number of 1's, and the parity is 0 if there are an even number of 1's. The weight of the parity check corresponds to the number of bits. For example, the above example with two bits is a weight-two parity check. The surface code\cite{litinski_game_2019}, a type of quantum error correction protocol, uses weight-four parity checks\cite{farhi_limit_1998}.

Parity-check operations enable us to detect errors in quantum systems without projecting and destroying quantum information. They are an example of a broader class of such operations called stabilizers\cite{gottesman_stabilizer_1997}. In the next sections, we will discuss these concepts in more detail.

\section{Introduction to the Bit-Flip Code: Detecting Single Errors} 
Before diving into error detection, let us first discuss at how the single-qubit state can be encoded quantum mechanically onto these code words. The idea is to start with three qubits. Qubit 1 is prepared in the superposition state $ \alpha $ 0 plus $ \beta $ 1. qubits 2 and 3 are initialized in state 0. As we have done in previous sections, we will show the mathematical steps as we step through the quantum circuit indicated by the vertical line\cite{laforest_mathematics_nodate}. The resulting logical qubit state is realized by implementing two CNOT gates, as shown, with qubits 2 and 3 being the target qubits conditioned on the state of qubit 1. The truth table for the CNOT gate is shown. the state of qubit J is flipped if qubit 1 is in state 1. Otherwise, it is left alone. This is essentially implementing an XOR operation. So, let us take them one gate at a time. we will start with a CNOT gate applied to qubits 1 and 2 with qubit 2 the target conditioned on the state of qubit 1. For the $ \alpha $ 0 component of qubit 1, qubit 2 remains in state 0. for the $ \beta $ 1 component of qubit 1, and qubit 2 will flip to state 1. we see that we have now associated state 0 of qubit 2 with the coefficient $ \alpha $ and state 1 of qubit 2 with the coefficient $ \beta $. Similarly, applying a CNOT gate to qubits 1 and 3 leads to the same result. The 0 and 1 states of qubit 3 are now associated with $ \alpha $ and $ \beta $, respectively. The result is an entangled state $ \alpha $ 000 plus $ \beta $ 111. we will drop the tensor products \cite{orus_practical_2014}, remove the subscripts Q1, Q2, and Q3. instead, just use single kets for all three qubits to simplify the notation going forward. so, we have encoded the single-qubit state using a three-qubit code. Now, how can we detect if a bit-flip occurred on one of these three qubits? The answer is to implement a type of majority voting scheme using a set of measurements called parity checks between neighboring qubits. Qubits in the same state have the same parity, while those in different states have different parity. The outputs of the parity measurements are called a syndrome. the syndrome is used to tell us if and where a single error occurred. we will realize a parity check on qubits 1 and 2 using the XOR gate and store the answer on an ancilla qubit, SA. Similarly, we will apply the XOR gate to qubits 2 and 3 and store the result on a second ancilla qubit, SB. we will then measure the ancilla qubits to get the values of SA and SB. For example, let us discuss at the parity measurement of qubits 1 and 2. The value SA is the exclusive OR of qubits 1 and 2. we can see that if both qubits are in state 0, or both are in state 1, then the value of SA is 0. That is, if the qubits are in the same state, meaning they have the same parity, then there is no error. However, if one of the qubits is in state 0, and the other is in state 1, it does not matter which is which then they have different parity. the value of SA is 1. we will do a similar XOR gate on qubits 2 and 3 to get the value of SB. SB takes on value 0 if qubits 2 and 3 have the same parity, indicating no error. Alternatively, SB will have value 1, indicating that qubits 2 and 3 have different parity, indicating that an error occurred on one of those two qubits. Now, we may have noticed something very interesting about performing parity measurements. this is very important to preserving quantum information\cite{nielsen_quantum_2011,mermin_quantum_2007,national_academies_of_sciences_quantum_2018}. When we measure parity, we never discuss anything about the specific states 0 or 1 or of the qubits 1, 2, or 3. The parity measurement never projects those qubits onto state 1 or state 0. Rather, we only discuss their parity. That is, whether the two qubits are in the same state, having the same parity. Alternatively, in different states, having different parity. Essentially, we have devised a measurement that indicates if an error occurred or not but never reveals, or projects out, the computational states of the qubits. That is really neat. it is the backbone of quantum error detection. So, we have a syndrome comprising two values SA and SB. Let us see how they are used to tell us where a single bit-flip error occurred. we will start with the encoded state $ \alpha $ 000 plus $ \beta $ 111. Now, let us assume that we have a single bit-flip error on qubit 1. The encoded state becomes $ \alpha $ 100 plus $ \beta $ 011. we perform a parity check on qubits 1 and 2. because they are in different states whether 10 or 01 SA equals 1. we then perform a parity check on qubits 2 and 3. because they are always in the same state either 00 or 11, we find SB equals 0. As before, the XOR operations are performed using sequential CNOT gates with the ancilla qubits SA and SB being the target qubits. For example, the state of SA is flipped conditioned on the state of qubit 1. then again on the state of qubit 2. similarly, for SB and qubits 2 and 3., we encourage us to work out the truth tables for SA and SB on our own. So, the syndrome SA equals 1 and SB equals 0 tells us that a single bit-flip error occurred on qubit 1. If instead, the error occurred on qubit 2, then we see that this error leads to having different parity for qubits 1 and 2, as well as for qubits 2 and 3. The resulting syndrome measurement then yields SA equals 1, and SB equals 1, indicating an error on qubit 2. an error on qubit 3 results from qubits 1 and 2 having the same parity, but qubits 2 and 3 having different parity. So, the syndrome, in this case, is SA equals 0, and SA equals 1, indicating an error on qubit 3. if no errors occurred, then both SA and SB equal 0. Thus, the syndrome measurement uniquely identifies if a single qubit error occurred. if so, on which qubit. 

How are parity checks used in error correction? To give an example, Let us consider a simple protocol the ``bit-flip code'' that uses redundancy to encode information. That is, instead of encoding information in a single physical qubit\cite{setia_superfast_2019}, we encode this information onto a single logical qubit comprising several physical qubits\cite{kapit_very_2016}. For example, Let us consider a mapping of one qubit on to three qubits,$  q_1, q_2, $ and $ q_3 $:
\begin{equation}\label{eq4_01}
\vert 0\rangle ~ \rightarrow ~ \vert 0\rangle _{q_1} \vert 0 \rangle _{q_2} \vert 0 \rangle _{q_3} \rightarrow \vert 000 \rangle \qquad \qquad \vert 1\rangle ~ \rightarrow ~ \vert 1\rangle _{q_1}\vert 1\rangle _{q_2}\vert 1\rangle _{q_3}\rightarrow \vert 111 \rangle
\end{equation}

Where we use a shorthand Dirac notation \cite{dirac_fundamental_1925, dirac_quantum_1926,dirac_mathematical_1978,dirac_basis_1929,dirac_quantum_1929,dirac_theory_1926} for the 000 and 111 states, these states are called codewords, and they are both chosen to have parity 0. This type of encoding can be used to detect a specific type of error a ``bit flip''. That is, if a single bit-flip error occurs, then the resulting state is no longer a codeword, and this can be detected.

To see how, Let us consider a single-qubit superposition state mapped onto the codewords of the three-qubit bit-flip code:
\begin{equation}\label{eq4_02}
\alpha \vert 0 \rangle + \beta \vert 1 \rangle ~ \rightarrow ~ \alpha \vert 000 \rangle + \beta \vert 111 \rangle .
\end{equation}

Suppose a quantum operation is executed on our encoded quantum state. Ideally, in the absence of errors, the operation may change the values of $ \alpha $ and $ \beta $, but it should not flip any of the bits within the codewords. However, in the presence of noise, an error may occur.

How can we use parity checks to detect if a single bit-flip error occurred? The concept is that single bit-flip errors will change the quantum states so that they are no longer codewords, and this is detectable through a change in parity. As we discuss in the accompanying section, in the case of the three-qubit bit-flip code, one can use two pairwise parity checks between qubits 1 and 2 $ (q_1 \oplus q_2) $, and between qubits 2 and 3 $ (q_2 \oplus q_3) $. We will store the results of the parity measurements on additional, ancillary qubits $ S_ A $ and $ S_ B $:
\begin{equation}\label{eq4_03}
S_ A=q_1 \oplus q_2,
\end{equation}

\begin{equation}\label{eq4_04}
S_ B = q_2 \oplus q_3.
\end{equation}

Together, the measured values of $ S_ A $ and $ S_ B $ form a syndrome, which can uniquely identify if and where an error occurred, provided there is at most one bit-flip. Because of this, the ancillary qubits are sometimes called syndrome qubits.

Importantly, when we measure the syndrome qubits, we only discuss the syndrome values the parity of the data qubits without projecting the data qubits individually onto values 0 or 1. For example, when a bit flip occurs on $ q_1 $, we never discuss if $ \vert 000 \rangle \rightarrow \vert 100 \rangle  $or $ \vert 111 \rangle \rightarrow \vert 011 \rangle $. We only become aware that the syndrome has changed. For both codewords, $ (S_ A,S_ B) = (0,0) \rightarrow (1,0) $. The syndrome is then used to identify the presence and location $ (q_1) $ of the error. Because we never projected out 0 or 1 of the data qubits, the quantum information is preserved. Said another way, the syndrome measurements project the logical states onto the codeword states (no error), or onto one of the error states (error) outside the codeword space with measurement values, in this case, the parity that distinguishes the various cases.

\begin{figure}[H] \centering{\includegraphics[scale=.47]{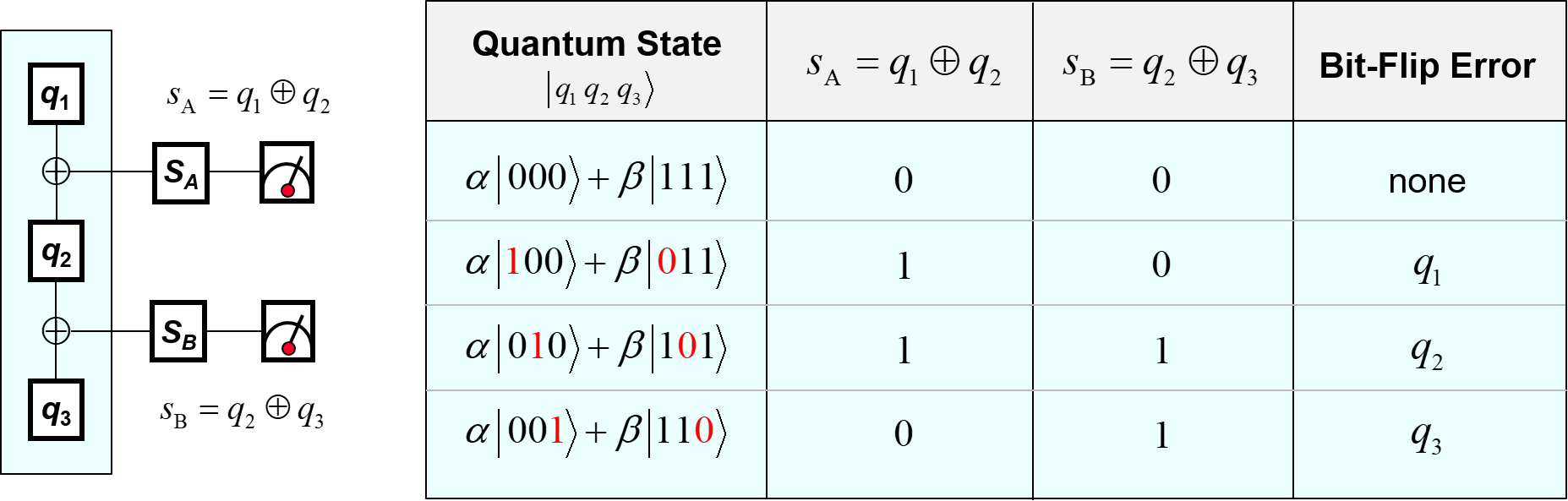}}\caption{Error detection with the three-qubit bit-flip code. The quantum state is encoded on to three qubits: $ q_1, q_2, $ and $ q_3 $. Parity checks between neighboring qubits are made and stored on the ancilla (syndrome) qubits $ S_ A $ and $ S_ B $. Measurements are then performed of the ancilla qubits, yielding the values $ s_ A  $and $ s_ B $. Given an arbitrary superposition state $ \alpha \vert 000 \rangle + \beta \vert 111 \rangle $, and assuming at most one bit-flip error (red text), the corresponding truth table for $ s_ A $ and $ s_ B $ is shown. Pairs of $ (s_ A,s_ B) $ values uniquely indicate the presence and location of a single error., Table of syndrome results for the three-qubit bit-flip error correction code with two additional ancilla qubits}\label{fig4_2}
\end{figure}

Finally, note that there is more than one way to realize the bit-flip code, also called a repetition code. In section presentations later this section, we will discuss another version of the bit-flip code (and related codes) that uses, for example, qubits $ q_2 $ and $ q_3 $ to store the syndrome, rather than using separate ancilla qubits. In that approach, the state of the physical qubit q1 is corrected, rather than the logically encoded state. While that approach uses fewer qubits, one needs to re-encode the state of qubit q1 onto the logical states after each syndrome measurement. Otherwise, many of the concepts and intuition are the same.

\section{Introduction to the Bit-Flip Code: Correcting Single Errors} 
So, now that we have detected a single error, what can we do? Clearly, if the syndrome is 0-0, we infer that no error occurred, and we do nothing. However, what about the other syndromes? Since the code indicates single bit-flip errors, we can take the result of the syndrome measurement and perform an X gate on the errant qubit to flip it back to the correct state. For example, if the syndrome is 1-0 indicating an error on qubit one, we apply an X gate to qubit one, flipping it back. Again, we never knew if qubit one was erroneously in state 0 or state 1. we just know that it was in the wrong state, whatever it was, and it needed to be flipped back to be in the correct state. Again, this lack of state-specific information means the quantum information has been preserved. Before we can proceed, we need to reset the ancilla qubits. In this case, S-A is flipped back to state 0 using an X gate. S-B, already in state 0, does not need to be flipped back. Next, let us assume that a bit-flip error occurs on qubit two. Here, the syndrome measurement yields 1-1, indicating qubit two has an error. So, we apply an X gate to qubit two to flip it back to the correct state. We apply X gates to both S-A and S-B to reset them to state 0-0. Finally, a syndrome 0-1 indicates a bit-flip error on qubit three, and so, we apply an X gate to qubit three to correct that error. We apply an X gate to ancilla qubit S-B to reset it back to state 0. so, to summarize, the bit flip code, although very simple, captures many key concepts of error correction. We started with a single qubit with an error probability p. we then redundantly encoded that qubit's state onto three qubits. The three-qubit bit-flip code can detect and correct for, at most, one-bit flip error. So, the logically encoded state is preserved unless two or more errors occur. Two errors happen with a probability c times p squared. Thus, we can define a threshold for this code here, and it is one over c that tells us at least how good the single qubits must be in order to achieve a net win when adding more together. We discuss that errors can be detected using parity measurements, which are combined to form a syndrome that tells us if and where an error occurred, without ever revealing or projecting out the quantum information. Based on the syndrome, we can correct for single errors and maintain the logical state. However, when two or more errors occurred, the syndrome will lead us to the wrong conclusion, and the scheme breaks down. Finally, we note that we can handle more errors by adding more redundancy. For example, using five qubits and four ancillas will correct for up to two-bit flip errors. Alternatively, if we want to improve the logical error rate, we can add more levels of recursion. That is, take each qubit in the level one logical encoding and represent each with its own logical encoding, and so on. Bit flips are not the only kind of error. In the remainder of this section and throughout the section, we will see a related code that can correct for phase flips, as well as more sophisticated codes that can correct for both types of error simultaneously. we will discuss the fault tolerance of particular codes, that is, the ability to handle errors wherever they may occur during the error detection and correction protocols. 
\section{Quantum Error Correction Codes: The Three-Qubit Bit Flip Error Correction Code} 
Now, we are going to go back to quantum error-correcting codes. In this section, we are going to do a quantum version of the repetition code. So, 0 goes to 000. 1 goes to 111. Well, let us try to quantumize it, or quantize it. 0 goes to 000. 1 goes to 111. we can write a circuit diagram for that. So, we will already know that this is not quantum cloning, because quantum cloning is impossible. So, $ \alpha $ 0 plus $ \beta $ 1 goes to $ \alpha $ 000, plus $ \beta $ 111. So, can we correct errors? Well, first, let us try to correct bit errors. So, what did this channel do? It corrected the one-bit error. Or not channel. What did that code do? It corrected the one-bit error. So, if we flipped one bit, we can get it back. So, let us start out by trying the $ \sigma $ x errors. $ \sigma $ x. Well, we have three bits. we can apply $ \sigma $ x on it. So, let us apply a $ \sigma $ x error to the encoded psi on the second bit. So, psi encoded is equal to what? Is equal to $ \alpha $ 010 plus $ \beta $ 101. So, can we correct this? Well, we can correct it—measure which bit is different. Now, we would like to measure which bit is different without measuring all three of these bits. Because if we measure all three of these bits, we will put a superposition into the code, $ \alpha $ zero plus $ \beta $ one. We would like the same superposition to come out of the code after we have decided it. So, how do we measure which bit is different? Well, actually, that is pretty easy. So, here is the encoded psi. We apply a control NOT from the first bit to the second bit, a control NOT from the first bit to the third bit. Back here, we apply a control NOT from the first to the second and control NOT to the first to the third. We get 0. So, $ \alpha $ 000, plus $ \beta $ 111 goes to $ \alpha $ 000 plus $ \beta $ 100. we measure two zeros. So, that is good. We can see that this decoding circuit is really the inverse of the encoding circuit. So, that is very good. Now, what happens if we made an error there? So, suppose we put $ \alpha $ 010, plus $ \beta $ 101 into that circuit. Psi error equals that. When we put it into that circuit, we get out $ \alpha $ 010 plus $ \beta $ 110. Because we are doing the control NOT from this guy to this guy, and from the first guy to the third guy. we measure 1, 0. our state is $ \alpha $ 0 plus $ \beta $ 1, because this is equal to $ \alpha $ 0 plus $ \beta $ 1 times 10. we measure these guys. So, what we measure tells us that our second bit is wrong, and we have recovered the first bit. So, suppose we put in $ \alpha $ 100 plus $ \beta $ 011. So, that is the error state. So, we got this by applying $ \sigma $ x to the first qubit. After circuit, we get $ \alpha $ 111 plus $ \beta $ 011 is equal to $ \alpha $ 1 plus $ \beta $ 0 times 11. we measure 11. So, what do we need to do to the state to fix it? If these measurement results are both ones, we apply x here. We should draw a classical stop in green. So, that is how we decode the qubit code. But, not everything is a $ \sigma $ x error. There are also $ \sigma $ z errors. So, what happens if we apply a $ \sigma $ z error to this code? Well, there are three things we can apply the $ \sigma $ z to. The first, second, or third qubit. What do we get? Well, here we get $ \alpha $ 000 minus $ \beta $ 111. Here we get $ \alpha $ 000 minus well, it is the same thing, right? And here we get the same thing again. all of these are encoded $ \alpha $ 0 minus $ \beta $ 1. So, what have we done? Well, we will call this the bit error-correcting code. 3 qubit bit error correction. If we have $ \sigma $ x, correct it. if we have a $ \sigma $ z, we have applied $ \sigma $ z to the encoded qubit.

The figure below shows the quantum circuit for an implementation of the bit-flip error correction code.
\begin{figure}[H] \centering{\includegraphics[scale=2]{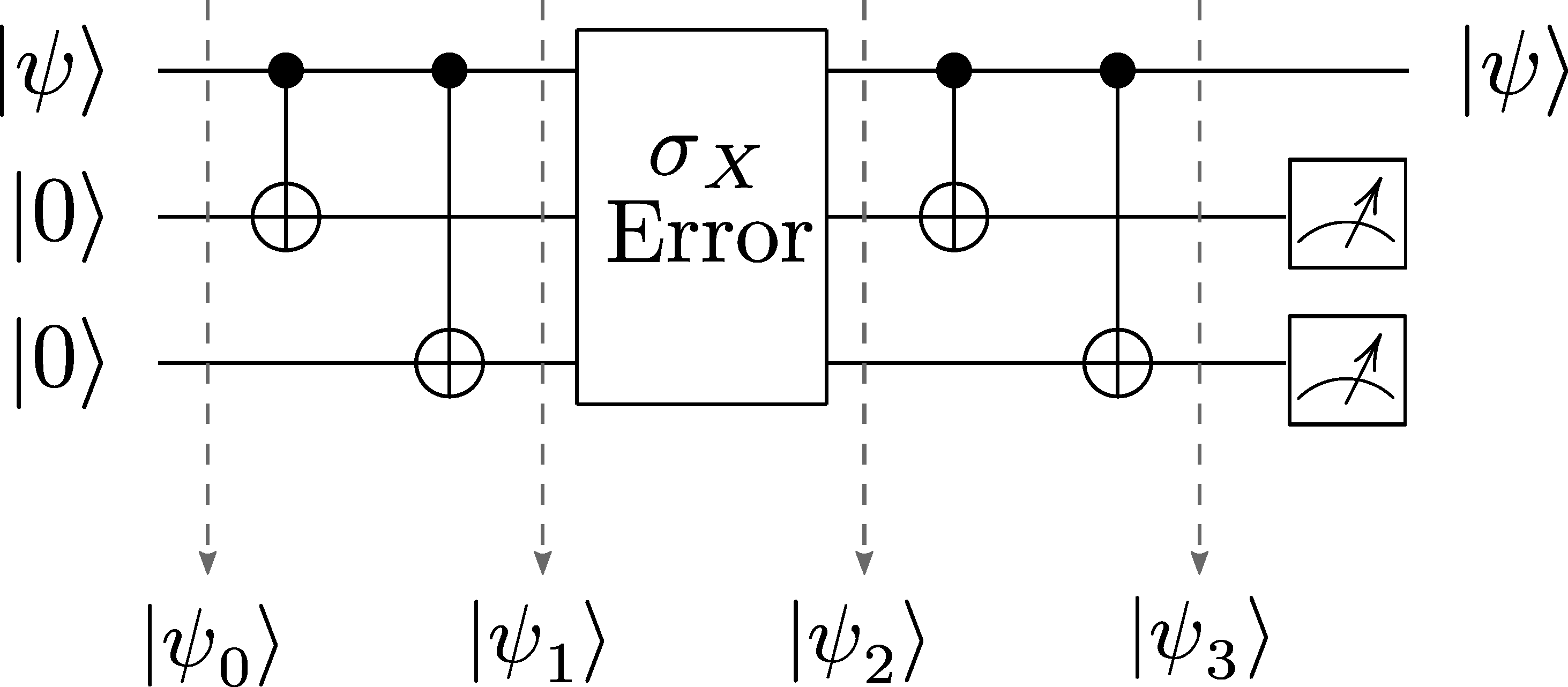}}\caption{Bit-flip}\label{fig4_3}
\end{figure}

1. The initial three-qubit state is
\begin{equation}\label{eq4_05}
\left\vert \psi _{0}\right\rangle =\left\vert \psi \right\rangle \left\vert 0\right\rangle \left\vert 0\right\rangle ,
\end{equation}

where the state of the qubit of interest is
\begin{equation}\label{eq4_06}
\left\vert \psi \right\rangle =\alpha \left\vert 0\right\rangle +\beta \left\vert 1\right\rangle .
\end{equation}

2. The first and second controlled-NOT (CNOT) gates \cite{chen_demonstration_2008} transform the state to
\begin{equation}\label{eq4_07}
\left\vert \psi _{0}\right\rangle \rightarrow \left\vert \psi _{1}\right\rangle =\alpha \left\vert 0\right\rangle \left\vert 0\right\rangle \left\vert 0\right\rangle +\beta \left\vert 1\right\rangle \left\vert 1\right\rangle \left\vert 1\right\rangle .
\end{equation}

3. First, Let us note that operationally a bit-flip error corresponds to an X-gate. This gate performs the following operation:
\begin{equation}\label{eq4_08}
\displaystyle X\left\vert 0\right\rangle    \displaystyle =    \displaystyle \left\vert 1\right\rangle ,          
\displaystyle X\left\vert 1\right\rangle    \displaystyle =    \displaystyle \left\vert 0\right\rangle .
\end{equation}

We will assume that the error occurs on at most one qubit.  The error can occur on the first, second, or third qubit, or it may not occur at all. The evolution of the system in these four cases is:

(a) There is no error on any of the qubits. In this case, the three-qubit state remains the same after $ E_{\text {bit}} $,
\begin{equation}\label{eq4_09}
\left\vert \psi _{1}\right\rangle \rightarrow \left\vert \psi _{2}^{a}\right\rangle =\alpha \left\vert 0\right\rangle \left\vert 0\right\rangle \left\vert 0\right\rangle +\beta \left\vert 1\right\rangle \left\vert 1\right\rangle \left\vert 1\right\rangle .
\end{equation}

(b) A bit-flip error occurs on the third qubit. In this case, the state of the system becomes:
\begin{equation}\label{eq4_10}
\left\vert \psi _{1}\right\rangle \rightarrow \left\vert \psi _{2}^{b}\right\rangle =\alpha \left\vert 0\right\rangle \left\vert 0\right\rangle \left\vert 1\right\rangle +\beta \left\vert 1\right\rangle \left\vert 1\right\rangle \left\vert 0\right\rangle .
\end{equation}

(c) A bit-flip error occurs on the second qubit. In this case, the state of the system becomes:
\begin{equation}\label{eq4_11}
\left\vert \psi _{1}\right\rangle \rightarrow \left\vert \psi _{2}^{c}\right\rangle =\alpha \left\vert 0\right\rangle \left\vert 1\right\rangle \left\vert 0\right\rangle +\beta \left\vert 1\right\rangle \left\vert 0\right\rangle \left\vert 1\right\rangle .
\end{equation}

(d) A bit-flip error occurs on the first qubit. In this case, the state of the system becomes:
\begin{equation}\label{eq4_12}
\left\vert \psi _{1}\right\rangle \rightarrow \left\vert \psi _{2}^{d}\right\rangle =\alpha \left\vert 1\right\rangle \left\vert 0\right\rangle \left\vert 0\right\rangle +\beta \left\vert 0\right\rangle \left\vert 1\right\rangle \left\vert 1\right\rangle .
\end{equation}

The third and fourth CNOT gates leave the system in state:
\begin{equation}\label{eq4_13}
\begin{split}
\displaystyle \left\vert \psi _{2}^{a}\right\rangle    \displaystyle \rightarrow    \displaystyle \left\vert \psi _{3}^{a}\right\rangle & =\alpha \left\vert 0\right\rangle \left\vert 0\right\rangle \left\vert 0\right\rangle +\beta \left\vert 1\right\rangle \left\vert 0\right\rangle \left\vert 0\right\rangle      \\     
\displaystyle \left\vert \psi _{2}^{b}\right\rangle    \displaystyle \rightarrow    \displaystyle \left\vert \psi _{3}^{b}\right\rangle & =\alpha \left\vert 0\right\rangle \left\vert 0\right\rangle \left\vert 1\right\rangle +\beta \left\vert 1\right\rangle \left\vert 0\right\rangle \left\vert 1\right\rangle \\ 
\displaystyle \left\vert \psi _{2}^{c}\right\rangle    \displaystyle \rightarrow    \displaystyle \left\vert \psi _{3}^{c}\right\rangle & =\alpha \left\vert 0\right\rangle \left\vert 1\right\rangle \left\vert 0\right\rangle +\beta \left\vert 1\right\rangle \left\vert 1\right\rangle \left\vert 0\right\rangle \\      
\displaystyle \left\vert \psi _{2}^{d}\right\rangle    \displaystyle \rightarrow    \displaystyle \left\vert \psi _{3}^{d}\right\rangle & =\alpha \left\vert 1\right\rangle \left\vert 1\right\rangle \left\vert 1\right\rangle +\beta \left\vert 0\right\rangle \left\vert 1\right\rangle \left\vert 1\right\rangle 
\end{split}    
\end{equation}

Finally, the four possible states before measuring the two ancillary qubits are:
\begin{equation}\label{eq4_14}
\begin{split}
\displaystyle \left\vert \psi _{3}^{a}\right\rangle    \displaystyle & =    \displaystyle \left( \alpha \left\vert 0\right\rangle +\beta \left\vert 1\right\rangle \right) \left\vert 0\right\rangle \left\vert 0\right\rangle \\          
\displaystyle \left\vert \psi _{3}^{b}\right\rangle    \displaystyle & =    \displaystyle \left( \alpha \left\vert 0\right\rangle +\beta \left\vert 1\right\rangle \right) \left\vert 0\right\rangle \left\vert 1\right\rangle \\          
\displaystyle \left\vert \psi _{3}^{c}\right\rangle    \displaystyle & =    \displaystyle \left( \alpha \left\vert 0\right\rangle +\beta \left\vert 1\right\rangle \right) \left\vert 1\right\rangle \left\vert 0\right\rangle \\          
\displaystyle \left\vert \psi _{3}^{d}\right\rangle    \displaystyle & =    \displaystyle \left( \alpha \left\vert 1\right\rangle +\beta \left\vert 0\right\rangle \right) \left\vert 1\right\rangle \left\vert 1\right\rangle .
\end{split}
\end{equation}

\begin{enumerate}[wide, labelwidth=!, labelindent=0pt]
\item  Bit-Flip Code I: Recovering the State; In which case(s) should a $ \sigma _{X} $ gate be applied to the first qubit to recover the initial state of qubit 1, $ \left\vert \psi_0 \right\rangle $?
\begin{itemize}
\item When the ancillary qubits are measured in $ \left\vert 0\right\rangle \left\vert 0\right\rangle $
\item When the ancillary qubits are measured in$  \left\vert 0\right\rangle \left\vert 1\right\rangle $
\item When the ancillary qubits are measured in $ \left\vert 1\right\rangle \left\vert 0\right\rangle $
\item When the ancillary qubits are measured in $ \left\vert 1\right\rangle \left\vert 1\right\rangle $
\end{itemize}
Solution:\\
If the ancillary qubits (qubits 2 and 3) are projected to $ \left\vert 0\right\rangle \left\vert 0\right\rangle $ (case a), then there was no error in any of the three qubits. If the ancillary qubits are projected to $ \left\vert 0\right\rangle \left\vert 1\right\rangle $ (case b), then there was bit-flip error in the third qubit. If the ancillary qubits are projected to $ \left\vert 1\right\rangle \left\vert 0\right\rangle $ (case c), then there was an error in the second qubit. In these three cases, there is no need to apply an extra operation to recover the initial state of the qubit of interest $ \left\vert \psi \right\rangle $. If the ancillary qubits are projected to $ \left\vert 1\right\rangle \left\vert 1\right\rangle $, then there was an error in the first qubit; and to recover the state $ \left\vert \psi \right\rangle $, an X gate has to be applied in the first qubit. Note that this implementation is predicated on the assumption of at most one error.

\section{Quantum Error Correction Codes: The Three-Qubit Phase Flip Error Correction Code} 
Suppose we have a dephasing channel and let us say that is 1 minus q, we apply a $ \sigma $ Z, probability Q, 1 minus q we apply nothing, probability Q we apply to $ \sigma $ Z. The probability of error is we get an error if we apply a $ \sigma $ Z to any qubit. We do not get an error if we apply $ \sigma $ Z to an even number of qubits. So, it is 3 Q, 1 minus Q squared plus Q cubed. We want to say this is three times the chance of $ \sigma $ Z error on an uncoded qubit. However, if we have a de-bidding channel, if we have a channel 1 minus Q $ \sigma $ X 1 minus Q identity, and Q $ \sigma $ x probability of error, well, we need at least two of these qubits to be wrong, to have an error in the originally encoded qubit. So, the probability of error is Q, 3 Q squared, 1 minus Q plus Q cubed. So, this is quadratically small if Q is small, and this is 3 times 3Q if Q is small. So, we have corrected the bit errors, and we made the phase errors even worse. Now, we probably know that Hadamard's interchange phase errors bit errors. So, intuitively this should mean that we can take this code and Hadamard it and get a code that corrects the bit phase errors but makes bit errors three times as worse. So, let us do it. The bit code, or the bit correcting code, was 0 goes to 0 0 0. 1 goes to 1 1 1. So, let us say this is the phase code. It is going to take H 0 to H cubed 0 0 0. H 1 to H cubed 1 1 1. this is just 0 plus 1, 1 over 2 goes to 1 over root 8, 0 0 0 plus 0 0 1 plus dot dot dot plus 1 1 1, and 1 over root 2 0 minus 1 goes to 1 over root 8 0 0 0 minus 0 0 1 through minus 1 1 1. this is minus 1, the number of 1's in this state. So, we worked this out for Simon's algorithm. This is what Hadamard cubed does. It takes up puts a minus 1 on any state with an odd number of ones. We can add these, and we can subtract these to get what the code does to 0 and 1. So, phase code 0 goes to 0 0 0 plus 0 1 1 plus 1 0 1 plus 1 1 0. 1 goes to 1/2 0 0 1 plus 0 1 0 plus 1 0 0 minus 1 1 plus 1 1. So, we take zero to the superposition of states with an even number of zeros, even the number of ones, and one to a superposition state with an odd number of ones. Because when we subtract this from this, we only get states with an odd number of ones. When we add these, we only get states with an even number of ones. So, we can correct this, and actually, it should be obvious that we can correct this because all we did was, we put a Hadamard in front of everything. So, just take our old correction circuit and add a whole bunch of Hadamard’s, and we get the new correction circuit with a phase code. However, we want to explain it in a different way. So, let us call this, let us call this a logical 0, and this a logical 1. Logical 0, logical 1, $ \sigma $ Z on the first qubit, a logical 0, $ \sigma $ Z on the first qubit, logical 1, $ \sigma $ Z on the second qubit, logical 0, are all orthogonal. Well, we can see that up here. We mean, we can note first clearly 0 and 1 are orthogonal because none of these basis states in the superposition of 1 appears in a superposition of 0. when we apply a $ \sigma $ Z to any of the bits, this is still true. So, any $ \sigma $ Z error on one is orthogonal to any other $ \sigma $ Z error on 0. further, if we apply $ \sigma $ Z error to 0, so, $ \sigma $ Z on first qubit of 0 logical 0 is equal to 1 over 2 0 0 0 plus 0 1 1 minus 1 0 1 minus 1 1 0. So, half the terms here are plus, half the terms here are minus, so, this is orthogonal to that. It is easy to check that all of these are orthogonal. So, project onto subspaces 0 L 1 L 0 $ \sigma $ Z 1 0 L $ \sigma $ Z 1 1 L. So, these are four orthogonal subspaces. So, there is a measurement that projects the state onto one of these four subspaces. In this case, if we get this subspace, we do nothing. If we get this subspace, we apply $ \sigma $ Z for the first qubit, and we get the next subspace, we apply $ \sigma $ Z to the second qubit. So, this shows us how to correct these errors.

\item  Phase-Flip Code I: Recovering the State; The figure below shows the quantum circuit for the implementation of the phase-flip error correction code.

\begin{figure}[H] \centering{\includegraphics[scale=1.8]{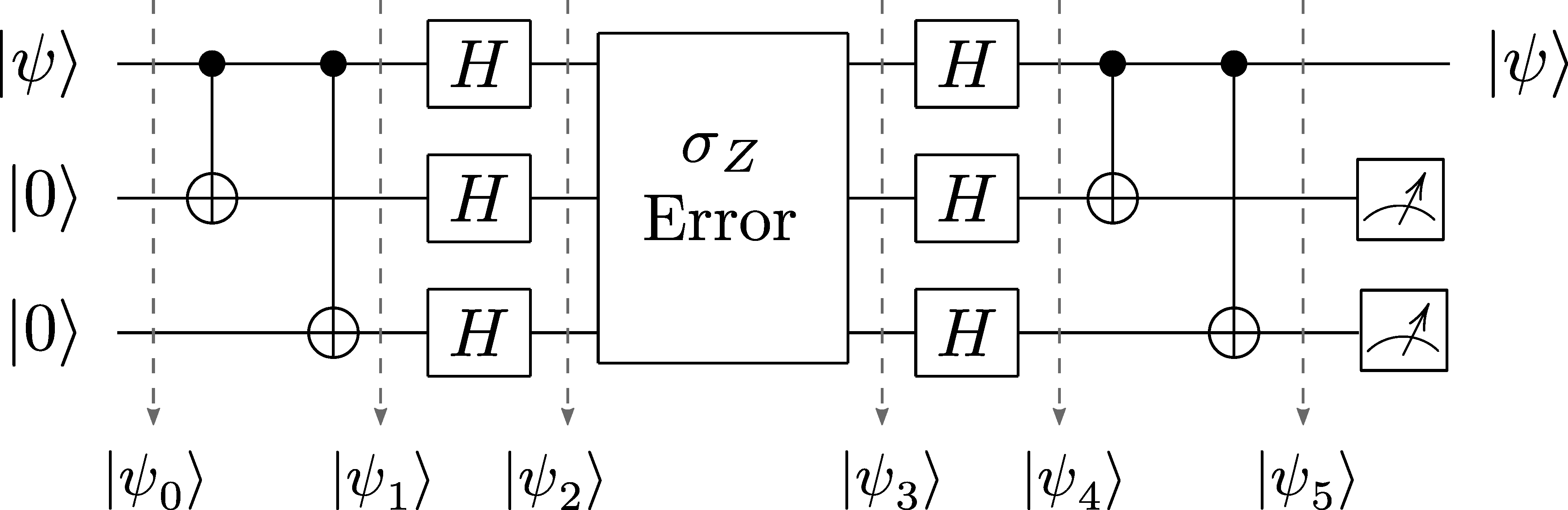}}\caption{Phase-Flip Code}\label{fig4_4}
\end{figure}

The initial three-qubit state is
\begin{equation}\label{eq4_15}
\left\vert \psi _{0}\right\rangle =\left\vert \psi \right\rangle \left\vert 0\right\rangle \left\vert 0\right\rangle ,
\end{equation}

where the state of the qubit of interest is
\begin{equation}\label{eq4_16}
\left\vert \psi \right\rangle =\alpha \left\vert 0\right\rangle +\beta \left\vert 1\right\rangle .
\end{equation}

The first and second CNOT gates transform the state to
\begin{equation}\label{eq4_17}
\left\vert \psi _{0}\right\rangle \rightarrow \left\vert \psi _{1}\right\rangle =\alpha \left\vert 0\right\rangle \left\vert 0\right\rangle \left\vert 0\right\rangle +\beta \left\vert 1\right\rangle \left\vert 1\right\rangle \left\vert 1\right\rangle .
\end{equation}

After a Hadamard gate in each of the qubits, the state becomes

\begin{equation}\label{eq4_18}
\left\vert \psi _{1}\right\rangle \rightarrow \left\vert \psi _{2}\right\rangle =\alpha \left\vert +\right\rangle \left\vert +\right\rangle \left\vert +\right\rangle +\beta \left\vert -\right\rangle \left\vert -\right\rangle \left\vert -\right\rangle ,
\end{equation}

where
\begin{equation}\label{eq4_19}
\left\vert \pm \right\rangle =\frac{\left\vert 0\right\rangle \pm \left\vert 1\right\rangle }{\sqrt{2}}.
\end{equation}

Then, the phase error occurs on at most one qubit. Note that a phase flip corresponds operationally to a $ \sigma _ Z $ gate. This gate implements the following transformation:

\begin{equation}\label{eq4_20}
\displaystyle \sigma _ Z\left\vert +\right\rangle    \displaystyle =    \displaystyle \frac{\left\vert 0\right\rangle -\left\vert 1\right\rangle }{\sqrt{2}},          
\displaystyle \sigma _ Z\left\vert -\right\rangle    \displaystyle =    \displaystyle \frac{\left\vert 0\right\rangle +\left\vert 1\right\rangle }{\sqrt{2}}.
\end{equation}
          
There would be four possible error cases depending on which qubit the error occurred if an error occurred at all.

(a) There is no error on any of the qubits. In this case, the three-qubit state remains the same,
\begin{equation}\label{eq4_21}
\left\vert \psi _{2}\right\rangle \rightarrow \left\vert \psi _{3}^{a}\right\rangle =\alpha \left\vert +\right\rangle \left\vert +\right\rangle \left\vert +\right\rangle +\beta \left\vert -\right\rangle \left\vert -\right\rangle \left\vert -\right\rangle .
\end{equation}

(b) A phase error occurs on the third qubit. In this case, the state of the system is transformed as:
\begin{equation}\label{eq4_22}
\left\vert \psi _{2}\right\rangle \rightarrow \left\vert \psi _{3}^{b}\right\rangle =\alpha \left\vert +\right\rangle \left\vert +\right\rangle \left\vert -\right\rangle +\beta \left\vert -\right\rangle \left\vert -\right\rangle \left\vert +\right\rangle .
\end{equation}

(c) A phase error occurs on the second qubit. In this case, the state of the system is transformed as:
\begin{equation}\label{eq4_23}
\left\vert \psi _{2}\right\rangle \rightarrow \left\vert \psi _{3}^{c}\right\rangle =\alpha \left\vert +\right\rangle \left\vert -\right\rangle \left\vert +\right\rangle +\beta \left\vert -\right\rangle \left\vert +\right\rangle \left\vert -\right\rangle .
\end{equation}

(d) A phase error occurs on the first qubit. In this case, the state of the system is transformed as:
\begin{equation}\label{eq4_24}
\left\vert \psi _{2}\right\rangle \rightarrow \left\vert \psi _{3}^{d}\right\rangle =\alpha \left\vert -\right\rangle \left\vert +\right\rangle \left\vert +\right\rangle +\beta \left\vert +\right\rangle \left\vert -\right\rangle \left\vert -\right\rangle .
\end{equation}

After applying a second set of Hadamard gate in each of the qubits, the state becomes:

\begin{equation}\label{eq4_25}
\begin{split}
\displaystyle \left\vert \psi _{3}^{a}\right\rangle    \displaystyle \rightarrow    \displaystyle \left\vert \psi _{4}^{a}\right\rangle & =\alpha \left\vert 0\right\rangle \left\vert 0\right\rangle \left\vert 0\right\rangle +\beta \left\vert 1\right\rangle \left\vert 1\right\rangle \left\vert 1\right\rangle \\          
\displaystyle \left\vert \psi _{3}^{b}\right\rangle    \displaystyle \rightarrow    \displaystyle \left\vert \psi _{4}^{b}\right\rangle & =\alpha \left\vert 0\right\rangle \left\vert 0\right\rangle \left\vert 1\right\rangle +\beta \left\vert 1\right\rangle \left\vert 1\right\rangle \left\vert 0\right\rangle \\     
\displaystyle \left\vert \psi _{3}^{c}\right\rangle    \displaystyle \rightarrow    \displaystyle \left\vert \psi _{4}^{c}\right\rangle & =\alpha \left\vert 0\right\rangle \left\vert 1\right\rangle \left\vert 0\right\rangle +\beta \left\vert 1\right\rangle \left\vert 0\right\rangle \left\vert 1\right\rangle \\     
\displaystyle \left\vert \psi _{3}^{d}\right\rangle    \displaystyle \rightarrow    \displaystyle \left\vert \psi _{4}^{d}\right\rangle & =\alpha \left\vert 1\right\rangle \left\vert 0\right\rangle \left\vert 0\right\rangle +\beta \left\vert 0\right\rangle \left\vert 1\right\rangle \left\vert 1\right\rangle \\
\end{split}
\end{equation}
          
The third and fourth CNOT gates transform the state to:

\begin{equation}\label{eq4_26}
\begin{split}
\displaystyle \left\vert \psi _{4}^{a}\right\rangle    \displaystyle \rightarrow    \displaystyle \left\vert \psi _{5}^{a}\right\rangle & =\alpha \left\vert 0\right\rangle \left\vert 0\right\rangle \left\vert 0\right\rangle +\beta \left\vert 1\right\rangle \left\vert 0\right\rangle \left\vert 0\right\rangle \\          
\displaystyle \left\vert \psi _{4}^{b}\right\rangle    \displaystyle \rightarrow    \displaystyle \left\vert \psi _{5}^{b}\right\rangle & =\alpha \left\vert 0\right\rangle \left\vert 0\right\rangle \left\vert 1\right\rangle +\beta \left\vert 1\right\rangle \left\vert 0\right\rangle \left\vert 1\right\rangle \\          
\displaystyle \left\vert \psi _{4}^{c}\right\rangle    \displaystyle \rightarrow    \displaystyle \left\vert \psi _{5}^{c}\right\rangle & =\alpha \left\vert 0\right\rangle \left\vert 1\right\rangle \left\vert 0\right\rangle +\beta \left\vert 1\right\rangle \left\vert 1\right\rangle \left\vert 0\right\rangle \\          
\displaystyle \left\vert \psi _{4}^{d}\right\rangle    \displaystyle \rightarrow    \displaystyle \left\vert \psi _{5}^{d}\right\rangle & =\alpha \left\vert 1\right\rangle \left\vert 1\right\rangle \left\vert 1\right\rangle +\beta \left\vert 0\right\rangle \left\vert 1\right\rangle \left\vert 1\right\rangle \\
\end{split}
\end{equation}
          
Finally, the four possible states before measuring the two ancillary qubits are:
\begin{equation}\label{eq4_27}
\begin{split}
\displaystyle \left\vert \psi _{5}^{a}\right\rangle    \displaystyle & =    \displaystyle \left( \alpha \left\vert 0\right\rangle +\beta \left\vert 1\right\rangle \right) \left\vert 0\right\rangle \left\vert 0\right\rangle \\          
\displaystyle \left\vert \psi _{5}^{b}\right\rangle    \displaystyle & =    \displaystyle \left( \alpha \left\vert 0\right\rangle +\beta \left\vert 1\right\rangle \right) \left\vert 0\right\rangle \left\vert 1\right\rangle \\          
\displaystyle \left\vert \psi _{5}^{c}\right\rangle    \displaystyle & =    \displaystyle \left( \alpha \left\vert 0\right\rangle +\beta \left\vert 1\right\rangle \right) \left\vert 1\right\rangle \left\vert 0\right\rangle \\          
\displaystyle \left\vert \psi _{5}^{d}\right\rangle    \displaystyle & =    \displaystyle \left( \alpha \left\vert 1\right\rangle +\beta \left\vert 0\right\rangle \right) \left\vert 1\right\rangle \left\vert 1\right\rangle 
\end{split}    
\end{equation}

\item  When should a $  \sigma _{X} $ gate be applied to qubit 1 to recover the initial state of qubit 1?$ \left\vert \psi \right\rangle $ ?
\begin{itemize}
\item When the ancillary qubits are measured in $ \left\vert 0\right\rangle \left\vert 0\right\rangle $
\item When the ancillary qubits are measured in $ \left\vert 0\right\rangle \left\vert 1\right\rangle $
\item When the ancillary qubits are measured in $ \left\vert 1\right\rangle \left\vert 0\right\rangle $
\item When the ancillary qubits are measured in $ \left\vert 1\right\rangle \left\vert 1\right\rangle $
\end{itemize}
Solution:\\

From the four final conditional states\\
$ \displaystyle \left\vert \psi _{5}^{a}\right\rangle    \displaystyle =    \displaystyle \left( \alpha \left\vert 0\right\rangle +\beta \left\vert 1\right\rangle \right) \left\vert 0\right\rangle \left\vert 0\right\rangle ,\\          
\displaystyle \left\vert \psi _{5}^{b}\right\rangle    \displaystyle =    \displaystyle \left( \alpha \left\vert 0\right\rangle +\beta \left\vert 1\right\rangle \right) \left\vert 0\right\rangle \left\vert 1\right\rangle ,    \\      
\displaystyle \left\vert \psi _{5}^{c}\right\rangle    \displaystyle =    \displaystyle \left( \alpha \left\vert 0\right\rangle +\beta \left\vert 1\right\rangle \right) \left\vert 1\right\rangle \left\vert 0\right\rangle ,     \\     
\displaystyle \left\vert \psi _{5}^{d}\right\rangle    \displaystyle =    \displaystyle \left( \alpha \left\vert 1\right\rangle +\beta \left\vert 0\right\rangle \right) \left\vert 1\right\rangle \left\vert 1\right\rangle , $\\          
we can see that if the ancillary qubits are projected to $ \left\vert 0\right\rangle \left\vert 0\right\rangle $ (case a), then there was no error in any of the three qubits.
If the ancillary qubits are projected to $ \left\vert 0\right\rangle \left\vert 1\right\rangle $ (case b), then there was a phase-flip error in the third qubit. If the ancillary qubits are projected to $ \left\vert 1\right\rangle \left\vert 0\right\rangle $ (case c), then there was an error in the second qubit. In these first-three cases, there is no need to apply an extra operation to recover the initial state $ \left\vert \psi \right\rangle $ of qubit 1.
However, if the ancillary qubits are projected to $ \left\vert 1\right\rangle \left\vert 1\right\rangle $, then there was an error on the first qubit. To recover state $ \left\vert \psi \right\rangle $, a $ \sigma_X $ gate has to be applied in the first qubit.

\section{Quantum Error Correction Codes: The Nine-Qubit Quantum Error Correction Code} 

we have a code that corrects bit errors, corrects that $ \sigma $ x errors, and makes $ \sigma $ z errors three times as likely. we have a code that corrects $ \sigma $ z errors and makes $ \sigma $ x errors three times as likely. we can easily check if we have a $ \sigma $ x error, it takes a logical 0 to logical 1 and vice versa. So, a $ \sigma $ x error on any of the three-bit qubits flips the logical 0 and logical 1. So, we apply a $ \sigma $ x to the logical qubit. So, we mean, this is we squeeze on the phase errors, we squeeze on the bit errors, the phase errors blow up. So, how do we fix this? Well, what we do is what is called concatenation. Well, what we do is combine the two codes. So, that they fix both bit errors and phases errors. in classical coding theory, it is called concatenating the two codes. So, first, what we do is we encode the code with one of these codes. then, we encode each of the qubits in the remaining code with the other. So, a zero we will just write it down goes to 1/2 to the 9/2 000 plus 111. Is 9/2 the wrong 9/2 is the wrong number. 1 over square root of 8 000 plus 111 000 plus 111. So, we are encoding one qubit into nine qubits. 1 goes to 1 over root 8 000 minus 111 000 minus 111 000 minus 111. So, to correct, first correct inner code. The inner code is just the bit error-correcting code, which means if there is a bit error, we correct it. Let us say we have a $ \sigma $ x error, what happens on, let us say, the fifth qubit? we know we had 000, we had the first of these. Then 010 plus or minus 101 and then the third of these, depending on whether the plus or minus depending on whether we have a logical 0 or logical 1. we know how to correct this. we asked which bit is different, and then we flip that bit. $ \sigma $ z error 5 error, 000 plus 111, so, these are we want to say that the fourth, fifth, and sixth bits go to 000 minus 111 on the fourth, fifth, sixth bits. Well, this is a logical 0, and this is a logical 1. So, really it is a logical error in the code. Well, not really, but we mean let us say this is 0 x and this is 1 x. So, we can think of a code that encodes a 0 like this and encodes a 1 like this. Now, what we have done is we have x goes to 0 x, 1 x, 0 x. Because we took the middle of these three triplets of qubits and we took it to its other thing, and if we call this 0 x and 1 x, this is just a bit error on the middle qubit. So, we can correct it. we correct it exactly the same way. we compare this state, this state, and this state and take the majority of them whether they have a plus or minus. we can write down a circuit for that, but we are not going to. Because it is rather complicated, and we not entirely sure how illuminating it is. However, we will make us do it on the homework. So, that is the nine-qubit code. So, so, far we have shown that the nine-qubit code can correct a $ \sigma $ x error, and it can correct the $ \sigma $ z error., in fact, it can correct a $ \sigma $ y error, because a $ \sigma $ y error is just a simultaneous $ \sigma $ x error and $ \sigma $ z error. the $ \sigma $ x error-correcting the $ \sigma $ x error here does not interfere with the $ \sigma $ z error, so, we can correct that later. So, the nine-qubit code corrects one error of any well, if $ \sigma $ x, $ \sigma $ y, $ \sigma $ z. But, there is lots and lots of errors that are not $ \sigma $ x, $ \sigma $ y, or $ \sigma $ z. we mean, we could apply a Hadamard gate to one of the qubits. Does it correct that as well? Well, there is a theorem. we do not want to get rid of the base code, so, let us erase this. If we can correct identity any code had better be able to correct the no error $ \sigma $ x, $ \sigma $ y, $ \sigma $ z on any qubit. Actually, we should not say the identity that is it is the identity of all the qubits. If we can $ \sigma $ x, $ \sigma $ y, or $ \sigma $ z on any single qubit, we can correct anyone qubit error. we mean, we can take our qubit and just remove it and add a 1 in its place, and that gets corrected. Alternatively, we can make a Hadamard gate on it, or a rotation of any kind, or measure it, and they all get corrected. So, we only need to correct $ \sigma $ x, $ \sigma $ y, sigma z errors. we are going to demonstrate this for the phase code where it is basically it is if we can correct the identity or if we can correct no errors, and if we can correct a phase error on any qubit, we can correct any phase error. we mean, if we can correct $ \sigma $ z on any qubit, we can correct any phase error on any qubit. this is very much related to the fact that this dephasing channel could also be done by making a small phase having a small phase error with probability 1/2 and opposite phase error with probability 1/2. 

\item  The Shor Code I: Implementation; The figure below shows the quantum circuit for the implementation of the Shor error correction code. This code is composed from the bit-flip (orange boxes), $ \sigma _ X $, and phase-flip (blue boxes), $ \sigma _ Z $, error correction codes.
The quantum gate in each red box is called a Toffoli gate. This conditional gate applies a $ \sigma _ X $ on the target qubit $ (\oplus) $ when both control qubits (solid black dots) are in the $ \left\vert 1\right\rangle $ state, and it leaves the state the same otherwise.

\begin{figure}[H] \centering{\includegraphics[scale=1.25]{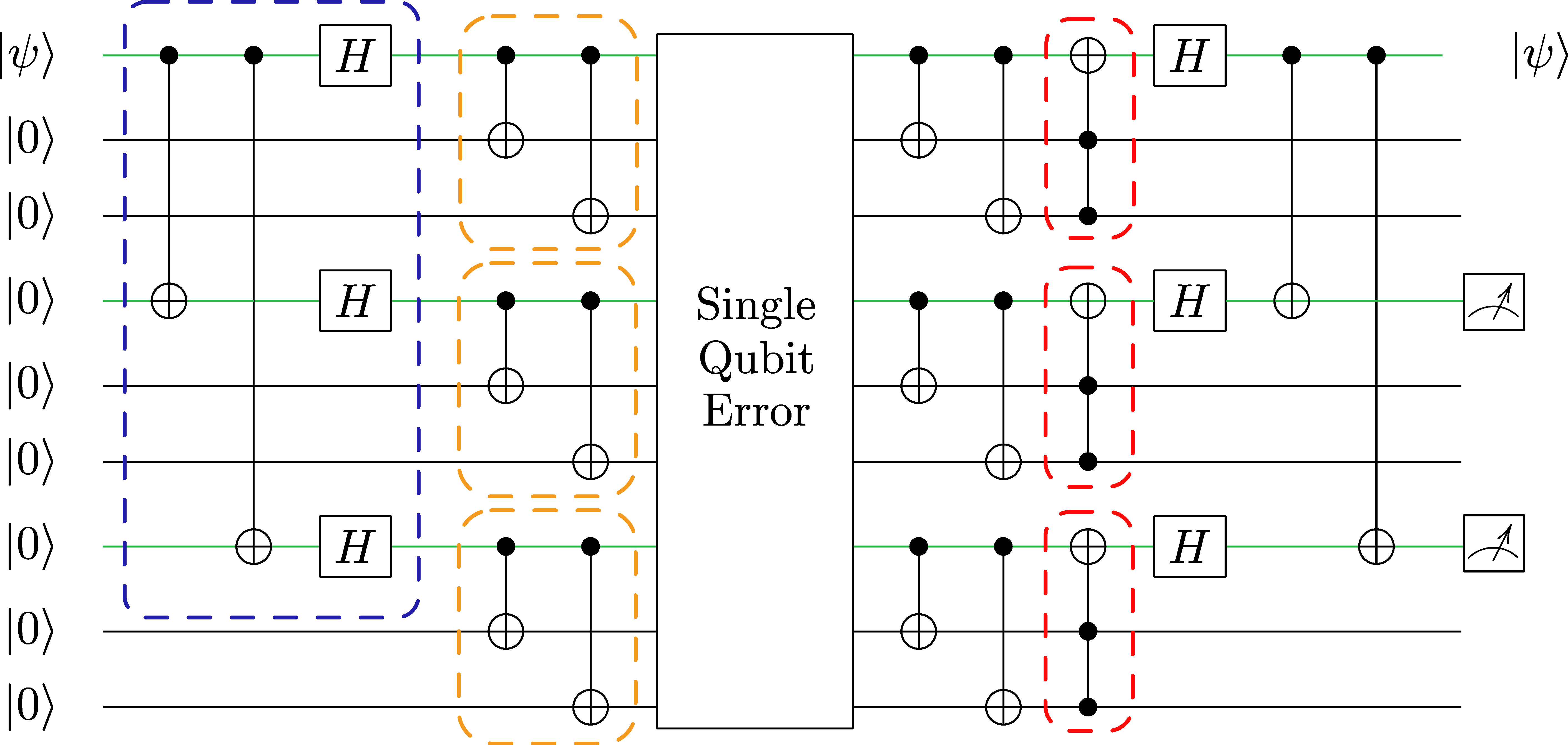}}\caption{Shor code}\label{fig4_5}
\end{figure}

Regarding the Shor quantum error correction code.
\begin{itemize}    
\item The code can correct at most one bit-flip error.
\item The code can correct a bit-flip and a phase-flip error, even if they occur on different qubits.
\item The circuit at the left of the error encodes $ \left\vert 0\right\rangle_1 $ into 
\begin{center}$ \lvert0_L\rangle=\left(\lvert000\rangle_{123}+\lvert111\rangle_{123}\right)\left(\lvert000\rangle_{456} +\lvert111\rangle_{456}\right)\left(\lvert000\rangle_{789}+\lvert111\rangle_{789}\right)/\sqrt{8}) $ \end{center}
\item Although the code can correct for single bit-flip errors occurring simultaneously on different blocks, [1,2,3], [4,5,6], and [7,8,9], it cannot correct simultaneous phase-flip errors on different blocks.
\end{itemize}
Solution:\\
The Shor code can correct both bit-flip and phase-flip errors. In fact, it can correct one bit-flip error in each one of the three blocks [1,2,3], [4,5,6], and [7,8,9].

\section{Computational Capacity: Communication Over Noise Wires}
 
This computational capacity idea that we will describe sets the stage for describing how we can build a classical computer out of noisy components. We will describe that in terms of a concept known as a threshold. Then we will describe the principles which make this possible for classical computation. Then we will try to generalize that to the quantum case. We want to begin this journey with us by describing a computational capacity. The first idea we want to share with us is capacity, which came out in the field of communication. This is about communication channels. It is a seminal result by Shannon, Claude Shannon, from the 1940s. The scenario was that one had a communication link between two parties, but the communication link was noisy. So, the question to ask is, what is the maximum rate at which the two parties can communicate with each other despite the noise on this communication channel? Moreover, it is a fascinating idea of this wire plus noise. We ask ourselves, at what rate can we communicate? So, let us think of this as the probability of error. It goes from 0 to 1 as a function of the rate at which we communicate over this wire. Think of this as being bits per second. We can use any other kind of measure of rate. We can imagine that if we start to speak faster and faster and faster and faster, then we probably cannot understand. Except we cannot talk very fast. However, we can still understand this way, yes? All right. However, there are others here who can speak much faster. The idea is that as we speak faster, the probability error goes up. We can believe that. We know, that means that if we speak at an intermediate rate, some of us will understand, and some of us will not. There will be a growing probability of error. This is quite reasonable, right? Moreover, what Shannon proved and showed the community that actually we can do much better than this. We can actually obtain asymptotically exponentially close to zero error for some rate of communication until we reach some magical point, called the capacity of the channel. This location reaching here is exponentially close to zero. This idea is one of the channel capacity. The reason we are able to do this is because of error correction codes. So, as long as the error rate is sufficiently low, we can, for example, repeat a message three times, repeat a message three times, repeat a message three times. As long as we do not speak too fast, if we caught it one of those or two of those three times, we could use, for example, majority voting to figure out what in the heck we meant, even though we were speaking fast. So, this is an idea about a discrete jump in the rate of a channel despite the error. So, this is pre-Shannon. This is Shannon.

\section{Computational Capacity: Computation with Noisy Gates} 
Jon von Neumann asked, what would happen if we ask the same kind of question, not about communication, but now about computation? What if we have gates plus noise? Moreover, von Neumann asked this in the 1950s. He has this beautiful paper that we really find amazing. We recommend that we read this article by him titled Probabilistic Logics and the Synthesis of. This is fascinating because von Neumann, at this time, he was working with vacuum tube classical computers. People did not yet know if he could build a large-scale computer, classical computer. So, his model for all of this was not the fact that we had computers, but rather the fact that we had biology. Biological organisms like ourselves, who are composed of, we know rather fallible pieces. We know we are, at least. However, we can largely talk and walk and communicate with us fairly reliably despite being built out of unreliable parts. So, von Neumann asked about the probabilistic logics and the Synthesis of reliable organisms from unreliable components. Could we use whatever biology does to make reliability emerge out of unreliable parts and do the same out of, in his case, unreliable vacuum tubes? Furthermore, we think this is a fascinating question? We can draw the same kind of graph. Now, this is the probability of error of a circuit as a function of the probability of error of a component. As the probability of error of the components grows, and we are not going to give us any reliable components, for example. We would like us, nevertheless, to build a reliable system despite everything in it being unreliable—the gates,  even the wires, but at least the gates. Every single gate in this construct, we are only allowed to use it with some failure probability. What von Neumann showed and argued is that we can have the same kind of capacity argument that we see over here. So, without doing any kind of strategic work, like error correction, we get a probability of error of a circuit, which grows as the probability of error of the component grows. However, if we construct our system appropriately, we can actually have something similar to Shannon's capacity. However, here we get a capacity, which is the computational capacity. This is the capacity of a set of noisy gates that compute some function with essentially no noise. It is fascinating. We do not use these ideas in today's computers because, largely, the semiconductors work with too much reliability. So, we do not need to assemble personal computers out of fallible personal computers. Although, we do at the largest scales of cloud computing synthesize reliable systems out of unreliable networks of computers\cite{dumitrescu_cloud_2018}. This is, for example, how Amazon Web Services and the Google Cloud Platform build their infrastructure out \cite{mcclean_openfermion_2019}. They assume that any a piece of it can fail and thus need to make the system reliable despite the system's components being unreliable. However, biology does this at a microscopic scale. Von Neumann asked the same question of the microscopic scale. It turns out that this is an essential idea to quantum computation because of the prevalence of noise at the level of gates in quantum computers. So, this is why we ask this question about fault tolerance. Let us give us a basic idea that von Neumann used. His basic idea was to assume that we have some circuit with the probability of error, and then we repeat it three times, for example. Then we correct. For example, we are using a majority voting gate. For now, let us assume that that is a perfect majority voting gate that does not fail. We can immediately see that the probability of error of this whole circuit is now something along the lines of 1 minus P error square. There are three choose two ways that we could fail and have the majority voting fail. So, this is very reasonable. This only communicates some of the principles behind it, however, because that is a flawed construction. We are not going to allow us a perfect component like a majority voting gate. Nevertheless, it gives us the idea that by using triple modular redundancy, we can start to increase the reliability of a system. So, issues, single-point failures like this one, we have to deal with that. We cannot allow such things to happen. This scheme also has no threshold. That is, there is not a single probability of failure of the gates, which we can say, would allow the system to scale to an arbitrary size. The idea here is that we might say that we want 1 minus P error square times 3 to be less than P error. Then we might say that this means something like P error is less than one third. That is a failure probability. Not 1 minus. So, this means that we have a reduction in error as long as the gate error probability is less than one third. However, then if we want to increase the size, it seems like that amount, that boundary continues to change. So, we are going to ask for something more. That thing will be what we will call a threshold. 

\item  Bit-Flip Code II: Before the Error; The figure below shows the quantum circuit for the implementation of the bit-flip quantum error correction code. What is the role of the controlled-NOT (CNOT) gates to the left of the error box (before the error occurs)?

\begin{figure}[H] \centering{\includegraphics[scale=2.4]{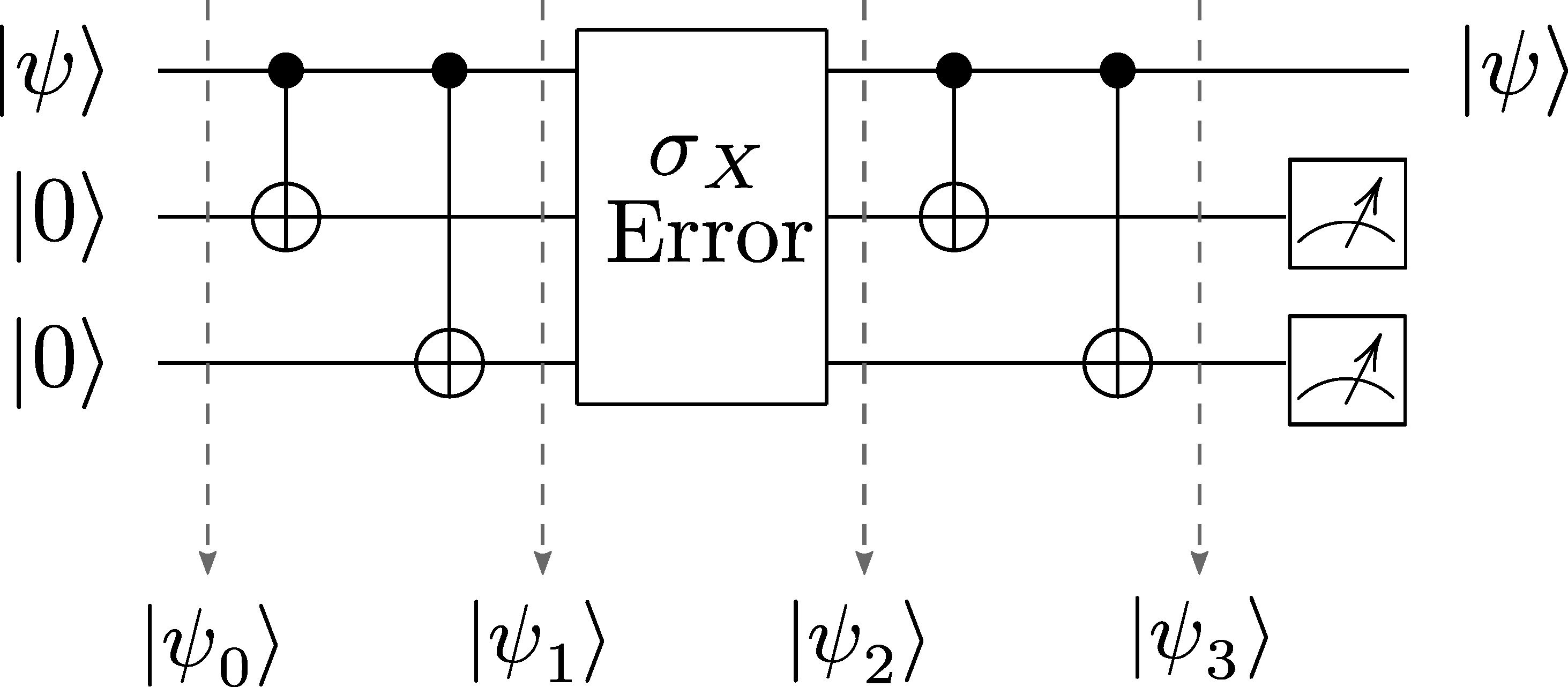}}\caption{Quantum circuit for a three-qubit bit-flip error correction code}\label{fig4_8}
\end{figure}

\begin{itemize}    
\item To entangle the first qubit with the second and third qubit.
\item To encode the single-qubit state $ \lvert 0\rangle $ in the three-qubit state $ \lvert 0\rangle \lvert 0\rangle \lvert 0\rangle $.
\item To encode the single-qubit state $ \lvert 1\rangle $ in the three-qubit state $ \lvert 1\rangle \lvert 1\rangle \lvert 1\rangle $.
\item All of the above
\end{itemize}

\item  Phase-Flip Code II: Before the error; The figure below shows the quantum circuit for implementation of the phase-flip quantum error correction code. What is the role of the three Hadamard gates to the left of the error box?
\begin{figure}[H] \centering{\includegraphics[scale=1.8]{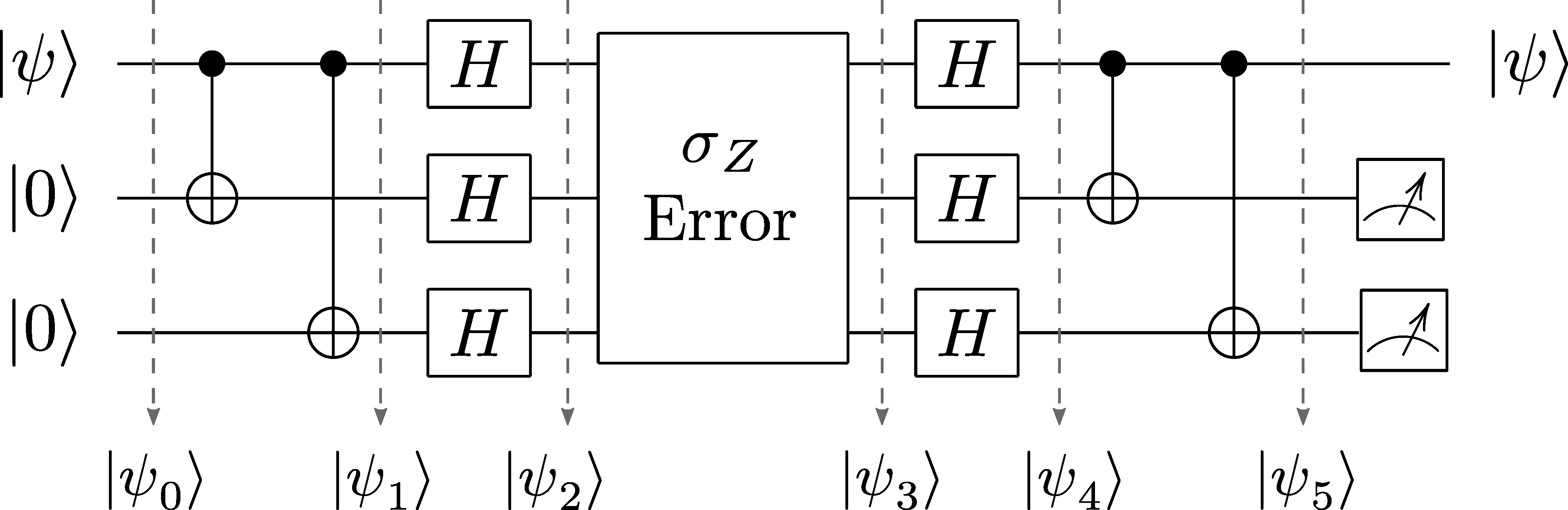}}\caption{Quantum circuit for a three-qubit phase-flip error correction code}\label{fig4_9}
\end{figure}

\begin{itemize}    
\item To entangle the first qubit with the second and third qubit.
\item To encode the single qubit state $ \lvert \psi \rangle $ in the three-qubit state $ \lvert 0\rangle \lvert 0\rangle \lvert 0\rangle $.
\item To encode the single qubit state $ \lvert \psi \rangle $ in the three-qubit state $ \lvert 1\rangle \lvert 1\rangle \lvert 1\rangle $.
\item To change the basis of each qubit from \\$ \left\{ \left\vert 0\right\rangle ,\left\vert 1\right\rangle \right\} $ to $ \left\{ \left(\left\vert 0\right\rangle +\left\vert 1\right\rangle \right)/\sqrt{2} ,\left(\left\vert 0\right\rangle -\left\vert 1\right\rangle \right)/\sqrt{2} \right\}. $
\end{itemize}
Solution:\\
A Hadamard gate is a single-qubit gate and therefore does not entangle two qubits. This also means that it can't encode a single-qubit state into a three-qubit entangled state.

Rather, the Hadamard gates perform a basis transformation. This transformation, in conjunction with the Hadamard gates that follow the error block and transform back to the original basis, enables phase flip errors to be treated in the same way as bit-flip errors.

\item  Relationship Between Bit-Flips and Phase-Flips; The bit-flip and phase-flip error correction codes require coupling two ancillary qubits to the system. Given the state of interest
\begin{equation}\label{eq4_28}
\left\vert \psi \right\rangle =\alpha \left\vert 0\right\rangle +\beta \left\vert 1\right\rangle ,
\end{equation}
the ancillary qubits are initialized in the $ \left\vert 0\right\rangle $ state, such that the initial three-qubit state is given by
\begin{equation}\label{eq4_29}
\displaystyle \left\vert \psi _{0}\right\rangle    \displaystyle =    \displaystyle \left\vert \psi \right\rangle \left\vert 0\right\rangle \left\vert 0\right\rangle ,          
\displaystyle =    \displaystyle \alpha \left\vert 0\right\rangle \left\vert 0\right\rangle \left\vert 0\right\rangle +\beta \left\vert 1\right\rangle \left\vert 0\right\rangle \left\vert 0\right\rangle
\end{equation}
          
This is done to find a possible error in the first qubit, given the measurement results in the ancillary qubits. Depending on the results of the measurement, one can know if there was an error and how to correct it without losing the quantum information represented by state $ \left\vert \psi \right\rangle $. These error correction codes can be divided into three sections.

\begin{figure}[H] \centering{\includegraphics[scale=1.8]{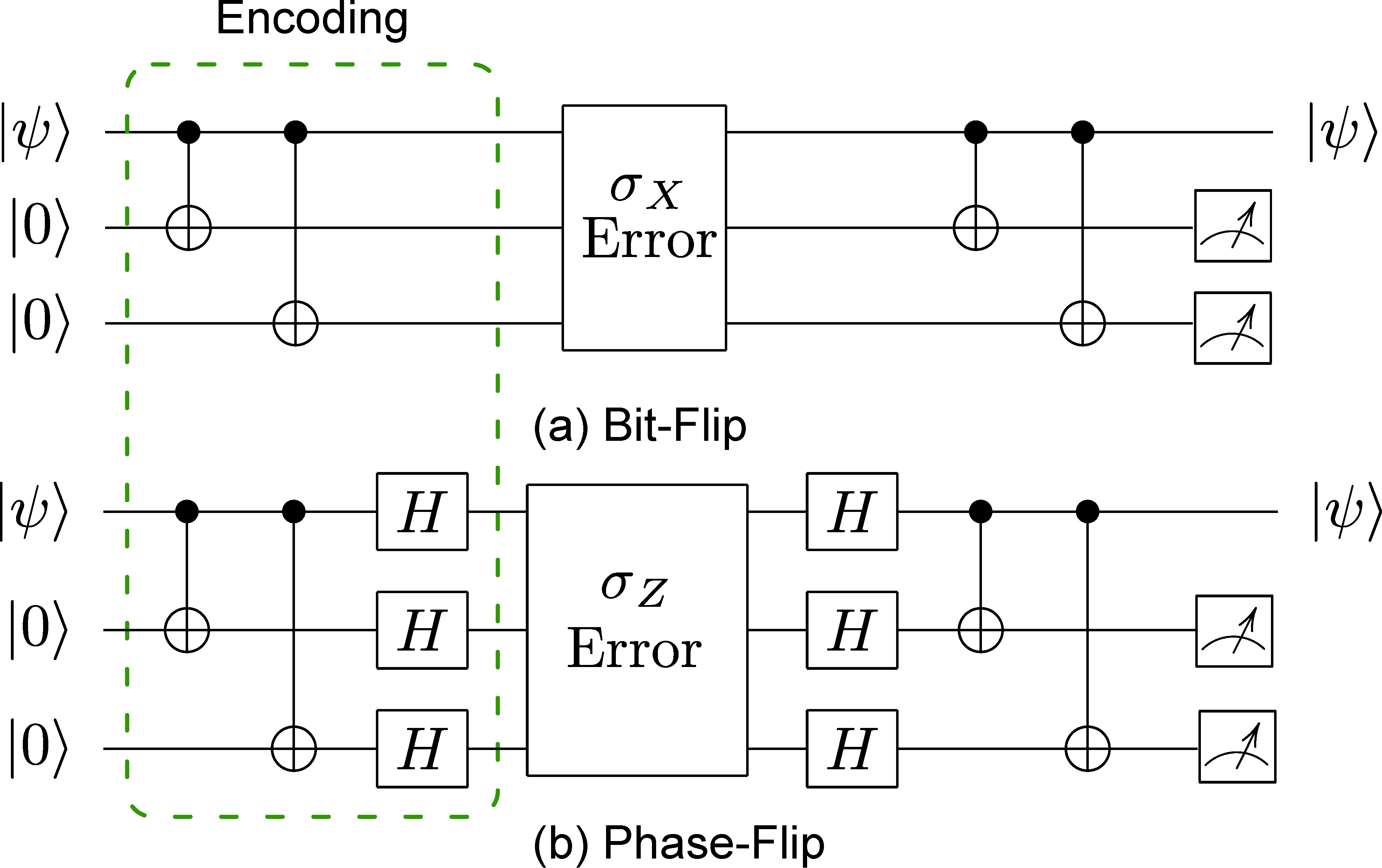}}\caption{Relation}\label{fig4_10}
\end{figure}

The figure shows the general quantum circuit for the bit-flip (a) and phase-flip (b) error correction codes. In figure (a), the encoding is performed using the first-two CNOT gates. The error box corresponds to a $ \sigma _{X} $ (bit-flip) error., and the recovery is made using two CNOT gates to achieve decoding, followed by ancilla measurement.

In figure (b), the encoding is performed using a combination of CNOT gates and a Hadamard gate on each qubit. The error box corresponds to a $ \sigma _{Z}  $(phase-flip) error., and the recovery is also made using Hadamard and CNOT gates to achieve decoding, followed by ancilla measurement.

A bit-flip error can be mathematically described by
\begin{equation}\label{eq4_30}
\displaystyle \sigma _{X}\left\vert 0\right\rangle    \displaystyle =    \displaystyle \left\vert 1\right\rangle ,          
\displaystyle \sigma _{X}\left\vert 1\right\rangle    \displaystyle =    \displaystyle \left\vert 0\right\rangle ,
\end{equation}
          
with $ \left\langle 0\right. \left\vert 1\right\rangle =0 $. The encoding takes the state $ \left\vert 0\right\rangle $ to $ \left\vert 0_{L}\right\rangle \equiv \left\vert 0\right\rangle \left\vert 0\right\rangle \left\vert 0\right\rangle $ and $ \left\vert 1\right\rangle $ to $ \left\vert 1_{L}\right\rangle \equiv \left\vert 1\right\rangle \left\vert 1\right\rangle \left\vert 1\right\rangle $.

A phase-flip error can be mathematically described by
\begin{equation}\label{eq4_31}
\displaystyle \sigma _{Z}\left\vert +\right\rangle    \displaystyle =    \displaystyle \left\vert -\right\rangle ,\\          
\displaystyle \sigma _{Z}\left\vert -\right\rangle    \displaystyle =    \displaystyle \left\vert +\right\rangle ,
\end{equation}     
with
\begin{equation}\label{eq4_32}
\left\vert \pm \right\rangle =\frac{\left\vert +\right\rangle \pm \left\vert -\right\rangle }{\sqrt{2}},
\end{equation}
and 
$ \left\langle +\right. \left\vert -\right\rangle =0. $ The encoding takes the state $ \left\vert 0\right\rangle $ to $ \left\vert 0_{L}\right\rangle \equiv \left\vert +\right\rangle \left\vert +\right\rangle \left\vert +\right\rangle $ and $ \left\vert 1\right\rangle $ to $ \left\vert 1_{L}\right\rangle \equiv \left\vert -\right\rangle \left\vert -\right\rangle \left\vert -\right\rangle $.

What is the relation between the bit-flip and phase-flip error correction codes?

\begin{itemize}
\item The phase-flip code is equivalent to the bit-flip code, since they both correct $ \sigma _{X} $ type of errors.

\item The phase-flip code is equivalent to the bit-flip code since they both correct $ \sigma _{Z} $ type of errors.

\item The phase-flip code is equivalent to the bit-flip code, since they both map
\begin{center}
$ \displaystyle \left\vert 0\right\rangle    \displaystyle \rightarrow    \displaystyle \left\vert 0_{L}\right\rangle \equiv \left\vert 0\right\rangle \left\vert 0\right\rangle \left\vert 0\right\rangle $ \\          
$ \displaystyle \left\vert 1\right\rangle    \displaystyle \rightarrow    \displaystyle \left\vert 1_{L}\right\rangle \equiv \left\vert 1\right\rangle \left\vert 1\right\rangle \left\vert 1\right\rangle $
\end{center}
         
\item The phase-flip code is equivalent to the bit-flip code, since they both map
\begin{center}
$ \displaystyle \left\vert 0\right\rangle    \displaystyle \rightarrow    \displaystyle \left\vert 0_{L}\right\rangle \equiv \left\vert e_{0}\right\rangle \left\vert e_{0}\right\rangle \left\vert e_{0}\right\rangle $ \\          
$ \displaystyle \left\vert 1\right\rangle    \displaystyle \rightarrow    \displaystyle \left\vert 1_{L}\right\rangle \equiv \left\vert e_{1}\right\rangle \left\vert e_{1}\right\rangle \left\vert e_{1}\right\rangle $
\end{center}
where $ \left\{ \left\vert e_{0}\right\rangle ,\left\vert e_{1}\right\rangle \right\} $ is a single-qubit basis $ \left\langle e_{0}\right. \left\vert e_{1}\right\rangle =0, $ where the basis is orthogonal to the direction of the error.
\end{itemize}

Solution:\\

The bit-flip and phase-flip codes map the $ \left\vert 0\right\rangle $ and $ \left\vert 1\right\rangle $ components of $ \left\vert \psi \right\rangle $ in a similar manner. Specifically, in both codes, the encoding part of the circuits maps the single qubit states $ \left\vert 0\right\rangle $ and $ \left\vert 1\right\rangle $ to a three-qubit logical $ \left\vert 0_ L\right\rangle $ and $ \left\vert 1_ L\right\rangle $,
\begin{center}
$ \displaystyle \left\vert 0\right\rangle    \displaystyle \rightarrow    \displaystyle \left\vert 0_{L}\right\rangle $ \\          
$ \displaystyle \left\vert 1\right\rangle    \displaystyle \rightarrow    \displaystyle \left\vert 1_{L}\right\rangle $
\end{center}          
Therefore, the difference between both codes is given by the basis in which the logical states $ \left\vert 0_{L}\right\rangle $ and $ \left\vert 1_{L}\right\rangle $ are defined. This is summarized on the table below.

\begin{table}[H]
\centering
\caption{The phase-flip code equivalence to the bit-flip code}
\label{tab:4_1:Table 4}
\begin{tabular}{|c|c|c|c|}\hline
Error type & Initial State & Encoding Basis & Encoded State \\\hline    
$ \sigma _{X} $ & $ \left\vert \psi \right\rangle \left\vert 0\right\rangle \left\vert 0\right\rangle $ & $ \left\{ \left\vert 0\right\rangle ,\left\vert 1\right\rangle \right\} $ & $ \alpha \left\vert 0\right\rangle \left\vert 0\right\rangle \left\vert 0\right\rangle +\beta \left\vert 1\right\rangle \left\vert 0\right\rangle \left\vert 0\right\rangle $ \\ \hline
$ \sigma _{Z} $ & $ \left\vert \psi \right\rangle \left\vert 0\right\rangle \left\vert 0\right\rangle $ & $ \left\{ \left\vert +\right\rangle ,\left\vert -\right\rangle \right\} $ & $ \alpha \left\vert +\right\rangle \left\vert +\right\rangle \left\vert +\right\rangle +\beta \left\vert -\right\rangle \left\vert -\right\rangle \left\vert -\right\rangle $ \\ \hline
\end{tabular}
\end{table}
Note that the direction of the error (X and Z) is orthogonal to the basis for the encoded state:$  0/1  $(Z-axis) and $ +/- $ (X-axis), respectively.

\item  The Shor Code II: Implementation; The Shor quantum error correction code can correct
\begin{itemize}
\item Single-qubit errors of the type $ \sigma _ X $
\item Single-qubit errors of the type $ \sigma _ Y $
\item Single-qubit errors of the type $ \sigma _ Z $
\item All of the above
\end{itemize}
Solution:\\
The Shor quantum error correction code is a combination of the bit-flip ($ \sigma _ X $) and phase-flip ($ \sigma _ Z $) error-correction codes. Since $ \sigma _ Y=\sigma _ Z\sigma _ X $, the Shor code can correct any single-qubit error, provided there is at most one error.
    
\item  The Shor Code III: The Role of Each Qubit; The figure below shows the quantum circuit for the implementation of the Shor error correction code. This code is composed from the bit-flip ($ \sigma _ X $) and phase-flip ($ \sigma _ Z $) codes. The ``double wires'' inside the red boxes represent the classical communication of the measurement results. The $ \sigma _ X $ gate only is applied when both measurements give 1 as a result.

\begin{figure}[H] \centering{\includegraphics[scale=1.2]{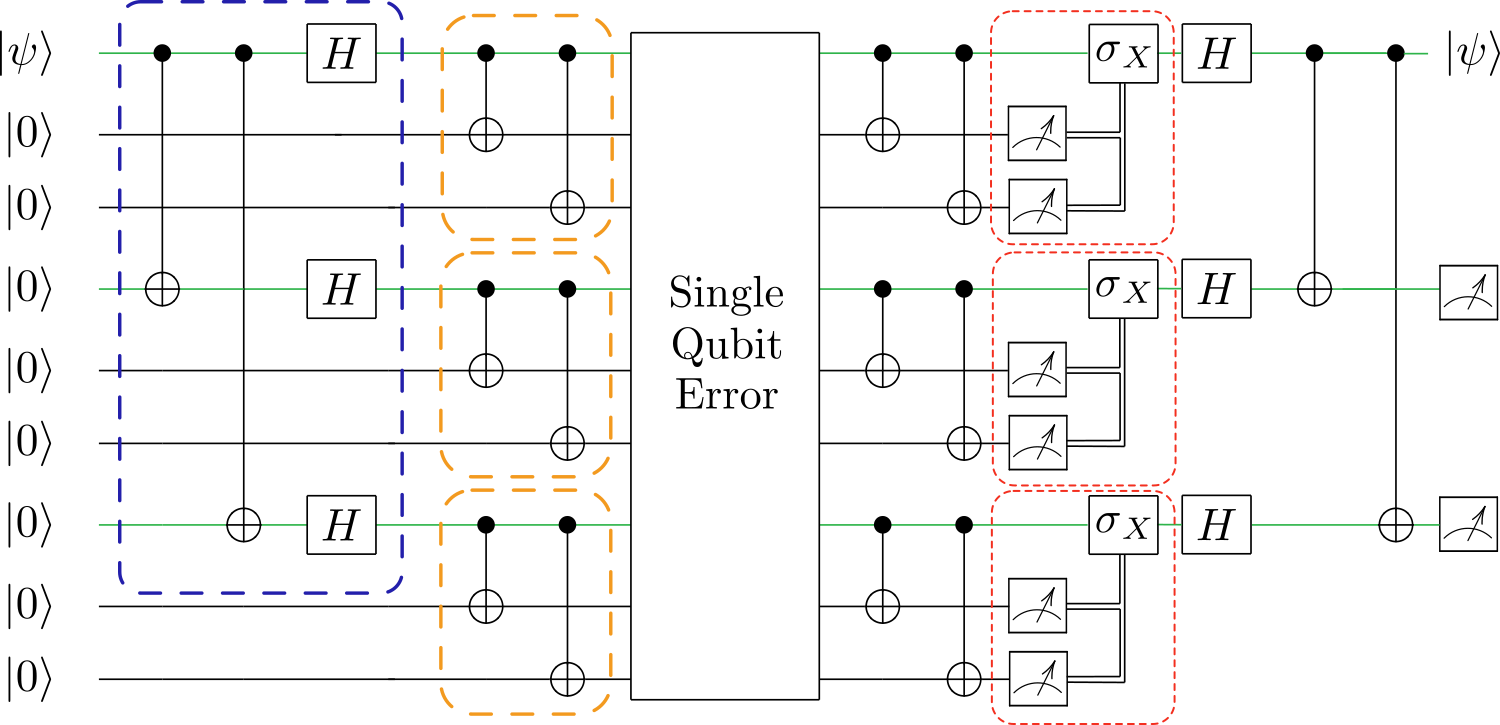}}\caption{Quantum circuit for a nine-qubit Shor error correction code}\label{fig4_11}
\end{figure}

How many measurements need to be done to correct an arbitrary error (X, Y, or Z) on state $ \left\vert \psi \right\rangle $?
\begin{itemize}    
\item Two measurements, qubits four and seven.
\item Three measurements, qubits one, four, and seven.
\item Six measurements, qubits two, three, five, six, eight, and nine.
\item Eight measurements, qubits two, three, four, five, six, seven, eight, and nine.
\end{itemize}
Solution:\\
Qubits four and seven are measured to detect if there was a phase-flip error in the first qubit. Qubits two and three, five and six, and eight and nine are measured to detect if there was a bit-flip error in the first qubit. To detect an arbitrary single-qubit error in the first qubit, all the ancillary qubits have to measure.

\section{How Can Reliable Classical and Quantum Machines be Built from Unreliable Components? Introduction}
In order to be useful, large-scale computers, whether classical computers or quantum computers need to operate reliably for extended periods of time in the presence of noise. However, we know that quantum computers are built from qubits and that qubits are highly susceptible to noise. Now in section three, we considered noisy intermediate-scale quantum algorithms that would operate on uncorrected systems of qubits, basically until too many qubits fail. Although many factors will contribute to the realizable circuit depth for such algorithms, ultimately, uncorrected quantum computers will be limited by the quality of the individual qubits from which they are built. This means their coherence times, the rate at which gates can be applied, and, most importantly, their gate fidelity. In contrast, for large-scale quantum computers addressing larger-scale problems, we will need to run algorithms for much longer than the lifetimes of any individual physical qubits. To extend the operational time of quantum computers built from such intrinsically faulty devices, we will need to implement quantum error correction. In section one, we discussed at several simple examples of error-correcting codes that can improve quantum system reliability through the introduction of qubit redundancy and overhead. We discuss that if qubits individually are good enough, then one can improve the overall system performance by encoding quantum information in larger numbers of such qubits. That is, by adding more qubits that are good enough, the overall error rate can be decreased. The question is, how do we determine the threshold for good enough? Furthermore, once we have such qubits, how can we devise an error-correcting code that can handle errors that will certainly at some point occur, no matter where or when they occur, even when those errors might happen during the error detection protocol itself? In this section, we will begin to address these and related questions by discussing at the threshold theorem and the principles of fault tolerance for both classical and quantum computers.
 
\section{ Threshold Theorem for Fault-Tolerant Computation, Threshold Theorem: Introduction}
What we are going to do first is to describe to us the threshold for classical computation. We will take von Neumann's core idea over here and try to explicitly tease out the ingredients that go into the fault tolerance of a classical computing system. Let us begin this journey by describing to us a theorem, which is going to define what we mean by a threshold for fault tolerance. This is a modern version of what von Neumann had described, and it will characterize what we wish the threshold to behave like in contrast to that argument given up there with the majority voting. What we will begin with is a family of circuits. For technical reasons, we will call this a uniform family. We will begin with a circuit that we wish to simulate. This circuit will have size n, so it will have error-free gates constructed in any which way we want. This describes the goal of the construction. So, we have a size, and we know that it is error-free. The theorem will state that we may simulate this error-free goal circuit with a certain probability of error bounded above by epsilon. Then, we need to say what it takes to simulate this, with what properties, what constraints. First, we will tell us how many gates are needed. That will be the order of poly log N over epsilon times N. This is an expression which means that the size of the simulating circuit and these are going to be gates which have some error, which we will describe in a moment must grow only logarithmically as 1 over the error desired. It must grow as the size of the circuit N. So, and this is very reasonable, this is actually a very aggressive scaling. This is very reasonable that it is proportional to N. then, the main idea of this threshold theorem is that the simulation is done with this number of gates. They are error-prone gates. They fail with some probability, P. the theorem only holds as long as this error probability P is less than some other value p threshold, Pth. We ask that this threshold is independent of N and epsilon. So, we want a capacity here that works for arbitrary-sized circuits. That is the independence of N., And we want it not to be a function of the error goal that we are discussing for. So, that means it is arbitrarily accurate in this sense. 

\item  Threshold Theorem I; Given a uniform circuit composed of N error-free gates, the Threshold Theorem stipulates that this circuit can be simulated by with an error probability less than $ \epsilon $ and using
\begin{equation}\label{eq4_35}
\mathcal{O} \left({{\mathsf{poly}}}\left(\log \frac{N}{\epsilon }\right)N\right)
\end{equation}
gates, with each gate having the same error probability P. This is valid as long as the probability P is smaller than a threshold probability $ P_{th} $.

The threshold probability $ P_{th} $ is a function that depends on:
\begin{itemize}    
\item The number of gates that composed the circuit, N.
\item The maximum error probability to simulate the error-free circuit, $ \epsilon $.
\item The error probability of the gates, P.
\end{itemize}
Solution:\\
The threshold probability  $ P_{th} $  is independent of the size of the circuit, implying that the circuit is scalable. It is also independent of the maximum error  $ \epsilon $  for which the error-free circuit can be simulated. Since the threshold probability sets a bound on how large the error probability  P  can be, while still being able to simulate the ideal circuit, it should not be a function of P.

\section{ Threshold Theorem: Proof Sketch}
This is a sketch of the proof. We are not going to prove it rigorously. However, we will get the main idea here. We will do this on classical circuits and classical gates. The main idea is to compute on encoded data. Do not decode the data as we are going along. How do we do that? Well, we give us a bit of an idea over here with this triple modular redundancy. Let us formalize that a little bit by constructing this notation. This would be a NAND gate where the output will be x and y, barred, inverted. Then another gate, which we will call the majority voting gate. The output of this will be the majority of the inputs. Each one of these gates fails with probability P. we will assume, for simplicity, that they are all identical probabilities. We can clearly work this out in the case where there are non-identical probabilities of failure. The idea for the construction is to encode zero as three zeroes and one as three ones. So, we have this notion of the physical states and the logical states. Logical zero and logical one will be just the triple redundancy code. Now we want to perform a NAND gate operation on the logical states. However, what we will actually have coming in we are going to switch to A and B from x and y so that we can use x and y for some other things later is that we will have three wires representing A coming in and three wires representing B coming in. We will have three NAND gates. Each one of the NAND gates takes two inputs. We will route them like this. The output of this, unfortunately, may have an error because there may be errors in the input. The way we correct for them is instead of doing a single majority voting gate as we have up there, we will have three majority voting gates. These three let us assume that we can now copy information from one wire to another. We will take inputs copied from each of the outputs of the NAND gates. Then we will have three outputs at the end. We claim that this output will be A and B inverted, where if there had been a single error in any of the inputs or if there had been a single gate failure, then we would have corrected that because of the majority voting. We can guarantee that there is, at most, one error in the output. Thus, the probability of failure of the output is going to be lower than P. What we have is a construction that has two parts. This first part is like a logical gate, a gate acting on the encoded data. The second part here is the error correction. We want to compute the probability of failure. We want to compute the probability of failure of this whole circuit construction. So, the output is incorrect only if two or more of the gates, the Ns and Ms fail. For simplicity, we are going to omit the discussion about the input errors from this. However, that also can be done. Let us assume what happens to analyze what happens when gates fail. So, if two or more gates fail, we can see, and this is a bound. So, if two gates happen to be good gates that fail, adding the same error, then the output is not going to be wrong. However, we can see that there are many ways in which, if two gates fail, the output definitely is wrong. Thus, let us bound that. The number of ways that two gates can fail is 6 choose 2 because there are six gates in this circuit. This is the number of possibilities. This is 6 times 5 divided by 2, which is 15. Thus, the probability of failure of a circuit is going to go like 15 P-squared. if the goal is to make this less than P, then we find that we could say that the threshold probability is 1 over 15. that is kind of remarkable because 1 over 15 is already better than, for example, one third. Now, there are some concerns about this. Again, we have to show us that we can make an arbitrarily sized circuit. Thus, far, we have done what we will call a level 1 encoding. So, the main idea that we are going to show we next is how we can recurse on this encoding and build a fault-tolerant circuit of arbitrary size. However, keep in mind that this argument about recursion is just one use for the proof. There are many other techniques by which we, in principle, could get to fault tolerance and to prove the theorem.

\item  Failure Probabilities; Consider a circuit comprising four faulty gates, each with a probability P of an error. The failure probability of the circuit will depend on the number of fault paths and the number of gate errors. Calculate the failure probability  to leading order in P  for each of the following cases:
(a) Exactly one of the four gates has an error.
(b) Exactly two of the four gates have an error.
(c) Exactly three of the four gates have an error.
The answers below are listed in order (a), (b), and (c).
\begin{itemize}
\item Case (a): P, case (b): $ P^2 $, and case (c) $ P^3 $.
\item Case (a): P, case (b): $ 15P^2 $, case (c)and $ 20P^3 $.
\item Case (a): 4P, case (b): $ 6P^2 $, and case (c): $ 4P^3 $.
\item Case (a): 4P, case (b): $ 15P^2 $, and case (c): $ 20P^3 $.
\end{itemize}
Solution:\\
Given a circuit composed of N gates, each with an error probability P, the probability that exactly j gates fail to leading order in P is given by:\\
$ P_{\text {fail}}=
\text {(number of fault paths)}\times P^{j} $,\\
where $ p^j (1-P)^{N-j}\approx P^j $ is the leading-order probability that exactly j gates fail, and the number of fault paths is\\
$ \binom {N}{j}=\frac{N\, !}{j\, !\, (N-j)\, !}. $\\
If the circuit has exactly one gate error, $ j=1 $, its failure probability is\\
$ P_{\text {fail}}=\frac{4\, !}{1\, !\, (4-1)\, !}\times P^{1}=4P $.
If the circuit has exactly two gate errors, $ j=2 $, its failure probability is\\
$ P_{\text {fail}}=\frac{4\, !}{2\, !\, (4-2)\, !}\times P^{2}=6P^2 $. \\
If the circuit has exactly three gate errors, j=3, its failure probability is\\
$ P_{\text {fail}}=\frac{4\, !}{3\, !\, (4-3)\, !}\times P^{3}=4P^3.$

\section{ Threshold Theorem: Recursion} 
So, let us recurse and see what happens if we do the same thing now recursively. What we want to do is to repeat the encoding now and let 0 be mapped to 000 000 000 and 1 be mapped to nine 1's. There are three groups in this. So, we can actually see what we have done is repeat and encode 000 by repeating that three times. Thus, this recursive step that we about to show we could be repeated again and again. Thus, we will get a system, which shows what happens when we scale to a large number of gates. This construction is very simple to imagine and is a little tedious to draw. So, what we will have is this will be A, and this will be B coming in. We will have three big blocks here. In each one of these three big blocks, we are going to have the same construction with six gates that we just showed us on the other board. Each one of these will output three wires and take six wires in. So, the way this works is we get part of A coming in over here, over here, over here, over here, over here, and so, forth. now, this is the encoded gate. We will subsequently also perform this is three blocks of Ms., And this is the error correction step. So, just like the construction that we discuss over there, what we are now going to do is to copy these three over in the appropriate ways. So, this one goes here, and it also goes here, it also goes here. This one goes here, and it also goes here. It also goes here, and so, on and so forth. Here is where we start to scribble. However, we get the idea. What we are going to get out is 123456789. these nine outputs are going to give we, in total, A and B negated. Good. Thus, this is a recursive structure. What we want to do is to calculate the probability of failure of this recursive structure. We want to write this formula out in a nice way for us. So, we can see that the output is incorrect only if two blocks of gates fail. So, this is the identical language that we used in the previous construction. There are 6 choose 2 ways of having blocks fail. Thus, the probability of failure goes as 15 times 15 P-squared quantity squared. However, let us write that in a nicer way. Let us get, now, a recursive formula for the error probability of this cascaded, concatenated data sequence. What we have up there is that Pfail goes as 15 P-squared squared. It is nice to write this as 15P to the fourth power, divide by 15. Because we claim that if we continue this recursion, then we can expand on this pattern that we have. At level one, we had that Pfail times C and if we define C as being the number of fault paths. In this case, above, it is 15. So, we have C times Pfail. This is the number C, which we moved over to the left-hand side is equal to C times P-squared. At level two, C Pfail is equal to CP to the 2 to the 2. At level three, if we kept on doing the same thing, we would find that CPfail is equal to C times P to the 2 to the 3. So, it would go square, fourth power, and eighth power, and then 16, 64. the nice thing about this is that we can write down a formula for the extent of how this goes at the case-level recursion. C times Pfail will be C P to the two to the k power. The idea of fault tolerance is that this is growing exponentially, or doubly exponentially, and thus, as long as P is less than 1 over C, then this probability of failure goes to 0 very rapidly as we increase the level of recursion doubly exponentially rapidly. This is part of the idea of this computational capacity that it goes so rapidly to zero.

\item  k=1 Level of Recursion; Consider an encoded circuit with only one level of recursion (k=1) that is composed of N faulty gates, each with a probability P. The circuit can correct for one error, and it will fail for two or more errors. To leading order in P, we are therefore concerned with the probability that exactly two gate errors occur. If C is the number of fault paths, and $ P_{th} $ is the threshold probability of the circuit, then
\begin{itemize}
\item The number of fault paths is at least $ C=\binom{N}{2}=\frac{N\,!}{2\,!(N-2)\,!}. $
\item The threshold probability is proportional to the number of fault paths.
\item If P is less than $ P_{th} $, then the net failure probability of the circuit is less than P.
\item The failure probability of the circuit is approximately given by the product of C and $ P^2 $
\end{itemize}

\section{Threshold Theorem: Reliable Classical Computation} 
Let us draw the probability of failure that we get from this formula as a function of the probability of failure of the gates. So, what we have is Pfail of the circuit as a function of the P error probability of the individual gates. If there is no error correction, it is just a straight line. If we have one level of recursion, it is this parabola that we are 15 p-squared. then we get a quartic and then the fourth power and so forth. So, it gets steeper and steeper as we reach this point. That is the threshold, Pth. So, remember, this is going as 1 over C., And this is a threshold which is independent of N and epsilon, exactly as we desire in the theorem that is given above here. This is really a remarkable property to be able to have. So, we have been describing this for classical circuits. We can actually go and make these classical circuit constructions. Let us give us some actual numbers. From von Neumann and his works in the 1950s, we have that the threshold probability is at least 0.073. there is a catch to this because von Neumann used three-input gates. His construction did not quite use all of the same kinds of strong assumptions that we require, such as no gates with zero error probability. However, this is what it largely translates into. Following that, it was quite a long time between then and the next result. This is from Hajck and Weller, from about 1991. they found a different construction, giving 0.167 about one over six. Again, there is a caveat to this. This was also using three in gates. We can see what we have done here is with NAND gates. It would be nice to know what the threshold for NAND gates is because NAND gates are universal for classical computation. We figure that if we are asking this question using three input gates, we are making a larger assumption about it because we have more capability with three input gates. So, in 1998, Evans and Pippenger, they came up with an interesting bound, which is approximately 0.089. Notice how this is worse than the three-input case bound here. However, this was for two input gates. Then similar to that, Evans and Pippenger again, in 2008, proved an interesting upper bound on Pth. This is fascinating. It is on k input. Their expression is one-half minus one over two square roots of k, like this. We can imagine why there is a one half in that expression immediately because we can just guess what the answer is. So, ideally, we would like this to be as high as possible, as low as possible because we would like to be able to have the lowest threshold to have to cross to be, we would like the highest possible error failure rate. Thus, we would like that. Would it not be great if we could compute everything with gates that failed with 75\% probability? So, that is not going to happen. Thus, we can imagine, well, how about a gate that fails with a 50\% probability? Well, then it is going just to output noise all the time. How about 0.499 probability? Well, it is close as k gets large here. So, but nobody can show that this is realizable. All right, so, this is a story with classical fault tolerance. What we want to take away from this message is that there is a process with a set of principles for fault tolerance. We would like to try now to abstract away those principles and try to apply them to quantum computation.

\section{ Principles of Fault Tolerance: Summary} 
What are the principles under which we were able to construct a theorem about fault tolerance? We want to abstract a series of ideas from this construction so that we can see if we can apply them to the quantum circuit case. We contend that one of the main ideas behind this theorem and construction is that we have arrived at this expression for a threshold, which reads as 1 over the number of fault paths. Here, we are defining the ideal fault paths as being the number of ways by which we can combine failure probabilities to make the circuit system fail. We are not defining it formally, but we have gotten the idea. However, one of the key things is that this is the number of fault paths in a universal gate construction. For example, we could not have made just an inverter and then expected to extrapolate and get a threshold probability, because the inverter by itself is not universal. So, we use a NAND gate. We know from classical Boolean algebra that any Boolean function can be constructed out of NAND gates, composed appropriately. So, we need to consider a family of universal gates or at least one gate from which any circuit can be constructed. The second idea is that this has to be in a fault-tolerant procedure\cite{chow_implementing_2014,corcoles_demonstration_2015,kandala_error_2019,colless_computation_2018}. The concept of a fault-tolerant procedure is something that we now want to be able to describe. Let us define this. This is essentially going to be the concept of how we control error propagation and minimize the number of fault paths. Let us reverse this logic. We are going to say a procedure is fault-tolerant if a single component failure, like one of the NAND gates or one of the majority voting gates, must not cascade and cause more errors that we cannot correct. So, we will say that a procedure is fault-tolerant if a single component failure causes at most one error in each encoded block of bits in the output. There are two main ideas here. First, the reason we are focusing on one error is that we are discussing only currently in the discussion at codes correcting one error. We can choose other higher rate codes that can correct for more errors, and then generalize this theorem. The reason we are thinking about each encoded block is that each one of the data blocks is a separate code right now. We could also conceive of encoding all of the data with multiple bits inside one code. That also changes this definition, but we can generalize it again fairly easily. So, for example, in the construction up here, this has one output block. However, we might have a different construction with two output blocks, and we would only want a maximum of one error in each one of those blocks separately. So, this is all about error propagation. 

\item  Fault-Tolerant Procedure; According to the section above, a procedure is a fault-tolerant if: 
\begin{itemize}    
\item It controls error propagation and minimizes the number of fault paths.
\item A failure causes at most one error in each encoded block of bits.
\item There is at most one fault path in which two gates can fail, meaning that the error-correcting code fails.
\item When a failure occurs, it does not cascade into more errors as the calculation continues.
\end{itemize}
Solution:\\
A procedure is a fault-tolerant if a single component (such as a majority gate) failure causes at most one error in each encoded block of bits in the output. In other words, if a failure occurs, it should not cascade into more errors as the calculation continues. The number of fault path sets the threshold probability, but it does not determine if a procedure is fault-tolerant or not.

\section{Principles of Fault Tolerance: Fault-Tolerance and Non-Fault-Tolerant Procedures} 
Let us describe some examples of fault-tolerant procedures and what are non-fault-tolerant procedures. Suppose we want to compute a NOT function. These will be poor ways of doing them, and these will be good or fault-tolerant ways of doing that. So, again, let us imagine that we encode in the triple redundancy code. So, we have a number, A, that comes in, and then, we perform three NOT gates. If we run through a single majority voting gate and then copy the output three times, this is no good because there is a single point of failure. This violates the idea here that a single component failure causes at most one error in the output. So, we found, as we discuss an example, that a good way of doing this fault-tolerantly is to have three majority voting gates, like so, and this, then, is a fault-tolerant procedure. If we have multiple bits do something and now let us start to use some of the quantum computing circuit terminology, and we want to compute a controlled-NOT gate, classically still, but using quantum circuit drawing, this is either 000 or 111 when it is encoded. So, clearly, if we wanted to, we could just perform a controlled-NOT gate by using the top bit because they are all supposed to be the same. Thus, if there were no errors, this would be the perfectly fine controlled-NOT gate between two blocks. However, clearly, if this wire failed, we would have another single-point failure, so it is not a fault-tolerant procedure. Instead, one of the canonical ideas behind fault tolerance is to separate gates off into different bits. This construction is called transversal gate constructions. It just means do our gates bitwise in parallel. This, then, allows errors only to propagate between the bits that are interacting together. We will use the same idea for the quantum circuit constructions that are fault-tolerant.

\item  Single Point of Failure; Which of the following sentences are correct about a ``single point of failure''? 

\begin{itemize}
\item It is a part of the circuit that, when a failure occurs, causes the entire process to fail.
\item It is challenging but possible to create a fault-tolerant procedure with a single point of failure.
\item When designing a fault-tolerant procedure, single-point failures must be avoided.
\end{itemize}
Solution:\\
It is impossible to have a fault-tolerant procedure with a single point of failure, since one error in such a part of the circuit cascades to more errors, causing the process to fail.

\section{ Circuit Size of Fault-Tolerant Procedures} 
One of the key things we have not shown we yet are the cost of fault tolerance. So, remember that the fault tolerance theorem has this order of polylog N over epsilon times N in the expression for the cost of the fault tolerance. Let us try to prove that now. How many erroneous error-prone gates do we require to construct an error-free circuit system? we want the overhead of this construction to be very low polylog 1 over epsilon. we have not shown that yet, but now we can do so, So, we discuss that if we have a circuit size d for the fault-free construction, using this recursive construction that we had to increase the size of the circuit. it goes approximately it has d to the k-th power. So, the circuit size, when we make it a fault-tolerant procedure, grows to d to the k, where k is the number of levels of recursion. Remember we showed we instead of 1 NAND gate, we went to 3, and then from 3 to 9. then after that, it would be 81, and so forth. So, this is the increase in that size that grows exponentially. So, what we need to do is to be able to compute the size of the number of levels of recursion needed. So, if the goal is to simulate an N gate circuit with the probability of failure bounded above by some epsilon remember that is the statement of the theorem then each gate we would like to simulate with the probability of failure N over epsilon divided by N. So, we can split this error probability up and proportion it to each one of the gates so, they will add up at worst using the triangle inequality. thus, if we have, again, let C be the number of fault paths, then we find that the equation we have is C times P fail is CP to the 2 to the k, as the scaling formula for the recursive construction. let us define Pth as 1 over C. Then, we can rewrite this equation in a really neat way. This is the way we like to remember the threshold theorem. P fail over Pth that is just this 1 over C over here is equal to P over Pth to the 2 to the k. think of this as the central equation of fault tolerance, at least using the recursive construction. this allows us to solve for k. So, the plugin that we would like P fail of a gate to be, well, less than epsilon divided by N. then we can plug this in here and solve for k, and we would get that k is equal to let us write this as an exponential 2 to the k is equal to the log of epsilon divided by N Pth divided by the log of P divided by Pth. So, this is fairly straightforward. we do not want through the algebra right here, but we want to show we the implication of having solved the formula here. By the way, there is a reason this is going to work out. The reason is that the circuit size is growing exponentially, but the failure probability of the circuit is going down doubly exponentially. So, this goes down much faster than this grows, even though this exponential cost looks very worrisome. So, we only need a very few levels of recursion, k, in order to be able to kill this error probability sufficiently rapidly with a small-sized circuit. So, the number of levels the recursion required is going to be fairly small. See all these logarithms over here. thus, we can solve and find the circuit size. So, lets actually work out that formula for we. we said it is d to the k. d to the k is the same thing as 2 to the k, just changing the base of the exponential. So, this is how we are going to rewrite this in a little better way by factoring out some things. By the way, the circuit size is actually N times d to the k. So, it is going to be N times log of N over epsilon Pth divided by log of Pth divided by P. we have inverted the top and bottom logarithm functions by using a minus sign it cancels to the log of base 2 of d. although this looks like a rather complex formula, this is actually N times a polynomial in the logarithm of 1 over epsilon N over epsilon. Because this is the key term here N over epsilon to some power. Everything else in there is not going to scale with N or epsilon. thus, this proves the theorem that we had set out to do. we have a logarithmically sized circuit required to become fault-tolerant as long as the probability of error is less than this threshold probability. This is really phenomenal., by the way, we believe we should probably build all of the future classical systems using this kind of idea. we really should be building the classical logic circuits out of very cheap, low threshold, deep subthreshold silicon chips which are built out of cheap fabrication facilities instead of billion-dollar, GDP size, large, perfect. So, instead of zero defect as a philosophy, allow there to be some probability of failure. just design the system in a clever way so that we get fault tolerance. Now, in the world of quantum computation\cite{langione_where_nodate}, this is not just a desirable philosophy. It is a necessary philosophy. Because we do not know any other way to reduce errors in quantum systems until they are tolerable. So, we now want to tell us about the story for fault tolerance for quantum computation.

\section{Threshold for Quantum Computation: Fault-Tolerant Quantum Procedures} 
we would like to know what is the maximum error that we can tolerate in the quantum gates and still build an arbitrarily sized, arbitrary low error quantum circuit out of those noisy gates. The ideas that we have seen for the principles of fault-tolerance here are going to apply almost directly to the quantum case, but there are some challenges. So, what we want to be able to do is to say that we have some threshold for quantum, which is one over the number of fault paths. However, again, this has to be in a universe gate, and we want to do this in a fault-tolerant procedure, which means that we want to compute on encoded data, which is now quantum. This idea of computing on encoded quantum data means, in this case, we will use stabilizer codes\cite{gottesman_stabilizer_1997}, and the gates are normalizer operations, as we have seen in the last few sections. However, because we need this to be a universal gate, this means more than just Clifford gates because of the Gottesman-Knill theorem, which says that all stabilizer circuits and Clifford gates can be efficiently simulated on a classical computer. They are not universal for quantum computation. So, we will need a gate that is outside of the Clifford set, and so, we can remind ourselves that the normalizable of a stabilizer is a set of all gates in the set that we are starting out with. For example, polys let us denote that as g such that ghg dagger is in the stabilizer for all h in the stabilizer set. This is the polys for this first discussion, but we may also generalize this and define some normalizer on some unitaries instead, where this becomes some arbitrary unitary instead. So, there are codes that have generalized normalizers that are outside of the poly group and outside of the Clifford group. 

The stabilizer formalism is a method used to design and implement quantum error detection and correction. A stabilizer is composed of a set of gates that leave a particular group of quantum states unaffected, and then the group is said to be ``stabilized''. A stabilizer applied to a quantum state that is not part of this group, for example, due to a single-qubit error in a logical qubit, is left altered and, hence, no longer ``stabilized''. This principle can be used to detect unwanted modifications of the stabilized group of quantum states.

The 3-qubit bit-flip code we introduced previously and, in particular, the parity measurements can be described within the stabilizer formalism. As we discussed previously, the logical encoding uses three physical qubits to form the codeword states $ \vert 000 \rangle $ and $ \vert 111 \rangle $. These can then be used to form a logically encoded superposition state.

\begin{equation}\label{eq4_36}
\vert 0\rangle _ L=\vert 000 \rangle \qquad \vert 1\rangle _ L=\vert 111 \rangle \qquad \rightarrow \qquad \alpha \vert 0 \rangle _ L+\beta \vert 1 \rangle _ L.
\end{equation}

Recall that the 3-qubit bit-flip code is composed of two sequential parity measurements. The parity operations are performed on qubits 1 and 2, and then on qubits 2 and 3. The results are stored on two ancilla qubits. Measuring the ancilla qubits provides two classical bits, the parity of each qubit pair, and these together form a syndrome. The syndrome values indicate the presence and location of up to one bit-flip error. Recall that a parity operation and measurement has the special property that it only indicates whether the two qubits are in the same state or in different states. It, therefore, does not destroy the quantum information.

Let us consider the ancilla qubit measurements in more detail. The ancilla qubit corresponding to qubits 1 and 2, for example, stores the parity of qubits 1 and 2., measuring the ancilla yields the parity, as described above. Is there another way we can think about this in the context of qubits 1 and 2?

In general, a qubit measured in the computational basis that is, along the Bloch sphere's z-axis, will output eigenvalues $ \pm 1 $ of the observable Z, depending on the state of the qubit. If the qubit is projected on the positive z-axis, a +1 results are indicating state $ \vert 0 \rangle $. In contrast, if the qubit is projected on the negative z-axis, a -1 results indicating $ \vert 1 \rangle $.

Now, let us consider the observable $ Z_1 Z_2 $ on qubits 1 and 2. Note, this does not represent two independent measurements of $ Z_1 $ and $ Z_2 $; that would clearly project out the qubit states. Rather, we can think of it as one measurement of a new observable comprising both $ Z_1 $ and $ Z_2 $. If both qubits are in-state $ \vert 0\rangle $, the measurement output is$  (+1)(+1)=1 $. And, if both qubits are in-state $ \vert 1\rangle $, the measurement output is $ (-1)(-1)=1 $. Similarly, we see that if the qubits are in different states, we get a measurement output of -1. Thus, we see that $ Z_1 Z_2  $is essentially the parity operation. Implementing an ancilla parity measurement is essentially equivalent to measuring the observable $ Z_1 Z_2 $. Moreover, the same is true for $ Z_2 Z_3 $. Consequently, the ``stabilizers'' of the 3-qubit bit-flip code are $ Z_1Z_2 $ and $ Z_2Z_3 $. These operations ``stabilize'' the logical qubit $ \alpha \vert 0 \rangle _ L+\beta \vert 1 \rangle _ L $.

The following table summarizes the expectation values of the stabilizers and shows that they are equivalent to the measurement of the ancilla qubits.

\begin{table}[H]
\centering
\caption{The expectation values of the stabilizers}
\label{tab:4_1:Table 1}
\resizebox{\textwidth}{!}{
\begin{tabular}{|c|c|c|c|c|}\hline
logical qubit &    $ \langle Z_1Z_2 \rangle \qquad \qquad $ &    $ \langle Z_2Z_3 \rangle \qquad \qquad $ &    ancillas &    Measurement \\\hline
$ \vert \psi \rangle _0=\alpha \vert 000\rangle +\beta \vert 111 \rangle $    & $ _0\langle \psi \vert Z_1Z_2\vert \psi \rangle _0=~ ~ 1 $ &    $ _0\langle \psi \vert Z_2Z_3\vert \psi \rangle _0=~ ~ 1 $ &    $ \vert 00\rangle _ A $    & $ ~ ~  1\qquad ~ 1 $  \\\hline
$ \vert \psi \rangle _1=\alpha \vert 100\rangle +\beta \vert 011 \rangle $    & $ _1\langle \psi \vert Z_1Z_2\vert \psi \rangle _1=-1 $ &    $ _1\langle \psi \vert Z_2Z_3\vert \psi \rangle _1=~ ~ 1 $ &    $ \vert 10\rangle _ A $    &$ -1\qquad ~ 1  $\\\hline
$ \vert \psi \rangle _2=\alpha \vert 010\rangle +\beta \vert 101 \rangle $    & $ _2\langle \psi \vert Z_1Z_2\vert \psi \rangle _2=-1 $ & $     _2\langle \psi \vert Z_2Z_3\vert \psi \rangle _2=-1 $ &    $ \vert 11\rangle _ A $    & $ -1\quad -1 $ \\\hline
$ \vert \psi \rangle _3=\alpha \vert 001\rangle +\beta \vert 110 \rangle $ &    $ _3\langle \psi \vert Z_1Z_2\vert \psi \rangle _3=~ ~ 1 $ &    $ _3\langle \psi \vert Z_2Z_3\vert \psi \rangle _3=-1 $ &    $ \vert 01\rangle _ A $    & $ ~ ~ 1\quad -1 $ \\\hline
\end{tabular}} 
\end{table}

The working principles of the 3-qubit bit-flip code can be captured with the stabilizers $ Z_1Z_2 $ and $ Z_2Z_3 $. Similarly, the 3-qubit phase-flip code is described with its two stabilizers $ X_1X_2 $ and $ X_2X_3 $. The reader is encouraged to show that these two stabilizers are an equivalent description of the 3-qubit phase-flip quantum error detection mechanism presented in an earlier text unit.

The utility of the stabilizer formalism becomes clearer for more complicated error correction codes. One example is the 5-qubit code, proposed by Raymond Laflamme, Cesar Miquel, Juan Pablo Paz, and Wojciech Zurek. The 5-qubit quantum error correction code requires the minimum number of physical qubits to correct one single arbitrary Pauli error (either a bit-flip, a phase-flip, or their combination)\cite{smart_experimental_2019}.

The stabilizers referred to as ``generators'' of the 5-qubit code are:
\begin{center}
$ \displaystyle g_1    \displaystyle =    \displaystyle XZZXI     $     \\
$ \displaystyle g_2    \displaystyle =    \displaystyle IXZZX     $      \\
$ \displaystyle g_3    \displaystyle =    \displaystyle XIXZZ     $      \\
$ \displaystyle g_4    \displaystyle =    \displaystyle ZXIXZ. $\\
\end{center}
          
The stabilized quantum state representing the logical qubits$  \vert 0\rangle _ L $ and $ \vert 1\rangle _ L, $ also referred to as logical code words $ \vert 0\rangle _ L $ and$  \vert 1\rangle _ L $, are then.

\begin{equation}\label{eq4_38}
\begin{split}
\displaystyle \vert 0 \rangle _ L    \displaystyle & =    \displaystyle \frac{1}{4}(\vert 00000\rangle +\vert 10010\rangle +\vert 01001\rangle +\vert 10100\rangle\\     & 
\displaystyle ~ +\vert 01010\rangle -\vert 11011\rangle -\vert 00110\rangle -\vert 11000 \rangle \\         & 
\displaystyle ~ -\vert 11101\rangle -\vert 00011\rangle -\vert 11110\rangle -\vert 01111 \rangle \\     &     
\displaystyle ~ -\vert 10001\rangle -\vert 01100\rangle -\vert 10111\rangle +\vert 00101 \rangle )    
\end{split}
\end{equation} 
\begin{equation}\label{eq4_38_1}
\begin{split} 
\displaystyle \vert 1 \rangle _ L    \displaystyle & =    \displaystyle \frac{1}{4}(\vert 11111\rangle +\vert 01101\rangle +\vert 10110\rangle +\vert 01011\rangle    \\     &       
\displaystyle ~ +\vert 10101\rangle -\vert 00100\rangle -\vert 11001\rangle -\vert 00111\rangle          \\     & 
\displaystyle ~ -\vert 00010\rangle -\vert 11100\rangle -\vert 00001\rangle -\vert 10000\rangle         \\     &  
\displaystyle ~ -\vert 01110\rangle -\vert 10011\rangle -\vert 01000\rangle +\vert 11010\rangle )
\end{split}
\end{equation}
      
Suppose the first qubit of quantum state $ \vert 0\rangle _ L $ experiences a bit bit-flip on the first qubit, $ X_1I_2I_3I_4I_5 $. The ``generator'' $ g_4 $ anti-commutes with the erroneous quantum operation and yields a -1 for the error state.

\begin{equation}\label{eq4_39}
\begin{split}
\displaystyle \vert 0 \rangle _{f}=X_1I_2I_3I_4I_5\vert 0 \rangle _ L    \displaystyle & =    \displaystyle \frac{1}{4}(\vert 10000\rangle +\vert 00010\rangle +\vert 11001\rangle +\vert 00100\rangle      \\     &     
\displaystyle ~ +\vert 11010\rangle -\vert 01011\rangle -\vert 10110\rangle -\vert 01000 \rangle      \\     &     
\displaystyle ~ -\vert 01101\rangle -\vert 10011\rangle -\vert 01110\rangle -\vert 11111 \rangle      \\     &     
\displaystyle ~ -\vert 00001\rangle -\vert 11100\rangle -\vert 00111\rangle +\vert 10101 \rangle )    
\end{split}
\end{equation}

\begin{equation}\label{eq4_39_1}
\begin{split}
\displaystyle g_1: \quad (XZZXI)(XIIII)\vert 0 \rangle _ L     \displaystyle & =    \displaystyle ~ ~ (XIIII)(XZZXI)\vert 0 \rangle _ L ~ \rightarrow ~ _ f\langle 0 \vert XZZXI \vert 0 \rangle _ f \\ & = +1
\end{split}
\end{equation}
\begin{equation}\label{eq4_39_2}
\begin{split}
\displaystyle g_2: \quad (IXZZX)(XIIII)\vert 0 \rangle _ L    \displaystyle & =    \displaystyle ~ ~ (XIIII)(IXZZX)\vert 0 \rangle _ L ~ \rightarrow ~ _ f\langle 0 \vert IXZZX \vert 0 \rangle _ f  \\ &= +1          
\end{split}
\end{equation}
\begin{equation}\label{eq4_39_3}
\begin{split}
\displaystyle g_3:\quad (XIXZZ)(XIIII)\vert 0 \rangle _ L    \displaystyle & =    \displaystyle ~ ~ (XIIII)(XIXZZ)\vert 0 \rangle _ L ~ \rightarrow ~ _ f\langle 0 \vert XIXZZ \vert 0 \rangle _ f \\& =+1         
\end{split}
\end{equation}
\begin{equation}\label{eq4_39_4}
\begin{split}
\displaystyle g_4:\quad (ZXIXZ)(XIIII)\vert 0 \rangle _ L    \displaystyle & =    \displaystyle -(XIIII)(ZXIXZ)\vert 0 \rangle _ L ~ \rightarrow ~ _ f\langle 0 \vert ZXIXZ \vert 0 \rangle _ f \\& =-1    
\end{split}
\end{equation}
          
Classical processing of the generator measurements the error syndrome reveals the erroneous physical qubit. Subsequently, the error can be corrected with a feed-forward counteracting quantum operation. The reader is encouraged to explore the working principles of the 5-qubit quantum error correction code in case of a phase-flip.

The stabilizer formalism enables the description and comparison of many quantum error correction codes beyond the 3-qubit bit-flip, 3-qubit phase flip, and 5-qubit codes referred to as stabilizer codes. The interested reader may read the following section on the stabilizer formalism and a more mathematical discussion of it\cite{rue_mathematical_nodate}.

\section{Stabilizers (Advanced)}

Stabilizer codes are an important class of quantum error correction codes based on the stabilizer formalism. The stabilizer formalism provides a general method for designing and implementing quantum error detection in support of error correction. A stabilizer is a configuration of operators that leave a particular group of quantum states untouched. However, for quantum states that are not part of this group, the stabilizer will effectively apply a ``label'' that identifies the state as being outside of the original set. This principle can be used to detect when changes, i.e., errors occur to codeword quantum states.

The stabilizer formalism is based on the Pauli group $ G_ n $ for n qubits. The Pauli group for a single qubit can be expressed as $ G_1\equiv \{ \pm I, \pm iI, \pm X, \pm iX, \pm Y, \pm iY, \pm Z, \pm iZ\} $. $ G_ n $ is composed of all n-fold tensor products of the Pauli matrices forming $ G_1 $.

To give an example, Let us consider the Bell state $ \vert \Phi ^+\rangle =1/\sqrt {2}(\vert 00 \rangle + \vert 11 \rangle ) $. Note that this may be viewed as qubits 1 and 2 from the discussion of the three-qubit bit-flip code. As we discuss previously, applying the operators $ X_1X_2 $ and $ Z_1Z_2 $ will leave the state $ \vert \Phi ^+\rangle $ unchanged:

\begin{equation}\label{eq4_40}
\begin{split}
\displaystyle X_1X_2\vert \Phi ^+\rangle =X_1X_2\frac{\vert 00 \rangle + \vert 11 \rangle }{\sqrt {2}}    \displaystyle =    \displaystyle ~ ~ \frac{\vert 00 \rangle + \vert 11 \rangle }{\sqrt {2}}=\vert \Phi ^+\rangle     \\     
\displaystyle Z_1Z_2\vert \Phi ^+\rangle ~ =~ Z_1Z_2\frac{\vert 00 \rangle + \vert 11 \rangle }{\sqrt {2}}    \displaystyle =    \displaystyle ~ ~ \frac{\vert 00 \rangle + \vert 11 \rangle }{\sqrt {2}}=\vert \Phi ^+\rangle
\end{split}
\end{equation}
          
The operators are thus said to ``stabilize'' the state.

Suppose S is a subspace of $ G_2 $ with $ S\equiv \{ I_1I_2,X_1X_2,Z_1Z_2\} $. Then, the vector space $ V_ S $ stabilized by S consists of $ \vert \Phi ^+\rangle $. The subspace S can be expressed in terms of its generators, a reduced set of operators that can be used to generate the elements of S. For example, the generators of S are $ \langle X_1X_2, Z_1Z_2 \rangle $, since $ I_1I_2=(X_1X_2)(X_1X_2) $ or $ I_1I_2=(Z_1Z_2)(Z_1Z_2) $. Thus, it is sufficient to define S solely by its generators, $ S \equiv \langle X_1X_2, Z_1Z_2 \rangle $, where we now use the $ \langle \ldots \rangle $ symbol to indicate that this lists only the generators of S. if a quantum state is stabilized by all generators of a group S, then all elements of group S including those derived from the generators will also stabilize this particular state.

There are two conditions for S to stabilize a non-trivial vector space, that is, a vector space spanned by non-zero vectors. First, -I cannot be an element of S, due to the fact that $ -I\vert \psi \rangle =-\vert \psi \rangle $ is only valid for the trivial case, $ \vert \psi \rangle =0 $. And, second is that the elements of S commute. Two operators (or, sets of operators) commute if their order of operation does not affect the outcome. For example, the stabilizers $ g_1 $ and $ g_2 $ commute if $ g_1g_2 - g_2g_1 = 0 \rightarrow g_1g_2=g_2g_1 $. To see why stabilizers must commute, consider an alternative case in which the two stabilizers happen to anti-commute, that is $ g_1g_2 + g_2g_1 = 0 \rightarrow g_1g_2=-g_2g_1 $. In this case, $  g_1g_2\vert \psi \rangle =-g_2g_1\vert \psi \rangle $, and so the stabilizers would label the state $ \vert \psi \rangle $ depending on the order in which they were applied, and this should not happen; the result of applying stabilizers should depend on the state, and not on the order of operation. If $ g_1 $ and $  g_2 $ were anti-communiting stabilizers, it would follow that $ -\vert \psi \rangle =\vert \psi \rangle $, which again can only be true for the trivial case.

Now, consider a bit-flip error $ I_1X_2 $, which leaves qubit 1 alone and flips the state of qubit 2. $ I_1X_2 $ commutes with stabilizer $ X_1X_2 $, but it anti-commutes with $ Z_1Z_2 $:

\begin{equation}\label{eq4_41}
\begin{split}
\displaystyle (X_1X_2)(I_1X_2)\frac{\vert 00 \rangle + \vert 11 \rangle }{\sqrt {2}}=~ ~ \frac{\vert 01 \rangle + \vert 10 \rangle }{\sqrt {2}} ~ ~    \displaystyle \\\&    \displaystyle ~ ~ (I_1X_2)(X_1X_2)\frac{\vert 00 \rangle + \vert 11 \rangle }{\sqrt {2}}=~ ~ \frac{\vert 01 \rangle + \vert 10 \rangle }{\sqrt {2}}     \\     
\displaystyle (Z_1Z_2)(I_1X_2)\frac{\vert 00 \rangle + \vert 11 \rangle }{\sqrt {2}}=-\frac{\vert 01 \rangle + \vert 10 \rangle }{\sqrt {2}}~ ~    \displaystyle \\\&    \displaystyle ~ ~ (I_1X_2)(Z_1Z_2)\frac{\vert 00 \rangle + \vert 11 \rangle }{\sqrt {2}}~ =~ ~ \frac{\vert 01 \rangle + \vert 10 \rangle }{\sqrt {2}}
\end{split}
\end{equation}
          
Thus, we see that applying the unitary bit-flip operator, an operation that takes a state within $ V_ S $, and puts it outside $ V_ S $, affects the results when applying the stabilizers of $ V_ S $ (as we would expect). More generally, applying a unitary gate $ U $ on a state vector $ \vert \psi \rangle $ consequently affects the generators of the associated stabilizer group. The new state $ U\vert \psi \rangle $ is an element in a new vector space $ U V_ S $, and it is stabilized by the generator $ UgU^\dagger $, which follows from $ U\vert \psi \rangle =Ug\vert \psi \rangle =UgU^\dagger U\vert \psi \rangle $.

The dynamics of a quantum state can be described by the evolution of the generators of the stabilizer group for a specific set of unitary quantum gates. Unitary gates mapping stabilizers to stabilizers are referred to as normalizers of $ G_ n $ and form the Clifford group. It can be shown that the CNOT, Hadamard, and phase gate can compose all examples of gates that are able to map elements of $ G_ n $ to $ G_ n $. The set of gates $ U $ complying with $ UG_ nU^\dagger =G_ n $ are called the normalizer gates$  N(G_ n) $.

\begin{equation}\label{eq4_42}
U_{CNOT}=\left(\begin{array}{cccc} 1&0&0&0\\ 0&1&0&0\\ 0&0&0&1\\ 0&0&1&0 \end{array}\right) \quad U_{Hadamard}=\frac{1}{\sqrt{2}}\left(\begin{array}{cc} 1&1\\ 1&-1 \end{array}\right) \quad U_{phase}=\left(\begin{array}{cc} 1&0\\ 0&i \end{array}\right) 
\end{equation}

For example a CNOT gate applied to $ \vert \Phi ^+\rangle $ and to the generator $ X_1X_2 $ alter the generators in the following way:

\begin{equation}\label{eq4_43}
\begin{split}
\displaystyle U_{CNOT}    \displaystyle & =    \displaystyle \left(\begin{array}{cccc} 1& 0& 0& 0\\ 0& 1& 0& 0\\ 0& 0& 0& 1\\ 0& 0& 1& 0 \end{array}\right)=U_{CNOT}^\dagger     \\     
\displaystyle X_1X_2    \displaystyle & =    \displaystyle \left(\begin{array}{cccc} 0& 0& 0& 1\\ 0& 0& 1& 0\\ 0& 1& 0& 0\\ 1& 0& 0& 0 \end{array}\right)          
\displaystyle \rightarrow    \displaystyle U_{CNOT}~ X_1X_2~ U_{CNOT}^\dagger =X_1I_2
\end{split}
\end{equation}
          
Not all unitary gates are normalizers of $ G_ n $. For example, the T gate, a gate often added to the Clifford gates to form universal quantum gate sets, is not a normalizer of $ G_ n $. Relatedly, the Gottesman-Knill theorem states that quantum circuits can be classically efficiently simulated if the three following conditions are true:\\
1. Quantum states are prepared in the computational basis. For example a single qubit in $ \vert 0 \rangle $.\\
2. Quantum operations are composed of elements of the Pauli group, CNOT, Hadamard, and phase gate.\\
3. Only measurements of Pauli observables are required. For example, a measurement in the computational basis for a single qubit and, thus, the observable Z with $ \langle 0 \vert Z\vert 0 \rangle \rightarrow 1 and \langle 1 \vert Z\vert 1 \rangle \rightarrow -1 $, is in this category.

The stabilizer formalism can entirely describe quantum circuits complying with the three conditions. A single quantum operation acting on an n-qubit state represented by an n-element stabilizer requires $ n^2 $ classical computational steps (due to conjugation). Consequently, a classical simulation of a quantum circuit with n qubits and m quantum operations requires $ \mathcal{O}(n^2m) $ classical computational steps. Note, since the stabilizer formalism does not fully describe universal gate sets, it cannot be used as a method to simulate universal quantum computation \cite{goto_universal_2016} with a classical computer efficiently. Nevertheless, the description of many quantum error correction and detection codes can be accomplished using the stabilizer formalism and, in particular, the normalizer gates.

Now, let us introduce the concept of how the stabilizer formalism can be used to describe quantum error detection and correction codes. Stabilizer codes are classified by the number of involved qubits n and generators n-k, abbreviated as [n,k] stabilizer codes. An [n,k] stabilizer code is composed of a vector space $ V_ S $, which represents the code space. In turn, the code space is stabilized by S and generated by $ S=\langle g_1, \cdots , g_{n-k}\rangle $.

Error detection is initiated by acquiring an error syndrome. The error syndrome is the summary of the measurements of all generators $ g_1, \cdots, g_{n-k} $ of the underlying stabilizer S. The error syndrome is subsequently classically processed to determine if an error occurred and, if so, on which qubit.

There are three possible outcomes of a stabilizer measurement applied to a code space that is corrupted by an error E represented as a quantum operation $ E\in G_ n $:\\
1. E is an element of S and maps one element to another element in the code space: the operation does not cause an actual error, and no recovery is necessary. \\
2. E anti-commutes with an element of the stabilizer S: the error E projects the quantum state to a subspace that is outside (orthogonal to) the code space. This is a detectable and potentially correctable error.\\
3. E commutes with all elements of S, but it is not an element of S: the error is not detectable, and hence recovery is not possible.

The third scenario is the most problematic one since error E goes undetected. The set of $ E\in G_ n $, such that E and g commute$  (Eg=gE) $ for all $ g\in S $, is referred to as the centralizer Z(S) of S in $ G_ n $ and identical to the normalizers N(S) of S.

The stabilizer formalism enables the description of a wide variety of quantum error correction codes, the stabilizer codes. A stabilizer code is able to detect errors $ E\in G_ n $ that are not part of the centralizer Z(S) or the stabilizer S itself. In the next section, we will introduce another important quantum error correction code, the 7-qubit Steane code, in detail, and we will see how efficient the stabilizer formalism captures the working principles of this more complex protocol.

\section{Fault-Tolerant Quantum Gates: 5-qubit and 7-qubit Codes}
Let us give some explicit examples so that we know what we are discussing in the context of quantum computation and all these gates. So, the 5-qubit code has these normalize $ XZZXI, IXZZX,$ $XIXZZ, ZXIXZ $. All we did is memorize this pattern, and then we cyclically shifted them to get the other three here. These are the stabilizers that generate the 5-qubit code, and there are some very useful normalizers. For example, we may define logical X on a code word of the 5-qubit code as being X five times. That is XXX, five X's, one X on each qubit. We can quite easily see that this commutes with every single generator, so that means that it is normalizable of this stabilizer group. Another case is, for example, Z. So, the Z bar on this is also equal to something simple. However, H, the Hadamard gate, cannot be done by just repeating H 5 times. This is fairly straightforward to see, because if we conjugate this it, turns X into Zs and Zs into Xs. Then, the pattern is different, and it does not commute with each of these five. So, the five-qubit code does not admit a simple way of doing universal gates on the data encoded inside it. We can also see this for the CNOT gate if we try to make that construction. On the other hand, there is another code in which we have seen the 7-qubit code, which has the stabilizers we, we, X. we, X, we. We are writing them in a funny way because it is easiest to remember this way. These are just binary numbers going from 1 through 7. So, 1, 2, 3, 4, 5, 6, 7. the generators of the 7-qubit code are these, plus the same with X replaced with Z. this is a CSS code, Calderbank-Shor-Steane code. It has the following beautiful properties. First, logical X is 7 Xs. Same thing for Z. However, we also have this nice thing, which is that a Hadamard is 7 Hadamard’s. That is obvious because the seven Hadamard’s just change these Xs to Zs. So, it flips this with this, and therefore, it stabilizes the code. In addition to this, if we want to do a controlled-NOT between two 7-qubit coded blocks, we can do it bitwise. This we are not going to show us, but we are going to make an assertion about it. All the way down to the last one over here. So, we say that the CNOT is transversal. A bitwise performance of controlled NOTs between each pair of bits in two different 7-qubit codeword blocks performs a controlled-NOT gate. This is such a beautiful feature that, in fact, we will make this assertion. It turns out that every CSS code has a transversal CNOT. Every code with a transversal CNOT is a CSS code. We encourage us to go and prove that for ourselves. It is a fun fact. Now, a catch. We cannot perform universal computation with Clifford H and X and Z., in fact, and we can also look at a logical S. Go work that out. Clifford is generated by H, S, and CNOT. There is an important theorem that has been proven, which is that, of all these quantum codes, the stabilizer codes are the most useful to work with, because they are the analog of classical linear codes. They have wonderful properties which give us large families. Unfortunately, there is this theorem from 2007 Eastin Knill, we think? That was also 2007. No stabilizer's code has a universal set of transversal gates. This is horrible. This means that we cannot get universal quantum computation by staying inside a single stabilizer code. This has meant that one has to use a variety of tricks to go in and out of the code to perform the computation or something else. in much of what we are going to discuss in the next few sections, we are going to find a series of tricks to be able to do that. Complicated and interesting codes that have multiple families together. Alternatively, using an idea called teleportation of gates\cite{podoshvedov_efficient_2019}. However, we are going to just stick with the assumption that we can look at the controlled-NOT gate, and then use that one assumption to derive a first-cut estimate of the threshold for quantum computation. So, there is this very strong caveat. Just keep that in mind. 

The 7-qubit Steane code \cite{steane_error_1996} is best understood from the perspective of the Hamming codes. Hamming codes are a family of classical error-correcting codes introduced by Richard Hamming in 1950 to deal with errors in punch card readers. Specifically, we will elaborate on the [7,4] Hamming code, which indicates that seven bits are used to communicate a four-bit message. The extra three bits in each logical codeword serve the purpose of redundancy and allow for any single-bit error to be detected and corrected.

With four bits of information, there need to be $ 2^4 $ distinct strings of length seven that can be used as logical codewords. These logical codewords can be found by multiplying any of the four-bit messages in the form of a binary column vector by the code generation matrix G given by:

\begin{equation*} G=\left(\begin{array}{cccc} 1 & 1 & 0 & 1 \\ 1 & 0 & 1 & 1 \\ 1 & 0 & 0 & 0 \\ 0 & 1 & 1 & 1 \\ 0 & 1 & 0 & 0 \\ 0 & 0 & 1 & 0 \\ 0 & 0 & 0 & 1 \\ \end{array}\right) \end{equation*}
which will result in a column vector of length seven.

Once an encoded message is transmitted, and potentially incurs errors, it can be analyzed by multiplying the seven-bit vector by the so-called parity-check matrix, given by:

\begin{equation*} H=\left(\begin{array}{ccccccc} 1 & 0 & 1 & 0 & 1 & 0 & 1 \\ 0 & 1 & 1 & 0 & 0 & 1 & 1 \\ 0 & 0 & 0 & 1 & 1 & 1 & 1 \\ \end{array}\right) \end{equation*}
which results in a three-bit vector, referred to as the syndrome vector. The syndrome vector can be used to determine which of the seven bits in the logical codeword have been flipped. Note that if the logical codeword is not corrupted, then the syndrome vector will be the zero vector for all logical codewords. It is important to remember that we are only considering bits, and so all matrix multiplication is modulo 2.

As an example, we can consider encoding (0 0 1 0) to find the seven-bit logical codeword.

\begin{equation*} \left(\begin{array}{cccc} 1 & 1 & 0 & 1 \\ 1 & 0 & 1 & 1 \\ 1 & 0 & 0 & 0 \\ 0 & 1 & 1 & 1 \\ 0 & 1 & 0 & 0 \\ 0 & 0 & 1 & 0 \\ 0 & 0 & 0 & 1 \\ \end{array}\right) \left(\begin{array}{c} 0\\ 0\\ 1\\ 0\\ \end{array}\right) = \left(\begin{array}{c} 0\\ 1\\ 0\\ 1\\ 0\\ 1\\ 0\\ \end{array}\right). \end{equation*}
We can first verify that multiplying the parity-check matrix into this logical codeword results in the zero vector.

\begin{equation*} H=\left(\begin{array}{ccccccc} 1 & 0 & 1 & 0 & 1 & 0 & 1 \\ 0 & 1 & 1 & 0 & 0 & 1 & 1 \\ 0 & 0 & 0 & 1 & 1 & 1 & 1 \\ \end{array}\right) \left(\begin{array}{c} 0\\ 1\\ 0\\ 1\\ 0\\ 1\\ 0\\ \end{array}\right) = \left(\begin{array}{c} 0 \mod 2\\ 2 \mod 2\\ 2 \mod 2\\ \end{array}\right) = \left(\begin{array}{c} 0\\ 0\\ 0\\ \end{array}\right). \end{equation*}
If instead, during transmission the first bit of the logical codeword is flipped from $ 0\rightarrow 1 $, then the syndrome vector would become

\begin{equation*} H=\left(\begin{array}{ccccccc} 1 & 0 & 1 & 0 & 1 & 0 & 1 \\ 0 & 1 & 1 & 0 & 0 & 1 & 1 \\ 0 & 0 & 0 & 1 & 1 & 1 & 1 \\ \end{array}\right) \left(\begin{array}{c} 1\\ 1\\ 0\\ 1\\ 0\\ 1\\ 0\\ \end{array}\right) = \left(\begin{array}{c} 1 \mod 2\\ 2 \mod 2\\ 2 \mod 2\\ \end{array}\right) = \left(\begin{array}{c} 1\\ 0\\ 0\\ \end{array}\right), \end{equation*}
which uniquely determines in which element of the length-seven logical codeword the error occurred (in this case, the first element). As we move to the quantum case, it is important to recognize that the syndrome vector only reveals which bit was flipped, and not the value of that bit.

The 7-qubit Steane code was introduced in 1996 by Andrew Steane of Oxford University and extended the Hamming code to the quantum realm. Since errors in quantum systems are more complicated than in classical systems, due to the existence of phase errors, the 7-qubit Steane code uses seven qubits to store a single qubit, rather than seven classical bits to store four classical bits as in the Hamming code. Similar to the classical notation for error correction codes such as the [7,4] Hamming code with seven bits encoding the information of four bits, there exists a quantum notation to summarize quantum codes. For example, the 7-qubit Steane code is written [[7,1,3]] (the double brackets indicate a quantum error correction code). The first value indicates the number of physical qubits, the second the number of encoded logical qubits, and the third the number of qubits containing information on the errors.

The Steane code is capable of correcting for a single arbitrary error, and it is based on a logical encoding with the following seven-qubit states:
\begin{equation}\label{eq4_44}
\begin{split}
\displaystyle \vert 0 \rangle _ L    \displaystyle =    \displaystyle \frac{1}{\sqrt {8}}(\vert 0000000\rangle +\vert 0001111\rangle +\vert 0110011\rangle +\vert 0111100\rangle    \\      
\displaystyle +    \displaystyle \vert 1010101\rangle +\vert 1011010\rangle +\vert 1100110\rangle + \vert 11101001 \rangle )     \\     
\displaystyle \vert 1 \rangle _ L    \displaystyle =    \displaystyle \frac{1}{\sqrt {8}}(\vert 1111111\rangle +\vert 1110000\rangle +\vert 1001100\rangle +\vert 1000011\rangle    \\      
\displaystyle +    \displaystyle \vert 0101010\rangle +\vert 0100101\rangle +\vert 0011001\rangle + \vert 00010110\rangle )
\end{split}
\end{equation}
          
Where the subscript L stands for logical codeword state. Each of these seven-qubit states corresponds to the form of the logical codeword in the [7,4] Hamming code, and hence it is possible to perform a quantum computation over the entire superposition to determine the syndrome vector and to correct a bit-flip error in this basis.

In the stabilizer formalism the 7-qubit Steane code can be expressed with 6 generators as:
\begin{equation}\label{eq4_45}
\begin{split}
\displaystyle g_1    \displaystyle =    \displaystyle IIIXXXX          \\
\displaystyle g_2    \displaystyle =    \displaystyle IXXIIXX          \\
\displaystyle g_3    \displaystyle =    \displaystyle XIXIXIX          \\
\displaystyle g_4    \displaystyle =    \displaystyle IIIZZZZ          \\
\displaystyle g_5    \displaystyle =    \displaystyle IZZIIZZ          \\
\displaystyle g_6    \displaystyle =    \displaystyle ZIZIZIZ
\end{split}
\end{equation}
          
To represent the code space. We encourage the reader to check a few qubit states and experience the ``stabilizing'' effects of the generators or combinations of them.

\begin{figure}[H] \centering{\includegraphics[scale=.52]{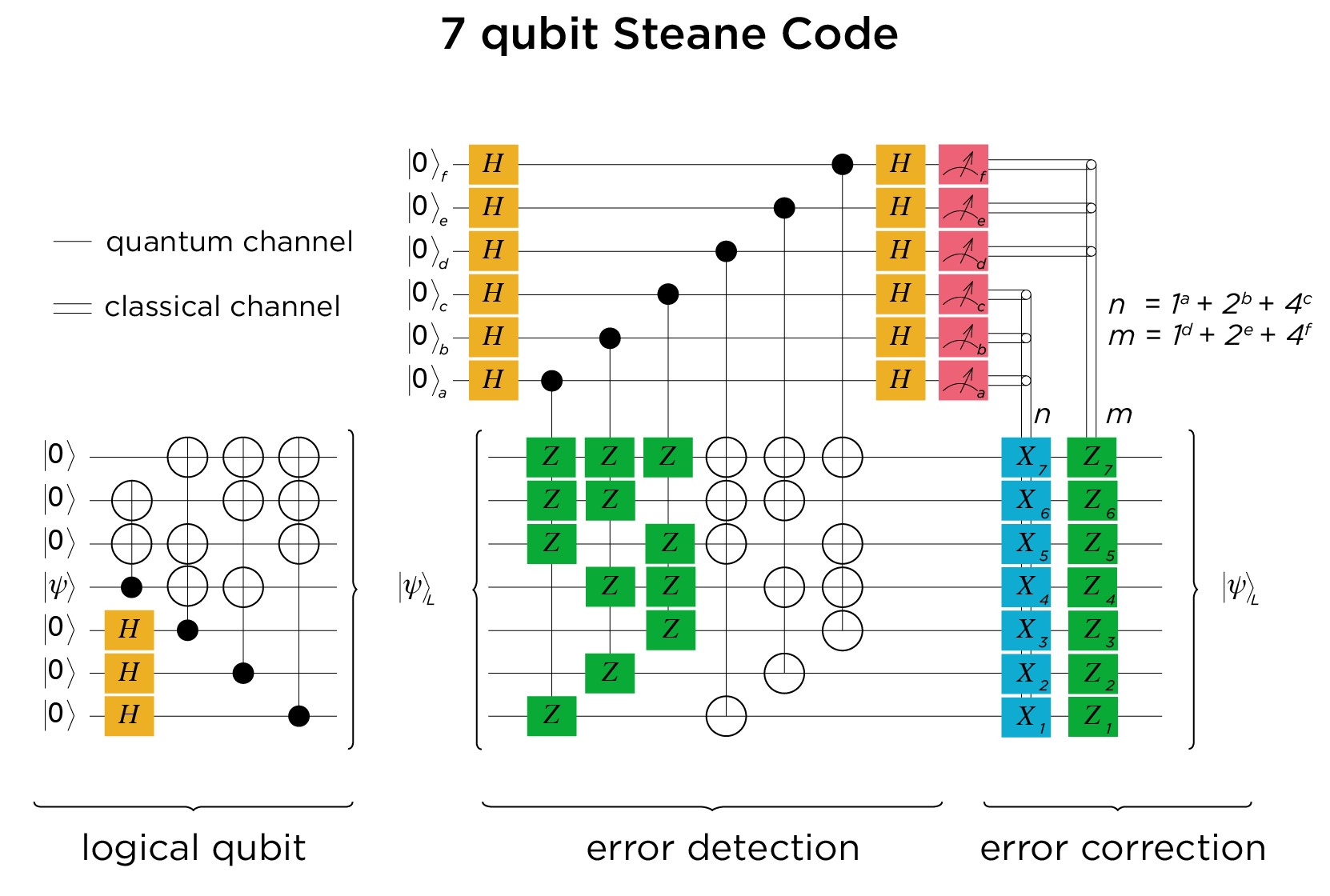}}\caption{Steane code}\label{fig4_12}
\end{figure}

The figure illustrates a possible implementation of the 7-qubit Steane code. The required logical qubit spans the code space, and a potential error is subsequently detected through a stabilizer measurement and corrected. The 7-qubit Steane code can, at most correct, a single bit-flip and phase-flip error, encoded in n and m, respectively, if the bit-flip and phase-flip errors happened to two different qubits.

So that this procedure can be used within a larger quantum computation, it is important that the detection and correction operations should not reveal any information about the encoded qubit. The seven qubits are therefore augmented with three ancilla qubits into which the syndrome will be encoded. As we noted above, the syndrome vector itself only indicates which qubit is faulty and not its value, and hence no quantum information is revealed.

So far we have only corrected bit-flip errors in the qubit basis. However, we also need to characterize and correct any phase-flip errors. This can be achieved simply by rotating the basis of each qubit, such that phase-flip errors become bit-flip errors, and repeating the same process. More specifically, consider that a phase flip, which takes $ \vert +\rangle = \frac{1}{\sqrt {2}}(\vert 0 \rangle +\vert 1\rangle ) $ to $ \vert -\rangle = \frac{1}{\sqrt {2}}(\vert 0 \rangle -\vert 1\rangle ) $, is equivalent to a bit flip error in the $ \{ +,-\} $ basis.

As an instructive example, we will consider the case of a bit-flip error in the first qubit of the states used to encode $ \vert \psi \rangle = a\vert 0\rangle _ L+b\vert 1\rangle _ L, $ where $ \vert a\vert ^{2}+\vert b\vert ^{2}=1 $. After the bit-flip error the state becomes $ \vert \psi '\rangle = E_{b1}\vert \psi \rangle $, where $ E_{b1} $ is an error operator that flips the first bit of the encoding states.

The system is now combined with three ancilla qubits in the $ \vert 000\rangle $ state, and an execution of the Steane code circuit results in:

\begin{equation*} \vert \psi'\rangle \vert 000\rangle \rightarrow \vert \psi'\rangle \vert S(\psi')\rangle \end{equation*}
where $ S(\psi ') $ represents the syndrome vector of $ \psi ' $.

As we discuss in the classical example above, the syndrome vector corresponding to a bit flip on the first bit of a Hamming code is given by

\begin{equation*} S(\psi')=\left(\begin{array}{c} 1\\ 0\\ 0\\ \end{array}\right). \end{equation*}
From this, we know that $ \vert \psi \rangle $ can be recovered by applying a bit flip to the first qubit. Note that no information about a and b is revealed throughout this process.

\item  7-qubit Codes; Consider the seven-qubit code described in the section above. Which of the following statements are correct?

\begin{itemize}
\item Three generators of the stabilizers of this code are IIIZZZZ, IZZIIZZ, and ZIZIZIZ.
\item It is possible to define the transversal Hadamard gate $ \overline{H}=HHHHHHH. $
\item The transversal Controlled-NOT (CNOT) gate is given by the combination of seven controlled gates between qubits 1 and 8, 2 and 9, 3, and 10.
\item The gate $ \overline{Z}=ZZZZZZZ $ cannot be considered a normalizer.
\end{itemize}
Solution:\\
The logical gate $ \overline{Z}=ZZZZZZZ $ can be considered a normalizer since it commutes with all generators of the stabilizer group.

\section{Fault-Tolerant Quantum Measurement of Error Syndromes}
Probably the most complicated part of the threshold for fault tolerance is for quantum computation, the methods that one has to use to perform the fault-tolerant measurement. We want to illustrate this challenge for us, and then we will see why this ends up dominating the cost for fault-tolerance and the threshold argument. So, the problem with fault-tolerant measurement is that in order to perform quantum error correction, we have to measure the stabilizers. So, recall that the quantum error correction procedure is to measure stabilizers the generators and then perform the recovery operation classically conditioned on these stabilizer measurement results. However, how do we measure stabilizers? Well, for example, if we look at this seven-qubit code up here, we have a number of generators that have Z's in them. In fact, we have four Zs, just like we have four X's here. So, we need to do four Control-Z gates, and the construction looks something like this. We start out with a single qubit. We perform a Hadamard, and then we have the seven-qubit data down here, and we perform controlled Z's on those qubits for which the generator has a Z. This is just an example, and they move around. Depending on where the Z's are in the generator, one then performs a Hadamard and a measurement. Clearly, this is not a good construction, because we have a single point of failure. If this qubit has an error in it, suddenly that error might propagate to four of the qubits down here. So, we need some fault-tolerant procedure for the measurement of the syndromes of the quantum error correction code. This is one of the largest hurdles for fault tolerance. So, let us tell us a story. Peter Shor invented the quantum factoring algorithm \cite{shor_polynomial-time_1997}. There was a lot of criticism that the algorithm would never be realistic because it relied on the perfect functioning of quantum computers. So, Peter Shor went away, and he wrote a second paper in which he largely invented the field of quantum error correction. This is a nine-qubit code, and what we have just discussed about in the last few sections. Then the community rallied again and said, but despite error correction, there is no plausible way we can realize Shor's algorithm \cite{shor_polynomial-time_1997} because errors will propagate too fast, and there is no way to build a fault-tolerant quantum computer. Error correction alone is not good enough. We need a code on which we can compute. So, Peter Shor went away, and then he came up with a fault-tolerant construction for quantum computation, at which point he got hired by MIT. So, he had three major results. Now we are essentially going to tell us about that third result. If we want, this is like the third Shor theorem. That is that we can perform this measurement procedure fault-tolerantly using a trick. This is a trick that Peter Shor invented. We measure this syndrome fault-tolerantly by first constructing something else which helps us an ancilla state. What we will do is we will construct using a quantum circuit. For example, this circuit. A four-qubit cat state. This is 0000 plus 1111 divided by root 2. we will construct this painstakingly well. We will explain more in a moment if we have time. Then we have the data qubits down here. We then perform the desired controlled Z measurements. We are just going to use a dot here to represent Z on whatever qubits we need to measure the stabilizers of. Just imagine that these are the four qubits. This is the ancilla. These are the data qubits. Notice how now this is transversal. They have split apart. However, because we have guaranteed that we are either all zero or all one, they actually do the same thing. So, that the effect of this measurement bumps a phase back up here, and it changes the parity once we perform the decoding of this cat state. So, we can decode by reversing that initial circuit, and then measure. We will measure all four of these and then look at those properties. What will it turn out is that the parity of the output over here will tell us the measurement result we want, while not causing a single error to propagate over here. Now, this is not very fault-tolerant, and this is not very fault-tolerant. However, we can do this part and then check those results. That is the key idea behind the fault tolerance.

\section{Crude Estimate of the Threshold for Reliable Quantum Computation}
Putting together everything we have seen to give us one final thing, which is an estimate of the fault tolerance. We had PTH for classical circuits, something like 1 over 6 or  0.089. What is it for a quantum computer? If we use this qubit code, we can discuss at the number of gates we need and their purpose. We will discuss at the gates required of the basic fault procedure, let us say for a controlled-NOT gate. So, here, we are going to need seven CNOTs for the gate. Then we need to do quantum error correction. Each of the steps of the quantum error correction has six syndrome operations. If we look up there, we have three X's and three Z's. So, add them up. That gives us 6. then we are going to have to measure four; each one has four gates, four non-zero syndrome qubits. then, we have to repeat this measurement three times because the measurement procedure itself is faulty. So, we have to be able to majority vote classically on the output of that. So, we multiply all these things up together. This gives us 72. we can see that the error correction far outweighs the actual steps we need for the gate. then we sum all of these, and this is going to give us 77 gates. We can tolerate 77 choose two, and therefore, if we take this as the number of fault paths possible, then we get something like around 3,081. So, the threshold is something like 01 over 3,081, which is something around 10 to the minus 4.And so, we say this is something around the threshold for fault-tolerant quantum computation. Questions? 79. Thank we. Other questions? Thank we for the fault-tolerant check. 

In this section, we provide a table comparing several simple codes. It is important to note here that there are many more quantum error correction codes and configurations than are summarized here.

For a quantum error correction code to be effective, it is important that its implementation is ``fault-tolerant'' to prevent error propagation. To this end, the ancillas assisting the quantum error detection can be implemented such that errors are unable to propagate and further corrupt the logical qubit. After implementing the classical processing to interpret the error syndrome and any subsequent error correction, the logical qubit is ideally free of errors. Such implementations are referred to as fault-tolerant since they enable the successful completion of a computational task even in the presence of errors.

\begin{figure}[H] \centering{\includegraphics[scale=.8]{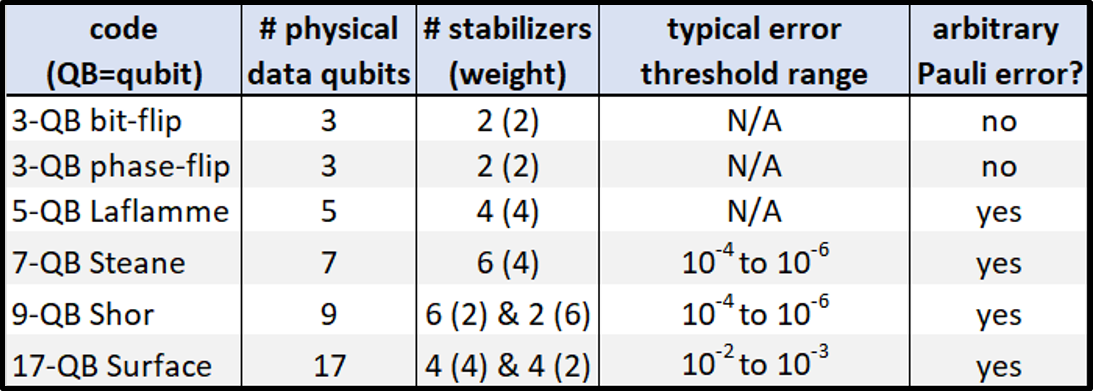}}\caption{Table: A basic comparison of quantum error correction codes. The stabilizers and weights provide an estimate on the number of ancilla qubits required to achieve fault-tolerance. The estimated error threshold is an indicative range and depends on the architecture used \cite{smith_practical_2017}. Thresholds are not provided (N/A) for the first three codes.}\label{fig4_13}
\end{figure}

The stabilizers describing a quantum error correction code can help to estimate the number of required ancilla qubits to achieve a fault-tolerant implementation. The number of stabilizers and their number of non-identity gates referred to as the stabilizer's weight, provide an estimate on how many interactions between the physical qubits and ancilla qubits are required to generate the error syndrome. Therefore, the number of ancilla qubits required for a fault-tolerant implementation of a quantum error correction code can be approximated by multiplying the number of stabilizers by their respective weights.

To compare the different fault-tolerant quantum error detection and correction methods, an important figure of merit is the physical qubit error rate. A quantum processor with an individual physical qubit error rate below a certain error threshold can, in principle, achieve fault-tolerance using quantum error correction codes. In general, the exact threshold depends on the architectural implementation of quantum error correction code, for example, the arrangement and connectivity of the qubits, the available gate set.

The table is organized to reflect an approximation of the classical effort required to detect errors. The complexity of the classical scheme increases from the top of the table to the bottom. In the next section, we will discuss the surface code in detail. Although it requires only nearest-neighbor couplings between qubits and has a relatively lenient threshold, making it a leading candidate for implementing error correction with today's available qubits, it requires a relatively large classical-computing overhead to interpret the syndrome measurements.

\item  Fault-Tolerant Procedure for a CNOT Gate; In a seven-qubit code, the procedure for a fault-tolerant CNOT gate requires:
\begin{itemize}    
\item Seven Controlled-NOT (CNOT) gates to generate a transversal CNOT.
\item Six syndrome operations for quantum error correction.
\item Four measurements per syndrome for quantum error correction.
\item Only two repetitions of each measurement to apply a majority gate.
\end{itemize}
Solution:\\
Since the measurement procedure is faulty, it is necessary to perform a majority vote, which needs three inputs and returns one output. This means that each measurement has to be repeated three times.

\item  Threshold Theorem II; A circuit constructed of unreliable gates can still simulate a uniform circuit composed of error-free gates. However, this simulation has to accept an error within a specified error tolerance.
Which of the following statements is correct about the number of gates required to simulate the ideal circuit: 
\begin{itemize}    
\item It grows with the number of gates of the ideal circuit.
\item It decreases with the threshold probability.
\item It scales as the logarithm of the inverse of the desired error tolerance.
\item It is greater than the number of error-free gates in the ideal circuit.
\end{itemize}
Solution:\\
If the ideal circuit is composed of N error-free gates, and the error tolerance is given by a number $ \epsilon $, the simulation circuit will be composed of\\
$ \mathcal{O} \left({{\mathsf{poly}}}\left(\log \frac{N}{\epsilon }\right)N\right) $ gates.

The larger the ideal circuit, the more gates it takes to simulate it. The smaller the error tolerance $ \epsilon $, the more gates it takes to reach such a level of precision. The threshold probability does not provide any information about how many gates are necessary to simulate the ideal circuit.

\item  Comparing Error-Correcting Circuits; Consider a given error correcting circuit A with a threshold probability larger than the threshold probability of another error correcting circuit B. Which of the following sentences is correct?

\begin{itemize}
\item A can tolerate working with gates with a larger error probability than B.
\item A can operate with gates with lower error probabilities than B. Those gates would not work on B.
\item There is no way that B can achieve the same level of error tolerance as A.

\end{itemize}
Solution:\\
A can tolerate working with gates with a larger error probability than B. However, both A and B can operate with arbitrarily low error probabilities. Provided that the single gate error probability is smaller than the threshold probability of B, A and B can simulate an ideal error-free circuit with the same level of precision.

\item  Threshold Theorem Equation; The threshold theorem can be summarized by the following equation,
\begin{equation}\label{eq4_46}
\frac{P_{fail}}{P_{th}}=\left(\frac{P}{P_{th}}\right)^{2^ k}.
\end{equation}
Which of the definitions below is correct? 
\begin{itemize}
\item $ P_{fail} $ is the probability of failure of the fault-tolerant procedure for a single gate.
\item P is the failure probability of an individual gate.
\item $ P_{th} $ is the threshold probability that sets the maximum value of the individual-gate failure probability for which the procedure can simulate, with arbitrary precision and error-free circuit.
\item k is the maximum number of times that the entire procedure can be performed.
\end{itemize}
Solution:\\
The term $ k $ is the level of recursions, that is, the number of times that a set of blocks is made into one larger block. For example, if a circuit is composed of eight gates, then the first level of recursion, $ k=1 $, is given by the same circuit, the second level, $ k=2 $, is given by eight blocks of eight gates each, the third level of recursion, $ k=3 $, is given by eight blocks each block composed of eight blocks composed of eight gates, and so for.

\item 5-qubit and 7-qubit Codes; Consider the section ``5-qubit and 7-qubit Codes''. Which of the following statements are correct?
\begin{itemize}
\item The Hadamard (H) gate, the S gate, and the Controlled-NOT (CNOT) gate are all examples of Clifford gates.
\item It is not possible to do universal quantum computation by operating solely within a single stabilizer code.
\item A transversal gate that commutes with every generator of the stabilizer group S is an element of the normalizer of S.
\item It is not possible to define a transversal Controlled-NOT (CNOT) gate for 5-qubit codes.
\end{itemize}

\item  Fault-Tolerance Elements; Which of the following are not required to realize fault-tolerant computation? 
\begin{itemize}
\item Fault-tolerant state preparation.
\item Fault-tolerant gates.
\item Fault-tolerant measurements.
\item They are all necessary elements of fault-tolerant computation.
\end{itemize}
Solution:\\
Initialization, operation, and measurement must all be fault-tolerant.

\section{Quantum Computation vs Analog Computation} 
In the early days of quantum computing, after Shor’s factoring algorithm came out, there were arguments on this, but there is something called Usenet, which was like an early Twitter. There were arguments on Usenet about whether quantum computers could ever be realized, because of noise. They said, well, these look very powerful, but we have already known that in theory, analog computers are very powerful. These amplitudes of quantum computers, they just look like the real numbers that we see in analog quantum computers. Thus, we should not be surprised that this model is more powerful, but we know that analog computers cannot be realized, at least not to very high precision, because if the voltage if all of the voltages are legal values right, in a digital computer, it is either 0 volts or 5 volts. Everything else is illegal. So, if the computer sees 4.9, it knows it should bump it up to 5. However, in an analog computer, if 3.7 is a legal voltage, and 3.70001 is also a legal voltage, there is nothing the computer can do to know if there is a little drift of voltage, if that was supposed to happen, or if it was something that should be corrected. So, analog computing is fundamentally not error correctable. In some cases, that is fine. We know, if our stereo does not produce exactly the right sound, then that is fine. However, if we want to use these real numbers for some complicated math calculations, it is not so good. Thus, there was that argument given that quantum computers suffered the same limitation as analog computers; namely, we cannot error correct amplitudes. Indeed, if we have some error that looks like e to the i $ \theta $ z, and just as a notation, by the way, we going to write x equals $ \sigma $ x, y equals $ \sigma $ y. From now on, we are going to use this kind of quantum computing notation for the poly matrices. So, we have this error, in the e to the i $ \theta $ z, $ \theta $ can continuously vary. It could be very, very small. We know, we cannot distinguish all those different errors. So, how can we correct them? That was a compelling argument against quantum computers before quantum error-correcting codes came along. We think that it was this Usenet discussion that was part of what pushed Peter Shor and others to develop quantum error-correcting codes. The answer we have is that e to the i $ \theta $ z is a linear combination of i and z. So, if we have a code that can correct i and z, then we can also correct at no extra cost, all linear combinations of them. So, we can correct e to the i $ \theta $ z, for all values of z. So, it is a remarkable and crucial feature of quantum error correction that the space of errors they correct is a linear subspace. 

We have discussed several examples of quantum error correction protocols. These protocols generally use syndrome measurements to detect if and where an error occurred. Detected errors can then be corrected using a ``feed-forward'' approach, in which pulses are applied to the errant qubits to correct the error. Because these protocols actively look for and correct errors, this approach is referred to as active error correction.

Implementing active error correction, particularly in a fault-tolerant manner, is resource-intensive, requiring significant overhead in terms of ancilla qubits, quantum operations, and measurements. The number of overhead decreases as the physical qubit error rates are reduced, and so it behooves us to use every tool available to mitigate errors in physical qubits before using them in logical qubits.

Passive error mitigation is one means to reduce physical qubit error rates\cite{kandala_error_2019}. In this case, specific pulse sequences are designed to either mitigate systematic qubit errors (e.g., under-rotation or over-rotation of the Bloch vector due to a systematic miscalibration), or to dynamically ``undo'' coherent errors (e.g., coherent dephasing errors). The pulses used are generally of the same type, e.g., $ \pi $-pulses, like those used in the execution of quantum algorithms \cite{de_ridder_quantum_2019}. However, their purpose is to counteract typical or likely errors. It is ``passive'' in the sense that we take preventative measures to counteract certain systematic or stochastic noise that we know are likely to exist in our system. However, it is not ``active,'' because we do not identify and correct specific errors.

Note that passive error mitigation can only address certain types of systematic and coherent stochastic errors; it will not mitigate all errors., for those errors that do remain, active error correction is required to fix them. However, by implementing passive error mitigation protocols, we can reduce the physical qubit error rate, and this translates to reduced overhead requirements when implementing active error correction protocols.

\section{Correcting Systematic Control Errors} 
A remarkable fact about quantum systems is that errors can be corrected simply by changing the control sequence as long as the errors are systematic, instead of random \cite{bardin_28nm_2019}. How this works can be illustrated by recalling the dynamics of single qubits. For example, coherent light applied to a two-level ion causes evolution under this Hamiltonian \cite{bravyi_tapering_2017}, where two degrees of freedom stand out, the laser amplitude and the laser phase, $ \phi $. Combined with the time applied, we get the angle of rotation affected by the laser pulse $ \theta $. Illustrated on the block sphere, a pulse with $ \theta $ equal to and $ \phi $ equal to 0 gives this quantum NOT gate. The problem is that these physical control prime may have systematic errors. Specifically, it is typical that the amplitude and phase may have some fixed offset error epsilon in all pulses applied. Let us consider just amplitude errors, which result in qubit over or under rotation. Denote the actual rotations implemented as M. Without this error, this would be the ideal rotation R of $ \theta $, but the error epsilon indicates an extra rotation that is undesired. So, the implemented gate has an overall error of order epsilon. Keep in mind that this error is unknown, but fixed. The effect of this error can be quantified with a quantum response function, which we represent here as the transition probability between 0 and 1 for the $ \pi $ pulse as implemented. Plotting this response as a function of epsilon, we see that it is just a cosine squared function. Here is a short animation sequence illustrating the error. These pulses are like driving a car on a sphere. On the left is Schrodinger, who controls his velocity very accurately. On the right is his cousin, Guido, who has a lead foot, and always pushes his pedal 15\% more than he should, although we do not know that it is 15\% versus 17\%. Starting the animation, we see Schrodinger drive precisely from start to the South Pole, whereas Guido overshoots by quite a bit. This error could possibly be compensated by calibration, but there is a more robust procedure, which can remove the error without knowing how large it is. This solution is known as composite pulse sequences or composite quantum gates, and it replaces a single pulse with multiple rotation pulses about different axes of rotation. For example, one of the most famous early pulses from nuclear magnetic resonance pioneer, Malcolm Levitt, is these three pulses 90, 180, 90 sequences about axes x, y, and x, which effects a robust $ \pi $ pulse with error epsilon squared instead of epsilon. Its quantum response function has this flat region for small epsilon, indicating its leading order insensitivity to error, which compares favorably to the single pulse case. Let us return to Schrodinger and Guido driving on the block sphere and see what happens with this 90, 180, 90 sequence. We see Schrodinger reaches the y-axis exactly after the first 90, but Guido overshoots. Due to the overshoot, Guido rotates backward when applying the middle 180-degree pulse. Whereas Schrodinger just rotates in place. The final 90 brings Schrodinger exactly to the South Pole, and this time, Guido gets very close because of the backtracking. Can one do better? Here is a 90, 360, 90 sequence, which involves rotations about some unusual axis, x and then an axis at 120 degrees in the XY plane, and then x, which is the desired $ \pi $ rotation up to epsilon cubed. This sequence has a quantum response function shown here in red, much flatter with respect to error than the 90, 180, 90, and the single pulse. On the block sphere, here are Schrodinger and Guido driving the 90, 360, 90 sequences. First, we have the 90, which Guido overshoots, but Schrodinger does not. The 360 takes Schrodinger back to the y-axis, but causes Guido to backtrack, compensating for his overshoot. Finally, the third pulse, a 90, brings both to the South Pole. Can one do even better? Here is a four-pulse sequence, known as BB1 with three unusual axes of rotation defined by this arc cosine expression, which has even less error. Indeed, its quantum response, shown here in green, is even flatter than all previous pulse sequences we have discussed at BB1 is also notable in that it performs a high-fidelity $ \pi $ pulse no matter what the initial state is\cite{harty_high-fidelity_2014,ballance_high-fidelity_2016,wang_high-fidelity_2020,gaebler_high-fidelity_2016}, whereas some other sequences only work when the initial state is at the North Pole. We have seen three examples of how sequences of pulses can correct amplitude errors without knowing the error. With fidelity, which increases with pulse sequence link, the ability to correct systematic errors using gate sequences is very useful in practical quantum computation.

\item  Composite Pulse Sequences; The use of composite pulse sequences can help reduce the system response to systematic errors in single quantum gates without specific information about how large the errors are. In this procedure, a single pulse is replaced with multiple pulses with different axes of rotation, in general.

Let us recall that an ideal qubit rotation about the $ \hat{n} $-axis at an angle $ \theta $ is given by
\begin{equation}\label{eq4_47}
R_{\hat{n}}(\theta )=\exp \left(-i\frac{\theta }{2}\hat{n}\cdot \vec{\sigma }\right),
\end{equation}

with $ \hat{n}\cdot \vec{\sigma }=n_ xX+n_ yY+n_ zZ $, where $ n_ x $, $ n_ y $, and $ n_ z $, are the x, y, and z components of $ \hat{n} $.

Similarly, a non-ideal qubit rotation about the $ \hat{n} $-axis at an angle $ \theta $ is given by
\begin{equation}\label{eq4_48}
M_{\hat{n}}(\theta )=\exp \left(-i\frac{\theta (1+\epsilon )}{2}\hat{n}\cdot \vec{\sigma }\right),
\end{equation}

Where $ \epsilon $ is a systematic rotation error.

According to the section above, which of the following pulse sequences implements a $ \pi $-pulse about the x-axis with the smallest net error in the system response?
\begin{itemize}
\item The sequence of two pulses, 90-180, implemented as one rotation about the x-axis and one about the y-axis,\\
\begin{center}
$ T=M_ y\left(\frac{\pi }{2}\right)M_ x(\pi ) $.
\end{center}
\item The sequence of three pulses, 90-180-90, implemented as two rotations about the x-axis and one about the y-axis,\\
\begin{center}
$ U=M_ x\left(\frac{\pi }{2}\right)M_ y(\pi )M_ x\left(\frac{\pi }{2}\right) $.
\end{center}
\item The sequence of three pulses, 90-360-90, implemented as two rotations about the x-axis and one about the an unusual axis represented by $ \frac{2\pi }{3}$,\\
\begin{center}
$ V=M_ x\left(\frac{\pi }{2}\right)M_{\frac{2\pi }{3}}(2\pi )M_ x\left(\frac{\pi }{2}\right) $.
\end{center}

\item The sequence of four pulses called $ ``BB1_{\pi  }'' $ given by\\
\begin{center}
$ B(\pi )=M_{\hat{n_\phi }}(\pi )M_{\hat{n_{3\phi }}}(2\pi )M_{\hat{n_\phi }}(\pi )M_ x(\pi ) $,
\end{center}
where $ \hat{n_\phi }=(\cos \phi ,\sin \phi ,0) $ with $ \phi =\cos ^{-1}\left(-\frac{1}{4}\right) $. 
\end{itemize}
Solution:\\
The sequence of pulses
\begin{center}
$ U=M_ x(\frac{\pi }{2})M_ y(\pi )M_ x(\frac{\pi }{2}) $
\end{center}
implements $ R_ x(\pi ) $ with error $ O(\epsilon ^2) $. The sequence
\begin{center}
$ V=M_ x(\frac{\pi }{2})M_{\frac{2\pi }{3}}(2\pi )M_ x(\frac{\pi }{2}) $
\end{center}
is better, as it implements $ R_ x(\pi ) $ with error $ O(\epsilon ^3) $. The sequence
\begin{center}
$ B(\pi )=M_{\hat{n_\phi }}(\pi )M_{\hat{n_{3\phi }}}(2\pi )M_{\hat{n_\phi }}(\pi )M_ x(\pi ) $
\end{center}
is the best, as it implements $ R_ x(\pi ) $ with error $ O(\epsilon ^4) $.

\section{Dynamical Decoupling: Introduction}
In section three, we discuss that noise could be categorized into two primary types systematic noise and stochastic noise. In this section, we will focus on error mitigation strategies that address stochastic noise. Recall that stochastic noise is the random fluctuation of a parameter that is coupled to the qubit, and it can cause decoherence in a couple of different ways. As we may remember from earlier sections, a qubit in state 1 can relax to the ground state by losing energy to its environment with a rate of $ \gamma $ 1, which is 1 over the relaxation time, T1. we will assume here that the qubit only relaxes to its ground state, and that it is rarely excited out of the ground state by its environment. Then, there is also decoherence of the block vector on the equator, as characterized by the decoherence rate $ \gamma $ 2, which is 1 over the T2 time. On the equator, two things can go wrong. First, the qubit could simply relax back to the ground state via a T1 process. clearly, this is a phase-breaking event, since once the block vector has relaxed to the North Pole, there is no way to tell which direction it had been pointing when it was on the equator. second, the block vector can diffuse around the equator, and this is called dephasing. The decoherence time, T2, is a combination of both the dephasing time, T $ \phi $, and the relaxation time, T1. note that if there is no dephasing, such that T $ \phi $ becomes very large, then T2 equals twice T1 and is completely limited by relaxation. While energy relaxation processes are generally irreversible, dephasing processes may, in fact, be coherent, and therefore reversible. as we will see, coherent dephasing errors can be mitigated using dynamical decoupling pulse sequences, a type of passive error suppression\cite{klimov_fluctuations_2018}. To gain an intuition for dynamical decoupling, let us first consider a lacrosse player. If we have ever played lacrosse or watched a lacrosse game, we know that we need to run down the lacrosse field with a lacrosse ball in a basket that is at the end of our lacrosse stick. as we run, our body moves up and down, left and right, and this shakes the stick. This is the noise. Now, if we just hold the stick in front of us as we run, then we will find that the ball will very easily fall out of the basket due to the noise associated with our running. So, lacrosse players do not do that. Instead, they cradle the stick as they run down the field, back and forth, back and forth. this keeps the ball in the basket. importantly, they do this out of habit, without even thinking about it. They are not reacting to something specific. They are just continuously cradling the ball. by doing this, they decouple the ball from the noise generated by their running. this is an example of dynamical error suppression. Now, cradling addresses running noise, but it does not prevent a defender from running up behind us and knocking the ball out of our basket and onto the ground. Those kinds of errors still occur. when that happens, we have to stop, find out where the ball went, go over, and physically pick it up again. that would be analogous to active error correction. so, lacrosse cradling is one classic example of dynamical error suppression. Let us now see how it works with a qubit. we will take as an example a superconducting flux qubit\cite{xu_coherent_2016}. The flux qubit shown here is a superconducting loop interrupted by 4 Josephson junctions \cite{abdo_active_2019,bergeal_analog_2010,abdo_josephson_2011,}. Threading a magnetic flux through the loop sets the qubit energy level E01. However, due to local magnetic field fluctuations in the qubit's environment, the energy levels also fluctuate, and this leads to qubit dephasing. To see this, let us take a qubit to be in a superposition state $ \alpha $ 0 plus $ \beta $ 1. Because state 1 is at higher energy than state 0, it accrues phase at a faster rate. we can write this as $ \alpha $ 0 plus $ \beta $ e to the i $ \phi $ of t state 1, where $ \phi $ of t is equal to the qubit energy E0 divided by the reduced Planck's constant, h bar, times the time, t. This is called free evolution because the phase of the qubit evolves solely due to the energy level difference and not due to any external driving. Now, in the absence of noise, E01 is fixed, and so, the phase accrual is deterministic, meaning that at any time t in the future, we can tell we precisely the value of the phase. However, in the presence of noise, such as low-frequency flux noise due to the local magnetic environment of the qubit, the energy level separation will fluctuate. There is now an average value of the energy, E01 bar, corresponding to an average phase, $ \phi $ bar. However, due to the stochastic noise, there is also fluctuating energy, E01 tilde, which fluctuates in time. this corresponds to a fluctuating phase, $ \phi $ tilde of t. it is this stochastic phase which leads to dephasing. we can write an expression for the resulting dephasing in an intuitive form by looking at the ensemble average of the dephasing. Phase accrues in an exponent, as e to the i $ \phi $, and the phase at time t is the time integral of the energy fluctuation, E01 tilde. In taking the ensemble average, we will assume that the noise is Gaussian-distributed, that is, arising from a large number of weakly coupled magnetic noise sources, and this leads to the following intuitive result. First off, we have the change in energy due to the fluctuations, which we will assume here are fluctuations in the magnetic flux, $ \phi $. This is the sensitivity of the qubit energy to fluctuations in magnetic flux, and it is squared. Next is the strength of the magnetic flux noise that drives the energy fluctuations. This is the integral over frequency of the noise power spectral density, and it has units of $ \phi $ squared. By itself, this would simply be the variance of the noise process. importantly, the noise power spectral density is shaped by a term called the filter function. The filter function depends on the pulse sequence that we apply to the qubit, and it serves to filter, or window, the noise. So, to summarize, the noise seen by a qubit will drive fluctuations in the energy levels, and this leads to dephasing. However, the noise spectrum itself can be shaped by a filter function related to the pulses we apply to the qubit. in the next section, we will discuss at a few specific pulse sequences, their corresponding filter functions, and how they dynamically decouple the qubit from its noisy environment.

\item  Stochastic Noise; Noise can be categorized into two primary types: systemic noise and stochastic noise. Regarding stochastic noise, which of the following statements is correct? 
\begin{itemize}    
\item Stochastic noise is given by a fixed and unknown offset error $ \epsilon $.
\item A qubit in the excited state $ \lvert 1\rangle $ may spontaneously relax to the ground state $ \lvert 0\rangle $ by stochastically emitting energy to its environment.
\item When the Bloch vector of a qubit diffuses around the equator, this is called dephasing.
\item Dephasing processes are always irreversible.
\end{itemize}
Solution:\\
Stochastic noise is given by the random fluctuation of a parameter that is coupled to a qubit. Dephasing processes can be coherent and, therefore, reversible.

\section{Dynamical Decoupling: Free Evolution}

In the last section, we considered a superconducting flux qubit \cite{yan_flux_2016, gambetta_building_2017}, in the presence of magnetic flux noise, and asked how we could potentially dynamically mitigate the resulting qubit dephasing. To gain some intuition, we consider dynamical decoupling in the context of lacrosse. Where a lacrosse player can use cradling to effectively decouple the lacrosse ball, in the basket of a lacrosse stick, from noise due to running down the field. we then discussed at how environmental noise drives fluctuations in the qubit energy levels and calculated the resulting phase decay function. In doing so, we discuss that the power spectrum of the noise that drives these fluctuations could be shaped by a filter function that is related to the pulses we intentionally apply to the qubit. To see how this works in practice, let us discuss at a few examples. we will start with a Ramsey experiment. Here a $ \pi/2 $ pulse along the y-axis of the block sphere brings the qubit block vector down to the equator, pointed along the x-axis. we then wait, while the qubit undergoes free evolution. The simulation shown here represents an ensemble of instances of the qubit in the presence of low frequency 1 over f noise, taken to be quasi-static. That is, the noise is essentially fixed for any given instance, and it is different for different instances according to the sum probability distribution. we also operate in a rotating frame, such that the average phase, in the absence of noise, always points along the x-axis. However, in the presence of noise, the block vector rotates clockwise or counterclockwise around the equator. at different speeds, depending on the specific size of the noise offset for each instance. we then use a second $ \pi/2 $ pulses to rotate back to the z-axis. ideally, in the absence of noise, the qubit should now align with state one at the south pole. However, clearly, due to the noise, the block vector is largely depolarized. The way to understand this is as follows. The magnetic flux noise spectral density is a 1 over f-type noise, which is large at low frequencies and decreases as 1 over frequency. During the free evolution period, we do nothing. we do not apply any external control pulses, which we represent here as n equals 0 where n will count the number of pulses we apply. However, here, n equals 0, and the qubit it evolves in the gray region, for a time tao. During this period of time, since we assume the noise is quasi-static, the qubit energy levels are essentially fixed for each instance and vary from instance to instance. For a given instance, the gray region represents the energy versus time. it is a rectangle. the Fourier transform of a rectangle is a sinc function centered at zero frequency. Which is to say, applying no pulses leads to a filter function that is highly peaked at zero frequency. That is, the filter passes noise at low frequencies. for 1 over f noise, this is where the noise is the strongest. Thus, this is the worst strategy for noise suppression of 1 over f-type noise. Referring back to the lacrosse example, this is analogous to running down the field without cradling the ball. However, we could improve the situation by performing a spin-echo experiment. In a spin echo, a $ \pi $ pulse is applied during the free evolution period. The $ \pi $ pulse inverts the phase diffusion along the equator, such that spins that were moving away from the x-axis before the pulse are now moving towards it after the pulse. for a pulse that is located precisely in the middle at the free evolution period, the block vector refocuses and coalesces right when the second $ \pi/2 $ pulse occurs, taking the block vector down to the south pole, and state 1. clearly, the polarization is significantly improved by adding this single refocusing pulse. So, what did we do? Well, adding the $ \pi $ pulse now divides the evolution into half periods of free evolution. The first is represented by the gray rectangle, with a plus sign. the second with a minus sign, due to the inversion of the system by the $ \pi $ pulse. The filter function corresponding to these two rectangle functions is shown in green. we see that its passband is now peaked at a higher frequency. Since the 1 over f noise decreases with frequency, the spin-echo filter function effectively lets through less noise. Thus, the dephasing is reduced. Adding the $ \pi $ pulse is an example of dynamically decoupling the qubit from its flux noise environment. In the context of the lacrosse example, this is cradling while running. as we might expect, the situation is further improved with an additional $ \pi $ pulses. Adding two $ \pi $ pulses break the free evolution into three distinct periods, raising the filter passband to even higher frequencies and further reducing the dephasing due to 1 over f noise. Adding equally spaced $ \pi $ pulses is called a CPMG sequence, named after Carr, Purcell, Meiboom, and Gill, who collectedly developed it. As the number of $ \pi $ pulses, n, is increased, the filter function has shifted to higher and higher frequencies. this reduces the dephasing rate more and more. as the dephasing is improved, the decoherence here in the time, t2, approaches the ideal value of twice t1. Finally, one can use pulse sequences and their corresponding narrow bandpass filters to sample an unknown noise power spectral density, as seen by the qubit. This is called noise spectroscopy. it essentially uses the qubit and the applied pulses as a noise spectrometer. Changing the number of pulses shifts the narrow filter around in frequency. by measuring the dephasing response of the qubit, one can reconstruct the noise spectral density causing the dephasing. Then, with knowledge of the specific form of the noise spectral density, one can, in turn, design a pulse sequence that corresponds to a noise filter that is optimal within certain specified constraints.

\item  CPMG Sequences; The effect of certain types of stochastic noise during free evolution can be reduced by applying $ \pi $ pulses about the x-axis (or y-axis). In particular, adding equally spaced $ \pi $-pulses is called a CP or CPMG sequence, depending on the rotation axis relative to the orientation of the Bloch vector on the equator (or, similarly, of the initial $ \pi /2 $ pulse used to reach the equator).
Which of the following statements describes the effect of this pulse sequence on the system? 

\begin{itemize}
\item It results in a filter function that windows the noise power spectral density.
\item It reduces the amplitude of the noise in the environment.
\item It generally reduces dephasing for environments with noise power spectra that decrease as frequency increases.
\end{itemize}
Solution:\\
Applying pulses to the qubit does not generally reduce the amplitude of the noise in the environment. Rather, the pulse sequences correspond to filter functions in the frequency domain that will window the environmental noise spectral density. For noise power spectra that reduce with frequency, for example, the $ 1/f $ noise discussed, adding more pulses will push the filter centroid to higher frequencies and, therefore, result in less noise as seen by the qubit.

\section{Dynamical Decoupling: Driven Evolution}
In the last section, we discussed at several examples of qubit dephasing during free evolution. We discuss that by applying $ \pi $ pulses during the free evolution period, we could effectively decouple the qubit from its noisy environment. In this section, we will discuss at dynamical error suppression during driven evolution and, in particular, a rotary echo, the driven evolution analog to the spin echo. To begin, let us discuss at a typical example of driven evolution called a rabi oscillation. The qubit starts in its ground state, and we then apply a continuous pulse to the qubit along the x-axis. Now ideally, the qubit would oscillate sinusoidal from the North Pole to the South Pole and back again, rotating around the driving field represented by the red arrow. The oscillation frequency is proportional to the driving field amplitude. Thus, in the absence of noise, the rotation occurs at a fixed frequency. However, as we discuss in section three, if the amplitude of the driving field is noisy, the qubit oscillation amplitude decays with time. This is illustrated on the block sphere by an ensemble of instances of the Rabi oscillations, assuming quasi-static noise on the driving field amplitude. For a given instantiation, the driving amplitude is fixed, but that amplitude changes with the noise for each different instance. The result is an ensemble of frequencies, which, as we showed in section three, leads to a decay of the resulting Rabi oscillation. This situation can be mitigated by using a rotary echo. In a rotary echo experiment, the driving field is reversed halfway through the driven evolution period. Again, assuming quasi-static noise, each instance drives the qubit at a given rate. Reversing the drive, no matter what amplitude the instances, will undo the unwanted over or under rotation, and this leads to a refocusing or revival of the block vector back at the North Pole. Just as dynamical decoupling during free evolution could be improved by moving from a single pulse spin echo to a multi-pulse CPMG protocol, for noise that decreases with frequency, a single reversal rotary echo can be extended to a multiple field reversal analog to CPMG. As the number of reversals increases, the corresponding filter function moves to higher and higher frequencies. In this case, the filter function shapes the noise spectral density of the driving field amplitude fluctuations. Dynamical decoupling during free evolution and driven evolution are examples of error suppression techniques that are called open loop. That is, they do not feed-forward to apply certain pulses based on measurement inputs, as is done with error correction. Rather, these are pulses or pulse sequences that are simply added to or built into the gates used to implement quantum logic. In doing so, the error rates due to coherent reversible errors are reduced. This, in turn, is quite useful. By using the same types of pulses that we already use to implement quantum logic, such as $ \pi $ pulses, we can reduce the effective error rates of physical qubits and thereby help to reduce the overhead required to implement full-blown error correction. 

Quantum systems interacting with their environment are referred to as open quantum systems. We need open systems in order to apply external controls pulses that enable us to implement quantum control. However, the environment also introduces noise that acts to decohere the quantum system. Ideally, we would like a means to faithfully apply the control pulses we want to reach our qubit, while, at the same time, decouple the qubit to the extent possible from the noisy environment. Although errors can be corrected with active quantum error correction protocols, as we have discussed, active error correction is relatively expensive. Dynamical decoupling pulse sequences act to decoupling the qubit from its noisy environment and thereby mitigate errors. Although these pulse sequences cannot mitigate every type of error, they do reduce the physical qubit error rate, and this, in turn, reduces the level of overhead required to implement active error correction for those errors that remain.

In a previous section, we were introduced to the spin echo, also known as the Hahn echo, proposed by Erwin Hahn in 1950. A spin-echo comprises a single $ \pi $-pulse applied to the qubit, and it mitigates dephasing noise. As introduced, the pulse sequence for a spin echo is $ (\pi /2)_ y\xrightarrow {\tau /2}\pi _ y\xrightarrow {\tau /2}(\pi /2)_ y $, where the $ \pi /2 $-pulses are used to bring the qubit Bloch vector from the z-axis to the equator where dephasing during free evolution occurs for a duration of time $ \tau $ and back again to project the resulting dephasing of the Bloch vector back onto the measurement axis (z-axis). The $ \pi $-pulse is used to refocus the Bloch vector dephasing during the free evolution period. Note that in this example, and in the absence of noise, the $ \pi $-pulse is applied to an axis perpendicular to the Bloch vector on the equator.

The extension of the above spin-echo sequence from a single $ \pi $-pulse to multiple $ \pi $-pulses is called a CP pulse sequence, named after Herman Carr and Edward Purcell. If we count the number N of $ \pi $ pulses, the above spin-echo sequence is identical to a CP pulse sequence with N=1. However, with the CP sequence, the number N can, in general, be larger, and as we have seen, adding more $ \pi $-pulses tends to reduce dephasing due to environmental noise spectra that decrease with frequency. As with the spin-echo above, the CP sequence applies the $ \pi $-pulses to an axis perpendicular to the Bloch vector, effectively toggling the Bloch vector between opposite sides of the Bloch sphere. It turns out that this approach is relatively sensitive to pulse errors, and these errors accumulate over the course of the pulse sequence.

In the section, we discussed a more robust gate sequence called the CPMG sequence, named after Herman Carr, Edward Purcell, Saul Meiboom, and David Gill. Meiboom and Gill pointed out that applying the $ \pi $ pulses co-aligned with the Bloch vector rather than perpendicular to it has the advantage that the overall approach becomes less sensitive to pulse errors. For example, a CPMG sequence with two $ \pi $-pulses would be $ (\pi /2)_ y\xrightarrow {\tau /4}\pi _ x\xrightarrow {\tau /2}\pi _ x\xrightarrow {\tau /4}(\pi /2)_ y $. The CPMG sequence is still employed today, due to the simplicity and robustness of its pulse sequence.

The CP and CPMG sequences are examples of a broader class of sequences called dynamical decoupling sequences. These sequences serve to mitigate coherent errors by effectively decoupling the system from its noisy environment. As we discussed, these pulse sequences in the time domain can be viewed as noise filters in the frequency domain. The shape and frequency passband of the filter depends on the number of pulses and the duration of the sequence. Provided the noise-induced dephasing dynamics are slow over the duration of the pulse sequence, dynamical decoupling can mitigate errors by ``refocusing'' the errant dynamics. The primary limitation of dynamical decoupling sequences is the accumulation of pulse-related errors. The pulses themselves are not perfect, and this leads to imperfections in the implementation of the sequence. As we discussed above with the CP and CPMG example, different sequences can have different levels of sensitivity to pulse errors.

To close, we will give two more examples of pulses sequences that are an extension of CPMG. The first sequence, XY-4, comprises four $ \pi $-rotations alternating between the x and y-axes. The combination of rotations around two orthogonal axis reduces the sensitivity to pulse imperfections, and it is particularly useful in cases where the Bloch vector orientation is not known. XY-4 can be further improved by adding an additional 4 $ \pi $-rotations, but in the reverse order. The resulting decoupling sequence is referred to as XY-8.
\begin{equation}\label{eq4_53}
\displaystyle XY-4    \displaystyle \equiv    \displaystyle (\pi /2)_ y\xrightarrow {\tau /8}\pi _ y\xrightarrow {\tau /4}\pi _ x\xrightarrow {\tau /4}\pi _ y\xrightarrow {\tau /4}\pi _ x\xrightarrow {\tau /8}(\pi /2)_ y    
\end{equation} 
\begin{equation}\label{eq4_53_1}
\begin{split}
\displaystyle XY-8    \displaystyle &\equiv    \displaystyle (\pi /2)_ y\xrightarrow {\tau /16}\pi _ y\xrightarrow {\tau /8}\pi _ x\xrightarrow {\tau /8}\pi _ y\xrightarrow {\tau /8}\pi _ x\xrightarrow {\tau /8}\pi _ x\\& \xrightarrow {\tau /8}\pi _ y\xrightarrow {\tau /8}\pi _ x\xrightarrow {\tau /8}\pi _ y\xrightarrow {\tau /16}(\pi /2)_ y
\end{split}    
\end{equation}
      
The advantage of XY-4 over CPMG and CP is its robustness to pulse related errors. Suppose a qubit is initialized in its ground state, pointing to the Bloch sphere's north pole. Then, the first $ \pi /2 $ pulse rotates the qubit around the y-axis. The resulting qubit orientation is ideally aligned with the x-axis. Now, suppose there is a constant over-rotation of each pulse. For a simple CP sequence, the over-rotation of each pulse accumulates and thus may induce more errors than mitigated. In contrast, the CPMG sequence is able to counteract the over-rotation of the $ \pi /2_ y $ pulse due to the $ \pi _ x $ rotation. The $ \pi _ x  $rotation causes the over-rotation of the initial and final $ \pi /2_ y $ rotation to compensate almost entirely. Nevertheless, the over-rotation of the $ \pi _ x $ remains. Therefore, the CPMG sequence is more robust than the CP sequence but is not entirely free of any pulse related errors. The XY-4 pulse sequence can help compensate constant imperfect rotations around the x-axis as well. We encourage we to review the rotations comprising the XY-4 pulse sequence and show that a constant over or under-rotation can be largely mitigated. XY-8 and more complicated sequences can reduce pulse related errors even further and are hence even more robust to pulse related errors. The trade-off is that more pulses within a fixed amount of time push the pulses closer together. At some point, increasing the duration of the sequence may be required to accommodate all of the pulses. In turn, longer pulse sequences take more time, and the system is still susceptible to incoherent errors that dynamical decoupling cannot protect against; these errors will still occur and will become more likely as the overall pulse length increases.

\begin{figure}[H] \centering{\includegraphics[scale=.18]{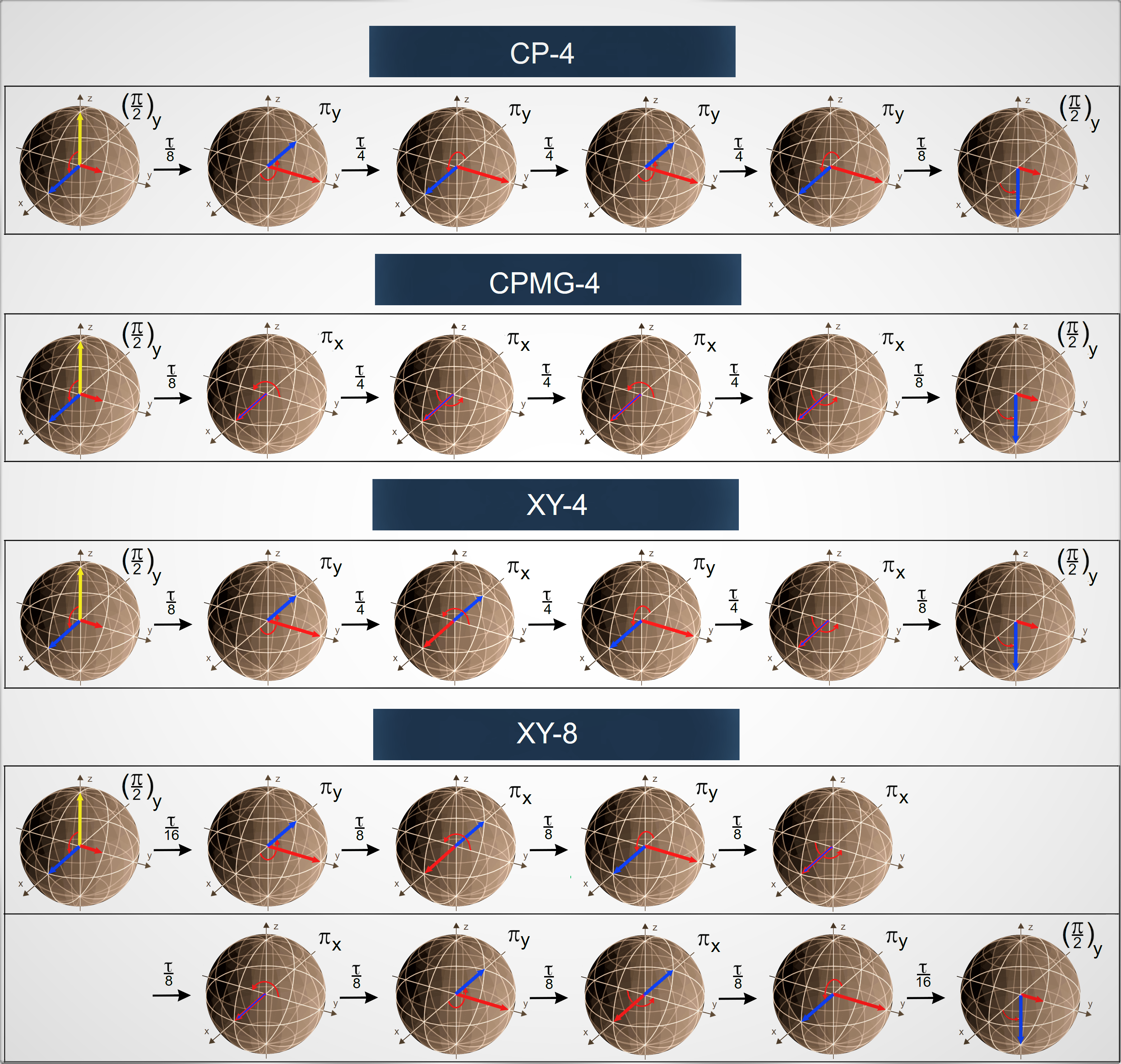}}\caption{Dynamical decoupling sequences: Shown are CP with four $ \pi $ pulses, CPMG with four $ \pi $ pulses, XY-4, and XY-8. All dynamical decoupling sequences start with the qubit in the ground-state, indicated with the yellow arrow pointing to the north pole on the Bloch sphere. The red arrow indicates the applied field used to implement the pulses. The blue arrow represents the Bloch vector as it rotates around the Bloch sphere.}\label{fig4_14}
\end{figure}

\item  Rabi Oscillations; A Rabi oscillation is a driven rotation of the Bloch vector. It is the driven-evolution analog to the oscillations observed during the free evolution of a frequency-detuned Ramsey experiment. Consider a Rabi experiment and a Ramsey experiment, where all pulses are applied along the x-axis. For the Ramsey experiment, the frequency detuning is defined as the difference between the driving frequency of the $ \pi/2 $ pulses and the qubit frequency. Assume we observe the experiments in a reference frame rotating at the qubit frequency.
What are the similarities and differences between these two types of experiments?

\begin{itemize}
\item The Rabi frequency is proportional to the driving amplitude, whereas the Ramsey frequency is proportional to the frequency detuning.
\item The Bloch vector in both cases rotates around the z-axis of the Bloch sphere.
\item Decay of the oscillations in both cases can be caused by stochastic noise.
\end{itemize}
Solution:\\
During Rabi oscillation, the state of the qubit rotates around the driving-field axis, in this case, the x-axis of the Bloch sphere.

\section{Fault-Tolerant Quantum Computation with Superconducting Qubits: The Surface Code and How it Works} 

Let us say we want to build a reliable quantum computer. We are going to need a way to detect and handle errors. The circuit churn is a generic way to do this, test the parity of a set of qubits, and note when the value of the parity changes. A change in value is called a detection event. Detection events indicate the nearby presence of errors. The surface code uses this idea in two dimensions\cite{jones_layered_2012}. The corners of every square correspond to data qubits we wish to protect. The squares and half-circles correspond to measure qubits that perform parity checks on the data qubits they touch. Every data qubit is touched by at least one white shape. If we perform parity checks on just these white shapes, we can detect all bit-flip errors. Quantum computers can also suffer phase flips, and a simple way of coping with this is to apply Hadamard gates to data qubits to convert phase flips into bit flips. One can then perform parity checks on the dark shapes to detect these, then finish with Hadamard gates on all data qubits again. This, when repeated, is the surface code approach to error detection. To first order, these checks are sufficient to detect all errors. Other effects such as leakage, dynamic loss of gate calibration \cite{motzoi_simple_2009,muller_interacting_2015}, and power outages can and must be handled using other techniques, such as periodic resetting, continuous calibration, and batteries and backup generators. So, why detect errors in this manner? Firstly, only nearest-neighbor interactions are required making this approach well-suited to many solid site technologies. Secondly, only about a half dozen gates are required to generate a bit of information about local errors. This means these bits are very likely to be reliable. This means a high error rate of order 1\% is, in principle, tolerable. No other error detection scheme tolerates as much error when restricted to nearest-neighbor interactions. In practice, the lower the error rate of physical quantum gates, the better error detection works. For the surface code, a gate error right around 10 to the minus 3 leads to rapid suppression of error as the size of the code is increased. While this low level of error has not been achieved in a scalable system yet, it seems tantalizingly within reach. Multiple groups around the world are racing towards this goal. 

\section{Fault-Tolerant Quantum Computation with Superconducting Qubits: Implementing Gates on the Surface Code} 

In the first two of the previous three sections, we introduced surface code logical qubits measuring multi-body logical operators and constructing both simple quantum gates and more complex computations, such as state distillation. The previous section presented a discussion of the challenges of implementing the surface code, using superconducting quantum devices. Ultimately, all of these challenges boiled down to the simple fact that we cannot currently execute physical gates with sufficiently low error to successfully implement the server's code. No meaningful scale-up can occur until error rates can be made sufficiently low to do this and kept sufficiently low as the system is scaled. In this section, we turn the attention to the classical processing associated with a quantum computer using the surface code. It is reasonable to expect that a superconducting quantum computer could execute a single round of surface code error detection in approximately one microsecond. Approximately half of the qubits in the quantum computer will be measured every round, each generating a classical bit of data. A large-scale superconducting computer with 100,000 to one million physical qubits, enough to perform quantum chemistry calculations \cite{reiher_elucidating_2017,hempel_quantum_2018,bassman_towards_2020} that would be intractable on a classical supercomputer, would, therefore, generate of order 100 billion to one trillion classical bits of information post-second. At a physical layer rate of approximately 0.1\%, roughly 1 in 20 of these bits will indicate the detection of nearby quantum errors. The most widely studied method of handling these detection events is to approximate quantum errors as only forming chains that can lie along with certain space-time directions\cite{paini_approximate_2019}, as shown on the screen. This image is intended to represent 2D hardware in the horizontal plane and time running vertically. Each layer of the structure represents a round of error detection. Each great cylinder is a directional long which a chain of errors can form. Red dots represent detection events, generated from the classical input bits. The 3D cylinder structure can be pre-computed classically. We call this structure a lattice. Given a lattice with detection events, we can choose paths connecting detection events in pairs or connecting them to the nearest boundary such that the total path length is minimal. The algorithm that does this is called minimum weight perfect matching. Choosing which pairs of detection events to match can be done by treating the cylinders like water pipes with water flowing more quickly along fatter pipes. Pouring water into a detection event creates an exploratory region. When two exploratory regions touch only each other, the associated detection events can be matched. This relatively simple operation is well-suited to implementation in an FPGA\cite{chen_quantum_2013}. More complicated cases with multiple colliding regions would be handled in CPUs. We have found that a single core of a modern CPU is sufficient to handle the output of approximately 100 physical qubits. This still leaves one planning a 1,000 to 10,000 core machine just to handle the error correction. It is not enough to use a standard operating system, which can pause processing on any core for multiple milliseconds. Instead, a custom real-time operating system is required and very likely custom hardware for CPU to CPU communication, to achieve very low latency processing. Going back to state distillation and recalling that time runs vertically, processing must be low latency. Ideally, sub 10 micro-second as every green box in the structure is included or not based on decisions made on corrective measurement results. If classical processing were high latency, the quantum computation itself would need to be slowed down by inserting additional rounds of quantum error detection. In summary, an error correction system for a large-scale quantum computer, one capable of performing classically intractable quantum chemistry calculations\cite{motta_low_2018,babbush_low-depth_2018,bassman_towards_2020,hempel_quantum_2018,mccaskey_quantum_2019,lanyon_towards_2010,li_efficient_2017}, is envisaged to consist of thousands of CPUs and FPGAs connected in a custom manner to achieve sub-10-microsecond latency processing. This level of performance is required to avoid significant additional error detection overhead. Research towards such a system is underway.
 
\section{Fault-Tolerant Quantum Computation with Superconducting Qubits: Challenges with Large-Scale Systems} 
we have discussed detecting bit-flip errors using the white shapes of the surface code and phase flip errors using the dark shapes. Long chains of bit flip, as shown in red, or phase flips, as shown in green, are, however, undetectable. Studying the red line, we can see that there are two-bit flips touching each adjacent white shape, so, no local parity change and no detection. This gives us the opportunity to store and protect data. If we start all data qubits in zero then start detecting errors, we will create and protect a logical zero state. If instead, after starting all data qubits in zero, we bit flip those along the red line then start detecting errors, we will create and protect a logical one state. Logical plus and minus can be prepared similarly, starting with old data qubits in plus and phase flipping along the green line. With logical initialization out of the way, we move on to a logical measurement. A logical measurement in the z basis corresponds to measuring each individual qubit in the z basis. After the many results of this transversal measurement are corrected, the overall logical measurement will be the parity of the measurements along the green line. Similarly, logical x-basis measurement is implemented using transversal x-basis measurements. we are not restricted to measuring individual logical qubits in the x or z basis. we can also measure multi-logical qubit operators such as logical zz, as shown. Logical zz corresponds to the parity of the two logical qubits. The large-scale fault-tolerant version of what one of the white half circles does to a pair of physical qubits. Indeed, the logical parity is measured by combining the parity of many white shapes. The parity of all the white shapes marked with blue dots gives us the parity of the two green lines. With a bit of extra effort, we can also measure logical operators like x, x z. we will notice that some shapes are half white and half dark. This just means the conversion from bit flip detection to phase flip detection using Hadamard gates is done at different times on different qubits around the shape. Again, the multi-logical measurement reduces to taking the parity of a large number of local parities indicated with blue dots. What can we do with multi-body logical measurements? It turns out a lot. Take, for example, the familiar CNOT gate. This can be broken into four gates, initialization to 0, measure xx, measure zz, then measure in the x-basis. All of these gates touch the central ancilla qubit with the CNOT control and target, top and bottom, respectively. Three additional single qubits can be seen, but these would be implemented in classical software\cite{larose_overview_2019}. The logical upright measurement approach can also be readily extended to give single control multi-target CNOT gates. For most of the remainder of this section, we will focus on explaining how to implement a logical T gate. T gates consume T states. So, really, we need to explain how to repair T states. The first stage of preparing a T state is an injection, which, using a method invented by Ying Li, involves starting with a single physical qubit and then turning on error detection in a particular way. If errors are detected early on, which happens about 50\% of the time at gate error rates around 10 to the minus 3, the injection files and must be repeated. This means that to have a high probability of success of injecting a T state at a particular place in a quantum computer at a particular time during an algorithm, we need enough space and time for multiple injection attempts. To keep images as simple as possible, an injection attempt will be represented by a red cuboid. The displayed image covers enough space and time up to 20 attempts. When an attempt succeeds, that logical qubit would be expanded to the full available space for more robust storage. Even when successful, the error rate of the logical T state is approximately the error rate of a single two cubic gate, so, not good enough for fault-tolerant quantum computation. Injected T states must be distilled to lower the error rate. There are many ways to distill T states. we will focus on 15 to 1 state distillation, meaning 15 injected T states are required to produce a single better T state. The circuit achieving this at first glance consists of many single controlled multi-target CNOT gates, which is great since we know how to do these with low overhead. However, the situation is really even better than this. The combination of plus state initialization followed by single control multi-target CNOT is equivalent to multi-qubit logical measurement. That is, we can replace every plus initialization with zero initialization, and each of the five columns of CNOTs with an 8-bodied logical x measurement. we are probably wondering what these five strange 3D objects are. Each one represents an 8-body logical x measurement. They are ordered left to right in exactly the same manner as the five columns of CNOT gates in the previous image. The central rectangle in each represents an ancilla patch performing the 8-body logical x measurement. The 16 cubes in each represent the 16 logical qubits in the quantum circuit. Those on the left represent the bottom eight qubits. Those on the right, the top eight in reverse order. So, the top qubit in the circuit is the bottom right cube. we should notice that not all of the cubes are interacting with anything in the left-most 3D object. Essentially, this means we are initializing logical qubits before we actually need them, which is pointless since it just provides extra time for errors to occur. Stacking these five objects one on top of another to represent time running vertically and removing unnecessary cubes and adding in state injection and a few extra pieces to perform the actual T gates at the end of the circuit, we get a nice engine-like picture of essentially 3D quantum assembly code designed to produce distilled T states. The distilled output of this engine would have an error rate of approximately 35P cubed, where P is the two-qubit physical gate error rate. For many algorithms, a single level of distillation would be insufficient. we would need to pipe the output of one level into a second level. While this might look very complicated, note that it is really just a bunch of multi-qubit x measurements and T gates that collectively output two low error T states. The two outputs correspond to the two white cubes closest to the front of the image. To give a sense of the scale of the image, if two-qubit gates had an error rate of 10 to the minus 3, approximately 150,000 physical qubits would be required. if a single round of error detection took one microsecond, the two outputs would be space around 200 microseconds apart. The quantum equivalent of an end gate requires four T states, so, nearly a millisecond to execute. Think about that the next time we hear someone say that quantum computers are fast. Fortunately, problems that exist with quantum computers have an exponential advantage. So, even this huge constant factor disadvantage can be overcome. Note also that if we can get the two-qubit gate error rate down to 10 to the minus 4, fewer than 15,000 qubits would be required, and the rate of distillation will be doubled. Given many qubits, we could also use many T state factories to achieve whatever distillation rate we desired to achieve faster quantum computation. To finish with a less complicated figure, what we see now is an example of how one might divide up a large square quantum computer into dark patches that represent protected logical qubits, a large open rectangle representing a T state factory, and white squares representing parts permitting logical qubits to be moved around and interacted. Note that the factory is a large but not dominant source of overhead. Not so, long ago, it was frequently necessary to devote 90\% or more of the area of a quantum computer to T state production to achieve a reasonable quantum algorithm run times. Then new visibly lower fraction reflects significant improvements in distillation and a shift in focus to quantum chemistry algorithms where massive overhead reduction has been achieved in recent years. 

\item  Surface Code Introduction; The surface code is a type of quantum error correction code. Which of the following are true for the surface code 
\begin{itemize}    
\item The surface code relies solely on nearest-neighbor interactions between qubits.
\item The surface code can detect and correct for both bit-flip and phase-flip errors.
\item The surface code error threshold is around 1\%.
\item The surface code is a leading candidate for solid-state quantum error correction implementations
\end{itemize}
Solution:\\
Due to the nearest-neighbor interactions, a relatively lenient threshold around 1\%, and the ability to correct for both bit-flip and phase-flip errors, the surface code is a leading candidate for solid-state implementations of quantum error correction.

\item  BitFlip Events; Referring to the surface-code qubit array shown in the section, assume that all data qubits are initialized in-state $ \vert 0 \rangle $. If each of the data qubits along a vertical line from the bottom edge to the top edge of the array are flipped to state $ \vert 1 \rangle $, then which of the following are true.
\begin{itemize}
\item These bit-flip events are detectable using the measure qubits since the surface code can detect bit-flip errors.
\item These bit-flip events are not detectable since each measure qubit touches two data qubits, and therefore there is no change in parity.
\item Because the vertical line of bit flips is not detectable as an error, and it can be used to encode logical $ \vert 1 \rangle $.
\item Flipping the vertical line of data qubits back to $ \vert 0 \rangle $ is also not detectable using the measure qubits, and this state is an encoding of logical $ \vert 0 \rangle $.
\end{itemize}
Solution:\\
A vertical line of data qubits that connects the bottom edge to the top edge of the qubit array has the property that each measure qubit next to the line will touch two data qubits. Therefore, if all data qubits in the line switch state, there is no detectable change in parity by the measure qubits. The same concept is true for a horizontal line connecting the left and right edges of the array. Note that if the line of data qubits stops short of an edge, then the measure qubit at the end of the line will touch only one data qubit and detect the bit flip as a change in parity. In this way, lines of data qubits connecting edges of the array are unique.

\item  Quantum Response Function; Consider an ideal qubit rotation about the x-axis at an angle $ \theta, R_ x(\theta ) $. If the angle of the rotation $ \theta $ is affected by a systematic error$  \epsilon $, then 
\begin{itemize}    
\item The implemented gate can be written as $ M_ x(\theta )=R_ x(\theta (1+\epsilon )) $.
\item The quantum response function for a $ \pi _ x $ pulse is given by the probability of transition between the ground and excited states when a $ \pi  $ pulse about the x-axis is applied to the system.
\item The quantum response function quantifies the effect of the systematic errors $ \epsilon $.
\item When there is no systematic error, $ \epsilon =0 $, the quantum response is equal to zero.
\end{itemize}
Solution:\\
The quantum response function quantifies the effect of systematic errors, and, for a $ \pi _ x $ pulse, it is given by the probability of transition between the ground $ \lvert 0 \rangle $ and exited $ \lvert 1 \rangle $ states when a $ \pi $ pulse on the x-axis is implemented, $ \left\vert \langle 1\rvert M_ x(\pi )\lvert 0\rangle \right\vert ^2 $, where the implemented gate can be written as,
\begin{center}
$ \displaystyle M_ x(\pi )    \displaystyle =    \displaystyle R_ x(\pi (1+\epsilon )) $\\         
$ \displaystyle =    \displaystyle \cos \left(\pi (1+\epsilon )\right)I-i\sin \left(\pi (1+\epsilon )\right)X $.
\end{center}
When the systematic error is zero, $ \epsilon =0 $, the quantum response is maximal and equal to one.

\item  Magnetic Flux Noise; In a superconducting flux qubit, the energy level separation between the ground and excited state, $ \lvert 0\rangle $, and $  \lvert 1\rangle $, can fluctuate due to local magnetic field fluctuations in the qubit environment., the energy difference determines the rate at which relative phase accrues between the components $ \lvert 0 \rangle $ and $ \lvert 1 \rangle $.

Which of the statements regarding the impact of magnetic flux noise in a flux qubit is correct?
\begin{itemize}
\item Fluctuations in the magnetic field translate to qubit energy fluctuations.
\item Fluctuations in the magnetic field translate to fluctuations of the relative phase accrual rate during free evolution.
\item Fluctuations of the relative phase during free evolution lead to dephasing.
\item All of the above.
\end{itemize}
Solution:\\
An external magnetic field sets the energy difference between the states of a flux qubit. Thus, fluctuations in the magnetic field translate to qubit energy fluctuations. The oscillation frequency during the free evolution of a qubit in the equator of the Bloch sphere is set by the energy difference between the states of a flux qubit. This is why fluctuations in the magnetic flux translate to fluctuations of oscillation frequency during free evolution. Fluctuations of oscillation frequency during free evolution creates dephasing when averaging over an ensemble of qubits, or over several implementations of the experiment on one qubit.

\item  Ramsey Experiment; A Ramsey experiment is sensitive to the energy difference between the two levels of a qubit or an atom., recall that energy is proportional to frequency through Planck's constant, E=hf. Thus, a Ramsey experiment can form the basis for ultra-precise atomic clocks\cite{delehaye_single-ion_2018,huntemann_single-ion_2016} by monitoring the frequency of an atomic transition with stable energy levels. The current metrological definition of a second is 9.192.631.770 cycles of two particular states of a Caesium atom in a Ramsey experiment.

Consider a Ramsey experiment, in which the qubit (atom) is prepared in-state $ \lvert 0\rangle $. A $ \pi /2 $ pulse along the $ \sigma _ x $ axis brings the Bloch vector to the equator. In the qubit frame of reference (that is, not using a rotating reference frame, the Bloch vector will rotate around the equator at a frequency f. After free evolution for a time $ \tau $, a second $ \pi /2 $ pulse is applied along the $ \sigma _ x $ axis. A measurement is then made that records the projection of the Bloch vector onto the $ \sigma _ z $ axis.

Assume ideal pulses, no noise, and operation in the qubit frame. Which of the following statements is true? 
\begin{itemize}
\item If $ \tau = 0 $, the qubit will be measured in-state $ \lvert 1 \rangle $.
\item If $ \tau $ is such that the qubit rotates halfway around the equator, the qubit will be measured in-state $ \lvert 0 \rangle $.
\item If $ \tau $ is such that the qubit rotates one time around the equator, the qubit will be measured in-state $ \lvert 1 \rangle $.
\item The inverse of the time $ \tau $ that takes to qubit to make one rotation around the equator is the qubit frequency.
\end{itemize}
Solution:\\
The qubit is left to evolve around the equator for a given period of $ \tau $. The energy difference between the two levels of a qubit causes the qubit to rotate around the z-axis of the Bloch sphere. This energy is proportional to frequency, $ E=hf $, where h is Planck's constant. Thus, measuring the time $ \tau $ for a single rotation around the equator is a measure of the frequency of $ f=1/ \tau $.

\item  Spin Echo; A ``spin-echo'' is a procedure implemented by adding a $ \pi $ pulse during a free-evolution period. It is used to reduce the effect of certain types of stochastic noise. Which of the following descriptions is correct about spin echo?
\begin{itemize}
\item It refocuses incoherent evolution introduced by a noisy environment.
\item It mitigates coherent, low-frequency noise.
\item It generally works for noise spectra that increase with frequency.
\end{itemize}
Solution:\\
A spin-echo procedure can only correct for stochastic noise from the environment that produces a coherent (reversible) evolution of the qubit. It generally works for noise spectra that decrease with frequency, because adding a $ \pi $ pulse will push the filter function centroid out to higher frequencies.

\item  Rotary Echo; A rotary echo is a procedure used to reduce the effect of certain types of stochastic noise during driven evolution. Which of the following descriptions is true about a rotary echo?
\begin{itemize}
\item It reduces the effect of coherent, but stochastic, evolution introduced by the noise in the driving field.
\item By reversing the driving field at the mid-point of the driven evolution, this procedure refocuses the Bloch vector.
\item The procedure works for environmental noise spectral densities that increase with frequency.
\end{itemize}
Solution:\\
Rotary echo, like spin echo, is applicable to environmental noise spectral densities that decrease with frequency.

\begin{figure}[H] \centering{\includegraphics[scale=.8]{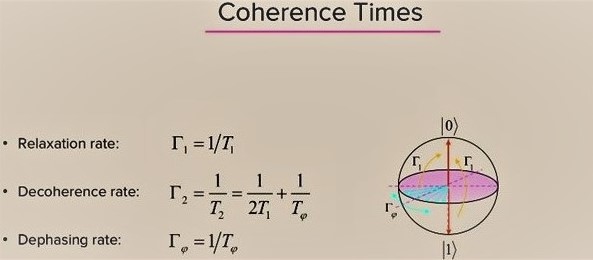}}\caption{Dynamical Decoupling During Free Evolution}\label{fig4_15}
\end{figure}

\begin{figure}[H] \centering{\includegraphics[scale=.11]{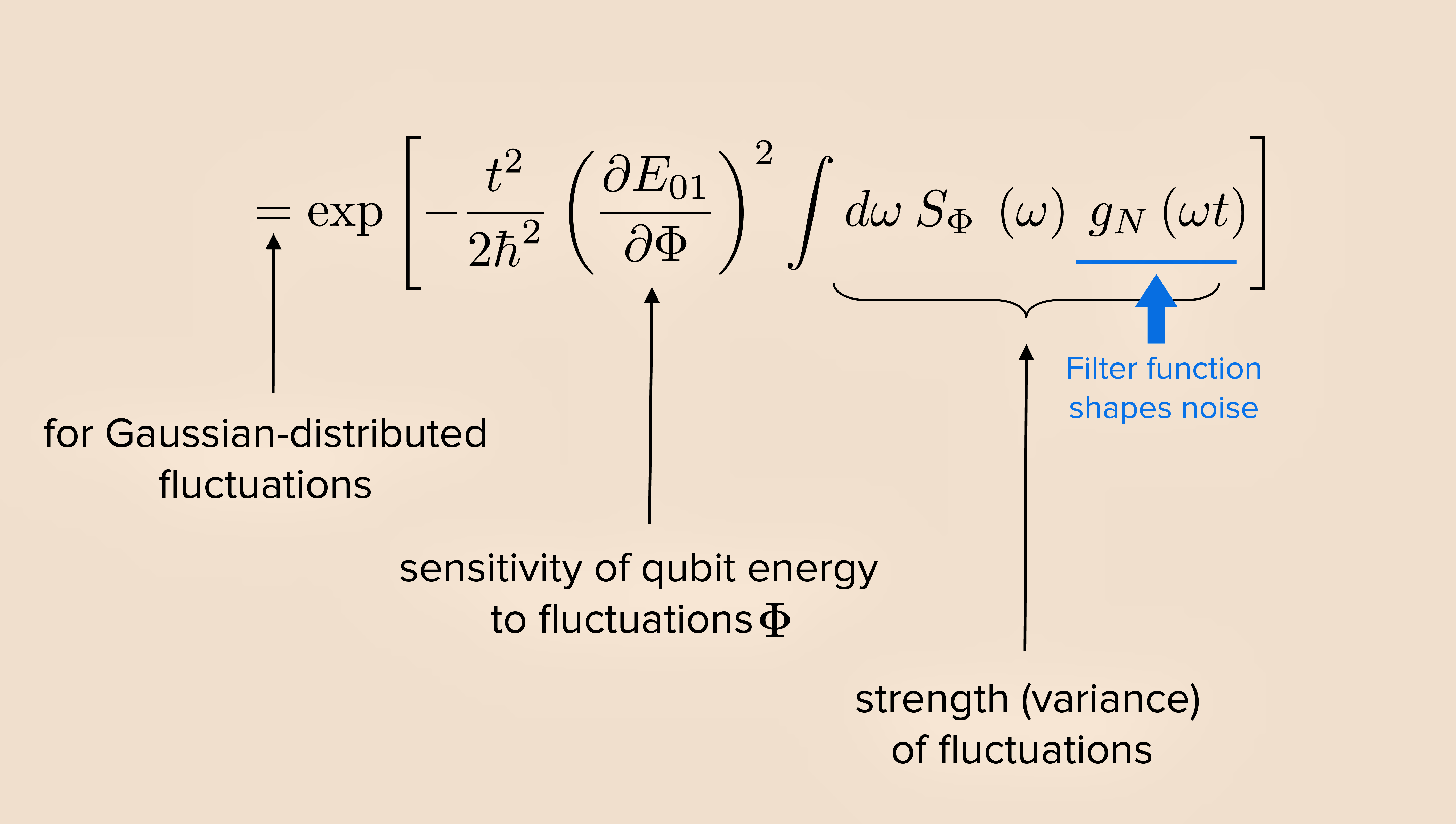}}\caption{Dynamical Decoupling During Free Evolution}\label{fig4_16}
\end{figure}

\begin{figure}[H] \centering{\includegraphics[scale=.20]{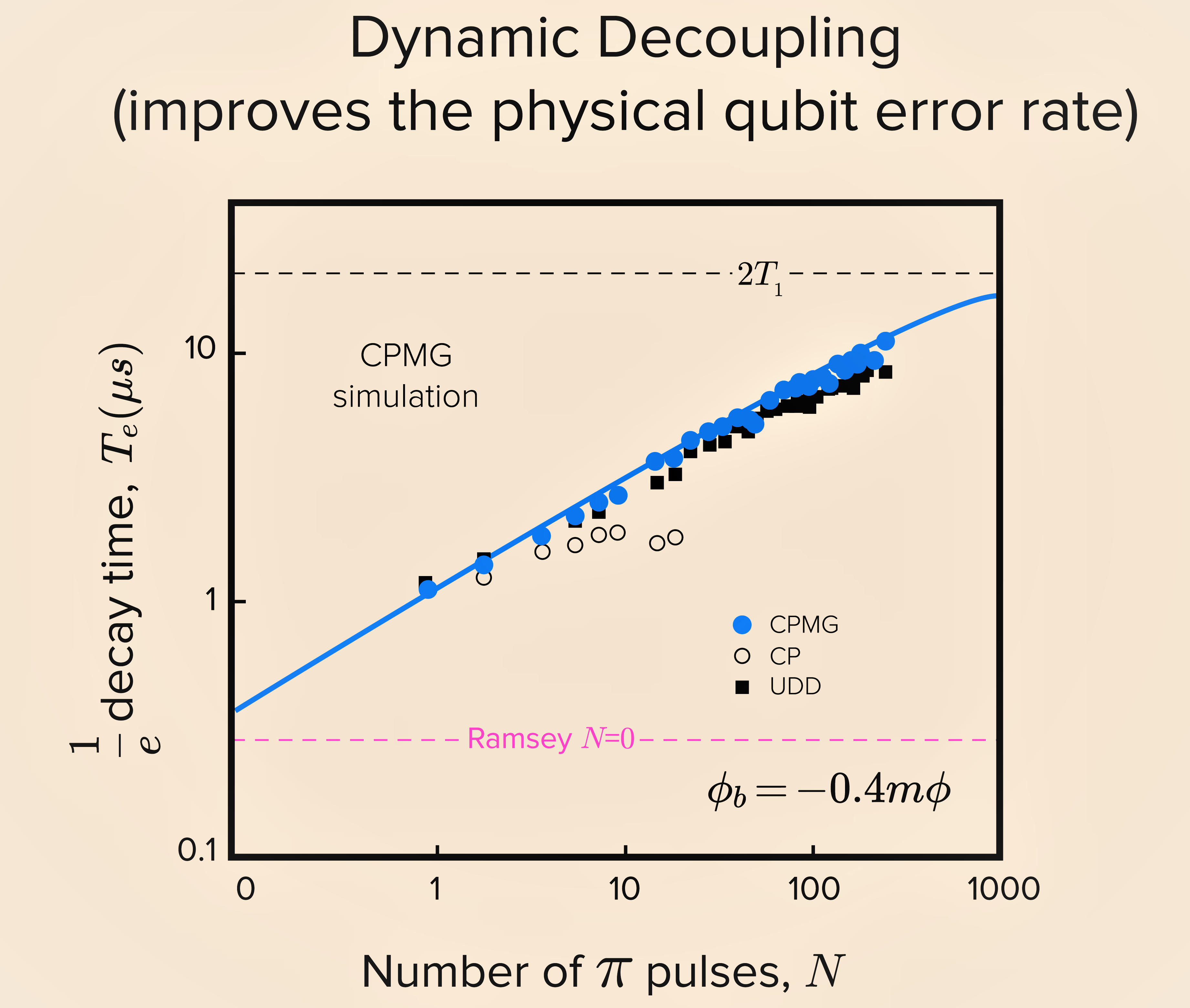}}\caption{Dynamical Decoupling During Driven Evolution}\label{fig4_17}
\end{figure}

\begin{figure}[H] \centering{\includegraphics[scale=.12]{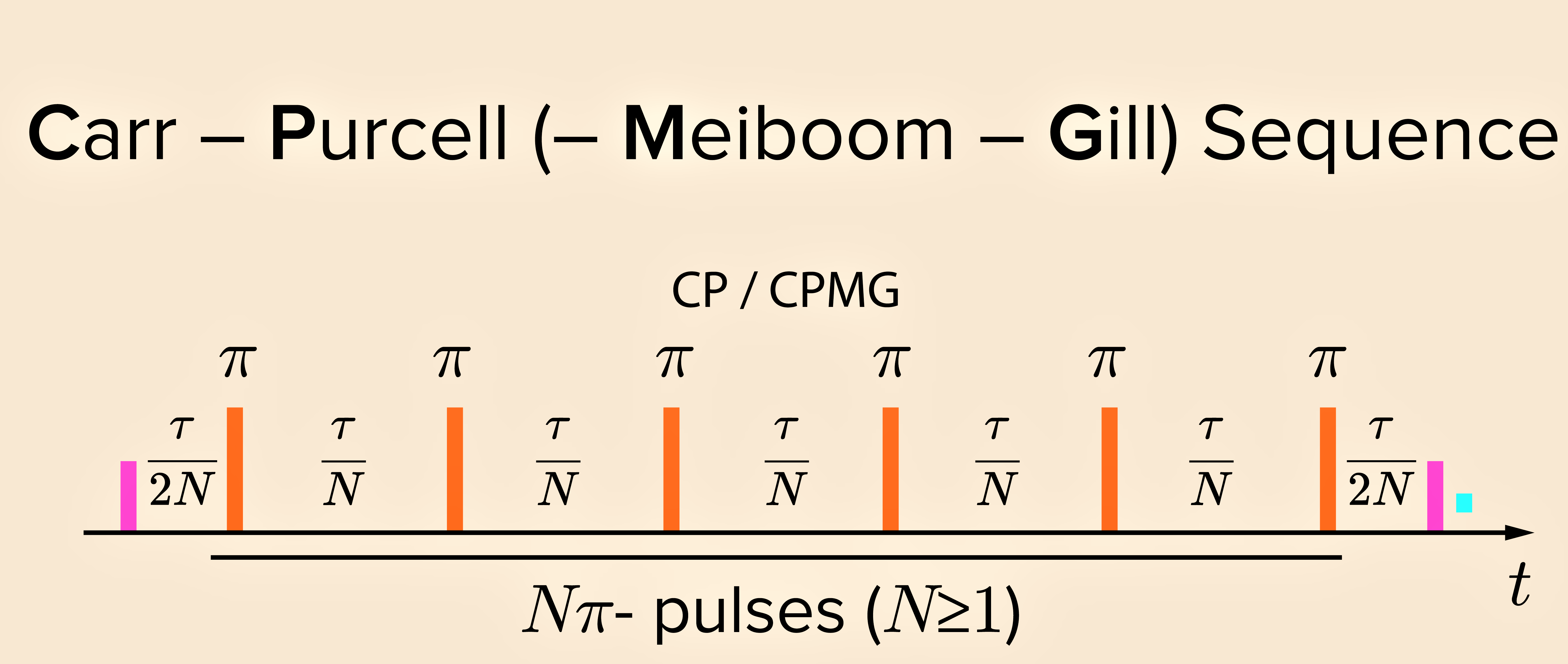}}\caption{Dynamical Decoupling During Driven Evolution}\label{fig4_18}
\end{figure}

\section{Computational Complexity and Quantum Supremacy: Introduction to Complexity Theory} 
Let us begin by describing the sort of problems that we might want to solve. In general, when we say that we want to compute something, what we mean usually is that we are computing a function. So, there is some function f that is input is a binary string, and whose output is a binary string. When we write 0, 1 star, what we mean is the set of strings of any length? So, 0, 1 star means the union over n 0, 1 to the n. so, this is a compact way of saying we have just taken a string of any length and output another binary string of any length. Because we can encode anything into bits, we can put any kind of standard function that we are interested in in this paradigm. So, this includes matrix multiplication, shortest path in a graph, integer factorization, the halting problem. This last one asks if we have some code and a given input, will that code ever terminate. All of these are functions that we can define on bit strings. Even though a matrix is described in terms of letting us say it is a matrix full of integers, we can write those integers. Alternatively, if they are rational, we can write those in terms of bits. They are real numbers. We can approximate them in terms of bits. All of these problems, even though their inputs may not look a priori like they are in terms of bits, we can always encode them in some way. Some of these are, much, much harder than others. So, when we are talking about complexity theory\cite{bernstein_quantum_1993}, the first task that we usually consider is computing functions. It is often helpful to simplify to the case where the answer is one bit. Partly to simplify some of the proofs, partly, it is just convenient. Thus, a lot of what we will discussing are describing to our conventions in the field of complexity theory where they could have defined it a different way. However, since the literature is mostly in terms of a particular convention, we want us to be familiar with it. So, the usual convention is to discuss a special case of functions called languages. What that means is just that the function has a 1-bit output? So, instead of outputting a string of any length, it just outputs a single bit. This has the same computational power as functions. We should not say the same. Nearly the same computational power as a general function, because we could always give this one more input that tells us which bit of the string we interested in. Then, by calling this 1-bit function repeatedly, extract all the information from a general function—for example, integer factorization. We take in a number. We output its prime factorization. That really seems like a function with many bits of output. However, we could ask we could say input a number, as well as another index i, and then tell the i-th a bit of the prime factorization. So, now that is just one bit. If we can have access to that function, we can call it repeatedly. We can read out all the bits of the prime factorization. There is a little bit of overhead in that we have to call this function many times. It is not the most efficient way to do it on a computer. However, there is not a big overhead in moving from the language picture to the function picture. Now, we might ask, why do we call it a language just because it a 1-bit output? Furthermore, the reason is, and this is also part of what makes it mathematically and a little bit nicer. If we have a 1-bit output, then we can describe the whole function by just specifying the set that gets mapped to 0, or the set that gets mapped to 1. So, let us define l to be f inverse of 1. In other words, the set of strings that gets mapped to 1. Thus, we say l is the language that describes this function. This is a subset of 0, 1 star. So, instead of discussing a function, we could just discuss a subset of strings corresponding to the inputs that give rise to one. We say that to decide on the language, and we should accept x in the language, or reject x, not in the language. So, we could think of this problem as instead of computing it bit, just equivalently, we can say it is determining membership in a set. If we are in the set l, we want to accept the input. If we are not on the set l, we want to reject the input—just different terminology for the same basic math. We want to output 1, which we call to accept, or 0, which we call reject. Thus, any general problem we can phrase is the yes/no problem. That is all that is going on. The language picture is, however, conventional. When we discuss complexity classes, we will mostly discuss it in terms of languages, although we can imagine the other versions.

\section{Classical Complexity Theory: Promise Problems}
So, these are the basic problems that we would like to solve. There are also some more complicated ones, which are often useful to discuss. So, one that comes up is promise problems. What this means is that there is instead of the input being all strings, we only need the input to be a subset of strings. So, we only need the correct answer when x belongs to some set p, which is some subset of strings. So, we could say that our input is promised to be in that set. Our algorithm only needs to run on strings satisfying the promise. When it is outside of that set, we do not need any guarantee on what our program does. The last type of problem we might consider is sampling problems. These are a generalization of functions. So, the function we can take in an input, and we are supposed to output a bit string. However, more generally, we might say, associated with every input x is some distribution that we would like to sample from. So, given input x, sample perhaps approximately from a probability distribution, d sub x. So, instead of having some output that is a deterministic string, now the goal is to output a sample from some distribution parameterized by x. So, we can see this is a significant generalization of functions, and also, in some cases, useful to consider. There are other types of problems we can consider as well, and our goal is to output a quantum state. However, we think this is probably a good first pass at the range of computational tasks that we might ask something to do. We guess we should say that promise problems modify all of the others, right? So, a promise problem could be a language, a function, or a sampling problem. Actually, we will mention one more, which is a relation. In this, there are multiple correct answers. So, we have sum r, which is a subset of zero one star given zero one star. Given x, the goal is to output y such that xy is in the relation. So, we can see this one example is we could view integer factorization also in this way. The goal is given a number, output any prime factor. Right? Thus, it could be multiple correct answers. So, these are the more advanced types of problems. Most things we can express just in terms of languages, but it can sometimes be convenient to look at these other types of problems. 

\section{Models of Computing: Circuits} 
So, let us discuss models of computing. This will be kind of like the problems that there are we going to give we a list of models of computing, but they are partially overlapping. These are attributes that can be combined with each other. So, the first one is one that we have seen already, which is the circuit model. So, in this, we have some bits, x1 through xn. if we have classical circuits, we combine them via gates like AND, and NOT, and OR. if it is a classical circuit so, classical circuits could have unbounded fan-in AND/OR fan-out. what this means is we could take the AND of many different bits. That could be a valid thing to do. then the output of that AND, we could feed into many, many other gates. There is, in principle, no restriction on this. Now we may prefer to consider special classes of classical circuits. we might prefer to consider ones only with bounded fan-in and fan-out. Fan-in means how many wires go in. Fan-out means how many wires go out. we might also prefer to consider reversible circuits. So, reversible is a special case. However, for classical circuits, we are not restricted to consider bounded fan-in, and out the same way, we are for quantum unitary circuits\cite{franson_limitations_2018}. So, compare, whereas for quantum circuits, the gates are unitary, and we start with some ancilla qubits in the zero state, we do not need ancillas in the classical case because we can kind of create more bits as we go along. The internal wires of the circuit are kind of like bits. So, if we have an AND and it has 100 outputs, those can be like the inputs to the next layer of the circuit. Whereas quantumly, if every gate is unitary, that means it has the same number of inputs and outputs. If we have control of NOT gate, two qubits go in, two qubits to go out, Toffoli gate, three qubits go in, three qubits go out. So, the gates are not going to create any extra data. we have to put them all in at the start with a bunch of ancillas initialized to the zero states. It is kind of the usual model of computing we have seen so far. So, as we are discussing these models of computing, we also want to kind of measure how complex they are. So, complexity theory is distinguished from computability theory. Computability theory says can we, yes or no, compute this function, or this relation, or this promise problem, whatever. Complexity theory says, given that we know we can compute it, how hard is it to compute it? How much time does it take? How much effort? To measure this for circuits, let us draw a model of a circuit again. Let us say we have a bunch of gates, some more gates. Then we measure. If it is a language, then we get the answer, which is, let us say, accept or reject, one or zero. So, how do we measure the complexity of such a circuit? So, the number of bits is called the width of the circuit. That is kind of how much memory it takes to run it. The amount of time it takes is called the depth. here, we imagine that the gates are in layers. If two gates happen on disjoint pairs of qubits, we can do them in parallel. Thus, in a one-time step, we could hit many different pairs of qubits with gates at the same time. the depth measures the number of layers in this kind of decomposition. then finally another measure is to say what is the number of gates. So, the size just counts the number of gates, which, in some cases, is also a measure of the amount of effort we have to put into the circuit. so, circuits can be measured in terms of their width, depth, and/or size as a way to measure their complexity. so, the question is what depth means in terms of the unitaries. Each one of these gates is a unitary operator. Size is the number of gates we apply. However, in this circuit, we see we did three gates all at the same time. So, in the circuit we have drawn, the depth would just be three. Here is one layer of gates. Here is another layer of gates. Here is one more layer of gates. The measurement we do not count. Some people would count it. However, we will choose not to count it. So, the circuit model is a very appealing one in many ways. we guess the good feature of the circuit model is that it is a very concrete model of computation. However, they have one big flaw, which is when we defined a function, we said the input could be any length, like matrix multiplication. That is just one function. It includes multiplying two by two matrices, or seven by seven matrices, or rectangular matrices. we do not want to define a new function for every size of the matrix. we just want to say matrix multiplication, and it encompasses all of those. Nevertheless, circuits are kind of funny. we can think of them on a classical circuit we could imagine actually soldering these things together. On a quantum circuit, this is just really a sequence of laser pulses, or magnetic fields, or whatever. However, they always have a fixed number of inputs. Any circuit does not have a variable-sized input. It has a definite number of bits that go in, a definite sequence of operations. Thus, the one big flaw of circuits is that they depend on the input size n. let us write n when we write the absolute value of x, and we mean the length of the string x. So, n is the number of bits of input. It is the length of the input x. So, for every value of n, we have to specify a circuit. Thus, when we tell we have a recipe for solving the problem, if we are going to tell we a circuit, that is only a recipe for one value of n. So, to fully specify how to solve a problem, we have to give we a circuit for every possible input length n. that is kind of an unsatisfactory feature of the model. In part, one reason why it is unsatisfactory is if we just say for each value n we want a circuit,  the width, and depth, and size should be some bounded functions of n. Nevertheless, that is it. we are allowing the circle to depend on n arbitrarily up to those limits. we could smuggle in information in that way. we could have it so that it could depend on n in a very hard to compute sort of way, in a way that would make this a somewhat unrealistic model of computing. So, that is kind of the one drawback of circuits. Let us just expand on this a little bit. Suppose we want to know does the nth Turing machine halt? So, we will come back to say in a minute what a Turing machine is. by halt, we mean if we just start it on a blank input, does it go into an infinite loop, or does it stop? Moreover, it turns out that there is no algorithm that will compute this for all inputs. However, we could smuggle it into something that could be solved on a circuit. we could just say, look at the length of the input. That tells n. we want to know does the nth Turing machine halt. because this model lets the circuit depend arbitrarily on n, the circuit could just do nothing, just ignore the input and just output zero or one immediately, and thereby would solve this problem. So, that sort of recognizes a flaw of the circuit model in that it gives us this uncontrolled dependence on n, which could give us in a sense too much computing power. There are ways around that which we will come back to in a minute. we end up still using circuits. we just have to kind of deal with this flaw. Another flaw, which is related, is that it has no loops. If we ever write code, and we write a program with no loops in it, our program is we would not want to write a long program that way. we are kind of only do that for a very simple program. Loops are a kind of an important feature of programming. No loops are also a pro. It means that there are no infinite loops. we know the circuit will always end after a certain amount of time. However, we give up some of the powerful features of programming that make this possible. so, that is one model of computing, which sort of gets at a lot of what computers do but misses out some of the important things, like being independent of the size of the input and having recursion and loops.
 
\section{Models of Computing: Random, Quantum, and Non-Determinism}
The basic models, we said, are circuits and Turing machines. These are good ways of describing deterministic computations. This is a section on quantum, so, let us discuss quantum, and even build up to that randomized computations. So, randomness is something that can go into the computing model. There are kind of two ways of doing it. We can either have some extra inputs, let us say r1 through r m are random bits. Alternatively, our gates or Turing machine rules can be random. For example, in that Turing machine table, we could say with 1/2 probability, if the current position is 0, then with 1/2 probability do this, and with 1/2 probability do that. Alternatively, we could add not just AND, OR, and NOT gates, but one extra gate called the ``random gate,'' that just outputs 0 or 1 randomly each time we use it. So, that is one way we could do it. Alternatively, we could have all our operations to be deterministic, but we push our randomness to the beginning. We say we are going to take in an extra input, which is a long string of random bits that could either be more inputs to the gates or more inputs to the Turing machine. We are going to use that as a source of randomness. These are basically equivalent. Thus, either one of these is a way of modeling randomness. Finally, let us discuss quantum. So, we kind of already described quantum circuits. We do not want to say too much more about it, because it is also the model we are used to seeing. So, quantum circuits have unitaries. They have ancillas in the zero state. They have some depth, some width, some size. The only drawback of them is this uncontrolled dependence on the input size. Just like classical circuits, they have the same problem. Thus, there are two ways around this. So, there are quantum Turing machines. They were defined in a paper about 20 years ago. However, almost nobody uses these. These are rarely used. Instead, one of the main preferred models is a classical Turing machine that outputs a quantum circuit. So, that is the compromise model we were discussing. For classical circuits, it does not make a whole lot of sense. We mean, there are cases in which it does, but often, if the Turing machine could output a circuit, it could just go on and compute the output of that circuit. Why bother outputting a circuit just to run it a minute later? However, here, we can have the advantage of a classical Turing machine. We can have a program with loops, recursion, a finite program for all input sizes. Then that could output a quantum circuit for each input length. Thus, then, this sort of keeps the benefits of the quantum circuit model, while using a classical Turing machine. So, when we discuss quantum circuits, formally, this is the computing model that we usually mean, that almost everybody means when they discuss quantum circuits being able to compute something. There is one last model, which is called the non-determinism. If we have discussed of NP, that is where this model is used. What it means is we can think of it as the classical computer takes the OR of two branches of a computation. So, if we go back to randomized and quantum, we can think of when we make a random choice, we are sort of averaging over two branches. If we keep making random choices, we are going to average over many, many branches. Then we are going to end up these branches will add up to some accept probability or add up to some reject probability. However, we are spreading probability among many branches. Quantumly, we can do something like a Hadamard that will put amplitude on what we could consider two different branches. Non-determinism is a physically unreasonable form of computation that can be useful for thinking about proof systems. There are some nice features of it that way, but it does not correspond to something we can compute in the real world. The idea is that our computer can go into two branches. Those branches can further branch. So, after n steps, we could have 2 to the n possible branches. Then, in the end, if any of those branches accept, the overall computer will accept. If all of them reject, then the overall computer will reject. That is what we mean by taking the OR of those branches. Accept corresponds to 1. So, if there are any 1's in the output, the OR is going to be 1. So, this is called non-determinism. We will discuss more this a little bit later. However, we could, for now, just think of it as another kind of wacky model of computation.

\item  Definition of a Language; Computation generates an output for a given input, directly analogous to a mathematical function. In a computer, the input and output are represented by strings of zeros and ones, denoted as $ \{0,1\}^* $. An algorithm calculates the action of a function f given an input string. To study the complexity of a function, that is, how hard it is to compute it is convenient to restrict the problem to functions with a single-bit output, zero or one, without loss of generality. Using this particular type of function, we can then introduce the definition of a Language. What is Language?
\begin{itemize}
\item It is the set of strings of zeros and ones $ (\{0,1\}^*) $ that when the input to a function f  results in an output equal to 1.
\item It is the inverse of the function f applied to output equal to 1.
\item It is a subset of all possible strings of zeros and ones $ (\{0,1\}^*) $.
Solution:\\
a) and b) are the same definition but said in different ways. Although c) is not a formal definition, it is true, and it can be inferred from a).
\end{itemize}

\item  The Circuit Model; There are different models of computing that depict how a given calculation is performed with a computer. Throughout, we have often represented our computation with circuits composed of wires and gates. Although this circuit model is concrete and intuitive, it has certain disadvantages. Which of the following are disadvantages of using the circuit model of computing
\begin{itemize}    
\item Circuits depend on the size of the input, i.e., the length of the input string of zeros and ones
\item Circuits might not halt, meaning that for certain problems, the program might continue running indefinitely.
\item Circuits have no analog of a ``software loop'', an important feature of programming.
\item All of the above.
\end{itemize}
Solution:\\
Since circuits have no loops, or recursive structures, that prevent completion of a task, they will generally run until they produce output and then stop. thus, circuits will not fall into infinite loops, which is one advantage of this model.

\section{Standard Classical Computational Complexity Classes: Deterministic Time} 

We have problems that we want to solve, we have models of computing, and the way that people put these together is they discuss complexity classes. These are, basically, classes of problems that all have similar complexity. So, one example, DTIME f of n stands for deterministic time. What this means, we say that a language belongs to DTIME of f of n, if there exists a Turing machine, m, such that m of x always halts with the correct answer. So, in other words, accept if x is in L, reject if x is not in L. So, so, far, that just means we have computed it correctly, but what about the time? Then we want to say in less than f of the size of x steps. So, for example, we might have DTIME of 100 n cubed, which means these are the languages that if the input is n bits, we can solve in at most 100 times n cubed steps.
 
\section{Complexity Classes: Polynomial Exponential and PSPACE}
 
This is a little bit precise. Usually, we do not care if the Turing machine takes 100 n cubed or 300 n cubed. We just want to know the scaling because, as we change from different memory access models, it might change the runtime a little bit. We do not care about that too much. We kind of want to know very crudely, is this poly time or exponential time. So, instead of just looking at D time for very specific functions, usually, we just discuss at these very crude ones so, P, probably the most famous complexity class, is the union of D time and the k for all values of k. So, k just ranges over all integers. Thus, this just means there is some polynomial of n, and we can solve the function in time that grows at most like a polynomial event. We could also discuss exponential time, which is the union over 2 to the n to the k. So, it means all the problems for which there is some finite k such that we can solve it in time 2 to the n to the k. So, in general, P, polynomial time, corresponds to the things that we think are relatively easy to compute, and we might object we might say, well, n or n squared seems like reasonable scaling. It was n to the 17 that does not sound very reasonable. That does not sound so easy to compute. N to the 100 already when n is 2, that is going to be out of reach. The reason why people study P is that, empirically, for problems that we are interested in, if we start with an algorithm, it takes time n to the 84, we tend to find further improvements. Usually, if people care about the problem, they find ways of doing it faster and bringing that number down. There is no law of nature that says it has to be that way. However, because it has often been true, people find P to be a useful organizing principle. Often m squared is not enough. We really want to be linear in n. It is not like P is the end of the story. However, it is a first pass at what is efficient or what is not. So, that is time. We might also consider space. So, we say the class space f of n is the set of languages that can be solved using at most f of n cells of tape. So, we can think of it as we have an infinite tape. However, if the Turing machine solves it without ever going beyond a certain region of tape, then we have not used that much. Thus, we say space of f of n means the space we need to solve the problem is bounded by f of n, the number of cells of tape. Again, we prefer to be vague about how much and just say P space means polynomial space. There is also log space, which is a little trickier to define because then we have to argue why we are not already using n just for the input. 

\section{Complexity Classes: NP (Nondeterministic Polynomial Time)}
 
There are a few other classes that are important. One of them is NP. So, NP is the set of problems that we can verify in polynomial time \cite{bremner_classical_2011}, and here, when we say verify, we mean only if the answer is yes. So, if the answer is yes, there should be a short we are of it, whereas if the answer is no, there does not have to be. Thus, an example of this would be the traveling salesman problem. Can we visit all of these cities without revisiting any city, just visiting each city once? Furthermore, do this where the total length of the route is less than 1,000 miles. If the answer is yes, we can prove that to us easily. We mean, there exists a short proof. It might be hard to come up with. However, there exists a proof, which we can verify easily. The proof is just, here is a list of cities. Then we have to count. Yes, do the miles add up to less than 1,000. If so, we verify that the answer to the original question about does there exist such a route is yes. If the answer is no, like there is no route under 1,000 miles, there might not be a succinct way to communicate that fact, to prove that fact. So, NP, we can think of it as a statement about non-determinism. Like, if a problem has a short proof, then one thing we could do is we could non-deterministically check all the proofs. Essentially, what we want to do is take the OR of all the possible proofs. However, it is also just a statement about what is provable. So, when we said at the beginning, we think we have erased it now complexity theory studies what we can compute quickly, but also what is the power of proof. We can think of NP as being a statement about proof complexity. The way we like to think of this that makes it simple to say that f is in NP if f of x can be expressed as a giant OR of g of XY where y should be a bit string whose length is polynomial in x. Sorry, the length of x. This is the length of x to the c. So, c if some constant. Y is, therefore, a polynomial length of x. g should be polynomial-time computable. So, the wasting of NP is we have something polytime commutable, but we take the OR of it over an exponentially large number of things. Thus, if f of x is equal to 1, the way we can demonstrate that to we is by exhibiting a y for which g of XY is 1. So, if f of x equals 1, then the proof of this is the y such that g of XY is equal to 1. If we give us such a y, then g is polytime commutable. We can quickly compute g of XY, verify that equals 1, and thereby be convinced that f of x is equal to 1. If f of x is 0, all that we need is that there is no such proof. If f of x is 0 and we try to convince us that it is 1 by giving some y, we check it. We say, hey, no, g of XY is 0. No matter which y we give we, g of XY will be 0. So, if f of x equals 0, then there is no such proof. 

In this section, we are evaluating the computational scaling of a classical or quantum processor from the perspective of a computer scientist. As we know, a quantum computer can outperform a classical computer for a limited set of computational problems. For many problems, a quantum computer performs no better than a classical computer. How can we differentiate these classes of problems? Computational problems are assigned to computational complexity classes\cite{aaronson_computational_2010,orus_tensor_2019} depending on how efficiently a classical or quantum computer can solve them. For example, a quantum computer can factor integers more efficiently than classical computers can use the best known classical algorithm\cite{smolin_pretending_2013}, but it may perform given tasks such as playing chess, proving mathematical theorems, solving the traveling salesman problem, or even multiplying two numbers no better than a classical computer.

Computational problems are classified based on the best algorithms known today. The different complexity classes arise due to the different scaling laws as a function of problem size for the physical resources and the time requirements for these best-known algorithms. For example, computing devices comprise a fixed number of physical memory units and a processor that can perform a certain number of computational steps per time unit. The physical memory of a computing device places an upper bound on the maximum, manageable problem complexity. The number of elementary computational steps required to solve a task can be used as a quantitative measure of time.

Suppose we have a computer that can compute the solutions to a fictitious problem with 2 variables of interest using 100 memory units and 4 computational steps. For 3 variables of interest, the computing device needs 1000 memory units and 8 computational steps. 4 variables require 10000 memory units and 16 computational steps. Obviously, the physical requirements for this fictitious problem scale unfavorably. The required memory is said to scale exponentially, $ 10^n $, with the number of variables n. In contrast, the number of computational steps scales as $ n^2 $. Computational problems for which a parameter scales as $ n^ x $, for a constant x, as n increases, are referred to as polynomial in n.

Here, we will focus on problems and complexity classes that show a polynomial physical memory scaling. The complexity class describing problems with polynomial memory requirements independent of the respective time complexity is referred to as PSPACE.

An example of a problem that scales polynomially in both time and memory is multiplication. The multiplication of two n-bit numbers requires $ n^2 $ computational steps to generate the solution.

\begin{table}[H]
\centering
\caption{The multiplication of two n-bit numbers requires $ n^2 $ computational steps}
\label{tab:4_1:Table 2}    
\resizebox{\textwidth}{!}{
\begin{tabular}{|c|c|c|c|c|}\hline
binary factors & $\quad N_1~$ & $\times$ & $ ~ N_2 $ & steps \\ \hline
{$N_1=N_2=001\rightarrow n=1$} & {$1\cdot 2^0$} & {$\times$} & {$1\cdot 2^0=00001$} & {$1^2=1$} \\ \hline
{$N_1=N_2=011\rightarrow n=2$} & {$1\cdot 2^0\times 1\cdot 2^0+1\cdot 2^0$} & {$\times$} & {$1\cdot 2^1\qquad \qquad$} &  \\ 
& {$ +1\cdot 2^1\times 1\cdot 2^0 + 1\cdot 2^1 $} & {$\times$} & {$1\cdot 2^1=01001$}  & {$2^2=4$} \\ \hline
{$N_1=N_2=100\rightarrow n=3$} & {$0\cdot 2^0\times 0\cdot 2^0+0\cdot 2^0\times 0\cdot 2^1+0\cdot 2^0$} & {$\times$} & {$1\cdot 2^2\qquad \qquad$} & \\
& {$+0\cdot 2^1\times 0\cdot 2^0+0\cdot 2^1\times 0\cdot 2^1+0\cdot 2^1 $} & {$\times$} & {$1\cdot 2^2\qquad \qquad$} & \\
&  {$ +1\cdot 2^2\times 0\cdot 2^0+1\cdot 2^2\times 0\cdot 2^1+1\cdot 2^2 $} & {$\times$} & {$1\cdot 2^2=10000$} & {$4^2=16$} \\\hline
\end{tabular}}
\end{table}
                                                       
Computational problems with algorithms able to compute solutions in polynomial time form the computational complexity class P. Problems in class P can be efficiently computed on a classical computer. Note that here, ``efficient'' does not necessarily mean ``on a human time scale;'' it only refers to the scaling law. For example, if a computer has to solve a large, polynomial problem, such as the multiplication of two numbers, each with $ n=10^{100} $ bits, it may still take an unrealistic number of computational steps, $ 10^{100^2} $, to finish in practice, despite being classified as efficient. Rather, the computation is efficient relative to another problem of a similar size but with an exponential time scaling, such as, for example, $ 2^{10^{100}} $.

Not every problem scales polynomially in time. For example, the traveling salesman problem scales exponentially with the number of cities. The problem is to evaluate the shortest route to visit multiple cities. Additional conditions are that each city is only visited once and that the traveling salesman returns to his original city at the end of his tour. For n cities, there are $ n!\propto n^{(n+0.5)} $ different options to check, and evaluating every possibility scales exponentially with n.

In general, a classical computer is unable to solve problems with exponential time requirements efficiently. However, once a solution is presented, it may only take polynomial computational time to confirm the solution. Factoring an integer into two prime numbers is an exponentially hard problem, in general. However, verifying the proposed solution involves a single multiplication, which is polynomial in time, as demonstrated above—problems with solutions that can be verified in polynomial computational time form the NP complexity class. NP stands for Non-deterministic Polynomial-time. Among the problems that can be efficiently evaluated are the problems in class P. If there exists an efficient algorithm to propose a solution to a problem, then they also exist an efficient algorithm to verify the solution.

The hardest computational problems in NP those for which the computation is exponential in time and, hence not in P are assigned to the NP-Complete complexity class\cite{botea_complexity_2018}. The traveling salesman is an NP-Complete problem\cite{lucas_ising_2014}. It is exponentially hard to find a solution, but it only takes polynomial time to evaluate the proposed solution. A proposed solution can be easily verified by depicting the proposed solution on a map and verifying that the conditions are not violated.

It is an open question if algorithms exist that are able to solve NP-Complete problems in polynomial time. If such algorithms exist, then NP would consequently collapse into P, meaning they are the same complexity classes, $ P=NP $. The search for such algorithms has continued for more than half a century. The inability to find such an algorithm leads most computer scientists to believe that, indeed, $ P\neq NP $.

\begin{figure}[H] \centering{\includegraphics[scale=.40]{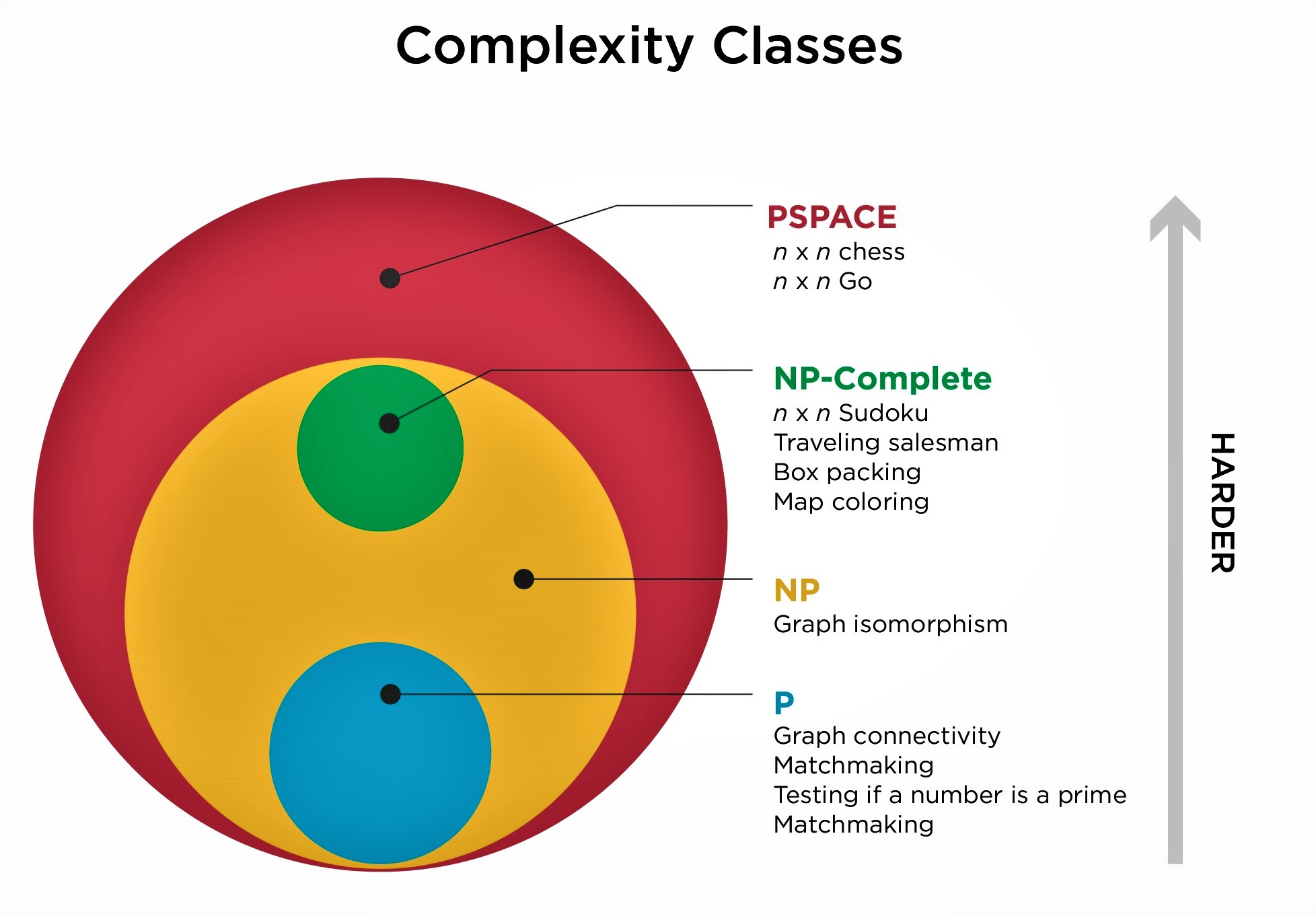}}\caption{Classical Computational Complexity Classes: PSPACE contains all classes and problems which require a polynomial amount of memory on a conventional computer independent of the number of necessary computational steps to find or verify a solution. P describes the class of classically efficient computable problems. Class NP contains problems that are efficiently verifiable. The hardest problems which are still efficiently verifiable form complexity class NP-Complete.}\label{fig4_19}
\end{figure}

Figure 1 summarizes the hierarchy of the four introduced complexity classes. Classical computers can efficiently solve computational tasks part of P. In the next section, and we will see how quantum computers perform and which problems it can solve efficiently and which ones it performs similarly to classical computers.

\item  DTIME; A complexity class is a set of problems with the same degree of computational complexity. To define certain complexity classes, it is helpful to think in terms of the time it would take to solve a given problem. This concept of ``how much time'' can be described mathematically using something called DTIME(f(n)), where f is a given function (not to be confused with the computed function f in section ) and n is the size of the input (previously denoted $ \lvert x\vert) $. Which of the following sentences is/are true about DTIME?
\begin{itemize}    
\item DTIME(f(n)) is a set of input strings of zeros and ones.
\item DTIME(f(n)) is the set of functions that can be computed in a deterministic time shorter than n, where n is the size of the input.
\item DTIME(f(n)) is the set of languages L that can be solved in deterministic time, equal to or smaller than f(n) steps.
\end{itemize}
Solution:\\
c) is the correct definition of DTIME, and a) is a true statement that can be inferred from c).

\item  Complexity Classes; Some complexity classes are defined by the scaling law for the amount of time it takes to compute a given set of problems. This time is characterized using the concept of DTIME(f(n)), where DTIME(f(n)) is the set of languages L that can be solved in a deterministic time, equal to or smaller than f(n) steps, where n is the length of the input string of bits. Which of the following are proper definitions of valid complexity classes?
\begin{itemize}    
\item EXPTIME complexity class: Exponential time is the union of all DTIME(f(n)), where $ f(n)=2^{n^k} $ and k runs over all positive integers.
\item P complexity class: Polynomial-time is the union of all DTIME(f(n)), where $ f(n)=n^k $ and k runs over all positive integers.
\item NP complexity class: Non-Polynomial time is the union of all DTIME(f(n)), where f(n) $ \neq n^k $ and k runs over all positive integers.
\end{itemize}
Solution:\\
In complexity theory, NP stands for non-deterministic polynomial time, and it is defined by the set of problems that can be solved in polynomial time by a non-deterministic Turing machine.

\section{BPP and BQP} 

The two more important classes, BPP and BQP \cite{aaronson_bqp_2009,aaronson_forrelation_2014,raz_oracle_2019}, these correspond to random and quantum polynomial time. So, the P or Q here means probabilistic or quantum, and the final P means polynomial time. Thus, that leaves only the B. So, the B stands for bounded error, right? Because if we are doing a deterministic computation, we are always thinking of the same answer. Thus, the correct behavior of the algorithm means it should be the right answer. Pretty clear. If our calculation is random, or quantum, then normally, the answer we get will not always be the same. However, we would like that it is the correct answer most of the time. Thus, here, let us say we are working with languages, so, the answer is a bit. So, we demand that so, here, correctness means that for all x, the probability that the algorithm on input x is equal to f of x is greater than or equal to 2/3. This is, we guess we should say, for f mapping to 0, 1, i.e., languages, although we guess it would work for functions also. However, it is sort of convenient to use it for languages. So, here, we are allowed to use random gates or quantum gates, and we should get the answer right most of the time. Alg means the algorithm. So, and let us say one more thing, which is, just like the others, these are classes of languages. So, just sort of as objects, we say that L belongs to BQP if there exists a quantum algorithm \cite{terhal_adaptive_2004}, we know, such that this is true. L here, we should think of as a subset of all the binary strings.
 
\section{BQP-Complete Problem: Quantum Circuit Evaluation} 

We discuss NP a lot because there are so many kinds of optimization and planning problems that are naturally NP-complete. Thus, it is historically something people have cared about a lot. A lot of the effort has been devoted to NP. But what about quantum computing? Can we do the same thing for BQP? Can we find a complete problem for a BQP? Can we say that this is BQP complete\cite{raz_oracle_2019}, and so, it is unlikely to have a poly-time solution? So, the complete natural problem for BQP, kind of analogous to circuit sat, is called ``Quantum Circuit Evaluation.'' Furthermore, the problem is the following so, the input is a circuit U. Let us say it is on n qubits. Let us say it starts in the 0 states, applies n squared gates, and at the end of those n squared gates, and it measures the first qubit. So, that is the format of the problem. The particular input will specify what those n squared gates are. We are promised that the probability of getting output 1 is either greater than 2/3 or less than 1/3. We want to determine which is the case. Because remember, for BQP, these are only problems where we get the answer right with probability at least 2/3. Thus, what that means is if the answer is yes, we get if the answer is 1, we get 1 with probability at least 2/3. If the true answer was 0, we should get output 1 with a probability of less than 1/3. So, we should always see outputs. We should never see things that are kind of in the middle. We should only see things that are more than 2/3 or less than 1/3. This condition here is called a promise. Remember, we discussed promise problems last time. We said they are not defined in all inputs. So, in this case, we only wanted to define this problem on the inputs that have this property. It is a very non-trivial restriction, and it might be hard to evaluate. However, promises do not have to be easy to evaluate. We just want to restrict attention to this and say, what that means is to solve quantum circuit evaluation, we only need to solve it on the inputs that have this property. Then we would be happy just to solve it in those cases. We claim that this is BQP-complete. So, first of all, we claim it is in BQP. Why is that? Well, if we are giving this input, we just run the circuit, and we measure. We use the output directly. We promised that the output has probably at least 2/3 of being one thing or the other. Thus, that output, with error probability less than 1/3, we can distinguish which is the case. If the way we have run it has introduced more errors, let us say from the theorem, we could just amplify. We could just repeat it a few times and drive our error probability below 1/3. So, it is in BQP. On the other hand, any problem in BQP can be expressed in this way. If we have something in BQP, practically the definition of BQP is to say there exists a circuit that does something like this. We might get nervous. We might say, well, BQP allowed to run polynomial like n to the 17. this only allows n squared gates. We had to pick something concrete, so we picked n squared. Does anyone see how to get around that problem? It shows us; we think some of the subtleties, we think, of poly-time reduction. Remember that is what we need for hardness. We need there to be a polynomial-time reduction from any problem in BQP to this one. The number of gates here is defined in terms of the input to this problem. So, we say, here, we have this many qubits and this many gates. Thus, we start with our original n, and then we might need to do n to the 10 gates. Then we have some new n prime, which is the length of this output, which could be like n to the 20. then we have got n prime qubits and n prime squared gates. That budget is more than enough to accomplish what we want. So, the answer, basically, is this poly-time reduction. Seems very innocent, but it can smuggle in polynomial overheads. The length can get polynomials larger. Thus, it is a very one way to think about this, and it is a very crude notion of complexity. As a practical matter, we care about the difference being n squared and n cubed. However, this crude notion of complexity will really just tell if it is polynomial versus exponential a little bit more than that. However, it is not going to distinguish between different kinds of polynomials. So, that is why we can say very crude things like this. It is enough to give we kind of a first-order answer. If we refine the same techniques, we can get a more refined answer, as well. 

\section{BQP and BPP: Quantum vs. Classical Complexity} 

Quantum circuit evaluation, the same thing we were over there. This is not in BPP unless BPP equals BQP. We could have put here P instead of BPP, remember BPP is randomized computation. Randomized seems like a slightly more, we know, closer to quantum. We should say that P is contained in BPP, is contained in BQP. So, just to give we sort of the cast of characters, P is the weakest. Everything else contains P., But these sorts of branches look like they may be incomparable. In other words, it looks like there may be problems in NP that are not in BQP and vice versa. BBP is a funny beast, so, it is conjectured. We mentioned that we believe that P is not equal to NP, but people also believe, we mean, the conventional wisdom sort of leans towards P being equal to BPP. There is no proof of this either. However, the intuition is, if we want to do random numbers on our computer, we have a pseudo-random number generator. Our computer just calls in a random number generator, but it is a pseudo-random number generator. It makes numbers, a sequence of numbers that are deterministic. However, the structure in them seems to be hard to extract by any program, kind of related to the intuition that we believe P is not equal to NP, we know, that there can be patterns that are present, which we cannot efficiently extract. So, we believe that the random number generator, even though there is information, theoretically, a pattern, it is too hard computationally to extract it. Thus, we can treat the output of the random number of our, sorry, our pseudo-random number generator as though it were true randomness, and our program will do just as well as if it had true randomness. So, that is the argument for why P is equal to BPP. There are some theorems that sort of move in that direction, but it is really conjecture. It is really not proven, but it is believed to be true., we believe that this is not equal to BQP, right? That is why we are interested in quantum computers because we think they can do things we cannot do with a randomized classical computer. 

Classical computers can efficiently solve problems in the computational complexity class P. There is a desire (and a question if it is possible) to develop a computational device able to solve exponentially hard problems in NP-Complete and even harder problems. With the complexity classes P, NP, NP-Complete\cite{vyskocil_embedding_2019}, and PSPACE defined, we will now discuss the performance of quantum computers, and where the class of problems a quantum computer can solve efficiently falls within the computational complexity class hierarchy.

Quantum computers outperform classical computers for specific problems, such as factoring large integer numbers, a problem that is believed to be part of NP. The advantage of a quantum computer arises due to its ability to exploit the problem's underlying mathematical structure differently than classical computers can. The quantum computer can determine the prime factors of a large integer number in polynomial time, which is exponentially faster than it takes classical computers to perform the same task.

Can a quantum computer solve NP-Complete problems? As the example of Shor's factoring algorithm suggests, quantum computers are able to outperform classical computers on problems that feature a mathematical structure that can be exploited by the working principles of quantum computers. NP-Complete problems do not seem to exhibit these sorts of favorable mathematical structures. There is no evidence to date that suggests quantum algorithms exist that could outperform classical algorithms for NP-Complete problems.

The classical search problem searching for a particular element in a list composed of n elements can take up to n trials to perform deterministically, and it takes on average $ n/2 $. Grover's quantum search algorithm finds the particular element in only $ \sqrt {n} $ steps \cite{grover_fast_1996}. While quantum processors outperform classical computers for some problems, the degree of quantum speedup may only be polynomial in certain cases\cite{moussa_function_2019}.

There exist other quantum algorithms able to generate minor speedups over the best known classical algorithms for problems in N and NP \cite{pednault_breaking_2018}. Still, there is no evidence that quantum computers can efficiently solve NP-Complete problems \cite{tang_quantum-inspired_2019}. There are proposals for quantum algorithms that may be able to verify a proposed solution in polynomial time for problems outside of NP. However, there is no evidence that quantum computers can outperform classical computers on problems outside of PSPACE. The problems quantum computers can solve efficiently form the computational complexity class BQP, with BQP being an abbreviation for Bounded-error Quantum Polynomial-time.

\begin{figure}[H] \centering{\includegraphics[scale=.32]{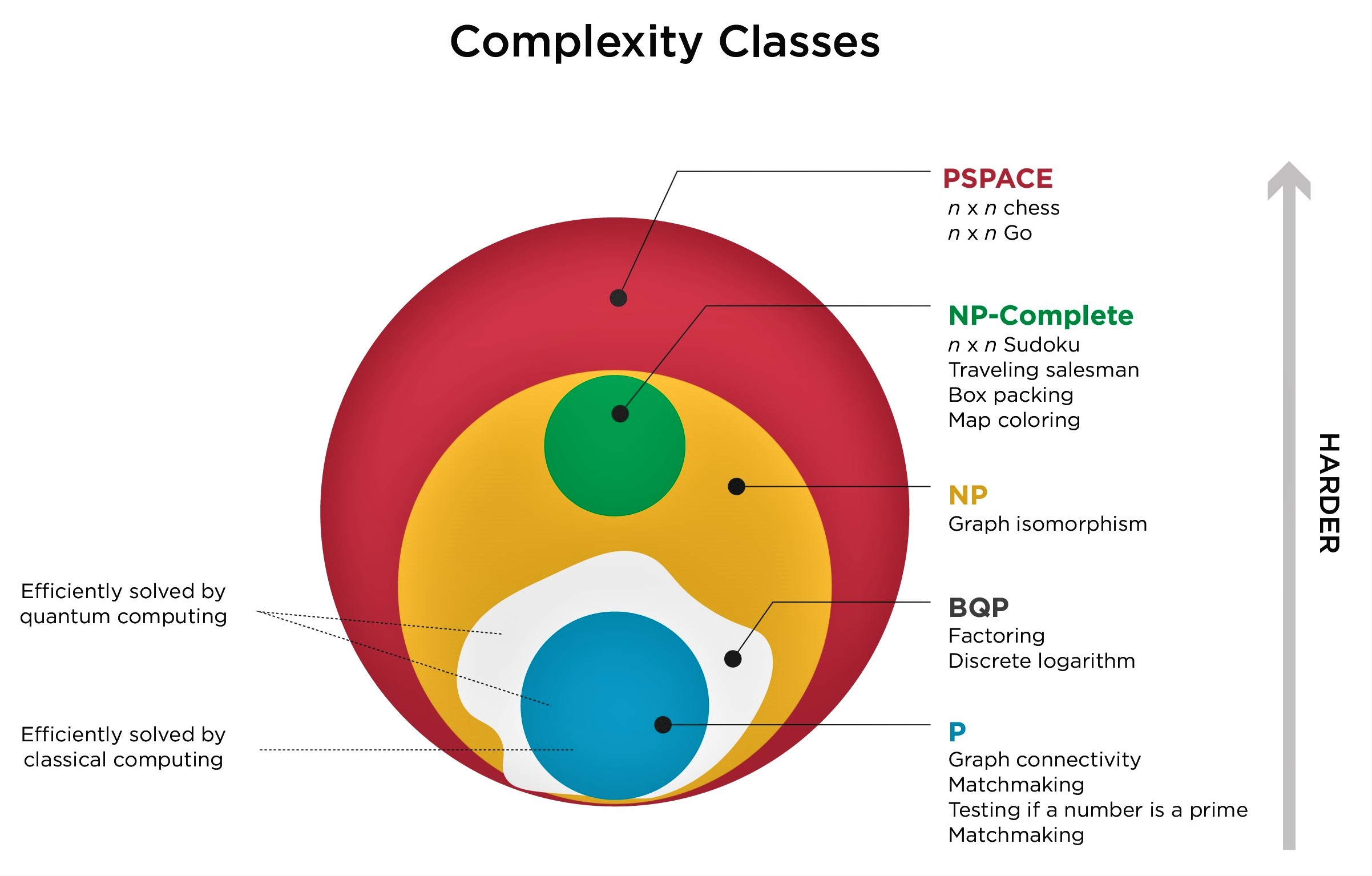}}\caption{Computational Complexity Classes: PSPACE contains all classes and problems which require a polynomial amount of memory on a conventional computer independent of the number of necessary computational steps to find or verify a solution. P describes the class of classically efficient computable problems. Problems are efficiently computable by a quantum computer form the complexity class BQP. Class NP contains problems that are efficiently verifiable. The hardest problems which are still efficiently verifiable form complexity class NP-Complete.}\label{fig4_20}
\end{figure}

Quantum computers are apparently more powerful than classical computers for certain problems. However, there is no formal proof that the hierarchy shown in Figure 1 is manifestly true. That is, there is no proof that an ``as-yet-undiscovered'' classical algorithm could not perform as well as a quantum algorithm for problems in NP where today there is a quantum advantage\cite{giri_review_2017}. The probability that a better classical algorithm exists is likely small, but it is not zero. Thus, quantum computers are able to outperform classical computers for certain tasks, but there is no formal proof that says it must be so.

\item  BPP and BQP; Two important complexity classes regarding random and quantum models of computation are BPP and BQP. Which of the following sentences are true about BPP and BQP:
\begin{itemize}    
\item BPP and BQP are defined by the set of languages that, after computation, lead to the wrong result more than 2/3 of the time in polynomial computational time.
\item BPP and BQP are defined by the set of languages that, after computation, lead to the wrong result of less than 1/3 of the time in polynomial computational time.
\item PP and BQP are defined by the set of languages that, after computation, lead to the correct result more than 2/3 of the time in exponential computational time.
\end{itemize}
Solution:\\
BPP and BQP stand for ``Bounded-error Probabilistic Polynomial'' time and ``Bounded-error Quantum Polynomial'' time, respectively. They represent the set of problems that can be computed in polynomial time with a bounded error of less than $ 1/3 $.

\item  Relation Between Complexity Classes; Complexity classes can overlap; they need not exclude one another. In fact, some complexity classes are entirely contained within larger complexity classes. Considering the complexity classes, we have studied so far  P, NP, BPP, and BQP, which of the following describe valid relationships between these complexity classes? (the notation $ ``x \subseteq y'' $ means ``all elements in x are also in y'').

\begin{itemize}
\item $ P \subseteq BQP \subseteq BPP \subseteq NP $
\item $ NP \subseteq BPP \subseteq BQP \subseteq P $
\item $ BPP \subseteq BQP \subseteq P \subseteq NP $
\item $ P \subseteq BPP \subseteq BQP $
\end{itemize}
Solution:\\
It is still an open question whether some of these complexity classes are in fact equal. One conjecture says that $ P=BPP $, and we might be familiar with the conjecture that $ P \neq NP $, the formal proof of which statement is one of the Millennium Prize Problems.

\section{How to Think About Quantum Supremacy: BQP and BPP - A Taste of Quantum Supremacy} 
This now gives us a way of arguing that a certain quant thing that a quantum computer can do cannot be done classically. So, this gives rise to what we could call quantum supremacy\cite{arute_quantum_2019,markov_quantum_2018}. If L is BQP-complete, then a quantum computer can solve it, and a classical computer cannot. So, then, a quantum computer can solve L. Classical computer-

\section{How to Think About Quantum Supremacy: Beyond NP - Counting Solutions} 
What we want to discuss about next is a way of going, in a sense, beyond NP. It may seem strange. We might think, well, look NP already takes exponential time. However, we want to discuss things were even to verify it takes exponential time. Thus, this is a way of separating NP from things even above it. We will see that it can be useful for reasoning about real-world devices, even devices that are weaker than a universal quantum computer. So, it shows we kind of the remarkable subtlety and flexibility of complexity theory\cite{bernstein_quantum_1993}. So, to do this, let us go back to three-set. So, the three-set is a Boolean formula. We put in input. We do these or's and these ands, and the answer is going to be either true or false. Let us say one or 0. So, let us say $ \phi $ of X is equal to 0 if unsatisfied. False one is satisfied, i.e., true. In other words, if all the clauses are true, we say the formula satisfied, $ \phi $ of X is 1. If not, the $ \phi $ of X is 0. This is $ \phi $ at a particular input. now, let us define the count of $ \phi $ to be the total number of satisfying assignments. So, in other words, we sum overall and bitstrings X, $ \phi $ of X. now, we can rephrase the basic problem of NP, in terms of this count function. We can say that an NP-complete task is to determine whether the count of $ \phi $ is either equal to 0 or greater than 0. So, remember, the three set problems ask, given this formula, does there exist an X that satisfies it? Furthermore, that is equivalent to saying, is the total number of solutions positive? Is it one or more versus it equal to 0? So, distinguishing these two cases is an empty NP-complete. $ \phi $ is given. $ \phi $ likes the input. So, we have a problem specified by our input, which is $ \phi $, and now we want to know the answer to this question. Is the count of the solution 0 or positive? The question is, is it NP-complete for arbitrary $ \phi $. the point is NP-completeness referred to as a language, or we could say a problem. So, the language is basically the set of all $ \phi $ that has one or more solutions. We could also think of this as a problem. The problem is given an arbitrary $ \phi $, determined if it has a satisfying ascent. Thus, problems are never defined for a fixed input. Because if so, there would not be hard. The answer would just be 0 or 1, and that would be the end of that. The key thing is we have to say this is the worst-case complexity, meaning that we have to be able to solve the problem, we demand to be able to solve it for any input $ \phi $. Just like for factoring. It is not hard to factor the number 1,372,000. Right? Any particular number is not hard because there is an answer. What is hard is factoring an arbitrary number. Just like what is hard is to compute the count of this for an arbitrary $ \phi $.
 
\section{How to Think About Quantum Supremacy: Beyond NP - Approximate and Exact Counting} 
So, this is what NP is. Let us now define some things that are kind of harder than this. So, one of them is there is a complexity class for it, but it is not the most natural one. So, we just in to call it the approximate counting problem which says given $ \phi $ at the threshold T, determine whether the count of $ \phi $ is either greater than 2T or less than T. So, now, we promise that one of these is the case, and now we are going to determine which one it is. So, T here, we would call it a threshold. So, we are promised there is some threshold. We either have more than let us say 2,000 solutions or less than 1,000. We want to determine which is the case. Alternatively, we have more than 2 to the 100 solutions or less than 2 to the 99 solutions, determine which is the case. It is clearly at least as hard as NP because we could take T to be like 1/2 or something, and that would include NP. However, we can choose T to be with it whatever we like, so it includes things that look quite different than NP. If we want to distinguish, there are more than 2 to the 1,000 or fewer than 2 to the 99 solutions, that do not look like something we can phrase as an NP problem. So, it looks a priori harder than NP. Then, we can do something else, which is yet harder than this, which is called exact counting. This actually corresponds to the complexity class known as sharp P. This is like sharp. It is kind of dumb. We say sharp like the musical note, but we really mean like the number sign, so it is also called number P. what this is is given $ \phi $ and T, is the count of $ \phi $ either greater than T or less than or equal to T minus 1? So, we are going to think of that as exact counting, except we try to phrase it actually, we sorry. We wrote sharp P. It should really be PP. The reason for that name, it is kind of a tangent, which we will not go on. The point is that it is the third one of these problems. The first one is, is the number 0 or positive? The second one is approximate to within a factor of 2. The third one nails it down exactly, figure out exactly is it above this threshold or below it? Are there more than 1,000,300,042 solutions, or are there fewer than that? Again, this does not look like it reduces to NP. It also does not look like it reduces to exact counting. If we can estimate it to within a factor of 2, that does not give the ability to nail it down on the nose, especially if the numbers we are discussing are exponentially large. Let us try to give us an example of this. So, let us take the Goldbach Conjecture. Here is a program and consider the following program. Does it take a purported proof of the Goldbach Conjecture that fits in a million lines? So, we have a proof written in some kind of formal reasoning system, that is no more than a million lines. It takes that proof, and it checks, does this purported proof actually prove the Goldbach Conjecture? Right? If it does, then it outputs 1. If it does not, it outputs 0. Now, most proofs do not even parse correctly, so they will just output 0. maybe the Goldbach Conjecture is false, or it has no proof in less than a million lines. In that case, it will always output 0. It will be, as we say, the trivial map. Nevertheless, maybe the Goldbach Conjecture is true and has a proof under a million lines, in which case this thing will not be trivial. It will sometimes output 1. So, distinguishing those cases sounds like a pretty hard problem. Right? It sounds like it would be surprising if a computer could do this, or if there was some like simple automated way of doing it in a very short amount of time. Because when we try to solve proofs, we have not really figured out a systematic way of doing it. We just use creativity. We bring in new ideas. We have never found like a foolproof way that will always tell us, is there a short proof or not? Thus, that is why the consensus is that such an algorithm does not exist, to tell, if we have a short little program, we want to tell if it is trivial or not, the consensus is that we cannot do it. 

\section{How to Think About Quantum Supremacy: Modern Argument of Quantum Supremacy}

Let us just tell us the high-level argument. The goal of quantum supremacy is to say, here is something that a quantum computer could do that a classical computer cannot do. By cannot, we have to mean relative to some assumption. So, the kind of the more modern approach is to say, assume that approximate counting is not equivalent to exact counting. In other words, if we can do approximate counting, there is no polynomial-time reduction from exact counting to approximate counting. So, we reason about things far above MP., But that is fine. We can still discuss them.
Moreover, they sure look different. It looks like if we can do approximate counting, there is no way with polynomial overhead to make it to exact counting. If we believe that, then that implies that there is no classical simulation of quantum computing\cite{chen_classical_2018}. So, that is kind of the more modern approach to showing quantum supremacy. There are kind of two advantages to it. One of them is that it applies even to non-universal quantum computing \cite{maslov_outlook_2019}. So, if we have a system of non-interacting photons, they have to be single photons going through a bunch of beam splitters\cite{bouland_generation_2014}, and then we measure them that has been called, sometimes, boson sampling \cite{neville_no_2017} that is not a universal architecture of quantum computing \cite{smith_practical_2017,linke_experimental_2017}.
However, even simulating this, it turns out, can be ruled out if we assume that approximate counting is not equal to exact counting. And then the second advantage is kind of more sociological, is that the assumption looks non-quantum. So, a lot of this started with Scott Aaronson, who had spent his time discussing to non-quantum computer scientists, who do not know what h-bar is. They are kind of fuzzy on complex numbers. They do not really believe in this whole quantum thing. Thus, when we say a quantum computer can do things that a classical computer cannot if we assume that BPP is not equal to BQP, they are not very impressed because we have just assumed that a quantum computer can do things that a classical computer cannot. So, what are we even saying here? However, if we say, assume that approximate counting is not equal to exact counting, we could say, well, we do not know about quantum mechanics. However, sure, that seems believable. And then we say, well, if we accept that, then o quantum computer can do things that the classic computer cannot, that is a more interesting statement. So, again, it comes down to the fact that complexity statements are somewhat sociological, and our belief in them depends on how plausible the assumption is and how widely it has been studied. Thus, that is why these quantum supremacy arguments are considered more plausible, and kind of the assumption is more desirable than the BPP versus BQP.

The performance of a computational task by a quantum computer that is beyond the capability of any classical computer is known as quantum supremacy. Such a demonstration would further confirm our understanding of quantum mechanics and contradict the extended Church-Turing thesis. The Church-Turing thesis states that quantum computers can be simulated by classical computers with polynomial overhead. More practically, a demonstration of quantum supremacy would increase our confidence in the eventual success of universal quantum computers.

While it is thought that existing quantum algorithms, such as Shor's factoring algorithm and Grover's search algorithm, could eventually be used to demonstrate quantum supremacy, the number of gates required for these demonstrations is far beyond the reach of current technology. Therefore, the focus of modern quantum supremacy research is on finding restricted models of quantum computation, which are not only realizable in the near term but ones for which we also have significant complexity-theoretic evidence that the classical simulation of these models is hard. It should be emphasized that the problems solved by these restricted models are, unlike Shor's factoring algorithm or Grover's search algorithm, generally not one of practical value beyond the demonstration of quantum supremacy\cite{kirke_application_2019}. Currently, the leading proposals for such a demonstration are boson sampling and random quantum circuits, each of which will be described below.

Boson sampling, originally proposed by Scott Aaronson and Alex Arkhipov, is a problem related to the simulation of photons traversing a linear-optical network. The network, usually generated at random, consists of m input and output ports and $ n\ll $ m photons injected into different input ports. The output ports are monitored by photon detectors, and the computational challenge is to sample the output photon distribution. This problem is of interest because the transition amplitudes of the photons in these experiments have been shown to be related to the permanent of a complex matrix, primarily used in combinatorics.

The permanent of a matrix M, with elements $ m_{i,j} $, is given by
\begin{equation}\label{eq4_54}
\mbox{per}(M)=\sum_{\sigma\in S_n}\prod_{i=1}^n m_{i,\sigma(i)}
\end{equation}

Where the $ \sigma \in S_ n $ means that the sum is over the symmetric group of size n, in practice, this means the sum is over all permutations of a list going from 1 to n.

As an example, consider a general $ 3\times 3 $ matrix; the sum would then be over lists with elements given by both the even permutations (1,2,3), (2,3,1), (3,1,2), and the odd permutations (2,1,3), (1,3,2) and (3,2,1). The first sum is then given by $ m_{1,1}m_{2,2}m_{3,3} $, the second sum by $ m_{1,2}m_{2,3}m_{3,1} $, the third by $ m_{1,3}m_{1,1},m_{3,2} $, and so on. Finding all of these terms results in the formula, for a general $ 3\times 3 $ matrix given by

$ \text{per}\left(\begin{array}{ccc} a&b&c \\ d&e&f \\ g&h&i  \end{array}\right)=aei + bfg + cdh + ceg + bdi + afh. $

This is closely related to the better-known matrix determinant which is given by

\begin{equation}\label{eq4_55}
det(M)=\sum_{\sigma \in S_n} \left( \text{sgn}(\sigma) \prod_{i=1}^n m_{i,\sigma_i}\right)
\end{equation}

Where sgn denotes the sign of the permutation. Using the same example as above, the determinant of a general $ 3\times 3 $ matrix becomes

$ \text{det}\left(\begin{array}{ccc} a&b&c \\ d&e&f \\ g&h&i  \end{array}\right)=aei + bfg + cdh - ceg - bdi - afh. $

Despite the similarities of these two definitions, the matrix permanent is thought to be significantly more difficult to compute as n increases. Understanding exactly why this is so is beyond the scope of this text unit. However, we can make the following simple argument for why we would expect this to be the case. While the above calculations follow directly from the definitions of the matrix permanent and determinant, we can also use an alternative formulation, more familiar to students of linear algebra, called Laplace's formula. Laplace's formula expands the calculation of the determinant along a single row or column and relies on the computation of the determinants of smaller sub-matrices.

For our matrix above, this would look like
\begin{equation}\label{eq4_56}
\left(\begin{array}{ccc}a&b&c\\d&e&f\\g&h&i\end{array}\right)=a\times\text{det}\left(\begin{array}{cc}e&f\\h&i\end{array}\right)-b\times\text{det}\left(\begin{array} {cc}d&f\\g&i\end{array}\right)+c\times\text{det}\left(\begin{array}{cc}d&e\\g&h\end{array}\right).
\end{equation}

An analogous formula exists for calculating the matrix permanent, but with the sign of all of the terms being positive.

The benefit of this formulation is that, since the matrix determinant is invariant to elementary row and column operations, a matrix can be reformulated such that a single row or column is mostly zeroed, significantly reducing the overall number of terms that need to be calculated. Surprisingly, the matrix permanent is not invariant to elementary row and column operations, and therefore this shortcut does not exist.

It is precisely the appearance of the matrix permanent in the calculation of detection statistics in boson sampling, which is used to estimate the computational complexity of the classical problem. Several experimental realizations of boson sampling have been achieved with nine modes and five photons. However, it is currently thought that at least 50 or more single photons would be required to reach quantum supremacy.

Random quantum circuits provide a model for quantum supremacy\cite{villalonga_flexible_2019}, which resides within the quantum circuit model. The most basic model referred to as Instantaneous Quantum Polynomial (IQP) circuits\cite{bremner_achieving_2017}, is based on the application of gates, which all commute with one another. One way this can be realized is, to begin with, a state where every qubit is zero, such as $ \vert 00...0\rangle $, and applying a Hadamard gate to each qubit. This is then followed by gates, which are only diagonal, and finally, Hadamard's are applied to each qubit one more time. Diagonal gates are required here because matrix multiplication is commutative for diagonal matrices, a fact that is easily proven.

The computational task is then to sample from the distribution of measurement outcomes for a random implementation of this circuit. Since these circuits are made up of a restricted class of gates, they are easier to implement physically and simpler to analyze theoretically than full-scale models of quantum computation\cite{nielsen_quantum_2011}. Unfortunately, it is thought that this simplicity causes IQP to become classically simulable when noise is present \cite{bremner_average-case_2016,bremner_classical_2011,calude_-quantizing_2007}. A related proposal for achieving quantum supremacy is to sample the output distribution of random circuits comprised of single and two-qubit gates, which in general do not commute.

\item  Relation Between Complexity Classes; According to the section, ``How to think about quantum supremacy: modern argument of quantum supremacy,'' which of the following are true statements about the modern concept of quantum supremacy?
\begin{itemize}    
\item Quantum supremacy implies a classical computer cannot simulate a quantum computer under certain assumptions.
\item The assumption ``approximate counting does not equal exact counting'' implies quantum supremacy.
\item The assumption ``approximate counting does not equal exact counting'' is viewed as more desirable than quantum supremacy arguments surrounding BQP and BPP, because it need not invoke quantum mechanics directly.
\item Quantum supremacy applies even to non-universal quantum computers.
\end{itemize}
Solution:\\
All of the above statements are true.

\section{Mitigating Errors in a Superconducting Qubit System with Composite Pulses}

In this IBM Q experience, we will apply the composite pulsing technique that we discussed during section 3. We strongly advise we to revisit the content on ``Correcting Systematic Control Errors.''

In the first section, we will discuss how to implement advanced single-qubit gates, and in the second section, we will discuss how to define new gates from the built-in ones, and in particular, how to define basic rotations. In the third section, we will review the basic theory about the composite pulse $ BB1(\pi) $ and will discuss how to define it in QASM\cite{noauthor_qiskitopenqasm_2020}. Last, in the fourth section, we will compare our IBM Q measurement results from the implementation of $ R_x(\pi)^{2N} $ and $ BB1(\pi)^{2N} $ to decide which one is the better implementation of a rotation about the x-axis at an angle $ \theta=\pi $ .

Advanced Gates

QASM codes \cite{cross_open_2017} that used the basic pre-defined single-qubit gates provided by the IBM Quantum Experience. We will need to go beyond these basic single-qubit gates in order to implement the composite pulse experiments.

In this first exercise, we will discuss how to use the advanced single-qubit gates that we will need for the subsequent experiments. We can gain access to these gates on the IBM Q website by clicking the ``Advanced'' checkbox, as shown below.

\begin{figure}[H] \centering{\includegraphics[scale=1]{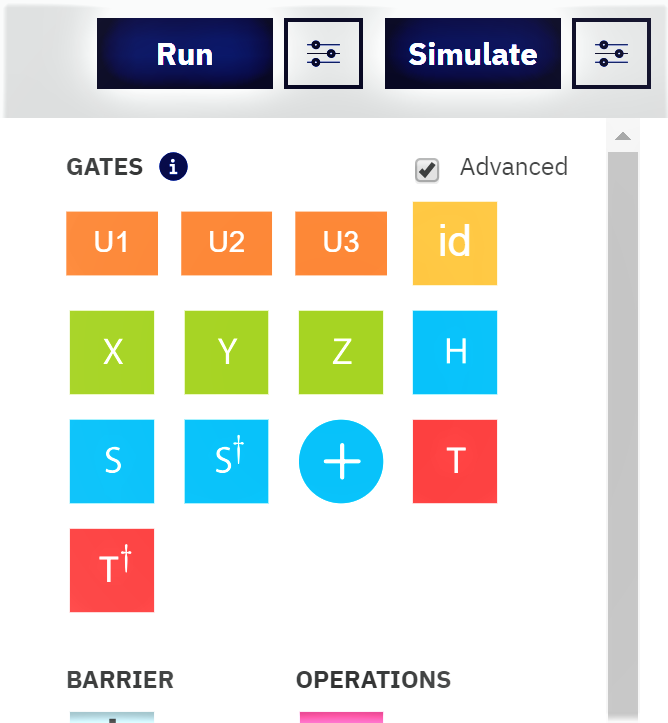}}\caption{Advance gates}\label{fig4_21}
\end{figure}

The table below shows the three pre-defined advanced single-qubit gates $ U_1(\lambda ), U_2(\phi ,\lambda ), $ and $ U_3(\theta ,\phi ,\lambda ) $. Using combinations of these three gates, it is possible to implement any single-qubit rotation.

\begin{table}[H]
\centering
\caption{Single-qubit gates $ U_1(\lambda ), U_2(\phi ,\lambda ), $ and $ U_3(\theta ,\phi ,\lambda ) $}
\label{tab:4_1:Table 3}
\resizebox{\textwidth}{!}{
    \begin{tabular}{|c|c|c|}\hline
QASM Code &    Matrix    &    Definition  \\\hline    
$ U_1(\lambda) $ &    
$ U_1(\lambda)=\left(\begin{array}{cc} 1 & 0\\ 0 & e^{i\lambda} \end{array} \right) $ &
One-parameter single-qubit phase gate  with zero duration (0-pulses).\\\hline
$ U_2(\phi,\lambda) $ &
$U_2(\phi,\lambda)=\frac{1}{\sqrt{2}}\left(\begin{array}{cc} 1 & -e^{i\lambda}\\ e^{i\phi} & e^{i(\phi+\lambda)} \end{array} \right) $ &
Two-parameter single-qubit gate with duration one unit of time (1-pulse).\\\hline
$ U_3(\theta,\phi,\lambda) $ &
$U_3(\theta,\phi,\lambda)=\frac{1}{\sqrt{2}}\left(\begin{array}{cc} \cos(\theta/2) & -e^{i\lambda}\sin(\theta/2)\\ e^{i\phi}\sin(\theta/2) & e^{i(\phi+\lambda)}\cos(\theta/2) \end{array} \right) $ &
Three-parameter single-qubit gate with duration 2 units of gate time (2-pulses).\\\hline
\end{tabular}}
\end{table}

Example 1:\\
The following QASM code applies a $ U_3(\theta ,\phi ,\lambda ) $ gate to the first qubit with $ \theta =\pi, \phi =\arccos(1/4) $, and $ \lambda =\pi /4 $,\\
\begin{lstlisting}
include ``qelib1.inc'';
qreg q[1];
creg c[1];
u3(pi, acos(1/4), pi/4) q[0]; 
measure q[0] -> c[0];
\end{lstlisting}
Example 2:\\
The following QASM code applies a $ U_2(\phi ,\lambda ) $ gate to the first qubit with $ \phi =3*\pi /4 $, and $ \lambda =5*\pi /4 $,\\
\begin{lstlisting}
qreg q[1];
creg c[1];
u2(3*pi/4, 5*pi/4) q[0];  
measure q[0] -> c[0];
\end{lstlisting}

\item  Advanced Gates; Run a QASM code that implements the following three gates in order: $ U_3(\theta,\phi,\lambda), U_2(\phi,\lambda), and U_1(\lambda) $. Use qubit q[0] and angles$  \theta=\pi, \phi=\pi/4 $, and $ \lambda=\pi/4 $.
\begin{lstlisting}
include ``qelib1.inc'';

qreg q[1];
creg c[1];
u3(pi, pi/4, pi/4) q[0]; 
measure q[0] -> c[0];
\end{lstlisting}
    
Gate Definition and Rotations\\
To implement the following exercises, we will need to define new gates using the built-in standard and advanced gates. A gate named name1 with parameters $ \alpha $, $ \beta $, and $ \gamma $ is defined as follows:\\
gate name1($ \alpha $,$ \beta $,$ \gamma $) a\\
{\\
    body a;\\
}\\
where a  indicates the qubit (within the definition of the new gate) on which the gate is acting, and the body is a single gate  such as u3($ \pi $,$ \beta $,$ \gamma $)  or a combination of gates.\\

Example 1: A rotation of an angle $ \theta $ about the x-axis, $ R_ x(\theta) $, and applied to qubit a is defined using the advanced gate u3($ \theta $,$ \beta $,$ \gamma $) by assigning the value of $ \beta $ to $ -\pi /2 $ and the value of $ \gamma $ to $ \pi /2 $.\\
gate rx($ \theta $) a\\
{\\
    u3(theta,$ -\pi/2 $,$ \pi/2 $) a;\\
}\\
\begin{lstlisting}
gate rx(theta) a
{
    u3(theta,-pi/2,pi/2) a;
}
\end{lstlisting}

Example 2: The sequence of single-qubit gates XYZ is defined using the basic built-in gates X, Y and Z in the following manner:\\
gate name a\\
{\\
    z a;      // In the sequence XYZ, Z is applied\\ first\\
    y a;\\
    x a;      // In the sequence XYZ, X is applied\\ last \\
}\\
Note that the basic built-in gates do not necessarily require arguments.\\

Example 3: In the code below, the gate id5 is defined as the multiplication of five identity gates IIIII, and it is implemented in the first qubit, in the following manner:\\
\begin{lstlisting}
include ``qelib1.inc'';
qreg q[1];
creg c[1];

gate id5 a
{ id a; id a; id a; id a; id a;
}

id5 q[0];
measure q[0] -> c[0];
\end{lstlisting}

On the IBM Q website, the code appears as follows:

\begin{figure}[H] \centering{\includegraphics[scale=.5]{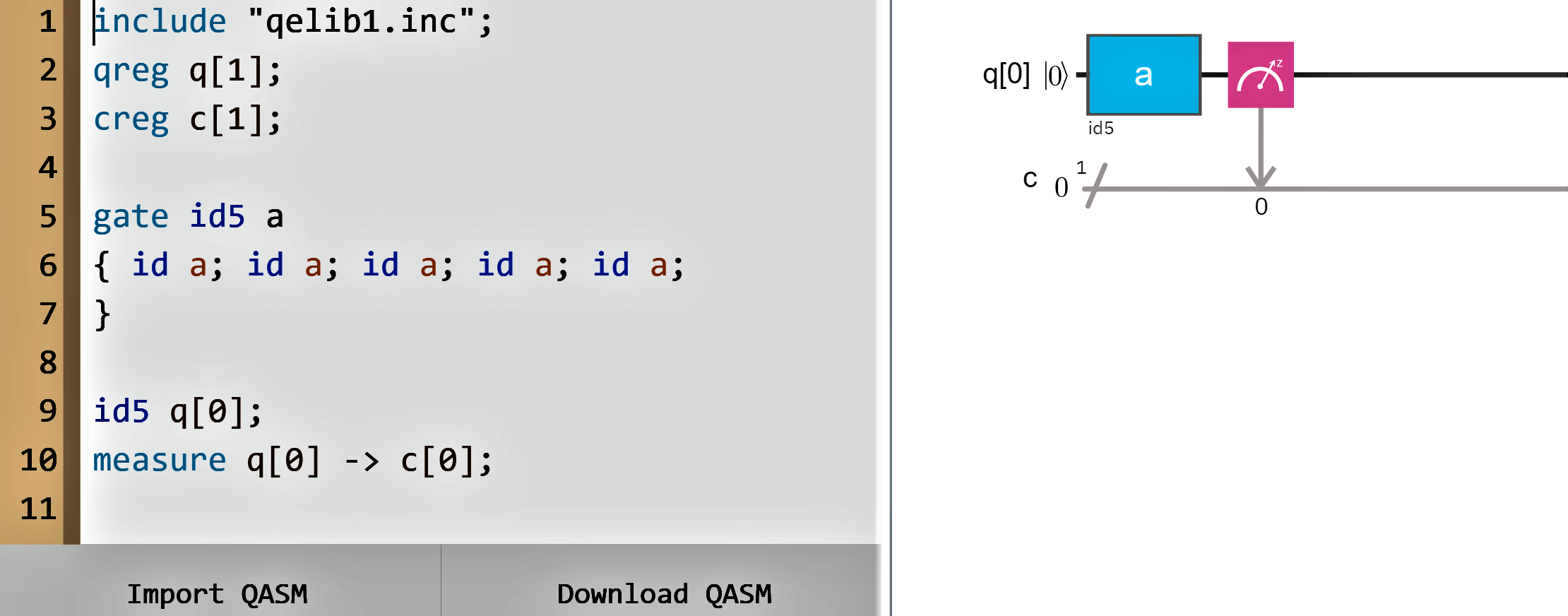}}\caption{gate id5}\label{fig4_22}
\end{figure}

\item  Gate Definition 1; Write the QASM code, which defines a gate called our gate. This gate implements a Z gate followed by an X gate, and it does this twice. Recall the order of these gates is read from right to left. We can represent the repeated concatenation of these gates as $ (XZ)^2 = XZXZ $. The state of a single qubit is transformed in the following manner:\\
$ \lvert 0\rangle\to \left(XZ\right)^2\lvert 0\rangle. $
Write the QASM codes that applies mygate to the first qubit.
\begin{lstlisting}
include ``qelib1.inc'';
qreg q[1];
creg c[1];

gate mygate a
{ z a; x a; z a; x a;
}

mygate q[0];
measure q[0] -> c[0];
\end{lstlisting}

\item  Gate Definition 2; Write a QASM code that defines a gate mygate2($ \alpha $) and then applies it to the first qubit with a parameter of $ \alpha = \pi $. This gate comprises the product of three $ U_3(\theta,\phi,\lambda) $ gates, and it transforms the state of the qubit as follows:

\begin{equation}\label{eq4_57}
\begin{split}
\lvert 0 \rangle\to  \\ & U_3\left(\pi,\alpha-\frac{\pi}{2},\alpha+\frac{\pi}{2}\right) U_3\left(\pi,3\alpha-\frac{\pi}{2},-3\alpha+\frac{\pi}{2}\right) U_3\left(\pi,\alpha+\frac{\pi}{2},\alpha-\frac{\pi}{2}\right)\lvert 0\rangle
\end{split}
\end{equation}
\begin{lstlisting}
include ``qelib1.inc'';
qreg q[1];
creg c[1];

gate mygate2(alpha) a
{   u3(pi, alpha + pi/2, alpha - pi/2) a;
    u3(pi, 3*alpha - pi/2, -3*alpha + pi/2) a;
    u3(pi, alpha - pi/2, alpha + pi/2) a;
}
mygate2(pi) q[0];
measure q[0] -> c[0];
\end{lstlisting}

Basic Gates \\

Single-qubit rotations around the x, y, and z axes are predefined as rx($ \theta $), ry($ \theta $), and rz($ \theta $) in the ``qelib1.inc'' file included in our QASM codes; as a result, they should work in any of the exercises on this site without needing to be defined again. While these gates are not yet supported on the IBM Q website, it is possible to implement them using the formalisms described above. The table below shows the matrix form of rotations around each axis, their corresponding advance gate matrix, and the QASM code we can use if rx($ \theta $), ry($ \theta $), and rz($ \theta $) are not available:
\begin{table}[H]
\centering
\caption{The matrix form of rotations around each axis, their corresponding advance gate matrix, and the QASM code we can use if rx($ \theta $), ry($ \theta $), and rz($ \theta $)}
\label{tab:4_1:Table 5}
\resizebox{\textwidth}{!}{
\begin{tabular}{|c|c|c|}\hline
Basic Rotation &    Advance Gate Matrix    &    QASM Equivalent Gate  \\\hline
$ R_{x}\left( \theta \right) =\left( \begin{array}{cc} \cos \frac{\theta }{2} & -i\sin \frac{\theta }{2} \\ -i\sin \frac{\theta }{2} & \cos \frac{\theta }{2}\end{array}\right) $ &    
$ U_{3}\left( \theta ,-\frac{\pi }{2},\frac{\pi }{2}\right) =\frac{1}{\sqrt{2}}\left( \begin{array}{cc} \cos \frac{\theta }{2} & -i\sin \frac{\theta }{2} \\ -i\sin \frac{\theta }{2} & \cos \frac{\theta }{2}\end{array}\right) $ &    
$ u3(theta,-pi/2,pi/2) $  \\\hline
$ R_{y}\left( \theta \right) =\left( \begin{array}{cc} \cos \frac{\theta }{2} & -\sin \frac{\theta }{2} \\ \sin \frac{\theta }{2} & \cos \frac{\theta }{2}\end{array}\right) $    &
$ U_{3}\left( \theta ,0,0 \right) =\frac{1}{\sqrt{2}}\left( \begin{array}{cc} \cos \frac{\theta }{2} & -\sin \frac{\theta }{2} \\ \sin \frac{\theta }{2} & \cos \frac{\theta }{2}\end{array} \right)     $ &
$ u3(theta,0,0) $ \\\hline
$ R_{z}\left( \theta \right) =\left( \begin{array}{cc} \exp {-i\frac{\theta }{2}} & 0 \\ 0 & \exp {i\frac{\theta }{2}} \end{array}\right) $ &    
$ U_1(\phi )= \exp {i\frac{\theta }{2} } \left(\begin{array}{cc} \exp {-i\frac{\theta }{2} } & 0\\ 0 & \exp {i\frac{\theta }{2} } \end{array} \right)     $&
$ u1(phi) $\\\hline
\end{tabular}}
\end{table}

Note that there are differences between the basic rotation matrices and their corresponding advanced gate matrices, but these differences correspond to a global phase, and therefore it is valid to consider the matrices as equivalent.

\item  Rotation About Y; Write a QASM code that applies a rotation of $ \theta=\frac{\pi}{2} $ about the y-axis to the first qubit. we can use either the predefined ry($ \theta $) gate or the equivalent advanced gate from the table above.
\begin{lstlisting}
include ``qelib1.inc'';
qreg q[1];
creg c[1];

ry(pi/2) q[0];  // Alternatively: u3(pi/2,0,0) q[0];

measure q[0] -> c[0];
\end{lstlisting}

Composite Pulse BB1($ \theta $)\\
Introduction\\
As we discussed there are different types of errors that may impact a desired operation. For example, the rotation of a single-qubit state about the $ \hat{n} $-axis at an angle $ \theta $,
\begin{equation}\label{eq4_59}
R_{\hat{n}}\left( \theta \right) =\exp \left( -i\frac{\theta }{2}\hat{n}\cdot \vec{\sigma }\right)
\end{equation}

may be affected by under and over-rotation errors, such that the rotation angle is no longer $ \theta $, but is instead given by $ \theta \left( 1+\epsilon \right) $, where $ \epsilon $ is an unknown, but fixed, parameter,
\begin{equation}\label{eq4_60}
R_{\hat{n}}\left( \theta \right) \rightarrow \tilde{R}_{\hat{n}}\left( \theta \left( 1+\epsilon \right) \right) =\exp \left( -i\frac{\theta \left( 1+\epsilon \right) }{2}\hat{n}\cdot \vec{\sigma }\right)
\end{equation}

Implementing the rotation using the $ BB1(\theta) $ composite pulse realizes the operation with $ \epsilon $ to higher order (see reference at the end). For example, whereas the average gate fidelity of the conventional $ R_{x}\left( \frac{\pi }{2}\right) $ rotation is quadratic in $ \epsilon $ for small $ \epsilon $, the net error for $ BB1(\pi/2) $ goes as $ \epsilon ^{6} $. The $ BB1(\theta) $ sequence is implemented using the product of four rotations about different axes and at different angles,
\begin{equation}\label{eq4_61}
BB1(\theta)=\tilde{R}_{\phi }\left( \pi \right) \tilde{R}_{3\phi }\left( 2\pi \right) \tilde{R}_{\phi }\left( \pi \right) \tilde{R}_{x}\left( \theta \right) 
\end{equation}
where\\
$ \tilde{R}_{\phi }\left( \theta \right) =\left( \begin{array}{cc} \cos \frac{\theta }{2} & -ie^{-i\phi }\sin \frac{\theta }{2} \\ -ie^{i\phi }\sin \frac{\theta }{2} & \cos \frac{\theta }{2}\end{array}\right)$\\
is a rotation about the $ \left( \cos \phi ,\sin \phi ,0\right) $-axis at an angle $ \theta $ with $ \phi =\arccos \left( -\theta /4\pi \right) $.\\

QASM Implementation\\

To write the $ BB1_{\theta } $ sequence in QASM, we will need the rotations $ \tilde{R}_{\phi }\left( \pi \right), \tilde{R}_{3\phi }\left( 2\pi \right) $ and $ \tilde{R}_{x}\left( \theta \right) $. These three operations can be implemented using the advanced gate,

\begin{equation}\label{eq4_62}
U_{3}\left( \theta ,\phi ,\lambda \right) =\frac{1}{\sqrt{2}}\left( \begin{array}{cc} \cos \frac{\theta }{2} & -e^{i\lambda }\sin \frac{\theta }{2} \\ e^{i\phi }\sin \frac{\theta }{2} & e^{i\left( \phi +\lambda \right) }\cos \frac{\theta }{2}\end{array}\right)
\end{equation}

A rotation of angle $ \theta=\pi $ about the $ \left( \cos \phi ,\sin \phi ,0\right) $-axis, \\
$ \tilde{R}_{\phi }\left( \pi \right) =\left( \begin{array}{cc} 0 & -ie^{-i\arccos \left( -\theta/4\pi \right) } \\ -ie^{i\arccos \left( -\theta/4\pi\right) } & 0\end{array}\right) $\\
can be implemented as,
\begin{equation}\label{eq4_63}
U_{3}\left( \pi ,\arccos \left( -\theta/4\pi\right) -\frac{\pi }{2},-\arccos \left( -\theta/4\pi\right) +\frac{\pi }{2}\right)
\end{equation}

A rotation of angle $ \theta=2\pi $ about the $ \left( \cos 3\phi ,\sin 3\phi ,0\right) $-axis,\\
$ \tilde{R}_{3\phi }\left( 2\pi \right) =\left( \begin{array}{cc} \cos \pi & -ie^{-3i\phi }\sin \pi \\ -ie^{3i\phi }\sin \pi & \cos \pi \end{array}\right), $\\
can be implemented as,
\begin{equation}\label{eq4_64}
U_{3}\left( 2\pi ,3\arccos \left( -\theta/4\pi\right) -\frac{\pi }{2},-3\arccos \left( -\theta/4\pi\right) +\frac{\pi }{2}\right)
\end{equation}

And, a rotation of angle $ \theta $ about the x-axis\\
$ \tilde{R}_{x}\left( \theta \right) =\left( \begin{array}{cc} \cos \frac{\theta }{2} & -i\sin \frac{\theta }{2} \\ -i\sin \frac{\theta }{2} & \cos \frac{\theta }{2}\end{array}\right) $\\
can be implemented as,

\begin{equation}\label{eq4_65}
U_{3}\left( \theta ,-\frac{\pi }{2},\frac{\pi }{2}\right) =\frac{1}{\sqrt{2}}\left( \begin{array}{cc} \cos \frac{\theta }{2} & -i\sin \frac{\theta }{2} \\ -i\sin \frac{\theta }{2} & \cos \frac{\theta }{2}\end{array}\right)
\end{equation}

QASM Code For BB1\\

Given the gate definitions above, we might naïvely expect to define a $ BB1(\theta) $ composite pulse in QASM as follows:

\begin{lstlisting}
gate bb1($ \theta $) a {
// Rotation about the x-axis at an angle $ \theta $:
u3($ \theta $, $ -\pi/2 $, $ \pi/2 $) a;
    
// Rotation about the (cos(phi),(sin(phi),0)-axis at an angle $ \pi $:
u3($ \pi $, acos(-$ \theta $/(4*$ \pi $)) - $ \pi/2 $, -acos(-$ \theta $/(4*$ \pi $)) + $ \pi/2 $) a;
    
// Rotation about the (cos(3phi),(sin(3phi),0)-axis at an angle 2pi:
u3(2*$ \pi $, 3*acos(-theta/(4*$ \pi $)) - $ \pi/2 $, -3*acos(-$ \theta $/(4*$ \pi $)) + $ \pi/2 $) a;
    
//Rotation about the (cos(phi),(sin(phi),0)-axis at an angle $ \pi $:
u3($ \pi $, acos(-$ \theta $/(4*$ \pi $)) - $ \pi/2 $, -acos(-$ \theta $/(4*$ \pi $)) + $ \pi/2 $) a;
}
\end{lstlisting}

If we were to try this on the IBMQ hardware, however, we might find that it does not perform as well as expected. The reason is subtle: while $ U_3 $ can describe a rotation of an arbitrary angle $ \theta $, the actual quantum computer does not have a physical gate capable of carrying out such a rotation in one step. In fact, all gates on the IBMQ hardware are constructed from a combination of:

(a) 90-degree rotations about the x-axis of the Bloch sphere (implemented using fixed-duration microwave pulses),\cite{randall_efficient_2015,paraoanu_microwave-induced_2006} and

(b) ``virtual'' rotations about the z-axis (implemented in the software by changing the reference phase of subsequent pulses, effectively changing the rotation axis in the X-Y plane).

In particular, $ U_3 $ is implemented using two 90-degree x pulses and three virtual z rotations \cite{larose_overview_2019}:

$ U_3(\theta, \phi, \lambda) = R_z(\phi + 3\pi) R_x\!\left(\tfrac{\pi}{2}\right) R_z(\theta + \pi) R_x\!\left(\tfrac{\pi}{2}\right) R_z(\lambda) $ \\

While the abstraction above is fine for quantum computing (which is what QASM was designed for!), it does mean that we must be careful when our aim is to execute specific pulse sequences like BB1. In this case, since all operations must eventually be translated to 90-degree hardware pulses, there are only two special cases of$  BB1(\theta) $ which can be implemented exactly on the IBMQ hardware: $ BB1(\pi) $ and$  BB1_{\pi/2} $. In QASM, these are:\\
\begin{lstlisting}
gate bb1pi a {
    u3(pi, -pi/2, pi/2) a;
    u3(pi, acos(-1/4) - pi/2, -acos(-1/4) + pi/2) a;
    u3(pi, 3*acos(-1/4) - pi/2, -3*acos(-1/4) + pi/2) a;
    u3(pi, 3*acos(-1/4) - pi/2, -3*acos(-1/4) + pi/2) a;
    u3(pi, acos(-1/4) - pi/2, -acos(-1/4) + pi/2) a;
}\\
gate bb1piover2 a {
    u2(-pi/2, pi/2) a; // Implements a single-pulse 90 degree rotation about the x-axis
    u3(pi, acos(-1/8) - pi/2, -acos(-1/8) + pi/2) a;
    u3(pi, 3*acos(-1/8) - pi/2, -3*acos(-1/8) + pi/2) a;
    u3(pi, 3*acos(-1/8) - pi/2, -3*acos(-1/8) + pi/2) a;
    u3(pi, acos(-1/8) - pi/2, -acos(-1/8) + pi/2) a;
}
\end{lstlisting}

Note that:\\
(a) the single $ U_3(2\pi, \dots) $ gate has been split into two applications of $ U_3(\pi, \dots) $, each of which is implemented with two 90-degree pulses on IBMQ hardware, and\\
(b) the x-rotation of $ \pi/2 $ is implemented using a different advanced gate $ U_2 $, which ensures that it is implemented in hardware with a single 90-degree pulse as desired.\\

\item  BB1 With $ \theta =\pi $; Run a QASM code that implements $ BB1(\theta ) $ with $ \theta =\pi $ using the gate definition above. Implement this sequence on the first qubit.
\begin{lstlisting}
include ``qelib1.inc'';
gate bb1pi a {
    u3(pi, -pi/2, pi/2) a;
    u3(pi, acos(-1/4) - pi/2, -acos(-1/4) + pi/2) a;
    u3(pi, 3*acos(-1/4) - pi/2, -3*acos(-1/4) + pi/2) a;
    u3(pi, 3*acos(-1/4) - pi/2, -3*acos(-1/4) + pi/2) a;
    u3(pi, acos(-1/4) - pi/2, -acos(-1/4) + pi/2) a;
}
qreg q[1];
creg c[1];

bb1pi q[0];
measure q[0] -> c[0];
\end{lstlisting}

\item  Rotation About x-Axis With At An Angle $ \theta =\pi $; Run a QASM code that implements a rotation about the x axis at an angle $ \theta=\pi $ using the gate definition above. Use $ U_3(\theta,\phi,\lambda) $ to implement the rotation on the first qubit.
\begin{lstlisting}
include ``qelib1.inc'';

qreg q[1];
creg c[1];

u3(pi,-pi/2,pi/2) q[0];
measure q[0] -> c[0];
\end{lstlisting}

BB1$(\pi) $ Versus Basic Rotation:\\

In this section, we will compare $ BB1(\theta ) $ with $ R_ x(\theta ) $, for the case $ \theta =\pi $. Ideally,  $ BB1(\pi ) $ and $ R_ x(\pi) $ implement an equivalent net rotation: they both rotate the state of a qubit an angle $ \theta=\pi $ about the x-axis. When they are applied to the initial state $ \lvert0\rangle $, they ideally will rotate this state to $ \lvert1\rangle $. When $ BB1(\pi ) $ and $  R_ x(\pi) $  are applied an even number of times, they  ideally rotate the state $ \lvert0\rangle $ back to $ \lvert0\rangle $. In reality, when one applies  $ BB1(\theta) $ and $ R_ x(\theta) $, due to noise and systematic imperfections, the measurements results are generally not the same.

\item  IBM Q Results: 2N=20, 40, 60, 80; Below, we will find the IBM Q measurement results of $ BB1(\pi )^{2N} $ and $ R_ x(\pi )^{2N} $ for 2N=20,40,60,80, alongside the usual simulation results (which do not suffer from systematic rotation errors and are therefore much closer to the ``ideal'' outcome). Use these results to respond which one of the following statements is correct.
\begin{itemize}    
\item The measured results from $ BB1(\pi) $ are always in better agreement with the idealized simulation results than the ones from the basic rotation$  R_x(\pi) $.
\item The measured results from $ BB1(\pi) $ are equivalent to the idealized simulation results, and in better agreement than the ones from the basic rotation $  R_x(\pi) $.
\item The measured results from $ BB1(\pi) $ are always in worse agreement with the idealized simulation results than the ones from the basic rotation $  R_x(\pi) $.
\end{itemize}
Solution:\\
From the two figures below, one can notice that the number of projections on the $ \lvert 0\rangle $ state is always greater for $ BB1(\pi )^{2N} $ than for $ R_ x(\pi )^{2N} $. For example, while the simulation result (blue) of $ BB1(\pi )^{40} $ is 994, and 989 for $ R_ x(\pi )^{40} $, the actual result (orange) obtained for $ BB1(\pi )^{40} $ is 786, and only 607 for $ R_ x(\pi )^{40} $.

The figure below shows the results from implementing $ BB1(\pi)^{2N} $ for 2N=20, 40, 60, 80. The vertical axis corresponds to the number of times the qubit was projected to state $ \lvert0\rangle $ from a total of 1024 measurements.

\begin{figure}[H] \centering{\includegraphics[scale=.5]{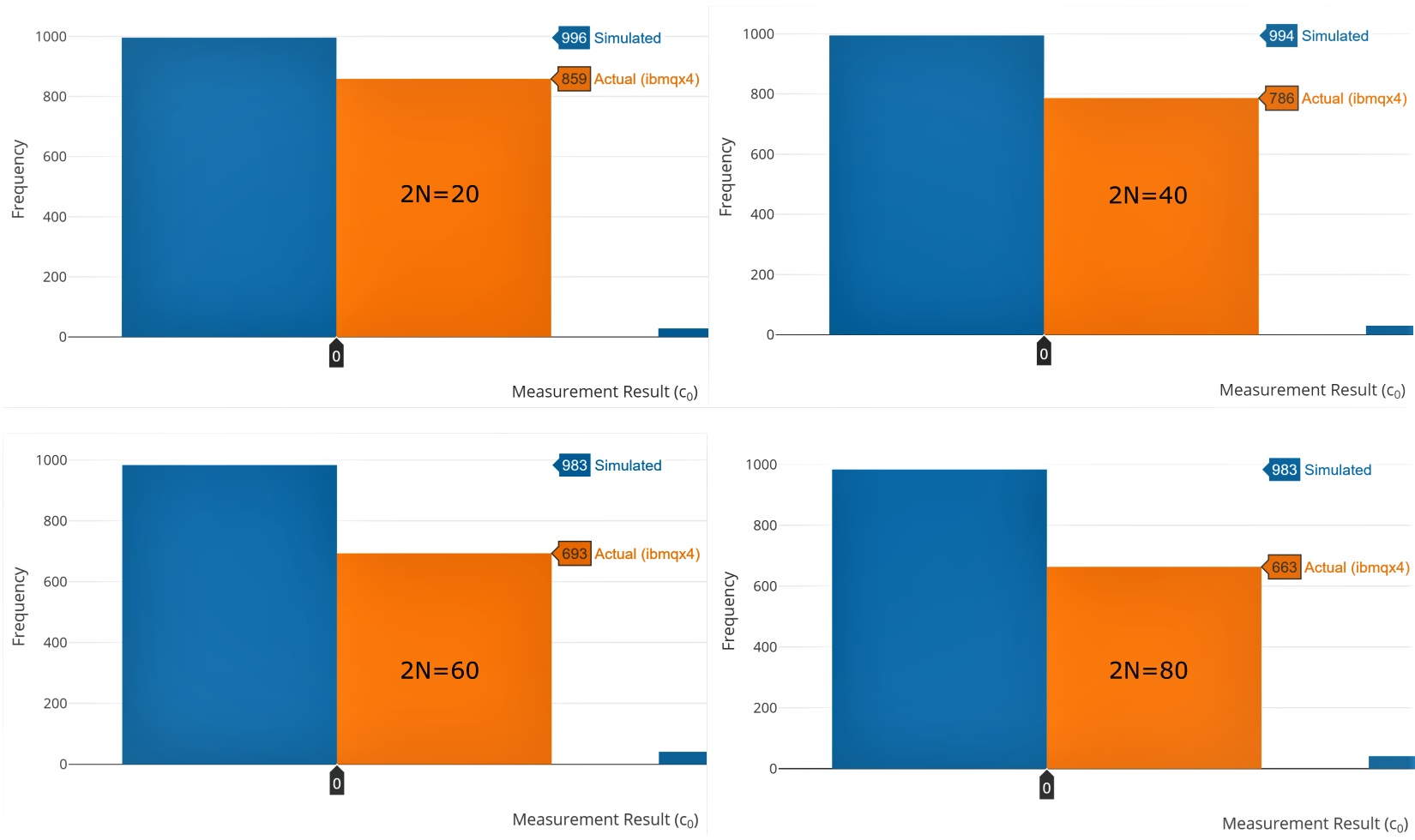}}\caption{$ BB1(\pi)^{2N} $ for 2N=20, 40, 60, 80}\label{fig4_23}
\end{figure}

The figure below shows the results from implementing $ R_x(\pi)^{2N} $ for 2N=20, 40, 60, 80. The vertical axis corresponds to the number of times the qubit was projected to state $ \lvert0\rangle $ from a total of 1024 measurements.

\begin{figure}[H] \centering{\includegraphics[scale=.5]{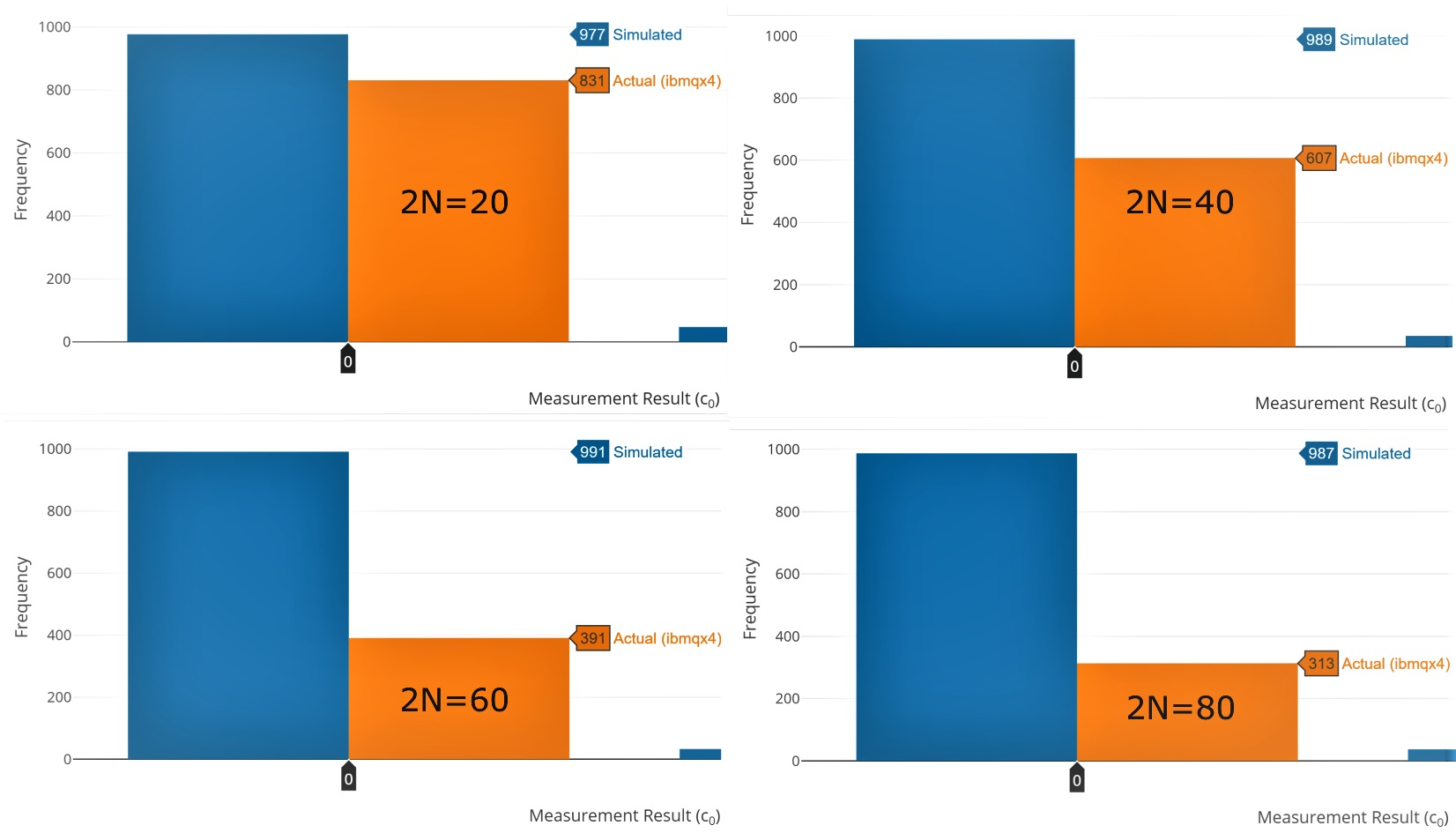}}\caption{$ R_x(\pi)^{2N} $ for 2N=20, 40, 60, 80}\label{fig4_24}
\end{figure}

Let us recall that the $  BB1(\pi) $ sequence pulse can be defined in QASM as
\begin{lstlisting}
gate bb1 pi a{

u3(pi, -pi/2, pi/2)a;
u3(pi, acos(-1/4)-pi/2, -acos(-1/4)+pi/2)a;
u3(pi, 3*acos(-1/4)-pi/2, -3*acos(-1/4)+pi/2)a;
u3(pi, 3*acos(-1/4)-pi/2, -3*acos(-1/4)+pi/2)a;
u3(pi, acos(-1/4)-pi/2, -acos(-1/4)+pi/2)a;
    
}
\end{lstlisting}
and that the rotation $ R_{x}\left( \theta \right) $ can be defined as 
\begin{lstlisting}
gate rx(theta) a {
    
    u3(theta,-pi/2,pi/2) a;  
}
\end{lstlisting}
However, we do not need to define rx($ \theta $) in the following assessments, since it was already defined in our platform.

For our convenience in the following exercises, we have defined a new command not present in standard QASM: ``repeat(N)''. This command repeats the indicated gate N times and applies it to the specified qubit. For example, 
\begin{lstlisting}
repeat(5) ry(pi/2) q[0]; 
\end{lstlisting}
applies five (5) y-axis rotations of $ \pi $ to the first qubit; that is, it is equivalent to writing
\begin{lstlisting}
ry(pi/2) q[0];
ry(pi/2) q[0];
ry(pi/2) q[0];
ry(pi/2) q[0];
ry(pi/2) q[0]; 
\end{lstlisting}
For some problems, we will also be able to leave the command as ``repeat(N)'', with N specified elsewhere. In these cases, our entire circuit will be run numerous times with a different N value each time, and in place of a histogram, we will obtain a plot with the IBM Q measurement results for each run. For example, 
\begin{lstlisting}
repeat(N) ry(pi/16) q[0];
\end{lstlisting}
produces the following plot
\begin{figure}[H] \centering{\includegraphics[scale=.4]{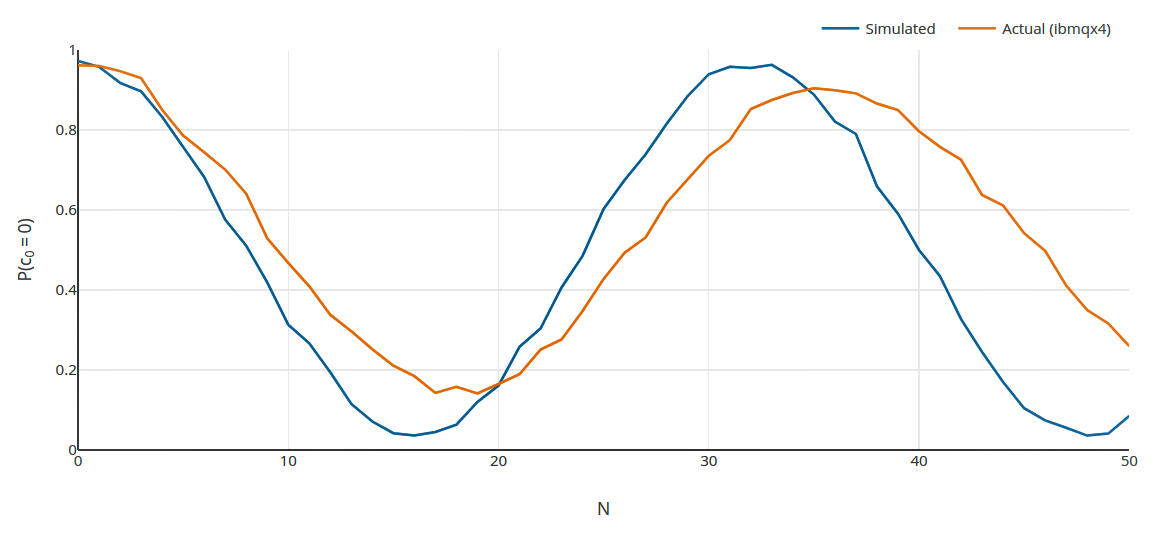}}\caption{QE Graph}\label{fig4_25}
\end{figure}

\item  BB1 With $ \theta=\pi $ And $ 2N=100 $; Run a QASM code that implements $ BB1(\theta)^{2N} $ with $ \theta=\pi $ and 2N=100 on the first qubit.
\begin{lstlisting}
include ``qelib1.inc'';

gate bb1pi a {
    u3(pi,-pi/2,pi/2) a;
    u3(pi, acos(-1/4) - pi/2, -acos(-1/4) + pi/2) a;
    u3(pi, 3*acos(-1/4) - pi/2, -3*acos(-1/4) + pi/2) a;
    u3(pi, 3*acos(-1/4) - pi/2, -3*acos(-1/4) + pi/2) a;
    u3(pi, acos(-1/4) - pi/2, -acos(-1/4) + pi/2) a;
}

qreg q[1];
creg c[1];

repeat(100) bb1pi q[0];

measure q -> c;
\end{lstlisting}

\item  Basic x-Axis Rotation With $ \theta=\pi $ And $ 2N=100 $; Run a QASM code that implements $ R_x(\theta)^{2N} $ with $ \theta=\pi $ and 2N=100 on the first qubit. we do not need to define rx(theta) since it was already defined for this activity.

\begin{lstlisting}
include ``qelib1.inc'';
qreg q[1];
creg c[1];

repeat(100) rx(pi) q[0];

measure q -> c;
\end{lstlisting}

\item  BB1 With $ \theta=\pi $ And $ 2N=120 $; Run a QASM code that implements $ BB1(\theta)^{2N} $ with $ \theta=\pi $ and 2N=120 on the first qubit.
\begin{lstlisting}
include ``qelib1.inc'';

gate bb1pi a {
    u3(pi,-pi/2,pi/2) a;
    u3(pi, acos(-1/4) - pi/2, -acos(-1/4) + pi/2) a;
    u3(pi, 3*acos(-1/4) - pi/2, -3*acos(-1/4) + pi/2) a;
    u3(pi, 3*acos(-1/4) - pi/2, -3*acos(-1/4) + pi/2) a;
    u3(pi, acos(-1/4) - pi/2, -acos(-1/4) + pi/2) a;
}

qreg q[1];
creg c[1];

repeat(120) bb1pi q[0];

measure q -> c;
\end{lstlisting}

\item  Basic x-Axis Rotation With $ \theta=\pi $ And $ 2N=120 $; Run a QASM code that implements $ R_x(\theta)^{2N} $ with $ \theta=\pi $ and 2N=120 on the first qubit. we do not need to define rx($ \theta $) since it was already defined for this activity.
\begin{lstlisting}
include ``qelib1.inc'';

qreg q[1];
creg c[1];

repeat(120) rx(pi) q[0];

measure q -> c;
\end{lstlisting}

\item  IBM Q Results: $ 2N=100, 120 $; Consider our IBM Q results from the last four assessments above for $ 2N=100, 120 $. Based on the results, one of the following statements is correct.
\begin{itemize}    
\item The measured results from $ BB1(\pi) $ are always in better agreement with the idealized simulation results than the results from the basic rotation $ R_x(\pi) $.
\item The measured results from $ BB1(\pi) $ are always in equal agreement with the idealized simulation results and are better than the results obtained from the basic rotation $ R_x(\pi) $.
\item The measured results from $ BB1(\pi) $ are always in worse agreement with the idealized simulation results, and the results from the basic rotation $ R_x(\pi) $ are in better agreement.
\end{itemize}

\item  Measurement Curve For BB1; Now that we analyzed the results of $ BB1(\pi )^{2N} $ and $ R_ x(\pi )^{2N}  $for 2N=20,40,60,80,100,120, we should have an idea about the behavior of both operations. Use the command ``repeat(N)'' to write a QASM code that implements B$ BB1(\pi )^{2N} $ and then outputs a plot of the number of projections on the $ \lvert0\rangle $ state versus N.

\begin{lstlisting}
include ``qelib1.inc'';

gate bb1pi a {
    u3(pi,-pi/2,pi/2) a;
    u3(pi, acos(-1/4) - pi/2, -acos(-1/4) + pi/2) a;
    u3(pi, 3*acos(-1/4) - pi/2, -3*acos(-1/4) + pi/2) a;
    u3(pi, 3*acos(-1/4) - pi/2, -3*acos(-1/4) + pi/2) a;
    u3(pi, acos(-1/4) - pi/2, -acos(-1/4) + pi/2) a;
}

qreg q[1];
creg c[1];

repeat(N) bb1pi q[0];
repeat(N) bb1pi q[0];

measure q -> c;
\end{lstlisting}

\item  Measurement Curve For Basic Rotation; Use the command ``repeat(N)'' to write a QASM code that implements $ R_ x(\pi )^{2N}  $ and then outputs a plot of the number of projections on the $ \lvert0\rangle $ state versus N. we do not need to define RX($ \theta $) since it was already defined in this activity.

\begin{lstlisting}
include ``qelib1.inc'';

qreg q[1];
creg c[1];

repeat(N) rx(pi) q[0];
repeat(N) rx(pi) q[0];

measure q -> c;
\end{lstlisting}
The actual results we obtain above will vary depending on day-to-day changes in IBM Q calibration and performance. For reference, here are results we obtained recently using the ibmqx4 computer\cite{shukla_complete_2018}; these were run using 8192 shots each to better illustrate the behavior. The results from $ BB1(\pi )^{2N} $ and $ R_ x(\pi )^{2N}  $ were overlaid on the same plot:

\begin{figure}[H] \centering{\includegraphics[scale=.9]{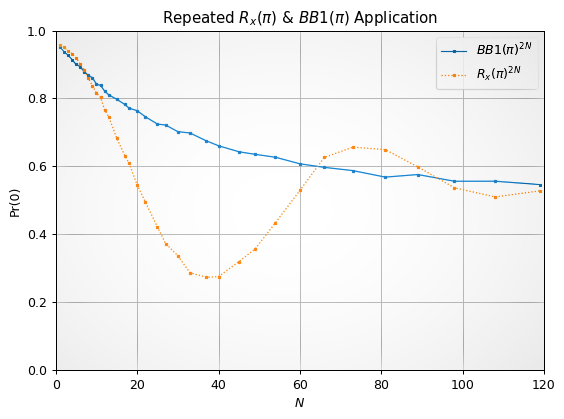}}\caption{$ BB1(\pi )^{2N} $ and $ R_ x(\pi )^{2N}  $}\label{fig4_26}
\end{figure}

Note that we expect an overall decay due to decoherence, which is not corrected by this composite pulse sequence. However, we also see that $ R_ x(\pi )^{2N}  $ exhibits oscillations; this due to accumulated systematic rotation errors. In contrast, $ BB1(\pi )^{2N} $ corrects for these systematic rotation errors and, thus, performs much better in general (no oscillations). While $ R_ x $ seemingly outperforms BB1 for values between N=65 and N=95, this is only an artifact related to it having accumulated an entire revolution's worth of rotation errors and has happened to start landing closer to the correct value as a result.

\item  Phase-Flip Code; The figure below shows the quantum circuit for implementation of the phase-flip quantum error correction code.
\begin{figure}[H] \centering{\includegraphics[scale=1.8]{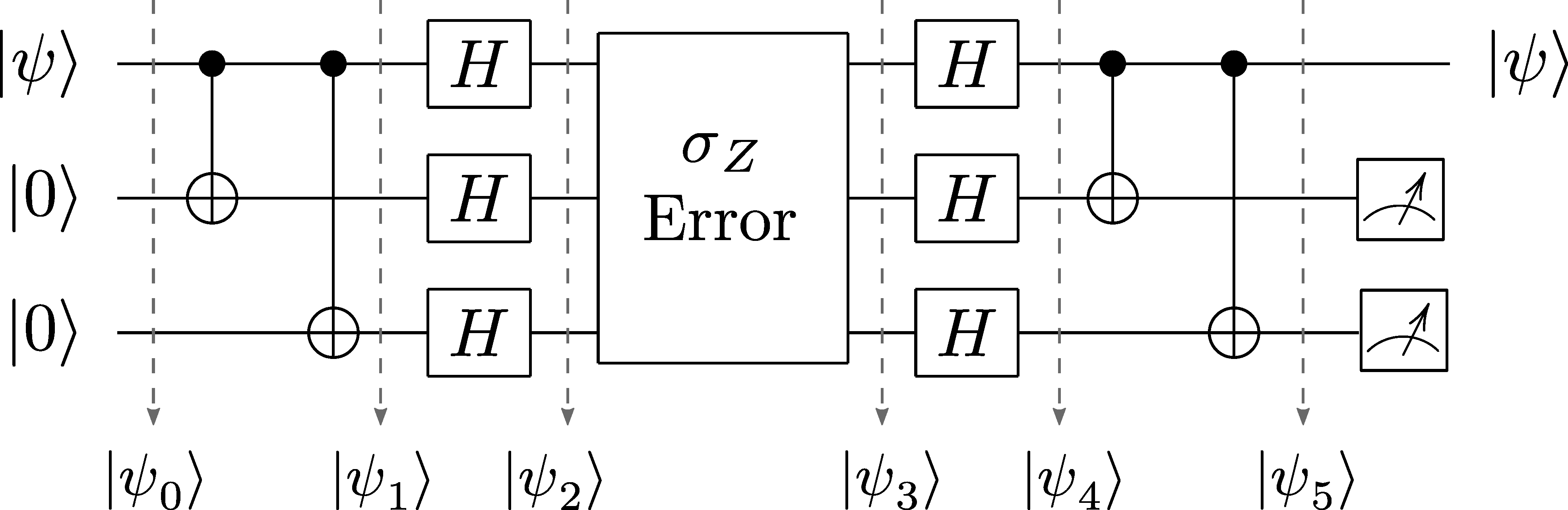}}\caption{Phase-Flip Code}\label{fig4_1}
\end{figure}
Which of the following statements is correct?
\begin{itemize}
\item The circuit creates identical copies of the state of the first qubit on the second qubit and the third qubits.
\item The circuit can correct three simultaneous $ \sigma_Z $ errors.
\item The circuit to the left of the error box encodes a physical qubit onto a logical qubit.
\item The circuit can correct errors of the type $ \sigma_Z  $without measuring the ancillary qubits.
\end{itemize}
	
\item  Shor Nine-Qubit Code; Which of the following statements about the Shor nine-qubit error correction code is correct?
\begin{itemize}
\item The code requires six ancilla qubits.
\item The code can only correct errors of the type $ \sigma _ X $
\item The code cannot correct for $ \sigma_Y $ errors.
\item The code is constructed from a combination of the bit-flip and phase-flip error-correction codes.
\end{itemize}
	
\item  Single Point of Failure; Which of the following sentences is correct about a ``single point of failure''?
\begin{itemize}
\item It is a part of the circuit that, when a failure occurs, causes the entire process to fail.
\item It is possible to create a fault-tolerant procedure with a single point of failure.
\item All circuits that have no single points of failure are, by definition, fault-tolerant.
\item They can only occur on quantum circuits.
\end{itemize}
	
\item  Fault-Tolerant Procedure; A procedure is fault-tolerant if:
\begin{itemize}
\item When errors occur, the procedure prevents these errors from propagating through the calculation.
\item A failure causes at most one error in each encoded block of bits.
\item There is at most one fault path in which two gates can fail.
\item When a failure occurs, it cascades into more errors as the calculation continues.
\end{itemize}
	
\item  Dynamical Error Suppression; In order to have a reliable quantum computer, we need to minimize systematic errors from the environment or from imperfect gate operations. Which one of the following is a platform-independent experimental procedure that reduces these types of errors?
\begin{itemize}
\item Replacing a single-qubit gate with an iteration of the same gate always reduces the intrinsic gate error.
\item Making redundant copies of a qubit always reduces the intrinsic gate error.
\item A sequence of pulses applied to a system can be designed to have a net effect that reduces the effect of environmental noise.
\item Adding more ancilla qubits always reduces the effect of environmental noise.
\end{itemize}
	
\item  Surface Codes; Which one of the following sentences does not describe the error correction code called the surface code?
\begin{itemize}
\item Surface codes can be implemented using a two-dimensional lattice of qubits.
\item In the surface code, one only needs to entangle nearest-neighbor qubits to create a logical qubit.
\item In the surface code, some qubits are used as data qubits, and some are used as syndrome qubits.
\item Surface codes can have an arbitrary large fault-tolerant threshold, allowing computation with arbitrarily large errors.
\end{itemize}
	
\item  Complexity Theory and Classes; Which of the following statements about complexity theory is correct?
\begin{itemize}
\item The time required to solve a polynomial-time problem of input size n grows as$  k^n $, where k is a positive constant that is problem-dependent.
\item Problems that are nondeterministic polynomial time (NP) are a subset of polynomial time.
\item The complexity class of a problem can generally be related to the input size of the problem, and the number of resources required to solve it.
\item All exponential-time problems can be reduced to polynomial-time problems with clever engineering.
\end{itemize}
	
\item  Complexity Classes for Quantum and Classical Computers; The quantum complexity class BQP (bounded quantum polynomial) includes the entirety of which of the following classical complexity classes?
\begin{itemize}
\item PSPACE
\item NP-complete
\item NP (nondeterministic polynomial time)
\item P (polynomial time)
\end{itemize}
\end{enumerate}

\bibliographystyle{IEEEtran}
\bibliography{Bib-Quantum}
\printindex

\printindex
\end{document}